\documentclass[letterpaper,11pt]{article}
\usepackage[utf8]{inputenc}
\usepackage[margin=1in]{geometry}
\usepackage[citestyle=numeric-comp]{biblatex}
\makeatletter
\newcommand{\removelatexerror}{\let\@latex@error\@gobble}
\makeatother

\usepackage[lined, noend, linesnumbered]{algorithm2e}
\SetKwBlock{Init}{}{}
\let\oldnl\nl% Store \nl in \oldnl
\newcommand{\nonl}{\renewcommand{\nl}{\let\nl\oldnl}}% Remove line number for one line
\SetAlCapSkip{1em}

\usepackage{paralist}
\usepackage{multicol}
\usepackage{caption}
\usepackage{mathtools}
\usepackage{amsthm}
\usepackage{amsmath}
\usepackage{amsfonts}
\usepackage{xcolor}
\usepackage{cleveref}

\usepackage[normalem]{ulem}
\usepackage[compact]{titlesec}
\usepackage{enumitem}
\setenumerate{label*=\arabic*.}
\setlist[itemize]{leftmargin=0pt}
\setlist[enumerate]{leftmargin=0pt}

\newtheoremstyle{compactthm}
  {2pt}         % Space above
  {2pt}         % Space below
  {\itshape}    % Body font (e.g., \itshape for theorems)
  {}            % Indent amount
  {\bfseries}   % Head font
  {.}           % Punctuation after head
  {.5em}        % Space after head
  {}            % Head spec
\theoremstyle{compactthm}

\newcommand{\rmv}[1]{}

\newcommand\red[1]{#1}
\newcommand\cyan[1]{#1}
\newcommand{\qH}[1]{\qedhere~$_{\text{#1}}$}

\newtheorem{papertheorem}{Theorem}
\newtheorem{paperdefinition}[papertheorem]{Definition}

\newtheoremstyle{sltheorem}
{}                % Space above
{}                % Space below
{\slshape}        % Theorem body font % (default is "\upshape")
{}                % Indent amount
{\bfseries}       % Theorem head font % (default is \mdseries)
{.}               % Punctuation after theorem head % default: no punctuation
{ }               % Space after theorem head
{}                % Theorem head spec
\theoremstyle{sltheorem}
\newtheorem{theorem}{Theorem}
\newtheorem{assumption}[theorem]{Assumption}
\newtheorem{scenario}[theorem]{Scenario}
\newtheorem{definition}[theorem]{Definition}
\newtheorem{observation}[theorem]{Observation}
\newtheorem{lemma}[theorem]{Lemma}

\newtheorem{proposition}[theorem]{Proposition}
\newtheorem{corollary}[theorem]{Corollary}

\numberwithin{theorem}{subsection}
\newtheorem*{lemmablank}{Lemma}
\newtheorem*{theoremblank}{Theorem}

\newtheorem{claimcustominner}{Claim}
\newenvironment{claimcustom}[1]{
    \renewcommand\theclaimcustominner{#1}
    \claimcustominner
}{\endclaimcustominner}

\newtheorem{corollaryinner}{Corollary}
\newenvironment{corollarycustom}[1]{
    \renewcommand\thecorollaryinner{#1}
    \corollaryinner
}{\endcorollaryinner}

\newcommand{\nocontentsline}[3]{}
\newcommand{\tocless}[2]{\bgroup\let\addcontentsline=\nocontentsline#1{#2}\egroup}

\newcommand{\old}{val_1}
\newcommand{\new}{val_2}

\newcommand{\Underline}[1]{\underline{\smash{#1}}}
\newcommand{\bemph}[1]{{\bfseries\itshape#1}}
\newcommand{\List}{\textbf{List}}

\newcommand{\CASop}{\text{CAS}}
\newcommand{\GCAS}{generalized-compare-and-swap}
\newcommand{\GCASop}{\text{GCAS}}
\newcommand{\FI}{fetch-and-increment}
\newcommand{\FIop}{\text{F\&I}}
\newcommand{\FA}{fetch-and-add}
\newcommand{\FAop}{\text{F\&A}}
\newcommand{\helpobject}[1]{C_{#1}}

\newcommand{\allocatecelloperation}{\text{AllocateCell}}
\newcommand{\freecelloperation}{\text{FreeCell}}

\newcommand{\highleveloperation}{DoHighLevelOp}
\newcommand{\doworkuntildone}{\text{DoLowLevelOp}}
\newcommand{\doaddcell}{DoAddCell}
\newcommand{\doapplyandcopyresponse}{\text{DoApply\&CopyResponse}}
\newcommand{\doremovecell}{DoRemoveCell}
\newcommand{\setrepositoryoperationresponse}{\text{SetResponse}}

\newcommand{\celluniverse}{\mathcal{C}}
\newcommand{\cellset}{Allocated} % used to S_C
\newcommand{\registerpointershort}{ptr}

\newcommand{\headobject}{H}

\newcommand{\clockobject}{C}
\newcommand{\announceobject}{A}
\newcommand{\linearizationobject}{L}
\newcommand{\stateobject}{S}

\newcommand{\timeshort}{t}
\newcommand{\operationshort}{hlo}
\newcommand{\statelong}{state}
\newcommand{\stateshort}{s}
\newcommand{\responselong}{resp}
\newcommand{\responseshort}{r}

\newcommand{\lastrepositoryoperationresponse}{response}
\newcommand{\result}{response}
\newcommand{\status}{status}
\newcommand{\nextlong}{next}
\newcommand{\acquisitions}{acquisitions}
\newcommand{\revocations}{revocations}
\newcommand{\view}{view}
\newcommand{\sealed}{sealed}

\newcommand{\uniquecellpointercontentlong}{ptr}
\newcommand{\cellpointerlong}{ptr}
\newcommand{\mycellpointershort}{ptr}

\newcommand{\uniquecellpointershort}{ptr}
\newcommand{\cellpointershort}{ptr}
\newcommand{\uniquecellpointercontentshort}{ptr}
\newcommand{\cellpointercontentshort}{ptr}

\newcommand{\repositoryoperationshort}{llo}
\newcommand{\uniquerepositoryoperationlong}{ullo}
\newcommand{\uniquerepositoryoperationshort}{ullo}
\newcommand{\myrepositoryoperationshort}{llo}
\newcommand{\myuniquerepositoryoperationshort}{ullo}
\newcommand{\nextuniquecellpointershort}{next\_\uniquecellpointershort{}}
\newcommand{\nextcellpointershort}{next\_\cellpointershort{}}
\newcommand{\currentuniquecellpointershort}{curr\_\uniquecellpointershort{}}
\newcommand{\currentcellpointershort}{curr\_\cellpointershort{}}
\newcommand{\previousuniquecellpointershort}{prev\_\uniquecellpointershort{}}
\newcommand{\previouscellpointershort}{prev\_\cellpointershort{}}

\newcommand{\targetcellpointershort}{target\_\cellpointershort{}}

\newcommand{\true}{\textsc{True}}
\newcommand{\false}{\textsc{False}}
\newcommand{\noop}{\textsc{Noop}}
\newcommand{\noopresponse}{\bot}
\newcommand{\nullconstant}{\textsc{Null}}
\newcommand{\arbitraryvalue}{*}
\newcommand{\unusedvalue}{-}
\newcommand{\genericobjecttype}{\mathcal{T}}
\newcommand{\genericobjecttransitionrelation}{apply}
\newcommand{\done}{\textsc{Done}}
\newcommand{\notdone}{\textsc{NotDone}}
\newcommand{\found}{\textsc{Found}}
\newcommand{\notfound}{\textsc{NotFound}}
\newcommand{\timechange}{\textsc{$\linearizationobject$-Changed}}
\newcommand{\addcell}{\textsc{AddCell}}
\newcommand{\doopandcopyresponse}{\textsc{Apply\&CopyResponse}}
\newcommand{\removecell}{\textsc{RemoveCell}}

\newcommand{\subhistory}[2]{#1\vert_{#2}}

\begin{document}

\title{Generalized Compare-and-Swap \\ and Space-Efficient Universal Constructions \\ for the Infinite-Arrival Model}

\author{Vassos Hadzilacos \and Myles Thiessen \and Sam Toueg}

\date{}

%\author{Vassos Hadzilacos}
% \affiliation{
%   \institution{University of Toronto}
%   \city{Toronto}
%   \country{Canada}
% }
% \email{vassos@cs.toronto.edu}

%\author{Myles Thiessen}
% \affiliation{
%   \institution{University of Toronto}
%   \city{Toronto}
%   \country{Canada}
% }
% \email{mthiessen@cs.toronto.edu}

%\author{Sam Toueg}
% \affiliation{
%   \institution{University of Toronto}
%   \city{Toronto}
%   \country{Canada}
% }
% \email{sam@cs.toronto.edu}

\maketitle

\begin{abstract}
    We introduce GCAS, a natural generalization of the well-known compare-and-swap (CAS) object.
Intuitively, GCAS just replaces the fixed equality test of CAS with a parametrized comparator chosen from $\{<, =, >\}$.
To showcase the utility of GCAS, we present two space-efficient wait-free universal constructions for systems where the number of participating processes is unknown and may be infinite (the \emph{infinite-arrival} model). 
The first has space-complexity linear in the number of processes that have participated so far, while the second has space-complexity linear in the point contention but assumes bounded concurrency.
To the best of our knowledge, these are the first wait-free universal constructions that achieve this space complexity in the infinite-arrival model.
To achieve space complexity linear in the point contention, our second universal construction uses a novel memory recycling scheme that works in the infinite-arrival model \red{with bounded concurrency}.
The ideas behind this recycling scheme could be of more general use.
\end{abstract}

\thispagestyle{empty}

\newpage

\setcounter{page}{1}

\LinesNotNumbered
\begin{figure}[t]
\DontPrintSemicolon
\removelatexerror
\begin{minipage}[t]{.15\columnwidth}
\end{minipage}
\hfill
\begin{minipage}[t]{.34\columnwidth}
\setlength{\algomargin}{0em}
\begin{algorithm}[H]
    \Init(\text{CAS($\old, \new$)}){
    
        $current \coloneqq O$
    
        \If{\textcolor{red}{$current = \old$}}{
            $O \coloneqq \new$
    
            \textbf{return} \true{} 
        }

        \vspace{1.5mm}
    
        \textbf{return} \false{}
        
    }
\end{algorithm}
\end{minipage}
\hfill
\begin{minipage}[t]{.34\columnwidth}
\setlength{\algomargin}{0em}
\begin{algorithm}[H]
    \Init(\text{GCAS($C, \old, \new$)}){
    
        $current \coloneqq O$
    
        \If{\textcolor{red}{$current\ C\ \old$}}{
            $O \coloneqq \new$
    
            \textbf{return} \true{} 
        }

        \vspace{1.5mm}
    
        \textbf{return} \false{}
    }    
\end{algorithm}
\end{minipage}
\hfill
\begin{minipage}[t]{.15\columnwidth}
\end{minipage}
\caption{CAS and GCAS operations.}
\label{casandgcas}
\end{figure}

\tocless\section{Introduction}\label{Introduction}

We propose a natural generalization of compare-and-swap (CAS), a fundamental object in shared memory systems, and show how it enables \emph{space-efficient}, wait-free universal constructions in systems where the number of participating processes is unknown and may be infinite (this is the \emph{infinite-arrival model} introduced by Merrit and Taubenfeld~\cite{merritt2000computing}).
This model encourages the design of \emph{adaptive} algorithms whose performance depends on the number of processes that \emph{actually} participate rather than the maximum number that \emph{could} participate.
We now describe~our~results.

A CAS object $O$ supports a CAS($\old$, $\new$) operation which compares the current value of~$O$ to $\old$ and, if equal, replaces it with $\new$; see \Cref{casandgcas} on the left.
The object also supports standard read and write operations.

We introduce \emph{generalized compare-and-swap} (\emph{GCAS}), a simple generalization of CAS that replaces the fixed equality test of CAS with a comparator $C \in \{ <, =, > \}$ supplied as a parameter. 
A GCAS object $O$ supports a GCAS($C$, $\old$, $\new$) operation which compares the current value~$v$ of $O$ to $\old$ using $C$ and, if $v$ $C$ $\old$, replaces it with $\new$; see \Cref{casandgcas} on the right.
Like CAS objects, GCAS objects also support standard read and write operations.
It is worth noting that GCAS should be implementable in hardware with minimal overhead compared to CAS (because  testing for inequality is not much harder than testing for equality).\footnote{In this paper we restrict the comparator $C$ of GCAS to be an equality or inequality test. We do so for two reasons:  (a)~these tests suffice for our universal constructions, and (b) this restriction minimizes the potential hardware overhead in implementing GCAS relative to CAS. More generally, $C$ could be any other binary comparator such as $\le$ or $\ge$, or even any function that takes two values and returns \textsc{True} or \textsc{False}. }

To showcase the utility of GCAS, we present two space-efficient wait-free universal constructions for the infinite-arrival model.
The space complexity of our first universal construction at any time $t$ is linear in the number of processes \emph{that have participated by time $t$}. To the best of our knowledge, this is the first universal construction to achieve this space complexity in the infinite-arrival model.

A drawback of our first construction is that once a process reserves memory, that memory remains allocated forever, even if the process later leaves the system.
Ideally, the space complexity at time~$t$ would be linear in the number of operations \emph{that are concurrent at time~$t$}, i.e., the point contention at time~$t$.
Our second universal construction achieves this space complexity, but under the assumption of an unknown upper bound on the maximum point contention (this is the infinite-arrival model \emph{with bounded concurrency}~\cite{aguilera2004pleasant, merritt2000computing}).
To the best of our knowledge, this is the first universal construction to attain this space complexity in this model.

We achieve this space complexity via a novel memory recycling scheme. Common approaches to memory recycling include \emph{reference counting} (e.g.,~\cite{detlefs2001lock, valois1995lock, herlihy2005nonblocking, anderson2021concurrent, nikolaev2019hyaline, nikolaev2021snapshot}) and \emph{hazard pointers} or related techniques (e.g.,~\cite{herlihy2002repeat, michael2002safe, michael2004hazard}), but, to the best of our knowledge, none of the existing schemes can be used to achieve our goals: some are non-blocking but not wait-free (e.g.,~\cite{detlefs2001lock, valois1995lock, herlihy2005nonblocking, nikolaev2019hyaline, nikolaev2021snapshot}), others do not work in the infinite-arrival model (e.g.,~\cite{herlihy2002repeat, michael2002safe, michael2004hazard, brown2015reclaiming, anderson2021concurrent}).
Our memory recycling scheme uses reference counters, with a key twist: each reference counter is decomposed into two counters, namely an \emph{acquisitions} and a \emph{revocations} counter, each stored at different locations; these are coalesced back into a single reference counter when its value is needed for recycling.

Our universal constructions leverage GCAS to achieve wait-freedom with a simple helping mechanism that prioritizes operations based on their timestamps.
Roughly speaking, to execute an operation $o$, a process $p$ obtains a timestamp $t$, and then it competes with other processes to have $o$ selected as the next operation to execute. To do so, $p$ tries to ``announce'' $o$ by applying a \mbox{GCAS($>$, $(t,o)$, $(t,o)$)} operation on a GCAS ``announcement'' object $A$:
if the timestamp $t$ of $o$ is smaller than the timestamp $t'$ of the operation $o'$ currently in $A$, this GCAS operation will replace $(t', o')$ with $(t,o)$ in $A$ (because $o$ has higher priority).\footnote{Throughout this paper we compare tuples in lexicographic order. In practice, this comparison can be achieved by reserving a field for each component of the tuple, concatenating these fields, and comparing the resulting bit strings.}
Eventually, the operation with the smallest timestamp will ``stick'' in $A$ and will be executed.
Once this operation is executed, however, it must be removed from $A$ \emph{even if it has a higher priority than any current and future operations.} So, if $p$ notices that the timestamped operation $(t',o')$ that is currently in $A$ has been executed, $p$ tries to replace it with its own operation $(t,o)$ by applying a GCAS($=$, $(t',o')$, $(t,o)$) operation on~$A$.

In summary, this paper makes the following four contributions:
\begin{compactitem}
    \item A natural generalization of the well-known compare-and-swap (CAS) object that replaces the fixed equality test with a parametrized comparator chosen from $\{<, =, >\}$.
    \item The first wait-free universal construction for the infinite-arrival model whose space complexity at time $t$ is linear in the number of processes that have participated by time $t$.
    \item The first wait-free universal construction for the infinite-arrival model with bounded concurrency whose space complexity at time $t$ is linear in the point contention at \mbox{time $t$.}
    \item A novel memory recycling scheme for the infinite-arrival model with bounded concurrency.
    The ideas behind this recycling scheme could be of more general use.
\end{compactitem}

It is worthwhile noting that our first universal construction guarantees more than wait-freedom: the step complexity of each operation is linear in the \emph{point contention}.

\textbf{Roadmap.}
In Section 2 we sketch our model.
In Section 3 we present a simple universal construction for the infinite-arrival model.
In Section 4 we describe our more space-efficient universal construction for the infinite-arrival model with bounded concurrency.
In Section 5 we discuss related work.
We conclude the paper with some remarks and open problems in Section 6.

\tocless\section{Model Sketch}\label{model}

We consider shared-memory systems where asynchronous processes may fail by crashing. In contrast to most work on shared-memory systems, which assumes a system with $n$ processes (the \emph{$n$-arrival} model), our system may have an infinite number of processes (the \emph{infinite-arrival} model).

\tocless\subsection{Objects, Implementations, and Runs}

Each object has a type that specifies how the object behaves when it is accessed sequentially.
We assume that
	the type $\mathcal{T}$ of object $O$ is given in the form of
	a (possibly nondeterministic) state-transition function $\textit{apply}_{\mathcal{T}}$:
	if $s$ is a state of $O$ and $o$ is an operation that can be applied to $O$,
	$\textit{apply}_{\mathcal{T}}(o, s)$ returns a pair of the form $(s',r)$,
	where~$s'$ is a possible new state of $O$ and $r$ is the corresponding response returned by $o$ when $o$ is applied to $O$ in state $s$.

An implementation of a \emph{target} object $O$ from a set of \emph{base} objects is
	a collection of procedures that specify how any process in the system
	can perform any operation of $O$ by applying operations to the base objects.
We only consider implementations that are \emph{linearizable}~\cite{herlihy1990linearizability} and \emph{wait-free}~\cite{herlihy1991wait}.
A universal construction from a set of base objects is an algorithm that takes the state-transition function $\textit{apply}_{\mathcal{T}}$ of an arbitrary type~$\mathcal{T}$ as input, and outputs an implementation of an object of type $\mathcal{T}$ from these base objects.

A run of an implementation of an object $O$ is a sequence of \emph{steps}, where each step is an invocation of an operation on $O$, an atomic operation on a base object, or a response from an operation on $O$.
Each step in a run $R$ has an associated ``time'' which is the sequence number of that step within $R$, i.e., the time of the $t$-th step in $R$ is $t$.
Furthermore, we say that a process $p$ \emph{has participated} at time $t$ in a run $R$ if $p$ has taken a step in $R$ before or at time~$t$.

\tocless\subsection{Concurrency}
The point contention at time $t$ in a run $R$ is
the number of \mbox{operations} that are pending at time~$t$. More precisely:

\begin{paperdefinition}[Point Contention]\label{def:point_contention}
    The \Underline{point contention at time~$t$ in a run $R$} is the number of \mbox{operations} that, by time $t$, have been invoked but have not responded.
\end{paperdefinition}

\begin{paperdefinition}[Bounded Concurrency]\label{def:bounded_concurrency}
A system has \Underline{bounded concurrency} if for every run~$R$ of the system there is a bound
$b_R \in \mathbb{N}$ such that the point contention at every time~$t$ in~$R$ is at most~$b_R$.
\end{paperdefinition}

We stress that in a system with bounded concurrency, processes do \emph{not} know the bound on the point contention (so they cannot use it); this is because the bound $b_R$ may be different in~each~run~$R$.

Recall that our first universal construction works in the infinite-arrival model.
This model does \emph{not} assume any bound on concurrency, i.e., there may be runs where the concurrency grows without bound (this is called the infinite-arrival model with \emph{unbounded concurrency} in~\cite{aguilera2004pleasant, merritt2000computing}).
Our second universal construction (which is more space efficient than our first) works in the infinite-arrival model with bounded concurrency.

\tocless\subsection{Memory Manager}

To support space efficiency, shared memory systems are often augmented with a memory manager that dynamically allocates and frees cells as instructed by processes.
For our purposes, a ``cell'' consists of a constant number of objects; i.e., it is a structure.
The memory manager
	maintains the
	set of currently unallocated cells.

When a process $p$ needs a new set of base objects, it asks the memory manager to allocate a new cell comprised of these objects.
The memory manager
	picks a cell $C$ that is not currently allocated,
	and
	returns
	a pointer to $C$ to $p$.
When a process determines that a cell $C$ is no longer needed \emph{by any process}, it asks the memory manager to free it for future reuse; the cell is no longer allocated.
We say that a base object is allocated by the memory manager if it resides in a cell that is currently allocated by the memory manager.

\setcounter{algocf}{0}
\LinesNumbered
\begin{algorithm}[t]
\DontPrintSemicolon
\footnotesize

% \nonl\textcolor{red}{Proof has to be fixed in four ways: $\noopresponse{}$ instead of $\perp$, $\helpobject{x}$ is a pointer instead of the object itself, some GCAS's were replaced with CAS's, ``response object" isn't a thing anymore, everything is cells.}

% \BlankLine

%\begin{multicols}{2}

\nonl\textbf{Statically allocated shared objects:}

\nonl\Init($\clockobject{}:$ A \FIop{} object, initially 1.){}

\nonl\Init($\announceobject{}:$ A \GCASop{} object with three fields:){
    \nonl$time:$ the timestamp $t$ of the operation $o$ in the field below, initially 0.

    \nonl$operation:$ the operation $o$ to execute, initially $\noop{}$. %(not an operation of type $\mathcal{T}$).

    \nonl$pointer:$ a pointer $\cellpointershort$ to the cell of the process that invoked $o$, initially a pointer $ptr_\noop{}$ to a dummy cell. %initialized to $(0, \noopresponse{})$.
}

\BlankLine

\nonl\Init($\stateobject{}:$ A \CASop{} object with four fields:){
    \nonl$time:$ the timestamp $t$ of the operation $o$ that resulted in the current object state $s$, initially 0.

    \nonl$state:$ the current object state $s$, initially the initial state of type $\mathcal{T}$.

    \nonl$response:$ the response $r$ of the operation $o$ that resulted in the current object state $s$, initially $\bot$.

    \nonl$pointer:$ a pointer $\cellpointershort$ to the cell of the process that invoked $o$, initially $ptr_\noop{}$.
}

\BlankLine

\nonl\textbf{Dynamically allocated shared objects:}

% \nonl\Init(The response object of process $p$, allocated when $p$ issues its first operation on $O$.
% It is a CAS object with two fields:){
\nonl\Init(Each cell contains a single \CASop{} object with two fields:){
    \nonl$time:$ the timestamp $t$ of the last operation $o$ invoked by the owner of this cell.

    \nonl$response:$ the response $r$ of the operation $o$ or $\nullconstant$. %(not a response of type $\mathcal{T}$).
}

\BlankLine

\nonl\textbf{Local persistent variable per process:}

\nonl\Init($ptr:$ a pointer to the cell of this process, initially $\nullconstant$.){}

%\end{multicols}

% \nonl\textbf{Code for process $p$:}

\BlankLine

\Init(\textbf{procedure} \text{DoOp($o$)}\label{line:op_start}\tcp*[f]{to perform an operation $o$ on the target object}){
    
    \lIf{$ptr = \nullconstant$}{$ptr \coloneqq \text{\textcolor{red}{\allocatecelloperation{}()}}$\label{line:semi-efficient-allocate-cell}\tcp*[f]{get a pointer to a cell from the memory manager}}

    $t \coloneqq \FIop(\clockobject{})$\label{line:time_assignment}\tcp*[f]{get a timestamp $t$ for $o$}

    $(*ptr) \coloneqq (t, \nullconstant)$\label{line:reset_help_struct}\tcp*[f]{initialize $o$'s response to $\nullconstant$}

    \While(\tcp*[f]{while $o$ is not done (i.e., its response is $\nullconstant$)}){$(*ptr) = (t, \nullconstant)$}{\label{line:loop_start}
        $(t^*, s^*, r^*, \cellpointershort^*) \coloneqq \stateobject{}$\label{line:g_r_query}\tcp*[f]{read the state of the target object}
        
        \CASop($(*\cellpointershort^*)$, $(t^*, \nullconstant)$, $(t^*, r^*)$)\label{line:help_pointer_cas}\tcp*[f]{copy the last operation's response into its cell}
    
        \GCASop($>$, $\announceobject{}$, $(t, o, ptr)$, $(t, o, ptr)$)\label{line:g_a_gcas}\tcp*[f]{if $o$ has higher priority, announce it}
    
        $(t', o', \cellpointershort') \coloneqq \announceobject{}$\label{line:g_a_query}\tcp*[f]{read the currently announced operation $o'$ to help}
    
        $(\hat{t}, \hat{r}) \coloneqq (*\cellpointershort')$\label{line:help_pointer_query}\tcp*[f]{read the response of $o'$ from its cell}

        \If(\tcp*[f]{if $o'$ is not done (i.e., its response is $\nullconstant$)}){$(\hat{t}, \hat{r}) = (t', \nullconstant)$}{\label{line:help}
            $(s', r') \coloneqq apply_{\mathcal{T}}(o', s^*)$\label{line:apply}\tcp*[f]{apply $o'$ to the target object state}

            \CASop($\stateobject{}$, $(t^*, s^*, r^*, \cellpointershort^*)$, $(t', s', r', \cellpointershort')$)\label{line:g_r_cas}\tcp*[f]{try to linearize $o'$ by changing $\stateobject{}$}
        }
        \Else(\tcp*[f]{if $o'$ is done (i.e., its response is not $\nullconstant$)}){
            \GCASop($=$, $\announceobject{}$, $(t', o', \cellpointershort')$, $(t, o, ptr)$)\label{line:g_a_cas}\tcp*[f]{and $o'$ is still in $\announceobject{}$, try to announce $o$}
        }
    }

    \textbf{return} $(*ptr).response$\label{line:op_done}\tcp*[f]{return the response of $o$ (found in $o$'s cell)}
}

\caption{\bf A simple \& space-efficient wait-free universal construction.}
\label{alg:wait-free-simple}
\end{algorithm}

We stress that if a process performs an operation on an object in a cell $C$ \emph{that has been freed and not yet reallocated}, the operation may return an incorrect value or may not return at all. This is because after the memory manager regains ownership of $C$ it may use $C$ arbitrarily—for example, it may assign $C$ to another application that accesses it in ways outside the control of~our~implementation.
In particular, this application could modify the contents of $C$ or change its formatting.

In systems with a memory manager, the base objects used by an implementation fall into two categories: \emph{statically allocated objects}, which exist and are known to all processes at the start of a run, and \emph{dynamically allocated objects}, which are currently allocated by the memory manager.
So we define the space complexity of an implementation as follows:

\begin{paperdefinition}[Space Complexity]\label{def:space_complexity}
    The \Underline{space complexity of an implementation at time $t$ of a run} is the number of statically allocated base objects plus the number of base objects dynamically allocated by the memory manager at time $t$ in that run.
\end{paperdefinition}

\tocless\section{A Simple \& Space-Efficient Universal Construction}\label{construction1}

We now describe a simple wait-free universal construction for the infinite-arrival model. Its space complexity at any time is linear in the number of processes that have participated by that time.
This universal construction, shown in \Cref{alg:wait-free-simple}, uses \GCASop{}, \CASop{}, and \FI{} (\FIop{}) objects to implement an object $O$ of an arbitrary type $\mathcal{T}$.\footnote{Since GCAS is a generalization of CAS, we can replace all CAS objects with GCAS objects. But, to highlight where the additional functionality of GCAS is~used, we opted to use CAS rather than GCAS whenever CAS is sufficient.}
In this construction, we leverage GCAS to implement the priority-based helping scheme outlined in Section 1.
\rmv{
In a nutshell, (1) a Fetch-and-Add object is used to timestamp operations, (2) a GCAS object $A$ is used to store the currently announced operation: the $>$ comparator is used to atomically compare the operation timestamps so that the operation with the lowest timestamp is eventually stored in $A$; and the $=$ comparator is used to remove from $A$ an operation that has already been applied to $O$, and (3) a CAS object is used to atomically update the state of $O$.
In this algorithm, each participating process gets a single cell from the memory manager: this cell stores the response of the last operation it invoked on $O$. So, the number of cells is linear in the number of participating processes.
}

When a process $p$ invokes its \emph{first} operation, it obtains from the memory manager
	a pointer to a cell consisting of a single CAS object,
	and assigns that pointer to a local variable $ptr$ (line~\ref{line:semi-efficient-allocate-cell}).
Thereafter, across all operations invoked by $p$,
    $ptr$ points to this cell.
This cell
	is used to store the response of each operation issued by $p$;
    since it is dedicated to $p$, we will call it \emph{$p$'s cell}.
In general it stores a pair~$(t,r)$, where
	$t$ is the timestamp of an operation $o$ on $O$ that $p$ has invoked and
	$r$ will eventually contain the response of $o$ (initially it is $\nullconstant$, indicating that $o$ is not done yet).

In addition to the cells that store the response of operations, this universal construction uses three statically allocated base objects:
\begin{compactitem}
    \item
    $C$ (for ``clock''): A \FIop{} object used to timestamp operations.
    \item $A$ (for ``announce''):
    A \GCASop{} object that processes use to announce the operations they wish to apply to $O$.
    It contains information about the oldest (highest priority) operation announced that has not yet
    been applied, namely
    a tuple~$(t,o,ptr)$,
    where $o$ is an operation, $t$ is its timestamp, and $ptr$ is a pointer to the cell
    of the process that invoked $o$.

    \item $S$ (for ``state''):
    A \CASop{} object that stores information about the state of the target object $O$.
    More precisely, it stores a tuple~$(t,s,r,ptr)$, where $t$ is the timestamp of the last operation~$o$ applied to $O$,
    $s$ is the state of $O$ after the application of $o$, $r$ is the response of $o$, and
    $ptr$ is a pointer to the cell of the process that invoked $o$.
\end{compactitem}

To perform an operation $o$, a process $p$
    first gets a timestamp $t$ for $o$ from the clock object $C$
    (line~\ref{line:time_assignment}).
Then, $p$ sets its response cell, which is pointed to by $ptr$, to $(t,\nullconstant)$ (line~\ref{line:reset_help_struct}).
Operation $o$ may be completed by $p$ itself or by a ``helper''. 
While $o$ is not done, i.e., while the response cell of $p$ still contains $(t,\nullconstant)$ (line~\ref{line:loop_start}):
    \begin{compactenum}
        \item $p$ reads the tuple $(t^*, s^*, r^*, ptr^*)$ currently in
        $S$
        (line~\ref{line:g_r_query}).
        \item $p$ ensures that the response of the last operation applied to~$O$ is copied into the response cell of the process that invoked it, i.e., $p$ ensures that $(t^*,r^*)$ is
        written in the cell pointed to by $ptr^*$,
        by applying a CAS operation on it (line~\ref{line:help_pointer_cas}).

        \item $p$ then tries to announce its own operation $o$ by applying a GCAS$(>, \ldots)$ operation on $A$
        	to write $(t,o,ptr)$ in it.
	This GCAS will succeed if $t$ is less than the timestamp of the operation presently in~$A$, i.e., if $o$ has higher priority (line~\ref{line:g_a_gcas}).

        \item Irrespective of whether this GCAS operation on $A$ was successful
        (i.e., whether $p$ succeeded in writing $(t,o,ptr)$ in $A$),
        $p$ now helps to execute \emph{whatever operation is currently in~$A$}. To~do~so, $p$ first reads from $A$ the tuple $(t',o',ptr')$ describing the operation to help (line~\ref{line:g_a_query}), and then it
        reads the response of $o'$ (in the cell pointed to by $ptr'$) to see whether $o'$ is already done (line~\ref{line:help_pointer_query}).
        \begin{compactenum}

            \item If $o'$ is not done,
            $p$
            applies $o'$ to the state $s^*$ of the target object $O$ that it read in step~(1),
            to get the new state~$s'$ of~$O$
            and the response $r'$
            of $o'$.
            It then attempts to linearize $o'$ by trying to replace the tuple it read from $S$ in step~(1) with $(t',s',r',ptr')$ (lines~\ref{line:help}-\ref{line:g_r_cas}).

            \item If $o'$ is done, $p$ tries to remove $o'$ from $A$ by replacing it with its own operation $o$.
            To do so, $p$ applies a GCAS($=, \ldots$) operation on $A$ to replace $(t',o',ptr')$ with $(t,o,ptr)$
            (line~\ref{line:g_a_cas}). 
        \end{compactenum}
    \end{compactenum}
When $p$ finds that $o$ is done,
    it returns the response of $o$,
	which is stored in the response cell of $p$ (line~\ref{line:op_done}).

\begin{papertheorem}\label{thm:algo_1}
    \Cref{alg:wait-free-simple} is a wait-free universal construction for the infinite-arrival model.
    Its space complexity at time $t$ is linear in the number of processes that have participated by time~$t$.
\end{papertheorem}

In fact, this universal construction guarantees more than wait-freedom: we prove that the step complexity of each operation is linear in the \emph{point contention} (at the time the invoking process gets a timestamp for this operation).
More precisely:

\begin{papertheorem}\label{thm:algo_1_step}
Suppose a process $p$ invokes an operation $o$ and executes \cref{line:time_assignment} within $o$.
Let $c$ be the point contention at this time.
Then, the number of steps that $p$ takes within $o$ is at most \mbox{linear in~$c$.}
\end{papertheorem}

\tocless\section{A More Space-Efficient Universal Construction}\label{construction2}

The space complexity of our first universal construction at time~$t$ is linear in the number of processes that have participated by $t$.
We now describe a universal construction whose space complexity at~$t$ is linear \emph{only} in the point contention at $t$ (Definition \ref{def:point_contention}).
It uses the same types of base objects as our first universal construction, except it also uses \FA{} (\FAop{}) in addition to \FI{}~(\FIop{}).
Both constructions work in the infinite-arrival model, but the second one requires the additional assumption of bounded concurrency (Definition \ref{def:bounded_concurrency}).
We first outline the main challenges in achieving this space complexity and how our universal construction solves them, and then present its pseudocode.

\tocless\subsection{Some Challenges and Their Solutions}
\label{sec:challenges_and_solutions}

In our first universal construction,
 the first time a process participates, it gets a new response cell from the memory manager and never frees it. In other words, this construction never recycles these cells.
To improve the space complexity, our second universal construction recycles cells, i.e., it \emph{frees} previously allocated cells.

\textbf{Cell recycling.}
One difficulty with
recycling is that before a process $p$ frees a cell $C$, it must be sure that no process will ever try to access an object $O$ within $C$ until $C$ is allocated again.
This is because if this were to happen, $O$ could misbehave: it could return a wrong value, or even not return~at~all.

A naive way to recycle cells with our first universal construction is as follows.
When a process $p$ invokes an operation $o$, it allocates a new cell $C$ to store the response of $o$.
Then, after $p$ finds the response of $o$ in $C$ (on line~\ref{line:op_done}), it immediately frees~$C$ (because $p$ no longer needs it).
The problem with this approach is that another process $q$ can now access $C$, even though $C$ is unallocated.
This occurs when $q$ reads a pointer to $C$ from the announce object $A$ on line~\ref{line:g_a_query} (before $C$ has been freed), goes to sleep, wakes up after $C$ has been freed, and then accesses $C$ on line~\ref{line:help_pointer_query}.

A common approach to enable the freeing of no-longer-needed cells
	is by using \emph{reference counters} (e.g.,~\cite{detlefs2001lock, valois1995lock, herlihy2005nonblocking, anderson2021concurrent, nikolaev2019hyaline, nikolaev2021snapshot}).
Intuitively, a reference counter for a cell $C$ stores the number of processes that currently have the right to access $C$.
A process acquires the right to access $C$ by incrementing the reference counter for $C$;
	and when it no longer needs to access $C$, the process decrements the reference counter for $C$.
A process that finds a cell's reference counter to be~0 can \mbox{free that cell.}

But where do we put the reference counter for a cell $C$?
If we put it in $C$ itself, then
    to acquire the right to access $C$ (by incrementing its reference counter)
	a process would have to access $C$ (where its reference counter is stored) ---
	a chicken-and-egg situation.
To solve this, we could try to put $C$'s reference counter outside of $C$.
But doing this raises another problem: when we recycle~$C$, we now must also recycle its reference counter; so we need a mechanism to recycle
the reference counters themselves ---
	a different kind of chicken-and-egg situation!

Our second universal construction solves this problem by (a)~threading the cells of the operations that are pending in a \emph{linked list}, and (b) \emph{splitting} the reference counter of each cell into two parts, each in a different location, as we now explain.
	
\textbf{Reference counter splitting.} At any time $t$, the reference counter for a response cell $C$ in the list is equal to
	the number of processes that have \emph{acquired} the right to access~$C$
	\emph{minus} the number of processes that have \emph{relinquished} that right by time $t$.
We store the reference counter for $C$ \emph{implicitly}
	by maintaining two separate counters:
	the \emph{acquisitions counter} for $C$, stored in the \emph{predecessor of $C$ in the list}; and
	the \emph{revocations counter} for $C$, stored \emph{in~$C$ itself}.
Note that the acquisitions counter for $C$ and the pointer to $C$, both of which reside in the predecessor of $C$, must be updated \emph{together} atomically.
So we store both of them in a CAS object called $next$ (in the predecessor of $C$).	
The revocations counter for $C$ is stored in a \FAop{} object, called $revocations$, in $C$ itself.

\textbf{List traversal.} To access any cell $C$ in the list, a process $p$ must first acquire the right to do so,
i.e., it must increment the acquisitions counter for~$C$.
Since this counter is located in the predecessor of~$C$, $p$ must \emph{traverse} the list to find
    (and acquire the right to access)
	the predecessor of $C$.
This traversal proceeds as follows.
Having obtained the right to access a cell~$C_i$ in the list,
	$p$ first obtains the right to access the next cell $C_{i + 1}$
	by incrementing the acquisitions counter for $C_{i + 1}$, which resides in~$C_i$
	($p$ does so by performing a successful CAS operation on the $next$ object of~$C_i$ because it contains the acquisitions counter for~$C_{i + 1}$).
After $p$ has acquired the right to access $C_{i + 1}$, it no longer needs access to $C_i$, so it relinquishes its right to access $C_i$.
It does so by incrementing $C_i$'s revocations counter,
    by performing a \FAop{} operation on the $revocations$ object
    of~$C_i$.
We note that the starting point of this traversal, i.e., the head of the list~$H=C_0$, is a \emph{statically} allocated cell that is never freed (so all processes always have the right to access $H$).

\textbf{Cell removal.} 
When an operation completes, its corresponding response cell is removed from the list.
To remove a cell $C$ from the list, a process $p$ must move the acquisitions counter for the \emph{successor}~$C^+$ of $C$, stored in $C$, to the \emph{predecessor} $C^-$ of $C$.
But the removal of $C$ and the move of the acquisitions counter for $C^+$ (from $C$ to $C^-$) must be done atomically to avoid the following bad scenario.
Before $p$ removes $C$ from the list, it reads the acquisitions counter for $C^+$ from $C$, say its value is~$a$.
Then, another process acquires the right to access $C^+$ by incrementing the acquisitions counter for $C^+$ stored in $C$; at this time, the number of acquisitions for $C^+$ is $a + 1$.
Now $p$ removes $C$ from the list and writes $a$ into $C^-$.
But the acquisitions counter for $C^+$, now stored in $C^-$, is incorrect: its value is $a$, but the true number of acquisitions for $C^+$ is $a + 1$!
To solve this problem, $p$ removes $C$ from the list as follows:
(1) it first \emph{freezes} the acquisitions counter for $C^+$,
(2) it then reads the acquisitions counter for $C^+$, say its value is $a$, and
(3) it finally removes $C$ from the list and writes $a$ into $C^-$; this last step is done atomically by doing a CAS operation on the $next$ object of $C^-$.
Process $p$ does step (1) by setting a $sealed$ flag in the $next$ object of $C$ (which contains the acquisitions counter for $C^+$).
Once this flag is set, the content of $next$ cannot change (the content of the $next$ object of $C$ is now \emph{sealed}).

\textbf{Cell freeing.} When a cell $C$ is removed from the list it cannot necessarily be freed yet.
This is because some process may still have the right to access $C$.
To determine when $C$ can be freed, we need to determine whether the reference counter for $C$ is zero.
This is done as follows.
When a process $p$ removes $C$ from the list, it computes the reference counter for $C$ by:
(1) reading the value $a$ of the acquisitions counter for $C$, which is stored in the predecessor of~$C$, and
(2)~subtracting $a$ from the revocations counter for $C$, which is stored in $C$, using a \FAop{} operation on the $revocations$ object of $C$.
Note that this subtraction changes the semantics of the $revocations$ object of $C$:
	it used to be the revocations counter for~$C$, it is now the \emph{negation} of the reference counter for~$C$.
This trick allows processes to relinquish their right to access $C$ in a \emph{uniform} way by incrementing the $revocations$ object of $C$ (irrespective of its current semantics).
We prove that the process that causes the reference counter for $C$ to become zero is the last process that had the right to access~$C$, so it can \mbox{safely free~$C$}.

\textbf{Wait-freedom.}
To achieve wait-freedom, we use a modified version of the priority-based helping mechanism of \Cref{alg:wait-free-simple}. But this is no longer sufficient here, because processes that are trying to use the list (e.g., traverse the list, add a cell, remove a cell, or change the content of a cell) may be prevented from doing so by other processes that are concurrently using the list. So we also need a mechanism to ensure that accessing the list is wait-free.
We now briefly elaborate on these two mechanisms.

To apply an operation $o$ on the target object $O$, a process $p$ first obtains from the memory manager a cell $C$ to store the response of $o$.
Then $p$ performs the following three ``low-level'' operations, possibly with the help of other processes, in that order:
 \begin{compactenum}
 \item \addcell{}: append $C$ to the end of the list;
 \item \doopandcopyresponse{}: apply $o$ to $O$ and then copy the response to $C$; and
 \item \removecell{}: remove $C$ from the list.
 \end{compactenum}
 
A process performs these three operations using a modified version of the priority-based helping mechanism of \Cref{alg:wait-free-simple}.
As before, process $p$ first obtains a timestamp for the operation that it wants to do from a \FIop{} object $C$,
	it tries to announce it by applying a \mbox{$\GCASop(>,\ldots)$} operation on the announce object $A$,
	and then tries to perform the operation~$o_A$ stored in~$A$.
Recall that in \Cref{alg:wait-free-simple}, $p$ tries to perform $o_A$ as follows:
(a) it first reads the state $s$ of $O$ from the state object $S$,
(b) it then applies $o_A$ to $s$ (using the state-transition function $apply_\mathcal{T}$ of the type $\mathcal{T}$ of ~$O$) to get the next state~$s'$ of $O$
	and the corresponding response $r'$, and
(c) it finally tries to write $(s',r')$ into the state object $S$ by doing a CAS operation on that object.
Note that, if this CAS is successful, step (3) does two things simultaneously:
it linearizes~$o_A$ \emph{and} changes the state of $O$ accordingly.
In contrast, our second universal construction separates these two \mbox{things, as follows.}

When $p$ tries to perform the operation $o_A$ stored in~$A$,
	it simply tries to write $o_A$ into a CAS object $L$.
We ensure that once an operation is written into $L$, \emph{it is not removed until it has taken effect.}
Thus, operations are linearized in the order
	they are written into $L$ (which is why this object is called $L$).
	
We now explain how processes perform the operations written into $L$.
Recall that in \Cref{alg:wait-free-simple}, before doing its own operation, a process $p$ \emph{copies the response} of the last operation that was linearized (into the appropriate cell).
In contrast, in our second universal construction, before doing its own operation, a process $p$ \emph{performs} the last operation that was linearized.
To do so, $p$ reads the operation~$o_L$ that is currently in $L$, and then:
 \begin{compactenum}
 \item If $o_L$ is an \addcell{} operation to add a cell $C$, $p$ traverses the list to append $C$ to the end of the list.
 \item If $o_L$ is an \doopandcopyresponse{} operation to apply an operation $o$ to the target object $O$, $p$ first applies $o$ to $O$ by performing steps (a), (b) and (c) above to the state object~$S$. Then $p$ traverses the list to find the appropriate cell and copies the response of $o$ into it.
 \item If $o_L$ is a \removecell{} operation to remove a cell $C$, $p$ traverses the list to find and remove $C$ from the list.
 \end{compactenum}

The priority-based helping mechanism described above, however, is not sufficient for performing the operation $o_L$ that is currently in~$L$ in a wait-free manner: as we see above, to perform $o_L$, a process~$p$ must traverse the list of cells, but this traversal could be impeded by concurrent processes that are also traversing the list.
To see this, recall that to reach the successor $C^+$ of a cell $C$ in the list, $p$ must acquire the right to access $C^+$ (by incrementing the acquisitions counter for $C^+$, which resides in $C$). 
But to do so $p$ must “win” a CAS operation on the $next$ object of $C$ which contains the acquisitions counter for $C^+$.
This is problematic because $p$ may keep losing its CAS operations on the $next$ object of $C$,
	so $p$ \emph{may get stuck at cell~$C$} while trying to traverse the list.
To avoid this, in our universal construction, $p$ periodically checks whether
	the operation $o_L$ that it is trying to perform is still in $L$ (recall that $p$ is traversing the list to perform $o_L$).
If $p$ sees that $o_L$ is no longer in $L$,
	it can be certain that~$o_L$ has already taken effect (because $o_L$ cannot be removed from $L$ until it has taken effect),
	and so $p$ \emph{bails out.}

This bail-out mechanism to achieve wait-freedom, however, works only under the assumption of bounded concurrency.
This is because, with \emph{un}bounded concurrency,
a process attempting to traverse the list may repeatedly lose its CAS operations on a $next$ object because there may be an unbounded stream of newly arriving processes, each of which wins a CAS operation on that $next$ object and then immediately crashes \emph{before changing~$L$}.

\textbf{Different incarnations of a cell.}
Recycling cells may also raise the following problem.
A process $p$ reads a $next$ object that contains a pointer to a cell $C$, but goes to sleep before acquiring the right to access $C$ (i.e., before incrementing the acquisitions counter for $C$, which also resides in this $next$ object).
Then $C$ gets recycled and reallocated; this is a new ``incarnation'' of $C$, and its content has changed.
Finally, $p$ wakes up and acquires the right to access (the new incarnation of) $C$,
    thinking that it has acquired the right to access the older incarnation of $C$ --- this is clearly problematic.
A simple solution to this problem is for processes to tag each pointer returned by the memory manager with a \emph{unique identifier} (which they can obtain by performing a \FIop{} operation):
    this creates ``unique pointers'' that are used in place of ``raw'' pointers everywhere (except for when a process needs a ``raw'' pointer to access a cell).
As it turns out, creating these unique pointers is not necessary: we show that the above scenario (and other problematic ones involving different incarnations of a cell) cannot occur in our universal construction.

\textbf{Space complexity.} We show that at any time $t$, our universal construction uses a number of cells that is linear in the point contention $c_t$ at time $t$. Intuitively, this follows from the following properties:

\begin{compactitem}

\item At any time, a process holds the right to access \emph{at most a small constant number $\alpha$ of cells.} This is ensured by relinquishing access to each cell as soon as it is no longer needed (for example, during list traversal, a process successively acquires and relinquishes cells as it traverses through the list).

\item When a process relinquishes the right to access a cell, it decrements the cell’s (implicit or explicit) reference counter. If the resulting value indicates that no process currently holds the right to access the cell, it frees the cell and returns it to the memory manager.

\item Before completing an operation on the target object, a process relinquishes the right to access every cell it acquired the right to access during that operation.

\end{compactitem}

The above properties ensure that the universal construction uses about $\alpha \cdot c_t$ cells (so about $3 \alpha \cdot c_t$ base objects) at any time $t$.

\begin{algorithm}
\DontPrintSemicolon
% \fontsize{6}{7}\selectfont
\tiny

\begin{multicols}{2}

% \nonl\Init(Each high-level operation begins by calling \text{\allocatecelloperation{}()} on the cell manager, which returns a pointer $\cellpointershort{}$ to a cell that is not currently allocated. Eventually, this cell is freed by calling \text{\freecelloperation{}($\cellpointershort{}$)}. This allows future \text{\allocatecelloperation{}()} calls to return $\cellpointershort{}$, enabling the cell to be reused by another high-level operation.){}

% \BlankLine

\nonl\textbf{Statically allocated shared objects:}

\nonl\Init($\headobject{}:$ A cell that serves as the head of a list of cells.){}

\nonl\Init($\clockobject{}:$ A \FIop{} object, initially 1.){}

\nonl\Init($\announceobject{}:$ A \GCASop{} object with two fields:){
    \nonl$\uniquerepositoryoperationlong{}:$ a unique low-level operation to linearize, initially $(0, \noop{})$.

    \nonl$\cellpointerlong{}:$ a pointer to a cell in which to store the response of $\uniquerepositoryoperationlong{}$, initially $\nullconstant{}$.
}

\BlankLine

\nonl\Init($\linearizationobject{}:$ A \CASop{} object with two fields:){

    \nonl$\uniquerepositoryoperationlong{}:$ the last unique low-level operation that was linearized, initially $(0, \noop{})$.

    \nonl$\cellpointerlong{}:$ a pointer to a cell in which to store the response of $\uniquerepositoryoperationlong{}$, initially $\nullconstant{}$.
}

\BlankLine

\nonl\Init($\stateobject{}:$ A \CASop{} object with three fields:){
    \nonl$\uniquerepositoryoperationlong{}:$ the unique \doopandcopyresponse{} operation that resulted in the current object state, initially $(0, \noop{})$.

    \nonl$\statelong{}:$ the current object state, initially the initial state of type~$\genericobjecttype{}$.

    \nonl$\responselong{}:$ the response of $\uniquerepositoryoperationlong{}$, initially $\noopresponse{}$.
}

\BlankLine

\BlankLine

\BlankLine

\nonl\textbf{Dynamically allocated shared objects:}

\nonl\Init(The response of each high-level operation is stored in a cell; a cell consists of the following objects){}

\nonl\Init(\textbf{type} cell:){

    % \nonl\Init($\lastwritelong{}:$ A \CAS{} object that stores the unique \writeregisteroperation{} operation that modified the value of the register this cell represents, initially \text{$(0, \langle \writecell{}, \arbitraryvalue{} \rangle)$}.){}

    % \BlankLine

    \nonl\Init($\lastrepositoryoperationresponse{}:$ A \CASop{} object with two fields:){
        \nonl$\uniquerepositoryoperationlong{}:$ the last unique low-level operation invoked by the owner of this cell, initially $(0, \noop{})$.

        \nonl$\responselong{}:$ the response of applying this low-level operation or $\nullconstant{}$, initially $\nullconstant{}$.
    }

    \BlankLine

    \nonl\Init($\revocations{}:$ A \FAop{} object that stores the number of times this cell was revoked, initially $0$.){}

    \nonl\Init($\nextlong{}:$ A \CASop{} object with four fields:){
        \nonl$\view{}:$ the number of times $\nextlong{}$ has changed in this incarnation of this cell, initially 0.
    
        \nonl$\sealed:$ $\true{}$ if this cell is in the process of being removed otherwise $\false{}$, initially $\false{}$.
    
        \nonl$\acquisitions{}:$ the number of times the cell pointed to by $\cellpointerlong{}$ was acquired, initially $0$.
    
        \nonl$\cellpointerlong{}:$ a pointer to a cell, initially $\nullconstant{}$.
    }
}

\end{multicols}

\BlankLine

\BlankLine

\BlankLine

\begin{multicols}{2}

\Init(\textbf{procedure} \text{\highleveloperation{}($\operationshort{}$)}){\label{line:ero:invocation_step}
    $\mycellpointershort{} \coloneqq \text{\textcolor{red}{\allocatecelloperation{}()}}$\label{line:ero:allocate_cell}

    \doworkuntildone{}($\addcell{}$, $\mycellpointershort{}$)\label{line:ero:low_level_add_cell}

    \doworkuntildone{}($\langle \doopandcopyresponse{}, \operationshort{} \rangle$, $\mycellpointershort{}$)\label{line:ero:low_level_apply_and_copy_response}

    $(\unusedvalue{}, \responselong{}) \coloneqq (*\mycellpointershort{}).\lastrepositoryoperationresponse{}$\label{line:ero:copy_response_out_of_cell}

    \doworkuntildone{}($\removecell{}$, $\mycellpointershort{}$)\label{line:ero:low_level_remove_cell}

    Relinquish($\mycellpointershort{}$)\label{line:ero:owner_relinquish}

    \textbf{return} $\responselong{}$\label{line:ero:response_step}
}

\BlankLine

\Init(\textbf{procedure} \text{\doworkuntildone{}($\myrepositoryoperationshort$, $\mycellpointershort{}$)}){
    $\timeshort{} \coloneqq \FIop{}(\clockobject{})$\label{line:ero:operation_timestamp}

    $\myuniquerepositoryoperationshort{} \coloneqq (\timeshort{}, \myrepositoryoperationshort)$\label{line:ero:create_unique_low_level_operation}

    $(*\mycellpointershort{}).\lastrepositoryoperationresponse{} \coloneqq (\myuniquerepositoryoperationshort{}, \nullconstant{})$\label{line:ero:do_work_initialize_response}

    \While{$(*\mycellpointershort{}).\lastrepositoryoperationresponse{} = (\myuniquerepositoryoperationshort{}, \nullconstant{})$}{\label{line:ero:do_work_while_loop}

    $(\uniquerepositoryoperationshort_\linearizationobject{}, \cellpointershort_\linearizationobject{}) \coloneqq \linearizationobject{}$\label{line:ero:linearization_read}

        \If{$\uniquerepositoryoperationshort_\linearizationobject{} = (\arbitraryvalue{}, \addcell{})$}{\label{line:ero:do_add_cell_condition}
            \doaddcell{}($\uniquerepositoryoperationshort_\linearizationobject{}$, $\cellpointershort_\linearizationobject{}$)\label{line:ero:do_add_cell}
        }
        \ElseIf{$\uniquerepositoryoperationshort_\linearizationobject{} = (\arbitraryvalue{}, \removecell{})$}{\label{line:ero:do_remove_cell_condition}
            \doremovecell{}($\uniquerepositoryoperationshort_\linearizationobject{}$, $\cellpointershort_\linearizationobject{}$)\label{line:ero:do_remove_cell}
        }
        \ElseIf{$\uniquerepositoryoperationshort_\linearizationobject{} = (\arbitraryvalue{}, \langle \doopandcopyresponse{}, \arbitraryvalue{}\rangle)$}{\label{line:ero:do_apply_and_copy_response_condition}
            \doapplyandcopyresponse{}($\uniquerepositoryoperationshort_\linearizationobject{}$, $\cellpointershort_\linearizationobject{}$)\label{line:ero:do_apply_and_copy_response}
        }
    
        \GCASop{}($>$, $\announceobject{}$, $(\myuniquerepositoryoperationshort{}, \mycellpointershort{})$, $(\myuniquerepositoryoperationshort{}, \mycellpointershort{})$)\label{line:ero:announce_gcas}
    
        $(\uniquerepositoryoperationshort_\announceobject{}, \cellpointershort_\announceobject{}) \coloneqq A$\label{line:ero:announce_read}

        $status \coloneqq \text{IsDone}(\uniquerepositoryoperationshort_\linearizationobject{}, \uniquerepositoryoperationshort_\announceobject{}, \cellpointershort_\announceobject{})$\label{line:ero:check_if_announce_is_done}
    
        \If{$status = \notdone$}{\label{line:ero:done_check}
            \CASop{}($\linearizationobject{}$, $(\uniquerepositoryoperationshort_\linearizationobject{}, \cellpointershort_\linearizationobject{})$, $(\uniquerepositoryoperationshort_\announceobject{}, \cellpointershort_\announceobject{})$)\label{line:ero:linearization_cas}
        }
        \ElseIf{$status = \done$}{\label{line:ero:not_done_check}
            \GCASop{}($=$, $A$, $(\uniquerepositoryoperationshort_\announceobject{}, \cellpointershort_\announceobject{})$, $(\myuniquerepositoryoperationshort{}, \mycellpointershort{})$)\label{line:ero:announce_cas}
        }
    }
}

\BlankLine

\Init(\textbf{procedure} \text{\doaddcell{}($\uniquerepositoryoperationshort_\linearizationobject{}$, $\cellpointershort_\linearizationobject{}$)}){\label{line:ero:add_cell_procedure}
    $\currentcellpointershort{} \coloneqq \&\headobject{}$\label{line:ero:add_cell_initial_current_pointer}

    \While{$\currentcellpointershort{} \neq \cellpointershort_\linearizationobject{}$}{\label{line:ero:add_cell_while_loop}    
        $(\status{}, \nextcellpointershort{}) \coloneqq \text{AcquireNext($\uniquerepositoryoperationshort_\linearizationobject{}$, $\currentcellpointershort{}$)}$\label{line:ero:add_cell_acquire_next}

        \lIf{$\status{} = \timechange{}$}{\textbf{break}}\label{line:ero:add_cell_acquire_next_l_changed}
        \ElseIf{$\status{} = \notfound{}$}{\label{line:ero:add_cell_acquire_next_not_found}
            $(\view{}, \unusedvalue{}, \unusedvalue{}, \unusedvalue{}) \coloneqq (*\currentcellpointershort{}).\nextlong{}$\label{line:ero:add_cell_read_end_of_list}

            \lIf{$\linearizationobject{}.\uniquerepositoryoperationlong{} \neq \uniquerepositoryoperationshort_\linearizationobject{}$}{\textbf{break}}\label{line:ero:add_cell_before_updating_end_of_list_linearization_changed_check}
        
            \CASop{}($(*\currentcellpointershort{}).\nextlong{}$, $(\view{}, \false{}, 0, \nullconstant{})$, $(\view{} + 1, \false{}, 1, \cellpointershort_\linearizationobject{})$)\label{line:ero:add_cell_to_list}

            \textbf{break}\label{line:ero:add_cell_not_found_break}
        }
        \ElseIf{$\status{} = \found{}$}{\label{line:ero:add_cell_acquire_next_found}
            Relinquish($\currentcellpointershort{}$)\label{line:ero:add_cell_traversal_relinquish}
        
            $\currentcellpointershort{} \coloneqq \nextcellpointershort{}$\label{line:ero:add_cell_update_current_pointer}
        }
    }

    Relinquish($\currentcellpointershort{}$)\label{line:ero:add_cell_final_relinquish}

    \setrepositoryoperationresponse{}($\uniquerepositoryoperationshort_\linearizationobject{}$, $\cellpointershort_\linearizationobject{}$, $\done{}$)\label{line:ero:add_cell_set_response}
}

\BlankLine

\Init(\textbf{procedure} \text{\doremovecell{}($\uniquerepositoryoperationshort_\linearizationobject{}$, $\cellpointershort_\linearizationobject{}$)}){\label{line:ero:do_remove_procedure}
    \setrepositoryoperationresponse{}($\uniquerepositoryoperationshort_\linearizationobject{}$, $\cellpointershort_\linearizationobject{}$, $\done{}$)\label{line:ero:remove_cell_set_response}

    $(\previouscellpointershort{}, \currentcellpointershort{}) \coloneqq (\nullconstant{}, \&\headobject{})$\label{line:ero:remove_cell_initialize_pointers}

    \While{$\currentcellpointershort{} \neq \cellpointershort_\linearizationobject{}$}{\label{line:ero:remove_cell_while_loop}
        $(\status{}, \nextcellpointershort{}) \coloneqq \text{AcquireNext($\uniquerepositoryoperationshort_\linearizationobject{}$, $\currentcellpointershort{}$)}$\label{line:ero:remove_cell_acquire_next}

        \If{$\status{} = \timechange{} \lor \status{} = \notfound{}$\label{line:ero:remove_cell_check_acquire_status}}{
            \textbf{goto} \cref{line:ero:remove_cell_prev_relinquish}\label{line:ero:exit_1}
        }
        \ElseIf{$\status{} = \found{}$}{\label{line:ero:remove_cell_acquire_next_found}
            Relinquish($\previouscellpointershort{}$)\label{line:ero:remove_cell_traversal_relinquish}
        
            $(\previouscellpointershort{}, \currentcellpointershort{}) \coloneqq (\currentcellpointershort{}, \nextcellpointershort{})$\label{line:ero:remove_cell_update_pointers}
        }
    }

    \While{$(*\cellpointershort_\linearizationobject{}).\nextlong{}.\sealed = \false$}{\label{line:ero:remove_cell_remove_seal_loop}
        $(\view{}, \sealed, a, \nextcellpointershort{}) \coloneqq (*\cellpointershort{}_\linearizationobject{}).\nextlong{}$\label{line:ero:remove_cell_read_pointer_to_remove_before_seal}

        \CASop{}($(*\cellpointershort{}_\linearizationobject{}).\nextlong{}$, $(\view{}, \sealed, a, \nextcellpointershort{})$, $(\view{} + 1, \true{}, a, \nextcellpointershort{})$)\label{line:ero:seal_cell}
    }

    $(\unusedvalue{}, \unusedvalue{}, a, \nextcellpointershort{}) \coloneqq (*\cellpointershort{}_\linearizationobject{}).\nextlong{}$\label{line:ero:remove_cell_read_pointer_to_remove}

    \Repeat{\upshape \CASop{}($(*\previouscellpointershort{}).\nextlong{}$, $(\view{}', \false{}, a', \cellpointershort_\linearizationobject{})$, $(\view{}' + 1, \false{}, a, \nextcellpointershort{})$)\label{line:ero:remove_cell_from_list}}{\label{line:ero:remove_cell_remove_repeat_loop}
    
        $(\view{}', \unusedvalue{}, a', \nextcellpointershort{}') \coloneqq (*\previouscellpointershort{}).\nextlong{}$\label{line:ero:remove_cell_read_previous_pointer}

        \If{$\linearizationobject{}.\uniquerepositoryoperationlong{} \neq \uniquerepositoryoperationshort_\linearizationobject{} \lor \nextcellpointershort{}' = \nextcellpointershort{}$\label{line:ero:remove_cell_before_removal_linearization_check}}{
            \textbf{goto} \cref{line:ero:remove_cell_prev_relinquish}\label{line:ero:exit_3}
        }
    }

    \FAop{}($(*\cellpointershort{}_\linearizationobject{}).\revocations{}$, $-a'$)\label{line:ero:copy_acquisitions_to_revocations}

    Relinquish($\previouscellpointershort{}$)\label{line:ero:remove_cell_prev_relinquish}

    Relinquish($\currentcellpointershort{}$)\label{line:ero:remove_cell_final_relinquish}
}

\BlankLine

\Init(\textbf{procedure} \text{\doapplyandcopyresponse{}($\uniquerepositoryoperationshort_\linearizationobject{}$, $\cellpointershort_\linearizationobject{}$)}){\label{line:ero:do_apply_and_copy_response_response}
    $(\uniquerepositoryoperationshort{}_\stateobject{}, \stateshort{}_\stateobject{}, \responseshort{}_\stateobject{}) \coloneqq \stateobject{}$\label{line:ero:state_read}

    \lIf{$\linearizationobject{}.\uniquerepositoryoperationlong{} \neq \uniquerepositoryoperationshort_\linearizationobject{}$}{\textbf{return}}\label{line:ero:state_linearization_check}

    \If{$\uniquerepositoryoperationshort{}_\stateobject{} \neq \uniquerepositoryoperationshort_\linearizationobject{}$}{\label{line:ero:check_if_already_applied}
        $(\unusedvalue{}, \langle \doopandcopyresponse{}, \operationshort{}_\linearizationobject{} \rangle) \coloneqq \uniquerepositoryoperationshort_\linearizationobject{}$\label{line:ero:unpack_operation_to_apply}
    
        $(\stateshort{}, \responseshort{}) \coloneqq \genericobjecttransitionrelation{}_{\genericobjecttype{}}(\operationshort{}_\linearizationobject{}, \stateshort{}_\stateobject{})$\label{line:ero:apply_op}
    
        \CASop{}($\stateobject{}$, $(\uniquerepositoryoperationshort{}_\stateobject{}, \stateshort{}_\stateobject{}, \responseshort{}_\stateobject{})$, $(\uniquerepositoryoperationshort_\linearizationobject{}, \stateshort{}, \responseshort{})$)\label{line:ero:state_cas}

        $\responseshort{}_\stateobject{} \coloneqq \stateobject{}.\responselong{}$\label{line:ero:apply_update_response}
    }

    \setrepositoryoperationresponse{}($\uniquerepositoryoperationshort_\linearizationobject{}$, $\cellpointershort_\linearizationobject{}$, $\responseshort{}_\stateobject{}$)\label{line:ero:apply_set_response}
}

\BlankLine

\Init(\textbf{procedure} \text{\setrepositoryoperationresponse{}($\uniquerepositoryoperationshort_\linearizationobject{}$, $\cellpointershort_\linearizationobject{}$, $\result{}$)}){\label{line:ero:set_repository_response_procedure}
    $\status{} \coloneqq \text{Acquire}(\uniquerepositoryoperationshort_\linearizationobject{}, \cellpointershort_\linearizationobject{})$\label{line:ero:set_response_acquire}
    
    \If{$\status{} = \found{}$}{\label{line:ero:set_response_found_check} 
    
        \CASop{}($(*\cellpointershort_\linearizationobject{}).\lastrepositoryoperationresponse{}$, $(\uniquerepositoryoperationshort_\linearizationobject{}, \nullconstant{})$, $(\uniquerepositoryoperationshort_\linearizationobject{}, \result{})$)\label{line:ero:responses_set_attempt}
    
        Relinquish($\cellpointershort_\linearizationobject{}$)\label{line:ero:set_response_relinquish}
    }
}

\BlankLine

\Init(\textbf{procedure} \text{IsDone($\uniquerepositoryoperationshort_\linearizationobject{}$, $\uniquerepositoryoperationshort_\announceobject{}$, $\cellpointershort_\announceobject{}$)}){
    $status \coloneqq \text{Acquire}(\uniquerepositoryoperationshort_\linearizationobject{}, \cellpointershort_\announceobject{})$\label{line:ero:announce_acquire}

    \lIf{$\status{} = \timechange{}$}{\textbf{return} $\timechange{}$}\label{line:ero:announce_acquire_l_changed_check}

    $response \coloneqq \done$\label{line:ero:done_initialization}

    \If{$\uniquerepositoryoperationshort_\announceobject{} = (\arbitraryvalue{}, \addcell{}) \land \status{} = \notfound{}$}{\label{line:ero:add_cell_done_check}
        $response \coloneqq \notdone$\label{line:ero:add_cell_done}
    }
    \ElseIf{$\uniquerepositoryoperationshort_\announceobject{} = (\arbitraryvalue{}, \removecell{}) \land \status{} = \found{}$}{\label{line:ero:remove_cell_done_check}
        $response \coloneqq \notdone$\label{line:ero:remove_cell_done}
    }
    \ElseIf{$\uniquerepositoryoperationshort_\announceobject{} = (\arbitraryvalue{}, \langle \doopandcopyresponse{}, \arbitraryvalue{}\rangle)$}{\label{line:ero:write_and_read_done_check}        
        \If{$(*\cellpointershort_\announceobject{}).\lastrepositoryoperationresponse{} = (\uniquerepositoryoperationshort_\announceobject{}, \nullconstant{})$}{\label{line:ero:announce_op_response_check}
            $response \coloneqq \notdone$\label{line:ero:apply_cell_done}
        }
    }

    \lIf{$\status{} = \found{}$}{Relinquish($\cellpointershort_\announceobject{}$)}\label{line:ero:announce_acquire_relinquish}

    \textbf{return} $response$

}

\BlankLine

\Init(\textbf{procedure} \text{Acquire($\uniquerepositoryoperationshort_\linearizationobject{}$, $\targetcellpointershort{}$)}){\label{line:ero:acquire_procedure}
    $\currentcellpointershort{} \coloneqq \&\headobject{}$\label{line:ero:acquire_initial_current_pointer}

    \While{$\currentcellpointershort{} \neq \targetcellpointershort{}$\label{line:ero:acquire_loop_until}}{
        $(\status{}, \nextcellpointershort{}) \coloneqq \text{AcquireNext($\uniquerepositoryoperationshort_\linearizationobject{}$, $\currentcellpointershort{}$)}$\label{line:ero:acquire_acquire_next}
    
        Relinquish($\currentcellpointershort{}$)\label{line:ero:acquire_relinquish}

        \If{$\status{} = \timechange{} \lor \status{} = \notfound{}$}{\label{line:ero:acquire_not_found_check}
            \textbf{return} $\status{}$\label{line:ero:acquire_early_exit}
        }
        \ElseIf{$\status{} = \found{}$}{\label{line:ero:acquire_found_check}
            $\currentcellpointershort{} \coloneqq \nextcellpointershort{}$\label{line:ero:acquire_update_current_unique_pointer}
        }
    }
    
    % \Repeat(\label{line:ero:acquire_loop}){$\currentcellpointershort{} = \targetcellpointershort{}$\label{line:ero:acquire_loop_until}}{
    %     $(\status{}, \nextcellpointershort{}) \coloneqq \text{AcquireNext($\uniquerepositoryoperationshort_\linearizationobject{}$, $\currentcellpointershort{}$)}$\label{line:ero:acquire_acquire_next}
    
    %     Relinquish($\currentcellpointershort{}$)\label{line:ero:acquire_relinquish}

    %     \If{$\status{} = \timechange{} \lor \status{} = \notfound{}$}{\label{line:ero:acquire_not_found_check}
    %         \textbf{return} $\status{}$\label{line:ero:acquire_early_exit}
    %     }
    %     \ElseIf{$\status{} = \found{}$}{\label{line:ero:acquire_found_check}
    %         $\currentcellpointershort{} \coloneqq \nextcellpointershort{}$\label{line:ero:acquire_update_current_unique_pointer}
    %     }
    % }

    \textbf{return} $\found{}$\label{line:ero:acquire_return}
}

\BlankLine

\Init(\textbf{procedure} \text{AcquireNext($\uniquerepositoryoperationshort_\linearizationobject{}$, $\currentcellpointershort{}$)}){\label{line:ero:acquire_next_procedure}

    \Repeat{\upshape \CASop{}($(*\currentcellpointershort{}).\nextlong{}$, $(\view{}, \false{}, a, \nextcellpointershort{})$, $(\view{} + 1, \false{}, a + 1, \nextcellpointershort{})$)\label{line:ero:acquire_next_cell}}{\label{line:ero:acquire_next_repeat_loop}
        $(\view{}, \unusedvalue{}, a, \nextcellpointershort{}) \coloneqq (*\currentcellpointershort{}).\nextlong{}$\label{line:ero:acquire_next_read_curr_unique_pointer}

        \If{$\linearizationobject{}.\uniquerepositoryoperationlong{} \neq \uniquerepositoryoperationshort_\linearizationobject{}$}{\label{line:ero:acquire_next_linearization_changed_check}
            \textbf{return} $(\timechange{}, \arbitraryvalue{})$\label{line:ero:acquire_next_linearization_changed_return}
        }

        \If{$\nextcellpointershort{} = \nullconstant{}$}{\label{line:ero:acquire_next_not_found_check}
            \textbf{return} $(\notfound{}, \arbitraryvalue{})$\label{line:ero:acquire_next_not_found_return}
        }
    }

    \textbf{return} $(\found{}, \nextcellpointershort{})$\label{line:ero:acquire_next_found_return}
}

\BlankLine

\Init(\textbf{procedure} \text{Relinquish($\currentcellpointershort{}$)}){\label{line:ero:relinquish}
    \lIf{$\currentcellpointershort{} = \nullconstant{} \lor \currentcellpointershort{} = \&\headobject{}$}{\textbf{return}}\label{line:ero:early_exit}

    $x \coloneqq $ \FAop{}($(*\currentcellpointershort{}).\revocations{}$, $1$)\label{line:ero:relinquish_revocations}

    \lIf{$x + 1 = 0$}{\textcolor{red}{\freecelloperation{}($\currentcellpointershort{}$)}}\label{line:ero:free_cell}
}

\end{multicols}
\caption{\bf A more space-efficient wait-free universal construction.}
\label{alg:efficient_algo}
\end{algorithm}

\tocless\subsection{Pseudocode Description}

The pseudocode of this universal construction is given in Algorithm~\ref{alg:efficient_algo}.
This algorithm uses the following statically allocated base objects: $H$ (for ``head''), $C$ (for ``clock''), $A$ (for ``announce''), $L$ (for ``linearize''), and $S$ (for ``state''). 
This algorithm also uses a list of dynamically allocated cells ($H$ is the head of this list).
Each cell consists of three base objects: $response$, $revocations$, and $next$.
The purpose of all the base objects was described in the previous section, and their type and content is given at the top of Algorithm~\ref{alg:efficient_algo}.
Note that $next$ is a CAS object with four fields: $view$, $sealed$, $acquisitions$, and $ptr$.
We already described the purpose of $sealed$, $acquisitions$, and $ptr$ in the previous section.
The $view$ field is a monotonically increasing counter that prevents ABA problems.\footnote{This field can be avoided by making $next$ an LL/SC object rather than a CAS object.}

\noindent$\bullet$~\textbf{\highleveloperation{}($hlo$)} is invoked by any process $p$ that 
wants to perform an operation $hlo$ on the target object (to differentiate $hlo$ from the three low-level operations that our universal construction does to perform $hlo$, we call $hlo$ a ``high-level'' operation).
In line~\ref{line:ero:allocate_cell} $p$ gets a pointer $ptr$ to a new cell from the memory manager.
Process $p$ then performs the operation $hlo$ on the target object by ensuring the following three low-level operations are performed in order: (1) add the cell pointed to by $ptr$ to the list of cells, (2) apply $hlo$ to the target object and copy its response into the cell pointed to by~$ptr$, and (3) remove this cell from the list.
This is done by invoking the $\doworkuntildone$ procedure with a first parameter of $\addcell$, $\langle \doopandcopyresponse, hlo \rangle$, and $\removecell{}$ on lines~\ref{line:ero:low_level_add_cell}, \ref{line:ero:low_level_apply_and_copy_response}, and~\ref{line:ero:low_level_remove_cell}, respectively.
After $p$ has finished these procedures, $hlo$ is done.
Hence, $p$ no longer needs the cell pointed to by $ptr$ (which was used to store the response of $hlo$), so it relinquishes its right to access it (line \ref{line:ero:owner_relinquish}).
Recall that the cell pointed to by $ptr$ is not necessarily recyclable yet because other processes may still have the right to access it.
Finally, $p$ returns the response of $hlo$ (line~\ref{line:ero:response_step}) that it read from the cell pointed to by $ptr$ on line~\ref{line:ero:copy_response_out_of_cell}.

\noindent$\bullet$~\textbf{\doworkuntildone($llo, ptr$)} performs the low-level operation $llo$
and ensures the response of this operation is stored
in the cell pointed to by $ptr$.
The high-level flow of this procedure is similar to \Cref{alg:wait-free-simple}.
Process $p$ first gets a unique timestamp $t$ (line~\ref{line:ero:operation_timestamp}) and forms the pair $(t, llo)$; we call the pair $(t, llo)$ a \emph{unique low-level operation} and denote it $ullo$ (line~\ref{line:ero:create_unique_low_level_operation}).
The timestamp $t$ is the priority of $llo$.
Process $p$ then sets the response of $ullo$ in the cell pointed to by $ptr$ to $\nullconstant$ (line~\ref{line:ero:do_work_initialize_response}), and enters the loop on line~\ref{line:ero:do_work_while_loop}.
$p$ exits this loop once the response of $ullo$ is not $\nullconstant$ (indicating $ullo$ has taken effect).

In each iteration of this loop: (1) $p$ reads the value ($ullo_L$, $ptr_L$) currently stored in $L$, and then invokes the \doaddcell{}, \doremovecell{}, or \doapplyandcopyresponse{} procedure, depending on the kind of operation $ullo_L$ is, to ensure $ullo_L$ takes effect and its corresponding response is written into the cell pointed to by $ptr_L$ (lines \ref{line:ero:do_add_cell_condition}-\ref{line:ero:do_apply_and_copy_response}); (2)~$p$ tries to announce its \emph{own} operation and pointer ($ullo$, $ptr$) by performing a $\GCASop(>,\ldots)$ operation on~$A$ (line~\ref{line:ero:announce_gcas});
(3)~regardless of the outcome of this GCAS operation, $p$
reads the low-level operation $ullo_A$ currently stored in $A$  (line \ref{line:ero:announce_read}),
and then it invokes the IsDone procedure to try to determine whether this operation has already taken effect
(line \ref{line:ero:check_if_announce_is_done}).
If IsDone returns $\notdone$, then $ullo_A$ has not yet taken effect and $p$ tries to store $(ullo_A, ptr_A)$ into $L$ by performing a \CASop{} operation on $L$ (line~\ref{line:ero:linearization_cas}).
If IsDone returns $\done$, then $ullo_A$ has taken effect, and in this case $p$ tries to remove $(ullo_A, ptr_A)$ from $A$ by replacing it with $(ullo, ptr)$ by performing $\GCASop(=,\ldots)$ operation on~$A$ (line~\ref{line:ero:announce_cas}).
Finally, if IsDone returns $\timechange$, then $p$ could not determine whether $ullo_A$ has taken effect or not; in this case, $p$ does nothing: it just proceeds to the next iteration of the loop.

\noindent$\bullet$~\textbf{\doaddcell{}($ullo_L$, $ptr_L$)} traverses the list of cells to find the last cell in the list, appends the cell $C$ pointed to by $ptr_L$ after it, and sets its response~to~\done{}.
To traverse the list, $p$ maintains a current pointer $curr\_ptr$ that points to some cell in the list ($curr\_ptr$ is initially a pointer to the head of the list~$\&H$).
While $curr\_ptr$ does not point to $C$ (i.e., $curr\_ptr \neq ptr_L$), $p$ tries to acquire the right to access the cell after the cell pointed to by $curr\_ptr$ by invoking the AcquireNext procedure (line~\ref{line:ero:add_cell_acquire_next}).
If the AcquireNext procedure returns $\found$ (line~\ref{line:ero:add_cell_acquire_next_found}), then $p$ continues the traversal: $p$ relinquishes its right to $curr\_ptr$ (line~\ref{line:ero:add_cell_traversal_relinquish}), updates $curr\_ptr$ to the next pointer (line~\ref{line:ero:add_cell_update_current_pointer}), and continues to the next iteration of the while loop.
If the AcquireNext procedure returns $\notfound{}$ (line~\ref{line:ero:add_cell_acquire_next_not_found}), $curr\_ptr$ points to the last cell $C_{last}$ in the list. Therefore, $p$ did not find $C$ in the list, and $p$ tries to add $C$ after $C_{last}$.
To do so, $p$ performs a CAS operation on $C_{last}$ to set its next pointer to $ptr_L$ (line~\ref{line:ero:add_cell_to_list}). We prove that, regardless of whether this CAS succeeds or not, $C$ is added to the list at the time of this CAS.
So, in all cases, $p$ exits the while loop after performing this CAS.
Note that the CAS that adds $C$ to the list sets the acquisitions counter for $C$ to be 1 to signify that the process that allocated $C$ has the right to access $C$ (this process will relinquish its right to access $C$ at the end of its operation on line~\ref{line:ero:owner_relinquish}).
After $p$ exits the while loop, $p$ first relinquishes its right to access the cell pointed to by $curr\_ptr$ (line~\ref{line:ero:add_cell_final_relinquish}). Then, $p$ invokes the $\setrepositoryoperationresponse{}$ procedure to write $\done$ into the $response$ object of $C$; this informs the process that allocated $C$ that $C$ has been added to the list (line~\ref{line:ero:add_cell_set_response}).

\noindent$\bullet$~\textbf{\setrepositoryoperationresponse{}($ullo_L$, $ptr_L$, $response$)} tries to set the response of $ullo_L$ in the cell $C$ pointed to by $ptr_L$ to $response$.
To do so, $p$ tries to acquire the right to access $C$ by invoking the Acquire procedure on line~\ref{line:ero:set_response_acquire}; if it is successful, then $p$ tries to write $response$ into the $response$ object of $C$ by performing a CAS operation on this object (line~\ref{line:ero:responses_set_attempt}).
We prove that by the time $p$ exits this procedure, the $response$ object of $C$ was set to $response$ (this is true even if $p$ was unsuccessful in acquiring the right to access $C$ or $p$ failed the CAS on line~\ref{line:ero:set_response_acquire}).
Note that, before exiting this procedure, if $p$ was successful in acquiring the right to access $C$, then $p$ relinquishes its right to access $C$ on line~\ref{line:ero:set_response_relinquish}.

\noindent$\bullet$~\textbf{AcquireNext($ullo_L$, $curr\_ptr$)} assumes that the process $p$ invoking it has the right to access the cell $C$ pointed to by $curr\_ptr$, and it returns one of the following: $(\found, next\_ptr)$, meaning $p$ has acquired the right to access the next cell in the list (i.e., the one pointed to by $next\_ptr$); $(\notfound{}, \arbitraryvalue)$, meaning there is no cell after the one pointed to by $curr\_ptr$ in the list (so $C$ is the last cell of the list); and $(\timechange, \arbitraryvalue)$, meaning the operation stored in $L$ is no~longer~$ullo_L$ (which implies that $ullo_L$ has already taken effect, so there is no needed to acquire the right to access the cell after $C$).
To acquire the right to access the next cell $C^+$ in the list, $p$ enters a repeat-until loop (line~\ref{line:ero:acquire_next_repeat_loop}) in which it repeatedly tries to increment the acquisitions counter for $C^+$, which is stored in $C$.
To do so, $p$ first reads the $next$ object of $C$ to learn the current number of acquisitions $a$ of $C^+$ and the pointer $next\_ptr$ to $C^+$ (line~\ref{line:ero:acquire_next_read_curr_unique_pointer}).
Then, $p$ performs a CAS operation on the $next$ object of $C$ to set the acquisitions counter for $C^+$ to $a + 1$ (line~\ref{line:ero:acquire_next_cell}).
If this CAS operation succeeds, $p$ has acquired the right to access $C^+$.
There are three ways that $p$ can exit this repeat-until loop: (1)~the CAS operation on line~\ref{line:ero:acquire_next_cell} succeeds, in which case $p$ returns $(\found, next\_ptr)$; (2)~$p$ finds that $C$ is the last cell in the list, so $C^+$ does not exist, in which case $p$ returns $(\notfound{}, \arbitraryvalue)$; or (3)~$p$ finds that $L$ no longer stores $ullo_L$, in which case $p$ returns $(\timechange, \arbitraryvalue)$.\footnote{This is the ``bail out'' mechanism described in Section 4.1. Recall that this is needed because $p$ could be stuck trying to acquire $C^+$ because its CAS operations fail forever.}

\noindent$\bullet$~\textbf{Acquire($ullo_L$, $target\_ptr$)} traverses the list of cells to find the cell pointed to by $target\_ptr$ and acquires the right to access it. This procedure returns one of the following: $\found$, meaning $p$ has acquired the right to access the cell pointed to by $target\_ptr$; $\notfound{}$, meaning the cell pointed to by $target\_ptr$ is not in the list; and $\timechange$, meaning the operation stored in $L$ is no longer $ullo_L$.
To determine whether the cell pointed to by $target\_ptr$ is in the list, $p$ searches for this cell by traversing the list, starting from the head $H$ of the list.
To do so, $p$ maintains a pointer $curr\_ptr$ to the current cell $C$ that it has reached in its traversal.
Process $p$ initializes $curr\_ptr$ to $\&H$, and then it
enters the while loop on line~\ref{line:ero:acquire_loop_until} which continues until $curr\_ptr = target\_ptr$, i.e., until $p$ finds the cell pointed to by $target\_ptr$; at this point $p$ has acquired the right to access it.
In each iteration of this loop, $p$ invokes the AcquireNext procedure to try to acquire the right to access the successor $C^+$ of the cell $C$ pointed to by $curr\_ptr$.
Then, irrespective of the result, it relinquishes the right to access $C$.
If $p$ fails in acquiring the right to access $C^+$, then $p$ returns $\timechange$ or $\notfound$ (depending on the reason why it failed).
Otherwise, (i.e., if $p$ succeeds in acquiring the right to $C^+$) $p$ sets $curr\_ptr$ to the pointer of $C^+$.
If $C^+$ is the cell pointed to by $target\_ptr$, then $p$ will exit the loop and return $\found$.

\noindent$\bullet$~\textbf{IsDone($ullo_L$, $ullo_A$, $ptr_A$)} checks whether $ullo_A$ has taken effect.
This procedure returns one of the following: $\notdone$, meaning~$ullo_A$ has not taken effect; $\done$, meaning $ullo_A$ has taken effect; and $\timechange$, meaning the operation stored in $L$ is no longer $ullo_L$.
Observe that this procedure \emph{cannot} simply check if $ullo_A$ has already taken effect by just reading the $response$ object of the cell $C_A$ pointed to by $ptr_A$.
This is because $p$ \emph{cannot} access any cell (including $C_A$) before acquiring the right to access it, i.e., incrementing the acquisitions counter for $C_A$ (which is located in the predecessor of $C_A$ in the list).
So, $p$ first tries to acquire the right to access $C_A$ by executing the Acquire procedure on line~\ref{line:ero:announce_acquire}; let $\status$ be its response.
If $\status = \timechange$, then IsDone returns $\timechange$.
Otherwise, $\status$ equals $\notfound$ or $\found$.
At this point, $p$ can determine whether $ullo_A$ has taken effect or not as follows: $ullo_A$ has \emph{not} taken effect if and only if $ullo_A$ is an $\addcell$ operation and $\status = \notfound$ (which means that the Acquire procedure did not find $C_A$ in the list); $ullo_A$ is a $\removecell{}$ operation and $\status = \found$ (which means that the Acquire procedure found $C_A$ in the list); or $ullo_A$ is an $\doopandcopyresponse$ operation, and the response in $C_A$ is still ($ullo_A$, $\nullconstant$).
If $p$ determines that $ullo_A$ has not taken effect, the IsDone procedure returns $\notdone$, and otherwise returns $\done$.
Before returning, however, $p$ relinquishes its right to access $C_A$ if the Acquire procedure found $C_A$ in the list, i.e., if $\status = \found$.

\noindent$\bullet$~\textbf{Relinquish($curr\_ptr$)} assumes that the process $p$ invoking it has the right to access the cell $C$ pointed to by $curr\_ptr$, and it is used by $p$ to relinquish its rights to access $C$.
To do so, $p$ increments the revocations object of $C$ by doing a fetch-and-increment operation on it.
Let $c$ be the value of this object immediately after this fetch-and-increment operation.
As we explained in Section 4.1, when $c > 0$, $c$ is the value of the \emph{revocations} counter for $C$; otherwise, $|c|$ is the value of the \emph{reference} counter for $C$.
We prove that if $c = 0$, this relinquish by $p$ is the last relinquish for $C$, and so $p$ can safely free $C$.\footnote{In the pseudocode, $c = x + 1$ because fetch-and-increment fetches the value of the object \emph{before} incrementing it.}

\noindent$\bullet$~\textbf{\doapplyandcopyresponse{}($ullo_L$, $ptr_L$)} is used by a process $p$ to: (1) apply the high-level operation $hlo_L$ stored in $ullo_L$ (where $ullo_L = (\arbitraryvalue, \langle \doopandcopyresponse, hlo_L \rangle)$) to the target object state stored in $S$, and (2) copy its response into the cell $C$ pointed to by $ptr_L$.
To do so, $p$ first reads the content $(ullo_S, s_S, r_S)$ of $S$ (line~\ref{line:ero:state_read}).
Recall that $ullo_S = (\arbitraryvalue, \langle \doopandcopyresponse, hlo_S \rangle)$ where $hlo_S$ is the last high-level operation to the target object that has taken effect, $s_S$ is the current state of the target object, and $r_S$ is the response of $hlo_S$. Then, $p$ checks if $ullo_L$ is still stored in $L$ (line~\ref{line:ero:state_linearization_check}); if $L$ has changed, then $hlo_L$ has already been applied to the target object and the response of $hlo_L$ has already been copied into $C$, so $p$ exits the procedure.
Otherwise, $ullo_L$ is still stored in $L$.
In this case, we prove that $hlo_L$ has been applied to the target object if and only if $ullo_S = ullo_L$.
So $p$ now checks whether $ullo_S \neq ullo_L$ (line~\ref{line:ero:check_if_already_applied}), and if so $p$ tries to apply $hlo_L$ to the target object.
To do so, $p$ determines the new state $s$ and response $r$ by applying $hlo_L$ to $s_S$ using the state-transition function $apply_{\mathcal{T}}$ of the type $\mathcal{T}$ of the target object (line~\ref{line:ero:apply_op}), and tries to write $(ullo_L, s, r)$ into $S$ using a CAS operation (line~\ref{line:ero:state_cas}).
We prove that regardless of whether this CAS is successful or not, $hlo_L$ has been applied to the target object.
What remains to be done is to copy the response from $S$ into $C$.
To do so, $p$ first reads the current response from $S$ (line~\ref{line:ero:apply_update_response}) ($p$ must re-read the response from $S$ because of non-determinism: the response that $p$ got from $\textit{apply}_{\mathcal{T}}$ may be different from the response that the process that succeeded in writing the response into $S$ received).
Then $p$ invokes the \setrepositoryoperationresponse{} procedure
    to copy this response into the cell $C$ pointed at by $ptr_L$.

\noindent$\bullet$~\textbf{\doremovecell{}($ullo_L$, $ptr_L$)} is used by a process $p$ to remove the cell $C$ pointed to by $ptr_L$ from the list if $p$ finds $C$ in the list.
Process $p$ does this in four stages: (1) $p$ traverses the list of cells to find $C$ and its predecessor $C^-$ (lines \ref{line:ero:remove_cell_initialize_pointers}-\ref{line:ero:remove_cell_update_pointers}); (2) if $p$ finds $C$, $p$ seals the acquisitions counter stored in $C$ to prevent any process from acquiring the right to access the successor $C^+$ of $C$ (lines \ref{line:ero:remove_cell_remove_seal_loop}-\ref{line:ero:seal_cell}); (3)~then it removes~$C$ from the list by setting the successor of $C^-$ to $C^+$ (lines \ref{line:ero:remove_cell_read_pointer_to_remove}-\ref{line:ero:remove_cell_from_list}); and (4) finally $p$ consolidates the acquisitions and revocations counter for $C$ into the reference counter for $C$ (line~\ref{line:ero:copy_acquisitions_to_revocations}).

To do stage (1), $p$ traverses the list of cells by maintaining a current pointer $curr\_ptr$ and a pointer to its predecessor $prev\_ptr$, which are initially $\&H$ and $\nullconstant$, respectively.
While $curr\_ptr$ does not point to $C$ (i.e., $curr\_ptr \neq ptr_L$), $p$ tries to acquire the right to access the cell after the cell pointed to by $curr\_ptr$ by invoking the AcquireNext procedure (line~\ref{line:ero:remove_cell_acquire_next}).
If the AcquireNext procedure returns $\found$ (line~\ref{line:ero:remove_cell_acquire_next_found}), then $p$ continues the traversal: $p$ relinquishes its right to $prev\_ptr$ (line~\ref{line:ero:remove_cell_traversal_relinquish}), updates $(prev\_ptr, curr\_ptr)$ to $curr\_ptr$ and the next pointer, respectively (line~\ref{line:ero:remove_cell_update_pointers}), and continues to the next iteration of the while loop.
If the AcquireNext procedure returns $\timechange$ or $\notfound{}$ (line~\ref{line:ero:remove_cell_check_acquire_status}), then $p$ skips stages (2), (3), and (4): $C$ has already been removed from the list.

To do stage (2), while the $next$ object of $C$ is not sealed, $p$ repeatedly performs a CAS operation on the $next$ object of $C$ to try to set its $sealed$ flag from $\false$ to $\true$ (line~\ref{line:ero:seal_cell}).
We prove that, even though these CAS operations can keep failing, the $next$ object of $C$ is eventually sealed, and so $p$ eventually exits this loop.

To do stage (3), $p$ repeatedly performs a CAS operation on the $next$ object of $C^-$ on line~\ref{line:ero:remove_cell_from_list} to try to: (a) copy the acquisitions counter for $C^+$ (which is in $C$) into $C^-$; and (b) set the successor of $C^-$ to $C^+$.
There are two ways $p$ can exit this loop.
First, if $p$ finds that $C$ has already been removed from the list (either because the successor of $C^-$ is $C^+$ or the operation $ullo_L$ to remove $C$ from the list is no longer stored in $L$), then $p$ immediately exits this loop.
Second, if $p$ performs a successful CAS operation on line~\ref{line:ero:remove_cell_from_list}, then $p$ is the process that removes $C$ from the list.

To do stage (4), $p$ must now compute the reference counter for $C$ from the acquisitions and revocations counters for $C$ (which are located in $C^-$ and $C$, respectively), and store it in $C$.
To do so, $p$ decrements the revocations counter for $C$ by the value of the acquisitions counter that $p$ previously read from $C^-$ (on line~\ref{line:ero:remove_cell_read_previous_pointer}).
Let $v$ be the value of the $revocations$ object of $C$ after this decrement.
Note that (a) $|v|$ now represents the \emph{reference} counter for $C$; and (b)~$v < 0$ (because $p$ still has the right to access $C$).

At this point (just after line~\ref{line:ero:copy_acquisitions_to_revocations}), $C$ has been removed from the list (and its reference counter was computed). The natural thing to do now would be to set the $response$ object of $C$ to $\done$ (by invoking $\setrepositoryoperationresponse{}(ullo_L, ptr_L, \done)$ between line~\ref{line:ero:copy_acquisitions_to_revocations} and \ref{line:ero:remove_cell_prev_relinquish}), \emph{but this does not work}.
To see why, consider the following scenario.
A process $p$ allocates a cell $C$ and completes the $\addcell$ and $\doopandcopyresponse$ operations and begins the $\removecell{}$ operation for $C$.
Process $p$ writes $(\removecell{},\&C)$ into the announce object $A$, and then goes to sleep.
A process $q$ now reads $(\removecell{},\&C)$ from $A$ and writes $(\removecell{},\&C)$ into $L$.
Then, $q$ reads $(\removecell{},\&C)$ from $L$ and invokes the \doremovecell{} procedure to try to remove $C$ from the list.
Process $q$ then successfully removes $C$ from the list, but crashes before setting $C$'s response to $\done$.
A process $r$ then invokes an operation on the target object, and begins the $\addcell$ operation for the cell it allocated.
Process $r$ then reads $(\removecell{},\&C)$ from $L$ and invokes the \doremovecell{} procedure to try to remove $C$ from the list.
During this procedure, when $r$ invokes the \setrepositoryoperationresponse{} procedure between line~\ref{line:ero:copy_acquisitions_to_revocations} and \ref{line:ero:remove_cell_prev_relinquish}, $r$ sees that $C$ has already been removed from the list (because $q$ removed it earlier).
Since $C$ is no longer in the list, the \setrepositoryoperationresponse{} procedure cannot acquire the right to access $C$.
Thus, it does \emph{not} set $C$'s response to $\done$ (to inform $p$ that $C$ has been removed).
So, $r$ exits the \doremovecell{} procedure \emph{without settings $C$'s response to $\done$}.\footnote{Notice that this argument does not depend on how $r$ tries to set $C$'s response to $\done$: once $C$ has been removed from the list, no newly arrived process can acquire the right to access $C$ in order to set its response to $\done$.}
Since no process other than $p$ can acquire the right to access $C$, no process other than $p$ can set the $response$ object of $C$ to $\done$.
Afterward, $r$ writes $\addcell$ into $A$, then into~$L$, and crashes.
At this point, $p$ wakes up and cannot determine that $(\removecell{},\&C)$ was written into~$L$.
This is because (1) $(\removecell{},\&C)$ is no longer in $L$ and (2) $p$ cannot determine whether $C$ has been removed from the list by traversing it to see if $C$ is still there (since new processes may arrive and traverse the list concurrently, impeding $p$'s traversal).
So, $p$ will not set the $response$ object of $C$ to $\done$.
Thus, this object remains \nullconstant{} forever, and so $p$ will loop forever in the loop on line~\ref{line:ero:do_work_while_loop}.

So instead of invoking $\setrepositoryoperationresponse{}(ullo_L, ptr_L, \done)$ between line~\ref{line:ero:copy_acquisitions_to_revocations} and \ref{line:ero:remove_cell_prev_relinquish}, to set the $response$ object of $C$ to $\done$, we do so in the \emph{first line} of the \doremovecell{} procedure.
In other words, we tell the process that wants to remove~$C$ from the list that $C$ has been removed from the list \emph{before we actually remove $C$ from the list!}
Although this does not look right, we prove that it does not affect the universal construction's correctness or asymptotic space complexity.
Roughly speaking, this is because the operation to remove $C$, namely $ullo_L$, is in $L$, and so no other operation can now occur unless this removal is done first.

\begin{papertheorem}\label{thm:algo_2}
    Algorithm~\ref{alg:efficient_algo} is a wait-free universal construction for the infinite-arrival model with bounded concurrency.
    Its space complexity at time $t$ is linear in the point contention at $t$.
\end{papertheorem}

\tocless\section{Related Work}\label{relatedwork}

Most object implementations in shared-memory systems do not work in the infinite-arrival model: they assume a system with $n$ processes, where $n$ is known to the processes (this is the \emph{$n$-arrival model} \cite{aguilera2004pleasant}).
Object implementations for these systems typically use this known $n$ in their code, and they use some number of base objects that depends on this $n$.
These base objects are statically allocated in \emph{every} run, even those in which fewer than $n$ processes actually participate.
This is clearly undesirable.

To avoid this, researchers have designed algorithms for systems where the number of processes is bounded but unknown (this is called the \emph{finite-arrival model} \cite{aguilera2004pleasant}), e.g., algorithms in~\cite{michael2004hazard, jayanti2005efficiently, aguilera2004pleasant}.
But such algorithms may not work in the infinite-arrival model, i.e., if an infinite number of processes may participate in a run. For example, as Aguilera pointed out in \cite{aguilera2004pleasant}, the simple naming algorithm in \cite[Figure 4]{aguilera2004pleasant} is not wait-free in these runs.

Researchers have also designed algorithms for the infinite-arrival model (e.g.,~\cite{merritt2000computing, gafni2001concurrency, aguilera2004pleasant, aspnes2002wait, afek2006common2, perrin2020extending, bedin2021wait, bedin2024wait, jayanti2005efficiently2}).
Existing algorithms for this model are typically not space-efficient, and some use infinitely many objects in every run (e.g.,~\cite{merritt2000computing, gafni2001concurrency, aguilera2004pleasant, aspnes2002wait, afek2006common2}).
% Researchers have also considered wait-free universal constructions for the infinite-arrival model~\cite{aspnes2002wait, perrin2020extending, bedin2021wait, bedin2024wait}.
% The universal constructions of \cite{aspnes2002wait} and \cite{perrin2020extending} were not designed to be space efficient.
In particular, the universal construction of \cite{aspnes2002wait} uses infinitely many base objects in every run because it relies on the collect algorithm of \cite{gafni2001concurrency}, and the universal construction of \cite{perrin2020extending} uses space linear in the number of \emph{operations} applied so far.
Observe that this can be much higher than the number of \emph{processes} that have participated so far, because each participating process can apply arbitrarily many operations.
So, the space complexity of \cite{perrin2020extending} can be much higher (and never less) than the space complexity of our first universal construction.
%Observe that this is higher than the space complexity of our first universal construction, whose space complexity at time $t$ is linear in the number of processes that have participated by time $t$.

Other work has focused on designing space-efficient algorithms for the infinite-arrival model. For example, the LL/SC implementation from CAS in~\cite{jayanti2005efficiently2} uses a number of base objects linear in the number of processes that have participated so far.

The universal constructions of~\cite{bedin2021wait, bedin2024wait} were also developed for the infinite-arrival model and aim to achieve space efficiency. However, the constructions of~\cite{bedin2021wait, bedin2024wait} rely on an external garbage collection mechanism that \emph{automatically frees} any object that “becomes inaccessible by any process in the system”~\cite{bedin2021wait}, even though no such mechanism is provided. Determining when an object becomes inaccessible and can be safely reclaimed is itself a difficult problem~\cite{brown2015reclaiming}. Indeed, to the best of our knowledge, there is no known automatic wait-free garbage collection mechanism for the infinite-arrival model.
In contrast, our universal constructions do not assume any garbage collection mechanism. In particular, in our second algorithm, processes \mbox{explicitly manage the recycling of objects.}

It is worth noting that the algorithms given ~\cite{bedin2021wait, bedin2024wait} are analyzed under a space complexity measure that
    accounts only for the space used \emph{at quiescent times}, i.e., only at times when no operations are executing.
But this measure provides no bounds on the space used in runs without quiescent times, i.e., runs in which at every moment at least one operation is executing.

% It is worth noting that the space complexity measure defined in~\cite{bedin2021wait, bedin2024wait} (which is used 
%     accounts only for the space used by algorithms \emph{at quiescent times}, i.e., only at times when no operations are executing.
% But this measure provides no bounds on the space that an algorithm takes in runs without quiescent times, i.e., runs in which at every moment at least one operation is being performed.\SMP{is in progress?}

%Since there exist runs with no quiescent times, this measure provides no bound for such executions.

\rmv{
Under our space-complexity measure and without assuming automatic garbage collection, the constructions of~\cite{bedin2021wait, bedin2024wait} use space linear in the number of \emph{operations} performed so far, which is often larger than the space complexity of our first universal construction.\MMP{If needed, we can remove.}
}

%There is a large body of work on memory management for shared memory systems.
%Closest to our paper are~\cite{detlefs2001lock, valois1995lock, herlihy2005nonblocking, herlihy2002repeat, michael2002safe, michael2004hazard, brown2015reclaiming}.
%These, however, are either not wait-free or don't work for the infinite-arrival model.

% There is a large body of work on memory management for shared-memory systems. The work most closely related to ours includes~\cite{detlefs2001lock, valois1995lock, herlihy2005nonblocking, herlihy2002repeat, michael2002safe, michael2004hazard, brown2015reclaiming}. However, these approaches are either not wait-free or do not apply to the infinite-arrival model.\MMP{group by problem: non-blocking, no infinite-arrival model, and optimistic accessing. We note that, in addition to these issues, some prior work (e.g., \cite{detlefs2001lock, valois1995lock}) makes the assumption that cells can always be accessed, even if they are unallocated. In contrast, our algorithms do not make such assumptions.}

\rmv{There is a large body of work on recycling memory for shared-memory systems, e.g.,~\cite{detlefs2001lock, valois1995lock, herlihy2005nonblocking, herlihy2002repeat, michael2002safe, michael2004hazard, brown2015reclaiming, anderson2021concurrent}.
Typically, memory is recycled using techniques such as \emph{reference counters} (e.g., \cite{detlefs2001lock, valois1995lock, herlihy2005nonblocking, anderson2021concurrent}) and \emph{hazard pointers} (and similar techniques)~\cite{herlihy2002repeat, michael2002safe, michael2004hazard}.
To the best of our knowledge, existing methods for reference counters or hazard pointers (or, more generally, for recycling memory) cannot be used in our setting for several reasons.
Some of these methods are only non-blocking but not wait-free~\cite{detlefs2001lock, valois1995lock, herlihy2005nonblocking}.
Others do not work in systems with infinitely many processes~\cite{herlihy2005nonblocking, herlihy2002repeat, michael2002safe, michael2004hazard, brown2015reclaiming, anderson2021concurrent}.
Some of them allow processes to access cells that are currently \emph{not} allocated~\cite{detlefs2001lock, valois1995lock}, which is typically not allowed.
}

\rmv{
There is extensive work on memory recycling in shared-memory systems (e.g., ~\cite{detlefs2001lock, valois1995lock, herlihy2005nonblocking, herlihy2002repeat, michael2002safe, michael2004hazard, brown2015reclaiming, anderson2021concurrent}).
Common approaches include reference counting (e.g., \cite{detlefs2001lock, valois1995lock, herlihy2005nonblocking, anderson2021concurrent}) and hazard pointers or related techniques (e.g., ~\cite{herlihy2002repeat, michael2002safe, michael2004hazard}).
To the best of our knowledge, none of the memory recycling methods described in the literature can be applied in our setting. Some are non-blocking but not wait-free (e.g., ~\cite{detlefs2001lock, valois1995lock, herlihy2005nonblocking}).
Others do not support infinitely many processes~(e.g.,~\cite{herlihy2005nonblocking, herlihy2002repeat, michael2002safe, michael2004hazard, brown2015reclaiming, anderson2021concurrent}).
Moreover, some allow processes to access cells that are not currently allocated (e.g., ~\cite{detlefs2001lock, valois1995lock}), which is typically not allowed.\MMP{We need to bring up the fact that many of these methods cannot even hope to achieve our complexity result because they use an array of length n or an array that is monotonically increasing.}
}
\rmv{
However, \cite{detlefs2001lock, valois1995lock, herlihy2005nonblocking} are only non-blocking (they are not wait-free) because they acquire the right to access a cell by repeatedly trying to win a CAS \emph{without aborting} (unlike the mechanism in \Cref{alg:efficient_algo}); 

and \cite{herlihy2002repeat, michael2002safe, michael2004hazard, brown2015reclaiming, anderson2021concurrent} do not work for the infinite-arrival model because they reclaim memory by scanning an array of length equal to the number of processes in the system (which would be infinite).
}

We note that the infinite-arrival model and its variants, including versions with bounded and unbounded concurrency, were introduced by Merritt and Taubenfeld in their seminal paper~\cite{merritt2000computing}.
The GCAS object and our first universal construction originally appeared in \cite{hadzilacos2024generalized2}.
This universal construction was inspired by the \emph{2-nonblocking} universal construction of \cite{chan2021differentiated}; in particular, as in our construction, processes compete on a single announce~object.

\rmv{
Finally, we note that the infinite-arrival model and several of its variants—including versions with bounded and unbounded concurrency—were introduced by Merritt and Taubenfeld in their seminal paper~\cite{merritt2000computing}.
}

% Some of them work only for the $n$-arrival or finite-arrival model~\cite{herlihy2002repeat, michael2002safe, michael2004hazard}, and the others are non-blocking~\cite{detlefs2001lock, valois1995lock, herlihy2005nonblocking}.

% \SMP{Mention all these references in related work, and explain why we cannot just adopt them. All of them were designed for the $n$-arrival model, not for the infinite-arrival model.
% Moreover, non-blocking rather than wait-free. Also mention first universal construction by herlihy also had recycling, but for n-arrival model only
% Pass the buck~\cite{herlihy2002repeat}, hazard pointers~\cite{michael2002safe}, both only work for the $n$-arrival model (they used a fixed array). Extension of hazard pointers works for the finite arrival model~\cite{michael2004hazard}}

\rmv{
In contrast, we do not assume an automatic garbage collection mechanism.
Our first algorithm does not attempt to recycle memory, so it does not use (or assume) a garbage collector.
Our second algorithm incorporates a novel garbage collection mechanism to free objects that become inaccessible; in fact, much of its complexity is due to this mechanism.
Furthermore, the space efficiency of these algorithms is with respect to a space complexity measure that is problematic: it measures the space used only at "quiescent" times, i.e., times when there are no operations being executed.
But there are runs with no quiescent times, and so this space complexity measure says nothing about them.
It's worthwhile noting that without the automatic garbage collection assumption and under our space complexity measure, the universal constructions of \cite{perrin2020extending, bedin2021wait, bedin2024wait} use space linear in the number of operations applied so far.
}

\tocless\section{Conclusion}\label{conclusion}
We introduced GCAS, a simple and natural generalization of CAS, and showed how it can be used to obtain two space-efficient, wait-free universal constructions in the infinite-arrival model. The first has space-complexity linear in the number of processes that have participated so far, the second has space-complexity linear in the point contention but assumes bounded concurrency.

A natural question is whether such universal constructions can be achieved using CAS instead of GCAS. Equivalently, can GCAS be implemented in a space-efficient manner using CAS in the infinite-arrival model?

If the answer is yes, then %substituting GCAS with 
plugging such an implementation into our algorithms would immediately yield space-efficient, wait-free universal constructions based on CAS.
If the answer is no, this would demonstrate that GCAS is strictly more powerful than CAS for at least one purpose: obtaining space-efficient, wait-free universal constructions in the infinite-arrival model.

We conclude with a final open question: can one achieve the space complexity of our second universal construction in the infinite-arrival model \emph{without assuming bounded concurrency?}

\printbibliography

\hfill\\

\clearpage

\appendix

\renewcommand{\contentsname}{Appendix Contents}

\tableofcontents

\newpage

% !TEX root = ../main.tex

\section{Model}
\label{sec:model}

In this section, we elaborate on the model given in Section 2 as needed for our proofs.
We consider an infinite arrival distributed system where possibly infinitely many asynchronous processes that may fail by crashing communicate via shared objects such as \GCAS{} and \FA{}.
In such systems, shared objects can be used to \emph{implement} other shared objects such that the implemented objects are linearizable and wait-free.

An implementation $\mathcal{A}$ of a \bemph{target} object $O$ from a set of \bemph{base} objects $\mathcal{B}$ is
	a collection of procedures that specify how any process in the system
	can perform any operation of $O$ by applying operations to the base objects in~$\mathcal{B}$.
	
A \bemph{configuration} of implementation $\mathcal{A}$ is
	a complete description of the state of the computation at some point in time
	during a run of the implementation.
Formally, it is a function that assigns
	a state to each process and
	a state to each shared object in $\mathcal{B}$ used by the implementation.
The state of a process consists of the values of its local variables and its program counter.
In our case the base objects used by the implementation
	include the objects contained in cells that can be allocated by the memory manager,
	\emph{even those that are not currently allocated}.
The \bemph{initial configuration} $C_0$ of the implementation $\mathcal{A}$ assigns
	to each process its initial state and
	to each object the initial state specified by the implementation.

Each process executes \bemph{steps}.
Formally, a step is a triple $(C,p,C')$,
	where $p$ is a process, $C,C'$ are configurations of the implementation,
	and $C'$ is obtained from $C$ by executing the instruction
	indicated by $p$'s program counter in $C$.
The instructions executed by $p$ are of the following types:
\begin{compactitem}[\noindent$\bullet$]
\item Invocation of an operation on the target object $O$.
\item Response of an operation on the target object $O$.
\item Computation involving only local variables of $p$ and at most \emph{one atomic} access to a base object.
% \item Atomic access to a base object;
%     this may involve the assignment of the return value of the object to a local variable of $p$.
%\VMP{Do we need to also mention AllocateCell and FreeCell instructions/steps?}\MMP{Since we model the memory manager as an object, it is ok to not mention it. Also need to mention that we update the local variables of the processes given the response of the object.}
\end{compactitem}
Accordingly we call the step $(C,p,C')$ involving the execution of such an instruction an
	\bemph{invocation}, \bemph{response}, or \bemph{implementation} step.%\MMP{I like this naming! But I will not have time to incorporate it into the proof before submission. I will do so eventually.}

An \bemph{implementation history} $\mathcal{I}$ of $\mathcal{A}$
	is a (finite or infinite) sequence of steps
	$$(C_0,p_1,C_1),(C_1,p_2,C_2),(C_2,p_3,C_3),\ldots.$$
That is, $\mathcal{I}$ describes a possible sequence of steps
	taken by processes during a run of $\mathcal{A}$,
	starting from its initial configuration.
For brevity, we write this implementation history as
	$$C_0,p_1,C_1,p_2,C_2,p_3,C_3,\ldots,$$
	but it is important to keep in mind that
	technically an implementation history is a sequence of \emph{steps}.
We note that, in the special case where $\mathcal{I}$ is an implementation history of $\mathcal{A}$ consisting of zero steps, $\mathcal{I} = C_0$ where $C_0$ is the initial configuration of $\mathcal{A}$.

We require that, for every process $p$,
	the subsequence of implementation history $\mathcal{I}$ consisting of the steps of $p$
	is composed of the repetition, zero or more times, of the following pattern:
\begin{compactenum}[\noindent(a)]
\item the invocation step for an operation $o$ of the target object $O$,
\item a sequence of implementation steps, and
\item the response step for $o$.
\end{compactenum}
where the last repetition may be a prefix of this pattern.
The sequence of implementation steps in~(b)
	is precisely as specified by the implementation $\mathcal{A}$
	for how $p$ is to perform operation $o$.

%In an implementation $\mathcal{A}$ of an object $O$, each process $p$ executes \emph{steps} of three kinds: the invocation by $p$ of an operation $o$ on $O$ (we call $p$ the \emph{owner} of $o$); the receipt by $p$ of a response $r$ from $O$; and the \emph{low-level steps}\VMP{Can we avoid this terminology?  Potential confusion with ``low-level operations'' in Algo~2.} required by $\mathcal{A}$ to perform $o$ such as performing operations on the base %\GCAS{} and \FA{}
%    objects.
%An \emph{implementation history} $\mathcal{I}$ of $\mathcal{A}$ is a sequence of invocation, response, and low-level steps such that for each process $p$, the subsequence of $\mathcal{I}$ involving only the steps of $p$, denoted as $\subhistory{\mathcal{I}}{p}$, follows the following pattern repeated zero or more times: an invocation step for some operation $o$, some number of low-level steps required by $\mathcal{A}$ to perform $o$, and a response step for $o$, where the last repetition
%    is a prefix of this pattern.
%%in $\subhistory{\mathcal{I}}{p}$ could be missing the response step.
An \bemph{operation execution} $opx$ of process $p$ in $\mathcal{I}$ consists of an invocation step of $p$ and all the subsequent steps of $p$ in $\mathcal{I}$ up to and including the next response step of $p$, if such a step exists.
If $opx$ ends with a response step, we say that $opx$ is \bemph{complete} in $\mathcal{I}$ and its invocation and response steps are \bemph{matching}; otherwise, we say that $opx$ is \bemph{incomplete} in $\mathcal{I}$.
Accordingly, the \bemph{point contention in $\mathcal{I}$} is the number of pending operations in $\mathcal{I}$.\footnote{Note that this is the same as \Cref{def:point_contention} except the time is fixed to be ``the end" of the implementation history.}

An \bemph{object history} $\mathcal{H}$ of $O$
	is a sequence of invocation and response steps, such that for each process~$p$, the subsequence of $\mathcal{H}$ involving only the steps of $p$, denoted $\subhistory{\mathcal{H}}{p}$, consists of an alternating sequence of invocation and response steps, starting with an invocation step.
An \bemph{operation execution} $opx$ of process $p$ in $\mathcal{H}$ is either a pair consisting of an invocation step of $p$ and the next response step of $p$, if such a step exists; or the last invocation step of $p$ in $\mathcal{H}$, if that step is not followed by a response step of $p$.
In the first case, we say that $opx$ is \bemph{complete} in $\mathcal{H}$ and the two steps of $opx$ are \bemph{matching}; in the second case, $opx$ is \bemph{incomplete} in $\mathcal{H}$.
A\cyan{n object} history~$\mathcal{H}$ is \bemph{complete} if all operation executions in $\mathcal{H}$ are complete.
A \bemph{completion} of $\mathcal{H}$ is a\cyan{n object} history~$\mathcal{H'}$ formed by removing the invocation step of, or adding a response step to, each incomplete operation execution in $\mathcal{H}$; thus $\mathcal{H'}$ is complete.
Two \cyan{object} histories $\mathcal{H}_1$ and $\mathcal{H}_2$ are \bemph{equivalent} if for all processes $p$, $\subhistory{\mathcal{H}_1}{p} = \subhistory{\mathcal{H}_2}{p}$.
A\cyan{n object} history $\mathcal{H}$ induces an irreflexive partial order $<_\mathcal{H}$ on operation executions: $opx_1 <_\mathcal{H} opx_2$ if $opx_1$'s response step occurs before $opx_2$'s invocation step in~$\mathcal{H}$.
A\cyan{n object} history $\mathcal{S}$ is \bemph{sequential} if $\mathcal{S}$ starts with an invocation step and each invocation step is immediately followed by its matching response step.

Each object has a type that specifies how the object behaves when it is accessed sequentially.
Formally, an \bemph{object of type $\mathcal{T}$} is specified by a tuple $(OP, RES, Q, \delta, s_0)$, where $OP$ is a set of operations, $RES$ is a set of responses, $Q$ is a set of states, $\delta \subseteq Q \times OP \times Q \times RES$ is a state transition relation, and $s_0\in Q$ is the initial state of $\mathcal{T}$.
A tuple $(s, o, s', r)$ in $\delta$ means that if type~$\mathcal{T}$ is in state $s$ when operation $o \in OP$ is invoked, then $\mathcal{T}$ can change its state to $s'$ and return the response $r$.
Note that $\delta$ is a relation as opposed to a function to capture non-determinism.
A sequential \cyan{object} history $\mathcal{S}$ is \bemph{legal with respect to $\mathcal{T}$} if the operation responses in $\mathcal{S}$ could be those received when applying these operations sequentially, in the order dictated by $\mathcal{S}$, on an object of type $\mathcal{T}$.
That is, $\mathcal{S} = invocation(opx_1, o_1), response(opx_1, r_1), \ldots$ is legal with respect to $\mathcal{T}$ if there are $s_1, s_2, \ldots$ in $Q$ such that $(s_{i-1}, o_i, s_i, r_i) \in \delta$ for all $i\geq1$, where $s_0$ is the initial state of~$\mathcal{T}$.

An implementation $\mathcal{A}$ of an object $O$ of type $\mathcal{T}$ should be \mbox{\bemph{linearizable with respect to $\mathcal{T}$}}: even when $O$ is accessed concurrently by processes that use $\mathcal{A}$, every operation on $O$ must appear to take effect instantaneously, at some point during its execution interval, according to type $\mathcal{T}$.
More precisely, \bemph{a\cyan{n object} history $\mathcal{H}$ of $O$ is linearizable with respect to $\mathcal{T}$} if there is a completion $\mathcal{H'}$ of~$\mathcal{H}$ that is equivalent to some sequential \cyan{object} history $\mathcal{S}$ that is legal with respect to $\mathcal{T}$ and $<_\mathcal{H'} \subseteq <_\mathcal{S}$. 
\bemph{An implementation history $\mathcal{I}$ of $\mathcal{A}$ is linearizable with respect to $\mathcal{T}$} if the \cyan{object} history~$\mathcal{H}$ obtained by removing all implementation steps from $\mathcal{I}$ is linearizable with respect to $\mathcal{T}$.
Finally, \cyan{implementation} $\mathcal{A}$ is \bemph{linearizable with respect~$\mathcal{T}$} if every implementation history $\mathcal{I}$ of $\mathcal{A}$ is linearizable with respect to type $\mathcal{T}$.

An implementation \cyan{$\mathcal{A}$} of an object is \bemph{wait-free} if, in any implementation history \cyan{of $\mathcal{A}$}, a process cannot invoke an operation and then take infinitely many steps without completing it.

\section{Proof of \Cref{alg:wait-free-simple}}

The goal of this section is to prove the step complexity of \Cref{alg:wait-free-simple} and that it is linearizable.
Throughout the proof of both algorithms, we use the notion of ``time" as a surrogate for the position of a step in any implementation history $\mathcal{I}$.
Specifically, if $\mathcal{I} = (C_0, p_1, C_1), (C_1, p_2, C_2), \ldots$, then time $t$ in $\mathcal{I}$ is the $t$th step in $\mathcal{I}$, i.e., $(C_{t - 1}, p_t, C_t)$.
When we talk about the value of an object or local variable at time $t$, we are referring to the value of this object or local variable in $C_t$.
Furthermore, when we talk about the operation that occurred at time $t$ (if any), we are referring to the operation that $p_t$ performed  $(C_{t - 1}, p_t, C_t)$.
So, when thinking about time $t$, we are always referring either to the action performed by $p_t$ or the state of $C_t$, and never $C_{t - 1}$.
In the few cases when we need to talk about the state of $C_{t - 1}$, we opt instead to talk about the state at time $t - 1$.

\subsection{Basic Facts}

Before we prove the step complexity of \Cref{alg:wait-free-simple} and that it is linearizable, we begin with some definitions and basic facts.
Throughout the proof of \Cref{alg:wait-free-simple}, we use the term ``operation" to refer to an operation execution.
For convenience, we would like the notion of a cell of a process to be well-defined in every implementation history of \Cref{alg:wait-free-simple} we consider in this proof.
To do so, we make the following assumption, which we note is without loss of generality for our purposes.

\begin{assumption}\label{assumption:cell_well_defined}
    Every process that participates in any implementation history of \Cref{alg:wait-free-simple} takes at least two steps.
\end{assumption}

This is assumed without loss of generality for our purposes because of the following.
Consider any implementation history $\mathcal{I}$ of \Cref{alg:wait-free-simple} in which some number of processes take a single step.
Let $\mathcal{I}^-$ be the subsequence of $\mathcal{I}$ where all steps by these processes are removed.
Since these processes just perform an invocation step and nothing else, it follows that $\mathcal{I}^-$ is an implementation history of \Cref{alg:wait-free-simple}.
Observe that any linearization of $\mathcal{I}^-$ is a linearization of $\mathcal{I}$.
This is because any completion of the object history of $\mathcal{I}^-$ (which is created by removing all of the implementation steps from $\mathcal{I}^-$) is a completion of the object history of $\mathcal{I}$ (because all operations invoked by processes that take a single step are not complete in $\mathcal{I}$, so we remove them in the completed object history).
Furthermore, an upper-bound on the step complexity of any operation in $\mathcal{I}^-$ is also an upper bound on the step complexity of any operation in $\mathcal{I}$ because the only operation in $\mathcal{I}$ not in $\mathcal{I}^-$ are ones which take a single step.
Finally, the space complexity of $\mathcal{I}^-$ and $\mathcal{I}$ is the same because all processes that take a single step in $\mathcal{I}$ do not perform any $\allocatecelloperation$ operations.

For convenience, we treat $\noop$ as an operation that occurred in \Cref{alg:wait-free-simple}.
So, when we consider any operation $o$, this is either (1) $\noop$ or (2) an operation execution.

Consider any implementation history $\mathcal{I}$ of \Cref{alg:wait-free-simple}.
All claims are with respect to $\mathcal{I}$.

\begin{assumption}\label{assumption:1}
    $\nullconstant$ is a value that differs from all possible responses to all operations of type $\mathcal{T}$ and $\noop$ is a value that differs from all operations of $\mathcal{T}$.
\end{assumption}

\begin{definition}
    The invocation and response steps for an operation $o$ are \cref{line:op_start} and \cref{line:op_done}, respectively.
\end{definition}

\begin{definition}
    We call a \CASop{} / \GCASop{} operation successful if its response is true and unsuccessful otherwise.
\end{definition}

\begin{definition}\label{def:op_state}
    For an operation $o \neq \noop$: $p(o)$ denotes the process that invoked $o$; $t(o)$ is equal to the response of the \FIop{} operation executed by $p(o)$ on \cref{line:time_assignment} within $o$ or $\infty$ if $p(o)$ has never executed \cref{line:time_assignment} within $o$; and we use $\helpobject{p(o)}$ to refer both to the response $p(o)$ received on its first execution of \cref{line:semi-efficient-allocate-cell} and the cell it points to (this is well-defined by \Cref{assumption:cell_well_defined}.).
    We sometimes say that $\helpobject{p(o)}$ is the cell of $p(o)$.
    For the special case of $\noop$, $p(\noop)$ is undefined, $t(\noop) = 0$, and $\helpobject{p(\noop)}$ equals $ptr_{\noop}$ which is a pointer to a dummy cell initialized to $(0, \bot)$.
\end{definition}

\begin{definition}[Complete at $T$]\label{def:complete}
    An operation $o$ is \emph{complete at time $T$} if $p(o)$ executed \cref{line:op_done} at some time $T' \leq T$ within $o$.
\end{definition}

\begin{definition}[Done at $T$]\label{def:done}
    An operation $o$ is \emph{done at time $T$} if $\helpobject{p(o)} = (t(o), r)$ and $r \neq \nullconstant$ at some time $T' \leq T$.
\end{definition}

We start with two observations about operations in general.
An immediate consequence of \Cref{def:done} is

\begin{observation}\label{observation:if_once_done_then_always_done}
    If operation $o$ is done at time $T$ then for all times $T' \geq T$ $o$ is done at $T'$.
\end{observation}

Since timestamp assignment is done using a \FI{} object (\cref{line:time_assignment}) that is initially 1, each operation is assigned a unique timestamp greater than or equal to 1. Hence

\begin{observation}\label{observation:unique_timestamps}
    For all operations $o$ and $o'$, if $t(o) \neq \infty$ and $t(o') \neq \infty$ then $o \neq o'$ if and only if $t(o) \neq t(o')$.
\end{observation}

The following three observations concern $\announceobject{}$ and $\stateobject{}$.
Since $\announceobject{}$ is initially $(t(\noop), \noop, \helpobject{\noop})$ and the new values passed on \cref{line:g_a_gcas} and \cref{line:g_a_cas} are always of the form $(t(o), o, \helpobject{p(o)})$, we have:

\begin{observation}\label{observation:g_a_is_well_formed}
    For all times $T$ there exists an operation $o$ such that $\announceobject{}$ equals $(t(o), o, \helpobject{p(o)})$ at $T$.
    In this case, we also say that ``$(t(o), o, \helpobject{p(o)})$ is stored in $\announceobject{}$" or, for brevity, ``$o$ is stored in $\announceobject{}$" at time $T$.
    Furthermore, we say ``an execution of \cref{line:g_a_gcas} or \cref{line:g_a_cas} is for operation $o$" to mean that it is of the form $\text{GCAS(}-, \announceobject{}, (t(o), o, \helpobject{p(o)}), (t(o), o, \helpobject{p(o)}))$.
\end{observation}

Since $t(o)$ is unique to $o$, and processes only write timestamps into $\announceobject{}$ they received on \cref{line:time_assignment}, we have: 

\begin{observation}\label{observation:p_o_writes_o_into_announce}
    Every execution of \cref{line:g_a_gcas} or \cref{line:g_a_cas} for operation $o$ is by $p(o)$ within $o$.
\end{observation}

Like with \Cref{observation:g_a_is_well_formed}, $\stateobject{}$ is initially $(t(\noop), s_0, \bot, \helpobject{\noop})$.
Since every new time and cell pointer written into $\stateobject{}$ on \cref{line:g_r_cas} is supplied from reading $\announceobject{}$, they are always $t(o)$ and $\helpobject{p(o)}$ by \Cref{observation:g_a_is_well_formed}.
Moreover, the response field is equal to the response from $apply_{\mathcal{T}}$ which is always not $\nullconstant$ by \Cref{assumption:1}.
Hence:

\begin{observation}\label{observation:g_r_is_well_formed}
    For all times $T$ there exists an operation $o$ such that $\stateobject{}$ equals $(t(o), -, r, \helpobject{p(o)})$ at $T$ for some response $r \neq \nullconstant$.
    In this case, we also say that ``$(t(o), -, r, \helpobject{p(o)})$ is stored in $\stateobject{}$" or, for brevity, ``$o$ is stored in $\stateobject{}$" at time $T$.
    Furthermore, we say ``an execution of \cref{line:g_r_cas} is for operation $o$" to mean that it is of the form $\text{CAS}(\stateobject{}, -, (t(o), -, r, \helpobject{p(o)}))$.
\end{observation}

The following two observations concern each process $p$'s cell $\helpobject{p}$.
By \Cref{observation:g_r_is_well_formed}:

\begin{observation}\label{observation:help_pointer_cas_is_well_formed}
    Each execution of \cref{line:help_pointer_cas} is of the form $\text{CAS}(\helpobject{p(o)}, (t(o), \nullconstant), (t(o), r))$ for some operation $o$ and response $r \neq \nullconstant$.
    Henceforth we abbreviate this as ``\cref{line:help_pointer_cas} is executed for $o$".
\end{observation}

Since the contents of $C_p$ only change on \cref{line:reset_help_struct} and \cref{line:help_pointer_cas} we have the following:

\begin{observation}\label{observation:only_owner_changes_times_of_help_struct}
    For all processes $p$ the following hold:
    \begin{compactitem}
        \item $\helpobject{p}.time$ only changes by $p$ executing \cref{line:reset_help_struct}.
        \item Every execution of \cref{line:reset_help_struct} by $p$ sets $\helpobject{p}.time$ to a unique value.
        \item $\helpobject{p}.time$ is monotonically increasing.
    \end{compactitem}
\end{observation}

The following observation concerns $\helpobject{\noop}$.
Since $\noop$ is a value that differs from all operations of $\mathcal{T}$ by \Cref{assumption:1}, DoOp($\noop$) is never invoked.
Thus, the contents of $\helpobject{\noop}$ can only change on \cref{line:help_pointer_cas}, but since $\helpobject{\noop}$ is initialized to $(0, \bot)$, all executions of \cref{line:help_pointer_cas} for $\noop$ will not change $\helpobject{\noop}$.
Hence:

\begin{observation}\label{observation:noop_help_pointer_never_changes}
    $\helpobject{\noop} = (0, \bot)$ at all times.
\end{observation}

We now prove some basic facts about \Cref{alg:wait-free-simple}.

\begin{lemma}\label{lemma:never_noop}
    Every CAS operation on $\stateobject{}$ on \cref{line:g_r_cas} is for an operation $o \neq \noop$.
\end{lemma}

\begin{proof}
    Suppose, for contradiction, that some CAS on $\stateobject{}$ on \cref{line:g_r_cas} is for $\noop$ and let $p$ be the process that executed this CAS.
    Thus, $p$ read $(0, \noop, \helpobject{\noop})$ from $\announceobject{}$ on \cref{line:g_a_query} and found the condition on \cref{line:help} to be true.
    Therefore, $\helpobject{\noop} = (0, \nullconstant) \neq (0, \bot)$ at some time.
    However, by \Cref{observation:noop_help_pointer_never_changes}, $\helpobject{\noop} = (0, \bot)$ at all times, a contradiction.
    \qH{\Cref{lemma:never_noop}}
\end{proof}

\begin{lemma}\label{lemma:once_response_is_not_null_only_owner_sets_to_null}
    Suppose $\helpobject{p} = (t, r)$ and $r \neq \nullconstant$ at time $T$.
    For all times $T' \geq T$ if $\helpobject{p}.time = t$ at $T'$ then $\helpobject{p}.response = r$ at $T'$.
\end{lemma}

\begin{proof}
    Suppose that $\helpobject{p} = (t, r)$ at time $T$ where $r \neq \nullconstant$, and $\helpobject{p}.time = t$ at time $T' \geq T$.
    Hence, since by \Cref{observation:only_owner_changes_times_of_help_struct} $\helpobject{p}.time$ is monotonically increasing, we have that $\helpobject{p}.time$ equals $t$ throughout $[T, T']$.
    Thus, since by \Cref{observation:only_owner_changes_times_of_help_struct} $\helpobject{p}.time$ only changes by $p$ executing \cref{line:reset_help_struct}, and every execution of \cref{line:reset_help_struct} by $p$ sets $\helpobject{p}.time$ to a unique value, it follows that \cref{line:reset_help_struct} has not been executed by $p$ throughout $[T, T']$.
    So, throughout $[T, T']$ the contents of $\helpobject{p}$ can only be changed by executions of \cref{line:help_pointer_cas}.
    However, since \cref{line:help_pointer_cas} only changes $\helpobject{p}.response$ if it equals $\nullconstant$, and $\helpobject{p}.response = r$ which is not $\nullconstant$ at $T$, all executions of \cref{line:help_pointer_cas} throughout $[T, T']$ are unsuccessful.
    Therefore, $\helpobject{p}.response = r$ throughout $[T, T']$ as wanted.
    \qH{\Cref{lemma:once_response_is_not_null_only_owner_sets_to_null}}
\end{proof}

\begin{lemma}\label{lemma:if_observed_done_wont_try_to_do}
    Suppose process $p$ executes \cref{line:g_a_query} at time $T$, operation $o$ is stored in $\announceobject{}$ at $T$, $o$ is done at $T$, and $p$ executes \cref{line:help} after $T$.
    Let the time of $p$'s next execution of \cref{line:help} after $T$ be $T'$.
    Then, $p$ finds the condition on \cref{line:help} to be false at $T'$.
\end{lemma}

\begin{proof}
    Since $o$ is done at $T$, by \Cref{def:done}, $\helpobject{p(o)} = (t(o), r)$ such that $r \neq \nullconstant$ at some time $T_1 \leq T$.
    Furthermore, since $p$ executes \cref{line:g_a_query} at time $T$, and operation $o$ is stored in $\announceobject{}$ at $T$, by \Cref{observation:g_a_is_well_formed}, $p$ reads $(t(o), o, \helpobject{p(o)})$ from $\announceobject{}$ on \cref{line:g_a_query} at $T$.
    Hence, since $p$ executes \cref{line:help} after $T$, we have that $p$ reads some value $(\hat{t}, \hat{r})$ from $\helpobject{p(o)}$ on its next execution of \cref{line:help_pointer_query} after $T$; say at time $T_2 > T$, so $T_2 \geq T_1$ (because $T_1 \leq T$).
    Thus, on $p$'s execution of \cref{line:help} at $T'$, we have that $p$ checks whether $(\hat{t}, \hat{r}) = (t(o), \nullconstant)$.
    If $\hat{t} = t(o)$, then since $\helpobject{p(o)} = (t(o), r)$ such that $r \neq \nullconstant$ at $T_1 \leq T_2$, by \Cref{lemma:once_response_is_not_null_only_owner_sets_to_null}, $\hat{r} = r \neq \nullconstant$, so $p$ finds the condition on \cref{line:help} to be false at $T'$.
    Otherwise, $\hat{t} \neq t(o)$, so $p$ finds the condition on \cref{line:help} to be false at $T'$.
    \qH{\Cref{lemma:if_observed_done_wont_try_to_do}}
\end{proof}

\begin{lemma}\label{lemma:complete_implies_done}
    If operation $o$ is complete at time $T$, then $o$ is done at some time $T' < T$.
\end{lemma}

\begin{proof}
    If operation $o$ is complete at time $T$, then by \Cref{def:complete} $p(o)$ executed \cref{line:op_done} at some time $T_1 \leq T$ within $o$.
    Hence, $p(o)$ found the condition on \cref{line:loop_start} to be false at some time $T_2 < T_1$ within $o$.
    Thus, since by \Cref{observation:only_owner_changes_times_of_help_struct} \cref{line:reset_help_struct} is the only step that changes $\helpobject{p(o)}.time$, we have that $\helpobject{p(o)}.time = t(o)$ at $T_2$.
    So, at $T_2$, $\helpobject{p(o)} = (t(o), r)$ for some $r \neq \nullconstant$.
    Therefore, by \Cref{def:done} $o$ is done at $T_2$, which is before $T$ as wanted.
    \qH{\Cref{lemma:complete_implies_done}}
\end{proof}

\begin{proposition}\label{lemma:help_struct_change_implies_done}
    If $\helpobject{p(o)} = (t(o), -)$ at time $T$ and $\helpobject{p(o)} = (t, -)$ for some $t \neq t(o)$ at time $T' > T$ then operation $o$ is done at some time $T^* < T'$.
\end{proposition}

\begin{proof}
    Since by \Cref{observation:only_owner_changes_times_of_help_struct} $\helpobject{p(o)}.time$ is monotonically increasing, $t \neq t(o)$, and $T' > T$, we have that $t(o) < t$.
    Hence, since by \Cref{observation:only_owner_changes_times_of_help_struct} \cref{line:reset_help_struct} is the only step that changes $\helpobject{p(o)}.time$, we have that at some time $T_1 \leq T'$, $p(o)$ executed \cref{line:reset_help_struct} within some operation $o'$ where $t = t(o')$.
    Thus, since $t(o) < t$, we have that $t(o) < t(o')$, and so $o' \neq o$.
    So, since $t(o) < t(o')$, and $p(o)$ invoked both $o$ and $o'$, by the monotonicity of the responses on \cref{line:time_assignment}, $p(o)$ completed $o$ at some time $T_2 < T_1$.
    Therefore, by \Cref{lemma:complete_implies_done} $o$ is done at some time $T^* < T_2 < T'$.
    \qH{\Cref{lemma:help_struct_change_implies_done}}
\end{proof}

\begin{lemma}\label{lemma:copy_help_struct_implies_done}
    If a process executes \cref{line:help_pointer_cas} for operation $o$ at time $T$ then $o$ is done at $T$.
\end{lemma}

\begin{proof}
    Suppose a process $p$ executes \cref{line:help_pointer_cas} for $o$ at some time $T$.
    If $o = \noop$ then by \Cref{observation:noop_help_pointer_never_changes} $\helpobject{\noop} = (0, \bot)$ at $T$.
    Hence by \Cref{def:done}, $o$ is done at $T$.
    
    Now suppose $o \neq \noop$.
    By \Cref{observation:help_pointer_cas_is_well_formed} $p$'s execution of \cref{line:help_pointer_cas} at time $T$ was of the form $\text{CAS}(\helpobject{p(o)}, (t(o), \nullconstant), (t(o), r))$ for some $r \neq \nullconstant$.
    If $p$'s \CASop{} is successful, then $\helpobject{p(o)} = (t(o), r)$ at time $T$, and so by \Cref{def:done} $o$ is done at $T$.
    If $p$'s \CASop{} is unsuccessful, then $\helpobject{p(o)} = (t', r')$ at time $T$ such that $t' \neq t(o)$ or $r' \neq \nullconstant$.
    If $t' = t(o)$, then $r' \neq \nullconstant$, so by \Cref{def:done} $o$ is done at $T$.
    Now suppose $t' \neq t(o)$.
    Hence, since $p$ executed \cref{line:help_pointer_cas} for $o$, we have that $o$ was stored in $\stateobject{}$ at the time $T_1 < T$ when $p$ executed \cref{line:g_r_query}  the same iteration of the loop.
    Thus, since $o \neq \noop$, it follows that some process $q$ executed a successful \CASop{} operation on \cref{line:g_r_cas} for $o$ at some time $T_2 < T_1$.
    So, $q$ read $(t(o), o, \helpobject{p(o)})$ from $\announceobject{}$ on its last execution of \cref{line:g_a_query} before $T_2$, and found the condition on \cref{line:help} to be true on its last execution of \cref{line:help} before $T_2$; say at time $T_3 < T_2$.
    Hence, $\helpobject{p(o)} = (t(o), \nullconstant)$ at the time of $q$'s last execution of \cref{line:help_pointer_query} before $T_3$; say at time $T_4 < T_3$.
    Since $\helpobject{p(o)} = (t(o), \nullconstant)$ at $T_4$ and $\helpobject{p(o)} = (t', r')$ where $t' \neq t(o)$ at $T > T_4$ (because $T_4 < T_3 < T_2 < T_1 < T$), by \Cref{lemma:help_struct_change_implies_done}, $o$ is done at some time $T_5 < T$.
    Therefore, by \Cref{observation:if_once_done_then_always_done}, $o$ is done at $T$ as wanted.
    \qH{\Cref{lemma:copy_help_struct_implies_done}}
\end{proof}

\begin{lemma}\label{lemma:non_null_help_struct_response}
    Suppose $\stateobject{}$ stores operations $o$ and $o'$ at times $T$ and $T' > T$, respectively.
    If $o \neq o'$ then $o$ is done at $T'$.
\end{lemma}

\begin{proof}
    Since $\stateobject{}$ stores $o$ at $T$ and $o' \neq o$ at $T' > T$, it follows that some process executed a successful \CASop{} operation on \cref{line:g_r_cas} between $T$ and $T'$.
    Let $T_1$ be the first such time, and let $p$ be the process that executed it.
    Hence, since $o$ is stored in $\stateobject{}$ at $T$, it follows that $o$ is stored in $\stateobject{}$ at the step before $T_1$.
    Thus, since the \CASop{} at $T_1$ is successful, it follows that $p$ read ($t(o)$, $\unusedvalue{}$, $\unusedvalue{}$, $\helpobject{p(o)}$) from $\stateobject{}$ on its last execution of \cref{line:g_r_query} before $T_1$, and so $p$ executed \cref{line:help_pointer_cas} for $o$ on its last execution of \cref{line:help_pointer_cas} before $T_1$; say at time $T_2$.
    Hence, by \Cref{lemma:copy_help_struct_implies_done}, $o$ is done at $T_2$.
    Therefore, since $T_2 < T_1 < T'$, by \Cref{observation:if_once_done_then_always_done} $o$ is done at $T'$ as wanted.
    \qH{\Cref{lemma:non_null_help_struct_response}}
\end{proof}

\begin{lemma}\label{lemma:done_implies_help_struct_form}
    If an operation $o$ is done at time $T$ then $\helpobject{p(o)} \neq (t(o), \nullconstant)$ from $T$ onwards.
\end{lemma}

\begin{proof}
    Since $o$ is done at $T$, by \Cref{def:done}, $\helpobject{p(o)} = (t(o), r)$ such that $r \neq \nullconstant$ at some time $T^* \leq T$.
    Hence, since by \Cref{observation:only_owner_changes_times_of_help_struct} $\helpobject{p(o)}.time$ is monotonically increasing, we have that $\helpobject{p(o)}.time$ equals some timestamp $t \geq t(o)$ at any time $T' \geq T^*$.
    If $t = t(o)$, then since $\helpobject{p(o)} = (t(o), r)$ such that $r \neq \nullconstant$ at $T^* \leq T'$ by \Cref{lemma:once_response_is_not_null_only_owner_sets_to_null} $\helpobject{p(o)} = (t(o), r)$ at $T'$.
    Otherwise, $\helpobject{p(o)} = (t, -)$ and $t > t(o)$ at $T'$.
    Hence, $\helpobject{p(o)} \neq (t(o), \nullconstant)$ at $T'$ and therefore from $T$ onwards as wanted.
    \qH{\Cref{lemma:done_implies_help_struct_form}}
\end{proof}

\begin{proposition}\label{lemma:double_execution_implies_concurrent}
    Suppose that:
    \begin{compactitem}
        \item At time $T^{\ref{line:g_r_cas}}_i$, process $p_i$ executes a successful \CASop{} operation on $\stateobject{}$ on \cref{line:g_r_cas} for \mbox{operation $o$.}
        \item At time $T^{\ref{line:g_r_cas}}_j > T^{\ref{line:g_r_cas}}_i$, process $p_j$ executes a \CASop{} operation on $\stateobject{}$ on \cref{line:g_r_cas} also for operation $o$.
        % This occurs in an iteration of the loop in line 4.
    \end{compactitem}
    Then the last reading of $\stateobject{}$ by $p_j$ on \cref{line:g_r_query} before $T^{\ref{line:g_r_cas}}_j$ occurs at some time $T^{\ref{line:g_r_query}}_j < T^{\ref{line:g_r_cas}}_i$.
\end{proposition}
% \hline
\begin{proof}
    Suppose, for contradiction, $p_j$'s last execution of \cref{line:g_r_query} before $T^{\ref{line:g_r_cas}}_j$ occurs at time $T^{\ref{line:g_r_query}}_j > T^{\ref{line:g_r_cas}}_i$.
    Let $T^{\ref{line:help_pointer_cas}}_j < T^{\ref{line:g_a_query}}_j < T^{\ref{line:help_pointer_query}}_j < T^{\ref{line:help}}_j$ be the times between $T^{\ref{line:g_r_query}}_j$ and $T^{\ref{line:g_r_cas}}_j$ when $p_j$ executed lines \ref{line:help_pointer_cas}, \ref{line:g_a_query}, \ref{line:help_pointer_query}, and \ref{line:help}, respectively.
    Since $p_j$ executed a \CASop{} operation on $\stateobject{}$ on \cref{line:g_r_cas} for operation $o$ at $T^{\ref{line:g_r_cas}}_j$, it read $(t(o), o, \helpobject{p(o)})$ on \cref{line:g_a_query} at $T^{\ref{line:g_a_query}}_j$.
    Hence, $p_j$ read from $\helpobject{p(o)}$ on \cref{line:help_pointer_query} at $T^{\ref{line:help_pointer_query}}_j$.
    Since $p_j$ executed \cref{line:g_r_cas} at $T^{\ref{line:g_r_cas}}_j$, it found the condition of \cref{line:help} to be true at $T^{\ref{line:help}}_j$.
    Thus, $\helpobject{p(o)} = (t(o), \nullconstant)$ at $T^{\ref{line:help_pointer_query}}_j$.

    We claim that $o$ is done at either $T^{\ref{line:g_r_query}}_j$ or $T^{\ref{line:help_pointer_cas}}_j$, which is before $T^{\ref{line:help_pointer_query}}_j$.
    There are two cases.
    \begin{enumerate}
        \item[] \hspace{0pt}\textbf{Case 1.} $o$ is stored in $\stateobject{}$ at $T^{\ref{line:g_r_query}}_j$.
        
        Hence, $p_j$'s \CASop{} operation on \cref{line:help_pointer_cas} at $T^{\ref{line:help_pointer_cas}}_j$ is for $o$.
        Thus, by \Cref{lemma:copy_help_struct_implies_done}, $o$ is done at $T^{\ref{line:help_pointer_cas}}_j$.
        
        \item[] \hspace{0pt}\textbf{Case 2.} $o' \neq o$ is stored in $\stateobject{}$ at $T^{\ref{line:g_r_query}}_j$.

        By assumption the \CASop{} on $\stateobject{}$ on \cref{line:g_r_cas} by $p_i$ at $T^{\ref{line:g_r_cas}}_i$ was for operation $o$ and is successful, so $\stateobject{}$ stores $o$ at $T^{\ref{line:g_r_cas}}_i$.
        Since $o' \neq o$ is stored in $\stateobject{}$ at time $T^{\ref{line:g_r_query}}_j > T^{\ref{line:g_r_cas}}_i$, by \Cref{lemma:non_null_help_struct_response} $o$ is done at $T^{\ref{line:g_r_query}}_j$.
    \end{enumerate}

    We now finish the proof.
    Since $o$ is done before $T^{\ref{line:help_pointer_query}}_j$, by \Cref{lemma:done_implies_help_struct_form}, $\helpobject{p(o)} \neq (t(o), \nullconstant)$ at $T^{\ref{line:help_pointer_query}}_j$.
    However, as established above, $\helpobject{p(o)} = (t(o), \nullconstant)$ at $T^{\ref{line:help_pointer_query}}_j$, a contradiction.
    \qH{\Cref{lemma:double_execution_implies_concurrent}}
\end{proof}

\begin{lemma}\label{claim:pairwise_distinct}
    Every successful \CASop{} operation on $\stateobject{}$ on \cref{line:g_r_cas} is for a different operation.
\end{lemma}

\begin{proof}
    Suppose, for contradiction, that processes $p_i$ and $p_j$ both execute successful \CASop{} operations on $\stateobject{}$ on \cref{line:g_r_cas} for the same operation $o$ at distinct times $T^{\ref{line:g_r_cas}}_i$ and $T^{\ref{line:g_r_cas}}_j$, respectively.
    Without loss of generality, assume $T^{\ref{line:g_r_cas}}_j$ is the minimum time when there are two successful \CASop{} operations on $\stateobject{}$ on \cref{line:g_r_cas} for the same operation.
    Hence, $T^{\ref{line:g_r_cas}}_i < T^{\ref{line:g_r_cas}}_j$.
    Let $T^{\ref{line:g_r_query}}_j$ be the time of $p_j$'s last execution of \cref{line:g_r_query} before $T^{\ref{line:g_r_cas}}_j$.
    Hence, by \Cref{lemma:double_execution_implies_concurrent}, $T^{\ref{line:g_r_query}}_j < T^{\ref{line:g_r_cas}}_i$.
    
    Let $o_1$ be the operation stored in $\stateobject{}$ at $T^{\ref{line:g_r_query}}_j$.
    We claim that $o_1 \neq o$.
    Suppose, for contradiction, that $o_1 = o$.
    Since $p_i$ and $p_j$ perform \CASop{} operations on $\stateobject{}$ on \cref{line:g_r_cas} for $o$, by \Cref{lemma:never_noop}, $o \neq \noop$.
    Hence, since $o_1 = o$, we have that $o_1 \neq \noop$.
    Thus, since $o_1$ is stored in $\stateobject{}$ at $T^{\ref{line:g_r_query}}_j$, it follows that there is a successful \CASop{} operation on~$\stateobject{}$ on \cref{line:g_r_cas} for $o_1$ (and hence $o$) before $T^{\ref{line:g_r_query}}_j$.
    Therefore, since $T^{\ref{line:g_r_query}}_j < T^{\ref{line:g_r_cas}}_i < T^{\ref{line:g_r_cas}}_j$, it follows that at time $T^{\ref{line:g_r_cas}}_i$ there are two successful \CASop{} operations on $\stateobject{}$ on \cref{line:g_r_cas} for $o$.
    However, $T^{\ref{line:g_r_cas}}_j$ is the minimum time when there are two successful \CASop{} operations on $\stateobject{}$ on \cref{line:g_r_cas} for the same operation, a contradiction.

    We now finish the proof of \Cref{claim:pairwise_distinct}.
    Since the \CASop{} operation on $\stateobject{}$ on \cref{line:g_r_cas} at $T_i^{\ref{line:g_r_cas}}$ is for $o$ and is successful, we have that $\stateobject{}$ stores $o$ at $T_i^{\ref{line:g_r_cas}}$.
    Hence, since $o_1$ is stored in $\stateobject{}$ at~$T^{\ref{line:g_r_query}}_j < T^{\ref{line:g_r_cas}}_i < T^{\ref{line:g_r_cas}}_j$, and $p_j$'s \CASop{} operation on $\stateobject{}$ on \cref{line:g_r_cas} at $T^{\ref{line:g_r_cas}}_j$ is successful, it follows that $\stateobject{}$ stores $o_1$ at the step before $T^{\ref{line:g_r_cas}}_j$.
    Thus, since as we proved above $o_1 \neq o$, it follows that there is a successful \CASop{} operation on $\stateobject{}$ on \cref{line:g_r_cas} for $o_1$ between $T^{\ref{line:g_r_cas}}_i$ and $T^{\ref{line:g_r_cas}}_j$.
    So, by \Cref{lemma:never_noop}, $o_1 \neq \noop$.
    Hence, since $o_1$ is stored in $\stateobject{}$ at~$T^{\ref{line:g_r_query}}_j$, it follows that there is a successful \CASop{} operation on $\stateobject{}$ on \cref{line:g_r_cas} for $o_1$ before $T^{\ref{line:g_r_query}}_j$. 
    Therefore, since $T^{\ref{line:g_r_query}}_j < T^{\ref{line:g_r_cas}}_i < T^{\ref{line:g_r_cas}}_j$, we have that there are two successful \CASop{} operations on $\stateobject{}$ on \cref{line:g_r_cas} for $o$ before $T^{\ref{line:g_r_cas}}_j$.
    However, $T^{\ref{line:g_r_cas}}_j$ is the minimum time when there are two successful \CASop{} operations on $\stateobject{}$ on \cref{line:g_r_cas} for the same operation, a contradiction.
    \qH{\Cref{claim:pairwise_distinct}}
\end{proof}

\subsection{Step Complexity}

In this section, we prove that the maximum number of steps a process takes to perform an operation~$o$ is adaptive to the point contention at the time it gets a timestamp for $o$.
More precisely:

\begin{theoremblank}
    Suppose a process $p$ invokes an operation $o$ and executes \cref{line:time_assignment} within $o$.
    Let $c$ be the point contention at this time.
    Then, the number of steps that $p$ takes within $o$ is at most \mbox{linear in $c$.}
\end{theoremblank}

Since if $p$ doesn't execute \cref{line:time_assignment} within $o$, it only takes two steps within $o$ ($p$ executes \cref{line:op_start} and~\ref{line:semi-efficient-allocate-cell} within $o$), and the point contention at any time is an integer, this theorem implies wait-freedom.

We now describe the high-level strategy for proving this theorem.
Since $p$ takes a constant number of steps before and after the loop on \cref{line:loop_start} within $o$, and $p$ takes a constant number of steps each iteration of the loop on \cref{line:loop_start} within $o$, it suffices to prove that $p$ performs at most linear in $c$ number of iterations of the loop on \cref{line:loop_start} within $o$.
Before we explain how we do this, we note that the reason $p$ continues to iterate in the loop is that it is either unable to announce $o$ into $\announceobject{}$ or it gets ``dislodged'' from $\announceobject{}$ by operations with higher priority (i.e., those with smaller timestamps).
So, to bound the number of iterations of the loop, we have to argue that eventually these high-priority operations are no longer in $\announceobject{}$, and they eventually stop dislodging $o$ from $\announceobject{}$.
We also note that there are at most $c$ of these operations: those that have a timestamp no larger than $t(o)$ and are pending at the time $p$ executes \cref{line:time_assignment} within $o$; denote them by $S(o)$.

To bound the number of iterations of the loop, we prove that every small constant number of iterations of the loop by $p$ within $o$, which we call a \Underline{period} (as defined formally later), $p$ can identify an operation $o'$ in $S(o)$.
The strategy is then that if $p$ performs too many periods, it can identify more operations in $S(o)$ than what are actually in $S(o)$, yielding a bound on the number of iterations.
For example, if $S(o)$ has five elements, and we were guaranteed that in every period $p$ could identify a unique operation in $S(o)$, then $p$ performs at most 5 periods because if it were to perform more, we could identify 6 elements in $S(o)$, contradicting the fact that $S(o)$ has five elements. 
The challenge with making this strategy work is that this identification may be redundant: $p$ performs many periods, but it always identifies the same operation in $S(o)$.
To avoid this, we require that $p$ identifies an operation $o' \in S(o)$ in any period with the following properties: (1) $o'$ is done during the period; (2) $o'$ is stored in $\announceobject{}$ during the period; and (3) $o'$ is dislodged from $\announceobject{}$ during the period after it was stored in $\announceobject{}$ and after it is done.
As a black box, these three properties let us prove that $p$ can identify the same operation in at most three periods.
This is because if $p$ identifies the same operation $o'$ in four different periods, then it would imply that $p(o')$ announced $o'$ in $\announceobject{}$ three times after $o'$ was done, but by the order in which operations happen in the loop, $p(o')$ will check whether $o'$ is done before announcing $o'$ for the third time, at which time it will see $o'$ is done, and stop trying to announce $o'$, contradicting the fact it announces $o'$ for a third time.
Since $p$ identifies a new operation in $S(o)$ at least every fourth period, we can conclude that $p$ performs at $3S(o)$ periods within $o$.
This is because if $p$ were to perform any more, it would imply $S(o)$ has more elements than it does.
So, since $S(o)$ has at most $c$ elements, we have $p$ performs at most $3c$ periods within $o$, and so $p$ performs at most linear in $c$ number of iterations within $o.$

The majority of the work is in proving that $p$ can identify an operation in each period with the above properties.
We prove this by considering various ``paths'' that $p$ can take during the period (e.g., in the first iteration of the loop, did $p$ execute \cref{line:g_r_cas} or \cref{line:g_a_cas} and was the \CASop{} operation successful or not) and identifying the desired operation in each case.

We now begin the proof.
We start by formalizing the concept of a period.

\begin{definition}\label{def:terminal_and_complete}
    We call an iteration $I$ of the loop on \cref{line:loop_start} by some process $p$ \Underline{terminal} if $p$ finds the condition on \cref{line:loop_start} to be false during $I$.
    Furthermore, we call $I$ \Underline{complete} if either (a) $I$ is terminal or (b) $p$ executes \cref{line:g_r_cas} or \cref{line:g_a_cas} during $I$.
\end{definition}

\begin{definition}\label{def:period}
    For each operation $o$, we call five consecutive complete and not terminal iterations of the loop on \cref{line:loop_start} by $p(o)$ within $o$ a \Underline{period} of $o$.
    We call a period $P$ of $o$ a \Underline{non-initial} period when the first iteration of $P$ is not the first iteration of the loop on \cref{line:loop_start} by $p(o)$ within~$o$.
    We call periods $P_1$ and $P_2$ of $o$ \Underline{distinct} when $P_1$ and $P_2$ are comprised of distinct iterations.
\end{definition}

We now define the set of high-priority operations that can compete with $o$.

\begin{definition}\label{def:set_of_interest}
    Let $S(o)$ be the set of operations that are pending at the time $p(o)$ executes \cref{line:time_assignment} within operation $o$, assuming it does, such that $\forall o' \in S(o)\ t(o') \leq t(o)$, and $\emptyset$ otherwise.
\end{definition}

We now define the properties of the operation we want to identify within each period.

\begin{definition}\label{def:useful_period}
    For a period $P$ of $o$ we say that $P$ is \Underline{useful} when the following are true.
    \begin{compactenum}
        \item There is an operation $o' \in S(o)$ that is done at some time $T_1$ during $P$.
        \item $o'$ is stored in $\announceobject{}$ at some time $T_2$ during $P$.
        \item Some operation $o^* \neq o'$ is stored in $\announceobject{}$ at some time $T_3$ during $P$ such that $T_3 > \max(T_1, T_2)$.
    \end{compactenum}
\end{definition}

The bulk of the work is to prove that every non-initial period is useful.
The next two lemmas motivate why we consider non-initial periods instead of periods.
Roughly speaking, by discarding the first iteration of the loop within an operation $o$, we don't need to worry about the case where a high-priority operation that is not in $S(o)$ blocks $p(o)$ from announcing $o$.
This is because the first iteration ensures that $p(o)$ gets them out of $\announceobject{}$ if they are there at all, and because they are not in $S(o)$, they are guaranteed not to be announced again (as is implied by the next claim).

\begin{proposition}\label{lemma:if_not_pending_when_timestamp_taken_then_completed}
    Consider any operation $o$ where $p(o)$ executes \cref{line:time_assignment} within $o$.
    For every operation $o'$ with $t(o') \leq t(o)$ if $o'$ is not pending at the time $p(o)$ executes \cref{line:time_assignment} within $o$, then $o'$ is complete at the time $p(o)$ executes \cref{line:time_assignment} within $o$.
\end{proposition}

\begin{proof}
    Suppose, for contradiction, there is an operation $o'$ with $t(o') \leq t(o)$ such that $o'$ is not pending and not complete at the time $p(o)$ executes \cref{line:time_assignment} within $o$.
    Hence, $o'$ was not invoked at the time $p(o)$ executes \cref{line:time_assignment} within $o$.
    Since $p(o)$ executes \cref{line:time_assignment} within $o$, by \Cref{def:op_state}, $t(o) \neq \infty$.
    Hence, since $t(o') \leq t(o)$, we have that $t(o') \neq \infty$.
    Thus, by \Cref{def:op_state}, $p(o')$ executes \cref{line:time_assignment} within $o'$.
    So, since $o'$ was not invoked at the time $p(o)$ executes \cref{line:time_assignment} within $o$, we have that $p(o')$ executes \cref{line:time_assignment} within $o'$ after $p(o)$ executes \cref{line:time_assignment} within $o$.
    Therefore, by \cref{line:time_assignment}, $t(o) < t(o')$.
    However, by assumption $t(o') \leq t(o)$, a contradiction.
    \qH{\Cref{lemma:if_not_pending_when_timestamp_taken_then_completed}}
\end{proof}

\begin{lemma}\label{lemma:any_inequality_gcas_after_first_iteration_ensures_operation_of_interest_is_in_announce}
    Consider any iteration $I$ of the loop on \cref{line:loop_start} by $p(o)$ within operation $o$ other than the first iteration.
    Some operation in $S(o)$ is stored in $\announceobject{}$ at the time $p(o)$ executes \cref{line:g_a_gcas} in $I$.
\end{lemma}

\begin{proof}
    Suppose, for contradiction, no operation in $S(o)$ is stored in $\announceobject{}$ at the time $p(o)$ executes \cref{line:g_a_gcas} in $I$; say time $T^{\ref{line:g_a_gcas}}$.
    By \Cref{observation:g_a_is_well_formed}, some operation $o'$ is stored in $\announceobject{}$ at time $T^{\ref{line:g_a_gcas}}$, i.e., $\announceobject{} = (t(o'), o', C_{p(o')})$ at $T^{\ref{line:g_a_gcas}}$.
    Since $p(o)$ executes \cref{line:g_a_gcas} in $I$ at time $T^{\ref{line:g_a_gcas}}$, we have that it is for $o$, so the \GCASop{} executed at time $T^{\ref{line:g_a_gcas}}$ is of the form $\GCASop{}(>, \announceobject{}, (t(o), o, C_{p(o)}), (t(o), o, C_{p(o)}))$.
    Hence, since $\announceobject{} = (t(o'), o', C_{p(o')})$ at $T^{\ref{line:g_a_gcas}}$, we have that $(t(o'), o', C_{p(o')}) \leq (t(o), o, C_{p(o)})$, and so $t(o') \leq t(o)$.
    Thus, if $o'$ is pending at the time $p(o)$ executes \cref{line:time_assignment} within $o$, by \Cref{def:set_of_interest}, $o' \in S(o)$, and so some operation in $S(o)$ is stored in $\announceobject{}$ at time $T^{\ref{line:g_a_gcas}}$, contradicting our initial assumption.
    So, $o'$ is not pending at the time $p(o)$ executes \cref{line:time_assignment} within $o$.
    Hence, since $t(o') \leq t(o)$, by \Cref{lemma:if_not_pending_when_timestamp_taken_then_completed}, $o'$ is complete at the time $p(o)$ executes \cref{line:time_assignment} within $o$.
    Since $I$ is any iteration of the loop on \cref{line:loop_start} by $p(o)$ within $o$ other than the first iteration, we have that $p(o)$ completed an iteration $I^-$ of the loop on \cref{line:loop_start} within $o$ before it began $I$.
    Hence, since $o'$ is complete at the time $p(o)$ executes \cref{line:time_assignment} within $o$, it follows that $o'$ is complete before $p(o)$ began $I^-$. 
    Suppose $o^*$ is the operation stored in $\announceobject{}$ at the time $p(o)$ executed \cref{line:g_a_query} in $I^-$; say at time $T^{\ref{line:g_a_query}}$, so $T^{\ref{line:g_a_query}} < T^{\ref{line:g_a_gcas}}$.
    There are two cases.

    \begin{itemize}
        \item[] \hspace{0pt}\textbf{Case 1.} $o^* \neq o'$.

        Hence, since $o^*$ is the operation stored in $\announceobject{}$ at $T^{\ref{line:g_a_query}}$, $o'$ is stored in $\announceobject{}$ at $T^{\ref{line:g_a_gcas}}$, and $T^{\ref{line:g_a_query}} < T^{\ref{line:g_a_gcas}}$, by \Cref{observation:p_o_writes_o_into_announce}, it follows that $p(o')$ executed \cref{line:g_a_gcas} or \cref{line:g_a_cas} for $o'$ within $o'$ after $T^{\ref{line:g_a_query}}$; say at time $T$.
        Thus, since $o'$ completed before $p(o)$ began $I^-$, and $T^{\ref{line:g_a_query}}$ is a time after $p(o)$ began $I^-$, by transitivity, $o'$ completed before $T$.
        Therefore, since $p(o')$ took the step at $T$ within $o'$, we have that $p(o')$ took a step within $o'$ after $o'$ completed, which is impossible.

        \item[] \hspace{0pt}\textbf{Case 2.} $o^* = o'$.

        Hence, since $o^*$ is the operation stored in $\announceobject{}$ at $T^{\ref{line:g_a_query}}$, we have that $o'$ is the operation stored in $\announceobject{}$ at $T^{\ref{line:g_a_query}}$.
        Since $o'$ is complete before $p(o)$ began $I^-$, by \Cref{lemma:complete_implies_done}, $o'$ is done before $p(o)$ began $I^-$.
        Hence, since $p(o)$ executed \cref{line:g_a_query} at time $T^{\ref{line:g_a_query}}$ in $I^-$, by \Cref{observation:if_once_done_then_always_done}, $o'$ is done at $T^{\ref{line:g_a_query}}$.
        Thus, since $p(o)$ executed \cref{line:g_a_query} at $T^{\ref{line:g_a_query}}$, $o'$ is the operation stored in $\announceobject{}$ at $T^{\ref{line:g_a_query}}$, $o'$ is done at $T^{\ref{line:g_a_query}}$, and $p(o)$ completes $I^-$, by \Cref{lemma:if_observed_done_wont_try_to_do}, $p(o)$ finds the condition on \cref{line:help} to be false during $I^-$.
        So, $p(o)$ executes \cref{line:g_a_cas} during $I^-$.
        Hence, since $o'$ is the operation stored in $\announceobject{}$ at $T^{\ref{line:g_a_query}}$, and $p(o)$ executed \cref{line:g_a_query} in $I^-$ at $T^{\ref{line:g_a_query}}$ within $o$, we have that $p(o)$'s execution of \cref{line:g_a_cas} during $I^-$ is of the form $\GCASop{}(=, \announceobject{}, (t(o'), o', C_{p(o')}), (t(o), o, C_{p(o)}))$.
        Thus, since $o'$ is complete at the time $p(o)$ executes \cref{line:time_assignment} within $o$, it follows that $o' \neq o$, and so regardless of whether $p(o)$'s execution of \cref{line:g_a_cas} during $I^-$ is successful, it follows that $o'$ is not stored in $\announceobject{}$ at the time of it.
        Thus, $o'$ is not stored in $\announceobject{}$ sometime during $I^-$.
        So, since $o'$ is stored in $\announceobject{}$ at $T^{\ref{line:g_a_gcas}}$, and $T^{\ref{line:g_a_gcas}}$ is the time of a step after $I^-$, by \Cref{observation:p_o_writes_o_into_announce}, it follows that $p(o')$ executed \cref{line:g_a_gcas} or \cref{line:g_a_cas} for $o'$ within $o'$ sometime after $I^-$ began.
        Therefore, since $o'$ completed before $p(o)$ began $I^-$, we have that $p(o')$ took a step within $o'$ after $o'$ completed, which is impossible.
        \qH{\Cref{lemma:any_inequality_gcas_after_first_iteration_ensures_operation_of_interest_is_in_announce}}
    \end{itemize}
\end{proof}

\begin{lemma}\label{lemma:after_competition_in_second_iteration_a_small_timestamp_op_in_announce_is_of_interest}
    Consider any iteration $I$ of the loop on \cref{line:loop_start} by $p(o)$ within operation $o$ other than the first iteration.
    If at any time after $p(o)$ executes \cref{line:g_a_gcas} in $I$ the operation $o'$ stored in $\announceobject{}$ has the property that $t(o') \leq t(o)$, then $o' \in S(o)$.
\end{lemma}

\begin{proof}
    Suppose, for contradiction, $o' \notin S(o)$.
    Since $t(o') \leq t(o)$, if $o'$ is pending at the time $p(o)$ executes \cref{line:time_assignment} within $o$, then by \Cref{def:set_of_interest} $o' \in S(o)$, contradicting $o' \notin S(o)$.
    Hence, it must be that $o'$ is not pending at the time $p(o)$ executes \cref{line:time_assignment} within $o$.
    Thus, since $t(o') \leq t(o)$, by \Cref{lemma:if_not_pending_when_timestamp_taken_then_completed}, $o'$ is complete at the time $p(o)$ executes \cref{line:time_assignment} within $o$.
    Let $T^{\ref{line:g_a_gcas}}$ be the time that $p(o)$ executes \cref{line:g_a_gcas} in $I$.
    Hence, since $o'$ is complete at the time $p(o)$ executes \cref{line:time_assignment} within $o$, and $p(o)$ executes \cref{line:time_assignment} within $o$ before $T^{\ref{line:g_a_gcas}}$, by transitivity, $o'$ is complete before $T^{\ref{line:g_a_gcas}}$.
    Furthermore, since $I$ is any iteration of the loop on \cref{line:loop_start} by $p(o)$ within $o$ other than the first iteration, by \Cref{lemma:any_inequality_gcas_after_first_iteration_ensures_operation_of_interest_is_in_announce}, some operation $o^* \in S(o)$ is stored in $\announceobject{}$ at $T^{\ref{line:g_a_gcas}}$.
    Thus, since $o' \notin S(o)$ is stored in $\announceobject{}$ at some time after $T^{\ref{line:g_a_gcas}}$, it follows that there is an execution of \cref{line:g_a_gcas} or \cref{line:g_a_cas} for $o'$ after $T^{\ref{line:g_a_gcas}}$.
    So, since $o'$ is complete before $T^{\ref{line:g_a_gcas}}$, by transitivity, there is an execution of \cref{line:g_a_gcas} or \cref{line:g_a_cas} for $o'$ after $o'$ is complete.
    Therefore, by \Cref{observation:p_o_writes_o_into_announce}, there is an execution of \cref{line:g_a_gcas} or \cref{line:g_a_cas} by $p(o')$ within $o'$ after $o'$ is complete, which is impossible, a contradiction.
    \qH{\Cref{lemma:after_competition_in_second_iteration_a_small_timestamp_op_in_announce_is_of_interest}}
\end{proof}

The goal of the next few claims is to prove that every non-initial period is useful.
The plan is to consider various ``paths'' that $p(o)$ can take during the period.

\begin{proposition}\label{lemma:read_low_priority_operation_from_announce}
    Consider any non-initial period $P$ of some operation $o$.
    If an operation $o' \notin S(o)$ is stored in $\announceobject{}$ at some time $T$ during $P$ such that $T$ is after the first time $p(o)$ executed \cref{line:g_a_gcas} during $P$, then $P$ is useful.    
\end{proposition}

\begin{proof}
    Let $T^{\ref{line:g_a_gcas}}$ be the first time $p(o)$ executed \cref{line:g_a_gcas} during $P$.
    Denote the iteration of the loop on \cref{line:loop_start} that $p(o)$ performed this execution of \cref{line:g_a_gcas} in as $I$.
    Since $P$ is a non-initial period, and $I$ is an iteration of $P$, by \Cref{def:period}, $I$ is not the first iteration of the loop on \cref{line:loop_start} by $p(o)$ within $o$.
    Hence, by \Cref{lemma:any_inequality_gcas_after_first_iteration_ensures_operation_of_interest_is_in_announce}, some operation $o^* \in S(o)$ is stored in $\announceobject{}$ at $T^{\ref{line:g_a_gcas}}$.
    Thus, since $o' \notin S(o)$ is stored in $\announceobject{}$ at $T$, and $T^{\ref{line:g_a_gcas}} < T$, we have that there is a non-empty finite sequence of $\GCASop{}$ operation on $\announceobject{}$ during $(T^{\ref{line:g_a_gcas}}, T]$, each of which is an execution of either \cref{line:g_a_gcas} or \cref{line:g_a_cas}.
    Let $e_1, \ldots e_n$ denote the sequence of $\GCASop{}$ operations on $\announceobject{}$ during $(T^{\ref{line:g_a_gcas}}, T]$.

    We claim that at least one of $e_1, \ldots e_n$ is a successful execution of \cref{line:g_a_cas}.
    Suppose, for contradiction, none of $e_1, \ldots e_n$ is a successful execution of \cref{line:g_a_cas}.
    Hence, since $e_1, \ldots e_n$ are executions of either \cref{line:g_a_gcas} or \cref{line:g_a_cas}, we have that each successful execution is on \cref{line:g_a_gcas}.
    Thus, since $o^* \in S(o)$ is stored in $\announceobject{}$ at $T^{\ref{line:g_a_gcas}}$, by \Cref{def:set_of_interest}, $t(o^*) \leq t(o)$, and so since $e_1, \ldots e_n$ is the sequence of $\GCASop{}$ operations on $\announceobject{}$ during $(T^{\ref{line:g_a_gcas}}, T]$, each successful execution in $e_1, \ldots e_n$ is for \cref{line:g_a_gcas}, and $o' \notin S(o)$ is stored in $\announceobject{}$ at $T$, it follows that $t(o') \leq t(o)$.
    Therefore, since $o'$ is an operation stored in $\announceobject{}$ after~$T^{\ref{line:g_a_gcas}}$, by \Cref{lemma:after_competition_in_second_iteration_a_small_timestamp_op_in_announce_is_of_interest}, $o' \in S(o)$.
    However, $o' \notin S(o)$, a contradiction.

    We now finish the proof of \Cref{lemma:read_low_priority_operation_from_announce}.
    Since at least one of $e_1, \ldots e_n$ is a successful execution of \cref{line:g_a_cas}, we have that there is a first $e_i$ in $e_1, \ldots e_n$ that is a successful execution of \cref{line:g_a_cas}.
    Hence, since $e_1, \ldots e_{i - 1}$ are executions of either \cref{line:g_a_gcas} or \cref{line:g_a_cas}, we have that each successful execution is on \cref{line:g_a_gcas}.
    Let $\hat{o}$ be the operation stored in $\announceobject{}$ at the step before $e_i$.
    Since $o^* \in S(o)$ is stored in $\announceobject{}$ at $T^{\ref{line:g_a_gcas}}$, by \Cref{def:set_of_interest}, $t(o^*) \leq t(o)$, and so since $e_1, \ldots e_{i - 1}$ is the sequence of $\GCASop{}$ operations on $\announceobject{}$ between $T^{\ref{line:g_a_gcas}}$ and the step before $e_i$, each successful execution in $e_1, \ldots e_{i - 1}$ is for \cref{line:g_a_gcas}, and $\hat{o}$ is stored in $\announceobject{}$ at at the step before $e_i$, it follows that $t(\hat{o}) \leq t(o)$.
    Hence, since $\hat{o}$ is an operation stored in $\announceobject{}$ after $T^{\ref{line:g_a_gcas}}$, by \Cref{lemma:after_competition_in_second_iteration_a_small_timestamp_op_in_announce_is_of_interest}, $\hat{o} \in S(o)$.
    Let $q$ be the process that executed~$e_i$.
    Since $e_i$ is a successful execution of \cref{line:g_a_cas}, and $\hat{o}$ is the operation stored in $\announceobject{}$ at the step before~$e_i$, by \Cref{observation:g_a_is_well_formed}, it follows that $e_i$ is of the form $\GCASop{}(=, \announceobject{}, (t(\hat{o}), \hat{o}, C_{p(\hat{o})}), \unusedvalue{})$.
    Hence, $q$ found the condition on \cref{line:help} to be false on its last execution of it before $e_i$, and so by \cref{line:help_pointer_query}, $q$ saw $C_{p(\hat{o})} \neq (t(\hat{o}), \nullconstant)$ before executing $e_i$.
    Thus, by \Cref{def:done}, $\hat{o}$ is done at before $e_i$, and so by \Cref{observation:if_once_done_then_always_done}, $\hat{o}$ is done at $e_i$.
    Lastly, since $\hat{o} \in S(o)$ is stored in $\announceobject{}$ at the step before $e_i$, $e_1, \ldots e_n$ is the sequence of $\GCASop{}$ operations on $\announceobject{}$ during $(T^{\ref{line:g_a_gcas}}, T]$, and $o' \notin S(o)$ is stored in $\announceobject{}$ at $T$, we have that $o' \neq \hat{o}$ is stored in $\announceobject{}$ at $T$.
    Therefore, since (1) $\hat{o} \in S(o)$ is done at $e_i$ which is in $P$, (2) $\hat{o}$ is stored in $\announceobject{}$ at the step before $e_i$ which is in $P$, and (3) $o' \neq \hat{o}$ is stored in $\announceobject{}$ at $T$ which is in $P$ and after $e_i$, by \Cref{def:useful_period}, $P$ is useful as wanted.
    \qH{\Cref{lemma:read_low_priority_operation_from_announce}}
\end{proof}

\begin{proposition}\label{lemma:read_done_operation_from_announce}
    Consider any non-initial period $P$ of some operation $o$.
    If $p(o)$ executes \cref{line:g_a_cas} during one of the first four iterations of $P$, then $P$ is useful.
\end{proposition}

\begin{proof}
    Consider any execution of \cref{line:g_a_cas} by $p(o)$ during any iteration $I$ of $P$ such that $I$ is one of the first four iterations of $P$.
    By \Cref{lemma:read_low_priority_operation_from_announce}, it suffices to consider the case where the operation $o'$ stored in $\announceobject{}$ at the time $p(o)$ executes \cref{line:g_a_query} during $I$ is in $S(o)$; say $T^{\ref{line:g_a_query}}$.
    Since $o'$ is stored in $\announceobject{}$ at $T^{\ref{line:g_a_query}}$, by \Cref{observation:g_a_is_well_formed}, $p(o)$ read $(t(o'), o', C_{p(o')})$ from $\announceobject{}$ on \cref{line:g_a_query} at $T^{\ref{line:g_a_query}}$.
    Hence, since $p(o)$ executed \cref{line:g_a_cas} during $I$, we have that $p(o)$ found the condition on \cref{line:help} to be false during $I$, and so $p(o)$ saw $C_{p(o')} \neq (t(o'), \nullconstant)$ on \cref{line:help_pointer_query} during $I$.
    Thus, by \Cref{def:done}, $o'$ is done at some time during $I$ and before $p(o)$ executed \cref{line:g_a_cas} during $I$.
    There are two cases.
    \begin{itemize}
        \item[] \hspace{0pt}\textbf{Case 1.} $o \neq o'$.

        Hence, $p(o)$ tries to write $(t(o), o, C_{p(o)}) \neq (t(o'), o', C_{p(o')})$ into $\announceobject{}$ on \cref{line:g_a_cas} during $I$.
        If $p(o)$ is successful, then an operation $o \neq o'$ is stored in $\announceobject{}$ at the time $p(o)$ executed \cref{line:g_a_cas} during $I$.
        Otherwise, if $p(o)$ is unsuccessful, since $p(o)$ read $(t(o'), o', C_{p(o')})$ from $\announceobject{}$ on \cref{line:g_a_query} during $I$, we have that $\announceobject{} \neq (t(o'), o', C_{p(o')})$ at the time $p(o)$ executed \cref{line:g_a_cas} during $I$.
        Hence, in all cases, some operation $o^* \neq o'$ is stored in $\announceobject{}$ at the time $p(o)$ executed \cref{line:g_a_cas} during $I$.
        Therefore, since (1) $o' \in S(o)$ is done at some time during $I$ (and thus $P$) before $p(o)$ executed \cref{line:g_a_cas} during $I$, (2) $o'$ is stored in $\announceobject{}$ at at the time $p(o)$ executes \cref{line:g_a_query} during $I$ which is in $P$, and (3) some operation $o^* \neq o'$ is stored in $\announceobject{}$ at the time $p(o)$ executed \cref{line:g_a_cas} during $I$ (and thus $P$) which is after the time identified in (1) and (2), by \Cref{def:useful_period}, $P$ is useful as wanted.

        \item[] \hspace{0pt}\textbf{Case 2.} $o = o'$.

        Since $I$ is one of the first four iterations of $P$, by \Cref{def:period}, there is a complete and not terminal iteration $I^+$ of the loop on \cref{line:loop_start} after $I$ by $p(o)$ within $o$.
        Hence, since $o'$ is done at some time during $I$ and before $p(o)$ executed \cref{line:g_a_cas} during $I$, and $o = o'$, by \Cref{observation:if_once_done_then_always_done}, $o$ is done at the time $p(o)$ executes \cref{line:loop_start} during $I^+$.
        Thus, by \Cref{lemma:done_implies_help_struct_form}, $\helpobject{p(o)} \neq (t(o), \nullconstant)$ at this time, and so $p(o)$ finds the condition on \cref{line:loop_start} to be false during $I^+$.
        Therefore, by \Cref{def:terminal_and_complete}, $I^+$ is terminal.
        However, $I^+$ is not terminal, a contradiction, so this case is impossible.
        \qH{\Cref{lemma:read_done_operation_from_announce}}
    \end{itemize}
\end{proof}

\begin{proposition}\label{lemma:read_not_done_operation_from_announce_and_successfully_helped}
    Consider any non-initial period $P$ of some operation $o$.
    If $p(o)$ executes \cref{line:g_r_cas} and the response is true in any of the first three iterations of $P$, then $P$ is useful.
\end{proposition}

\begin{proof}
    Consider any execution of \cref{line:g_r_cas} whose response is true during any iteration $I$ of $P$ such that $I$ is one of the first three iterations of $P$.
    By \Cref{lemma:read_low_priority_operation_from_announce}, it suffices to consider the case where the operation $o'$ stored in $\announceobject{}$ at the time $p(o)$ executes \cref{line:g_a_query} during $I$ is in $S(o)$; say $T^{\ref{line:g_a_query}}$.
    Since $o'$ is stored in $\announceobject{}$ at $T^{\ref{line:g_a_query}}$, by \Cref{observation:g_a_is_well_formed}, $p(o)$ read $(t(o'), o', C_{p(o')})$ from $\announceobject{}$ on \cref{line:g_a_query} at $T^{\ref{line:g_a_query}}$.
    Hence, since $p(o)$ executed \cref{line:g_r_cas} during $I$, it is of the form $\CASop{}(\stateobject{}, \unusedvalue{}, (t(o'), \unusedvalue{}, \unusedvalue{}, C_{p(o')}))$, so by \Cref{observation:g_r_is_well_formed}, it is for $o'$.
    Thus, since the response of this \CASop{} operation is true, we have that $o'$ is stored in $\stateobject{}$ during $I$; say at time $T$.
    Since $I$ is one of the first three iterations of $P$, by \Cref{def:period}, there is a complete and not terminal iterations of the loop on \cref{line:loop_start} by $p(o)$ within $o$ after $I$.
    Let $I^+$ be the next one after $I$.
    Since $I^+$ is complete and not terminal, $p(o)$ executes \cref{line:g_r_query} (resp. \cref{line:help_pointer_cas}) during $I^+$; say at time $T^{\ref{line:g_r_query}}$ (resp. $T^{\ref{line:help_pointer_cas}}$).
    Hence, since $I$ completes before $I^+$ begins, and $T$ is a time during $I$, we have that $T < T^{\ref{line:g_r_query}} < T^{\ref{line:help_pointer_cas}}$.
    Let $o^*$ be the operation stored in $\stateobject{}$ at $T^{\ref{line:g_r_query}}$.
    If $o' \neq o^*$, then since $o'$ is stored in $\stateobject{}$ at $T$, $o^*$ is stored in $\stateobject{}$ at $T^{\ref{line:g_r_query}}$, and $T < T^{\ref{line:g_r_query}}$, by \Cref{lemma:non_null_help_struct_response}, $o'$ is done at $T^{\ref{line:g_r_query}}$, and so since $T^{\ref{line:g_r_query}} < T^{\ref{line:help_pointer_cas}}$, by \Cref{observation:if_once_done_then_always_done}, $o'$ is done at $T^{\ref{line:help_pointer_cas}}$.
    If $o' = o^*$, then since $o^*$ is stored in $\stateobject{}$ at $T^{\ref{line:g_r_query}}$, we have that $p(o)$ executes \cref{line:help_pointer_cas} for $o'$ at $T^{\ref{line:help_pointer_cas}}$, and so by \Cref{lemma:copy_help_struct_implies_done}, $o'$ is done at $T^{\ref{line:help_pointer_cas}}$.
    Therefore, in all cases, $o'$ is done at $T^{\ref{line:help_pointer_cas}}$.
    There are two cases.

    \begin{itemize}
        \item[] \hspace{0pt}\textbf{Case 1.} $o'$ is not stored in $\announceobject{}$ at the time $p(o)$ executes \cref{line:g_a_query} during $I^+$.

        Therefore, since (1) $o' \in S(o)$ is done at $T^{\ref{line:help_pointer_cas}}$ during $I^+$ (and thus $P$), (2) $o'$ is stored in $\announceobject{}$ at $T^{\ref{line:g_a_query}}$ during $I$ (and thus $P$), and (3) $o'$ is not stored in $\announceobject{}$ at the time $p(o)$ executes \cref{line:g_a_query} during $I^+$ which is after the times in (1) and (2), by \Cref{def:useful_period}, $P$ is useful as wanted.

        \item[] \hspace{0pt}\textbf{Case 2.} $o'$ is stored in $\announceobject{}$ at the time $p(o)$ executes \cref{line:g_a_query} during $I^+$.

        Hence, $p(o)$ read $(t(o'), o', C_{p(o')})$ from $\announceobject{}$ at this time.
        Thus, since $o'$ is done at $T^{\ref{line:help_pointer_cas}}$, by \Cref{observation:if_once_done_then_always_done}, $o'$ is done at the time $p(o)$ executes \cref{line:help_pointer_query} during $I^+$.
        So, by \Cref{lemma:done_implies_help_struct_form} $C_{p(o')} \neq (t(o'), \nullconstant)$ at this time.
        Hence, $p(o)$ finds the condition on \cref{line:help} to be false during $I^+$, and so $p(o)$ executes \cref{line:g_a_cas} during $I^+$.
        Thus, since $I$ is one of the first three iterations of $P$, and $I^+$ is the iteration after $I$ in $P$, we have that $I^+$ is one of the first four iterations of $P$.
        Therefore, since $p(o)$ executes \cref{line:g_a_cas} during $I^+$, by \Cref{lemma:read_done_operation_from_announce}, $P$ is useful as wanted.
        \qH{\Cref{lemma:read_not_done_operation_from_announce_and_successfully_helped}}
    \end{itemize}
\end{proof}

\begin{proposition}\label{lemma:the_annoying_case}
    Consider any non-initial period $P$ of some operation $o$.
    If $p(o)$ executes \cref{line:g_r_cas} and the response is false in the second and third iteration of $P$, then $P$ is useful.
\end{proposition}

\begin{proof}
    Denote the first three iterations of $P$ as $I_1$, $I_2$, and $I_3$.
    Let $T^{\ref{line:g_r_query}}_2$ (resp. $T^{\ref{line:g_r_query}}_3$) be the time that $p(o)$ executed \cref{line:g_r_query} during $I_2$ (resp. $I_3$) and let $T^{\ref{line:g_r_cas}}_2$ (resp. $T^{\ref{line:g_r_cas}}_3$) be the time that $p(o)$ executed \cref{line:g_r_cas} during $I_2$ (resp. $I_3$).
    Since $p(o)$ executed \cref{line:g_r_cas} during $I_2$ (resp. $I_3$) and received a response of false, it follows that the value of $\stateobject{}$ changed between $T^{\ref{line:g_r_query}}_2$ and $T^{\ref{line:g_r_cas}}_2$ (resp. $T^{\ref{line:g_r_query}}_3$ and $T^{\ref{line:g_r_cas}}_3$).
    Hence, there is a successful execution of the \CASop{} operation on \cref{line:g_r_cas} between $T^{\ref{line:g_r_query}}_2$ and $T^{\ref{line:g_r_cas}}_2$ (resp. $T^{\ref{line:g_r_query}}_3$ and $T^{\ref{line:g_r_cas}}_3$).
    Let $e_2$ (resp. $e_3$) be a successful execution of the \CASop{} operation on \cref{line:g_r_cas} between $T^{\ref{line:g_r_query}}_2$ and $T^{\ref{line:g_r_cas}}_2$ (resp. $T^{\ref{line:g_r_query}}_3$ and $T^{\ref{line:g_r_cas}}_3$).
    Hence, since $T^{\ref{line:g_r_query}}_2$ and $T^{\ref{line:g_r_cas}}_2$ (resp. $T^{\ref{line:g_r_query}}_3$ and $T^{\ref{line:g_r_cas}}_3$) are during $I_2$ (resp. $I_3$), and $I_2$ is completed before $I_3$ begins, we have the following ordering:
    \begin{align*}
        T^{\ref{line:g_r_query}}_2 < e_2 < T^{\ref{line:g_r_cas}}_2 < T^{\ref{line:g_r_query}}_3 < e_3 < T^{\ref{line:g_r_cas}}_3.
    \end{align*}
    
    Let $q$ be the process that executed $e_3$ and let $T^{\ref{line:g_r_query}}_q$ be the time of $q$'s last execution of \cref{line:g_r_query} before $e_3$.
    We claim that $T^{\ref{line:g_r_query}}_q$ is after $e_2$.
    Suppose, for contradiction, $T^{\ref{line:g_r_query}}_q$ is before $e_2$.
    Let $o_3$ be the operation stored in $\stateobject{}$ at the step before $e_3$.
    Hence, since $e_3$ is successful, it follows that $q$ read $(t(o_3), \unusedvalue{}, \unusedvalue{}, C_{p(o_3))})$ from $\stateobject{}$ on \cref{line:g_r_query} at $T^{\ref{line:g_r_query}}_q$.
    Thus, by \Cref{observation:g_r_is_well_formed}, $o_3$ is stored in $\stateobject{}$ at $T^{\ref{line:g_r_query}}_q$.
    Since $T^{\ref{line:g_r_query}}_q$ is before $e_2$, $e_2$ is before $e_3$, and $o_3$ is the operation stored in $\stateobject{}$ at the step before $e_3$, we have that there is a successful \CASop{} operation on $\stateobject{}$ on \cref{line:g_r_cas} for $o_3$ strictly between $T^{\ref{line:g_r_query}}_q$ and $e_3$.
    Hence, by \Cref{lemma:never_noop}, $o_3 \neq \noop$.
    Thus, since $o_3$ is stored in $\stateobject{}$ at $T^{\ref{line:g_r_query}}_q$, we have that there is a successful \CASop{} operation on $\stateobject{}$ on \cref{line:g_r_cas} for $o_3$ before $T^{\ref{line:g_r_query}}_q$.
    Therefore, there are two successful \CASop{} operation on $\stateobject{}$ on \cref{line:g_r_cas} for $o_3$.
    However, by \Cref{claim:pairwise_distinct}, every \CASop{} operation on $\stateobject{}$ on \cref{line:g_r_cas} is for a different operation, a contradiction.

    We now finish the proof of \Cref{lemma:the_annoying_case}.
    Since $T^{\ref{line:g_r_query}}_q$ is after $e_2$, we have that $q$'s last execution of \cref{line:g_a_query} before $e_3$ is between $e_2$ and $e_3$; say at time $T^{\ref{line:g_a_query}}_q$.
    Let $T^{\ref{line:g_a_gcas}}_1$ be the time $p(o)$ executed \cref{line:g_a_gcas} during $I_1$.
    Hence, since $I_1$ completed before $I_2$ began, and $T^{\ref{line:g_r_query}}_2$ is during $I_2$, by transitivity, $T^{\ref{line:g_a_gcas}}_1 < T^{\ref{line:g_r_query}}_2$.
    Thus, since $T^{\ref{line:g_r_query}}_2 < e_2$, $e_2 < T^{\ref{line:g_a_query}}_q < e_3$, and $e_3 < T^{\ref{line:g_r_cas}}_3$, by transitivity, $T^{\ref{line:g_a_gcas}}_1 < T^{\ref{line:g_a_query}}_q < T^{\ref{line:g_r_cas}}_3$.
    So, since $T^{\ref{line:g_a_gcas}}_1$ is a time during $I_1$, $T^{\ref{line:g_r_cas}}_3$ is a time during $I_3$, and both $I_1$ and $I_3$ are during $P$, we have that $T^{\ref{line:g_a_query}}_q$ is a time during $P$.
    Let $o'$ be the operation stored in $\announceobject{}$ at $T^{\ref{line:g_a_query}}_q$.
    There are two cases.
    \begin{itemize}
        \item[] \hspace{0pt}\textbf{Case 1.} $o' \notin S(o)$.

        Since $o' \notin S(o)$ is stored in $\announceobject{}$ at time $T^{\ref{line:g_a_query}}_q$, which by above is during $P$, and $T^{\ref{line:g_a_query}}_q$ is after the first time $p(o)$ executed \cref{line:g_a_gcas} during $P$ (namely $T^{\ref{line:g_a_gcas}}_1$), by \Cref{lemma:read_low_priority_operation_from_announce}, $P$ is useful as wanted.
    
        \item[] \hspace{0pt}\textbf{Case 2.} $o' \in S(o)$.

        The argument is essentially the same as \Cref{lemma:read_not_done_operation_from_announce_and_successfully_helped}.
        Since $o'$ is the operation stored in $\announceobject{}$ at $T^{\ref{line:g_a_query}}_q$, by \Cref{observation:g_a_is_well_formed}, $q$ read $(t(o'), o', C_{p(o')})$ from $\announceobject{}$ on \cref{line:g_a_query} at $T^{\ref{line:g_a_query}}_q$.
        Hence, since  $T^{\ref{line:g_a_query}}_q$ is the time of $q$'s last execution of \cref{line:g_a_query} before $e_3$, we have that $e_3$ is of the form $\CASop{}(\stateobject{}, \unusedvalue{}, (t(o'), \unusedvalue{}, \unusedvalue{}, C_{p(o')}))$, so by \Cref{observation:g_r_is_well_formed}, it is for $o'$.
        Thus, since the response of this \CASop{} operation is true, we have that $o'$ is stored in $\stateobject{}$ at $e_3$.
        So, since $T^{\ref{line:g_r_query}}_3 < e_3 < T^{\ref{line:g_r_cas}}_3$, and $T^{\ref{line:g_r_query}}_3$ and $T^{\ref{line:g_r_cas}}_3$ are during $I_3$, we have that $o'$ is stored in $\stateobject{}$ during $I_3$; say at time $T$.
        Since $I_3$ is the third iteration of $P$, by \Cref{def:period}, there is a complete and not terminal iteration of the loop on \cref{line:loop_start} by $p(o)$ within $o$ after $I_3$.
        Let $I^+$ be the next one after $I_3$.
        Since $I^+$ is complete and not terminal, $p(o)$ executes \cref{line:g_r_query} (resp. \cref{line:help_pointer_cas}) during $I^+$; say at time $T^{\ref{line:g_r_query}}$ (resp. $T^{\ref{line:help_pointer_cas}}$).
        Hence, since $I_3$ completes before $I^+$ begins, and $T$ is a time during $I_3$, we have that $T < T^{\ref{line:g_r_query}} < T^{\ref{line:help_pointer_cas}}$.
        Let $o^*$ be the operation stored in $\stateobject{}$ at $T^{\ref{line:g_r_query}}$.
        If $o' \neq o^*$, then since $o'$ is stored in $\stateobject{}$ at $T$, $o^*$ is stored in $\stateobject{}$ at $T^{\ref{line:g_r_query}}$, and $T < T^{\ref{line:g_r_query}}$, by \Cref{lemma:non_null_help_struct_response}, $o'$ is done at $T^{\ref{line:g_r_query}}$, and so since $T^{\ref{line:g_r_query}} < T^{\ref{line:help_pointer_cas}}$, by \Cref{observation:if_once_done_then_always_done}, $o'$ is done at $T^{\ref{line:help_pointer_cas}}$.
        If $o' = o^*$, then since $o^*$ is stored in $\stateobject{}$ at $T^{\ref{line:g_r_query}}$, we have that $p(o)$ executes \cref{line:help_pointer_cas} for $o'$ at $T^{\ref{line:help_pointer_cas}}$, and so by \Cref{lemma:copy_help_struct_implies_done}, $o'$ is done at $T^{\ref{line:help_pointer_cas}}$.
        Therefore, in all cases, $o'$ is done at $T^{\ref{line:help_pointer_cas}}$.
        There are two cases.
    
        \begin{itemize}
            \item[] \hspace{0pt}\textbf{Case 2.1.} $o'$ is not stored in $\announceobject{}$ at the time $p(o)$ executes \cref{line:g_a_query} during $I^+$.

            Hence, since $T^{\ref{line:g_a_gcas}}_1 < T^{\ref{line:g_a_query}}_q < T^{\ref{line:g_r_cas}}_3$, $T^{\ref{line:g_r_cas}}_3$ is during $I_3$, and $I_3$ completed before $I^+$ began, by transitivity, the time $p(o)$ executes \cref{line:g_a_query} during $I^+$ is after $T^{\ref{line:g_a_query}}_q$.
            Therefore, since (1) $o' \in S(o)$ is done at $T^{\ref{line:help_pointer_cas}}$ during $I^+$ (and thus $P$), (2) $o'$ is stored in $\announceobject{}$ at $T^{\ref{line:g_a_query}}_q$ which by the above is during $P$, and (3) $o'$ is not stored in $\announceobject{}$ at the time $p(o)$ executes \cref{line:g_a_query} during $I^+$ which is after the times in (1) and (2), by \Cref{def:useful_period}, $P$ is useful as wanted.
    
            \item[] \hspace{0pt}\textbf{Case 2.2.} $o'$ is stored in $\announceobject{}$ at the time $p(o)$ executes \cref{line:g_a_query} during $I^+$.
    
            Hence, $p(o)$ read $(t(o'), o', C_{p(o')})$ from $\announceobject{}$ at this time.
            Thus, since $o'$ is done at $T^{\ref{line:help_pointer_cas}}$, by \Cref{observation:if_once_done_then_always_done}, $o'$ is done at the time $p(o)$ executes \cref{line:help_pointer_query} during $I^+$.
            So, by \Cref{lemma:done_implies_help_struct_form} $C_{p(o')} \neq (t(o'), \nullconstant)$ at this time.
            Hence, $p(o)$ finds the condition on \cref{line:help} to be false during $I^+$, and so $p(o)$ executes \cref{line:g_a_cas} during $I^+$.
            Thus, since $I_3$ is the third iteration of $P$, and $I^+$ is the iteration after $I_3$ in $P$, we have that $I^+$ is one of the first four iterations of $P$.
            Therefore, since $p(o)$ executes \cref{line:g_a_cas} during $I^+$, by \Cref{lemma:read_done_operation_from_announce}, $P$ is useful as wanted.
            \qH{\Cref{lemma:the_annoying_case}}
        \end{itemize}
    \end{itemize}
\end{proof}

\begin{lemma}\label{lemma:non_initial_periods_are_useful}
    Every non-initial period is useful.
\end{lemma}

\begin{proof}
    Consider any non-initial period $P$ of an operation $o$.
    If either: (1) $p(o)$ executes \cref{line:g_a_cas} during any of the first three iterations of $P$; (2) $p(o)$ executes \cref{line:g_r_cas} and the response is true in any of the first three iterations of $P$; or (3) $p(o)$ executes \cref{line:g_r_cas} and the response is false in each of the first three iterations of $P$, then by Propositions \ref{lemma:read_done_operation_from_announce}, \ref{lemma:read_not_done_operation_from_announce_and_successfully_helped}, and \ref{lemma:the_annoying_case}, respectively, $P$ is useful.
    So, it suffices to assume $P$ does not satisfy case (1), (2), or (3).
    We show that this is impossible, completing the proof.
    By \Cref{def:period} $P$ is five consecutive complete and not terminal iterations of the loop on \cref{line:loop_start} by $p(o)$ within $o$; denote the first three of them by $I_1$, $I_2$, and $I_3$.
    Since these iterations are complete and not terminal, by \Cref{def:terminal_and_complete}, $p(o)$ executes either \cref{line:g_r_cas} or \cref{line:g_a_cas} in each of them.
    However, since $P$ does not satisfy case (1), we have that $p(o)$ executed \cref{line:g_r_cas} during $I_1$, $I_2$, and $I_3$.
    Hence, since $P$ does not satisfy case (2), we have that when $p(o)$ executes \cref{line:g_r_cas} and receives a response of false in $I_1$, $I_2$, and $I_3$.
    Therefore, $P$ satisfies case (3).
    However, $P$ does not satisfy case (3), a contradiction.
    \qH{\Cref{lemma:non_initial_periods_are_useful}}
\end{proof}

We now bound the number of distinct non-initial periods per operation.
We start by proving that after four distinct non-initial periods, we have identified at least two operations.

\begin{proposition}\label{lemma:four_non_initial_periods_provide_at_least_two_done_operations}
    Consider four distinct non-initial periods $P_1$, $P_2$, $P_3$, and $P_4$ of an operation.
    By \Cref{lemma:non_initial_periods_are_useful}, they are useful, so by \Cref{def:useful_period}, there is an operation $o_1$, $o_2$, $o_3$, and $o_4$ that is done at some time in $P_1$, $P_2$, $P_3$, and $P_4$, respectively.
    Then, $|\{o_1, o_2, o_3, o_4\}| \geq 2$.
\end{proposition}

\begin{proof}
    Suppose, for contradiction, $|\{o_1, o_2, o_3, o_4\}| < 2$.
    Hence, $o_1 = o_2 = o_3 = o_4$; denote this operation by $o$.
    Without loss of generality, suppose $P_1, P_2, P_3, P_4$ is the order in which they occur.
    Since $P_1$ is useful, by \Cref{def:useful_period}, $o$ is done at some time $D$ during $P_1$.
    Furthermore, since $P_1$, $P_2$, $P_3$, and $P_4$ are useful, by \Cref{def:useful_period}, there is a time $T^{\announceobject{}}_2$, $T^{\announceobject{}}_3$, and $T^{\announceobject{}}_4$ during $P_2$, $P_3$, and $P_4$, respectively, where $o$ is stored in $\announceobject{}$, and there is a time $T^X_1$, $T^X_2$, and $T^X_3$ during $P_1$, $P_2$, and $P_3$, respectively, where $o$ is not stored in $\announceobject{}$ such that $T^X_1 > D$, $T^X_2 > T^{\announceobject{}}_2$, and $T^X_3 > T^{\announceobject{}}_3$.
    Hence, since $P_1, P_2, P_3, P_4$ are distinct, and this is the order in which they occur, it follows that
    \begin{align*}
        D < T^X_1 < T^{\announceobject{}}_2 < T^X_2 < T^{\announceobject{}}_3 < T^X_3 < T^{\announceobject{}}_4.
    \end{align*}
    Thus, for each $i \in [1..3]$, since $o$ is not stored in $\announceobject{}$ at $T^X_i$ and $o$ is stored in $\announceobject{}$ at $T^{\announceobject{}}_{i + 1}$, it follows that between $T^X_i$ and $T^{\announceobject{}}_{i + 1}$ there is an execution of \cref{line:g_a_gcas} or \cref{line:g_a_cas} for operation $o$; denote this execution as $e_i$.
    So, by \Cref{observation:p_o_writes_o_into_announce}, $e_i$ is performed by $p(o)$ within $o$.
    Since $e_1$, $e_2$, and $e_3$ are each an execution of \cref{line:g_a_gcas} or \cref{line:g_a_cas} by $p(o)$ within $o$, it follows that $p(o)$ executes \cref{line:loop_start} within $o$ some time strictly between $e_1$ and $e_3$; say at time $T$.
    Thus, since $T$ is after $e_1$, $e_1$ is after $T^X_1$, and $T^X_1$ is after $D$, by transitivity, $T$ is after $D$.
    So, since $o$ is done at $D$, by \Cref{lemma:done_implies_help_struct_form}, $\helpobject{p(o)} \neq (t(o), \nullconstant)$ at $T$.
    Since $T$ is the time of an execution of \cref{line:loop_start} by $p(o)$ within $o$, it follows that $p(o)$ checks whether $\helpobject{p(o)} = (t(o), \nullconstant)$ on \cref{line:loop_start} at time $T$.
    Therefore, since $\helpobject{p(o)} \neq (t(o), \nullconstant)$ at time $T$, we have that $p(o)$ finds the condition on \cref{line:loop_start} to be false at time $T$, and so $p(o)$ does not execute \cref{line:g_a_gcas} or \cref{line:g_a_cas} from $T$ onwards within $o$.
    However, since $T$ is strictly before $e_3$, and $e_3$ is an execution of \cref{line:g_a_gcas} or \cref{line:g_a_cas} by $p(o)$ within $o$, we have that $p(o)$ executes \cref{line:g_a_gcas} or \cref{line:g_a_cas} from $T$ onwards within $o$, a contradiction.
    \qH{\Cref{lemma:four_non_initial_periods_provide_at_least_two_done_operations}}
\end{proof}

\begin{lemma}\label{lemma:o_has_few_periods}
    Every operation $o$ has at most $3|S(o)|$ distinct non-initial periods. 
\end{lemma}

\begin{proof}
    Suppose, for contradiction, some operation $o$ has more than $3|S(o)|$ distinct non-initial periods.
    Hence, there are at least $3|S(o)| + 1$ distinct non-initial periods of $o$.
    Denote them as $P_1, P_2, \ldots, P_{3|S(o)| + 1}$.
    For each $i \in [1..3|S(o)| + 1]$, by \Cref{lemma:non_initial_periods_are_useful}, $P_i$ is useful, so by \Cref{def:useful_period}, some operation $o_i \in S(o)$ is done at some time during $P_i$.
    For each $n \in [0..|S(o)|]$ let $\mathcal{P}(n)$ be the predicate: $|O_n| \geq n + 1$ where $O_n = \{o_1, \ldots o_{3n + 1}\}$.
    We prove $\mathcal{P}(n)$ by induction on $n$.
    \begin{itemize}
        \item[] \hspace{0pt}\textbf{Base Case.} $n = 0$.

        Hence, $O_n = \{o_1\}$ and $n + 1 = 1$.
        Therefore, since $|\{o_1\}| \geq 1$, we have that $\mathcal{P}(0)$ holds.

        \item[] \hspace{0pt}\textbf{Inductive Case.} $\forall n \in [0..|S(o)|)\ \mathcal{P}(n) \implies \mathcal{P}(n + 1)$.

        Suppose for some $n \in [0..|S(o)|)$ $\mathcal{P}(n)$ holds.
        This is the inductive hypothesis.
        Suppose, for contradiction, $\mathcal{P}(n + 1)$ does not hold, so $|O_{n + 1}| < n + 2$.
        Since $\mathcal{P}(n)$ holds, we have that $|O_n| \geq n + 1$.
        Hence, since $O_n \subseteq O_{n + 1}$, we have that $|O_{n + 1}| \geq n + 1$, and so $n + 1 \leq |O_{n + 1}| < n + 2$.
        Thus, $|O_{n + 1}| = n + 1$.
        So, since $|O_n| \geq n + 1$ and $O_n \subseteq O_{n + 1}$, it follows that $O_n = O_{n + 1}$.
        Therefore, since $O_{n} \neq \emptyset$, and $O_{n + 1} = O_n \cup \{o_{3n + 2}, o_{3n + 3}, o_{3n + 4}\}$, we have that $o' = o_{3n + 2} = o_{3n + 3} = o_{3n + 4}$ for some $o' \in O_n$.
        However, by \Cref{lemma:four_non_initial_periods_provide_at_least_two_done_operations}, $|\{o', o_{3n + 2}, o_{3n + 3}, o_{3n + 4}\}| \geq 2$, a contradiction.
    \end{itemize}

    We now finish the proof of \Cref{lemma:o_has_few_periods}.
    Since for each $i \in [1..3|S(o)| + 1]$ $o_i \in S(o)$, we have that $O_{|S(o)|} \subseteq S(o)$.
    Furthermore, since $\mathcal{P}(|S(o)|)$ holds, we have that $|O_{|S(o)|}| \geq |S(o)| + 1$.
    Therefore, since $O_{|S(o)|} \subseteq S(o)$, we have that $|S(o)| \geq |S(o)| + 1$, which is impossible.
    \qH{\Cref{lemma:o_has_few_periods}}
\end{proof}

We now have all we need to prove the main result of this section.

\begin{theorem}[\Cref{thm:algo_1_step} restated]\label{thm:algo_1_time_complexity}
    Suppose a process $p$ invokes an operation $o$ and executes \cref{line:time_assignment} within $o$.
    Let~$c$ be the point contention at this time.
    Then, the number of steps that $p$ takes within $o$ is at most \mbox{linear in $c$.}
\end{theorem}

\begin{proof}
    By \Cref{def:op_state}, $p = p(o)$.
    Furthermore, by \Cref{def:set_of_interest} $|S(o)|$ is at most the number of pending operations at the time $p(o)$ executes \cref{line:time_assignment} within $o$, so $|S(o)| \leq c$.
    Hence, since by \Cref{lemma:o_has_few_periods}, $o$ has at most $3|S(o)|$ distinct non-initial periods, it follows that $o$ has at most $3|c|$ distinct non-initial periods.
    Thus, by \Cref{def:period}, $o$ has at most $3|c|$ distinct periods after the first iteration of the loop on \cref{line:loop_start} by $p(o)$ within $o$.
    So, by \Cref{def:period}, $p(o)$ completes at most $15|c|$ non-terminal iterations of the loop on \cref{line:loop_start} within $o$ after the first iteration of the loop on \cref{line:loop_start} by $p(o)$ within~$o$.
    Hence, by \Cref{def:terminal_and_complete}, $p(o)$ completes at most $15|c| + 2$ iterations of the loop on \cref{line:loop_start} within $o$ (the $2$ accounts for the first iteration and the terminal iteration).
    Therefore, since $p(o)$ takes a constant number of steps before entering (resp. after exiting) the loop on \cref{line:loop_start} within $o$, and $p(o)$ takes a constant number of steps during each iteration of the loop on \cref{line:loop_start} within $o$, it follows that $p(o)$ takes at most linear in $c$ steps within $o$ as wanted.
    \qH{\Cref{thm:algo_1_time_complexity}}
\end{proof}

\subsection{Linearizability}

In this section, we prove that \Cref{alg:wait-free-simple} is linearizable with respect to type $\mathcal{T}$.

\begin{theorem}\label{theorem:linearizable}
    \Cref{alg:wait-free-simple} is linearizable with respect to type $\mathcal{T}$.
\end{theorem}

\begin{proof}
    Consider any implementation history $\mathcal{I}$ of \Cref{alg:wait-free-simple}.
    Let $\mathcal{H}$ be the object history obtained by removing all implementation steps from $\mathcal{I}$.
    We must prove that $\mathcal{H}$ is linearizable with respect to $\mathcal{T}$; that is, we must prove that there is a completion~$\mathcal{H'}$ of~$\mathcal{H}$ that is equivalent to some sequential history $\mathcal{S}$ such that $\mathcal{S}$ is legal with respect to $\mathcal{T}$ and $<_{\mathcal{H'}} \subseteq <_{\mathcal{S}}$.
    Let
    \begin{align*}
        (t_1, s_1, r_1, -), (t_2, s_2, r_2, -), (t_3, s_3, r_3, -), \ldots
    \end{align*}
    be the sequence of values written into $\stateobject{}$ in $\mathcal{I}$.
    These values were written by the sequence of successful CAS executions on $\stateobject{}$ on \cref{line:g_r_cas} in $\mathcal{I}$.
    Observe that there is a unique operation $o_i$ on $O$ that has timestamp $t_i$ (see \Cref{observation:unique_timestamps}).
    So the sequence
    \begin{align*}
        Ops = o_1, o_2, o_3, \ldots
    \end{align*}
    is the sequence of operations stored in $\stateobject{}$ during $\mathcal{I}$ (see \Cref{observation:g_r_is_well_formed}).
    Using $Ops$, we define the completion~$\mathcal{H'}$ of $\mathcal{H}$ as follows.
    Consider any incomplete operation $o$ in $\mathcal{H}$.
    If $o$ is in $Ops$ and the first index it appears at is $i$, then a response step for~$o$ is appended at the end of $\mathcal{H'}$ with response $r_i$.
    Otherwise, $o$'s invocation step is removed in $\mathcal{H'}$.
    Also using~$Ops$, we define a sequential history $\mathcal{S}$ as the sequence
    \begin{align*}
        invocation(o_1), response(o_1, r_1), invocation(o_2), response(o_2, r_2), \ldots
    \end{align*}
    where $invocation(o_i)$ is the invocation step of operation $o_i$ (\cref{line:op_start}) and $response(o_i, r_i)$ is the response step of $o_i$ which returned the response $r_i$ (\cref{line:op_done}).
    The remainder of this proof will go as follows.
    \begin{compactitem}
        
        \item First we define the \emph{linearization point} $\ell(o_i)$ of $o_i$ to be the time when the successful CAS on \cref{line:g_r_cas} wrote $(t_i, s_i, r_i, -)$ in $\stateobject{}$.
        Since each $o_i$ appears exactly once in $Ops$, $\ell(o_i)$ is well-defined, and the operations in $Ops$ appear in increasing order of their linearization points.

        \item We will prove that (a) every complete operation in $\mathcal{H}$ is in $Ops$ and (b) the linearization point $\ell(o_i)$ of every operation $o_i$ in $Ops$ (whether complete in $\mathcal{I}$ or not) is after $o_i$'s invocation step in $\mathcal{I}$ and before $o_i$'s response step $\mathcal{I}$ if it exists.
        These two facts together imply that $<_{\mathcal{H'}} \subseteq <_{\mathcal{S}}$.

        \item We will then prove that for every $o_i$ in $Ops$, $(s_i, r_i) = apply_\mathcal{T}(o_i, s_{i - 1})$.
        This implies that $\mathcal{S}$ is legal with respect to $\mathcal{T}$.

        \item We will then prove that if $o_i$ is a complete operation in $\mathcal{I}$ then its response is $r_i$ in $\mathcal{I}$.

        \item Finally, we will prove that $\mathcal{H'}$ is equivalent to $\mathcal{S}$.
    \end{compactitem}

    For every operation $o$ in $Ops$, define $\ell(o)$ to be the time of the successful CAS on $\stateobject{}$ on \cref{line:g_r_cas} for $o$ in~$\mathcal{I}$.
    This is well defined by \Cref{claim:pairwise_distinct}.

    \begin{claimcustom}{\ref{theorem:linearizable}.1}\label{claim:linearization_points}
        (a) Every complete operation in $\mathcal{I}$ is in $Ops$.\\
        (b) The linearization point $\ell(o)$ of every operation $o$ in $Ops$ (whether complete in $\mathcal{I}$ or not) is after $o$'s invocation step in $\mathcal{I}$ and before $o$'s response step in $\mathcal{I}$ if it exists.
    \end{claimcustom}

    \begin{proof}
        For part (a) consider any complete operation $o$ in $\mathcal{I}$ and let $p = p(o)$.
        Therefore, $p$ found the response~$r$ to $o$ in $\helpobject{p}.response$ (see lines \ref{line:loop_start} and \ref{line:op_done}).
        Thus some process $q$ previously executed a successful CAS on~$\helpobject{p}$ on \cref{line:help_pointer_cas} that wrote $(t(o), r)$ in $\helpobject{p}$.
        Hence, $q$ previously read $(t(o), -, r, \helpobject{p})$ in $\stateobject{}$ on \cref{line:g_r_query}, say at time~$T^{\ref{line:g_r_query}}_q$.
        Thus a successful CAS on $\stateobject{}$ on \cref{line:g_r_cas} and wrote that value before~$T^{\ref{line:g_r_query}}_q$.
        Therefore, the complete operation $o$ is one of the operations in $Ops$, say $o_i$, which completes the proof of part (a).
        Furthermore, $o_i$'s linearization point $\ell(o_i)$ occurred before $T^{\ref{line:g_r_query}}_q$, and so before $o$'s response step, which is needed for part (b).

        To complete the proof of part (b) consider any operation $o$ in $\mathcal{I}$, whether complete or not.
        The linearization point $\ell(o)$ of $o$ is the time when some process $p$ executes a successful CAS on $\stateobject{}$ on \cref{line:g_r_cas} and writes $(t(o), -, -, -)$ in $\stateobject{}$.
        For this to happen $p$ must have previously read $(t(o), -, -)$ in $\announceobject{}$ on \cref{line:g_a_query}, which means that $p(o)$ executed a successful GCAS on $\announceobject{}$ on either \cref{line:g_a_gcas} or \cref{line:g_a_cas} within $o$.
        Therefore, the linearization point $\ell(o)$ occurs after $o$'s invocation step, which completes the proof of part (b).
        \qH{\Cref{claim:linearization_points}}
    \end{proof}

    \begin{claimcustom}{\ref{theorem:linearizable}.2}\label{claim:respect_real_time_order}
        $<_\mathcal{H'} \subseteq <_\mathcal{S}$.
    \end{claimcustom}

    \begin{proof}
        Consider any two operations $o$ and $o'$ in $\mathcal{H'}$ such that $o <_\mathcal{H'} o'$.
        Thus, $o$'s response step in $\mathcal{H'}$ is before $o'$'s invocation step in $\mathcal{H'}$.
        Hence, by the construction of $\mathcal{H'}$: $o$ is complete in $\mathcal{H}$, $o <_\mathcal{H} o'$, and $o'$ is in~$Ops$.
        Since $\mathcal{H}$ is the result of removing all implementation steps in $\mathcal{I}$, this implies that $o$ is complete in $\mathcal{I}$ and that $o$'s response step in $\mathcal{I}$ is before $o'$'s invocation step in $\mathcal{I}$.
        Since $o$ is complete in $\mathcal{I}$, \Cref{claim:linearization_points} (a) asserts that $o$ is in $Ops$.
        Thus by \Cref{claim:linearization_points} (b), $\ell(o)$ is before $o$'s response step in $\mathcal{I}$ (it exists since $o$ is complete in $\mathcal{I}$).
        Likewise, since $o'$ is in $Ops$, \Cref{claim:linearization_points} (b) states that $\ell(o')$ is after $o'$'s invocation step in~$\mathcal{I}$.
        Thus, $\ell(o)$ is before $\ell(o')$ in $\mathcal{I}$ and therefore $o$ is before $o'$ in $Ops$.
        Hence, by the construction of $\mathcal{S}$,~$o$'s response step is before $o'$'s invocation step in $\mathcal{S}$.
        Therefore, $o <_\mathcal{S} o'$ as wanted.
        \qH{\Cref{claim:respect_real_time_order}}
    \end{proof}

    \begin{claimcustom}{\ref{theorem:linearizable}.3}\label{claim:correct_execution}
        For every $o_i$ in $Ops$, $(s_i, r_i) = apply_\mathcal{T}(o_i, s_{i - 1})$, where $s_0$ is the initial state of type $\mathcal{T}$.
    \end{claimcustom}

    \begin{proof}
        By definition, $s_i$ and $r_i$ are the values written in $\stateobject{}.state$ and $\stateobject{}.response$, respectively, by the $i$th successful CAS on $\stateobject{}$ on \cref{line:g_r_cas} in $\mathcal{I}$.
        Let $q$ be the process that performed this CAS.
        Therefore, by \cref{line:apply}, $(s_i, r_i) = apply_{\mathcal{T}}(o', s^*)$, where $o'$ is the operation in $\announceobject{}.operation$ when $q$ last read $\announceobject{}$ on \cref{line:g_a_query} and $s^*$ is the state in $\stateobject{}.state$ when $q$ last read $\stateobject{}$ on \cref{line:g_r_query}.
        So the timestamp $t'$ in $\announceobject{}.time$ when $q$ read $\announceobject{}$ on \cref{line:g_a_query} is the timestamp $t_i$ that $q$ wrote into $\stateobject{}.time$ in the $i$-th successful CAS on $\stateobject{}$ on \cref{line:g_r_cas}; therefore $t' = t_i$, and $o' = o_i$.
        Furthermore, $s^*$ is the value in $\stateobject{}.state$ when $q$ executed the $i$-th successful CAS on $\stateobject{}$ on \cref{line:g_r_cas}: otherwise, that CAS would not be successful.
        Therefore, $s^*$ is the value written in $\stateobject{}.state$ by the $(i-1)$-th successful CAS on $\stateobject{}$ on \cref{line:g_r_cas}, or the initial state $s_0$ of $O$, if $i = 1$, so $s^* = s_{i - 1}$.
        Thus $(s_i, r_i) = apply_{\mathcal{T}}(o_i, s_{i - 1})$, as wanted.
        \qH{\Cref{claim:correct_execution}}
    \end{proof}

    By the definition of $apply_{\mathcal{T}}$ \Cref{claim:correct_execution} immediately implies:
    
    \begin{corollarycustom}{\ref{theorem:linearizable}.4}\label{corollary:correct_execution}
        For every $o_i$ in $Ops$,  $(s_{i - 1}, o_i, s_i, r_i) \in \delta$, where $\delta$ is the state-transition relation of type $\mathcal{T}$, and $s_0$ is the initial state of type $\mathcal{T}$, so $\mathcal{S}$ is legal with respect to $\mathcal{T}$.
    \end{corollarycustom}

    \begin{claimcustom}{\ref{theorem:linearizable}.5}\label{claim:correct_response}
        For every $o_i$ in $Ops$, if $o_i$ is complete in $\mathcal{I}$ then its response is $r_i$ in $\mathcal{I}$.
    \end{claimcustom}

    \begin{proof}
        Suppose that $o_i$'s response is $r$ in $\mathcal{I}$.
        Let $p = p(o_i)$.
        Since $o_i$ is complete and (by definition) $t(o_i) = t_i$, $p$ found $\helpobject{p} = (t_i, r)$ on \cref{line:loop_start}, for some $r \neq \nullconstant$, and returned $r$ on \cref{line:op_done}.
        Therefore, some process $q$ wrote $(t_i, r)$ in $\helpobject{p}$ by a successful CAS on \cref{line:help_pointer_cas}.
        This means that $q$ read $(t_i, s, r, \helpobject{p})$ from $\stateobject{}$ on \cref{line:g_r_query}, for some state $s$.
        This, in turn, implies that some process $q'$ wrote $(t_i, s, r, \helpobject{p})$ into $\stateobject{}$ via a successful CAS on \cref{line:g_r_cas}.
        By \Cref{claim:pairwise_distinct} there is only one successful CAS on $\stateobject{}$ on \cref{line:g_r_cas} in $\mathcal{I}$ for $o_i$.
        So, $s = s_i$ and $r = r_i$.
        Therefore, $o_i$ returns $r_i$ in $\mathcal{I}$.
        \qH{\Cref{claim:correct_response}}
    \end{proof}

    \begin{claimcustom}{\ref{theorem:linearizable}.6}\label{claim:equivalent}
        $\mathcal{H'}$ is equivalent to $\mathcal{S}$.
    \end{claimcustom}

    \begin{proof}
        We must prove that $\subhistory{\mathcal{H'}}{p} = \subhistory{\mathcal{S}}{p}$ for each process $p$.
        By the definition of $\mathcal{H'}$, $\mathcal{S}$, and Claims \ref{claim:pairwise_distinct} and \ref{claim:linearization_points} (a), there is a one-to-one mapping between steps of $\subhistory{\mathcal{H'}}{p}$ and those of $\subhistory{\mathcal{S}}{p}$.
        Furthermore, by definition the operations of these operation executions are the same in $\subhistory{\mathcal{H'}}{p}$ and those of $\subhistory{\mathcal{S}}{p}$, and by \Cref{claim:correct_response} it follows that their responses are the same.
        Since operations for each process $p$ appear sequentially in $\mathcal{I}$ from which $\mathcal{H'}$ is derived, $\subhistory{\mathcal{H'}}{p}$ is a sequential history.
        Thus $<_{\subhistory{\mathcal{H'}}{p}}$ is a total order over all operations in $\subhistory{\mathcal{H'}}{p}$.
        Likewise, since $\subhistory{\mathcal{S}}{p}$ is a sequential history, $<_{\subhistory{\mathcal{S}}{p}}$ is a total order over all operations in $\subhistory{\mathcal{S}}{p}$.
        Since (1) there is a one-to-one mapping between steps of $\subhistory{\mathcal{H'}}{p}$ and those of $\subhistory{\mathcal{S}}{p}$ for each process $p$, (2) $<_{\subhistory{\mathcal{H'}}{p}}$ is a total order over all operations in $\subhistory{\mathcal{H'}}{p}$, (3) $<_{\subhistory{\mathcal{S}}{p}}$ is a total order over all operations in $\subhistory{\mathcal{S}}{p}$, and (4) $<_{\subhistory{\mathcal{H'}}{p}} \subseteq <_{\subhistory{\mathcal{S}}{p}}$ by \Cref{claim:respect_real_time_order}, $\subhistory{\mathcal{H'}}{p} = \subhistory{\mathcal{S}}{p}$.
        \qH{\Cref{claim:equivalent}}
    \end{proof}

    By \Cref{claim:equivalent} $\mathcal{H'}$ is equivalent to $\mathcal{S}$, by \Cref{corollary:correct_execution} $\mathcal{S}$ is legal with respect to $\mathcal{T}$, and by \Cref{claim:respect_real_time_order} $<_{\mathcal{H'}} \subseteq <_{\mathcal{S}}$.
    Therefore, $\mathcal{H}$ is linearizable with respect to $\mathcal{T}$.
    \qH{\Cref{theorem:linearizable}}
\end{proof}

\begin{theorem}[\Cref{thm:algo_1} restated]\label{thm:algo_1_restated}
    \Cref{alg:wait-free-simple} is a wait-free universal construction for the infinite-arrival model.
    Its space complexity at time $t$ is linear in the number of processes that have participated by time~$t$.
\end{theorem}

\begin{proof}
    By \Cref{thm:algo_1_time_complexity} and \Cref{theorem:linearizable}, \Cref{alg:wait-free-simple} is a wait-free universal construction for the infinite-arrival model.
    The space complexity immediately follows from the observation that every process performs at most one $\allocatecelloperation$ operation.
    \qH{\Cref{thm:algo_1_restated}}
\end{proof}

\section{Proof of \Cref{alg:efficient_algo}}
\label{sec:algo_2_proof}

In this section, we prove that \Cref{alg:efficient_algo} is linearizable, wait-free, and has space complexity linear in the point contention.
The high-level strategy is to prove that \Cref{alg:efficient_algo} satisfies these properties assuming that the memory manager does not reuse freed cells (we call this version algorithm $\mathcal{B}$).
We then show that violation of any of these properties by \Cref{alg:efficient_algo} when the memory manager can reuse freed cells (we call this version algorithm $\mathcal{A}$) would imply the violation of that property of $\mathcal{B}$, contradicting the first result.
This is done by establishing a ``correctness-preserving mapping'' from implementation histories of $\mathcal{A}$ to $\mathcal{B}$.
A high-level description of this mapping is given at the start of \Cref{sec:mapping_a_to_b}.
Also, for convenience, we treat the memory manager as a base object, and define $\mathcal{A}$ and $\mathcal{B}$ below.

\begin{definition}[$\mathcal{A}$]\label{def:reduction:algorithm_a}
    Algorithm $\mathcal{A}$ is \Cref{alg:efficient_algo} using the memory manager given in \Cref{alg:cell_manager_specification}.
    For a step $(C, p, C')$, there are two details not specified in the pseudocode:
    \begin{compactenum}
        \item If during this step $p$ performs an $\allocatecelloperation$ operation on the memory manager whose response is $ptr$, then the state assigned to each object of the cell pointed to by $ptr$ in $C'$ is the initial state specified in \Cref{alg:efficient_algo}.
    
        \item If during this step $p$ performs an operation $o$ on an object of a cell whose pointer is not in the state assigned to the memory manager in $C$, then the response of $o$ is arbitrary.
    \end{compactenum}
\end{definition}

\LinesNotNumbered
\setlength{\algomargin}{1.5em}
\begin{algorithm}[t]
\DontPrintSemicolon
\small

\nonl\Init(Let $\celluniverse{}$ be an infinite set of pointers to unique cells.){}

\BlankLine

\nonl\textbf{State:}

\nonl\Init($\cellset{}:$ A set of pointers to cells, initially $\emptyset$.){}

\BlankLine

\begin{multicols}{2}

\Init(\text{\allocatecelloperation{}()}){
    $\registerpointershort{} \coloneqq \text{pick a pointer from $\celluniverse{} \setminus \cellset{}$}$

    $\cellset{} \coloneqq \cellset{} \cup \{\registerpointershort{}\}$

    \textbf{return} $\registerpointershort{}$
}

\BlankLine

\Init(\text{\freecelloperation{}{}($\registerpointershort{}$)}){

    $\cellset{} \coloneqq \cellset{} \setminus \{\registerpointershort{}\}$

    \textbf{return} $\done{}$
}

\end{multicols}
\caption{\bf Memory manager operations.}
\label{alg:cell_manager_specification}
\end{algorithm}

\setlength{\algomargin}{1.5em}
\begin{algorithm}[t]
\DontPrintSemicolon
\small

\nonl\textbf{State:}

\nonl\Init($\cellset:$ A set of pointers, initially $\emptyset$.){}

% \nonl\Init($F:$ The set of freed pointers, initially $\emptyset$.){}

\BlankLine

\begin{multicols}{2}

\Init(\text{\allocatecelloperation{}()}){
    $\registerpointershort{} \coloneqq \text{pick a pointer from $\celluniverse{} \setminus \cellset$}$

    $\cellset \coloneqq \cellset \cup \{\registerpointershort{}\}$

    \textbf{return} $\registerpointershort{}$
}

\BlankLine

\Init(\text{\freecelloperation{}{}($\registerpointershort{}$)}){
    % $F \coloneqq F \cup \{\registerpointershort{}\}$

    \textbf{return} $\done{}$
}

\end{multicols}

\caption{\bf Lazy memory manager operations.}\label{alg:lazy_cell_manager_specification}
\end{algorithm}

\begin{definition}[$\mathcal{B}$]\label{def:reduction:algorithm_b}
    Algorithm $\mathcal{B}$ is \Cref{alg:efficient_algo} using the ``lazy" memory manager given in \Cref{alg:lazy_cell_manager_specification}.\footnote{This memory manager is lazy in the sense that $\freecelloperation$ operations do nothing.}
    In contrast to algorithm $\mathcal{A}$, $\allocatecelloperation$
    operations do not change the state of objects of a cell, and every operation on an object of a cell respects the semantics of its type.
\end{definition}

\textbf{Roadmap.}
\Cref{sec:basic_facts_about_b} proves some basic facts about $\mathcal{B}$, and states some key-invariants that we prove in \Cref{sec:l_invariants}.
We then prove that $\mathcal{B}$ is linearizable, wait-free, and has space complexity linear in the point contention in Appendices \ref{sec:b_is_linearizable}, \ref{sec:b_is_wait_free}, and \ref{sec:b_memory_management_and_space_efficiency}, respectively.
Lastly, we prove that $\mathcal{A}$ has all these properties in \Cref{sec:mapping_a_to_b}.
We note that Appendices \ref{sec:l_invariants}-\ref{sec:mapping_a_to_b} are logically independent, but all they all depend on the definitions and basic facts given in \Cref{sec:basic_facts_about_b}.

\textbf{Conventions.}
We use the symbol $\arbitraryvalue$ to mean any value, and $\unusedvalue$ is used in the code to mean that that field is not needed.
A prefix $\mathcal{I}'$ of an implementation history $\mathcal{I}$ during $[t_1, t_2]$ means that if $\mathcal{I} = (C_0, p_1, C_1), (C_1, p_2, C_2), \ldots$, then $\mathcal{I}' = (C_0, p_1, C_1), \ldots, (C_{i - 1}, p_i, C_i)$ for some $i \in [t_1, t_2]$.
% Likewise, a prefix $\mathcal{I}'$ of $\mathcal{I}$ up to and including (resp. but excluding) some step $s$ means if $\mathcal{I} = (C_0, p_1, C_1), (C_1, p_2, C_2), \ldots, (C_{i - 1}, p_i, C_i), s, \ldots$, then $\mathcal{I}' = (C_0, p_1, C_1), \ldots, (C_{i - 1}, p_i, C_i), s$ (resp. $\mathcal{I}' = (C_0, p_1, C_1), \ldots, (C_{i - 1}, p_i, C_i)$).
We will often talk about an operation that occurs during some step $s$, e.g., a \CASop{} operation on some base object, and denote it by $o$, and then later refer to some step $s$ that happened before (resp. after) $o$, and use the notation $s < o$ (resp. $o < s$) to mean that the step number of $s$ is smaller (resp. larger) than the step number of $o$ in the implementation history $\mathcal{I}$ they both occur in.
Throughout the entire proof, every number we refer to is an integer.
Lastly, to help keep track of the important statements, the headwords (i.e., Claim, Lemma, Proposition, Theorem) in this appendix are used as follows.
Theorems are only for major properties of an algorithm (i.e., linearizability, and there are only four theorems for $\mathcal{B}$ and one for $\mathcal{A}$), lemmas are properties referenced outside of the subsubsection they are stated in, propositions are properties not referenced outside of the subsubsection they are stated in, and claims are properties stated and used inside a proof.

\subsection{Basic Facts About $\mathcal{B}$}
\label{sec:basic_facts_about_b}

In this section prove some basic facts about $\mathcal{B}$.
Throughout this section, $\mathcal{I}^\mathcal{B}$ refers to an arbitrary implementation history of $\mathcal{B}$, i.e., all statements that refer to $\mathcal{I}^\mathcal{B}$ begin with ``for every implementation history $\mathcal{I}^\mathcal{B}$ of $\mathcal{B}$", which is omitted for brevity.

\subsubsection{Assumptions, definitions, and observations}

We first state all assumptions used throughout the proof.

\begin{assumption}\label{assumption:ero:bounded_concurrency}
    The system has bounded concurrency (see \Cref{def:bounded_concurrency}).
\end{assumption}

\begin{assumption}\label{assumption:ero:head_and_null_not_in_cell_universe}
    $\nullconstant \notin \celluniverse{}$ and $\&\headobject \notin \celluniverse{} \cup \{\nullconstant\}$.
\end{assumption}

\begin{assumption}\label{assumption:operations_and_responses}
    $\nullconstant$ differs from all possible responses to all operations of $\mathcal{T}$.
\end{assumption}

\begin{definition}
    The invocation and response steps for an operation execution are lines \ref{line:ero:invocation_step} and~\ref{line:ero:response_step}, respectively.
\end{definition}

The purpose of the long definition that follows is to provide hopefully meaningful and evocative terminology (rather than referring to line numbers) for the steps that affect the states of the base objects.
This makes the statements of the claims that follow more natural and their proofs easier to follow.
% We now define almost all of the terminology we will use throughout the proof for $\mathcal{B}$.\footnote{A small amount of terminology is introduced in Appendices \ref{sec:b_is_wait_free} and \ref{sec:b_memory_management_and_space_efficiency}.}

\begin{definition}\label{def:ero:english}
We define the following terminology for steps in $\mathcal{I}^\mathcal{B}$.
\begin{compactitem}

    \item An execution of a \GCASop{} (resp. \CASop{}) operation is \bemph{successful} if it returns $\true$ and \bemph{unsuccessful} if it returns $\false$.

    \item Consider a \Underline{successful} execution of the \GCASop{} operation in \cref{line:ero:announce_gcas} or a  \Underline{successful} execution of the \CASop{} operation in \cref{line:ero:announce_cas}.
    (These lines are the only places in the algorithm that modify the contents of~$\announceobject$, see \Cref{observation:ero:where_objects_change}.)
    Such an operation writes into $\announceobject$ a value of the form $((\timeshort{},\myrepositoryoperationshort), \cellpointershort)$.
    We refer to these operations as \bemph{$A$-events for timestamp $\timeshort{}$} or \bemph{$A$-events for $\cellpointershort$}.
    Specifically,
    \begin{compactitem}
        \item If $\myrepositoryoperationshort=\addcell{}$, we say that this is an \bemph{$A$-add event for timestamp $\timeshort{}$} or an \bemph{$A$-add event for $ptr$}.
        \item If $\myrepositoryoperationshort=\langle \doopandcopyresponse{}, \arbitraryvalue \rangle$, we say that this is an \bemph{$A$-apply event for timestamp $\timeshort{}$} or an \bemph{$A$-apply event for $ptr$}.
        \item If $\myrepositoryoperationshort=\removecell{}$, we say that this is an \bemph{$A$-remove event for timestamp $\timeshort{}$} or an \bemph{$A$-remove event for $ptr$}.
    \end{compactitem}
    (As we will see in \Cref{lemma:ero:every_a_event_is_for_timestamp_other_than_zero} $\timeshort{} > 0$, in \Cref{lemma:ero:every_a_event_is_for_pointer_from_universe} $\cellpointershort \in \celluniverse{}$, and in \Cref{lemma:ero:every_a_event_is_add_apply_or_remove} every $A$-event is either an $A$-add, $A$-apply, or $A$-remove event.)
        
    \item Consider a \Underline{successful} execution of the \CASop{} operation in \cref{line:ero:linearization_cas}.
    (This line is the only place in the algorithm that modifies the contents of~$\linearizationobject{}$, see \Cref{observation:ero:where_objects_change}.)
    Such an operation writes into $\linearizationobject{}$ a value of the form $((\timeshort{},\myrepositoryoperationshort),\cellpointershort)$.
    We refer to these operations as \bemph{$L$-events for timestamp $\timeshort{}$} or \bemph{$L$-events for $\cellpointershort$}.
    Specifically,
    \begin{compactitem}
        \item If $\myrepositoryoperationshort=\addcell{}$, we say that this is an \bemph{$L$-add event for timestamp $\timeshort{}$} or an \bemph{$L$-add event for $ptr$}.
        \item If $\myrepositoryoperationshort=\langle \doopandcopyresponse{}, \arbitraryvalue \rangle$, we say that this is an \bemph{$L$-apply event for timestamp $\timeshort{}$} or an \bemph{$L$-apply event for $ptr$}.
        \item If $\myrepositoryoperationshort=\removecell{}$, we say that this is an \bemph{$L$-remove event for timestamp $\timeshort{}$} or an \bemph{$L$-remove event for $ptr$}.
    \end{compactitem}
    (As we will see in \Cref{lemma:ero:every_l_event_is_for_timestamp_other_than_zero} $\timeshort{} > 0$, in \Cref{lemma:ero:every_l_event_is_for_pointer_from_universe} $\cellpointershort \in \celluniverse{}$, and in \Cref{lemma:ero:every_l_event_is_add_apply_or_remove} every $L$-event is either an $L$-add, $L$-apply, or $L$-remove event.)

    \item Consider an (unsuccessful or successful) execution of the \CASop{} operation in \cref{line:ero:state_cas}.
    Observe that this \CASop{} operation occurs during an invocation of the $\doapplyandcopyresponse{}$ procedure with parameters $((\timeshort, \arbitraryvalue),\arbitraryvalue)$.
    This \CASop{} operation attempts to write into $\stateobject$ a value of the form $((\timeshort, \arbitraryvalue), \arbitraryvalue, \arbitraryvalue)$.
    We refer to these operations as \textbf{$S$-attempts for $\timeshort$}.
    (We use the word ``attempt'' as opposed to ``event'', to highlight the fact that, in contrast to $A$- and $L$-events, which refer to successful \CASop{} operations, $S$-attempts are not necessarily successful. As we will see in \Cref{lemma:ero:every_s_attempt_is_for_timestamp_other_than_zero} $\timeshort{} > 0$.)
	
    \item Consider an (unsuccessful or successful) execution of the \CASop{} operation in \cref{line:ero:add_cell_to_list}.
    Observe that this \CASop{} operation occurs during an invocation of the $\doaddcell$ procedure with parameters $(\arbitraryvalue,\cellpointershort)$.
    This \CASop{} operation attempts to append to the list the cell pointed to by $\cellpointershort$ by changing the $\nextlong$ field of the cell pointed to by some pointer $\currentcellpointershort{}$ to $(\arbitraryvalue,\arbitraryvalue,\arbitraryvalue, \cellpointershort)$.
    We call the execution of such a \CASop{} operation a \bemph{list-add attempt for $ptr$ after $cur\_ptr$}.
    (As we will see in \Cref{lemma:ero:every_list_add_seal_and_remove_attempt_is_for_ptr_from_universe} $\cellpointershort \in \celluniverse{}$ and in \Cref{lemma:ero:every_list_add_attempt_is_after_a_pointer_from_universe_or_head} $\currentcellpointershort{} \in \celluniverse{} \cup \{\&\headobject\}$.)

    \item Consider an (unsuccessful or successful) execution of the \CASop{} operation in \cref{line:ero:seal_cell}.
    Observe that this \CASop{} operation occurs during an invocation of the $\doremovecell$ procedure with parameters $(\arbitraryvalue, \cellpointershort)$.
    This \CASop{} operation attempts to seal the cell pointed to by $\cellpointershort$ by changing the $\nextlong.sealed$ field of the cell pointed to by $\cellpointershort$ from $\false$ to $\true$.
    We call the execution of such a \CASop{} operation a \bemph{list-seal attempt for $\cellpointershort{}$}.
    (As we will see in \Cref{lemma:ero:every_list_add_seal_and_remove_attempt_is_for_ptr_from_universe} $\cellpointershort \in \celluniverse{}$.)
        
    \item Consider an (unsuccessful or successful) execution of the \CASop{} operation in \cref{line:ero:remove_cell_from_list}.
    Observe that this \CASop{} operation occurs during an invocation of the $\doremovecell$ procedure with parameters $(\arbitraryvalue, \cellpointershort)$.
    This \CASop{} operation attempts to remove from the list the cell pointed to by $\cellpointershort$ by changing the $\nextlong$ field of the cell pointed to by $\previouscellpointershort{}$ to $(\arbitraryvalue,\arbitraryvalue,\arbitraryvalue,\nextcellpointershort{})$.
    We call the execution of such a \CASop{} operation a \bemph{list-remove attempt for $\cellpointershort{}$ between $\previouscellpointershort{}$ and $\nextcellpointershort{}$}.
    (As we will see in \Cref{lemma:ero:every_list_add_seal_and_remove_attempt_is_for_ptr_from_universe} $\cellpointershort \in \celluniverse{}$ and in \Cref{lemma:ero:every_list_remove_attempt_is_between_a_pointer_from_universe_or_header_and_a_pointer_from_universe_or_null} $\previouscellpointershort{} \in \celluniverse{} \cup \{\&\headobject\}$ and $\nextcellpointershort{} \in \celluniverse{} \cup \{\nullconstant\}$.)

    \item Consider an execution of the write operation in \cref{line:ero:do_work_initialize_response} for some $\cellpointershort$.
    Observe that this write operation occurs during an invocation of the $\doworkuntildone{}$ procedure with parameters $(\myrepositoryoperationshort, \cellpointershort)$.
    This write operation sets the value of $(*\cellpointershort).\lastrepositoryoperationresponse{} = ((\arbitraryvalue{}, \myrepositoryoperationshort), \nullconstant{})$.
    We refer to these operations as \bemph{response-reset events for $\cellpointershort$}.
    Specifically,
    \begin{compactitem}
        \item If $\myrepositoryoperationshort = \addcell{}$, we say that this is an \textbf{add-response-reset event for $\cellpointershort$}.
        \item If $\myrepositoryoperationshort = \langle \doopandcopyresponse{}, \arbitraryvalue \rangle$, we say that this is an \textbf{apply-response-reset event for $\cellpointershort$}.
        \item If $\myrepositoryoperationshort = \removecell{}$, we say that this is a \textbf{remove-response-reset event for $\cellpointershort$}.
    \end{compactitem}
    (We will see in \Cref{lemma:ero:every_response_reset_is_add_apply_or_remove} $\cellpointershort \in \celluniverse{}$ and every response-reset event is either an add-response-reset, apply-response-reset, or remove-response-reset event.)
    
    \item Consider an execution of the \CASop{} operation in \cref{line:ero:responses_set_attempt} for some $\cellpointershort$.
    Observe that this \CASop{} operation occurs during an invocation of the $\setrepositoryoperationresponse{}$ procedure with parameters $((\arbitraryvalue{}, \myrepositoryoperationshort), \cellpointershort, \result{})$.
    This \CASop{} operation attempts to set the value of $(*\cellpointershort).\lastrepositoryoperationresponse{} = ((\arbitraryvalue{}, \myrepositoryoperationshort), \result{})$.
    We refer to these operations as \bemph{response-set attempts for $\cellpointershort$ to $\result{}$}.
    Specifically,
    \begin{compactitem}
        \item If $\myrepositoryoperationshort = \addcell{}$, we say that this is an \textbf{add-response-set attempt for $\cellpointershort$ to $\result{}$}.
        \item If $\myrepositoryoperationshort = \langle \doopandcopyresponse{}, \arbitraryvalue \rangle$, we say that this is an \textbf{apply-response-set attempt for $\cellpointershort$ to $\result{}$}.
        \item If $\myrepositoryoperationshort = \removecell{}$, we say that this is a \textbf{remove-response-set attempt for $\cellpointershort$ to $\result{}$}.
    \end{compactitem}
    (We will see in \Cref{lemma:ero:every_response_set_attempt_is_for_pointer_from_universe} $\cellpointershort \in \celluniverse{}$ and every response-set attempt is either an add-response-set, apply-response-set, or remove-response-set attempt.)
    
    \item Consider an (unsuccessful or successful) execution of the \CASop{} operation in \cref{line:ero:acquire_next_cell}.
    Observe that this \CASop{} operation occurs during an invocation of the AcquireNext procedure with parameters $(\arbitraryvalue, \cellpointershort)$.
    This \CASop{} operation attempts to acquire the cell after the cell pointed to by $\cellpointershort$ by changing the $\nextlong$ field of the cell pointed to by $\cellpointershort$ from $(\arbitraryvalue, \arbitraryvalue, a, \nextcellpointershort{})$ to $(\arbitraryvalue, \arbitraryvalue, a + 1, \nextcellpointershort{})$.
    We call the execution of such a \CASop{} operation a \bemph{list-acquire-next attempt for $\nextcellpointershort{}$ after $\cellpointershort{}$}.(As we will see in \Cref{lemma:ero:list_acquire_next_attempt_for_pointer_from_universe_and_after_pointer_from_universe_or_head} $\nextcellpointershort{} \in \celluniverse{}$ and $\cellpointershort \in \celluniverse{} \cup \{\&\headobject\}$)

    \item Consider an execution of the \FAop{} operation in \cref{line:ero:copy_acquisitions_to_revocations}.
    Observe that this \FAop{} operation occurs during an invocation of the $\doremovecell$ procedure with parameters $(\arbitraryvalue,\cellpointershort)$.
    This \FAop{} operation copies the final number of acquires for $\cellpointershort$ into the cell that $\cellpointershort$ points to after $\cellpointershort$ was removed from the list.
    We call the execution of such a \FAop{} operation an \bemph{acquire-copy event for $\cellpointershort$}. (As we will see in \Cref{lemma:ero:every_acquire_copy_event_is_for_pointer_from_universe} $\cellpointershort \in \celluniverse{}$)

    \item Consider an execution of the \FAop{} operation in \cref{line:ero:relinquish_revocations}.
    Observe that this \FAop{} operation occurs during an invocation of the Relinquish procedure with parameters $(\arbitraryvalue,\cellpointershort)$.
    This \FAop{} operation increments the number of revocations of $\cellpointershort$ followed by freeing $\cellpointershort$ if it is no longer in use.
    We call the execution of such a \FAop{} operation a \bemph{revocation event for $\cellpointershort$}. (As we will see in \Cref{lemma:ero:revocation_event_is_for_pointer_from_universe} $\cellpointershort \in \celluniverse{}$)
\end{compactitem}

\end{definition}

The benefit of this terminology is that it captures with evocative words (rather than line numbers) the steps that change the values of the base objects.
We summarize this in the following observation.
In this observation, we use the terminology ``set'' to mean a step that sets the state of an object or one of its fields, and ``change'' to mean that a step can actually change the object or one of its fields.
This distinction is important because some steps set the state of an object but do not change some of its fields.

\begin{observation}\label{observation:ero:where_objects_change}
    The following are true for steps in $\mathcal{I}^\mathcal{B}$:
    
    \begin{compactitem}
        \item The only steps that \Underline{set} the value of $\announceobject$ are $A$-events.

        \item The only steps that \Underline{set} the value of $\linearizationobject{}$ are $L$-events.

        \item The only steps that \Underline{set} the value of $\stateobject$ are successful $S$-attempts.

        \item For every $\cellpointershort \in \celluniverse{} \cup \{\&\headobject\}$ the only steps that \Underline{set} the value of $(*\cellpointershort).\lastrepositoryoperationresponse{}$ are response-reset events for $\cellpointershort$ and successful response-set attempts for $\cellpointershort$.
        
        \item For every $\cellpointershort \in \celluniverse{} \cup \{\&\headobject\}$ the only steps that \Underline{set} the value of $(*\cellpointershort).\revocations$ are acquire-copy events for $\cellpointershort$ and revocation events for $\cellpointershort$.
        
        \item For every $\cellpointershort \in \celluniverse{} \cup \{\&\headobject\}$ the only steps that \Underline{set} the value of $(*\cellpointershort).\nextlong$ are successful list-add attempts after $\cellpointershort$, successful list-seal attempts for $\cellpointershort$, successful list-remove attempts between $\cellpointershort$ and some pointer, and successful list-acquire-next attempts after $\cellpointershort$.
        
        \item For every $\cellpointershort \in \celluniverse{} \cup \{\&\headobject\}$ the only steps that \Underline{change} the value of $(*\cellpointershort).\nextlong.\acquisitions$ are successful list-add attempts after $\cellpointershort$, successful list-remove attempts between $\cellpointershort$ and some pointer, and successful list-acquire-next-attempts after $\cellpointershort$.
        
        \item For every $\cellpointershort \in \celluniverse{} \cup \{\&\headobject\}$ the only step that \Underline{change} the value of $(*\cellpointershort).\nextlong.sealed$ are successful list-sealed attempts for $\cellpointershort$.
        
        \item For every $\cellpointershort \in \celluniverse{} \cup \{\&\headobject\}$ the only steps that \Underline{change} the value of $(*\cellpointershort).\nextlong.\uniquecellpointercontentlong{}$ are successful list-add attempts after $\cellpointershort$ and successful list-remove attempts between $\cellpointershort$ and some pointer.
    \end{compactitem}
\end{observation}

In addition to the above terminology, a central concept throughout the proof is the ``shape" of the list.
As we will see, the shape of the list at some time is determined by the sequence of $L$-events up to and including that time.
We formalize below what the ``shape" of the list should be. 

\begin{definition}\label{def:ero:logical_list}
Consider any \Underline{finite} implementation history $\mathcal{I}$ of $\mathcal{B}$.
\begin{compactitem}
    \item Let $e_1,e_2,\ldots,e_n$ be the (possibly empty) subsequence of $\mathcal{I}$ consisting of the $L$-add events for some pointer for which there are no subsequent $L$-remove events for \Underline{that} pointer in $\mathcal{I}$.
    Let $e_i$ be an $L$-add event for pointer $\cellpointershort_i$.
    We let $\List(\mathcal{I}) = \&\headobject,\cellpointershort_1,\cellpointershort_2,\ldots,\cellpointershort_n,\nullconstant$.

    \item We say that the list of cells \bemph{conforms to} $\cellpointershort_0,\cellpointershort_1,\ldots,\cellpointershort_n,\cellpointershort_{n+1}$ \textbf{in $\mathcal{I}$} if, at the end of $\mathcal{I}$, for all $i\in[0..n]$, $(*\cellpointershort_i).\nextlong.\uniquecellpointercontentlong{}= \cellpointershort_{i+1}$ (assuming $\cellpointershort_i \in \celluniverse{} \cup \{\&\headobject\}$).
\end{compactitem}
\end{definition}

As we will see in \Cref{lemma:ero:conditional_classification_lemma}, $\List(\mathcal{I})$ is essentially the ``shape" of the list at the end of $\mathcal{I}$, or more precisely, the list of cells conforms to $\List(\mathcal{I})$ in $\mathcal{I}$.
We note that we use the word ``essentially" here to disregard the fact that there is some lag between the moment an $L$-add (resp. $L$-remove) event for $\cellpointershort$ occurs and the time $\cellpointershort$ is added (resp. removed) from the list.
See \Cref{lemma:ero:conditional_classification_lemma} for a precise description of the conditions when this lag does and does not occur.

The high-level strategy for proving this fact, and many useful and natural facts about the algorithm, are stated and proved conditionally on the invariants holding.
The reason for this is that these statements are needed to \emph{prove the invariants themselves}.
This is not circular because we prove the invariants by induction: we
consider any implementation history $\mathcal{I}$ of $n + 1$ steps and assume that the
invariants hold for the prefix $\mathcal{I}'$ of it up to and including the $n$th step and
show that the invariants hold for $\mathcal{I}$. 
In our proof that the invariants hold for $\mathcal{I}$, we invoke the conditional facts for $\mathcal{I}'$, whose assumptions hold because the invariants hold for $\mathcal{I}'$ by the inductive hypothesis.

\begin{definition}
We define the following four invariants.

\medskip\noindent
{\normalfont\textbf{Invariant $P(\mathcal{I}^\mathcal{B})$:}}\/
For each $\cellpointershort$, there is at most one $L$-add event for $\cellpointershort$,
	at most one $L$-apply event for $\cellpointershort$, and
	at most one $L$-remove event for $\cellpointershort$ in $\mathcal{I}^\mathcal{B}$.

\medskip\noindent
{\normalfont\textbf{Invariant $Q(\mathcal{I}^\mathcal{B})$:}}\/
All of the following are true:

\begin{compactenum}
    \item Every list-add attempt for some $\cellpointershort$ after some $\currentcellpointershort{}$ in $\mathcal{I}^\mathcal{B}$ is preceded by a unique $L$-add event for $\cellpointershort$; furthermore, if $\mathcal{I}$ is the prefix of $\mathcal{I}^\mathcal{B}$ up to but excluding that $L$-add event, $\currentcellpointershort{}$ is the second last pointer in $\List(\mathcal{I})$ --- i.e., the one preceding $\nullconstant$.

    \item Every list-remove attempt for some $\cellpointershort$ between some $\previouscellpointershort{}$ and some $\nextuniquecellpointershort{}$ in $\mathcal{I}^\mathcal{B}$ is preceded by a unique $L$-remove event for $\cellpointershort$; furthermore, if $\mathcal{I}$ is the prefix of $\mathcal{I}^\mathcal{B}$ up to but excluding that $L$-remove event, $\cellpointershort$ is in $\List(\mathcal{I})$ exactly once and $\previouscellpointershort{}$ and $\nextcellpointershort{}$ are the pointers preceding and succeeding $\cellpointershort$ in $\List(\mathcal{I})$.
\end{compactenum}

\smallskip\noindent
{\normalfont\textbf{Invariant $R(\mathcal{I}^\mathcal{B})$:}}\/
All of the following are true for any two successive $L$-events $e$ and $e'$ in $\mathcal{I}^\mathcal{B}$:

\begin{compactenum}
    \item If $e$ is an $L$-add event for $\cellpointershort$, then the interval between $e$ and $e'$ contains one \Underline{successful} list-add attempt for $\cellpointershort$ and no other \Underline{successful} list-add or list-remove attempt for any pointer.
    
    \item If $e$ is an $L$-apply event, then the interval between $e$ and $e'$ contains no \Underline{successful} list-add or list-remove attempts for any pointer.   
    
    \item If $e$ is an $L$-remove event for $\cellpointershort$, the interval between $e$ and $e'$ contains one \Underline{successful} list-remove attempt for $\cellpointershort$ and no other \Underline{successful} list-remove or list-add attempt for any pointer.
\end{compactenum}

\medskip\noindent
{\normalfont\textbf{Invariant $O(\mathcal{I}^\mathcal{B})$:}}\/
All of the following are true for any two successive $L$-events $e$ and $e'$ in $\mathcal{I}^\mathcal{B}$:
\begin{compactitem}
    \item If $e$ is an $L$-add or $L$-remove event, then the interval between $e$ and $e'$ contains no successful $S$-attempts.
    \item If $e$ is an $L$-apply event for a timestamp $\timeshort{}$, then between $e$ and $e'$ there is one successful $S$-attempt for $\timeshort{}$ and no other successful $S$-attempts for any timestamp.
\end{compactitem}

\end{definition}

We now record some basic observations that are useful throughout the proof.

\begin{observation}\label{observation:ero:invariants_hold_for_prefixes}
    For every prefix $\mathcal{I}$ of $\mathcal{I}^\mathcal{B}$ if $P(\mathcal{I}^\mathcal{B})$, $Q(\mathcal{I}^\mathcal{B})$, $R(\mathcal{I}^\mathcal{B})$, or $O(\mathcal{I}^\mathcal{B})$ holds, then $P(\mathcal{I})$, $Q(\mathcal{I})$, $R(\mathcal{I})$, or $O(\mathcal{I})$ holds, respectively.
\end{observation}

Because of this observation, for brevity, when we are satisfying the conditions of a lemma regarding some prefix $\mathcal{I}$ of $\mathcal{I}^\mathcal{B}$, we will satisfy the condition of the lemma that $X(\mathcal{I})$ holds, where $X$ is one of the invariants, by stating that $X(\mathcal{I}^\mathcal{B})$ holds.

\begin{observation}\label{observation:ero:views_are_monotonic}
    For every $\cellpointershort \in \celluniverse{} \cup \{\&\headobject\}$, $(*\cellpointershort).\nextlong.\view$ is monotonically increasing and is greater than or equal to $0$.
\end{observation}

\begin{observation}\label{observation:ero:operation_timestamp_is_unique}
    Every timestamp returned on \cref{line:ero:operation_timestamp} is unique and is an integer larger than~0.
\end{observation}

\begin{observation}\label{observation:ero:response_set_attempt_invocation}
    Consider any response-set attempt $a$ during an invocation $I$ of the \setrepositoryoperationresponse{} procedure.
    The following are true.
    \begin{compactenum}
        \item $a$ is an add-response-set attempt if and only if $I$ was invoked during an invocation of the \doaddcell{} procedure.
        \item $a$ is a remove-response-set attempt if and only if $I$ was invoked during an invocation of the \doremovecell{} procedure.
        \item $a$ is an apply-response-set attempt if and only if $I$ was invoked during an invocation of the \doapplyandcopyresponse{} procedure.
    \end{compactenum}
\end{observation}

\subsubsection{$A$-events, $L$-events, $S$-attempts, and list-attempts}

We start with some facts about $A$-events.

\begin{proposition}\label{lemma:ero:every_a_event_is_for_timestamp_other_than_zero}
    Every $A$-event in $\mathcal{I}^\mathcal{B}$ is for a timestamp larger than $0$.
\end{proposition}

\begin{proof}
    Consider any $A$-event $e$ for some timestamp $\timeshort{}$ executed by some process $p$.
    Hence, by \Cref{def:ero:english}, $e$ set $\announceobject$ to $((\timeshort{}, \arbitraryvalue), \arbitraryvalue)$, so by \Cref{def:ero:english} $\timeshort{}$ is the response $p$ received on \cref{line:ero:operation_timestamp} during the invocation of the \doworkuntildone{} procedure that $p$ executed $e$ during.
    Therefore, by \Cref{observation:ero:operation_timestamp_is_unique} $\timeshort{} > 0$, and so $e$ is for a timestamp larger than $0$ as wanted.
    \qH{\Cref{lemma:ero:every_a_event_is_for_timestamp_other_than_zero}}
\end{proof}

\begin{lemma}\label{lemma:ero:every_a_event_is_for_pointer_from_universe}
    Every $A$-event in $\mathcal{I}^\mathcal{B}$ is for some pointer in $\celluniverse{}$.
\end{lemma}

\begin{proof}
    Consider any $A$-event $e$ for some $\cellpointershort$ executed by some process $p$.
    Hence, by \Cref{def:ero:english}, $p$ executed $e$ during some invocation $I$ of the \doworkuntildone{} procedure with a second parameter of $\cellpointershort$.
    Thus, since this procedure is only invoked on lines \ref{line:ero:low_level_add_cell}, \ref{line:ero:low_level_apply_and_copy_response}, and \ref{line:ero:low_level_remove_cell}, $p$ performed an $\allocatecelloperation$ whose response is $\cellpointershort$ before invoking $I$.
    Therefore, by \Cref{alg:lazy_cell_manager_specification}, $\cellpointershort \in \celluniverse{}$.
    \qH{\Cref{lemma:ero:every_a_event_is_for_pointer_from_universe}}
\end{proof}

\begin{lemma}\label{lemma:ero:every_a_event_is_add_apply_or_remove}
    Every $A$-event in $\mathcal{I}^\mathcal{B}$ is either an $A$-add, $A$-apply, or $A$-remove event.
\end{lemma}

\begin{proof}
    Let $p$ be a process that executed an $A$-event $e$ which sets $\announceobject.\uniquerepositoryoperationshort = (\arbitraryvalue, \myrepositoryoperationshort)$.
    By \Cref{def:ero:english}, $e$ is an execution of \cref{line:ero:announce_gcas} or \cref{line:ero:announce_cas} and so $p$ executed $e$ during some invocation $I$ of the \doworkuntildone{} procedure.
    Hence, since $e$ set $\announceobject.\uniquerepositoryoperationshort = (\arbitraryvalue, \myrepositoryoperationshort)$, it follows that the first parameter of $I$ is $\myrepositoryoperationshort$.
    Thus, since the \doworkuntildone{} procedure is only invoked on lines \ref{line:ero:low_level_add_cell}, \ref{line:ero:low_level_apply_and_copy_response}, and \ref{line:ero:low_level_remove_cell}, we have that $\myrepositoryoperationshort$ is either $\addcell$, $\langle \doopandcopyresponse{}, \arbitraryvalue \rangle$, or $\removecell$.
    Therefore, by \Cref{def:ero:english}, $e$ is either an $A$-add, $A$-apply, or $A$-remove event as wanted.
    \qH{\Cref{lemma:ero:every_a_event_is_add_apply_or_remove}}
\end{proof}

\begin{proposition}\label{lemma:ero:matching_timestamp_in_a_implies_matching_value_in_a}
    If the left field of $\announceobject{}.\uniquerepositoryoperationlong{}$ is the same at times $T$ and $T'$ in $\mathcal{I}^\mathcal{B}$, then the value of $\announceobject{}$ is the same at $T$ and $T'$.
\end{proposition}

\begin{proof}
    Suppose, for contradiction, that the left field of $\announceobject{}.\uniquerepositoryoperationlong{}$ is the same at $T$ and $T'$ and the value of $\announceobject{}$ is different at $T$ and $T'$.
    Suppose $\announceobject{} = ((\timeshort{}, \myrepositoryoperationshort), \cellpointershort)$ at $T$
    and suppose $\announceobject{} = ((\timeshort{}, \myrepositoryoperationshort'), \cellpointershort')$ at $T'$ such that $(\myrepositoryoperationshort, \cellpointershort) \neq (\myrepositoryoperationshort', \cellpointershort')$.
    Without loss of generality, suppose $T < T'$.
    Hence, the value of $\announceobject{}$ was set to $((\timeshort{}, \myrepositoryoperationshort'), \cellpointershort')$.
    Thus, by \Cref{observation:ero:where_objects_change}, an $A$-event $e'$ set $\announceobject{}.\uniquerepositoryoperationlong{}$ to $((\timeshort{}, \myrepositoryoperationshort'), \cellpointershort')$.
    So, by \Cref{lemma:ero:every_a_event_is_for_timestamp_other_than_zero} $\timeshort{} \neq 0$.
    Hence, since $\announceobject{}$ is initially $((0, \noop), \nullconstant)$, the value of $\announceobject{}$ at $T$ is not the initial value, and so $\announceobject{}$ was set to $((\timeshort{}, \myrepositoryoperationshort), \cellpointershort)$.
    Thus, by \Cref{observation:ero:where_objects_change}, an $A$-event $e$ set $\announceobject{}.\uniquerepositoryoperationlong{}$ to $((\timeshort{}, \myrepositoryoperationshort), \cellpointershort)$.
    Let $p$ (resp. $p'$) be the process that executed $e$ (resp. $e'$).
    Since $e$ and $e'$ both set $\announceobject.\uniquerepositoryoperationlong{} = (\timeshort{}, \arbitraryvalue)$, both $p$ and $p'$ received $\timeshort{}$ on \cref{line:ero:operation_timestamp} during some invocation $I$ (resp. $I'$) of the \doworkuntildone{} procedure.
    Hence, since by \Cref{observation:ero:operation_timestamp_is_unique} every response on \cref{line:ero:operation_timestamp} is unique, we have that $p = p'$ and $I = I'$.
    Since $p$ (resp. $p'$) set $\announceobject$ to $((\timeshort{}, \myrepositoryoperationshort), \cellpointershort)$ (resp. $((\timeshort{}, \myrepositoryoperationshort'), \cellpointershort')$), we have that the parameters of $I$ (resp. $I'$) are $(\myrepositoryoperationshort, \cellpointershort)$ (resp. $(\myrepositoryoperationshort', \cellpointershort')$).
    Therefore, since $I = I'$, we have that $(\myrepositoryoperationshort, \cellpointershort) = (\myrepositoryoperationshort', \cellpointershort')$.
    However, $(\myrepositoryoperationshort, \cellpointershort) \neq (\myrepositoryoperationshort', \cellpointershort')$, a contradiction.
    \qH{\Cref{lemma:ero:matching_timestamp_in_a_implies_matching_value_in_a}}
\end{proof}

\begin{lemma}\label{lemma:ero:try_to_set_a_to_same_timestamp_by_same_process}
    Let $e_1$ and $e_2$ be two executions of either \cref{line:ero:announce_gcas} or \ref{line:ero:announce_cas} that try to set $\announceobject = ((\timeshort{}, \arbitraryvalue), \arbitraryvalue)$ in $\mathcal{I}^\mathcal{B}$.
    Then, $e_1$ and $e_2$ are executed by the same process $p$ during the same invocation $I$ of the \doworkuntildone{} procedure such that $p$ received $\timeshort$ as a response on \cref{line:ero:operation_timestamp} during $I$.
\end{lemma}

\begin{proof}
    Suppose $e_1$ and $e_2$ are performed by processes $p_1$ and $p_2$ during invocations $I_1$ and $I_2$ of the \doworkuntildone{} procedure, respectively.
    Since $p_1$ and $p_2$ both try to set $\announceobject = ((\timeshort, \arbitraryvalue), \arbitraryvalue)$ during $I_1$ and $I_2$, we have that $p_1$ and $p_2$ both received $\timeshort$ as a response on \cref{line:ero:operation_timestamp} during $I_1$ and $I_2$, respectively.
    Therefore, since by \Cref{observation:ero:operation_timestamp_is_unique} the responses on \cref{line:ero:operation_timestamp} are unique, we have that $p_1 = p_2 = p$, $I_1 = I_2 = I$, and $p$ received $\timeshort$ as a response on \cref{line:ero:operation_timestamp} during $I$ as wanted.
    \qH{\Cref{lemma:ero:try_to_set_a_to_same_timestamp_by_same_process}}
\end{proof}

\Cref{lemma:ero:try_to_set_a_to_same_timestamp_by_same_process} implies the following.

\begin{corollary}\label{lemma:ero:try_to_set_a_to_same_value_by_same_process}
    Let $e_1$ and $e_2$ be two executions of either \cref{line:ero:announce_gcas} or \ref{line:ero:announce_cas} that try to set $\announceobject = v$ in $\mathcal{I}^\mathcal{B}$.
    Then, $e_1$ and $e_2$ are executed by the same process during the same invocation of the \doworkuntildone{} procedure.
\end{corollary}

\begin{proposition}\label{lemma:ero:a_events_for_same_pointer_are_by_same_process_in_same_opx}
    Consider any $A$-events $e_1$ and $e_2$ for $\cellpointershort{}$ in $\mathcal{I}^\mathcal{B}$.
    Let $I_1$ and $I_2$ be the invocations of the \doworkuntildone{} procedure that $e_1$ and $e_2$ were executed in, respectively.
    Then, $I_1$ and $I_2$ were invoked by the same processes and invoked during the same invocation of the \highleveloperation{} procedure.
\end{proposition}

\begin{proof}
    Let $p_1$ and $p_2$ be the processes that executed $e_1$ and $e_2$, respectively.
    Hence, since $e_1$ and $e_2$ are $A$-events for $\cellpointershort{}$ during $I_1$ and $I_2$, respectively, by \Cref{def:ero:english}, the second parameter of $I_1$ and $I_2$ is $\cellpointershort{}$.
    Thus, $p_1$ and $p_2$ received $\cellpointershort{}$ as response on \cref{line:ero:allocate_cell}.
    Therefore, since by \Cref{alg:lazy_cell_manager_specification} the responses on \cref{line:ero:allocate_cell} are unique in $\mathcal{B}$, we have that $p_1 = p_2$ and $I_1$ and $I_2$ were invoked during the same invocation of the \highleveloperation{} procedure.
    \qH{\Cref{lemma:ero:a_events_for_same_pointer_are_by_same_process_in_same_opx}}
\end{proof}

\begin{lemma}\label{lemma:ero:no_apply_announce_after_remove_announce}
    Consider any $A$-remove event $e$ for $\cellpointershort$ in $\mathcal{I}^\mathcal{B}$.
    Then, there are no $A$-apply events for $\cellpointershort$ from $e$ onwards in $\mathcal{I}^\mathcal{B}$.
\end{lemma}

\begin{proof}
    Suppose, for contradiction, there is an $A$-remove event $e$ for $\cellpointershort$ in $\mathcal{I}^\mathcal{B}$ and there is an $A$-apply event $e'$ for $\cellpointershort$ after $e$ in $\mathcal{I}^\mathcal{B}$.
    Hence, by \Cref{lemma:ero:a_events_for_same_pointer_are_by_same_process_in_same_opx}, the same process $p$ executed $e$ and $e'$ and did so during the same invocation $I$ of the \highleveloperation{} procedure.
    By \Cref{def:ero:english}, $p$ executed $e$ during an invocation of the \doworkuntildone{} procedure invoked on \cref{line:ero:low_level_remove_cell} during $I$, and $p$ executed $e'$ during an invocation of the \doworkuntildone{} procedure invoked on \cref{line:ero:low_level_apply_and_copy_response} during $I$.
    Therefore, $e' < e$.
    However, by assumption $e < e'$, a contradiction.
    \qH{\Cref{lemma:ero:no_apply_announce_after_remove_announce}}
\end{proof}

\begin{lemma}\label{lemma:ero:no_add_announce_after_remove_announce}
    Consider any $A$-remove event $e$ for $\cellpointershort$ in $\mathcal{I}^\mathcal{B}$.
    Then, there are no $A$-add events for $\cellpointershort$ from $e$ onwards in $\mathcal{I}^\mathcal{B}$.
\end{lemma}

\begin{proof}
    By essentially the same argument as \Cref{lemma:ero:no_apply_announce_after_remove_announce}, which is provided below for completeness.
    Suppose, for contradiction, there is an $A$-remove event $e$ for $\cellpointershort$ in $\mathcal{I}^\mathcal{B}$ and there is an $A$-add event $e'$ for $\cellpointershort$ after $e$ in $\mathcal{I}^\mathcal{B}$.
    Hence, by \Cref{lemma:ero:a_events_for_same_pointer_are_by_same_process_in_same_opx}, the same process $p$ executed $e$ and $e'$ and did so during the same invocation $I$ of the \highleveloperation{} procedure.
    By \Cref{def:ero:english}, $p$ executed $e$ during an invocation of the \doworkuntildone{} procedure invoked on \cref{line:ero:low_level_remove_cell} during $I$, and $p$ executed $e'$ during an invocation of the \doworkuntildone{} procedure invoked on \cref{line:ero:low_level_add_cell} during $I$.
    Therefore, $e' < e$.
    However, by assumption $e < e'$, a contradiction.
    \qH{\Cref{lemma:ero:no_add_announce_after_remove_announce}}
\end{proof}

\begin{proposition}\label{lemma:ero:a_events_of_same_type_for_same_pointer_set_same_value}
    Consider any $A$-$X$ events $e_1$ and $e_2$ for $\cellpointershort{}$ where $X \in \{\text{add}, \text{apply}, \text{remove}\}$ in $\mathcal{I}^\mathcal{B}$.
    Then, $e_1$ and $e_2$ set $\announceobject$ to the same value.
\end{proposition}

\begin{proof}
    Suppose $e_1$ set $\announceobject = v_1$ and $e_2$ set $\announceobject = v_2$.
    Since $e_1$ and $e_2$ are $A$-$X$ events for $\cellpointershort{}$, by \Cref{def:ero:english}, $v_1 = ((\timeshort_1, \repositoryoperationshort), \cellpointershort{})$ and $v_2 = ((\timeshort_2, \repositoryoperationshort), \cellpointershort{})$.
    Let $p_1$ and $p_2$ be the processes that executed $e_1$ and $e_2$, respectively, and let $I_1$ and $I_2$ be the invocation of the \doworkuntildone{} procedure that $p_1$ and $p_2$ executed $e_1$ and $p_2$ during, respectively.
    Since $e_1$ set $\announceobject = ((\timeshort_1, \repositoryoperationshort), \cellpointershort{})$ and $e_2$ set $\announceobject{} = ((\timeshort_2, \repositoryoperationshort), \cellpointershort{})$, we have that the parameters of $I_1$ and $I_2$ are $(\repositoryoperationshort, \cellpointershort{})$ and $(\repositoryoperationshort, \cellpointershort{})$, respectively, and $p_1$ and $p_2$ received $\timeshort_1$ and $\timeshort_2$ as a response on \cref{line:ero:operation_timestamp} during $I_1$ and $I_2$, respectively.
    Hence, since by \Cref{lemma:ero:a_events_for_same_pointer_are_by_same_process_in_same_opx} $p_1 = p_2$ and $I_1$ and $I_2$ were invoked during the same invocation $I$ of the \highleveloperation{} procedure, and the \doworkuntildone{} procedure is only invoked with the same first parameter on a single line during $I$, we have that $I_1 = I_2$.
    Thus, since $p_1$ and $p_2$ received $\timeshort_1$ and $\timeshort_2$ as a response on \cref{line:ero:operation_timestamp} during $I_1$ and $I_2$, respectively, we have that $\timeshort_1 = \timeshort_2$.
    Therefore, it follows that $v_1 = v_2$ as wanted.
    \qH{\Cref{lemma:ero:a_events_of_same_type_for_same_pointer_set_same_value}}
\end{proof}

We now prove some facts about $L$-events.

\begin{lemma}\label{lemma:ero:l_events_have_corresponding_a_events}
    Consider an $L$-event $e$ which set $\linearizationobject{}$ to $v$ in $\mathcal{I}^\mathcal{B}$.
    Then, there is an $A$-event which set $\announceobject{}$ to $v$ before $e$.
    We call the last $A$-event which set $\announceobject$ to $v$ before $e$, $e$'s corresponding $A$-event.
\end{lemma}

\begin{proof}
    Consider any $L$-event $e$ which set $\linearizationobject{}$ to some value $v$.
    Let $p$ be the process that executed $e$.
    Since $e$ set $\linearizationobject{}$ to $v$, $p$ read $v$ from $\announceobject$ on its last execution of \cref{line:ero:announce_read} before $e$; say at time $T^{\ref{line:ero:announce_read}}$.
    To prove that there is an $A$-event which sets $\announceobject{}$ to $v$ before $e$, it suffices to prove that $v$ does not equal the initial value of $\announceobject$.
    Suppose, for contradiction, $v$ is the initial value of $\announceobject$, i.e., $v = ((0, \noop), \nullconstant)$.
    Since $p$ executed $e$ after reading $v$ from $\announceobject$ at $T^{\ref{line:ero:announce_read}}$, $p$ found the condition \cref{line:ero:done_check} to be true between $T^{\ref{line:ero:announce_read}}$ and $e$; say at time $T^{\ref{line:ero:done_check}}$.
    Hence, $\status = \notdone$ at $T^{\ref{line:ero:done_check}}$.
    Let $I$ be $p$'s invocation of the IsDone procedure on \cref{line:ero:check_if_announce_is_done} between $T^{\ref{line:ero:announce_read}}$ and $T^{\ref{line:ero:done_check}}$.
    Hence, the response of $I$ is $\notdone$.
    Thus, $p$ found the condition on \cref{line:ero:add_cell_done}, \cref{line:ero:remove_cell_done}, or \cref{line:ero:apply_cell_done} to be true during $I$.
    Therefore, since $p$ read $((0, \noop), \nullconstant)$ from $\announceobject$ at $T^{\ref{line:ero:announce_read}}$, the parameters of $I$ are $(\arbitraryvalue, (0, \noop), \nullconstant)$, and so $\noop$ equals either $\addcell$, $\removecell$, or $\langle \doopandcopyresponse{}, \arbitraryvalue \rangle$, a contradiction.
    \qH{\Cref{lemma:ero:l_events_have_corresponding_a_events}}
\end{proof}

By \Cref{def:ero:english}, this implies the following.

\begin{corollary}\label{lemma:ero:l_event_corresponding_a_event_type_and_pointer_relations}
    Consider any $L$-event $e$ in $\mathcal{I}^\mathcal{B}$ and its corresponding $A$-event $e'$.
    \begin{compactitem}
        \item $e$ is an $L$-add event if and only if $e'$ is an $A$-add event.
        \item $e$ is an $L$-apply event if and only if $e'$ is an $A$-apply event.
        \item $e$ is an $L$-remove event if and only if $e'$ is an $A$-remove event.
        \item $e$ is for timestamp $\timeshort{}$ if and only if $e'$ is for timestamp $\timeshort{}$.
        \item $e$ is for $\cellpointershort$ if and only if $e'$ is for $\cellpointershort$.
    \end{compactitem}
\end{corollary}

Using this corollary, we can now port basic facts about $A$-events to $L$-events.

\begin{lemma}\label{lemma:ero:a_and_l_events_of_same_type_for_same_pointer_set_same_value}
    Consider any $A$-$X$ event $e_A$ for some $\cellpointershort{}$, and an $L$-$X$ event $e_L$ for $\cellpointershort{}$ where $X \in \{\text{add}, \text{apply}, \text{remove}\}$ in $\mathcal{I}^\mathcal{B}$.
    Suppose $e_A$ set $\announceobject = v_A$ and $e_L$ set $\linearizationobject{} = v_L$.
    Then, $v_A = v_L$.
\end{lemma}

\begin{proof}
    Since $e_L$ set $\linearizationobject{} = v_L$, by \Cref{lemma:ero:l_events_have_corresponding_a_events}, there is an $A$-event $e$ that sets $\announceobject = v_L$.
    Furthermore, since $e_L$ is an $L$-$X$ event for $\cellpointershort{}$, by \Cref{lemma:ero:l_event_corresponding_a_event_type_and_pointer_relations}, $e$ is an $A$-$X$ event for $\cellpointershort{}$.
    Therefore, since $e_A$ and $e$ are $A$-$X$ events for $\cellpointershort{}$, $e_A$ set  $\announceobject = v_A$, and $e$ set $\announceobject = v_L$, by \Cref{lemma:ero:a_events_of_same_type_for_same_pointer_set_same_value}, $v_A = v_L$ as wanted.
    \qH{\Cref{lemma:ero:a_and_l_events_of_same_type_for_same_pointer_set_same_value}}
\end{proof}

\begin{lemma}\label{lemma:ero:every_l_event_is_for_timestamp_other_than_zero}
    Every $L$-event in $\mathcal{I}^\mathcal{B}$ is for a timestamp larger than $0$.
\end{lemma}

\begin{proof}
    Consider any $L$-event $e$ for some timestamp $\timeshort{}$.
    Let $e'$ be $e$'s corresponding $A$-event (see \Cref{lemma:ero:l_events_have_corresponding_a_events}).
    By \Cref{lemma:ero:l_event_corresponding_a_event_type_and_pointer_relations}, $e'$ is for timestamp $\timeshort{}$.
    Therefore, by \Cref{lemma:ero:every_a_event_is_for_timestamp_other_than_zero}, $\timeshort{}$ is larger than $0$ as wanted.
    \qH{\Cref{lemma:ero:every_l_event_is_for_timestamp_other_than_zero}}
\end{proof}

This implies the following.

\begin{corollary}\label{lemma:ero:l_events_set_llo_to_a_value_different_from_initial}
    Every $L$-event in $\mathcal{I}^\mathcal{B}$ sets $\linearizationobject{}.\uniquerepositoryoperationlong \neq (0, \noop)$.
\end{corollary}

\begin{lemma}\label{lemma:ero:every_l_event_is_for_pointer_from_universe}
    Every $L$-event in $\mathcal{I}^\mathcal{B}$ is for some pointer in $\celluniverse{}$.
\end{lemma}

\begin{proof}
    Consider any $L$-event $e$ for some $\cellpointershort$.
    Let $e'$ be $e$'s corresponding $A$-event (see \Cref{lemma:ero:l_events_have_corresponding_a_events}).
    Therefore, by \Cref{lemma:ero:l_event_corresponding_a_event_type_and_pointer_relations}, $e'$ is for $\cellpointershort$, so by \Cref{lemma:ero:every_a_event_is_for_pointer_from_universe}, $\cellpointershort \in \celluniverse{}$.
    \qH{\Cref{lemma:ero:every_l_event_is_for_pointer_from_universe}}
\end{proof}

\begin{lemma}\label{lemma:ero:every_l_event_is_add_apply_or_remove}
    Every $L$-event in $\mathcal{I}^\mathcal{B}$ is either an $L$-add, $L$-apply, or $L$-remove event.
\end{lemma}

\begin{proof}
    Consider any $L$-event $e$.
    Let $e'$ be $e$'s corresponding $A$-event defined by \Cref{lemma:ero:l_events_have_corresponding_a_events}.
    Therefore, by \Cref{lemma:ero:every_a_event_is_add_apply_or_remove} $e'$ is either an $A$-add, $A$-apply, or $A$-remove event, and so by \Cref{lemma:ero:l_event_corresponding_a_event_type_and_pointer_relations} $e$ is either an $L$-add, $L$-apply, or $L$-remove event.
    \qH{\Cref{lemma:ero:every_l_event_is_add_apply_or_remove}}
\end{proof}

We now show that invariant $P$ implies the uniqueness of $L$-events in $\mathcal{I}^\mathcal{B}$ in different senses.

\begin{lemma}\label{lemma:ero:p_implies_unique_values_in_linearization}
    If $P(\mathcal{I}^\mathcal{B})$ holds, then every $L$-event in $\mathcal{I}^\mathcal{B}$ sets $\linearizationobject{}$ to a unique value.
\end{lemma}

\begin{proof}
    Suppose, for contradiction, $P(\mathcal{I}^\mathcal{B})$ holds and there are two $L$-events in $\mathcal{I}^\mathcal{B}$, say $e_1$ and $e_2$, that set $\linearizationobject{}$ to the same value $((\arbitraryvalue, \myrepositoryoperationshort), \cellpointershort)$.
    Hence, by \Cref{def:ero:english}, $e_1$ and $e_2$ are $L$-events for $\cellpointershort$, so by \Cref{lemma:ero:every_l_event_is_for_pointer_from_universe} $\cellpointershort \in \celluniverse{}$.
    Furthermore, by \Cref{lemma:ero:every_l_event_is_add_apply_or_remove}, $\myrepositoryoperationshort$ is either $\addcell$, $\langle \doopandcopyresponse{}, \arbitraryvalue \rangle$, or $\removecell$.
    Thus, by \Cref{def:ero:english}: if $\myrepositoryoperationshort = \addcell$, then $e_1$ and $e_2$ are both $L$-add events for $\cellpointershort$; if $\myrepositoryoperationshort = \langle \doopandcopyresponse{}, \arbitraryvalue \rangle$, then $e_1$ and $e_2$ are both $L$-apply events for $\cellpointershort$; and if $\myrepositoryoperationshort = \removecell$, then $e_1$ and $e_2$ are both $L$-remove events for $\cellpointershort$.
    Therefore, for some $\cellpointershort \in \celluniverse{}$, in $\mathcal{I}^\mathcal{B}$ there is either two $L$-add events for $\cellpointershort$, two $L$-apply events for $\cellpointershort$, or two $L$-remove events for $\cellpointershort$.
    However, since $\cellpointershort \in \celluniverse{}$, by $P(\mathcal{I}^\mathcal{B})$, in $\mathcal{I}^\mathcal{B}$ there is at most one $L$-add event for $\cellpointershort$, at most one $L$-apply event for $\cellpointershort$, and at most one $L$-remove event for $\cellpointershort$, a contradiction.
    \qH{\Cref{lemma:ero:p_implies_unique_values_in_linearization}}
\end{proof}

\begin{proposition}\label{lemma:ero:matching_timestamp_implies_matching_low_level_op}
    If the left field of $\linearizationobject{}.\uniquerepositoryoperationlong{}$ is the same at times $T$ and $T'$ in $\mathcal{I}^\mathcal{B}$, then the value of $\linearizationobject{}$ is the same at $T$ and $T'$.
\end{proposition}

\begin{proof}
    Suppose, for contradiction, that the left field of $\linearizationobject{}.\uniquerepositoryoperationlong{}$ is the same at $T$ and $T'$ and the value of $\linearizationobject{}$ is different at $T$ and $T'$.
    Suppose $\linearizationobject{} = ((\timeshort{}, \myrepositoryoperationshort), \cellpointershort)$ at $T$
    and suppose $\linearizationobject{} = ((\timeshort{}, \myrepositoryoperationshort'), \cellpointershort')$ at $T'$  such that $(\myrepositoryoperationshort, \cellpointershort) \neq (\myrepositoryoperationshort', \cellpointershort')$.
    Without loss of generality, suppose $T < T'$.
    Hence, the value of $\linearizationobject{}$ was set to $((\timeshort{}, \myrepositoryoperationshort'), \cellpointershort')$ sometime during $(T, T']$.
    Thus, by \Cref{observation:ero:where_objects_change}, an $L$-event $e'$ set $\linearizationobject{}$ to $((\timeshort{}, \myrepositoryoperationshort'), \cellpointershort')$.
    So, by \Cref{lemma:ero:l_events_have_corresponding_a_events}, $\announceobject{} = ((\timeshort{}, \myrepositoryoperationshort'), \cellpointershort')$ at some time $T'_A$.
    Furthermore, by \Cref{lemma:ero:every_l_event_is_for_timestamp_other_than_zero}, $\timeshort{} \neq 0$.
    Hence, since $\linearizationobject{}$ is initially $((0, \noop), \nullconstant)$, the value of $\linearizationobject{}$ at $T$ is not the initial value, and so $\linearizationobject{}$ was set to $((\timeshort{}, \myrepositoryoperationshort), \cellpointershort)$.
    Thus, by \Cref{observation:ero:where_objects_change}, an $L$-event set $\linearizationobject{}$ to $((\timeshort{}, \myrepositoryoperationshort), \cellpointershort)$.
    So, by \Cref{lemma:ero:l_events_have_corresponding_a_events}, $\announceobject{} = ((\timeshort{}, \myrepositoryoperationshort), \cellpointershort)$ at some time $T_A$.
    Therefore, since $\announceobject{} = ((\timeshort{}, \myrepositoryoperationshort), \cellpointershort)$ at $T_A$ and $\announceobject{} = ((\timeshort{}, \myrepositoryoperationshort'), \cellpointershort')$ at $T'_A$, by \Cref{lemma:ero:matching_timestamp_in_a_implies_matching_value_in_a}, $(\myrepositoryoperationshort, \cellpointershort) = (\myrepositoryoperationshort', \cellpointershort')$.
    However, $(\myrepositoryoperationshort, \cellpointershort) \neq (\myrepositoryoperationshort', \cellpointershort')$, a contradiction.
    \qH{\Cref{lemma:ero:matching_timestamp_implies_matching_low_level_op}}
\end{proof}

\begin{lemma}\label{lemma:ero:p_implies_unique_low_level_operations_in_linearization}
    If $P(\mathcal{I}^\mathcal{B})$ holds, then every $L$-event in $\mathcal{I}^\mathcal{B}$ sets $\linearizationobject{}.\uniquerepositoryoperationlong{}$ to a unique value.
\end{lemma}

\begin{proof}
    Suppose, for contradiction, $P(\mathcal{I}^\mathcal{B})$ holds and there are two $L$-events $e_1$ and $e_2$ in $\mathcal{I}^\mathcal{B}$ that set the value of $\linearizationobject{}.\uniquerepositoryoperationlong{}$ to the same value.
    Since $e_1$ and $e_2$ are in $\mathcal{I}^\mathcal{B}$ and by assumption $P(\mathcal{I}^\mathcal{B})$ holds, by \Cref{lemma:ero:p_implies_unique_values_in_linearization}, $e_1$ and $e_2$ set $\linearizationobject{}$ to different values.
    Therefore, since $e_1$ and $e_2$ set $\linearizationobject{}.\uniquerepositoryoperationlong{}$ to the same value, it follows that $e_1$ and $e_2$ set $\linearizationobject{}.\cellpointerlong{}$ to different values.
    However, since $\linearizationobject{}.\uniquerepositoryoperationlong{}$ is the same at $e_1$ and $e_2$, by \Cref{lemma:ero:matching_timestamp_implies_matching_low_level_op}, $\linearizationobject{}.\cellpointerlong{}$ is the same at $e_1$ and $e_2$, so $e_1$ and $e_2$ set  $\linearizationobject{}.\cellpointerlong{}$ to the same value, a contradiction.
    \qH{\Cref{lemma:ero:p_implies_unique_low_level_operations_in_linearization}}
\end{proof}

\begin{lemma}\label{lemma:ero:every_l_event_has_a_unique_timestamp}
    If $P(\mathcal{I}^\mathcal{B})$ holds, then every $L$-event in $\mathcal{I}^\mathcal{B}$ is for a unique timestamp.
\end{lemma}

\begin{proof}
    Suppose, for contradiction, $P(\mathcal{I}^\mathcal{B})$ holds and there are two $L$-events $e_1$ and $e_2$ in $\mathcal{I}^\mathcal{B}$ for the same timestamp $\timeshort$.
    Let $((\timeshort{}, \myrepositoryoperationshort_1), \cellpointershort_1)$ and $((\timeshort{}, \myrepositoryoperationshort_2), \cellpointershort_2)$ be the values that $e_1$ and $e_2$ write into $\linearizationobject$, respectively.
    Since $e_1$ and $e_2$ set $\linearizationobject{}.\uniquerepositoryoperationlong{}$ to $(\timeshort{}, \myrepositoryoperationshort_1)$ and $(\timeshort{}, \myrepositoryoperationshort_2)$, respectively, by \Cref{lemma:ero:matching_timestamp_implies_matching_low_level_op}, $\myrepositoryoperationshort_1 = \myrepositoryoperationshort_2$.
    Therefore, $e_1$ and $e_2$ set $\linearizationobject{}.\uniquerepositoryoperationlong{}$ to the same value.
    However, since $P(\mathcal{I}^\mathcal{B})$ holds, by \Cref{lemma:ero:p_implies_unique_low_level_operations_in_linearization}, every $L$-event in $\mathcal{I}^\mathcal{B}$ sets $\linearizationobject{}.\uniquerepositoryoperationlong{}$ to a unique value, a contradiction.
    \qH{\Cref{lemma:ero:every_l_event_has_a_unique_timestamp}}
\end{proof}

We now define the concept of a corresponding $L$-event for invocations of the \doaddcell{}, \doremovecell{}, and \doapplyandcopyresponse{} procedures.

\begin{proposition}\label{lemma:ero:l_add_event_for_do_add_cell_parameters}
    Let $I$ be any invocation of the \doaddcell{} procedure in $\mathcal{I}^\mathcal{B}$ with parameters $(\uniquerepositoryoperationshort_\linearizationobject{}, \cellpointershort_\linearizationobject{})$.
    There is an $L$-add event for $\cellpointershort_\linearizationobject{}$ before $I$ was invoked that set $\linearizationobject{} = (\uniquerepositoryoperationshort_\linearizationobject{}, \cellpointershort_\linearizationobject{})$.
\end{proposition}

\begin{proof}
    Let $p$ be the process that invoked $I$.
    Hence, $p$ read $(\uniquerepositoryoperationshort_\linearizationobject{}, \cellpointershort_\linearizationobject{})$ from $\linearizationobject{}$ on \cref{line:ero:linearization_read} before invoking $I$, and so by the condition on \cref{line:ero:do_add_cell_condition} $\uniquerepositoryoperationshort_\linearizationobject{} = (\arbitraryvalue, \addcell)$.
    Thus, $\uniquerepositoryoperationshort_\linearizationobject{} \neq (0, \noop)$, and so since $\linearizationobject{}$ is initially $((0, \noop), \nullconstant)$, we have that $\linearizationobject{}$ was set to $(\uniquerepositoryoperationshort_\linearizationobject{}, \cellpointershort_\linearizationobject{})$ before $I$ was invoked.
    So, by \Cref{observation:ero:where_objects_change}, there is an $L$-event $e$ for $\cellpointershort_\linearizationobject{}$ that set $\linearizationobject{} = (\uniquerepositoryoperationshort_\linearizationobject{}, \cellpointershort_\linearizationobject{})$ before $I$ was invoked.
    Therefore, since $\uniquerepositoryoperationshort_\linearizationobject{} = (\arbitraryvalue, \addcell)$, by \Cref{def:ero:english}, $e$ is an $L$-add event for $\cellpointershort_\linearizationobject{}$ as wanted.
    \qH{\Cref{lemma:ero:l_add_event_for_do_add_cell_parameters}}
\end{proof}

\begin{proposition}\label{lemma:ero:l_remove_event_for_do_remove_cell_parameters}
    Let $I$ be any invocation of the \doremovecell{} procedure in $\mathcal{I}^\mathcal{B}$ with parameters $(\uniquerepositoryoperationshort_\linearizationobject{}, \cellpointershort_\linearizationobject{})$.
    There is an $L$-remove event for $\cellpointershort_\linearizationobject{}$ before $I$ was invoked that set $\linearizationobject{} = (\uniquerepositoryoperationshort_\linearizationobject{}, \cellpointershort_\linearizationobject{})$.
\end{proposition}

\begin{proof}
    By essentially the same argument as \Cref{lemma:ero:l_add_event_for_do_add_cell_parameters}, which is provided below for completeness.
    Let $p$ be the process that invoked $I$.
    Hence, $p$ read $(\uniquerepositoryoperationshort_\linearizationobject{}, \cellpointershort_\linearizationobject{})$ from $\linearizationobject{}$ on \cref{line:ero:linearization_read} before invoking $I$, and so by the condition on \cref{line:ero:do_remove_cell_condition} $\uniquerepositoryoperationshort_\linearizationobject{} = (\arbitraryvalue, \removecell)$.
    Thus, $\uniquerepositoryoperationshort_\linearizationobject{} \neq (0, \noop)$, and so since $\linearizationobject{}$ is initially $((0, \noop), \nullconstant)$, we have that $\linearizationobject{}$ was set to $(\uniquerepositoryoperationshort_\linearizationobject{}, \cellpointershort_\linearizationobject{})$ before $I$ was invoked.
    So, by \Cref{observation:ero:where_objects_change}, there is an $L$-event $e$ for $\cellpointershort_\linearizationobject{}$ that set $\linearizationobject{} = (\uniquerepositoryoperationshort_\linearizationobject{}, \cellpointershort_\linearizationobject{})$ before $I$ was invoked.
    Therefore, since $\uniquerepositoryoperationshort_\linearizationobject{} = (\arbitraryvalue, \removecell)$, by \Cref{def:ero:english}, $e$ is an $L$-remove event for $\cellpointershort_\linearizationobject{}$ as wanted.
    \qH{\Cref{lemma:ero:l_remove_event_for_do_remove_cell_parameters}}
\end{proof}

\begin{proposition}\label{lemma:ero:l_apply_event_for_do_apply_parameters}
    Let $I$ be any invocation of the \doapplyandcopyresponse{} procedure in $\mathcal{I}^\mathcal{B}$ with parameters $(\uniquerepositoryoperationshort_\linearizationobject{}, \cellpointershort_\linearizationobject{})$.
    There is an $L$-apply event for $\cellpointershort_\linearizationobject{}$ before $I$ was invoked that set $\linearizationobject{} = (\uniquerepositoryoperationshort_\linearizationobject{}, \cellpointershort_\linearizationobject{})$.
\end{proposition}

\begin{proof}
    By essentially the same argument as \Cref{lemma:ero:l_add_event_for_do_add_cell_parameters}, which is provided below for completeness.
    Let $p$ be the process that invoked $I$.
    Hence, $p$ read $(\uniquerepositoryoperationshort_\linearizationobject{}, \cellpointershort_\linearizationobject{})$ from $\linearizationobject{}$ on \cref{line:ero:linearization_read} before invoking $I$, and so by the condition on \cref{line:ero:do_apply_and_copy_response_condition} $\uniquerepositoryoperationshort_\linearizationobject{} = (\arbitraryvalue, \langle \doopandcopyresponse{}, \arbitraryvalue \rangle)$.
    Thus, $\uniquerepositoryoperationshort_\linearizationobject{} \neq (0, \noop)$, and so since $\linearizationobject{}$ is initially $((0, \noop), \nullconstant)$, we have that $\linearizationobject{}$ was set to $(\uniquerepositoryoperationshort_\linearizationobject{}, \cellpointershort_\linearizationobject{})$ before $I$ was invoked.
    So, by \Cref{observation:ero:where_objects_change}, there is an $L$-event $e$ for $\cellpointershort_\linearizationobject{}$ that set $\linearizationobject{} = (\uniquerepositoryoperationshort_\linearizationobject{}, \cellpointershort_\linearizationobject{})$ before $I$ was invoked.
    Therefore, since $\uniquerepositoryoperationshort_\linearizationobject{} = (\arbitraryvalue, \langle \doopandcopyresponse{}, \arbitraryvalue \rangle)$, by \Cref{def:ero:english}, $e$ is an $L$-apply event for $\cellpointershort_\linearizationobject{}$ as wanted.
    \qH{\Cref{lemma:ero:l_apply_event_for_do_apply_parameters}}
\end{proof}

By Lemmas \ref{lemma:ero:l_add_event_for_do_add_cell_parameters}, \ref{lemma:ero:l_remove_event_for_do_remove_cell_parameters}, and \ref{lemma:ero:l_apply_event_for_do_apply_parameters} we have the following.

\begin{corollary}\label{lemma:ero:l_event_corresponding_to_do_low_level_op}
    Let $I$ be any invocation of the \doaddcell{}, \doapplyandcopyresponse{}, or \doremovecell{} procedure in $\mathcal{I}^\mathcal{B}$ with parameters $(\uniquerepositoryoperationshort_\linearizationobject{}, \cellpointershort_\linearizationobject{})$.
    The following are true.
    \begin{compactitem}
        \item There is an $L$-event $e$ for $\cellpointershort_\linearizationobject{}$ before $I$ was invoked that set $\linearizationobject{} = (\uniquerepositoryoperationshort_\linearizationobject{}, \cellpointershort_\linearizationobject{})$.
        \item $I$ is an invocation of \doaddcell{} if and only if $e$ is an $L$-add event.
        \item $I$ is an invocation of \doremovecell{} if and only if $e$ is an $L$-remove event.
        \item $I$ is an invocation of \doapplyandcopyresponse{} if and only if $e$ is an $L$-apply event.
    \end{compactitem}
    We call $e$ the \textbf{corresponding $L$-event of $I$}. For convenience, we sometimes call $e$ the corresponding $L$-event of a step during $I$ performed by the process that invoked $I$. 
\end{corollary}

We now prove that $L$-events appear in the expected order for a given pointer.

\begin{lemma}\label{lemma:ero:apply_events_are_preceeded_by_add_events}
    Every $L$-apply event for $\cellpointershort$ in $\mathcal{I}^\mathcal{B}$ is preceded by an $L$-add event for $\cellpointershort$. 
\end{lemma}

\begin{proof}
    Consider any $L$-apply event $e$ for $\cellpointershort$ in $\mathcal{I}^\mathcal{B}$.
    Hence, by \Cref{lemma:ero:every_l_event_is_for_pointer_from_universe}, $\cellpointershort \in \celluniverse{}$.
    Let $p$ be the process that executed $e$.
    We prove two intermediate claims.
	
    \begin{claimcustom}{\ref{lemma:ero:apply_events_are_preceeded_by_add_events}.1}\label{lemma:ero:apply_events_are_preceeded_by_add_events:first_claim}
       Some process $q$ found the condition on \cref{line:ero:do_work_while_loop} to be false at some time $T^{\ref{line:ero:do_work_while_loop}}_q < e$ during some invocation of the \doworkuntildone{} procedure with parameters $(\addcell, \cellpointershort)$.
    \end{claimcustom}

    \begin{proof}
        Since $e$ is an $L$-apply event for $\cellpointershort$, by \Cref{lemma:ero:l_event_corresponding_a_event_type_and_pointer_relations}, there is an $A$-apply event $e'$ for $\cellpointershort$ before $e$; say by process $q$.
        Thus, by \Cref{def:ero:english}, $q$ executed $e'$ during an invocation $I$ of the \doworkuntildone{} procedure with parameters $(\langle \doopandcopyresponse{}, \arbitraryvalue\rangle, \cellpointershort)$.
        Let $I^{hlo}$ be the invocation of the \highleveloperation{} procedure that $q$ invoked $I$ during.
        So, $q$ invoked $I$ on \cref{line:ero:low_level_apply_and_copy_response} during $I^{hlo}$.
        Hence, before $q$ invoked $I$, $q$ invoked and exited the \doworkuntildone{} procedure on \cref{line:ero:low_level_add_cell} during $I^{hlo}$; let $I'$ denote this invocation.
        Thus, since $I$ has parameters $(\arbitraryvalue, \cellpointershort)$, it follows that $I'$ has parameters $(\addcell, \cellpointershort)$.
        Since $q$ exited $I'$, we have that $q$ found the condition on \cref{line:ero:do_work_while_loop} to be false during $I'$; say at time $T^{\ref{line:ero:do_work_while_loop}}_q$.
        Therefore, since $q$ executed the step at $T^{\ref{line:ero:do_work_while_loop}}_q$ during $I'$, $q$ exited $I'$ before invoking $I$, $q$ executed $e'$ during $I$, and $e' < e$, by transitivity, $T^{\ref{line:ero:do_work_while_loop}}_q < e$.
        \qH{\Cref{lemma:ero:apply_events_are_preceeded_by_add_events:first_claim}}
    \end{proof}
    
    \begin{claimcustom}{\ref{lemma:ero:apply_events_are_preceeded_by_add_events}.2}\label{lemma:ero:apply_events_are_preceeded_by_add_events:second_claim}
    	Let $T^{\ref{line:ero:do_work_initialize_response}}_q$ be the time of $q$'s last execution of \cref{line:ero:do_work_initialize_response} before $T^{\ref{line:ero:do_work_while_loop}}_q$.
        Then, there is a successful add-response-set attempt for $\cellpointershort{}$ during $(T^{\ref{line:ero:do_work_initialize_response}}_q, T^{\ref{line:ero:do_work_while_loop}}_q)$.
    \end{claimcustom}

    \begin{proof}
        Since $q$ executed the step at $T^{\ref{line:ero:do_work_while_loop}}_q$ during some invocation $I_q$ of the \doworkuntildone{} procedure with parameters $(\addcell, \cellpointershort)$ and $q$'s step at  $T^{\ref{line:ero:do_work_initialize_response}}_q$ is its last execution of \cref{line:ero:do_work_initialize_response} before $T^{\ref{line:ero:do_work_while_loop}}_q$, we have that $q$'s step at $T^{\ref{line:ero:do_work_initialize_response}}_q$ is during $I_q$.
        Hence, since $I_q$'s parameters are $(\addcell, \cellpointershort)$ and $T^{\ref{line:ero:do_work_initialize_response}}_q$ is during $I_q$, we have that $q$ set the value of $(*\cellpointershort).\lastrepositoryoperationresponse{} = ((\arbitraryvalue, \addcell), \nullconstant)$ at $T^{\ref{line:ero:do_work_initialize_response}}_q$.
        Since $q$ found the condition on \cref{line:ero:do_work_while_loop} to be false during $I_q$ at  $T^{\ref{line:ero:do_work_while_loop}}_q$, we have that $(*\cellpointershort).\lastrepositoryoperationresponse{} \neq ((\arbitraryvalue{}, \addcell), \nullconstant)$ at $T^{\ref{line:ero:do_work_while_loop}}_q$.
        Hence, since $(*\cellpointershort).\lastrepositoryoperationresponse{} = ((\arbitraryvalue{}, \addcell), \nullconstant)$ at $T^{\ref{line:ero:do_work_initialize_response}}_q$, and $T^{\ref{line:ero:do_work_initialize_response}}_q < T^{\ref{line:ero:do_work_while_loop}}_q$, the value of $(*\cellpointershort).\lastrepositoryoperationresponse{}$ changed during $(T^{\ref{line:ero:do_work_initialize_response}}_q, T^{\ref{line:ero:do_work_while_loop}}_q)$.
        Thus, by \Cref{observation:ero:where_objects_change}, a response-reset event for $\cellpointershort$ or a successful response-set attempt for $\cellpointershort$ changed $(*\cellpointershort).\lastrepositoryoperationresponse{}$ during $(T^{\ref{line:ero:do_work_initialize_response}}_q, T^{\ref{line:ero:do_work_while_loop}}_q)$.
        Let $a$ be the first step that changed $(*\cellpointershort).\lastrepositoryoperationresponse{}$ during $(T^{\ref{line:ero:do_work_initialize_response}}_q, T^{\ref{line:ero:do_work_while_loop}}_q)$.
        \begin{itemize}
            \item[] \hspace{0pt}\textbf{Case 1.} $a$ is a response-reset event for $\cellpointershort$.

            Hence, by \Cref{def:ero:english}, $a$ set $(*\cellpointershort).\lastrepositoryoperationresponse{} = (\arbitraryvalue, \nullconstant)$ on \cref{line:ero:do_work_initialize_response}.
            Thus, some process $r$ executed $a$ during an invocation $I_r$ of the \doworkuntildone{} procedure with parameters $(\arbitraryvalue, \cellpointershort)$.
            Hence, $r$ received $\cellpointershort$ as a response on \cref{line:ero:allocate_cell} during the invocation $I^{hlo}_r$ of the \highleveloperation{} procedure in which $r$ invoked $I_r$ during.
            Let $I^{hlo}_q$ be the invocation of the \highleveloperation{} procedure in which $q$ invoked $I_q$ during.
            Since the second parameter of $I_q$ is $\cellpointershort$, it follows that $q$ received $\cellpointershort$ as a response on \cref{line:ero:allocate_cell} during $I^{hlo}_q$.
            Thus, since by \Cref{alg:lazy_cell_manager_specification} the response of every \allocatecelloperation{} operation is unique, we have that $q = r$ and $I^{hlo}_q = I^{hlo}_r$.
            Therefore, since $a$ is an execution of \cref{line:ero:do_work_initialize_response} and is during $(T^{\ref{line:ero:do_work_initialize_response}}_q, T^{\ref{line:ero:do_work_while_loop}}_q)$, $q$ executed \cref{line:ero:do_work_initialize_response} during the loop on \cref{line:ero:do_work_while_loop} during $I_q$.
            However, $q$ must exit the loop on \cref{line:ero:do_work_while_loop} to execute \cref{line:ero:do_work_initialize_response}, a contradiction, so this case is impossible.

            \item[] \hspace{0pt}\textbf{Case 2.} $a$ is a successful response-set attempt for $\cellpointershort$.

            Hence, by \Cref{def:ero:english}, $a$ is an execution of \cref{line:ero:responses_set_attempt}.
            Thus, since by definition $a$ is the first step that changed $(*\cellpointershort).\lastrepositoryoperationresponse{}$ during $(T^{\ref{line:ero:do_work_initialize_response}}_q, T^{\ref{line:ero:do_work_while_loop}}_q)$, and as established above $(*\cellpointershort).\lastrepositoryoperationresponse{} = ((\arbitraryvalue{}, \addcell), \nullconstant)$ at $T^{\ref{line:ero:do_work_initialize_response}}_q$, it follows that the first parameter of $a$ is $((\arbitraryvalue{}, \addcell), \nullconstant)$ (if it was anything else $a$ would not be successful).
            Hence, by \Cref{def:ero:english}, $a$ is an add-response-set attempt for $\cellpointershort$.
            Therefore, since $a$ is successful and is during $(T^{\ref{line:ero:do_work_initialize_response}}_q, T^{\ref{line:ero:do_work_while_loop}}_q)$, we have that $a$ is a successful add-response-set attempt for $\cellpointershort$ during $(T^{\ref{line:ero:do_work_initialize_response}}_q, T^{\ref{line:ero:do_work_while_loop}}_q)$ as required.
            \qH{\Cref{lemma:ero:apply_events_are_preceeded_by_add_events:second_claim}}
        \end{itemize}
    \end{proof}
    
    We now return to the proof of \Cref{lemma:ero:apply_events_are_preceeded_by_add_events}.
    Let $a$ be the successful add-response-set attempt for $\cellpointershort{}$ during $(T^{\ref{line:ero:do_work_initialize_response}}_q, T^{\ref{line:ero:do_work_while_loop}}_q)$ identified by \Cref{lemma:ero:apply_events_are_preceeded_by_add_events:second_claim} and let $r$ be the process that executed $a$.
	Since by \Cref{lemma:ero:apply_events_are_preceeded_by_add_events:first_claim} $T^{\ref{line:ero:do_work_while_loop}}_q < e$, by transitivity, $a < e$.
    Furthermore, since $a$ is an add-response-set attempt for $\cellpointershort$, by \Cref{def:ero:english}, $a$ was executed during an invocation $I$ of the \setrepositoryoperationresponse{} procedure with parameters $((\arbitraryvalue, \addcell), \cellpointershort, \arbitraryvalue)$.
    Hence, by \Cref{observation:ero:response_set_attempt_invocation}, $I$ was invoked during an invocation $I'$ of the \doaddcell{} procedure.
    Thus, since $I$ has parameters $((\arbitraryvalue, \addcell), \cellpointershort, \arbitraryvalue)$, it follows that $I'$ has parameters $((\arbitraryvalue, \addcell), \cellpointershort)$.
    So, by \Cref{lemma:ero:l_event_corresponding_to_do_low_level_op}, there is an $L$-add $e'$ for $\cellpointershort$ before $I'$ was invoked.
    Hence, since $I$ was invoked during $I'$, $a$ was executed during $I$, and $a < e$, by transitivity, $e' < e$.
    Therefore, there is an $L$-add event for $\cellpointershort$ before $e$.
    \qH{\Cref{lemma:ero:apply_events_are_preceeded_by_add_events}}
\end{proof}

\begin{lemma}\label{lemma:ero:remove_events_are_preceeded_by_apply_events}
    Every $L$-remove event for $\cellpointershort$ in $\mathcal{I}^\mathcal{B}$ is preceded by an $L$-apply event for $\cellpointershort$. 
\end{lemma}

\begin{proof}
    By essentially the same argument as \Cref{lemma:ero:apply_events_are_preceeded_by_add_events}, which we provide below for completeness.
    Consider any $L$-remove event $e$ for $\cellpointershort$ in $\mathcal{I}^\mathcal{B}$.
    Hence, by \Cref{lemma:ero:every_l_event_is_for_pointer_from_universe}, $\cellpointershort \in \celluniverse{}$.
    Let $p$ be the process that executed $e$.
    We prove two intermediate claims.
	
    \begin{claimcustom}{\ref{lemma:ero:remove_events_are_preceeded_by_apply_events}.1}\label{lemma:ero:remove_events_are_preceeded_by_apply_events:first_claim}
       Some process $q$ found the condition on \cref{line:ero:do_work_while_loop} to be false at some time $T^{\ref{line:ero:do_work_while_loop}}_q < e$ during some invocation of the \doworkuntildone{} procedure with parameters\\
       $(\langle \doopandcopyresponse{}, \arbitraryvalue \rangle, \cellpointershort)$.
    \end{claimcustom}

    \begin{proof}
        Since $e$ is an $L$-remove event for $\cellpointershort$, by \Cref{lemma:ero:l_event_corresponding_a_event_type_and_pointer_relations}, there is an $A$-remove event $e'$ for $\cellpointershort$ before $e$; say by process $q$.
        Thus, by \Cref{def:ero:english}, $q$ executed $e'$ during an invocation $I$ of the \doworkuntildone{} procedure with parameters $(\removecell, \cellpointershort)$.
        Let $I^{hlo}$ be the invocation of the \highleveloperation{} procedure that $q$ invoked $I$ during.
        So, $q$ invoked $I$ on \cref{line:ero:low_level_remove_cell} during $I^{hlo}$.
        Hence, before $q$ invoked $I$, $q$ invoked and exited the \doworkuntildone{} procedure on \cref{line:ero:low_level_apply_and_copy_response} during $I^{hlo}$; let $I'$ denote this invocation.
        Thus, since $I$ has parameters $(\arbitraryvalue, \cellpointershort)$, it follows that $I'$ has parameters $(\langle \doopandcopyresponse{}, \arbitraryvalue \rangle, \cellpointershort)$.
        Since $q$ exited $I'$, we have that $q$ found the condition on \cref{line:ero:do_work_while_loop} to be false during $I'$; say at time $T^{\ref{line:ero:do_work_while_loop}}_q$.
        Therefore, since $q$ executed the step at $T^{\ref{line:ero:do_work_while_loop}}_q$ during $I'$, $q$ exited $I'$ before invoking $I$, $q$ executed $e'$ during $I$, and $e' < e$, by transitivity, $T^{\ref{line:ero:do_work_while_loop}}_q < e$.
        \qH{\Cref{lemma:ero:remove_events_are_preceeded_by_apply_events:first_claim}}
    \end{proof}
    
    \begin{claimcustom}{\ref{lemma:ero:remove_events_are_preceeded_by_apply_events}.2}\label{lemma:ero:remove_events_are_preceeded_by_apply_events:second_claim}
    	Let $T^{\ref{line:ero:do_work_initialize_response}}_q$ be the time of $q$'s last execution of \cref{line:ero:do_work_initialize_response} before $T^{\ref{line:ero:do_work_while_loop}}_q$.
        Then, there is a successful apply-response-set attempt for $\cellpointershort{}$ during $(T^{\ref{line:ero:do_work_initialize_response}}_q, T^{\ref{line:ero:do_work_while_loop}}_q)$.
    \end{claimcustom}

    \begin{proof}
        Since $q$ executed the step at $T^{\ref{line:ero:do_work_while_loop}}_q$ during some invocation $I_q$ of the \doworkuntildone{} procedure with parameters $(\langle \doopandcopyresponse{}, \arbitraryvalue \rangle, \cellpointershort)$ and $q$'s step at $T^{\ref{line:ero:do_work_initialize_response}}_q$ is its last execution of \cref{line:ero:do_work_initialize_response} before $T^{\ref{line:ero:do_work_while_loop}}_q$, we have that $q$'s step at $T^{\ref{line:ero:do_work_initialize_response}}_q$ is during $I_q$.
        Hence, since $I_q$'s parameters are $(\langle \doopandcopyresponse{}, \arbitraryvalue \rangle, \cellpointershort)$ and $T^{\ref{line:ero:do_work_initialize_response}}_q$ is during $I_q$, we have that $q$ set the value of $(*\cellpointershort).\lastrepositoryoperationresponse{} = ((\arbitraryvalue, \langle \doopandcopyresponse{}, \arbitraryvalue \rangle), \nullconstant)$ at $T^{\ref{line:ero:do_work_initialize_response}}_q$.
        Since $q$ found the condition on \cref{line:ero:do_work_while_loop} to be false at  $T^{\ref{line:ero:do_work_while_loop}}_q$, we have that $(*\cellpointershort).\lastrepositoryoperationresponse{} \neq ((\arbitraryvalue{}, \langle \doopandcopyresponse{}, \arbitraryvalue \rangle), \nullconstant)$ at $T^{\ref{line:ero:do_work_while_loop}}_q$.
        Hence, since $(*\cellpointershort).\lastrepositoryoperationresponse{} = ((\arbitraryvalue{}, \langle \doopandcopyresponse{}, \arbitraryvalue \rangle), \nullconstant)$ at $T^{\ref{line:ero:do_work_initialize_response}}_q$, and $T^{\ref{line:ero:do_work_initialize_response}}_q < T^{\ref{line:ero:do_work_while_loop}}_q$, the value of $(*\cellpointershort).\lastrepositoryoperationresponse{}$ changed during $(T^{\ref{line:ero:do_work_initialize_response}}_q, T^{\ref{line:ero:do_work_while_loop}}_q)$.
        Thus, by \Cref{observation:ero:where_objects_change}, a response-reset event for $\cellpointershort$ or a successful response-set attempt for $\cellpointershort$ changed $(*\cellpointershort).\lastrepositoryoperationresponse{}$ during $(T^{\ref{line:ero:do_work_initialize_response}}_q, T^{\ref{line:ero:do_work_while_loop}}_q)$.
        Let $a$ be the first step that changed $(*\cellpointershort).\lastrepositoryoperationresponse{}$ during $(T^{\ref{line:ero:do_work_initialize_response}}_q, T^{\ref{line:ero:do_work_while_loop}}_q)$.
        \begin{itemize}
            \item[] \hspace{0pt}\textbf{Case 1.} $a$ is a response-reset event for $\cellpointershort$.

            Hence, by \Cref{def:ero:english}, $a$ set $(*\cellpointershort).\lastrepositoryoperationresponse{} = (\arbitraryvalue, \nullconstant)$ on \cref{line:ero:do_work_initialize_response}.
            Thus, some process $r$ executed $a$ during an invocation $I_r$ of the \doworkuntildone{} procedure with parameters $(\arbitraryvalue, \cellpointershort)$.
            Hence, $r$ received $\cellpointershort$ as a response on \cref{line:ero:allocate_cell} during the invocation $I^{hlo}_r$ of the \highleveloperation{} procedure in which $r$ invoked $I_r$ during.
            Let $I^{hlo}_q$ be the invocation of the \highleveloperation{} procedure in which $q$ invoked $I_q$ during.
            Since the second parameter of $I_q$ is $\cellpointershort$, it follows that $q$ received $\cellpointershort$ as a response on \cref{line:ero:allocate_cell} during $I^{hlo}_q$.
            Thus, since by \Cref{alg:lazy_cell_manager_specification} the response of every \allocatecelloperation{} operation is unique, we have that $q = r$ and $I^{hlo}_q = I^{hlo}_r$.
            Therefore, since $a$ is an execution of \cref{line:ero:do_work_initialize_response} and is during $(T^{\ref{line:ero:do_work_initialize_response}}_q, T^{\ref{line:ero:do_work_while_loop}}_q)$, $q$ executed \cref{line:ero:do_work_initialize_response} during the loop on \cref{line:ero:do_work_while_loop} during $I_q$.
            However, $q$ must exit the loop on \cref{line:ero:do_work_while_loop} to execute \cref{line:ero:do_work_initialize_response}, a contradiction, so this case is impossible.

            \item[] \hspace{0pt}\textbf{Case 2.} $a$ is a successful response-set attempt for $\cellpointershort$.

            Hence, by \Cref{def:ero:english}, $a$ is an execution of \cref{line:ero:responses_set_attempt}.
            Thus, since by definition $a$ is the first step that changed $(*\cellpointershort).\lastrepositoryoperationresponse{}$ during $(T^{\ref{line:ero:do_work_initialize_response}}_q, T^{\ref{line:ero:do_work_while_loop}}_q)$, and as established above $(*\cellpointershort).\lastrepositoryoperationresponse{} = ((\arbitraryvalue{}, \langle \doopandcopyresponse{}, \arbitraryvalue \rangle), \nullconstant)$ at $T^{\ref{line:ero:do_work_initialize_response}}_q$, it follows that the first parameter of the \CASop{} operation $a$ performs is $((\arbitraryvalue{}, \langle \doopandcopyresponse{}, \arbitraryvalue \rangle), \nullconstant)$ (if it was anything else $a$ would not be successful).
            Hence, by \Cref{def:ero:english}, $a$ is an apply-response-set attempt for $\cellpointershort$.
            Therefore, since $a$ is successful and is during $(T^{\ref{line:ero:do_work_initialize_response}}_q, T^{\ref{line:ero:do_work_while_loop}}_q)$, we have that $a$ is a successful apply-response-set attempt for $\cellpointershort$ during $(T^{\ref{line:ero:do_work_initialize_response}}_q, T^{\ref{line:ero:do_work_while_loop}}_q)$ as required.
            \qH{\Cref{lemma:ero:remove_events_are_preceeded_by_apply_events:second_claim}}
        \end{itemize}
    \end{proof}
    
    We now return to the proof of \Cref{lemma:ero:remove_events_are_preceeded_by_apply_events}.
    Let $a$ be the successful apply-response-set attempt for $\cellpointershort{}$ during $(T^{\ref{line:ero:do_work_initialize_response}}_q, T^{\ref{line:ero:do_work_while_loop}}_q)$ identified by \Cref{lemma:ero:remove_events_are_preceeded_by_apply_events:second_claim} and let $r$ be the process that executed $a$.
	Since by \Cref{lemma:ero:remove_events_are_preceeded_by_apply_events:first_claim} $T^{\ref{line:ero:do_work_while_loop}}_q < e$, by transitivity, $a < e$.
    Furthermore, since $a$ is an apply-response-set attempt for $\cellpointershort$, by \Cref{def:ero:english}, $a$ was executed during an invocation $I$ of the \setrepositoryoperationresponse{} procedure with parameters $((\arbitraryvalue, \langle \doopandcopyresponse{}, \arbitraryvalue \rangle), \cellpointershort, \arbitraryvalue)$.
    Hence, by \Cref{observation:ero:response_set_attempt_invocation}, $I$ was invoked during an invocation $I'$ of the \doapplyandcopyresponse{} procedure.
    Thus, since $I$ has parameters $(\arbitraryvalue, \cellpointershort, \arbitraryvalue)$, it follows that $I'$ has parameters $(\arbitraryvalue, \cellpointershort)$.
    So, since $I'$ is an invocation of the \doapplyandcopyresponse{} procedure, by \Cref{lemma:ero:l_event_corresponding_to_do_low_level_op}, there is an $L$-apply event $e'$ for $\cellpointershort$ before $I'$ was invoked.
    Hence, since $I$ was invoked during $I'$, $a$ was executed during $I$, and $a < e$, by transitivity, $e' < e$.
    Therefore, there is an $L$-apply event for $\cellpointershort$ before $e$.
    \qH{\Cref{lemma:ero:remove_events_are_preceeded_by_apply_events}}
\end{proof}

\Cref{lemma:ero:apply_events_are_preceeded_by_add_events} and \Cref{lemma:ero:remove_events_are_preceeded_by_apply_events} imply the following.

\begin{corollary}\label{lemma:ero:remove_events_are_preceeded_by_add_events}
    Every $L$-remove event for $\cellpointershort$ in $\mathcal{I}^\mathcal{B}$ is preceded by an $L$-add event for $\cellpointershort$. 
\end{corollary}

We now prove two properties about successive $L$-events.

\begin{lemma}\label{lemma:ero:successive_l_event_read_previous_l_event_value}
    For every two successive $L$-events $e$ and $e'$ in  $\mathcal{I}^\mathcal{B}$, the process that executed $e'$ read the value that $e$ set $\linearizationobject{}$ to on its last execution of \cref{line:ero:linearization_read} \mbox{before $e'$.}
\end{lemma}

\begin{proof}
    Suppose, for contradiction, the process $p$ that executed $e'$ read a different value $v'$ on its last execution of \cref{line:ero:linearization_read} before $e'$ such that $v' \neq v$ where $e$ set $\linearizationobject{}$ to $v$.
    Since $p$ read $v'$ from $\linearizationobject{}$ on its last execution of \cref{line:ero:linearization_read} before $e'$ and $e'$ is a successful \CASop{} operation on \cref{line:ero:linearization_cas}, the value of $\linearizationobject{}$ at the step before $e'$ is $v'$.
    Hence, since $e < e'$, and $e$ set $\linearizationobject{}$ to $v \neq v'$, the value stored in $\linearizationobject{}$ changed during $(e, e')$.
    Therefore, by \Cref{observation:ero:where_objects_change}, there is an $L$-event during $(e, e')$.
    However, this contradicts the fact that $e$ and $e'$ are successive $L$-events.
    \qH{\Cref{lemma:ero:successive_l_event_read_previous_l_event_value}}
\end{proof}

\begin{lemma}\label{lemma:ero:successive_l_event_read_previous_l_event_value_after_it_happened}
    Let $\mathcal{I}^\mathcal{B}_+$ be any implementation history of $\mathcal{B}$ such that $\mathcal{I}^\mathcal{B}$ is a prefix of $\mathcal{I}^\mathcal{B}_+$.
    Furthermore, let $e$ and $e'$ be any successive $L$-events in $\mathcal{I}^\mathcal{B}_+$ such that $e$ is in $\mathcal{I}^\mathcal{B}$.
    Lastly, let $p$ be the process that executed $e'$.
    If $P(\mathcal{I}^\mathcal{B})$ holds, then $p$'s last execution of \cref{line:ero:linearization_read} before $e'$ is after $e$.
\end{lemma}

\begin{proof}
    Suppose, for contradiction, $T^{\ref{line:ero:linearization_read}} < e$ where $T^{\ref{line:ero:linearization_read}}$ is the time of $p$'s last execution of \cref{line:ero:linearization_read} before $e'$.
    Let $v$ be the value $e$ sets $\linearizationobject{}$ to.
    Hence, by \Cref{lemma:ero:l_events_set_llo_to_a_value_different_from_initial}, $v$ is not the initial value of $\linearizationobject{}$.
    Since $e$ and $e'$ are successive $L$-events, by \Cref{lemma:ero:successive_l_event_read_previous_l_event_value}, $p$ read $v$ on \cref{line:ero:linearization_read} at $T^{\ref{line:ero:linearization_read}}$.
    Hence, since $v$ is not the initial value of $\linearizationobject{}$ and $T^{\ref{line:ero:linearization_read}} < e$, we have that some step before $T^{\ref{line:ero:linearization_read}}$ set $\linearizationobject{}$ to $v$.
    Thus, by \Cref{observation:ero:where_objects_change}, some $L$-event $e^*$ before $T^{\ref{line:ero:linearization_read}}$ set $\linearizationobject{}$ to $v$.
    Hence, since $T^{\ref{line:ero:linearization_read}} < e$, by transitivity, $e^* < e$, and so $e^* \neq e$.
    Furthermore, since $e$ is in $\mathcal{I}^\mathcal{B}$ and $e^* < e$, we have that $e^*$ is in $\mathcal{I}^\mathcal{B}$.
    Therefore, $e$ and $e^*$ are two $L$-events in $\mathcal{I}^\mathcal{B}$ that set $\linearizationobject{}$ to $v$.
    However, since by assumption $P(\mathcal{I}^\mathcal{B})$ holds, by \Cref{lemma:ero:p_implies_unique_values_in_linearization}, every $L$-event in $\mathcal{I}^\mathcal{B}$ sets $\linearizationobject{}$ to a unique value, a contradiction.
    \qH{\Cref{lemma:ero:successive_l_event_read_previous_l_event_value_after_it_happened}}
\end{proof}

We now prove some facts about $S$-attempts.

\begin{lemma}\label{lemma:ero:s_attempts_have_corresponding_l_events}
    Consider an $S$-attempt $a$ which attempts to set $\stateobject.\uniquerepositoryoperationlong{}$ to some value $v$ in $\mathcal{I}^\mathcal{B}$ during some invocation $I$ of the \doapplyandcopyresponse{} procedure.
    Then, $a$'s corresponding $L$-event is an $L$-apply event which set $\linearizationobject{}.\uniquerepositoryoperationlong{}$ to $v$ before $I$ was invoked.
    % We call the last $L$-apply event which set $\linearizationobject{}.\uniquerepositoryoperationlong{}$ to $v$ before $a$, $a$'s corresponding $L$-event.
\end{lemma}

\begin{proof}
    %Consider any $S$-attempt $a$ which attempts to set $\stateobject.\uniquerepositoryoperationlong{}$ to some value $v$.
    Since $a$ tries to set $\stateobject.\uniquerepositoryoperationlong{}$ to $v$, by \Cref{def:ero:english}, the process that executed $a$ did so during an invocation of the \doapplyandcopyresponse{} procedure with a first parameter of $v$.
    Therefore, by \Cref{lemma:ero:l_event_corresponding_to_do_low_level_op}, the claim follows.
    \qH{\Cref{lemma:ero:s_attempts_have_corresponding_l_events}}
\end{proof}

By \Cref{def:ero:english}, this implies the following.

\begin{corollary}\label{corollary:ero:s_attempts_and_corresponding_l_events_are_for_matching_timestamps}
    Consider a $S$-attempt $a$ in $\mathcal{I}^\mathcal{B}$ and let $e$ be its corresponding $L$-event.
    Then, $a$ is for timestamp $\timeshort{}$ if and only if $e$ is for timestamp $\timeshort{}$.
\end{corollary}

\begin{lemma}\label{lemma:ero:every_s_attempt_is_for_timestamp_other_than_zero}
    Every $S$-attempt in $\mathcal{I}^\mathcal{B}$ is for a timestamp larger than $0$.
\end{lemma}

\begin{proof}
    Consider any $S$-attempt $a$ for some timestamp $\timeshort{}$.
    Let $e$ be $a$'s corresponding $L$-event.
    By \Cref{corollary:ero:s_attempts_and_corresponding_l_events_are_for_matching_timestamps} $e$ is for timestamp $\timeshort{}$.
    Hence, by \Cref{lemma:ero:every_l_event_is_for_timestamp_other_than_zero} $\timeshort{} > 0$ as wanted.
    \qH{\Cref{lemma:ero:every_s_attempt_is_for_timestamp_other_than_zero}}
\end{proof}

We now prove some facts about list-attempts.

\begin{lemma}\label{lemma:ero:every_list_add_seal_and_remove_attempt_is_for_ptr_from_universe}
    Every list-add, list-seal, and list-remove attempt in $\mathcal{I}^\mathcal{B}$ is for some pointer in $\celluniverse{}$.
\end{lemma}

\begin{proof}
    Consider any list-add attempt $a$ for some $\cellpointershort$.
    Hence, by \Cref{def:ero:english}, $a$ occurred during an invocation of the \doaddcell{} procedure with parameters $(\arbitraryvalue, \cellpointershort)$.
    Thus, by \Cref{lemma:ero:l_event_corresponding_to_do_low_level_op}, there is an $L$-add event for $\cellpointershort$ in $\mathcal{I}^\mathcal{B}$.
    Therefore, by \Cref{lemma:ero:every_l_event_is_for_pointer_from_universe}, $\cellpointershort \in \celluniverse{}$.
    Now, consider any list-seal or list-remove attempt $a$ for some $\cellpointershort$.
    Hence, by \Cref{def:ero:english}, $a$ occurred during an invocation of the \doremovecell{} procedure with parameters $(\arbitraryvalue, \cellpointershort)$.
    Thus, by \Cref{lemma:ero:l_event_corresponding_to_do_low_level_op}, there is an $L$-remove event for $\cellpointershort$ in $\mathcal{I}^\mathcal{B}$.
    Therefore, by \Cref{lemma:ero:every_l_event_is_for_pointer_from_universe}, $\cellpointershort \in \celluniverse{}$.
    \qH{\Cref{lemma:ero:every_list_add_seal_and_remove_attempt_is_for_ptr_from_universe}}
\end{proof}

The next few statements show that pointer fields and variables contain legitimate pointer values.

\begin{lemma}\label{lemma:ero:next_pointer_is_always_from_universe_or_null}
    For every $\cellpointershort \in \celluniverse{} \cup \{\&\headobject\}$, if $(*\cellpointershort).\nextlong.\uniquecellpointercontentlong{} = \nextcellpointershort{}$ at any time in $\mathcal{I}^\mathcal{B}$, then $\nextcellpointershort{} \in \celluniverse{} \cup \{\nullconstant\}$.
\end{lemma}

\begin{proof}
    Suppose, for contradiction, there exists a $\cellpointershort \in \celluniverse{} \cup \{\&\headobject\}$ such that at some time $T$ in $\mathcal{I}^\mathcal{B}$ $(*\cellpointershort).\nextlong.\uniquecellpointercontentlong{} = \nextcellpointershort{}$ for some $\nextcellpointershort{} \notin \celluniverse{} \cup \{\nullconstant\}$.
    Without loss of generality, suppose $T$ is the first time the lemma is violated for \Underline{any} pointer in $\celluniverse{} \cup \{\&\headobject\}$.
    Since $(*\cellpointershort).\nextlong.\uniquecellpointercontentlong{}$ is initially $\nullconstant$, $(*\cellpointershort).\nextlong.\uniquecellpointercontentlong{} = \nextcellpointershort{} \neq \nullconstant$ at $T$, and $T$ is the first time the lemma is violated in $\mathcal{I}^\mathcal{B}$, it follows that the step at $T$ sets the value of $(*\cellpointershort).\nextlong.\uniquecellpointercontentlong{}$ to $\nextcellpointershort{}$.
    Hence, by \Cref{observation:ero:where_objects_change}, the step at $T$ is either a successful list-add attempt for $\nextcellpointershort{}$ after $\cellpointershort$ or a successful list-remove attempt between $\cellpointershort$ and $\nextcellpointershort{}$.
    We consider each case separately.
    \begin{itemize}
        \item[] \hspace{0pt}\textbf{Case 1.} The step at $T$ is a successful list-add attempt for $\nextcellpointershort{}$ after $\cellpointershort$.

        Therefore, by \Cref{lemma:ero:every_list_add_seal_and_remove_attempt_is_for_ptr_from_universe}, $\nextcellpointershort{} \in \celluniverse{}$, contradicting $\nextcellpointershort{} \notin \celluniverse{} \cup \{\nullconstant\}$.

        \item[] \hspace{0pt}\textbf{Case 2.} The step at $T$ is a successful list-remove attempt between $\cellpointershort$ and $\nextcellpointershort{}$.

        Suppose this step is for $\cellpointershort'$.
        Hence, by \Cref{lemma:ero:every_list_add_seal_and_remove_attempt_is_for_ptr_from_universe}, $\cellpointershort' \in \celluniverse{}$.
        Let $p$ be the process that performed the step at $T$.
        Hence, $p$ read $\nextcellpointershort{}$ from $(*\cellpointershort').\nextlong.\uniquecellpointercontentlong{}$ on its last execution of \cref{line:ero:remove_cell_read_pointer_to_remove} before $T$.
        Therefore, since $\cellpointershort' \in \celluniverse{}$ and $p$ read $\nextcellpointershort{}$ from $(*\cellpointershort').\nextlong.\uniquecellpointercontentlong{}$ before $T$, by the minimality of $T$, $\nextcellpointershort{} \in \celluniverse{} \cup \{\nullconstant\}$.
        However, $\nextcellpointershort{} \notin \celluniverse{} \cup \{\nullconstant\}$, a contradiction.
        \qH{\Cref{lemma:ero:next_pointer_is_always_from_universe_or_null}}
    \end{itemize}
\end{proof}

\begin{lemma}\label{lemma:ero:add_cell_current_pointer_always_in_universe_or_head}
    Let $I$ be any invocation of the \doaddcell{} procedure in $\mathcal{I}^\mathcal{B}$ by some process $p$ and let $T^{\ref{line:ero:add_cell_initial_current_pointer}}$ be the time $p$ executed \cref{line:ero:add_cell_initial_current_pointer} during $I$ (assuming $p$ does).
    At all times at or after $T^{\ref{line:ero:add_cell_initial_current_pointer}}$ and before $I$ returns, the value of the local variable $\currentuniquecellpointershort{}$ in $I$ is in $\celluniverse{} \cup \{\&\headobject\}$.
\end{lemma}

\begin{proof}
    Suppose, for contradiction, there is a time $T$ at or after $T^{\ref{line:ero:add_cell_initial_current_pointer}}$ and before $I$ returns (if it ever does) such that the value of the local variable $\currentuniquecellpointershort{}$ in $I$ is $\cellpointershort \notin \celluniverse{} \cup \{\&\headobject\}$.
    Without loss of generality, suppose $T$ is the first such time.
    Since $p$ executed \cref{line:ero:add_cell_initial_current_pointer} at $T^{\ref{line:ero:add_cell_initial_current_pointer}}$ during $I$, the value of $\currentuniquecellpointershort{}$ is $\&\headobject$ at $T^{\ref{line:ero:add_cell_initial_current_pointer}}$.
    Hence, since the value of $\currentuniquecellpointershort{}$ is $\cellpointershort \notin \celluniverse{} \cup \{\&\headobject\}$ at $T \geq T^{\ref{line:ero:add_cell_initial_current_pointer}}$, it follows that the value of $\currentuniquecellpointershort{}$ was set to $\cellpointershort$ at $T$.
    Thus, since the value of $\currentuniquecellpointershort{}$ only changes on \cref{line:ero:add_cell_update_current_pointer} after $T^{\ref{line:ero:add_cell_initial_current_pointer}}$ during $I$, we have that $p$ set $\currentuniquecellpointershort{}$ to $\cellpointershort$ by executing \cref{line:ero:add_cell_update_current_pointer} at $T$.
    So, the value of the local variable $\nextuniquecellpointershort{}$ in $I$ is $\cellpointershort$ at $T$.
    Therefore, the response of the invocation $I'$ of the AcquireNext procedure on \cref{line:ero:add_cell_acquire_next} during the same iteration of the while loop on \cref{line:ero:add_cell_while_loop} as $T$ is $(\found, \cellpointershort)$.
    Let $T^{\ref{line:ero:add_cell_acquire_next}}$ be the time $p$ invoked $I'$, and let $\cellpointershort'$ be the second parameter of $I'$.
    Hence, the value of $\currentuniquecellpointershort{}$ is $\cellpointershort'$ at $T^{\ref{line:ero:add_cell_acquire_next}}$.
    Thus, since $p$ invoked $I'$ at $T^{\ref{line:ero:add_cell_acquire_next}}$ strictly before $T$, by the minimality of $T$, $\cellpointershort' \in \celluniverse{} \cup \{\&\headobject\}$.
    Since the second parameter of $I'$ is $\cellpointershort'$, and the response of $I'$ is $(\found, \cellpointershort)$, it follows that $p$ read $\cellpointershort$ from $(*\cellpointershort').\nextlong.\uniquecellpointercontentlong{}$ on the last execution of \cref{line:ero:acquire_next_read_curr_unique_pointer} during $I'$; say at time $T^{\ref{line:ero:acquire_next_read_curr_unique_pointer}}$.
    Hence, since $\cellpointershort' \in \celluniverse{} \cup \{\&\headobject\}$, by \Cref{lemma:ero:next_pointer_is_always_from_universe_or_null}, $\cellpointershort \in \celluniverse{} \cup \{\nullconstant\}$.
    Therefore, since $\cellpointershort \notin \celluniverse{} \cup \{\&\headobject\}$, we have that $\cellpointershort = \nullconstant$.
    However, since $p$ exited $I'$ with response $(\found, \cellpointershort)$, we have that $p$ found the clause on \cref{line:ero:acquire_next_not_found_check} to be false on its last execution of \cref{line:ero:acquire_next_not_found_check} during $I'$, and since $p$ read $\cellpointershort$ from $(*\cellpointershort').\nextlong.\uniquecellpointercontentlong{}$ on its last execution of \cref{line:ero:acquire_next_read_curr_unique_pointer} during $I'$, this implies that $\cellpointershort \neq \nullconstant$, a contradiction.    
    \qH{\Cref{lemma:ero:add_cell_current_pointer_always_in_universe_or_head}}
\end{proof}

\begin{lemma}\label{lemma:ero:every_list_add_attempt_is_after_a_pointer_from_universe_or_head}
    Every list-add attempt in $\mathcal{I}^\mathcal{B}$ is after some pointer in $\celluniverse{} \cup \{\&\headobject\}$.
\end{lemma}

\begin{proof}
    Consider any list-add attempt $a$ for any pointer after some pointer $\cellpointershort$ by some process $p$.
    Let $I$ be the invocation of the \doaddcell{} procedure that $p$ executed $a$ during.
    Hence, $p$ executed \cref{line:ero:add_cell_initial_current_pointer} before $a$ during $I$.
    Furthermore, since $a$ is a list-add attempt after $\cellpointershort$, the value of the local variable $\currentuniquecellpointershort{}$ in $I$ is $\cellpointershort$ at $a$.
    Therefore, by \Cref{lemma:ero:add_cell_current_pointer_always_in_universe_or_head} $\cellpointershort \in \celluniverse{} \cup \{\&\headobject\}$, and so $a$ is after some pointer in $\celluniverse{} \cup \{\&\headobject\}$ as wanted.
    \qH{\Cref{lemma:ero:every_list_add_attempt_is_after_a_pointer_from_universe_or_head}}
\end{proof}

Below, we bound the domain of values that a list-remove attempt can be between.

\begin{lemma}\label{lemma:ero:remove_cell_current_pointer_always_in_universe_or_head}
    Let $I$ be any invocation of the \doremovecell{} procedure in $\mathcal{I}^\mathcal{B}$ by some process $p$ and let $T^{\ref{line:ero:remove_cell_initialize_pointers}}$ be the time $p$ executed \cref{line:ero:remove_cell_initialize_pointers} during $I$ (assuming $p$ does).
    At all times at or after $T^{\ref{line:ero:remove_cell_initialize_pointers}}$ and before $I$ returns, the value of the local variable $\currentuniquecellpointershort{}$ in $I$ is in $\celluniverse{} \cup \{\&\headobject\}$.
\end{lemma}

\begin{proof}
    By essentially the same argument as \Cref{lemma:ero:add_cell_current_pointer_always_in_universe_or_head}, which we provide below for completeness.
    Suppose, for contradiction, there is a time $T$ at or after $T^{\ref{line:ero:remove_cell_initialize_pointers}}$ and before $I$ returns (if it ever does) such that the value of the local variable $\currentuniquecellpointershort{}$ in $I$ is $\cellpointershort \notin \celluniverse{} \cup \{\&\headobject\}$.
    Without loss of generality, suppose $T$ is the first such time.
    Since $p$ executed \cref{line:ero:remove_cell_initialize_pointers} at $T^{\ref{line:ero:remove_cell_initialize_pointers}}$ during $I$, the value of $\currentuniquecellpointershort{}$ is $\&\headobject$ at $T^{\ref{line:ero:remove_cell_initialize_pointers}}$.
    Hence, since the value of $\currentuniquecellpointershort{}$ is $\cellpointershort \notin \celluniverse{} \cup \{\&\headobject\}$ at $T \geq T^{\ref{line:ero:remove_cell_initialize_pointers}}$, it follows that the value of $\currentuniquecellpointershort{}$ was set to $\cellpointershort$ at $T$.
    Thus, since the value of $\currentuniquecellpointershort{}$ only changes on \cref{line:ero:remove_cell_update_pointers} after $T^{\ref{line:ero:remove_cell_initialize_pointers}}$ during $I$, we have that $p$ set $\currentuniquecellpointershort{}$ to $\cellpointershort$ by executing \cref{line:ero:remove_cell_update_pointers} at $T$.
    So, the value of the local variable $\nextuniquecellpointershort{}$ in $I$ is $\cellpointershort$ at $T$.
    Therefore, the response of the invocation $I'$ of the AcquireNext procedure on \cref{line:ero:remove_cell_acquire_next} during the same iteration of the while loop on \cref{line:ero:remove_cell_while_loop} as $T$ is $(\found, \cellpointershort)$.
    Let $T^{\ref{line:ero:remove_cell_acquire_next}}$ be the time $p$ invoked $I'$, and let $\cellpointershort'$ be the second parameter of $I'$.
    Hence, the value of $\currentuniquecellpointershort{}$ is $\cellpointershort'$ at $T^{\ref{line:ero:remove_cell_acquire_next}}$.
    Thus, since $p$ invoked $I'$ at $T^{\ref{line:ero:remove_cell_acquire_next}}$ strictly before $T$, by the minimality of $T$, $\cellpointershort' \in \celluniverse{} \cup \{\&\headobject\}$.
    Since the second parameter of $I'$ is $\cellpointershort'$, and the response of $I'$ is $(\found, \cellpointershort)$, it follows that $p$ read $\cellpointershort$ from $(*\cellpointershort').\nextlong.\uniquecellpointercontentlong{}$ on the last execution of \cref{line:ero:acquire_next_read_curr_unique_pointer} during $I'$; say at time $T^{\ref{line:ero:acquire_next_read_curr_unique_pointer}}$.
    Hence, since $\cellpointershort' \in \celluniverse{} \cup \{\&\headobject\}$, by \Cref{lemma:ero:next_pointer_is_always_from_universe_or_null}, $\cellpointershort \in \celluniverse{} \cup \{\nullconstant\}$.
    Therefore, since $\cellpointershort \notin \celluniverse{} \cup \{\&\headobject\}$, we have that $\cellpointershort = \nullconstant$.
    However, since $p$ exited $I'$ with response $(\found, \cellpointershort)$, we have that $p$ found the clause on \cref{line:ero:acquire_next_not_found_check} to be false on its last execution of \cref{line:ero:acquire_next_not_found_check} during $I'$, and since $p$ read $\cellpointershort$ from $(*\cellpointershort').\nextlong.\uniquecellpointercontentlong{}$ on its last execution of \cref{line:ero:acquire_next_read_curr_unique_pointer} during $I'$, this implies that $\cellpointershort \neq \nullconstant$, a contradiction.
    \qH{\Cref{lemma:ero:remove_cell_current_pointer_always_in_universe_or_head}}
\end{proof}

\begin{lemma}\label{lemma:ero:remove_cell_previous_pointer_always_in_universe_or_head}
    Let $I$ be any invocation of the \doremovecell{} procedure in $\mathcal{I}^\mathcal{B}$ by some process $p$ and let $T^{\ref{line:ero:remove_cell_update_pointers}}$ be the first time $p$ executed \cref{line:ero:remove_cell_update_pointers} during $I$ (assuming $p$ does).
    At all times at or after $T^{\ref{line:ero:remove_cell_update_pointers}}$ and before $I$ returns, the value of the local variable $\previousuniquecellpointershort{}$ in $I$ is in $\celluniverse{} \cup \{\&\headobject\}$.
\end{lemma}

\begin{proof}
    Suppose $I$ is an invocation of the \doremovecell{} procedure by some process $p$ and $T^{\ref{line:ero:remove_cell_update_pointers}}$ is the first time $p$ executed \cref{line:ero:remove_cell_update_pointers} during $I$.
    Hence, $p$ executed \cref{line:ero:remove_cell_initialize_pointers} before $T^{\ref{line:ero:remove_cell_update_pointers}}$; say at time $T^{\ref{line:ero:remove_cell_initialize_pointers}} < T^{\ref{line:ero:remove_cell_update_pointers}}$.
    Thus, by \Cref{lemma:ero:remove_cell_current_pointer_always_in_universe_or_head} at all times at or after $T^{\ref{line:ero:remove_cell_initialize_pointers}}$ and before $I$ returns, the value of the local variable $\currentuniquecellpointershort{}$ in $I$ is in $\celluniverse{} \cup \{\&\headobject\}$.
    Therefore, since every time $p$ sets the value of $\previouscellpointershort{}$ at or after $T^{\ref{line:ero:remove_cell_update_pointers}}$ during $I$ it is to the value of $\currentcellpointershort{}$ (see \cref{line:ero:remove_cell_update_pointers}) the claim follows.
    \qH{\Cref{lemma:ero:remove_cell_previous_pointer_always_in_universe_or_head}}
\end{proof}

This implies the following.

\begin{corollary}\label{lemma:ero:remove_cell_previous_pointer_always_in_universe_or_head_or_null}
    Let $I$ be any invocation of the \doremovecell{} procedure in $\mathcal{I}^\mathcal{B}$ by some process $p$ and let $T^{\ref{line:ero:remove_cell_initialize_pointers}}$ be the time $p$ executed \cref{line:ero:remove_cell_initialize_pointers} during $I$ (assuming $p$ does).
    At all times at or after $T^{\ref{line:ero:remove_cell_initialize_pointers}}$ and before $I$ returns, the value of the local variable $\previousuniquecellpointershort{}$ in $I$ is in $\celluniverse{} \cup \{\&\headobject, \nullconstant\}$.
\end{corollary}

\begin{lemma}\label{lemma:ero:every_list_remove_attempt_is_between_a_pointer_from_universe_or_header_and_a_pointer_from_universe_or_null}
    Every list-remove attempt in $\mathcal{I}^\mathcal{B}$ is between some pointer in $\celluniverse{} \cup \{\&\headobject\}$ and some pointer in $\celluniverse{} \cup \{\nullconstant\}$.
\end{lemma}

\begin{proof}
    Consider any list-remove attempt $a$ for some $\cellpointershort$ between some $\previouscellpointershort{}$ and some $\nextcellpointershort{}$.
    Hence, by \Cref{lemma:ero:every_list_add_seal_and_remove_attempt_is_for_ptr_from_universe}, $\cellpointershort \in \celluniverse{}$, so by \Cref{assumption:ero:head_and_null_not_in_cell_universe} $\cellpointershort \neq \&\headobject$.
    Furthermore, by \Cref{def:ero:english}, some process $p$ executed $a$ during some invocation $I$ of the \doremovecell{} procedure with parameters $(\arbitraryvalue, \cellpointershort)$.
    Since the value of the local variable $\currentcellpointershort{}$ is initially $\&\headobject$ in $I$, and $\cellpointershort \neq \&\headobject$, we have that $p$ found the condition on \cref{line:ero:remove_cell_while_loop} to be true at least once in $I$.
    Hence, since $p$ executes \cref{line:ero:remove_cell_from_list} in $I$ (because $p$ executed $a$ during $I$), we have that $p$ executes \cref{line:ero:remove_cell_update_pointers} at least once in $I$; say at time $T^{\ref{line:ero:remove_cell_update_pointers}}$.
    Thus, by \Cref{lemma:ero:remove_cell_previous_pointer_always_in_universe_or_head}, the value of the local variable $\previousuniquecellpointershort{}$ in $I$ is in $\celluniverse{} \cup \{\&\headobject\}$ at all times from $T^{\ref{line:ero:remove_cell_update_pointers}}$ until $I$ returns (if ever).
    Therefore, since $a$ is a list-remove attempt between $\previouscellpointershort{}$ and $\nextcellpointershort{}$, and $p$ executed $a$ during $I$, we have that $\previouscellpointershort{}$ is the value of the local variable $\previouscellpointershort{}$ at $a$ during $I$, and since $a$ is after $T^{\ref{line:ero:remove_cell_update_pointers}}$, we have that $\previouscellpointershort{} \in \celluniverse{} \cup \{\&\headobject\}$ as wanted.
    Furthermore, since $a$ is a list-remove attempt for $\cellpointershort$ between $\previouscellpointershort{}$ and $\nextcellpointershort{}$, we have that $p$ read $\nextcellpointershort{}$ from $(*\cellpointershort).\nextlong.\uniquecellpointercontentlong{}$ on its last execution of \cref{line:ero:remove_cell_read_pointer_to_remove} before executing $a$ in $I$, and so since $\cellpointershort \neq \&\headobject$, by \Cref{lemma:ero:next_pointer_is_always_from_universe_or_null}, $\nextcellpointershort{} \in \celluniverse{} \cup \{\nullconstant\}$.    \qH{\Cref{lemma:ero:every_list_remove_attempt_is_between_a_pointer_from_universe_or_header_and_a_pointer_from_universe_or_null}}
\end{proof}

We now prove that invariants $P$ and $Q$ imply some useful facts about list-add and list-remove attempts.

\begin{lemma}\label{lemma:ero:list_add_attempt_for_some_pointer_are_after_same_pointer}
    Let $a_1$ and $a_2$ be two list-add attempts for some $\cellpointershort$ in $\mathcal{I}^\mathcal{B}$.
    If $Q(\mathcal{I}^\mathcal{B})$ holds, then $a_1$ and $a_2$ are after the same $\currentcellpointershort{}$.
\end{lemma}

\begin{proof}
    Suppose $a_1$ (resp. $a_2$) is after $\currentcellpointershort{}_1$ (resp. $\currentcellpointershort{}_2$).
    Hence, since $a_1$ (resp. $a_2$) is a list-add attempts for $\cellpointershort$, by $Q(\mathcal{I}^\mathcal{B})$, there is a unique $L$-add event $e$ for $\cellpointershort$ before $a_1$ (resp. $a_2$), thus $e$ is the same for $a_1$ and $a_2$, and if $\mathcal{I}$ is the prefix of $\mathcal{I}^\mathcal{B}$ up to but excluding $e$, $\currentcellpointershort{}_1$ (resp. $\currentcellpointershort{}_2$) is the second last pointer in $\List(\mathcal{I})$.
    Therefore, $\currentcellpointershort{}_1 = \currentcellpointershort{}_2$.
    \qH{\Cref{lemma:ero:list_add_attempt_for_some_pointer_are_after_same_pointer}}
\end{proof}

\begin{lemma}\label{lemma:ero:list_add_attempt_has_different_after}
    Let $a$ be a list-add attempt for some $\cellpointershort$ after some $\currentcellpointershort{}$ in $\mathcal{I}^\mathcal{B}$.
    If $P(\mathcal{I}^\mathcal{B})$ and $Q(\mathcal{I}^\mathcal{B})$ hold, then $\currentcellpointershort{} \neq \cellpointershort$.
\end{lemma}

\begin{proof}
    Suppose, for contradiction, $\currentcellpointershort{} = \cellpointershort$.
    Since $a$ is a list-add attempt for $\cellpointershort$, by \Cref{lemma:ero:every_list_add_seal_and_remove_attempt_is_for_ptr_from_universe} $\cellpointershort \in \celluniverse{}$.
    Furthermore, by $Q(\mathcal{I}^\mathcal{B})$, there is a unique $L$-add event $e$ for $\cellpointershort$ before $a$, and if $\mathcal{I}$ is the prefix of $\mathcal{I}^\mathcal{B}$ up to but excluding $e$, then $\currentcellpointershort{}$ is the second last pointer in $\List(\mathcal{I})$.
    Hence, since $\currentcellpointershort{} = \cellpointershort$, we have that $\cellpointershort$ is the second last pointer in $\List(\mathcal{I})$.
    Thus, since $\cellpointershort \in \celluniverse{}$, by \Cref{assumption:ero:head_and_null_not_in_cell_universe} $\cellpointershort \neq \&\headobject$, and so by the definition of $\List(\mathcal{I})$ (see \Cref{def:ero:logical_list}), there is an $L$-add event $e'$ for $\cellpointershort$ in $\mathcal{I}$.
    Hence, since $e$ is an $L$-add event for $\cellpointershort$, and $\mathcal{I}$ is the prefix of $\mathcal{I}^\mathcal{B}$ up to but excluding $e$, it follows that $e \neq e'$.
    Therefore, since both $e$ and $e'$ are $L$-add events for $\cellpointershort$ in $\mathcal{I}^\mathcal{B}$, there are two $L$-add events for $\cellpointershort$ in $\mathcal{I}^\mathcal{B}$.
    However, since $\cellpointershort \in \celluniverse{}$, by $P(\mathcal{I}^\mathcal{B})$, there is at most one $L$-add event for $\cellpointershort$ in $\mathcal{I}^\mathcal{B}$, a contradiction.
    \qH{\Cref{lemma:ero:list_add_attempt_has_different_after}}
\end{proof}

\begin{lemma}\label{lemma:ero:list_remove_attempt_for_some_pointer_are_between_same_pointer}
    Let $a_1$ and $a_2$ be two list-remove attempts for some $\cellpointershort$ in $\mathcal{I}^\mathcal{B}$.
    If $Q(\mathcal{I}^\mathcal{B})$ holds, then $a_1$ and $a_2$ are between the same $\previouscellpointershort{}$ and $\nextcellpointershort{}$.
\end{lemma}

\begin{proof}
    Suppose $a_1$ (resp. $a_2$) is between $\previouscellpointershort{}_1$ (resp. $\previouscellpointershort{}_2$) and $\nextcellpointershort{}_1$ (resp. $\nextcellpointershort{}_2$).
    Hence, since $a_1$ (resp. $a_2$) is a list-remove attempts for $\cellpointershort$, by $Q(\mathcal{I}^\mathcal{B})$, there is a unique $L$-remove event $e$ for $\cellpointershort$ before $a_1$ (resp. $a_2$), thus $e$ is the same for $a_1$ and $a_2$, and if $\mathcal{I}$ is the prefix of $\mathcal{I}^\mathcal{B}$ up to but excluding $e$, $\cellpointershort$ is in $\List(\mathcal{I})$ exactly once, and $\previouscellpointershort{}_1$ (resp. $\previouscellpointershort{}_2$) and $\nextcellpointershort{}_1$ (resp. $\nextcellpointershort{}_2$) are the pointers preceding and succeeding $\cellpointershort$ in $\List(\mathcal{I})$.
    Therefore $\previouscellpointershort{}_1 = \previouscellpointershort{}_2$ and $\nextcellpointershort{}_1 = \nextcellpointershort{}_2$ as wanted.
    \qH{\Cref{lemma:ero:list_remove_attempt_for_some_pointer_are_between_same_pointer}}
\end{proof}

\begin{lemma}\label{lemma:ero:list_remove_attempt_has_different_next}
    Let $a$ be a list-remove attempt for some $\cellpointershort$ between some $\previouscellpointershort{}$ and some $\nextcellpointershort{}$ in $\mathcal{I}^\mathcal{B}$.
    If $P(\mathcal{I}^\mathcal{B})$ and $Q(\mathcal{I}^\mathcal{B})$ hold, then $\previouscellpointershort{}$, $\cellpointershort$, and $\nextcellpointershort{}$ are distinct.
\end{lemma}

\begin{proof}
    Suppose, for contradiction, either $\previouscellpointershort{} = \cellpointershort$, $\cellpointershort = \nextcellpointershort{}$, or $\previouscellpointershort{} = \nextcellpointershort{}$.
    Since $a$ is a list-remove attempt for $\cellpointershort$ between $\previouscellpointershort{}$ and $\nextcellpointershort{}$, by \Cref{lemma:ero:every_list_add_seal_and_remove_attempt_is_for_ptr_from_universe} $\cellpointershort \in \celluniverse{}$ and by \Cref{lemma:ero:every_list_remove_attempt_is_between_a_pointer_from_universe_or_header_and_a_pointer_from_universe_or_null} $\previouscellpointershort{} \in \celluniverse{} \cup \{\&\headobject\}$ and $\nextcellpointershort{} \in \celluniverse{} \cup \{\nullconstant\}$.
    Hence, since $a$ is in $\mathcal{I}^\mathcal{B}$, by $Q(\mathcal{I}^\mathcal{B})$, there is a unique $L$-remove event $e$ for $\cellpointershort$ before $a$, and if $\mathcal{I}$ is the prefix of $\mathcal{I}^\mathcal{B}$ up to but excluding $e$, then $\cellpointershort \in \List(\mathcal{I})$ exactly once and $\previouscellpointershort{}$ and $\nextcellpointershort{}$ are the pointers preceding and succeeding $\cellpointershort$ in $\List(\mathcal{I})$.

    We prove that some $\cellpointershort^* \in \celluniverse{}$ occurs twice in $\List(\mathcal{I})$.
    There are two cases.
    \begin{itemize}
        \item[] \hspace{0pt}\textbf{Case 1.} $\previouscellpointershort{} = \cellpointershort$ or $\cellpointershort = \nextcellpointershort{}$.

        Hence, since $\previouscellpointershort{}$ is the pointer preceding $\cellpointershort$ in $\List(\mathcal{I})$, and $\nextcellpointershort{}$ is the pointer succeeding $\cellpointershort$ in $\List(\mathcal{I})$, we have that $\cellpointershort$ occurs twice in $\List(\mathcal{I})$ and is in $\celluniverse{}$.

        \item[] \hspace{0pt}\textbf{Case 2.} $\previouscellpointershort{} = \nextcellpointershort{}$.

        Since $\previouscellpointershort{} \in \celluniverse{} \cup \{\&\headobject\}$, $\nextcellpointershort{} \in \celluniverse{} \cup \{\nullconstant\}$, and by \Cref{assumption:ero:head_and_null_not_in_cell_universe} $\&\headobject \neq \nullconstant$, we have that $\previouscellpointershort{} \in \celluniverse{}$.
        Hence, since $\previouscellpointershort{}$ is the pointer preceding $\cellpointershort$ in $\List(\mathcal{I})$, and $\nextcellpointershort{}$ is the pointer succeeding $\cellpointershort$ in $\List(\mathcal{I})$, we have that $\previouscellpointershort{}$ occurs twice in $\List(\mathcal{I})$ and is in $\celluniverse{}$.
    \end{itemize}

    We now finish the proof of \Cref{lemma:ero:list_remove_attempt_has_different_next}.
    Since $\cellpointershort^* \in \celluniverse{}$, by \Cref{assumption:ero:head_and_null_not_in_cell_universe}, $\cellpointershort^* \neq \&\headobject$ and $\cellpointershort^* \neq \nullconstant$.
    Thus, since $\cellpointershort^*$ occurs twice in $\List(\mathcal{I})$, by the definition of $\List(\mathcal{I})$ (see \Cref{def:ero:logical_list}), there are two $L$-add events for $\cellpointershort^*$ in $\mathcal{I}$.
    Therefore, since $\mathcal{I}$ is a prefix of $\mathcal{I}^\mathcal{B}$, there are two $L$-add events for $\cellpointershort^*$ in $\mathcal{I}^\mathcal{B}$.
    However, since $\cellpointershort^* \in \celluniverse{}$, by $P(\mathcal{I}^\mathcal{B})$, there is at most one $L$-add event for $\cellpointershort^*$ in $\mathcal{I}^\mathcal{B}$, a contradiction.
    \qH{\Cref{lemma:ero:list_remove_attempt_has_different_next}}
\end{proof}

Lastly, we prove some facts relating list-add and list-remove attempts with list-seal attempts.

\begin{lemma}\label{lemma:ero:list_seal_before_list_remove}
    Consider any list-remove attempt $a_{remove}$ for $\cellpointershort$ in $\mathcal{I}^\mathcal{B}$.
    Let $p$ be the process that executed $a_{remove}$ and let $T^{\ref{line:ero:remove_cell_read_pointer_to_remove}}$ be the time of $p$'s last execution of \cref{line:ero:remove_cell_read_pointer_to_remove} before $a_{remove}$.
    There is a successful list-seal attempt for $\cellpointershort$ before $T^{\ref{line:ero:remove_cell_read_pointer_to_remove}}$ in $\mathcal{I}^\mathcal{B}$.
\end{lemma}

\begin{proof}
    By \Cref{lemma:ero:every_list_add_seal_and_remove_attempt_is_for_ptr_from_universe}, $\cellpointershort \in \celluniverse$.
    Let $I$ be the invocation of the \doremovecell{} procedure that $p$ executed $a_{remove}$ during.
    Since $a_{remove}$ is a list-remove attempt for $\cellpointershort$ during $I$, by \Cref{def:ero:english}, the second parameter of $I$ is $\cellpointershort$.
    Furthermore, since $p$ executed $a_{remove}$ during $I$, $p$ found the condition on \cref{line:ero:remove_cell_remove_seal_loop} to be false before $T^{\ref{line:ero:remove_cell_read_pointer_to_remove}}$ during $I$.
    Hence, since the second parameter of $I$ is $\cellpointershort$, we have that $(*\cellpointershort).\nextlong{}.sealed \neq \false$ at some time $T$ before $T^{\ref{line:ero:remove_cell_read_pointer_to_remove}}$.
    Thus, since $\cellpointershort \in \celluniverse$, we have that $(*\cellpointershort).\nextlong{}.sealed$ is initially $\false$, and so $(*\cellpointershort).\nextlong{}.sealed$ was changed before $T$.
    Hence, by \Cref{observation:ero:where_objects_change}, there is a successful list-seal attempt for $\cellpointershort$ before $T$.
    Therefore, since $T < T^{\ref{line:ero:remove_cell_read_pointer_to_remove}}$, there is a successful list-seal attempt for $\cellpointershort$ before $T^{\ref{line:ero:remove_cell_read_pointer_to_remove}}$ in $\mathcal{I}^\mathcal{B}$.
    \qH{\Cref{lemma:ero:list_seal_before_list_remove}}
\end{proof}

\begin{lemma}\label{lemma:ero:sealed_is_forever}
    Consider any successful list-seal attempt $a_{seal}$ for $\cellpointershort$ in $\mathcal{I}^\mathcal{B}$.
    By \Cref{lemma:ero:every_list_add_seal_and_remove_attempt_is_for_ptr_from_universe}, $\currentcellpointershort \in \celluniverse$.
    Then, from $a_{seal}$ onwards in $\mathcal{I}^\mathcal{B}$ $(*\currentcellpointershort).\nextlong{}.sealed = \true{}$.
\end{lemma}

\begin{proof}
    By \Cref{def:ero:english}, $a_{seal}$ sets $(*\currentcellpointershort).\nextlong{}.sealed = \true$.
    Hence, since $\currentcellpointershort \in \celluniverse$, by \Cref{observation:ero:where_objects_change}, the only steps that change the value of $(*\currentcellpointershort).\nextlong{}.sealed$ are successful list-sealed attempts for $\currentcellpointershort$, and since every successful list-sealed attempt for $\currentcellpointershort$ sets $(*\currentcellpointershort).\nextlong{}.sealed = \true{}$, the claim follows.
    \qH{\Cref{lemma:ero:sealed_is_forever}}
\end{proof}

\begin{lemma}\label{lemma:ero:no_list_seal_before_add}
    Consider any successful list-add attempt $a_{add}$ after $\currentcellpointershort$ in $\mathcal{I}^\mathcal{B}$.
    There are no successful list-seal attempts for $\currentcellpointershort$ before $a_{add}$ in $\mathcal{I}^\mathcal{B}$.
\end{lemma}

\begin{proof}
    Suppose, for contradiction, there is a successful list-seal attempt $a_{seal}$ for $\currentcellpointershort$ before $a_{add}$ in $\mathcal{I}^\mathcal{B}$.
    By \Cref{lemma:ero:sealed_is_forever}, from $a_{seal}$ onwards in $\mathcal{I}^\mathcal{B}$ $(*\currentcellpointershort).\nextlong{}.sealed = \true{}$.
    Therefore, since $a_{seal} < a_{add}$, we have that $(*\currentcellpointershort).\nextlong{}.sealed = \true{}$ at $a_{add}$.
    However, since $a_{add}$ is a successful list-add attempt after $\currentcellpointershort$, it follows that $(*\currentcellpointershort).\nextlong{}.sealed = \false{}$ at $a_{add}$, a contradiction.
    \qH{\Cref{lemma:ero:no_list_seal_before_add}}
\end{proof}

\begin{lemma}\label{lemma:ero:no_list_seal_before_remove}
    Consider any successful list-remove attempt $a_{remove}$ between $\currentcellpointershort$ and some pointer in $\mathcal{I}^\mathcal{B}$.
    There are no successful list-seal attempts for $\currentcellpointershort$ before $a_{remove}$ in $\mathcal{I}^\mathcal{B}$.
\end{lemma}

\begin{proof}
    Suppose, for contradiction, there is a successful list-seal attempt $a_{seal}$ for $\currentcellpointershort$ before $a_{remove}$ in $\mathcal{I}^\mathcal{B}$.
    By \Cref{lemma:ero:sealed_is_forever}, from $a_{seal}$ onwards in $\mathcal{I}^\mathcal{B}$ $(*\currentcellpointershort).\nextlong{}.sealed = \true{}$.
    Therefore, since $a_{seal} < a_{remove}$, we have that $(*\currentcellpointershort).\nextlong{}.sealed = \true{}$ at $a_{remove}$.
    However, since $a_{remove}$ is a successful list-remove attempt between $\currentcellpointershort$ and some pointer, it follows that $(*\currentcellpointershort).\nextlong{}.sealed = \false{}$ at $a_{remove}$, a contradiction.
    \qH{\Cref{lemma:ero:no_list_seal_before_add}}
\end{proof}

\subsubsection{Response-reset events and response-set attempts}

We now prove some facts about response-reset events.

\begin{lemma}\label{lemma:ero:every_response_reset_is_add_apply_or_remove}
    Every response-reset event in $\mathcal{I}^\mathcal{B}$ is for some pointer in $\celluniverse{}$ and is either an add-response-reset, apply-response-reset, or remove-response-reset event.
\end{lemma}

\begin{proof}
    Consider any response-reset event $e$ for some $\cellpointershort$ by some process $p$.
    By \Cref{def:ero:english}, $e$ is an execution of \cref{line:ero:do_work_initialize_response} during some invocation $I$ of the \doworkuntildone{} procedure with a second parameter of $\cellpointershort$.
    Thus, since this procedure is only invoked on lines \ref{line:ero:low_level_add_cell}, \ref{line:ero:low_level_apply_and_copy_response}, and \ref{line:ero:low_level_remove_cell}, $p$ performed an $\allocatecelloperation$ whose response is $\cellpointershort$ before invoking $I$.
    Therefore, by \Cref{alg:lazy_cell_manager_specification}, $\cellpointershort \in \celluniverse{}$.
    Now suppose $e$ sets the value of $(*\cellpointershort).\lastrepositoryoperationresponse{} = (\myuniquerepositoryoperationshort{}, \nullconstant)$.
    By \cref{line:ero:create_unique_low_level_operation}, $\myuniquerepositoryoperationshort{} = (\arbitraryvalue, \myrepositoryoperationshort)$ where $\myrepositoryoperationshort$ is the first parameter of $I$.
    Hence, $e$ set $(*\cellpointershort).\lastrepositoryoperationresponse{} = ((\arbitraryvalue, \myrepositoryoperationshort), \nullconstant)$.
    Since \doworkuntildone{} is only invoked on lines \ref{line:ero:low_level_add_cell}, \ref{line:ero:low_level_apply_and_copy_response}, and \ref{line:ero:low_level_remove_cell}, $\myrepositoryoperationshort$ is either $\addcell$, $\langle \doopandcopyresponse{}, \arbitraryvalue \rangle$, or $\removecell$.
    Therefore, by \Cref{def:ero:english}, $e$ is either an add-response-reset, apply-response-reset, or remove-response-reset event as wanted.
    \qH{\Cref{lemma:ero:every_response_reset_is_add_apply_or_remove}}
\end{proof}

\begin{proposition}\label{lemma:ero:at_most_one_add_reset_per_pointer}
    There is at most one add-response-reset event for $\cellpointershort$ in $\mathcal{I}^\mathcal{B}$.
\end{proposition}

\begin{proof}
    Suppose, for contradiction, there is more than one add-response-reset event for some $\cellpointershort$.
    Let $e_1$ and $e_2$ be two add-response-reset events for $\cellpointershort$ and let $p_1$ (resp. $p_2$) be the process that executed $e_1$ (resp. $e_2$).
    Hence, by \Cref{def:ero:english}, $p_1$ and $p_2$ executed $e_1$ and $e_2$ in an invocation of the \doworkuntildone{} procedure with parameters $(\addcell, \cellpointershort)$.
    Let $I_1$ (resp $I_2$) be the invocation the \doworkuntildone{} procedure that $p_1$ (resp. $p_2$) executed $e_1$ (resp. $e_2$) during.
    Furthermore, let $I^{hlo}_1$ (resp. $I^{hlo}_2$) be the invocation of the \highleveloperation{} procedure that $p_1$ (resp. $p_2$) invoked $I_1$ (resp. $I_2$) during.
    Since the the second parameter of $I_1$ and $I_2$ is $\cellpointershort$, it follows that $p_1$ and $p_2$ both performed an $\allocatecelloperation$ operation on \cref{line:ero:allocate_cell} whose response is $\cellpointershort$ during $I^{hlo}_1$ and $I^{hlo}_2$, and since by \Cref{alg:lazy_cell_manager_specification} the response of every $\allocatecelloperation$ operation is unique, we have that $p_1 = p_2$ and $I^{hlo}_1 = I^{hlo}_2$.
    Hence, since the first parameter of $I_1$ and $I_2$ is $\addcell$, $I_1$ and $I_2$ are invoked by the same process, and $I_1$ and $I_2$ are both invoked during the same invocation of the \highleveloperation{} procedure, it follows that $I_1 = I_2$.
    Thus, $e_1$ and $e_2$ are performed during the same invocation of the \doworkuntildone{} procedure.
    Therefore, since there is at most one response-reset event per invocation of the \doworkuntildone{} procedure (because \cref{line:ero:do_work_initialize_response} is executed at most once), we have that $e_1 = e_2$.
    However, $e_1 \neq e_2$, a contradiction.
    \qH{\Cref{lemma:ero:at_most_one_add_reset_per_pointer}}
\end{proof}

\begin{proposition}\label{lemma:ero:at_most_one_apply_reset_per_pointer}
    There is at most one apply-response-reset event for $\cellpointershort$ in $\mathcal{I}^\mathcal{B}$.
\end{proposition}

\begin{proof}
    By essentially the same argument as \Cref{lemma:ero:at_most_one_add_reset_per_pointer}, which we provide below for completeness.
    Suppose, for contradiction, there is more than one apply-response-reset event for some $\cellpointershort$.
    Let $e_1$ and $e_2$ be two apply-response-reset events for $\cellpointershort$ and let $p_1$ (resp. $p_2$) be the process that executed $e_1$ (resp. $e_2$).
    Hence, by \Cref{def:ero:english}, $p_1$ and $p_2$ executed $e_1$ and $e_2$ in an invocation of the \doworkuntildone{} procedure with parameters $(\langle \doopandcopyresponse{}, \arbitraryvalue \rangle, \cellpointershort)$.
    Let $I_1$ (resp $I_2$) be the invocation the \doworkuntildone{} procedure that $p_1$ (resp. $p_2$) executed $e_1$ (resp. $e_2$) during.
    Furthermore, let $I^{hlo}_1$ (resp. $I^{hlo}_2$) be the invocation of the \highleveloperation{} procedure that $p_1$ (resp. $p_2$) invoked $I_1$ (resp. $I_2$) during.
    Since the the second parameter of $I_1$ and $I_2$ is $\cellpointershort$, it follows that $p_1$ and $p_2$ both performed an $\allocatecelloperation$ operation on \cref{line:ero:allocate_cell} whose response is $\cellpointershort$ during $I^{hlo}_1$ and $I^{hlo}_2$, and since by \Cref{alg:lazy_cell_manager_specification} the response of every $\allocatecelloperation$ operation is unique, we have that $p_1 = p_2$ and $I^{hlo}_1 = I^{hlo}_2$.
    Hence, since the first parameter of $I_1$ and $I_2$ is $\langle \doopandcopyresponse{}, \arbitraryvalue \rangle$, $I_1$ and $I_2$ are invoked by the same process, and $I_1$ and $I_2$ are both invoked during the same invocation of the \highleveloperation{} procedure, it follows that $I_1 = I_2$.
    Thus, $e_1$ and $e_2$ are performed during the same invocation of the \doworkuntildone{} procedure.
    Therefore, since there is at most one response-reset event per invocation of the \doworkuntildone{} procedure (because \cref{line:ero:do_work_initialize_response} is executed at most once), we have that $e_1 = e_2$.
    However, $e_1 \neq e_2$, a contradiction.
    \qH{\Cref{lemma:ero:at_most_one_apply_reset_per_pointer}}
\end{proof}

\begin{proposition}\label{lemma:ero:at_most_one_remove_reset_per_pointer}
    There is at most one remove-response-reset event for $\cellpointershort$ in $\mathcal{I}^\mathcal{B}$.
\end{proposition}

\begin{proof}
    By essentially the same argument as \Cref{lemma:ero:at_most_one_add_reset_per_pointer}, which we provide below for completeness.
    Suppose, for contradiction, there is more than one remove-response-reset event for some $\cellpointershort$.
    Let $e_1$ and $e_2$ be two remove-response-reset events for $\cellpointershort$ and let $p_1$ (resp. $p_2$) be the process that executed $e_1$ (resp. $e_2$).
    Hence, by \Cref{def:ero:english}, $p_1$ and $p_2$ executed $e_1$ and $e_2$ in an invocation of the \doworkuntildone{} procedure with parameters $(\removecell, \cellpointershort)$.
    Let $I_1$ (resp $I_2$) be the invocation the \doworkuntildone{} procedure that $p_1$ (resp. $p_2$) executed $e_1$ (resp. $e_2$) during.
    Furthermore, let $I^{hlo}_1$ (resp. $I^{hlo}_2$) be the invocation of the \highleveloperation{} procedure that $p_1$ (resp. $p_2$) invoked $I_1$ (resp. $I_2$) during.
    Since the the second parameter of $I_1$ and $I_2$ is $\cellpointershort$, it follows that $p_1$ and $p_2$ both performed an $\allocatecelloperation$ operation on \cref{line:ero:allocate_cell} whose response is $\cellpointershort$ during $I^{hlo}_1$ and $I^{hlo}_2$, and since by \Cref{alg:lazy_cell_manager_specification} the response of every $\allocatecelloperation$ operation is unique, we have that $p_1 = p_2$ and $I^{hlo}_1 = I^{hlo}_2$.
    Hence, since the first parameter of $I_1$ and $I_2$ is $\removecell$, $I_1$ and $I_2$ are invoked by the same process, and $I_1$ and $I_2$ are both invoked during the same invocation of the \highleveloperation{} procedure, it follows that $I_1 = I_2$.
    Thus, $e_1$ and $e_2$ are performed during the same invocation of the \doworkuntildone{} procedure.
    Therefore, since there is at most one response-reset event per invocation of the \doworkuntildone{} procedure (because \cref{line:ero:do_work_initialize_response} is executed at most once), we have that $e_1 = e_2$.
    However, $e_1 \neq e_2$, a contradiction.
    \qH{\Cref{lemma:ero:at_most_one_remove_reset_per_pointer}}
\end{proof}

By Propositions \ref{lemma:ero:at_most_one_add_reset_per_pointer}, \ref{lemma:ero:at_most_one_apply_reset_per_pointer}, and \ref{lemma:ero:at_most_one_remove_reset_per_pointer} we have the following.

\begin{corollary}\label{lemma:ero:at_most_one_reset_per_type_and_pointer}
    There is at most one add-response-reset, apply-response-reset, and remove-response-reset event for $\cellpointershort$ in $\mathcal{I}^\mathcal{B}$.
\end{corollary}

\begin{proposition}\label{lemma:ero:every_l_add_event_is_preceeded_by_unique_add_response_reset}
    Let $e$ be an $L$-add event for $\cellpointershort$ in $\mathcal{I}^\mathcal{B}$ that set $\linearizationobject{}.\uniquerepositoryoperationlong{} = \uniquerepositoryoperationshort$.
    There is exactly one add-response-reset event for $\cellpointershort$ before $e$, and it set $(*\cellpointershort).\lastrepositoryoperationresponse{} = (\uniquerepositoryoperationshort, \nullconstant)$.
\end{proposition}

\begin{proof}
    By \Cref{lemma:ero:at_most_one_reset_per_type_and_pointer}, there is at most one add-response-reset event for $\cellpointershort$ before $e$, so it suffices to prove that there is at least one before $e$.
    Since $e$ is an $L$-add event for $\cellpointershort$ that set $\linearizationobject{}.\uniquerepositoryoperationlong{} = \uniquerepositoryoperationshort$, by \Cref{def:ero:english},  $\uniquerepositoryoperationshort = (\arbitraryvalue, \addcell)$ and $e$ set $\linearizationobject{}$ to $(\uniquerepositoryoperationshort, \cellpointershort)$.
    Hence, by \Cref{lemma:ero:l_events_have_corresponding_a_events}, there is an $A$-event $e'$ which set $\announceobject$ to $(\uniquerepositoryoperationshort, \cellpointershort)$ before $e$.
    Thus, by \Cref{def:ero:english}, $e'$ is an execution of either \cref{line:ero:announce_gcas} or \ref{line:ero:announce_cas}, and so the process that executed $e'$ set $(*\cellpointershort).\lastrepositoryoperationresponse{} = (\uniquerepositoryoperationshort, \nullconstant)$ on \cref{line:ero:do_work_initialize_response} before $e'$ (and thus $e$ by transitivity).
    Therefore, since $\uniquerepositoryoperationshort = (\arbitraryvalue, \addcell)$, by \Cref{def:ero:english}, this is an add-response-reset event for $\cellpointershort$ before $e$ as wanted.
    \qH{\Cref{lemma:ero:every_l_add_event_is_preceeded_by_unique_add_response_reset}}
\end{proof}

\begin{proposition}\label{lemma:ero:every_l_apply_event_is_preceeded_by_unique_apply_response_reset}
    Let $e$ be an $L$-apply event for $\cellpointershort$ in $\mathcal{I}^\mathcal{B}$ that set $\linearizationobject{}.\uniquerepositoryoperationlong{} = \uniquerepositoryoperationshort$.
    There is exactly one apply-response-reset event for $\cellpointershort$ before $e$, and it set $(*\cellpointershort).\lastrepositoryoperationresponse{} = (\uniquerepositoryoperationshort, \nullconstant)$.
\end{proposition}

\begin{proof}
    By essentially the same argument as \Cref{lemma:ero:every_l_add_event_is_preceeded_by_unique_add_response_reset}, which we provide below for completeness.
    By \Cref{lemma:ero:at_most_one_reset_per_type_and_pointer}, there is at most one apply-response-reset event for $\cellpointershort$ before $e$, so it suffices to prove that there is at least one before $e$.
    Since $e$ is an $L$-apply event for $\cellpointershort$ that set $\linearizationobject{}.\uniquerepositoryoperationlong{} = \uniquerepositoryoperationshort$, by \Cref{def:ero:english},  $\uniquerepositoryoperationshort = (\arbitraryvalue, \langle \doopandcopyresponse{}, \arbitraryvalue \rangle)$ and $e$ set $\linearizationobject{}$ to $(\uniquerepositoryoperationshort, \cellpointershort)$.
    Hence, by \Cref{lemma:ero:l_events_have_corresponding_a_events}, there is an $A$-event $e'$ which set $\announceobject$ to $(\uniquerepositoryoperationshort, \cellpointershort)$ before $e$.
    Thus, by \Cref{def:ero:english}, $e'$ is an execution of either \cref{line:ero:announce_gcas} or \ref{line:ero:announce_cas}, and so the process that executed $e'$ set $(*\cellpointershort).\lastrepositoryoperationresponse{} = (\uniquerepositoryoperationshort, \nullconstant)$ on \cref{line:ero:do_work_initialize_response} before $e'$ (and thus $e$ by transitivity).
    Therefore, since $\uniquerepositoryoperationshort = (\arbitraryvalue, \langle \doopandcopyresponse{}, \arbitraryvalue \rangle)$, by \Cref{def:ero:english}, this is an apply-response-reset event for $\cellpointershort$ before $e$ as wanted.
    \qH{\Cref{lemma:ero:every_l_apply_event_is_preceeded_by_unique_apply_response_reset}}
\end{proof}

\begin{proposition}\label{lemma:ero:every_l_remove_event_is_preceeded_by_unique_remove_response_reset}
    Let $e$ be an $L$-remove event for $\cellpointershort$ in $\mathcal{I}^\mathcal{B}$ that set $\linearizationobject{}.\uniquerepositoryoperationlong{} = \uniquerepositoryoperationshort$.
    There is exactly one remove-response-reset event for $\cellpointershort$ before $e$, and it set $(*\cellpointershort).\lastrepositoryoperationresponse{} = (\uniquerepositoryoperationshort, \nullconstant)$.
\end{proposition}

\begin{proof}
    By essentially the same argument as \Cref{lemma:ero:every_l_add_event_is_preceeded_by_unique_add_response_reset}, which we provide below for completeness.
    By \Cref{lemma:ero:at_most_one_reset_per_type_and_pointer}, there is at most one remove-response-reset event for $\cellpointershort$ before $e$, so it suffices to prove that there is at least one before $e$.
    Since $e$ is an $L$-remove event for $\cellpointershort$ that set $\linearizationobject{}.\uniquerepositoryoperationlong{} = \uniquerepositoryoperationshort$, by \Cref{def:ero:english},  $\uniquerepositoryoperationshort = (\arbitraryvalue, \removecell)$ and $e$ set $\linearizationobject{}$ to $(\uniquerepositoryoperationshort, \cellpointershort)$.
    Hence, by \Cref{lemma:ero:l_events_have_corresponding_a_events}, there is an $A$-event $e'$ which set $\announceobject$ to $(\uniquerepositoryoperationshort, \cellpointershort)$ before $e$.
    Thus, by \Cref{def:ero:english}, $e'$ is an execution of either \cref{line:ero:announce_gcas} or \ref{line:ero:announce_cas}, and so the process that executed $e'$ set $(*\cellpointershort).\lastrepositoryoperationresponse{} = (\uniquerepositoryoperationshort, \nullconstant)$ on \cref{line:ero:do_work_initialize_response} before $e'$ (and thus $e$ by transitivity).
    Therefore, since $\uniquerepositoryoperationshort = (\arbitraryvalue, \removecell)$, by \Cref{def:ero:english}, this is a remove-response-reset event for $\cellpointershort$ before $e$ as wanted.
    \qH{\Cref{lemma:ero:every_l_remove_event_is_preceeded_by_unique_remove_response_reset}}
\end{proof}

By Propositions \ref{lemma:ero:every_l_add_event_is_preceeded_by_unique_add_response_reset}, \ref{lemma:ero:every_l_apply_event_is_preceeded_by_unique_apply_response_reset}, and \ref{lemma:ero:every_l_remove_event_is_preceeded_by_unique_remove_response_reset} we have the following.

\begin{corollary}\label{lemma:ero:every_l_event_is_preceeded_by_unique_response_reset}
    Let $e$ be an $L$-$X$ event for $\cellpointershort$ in $\mathcal{I}^\mathcal{B}$ where $X \in \{\text{add}, \text{apply}, \text{remove}\}$ that set $\linearizationobject{}.\uniquerepositoryoperationlong{} = \uniquerepositoryoperationshort$.
    There is exactly one $X$-response-reset event for $\cellpointershort$ before $e$, and it set\\\mbox{$(*\cellpointershort).\lastrepositoryoperationresponse{} = (\uniquerepositoryoperationshort, \nullconstant)$.}
\end{corollary}

We now prove some facts about response-set attempts.

\begin{lemma}\label{lemma:ero:l_event_corresponding_to_set_response}
    Let $I$ be any invocation of the \setrepositoryoperationresponse{} procedure in $\mathcal{I}^\mathcal{B}$ with parameters $(\uniquerepositoryoperationshort, \cellpointershort, \arbitraryvalue)$.
    Then, there is an $L$-event $e$ before $I$ was invoked that set $\linearizationobject{} = (\uniquerepositoryoperationshort, \cellpointershort)$.
\end{lemma}

\begin{proof}
    Since some process $p$ invoked $I$ on either \cref{line:ero:add_cell_set_response}, \ref{line:ero:remove_cell_set_response}, or \ref{line:ero:apply_set_response}, we have that $p$ invoked $I$ during an invocation $I'$ of either the \doaddcell{}, \doremovecell{}, or \doapplyandcopyresponse{} procedure.
    Hence, since the parameters of $I$ are $(\uniquerepositoryoperationshort, \cellpointershort, \arbitraryvalue)$, it follows that the parameters of $I'$ are $(\uniquerepositoryoperationshort, \cellpointershort)$.
    Thus, by \Cref{lemma:ero:l_event_corresponding_to_do_low_level_op}, there is an $L$-event $e$ before $I'$ was invoked that set $\linearizationobject{}$ to $(\uniquerepositoryoperationshort, \cellpointershort)$.
    Therefore, since $I'$ was invoked before $I$, the claim follows.
    \qH{\Cref{lemma:ero:l_event_corresponding_to_set_response}}
\end{proof}

\begin{proposition}\label{lemma:ero:every_response_set_attempt_is_for_pointer_from_universe_helper}
    Every response-set attempt in $\mathcal{I}^\mathcal{B}$ is for some pointer in $\celluniverse{}$.
\end{proposition}

\begin{proof}
    Consider any response-set attempt $a$ for $\cellpointershort$ in $\mathcal{I}^\mathcal{B}$.
    Hence, by \Cref{def:ero:english}, $a$ was executed during some invocation $I$ of the \setrepositoryoperationresponse{} procedure with parameters $(\arbitraryvalue, \cellpointershort, \arbitraryvalue)$.
    Thus, by \Cref{lemma:ero:l_event_corresponding_to_set_response}, there is an $L$-event for $\cellpointershort$.
    Therefore, by \Cref{lemma:ero:every_l_event_is_for_pointer_from_universe}, $\cellpointershort \in \celluniverse{}$ as wanted.
    \qH{\Cref{lemma:ero:every_response_set_attempt_is_for_pointer_from_universe_helper}}
\end{proof}

\begin{lemma}\label{lemma:ero:response_set_attempts_have_corresponding_l_events}
    Consider any response-set attempt $a$ for $\cellpointershort$, so by \Cref{lemma:ero:every_response_set_attempt_is_for_pointer_from_universe_helper} $\cellpointershort \in \celluniverse{}$, which tries to set $(*\cellpointershort).\lastrepositoryoperationresponse{}.\uniquerepositoryoperationlong{}$ to some value $v$ in $\mathcal{I}^\mathcal{B}$.
    Then, there is an $L$-event which set $\linearizationobject{}$ to $(v, \cellpointershort)$ before $a$.
    We call the last $L$-event which set $\linearizationobject{}$ to $(v, \cellpointershort)$ before $a$, $a$'s corresponding $L$-event.
\end{lemma}

\begin{proof}
    Consider any response-set attempt $a$ for $\cellpointershort$ by some process $p$ which attempts to set the value of $(*\cellpointershort).\lastrepositoryoperationresponse{}.\uniquerepositoryoperationlong{}$ to $v$.
    Hence, by \Cref{def:ero:english}, $p$ executed $a$ during some invocation of the \setrepositoryoperationresponse{} procedure with parameters $(v, \cellpointershort, \arbitraryvalue)$.
    Thus, by \Cref{lemma:ero:l_event_corresponding_to_set_response}, there is an $L$-event before this invocation was invoked that set $\linearizationobject{} = (v, \cellpointershort)$.
    \qH{\Cref{lemma:ero:response_set_attempts_have_corresponding_l_events}}
\end{proof}

By \Cref{def:ero:english}, this implies the following.

\begin{corollary}\label{corollary:ero:response_set_attempts_and_corresponding_l_events_are_matching}
    Consider a response-set attempt $a$ for $\cellpointershort$ in $\mathcal{I}^\mathcal{B}$ and let $e$ be its corresponding $L$-event.
    The following are true.
    \begin{compactitem}
        \item $e$ is for $\cellpointershort$.
        \item $a$ is an add-response-set attempt if and only if $e$ is an $L$-add event.
        \item $a$ is an apply-response-set attempt if and only if $e$ is an $L$-apply event.
        \item $a$ is an remove-response-set attempt if and only if $e$ is an $L$-remove event.
    \end{compactitem}
\end{corollary}

\begin{lemma}\label{lemma:ero:every_response_set_attempt_is_for_pointer_from_universe}
    Every response-set attempt in $\mathcal{I}^\mathcal{B}$ is for some pointer in $\celluniverse{}$ and is either an add-response-set, apply-response-set, or remove-response-set attempt.
\end{lemma}

\begin{proof}
    Consider any response-set attempt $a$ for $\cellpointershort$ in $\mathcal{I}^\mathcal{B}$ and let $e$ be its corresponding $L$-event (see \Cref{lemma:ero:response_set_attempts_have_corresponding_l_events}).
    Hence, by \Cref{corollary:ero:response_set_attempts_and_corresponding_l_events_are_matching}, $e$ is for $\cellpointershort$, and so by \Cref{lemma:ero:every_l_event_is_for_pointer_from_universe} $\cellpointershort \in \celluniverse{}$.
    Furthermore, since by \Cref{lemma:ero:every_l_event_is_add_apply_or_remove} $e$ is either an $L$-add, $L$-apply, or $L$-remove event, by \Cref{corollary:ero:response_set_attempts_and_corresponding_l_events_are_matching}, $a$ is either an add-response-set, apply-response-set, or remove-response-set attempt.
    \qH{\Cref{lemma:ero:every_response_set_attempt_is_for_pointer_from_universe}}
\end{proof}

\begin{lemma}\label{lemma:ero:every_successful_response_set_attempt_is_preceeded_by_a_reset_event}
    If there is a $X$-response-set attempt for $\cellpointershort$ in $\mathcal{I}^\mathcal{B}$ where $X \in \{\text{add}, \text{apply}, \text{remove}\}$, then there is a $X$-response-reset event for $\cellpointershort$ beforehand.
\end{lemma}

\begin{proof}
    Consider any $X$-response-set attempt $a$ for $\cellpointershort$ in $\mathcal{I}^\mathcal{B}$.
    Hence, by \Cref{corollary:ero:response_set_attempts_and_corresponding_l_events_are_matching}, there is an $L$-$X$ event $e$ for $\cellpointershort$ before $a$.
    Thus, by \Cref{lemma:ero:every_l_event_is_preceeded_by_unique_response_reset}, there is an $X$-response-reset event for $\cellpointershort$ before $e$.
    Therefore, since $e$ is before $a$, the claim follows.
    \qH{\Cref{lemma:ero:every_successful_response_set_attempt_is_preceeded_by_a_reset_event}}
\end{proof}

\begin{lemma}\label{lemma:ero:every_apply_response_set_attempt_is_to_a_not_null_response}
    Every response-set attempt in $\mathcal{I}^\mathcal{B}$ is to a response other than $\nullconstant$.
\end{lemma}

\begin{proof}
    Consider any response-set attempt $a$ to response $r$ by some process $p$.
    Hence, by \Cref{def:ero:english}, $p$ executed $a$ during some invocation $I$ of the \setrepositoryoperationresponse{} procedure with parameters $(\arbitraryvalue, \arbitraryvalue, r)$.
    Since $p$ invoked $I$ on either \cref{line:ero:add_cell_set_response}, \ref{line:ero:remove_cell_set_response}, or \ref{line:ero:apply_set_response}, we have that $r$ is $\done$ or the value of $\stateobject.\responselong$.    
    Hence, since $\stateobject.\responselong$ is initially $\noopresponse$ and is only set to the right field of the response of $apply_\mathcal{T}$, we have that $r$ is either $\done$, $\noopresponse$, or some response to an operation of type $\mathcal{T}$.
    Therefore, by \Cref{assumption:operations_and_responses}, $r \neq \nullconstant$ as wanted.
    \qH{\Cref{lemma:ero:every_apply_response_set_attempt_is_to_a_not_null_response}}
\end{proof}

\begin{proposition}\label{lemma:ero:at_most_one_add_response_set_per_pointer}
    There is at most one successful add-response-set attempt for $\cellpointershort$ in $\mathcal{I}^\mathcal{B}$.
\end{proposition}

\begin{proof}
    Suppose, for contradiction, there is more than one successful add-response-set attempt for $\cellpointershort$.
    Hence, by \Cref{lemma:ero:every_response_set_attempt_is_for_pointer_from_universe}, $\cellpointershort \in \celluniverse{}$.
    Let $a_1$ and $a_2$ be two successful add-response-set attempts for $\cellpointershort$.
    Without loss of generality, suppose $a_1 < a_2$.
    Hence, since $\cellpointershort \in \celluniverse{}$, by \Cref{lemma:ero:every_apply_response_set_attempt_is_to_a_not_null_response}, $a_1$ sets $(*\cellpointershort).\lastrepositoryoperationresponse{} \neq (\arbitraryvalue, \nullconstant)$.
    Furthermore, since $a_2$ is a successful add-response-set attempt, by \Cref{def:ero:english}, $(*\cellpointershort).\lastrepositoryoperationresponse{} = (\myuniquerepositoryoperationshort{}, \nullconstant)$ at the step before $a_2$ where $\myuniquerepositoryoperationshort{} = (\arbitraryvalue, \addcell)$.
    Hence, since $a_1 < a_2$, we have that $(*\cellpointershort).\lastrepositoryoperationresponse{}$ was set to $(\uniquerepositoryoperationshort, \nullconstant)$ between $a_1$ and $a_2$.
    Thus, by \Cref{observation:ero:where_objects_change}, either a response-reset event or a successful response-set attempt for $\cellpointershort$ set $(*\cellpointershort).\lastrepositoryoperationresponse{} = (\uniquerepositoryoperationshort, \nullconstant)$ between $a_1$ and $a_2$.
    However, since by \Cref{lemma:ero:every_apply_response_set_attempt_is_to_a_not_null_response} every successful response-set attempt for $\cellpointershort$ sets the value of $(*\cellpointershort).\lastrepositoryoperationresponse{} \neq (\arbitraryvalue, \nullconstant)$, we have that there is a response-reset event $e_2$ for $\cellpointershort$ between $a_1$ and $a_2$ that set $(*\cellpointershort).\lastrepositoryoperationresponse{} = (\uniquerepositoryoperationshort, \nullconstant)$.
    Hence, since $\myuniquerepositoryoperationshort{} = (\arbitraryvalue, \addcell)$, by \Cref{def:ero:english}, $e_2$ is an add-response-reset event for $\cellpointershort$.
    Since $a_1$ is an add-response-set attempt for $\cellpointershort$, by \Cref{lemma:ero:every_successful_response_set_attempt_is_preceeded_by_a_reset_event}, there is an add-response-reset event $e_1$ for $\cellpointershort$ before $a_1$.
    Therefore, since $e_1$ is before $a_1$, and $e_2$ is between $a_1$ and $a_2$, it follows that there are two add-response-reset events for $\cellpointershort$ in $\mathcal{I}^\mathcal{B}$.
    However, by \Cref{lemma:ero:at_most_one_reset_per_type_and_pointer}, there is at most one add-response-reset event for $\cellpointershort$ in $\mathcal{I}^\mathcal{B}$, a contradiction.
    \qH{\Cref{lemma:ero:at_most_one_add_response_set_per_pointer}}
\end{proof}

\begin{proposition}\label{lemma:ero:at_most_one_apply_response_set_per_pointer}
    There is at most one successful apply-response-set attempt for $\cellpointershort$ in $\mathcal{I}^\mathcal{B}$.
\end{proposition}

\begin{proof}
    By essentially the same argument as \Cref{lemma:ero:at_most_one_add_response_set_per_pointer}, which we provide below for completeness.
    Suppose, for contradiction, there is more than one successful apply-response-set attempt for $\cellpointershort$.
    Hence, by \Cref{lemma:ero:every_response_set_attempt_is_for_pointer_from_universe}, $\cellpointershort \in \celluniverse{}$.
    Let $a_1$ and $a_2$ be two successful apply-response-set attempts for $\cellpointershort$.
    Without loss of generality, suppose $a_1 < a_2$.
    Hence, since $\cellpointershort \in \celluniverse{}$, by \Cref{lemma:ero:every_apply_response_set_attempt_is_to_a_not_null_response}, $a_1$ sets $(*\cellpointershort).\lastrepositoryoperationresponse{} \neq (\arbitraryvalue, \nullconstant)$.
    Furthermore, since $a_2$ is a successful apply-response-set attempt, by \Cref{def:ero:english}, $(*\cellpointershort).\lastrepositoryoperationresponse{} = (\myuniquerepositoryoperationshort{}, \nullconstant)$ at the step before $a_2$ where $\myuniquerepositoryoperationshort{} = (\arbitraryvalue, \langle \doopandcopyresponse{}, \arbitraryvalue \rangle)$.
    Hence, since $a_1 < a_2$, we have that $(*\cellpointershort).\lastrepositoryoperationresponse{}$ was set to $(\uniquerepositoryoperationshort, \nullconstant)$ between $a_1$ and $a_2$.
    Thus, by \Cref{observation:ero:where_objects_change}, either a response-reset event or a successful response-set attempt for $\cellpointershort$ set $(*\cellpointershort).\lastrepositoryoperationresponse{} = (\uniquerepositoryoperationshort, \nullconstant)$ between $a_1$ and $a_2$.
    However, since by \Cref{lemma:ero:every_apply_response_set_attempt_is_to_a_not_null_response} every successful response-set attempt for $\cellpointershort$ sets the value of $(*\cellpointershort).\lastrepositoryoperationresponse{} \neq (\arbitraryvalue, \nullconstant)$, we have that there is a response-reset event $e_2$ for $\cellpointershort$ between $a_1$ and $a_2$ that set $(*\cellpointershort).\lastrepositoryoperationresponse{} = (\uniquerepositoryoperationshort, \nullconstant)$.
    Hence, since $\myuniquerepositoryoperationshort{} = (\arbitraryvalue, \langle \doopandcopyresponse{}, \arbitraryvalue \rangle)$, by \Cref{def:ero:english}, $e_2$ is an apply-response-reset event for $\cellpointershort$.
    Since $a_1$ is an apply-response-set attempt for $\cellpointershort$, by \Cref{lemma:ero:every_successful_response_set_attempt_is_preceeded_by_a_reset_event}, there is an apply-response-reset event $e_1$ for $\cellpointershort$ before $a_1$.
    Therefore, since $e_1$ is before $a_1$, and $e_2$ is between $a_1$ and $a_2$, it follows that there are two apply-response-reset events for $\cellpointershort$ in $\mathcal{I}^\mathcal{B}$.
    However, by \Cref{lemma:ero:at_most_one_reset_per_type_and_pointer}, there is at most one apply-response-reset event for $\cellpointershort$ in $\mathcal{I}^\mathcal{B}$, a contradiction.
    \qH{\Cref{lemma:ero:at_most_one_apply_response_set_per_pointer}}
\end{proof}

\begin{proposition}\label{lemma:ero:at_most_one_remove_response_set_per_pointer}
    There is at most one successful remove-response-set attempt for $\cellpointershort$ in $\mathcal{I}^\mathcal{B}$.
\end{proposition}

\begin{proof}
    By essentially the same argument as \Cref{lemma:ero:at_most_one_add_response_set_per_pointer}, which we provide below for completeness.
    Suppose, for contradiction, there is more than one successful remove-response-set attempt for $\cellpointershort$.
    Hence, by \Cref{lemma:ero:every_response_set_attempt_is_for_pointer_from_universe}, $\cellpointershort \in \celluniverse{}$.
    Let $a_1$ and $a_2$ be two successful remove-response-set attempts for $\cellpointershort$.
    Without loss of generality, suppose $a_1 < a_2$.
    Hence, since $\cellpointershort \in \celluniverse{}$, by \Cref{lemma:ero:every_apply_response_set_attempt_is_to_a_not_null_response}, $a_1$ sets $(*\cellpointershort).\lastrepositoryoperationresponse{} \neq (\arbitraryvalue, \nullconstant)$.
    Furthermore, since $a_2$ is a successful remove-response-set attempt, by \Cref{def:ero:english}, $(*\cellpointershort).\lastrepositoryoperationresponse{} = (\myuniquerepositoryoperationshort{}, \nullconstant)$ at the step before $a_2$ where $\myuniquerepositoryoperationshort{} = (\arbitraryvalue, \removecell)$.
    Hence, since $a_1 < a_2$, we have that $(*\cellpointershort).\lastrepositoryoperationresponse{}$ was set to $(\uniquerepositoryoperationshort, \nullconstant)$ between $a_1$ and $a_2$.
    Thus, by \Cref{observation:ero:where_objects_change}, either a response-reset event or a successful response-set attempt for $\cellpointershort$ set $(*\cellpointershort).\lastrepositoryoperationresponse{} = (\uniquerepositoryoperationshort, \nullconstant)$ between $a_1$ and $a_2$.
    However, since by \Cref{lemma:ero:every_apply_response_set_attempt_is_to_a_not_null_response} every successful response-set attempt for $\cellpointershort$ sets the value of $(*\cellpointershort).\lastrepositoryoperationresponse{} \neq (\arbitraryvalue, \nullconstant)$, we have that there is a response-reset event $e_2$ for $\cellpointershort$ between $a_1$ and $a_2$ that set $(*\cellpointershort).\lastrepositoryoperationresponse{} = (\uniquerepositoryoperationshort, \nullconstant)$.
    Hence, since $\myuniquerepositoryoperationshort{} = (\arbitraryvalue, \removecell)$, by \Cref{def:ero:english}, $e_2$ is a remove-response-reset event for $\cellpointershort$.
    Since $a_1$ is a remove-response-set attempt for $\cellpointershort$, by \Cref{lemma:ero:every_successful_response_set_attempt_is_preceeded_by_a_reset_event}, there is a remove-response-reset event $e_1$ for $\cellpointershort$ before $a_1$.
    Therefore, since $e_1$ is before $a_1$, and $e_2$ is between $a_1$ and $a_2$, it follows that there are two remove-response-reset events for $\cellpointershort$ in $\mathcal{I}^\mathcal{B}$.
    However, by \Cref{lemma:ero:at_most_one_reset_per_type_and_pointer}, there is at most one remove-response-reset event for $\cellpointershort$ in $\mathcal{I}^\mathcal{B}$, a contradiction.
    \qH{\Cref{lemma:ero:at_most_one_remove_response_set_per_pointer}}
\end{proof}

By Propositions \ref{lemma:ero:at_most_one_add_response_set_per_pointer}, \ref{lemma:ero:at_most_one_apply_response_set_per_pointer}, and \ref{lemma:ero:at_most_one_remove_response_set_per_pointer} we have the following.

\begin{corollary}\label{lemma:ero:at_most_one_response_set_per_type_and_pointer}
    There is at most one successful add-response-set, apply-response-set, and remove-response-set attempt for $\cellpointershort$ in $\mathcal{I}^\mathcal{B}$.
\end{corollary}

\begin{lemma}\label{lemma:ero:once_response_not_null_for_add_never_null_for_add_again}
    For every $\cellpointershort{} \in \celluniverse$, if there is a successful add-response-set attempt $a$ for $\cellpointershort$ during $\mathcal{I}^\mathcal{B}$, then from $a$ onwards $(*\cellpointershort{}).\lastrepositoryoperationresponse{} \neq ((\arbitraryvalue, \addcell), \nullconstant)$. 
\end{lemma}

\begin{proof}
    Suppose, for contradiction, there is a successful add-response-set attempt $a$ for $\cellpointershort$ during $\mathcal{I}^\mathcal{B}$ and $(*\cellpointershort{}).\lastrepositoryoperationresponse{} = ((\arbitraryvalue, \addcell), \nullconstant)$ at or after $a$.
    Let $a$ be a successful add-response-set attempt to $r$.
    By \Cref{lemma:ero:every_apply_response_set_attempt_is_to_a_not_null_response}, $r \neq \nullconstant$.
    Hence, since $a$ is a successful add-response-set attempt for $\cellpointershort{}$ to $r$, we have that $(*\cellpointershort{}).\lastrepositoryoperationresponse{} \neq ((\arbitraryvalue, \addcell), \nullconstant)$ at $a$.
    Thus, since $(*\cellpointershort{}).\lastrepositoryoperationresponse{} = ((\arbitraryvalue, \addcell), \nullconstant)$ at or after $a$, we have $(*\cellpointershort{}).\lastrepositoryoperationresponse{}$ was set to $((\arbitraryvalue, \addcell), \nullconstant)$ after $a$.
    Hence, by \Cref{observation:ero:where_objects_change}, either an add-response-reset event for $\cellpointershort$ or a successful add-response-set attempt for $\cellpointershort$ set $(*\cellpointershort).\lastrepositoryoperationresponse{}$ to $((\arbitraryvalue, \addcell), \nullconstant)$ after $a$.
    Since by \Cref{lemma:ero:every_apply_response_set_attempt_is_to_a_not_null_response}, every add-response-set attempt does not set the response to $\nullconstant$, it cannot be a successful add-response-set attempt that set the value of $(*\cellpointershort).\lastrepositoryoperationresponse{}$ to $((\arbitraryvalue, \addcell), \nullconstant)$ after $a$.
    Hence, there is an add-response-reset event for $\cellpointershort$ after $a$.
    Thus, since $a$ is a successful add-response-set attempt for $\cellpointershort$, by \Cref{lemma:ero:every_successful_response_set_attempt_is_preceeded_by_a_reset_event}, there is an add-response-reset event for $\cellpointershort$ before $a$.
    Therefore, since there is an add-response-reset event for $\cellpointershort$ after $a$, there are two add-response-reset events for $\cellpointershort$ in $\mathcal{I}^\mathcal{B}$.
    However, by \Cref{lemma:ero:at_most_one_reset_per_type_and_pointer}, there is at most one add-response-reset event for $\cellpointershort$, a contradiction.
    \qH{\Cref{lemma:ero:once_response_not_null_for_add_never_null_for_add_again}}
\end{proof}

\begin{lemma}\label{lemma:ero:once_response_not_null_for_apply_never_null_for_apply_again}
    For every $\cellpointershort{} \in \celluniverse$, if there is a successful apply-response-set attempt $a$ for $\cellpointershort$ during $\mathcal{I}^\mathcal{B}$, then from $a$ onwards $(*\cellpointershort{}).\lastrepositoryoperationresponse{} \neq ((\arbitraryvalue, \langle \doopandcopyresponse, \arbitraryvalue\rangle), \nullconstant)$.
\end{lemma}

\begin{proof}
    By essentially the same argument as \Cref{lemma:ero:once_response_not_null_for_add_never_null_for_add_again}, which we provide below for completeness.
    Suppose, for contradiction, there is a successful apply-response-set attempt $a$ for $\cellpointershort$ during $\mathcal{I}^\mathcal{B}$ and $(*\cellpointershort{}).\lastrepositoryoperationresponse{} = ((\arbitraryvalue, \langle \doopandcopyresponse{}, \arbitraryvalue \rangle), \nullconstant)$ at or after $a$.
    Let $a$ be a successful apply-response-set attempt to $r$.
    By \Cref{lemma:ero:every_apply_response_set_attempt_is_to_a_not_null_response}, $r \neq \nullconstant$, and so $(*\cellpointershort{}).\lastrepositoryoperationresponse{} \neq ((\arbitraryvalue, \langle \doopandcopyresponse{}, \arbitraryvalue \rangle), \nullconstant)$ at $a$.
    Thus, since by assumption $(*\cellpointershort{}).\lastrepositoryoperationresponse{} = ((\arbitraryvalue, \langle \doopandcopyresponse{}, \arbitraryvalue \rangle), \nullconstant)$ at or after $a$, we have $(*\cellpointershort{}).\lastrepositoryoperationresponse{}$ was set to the value $((\arbitraryvalue, \langle \doopandcopyresponse{}, \arbitraryvalue \rangle), \nullconstant)$ after $a$.
    Hence, by \Cref{observation:ero:where_objects_change}, either an apply-response-reset event for $\cellpointershort$ or a successful apply-response-set attempt for $\cellpointershort$ set $(*\cellpointershort).\lastrepositoryoperationresponse{}$ to $((\arbitraryvalue, \langle \doopandcopyresponse{}, \arbitraryvalue \rangle), \nullconstant)$ after $a$.
    Since by \Cref{lemma:ero:every_apply_response_set_attempt_is_to_a_not_null_response}, every apply-response-set attempt does not set the response to $\nullconstant$, it cannot be a successful apply-response-set attempt that set the value of $(*\cellpointershort).\lastrepositoryoperationresponse{}$ to $((\arbitraryvalue, \langle \doopandcopyresponse{}, \arbitraryvalue \rangle), \nullconstant)$ after $a$.
    Hence, there is an apply-response-reset event for $\cellpointershort$ after $a$.
    Thus, since $a$ is a successful apply-response-set attempt for $\cellpointershort$, by \Cref{lemma:ero:every_successful_response_set_attempt_is_preceeded_by_a_reset_event}, there is an apply-response-reset event for $\cellpointershort$ before $a$.
    Therefore, since there is an apply-response-reset event for $\cellpointershort$ after $a$, there are two apply-response-reset events for $\cellpointershort$ in $\mathcal{I}^\mathcal{B}$.
    However, by \Cref{lemma:ero:at_most_one_reset_per_type_and_pointer}, there is at most one apply-response-reset event for $\cellpointershort$, a contradiction.
    \qH{\Cref{lemma:ero:once_response_not_null_for_apply_never_null_for_apply_again}}
\end{proof}

\begin{lemma}\label{lemma:ero:once_response_not_null_for_remove_never_null_for_remove_again}
    For every $\cellpointershort{} \in \celluniverse$, if there is a successful remove-response-set attempt $a$ for $\cellpointershort$ during $\mathcal{I}^\mathcal{B}$, then from $a$ onwards $(*\cellpointershort{}).\lastrepositoryoperationresponse{} \neq ((\arbitraryvalue, \removecell), \nullconstant)$. 
\end{lemma}

\begin{proof}
    By essentially the same argument as \Cref{lemma:ero:once_response_not_null_for_add_never_null_for_add_again}, which we provide below for completeness.
    Suppose, for contradiction, there is a successful remove-response-set attempt $a$ for $\cellpointershort$ during $\mathcal{I}^\mathcal{B}$ and $(*\cellpointershort{}).\lastrepositoryoperationresponse{} = ((\arbitraryvalue, \removecell), \nullconstant)$ at or after $a$.
    Let $a$ be a successful remove-response-set attempt to $r$.
    By \Cref{lemma:ero:every_apply_response_set_attempt_is_to_a_not_null_response}, $r \neq \nullconstant$.
    Hence, since $a$ is a successful remove-response-set attempt for $\cellpointershort{}$ to $r$, we have that $(*\cellpointershort{}).\lastrepositoryoperationresponse{} \neq ((\arbitraryvalue, \removecell), \nullconstant)$ at $a$.
    Thus, since $(*\cellpointershort{}).\lastrepositoryoperationresponse{} = ((\arbitraryvalue, \removecell), \nullconstant)$ at or after $a$, we have $(*\cellpointershort{}).\lastrepositoryoperationresponse{}$ was set to $((\arbitraryvalue, \removecell), \nullconstant)$ after $a$.
    Hence, by \Cref{observation:ero:where_objects_change}, either a remove-response-reset event for $\cellpointershort$ or a successful remove-response-set attempt for $\cellpointershort$ set $(*\cellpointershort).\lastrepositoryoperationresponse{}$ to $((\arbitraryvalue, \removecell), \nullconstant)$ after $a$.
    Since by \Cref{lemma:ero:every_apply_response_set_attempt_is_to_a_not_null_response}, every remove-response-set attempt does not set the response to $\nullconstant$, it cannot be a successful remove-response-set attempt that set the value of $(*\cellpointershort).\lastrepositoryoperationresponse{}$ to $((\arbitraryvalue, \addcell), \nullconstant)$ after $a$.
    Hence, there is a remove-response-reset event for $\cellpointershort$ after $a$.
    Thus, since $a$ is a successful remove-response-set attempt for $\cellpointershort$, by \Cref{lemma:ero:every_successful_response_set_attempt_is_preceeded_by_a_reset_event}, there is a remove-response-reset event for $\cellpointershort$ before $a$.
    Therefore, since there is a remove-response-reset event for $\cellpointershort$ after $a$, there are two remove-response-reset events for $\cellpointershort$ in $\mathcal{I}^\mathcal{B}$.
    However, by \Cref{lemma:ero:at_most_one_reset_per_type_and_pointer}, there is at most one remove-response-reset event for $\cellpointershort$, a contradiction.
    \qH{\Cref{lemma:ero:once_response_not_null_for_remove_never_null_for_remove_again}}
\end{proof}

\subsubsection{List-acquire-next attempts, acquire-copy events, and revocation events}

We now prove some facts about list-acquire-next attempts.

\begin{lemma}\label{lemma:ero:no_list_seal_before_acquire}
    Consider any successful list-acquire-next attempt $a_{acquire}$ after $\currentcellpointershort$ in $\mathcal{I}^\mathcal{B}$.
    There are no successful list-seal attempts for $\currentcellpointershort$ before $a_{acquire}$ in $\mathcal{I}^\mathcal{B}$.
\end{lemma}

\begin{proof}
    Suppose, for contradiction, there is a successful list-seal attempt $a_{seal}$ for $\currentcellpointershort$ before $a_{acquire}$ in $\mathcal{I}^\mathcal{B}$.
    By \Cref{lemma:ero:sealed_is_forever}, from $a_{seal}$ onwards in $\mathcal{I}^\mathcal{B}$ $(*\currentcellpointershort).\nextlong{}.sealed = \true{}$.
    Therefore, since $a_{seal} < a_{acquire}$, we have that $(*\currentcellpointershort).\nextlong{}.sealed = \true{}$ at $a_{acquire}$.
    However, since $a_{acquire}$ is a successful list-acquire-next attempt after $\currentcellpointershort$, it follows that $(*\currentcellpointershort).\nextlong{}.sealed = \false{}$ at $a_{acquire}$, a contradiction.
    \qH{\Cref{lemma:ero:no_list_seal_before_acquire}}
\end{proof}

\begin{lemma}\label{lemma:ero:no_list_remove_before_acquire}
    Consider any successful list-acquire-next attempt $a_{acquire}$ after $\currentcellpointershort$ in $\mathcal{I}^\mathcal{B}$.
    There are no successful list-remove attempts for $\currentcellpointershort$ before $a_{acquire}$ in $\mathcal{I}^\mathcal{B}$.
\end{lemma}

\begin{proof}
    Suppose, for contradiction, there is a successful list-remove attempt $a_{remove}$ for $\currentcellpointershort$ before $a_{acquire}$ in $\mathcal{I}^\mathcal{B}$.
    Let $p$ be the process that executed $a_{remove}$ and let $T^{\ref{line:ero:remove_cell_read_pointer_to_remove}}$ be the time of $p$'s last execution of \cref{line:ero:remove_cell_read_pointer_to_remove} before $a_{remove}$.
    Hence, by \Cref{lemma:ero:list_seal_before_list_remove}, there is a successful list-seal attempt for $\currentcellpointershort$ before $T^{\ref{line:ero:remove_cell_read_pointer_to_remove}}$ in $\mathcal{I}^\mathcal{B}$.
    Therefore, since $T^{\ref{line:ero:remove_cell_read_pointer_to_remove}} < a_{remove}$ and $a_{remove} < a_{acquire}$, by transitivity, there is a successful list-seal attempt for $\currentcellpointershort$ before $a_{acquire}$ in $\mathcal{I}^\mathcal{B}$.
    However, by \Cref{lemma:ero:no_list_seal_before_acquire}, there are no successful list-seal attempts for $\currentcellpointershort$ before $a_{acquire}$ in $\mathcal{I}^\mathcal{B}$, a contradiction.
    \qH{\Cref{lemma:ero:no_list_remove_before_acquire}}
\end{proof}

\begin{lemma}\label{lemma:ero:l_remove_event_is_last_before_acquire}
    Consider any successful list-acquire-next attempt $a_{acquire}$ after $\currentcellpointershort$ \mbox{in $\mathcal{I}^\mathcal{B}$.}
    If $R(\mathcal{I}^\mathcal{B})$ holds, and there is an $L$-remove event for $\currentcellpointershort$ before $a_{acquire}$ in $\mathcal{I}^\mathcal{B}$, then it is the last $L$-event before $a_{acquire}$ in $\mathcal{I}^\mathcal{B}$.
\end{lemma}

\begin{proof}
    Suppose, for contradiction, there is an $L$-remove event $e$ for $\currentcellpointershort$ before $a_{acquire}$ in $\mathcal{I}^\mathcal{B}$ and $e$ is not the last $L$-event before $a_{acquire}$ in $\mathcal{I}^\mathcal{B}$.
    Hence, there is an $L$-event after $e$ but before $a_{acquire}$ in $\mathcal{I}^\mathcal{B}$.
    Let $e'$ be the next $L$-event after $e$ in $\mathcal{I}^\mathcal{B}$, so $e' < a_{acquire}$.
    Hence, $e$ and $e'$ are successive $L$-events in $\mathcal{I}^\mathcal{B}$.
    Thus, since $e$ is an $L$-remove event for $\currentcellpointershort$ and $R(\mathcal{I}^\mathcal{B})$ holds, we have that there is a successful list-remove attempt for $\currentcellpointershort$ before $e'$ in $\mathcal{I}^\mathcal{B}$.
    Therefore, since $e' < a_{acquire}$, we have that there is a successful list-remove attempt for $\currentcellpointershort$ before $a_{acquire}$ in $\mathcal{I}^\mathcal{B}$.
    However, by \Cref{lemma:ero:no_list_remove_before_acquire}, there are no successful list-remove attempts for $\currentcellpointershort$ before $a_{acquire}$ in $\mathcal{I}^\mathcal{B}$, a contradiction.
    \qH{\Cref{lemma:ero:l_remove_event_is_last_before_acquire}}
\end{proof}

\begin{proposition}\label{lemma:ero:aquire_current_pointer_always_in_universe_or_head}
    Let $I$ be any invocation of the Acquire procedure in $\mathcal{I}^\mathcal{B}$ by some process $p$ and let $T^{\ref{line:ero:acquire_initial_current_pointer}}$ be the time $p$ executed \cref{line:ero:acquire_initial_current_pointer} during $I$ (assuming $p$ does).
    At all times at or after $T^{\ref{line:ero:acquire_initial_current_pointer}}$ and before $I$ returns, the value of the local variable $\currentuniquecellpointershort{}$ in $I$ is in $\celluniverse{} \cup \{\&\headobject\}$.
\end{proposition}

\begin{proof}
    By essentially the same argument as \Cref{lemma:ero:add_cell_current_pointer_always_in_universe_or_head}, which we provide below for completeness.
    Suppose, for contradiction, there is a time $T$ at or after $T^{\ref{line:ero:acquire_initial_current_pointer}}$ and before $I$ returns (if it ever does) such that the value of the local variable $\currentuniquecellpointershort{}$ in $I$ is $\cellpointershort \notin \celluniverse{} \cup \{\&\headobject\}$.
    Without loss of generality, suppose $T$ is the first such time.
    Since $p$ executed \cref{line:ero:acquire_initial_current_pointer} at $T^{\ref{line:ero:acquire_initial_current_pointer}}$ during $I$, the value of $\currentuniquecellpointershort{}$ is $\&\headobject$ at $T^{\ref{line:ero:acquire_initial_current_pointer}}$.
    Hence, since the value of $\currentuniquecellpointershort{}$ is $\cellpointershort \notin \celluniverse{} \cup \{\&\headobject\}$ at $T \geq T^{\ref{line:ero:acquire_initial_current_pointer}}$, it follows that the value of $\currentuniquecellpointershort{}$ was set to $\cellpointershort$ at $T$.
    Thus, since the value of $\currentuniquecellpointershort{}$ only changes on \cref{line:ero:acquire_update_current_unique_pointer} after $T^{\ref{line:ero:acquire_initial_current_pointer}}$ during $I$, we have that $p$ set $\currentuniquecellpointershort{}$ to $\cellpointershort$ by executing \cref{line:ero:acquire_update_current_unique_pointer} at $T$.
    So, the value of the local variable $\nextuniquecellpointershort{}$ in $I$ is $\cellpointershort$ at $T$.
    Therefore, the response of the invocation $I'$ of the AcquireNext procedure on \cref{line:ero:acquire_acquire_next} during the same iteration of the while loop on \cref{line:ero:acquire_loop_until} as $T$ is $(\found, \cellpointershort)$.
    Let $T^{\ref{line:ero:acquire_acquire_next}}$ be the time $p$ invoked $I'$, and let $\cellpointershort'$ be the second parameter of $I'$.
    Hence, the value of $\currentuniquecellpointershort{}$ is $\cellpointershort'$ at $T^{\ref{line:ero:acquire_acquire_next}}$.
    Thus, since $p$ invoked $I'$ at $T^{\ref{line:ero:acquire_acquire_next}}$ strictly before $T$, by the minimality of $T$, $\cellpointershort' \in \celluniverse{} \cup \{\&\headobject\}$.
    Since the second parameter of $I'$ is $\cellpointershort'$, and the response of $I'$ is $(\found, \cellpointershort)$, it follows that $p$ read $\cellpointershort$ from $(*\cellpointershort').\nextlong.\uniquecellpointercontentlong{}$ on the last execution of \cref{line:ero:acquire_next_read_curr_unique_pointer} during $I'$; say at time $T^{\ref{line:ero:acquire_next_read_curr_unique_pointer}}$.
    Hence, since $\cellpointershort' \in \celluniverse{} \cup \{\&\headobject\}$, by \Cref{lemma:ero:next_pointer_is_always_from_universe_or_null}, $\cellpointershort \in \celluniverse{} \cup \{\nullconstant\}$.
    Therefore, since $\cellpointershort \notin \celluniverse{} \cup \{\&\headobject\}$, we have that $\cellpointershort = \nullconstant$.
    However, since $p$ exited $I'$ with response $(\found, \cellpointershort)$, we have that $p$ found the clause on \cref{line:ero:acquire_next_not_found_check} to be false on its last execution of \cref{line:ero:acquire_next_not_found_check} during $I'$, and since $p$ read $\cellpointershort$ from $(*\cellpointershort').\nextlong.\uniquecellpointercontentlong{}$ on its last execution of \cref{line:ero:acquire_next_read_curr_unique_pointer} during $I'$, this implies that $\cellpointershort \neq \nullconstant$, a contradiction.
    \qH{\Cref{lemma:ero:aquire_current_pointer_always_in_universe_or_head}}
\end{proof}

Since the AcquireNext procedure is invoked only on lines \ref{line:ero:add_cell_acquire_next}, \ref{line:ero:remove_cell_acquire_next}, and \ref{line:ero:acquire_acquire_next}, Propositions \ref{lemma:ero:add_cell_current_pointer_always_in_universe_or_head}, \ref{lemma:ero:remove_cell_current_pointer_always_in_universe_or_head}, and \ref{lemma:ero:aquire_current_pointer_always_in_universe_or_head} imply the following.

\begin{corollary}\label{lemma:ero:acquire_next_is_for_pointer_from_universe_or_head}
    Consider any invocation of the AcquireNext procedure in $\mathcal{I}^\mathcal{B}$ and let $\currentcellpointershort{}$ be its second parameter.
    Then, $\currentcellpointershort{} \in \celluniverse \cup \{\&\headobject\}$.
\end{corollary}

\begin{lemma}\label{lemma:ero:list_acquire_next_attempt_for_pointer_from_universe_and_after_pointer_from_universe_or_head}
    Consider any list-acquire-next attempt for $\nextcellpointershort{}$ after $\currentcellpointershort{}$ in $\mathcal{I}^\mathcal{B}$.
    Then, $\nextcellpointershort{} \in \celluniverse$ and $\currentcellpointershort{} \in \celluniverse \cup \{\&\headobject\}$.
\end{lemma}

\begin{proof}
    Consider any list-acquire-next attempt $a$ for $\nextcellpointershort{}$ after $\currentcellpointershort{}$ by some process $p$.
    By \Cref{def:ero:english}, $p$ performed $a$ during some invocation $I$ of the AcquireNext procedure with parameters $(\arbitraryvalue, \currentcellpointershort{})$.
    Hence, by \Cref{lemma:ero:acquire_next_is_for_pointer_from_universe_or_head}, $\currentcellpointershort{} \in \celluniverse \cup \{\&\headobject\}$.
    Thus, since $a$ is for $\nextcellpointershort{}$, we have that $p$ read $\nextcellpointershort{}$ from $(*\currentcellpointershort{}).\nextlong.\uniquecellpointercontentlong{}$ on $p$'s last execution of \cref{line:ero:acquire_next_read_curr_unique_pointer} during $I$; say at time $T^{\ref{line:ero:acquire_next_read_curr_unique_pointer}}$.
    Hence, since $\currentcellpointershort{} \in \celluniverse \cup \{\&\headobject\}$, by \Cref{lemma:ero:next_pointer_is_always_from_universe_or_null}, $\nextcellpointershort{} \in \celluniverse \cup \{\nullconstant\}$.
    If $\nextcellpointershort{} \in \celluniverse$, then we are done, so suppose $\nextcellpointershort{} = \nullconstant$.
    Since $a$ is a list-acquire-next attempt for $\nextcellpointershort{}$, we have that $p$ found the condition on \cref{line:ero:acquire_next_linearization_changed_check} to be false between $T^{\ref{line:ero:acquire_next_read_curr_unique_pointer}}$ and $a$.
    Hence, $\nextcellpointershort{} \neq \nullconstant$, contradicting the fact that $\nextcellpointershort{} = \nullconstant$, so this case impossible.
    Therefore, $\nextcellpointershort{} \in \celluniverse{}$ as wanted.
    \qH{\Cref{lemma:ero:list_acquire_next_attempt_for_pointer_from_universe_and_after_pointer_from_universe_or_head}}
\end{proof}

\begin{lemma}\label{lemma:ero:acquire_next_found_response_is_from_universe}
    Consider any invocation of the AcquireNext procedure in $\mathcal{I}^\mathcal{B}$ whose response is $(\found, \nextcellpointershort{})$.
    Then, $\nextcellpointershort{} \in \celluniverse$.
\end{lemma}

\begin{proof}
    Consider any invocation $I$ of the AcquireNext procedure whose response is $(\found, \nextcellpointershort{})$ by some process $p$.
    Hence, $p$ exited $I$ on \cref{line:ero:acquire_next_found_return}, and so $p$ performed a successful list-acquire-next attempt for $\nextcellpointershort{}$ on \cref{line:ero:acquire_next_cell}.
    Therefore, by \Cref{lemma:ero:list_acquire_next_attempt_for_pointer_from_universe_and_after_pointer_from_universe_or_head}, $\nextcellpointershort{} \in \celluniverse$.
    \qH{\Cref{lemma:ero:acquire_next_found_response_is_from_universe}}
\end{proof}

\begin{lemma}\label{lemma:ero:acquire_second_parameter_is_pointer_or_null}
    Consider any invocation $I$ of the Acquire procedure in $\mathcal{I}^\mathcal{B}$.
    The second parameter of $I$ is in $\celluniverse{} \cup \{\nullconstant\}$.
\end{lemma}

\begin{proof}
    Observe that $I$ is invoked on either \cref{line:ero:set_response_acquire} or \cref{line:ero:announce_acquire}.
    Hence, the second parameter of $I$ was read from either $\linearizationobject{}.\cellpointerlong$ or $\announceobject.\cellpointerlong$.
    Since $\linearizationobject{}.\cellpointerlong$ (resp. $\announceobject.\cellpointerlong$) is initially $\nullconstant$, by \Cref{observation:ero:where_objects_change} only $L$-events (resp. $A$-events) change the value of $\linearizationobject{}.\cellpointerlong$ (resp. $\announceobject.\cellpointerlong$), and by \Cref{lemma:ero:every_l_event_is_for_pointer_from_universe} (resp. \Cref{lemma:ero:every_a_event_is_for_pointer_from_universe}) every $L$-event (resp. $A$-event) is for a pointer from $\celluniverse{}$, the lemma follows.
    \qH{\Cref{lemma:ero:acquire_second_parameter_is_pointer_or_null}}
\end{proof}

Since $\&\headobject \notin \celluniverse{} \cup \{\nullconstant\}$ (\Cref{assumption:ero:head_and_null_not_in_cell_universe}), the second parameter of $I$ is not $\&\headobject$, which implies the following.

\begin{corollary}\label{lemma:ero:exit_acquire_implies_executing_105}
    Consider any invocation $I$ of the Acquire procedure that exits in $\mathcal{I}^\mathcal{B}$.
    The process that invoked $I$ executed \cref{line:ero:acquire_next_linearization_changed_check} at least once during $I$.
\end{corollary}

\begin{lemma}\label{lemma:ero:before_any_list_acquire_is_a_successful_list_add_or_list_remove_attempt}
    Consider any list-acquire-next attempt $a_{acquire}$ for $\cellpointershort$ after $\currentcellpointershort$ in $\mathcal{I}^\mathcal{B}$.
    There is either a successful list-add attempt for $\cellpointershort$ after $\currentcellpointershort$ or a successful list-remove attempt between $\currentcellpointershort$ and $\cellpointershort$ before $a_{acquire}$ in $\mathcal{I}^\mathcal{B}$.
\end{lemma}

\begin{proof}
    Let $p$ be the process that executed $a_{acquire}$.
    Since $a_{acquire}$ is a list-acquire-next attempt for $\cellpointershort$ after $\currentcellpointershort$, by \Cref{lemma:ero:list_acquire_next_attempt_for_pointer_from_universe_and_after_pointer_from_universe_or_head}, $\cellpointershort \in \celluniverse$ and $\currentcellpointershort \in \celluniverse \cup \{\&\headobject{}\}$.
    Furthermore, $p$ read $\cellpointershort$ from $(*\currentcellpointershort).\nextlong{}.\cellpointerlong$ on its last execution of \cref{line:ero:acquire_next_read_curr_unique_pointer} before $a_{acquire}$; say at time $T^{\ref{line:ero:acquire_next_read_curr_unique_pointer}}$.
    Hence, since $\cellpointershort \in \celluniverse$, by \Cref{assumption:ero:head_and_null_not_in_cell_universe}, $\cellpointershort \neq \nullconstant$.
    Thus, since $\currentcellpointershort \in \celluniverse \cup \{\&\headobject{}\}$, we have that $(*\currentcellpointershort).\nextlong{}.\cellpointerlong$ is initially $\nullconstant$, and so since $p$ read $\cellpointershort$ from $(*\currentcellpointershort).\nextlong{}.\cellpointerlong$ at $T^{\ref{line:ero:acquire_next_read_curr_unique_pointer}}$, it follows that $(*\currentcellpointershort).\nextlong{}.\cellpointerlong$ was set to $\cellpointershort$ before $T^{\ref{line:ero:acquire_next_read_curr_unique_pointer}}$.
    Therefore, since $\currentcellpointershort \in \celluniverse \cup \{\&\headobject{}\}$, by \Cref{observation:ero:where_objects_change}, there is either a successful list-add attempt for $\cellpointershort$ after $\currentcellpointershort$ or a successful list-remove attempt between $\currentcellpointershort$ and $\cellpointershort$ before $T^{\ref{line:ero:acquire_next_read_curr_unique_pointer}}$ (and thus $a_{acquire}$).
    \qH{\Cref{lemma:ero:before_any_list_acquire_is_a_successful_list_add_or_list_remove_attempt}}
\end{proof}

We now prove some facts about acquire-copy events.

\begin{lemma}\label{lemma:ero:before_acquire_copy_is_successful}
    Consider any acquire-copy event $e$ for $\cellpointershort$ in $\mathcal{I}^\mathcal{B}$.
    Let $p$ be the process that executed $e$ and let $I$ be the invocation of the \doremovecell{} procedure that $e$ was executed during.
    Then, $p$ performed a successful list-remove attempt for $\cellpointershort$ before $e$ during $I$.
\end{lemma}

\begin{proof}
    Since $e$ is an acquire-copy event for $\cellpointershort$ during $I$, by \Cref{def:ero:english}, the second parameter of $I$ is $\cellpointershort$.
    Furthermore, it follows that $p$ executed a successful \CASop{} on \cref{line:ero:remove_cell_from_list} before $e$ during $I$.
    Let $a$ denote this successful \CASop{} on \cref{line:ero:remove_cell_acquire_next} before $e$ during $I$.
    Therefore, since the second parameter of $I$ is $\cellpointershort$, by \Cref{def:ero:english}, $a$ a successful list-remove attempt for $\cellpointershort$.
    \qH{\Cref{lemma:ero:before_acquire_copy_is_successful}}
\end{proof}

\begin{lemma}\label{lemma:ero:every_acquire_copy_event_is_for_pointer_from_universe}
    Every acquire-copy event in $\mathcal{I}^\mathcal{B}$ is for a pointer in $\celluniverse$.
\end{lemma}

\begin{proof}
    By \Cref{lemma:ero:before_acquire_copy_is_successful}, the process that performs any acquire-copy event for some $\cellpointershort$ previously performs a successful list-remove attempt for $\cellpointershort$.
    Thus, by \Cref{lemma:ero:every_list_add_seal_and_remove_attempt_is_for_ptr_from_universe}, $\cellpointershort \in \celluniverse$.
    \qH{\Cref{lemma:ero:every_acquire_copy_event_is_for_pointer_from_universe}}
\end{proof}

We now prove some facts about revocation events.

\begin{lemma}\label{lemma:ero:revocation_event_is_for_pointer_from_universe}
    Every revocation event in $\mathcal{I}^\mathcal{B}$ is for a pointer in $\celluniverse$.
\end{lemma}

\begin{proof}
    Consider any revocation event $e$ for some $\cellpointershort$ in $\mathcal{I}^\mathcal{B}$.
    Hence, by \Cref{def:ero:english}, $e$ was performed by some process $p$ during an invocation $I$ of the Relinquish procedure with parameters $\cellpointershort$.
    Thus, since $e$ is an execution of \cref{line:ero:relinquish_revocations}, we have that $p$ found the condition on \cref{line:ero:early_exit} to be false during $I$, so $\cellpointershort \neq \nullconstant$ and $\cellpointershort \neq \&\headobject{}$.
    Observe that $I$ could be invoked on line \ref{line:ero:owner_relinquish}, \ref{line:ero:add_cell_traversal_relinquish}, \ref{line:ero:add_cell_final_relinquish}, \ref{line:ero:remove_cell_traversal_relinquish}, \ref{line:ero:remove_cell_prev_relinquish}, \ref{line:ero:remove_cell_final_relinquish}, \ref{line:ero:set_response_relinquish}, \ref{line:ero:announce_acquire_relinquish}, \ref{line:ero:acquire_relinquish}.
    We consider each case.

    \begin{itemize}
        \item[] \hspace{0pt}\textbf{Case 1.} $I$ was invoked on \cref{line:ero:owner_relinquish}.

        Hence, since $I$'s parameter is $\cellpointershort$, $p$ received $\cellpointershort$ as a response to an $\allocatecelloperation{}$ operation on \cref{line:ero:allocate_cell}.
        Therefore, by \Cref{alg:lazy_cell_manager_specification}, $\cellpointershort \in \celluniverse$.

        \item[] \hspace{0pt}\textbf{Case 2.} $I$ was invoked on \cref{line:ero:add_cell_traversal_relinquish} or \ref{line:ero:add_cell_final_relinquish}.

        Hence, since $I$'s parameter is $\cellpointershort$, we have that $\cellpointershort$ is the value of $p$'s local variable $\currentuniquecellpointershort$ during an invocation of the \doaddcell{} procedure.
        Thus, by \Cref{lemma:ero:add_cell_current_pointer_always_in_universe_or_head}, $\cellpointershort \in \celluniverse{} \cup \{\&\headobject\}$.
        Therefore, since $\cellpointershort \neq \&\headobject$, we have that $\cellpointershort \in \celluniverse{}$ as wanted.

        \item[] \hspace{0pt}\textbf{Case 3.} $I$ was invoked on \cref{line:ero:remove_cell_traversal_relinquish} or \ref{line:ero:remove_cell_prev_relinquish}.

        Hence, since $I$'s parameter is $\cellpointershort$, we have that $\cellpointershort$ is the value of $p$'s local variable $\previousuniquecellpointershort$ during an invocation of the \doremovecell{} procedure.
        Thus, by \Cref{lemma:ero:remove_cell_previous_pointer_always_in_universe_or_head_or_null}, $\cellpointershort \in \celluniverse{} \cup \{\&\headobject\}$.
        Therefore, since $\cellpointershort \neq \nullconstant$ and $\cellpointershort \neq \&\headobject$, we have that $\cellpointershort \in \celluniverse{}$ as wanted.

        \item[] \hspace{0pt}\textbf{Case 4.} $I$ was invoked on \cref{line:ero:remove_cell_final_relinquish}.

        Hence, since $I$'s parameter is $\cellpointershort$, we have that $\cellpointershort$ is the value of $p$'s local variable $\currentuniquecellpointershort$ during an invocation of the \doremovecell{} procedure.
        Thus, by \Cref{lemma:ero:remove_cell_current_pointer_always_in_universe_or_head}, $\cellpointershort \in \celluniverse{} \cup \{\&\headobject\}$.
        Therefore, since $\cellpointershort \neq \&\headobject$, we have that $\cellpointershort \in \celluniverse{}$ as wanted.

        \item[] \hspace{0pt}\textbf{Case 5.} $I$ was invoked on \cref{line:ero:set_response_relinquish}.

        Hence, since $I$'s parameter is $\cellpointershort$, we have that $\cellpointershort$ is the second parameter of some invocation of the \setrepositoryoperationresponse{} procedure.
        Thus, by \Cref{lemma:ero:l_event_corresponding_to_set_response}, there is an $L$-event for $\cellpointershort$ in $\mathcal{I}^\mathcal{B}$, and so by \Cref{lemma:ero:every_l_event_is_for_pointer_from_universe} $\cellpointershort \in \celluniverse{}$ as wanted.

        \item[] \hspace{0pt}\textbf{Case 6.} $I$ was invoked on \cref{line:ero:announce_acquire_relinquish}.

        Hence, since $I$'s parameter is $\cellpointershort$, we have that $p$ read $\cellpointershort$ from $\announceobject.\cellpointerlong$.
        Since $\announceobject.\cellpointerlong$ is initially $\nullconstant$, and $\cellpointershort \neq \nullconstant$, we have that $\announceobject.\cellpointerlong$ was set to $\cellpointershort$.
        Hence, by \Cref{observation:ero:where_objects_change}, some $A$-event $e_A$ set $\announceobject.\cellpointerlong = \cellpointershort$, and so by \Cref{def:ero:english}, $e_A$ is an $A$-event for $\cellpointershort$.
        Therefore, by \Cref{lemma:ero:every_a_event_is_for_pointer_from_universe}, $\cellpointershort \in \celluniverse$ as wanted.

        \item[] \hspace{0pt}\textbf{Case 7.} $I$ was invoked on \cref{line:ero:acquire_relinquish}.

        Hence, since $I$'s parameter is $\cellpointershort$, we have that $\cellpointershort$ is the value of $p$'s local variable $\currentuniquecellpointershort$ during an invocation of the Acquire procedure.
        Thus, by \Cref{lemma:ero:aquire_current_pointer_always_in_universe_or_head}, $\cellpointershort \in \celluniverse{} \cup \{\&\headobject\}$.
        Therefore, since $\cellpointershort \neq \&\headobject$, we have that $\cellpointershort \in \celluniverse{}$ as wanted.
        \qH{\Cref{lemma:ero:revocation_event_is_for_pointer_from_universe}}
    \end{itemize}
\end{proof}

Since every $\freecelloperation(\cellpointershort)$ operation is preceded by a revocation event for $\cellpointershort$, \Cref{lemma:ero:revocation_event_is_for_pointer_from_universe} implies the following.

\begin{corollary}\label{lemma:ero:free_cell_operations_are_from_universe}
    The input of every $\freecelloperation$ in $\mathcal{I}^\mathcal{B}$ is in $\celluniverse{}$.
\end{corollary}

We are now ready to prove that every line of $\mathcal{B}$ that tries to perform an operation on an object of a cell actually does.
In other words, in $\mathcal{B}$, no step de-references a value that isn't in $\celluniverse{} \cup \{\&\headobject\}$.

\begin{lemma}\label{lemma:ero:b_never_performs_an_op_on_a_bad_pointer}
    Every execution of line \ref{line:ero:copy_response_out_of_cell}, \ref{line:ero:do_work_initialize_response}, \ref{line:ero:do_work_while_loop}, \ref{line:ero:add_cell_read_end_of_list}, \ref{line:ero:add_cell_to_list}, \ref{line:ero:remove_cell_remove_seal_loop}, \ref{line:ero:remove_cell_read_pointer_to_remove_before_seal}, \ref{line:ero:seal_cell}, \ref{line:ero:remove_cell_read_pointer_to_remove}, \ref{line:ero:remove_cell_read_previous_pointer}, \ref{line:ero:remove_cell_from_list}, \ref{line:ero:copy_acquisitions_to_revocations}, \ref{line:ero:responses_set_attempt}, \ref{line:ero:announce_op_response_check}, \ref{line:ero:acquire_next_read_curr_unique_pointer}, \ref{line:ero:acquire_next_cell}, and \ref{line:ero:relinquish_revocations} in $\mathcal{I}^\mathcal{B}$ performs an operation on an object of a cell whose pointer is in $\celluniverse{} \cup \{\&\headobject\}$.
\end{lemma}

\begin{proof}
    The claim is trivial for lines \ref{line:ero:copy_response_out_of_cell}, \ref{line:ero:do_work_initialize_response}, and \ref{line:ero:do_work_while_loop}.
    For \cref{line:ero:add_cell_read_end_of_list} the claim follows from \Cref{lemma:ero:add_cell_current_pointer_always_in_universe_or_head}.
    For \cref{line:ero:add_cell_to_list} the claim follows from \Cref{lemma:ero:every_list_add_attempt_is_after_a_pointer_from_universe_or_head}.
    For lines \ref{line:ero:remove_cell_remove_seal_loop}, \ref{line:ero:remove_cell_read_pointer_to_remove_before_seal}, \ref{line:ero:seal_cell}, and \ref{line:ero:remove_cell_read_pointer_to_remove} the claim follows from \Cref{lemma:ero:l_event_corresponding_to_do_low_level_op}.
    For lines \ref{line:ero:remove_cell_read_previous_pointer} and \ref{line:ero:remove_cell_from_list}, let $I$ be the invocation of the \doremovecell{} procedure that either is executed in, and let $\cellpointershort_\linearizationobject{}$ be the second parameter of $I$.
    Hence, by \Cref{lemma:ero:l_event_corresponding_to_do_low_level_op}, $\cellpointershort_\linearizationobject{} \in \celluniverse{}$, and so by \Cref{assumption:ero:head_and_null_not_in_cell_universe}, $\cellpointershort_\linearizationobject{} \neq \&\headobject$.
    Thus, the process that invoked $I$ found the condition on \cref{line:ero:remove_cell_while_loop} to be false its first time in $I$, and so it executed \cref{line:ero:remove_cell_update_pointers} at least once during $I$ (since it executed line \ref{line:ero:remove_cell_read_previous_pointer} or \ref{line:ero:remove_cell_from_list} in $I$).
    So, the claim follows from \Cref{lemma:ero:remove_cell_previous_pointer_always_in_universe_or_head}.
    For \cref{line:ero:copy_acquisitions_to_revocations} the claim follows from \Cref{lemma:ero:every_acquire_copy_event_is_for_pointer_from_universe}.
    For \cref{line:ero:responses_set_attempt} the claim follows from \Cref{lemma:ero:every_response_set_attempt_is_for_pointer_from_universe}.
    For \cref{line:ero:announce_op_response_check}, observe that $(\uniquerepositoryoperationshort_\announceobject{}, \cellpointershort_\announceobject{})$ was read from $\announceobject$, which is initially $((0, \noop), \nullconstant)$, and since $\uniquerepositoryoperationshort_\announceobject{} = (\arbitraryvalue{}, \langle \doopandcopyresponse{}, \arbitraryvalue{}\rangle)$ by the condition on \cref{line:ero:write_and_read_done_check}, we have that $\announceobject$ was set to $(\uniquerepositoryoperationshort_\announceobject{}, \cellpointershort_\announceobject{})$, and so by \Cref{observation:ero:where_objects_change}, some $A$-event set $\announceobject$ to $(\uniquerepositoryoperationshort_\announceobject{}, \cellpointershort_\announceobject{})$, and thus by \Cref{lemma:ero:every_a_event_is_for_pointer_from_universe}, $\cellpointershort_\announceobject \in \celluniverse{}$.
    For lines \ref{line:ero:acquire_next_read_curr_unique_pointer} and \ref{line:ero:acquire_next_cell} the claim follows from \Cref{lemma:ero:acquire_next_is_for_pointer_from_universe_or_head}.
    Finally, for \cref{line:ero:relinquish_revocations}, the claim follows from \Cref{lemma:ero:revocation_event_is_for_pointer_from_universe}.
    \qH{\Cref{lemma:ero:b_never_performs_an_op_on_a_bad_pointer}}
\end{proof}

% \textcolor{red}{At this point, we know that every step in $\mathcal{B}$ either performs no operation on a shared object, an operation on a static object, or an operation on an object of some cell in $\celluniverse{}$. It seems like a model-level issue because we technically don't care about this.}

\subsubsection{The \doworkuntildone{} procedure}

We now prove that if an invocation of the \doworkuntildone{} procedure exits, then the low-level operation it was trying to do is ``done" in the sense described below.

\begin{lemma}\label{lemma:ero:at_most_one_response_reset_for_ptr_during_low_level_op_never_exits}
    Consider any invocation $I$ of the \doworkuntildone{} procedure with a second parameter of $\cellpointershort{}$ that never exits in $\mathcal{I}^\mathcal{B}$.
    Let $T^{\ref{line:ero:do_work_initialize_response}}$ be the time \cref{line:ero:do_work_initialize_response} is executed during $I$.
    Then, from $T^{\ref{line:ero:do_work_initialize_response}}$ onwards in $\mathcal{I}^\mathcal{B}$ there are no response-reset events for $\cellpointershort$.
\end{lemma}

\begin{proof}
    Suppose, for contradiction, there is a response-reset event $e$ for $\cellpointershort$ after $T^{\ref{line:ero:do_work_initialize_response}}$.
    Let $p$ be the process that invoked $I$ and let $q$ be the process that executed $e$.
    Let $I'$ be the invocation of the \doworkuntildone{} procedure that $q$ executed $e$ during.
    Since $e$ is a response-reset event for $\cellpointershort$, the second parameter of $I'$ is $\cellpointershort{}$.
    Since the second parameter of $I$ (resp. $I'$) is $\cellpointershort{}$, it follows that $p$ (resp. $q$) received $\cellpointershort$ as a response on \cref{line:ero:allocate_cell}.
    Hence, since by \Cref{alg:lazy_cell_manager_specification} every response on \cref{line:ero:allocate_cell} is unique, we have that $p = q$.
    Therefore, since $e$ is after $T^{\ref{line:ero:do_work_initialize_response}}$, $p$ never exits $I$ after $T^{\ref{line:ero:do_work_initialize_response}}$, and $p$ executes \cref{line:ero:do_work_initialize_response} at $T^{\ref{line:ero:do_work_initialize_response}}$, $p$ executes \cref{line:ero:do_work_initialize_response} twice during $I$.
    However, there is at most one execution of \cref{line:ero:do_work_initialize_response} per invocation of the \doworkuntildone{} procedure, a contradiction.
    \qH{\Cref{lemma:ero:at_most_one_response_reset_for_ptr_during_low_level_op_never_exits}}
\end{proof}

\begin{proposition}\label{lemma:ero:at_most_one_response_reset_for_ptr_during_low_level_op}
    Consider any invocation $I$ of the \doworkuntildone{} procedure with a second parameter of $\cellpointershort$ that exits at some time $T_e$ in $\mathcal{I}^\mathcal{B}$.
    Let $T^{\ref{line:ero:do_work_initialize_response}}$ be the time \cref{line:ero:do_work_initialize_response} is executed during $I$.
    Then, between $T^{\ref{line:ero:do_work_initialize_response}}$ and $T_e$, there are no response-reset events for $\cellpointershort$.
\end{proposition}

\begin{proof}
    Let $\mathcal{I}$ be the prefix of $\mathcal{I}^\mathcal{B}$ up to but excluding the last step of $I$, so $I$ never exits in $\mathcal{I}$.
    By plugging in $\mathcal{I}$ for $\mathcal{I}^\mathcal{B}$ in \Cref{lemma:ero:at_most_one_response_reset_for_ptr_during_low_level_op_never_exits} the claim follows.
    \qH{\Cref{lemma:ero:at_most_one_response_reset_for_ptr_during_low_level_op}}
\end{proof}

\begin{lemma}\label{lemma:ero:successful_add_response_set_before_add_low_level_exits}
    Consider any invocation $I$ of the \doworkuntildone{} procedure with parameters $(\addcell, \cellpointershort{})$ that exits at some time $T_e$ in $\mathcal{I}^\mathcal{B}$.
    Let $T^{\ref{line:ero:do_work_initialize_response}}$ be the time \cref{line:ero:do_work_initialize_response} is executed during $I$.
    Then, between $T^{\ref{line:ero:do_work_initialize_response}}$ and $T_e$, there is a successful add-response-set attempt for $\cellpointershort$.
\end{lemma}
 
\begin{proof}
    Let $p$ be the process that executed $I$.
    Hence, $p$ received $\cellpointershort$ as a response on \cref{line:ero:allocate_cell}, and so by \Cref{alg:lazy_cell_manager_specification} $\cellpointershort \in \celluniverse$.
    Since $p$ exited $I$, we have that $p$ finds the condition on \cref{line:ero:do_work_while_loop} to be false at some time during $I$; say at time $T^{\ref{line:ero:do_work_while_loop}}$.
    Hence, since $p$ set $(*\cellpointershort).\lastrepositoryoperationresponse{} = (\myuniquerepositoryoperationshort{}, \nullconstant{})$ at $T^{\ref{line:ero:do_work_initialize_response}}$ during $I$, and then later found \cref{line:ero:do_work_while_loop} to be false at $T^{\ref{line:ero:do_work_while_loop}}$ during $I$, we have that between $T^{\ref{line:ero:do_work_initialize_response}}$ and $T^{\ref{line:ero:do_work_while_loop}}$, the value of $(*\cellpointershort).\lastrepositoryoperationresponse{}$ changed.
    Thus, since $\cellpointershort \in \celluniverse$, by \Cref{observation:ero:where_objects_change}, there is either a response-reset event for $\cellpointershort$ or a successful response-set attempt for $\cellpointershort$ between $T^{\ref{line:ero:do_work_initialize_response}}$ and $T^{\ref{line:ero:do_work_while_loop}}$.
    Hence, since $T^{\ref{line:ero:do_work_while_loop}}$ is during $I$ and $I$ exits at $T_e$, by transitivity, $T^{\ref{line:ero:do_work_while_loop}} < T_e$, and so there is either a response-reset event for $\cellpointershort$ or a successful response-set attempt for $\cellpointershort$ between $T^{\ref{line:ero:do_work_initialize_response}}$ and $T_e$.
    However, since by \Cref{lemma:ero:at_most_one_response_reset_for_ptr_during_low_level_op} there are no response-reset events for $\cellpointershort$ between $T^{\ref{line:ero:do_work_initialize_response}}$ and $T_e$, we have that there is a successful response-set attempt for $\cellpointershort$ between $T^{\ref{line:ero:do_work_initialize_response}}$ and $T_e$.
    Let $a$ be the first successful response-set attempt for $\cellpointershort$ between $T^{\ref{line:ero:do_work_initialize_response}}$ and $T_e$.
    Hence, since $(*\cellpointershort).\lastrepositoryoperationresponse{} = (\myuniquerepositoryoperationshort{}, \nullconstant{})$ at $T^{\ref{line:ero:do_work_initialize_response}}$, and $a$ is a successful execution of \cref{line:ero:responses_set_attempt}, we have that $a$ is of the form \CASop{}($(*\cellpointershort).\lastrepositoryoperationresponse{}$, $(\myuniquerepositoryoperationshort{}, \nullconstant{})$, $\arbitraryvalue$).
    Since $(*\cellpointershort).\lastrepositoryoperationresponse{} = (\myuniquerepositoryoperationshort{}, \nullconstant{})$ at $T^{\ref{line:ero:do_work_initialize_response}}$ during $I$ and $I$'s first parameter is $\addcell$, we have that $\myuniquerepositoryoperationshort{} = (\arbitraryvalue, \addcell)$.
    Hence, $a$ is of the form \CASop{}($(*\cellpointershort).\lastrepositoryoperationresponse{}$, $((\arbitraryvalue, \addcell), \nullconstant{})$, $\arbitraryvalue$), and so by \Cref{def:ero:english}, $a$ is a successful add-response-set attempt for $\cellpointershort$.
    Therefore, since $a$ is between $T^{\ref{line:ero:do_work_initialize_response}}$ and $T_e$, we have there is a successful add-response-set attempt for $\cellpointershort$ between $T^{\ref{line:ero:do_work_initialize_response}}$ and $T_e$ as wanted.
    \qH{\Cref{lemma:ero:successful_add_response_set_before_add_low_level_exits}}
\end{proof}

\begin{lemma}\label{lemma:ero:successful_apply_response_set_before_apply_low_level_exits}
    Consider any invocation $I$ of the \doworkuntildone{} procedure with parameters $(\langle \doopandcopyresponse{}, \arbitraryvalue \rangle, \cellpointershort)$ that exits at some time $T_e$ in $\mathcal{I}^\mathcal{B}$.
    Let $T^{\ref{line:ero:do_work_initialize_response}}$ be the time \cref{line:ero:do_work_initialize_response} is executed during $I$.
    Then, between $T^{\ref{line:ero:do_work_initialize_response}}$ and $T_e$, there is a successful apply-response-set attempt for $\cellpointershort$.
\end{lemma}

\begin{proof}
    By essentially the same argument as \Cref{lemma:ero:successful_add_response_set_before_add_low_level_exits}, which we provide below for completeness.
    Let $p$ be the process that executed $I$.
    Hence, $p$ received $\cellpointershort$ as a response on \cref{line:ero:allocate_cell}, and so by \Cref{alg:lazy_cell_manager_specification} $\cellpointershort \in \celluniverse$.
    Since $p$ exited $I$, we have that $p$ finds the condition on \cref{line:ero:do_work_while_loop} to be false at some time during $I$; say at time $T^{\ref{line:ero:do_work_while_loop}}$.
    Hence, since $p$ set $(*\cellpointershort).\lastrepositoryoperationresponse{} = (\myuniquerepositoryoperationshort{}, \nullconstant{})$ at $T^{\ref{line:ero:do_work_initialize_response}}$ during $I$, and then later found \cref{line:ero:do_work_while_loop} to be false at $T^{\ref{line:ero:do_work_while_loop}}$ during $I$, we have that between $T^{\ref{line:ero:do_work_initialize_response}}$ and $T^{\ref{line:ero:do_work_while_loop}}$, the value of $(*\cellpointershort).\lastrepositoryoperationresponse{}$ changed.
    Thus, since $\cellpointershort \in \celluniverse$, by \Cref{observation:ero:where_objects_change}, there is either a response-reset event for $\cellpointershort$ or a successful response-set attempt for $\cellpointershort$ between $T^{\ref{line:ero:do_work_initialize_response}}$ and $T^{\ref{line:ero:do_work_while_loop}}$.
    Hence, since $T^{\ref{line:ero:do_work_while_loop}}$ is during $I$ and $I$ exits at $T_e$, by transitivity, $T^{\ref{line:ero:do_work_while_loop}} < T_e$, and so there is either a response-reset event for $\cellpointershort$ or a successful response-set attempt for $\cellpointershort$ between $T^{\ref{line:ero:do_work_initialize_response}}$ and $T_e$.
    However, since by \Cref{lemma:ero:at_most_one_response_reset_for_ptr_during_low_level_op} there are no response-reset events for $\cellpointershort$ between $T^{\ref{line:ero:do_work_initialize_response}}$ and $T_e$, we have that there is a successful response-set attempt for $\cellpointershort$ between $T^{\ref{line:ero:do_work_initialize_response}}$ and $T_e$.
    Let $a$ be the first successful response-set attempt for $\cellpointershort$ between $T^{\ref{line:ero:do_work_initialize_response}}$ and $T_e$.
    Hence, since $(*\cellpointershort).\lastrepositoryoperationresponse{} = (\myuniquerepositoryoperationshort{}, \nullconstant{})$ at $T^{\ref{line:ero:do_work_initialize_response}}$, and $a$ is a successful execution of \cref{line:ero:responses_set_attempt}, we have that $a$ is of the form \CASop{}($(*\cellpointershort).\lastrepositoryoperationresponse{}$, $(\myuniquerepositoryoperationshort{}, \nullconstant{})$, $\arbitraryvalue$).
    Since $(*\cellpointershort).\lastrepositoryoperationresponse{} = (\myuniquerepositoryoperationshort{}, \nullconstant{})$ at $T^{\ref{line:ero:do_work_initialize_response}}$ during $I$ and $I$'s first parameter is $\langle \doopandcopyresponse{}, \arbitraryvalue \rangle$, we have that $\myuniquerepositoryoperationshort{} = (\arbitraryvalue, \langle \doopandcopyresponse{}, \arbitraryvalue \rangle)$.
    Hence, $a$ is of the form \CASop{}($(*\cellpointershort).\lastrepositoryoperationresponse{}$, $((\arbitraryvalue, \langle \doopandcopyresponse{}, \arbitraryvalue \rangle), \nullconstant{})$, $\arbitraryvalue$), and so by \Cref{def:ero:english}, $a$ is a successful apply-response-set attempt for $\cellpointershort$.
    Therefore, since $a$ is between $T^{\ref{line:ero:do_work_initialize_response}}$ and $T_e$, we have there is a successful apply-response-set attempt for $\cellpointershort$ between $T^{\ref{line:ero:do_work_initialize_response}}$ and $T_e$ as wanted.
    \qH{\Cref{lemma:ero:successful_apply_response_set_before_apply_low_level_exits}}
\end{proof}

\begin{lemma}\label{lemma:ero:successful_remove_response_set_before_remove_low_level_exits}
    Consider any invocation $I$ of the \doworkuntildone{} procedure with parameters $(\removecell, \cellpointershort)$ that exits at some time $T_e$ in $\mathcal{I}^\mathcal{B}$.
    Let $T^{\ref{line:ero:do_work_initialize_response}}$ be the time \cref{line:ero:do_work_initialize_response} is executed during $I$.
    Then, between $T^{\ref{line:ero:do_work_initialize_response}}$ and $T_e$, there is a successful remove-response-set attempt for $\cellpointershort$.
\end{lemma}

\begin{proof}
    By essentially the same argument as \Cref{lemma:ero:successful_add_response_set_before_add_low_level_exits}, which we provide below for completeness.
    Let $p$ be the process that executed $I$.
    Hence, $p$ received $\cellpointershort$ as a response on \cref{line:ero:allocate_cell}, and so by \Cref{alg:lazy_cell_manager_specification} $\cellpointershort \in \celluniverse$.
    Since $p$ exited $I$, we have that $p$ finds the condition on \cref{line:ero:do_work_while_loop} to be false at some time during $I$; say at time $T^{\ref{line:ero:do_work_while_loop}}$.
    Hence, since $p$ set $(*\cellpointershort).\lastrepositoryoperationresponse{} = (\myuniquerepositoryoperationshort{}, \nullconstant{})$ at $T^{\ref{line:ero:do_work_initialize_response}}$ during $I$, and then later found \cref{line:ero:do_work_while_loop} to be false at $T^{\ref{line:ero:do_work_while_loop}}$ during $I$, we have that between $T^{\ref{line:ero:do_work_initialize_response}}$ and $T^{\ref{line:ero:do_work_while_loop}}$, the value of $(*\cellpointershort).\lastrepositoryoperationresponse{}$ changed.
    Thus, since $\cellpointershort \in \celluniverse$, by \Cref{observation:ero:where_objects_change}, there is either a response-reset event for $\cellpointershort$ or a successful response-set attempt for $\cellpointershort$ between $T^{\ref{line:ero:do_work_initialize_response}}$ and $T^{\ref{line:ero:do_work_while_loop}}$.
    Hence, since $T^{\ref{line:ero:do_work_while_loop}}$ is during $I$ and $I$ exits at $T_e$, by transitivity, $T^{\ref{line:ero:do_work_while_loop}} < T_e$, and so there is either a response-reset event for $\cellpointershort$ or a successful response-set attempt for $\cellpointershort$ between $T^{\ref{line:ero:do_work_initialize_response}}$ and $T_e$.
    However, since by \Cref{lemma:ero:at_most_one_response_reset_for_ptr_during_low_level_op} there are no response-reset events for $\cellpointershort$ between $T^{\ref{line:ero:do_work_initialize_response}}$ and $T_e$, we have that there is a successful response-set attempt for $\cellpointershort$ between $T^{\ref{line:ero:do_work_initialize_response}}$ and $T_e$.
    Let $a$ be the first successful response-set attempt for $\cellpointershort$ between $T^{\ref{line:ero:do_work_initialize_response}}$ and $T_e$.
    Hence, since $(*\cellpointershort).\lastrepositoryoperationresponse{} = (\myuniquerepositoryoperationshort{}, \nullconstant{})$ at $T^{\ref{line:ero:do_work_initialize_response}}$, and $a$ is a successful execution of \cref{line:ero:responses_set_attempt}, we have that $a$ is of the form \CASop{}($(*\cellpointershort).\lastrepositoryoperationresponse{}$, $(\myuniquerepositoryoperationshort{}, \nullconstant{})$, $\arbitraryvalue$).
    Since $(*\cellpointershort).\lastrepositoryoperationresponse{} = (\myuniquerepositoryoperationshort{}, \nullconstant{})$ at $T^{\ref{line:ero:do_work_initialize_response}}$ during $I$ and $I$'s first parameter is $\removecell$, we have that $\myuniquerepositoryoperationshort{} = (\arbitraryvalue, \removecell)$.
    Hence, $a$ is of the form \CASop{}($(*\cellpointershort).\lastrepositoryoperationresponse{}$, $((\arbitraryvalue, \removecell), \nullconstant{})$, $\arbitraryvalue$), and so by \Cref{def:ero:english}, $a$ is a successful remove-response-set attempt for $\cellpointershort$.
    Therefore, since $a$ is between $T^{\ref{line:ero:do_work_initialize_response}}$ and $T_e$, we have there is a successful remove-response-set attempt for $\cellpointershort$ between $T^{\ref{line:ero:do_work_initialize_response}}$ and $T_e$ as wanted.
    \qH{\Cref{lemma:ero:successful_remove_response_set_before_remove_low_level_exits}}
\end{proof}

\begin{lemma}\label{lemma:ero:l_add_event_before_add_low_level_exits}
    Consider any invocation $I$ of the \doworkuntildone{} procedure with parameters $(\addcell, \cellpointershort)$ that is invoked at time $T_b$ and exits at some time $T_e$ in $\mathcal{I}^\mathcal{B}$.
    If $P(\mathcal{I}^\mathcal{B})$ holds, then there is an $L$-add event for $\cellpointershort$ between $T_b$ and $T_e$.
\end{lemma}

\begin{proof}
    Let $p$ be the process that invoked $I$.
    We first prove that there is an $L$-add event for $\cellpointershort$ before $T_e$.
    Let $T^{\ref{line:ero:do_work_initialize_response}}$ be the time \cref{line:ero:do_work_initialize_response} is executed during $I$.
    By \Cref{lemma:ero:successful_add_response_set_before_add_low_level_exits}, there is a successful add-response-set attempt for $\cellpointershort$ between $T^{\ref{line:ero:do_work_initialize_response}}$ and $T_e$.
    Let $a$ be this successful add-response-set attempt.
    Hence, by \Cref{corollary:ero:response_set_attempts_and_corresponding_l_events_are_matching}, there is an $L$-add event $e$ for $\cellpointershort$ before $a$ (and thus $T_e$ since $a < T_e$).
    Thus, by $P(\mathcal{I}^\mathcal{B})$, $e$ is the only $L$-add event for $\cellpointershort$ in $\mathcal{I}^\mathcal{B}$.
    We now prove that $e$ is after $T_b$, which completes the proof.
    Suppose, for contradiction, $e < T_b$.
    Since $e$ is an $L$-add event for $\cellpointershort$, by \Cref{def:ero:english}, $e$ set $\linearizationobject{}$ to a value of the form $((\arbitraryvalue, \addcell), \cellpointershort)$.
    Hence, by \Cref{lemma:ero:l_events_have_corresponding_a_events}, there is an $A$-event $e'$ before $e$ which set $\announceobject$ to the same value.
    Thus, by \Cref{def:ero:english}, $e'$ is an $A$-add event for $\cellpointershort$.
    Let $q$ be the process that executed $e'$.
    Since $e'$ is an $A$-add event for $\cellpointershort$, we have that $q$ executed $e'$ during an invocation $I'$ of the \doworkuntildone{} procedure with parameters of the form $(\addcell, \cellpointershort)$.
    Hence, since $e'$ is before $e$, and by assumption $e < T_b$, we have that $I'$ was invoked before $I$ was invoked, and so $I' \neq I$.
    Therefore, since $I$ and $I'$ are both invocations of the \doworkuntildone{} procedure with parameters of the form $(\addcell, \cellpointershort)$, we have that there are two invocations of the \doworkuntildone{} procedure with parameters of the form $(\addcell, \cellpointershort)$ in $\mathcal{I}^\mathcal{B}$.
    However, since the response on \cref{line:ero:allocate_cell} is unique, there is at most one invocation of the \doworkuntildone{} procedure with these parameters in $\mathcal{I}^\mathcal{B}$, a contradiction.
    \qH{\Cref{lemma:ero:l_add_event_before_add_low_level_exits}}
\end{proof}

\begin{lemma}\label{lemma:ero:l_apply_event_before_apply_low_level_exits}
    Consider any invocation $I$ of the \doworkuntildone{} procedure with parameters $(\langle \doopandcopyresponse{}, \arbitraryvalue \rangle, \cellpointershort)$ that is invoked at time $T_b$ and exits at some time $T_e$ in $\mathcal{I}^\mathcal{B}$.
    If $P(\mathcal{I}^\mathcal{B})$ holds, then there is an $L$-apply event for $\cellpointershort$ between $T_b$ and $T_e$.
\end{lemma}

\begin{proof}
    By essentially the same argument as \Cref{lemma:ero:l_add_event_before_add_low_level_exits}, which we provide below for completeness.
    Let $p$ be the process that invoked $I$.
    We first prove that there is an $L$-apply event for $\cellpointershort$ before $T_e$.
    Let $T^{\ref{line:ero:do_work_initialize_response}}$ be the time \cref{line:ero:do_work_initialize_response} is executed during $I$.
    By \Cref{lemma:ero:successful_apply_response_set_before_apply_low_level_exits}, there is a successful apply-response-set attempt for $\cellpointershort$ between $T^{\ref{line:ero:do_work_initialize_response}}$ and $T_e$.
    Let $a$ be this successful apply-response-set attempt.
    Hence, by \Cref{corollary:ero:response_set_attempts_and_corresponding_l_events_are_matching}, there is an $L$-apply event $e$ for $\cellpointershort$ before $a$ (and thus $T_e$ since $a < T_e$).
    Thus, by $P(\mathcal{I}^\mathcal{B})$, $e$ is the only $L$-apply event for $\cellpointershort$ in $\mathcal{I}^\mathcal{B}$.
    We now prove that $e$ is after $T_b$, which completes the proof.
    Suppose, for contradiction, $e < T_b$.
    Since $e$ is an $L$-apply event for $\cellpointershort$, by \Cref{def:ero:english}, $e$ set $\linearizationobject{}$ to $((\arbitraryvalue, \langle \doopandcopyresponse{}, \arbitraryvalue \rangle), \cellpointershort)$.
    Hence, by \Cref{lemma:ero:l_events_have_corresponding_a_events}, there is an $A$-event $e'$ before $e$ which set $\announceobject$ to the same value.
    Thus, by \Cref{def:ero:english}, $e'$ is an $A$-apply event for $\cellpointershort$.
    Let $q$ be the process that executed $e'$.
    Since $e'$ is an $A$-apply event for $\cellpointershort$, we have that $q$ executed $e'$ during an invocation $I'$ of the \doworkuntildone{} procedure with parameters $(\langle \doopandcopyresponse{}, \arbitraryvalue \rangle, \cellpointershort)$.
    Hence, since $e'$ is before $e$, and by assumption $e < T_b$, we have that $I'$ was invoked before $I$ was invoked, and so $I' \neq I$.
    Therefore, since $I$ and $I'$ are both invocations of the \doworkuntildone{} procedure with parameters $(\langle \doopandcopyresponse{}, \arbitraryvalue \rangle, \cellpointershort)$, we have that there are two invocations of the \doworkuntildone{} procedure with parameters $(\langle \doopandcopyresponse{}, \arbitraryvalue \rangle, \cellpointershort)$ in $\mathcal{I}^\mathcal{B}$.
    However, since the response on \cref{line:ero:allocate_cell} is unique, there is at most one invocation of the \doworkuntildone{} procedure with these parameters in $\mathcal{I}^\mathcal{B}$, a contradiction.
    \qH{\Cref{lemma:ero:l_apply_event_before_apply_low_level_exits}}
\end{proof}

\begin{lemma}\label{lemma:ero:l_remove_event_before_remove_low_level_exits}
    Consider any invocation $I$ of the \doworkuntildone{} procedure with parameters $(\removecell, \cellpointershort)$ that is invoked at time $T_b$ and exits at some time $T_e$ in $\mathcal{I}^\mathcal{B}$.
    If $P(\mathcal{I}^\mathcal{B})$ holds, then there is an $L$-remove event for $\cellpointershort$ between $T_b$ and $T_e$.
\end{lemma}

\begin{proof}
    By essentially the same argument as \Cref{lemma:ero:l_add_event_before_add_low_level_exits}, which we provide below for completeness.
    Let $p$ be the process that invoked $I$.
    We first prove that there is an $L$-remove event for $\cellpointershort$ before $T_e$.
    Let $T^{\ref{line:ero:do_work_initialize_response}}$ be the time \cref{line:ero:do_work_initialize_response} is executed during $I$.
    By \Cref{lemma:ero:successful_remove_response_set_before_remove_low_level_exits}, there is a successful remove-response-set attempt for $\cellpointershort$ between $T^{\ref{line:ero:do_work_initialize_response}}$ and $T_e$.
    Let $a$ be this successful remove-response-set attempt.
    Hence, by \Cref{corollary:ero:response_set_attempts_and_corresponding_l_events_are_matching}, there is an $L$-remove event $e$ for $\cellpointershort$ before $a$ (and thus $T_e$ since $a < T_e$).
    Thus, by $P(\mathcal{I}^\mathcal{B})$, $e$ is the only $L$-remove event for $\cellpointershort$ in $\mathcal{I}^\mathcal{B}$.
    We now prove that $e$ is after $T_b$, which completes the proof.
    Suppose, for contradiction, $e < T_b$.
    Since $e$ is an $L$-remove event for $\cellpointershort$, by \Cref{def:ero:english}, $e$ set $\linearizationobject{}$ to $((\arbitraryvalue, \removecell), \cellpointershort)$.
    Hence, by \Cref{lemma:ero:l_events_have_corresponding_a_events}, there is an $A$-event $e'$ before $e$ which set $\announceobject$ to the same value.
    Thus, by \Cref{def:ero:english}, $e'$ is an $A$-remove event for $\cellpointershort$.
    Let $q$ be the process that executed $e'$.
    Since $e'$ is an $A$-remove event for $\cellpointershort$, we have that $q$ executed $e'$ during an invocation $I'$ of the \doworkuntildone{} procedure with parameters of the form $(\removecell, \cellpointershort)$.
    Hence, since $e'$ is before $e$, and by assumption $e < T_b$, we have that $I'$ was invoked before $I$ was invoked, and so $I' \neq I$.
    Therefore, since $I$ and $I'$ are both invocations of the \doworkuntildone{} procedure with parameters of the form $(\removecell, \cellpointershort)$, we have that there are two invocations of the \doworkuntildone{} procedure with parameters of the form $(\removecell, \cellpointershort)$ in $\mathcal{I}^\mathcal{B}$.
    However, since the response on \cref{line:ero:allocate_cell} is unique, there is at most one invocation of the \doworkuntildone{} procedure with these parameters in $\mathcal{I}^\mathcal{B}$, a contradiction.
    \qH{\Cref{lemma:ero:l_remove_event_before_remove_low_level_exits}}
\end{proof}

\begin{lemma}\label{lemma:ero:l_apply_event_before_apply_low_level_exits_is_for_correct_timestamp}
    Consider any invocation $I$ of the \doworkuntildone{} procedure with parameters of the form $(\langle \doopandcopyresponse{}, \arbitraryvalue \rangle, \cellpointershort)$ that is invoked at time $T_b$ and exits at some time $T_e$ in $\mathcal{I}^\mathcal{B}$.
    Suppose $P(\mathcal{I}^\mathcal{B})$ holds.
    Let $\timeshort{}$ be the response on \cref{line:ero:operation_timestamp} during $I$ and let $e$ be the $L$-apply event for $\cellpointershort$ between $T_b$ and $T_e$ identified by \Cref{lemma:ero:l_apply_event_before_apply_low_level_exits}.
    Then, $e$ is for timestamp $\timeshort{}$.
\end{lemma}

\begin{proof}
    Suppose, for contradiction $e$ is for a timestamp $\timeshort{}' \neq \timeshort{}$.
    Hence, since $e$ is an $L$-apply event for $\cellpointershort$, by \Cref{def:ero:english}, $e$ set $\linearizationobject{}$ to $((\timeshort{}', \langle \doopandcopyresponse{}, \arbitraryvalue \rangle), \cellpointershort)$.
    Thus, by \Cref{lemma:ero:l_events_have_corresponding_a_events}, there is an $A$-event $e'$ which set $\announceobject$ to $((\timeshort{}', \langle \doopandcopyresponse{}, \arbitraryvalue \rangle), \cellpointershort)$.
    Hence, $e'$ is invoked during an invocation $I'$ of the \doworkuntildone{} procedure with parameters $(\langle \doopandcopyresponse{}, \arbitraryvalue \rangle, \cellpointershort)$ whose response on \cref{line:ero:operation_timestamp} is $\timeshort{}'$.
    Thus, since $I$ is an invocation of the \doworkuntildone{} procedure with $(\langle \doopandcopyresponse{}, \arbitraryvalue \rangle, \cellpointershort)$ whose response on \cref{line:ero:operation_timestamp} is $\timeshort{}$, and $\timeshort{}' \neq \timeshort{}$, we have that $I \neq I'$.
    Therefore, there are two invocations of the \doworkuntildone{} procedure with parameters of the form $(\langle \doopandcopyresponse{}, \arbitraryvalue \rangle, \cellpointershort)$ in $\mathcal{I}^\mathcal{B}$.
    However, since the response on \cref{line:ero:allocate_cell} is unique, there is at most one invocation of the \doworkuntildone{} procedure with parameters of the form $(\langle \doopandcopyresponse{}, \arbitraryvalue \rangle, \cellpointershort)$ in $\mathcal{I}^\mathcal{B}$, a contradiction.
    \qH{\Cref{lemma:ero:l_apply_event_before_apply_low_level_exits_is_for_correct_timestamp}}
\end{proof}

Another consequence of the fact that once a \doworkuntildone{} procedure exits there is a successful response-set attempt is that if there is an unsuccessful response-set attempt, then there must be a successful one beforehand.

\begin{lemma}\label{lemma:ero:unsuccessful_add_response_set_implies_successful_add_response_set}
    Consider any unsuccessful add-response-set attempt for $\cellpointershort$ in $\mathcal{I}^\mathcal{B}$.
    Then there is a successful add-response-set attempt for $\cellpointershort$ beforehand.
\end{lemma}

\begin{proof}
    Consider any unsuccessful add-response-set attempt $a$ for $\cellpointershort$ in $\mathcal{I}^\mathcal{B}$.
    Hence, by \Cref{lemma:ero:every_response_set_attempt_is_for_pointer_from_universe} $\cellpointershort \in \celluniverse{}$.
    Suppose $a$ tried to set $(*\cellpointershort).\lastrepositoryoperationresponse{}.\uniquerepositoryoperationlong{} = \uniquerepositoryoperationshort$.
    Then, by \Cref{lemma:ero:response_set_attempts_have_corresponding_l_events}, there is an $L$-event $e$ before $a$ which set $\linearizationobject{} = (\uniquerepositoryoperationshort, \cellpointershort)$, and so $e$ is an $L$-add event for $\cellpointershort$.
    Since $a$ tries to set $(*\cellpointershort).\lastrepositoryoperationresponse{}.\uniquerepositoryoperationlong{} = \uniquerepositoryoperationshort$, and is unsuccessful, we have that $(*\cellpointershort{}).\lastrepositoryoperationresponse{} \neq (\uniquerepositoryoperationshort{}, \nullconstant)$ at the step before $a$.
    Since $e$ is an $L$-add event for $\cellpointershort{}$ that set $\linearizationobject{}.\uniquerepositoryoperationlong{} = \uniquerepositoryoperationshort{}$, by \Cref{lemma:ero:every_l_event_is_preceeded_by_unique_response_reset}, there is exactly one add-response-reset event $e_r$ for $\cellpointershort$ before $e$ that set $(*\cellpointershort).\lastrepositoryoperationresponse{} = (\uniquerepositoryoperationshort{}, \nullconstant)$.
    Hence, since $e < a$, and $(*\cellpointershort{}).\lastrepositoryoperationresponse{} \neq (\uniquerepositoryoperationshort{}, \nullconstant)$ at the step before $a$, we have that the value of $(*\cellpointershort).\lastrepositoryoperationresponse{}$ changed during $(e_r, a)$.
    Let $e_c$ be the first step during $(e_r, a)$ that changed the value of $(*\cellpointershort).\lastrepositoryoperationresponse{}$.
    Thus, since $\cellpointershort \in \celluniverse{}$, by \Cref{observation:ero:where_objects_change}, $e_c$ is either a response-reset event for $\cellpointershort$ or a successful response-set attempt for $\cellpointershort$.
    We consider each case separately.
    \begin{itemize}
        \item[] \hspace{0pt}\textbf{Case 1.} $e_c$ is a response-reset event for $\cellpointershort$.

        Hence, by \Cref{def:ero:english}, $e_c$ set $(*\cellpointershort).\lastrepositoryoperationresponse{} = (\arbitraryvalue, \nullconstant)$ on \cref{line:ero:do_work_initialize_response}.
        Let $p_r$ (resp. $p_c$) be the process that executed $e_r$ (resp. $e_c$).
        Since $e_r$ is an add-response-reset event for $\cellpointershort$ and $e_c$ is a response-reset event for $\cellpointershort$, $p_r$ (resp. $p_c$) performed $e_r$ (resp. $e_c$) during an invocation $I_r$ (resp. $I_c$) of the \doworkuntildone{} procedure whose second parameter is $\cellpointershort$.
        Let $I^{hlo}_r$ (resp. $I^{hlo}_c$) be the invocation of the \highleveloperation{} procedure that $p_r$ (resp. $p_c$) invoked $I_r$ (resp. $I_c$) during.
        Since the second parameter of $I_r$ (resp. $I_c$) is $\cellpointershort$, it follows that $p_r$ (resp. $p_c$) received $\cellpointershort$ as a response on \cref{line:ero:allocate_cell} during $I^{hlo}_r$ (resp. $I^{hlo}_c$).
        Hence, since by \Cref{alg:lazy_cell_manager_specification} every $\allocatecelloperation$ operation returns a unique response, we have that $p_r = p_c$ and $I^{hlo}_r = I^{hlo}_c$.
        Let $p_*$ be this process.
        Since $e_r$ (resp. $e_c$) is an execution of \cref{line:ero:do_work_initialize_response} during $I_r$ (resp. $I_c$), and there is at most one execution of \cref{line:ero:do_work_initialize_response} per invocation of the \doworkuntildone{} procedure, we have that $I_r \neq I_c$.
        Hence, since $e_r < e_c$, and $p_*$ invoked both $I_r$ and $I_c$, it follows that $I_r$ exited before $I_c$ was invoked. 
        Since $p_*$ executed $e_r$ during $I_r$, and $e_r$ is an add-response-reset event for $\cellpointershort{}$, it follows that the parameters are $I_r$ are $(\addcell, \cellpointershort{})$.
        Hence, by \Cref{lemma:ero:successful_add_response_set_before_add_low_level_exits}, there is a successful add-response-set attempt for $\cellpointershort{}$ before $I_r$ exits.
        Therefore, since $I_r$ exits before $I_c$ is invoked, $p_*$ executes $e_c$ during $I_c$, and $e_c < a$, by transitivity, there is a successful add-response-set attempt for $\cellpointershort{}$ before $a$.
        
        \item[] \hspace{0pt}\textbf{Case 2.} $e_c$ is a successful response-set attempt for $\cellpointershort$.

        Since $e_r$ set $(*\cellpointershort).\lastrepositoryoperationresponse{} = (\uniquerepositoryoperationshort{}, \nullconstant)$, and the first time it changes after $e_r$ is $e_c$, it follows that $(*\cellpointershort).\lastrepositoryoperationresponse{} = (\uniquerepositoryoperationshort{}, \nullconstant)$ throughout $[e_r, e_c)$.
        Hence, since $e_c$ is a successful response-set attempt for $\cellpointershort$, by \Cref{def:ero:english}, $e_c$ is a \CASop{} operation on \cref{line:ero:responses_set_attempt}.
        Thus, since $e_c$ is successful and $(*\cellpointershort).\lastrepositoryoperationresponse{} = (\uniquerepositoryoperationshort{}, \nullconstant)$ throughout $[e_r, e_c)$, the first parameter of $e_c$ is $(\uniquerepositoryoperationshort{}, \nullconstant)$.
        Hence, since $\uniquerepositoryoperationshort{} = (\arbitraryvalue, \addcell)$ (because $e$ set $\linearizationobject{}.\uniquerepositoryoperationlong{} = \uniquerepositoryoperationshort{}$ and $e$ is an $L$-add event), by \Cref{def:ero:english}, $e_c$ is a successful add-response-set attempt for $\cellpointershort$.
        Therefore, since $e_c < a$, there is a successful add-response-set attempt for $\cellpointershort$ before $a$.
        \qH{\Cref{lemma:ero:unsuccessful_add_response_set_implies_successful_add_response_set}}
    \end{itemize}
\end{proof}

\begin{lemma}\label{lemma:ero:unsuccessful_apply_response_set_implies_successful_apply_response_set}
    Consider any unsuccessful apply-response-set attempt for $\cellpointershort$ in $\mathcal{I}^\mathcal{B}$.
    Then there is a successful apply-response-set attempt for $\cellpointershort$ beforehand.
\end{lemma}

\begin{proof}
    By essentially the same argument as \Cref{lemma:ero:unsuccessful_add_response_set_implies_successful_add_response_set}, which we provide below for completeness.
    Consider any unsuccessful apply-response-set attempt $a$ for $\cellpointershort$ in $\mathcal{I}^\mathcal{B}$.
    Hence, by \Cref{lemma:ero:every_response_set_attempt_is_for_pointer_from_universe} $\cellpointershort \in \celluniverse{}$.
    Suppose $a$ tried to set $(*\cellpointershort).\lastrepositoryoperationresponse{}.\uniquerepositoryoperationlong{} = \uniquerepositoryoperationshort$.
    Then, by \Cref{lemma:ero:response_set_attempts_have_corresponding_l_events}, there is an $L$-event $e$ before $a$ which set $\linearizationobject{} = (\uniquerepositoryoperationshort, \cellpointershort)$, and so $e$ is an $L$-apply event for $\cellpointershort$.
    Since $a$ tries to set $(*\cellpointershort).\lastrepositoryoperationresponse{}.\uniquerepositoryoperationlong{} = \uniquerepositoryoperationshort$, and is unsuccessful, we have that $(*\cellpointershort{}).\lastrepositoryoperationresponse{} \neq (\uniquerepositoryoperationshort{}, \nullconstant)$ at the step before $a$.
    Since $e$ is an $L$-apply event for $\cellpointershort{}$ that set $\linearizationobject{}.\uniquerepositoryoperationlong{} = \uniquerepositoryoperationshort{}$, by \Cref{lemma:ero:every_l_event_is_preceeded_by_unique_response_reset}, there is exactly one apply-response-reset event $e_r$ for $\cellpointershort$ before $e$ that set $(*\cellpointershort).\lastrepositoryoperationresponse{} = (\uniquerepositoryoperationshort{}, \nullconstant)$.
    Hence, since $e < a$, and $(*\cellpointershort{}).\lastrepositoryoperationresponse{} \neq (\uniquerepositoryoperationshort{}, \nullconstant)$ at the step before $a$, we have that the value of $(*\cellpointershort).\lastrepositoryoperationresponse{}$ changed during $(e_r, a)$.
    Let $e_c$ be the first step during $(e_r, a)$ that changed the value of $(*\cellpointershort).\lastrepositoryoperationresponse{}$.
    Thus, since $\cellpointershort \in \celluniverse{}$, by \Cref{observation:ero:where_objects_change}, $e_c$ is either a response-reset event for $\cellpointershort$ or a successful response-set attempt for $\cellpointershort$.
    We consider each case separately.
    \begin{itemize}
        \item[] \hspace{0pt}\textbf{Case 1.} $e_c$ is a response-reset event for $\cellpointershort$.

        Hence, by \Cref{def:ero:english}, $e_c$ set $(*\cellpointershort).\lastrepositoryoperationresponse{} = (\arbitraryvalue, \nullconstant)$ on \cref{line:ero:do_work_initialize_response}.
        Let $p_r$ (resp. $p_c$) be the process that executed $e_r$ (resp. $e_c$).
        Since $e_r$ is an apply-response-reset event for $\cellpointershort$ and $e_c$ is a response-reset event for $\cellpointershort$, $p_r$ (resp. $p_c$) performed $e_r$ (resp. $e_c$) during an invocation $I_r$ (resp. $I_c$) of the \doworkuntildone{} procedure whose second parameter is $\cellpointershort$.
        Let $I^{hlo}_r$ (resp. $I^{hlo}_c$) be the invocation of the \highleveloperation{} procedure that $p_r$ (resp. $p_c$) invoked $I_r$ (resp. $I_c$) during.
        Since the second parameter of $I_r$ (resp. $I_c$) is $\cellpointershort$, it follows that $p_r$ (resp. $p_c$) received $\cellpointershort$ as a response on \cref{line:ero:allocate_cell} during $I^{hlo}_r$ (resp. $I^{hlo}_c$).
        Hence, since by \Cref{alg:lazy_cell_manager_specification} every $\allocatecelloperation$ operation returns a unique response, we have that $p_r = p_c$ and $I^{hlo}_r = I^{hlo}_c$.
        Let $p_*$ be this process.
        Since $e_r$ (resp. $e_c$) is an execution of \cref{line:ero:do_work_initialize_response} during $I_r$ (resp. $I_c$), and there is at most one execution of \cref{line:ero:do_work_initialize_response} per invocation of the \doworkuntildone{} procedure, we have that $I_r \neq I_c$.
        Hence, since $e_r < e_c$, and $p_*$ invoked both $I_r$ and $I_c$, it follows that $I_r$ exited before $I_c$ was invoked. 
        Since $p_*$ executed $e_r$ during $I_r$, and $e_r$ is an apply-response-reset event for $\cellpointershort{}$, it follows that the parameters are $I_r$ are $(\langle \doopandcopyresponse{}, \arbitraryvalue \rangle, \cellpointershort{})$.
        Hence, by \Cref{lemma:ero:successful_apply_response_set_before_apply_low_level_exits}, there is a successful apply-response-set attempt for $\cellpointershort{}$ before $I_r$ exits.
        Therefore, since $I_r$ exits before $I_c$ is invoked, $p_*$ executes $e_c$ during $I_c$, and $e_c < a$, by transitivity, there is a successful apply-response-set attempt for $\cellpointershort{}$ before $a$.
        
        \item[] \hspace{0pt}\textbf{Case 2.} $e_c$ is a successful response-set attempt for $\cellpointershort$.

        Since $e_r$ set $(*\cellpointershort).\lastrepositoryoperationresponse{} = (\uniquerepositoryoperationshort{}, \nullconstant)$, and the first time it changes after $e_r$ is $e_c$, it follows that $(*\cellpointershort).\lastrepositoryoperationresponse{} = (\uniquerepositoryoperationshort{}, \nullconstant)$ throughout $[e_r, e_c)$.
        Hence, since $e_c$ is a successful response-set attempt for $\cellpointershort$, by \Cref{def:ero:english}, $e_c$ is a \CASop{} operation on \cref{line:ero:responses_set_attempt}.
        Thus, since $e_c$ is successful and $(*\cellpointershort).\lastrepositoryoperationresponse{} = (\uniquerepositoryoperationshort{}, \nullconstant)$ throughout $[e_r, e_c)$, the first parameter of $e_c$ is $(\uniquerepositoryoperationshort{}, \nullconstant)$.
        Hence, since $\uniquerepositoryoperationshort{} = (\arbitraryvalue, \langle \doopandcopyresponse{}, \arbitraryvalue \rangle)$ (because $e$ set $\linearizationobject{}.\uniquerepositoryoperationlong{} = \uniquerepositoryoperationshort{}$ and $e$ is an $L$-apply event), by \Cref{def:ero:english}, $e_c$ is a successful apply-response-set attempt for $\cellpointershort$.
        Therefore, since $e_c < a$, there is a successful apply-response-set attempt for $\cellpointershort$ before $a$.
        \qH{\Cref{lemma:ero:unsuccessful_apply_response_set_implies_successful_apply_response_set}}
    \end{itemize}
\end{proof}

\begin{lemma}\label{lemma:ero:unsuccessful_remove_response_set_implies_successful_remove_response_set}
    Consider any unsuccessful remove-response-set attempt for $\cellpointershort$ in $\mathcal{I}^\mathcal{B}$.
    Then there is a successful remove-response-set attempt for $\cellpointershort$ beforehand.
\end{lemma}

\begin{proof}
    By essentially the same argument as \Cref{lemma:ero:unsuccessful_add_response_set_implies_successful_add_response_set}, which we provide below for completeness.
    Consider any unsuccessful remove-response-set attempt $a$ for $\cellpointershort$ in $\mathcal{I}^\mathcal{B}$.
    Hence, by \Cref{lemma:ero:every_response_set_attempt_is_for_pointer_from_universe} $\cellpointershort \in \celluniverse{}$.
    Suppose $a$ tried to set $(*\cellpointershort).\lastrepositoryoperationresponse{}.\uniquerepositoryoperationlong{} = \uniquerepositoryoperationshort$.
    Then, by \Cref{lemma:ero:response_set_attempts_have_corresponding_l_events}, there is an $L$-event $e$ before $a$ which set $\linearizationobject{} = (\uniquerepositoryoperationshort, \cellpointershort)$, and so $e$ is an $L$-remove event for $\cellpointershort$.
    Since $a$ tries to set $(*\cellpointershort).\lastrepositoryoperationresponse{}.\uniquerepositoryoperationlong{} = \uniquerepositoryoperationshort$, and is unsuccessful, we have that $(*\cellpointershort{}).\lastrepositoryoperationresponse{} \neq (\uniquerepositoryoperationshort{}, \nullconstant)$ at the step before $a$.
    Since $e$ is an $L$-remove event for $\cellpointershort{}$ that set $\linearizationobject{}.\uniquerepositoryoperationlong{} = \uniquerepositoryoperationshort{}$, by \Cref{lemma:ero:every_l_event_is_preceeded_by_unique_response_reset}, there is exactly one remove-response-reset event $e_r$ for $\cellpointershort$ before $e$ that set $(*\cellpointershort).\lastrepositoryoperationresponse{} = (\uniquerepositoryoperationshort{}, \nullconstant)$.
    Hence, since $e < a$, and $(*\cellpointershort{}).\lastrepositoryoperationresponse{} \neq (\uniquerepositoryoperationshort{}, \nullconstant)$ at the step before $a$, we have that the value of $(*\cellpointershort).\lastrepositoryoperationresponse{}$ changed during $(e_r, a)$.
    Let $e_c$ be the first step during $(e_r, a)$ that changed the value of $(*\cellpointershort).\lastrepositoryoperationresponse{}$.
    Thus, since $\cellpointershort \in \celluniverse{}$, by \Cref{observation:ero:where_objects_change}, $e_c$ is either a response-reset event for $\cellpointershort$ or a successful response-set attempt for $\cellpointershort$.
    We consider each case separately.
    \begin{itemize}
        \item[] \hspace{0pt}\textbf{Case 1.} $e_c$ is a response-reset event for $\cellpointershort$.

        Hence, by \Cref{def:ero:english}, $e_c$ set $(*\cellpointershort).\lastrepositoryoperationresponse{} = (\arbitraryvalue, \nullconstant)$ on \cref{line:ero:do_work_initialize_response}.
        Let $p_r$ (resp. $p_c$) be the process that executed $e_r$ (resp. $e_c$).
        Since $e_r$ is an remove-response-reset event for $\cellpointershort$ and $e_c$ is a response-reset event for $\cellpointershort$, $p_r$ (resp. $p_c$) performed $e_r$ (resp. $e_c$) during an invocation $I_r$ (resp. $I_c$) of the \doworkuntildone{} procedure whose second parameter is $\cellpointershort$.
        Let $I^{hlo}_r$ (resp. $I^{hlo}_c$) be the invocation of the \highleveloperation{} procedure that $p_r$ (resp. $p_c$) invoked $I_r$ (resp. $I_c$) during.
        Since the second parameter of $I_r$ (resp. $I_c$) is $\cellpointershort$, it follows that $p_r$ (resp. $p_c$) received $\cellpointershort$ as a response on \cref{line:ero:allocate_cell} during $I^{hlo}_r$ (resp. $I^{hlo}_c$).
        Hence, since by \Cref{alg:lazy_cell_manager_specification} every $\allocatecelloperation$ operation returns a unique response, we have that $p_r = p_c$ and $I^{hlo}_r = I^{hlo}_c$.
        Let $p_*$ be this process.
        Since $e_r$ (resp. $e_c$) is an execution of \cref{line:ero:do_work_initialize_response} during $I_r$ (resp. $I_c$), and there is at most one execution of \cref{line:ero:do_work_initialize_response} per invocation of the \doworkuntildone{} procedure, we have that $I_r \neq I_c$.
        Hence, since $e_r < e_c$, and $p_*$ invoked both $I_r$ and $I_c$, it follows that $I_r$ exited before $I_c$ was invoked. 
        Since $p_*$ executed $e_r$ during $I_r$, and $e_r$ is an remove-response-reset event for $\cellpointershort{}$, it follows that the parameters are $I_r$ are $(\removecell, \cellpointershort{})$.
        Hence, by \Cref{lemma:ero:successful_remove_response_set_before_remove_low_level_exits}, there is a successful remove-response-set attempt for $\cellpointershort{}$ before $I_r$ exits.
        Therefore, since $I_r$ exits before $I_c$ is invoked, $p_*$ executes $e_c$ during $I_c$, and $e_c < a$, by transitivity, there is a successful remove-response-set attempt for $\cellpointershort{}$ before $a$.
        
        \item[] \hspace{0pt}\textbf{Case 2.} $e_c$ is a successful response-set attempt for $\cellpointershort$.

        Since $e_r$ set $(*\cellpointershort).\lastrepositoryoperationresponse{} = (\uniquerepositoryoperationshort{}, \nullconstant)$, and the first time it changes after $e_r$ is $e_c$, it follows that $(*\cellpointershort).\lastrepositoryoperationresponse{} = (\uniquerepositoryoperationshort{}, \nullconstant)$ throughout $[e_r, e_c)$.
        Hence, since $e_c$ is a successful response-set attempt for $\cellpointershort$, by \Cref{def:ero:english}, $e_c$ is a \CASop{} operation on \cref{line:ero:responses_set_attempt}.
        Thus, since $e_c$ is successful and $(*\cellpointershort).\lastrepositoryoperationresponse{} = (\uniquerepositoryoperationshort{}, \nullconstant)$ throughout $[e_r, e_c)$, the first parameter of $e_c$ is $(\uniquerepositoryoperationshort{}, \nullconstant)$.
        Hence, since $\uniquerepositoryoperationshort{} = (\arbitraryvalue, \removecell)$ (because $e$ set $\linearizationobject{}.\uniquerepositoryoperationlong{} = \uniquerepositoryoperationshort{}$ and $e$ is an $L$-remove event), by \Cref{def:ero:english}, $e_c$ is a successful remove-response-set attempt for $\cellpointershort$.
        Therefore, since $e_c < a$, there is a successful remove-response-set attempt for $\cellpointershort$ before $a$.
        \qH{\Cref{lemma:ero:unsuccessful_remove_response_set_implies_successful_remove_response_set}}
    \end{itemize}
\end{proof}

Lastly, we prove a basic property of the IsDone procedure.

\begin{lemma}\label{lemma:ero:if_announce_acquire_does_not_return_time_change_then_no_l_events}
    Consider any process $p$ and any iteration $I$ of the loop on \cref{line:ero:do_work_while_loop} such that during $I$ $p$ exits some invocation $I'$ of the Acquire procedure on \cref{line:ero:announce_acquire} with response $\status$ during $\mathcal{I}^\mathcal{B}$.
    Let $T^{\ref{line:ero:linearization_read}}$ be the time $p$ executes \cref{line:ero:linearization_read} during $I$ and let $T^{\ref{line:ero:acquire_next_linearization_changed_check}}$ be the last time $p$ executes \cref{line:ero:acquire_next_linearization_changed_check} during $I'$.
    Recall that $T^{\ref{line:ero:acquire_next_linearization_changed_check}}$ is well-defined by \Cref{lemma:ero:exit_acquire_implies_executing_105}.
    If $P(\mathcal{I}^\mathcal{B})$ holds and $\status \neq \timechange$, then there are no $L$-events throughout $[T^{\ref{line:ero:linearization_read}}, T^{\ref{line:ero:acquire_next_linearization_changed_check}}]$ during $\mathcal{I}^\mathcal{B}$.
\end{lemma}

\begin{proof}
    Suppose $p$ read $\uniquerepositoryoperationshort_\linearizationobject{}$ from $\linearizationobject{}.\uniquerepositoryoperationlong$ on \cref{line:ero:linearization_read} during $I$ at time $T^{\ref{line:ero:linearization_read}}$.
    Hence, since $I'$ is invoked during $I$, we have that the first parameter of $I'$ is $\uniquerepositoryoperationshort_\linearizationobject{}$.

    We first prove that $\linearizationobject{}.\uniquerepositoryoperationlong = \uniquerepositoryoperationshort_\linearizationobject{}$ at $T^{\ref{line:ero:acquire_next_linearization_changed_check}}$ (*).
    Since, by assumption, the response of $I'$ is not $\timechange$, we have that the response of every invocation of the AcquireNext procedure during $I'$ is also not $\timechange$ (otherwise $I'$'s response would be $\timechange$).
    Hence, $p$ finds the condition on \cref{line:ero:acquire_next_linearization_changed_check} to be false on every execution of \cref{line:ero:acquire_next_linearization_changed_check} during $I'$.
    Thus, since the first parameter of $I'$ is $\uniquerepositoryoperationshort_\linearizationobject{}$, the first parameter of every invocation of the AcquireNext procedure during $I'$ is also $\uniquerepositoryoperationshort_\linearizationobject{}$, and so at the time of every execution of \cref{line:ero:acquire_next_linearization_changed_check} during $I'$ $p$ finds $\linearizationobject{}.\uniquerepositoryoperationlong = \uniquerepositoryoperationshort_\linearizationobject{}$.
    Therefore, $\linearizationobject{}.\uniquerepositoryoperationlong = \uniquerepositoryoperationshort_\linearizationobject{}$ at $T^{\ref{line:ero:acquire_next_linearization_changed_check}}$.

    We now finish the proof of \Cref{lemma:ero:if_announce_acquire_does_not_return_time_change_then_no_l_events}.
    Suppose, for contradiction, there is an $L$-event during $[T^{\ref{line:ero:linearization_read}}, T^{\ref{line:ero:acquire_next_linearization_changed_check}}]$.
    Let $e$ be the last $L$-event in $[T^{\ref{line:ero:linearization_read}}, T^{\ref{line:ero:acquire_next_linearization_changed_check}}]$.
    Hence, since by (*) $\linearizationobject{}.\uniquerepositoryoperationlong = \uniquerepositoryoperationshort_\linearizationobject{}$ at $T^{\ref{line:ero:acquire_next_linearization_changed_check}}$, we have that $e$ set $\linearizationobject{}.\uniquerepositoryoperationlong = \uniquerepositoryoperationshort_\linearizationobject{}$, and so by \Cref{lemma:ero:l_events_set_llo_to_a_value_different_from_initial}, $\uniquerepositoryoperationshort_\linearizationobject{} \neq (0, \noop)$.
    Thus, since $p$ read $\uniquerepositoryoperationshort_\linearizationobject{}$ from $\linearizationobject{}.\uniquerepositoryoperationlong$ at $T^{\ref{line:ero:linearization_read}}$, and $\linearizationobject{}.\uniquerepositoryoperationlong$ is initially $(0, \noop)$, we have that $\linearizationobject{}.\uniquerepositoryoperationlong$ was set to $\uniquerepositoryoperationshort_\linearizationobject{}$ before $T^{\ref{line:ero:linearization_read}}$.
    So, by \Cref{observation:ero:where_objects_change}, there is an $L$-event $e'$ before $T^{\ref{line:ero:linearization_read}}$ that set $\linearizationobject{}.\uniquerepositoryoperationlong = \uniquerepositoryoperationshort_\linearizationobject{}$.
    Since $e'$ is before $T^{\ref{line:ero:linearization_read}}$ and $e$ is during $[T^{\ref{line:ero:linearization_read}}, T^{\ref{line:ero:acquire_next_linearization_changed_check}}]$, we have that $e' < e$, and so $e' \neq e$. 
    Therefore, there are two $L$-events that set $\linearizationobject{}.\uniquerepositoryoperationlong = \uniquerepositoryoperationshort_\linearizationobject{}$ in $\mathcal{I}^\mathcal{B}$.
    However, since $P(\mathcal{I}^\mathcal{B})$ holds, by \Cref{lemma:ero:p_implies_unique_low_level_operations_in_linearization}, every $L$-event in $\mathcal{I}^\mathcal{B}$ sets the value of $\linearizationobject{}.\uniquerepositoryoperationlong$ to a unique value, a contradiction.
    \qH{\Cref{lemma:ero:if_announce_acquire_does_not_return_time_change_then_no_l_events}}
\end{proof}

\subsubsection{Properties of $\List$}

We now prove some facts about $\List$ (see \Cref{def:ero:logical_list}).

\begin{lemma}\label{lemma:ero:every_list_sequence_is_from_universe}
    Let $\mathcal{I}$ be a finite implementation history of $\mathcal{B}$ and let $\List(\mathcal{I}) = \cellpointershort_0, \ldots, \cellpointershort_{n + 1}$ for some integer $n \geq 0$.
    Then, $\cellpointershort_0 = \&\headobject$, for every $i \in [1..n]$ $\cellpointershort_i \in \celluniverse{}$, and $\cellpointershort_{n + 1} = \nullconstant$.
\end{lemma}

\begin{proof}
    Since $\List(\mathcal{I}) = \cellpointershort_0, \ldots, \cellpointershort_{n + 1}$, by \Cref{def:ero:logical_list}, $\cellpointershort_0 = \&\headobject$, $\cellpointershort_{n + 1} = \nullconstant$, and for every $i \in [1..n]$, there is an $L$-add event for $\cellpointershort_i$ in $\mathcal{I}$, so by \Cref{lemma:ero:every_l_event_is_for_pointer_from_universe} $\cellpointershort_i \in \celluniverse{}$.
    \qH{\Cref{lemma:ero:every_list_sequence_is_from_universe}}
\end{proof}

\begin{lemma}\label{lemma:ero:pointers_in_list_are_unique}
    Let $\mathcal{I}$ be a finite implementation history of $\mathcal{B}$ and let $\List(\mathcal{I}) = \cellpointershort_0, \ldots, \cellpointershort_{n + 1}$ for some integer $n \geq 0$.
    If $P(\mathcal{I})$ holds, then for every $i, j \in [0..n+1]$, if $i \neq j$, then $\cellpointershort_i \neq \cellpointershort_j$.
\end{lemma}

\begin{proof}
    Suppose, for contradiction, for some $i, j \in [0..n+1]$ $i \neq j$ and $\cellpointershort_i = \cellpointershort_j$.
    \begin{itemize}
        \item[] \hspace{0pt}\textbf{Case 1.} $i = 0$ or $j = 0$.

        Hence, since by assumption $i \neq j$, either $i$ or $j$ is not zero.
        Without loss of generality, suppose $i = 0$ and $j \neq 0$.
        Hence, by \Cref{def:ero:logical_list}, $\cellpointershort_i = \&\headobject$.
        Thus, since by assumption $\cellpointershort_i = \cellpointershort_j$, we have that $\cellpointershort_j = \&\headobject$.
        Since $j \neq 0$, $j \in [1..n +1]$.
        Hence, by \Cref{lemma:ero:every_list_sequence_is_from_universe}, $\cellpointershort_j \in \celluniverse{} \cup \{\nullconstant\}$.
        Therefore, by \Cref{assumption:ero:head_and_null_not_in_cell_universe}, $\cellpointershort_j \neq \&\headobject$.
        However, $\cellpointershort_j = \&\headobject$, a contradiction.

        \item[] \hspace{0pt}\textbf{Case 2.} $i \neq 0$ and $j \neq 0$.

        Hence, since $i, j \in [0..n +1]$ and by assumption $i \neq j$, either $i, j \in (0..n + 1)$ or one of $i$ and $j$ is in $(0..n+1)$ and the other equals $n + 1$.
        We consider each case separately.

        \begin{itemize}
            \item[] \hspace{0pt}\textbf{Case 2.1.} $i \in (0..n + 1)$ and $j \in (0..n+1)$.

            Hence, $i, j \in [1..n]$, and since $\List(\mathcal{I}) = \cellpointershort_0, \ldots, \cellpointershort_{n + 1}$, by \Cref{def:ero:logical_list}, there is an $L$-add event for $\cellpointershort_i$ in $\mathcal{I}$, an $L$-add event for $\cellpointershort_j$ in $\mathcal{I}$, and so by \Cref{lemma:ero:every_l_event_is_for_pointer_from_universe} $\cellpointershort_i, \cellpointershort_j \in \celluniverse{}$.
            Therefore, since by assumption $\cellpointershort_i = \cellpointershort_j$, there are two $L$-add events in $\mathcal{I}$ for the same pointer in $\celluniverse{}$.
            However, by $P(\mathcal{I})$, there is at most one $L$-add event in $\mathcal{I}$ for every pointer in $\celluniverse{}$, a contradiction.

            \item[] \hspace{0pt}\textbf{Case 2.2.} One of $i$ and $j$ is in $(0..n+1)$ and the other equals $n + 1$.

            Without loss of generality suppose $i \in (0..n+1)$ and $j = n + 1$.
            Hence, $i \in [1..n]$, and since $\List(\mathcal{I}) = \cellpointershort_0, \ldots, \cellpointershort_{n + 1}$, by \Cref{def:ero:logical_list}, there is an $L$-add event for $\cellpointershort_i$ in $\mathcal{I}$,  and so by \Cref{lemma:ero:every_l_event_is_for_pointer_from_universe} $\cellpointershort_i \in \celluniverse{}$.
            Thus, by \Cref{assumption:ero:head_and_null_not_in_cell_universe}, $\cellpointershort_i \neq \nullconstant$.
            Therefore, since by assumption $\cellpointershort_i = \cellpointershort_j$, $\cellpointershort_j \neq \nullconstant$.
            However, since $j = n + 1$, by \Cref{lemma:ero:every_list_sequence_is_from_universe}, $\cellpointershort_j = \nullconstant$, a contradiction.
            \qH{\Cref{lemma:ero:pointers_in_list_are_unique}}
        \end{itemize}
    \end{itemize}
\end{proof}

\begin{lemma}\label{lemma:ero:list_neighbours_are_the_same_if_they_contained_in_an_extended_list}
    Consider any finite implementation histories $\mathcal{I}_1$ and $\mathcal{I}_2$ of $\mathcal{B}$ such that $\mathcal{I}_1$ is a prefix of $\mathcal{I}_2$.
    Suppose some $\currentcellpointershort \in \celluniverse \cup \{\&\headobject{}\}$ is in $\List(\mathcal{I}_1)$ and it appears immediately before some $\cellpointershort \in \celluniverse$ in $\List(\mathcal{I}_1)$.
    If $P(\mathcal{I}_2)$ holds, and both $\currentcellpointershort$ and $\cellpointershort$ are in $\List(\mathcal{I}_2)$, then $\currentcellpointershort$ appears immediately before $\cellpointershort$ in $\List(\mathcal{I}_2)$.
\end{lemma}

\begin{proof}
    Suppose $\currentcellpointershort$ and $\cellpointershort$ are in $\List(\mathcal{I}_2)$.
    There are two cases.

    \begin{itemize}
        \item[] \hspace{0pt}\textbf{Case 1.} $\currentcellpointershort = \&\headobject{}$.

        Since $\cellpointershort \in \celluniverse$, by \Cref{assumption:ero:head_and_null_not_in_cell_universe}, $\cellpointershort \neq \&\headobject{}$ and $\cellpointershort \neq \nullconstant$.
        Hence, since $\currentcellpointershort = \&\headobject{}$ and it appears immediately before $\cellpointershort$ in $\List(\mathcal{I}_1)$, by \Cref{def:ero:logical_list}, there is an $L$-add event $e_1$ for $\cellpointershort$ in $\mathcal{I}_1$ such that for every $L$-event $e$ before $e_1$ in $\mathcal{I}_1$, if $e$ is an $L$-add event for $\cellpointershort'$, then there is an $L$-remove event for $\cellpointershort'$ after $e$ in $\mathcal{I}_1$.
        Furthermore, since $\cellpointershort$ is in $\List(\mathcal{I}_2)$, by \Cref{def:ero:logical_list}, there is an $L$-add event $e_2$ for $\cellpointershort$ in $\mathcal{I}_2$ such that there is no $L$-remove event for $\cellpointershort$ after $e_2$ in $\mathcal{I}_2$.
        Since $e_1$ is in $\mathcal{I}_1$ and $\mathcal{I}_1$ is a prefix of $\mathcal{I}_2$, we have that $e_1$ is in $\mathcal{I}_2$.
        Hence, by $P(\mathcal{I}_2)$, there is at most one $L$-add event for $\cellpointershort$ in $\mathcal{I}_2$, and so $e_1 = e_2$.
        Thus, since for every $L$-event $e$ before $e_1$ in $\mathcal{I}_1$, if $e$ is an $L$-add event for $\cellpointershort'$, then there is an $L$-remove event for $\cellpointershort'$ after $e$ in $\mathcal{I}_1$, and $\mathcal{I}_1$ is a prefix of $\mathcal{I}_2$, we have that for every $L$-event $e$ before $e_2$ in $\mathcal{I}_2$, if $e$ is an $L$-add event for $\cellpointershort'$, then there is an $L$-remove event for $\cellpointershort'$ after $e$ in $\mathcal{I}_2$.
        Hence, $e_2$ is the first $L$-add event for a pointer without a subsequent $L$-remove event for that pointer in $\mathcal{I}_2$.
        Therefore, since $e_2$ is an $L$-add event for $\cellpointershort$, by \Cref{def:ero:logical_list}, the first two elements of $\List(\mathcal{I}_2)$ are $\&\headobject{}$ and $\cellpointershort$, and so $\currentcellpointershort$ appears immediately before $\cellpointershort$ in $\List(\mathcal{I}_2)$ as required.  

        \item[] \hspace{0pt}\textbf{Case 2.} $\currentcellpointershort \in \celluniverse$.

        Since $\currentcellpointershort \in \celluniverse$ and $\cellpointershort \in \celluniverse$, by \Cref{assumption:ero:head_and_null_not_in_cell_universe}, $\cellpointershort \neq \&\headobject{}$, $\currentcellpointershort \neq \&\headobject{}$, $\cellpointershort \neq \nullconstant$, and $\currentcellpointershort \neq \nullconstant$.
        Hence, since $\currentcellpointershort$ appears immediately before $\cellpointershort$ in $\List(\mathcal{I}_1)$, by \Cref{def:ero:logical_list}, there is an $L$-add event $e_1$ for $\currentcellpointershort$ in $\mathcal{I}_1$ and an $L$-add event $e'_1$ for $\cellpointershort$ in $\mathcal{I}_1$ such that $e_1 < e'_1$ and for every $L$-event $e$ between $e_1$ and $e'_1$ in $\mathcal{I}_1$, if $e$ is an $L$-add event for $\cellpointershort'$, then there is an $L$-remove event for $\cellpointershort'$ after $e$ in $\mathcal{I}_1$.
        Furthermore, since $\currentcellpointershort$ and $\cellpointershort$ are in $\List(\mathcal{I}_2)$, by \Cref{def:ero:logical_list}, there is an $L$-add event $e_2$ for $\currentcellpointershort$ in $\mathcal{I}_2$ and an $L$-add event $e'_2$ for $\cellpointershort$ in $\mathcal{I}_2$ such that: (1) $e_2 < e'_2$; (2) there is no $L$-remove event for $\currentcellpointershort$ after $e_2$ in $\mathcal{I}_2$; and (3) there is no $L$-remove event for $\cellpointershort$ after $e'_2$ in $\mathcal{I}_2$.
        Since $e_1$ and $e'_1$ are in $\mathcal{I}_1$ and $\mathcal{I}_1$ is a prefix of $\mathcal{I}_2$, we have that $e_1$ and $e'_1$ are in $\mathcal{I}_2$.
        Hence, by $P(\mathcal{I}_2)$, there is at most one $L$-add event for $\currentcellpointershort$ (resp. $\cellpointershort$) in $\mathcal{I}_2$, and so $e_1 = e_2$ (resp. $e'_1 = e'_2$).
        Thus, since for every $L$-event $e$ between $e_1$ and $e'_1$ in $\mathcal{I}_1$, if $e$ is an $L$-add event for $\cellpointershort'$, then there is an $L$-remove event for $\cellpointershort'$ after $e$ in $\mathcal{I}_1$, and $\mathcal{I}_1$ is a prefix of $\mathcal{I}_2$, we have that for every $L$-event $e$ between $e_2$ and $e'_2$ in $\mathcal{I}_2$, if $e$ is an $L$-add event for $\cellpointershort'$, then there is an $L$-remove event for $\cellpointershort'$ after $e$ in $\mathcal{I}_2$.
        Hence, $e_2$ and $e'_2$ are successive $L$-add events for a pointer without a subsequent $L$-remove event for that pointer in $\mathcal{I}_2$.
        Therefore, since $e_2$ (resp. $e'_2$) is an $L$-add event for $\currentcellpointershort$ (resp. $\cellpointershort$), by \Cref{def:ero:logical_list}, $\currentcellpointershort$ and $\cellpointershort$ are successive elements of $\List(\mathcal{I}_2)$, and so $\currentcellpointershort$ appears immediately before $\cellpointershort$ in $\List(\mathcal{I}_2)$ as required.
        \qH{\Cref{lemma:ero:list_neighbours_are_the_same_if_they_contained_in_an_extended_list}}
    \end{itemize}
\end{proof}

\begin{lemma}\label{lemma:ero:if_in_list_then_no_l_remove_events_for_in_history}
    Suppose $P(\mathcal{I}^\mathcal{B})$ holds.
    For every finite prefix $\mathcal{I}$ of $\mathcal{I}^\mathcal{B}$ if $\cellpointershort \in \List(\mathcal{I})$, then there are no $L$-remove events for $\cellpointershort$ in $\mathcal{I}$ and there are no list-remove attempts for $\cellpointershort$ in $\mathcal{I}$.
\end{lemma}

\begin{proof}
    Suppose $\cellpointershort \in \List(\mathcal{I})$.
    We first prove that there are no $L$-remove events for $\cellpointershort$ in $\mathcal{I}$.
    Suppose, for contradiction, there is an $L$-remove event $e_{remove}$ for $\cellpointershort$ in $\mathcal{I}$.
    Hence, by \Cref{lemma:ero:every_l_event_is_for_pointer_from_universe}, $\cellpointershort \in \celluniverse$, and so by \Cref{assumption:ero:head_and_null_not_in_cell_universe}, $\cellpointershort \neq \&\headobject{}$ and $\cellpointershort \neq \nullconstant$.
    Thus, since $\cellpointershort \in \List(\mathcal{I})$, by \Cref{def:ero:logical_list}, there is an $e_{add}$ event for $\cellpointershort$ in $\mathcal{I}$ such that from $e_{add}$ onwards in $\mathcal{I}$ there are no $L$-remove events for $\cellpointershort$.
    Hence, since $e_{remove}$ is an $L$-remove event for $\cellpointershort$ in $\mathcal{I}$, we have that $e_{remove} \leq e_{add}$.
    Thus, by \Cref{lemma:ero:remove_events_are_preceeded_by_add_events}, there is a $L$-add event $e$ for $\cellpointershort$ before $e_{remove}$ in $\mathcal{I}$.
    Hence, since $e_{remove} \leq e_{add}$, by transitivity, $e < e_{add}$, and so $e \neq e_{add}$.
    Therefore, there are two $L$-add events for $\cellpointershort$ in $\mathcal{I}$.
    However, by $P(\mathcal{I}^\mathcal{B})$, there is at most one $L$-add event for $\cellpointershort$ in $\mathcal{I}$, a contradiction.

    We now prove there are no list-remove attempts for $\cellpointershort$ in $\mathcal{I}$.
    Suppose, for contradiction, there is a list-remove attempt $a$ for $\cellpointershort$ in $\mathcal{I}$.
    Hence, by \Cref{lemma:ero:l_event_corresponding_to_do_low_level_op} there is a $L$-remove event $e$ for $\cellpointershort$ before $a$.
    Therefore, since $a$ is in $\mathcal{I}$, we have that $e$ is in $\mathcal{I}$, and so there is an $L$-remove event for $\cellpointershort$ in $\mathcal{I}$.
    However, there are no $L$-remove events for $\cellpointershort$ in $\mathcal{I}$, a contradiction.
    \qH{\Cref{lemma:ero:if_in_list_then_no_l_remove_events_for_in_history}}
\end{proof}

\begin{lemma}\label{lemma:ero:l_add_for_ptr_is_in_exclude_list}
    Consider any $L$-add event $e$ for $\cellpointershort$ in $\mathcal{I}^\mathcal{B}$ and let $\mathcal{I}^{exclude}_e$ be the prefix of $\mathcal{I}^\mathcal{B}$ up to but excluding $e$.
    If $P(\mathcal{I}^\mathcal{B})$ holds, then, $\cellpointershort \notin \List(\mathcal{I}^{exclude}_e)$.
\end{lemma}

\begin{proof}
    Suppose, for contradiction, $\cellpointershort \in \List(\mathcal{I}^{exclude}_e)$.
    Since $e$ is a $L$-event for $\cellpointershort$, we have that by \Cref{lemma:ero:every_l_event_is_for_pointer_from_universe}, $\cellpointershort \in \celluniverse{}$, so by \Cref{assumption:ero:head_and_null_not_in_cell_universe}, $\cellpointershort \neq \&\headobject$ and $\cellpointershort \neq \nullconstant$.
    Hence, since $\cellpointershort \in \List(\mathcal{I}^{exclude}_e)$, by \Cref{def:ero:english}, there is an $L$-add event $e'$ for $\cellpointershort$ in $\mathcal{I}^{exclude}_e$.
    Thus, since $\mathcal{I}^{exclude}_e$ is the prefix of $\mathcal{I}^\mathcal{B}$ up to but excluding $e$, we have that $e'$ is before $e$ in $\mathcal{I}^\mathcal{B}$, and so $e' \neq e$.
    Therefore, since $e$ and $e'$ are $L$-add events for $\cellpointershort$ in $\mathcal{I}^\mathcal{B}$, there are two $L$-add events for $\cellpointershort$ in $\mathcal{I}^\mathcal{B}$.
    However, by $P(\mathcal{I}^\mathcal{B})$, there is at most one $L$-add event for $\cellpointershort$ in $\mathcal{I}^\mathcal{B}$, a contradiction.
    \qH{\Cref{lemma:ero:l_add_for_ptr_is_in_exclude_list}}
\end{proof}

\begin{lemma}\label{lemma:ero:l_remove_for_ptr_is_in_exclude_list}
    Consider any $L$-remove event $e$ for $\cellpointershort$ in $\mathcal{I}^\mathcal{B}$ and let $\mathcal{I}^{exclude}_e$ be the prefix of $\mathcal{I}^\mathcal{B}$ up to but excluding $e$.
    If $P(\mathcal{I}^\mathcal{B})$ holds, then $\cellpointershort \in \List(\mathcal{I}^{exclude}_e)$.
\end{lemma}

\begin{proof}
    Suppose, for contradiction, $\cellpointershort \notin \List(\mathcal{I}^{exclude}_e)$.
    Hence, since $\cellpointershort \notin \List(\mathcal{I}^{exclude}_e)$, by \Cref{def:ero:logical_list}, either there is an $L$-add event for $\cellpointershort$ with a subsequent $L$-remove event for $\cellpointershort$ in $\mathcal{I}^{exclude}_e$ or there is no $L$-add event for $\cellpointershort$ in $\mathcal{I}^{exclude}_e$.
    Thus, since $\mathcal{I}^{exclude}_e$ is the prefix of $\mathcal{I}^\mathcal{B}$ up to but excluding $e$, either there is an $L$-add event for $\cellpointershort$ with a subsequent $L$-remove event for $\cellpointershort$ before $e$ in $\mathcal{I}^\mathcal{B}$ or there is no $L$-add event for $\cellpointershort$ before $e$ in $\mathcal{I}^\mathcal{B}$.
    \begin{itemize}
        \item[] \hspace{0pt}\textbf{Case 1.} There is an $L$-add event for $\cellpointershort$ with a subsequent $L$-remove event for $\cellpointershort$ before $e$ in $\mathcal{I}^\mathcal{B}$.

        Hence, since $e$ is an $L$-remove event for $\cellpointershort$, there are two $L$-remove events for $\cellpointershort$ in $\mathcal{I}^\mathcal{B}$.
        However, by $P(\mathcal{I}^\mathcal{B})$, there is at most one $L$-remove event for $\cellpointershort$ in $\mathcal{I}^\mathcal{B}$, a contradiction.

        \item[] \hspace{0pt}\textbf{Case 2.} There is no $L$-add event for $\cellpointershort$ before $e$ in $\mathcal{I}^\mathcal{B}$.

        However, since $e$ is an $L$-remove event for $\cellpointershort$ in $\mathcal{I}^\mathcal{B}$, by \Cref{lemma:ero:remove_events_are_preceeded_by_add_events}, there is an $L$-add for $\cellpointershort$ before $e$ in $\mathcal{I}^\mathcal{B}$, a contradiction.
        \qH{\Cref{lemma:ero:l_remove_for_ptr_is_in_exclude_list}}
    \end{itemize}
\end{proof}

\begin{lemma}\label{lemma:ero:l_remove_for_ptr_is_uniquely_in_exclude_list}
    Consider any $L$-remove event $e$ for $\cellpointershort$ in $\mathcal{I}^\mathcal{B}$ and let $\mathcal{I}^{exclude}_e$ be the prefix of $\mathcal{I}^\mathcal{B}$ up to but excluding $e$.
    Let $\List(\mathcal{I}^{exclude}_{e}) = \cellpointershort_0, \ldots, \cellpointershort_{n + 1}$ for some integer $n \geq 0$.
    If $P(\mathcal{I}^\mathcal{B})$ holds, then there is exactly one $i \in [1..n]$ such that $\cellpointershort_i = \cellpointershort$.
\end{lemma}

\begin{proof}
    We first prove $i$ exists.
    Since $e$ is an $L$-event for $\cellpointershort$, by \Cref{lemma:ero:every_l_event_is_for_pointer_from_universe}, $\cellpointershort \in \celluniverse{}$, and so by \Cref{assumption:ero:head_and_null_not_in_cell_universe} $\cellpointershort \neq \&\headobject$ and $\cellpointershort \neq \nullconstant$.
    Furthermore, by \Cref{lemma:ero:every_list_sequence_is_from_universe}, $\cellpointershort_0 = \&\headobject$, and $\cellpointershort_{n + 1} = \nullconstant$.
    Hence, since $P(\mathcal{I}^\mathcal{B})$ holds, by\Cref{lemma:ero:l_remove_for_ptr_is_in_exclude_list} $\cellpointershort \in \List(\mathcal{I}^{exclude}_{e})$, and so since $\List(\mathcal{I}^{exclude}_{e}) = \&\headobject, \cellpointershort_1, \ldots, \cellpointershort_{n}, \nullconstant$, there is at least one $i \in [1..n]$ such that $\cellpointershort_i = \cellpointershort$.
    We now prove that $i$ is unique.
    Suppose, for contradiction, there exists $i, j \in [1..n]$ such that $i \neq j$ and $\cellpointershort_i = \cellpointershort_j = \cellpointershort$.
    However, since $\mathcal{I}^{exclude}_{e_b}$ is a prefix of $\mathcal{I}$, $\List(\mathcal{I}^{exclude}_{e_b}) = \cellpointershort_0, \ldots, \cellpointershort_{n + 1}$, by assumption $P(\mathcal{I})$ holds, and $i \neq j$, by \Cref{lemma:ero:pointers_in_list_are_unique} $\cellpointershort_i \neq \cellpointershort_j$, a contradiction.
    \qH{\Cref{lemma:ero:l_remove_for_ptr_is_uniquely_in_exclude_list}}
\end{proof}

\subsubsection{Miscellaneous}
We finish the basic facts section with some miscellaneous properties. 

\begin{lemma}\label{lemma:ero:list_add_before_list_remove}
    Consider any list-remove attempt $a_{remove}$ for $\cellpointershort$ in $\mathcal{I}^\mathcal{B}$.
    If $R(\mathcal{I}^\mathcal{B})$ holds, then there is a successful list-add attempt for $\cellpointershort$ before $a_{remove}$'s corresponding $L$-event in $\mathcal{I}^\mathcal{B}$.
\end{lemma}

\begin{proof}
    Since $a_{remove}$ is a list-remove attempt for $\cellpointershort$, by \Cref{lemma:ero:l_event_corresponding_to_do_low_level_op}, there is a $L$-remove event $e_{remove}$ for $\cellpointershort$ before $a_{remove}$ in $\mathcal{I}^\mathcal{B}$.
    Hence, by \Cref{lemma:ero:remove_events_are_preceeded_by_add_events}, there is a $L$-add event $e_{add}$ for $\cellpointershort$ before $e_{remove}$ in $\mathcal{I}^\mathcal{B}$.
    Thus, there is an $L$-event after $e_{add}$ in $\mathcal{I}^\mathcal{B}$.
    Let $e$ be the next $L$-event after $e_{add}$ in $\mathcal{I}^\mathcal{B}$, so $e \leq e_{remove}$.
    Since $e_{add}$ is an $L$-add event for $\cellpointershort$, and $e_{add}$ and $e$ are successive $L$-events in $\mathcal{I}^\mathcal{B}$, by $R(\mathcal{I}^\mathcal{B})$, there is a successful list-add attempt $a_{add}$ for $\cellpointershort$ before $e$.
    Therefore, since $a_{add} < e$ and $e \leq e_{remove}$, by transitivity, $a_{add} < e_{remove}$, and so there is a successful list-add attempt for $\cellpointershort$ before $a_{remove}$'s corresponding $L$-event in $\mathcal{I}^\mathcal{B}$ as wanted.
    \qH{\Cref{lemma:ero:list_add_before_list_remove}}
\end{proof}

\begin{proposition}\label{lemma:ero:if_next_pointer_gets_set_to_null_a_list_remove_did_it}
    For every $\cellpointershort \in \celluniverse{} \cup \{\&\headobject\}$, if $(*\cellpointershort).\nextlong.\uniquecellpointercontentlong{} \neq \nullconstant$ at time $T$ in $\mathcal{I}^\mathcal{B}$, and $(*\cellpointershort).\nextlong.\uniquecellpointercontentlong{} = \nullconstant$ at time $T' > T$ in $\mathcal{I}^\mathcal{B}$, then there is a successful list-remove attempt between $\cellpointershort$ and $\nullconstant$ during $(T, T']$.
\end{proposition}

\begin{proof}
    Since $(*\cellpointershort).\nextlong.\uniquecellpointercontentlong{}$ does not equal $\nullconstant$ at $T$ and equals $\nullconstant$ at $T'$ where $T' > T$, it follows that $(*\cellpointershort).\nextlong.\uniquecellpointercontentlong{}$ was set to $\nullconstant$ during $(T, T']$.
    Hence, by \Cref{observation:ero:where_objects_change}, there is either a successful list-add attempt for $\nullconstant$ after $\cellpointershort$ or a successful list-remove attempt between $\cellpointershort$ and $\nullconstant$ during $(T, T']$.
    Therefore, since by \Cref{lemma:ero:every_list_add_seal_and_remove_attempt_is_for_ptr_from_universe} every list-add attempt is for a pointer in $\celluniverse{}$, and by \Cref{assumption:ero:head_and_null_not_in_cell_universe} every pointer in $\celluniverse{}$ is not $\nullconstant$, the former is impossible, so the latter is the only possibility as wanted.
    \qH{\Cref{lemma:ero:if_next_pointer_gets_set_to_null_a_list_remove_did_it}}
\end{proof}

\begin{lemma}\label{lemma:ero:two_successful_list_add_attempts_after_same_pointer_implies_a_removal}
    Consider two successful list-add attempts after $\currentcellpointershort{}$ in $\mathcal{I}^\mathcal{B}$ denoted by $a_1$ and $a_2$ such that $a_1 < a_2$.
    Then there is a successful list-remove attempt between $\currentcellpointershort{}$ and $\nullconstant$ between $a_1$ and $a_2$ in $\mathcal{I}^\mathcal{B}$.
\end{lemma}

\begin{proof}
    Suppose $a_1$ is for $\cellpointershort$, so by \Cref{lemma:ero:every_list_add_seal_and_remove_attempt_is_for_ptr_from_universe} $\cellpointershort \in \celluniverse{}$.
    Since $a_1$ is after $\currentcellpointershort{}$, by \Cref{lemma:ero:every_list_add_attempt_is_after_a_pointer_from_universe_or_head}, $\currentcellpointershort{} \in \celluniverse{} \cup \{\&\headobject\}$.
    Hence, since $a_1$ is for $\cellpointershort$ $(*\currentcellpointershort{}).\nextlong.\uniquecellpointercontentlong{} = \cellpointershort$ at $a_1$.
    Thus, since $\cellpointershort \in \celluniverse{}$, by \Cref{assumption:ero:head_and_null_not_in_cell_universe}, $\cellpointershort \neq \nullconstant$, and so $(*\currentcellpointershort{}).\nextlong \neq \nullconstant$ at $a_1$.
    Let $s$ be the step before $a_2$ in $\mathcal{I}^\mathcal{B}$.
    Hence, $a_1 \leq s$ (since $a_1 < a_2$).
    Furthermore, since $a_2$ is a successful list-add attempt after $\currentcellpointershort{}$, by \Cref{def:ero:english}, $(*\currentcellpointershort{}).\nextlong.\uniquecellpointercontentlong{} = \nullconstant$ at $s$.
    Thus, since $(*\currentcellpointershort{}).\nextlong \neq \nullconstant$ at $a_1$, we have that $a_1 \neq s$, and thus $a_1 < s$ (since $a_1 \leq s$).
    So, since $\currentcellpointershort{} \in \celluniverse{} \cup \{\&\headobject\}$, $(*\currentcellpointershort{}).\nextlong \neq \nullconstant$ at $a_1$, $(*\currentcellpointershort{}).\nextlong.\uniquecellpointercontentlong{} = \nullconstant$ at $s$, and $a_1 < s$, by \Cref{lemma:ero:if_next_pointer_gets_set_to_null_a_list_remove_did_it}, there is a successful list-remove attempt between $\currentcellpointershort{}$ and $\nullconstant$ during $(a_1, s]$.
    Therefore, since $s < a_2$, the claim follows.
    \qH{\Cref{lemma:ero:two_successful_list_add_attempts_after_same_pointer_implies_a_removal}}
\end{proof}

\begin{lemma}\label{lemma:ero:if_p_reads_last_l_event_then_announce_acquire_never_fails}
    Suppose there is a last $L$-event in $\mathcal{I}^\mathcal{B}$; say $e_{last}$.
    Consider any process $p$ and iteration $I$ of the loop on \cref{line:ero:do_work_while_loop} by $p$ during $\mathcal{I}^\mathcal{B}$ such that $p$'s execution of \cref{line:ero:linearization_read} during $I$ is after $e_{last}$.
    Then, $p$ does not find the condition on \cref{line:ero:announce_acquire_l_changed_check} to be true during $I$.
\end{lemma}

\begin{proof}
    Suppose, for contradiction, $p$ finds the condition on \cref{line:ero:announce_acquire_l_changed_check} to be true during $I$.
    Suppose $e_{last}$ set $\linearizationobject{} = (\uniquerepositoryoperationshort, \uniquecellpointershort)$.
    Hence, since by assumption $e_{last}$ is the last $L$-event in $\mathcal{I}^\mathcal{B}$, by \Cref{observation:ero:where_objects_change}, $\linearizationobject{}  = (\uniquerepositoryoperationshort, \uniquecellpointershort)$ from $e_{last}$ onwards in $\mathcal{I}^\mathcal{B}$.
    Thus, since $p$'s execution of \cref{line:ero:linearization_read} during $I$ is after $e_{last}$, $p$ reads $(\uniquerepositoryoperationshort, \uniquecellpointershort)$ from $\linearizationobject{}$ on \cref{line:ero:linearization_read} during $I$.
    Hence, the first parameter of the Acquire procedure on \cref{line:ero:announce_acquire} during $I$ is $\uniquerepositoryoperationshort$.
    Let $I'$ denote this invocation of the Acquire procedure.
    Since $p$ finds the condition on \cref{line:ero:announce_acquire_l_changed_check} to be true during $I$, we have that the response of $I'$ is $\timechange$.
    Hence, $p$ found the condition on \cref{line:ero:acquire_next_linearization_changed_check} to be true during an invocation $I^*$ of the AcquireNext procedure invoked during $I'$; say at time $T^{\ref{line:ero:acquire_next_linearization_changed_check}}$.
    Since $I^*$ is invoked during $I'$ and $I'$'s first parameter is $\uniquerepositoryoperationshort$, we have that $I^*$'s first parameter is $\uniquerepositoryoperationshort$.
    Furthermore, since $p$'s execution of \cref{line:ero:linearization_read} during $I$ is after $e_{last}$, $I'$ was invoked on \cref{line:ero:announce_acquire} during $I$, $I^*$ was invoked during $I'$, and $T^{\ref{line:ero:acquire_next_linearization_changed_check}}$ is during $I^*$, by transitivity, $e_{last} < T^{\ref{line:ero:acquire_next_linearization_changed_check}}$.
    Therefore, since $I^*$'s first parameter is $\uniquerepositoryoperationshort$, and $p$ finds the condition on \cref{line:ero:acquire_next_linearization_changed_check} to be true during $I^*$ at $T^{\ref{line:ero:acquire_next_linearization_changed_check}}$ which is after $e_{last}$, we have that $\linearizationobject{}.\uniquerepositoryoperationlong \neq \uniquerepositoryoperationshort$ after $e_{last}$.
    However, $\linearizationobject{} = (\uniquerepositoryoperationshort, \uniquecellpointershort)$ from $e_{last}$ onwards in $\mathcal{I}^\mathcal{B}$, a contradiction.
    \qH{\Cref{lemma:ero:if_p_reads_last_l_event_then_announce_acquire_never_fails}}
\end{proof}

\begin{lemma}\label{lemma:ero:if_next_upointer_is_initial_then_acquisitions_is_zero}
    For every $\cellpointershort \in \celluniverse{} \cup \{\&\headobject\}$, if $(*\cellpointershort).\nextlong = (\arbitraryvalue, \arbitraryvalue, acq, \nullconstant)$ at any time in $\mathcal{I}^\mathcal{B}$, then $acq = 0$.
\end{lemma}

\begin{proof}
    Suppose, for contradiction, there exists a $\cellpointershort \in \celluniverse{} \cup \{\&\headobject\}$ such that at some time $T$ in $\mathcal{I}^\mathcal{B}$ $(*\cellpointershort).\nextlong = (\arbitraryvalue, \arbitraryvalue, acq, \nullconstant)$ for some $acq \neq 0$.
    Without loss of generality, suppose $T$ is the first time the lemma is violated for \Underline{any} pointer in $\celluniverse{} \cup \{\&\headobject\}$.
    Since $(*\cellpointershort).\nextlong.\acquisitions$ is initially $0$, $(*\cellpointershort).\nextlong.\acquisitions = acq \neq 0$ at $T$, and $T$ is the first such time in $\mathcal{I}^\mathcal{B}$, it follows that the step at $T$ sets the value of $(*\cellpointershort).\nextlong$ to $(\arbitraryvalue, \arbitraryvalue, acq, \nullconstant)$ at $T$.
    Hence, by \Cref{observation:ero:where_objects_change}, the step at $T$ is either a successful list-add attempt after $\cellpointershort$, a successful list-seal attempt for $\cellpointershort$, a successful list-remove attempt between $\cellpointershort$ and $\nullconstant$, or a successful list-acquire-next attempt after $\cellpointershort$.
    Denote by $a$ this attempt at $T$ and so $a$ set $(*\cellpointershort).\nextlong$ to $(\arbitraryvalue, \arbitraryvalue, acq, \nullconstant)$ (*).

    \begin{itemize}
        \item[] \hspace{0pt}\textbf{Case 1.} $a$ is a successful list-add attempt after $\cellpointershort$.

        Hence, by \Cref{lemma:ero:every_list_add_seal_and_remove_attempt_is_for_ptr_from_universe}, $a$ is for some $\cellpointershort' \in \celluniverse{}$.
        Thus, since $a$ is successful, by \Cref{def:ero:english} $a$ set $(*\cellpointershort).\nextlong.\uniquecellpointercontentlong{} = \cellpointershort'$.
        Therefore, since $\cellpointershort' \in \celluniverse{}$, by \Cref{assumption:ero:head_and_null_not_in_cell_universe}, $\cellpointershort' \neq \nullconstant$, and so $a$ set $(*\cellpointershort).\nextlong.\uniquecellpointercontentlong{} \neq \nullconstant$.
        However, by (*) $a$ set $(*\cellpointershort).\nextlong.\uniquecellpointercontentlong{} = \nullconstant$, a contradiction.

        \item[] \hspace{0pt}\textbf{Case 2.} $a$ is a successful list-remove attempt between $\cellpointershort$ and $\nullconstant$

        Let $p$ be the process that executed $a$ and suppose $a$ is for some $\cellpointershort'$.
        Hence, by \Cref{lemma:ero:every_list_add_seal_and_remove_attempt_is_for_ptr_from_universe}, $\cellpointershort' \in \celluniverse{}$.
        Since $a$ is a successful list-remove attempt for $\cellpointershort'$ between $\cellpointershort$ and $\nullconstant$ and $a$ set $(*\cellpointershort).\nextlong$ to $(\arbitraryvalue, \arbitraryvalue, acq, \nullconstant)$, $p$ read $(\arbitraryvalue, \arbitraryvalue, acq, \nullconstant)$ from $(*\cellpointershort').\nextlong$ on its last execution of \cref{line:ero:remove_cell_read_pointer_to_remove} before $a$.
        Therefore, for some $\cellpointershort' \in \celluniverse{}$, $(*\cellpointershort').\nextlong = (\arbitraryvalue, \arbitraryvalue, acq, \nullconstant)$ at some time before $T$, and so by the minimality of $T$, $acq = 0$.
        However, by assumption $acq \neq 0$, a contradiction.

        \item[] \hspace{0pt}\textbf{Case 3.} $a$ is a successful list-seal attempt for $\cellpointershort$.

        Let $p$ be the process that executed $a$.
        Since $a$ is a successful list-seal attempt for $\cellpointershort$ and $a$ set $(*\cellpointershort).\nextlong$ to $(\arbitraryvalue, \arbitraryvalue, acq, \nullconstant)$, we have that $p$ read $(\arbitraryvalue, \arbitraryvalue, acq, \nullconstant)$ from $(*\cellpointershort').\nextlong$ on its last execution of \cref{line:ero:remove_cell_read_pointer_to_remove_before_seal} before $a$.
        Therefore, $(*\cellpointershort).\nextlong = (\arbitraryvalue, \arbitraryvalue, acq, \nullconstant)$ at some time before $T$, and so by the minimality of $T$, $acq = 0$.
        However, by assumption $acq \neq 0$, a contradiction.

        \item[] \hspace{0pt}\textbf{Case 4.} $a$ is a successful list-acquire-next attempt after $\cellpointershort$.

        Let $p$ be the process that executed $a$.
        Since $a$ set $(*\cellpointershort).\nextlong$ to $(\arbitraryvalue, \arbitraryvalue, acq, \nullconstant)$, we have that  $p$ read $(\arbitraryvalue, \arbitraryvalue, \arbitraryvalue, \nullconstant)$ from $(*\cellpointershort).\nextlong$ on its last execution of \cref{line:ero:acquire_next_read_curr_unique_pointer} before $a$. 
        Therefore, $p$ found the condition on \cref{line:ero:acquire_next_not_found_check} to be true between this time and $a$.
        However, since $p$ executed $a$, it must have found the condition on \cref{line:ero:acquire_next_not_found_check} to be false between this time and $a$, a contradiction.
        \qH{\Cref{lemma:ero:if_next_upointer_is_initial_then_acquisitions_is_zero}}
    \end{itemize}
\end{proof}

\begin{lemma}\label{lemma:ero:acquisition_counter_always_non_negative}
    For every $\cellpointershort \in \celluniverse{} \cup \{\&\headobject\}$, $(*\cellpointershort).\nextlong.\acquisitions \geq 0$ at all times in $\mathcal{I}^\mathcal{B}$.
\end{lemma}

\begin{proof}
    Suppose, for contradiction, there is a $\cellpointershort \in \celluniverse{} \cup \{\&\headobject\}$ such that $(*\cellpointershort).\nextlong.\acquisitions < 0$ at some time $T$ in $\mathcal{I}^\mathcal{B}$.
    Without loss of generality, suppose $T$ is the first time in $\mathcal{I}^\mathcal{B}$ the lemma is violated, i.e., for every $\cellpointershort \in \celluniverse{} \cup \{\&\headobject\}$, $(*\cellpointershort).\nextlong.\acquisitions \geq 0$ at all times before $T$ in $\mathcal{I}^\mathcal{B}$.
    Since $\cellpointershort \in \celluniverse{} \cup \{\&\headobject\}$ $(*\cellpointershort).\nextlong.\acquisitions$ is initially $0$.
    Hence, the step at time $T$ set $(*\cellpointershort).\nextlong.\acquisitions < 0$.
    Thus, by \Cref{observation:ero:where_objects_change}, the step at $T$ is either a successful list-add attempt after $\cellpointershort$, a successful list-remove attempt between $\cellpointershort$ and some pointer, or a successful list-acquire-next attempt after $\cellpointershort$.
    In the first and last case, by \cref{line:ero:add_cell_to_list} and \cref{line:ero:acquire_next_cell}, the value of $(*\cellpointershort).\nextlong.\acquisitions$ is one larger at time $T$ than $T - 1$, so $(*\cellpointershort).\nextlong.\acquisitions < 0$ at $T - 1$, a contradiction to the minimality of $T$.
    Now consider the second case.
    Let the step at $T$ be a successful list-remove attempt for $\cellpointershort_\linearizationobject{}$, so by \Cref{lemma:ero:every_list_add_seal_and_remove_attempt_is_for_ptr_from_universe} $\cellpointershort_\linearizationobject \in \celluniverse{}$.
    By \cref{line:ero:remove_cell_from_list}, the step at $T$ set $(*\cellpointershort).\nextlong.\acquisitions = a$ where $(*\cellpointershort_\linearizationobject{}).\nextlong.\acquisitions = a$ at some time before $T$.
    Therefore, since $\cellpointershort_\linearizationobject{} \in \celluniverse{}$, by the minimality of $T$, $a \geq 0$.
    However, since $(*\cellpointershort).\nextlong.\acquisitions < 0$ at $T$, and $(*\cellpointershort).\nextlong.\acquisitions = a$ at $T$, we have that $a < 0$, a contradiction.
    \qH{\Cref{lemma:ero:acquisition_counter_always_non_negative}}
\end{proof}

\begin{lemma}\label{lemma:ero:free_cell_operations_are_preceeded_by_successful_list_remove_attempt}
    Every $\freecelloperation(\cellpointershort)$ operation in $\mathcal{I}^\mathcal{B}$ is preceded by a successful list-remove attempt for $\cellpointershort$.
\end{lemma}

\begin{proof}
    Consider any $\freecelloperation(\cellpointershort)$ operation $o$ in $\mathcal{I}^\mathcal{B}$ by some process $p$, so by \Cref{lemma:ero:free_cell_operations_are_from_universe} $\cellpointershort \in \celluniverse{}$.
    Since $p$ performed $o$, it did so on \cref{line:ero:free_cell}, and thus $p$ performed a revocation event for $\cellpointershort$ whose response is $-1$ before $o$.
    Hence, since $\cellpointershort \in \celluniverse$ (a) by \Cref{observation:ero:where_objects_change} the only steps that change the value of $(*\cellpointershort).\revocations$ are acquire-copy events for $\cellpointershort$ and revocation events for $\cellpointershort$, (b) each revocation event for $\cellpointershort$ increases the value of $(*\cellpointershort).\revocations$ by 1, and (c) $(*\cellpointershort).\revocations$ is initially 0, we have that there is an acquire-copy event for $\cellpointershort$ before $o$.
    Therefore, by \Cref{lemma:ero:before_acquire_copy_is_successful}, there is a successful list-remove attempt for $\cellpointershort$ before $o$ as wanted.
    \qH{\Cref{lemma:ero:free_cell_operations_are_preceeded_by_successful_list_remove_attempt}}
\end{proof}

Since by \Cref{lemma:ero:l_event_corresponding_to_do_low_level_op} every list-remove attempt for $\cellpointershort$ is preceded by an $L$-remove event for $\cellpointershort$, \Cref{lemma:ero:free_cell_operations_are_preceeded_by_successful_list_remove_attempt} implies the following.

\begin{corollary}\label{lemma:ero:free_cell_operations_are_preceeded_by_remove_event}
    Every $\freecelloperation(\cellpointershort)$ operation in $\mathcal{I}^\mathcal{B}$ is preceded by an $L$-remove event for $\cellpointershort$.
\end{corollary}

\begin{lemma}\label{lemma:ero:if_next_pointer_not_null_there_is_a_l_event_for_it}
    Suppose $Q(\mathcal{I}^\mathcal{B})$ holds.
    For every $\cellpointershort \in \celluniverse{} \cup \{\&\headobject\}$, if $(*\cellpointershort).\nextlong.\cellpointerlong = \nextcellpointershort{}$ at some time $T$ in $\mathcal{I}^\mathcal{B}$ where $\nextcellpointershort{} \in \celluniverse{}$, then there is an $L$-event for $\cellpointershort$ before $T$ in $\mathcal{I}^\mathcal{B}$. 
\end{lemma}

\begin{proof}
    Since $\cellpointershort \in \celluniverse \cup \{\&\headobject\}$, we have that $(*\cellpointershort).\nextlong.\cellpointerlong$ is initially $\nullconstant$, and since $\nextcellpointershort{} \in \celluniverse$, by \Cref{assumption:ero:head_and_null_not_in_cell_universe},  $\nextcellpointershort{} \neq \&\headobject$ and $\nextcellpointershort{} \neq \nullconstant$, and so $(*\cellpointershort).\nextlong.\cellpointerlong$ was set to $\nextcellpointershort{}$ before $T$ in $\mathcal{I}^\mathcal{B}$.
    Thus, by \Cref{observation:ero:where_objects_change}, there is either a successful list-add attempt for $\nextcellpointershort{}$ after $\cellpointershort$ or a successful list-remove attempt between $\cellpointershort$ and $\nextcellpointershort{}$ before $T$ in $\mathcal{I}^\mathcal{B}$.
    Let $a$ denote this successful list attempt.
    If $a$ is a successful list-add attempt for $\cellpointershort$, by \Cref{lemma:ero:l_event_corresponding_to_do_low_level_op}, there is an $L$-event for $\cellpointershort$ before $a$.
    Furthermore, if $a$ is a successful list-remove attempt between $\cellpointershort$ and $\nextcellpointershort{}$, by $Q(\mathcal{I}^\mathcal{B})$, $\cellpointershort \in \List(\mathcal{I})$ where $\mathcal{I}$ is a prefix of $\mathcal{I}^\mathcal{B}$ before $a$.
    Hence, since $\cellpointershort \neq \nullconstant$ and $\cellpointershort \neq \&\headobject$, by \Cref{def:ero:logical_list}, there is an $L$-event for $\cellpointershort$ in $\mathcal{I}$.
    Therefore, in all cases, there is an $L$-event for $\cellpointershort$ before $a$ (and thus $T$) in $\mathcal{I}^\mathcal{B}$.
    \qH{\Cref{lemma:ero:if_next_pointer_not_null_there_is_a_l_event_for_it}}
\end{proof}

\subsection{The $L$-Invariants Hold}
\label{sec:l_invariants}

The high-level strategy for proving many facts about $\mathcal{B}$ is to do so under the assumption that the $L$-invariants hold.
We already saw some basic examples of this in the last section.
The main goal of this section is to prove that the $L$-invariants hold for every implementation history of $\mathcal{B}$.
The high-level strategy for doing so is as follows.
First, we will prove that $\List(\mathcal{I})$ is essentially the ``shape" of the list at the end of $\mathcal{I}$ where $\mathcal{I}$ is a finite implementation history of $\mathcal{B}$.
Then, using this fact, we prove that the \doaddcell{}, \doapplyandcopyresponse{}, and \doremovecell{} procedures have the intended effect: if a process exits any of these procedures, then the task it wanted to complete has been completed (but not necessarily by itself).
For example, if a process exits the \doaddcell{} procedure with parameters $(\arbitraryvalue, \cellpointershort)$, then by the time this procedure exits, $\cellpointershort$ has been added to the list, i.e., there is a successful list-add attempt for $\cellpointershort$.
We then use these facts to prove that the IsDone procedure has the intended effect in the sense that its response informs the invoking process whether the inputted low-level operation has been written into $\linearizationobject{}$.
For example, if a process invokes the IsDone procedure with parameters $(\arbitraryvalue, (\arbitraryvalue, \addcell), \cellpointershort)$ and receives response $\notdone$, then it knows that there has yet to be an $L$-add event for $\cellpointershort$ and conversely if it receives response $\done$, then it knows that there is an $L$-add event for $\cellpointershort$ (note that the exact time when these are true is delicate).
Finally, these facts let us prove inductively that the $L$-inva3riants hold for any implementation history of $\mathcal{B}$.
Throughout this section, $\mathcal{I}^\mathcal{B}$ refers to an arbitrary implementation history of $\mathcal{B}$, i.e., all statements that refer to $\mathcal{I}^\mathcal{B}$ begin with ``for every implementation history $\mathcal{I}^\mathcal{B}$ of $\mathcal{B}$" which is omitted for brevity.

\subsubsection{The sequence of $L$-events determines the shape of the list}

We start by proving that the $\List(\mathcal{I})$ is essentially the ``shape" of the list at the end of $\mathcal{I}$.
Given $P(\mathcal{I})$, $Q(\mathcal{I})$, and $R(\mathcal{I})$ hold, this is mostly a matter of capturing the ``lag" between the time of an $L$-event and the time it ``takes effect" (i.e., the relevant cell is added to or removed from the list), and carefully applying these invariants along with the definition of $\List(\mathcal{I})$.
The one difficulty is dealing with the period after the last $L$-event in $\mathcal{I}$ because invariant $R$ only tells us how the list changes between successive $L$-events.
So, the first step is to ``extend" $R$ beyond the last $L$-event $e_{last}$ in $\mathcal{I}$ to state that there is either (1) at most one successful list-add attempt for $\cellpointershort$ and no other successful list-add or list-remove attempts for any pointer (if $e_{last}$ is an $L$-add event for $\cellpointershort$), (2) zero successful list-add or list-remove attempts for any pointer (if $e_{last}$ is an $L$-apply event), or (3) at most one successful list-remove attempt for $\cellpointershort$ and no other successful list-add or list-remove attempts for any pointer (if $e_{last}$ is an $L$-remove event for $\cellpointershort$).
We accomplish this goal via the next four lemmas.

We start by showing that the success of a list-add / list-remove attempt implies the non-existence of an $L$-event between its corresponding $L$-event and it.
As we will see shortly, this is the key fact that lets us ``extend" $R$ beyond the last $L$-event within $\mathcal{I}$.

\begin{proposition}\label{lemma:ero:any_list_attempt_outside_its_window_is_unsuccessful}
    Consider a list-add or list-remove attempt $a$ in $\mathcal{I}^\mathcal{B}$ and let $e_b$ be its corresponding $L$-event (see \Cref{lemma:ero:l_event_corresponding_to_do_low_level_op}).
    Suppose there is an $L$-event after $e_b$ in $\mathcal{I}^\mathcal{B}$ and that $P(\mathcal{I}^\mathcal{B})$, $Q(\mathcal{I}^\mathcal{B})$, and $R(\mathcal{I}^\mathcal{B})$ hold.
    Let $e_a$ be the next $L$-event after $e_b$ in $\mathcal{I}^\mathcal{B}$.
    If $e_a < a$, then $a$ is unsuccessful.
\end{proposition}

\begin{proof}
    Suppose, for contradiction, $e_a < a$, and $a$ is successful.
    Let $p$ be the process that executed $a$.
    If $a$ is a list-add attempt, suppose it is for some $\cellpointershort$ after some $\currentcellpointershort{}$ and otherwise, suppose it is for $\cellpointershort$ between $\currentcellpointershort{}$ and some $\nextcellpointershort{}$.
    Hence, by Lemmas \ref{lemma:ero:every_list_add_seal_and_remove_attempt_is_for_ptr_from_universe}, \ref{lemma:ero:every_list_add_attempt_is_after_a_pointer_from_universe_or_head}, and \ref{lemma:ero:every_list_remove_attempt_is_between_a_pointer_from_universe_or_header_and_a_pointer_from_universe_or_null}, $\cellpointershort \in \celluniverse{}$, $\currentcellpointershort{} \in \celluniverse{} \cup \{\&\headobject\}$, and $\nextcellpointershort{} \in \celluniverse{} \cup \{\nullconstant\}$.

    \begin{claimcustom}{\ref{lemma:ero:any_list_attempt_outside_its_window_is_unsuccessful}.1}\label{lemma:ero:any_list_attempt_outside_its_window_is_unsuccessful_one_claim}
        If $a$ is a list-add attempt, then $T$ is the time of $p$'s last execution of \cref{line:ero:add_cell_read_end_of_list} before $a$; otherwise, $T$ is the time of $p$'s last execution of \cref{line:ero:remove_cell_read_previous_pointer} before $a$.
        Then, $T \in (e_b, e_a)$.
    \end{claimcustom}

    \begin{proof}
        Let $I$ be the invocation of the \doaddcell{} or \doremovecell{} procedure that $p$ executed $a$ during.
        Hence, $T$ is the time of a step that $p$ performed during $I$, and by \Cref{lemma:ero:l_event_corresponding_to_do_low_level_op} $e_b$ is before $p$ invoked $I$.
        Therefore, $e_b < T$.
        So, what remains is to prove that $T < e_a$.
        
        Suppose, for contradiction, $e_a \leq T$.
        If $a$ is a list-add attempt, then $T'$ is the time of $p$'s execution of \cref{line:ero:add_cell_before_updating_end_of_list_linearization_changed_check} between $T$ and $a$; otherwise, $T'$ is the time of $p$'s execution of \cref{line:ero:remove_cell_before_removal_linearization_check} between $T$ and $a$.
        Hence, $T < T'$.
        Since $e_b < e_a$, $e_a \leq T$, and $T < T'$, by transitivity, $e_b < e_a < T'$.
        Furthermore, since $p$ executed $a$, it follows that the condition on \cref{line:ero:add_cell_before_updating_end_of_list_linearization_changed_check} or \ref{line:ero:remove_cell_before_removal_linearization_check} is false at $T'$.
        Hence, $\linearizationobject{}.\uniquerepositoryoperationlong{} = \uniquerepositoryoperationshort{}_\linearizationobject{}$ at $T'$ where $\uniquerepositoryoperationshort{}_\linearizationobject{}$ is the first parameter of $I$.
        Thus, since $e_b$ is $a$'s corresponding $L$-event, and $p$ executed $a$ during $I$, by \Cref{lemma:ero:l_event_corresponding_to_do_low_level_op}, $e_b$ set $\linearizationobject{}.\uniquerepositoryoperationlong = \uniquerepositoryoperationshort{}_\linearizationobject{}$.
        Furthermore, since $e_b < T'$, by \Cref{observation:ero:where_objects_change}, the last $L$-event before $T'$ set $\linearizationobject{}.\uniquerepositoryoperationlong = \uniquerepositoryoperationshort{}_\linearizationobject{}$, and since $e_b < e_a < T'$ there is an $L$-event during $(e_b, T')$ that set $\linearizationobject{}.\uniquerepositoryoperationlong = \uniquerepositoryoperationshort{}_\linearizationobject{}$.
        Therefore, since $e_b$ set $\linearizationobject{}.\uniquerepositoryoperationlong = \uniquerepositoryoperationshort{}_\linearizationobject{}$, we have that two $L$-events in $\mathcal{I}^\mathcal{B}$ set $\linearizationobject{}.\uniquerepositoryoperationlong$ to the same value.
        However, since $P(\mathcal{I}^\mathcal{B})$ holds, by \Cref{lemma:ero:p_implies_unique_low_level_operations_in_linearization}, every $L$-event in $\mathcal{I}^\mathcal{B}$ sets $\linearizationobject{}.\uniquerepositoryoperationlong$ to a unique value, contradiction.
        \qH{\Cref{lemma:ero:any_list_attempt_outside_its_window_is_unsuccessful_one_claim}}
    \end{proof}

    \begin{claimcustom}{\ref{lemma:ero:any_list_attempt_outside_its_window_is_unsuccessful}.2}\label{lemma:ero:any_list_attempt_outside_its_window_is_unsuccessful_two_claim}
        During $(T, a)$, there is either a successful list-add attempt after $\currentcellpointershort{}$ or there is a successful list-remove attempt between $\currentcellpointershort{}$ and some pointer.
    \end{claimcustom}

    \begin{proof}
        There are two cases.
        \begin{enumerate}
            \item[] \hspace{0pt}\textbf{Case 1.} $a$ is a list-add attempt for $\cellpointershort$ after $\currentcellpointershort{}$.

            Let $I$ be the invocation of the \doaddcell{} procedure that $p$ executed $a$ during.
            Since $a$ is a list-add attempt for $\cellpointershort$, by \Cref{def:ero:english}, the second parameter of $I$ is $\cellpointershort$.
            Hence, since $e_b$ is $a$'s corresponding $L$-event, by \Cref{lemma:ero:l_event_corresponding_to_do_low_level_op}, $e_b$ is an $L$-add event for $\cellpointershort$.
            Since by assumption $e_b$ and $e_a$ are in $\mathcal{I}^\mathcal{B}$ and $e_a$ is defined as the next $L$-event after $e_b$ in $\mathcal{I}^\mathcal{B}$, $e_b$ and $e_a$ are successive $L$-events in $\mathcal{I}^\mathcal{B}$.
            Therefore, since $e_b$ is an $L$-add event for $\cellpointershort$, by $R(\mathcal{I}^\mathcal{B})$, there is one successful list-add attempt for $\cellpointershort$ during $(e_b, e_a)$; say $a'$.
            Since $a' < e_a$ and $e_a < a$, by transitivity, $a' < a$.
            Furthermore, since $a$ and $a'$ are both list-add attempts for $\cellpointershort$ in $\mathcal{I}^\mathcal{B}$, $a$ is after $\currentcellpointershort{}$, and $Q(\mathcal{I}^\mathcal{B})$ holds, by \Cref{lemma:ero:list_add_attempt_for_some_pointer_are_after_same_pointer}, $a'$ is after $\currentcellpointershort{}$.
            There are two cases.

            \begin{enumerate}
                \item[] \hspace{0pt}\textbf{Case 1.1.} $T < a'$.

                Hence, since $a' < a$, we have that $a' \in (T, a)$.
                Therefore, since $a'$ is a successful list-add attempt after $\currentcellpointershort{}$, the claim follows.

                \item[] \hspace{0pt}\textbf{Case 1.2.} $a' < T$.
                
                Hence, since by definition $T < a$, by transitivity, $a' < a$.
                Since $a'$ and $a$ are two successful list-add attempts after $\currentcellpointershort{}$ in $\mathcal{I}^\mathcal{B}$ such that $a' < a$, by \Cref{lemma:ero:two_successful_list_add_attempts_after_same_pointer_implies_a_removal}, there is a successful list-remove attempt $a^*$ between $\currentcellpointershort{}$ and $\nullconstant$ between $a'$ and $a$ in $\mathcal{I}^\mathcal{B}$.
                There are two more cases.

                \begin{enumerate}
                    \item[] \hspace{0pt}\textbf{Case 1.2.1.} $T < a^*$.

                    Hence, since $a^* < a$, we have that $a^* \in (T, a)$.
                    Therefore, since $a^*$ is a successful list-remove attempt between $\currentcellpointershort{}$ and some pointer, the claim follows.

                    \item[] \hspace{0pt}\textbf{Case 1.2.2.} $a^* < T$.

                    Hence, since $a' < a^*$, we have that $a^* \in (a', T)$.
                    Thus, since $e_b < a'$ and by \Cref{lemma:ero:any_list_attempt_outside_its_window_is_unsuccessful_one_claim} $T < e_a$, by transitivity, $a^* \in (e_b, e_a)$.
                    Therefore, there is a successful list-remove attempt during $(e_b, e_a)$.
                    However, since $e_b$ and $e_a$ are successive $L$-events in $\mathcal{I}^\mathcal{B}$ and $e_b$ is an $L$-add event, by $R(\mathcal{I}^\mathcal{B})$, there are no successful list-remove attempts during $(e_b, e_a)$, a contradiction, so this case is impossible.
                \end{enumerate}
            \end{enumerate}
                  
            \item[] \hspace{0pt}\textbf{Case 2.} $a$ is a list-remove attempt for $\cellpointershort$ between $\currentcellpointershort{}$ and $\nextcellpointershort{}$.

            Let $I$ be the invocation of the \doremovecell{} procedure that $p$ executed $a$ during.
            Since $a$ is a list-remove attempt for $\cellpointershort$, by \Cref{def:ero:english}, the second parameter of $I$ is $\cellpointershort$.
            Hence, since $e_b$ is $a$'s corresponding $L$-event, by \Cref{lemma:ero:l_event_corresponding_to_do_low_level_op}, $e_b$ is an $L$-remove event for $\cellpointershort$.
            Since by assumption $e_b$ and $e_a$ are in $\mathcal{I}^\mathcal{B}$ and $e_a$ is defined as the next $L$-event after $e_b$ in $\mathcal{I}^\mathcal{B}$, $e_b$ and $e_a$ are successive $L$-events in $\mathcal{I}^\mathcal{B}$.
            Therefore, since $e_b$ is an $L$-remove event for $\cellpointershort$, by $R(\mathcal{I}^\mathcal{B})$, there is one successful list-remove attempt for $\cellpointershort$ during $(e_b, e_a)$; say $a'$.
            Since $a' < e_a$ and $e_a < a$, by transitivity, $a' < a$.
            Furthermore, since $a$ and $a'$ are both list-remove attempts for $\cellpointershort$ in $\mathcal{I}^\mathcal{B}$, $a$ is between $\currentcellpointershort{}$ and $\nextcellpointershort{}$, and $Q(\mathcal{I}^\mathcal{B})$ holds, by \Cref{lemma:ero:list_remove_attempt_for_some_pointer_are_between_same_pointer}, $a'$ is between $\currentcellpointershort{}$ and $\nextcellpointershort{}$.
            There are two cases.

            \begin{enumerate}
                \item[] \hspace{0pt}\textbf{Case 2.1.} $T < a'$.

                Hence, since $a' < a$, we have that $a' \in (T, a)$.
                Therefore, since $a'$ is a successful list-remove attempt between $\currentcellpointershort{}$ and some pointer, the claim follows.

                \item[] \hspace{0pt}\textbf{Case 2.2.} $a' < T$.

                Since $e_b$ and $e_a$ are successive $L$-events in $\mathcal{I}^\mathcal{B}$ and $e_b$ is an $L$-remove event for $\cellpointershort$, by $R(\mathcal{I}^\mathcal{B})$, there are no successful list-add or list-remove attempts during $(e_b, e_a)$ other than $a'$.
                Hence, since $a'$ is a list-remove attempt for $\cellpointershort$ between $\currentcellpointershort{}$ and $\nextcellpointershort{}$, by \Cref{observation:ero:where_objects_change}, $(*\currentcellpointershort{}).\nextlong.\uniquecellpointercontentlong{}$ is set exactly once during $(e_b, e_a)$ (at $a'$), so $(*\currentcellpointershort{}).\nextlong.\uniquecellpointercontentlong{}$ is unchanged during $(a', e_a)$.
                
                We now deduce the values that $p$ read from $\cellpointershort$ at the time of $p$'s last execution of \cref{line:ero:remove_cell_read_pointer_to_remove} before $a$, say $T^{\ref{line:ero:remove_cell_read_pointer_to_remove}}$, and from $\currentcellpointershort{}$ at $T$.
                First $\cellpointershort$ at $T^{\ref{line:ero:remove_cell_read_pointer_to_remove}}$.
                Since $a$ is a list-remove attempt for $\cellpointershort$ between $\currentcellpointershort{}$ and $\nextcellpointershort{}$, we have that $p$ read $\nextcellpointershort{}$ from $(*\cellpointershort{}).\nextlong.\uniquecellpointercontentlong{}$ on \cref{line:ero:remove_cell_read_pointer_to_remove} at $T^{\ref{line:ero:remove_cell_read_pointer_to_remove}}$.
                Now $\currentcellpointershort{}$ at $T$.
                Since $a'$ is a successful list-remove attempt between $\currentcellpointershort{}$ and $\nextcellpointershort{}$, we have that $(*\currentcellpointershort{}).\nextlong.\uniquecellpointercontentlong{} = \nextcellpointershort{}$ at $a'$.
                Hence, since $(*\currentcellpointershort{}).\nextlong.\uniquecellpointercontentlong{}$ is unchanged during $(a', e_a)$, we have that $(*\currentcellpointershort{}).\nextlong.\uniquecellpointercontentlong{} = \nextcellpointershort{}$ throughout $(a', e_a)$.
                Furthermore, since $a' < T$ and by \Cref{lemma:ero:any_list_attempt_outside_its_window_is_unsuccessful_one_claim} $T < e_a$, we have that $T \in (a', e_a)$.
                Hence, $(*\currentcellpointershort{}).\nextlong.\uniquecellpointercontentlong{} = \nextcellpointershort{}$ at $T$.
                Therefore, since $p$ executes $a$, $a$ is a list-remove attempt for $\cellpointershort$ between $\currentcellpointershort{}$ and $\nextcellpointershort{}$, and $T$ is the time of $p$'s last execution of \cref{line:ero:remove_cell_read_previous_pointer} before $a$, we have that $p$ read $\nextcellpointershort{}$ from $(*\currentcellpointershort{}).\nextlong.\uniquecellpointercontentlong{}$ at $T$.

                We now complete the proof of Case 2.2.
                Since $T$ is the time of $p$'s last execution of \cref{line:ero:remove_cell_read_previous_pointer} before $a$, we have that $p$ performed the following steps back-to-back: (1) \cref{line:ero:remove_cell_read_previous_pointer} at $T$; (2) \cref{line:ero:remove_cell_before_removal_linearization_check} at some time $T^{\ref{line:ero:remove_cell_before_removal_linearization_check}}$; and (3) \cref{line:ero:remove_cell_from_list} at $a$.
                Therefore, since $p$ read $\nextcellpointershort{}$ from $(*\cellpointershort{}).\nextlong.\uniquecellpointercontentlong{}$ at $T^{\ref{line:ero:remove_cell_read_pointer_to_remove}}$, and $p$ read $\nextcellpointershort{}$ from $(*\currentcellpointershort{}).\nextlong.\uniquecellpointercontentlong{}$ at $T$, it follows that $p$ found the second clause of \cref{line:ero:remove_cell_before_removal_linearization_check} to be true at $T^{\ref{line:ero:remove_cell_before_removal_linearization_check}}$.
                However, since $p$ executed $a$ after executing \cref{line:ero:remove_cell_before_removal_linearization_check} at $T^{\ref{line:ero:remove_cell_before_removal_linearization_check}}$, $p$ found the second clause of \cref{line:ero:remove_cell_before_removal_linearization_check} to be false at $T^{\ref{line:ero:remove_cell_before_removal_linearization_check}}$, a contradiction, so this case is impossible.
                \qH{\Cref{lemma:ero:any_list_attempt_outside_its_window_is_unsuccessful_two_claim}}
            \end{enumerate}
        \end{enumerate}
    \end{proof}

    We now return to the proof of \Cref{lemma:ero:any_list_attempt_outside_its_window_is_unsuccessful}.
    Let $a'$ be the attempt identified in \Cref{lemma:ero:any_list_attempt_outside_its_window_is_unsuccessful_two_claim}.
    By \Cref{def:ero:english}, $a'$ is of the form \CASop{}$((*\currentcellpointershort{}).\nextlong, (v, \arbitraryvalue, \arbitraryvalue, \arbitraryvalue), (v + 1, \arbitraryvalue, \arbitraryvalue, \arbitraryvalue))$ for some $v$.
    Thus, since $a'$ is successful, $(*\currentcellpointershort{}).\nextlong.\view$ equals $v$ at the step before $a'$, and $v + 1$ at $a'$.
    Hence, since $\currentcellpointershort{} \in \celluniverse{} \cup \{\&\headobject\}$, by \Cref{observation:ero:views_are_monotonic}, $(*\currentcellpointershort{}).\nextlong.\view$ is monotonically increasing, and since $a' \in (T, a)$, it follows that (1) $(*\currentcellpointershort{}).\nextlong.\view \leq v$ at $T$, and (2) $(*\currentcellpointershort{}).\nextlong.\view > v$ at the step before $a$.
    Since $a$ is either a successful list-add attempt after $\currentcellpointershort{}$ or $a$ is a successful list-remove attempt between $\currentcellpointershort{}$ and some pointer, by \Cref{def:ero:english} $a$ is of the form \CASop{}$((*\currentcellpointershort{}).\nextlong, (v', \arbitraryvalue, \arbitraryvalue, \arbitraryvalue), (v' + 1, \arbitraryvalue, \arbitraryvalue, \arbitraryvalue))$ for some $v'$.
    Hence, by the definition of $T$, $p$ read $(*\currentcellpointershort{}).\nextlong.\view = v'$ at $T$. 
    Thus, by (1), $v' \leq v$.
    Therefore, since $a$ is successful, $(*\currentcellpointershort{}).\nextlong.\view = v'$ at the step before $a$, and since $v' \leq v$, $(*\currentcellpointershort{}).\nextlong.\view \leq v$ at the step before $a$.
    However, this contradicts (2).
    \qH{\Cref{lemma:ero:any_list_attempt_outside_its_window_is_unsuccessful}}
\end{proof}

This implies the following.

\begin{corollary}\label{lemma:ero:any_list_attempt_outside_its_window_is_unsuccessful_alternate_statement}
    Consider a successful list-add or list-remove attempt $a$ in $\mathcal{I}^\mathcal{B}$ and let $e_b$ be its corresponding $L$-event (see \Cref{lemma:ero:l_event_corresponding_to_do_low_level_op}).
    If $P(\mathcal{I}^\mathcal{B})$, $Q(\mathcal{I}^\mathcal{B})$, and $R(\mathcal{I}^\mathcal{B})$ hold, then there are no $L$-events during $(e_b, a)$ in $\mathcal{I}^\mathcal{B}$.
    Equivalently, $e_b$ is the last $L$-event before $a$ in $\mathcal{I}^\mathcal{B}$.
\end{corollary}

We now ``extend" $R$ beyond the last $L$-event.
We do so based on the type of the last $L$-event.

\begin{lemma}\label{lemma:ero:1_of_r_safety_holds}
    Suppose $\mathcal{I}^\mathcal{B}$ has a last $L$-event denoted by $e$, and $P(\mathcal{I}^\mathcal{B})$, $Q(\mathcal{I}^\mathcal{B})$, and $R(\mathcal{I}^\mathcal{B})$ hold.
    If $e$ is an $L$-add event for $\cellpointershort$, then from $e$ onwards in $\mathcal{I}^\mathcal{B}$, there is at most one successful list-add attempt for $\cellpointershort$ and no other successful list-add or list-remove attempts for any pointer.
\end{lemma}

\begin{proof}
    Suppose, for contradiction, $e$ is an $L$-add event for $\cellpointershort$, and at or after $e$ in $\mathcal{I}^\mathcal{B}$ there is either at least two successful list-add attempts for $\cellpointershort$, one or more successful list-add attempts for a pointer other than $\cellpointershort$, or one or more successful list-remove attempts.
    We consider each case separately.
    Note that since $e$ is an $L$-add event for $\cellpointershort$, by \Cref{lemma:ero:every_l_event_is_for_pointer_from_universe}, $\cellpointershort \in \celluniverse{}$.

    \begin{enumerate}
        \item[] \hspace{0pt}\textbf{Case 1.} There are at least two successful list-add attempts for $\cellpointershort$ at or after $e$ in $\mathcal{I}^\mathcal{B}$.

        Let $a_1$ and $a_2$ be the first two successful list-add attempts for $\cellpointershort$ at or after $e$ in $\mathcal{I}^\mathcal{B}$ such that $a_1 < a_2$.
        Since $a_1$ and $a_2$ are in $\mathcal{I}^\mathcal{B}$ and $Q(\mathcal{I}^\mathcal{B})$ holds, by \Cref{lemma:ero:list_add_attempt_for_some_pointer_are_after_same_pointer}, $a_1$ and $a_2$ are after the same $\currentcellpointershort{}$.
        Hence, $a_1$ and $a_2$ are two successful list-add attempts after $\currentcellpointershort{}$ in $\mathcal{I}^\mathcal{B}$ such that $a_1 < a_2$, so by \Cref{lemma:ero:two_successful_list_add_attempts_after_same_pointer_implies_a_removal}, there is a successful list-remove attempt between $\currentcellpointershort{}$ and $\nullconstant$ between $a_1$ and $a_2$ in $\mathcal{I}^\mathcal{B}$.
        Therefore, since $a_1$ is at or after $e$ in $\mathcal{I}^\mathcal{B}$, there is a successful list-remove attempt at or after $e$ in $\mathcal{I}^\mathcal{B}$, so this case reduces to the third case.

        \item[] \hspace{0pt}\textbf{Case 2.} There is a successful list-add attempt $a$ for $\cellpointershort' \neq \cellpointershort$ at or after $e$ in $\mathcal{I}^\mathcal{B}$.

        Hence, by \Cref{def:ero:english}, $a$ was executed during an invocation of the \doaddcell{} procedure whose parameters are $(\arbitraryvalue, \cellpointershort')$, and so by \Cref{lemma:ero:l_event_corresponding_to_do_low_level_op}, $a$'s corresponding event $e_b$ is an $L$-event for $\cellpointershort'$.
        Thus, since $e$ is for $\cellpointershort$ and $\cellpointershort' \neq \cellpointershort$, we have that $e_b \neq e$.
        Therefore, since $e$ is the last $L$-event in $\mathcal{I}^\mathcal{B}$, we have that $e_b < e$, and since $e < a$ (because $e \leq a$ and $e \neq a$ since $e$ is an $L$-event and $a$ is a list-add attempt), we have that $e \in (e_b, a)$.
        However, since $P(\mathcal{I}^\mathcal{B})$, $Q(\mathcal{I}^\mathcal{B})$, and $R(\mathcal{I}^\mathcal{B})$ hold, and $a$ is a successful list-add attempt in $\mathcal{I}^\mathcal{B}$ whose corresponding $L$-event is $e_b$, by \Cref{lemma:ero:any_list_attempt_outside_its_window_is_unsuccessful_alternate_statement}, there are no $L$-events during $(e_b, a)$ in $\mathcal{I}^\mathcal{B}$, a contradiction.

        \item[] \hspace{0pt}\textbf{Case 3.} There is a successful list-remove attempt $a$ at or after $e$ in $\mathcal{I}^\mathcal{B}$.

        By \Cref{def:ero:english}, $a$ was executed during some invocation of the \doremovecell{} procedure, and so by \Cref{lemma:ero:l_event_corresponding_to_do_low_level_op}, $a$'s corresponding event $e_b$ is an $L$-remove event.
        Thus, since $e$ is an $L$-add event, we have that $e_b \neq e$.
        Therefore, since $e$ is the last $L$-event in $\mathcal{I}^\mathcal{B}$, we have that $e_b < e$, and since $e < a$ (because $e \leq a$ and $e \neq a$ since $e$ is an $L$-event and $a$ is a list-remove attempt), we have that $e \in (e_b, a)$.
        However, since $P(\mathcal{I}^\mathcal{B})$, $Q(\mathcal{I}^\mathcal{B})$, and $R(\mathcal{I}^\mathcal{B})$ hold, and $a$ is a successful list-remove attempt in $\mathcal{I}^\mathcal{B}$ whose corresponding $L$-event is $e_b$, by \Cref{lemma:ero:any_list_attempt_outside_its_window_is_unsuccessful_alternate_statement}, there are no $L$-events during $(e_b, a)$ in $\mathcal{I}^\mathcal{B}$, a contradiction.
        \qH{\Cref{lemma:ero:1_of_r_safety_holds}}
    \end{enumerate}
\end{proof}

\begin{lemma}\label{lemma:ero:2_of_r_safety_holds}
    Suppose $\mathcal{I}^\mathcal{B}$ has a last $L$-event denoted as $e$, and $P(\mathcal{I}^\mathcal{B})$, $Q(\mathcal{I}^\mathcal{B})$, and $R(\mathcal{I}^\mathcal{B})$ hold.
    If $e$ is an $L$-apply event, then from $e$ onwards in $\mathcal{I}^\mathcal{B}$ there are no successful list-add or list-remove attempts.  
\end{lemma}

\begin{proof}
    Let $a$ be any list-add or list-remove attempts at or after $e$ in $\mathcal{I}^\mathcal{B}$.
    Hence, by \Cref{def:ero:english}, $a$ was executed during some invocation of the \doaddcell{} or \doremovecell{} procedure.
    Thus, by \Cref{lemma:ero:l_event_corresponding_to_do_low_level_op}, $a$'s corresponding event $e_b$ is either an $L$-add or $L$-remove event.
    So, since $e$ is an $L$-apply event, it follows that $e_b \neq e$.
    Therefore, since $e$ is the last $L$-event in $\mathcal{I}^\mathcal{B}$, we have that $e_b < e$, and since $e < a$ (because $e \leq a$ and $e \neq a$ since $e$ is an $L$-event and $a$ is a list-add or list-remove attempt), we have that $e \in (e_b, a)$.
    However, since $P(\mathcal{I}^\mathcal{B})$, $Q(\mathcal{I}^\mathcal{B})$, and $R(\mathcal{I}^\mathcal{B})$ hold, and $a$ is a successful list-add or list-remove attempt in $\mathcal{I}^\mathcal{B}$ whose corresponding $L$-event is $e_b$, by \Cref{lemma:ero:any_list_attempt_outside_its_window_is_unsuccessful_alternate_statement}, there are no $L$-events during $(e_b, a)$ in $\mathcal{I}^\mathcal{B}$, a contradiction.
    \qH{\Cref{lemma:ero:2_of_r_safety_holds}}
\end{proof}

\begin{lemma}\label{lemma:ero:3_of_r_safety_holds}
    Suppose $\mathcal{I}^\mathcal{B}$ has a last $L$-event denoted as $e$, and $P(\mathcal{I}^\mathcal{B})$, $Q(\mathcal{I}^\mathcal{B})$, and $R(\mathcal{I}^\mathcal{B})$ hold.
    If $e$ is an $L$-remove event for $\cellpointershort$, then from $e$ onwards in $\mathcal{I}^\mathcal{B}$ there is at most one successful list-remove attempt for $\cellpointershort$ and no other successful list-remove or list-add attempt for any pointer.
\end{lemma}

\begin{proof}
    Suppose, for contradiction, $e$ is an $L$-remove event for $\cellpointershort$, and at or after $e$ in $\mathcal{I}^\mathcal{B}$ there is either at least two successful list-remove attempts for $\cellpointershort$, one or more successful list-remove attempts for a pointer other than $\cellpointershort$, or one or more successful list-add attempts.
    We consider each case separately.
    Note that since $e$ is an $L$-remove event for $\cellpointershort$, by \Cref{lemma:ero:every_l_event_is_for_pointer_from_universe}, $\cellpointershort \in \celluniverse{}$.

    \begin{enumerate}
        \item[] \hspace{0pt}\textbf{Case 1.} There are at least two successful list-remove attempts for $\cellpointershort$ at or after $e$ in $\mathcal{I}^\mathcal{B}$.

        Let $a_1$ and $a_2$ be the first two successful list-remove attempts for $\cellpointershort$ at or after $e$ in $\mathcal{I}^\mathcal{B}$ such that $a_1 < a_2$.
        Since $a_1$ and $a_2$ are in $\mathcal{I}^\mathcal{B}$ and $Q(\mathcal{I}^\mathcal{B})$ holds, by \Cref{lemma:ero:list_remove_attempt_for_some_pointer_are_between_same_pointer}, $a_1$ and $a_2$ are successful list-remove attempts for $\cellpointershort$ between the same $\previouscellpointershort{}$ and $\nextcellpointershort{}$, and by \Cref{lemma:ero:every_list_remove_attempt_is_between_a_pointer_from_universe_or_header_and_a_pointer_from_universe_or_null} $\previouscellpointershort{} \in \celluniverse{} \cup \{\&\headobject\}$.
        Furthermore, since $P(\mathcal{I}^\mathcal{B})$ and $Q(\mathcal{I}^\mathcal{B})$ hold, by \Cref{lemma:ero:list_remove_attempt_has_different_next}, $\nextcellpointershort{} \neq \cellpointershort$.
        Since $a_1$ and $a_2$ are both successful list-remove attempts for $\cellpointershort$ between $\previouscellpointershort{}$ and $\nextcellpointershort{}$, and $\previouscellpointershort{} \in \celluniverse{} \cup \{\&\headobject\}$, by \Cref{def:ero:english}, $(*\previouscellpointershort{}).\nextlong.\uniquecellpointercontentlong{} = \cellpointershort$ at the step before both $a_1$ and $a_2$ and $(*\previouscellpointershort{}).\nextlong.\uniquecellpointercontentlong{} = \nextcellpointershort{}$ at $a_1$ and $a_2$.
        Hence, since $\nextcellpointershort \neq \cellpointershort$ and $a_1 < a_2$, the value of $(*\previouscellpointershort{}).\nextlong.\uniquecellpointercontentlong{}$ was set to $\cellpointershort$ during $(a_1, a_2)$.
        Thus, by \Cref{observation:ero:where_objects_change}, there is either a successful list-add attempt for $\cellpointershort$ after $\previouscellpointershort{}$ or a successful list-remove attempt for some $\cellpointershort'$ between $\previouscellpointershort{}$ and $\cellpointershort{}$ during $(a_1, a_2)$.
        Since $a_1$ is at or after $e$ in $\mathcal{I}^\mathcal{B}$, by transitivity, there is either a successful list-add attempt for $\cellpointershort$ or a successful list-remove attempt for $\cellpointershort'$ at or after $e$ in $\mathcal{I}^\mathcal{B}$.
        If the former, then this case reduces to the third.
        If the latter, then by the minimality of $a_2$ $\cellpointershort' \neq \cellpointershort$, so this case reduces to the second.

        \item[] \hspace{0pt}\textbf{Case 2.} There is a successful list-remove attempt $a$ for $\cellpointershort' \neq \cellpointershort$ at or after $e$ in $\mathcal{I}^\mathcal{B}$.

        Hence, by \Cref{def:ero:english}, $a$ was executed during an invocation of the \doremovecell{} procedure whose parameters are $(\arbitraryvalue, \cellpointershort')$, and so by \Cref{lemma:ero:l_event_corresponding_to_do_low_level_op}, $a$'s corresponding event $e_b$ is an $L$-event for $\cellpointershort'$.
        Thus, since $e$ is for $\cellpointershort$ and $\cellpointershort' \neq \cellpointershort$, we have that $e_b \neq e$.
        Therefore, since $e$ is the last $L$-event in $\mathcal{I}^\mathcal{B}$, we have that $e_b < e$, and since $e < a$ (because $e \leq a$ and $e \neq a$ since $e$ is an $L$-event and $a$ is a list-remove attempt), we have that $e \in (e_b, a)$.
        However, since $P(\mathcal{I}^\mathcal{B})$, $Q(\mathcal{I}^\mathcal{B})$, and $R(\mathcal{I}^\mathcal{B})$ hold, and $a$ is a successful list-remove attempt in $\mathcal{I}^\mathcal{B}$ whose corresponding $L$-event is $e_b$, by \Cref{lemma:ero:any_list_attempt_outside_its_window_is_unsuccessful_alternate_statement}, there are no $L$-events during $(e_b, a)$ in $\mathcal{I}^\mathcal{B}$, a contradiction.

        \item[] \hspace{0pt}\textbf{Case 3.} There is a successful list-add attempt $a$ at or after $e$ in $\mathcal{I}^\mathcal{B}$.

        By \Cref{def:ero:english}, $a$ was executed during some invocation of the \doaddcell{} procedure, and so by \Cref{lemma:ero:l_event_corresponding_to_do_low_level_op}, $a$'s corresponding event $e_b$ is an $L$-add event.
        Thus, since $e$ is an $L$-remove event, we have that $e_b \neq e$.
        Therefore, since $e$ is the last $L$-event in $\mathcal{I}^\mathcal{B}$, we have that $e_b < e$, and since $e < a$ (because $e \leq a$ and $e \neq a$ since $e$ is an $L$-event and $a$ is a list-add attempt), we have that $e \in (e_b, a)$.
        However, since $P(\mathcal{I}^\mathcal{B})$, $Q(\mathcal{I}^\mathcal{B})$, and $R(\mathcal{I}^\mathcal{B})$ hold, and $a$ is a successful list-add attempt in $\mathcal{I}^\mathcal{B}$ whose corresponding $L$-event is $e_b$, by \Cref{lemma:ero:any_list_attempt_outside_its_window_is_unsuccessful_alternate_statement}, there are no $L$-events during $(e_b, a)$ in $\mathcal{I}^\mathcal{B}$, a contradiction.
        \qH{\Cref{lemma:ero:3_of_r_safety_holds}}
    \end{enumerate}
\end{proof}

We are now ready to prove that $\List(\mathcal{I})$ is essentially the ``shape" of the list at the end of $\mathcal{I}$.

\begin{lemma}\label{lemma:ero:conditional_classification_lemma}
    For every \Underline{finite} implementation history $\mathcal{I}$ of $\mathcal{B}$, if $P(\mathcal{I})$, $Q(\mathcal{I})$, and $R(\mathcal{I})$ hold, then the list of cells conforms to either one of two sequences in $\mathcal{I}$:
    \begin{compactenum}
        \item if $\mathcal{I}$ has at least one $L$-event, the last $L$-event in $\mathcal{I}$ denoted by $e$ is a $L$-add or $L$-remove event, and from $e$ onwards in $\mathcal{I}$ there are no successful list-add or list-remove attempts, then the list of cells conforms to $\List(\mathcal{I}^{exclude}_e)$ in $\mathcal{I}$ where $\mathcal{I}^{exclude}_e$ is the prefix of $\mathcal{I}$ up to but excluding $e$;
        \item otherwise, the list of cells conforms to $\List(\mathcal{I})$ in $\mathcal{I}$.
    \end{compactenum}
\end{lemma}

\begin{proof}
    Let $\mathcal{P}(n)$ be the predicate: for every implementation history $\mathcal{I}$ of $\mathcal{B}$ of $n$ steps, if $P(\mathcal{I})$, $Q(\mathcal{I})$, and $R(\mathcal{I})$ hold, then the list of cells conforms to either one of two sequences in $\mathcal{I}$ as described in the statement of the lemma.
    We prove $\mathcal{P}(n)$ by induction on $n$.

    \begin{itemize}
        \item[] \hspace{0pt}\textbf{Base Case.} $\mathcal{P}(0)$.

        Let $\mathcal{I}_0$ be an implementation history of $\mathcal{B}$ of zero steps.
        Hence, there are zero $L$-events in $\mathcal{I}_0$.
        Thus, we must prove that the list of cells conforms to $\List(\mathcal{I})$ in $\mathcal{I}_0$.
        Since there are zero $L$-events in $\mathcal{I}_0$, by \Cref{def:ero:logical_list} $\List(\mathcal{I}) = \&\headobject, \nullconstant$, and so we must prove that $\headobject.\nextlong.\uniquecellpointercontentlong{} = \nullconstant$ at the end of $\mathcal{I}_0$.
        Since the end of $\mathcal{I}_0$ is the initial configuration, this follows from the initialization of $\headobject$.

        \item[] \hspace{0pt}\textbf{Inductive Case.} $\forall n\ \mathcal{P}(n) \implies \mathcal{P}(n + 1)$.

        Suppose for some $n \geq 0$ $\mathcal{P}(n)$ holds.
        To prove that $\mathcal{P}(n + 1)$ holds, consider any implementation history $\mathcal{I}_{n + 1}$ of $\mathcal{B}$ of $n + 1$ steps, and suppose that $P(\mathcal{I}_{n + 1})$, $Q(\mathcal{I}_{n + 1})$, and $R(\mathcal{I}_{n + 1})$ hold. 
        Let $s$ be the last step in $\mathcal{I}_{n + 1}$, and let $\mathcal{I}_n$ be the prefix of $\mathcal{I}_{n + 1}$ up to and including the $n$th step, so $s$ is the only step in $\mathcal{I}_{n + 1}$ not in $\mathcal{I}_n$.
        There are two cases.

        \begin{itemize}
            \item[] \hspace{0pt}\textbf{Case 1.} $s$ is not a successful list-add or list-remove attempt.

            We start with a claim.

            \begin{claimcustom}{\ref{lemma:ero:conditional_classification_lemma}.1}\label{lemma:ero:conditional_classification_lemma:zero_claim}
                If the list of cells conforms to $\cellpointershort_0, \ldots, \cellpointershort_{m + 1}$ in $\mathcal{I}_n$, then the list of cells conforms to $\cellpointershort_0, \ldots, \cellpointershort_{m + 1}$ in $\mathcal{I}_{n + 1}$.
            \end{claimcustom}

            \begin{proof}
                Since by assumption the list of cells conforms to $\cellpointershort_0, \ldots, \cellpointershort_{m + 1}$ in $\mathcal{I}_n$, by \Cref{def:ero:logical_list}, at the end of $\mathcal{I}_n$, for all $i\in[0..m]$, $(*\cellpointershort_i).\nextlong.\uniquecellpointercontentlong{}= \cellpointershort_{i+1}$ and $\cellpointershort_i \in \celluniverse{} \cup \{\&\headobject\}$.
                Hence, since $s$ is the only step in $\mathcal{I}_{n + 1}$ not in $\mathcal{I}_n$, $s$ is not a successful list-add or list-remove attempt, and by \Cref{observation:ero:where_objects_change} for all $i\in[0..m]$ $(*\cellpointershort_i).\nextlong.\uniquecellpointercontentlong{}$ only changes as the result of a successful list-add or list-remove attempt, at the end of $\mathcal{I}_{n + 1}$, for all $i\in[0..m]$, $(*\cellpointershort_i).\nextlong.\uniquecellpointercontentlong{}= \cellpointershort_{i+1}$.
                Therefore, by \Cref{def:ero:logical_list}, the list of cells conforms to $\cellpointershort_0, \ldots, \cellpointershort_{m + 1}$ in $\mathcal{I}_{n + 1}$ as wanted.
                \qH{\Cref{lemma:ero:conditional_classification_lemma:zero_claim}}
            \end{proof}
            
            We now return to the proof of Case 1.
            Since $\mathcal{I}_n$ is an implementation history of $\mathcal{B}$ of $n$ steps, and by assumption $\mathcal{P}(n)$ holds, there are two cases.

            \begin{itemize}
                \item[] \hspace{0pt}\textbf{Case 1.1.} $\mathcal{I}_n$ has at least one $L$-event, the last $L$-event in $\mathcal{I}_n$ denoted by $e$ is a $L$-add or $L$-remove event, from $e$ onwards in $\mathcal{I}_n$ there are no successful list-add or list-remove attempts, and the list of cells conforms to $\List(\mathcal{I}^{exclude}_{e})$ in $\mathcal{I}_n$ where $\mathcal{I}^{exclude}_{e}$ is the prefix of $\mathcal{I}_n$ up to but excluding $e$.

                We start with two claims.

                \begin{claimcustom}{\ref{lemma:ero:conditional_classification_lemma}.2}\label{lemma:ero:conditional_classification_lemma:first_claim}
                    $e$ is the last $L$-event in $\mathcal{I}_{n + 1}$.
                \end{claimcustom}

                \begin{proof}
                    Since $\mathcal{I}_n$ is a prefix of $\mathcal{I}_{n + 1}$ and $e$ is in $\mathcal{I}_n$, we have that $e$ is in $\mathcal{I}_{n + 1}$, and so $\mathcal{I}_{n + 1}$ has at least one $L$-event.
                    Now suppose, for contradiction, that $e$ is not the last $L$-event in $\mathcal{I}_{n + 1}$.
                    Hence, since $e$ is in $\mathcal{I}_{n + 1}$, and $s$ is the only step in $\mathcal{I}_{n + 1}$ not in $\mathcal{I}_n$, it follows that $s$ is an $L$-event, and so $e$ and $s$ are successive $L$-events in $\mathcal{I}_{n + 1}$.
                    Thus, since by Case 1.1 $e$ is either an $L$-add or $L$-remove event, by $R(\mathcal{I}_{n + 1})$, there is either a successful list-add attempt or a successful list-remove attempt between $e$ and $s$.
                    Therefore, since $\mathcal{I}_{n}$ is a prefix of $\mathcal{I}_{n + 1}$ up to $s$, we have that from $e$ onwards in $\mathcal{I}_n$ there is either a successful list-add or a successful list-remove attempt.
                    However, by Case 1.1 from $e$ onwards in $\mathcal{I}_n$ there are no successful list-add or list-remove attempts, a contradiction.
                    \qH{\Cref{lemma:ero:conditional_classification_lemma:first_claim}}
                \end{proof}

                \begin{claimcustom}{\ref{lemma:ero:conditional_classification_lemma}.3}\label{lemma:ero:conditional_classification_lemma:case_1_1_reduction}
                    If the list of cells conforms to $\List(\mathcal{I}^{exclude}_{e})$ in $\mathcal{I}_{n + 1}$, then $\mathcal{P}(n + 1)$ holds.
                \end{claimcustom}

                \begin{proof}
                    This follows from three facts.
                    (1) By \Cref{lemma:ero:conditional_classification_lemma:first_claim} $\mathcal{I}_{n + 1}$ has at least one $L$-event.
                    (2) Since by Case 1.1 $e$ is a $L$-add or $L$-remove event, and by \Cref{lemma:ero:conditional_classification_lemma:first_claim} $e$ is the last $L$-event of $\mathcal{I}_{n + 1}$, the last $L$-event in $\mathcal{I}_{n + 1}$ is an $L$-add or $L$-remove event.
                    (3) Since by Case 1.1 from $e$ onwards in $\mathcal{I}_n$ there are no successful list-add or list-remove attempts, by Case 1 $s$ is not a successful list-add or list-remove attempt, and $s$ is the only step in $\mathcal{I}_{n + 1}$ not in $\mathcal{I}_n$, it follows that from $e$ onwards in $\mathcal{I}_{n + 1}$ there are no successful list-add or list-remove attempts.
                    Therefore, $\mathcal{P}(n + 1)$ requires that the list of cells conforms to $\List(\mathcal{I}^{exclude}_{e})$ in $\mathcal{I}_{n + 1}$ as wanted.
                    \qH{\Cref{lemma:ero:conditional_classification_lemma:case_1_1_reduction}}
                \end{proof}

                We now finish the proof of Case 1.1.
                Since by Case 1.1 the list of cells conforms to $\List(\mathcal{I}^{exclude}_{e})$ in $\mathcal{I}_n$, by \Cref{lemma:ero:conditional_classification_lemma:zero_claim}, the list of cells conforms to $\List(\mathcal{I}^{exclude}_{e})$ in $\mathcal{I}_{n + 1}$.
                Therefore, by \Cref{lemma:ero:conditional_classification_lemma:case_1_1_reduction}, $\mathcal{P}(n + 1)$ holds as wanted.

                \item[] \hspace{0pt}\textbf{Case 1.2.} Either (1) $\mathcal{I}_n$ as zero $L$-events; (2) the last $L$-event in $\mathcal{I}_n$ is not an $L$-add or $L$-remove event; or (3) from the last $L$-event in $\mathcal{I}_n$ onwards in $\mathcal{I}_n$, there is a successful list-add or list-remove attempt.
                In any case, the list of cells conforms to $\List(\mathcal{I}_n)$ in $\mathcal{I}_n$.

                The core of the proof of Case 1.2 is the following claim.

                \begin{claimcustom}{\ref{lemma:ero:conditional_classification_lemma}.4}\label{lemma:ero:conditional_classification_lemma:case_1_2_reduction}
                    If the list of cells conforms to $\List(\mathcal{I}_n)$ in $\mathcal{I}_{n + 1}$, then $\mathcal{P}(n + 1)$ holds.
                \end{claimcustom}

                \begin{proof}
                    Since $s$ is either not an $L$-event or it is an $L$-event, and by \Cref{lemma:ero:every_l_event_is_add_apply_or_remove}, if $s$ is an $L$-event then it is either an $L$-add, $L$-apply, or $L$-remove event, it follows that there are three cases.
    
                    \begin{itemize}
                        \item[] \hspace{0pt}\textbf{Case A.} $s$ is not an $L$-event.
    
                        We first reduce the task of proving $\mathcal{P}(n + 1)$ to proving that the list of cells conforms to $\List(\mathcal{I}_{n + 1})$ in $\mathcal{I}_{n + 1}$.\footnote{Notice that here we are referring to $\List(\mathcal{I}_{n + 1})$ whereas the claim refers to $\List(\mathcal{I}_{n})$.}
                        First, suppose (1) is true.
                        Since $\mathcal{I}_n$ has zero $L$-events and $s$ is the only step in $\mathcal{I}_{n + 1}$ not in $\mathcal{I}_n$, there are zero $L$-events in $\mathcal{I}_{n + 1}$.
                        Thus, $\mathcal{P}(n + 1)$ requires that the list of cells conforms to $\List(\mathcal{I}_{n + 1})$ in $\mathcal{I}_{n + 1}$.
                        Now suppose (2) is true.
                        Since $s$ is the only step in $\mathcal{I}_{n + 1}$ not in $\mathcal{I}_n$ and $s$ is not an $L$-event, the last $L$-event is the same in $\mathcal{I}_n$ and $\mathcal{I}_{n + 1}$, and since the last $L$-event in $\mathcal{I}_n$ is not an $L$-add or $L$-remove event, the last $L$-event in $\mathcal{I}_{n + 1}$ is not an $L$-add or $L$-remove event.
                        Thus, $\mathcal{P}(n + 1)$ requires that the list of cells conforms to $\List(\mathcal{I}_{n + 1})$ in $\mathcal{I}_{n + 1}$.
                        Finally, suppose (3) is true.
                        By the same argument above, the last $L$-event is the same in $\mathcal{I}_n$ and $\mathcal{I}_{n + 1}$, and since from the last $L$-event in $\mathcal{I}_n$ onwards in $\mathcal{I}_n$ there is a successful list-add or list-remove attempt, it follows that from the last $L$-event in $\mathcal{I}_{n + 1}$ onwards in $\mathcal{I}_{n + 1}$ there is a successful list-add or list-remove attempt.
                        Thus, $\mathcal{P}(n + 1)$ requires that the list of cells conforms to $\List(\mathcal{I}_{n + 1})$ in $\mathcal{I}_{n + 1}$.
                        Therefore, in all cases, $\mathcal{P}(n + 1)$ requires that the list of cells conforms to $\List(\mathcal{I}_{n + 1})$ in $\mathcal{I}_{n + 1}$.
                        
                        Since $s$ is the only step in $\mathcal{I}_{n + 1}$ not in $\mathcal{I}_n$ and $s$ is not an $L$-event, by \Cref{def:ero:logical_list}, $\List(\mathcal{I}_n) = \List(\mathcal{I}_{n + 1})$, so if the list of cells conforms to $\List(\mathcal{I}_{n})$ in $\mathcal{I}_{n + 1}$, then $\mathcal{P}(n + 1)$ holds as wanted.
                        
                        \item[] \hspace{0pt}\textbf{Case B.} $s$ is an $L$-add or $L$-remove event.
    
                        Hence, $\mathcal{I}_{n + 1}$ has at least one $L$-event, the last $L$-event in $\mathcal{I}_{n + 1}$, $s$, is an $L$-add or $L$-remove event, and from $s$ onwards in $\mathcal{I}_{n + 1}$ there are no successful list-add or list-remove attempts, so $\mathcal{P}(n + 1)$ requires that the list of cells conforms to $\List(\mathcal{I}^{exclude}_s)$ in $\mathcal{I}_{n + 1}$ where $\mathcal{I}^{exclude}_s$ is the prefix of $\mathcal{I}_{n + 1}$ up to but excluding $s$.
                        Since $s$ is the only step in $\mathcal{I}_{n + 1}$ not in $\mathcal{I}_n$, we have that $\mathcal{I}^{exclude}_s = \mathcal{I}_n$, and so if the list of cells conforms to $\List(\mathcal{I}_{n})$ in $\mathcal{I}_{n + 1}$, then $\mathcal{P}(n + 1)$ holds as wanted.
    
                        \item[] \hspace{0pt}\textbf{Case C.} $s$ is an $L$-apply event.
    
                        Hence, the last $L$-event in $\mathcal{I}_{n + 1}$, $s$, is not an $L$-add or $L$-remove event, so $\mathcal{P}(n + 1)$ requires that the list of cells conforms to $\List(\mathcal{I}_{n + 1})$ in $\mathcal{I}_{n + 1}$.
                        Since $s$ is an $L$-apply event, by \Cref{def:ero:logical_list}, $\List(\mathcal{I}^{exclude}_s) = \List(\mathcal{I}_{n + 1})$  where $\mathcal{I}^{exclude}_s$ is the prefix of $\mathcal{I}_{n + 1}$ up to but excluding $s$, so $\mathcal{P}(n + 1)$ requires that the list of cells conforms to $\List(\mathcal{I}^{exclude}_s)$ in $\mathcal{I}_{n + 1}$.
                        Since $s$ is the only step in $\mathcal{I}_{n + 1}$ not in $\mathcal{I}_n$, we have that $\mathcal{I}^{exclude}_s = \mathcal{I}_n$, and so if the list of cells conforms to $\List(\mathcal{I}_{n})$ in $\mathcal{I}_{n + 1}$, then $\mathcal{P}(n + 1)$ holds as wanted.
                        \qH{\Cref{lemma:ero:conditional_classification_lemma:case_1_2_reduction}}
                    \end{itemize}
                \end{proof}

                We now finish the proof of Case 1.2.
                Since by Case 1.2 the list of cells conforms to $\List(\mathcal{I}_n)$ in $\mathcal{I}_n$, by \Cref{lemma:ero:conditional_classification_lemma:zero_claim}, the list of cells conforms to $\List(\mathcal{I}_n)$ in $\mathcal{I}_{n + 1}$.
                Therefore, by \Cref{lemma:ero:conditional_classification_lemma:case_1_2_reduction}, $\mathcal{P}(n + 1)$ holds as wanted.
            \end{itemize}

            \item[] \hspace{0pt}\textbf{Case 2.} $s$ is a successful list-add or list-remove attempt.

            We start with a few claims.

            \begin{claimcustom}{\ref{lemma:ero:conditional_classification_lemma}.5}\label{lemma:ero:conditional_classification_lemma:third_claim}
                Let $e$ be the corresponding $L$-event of $s$.
                Then, $e$ is the last $L$-event in $\mathcal{I}_{n + 1}$.
            \end{claimcustom}

            \begin{proof}
                Since $s$ is a successful list-add or list-remove attempt in $\mathcal{I}_{n + 1}$, and by assumption $P(\mathcal{I}_{n + 1})$, $Q(\mathcal{I}_{n + 1})$, $R(\mathcal{I}_{n + 1})$ hold, by \Cref{lemma:ero:any_list_attempt_outside_its_window_is_unsuccessful_alternate_statement}, $e$ is the last $L$-event before $s$ in $\mathcal{I}_{n + 1}$.
                Therefore, since $s$ is the last step in $\mathcal{I}_{n + 1}$ and is not an $L$-event, the claim follows.
                \qH{\Cref{lemma:ero:conditional_classification_lemma:third_claim}}
            \end{proof}

            \begin{claimcustom}{\ref{lemma:ero:conditional_classification_lemma}.6}\label{lemma:ero:conditional_classification_lemma:case_2_reduction}
                If the list of cells conforms to $\List(\mathcal{I}_{n + 1})$ in $\mathcal{I}_{n + 1}$, then $\mathcal{P}(n + 1)$ holds.
            \end{claimcustom}

            \begin{proof}
                Since $s$ is the last step of $\mathcal{I}_{n + 1}$, $s$ is a successful list-add or list-remove attempt, and by \Cref{lemma:ero:conditional_classification_lemma:third_claim} $e$ is a last $L$-event in $\mathcal{I}_{n + 1}$, from $e$ onwards in $\mathcal{I}_{n + 1}$ there is a successful list-add or list-remove attempt.
                Therefore, the claim follows from the definition of $\mathcal{P}(n + 1)$.
                \qH{\Cref{lemma:ero:conditional_classification_lemma:case_2_reduction}}
            \end{proof}

            \begin{claimcustom}{\ref{lemma:ero:conditional_classification_lemma}.7}\label{lemma:ero:conditional_classification_lemma:second_claim}
                The following properties hold regarding $\mathcal{I}_n$.
                \begin{compactenum}
                    \item $\mathcal{I}_n$ has at least one $L$-event.
                    \item $e$ is the last $L$-event in $\mathcal{I}_n$ and it is an $L$-add or $L$-remove event.
                    \item From $e$ onwards in $\mathcal{I}_n$ there are no successful list-add or list-remove attempts.
                    \item The list of cells conforms to $\List(\mathcal{I}^{exclude}_e)$ in $\mathcal{I}_n$ where $\mathcal{I}^{exclude}_e$ is the prefix of $\mathcal{I}_n$ up to but excluding $e$.
                \end{compactenum}
            \end{claimcustom}

            \begin{proof}
                First 1.
                Since $e$ is the corresponding $L$-event of $s$, by \Cref{lemma:ero:l_event_corresponding_to_do_low_level_op}, $e < s$.
                Hence, since $s$ is the only step in $\mathcal{I}_{n + 1}$ not in $\mathcal{I}_n$, we have that $e$ is in $\mathcal{I}_n$.
                Therefore, $\mathcal{I}_n$ has at least one $L$-event.

                Now 2.
                Since by 1 $e$ is in $\mathcal{I}_n$, by \Cref{lemma:ero:conditional_classification_lemma:third_claim} $e$ is the last $L$-event in $\mathcal{I}_{n + 1}$, and $\mathcal{I}_n$ is a prefix of $\mathcal{I}_{n + 1}$, we have that $e$ is the last $L$-event in $\mathcal{I}_n$.
                Furthermore, since $e$ is the corresponding $L$-event of $s$ and $s$ is a successful list-add or list-remove attempt, by \Cref{lemma:ero:l_event_corresponding_to_do_low_level_op}, $e$ is either an $L$-add or $L$-remove event.

                Now 3.
                Since by \Cref{lemma:ero:conditional_classification_lemma:third_claim} $e$ is the last $L$-event in $\mathcal{I}_{n + 1}$, the last step of $\mathcal{I}_{n + 1}$ is a successful list-add or list-remove attempt, and $P(\mathcal{I}_{n + 1})$, $Q(\mathcal{I}_{n + 1})$, and $R(\mathcal{I}_{n + 1})$ hold, by Lemmas \ref{lemma:ero:1_of_r_safety_holds} and \ref{lemma:ero:3_of_r_safety_holds}, $s$ is the only successful list-add or list-remove attempt from $e$ onwards in $\mathcal{I}_{n + 1}$.
                Therefore, since $s$ is the only step in $\mathcal{I}_{n + 1}$ not in $\mathcal{I}_n$, it follows that from $e$ onwards in $\mathcal{I}_n$ there are no successful list-add or list-remove attempts.

                Lastly 4.
                Since $\mathcal{P}(n)$ holds and $\mathcal{I}_n$ is a finite implementation history of $\mathcal{B}$ of $n$ steps, by 1, 2, and 3, the list of cells conforms to $\List(\mathcal{I}^{exclude}_e)$ in $\mathcal{I}_n$ as wanted.
                \qH{\Cref{lemma:ero:conditional_classification_lemma:second_claim}}
            \end{proof}

            We now return to the proof of Case 2.
            By \Cref{lemma:ero:conditional_classification_lemma:case_2_reduction}, it suffices to prove that the list of cells conforms to $\List(\mathcal{I}_{n + 1})$ in $\mathcal{I}_{n + 1}$.
            Let $\List(\mathcal{I}^{exclude}_{e}) = \cellpointershort_0, \ldots, \cellpointershort_m + 1$ for some $m$.
            Since $P(\mathcal{I}_{n + 1})$ holds, by \Cref{lemma:ero:every_list_sequence_is_from_universe}, $\cellpointershort_0 = \&\headobject$, for every $i \in [1..m]\ \cellpointershort_i \in \celluniverse{}$, and $\cellpointershort_{m + 1} = \nullconstant$.
            Also, note that since $\mathcal{I}^{exclude}_e$ is the prefix of $\mathcal{I}_n$ up to but excluding $e$, and $\mathcal{I}_n$ is a prefix of $\mathcal{I}_{n + 1}$, we have that $\mathcal{I}^{exclude}_e$ is also the prefix of $\mathcal{I}_{n + 1}$ up to but excluding $e$.
  
            \begin{itemize}
            
                \item[] \hspace{0pt}\textbf{Case 2.1.} $s$ is a list-add attempt for some $\cellpointershort$ after some $\currentcellpointershort{}$.

                Hence, by \Cref{lemma:ero:every_list_add_seal_and_remove_attempt_is_for_ptr_from_universe} $\cellpointershort \in \celluniverse{}$ and by \Cref{lemma:ero:every_list_add_attempt_is_after_a_pointer_from_universe_or_head} $\currentcellpointershort{} \in \celluniverse{} \cup \{\&\headobject\}$.
                Thus, since $s$ is a list-add attempt for $\cellpointershort$, by \Cref{def:ero:english}, $s$ was executed during an invocation of the \doaddcell{} procedure whose second parameter was $\cellpointershort$, so by \Cref{lemma:ero:l_event_corresponding_to_do_low_level_op}, its corresponding $L$-event is before $s$ and is an $L$-add event for $\cellpointershort$.
                So, since by \Cref{lemma:ero:conditional_classification_lemma:third_claim} $e$ is the corresponding $L$-event of $s$, we have that $e$ is an $L$-add event for $\cellpointershort$.
                Therefore, since $e < s$, by $Q(\mathcal{I}_{n + 1})$, $e$ is the unique $L$-add event that precedes $s$, and since $\mathcal{I}^{exclude}_{e}$ is the prefix of $\mathcal{I}_{n + 1}$ up to but excluding $e$, $\currentcellpointershort{}$ is the second last pointer in $\List(\mathcal{I}^{exclude}_{e})$.
                Since $\List(\mathcal{I}^{exclude}_{e}) = \cellpointershort_0, \ldots, \cellpointershort_m + 1$, we have that $\currentcellpointershort{} = \cellpointershort_m$, so $s$ is a successful list-add attempt for $\cellpointershort$ after $\cellpointershort_m$.
                Furthermore, since by \Cref{lemma:ero:conditional_classification_lemma:third_claim} $e$ is the last $L$-event in $\mathcal{I}_{n + 1}$, and $\mathcal{I}^{exclude}_{e}$ is the prefix of $\mathcal{I}_{n + 1}$ up to but excluding $e$, we have that $e$ is the only $L$-event in $\mathcal{I}_{n + 1}$ not in $\mathcal{I}^{exclude}_{e}$, and $e$ is after all $L$-events in $\mathcal{I}^{exclude}_{e}$.
                Hence, since $e$ is an $L$-add event for $\cellpointershort$, and $\List(\mathcal{I}^{exclude}_{e}) = \&\headobject, \cellpointershort_1, \ldots, \cellpointershort_m, \nullconstant$, by \Cref{def:ero:logical_list}, $\List(\mathcal{I}_{n + 1}) = \&\headobject, \cellpointershort_1, \ldots, \cellpointershort_m, \cellpointershort, \nullconstant$.
                Therefore, to prove that the list of cells conforms to $\List(\mathcal{I}_{n + 1})$ in $\mathcal{I}_{n + 1}$, by \Cref{def:ero:logical_list}, we must prove that the following properties hold at the end of $\mathcal{I}_{n + 1}$: (1) for all $i \in [0..m)$ $(*\cellpointershort_i).\nextlong.\uniquecellpointercontentlong{} = \cellpointershort_{i + 1}$; (2) $(*\cellpointershort_m).\nextlong.\uniquecellpointercontentlong{} = \cellpointershort$; and (3) $(*\cellpointershort).\nextlong.\uniquecellpointercontentlong{} = \nullconstant$.
                The following claims prove these properties, completing Case 2.1.

                \begin{claimcustom}{\ref{lemma:ero:conditional_classification_lemma}.8}\label{lemma:ero:conditional_classification_lemma:fourth_claim}
                    For all $i \in [0..m)$, $(*\cellpointershort_i).\nextlong.\uniquecellpointercontentlong{} = \cellpointershort_{i + 1}$ at the end of $\mathcal{I}_{n + 1}$.
                \end{claimcustom}

                \begin{proof}
                    Since by \Cref{lemma:ero:conditional_classification_lemma:second_claim} the list of cells conforms to $\List(\mathcal{I}^{exclude}_{e})$ in $\mathcal{I}_n$, and $\List(\mathcal{I}^{exclude}_{e}) = \cellpointershort_0, \ldots, \cellpointershort_{m + 1}$, by \Cref{def:ero:logical_list}, at the end of $\mathcal{I}_n$, for all $i\in[0..m)$, $(*\cellpointershort_i).\nextlong.\uniquecellpointercontentlong{}= \cellpointershort_{i+1}$.
                    Hence, since $s$ is the only step in $\mathcal{I}_{n + 1}$ not in $\mathcal{I}_n$, it suffices to show that $s$ does not change the value of $(*\cellpointershort_i).\nextlong.\uniquecellpointercontentlong{}$ for all $i \in [0..m)$.
                    This follows from three facts: (1) $s$ is a successful list-add attempt for $\cellpointershort$ after $\cellpointershort_m$; (2) since for all $i\in[0..m)$ $\cellpointershort_i \in \celluniverse{} \cup \{\&\headobject\}$, by \Cref{observation:ero:where_objects_change}, $(*\cellpointershort_i).\nextlong.\uniquecellpointercontentlong{}$ only changes as the result of a successful list-add attempt after $\cellpointershort_i$ or list-remove attempt between $\cellpointershort_i$ and some pointer; and (3) since $\mathcal{I}^{exclude}_{e}$ is a prefix of $\mathcal{I}_{n + 1}$, $\List(\mathcal{I}^{exclude}_{e}) = \cellpointershort_0, \ldots, \cellpointershort_{m + 1}$, $P(\mathcal{I}_{n + 1})$ holds, and for all $i \in [0..m)$ $i \neq m$, by \Cref{lemma:ero:pointers_in_list_are_unique}, $\cellpointershort_i \neq \cellpointershort_m$.
                    \qH{\Cref{lemma:ero:conditional_classification_lemma:fourth_claim}}
                \end{proof}

                \begin{claimcustom}{\ref{lemma:ero:conditional_classification_lemma}.9}\label{lemma:ero:conditional_classification_lemma:fifth_claim}
                    $(*\cellpointershort_m).\nextlong.\uniquecellpointercontentlong{} = \cellpointershort$ at the end of $\mathcal{I}_{n + 1}$.
                \end{claimcustom}

                \begin{proof}
                    $s$ is a successful list-add attempt for $\cellpointershort$ after $\cellpointershort_m$.
                    \qH{\Cref{lemma:ero:conditional_classification_lemma:fifth_claim}}
                \end{proof}

                \begin{claimcustom}{\ref{lemma:ero:conditional_classification_lemma}.10}\label{lemma:ero:conditional_classification_lemma:sixth_claim}
                    $(*\cellpointershort).\nextlong.\uniquecellpointercontentlong{} = \nullconstant$ at the end of $\mathcal{I}_{n + 1}$.
                \end{claimcustom}

                \begin{proof}
                    Suppose, for contradiction, $(*\cellpointershort).\nextlong.\uniquecellpointercontentlong{} \neq \nullconstant$ at the end of $\mathcal{I}_{n + 1}$.
                    Hence, since $\cellpointershort \in \celluniverse{}$,  $(*\cellpointershort).\nextlong.\uniquecellpointercontentlong{}$ is initialized to $\nullconstant$, and so it was changed in $\mathcal{I}_{n + 1}$.
                    Thus, by \Cref{observation:ero:where_objects_change}, there is a successful list-add attempt after $\cellpointershort$ or a successful list-remove attempt between $\cellpointershort$ and some pointer in $\mathcal{I}_{n + 1}$.
                    Let $a$ be this successful list attempt.
                    Hence, by $Q(\mathcal{I}_{n + 1})$, $\cellpointershort\in\List(\mathcal{I}^{exclude}_{e'})$ where $e'$ is the unique $L$-event preceding $a$ in $\mathcal{I}_{n + 1}$ for the same pointer as $a$, and $\mathcal{I}^{exclude}_{e'}$ is the prefix of $\mathcal{I}_{n + 1}$ up to but excluding $e'$.
                    Thus, since by \Cref{lemma:ero:conditional_classification_lemma:third_claim} $e$ is the last $L$-event in $\mathcal{I}_{n + 1}$, we have that $e' \leq e$.
                    Furthermore, since $\cellpointershort \in \celluniverse{}$, by \Cref{assumption:ero:head_and_null_not_in_cell_universe}, $\cellpointershort \neq \&\headobject$ and $\cellpointershort \neq \nullconstant$, and so since $\cellpointershort\in\List(\mathcal{I}^{exclude}_{e'})$, by \Cref{def:ero:logical_list}, there is a $L$-add event for $\cellpointershort$ in $\mathcal{I}^{exclude}_{e'}$.
                    Hence, since $\mathcal{I}^{exclude}_{e'}$ is the prefix of $\mathcal{I}_{n + 1}$ up to but excluding $e'$, there is a $L$-add event for $\cellpointershort$ before $e'$ in $\mathcal{I}_{n + 1}$.
                    Thus, since $e' \leq e$, there is a $L$-add event for $\cellpointershort$ before $e$ in $\mathcal{I}_{n + 1}$.
                    Therefore, since $e$ is an $L$-add for $\cellpointershort$, it follows that there are two $L$-add events for $\cellpointershort$ in $\mathcal{I}_{n + 1}$.
                    However, by $P(\mathcal{I}_{n + 1})$, there is at most one $L$-add event for $\cellpointershort$ in $\mathcal{I}_{n + 1}$, a contradiction.
                    \qH{\Cref{lemma:ero:conditional_classification_lemma:sixth_claim}}
                \end{proof}

                \item[] \hspace{0pt}\textbf{Case 2.2.} $s$ is a list-remove attempt for some $\cellpointershort$ between some $\previouscellpointershort{}$ and some $\nextcellpointershort{}$.

                Hence, by \Cref{lemma:ero:every_list_add_seal_and_remove_attempt_is_for_ptr_from_universe} $\cellpointershort \in \celluniverse{}$ and by \Cref{lemma:ero:every_list_remove_attempt_is_between_a_pointer_from_universe_or_header_and_a_pointer_from_universe_or_null} $\previouscellpointershort{} \in \celluniverse{} \cup \{\&\headobject\}$ and $\nextcellpointershort{} \in \celluniverse{} \cup \{\nullconstant\}$.
                Thus, since $s$ is a list-remove attempt for $\cellpointershort$, by \Cref{def:ero:english}, $s$ was executed during an invocation of the \doremovecell{} procedure whose second parameter was $\cellpointershort$, so by \Cref{lemma:ero:l_event_corresponding_to_do_low_level_op}, its corresponding $L$-event is before $s$ and is an $L$-remove event for $\cellpointershort$.
                So, since by \Cref{lemma:ero:conditional_classification_lemma:third_claim} $e$ is the corresponding $L$-event of $s$, we have that $e$ is an $L$-remove event for $\cellpointershort$.
                Therefore, since $e < s$, by $Q(\mathcal{I}_{n + 1})$, $e$ is the unique $L$-remove event that precedes $s$, and since $\mathcal{I}^{exclude}_{e}$ is the prefix of $\mathcal{I}_{n + 1}$ up to but excluding $e$, $\cellpointershort$ is in $\List(\mathcal{I}^{exclude}_{e})$ exactly once and $\previouscellpointershort{}$ and $\nextcellpointershort{}$ are the pointers preceding and succeeding $\cellpointershort$ in $\List(\mathcal{I}^{exclude}_{e})$.
                Since $\cellpointershort \in \celluniverse{}$, by \Cref{assumption:ero:head_and_null_not_in_cell_universe} $\cellpointershort \neq \&\headobject$ and $\cellpointershort \neq \nullconstant$, and so since $\List(\mathcal{I}^{exclude}_{e}) = \&\headobject, \cellpointershort_1, \ldots, \cellpointershort_m, \nullconstant$, we have that $\cellpointershort = \cellpointershort_i$ for a unique $i \in [1..m]$, $\previouscellpointershort{} = \cellpointershort_{i - 1}$, and $\nextcellpointershort{} = \cellpointershort_{i + 1}$, so $s$ is a successful list-remove attempt for $\cellpointershort_i$ between $\cellpointershort_{i - 1}$ and $\cellpointershort_{i + 1}$.
                Furthermore, since by \Cref{lemma:ero:conditional_classification_lemma:third_claim} $e$ is the last $L$-event in $\mathcal{I}_{n + 1}$, and $\mathcal{I}^{exclude}_{e}$ is the prefix of $\mathcal{I}_{n + 1}$ up to but excluding $e$, we have that $e$ is the only $L$-event in $\mathcal{I}_{n + 1}$ not in $\mathcal{I}^{exclude}_{e}$, and $e$ is after all $L$-events in $\mathcal{I}^{exclude}_{e}$.
                Hence, since $e$ is an $L$-remove event for $\cellpointershort$, $\cellpointershort = \cellpointershort_i$ for a unique $i \in [1..m]$, and $\List(\mathcal{I}^{exclude}_{e}) = \&\headobject, \cellpointershort_1, \ldots, \cellpointershort_m, \nullconstant$, by \Cref{def:ero:logical_list}, $\List(\mathcal{I}_{n + 1}) = \&\headobject, \cellpointershort_1, \ldots, \cellpointershort_{i - 1}, \cellpointershort_{i + 1}, \ldots, \cellpointershort_m, \nullconstant$.
                Therefore, to prove that the list of cells conforms to $\List(\mathcal{I}_{n + 1})$ in $\mathcal{I}_{n + 1}$, by \Cref{def:ero:logical_list}, we must prove that the following hold at the end of $\mathcal{I}_{n + 1}$: (1) for all $j \in [0..i-1) \cup [i+1..m]$, $(*\cellpointershort_j).\nextlong.\uniquecellpointercontentlong{} = \cellpointershort_{j + 1}$; and (2) $(*\cellpointershort_{i - 1}).\nextlong.\uniquecellpointercontentlong{} = \cellpointershort_{i + 1}$.
                The following claims prove this, completing Case 2.2.

                \begin{claimcustom}{\ref{lemma:ero:conditional_classification_lemma}.11}\label{lemma:ero:conditional_classification_lemma:seventh_claim}
                    For all $j \in [0..i-1) \cup [i+1..m]$, $(*\cellpointershort_j).\nextlong.\uniquecellpointercontentlong{} = \cellpointershort_{j + 1}$ at the end of $\mathcal{I}_{n + 1}$.
                \end{claimcustom}

                \begin{proof}
                    Since by \Cref{lemma:ero:conditional_classification_lemma:second_claim} the list of cells conforms to $\List(\mathcal{I}^{exclude}_{e})$ in $\mathcal{I}_n$, and $\List(\mathcal{I}^{exclude}_{e}) = \cellpointershort_0, \ldots, \cellpointershort_{m + 1}$, by \Cref{def:ero:logical_list}, for all $j\in[0..i-1) \cup [i+1..m]$ $(*\cellpointershort_j).\nextlong.\uniquecellpointercontentlong{}=\cellpointershort_{j+1}$ at the end of $\mathcal{I}_n$.
                    Hence, since $s$ is the only step in $\mathcal{I}_{n + 1}$ not in $\mathcal{I}_n$, it suffices to show that $s$ does not change $(*\cellpointershort_j).\nextlong.\uniquecellpointercontentlong{}$ for all $j \in [0..i-1) \cup [i+1..m]$.
                    This follows from three facts: (1) $s$ is a successful list-remove attempt for $\cellpointershort_i$ between $\cellpointershort_{i - 1}$ and $\cellpointershort_{i + 1}$; (2) since for all $j \in [0..i-1) \cup [i+1..m]$ $\cellpointershort_j \in \celluniverse{} \cup \{\&\headobject\}$, by \Cref{observation:ero:where_objects_change}, $(*\cellpointershort_j).\nextlong.\uniquecellpointercontentlong{}$ only changes as the result of a successful list-add for some pointer after $\cellpointershort_j$ or list-remove attempt for some pointer between $\cellpointershort_j$ and some pointer; and (3) since $\mathcal{I}^{exclude}_{e}$ is a prefix of $\mathcal{I}_{n + 1}$, $\List(\mathcal{I}^{exclude}_{e}) = \cellpointershort_0, \ldots, \cellpointershort_{m + 1}$, $P(\mathcal{I}_{n + 1})$ holds, and for all $j \in [0..i-1) \cup [i+1..m]$ $j \neq i-1$, by \Cref{lemma:ero:pointers_in_list_are_unique}, $\cellpointershort_j \neq \cellpointershort_{i - 1}$.
                    \qH{\Cref{lemma:ero:conditional_classification_lemma:seventh_claim}}
                \end{proof}

                \begin{claimcustom}{\ref{lemma:ero:conditional_classification_lemma}.12}\label{lemma:ero:conditional_classification_lemma:eigth_claim}
                    $(*\cellpointershort_{i - 1}).\nextlong.\uniquecellpointercontentlong{} = \cellpointershort_{i + 1}$ at the end of $\mathcal{I}_{n + 1}$.
                \end{claimcustom}

                \begin{proof}
                    $s$ is a successful list-remove attempt between $\cellpointershort_{i - 1}$ and $\cellpointershort_{i + 1}$.
                    \qH{\Cref{lemma:ero:conditional_classification_lemma:eigth_claim}}
                \end{proof}
            \end{itemize}
        \end{itemize}
    \end{itemize}
    \qH{\Cref{lemma:ero:conditional_classification_lemma}}
\end{proof}

\subsubsection{Each successful $S$-attempt is immediately after its corresponding $L$-event}

In the last section, we proved, roughly speaking, that successful list-attempts are immediately after their corresponding $L$-event.
We now prove analogous facts for $S$-attempts.
These facts are useful because they let us prove the invariant $O$, and subsequently define the linearization points for~$\mathcal{B}$.

\begin{lemma}\label{lemma:ero:any_s_attempt_outside_its_window_is_unsuccessful}
    Consider an $S$-attempt $a$ in $\mathcal{I}^\mathcal{B}$ and let $e_b$ be its corresponding $L$-event (see \Cref{lemma:ero:l_event_corresponding_to_do_low_level_op}).
    Suppose there is a $L$-event after $e_b$ in $\mathcal{I}^\mathcal{B}$ and that $P(\mathcal{I}^\mathcal{B})$ and $O(\mathcal{I}^\mathcal{B})$ holds.
    Let $e_a$ be the next $L$-event after $e_b$ in $\mathcal{I}^\mathcal{B}$.
    If $e_a < a$, then $a$ is unsuccessful. 
\end{lemma}

\begin{proof}
    Suppose, for contradiction, $e_a < a$ and $a$ is successful.
    Without loss of generality, suppose $a$ is the first such $S$-attempt in $\mathcal{I}^\mathcal{B}$.
    More precisely, for every $S$-attempt $a'$ before $a$ in $\mathcal{I}^\mathcal{B}$, if $e'_b$ is $a'$'s corresponding $L$-event, $e'_a$ is the next $L$-event after $e'_b$ in $\mathcal{I}^\mathcal{B}$, and $e'_a < a'$, then $a'$ is unsuccessful.
    Let $p$ be the process that executed $a$, and let $T^{\ref{line:ero:state_read}}$ be the last time $p$ executed \cref{line:ero:state_read} before $a$.
    Hence, by \Cref{def:ero:english}, $p$ executed $T^{\ref{line:ero:state_read}}$ and $a$ during the same invocation $I$ of the \doapplyandcopyresponse{} procedure.
    Let the first parameter of $I$ be $\uniquerepositoryoperationshort{}_\linearizationobject{}$.
    Hence, since $e_b$ is $a$'s corresponding $L$-event, by \Cref{lemma:ero:l_event_corresponding_to_do_low_level_op}, $e_b$ set $\linearizationobject{}.\uniquerepositoryoperationlong{} = \uniquerepositoryoperationshort{}_\linearizationobject{}$, $e_b$ is an $L$-apply event, and $e_b$ is before $I$ was invoked.
    Thus, by $O(\mathcal{I}^\mathcal{B})$, there is exactly one successful $S$-attempt between $e_b$ and $e_a$; say $a'$.
    We now prove two simple facts about $T^{\ref{line:ero:state_read}}$ and $a'$.
    
    \begin{claimcustom}{\ref{lemma:ero:any_s_attempt_outside_its_window_is_unsuccessful}.1}\label{lemma:ero:any_s_attempt_outside_its_window_is_unsuccessful:claim_one}
        $T^{\ref{line:ero:state_read}} \in (e_b, e_a)$.
    \end{claimcustom}

    \begin{proof}
        Since $e_b$ is before $I$ was invoked and $T^{\ref{line:ero:state_read}}$ is during $I$, we have that $e_b < T^{\ref{line:ero:state_read}}$, so it suffices to prove that $T^{\ref{line:ero:state_read}} < e_a$.
        Suppose, for contradiction, that $e_a < T^{\ref{line:ero:state_read}}$.
        Let $T^{\ref{line:ero:state_linearization_check}}$ be the time of $p$'s execution of \cref{line:ero:state_linearization_check} between $T^{\ref{line:ero:state_read}}$ and $a$.
        Since $e_b < e_a$, $e_a < T^{\ref{line:ero:state_read}}$, and $T^{\ref{line:ero:state_read}} < T^{\ref{line:ero:state_linearization_check}}$, by transitivity, $e_b < e_a < T^{\ref{line:ero:state_linearization_check}}$.
        Furthermore, since $p$ executed $a$, we have that $p$ found the condition on \cref{line:ero:state_linearization_check} to be false at $T^{\ref{line:ero:state_linearization_check}}$.
        Hence, since the first parameter of $I$ is $\uniquerepositoryoperationshort{}_\linearizationobject{}$, we have that $\linearizationobject{}.\uniquerepositoryoperationlong{} = \uniquerepositoryoperationshort{}_\linearizationobject{}$ at $T^{\ref{line:ero:state_linearization_check}}$.
        Thus, since $e_b$ set $\linearizationobject{}.\uniquerepositoryoperationlong{} = \uniquerepositoryoperationshort{}_\linearizationobject{}$, the value of $\linearizationobject{}.\uniquerepositoryoperationlong{}$ is the same at $e_b$ and $T^{\ref{line:ero:state_linearization_check}}$.
        So, since $e_b < e_a < T^{\ref{line:ero:state_linearization_check}}$, it follows that that $\linearizationobject{}.\uniquerepositoryoperationlong{}$ was set to $\uniquerepositoryoperationshort{}_\linearizationobject{}$ during $(e_b, T^{\ref{line:ero:state_linearization_check}})$, and thus by \Cref{observation:ero:where_objects_change}, there is an $L$-event during $(e_b, T^{\ref{line:ero:state_linearization_check}})$ that set $\linearizationobject{}.\uniquerepositoryoperationlong{} = \uniquerepositoryoperationshort{}_\linearizationobject{}$.
        Therefore, since $e_b$ set $\linearizationobject{}.\uniquerepositoryoperationlong{} = \uniquerepositoryoperationshort{}_\linearizationobject{}$, there are two $L$-events in $\mathcal{I}^\mathcal{B}$ that set $\linearizationobject{}.\uniquerepositoryoperationlong{} = \uniquerepositoryoperationshort{}_\linearizationobject{}$.
        However, since $P(\mathcal{I}^\mathcal{B})$ holds, by \Cref{lemma:ero:p_implies_unique_low_level_operations_in_linearization}, every $L$-event in $\mathcal{I}^\mathcal{B}$ sets $\linearizationobject{}.\uniquerepositoryoperationlong{}$ to a unique value, a contradiction.
        \qH{\Cref{lemma:ero:any_s_attempt_outside_its_window_is_unsuccessful:claim_one}}
    \end{proof}

    \begin{claimcustom}{\ref{lemma:ero:any_s_attempt_outside_its_window_is_unsuccessful}.2}\label{lemma:ero:any_s_attempt_outside_its_window_is_unsuccessful:claim_two}
        $\stateobject.\uniquerepositoryoperationlong{} = \uniquerepositoryoperationshort{}_\linearizationobject{}$ at $a'$.
    \end{claimcustom}

    \begin{proof}
        Suppose $q$ executed $a'$. 
        Hence, by \Cref{def:ero:english}, $q$ executed $a'$ during some invocation $I'$ of the \doapplyandcopyresponse{} procedure.
        By $O(\mathcal{I}^\mathcal{B})$, $e_b$ and $a'$ are for the same timestamp; say $\timeshort{}$.
        Hence, by \Cref{def:ero:english}, the first parameter of $I'$ is some $\uniquerepositoryoperationshort{} = (\timeshort{}, \arbitraryvalue)$.
        Thus, by \Cref{lemma:ero:l_event_corresponding_to_do_low_level_op}, there is an $L$-event $e'_b$ that set $\linearizationobject{}.\uniquerepositoryoperationlong{} = \uniquerepositoryoperationshort{}$, and so by \Cref{def:ero:english}, $e'_b$ is for timestamp $t$.
        So, since $e_b$ is for timestamp $t$, and $P(\mathcal{I}^\mathcal{B})$ holds, by \Cref{lemma:ero:every_l_event_has_a_unique_timestamp}, $e_b = e'_b$.
        Hence, since $e'_b$ set $\linearizationobject{}.\uniquerepositoryoperationlong{} = \uniquerepositoryoperationshort{}$ and $e_b$ set $\linearizationobject{}.\uniquerepositoryoperationlong{} = \uniquerepositoryoperationshort{}_\linearizationobject{}$, we have that $ \uniquerepositoryoperationshort{} = \uniquerepositoryoperationshort{}_\linearizationobject{}$.
        Thus, the first parameter of $I'$ is $\uniquerepositoryoperationshort{}_\linearizationobject{}$.
        Therefore, since $a'$ is successful, we have that $\stateobject.\uniquerepositoryoperationlong{} = \uniquerepositoryoperationshort{}_\linearizationobject{}$ at $a'$ as wanted.
        \qH{\Cref{lemma:ero:any_s_attempt_outside_its_window_is_unsuccessful:claim_two}}
    \end{proof}

    We now identify a successful $S$-attempt that contradicts the minimality of $a$.

    \begin{claimcustom}{\ref{lemma:ero:any_s_attempt_outside_its_window_is_unsuccessful}.3}\label{lemma:ero:any_s_attempt_outside_its_window_is_unsuccessful:claim_three}
        The corresponding $L$-event of a successful $S$-attempt during $(a', a)$ is before $e_b$.
    \end{claimcustom}

    \begin{proof}
        There are two cases.
        \begin{itemize}
            \item[] \hspace{0pt}\textbf{Case 1.} $T^{\ref{line:ero:state_read}} < a'$.
    
            Since by \Cref{lemma:ero:any_s_attempt_outside_its_window_is_unsuccessful:claim_one} $T^{\ref{line:ero:state_read}} \in (e_b, e_a)$, and $a' \in (e_b, e_a)$, by transitivity, $e_b < T^{\ref{line:ero:state_read}} < a' < e_a$.
            Suppose $p$ read $\uniquerepositoryoperationshort{}$ from $\stateobject.\uniquerepositoryoperationlong{}$ at $T^{\ref{line:ero:state_read}}$.
            Since $p$ executed $T^{\ref{line:ero:state_read}}$ and $a$ during $I$, we have that $p$ found the condition on \cref{line:ero:check_if_already_applied} to be true during $I$.
            Hence, since $I$'s first parameter is $\uniquerepositoryoperationshort{}_\linearizationobject{}$, we have that $\uniquerepositoryoperationshort{} \neq \uniquerepositoryoperationshort{}_\linearizationobject{}$.
            Furthermore, since $p$ executed $a$ during $I$, $p$ read $\uniquerepositoryoperationshort{}$ from $\stateobject.\uniquerepositoryoperationlong{}$ at $T^{\ref{line:ero:state_read}}$, and by assumption $a$ is successful, we have that $\stateobject.\uniquerepositoryoperationlong{} = \uniquerepositoryoperationshort{}$ at the step before $a$.
            Hence, since $a' < a$ (because $a' < e_a$ and $e_a < a$), by \Cref{lemma:ero:any_s_attempt_outside_its_window_is_unsuccessful:claim_two} $\stateobject.\uniquerepositoryoperationlong{} = \uniquerepositoryoperationshort{}_\linearizationobject{}$ at $a'$, and $\uniquerepositoryoperationshort{} \neq \uniquerepositoryoperationshort{}_\linearizationobject{}$, we have that some step set the value of $\stateobject.\uniquerepositoryoperationlong{} = \uniquerepositoryoperationshort{}$ during $(a', a)$.
            Thus, by \Cref{observation:ero:where_objects_change}, some successful $S$-attempt $a^*$ set the value of $\stateobject.\uniquerepositoryoperationlong{} = \uniquerepositoryoperationshort{}$ during $(a', a)$.
            So, by \Cref{lemma:ero:every_s_attempt_is_for_timestamp_other_than_zero}, $a^*$ is for a timestamp larger than $0$, and so $\uniquerepositoryoperationshort{} \neq (0, \arbitraryvalue)$.
            Thus, since $p$ read $\uniquerepositoryoperationshort{}$ from $\stateobject.\uniquerepositoryoperationlong{}$ at $T^{\ref{line:ero:state_read}}$ and the initial value of $\stateobject.\uniquerepositoryoperationlong{} = (0, \noop)$, we have that $\stateobject.\uniquerepositoryoperationlong{}$ was set to $\uniquerepositoryoperationshort{}$ before $T^{\ref{line:ero:state_read}}$.
            Therefore, by \Cref{observation:ero:where_objects_change}, some successful $S$-attempt $\hat{a}$ set $\stateobject.\uniquerepositoryoperationlong{} = \uniquerepositoryoperationshort{}$ before $T^{\ref{line:ero:state_read}}$.
    
            Let $q$ be the process that executed $a^*$, and let $e^*_b$ be $a^*$'s corresponding $L$-event.
            Since $a^*$ set $\stateobject.\uniquerepositoryoperationlong{} = \uniquerepositoryoperationshort{}$, by \Cref{lemma:ero:l_event_corresponding_to_do_low_level_op}, $e^*_b$ set $\linearizationobject{}.\uniquerepositoryoperationlong{} = \uniquerepositoryoperationshort{}$.
            Thus, since $e_b$ set $\linearizationobject{}.\uniquerepositoryoperationlong{} = \uniquerepositoryoperationshort{}_\linearizationobject{}$ and $\uniquerepositoryoperationshort{} \neq \uniquerepositoryoperationshort{}_\linearizationobject{}$, we have that $e'_b \neq e_b$.
            We now prove that $e^*_b \leq e_b$.
            Let $r$ be the process that executed $\hat{a}$, and let $\hat{e_b}$ be $\hat{a}$'s corresponding $L$-event.
            Since $\hat{a}$ set $\stateobject.\uniquerepositoryoperationlong{} = \uniquerepositoryoperationshort{}$, by \Cref{lemma:ero:s_attempts_have_corresponding_l_events}, $\hat{e_b}$ set $\linearizationobject{}.\uniquerepositoryoperationlong{} = \uniquerepositoryoperationshort{}$.
            Hence, since $e^*_b$ and $\hat{e_b}$ are two $L$-events in $\mathcal{I}^\mathcal{B}$ that set $\linearizationobject{}.\uniquerepositoryoperationlong{} = \uniquerepositoryoperationshort{}$ and $P(\mathcal{I}^\mathcal{B})$ holds, by \Cref{lemma:ero:p_implies_unique_low_level_operations_in_linearization}, $e^*_b = \hat{e_b}$.
            Since by \Cref{lemma:ero:any_s_attempt_outside_its_window_is_unsuccessful:claim_one} $T^{\ref{line:ero:state_read}} \in (e_b, e_a)$, and $e_b$ and $e_a$ are successive $L$-events, $e_b$ is the last $L$-event before $T^{\ref{line:ero:state_read}}$.
            Furthermore, since $\hat{e_b} < \hat{a}$ and $\hat{a} < T^{\ref{line:ero:state_read}}$, by transitivity, $\hat{e_b} < T^{\ref{line:ero:state_read}}$, and since $e^*_b = \hat{e_b}$, we have that $e^*_b < T^{\ref{line:ero:state_read}}$.
            Therefore, since $e_b$ is the last $L$-event before $T^{\ref{line:ero:state_read}}$, and $e^*_b$ is an $L$-event before $T^{\ref{line:ero:state_read}}$, we have that $e^*_b \leq e_b$ as wanted.
            Since $e^*_b \neq e_b$ and $e^*_b \leq e_b$, we have that $e^*_b < e_b$.
            Therefore, there is a successful $S$-attempt during $(a', a)$ (namely $a^*$) whose corresponding $L$-event (namely $e^*_b$) is before $e_b$ as wanted.
            
            \item[] \hspace{0pt}\textbf{Case 2.} $a' < T^{\ref{line:ero:state_read}}$.
    
            Since by \Cref{lemma:ero:any_s_attempt_outside_its_window_is_unsuccessful:claim_one} $T^{\ref{line:ero:state_read}} \in (e_b, e_a)$ and $a' \in (e_b, e_a)$, by transitivity, $e_b < a' < T^{\ref{line:ero:state_read}} < e_a$.
            Hence, since $a'$ is the only successful $S$-attempt during $(e_b, e_a)$, and by \Cref{lemma:ero:any_s_attempt_outside_its_window_is_unsuccessful:claim_two} $\stateobject.\uniquerepositoryoperationlong{} = \uniquerepositoryoperationshort{}_\linearizationobject{}$ at $a'$, we have that $\stateobject.\uniquerepositoryoperationlong{} = \uniquerepositoryoperationshort{}_\linearizationobject{}$ throughout $(a', e_a)$.
            Thus, since $T^{\ref{line:ero:state_read}} \in (a', e_a)$, we have that $p$ read $\uniquerepositoryoperationshort{}_\linearizationobject{}$ from $\stateobject.\uniquerepositoryoperationlong{}$ on \cref{line:ero:state_read} at $T^{\ref{line:ero:state_read}}$.
            Hence, since $I$'s first parameter is $\uniquerepositoryoperationshort{}_\linearizationobject{}$, we have that $p$ finds the condition on \cref{line:ero:check_if_already_applied} to be false during $I$.
            Therefore, $p$ does not execute \cref{line:ero:state_cas} during $I$.
            However, $p$ executes $a$ during $I$, a contradiction, so this case is impossible.
            \qH{\Cref{lemma:ero:any_s_attempt_outside_its_window_is_unsuccessful:claim_three}}
        \end{itemize}
    \end{proof}

    We now finish the proof of \Cref{lemma:ero:any_s_attempt_outside_its_window_is_unsuccessful}.
    Let $a^*$ be the successful $S$-attempt identified by \Cref{lemma:ero:any_s_attempt_outside_its_window_is_unsuccessful:claim_three} and let $e^*_b$ be its corresponding $L$-event.
    Hence, $a^*$ is during $(a', a)$ and $e^*_b < e_b$.
    Since $e^*_b < e_b$, it follows that there is an $L$-event after $e^*_b$.
    Let $e^*_a$ be the next $L$-event after $e^*_b$.
    Hence, since $e^*_b < e_b$, we have that $e^*_a \leq e_b$.
    Thus, since $e_b < a'$ (because $a' \in (e_b, e_a)$) and $a' < a^*$ (because $a^* \in (a', a)$), by transitivity, $e^*_a < a^*$.
    Therefore, since $a^* < a$, we have shown that $a^*$ is a successful $S$-attempt before $a$ such that the next $L$-event after $a^*$'s corresponding $L$-event is before $a^*$.
    However, by the minimality of $a$, $a^*$ must be unsuccessful, a contradiction.
    \qH{\Cref{lemma:ero:any_s_attempt_outside_its_window_is_unsuccessful}}
\end{proof}

This implies the following.

\begin{corollary}\label{lemma:ero:any_s_attempt_outside_its_window_is_unsuccessful_alternate_statement}
    Consider a successful $S$-attempt $a$ in $\mathcal{I}^\mathcal{B}$ and let $e_b$ be its corresponding $L$-event (see \Cref{lemma:ero:l_event_corresponding_to_do_low_level_op}).
    If $P(\mathcal{I}^\mathcal{B})$ and $O(\mathcal{I}^\mathcal{B})$ hold, then there are no $L$-events during $(e_b, a)$ in $\mathcal{I}^\mathcal{B}$.
\end{corollary}

We now ``extend" $O$ beyond the last $L$-event.
We do so based on the type of the last $L$-event.

\begin{lemma}\label{lemma:ero:1_o_safety_holds}
    Suppose $\mathcal{I}^\mathcal{B}$ has a last $L$-event denoted as $e$ and $P(\mathcal{I}^\mathcal{B})$ and $O(\mathcal{I}^\mathcal{B})$ hold.
    If $e$ is an $L$-apply event for a timestamp $\timeshort{}$, then from $e$ onwards in $\mathcal{I}^\mathcal{B}$ there is at most one successful $S$-attempt for $\timeshort{}$ and no other successful $S$-attempts for any timestamp.
\end{lemma}

\begin{proof}
    Suppose, for contradiction, $e$ is a $L$-apply event for timestamp $\timeshort{}$, so by \Cref{lemma:ero:every_l_event_is_for_timestamp_other_than_zero} $\timeshort{} \neq 0$, and at or after $e$ in $\mathcal{I}^\mathcal{B}$ there is either at least two successful $S$-attempts for $\timeshort{}$, or one or more successful $S$-attempts for a timestamp other than $\timeshort{}$.
    We consider each case separately.

    \begin{itemize}
        \item[] \hspace{0pt}\textbf{Case 1.} There are at least two successful $S$-attempts for $\timeshort{}$ at or after $e$ in $\mathcal{I}^\mathcal{B}$.

        Let $a_1$ and $a_2$ be the first two successful $S$-attempts for $\timeshort{}$ at or after $e$ in $\mathcal{I}^\mathcal{B}$.
        Without loss of generality, suppose $a_1 < a_2$.
        Let $p_1$ (resp. $p_2$) be the process that executed $a_1$ (resp. $a_2$).
        Furthermore, let $e_1$ (resp. $e_2$) be $a_1$'s (resp. $a_2$'s) corresponding $L$-event (see \Cref{lemma:ero:l_event_corresponding_to_do_low_level_op}).
        Since $a_1$ and $a_2$ are for timestamp $\timeshort{}$, by \Cref{corollary:ero:s_attempts_and_corresponding_l_events_are_for_matching_timestamps}, $e_1$ and $e_2$ are for timestamp $\timeshort{}$.
        Since $e$, $e_1$, and $e_2$ are all $L$-events in $\mathcal{I}^\mathcal{B}$ for timestamp $\timeshort{}$, and $P(\mathcal{I}^\mathcal{B})$ holds, by \Cref{lemma:ero:every_l_event_has_a_unique_timestamp}, $e = e_1 = e_2$, so by \Cref{lemma:ero:s_attempts_have_corresponding_l_events} $a_1$ and $a_2$ set $\stateobject.\uniquerepositoryoperationlong{}$ to the same value; say $\uniquerepositoryoperationshort{}_\linearizationobject{}$.
        So, by \Cref{lemma:ero:s_attempts_have_corresponding_l_events}, $e$ set $\linearizationobject{}.\uniquerepositoryoperationlong{} = \uniquerepositoryoperationshort{}_\linearizationobject{}$.
        Let $I_1$ and $I_2$ be the invocations of the \doapplyandcopyresponse{} procedure that $a_1$ and $a_2$ are executed during, respectively, so the first parameter of $I_1$ and $I_2$ is $\uniquerepositoryoperationshort{}_\linearizationobject{}$.
        Since $p_2$ executed $a_2$ during $I_2$, $p_2$ found the condition on \cref{line:ero:check_if_already_applied} to be true during $I_2$.
        Hence, $p_2$ read $\stateobject.\uniquerepositoryoperationlong{} = \uniquerepositoryoperationshort{}$ on \cref{line:ero:state_read} during $I_2$, say at time $T^{\ref{line:ero:state_read}}$, such that $\uniquerepositoryoperationshort{} \neq \uniquerepositoryoperationshort{}_\linearizationobject{}$.
        
        We now prove that there is a successful $S$-attempt that set $\stateobject.\uniquerepositoryoperationlong{} = \uniquerepositoryoperationshort{}$ between $a_1$ and $a_2$ (*).
        There are two cases.
    
        \begin{itemize}
            \item[] \hspace{0pt}\textbf{Case 1.1.} $T^{\ref{line:ero:state_read}} < a_1$.
    
            Hence, since $a_1 < a_2$, by transitivity, $T^{\ref{line:ero:state_read}} < a_1 < a_2$.
            Thus, since $\stateobject.\uniquerepositoryoperationlong{} = \uniquerepositoryoperationshort{}$ at $T^{\ref{line:ero:state_read}}$, $\stateobject.\uniquerepositoryoperationlong{} = \uniquerepositoryoperationshort{}_\linearizationobject{}$ at $a_1$, $\uniquerepositoryoperationshort{} \neq \uniquerepositoryoperationshort{}_\linearizationobject{}$, and $a_2$ is successful, we have that $\stateobject.\uniquerepositoryoperationlong{}$ was set to $\uniquerepositoryoperationshort{}$ between $a_1$ and $a_2$.
            Hence, by \Cref{observation:ero:where_objects_change}, (*) follows.
    
            \item[] \hspace{0pt}\textbf{Case 1.2.} $a_1 < T^{\ref{line:ero:state_read}}$.

            Hence, since $T^{\ref{line:ero:state_read}} < a_2$, by transitivity, $a_1 < T^{\ref{line:ero:state_read}} < a_2$.
            Since $a_1$ set $\stateobject.\uniquerepositoryoperationlong{} = \uniquerepositoryoperationshort{}_\linearizationobject{}$, $\stateobject.\uniquerepositoryoperationlong{} = \uniquerepositoryoperationshort{}$ at $T^{\ref{line:ero:state_read}}$, and $\uniquerepositoryoperationshort{} \neq \uniquerepositoryoperationshort{}_\linearizationobject{}$, we have that $\stateobject.\uniquerepositoryoperationlong{}$ was set to $\uniquerepositoryoperationshort{}$ between $a_1$ and $T^{\ref{line:ero:state_read}}$.
            Hence, by \Cref{observation:ero:where_objects_change}, there is a successful $S$-attempt that set $\stateobject.\uniquerepositoryoperationlong{} = \uniquerepositoryoperationshort{}$ between $a_1$ and $T^{\ref{line:ero:state_read}}$.
            Therefore, since $T^{\ref{line:ero:state_read}} < a_2$, (*) follows.
        \end{itemize}
    
        We now finish the proof of Case 1.
        Let $a'$ be the successful $S$-attempt that set $\stateobject.\uniquerepositoryoperationlong{} = \uniquerepositoryoperationshort{}$ between $a_1$ and $a_2$ identified by (*).
        Let $e'$ be $a'$'s corresponding $L$-event, so by \Cref{lemma:ero:s_attempts_have_corresponding_l_events}, $e'$ set $\linearizationobject{}.\uniquerepositoryoperationlong{} = \uniquerepositoryoperationshort{}$.
        Hence, since $e$ set $\linearizationobject{}.\uniquerepositoryoperationlong{} = \uniquerepositoryoperationshort{}_\linearizationobject{}$, and $\uniquerepositoryoperationshort{} \neq \uniquerepositoryoperationshort{}_\linearizationobject{}$, we have that $e \neq e'$.
        Thus, since $e$ and $e'$ are two different $L$-events in $\mathcal{I}^\mathcal{B}$, $e$ is for timestamp $\timeshort{}$, and $P(\mathcal{I}^\mathcal{B})$ holds, by \Cref{lemma:ero:every_l_event_has_a_unique_timestamp}, $e'$ is for a timestamp $\timeshort{}' \neq \timeshort{}$.
        Hence, since $e'$ is $a'$'s corresponding $L$-event, and $e'$ is for timestamp $\timeshort{}'$, by \Cref{corollary:ero:s_attempts_and_corresponding_l_events_are_for_matching_timestamps}, $a'$ is for timestamp $\timeshort{}'$.
        Thus, there is a successful $S$-attempt for a timestamp other than $\timeshort{}$ between $a_1$ and $a_2$.
        So, since $a_1$ is at or after $e$ in $\mathcal{I}^\mathcal{B}$, we have that there is a successful $S$-attempt for a timestamp other than $\timeshort{}$ at or after $e$ in $\mathcal{I}^\mathcal{B}$.
        Therefore, this case reduces to the next one.

        \item[] \hspace{0pt}\textbf{Case 2.} There is a successful $S$-attempt $a$ for a timestamp other than $\timeshort{}$ at or after $e$ in $\mathcal{I}^\mathcal{B}$.

        We will apply \Cref{lemma:ero:any_s_attempt_outside_its_window_is_unsuccessful}.
        Suppose $e_b$ is $a$'s corresponding $L$-event and suppose $a$ is for timestamp $\timeshort{}' \neq \timeshort{}$.
        Hence, by \Cref{corollary:ero:s_attempts_and_corresponding_l_events_are_for_matching_timestamps}, $e_b$ is for timestamp $\timeshort{}'$.
        Thus, since $\timeshort{}' \neq \timeshort{}$ and $e$ is for timestamp $\timeshort{}$, we have that $e_b \neq e$.
        So, since $e_b < a$, $a$ is in $\mathcal{I}^\mathcal{B}$, and $e$ is the last $L$-event in $\mathcal{I}^\mathcal{B}$, we have that $e_b < e$.
        Hence, there is a next $L$-event after $e_b$ in $\mathcal{I}^\mathcal{B}$; say $e_a$.
        Thus, $e_a \leq e$, and since $e < a$, by transitivity, $e_a < a$.
        Therefore, since $P(\mathcal{I}^\mathcal{B})$ and $O(\mathcal{I}^\mathcal{B})$ hold, by \Cref{lemma:ero:any_s_attempt_outside_its_window_is_unsuccessful}, $a$ is unsuccessful.
        However, $a$ was assumed to be successful, a contradiction.
        \qH{\Cref{lemma:ero:1_o_safety_holds}}
    \end{itemize}
\end{proof}

\begin{lemma}\label{lemma:ero:2_o_safety_holds}
    Suppose $\mathcal{I}^\mathcal{B}$ has a last $L$-event denoted as $e$ and $P(\mathcal{I}^\mathcal{B})$ and $O(\mathcal{I}^\mathcal{B})$ hold.
    If $e$ is an $L$-add or $L$-remove event, then from $e$ onwards in $\mathcal{I}^\mathcal{B}$ there are no successful $S$-attempts.
\end{lemma}

\begin{proof}
    Suppose, for contradiction, $e$ is a $L$-add or $L$-remove event and at or after $e$ in $\mathcal{I}^\mathcal{B}$ there is a successful $S$-attempt $a$.
    We will apply \Cref{lemma:ero:any_s_attempt_outside_its_window_is_unsuccessful}.
    Suppose $e_b$ is $a$'s corresponding $L$-event, so by \Cref{lemma:ero:l_event_corresponding_to_do_low_level_op}, $e_b$ is an $L$-apply event.
    Hence, since $e$ is an $L$-add or $L$-remove event, it follows that $e_b \neq e$.
    Thus, since $e_b < a$, $a$ is in $\mathcal{I}^\mathcal{B}$, and $e$ is the last $L$-event in $\mathcal{I}^\mathcal{B}$, we have that $e_b < e$.
    Hence, there is a next $L$-event after $e_b$ in $\mathcal{I}^\mathcal{B}$; say $e_a$.
    Thus, $e_a \leq e$, and since $e < a$, by transitivity, $e_a < a$.
    Therefore, since $P(\mathcal{I}^\mathcal{B})$ and $O(\mathcal{I}^\mathcal{B})$ hold, by \Cref{lemma:ero:any_s_attempt_outside_its_window_is_unsuccessful}, $a$ is unsuccessful.
    However, $a$ was assumed to be successful, a contradiction.
    \qH{\Cref{lemma:ero:2_o_safety_holds}}
\end{proof}

\subsubsection{The AcquireNext procedure has the intended effect}

We are now ready to prove that various procedures have the intended effect: once they exit, what they were trying to do is done.

\begin{lemma}\label{lemma:ero:acquire_next_response_classification_weak}
    Suppose some process $p$ exited some invocation $I$ of the AcquireNext procedure with parameters $(\uniquerepositoryoperationshort_\linearizationobject{}, \currentcellpointershort)$ in $\mathcal{I}^\mathcal{B}$.
    By \Cref{lemma:ero:acquire_next_is_for_pointer_from_universe_or_head}, $\currentcellpointershort \in \celluniverse{} \cup \{\&\headobject\}$.
    Let $T^{\ref{line:ero:acquire_next_read_curr_unique_pointer}}$ (resp. $T^{\ref{line:ero:acquire_next_linearization_changed_check}}$) be the last time $p$ executed \cref{line:ero:acquire_next_read_curr_unique_pointer} (resp. \cref{line:ero:acquire_next_linearization_changed_check}) during $I$.
    If $\linearizationobject{}.\uniquerepositoryoperationlong{} = \uniquerepositoryoperationshort_\linearizationobject{}$ at $T^{\ref{line:ero:acquire_next_linearization_changed_check}}$, then 
    \begin{compactenum}
        \item if $(*\currentcellpointershort).\nextlong.\uniquecellpointercontentlong{} = \nullconstant$ at $T^{\ref{line:ero:acquire_next_read_curr_unique_pointer}}$, then $I$'s response is $(\notfound, \arbitraryvalue)$; otherwise
        \item $I$'s response is $(\found, \nextcellpointershort{})$ where $(*\currentcellpointershort).\nextlong.\uniquecellpointercontentlong{} = \nextcellpointershort{}$ at $T^{\ref{line:ero:acquire_next_read_curr_unique_pointer}}$.
    \end{compactenum} 
\end{lemma}

\begin{proof}
    Since $\linearizationobject{}.\uniquerepositoryoperationlong{} = \uniquerepositoryoperationshort_\linearizationobject{}$ at $T^{\ref{line:ero:acquire_next_linearization_changed_check}}$, it follows that $p$ finds the condition on \cref{line:ero:acquire_next_linearization_changed_check} to be false at $T^{\ref{line:ero:acquire_next_linearization_changed_check}}$.
    Hence, since $p$ exits $I$, $p$ executes \cref{line:ero:acquire_next_not_found_check} after $T^{\ref{line:ero:acquire_next_linearization_changed_check}}$ during $I$.
    Let $T^{\ref{line:ero:acquire_next_not_found_check}}$ be the time of $p$'s next execution of \cref{line:ero:acquire_next_not_found_check} after $T^{\ref{line:ero:acquire_next_linearization_changed_check}}$.
    By definition $T^{\ref{line:ero:acquire_next_read_curr_unique_pointer}} < T^{\ref{line:ero:acquire_next_linearization_changed_check}} < T^{\ref{line:ero:acquire_next_not_found_check}}$, and $T^{\ref{line:ero:acquire_next_read_curr_unique_pointer}}$, $T^{\ref{line:ero:acquire_next_linearization_changed_check}}$, and $T^{\ref{line:ero:acquire_next_not_found_check}}$ occur within the last iteration $I'$ of the loop on \cref{line:ero:acquire_next_repeat_loop} during $I$.
    There are two cases.
    \begin{itemize}
        \item[] \hspace{0pt}\textbf{Case 1.} $(*\currentcellpointershort).\nextlong.\uniquecellpointercontentlong{} = \nullconstant$ at $T^{\ref{line:ero:acquire_next_read_curr_unique_pointer}}$.

        Hence, $p$ finds the condition on \cref{line:ero:acquire_next_not_found_check} to be true at $T^{\ref{line:ero:acquire_next_not_found_check}}$.
        Therefore, since $p$ exits $I$, $p$ exits on \cref{line:ero:acquire_next_not_found_return} with response $(\notfound, \arbitraryvalue)$ as required.

        \item[] \hspace{0pt}\textbf{Case 2.} $(*\currentcellpointershort).\nextlong.\uniquecellpointercontentlong{} \neq \nullconstant$ at $T^{\ref{line:ero:acquire_next_read_curr_unique_pointer}}$.

        Let $(*\currentcellpointershort).\nextlong.\uniquecellpointercontentlong{} = \nextcellpointershort{}$ at $T^{\ref{line:ero:acquire_next_read_curr_unique_pointer}}$.
        Hence, $\nextuniquecellpointershort{} \neq \nullconstant$.
        Thus, $p$ finds the condition on \cref{line:ero:acquire_next_not_found_check} to be false at $T^{\ref{line:ero:acquire_next_not_found_check}}$.
        Since (1) $T^{\ref{line:ero:acquire_next_read_curr_unique_pointer}}$, $T^{\ref{line:ero:acquire_next_linearization_changed_check}}$, and $T^{\ref{line:ero:acquire_next_not_found_check}}$ are within $I'$, (2) $I'$ is the last iteration of the loop on \cref{line:ero:acquire_next_repeat_loop} during $I$, (3) $p$ exits $I$, (4) $p$ finds the condition on \cref{line:ero:acquire_next_linearization_changed_check} to be false at $T^{\ref{line:ero:acquire_next_linearization_changed_check}}$, and (5) $p$ finds the condition on \cref{line:ero:acquire_next_not_found_check} to be false at $T^{\ref{line:ero:acquire_next_not_found_check}}$, we have that $p$ finds the condition on \cref{line:ero:acquire_next_cell} to be true during $I'$.
        Therefore, since $p$ exits $I$, we have that $p$ exits $I$ on \cref{line:ero:acquire_next_found_return} with the response $(\found, \nextcellpointershort{})$ as required.
        \qH{\Cref{lemma:ero:acquire_next_response_classification_weak}}
    \end{itemize}
\end{proof}

\subsubsection{The Acquire procedure has the intended effect}

\begin{lemma}\label{lemma:ero:under_certain_conditions_acquire_returns_found_weak}
    Suppose some process $p$ exited some invocation $I$ of the Acquire procedure in $\mathcal{I}^\mathcal{B}$ with parameters $(\uniquerepositoryoperationshort_\linearizationobject{}$, $\targetcellpointershort)$ for some $\targetcellpointershort \in \celluniverse{}$ and returns response $\status$.
    Let $T_b$ be the time $p$ invoked $I$, and let $T^{\ref{line:ero:acquire_next_linearization_changed_check}}$ be the last time $p$ executes \cref{line:ero:acquire_next_linearization_changed_check} during $I$.
    Recall that $T^{\ref{line:ero:acquire_next_linearization_changed_check}}$ is well-defined by \Cref{lemma:ero:exit_acquire_implies_executing_105}.
    If $P(\mathcal{I}^\mathcal{B})$ holds and for some finite prefix $\mathcal{I}$ of $\mathcal{I}^\mathcal{B}$ the following two conditions hold for every prefix $\mathcal{I}'$ of $\mathcal{I}^\mathcal{B}$ during $[T_b, T^{\ref{line:ero:acquire_next_linearization_changed_check}}]$: 
    \begin{compactitem}
        \item $\linearizationobject{}.\uniquerepositoryoperationlong{} = \uniquerepositoryoperationshort_\linearizationobject{}$ at the end of $\mathcal{I}'$; and
        \item the list of cells conforms to $\List(\mathcal{I})$ in $\mathcal{I}'$
    \end{compactitem}
    then
    \begin{compactenum}
        \item if $\targetcellpointershort \in \List(\mathcal{I})$, then $\status = \found{}$; and
        \item if $\targetcellpointershort \not\in \List(\mathcal{I})$, then $\status = \notfound{}$.
    \end{compactenum}
\end{lemma}

\begin{proof}
    We start by showing that our first assumption implies the following claim.
    As we will see, this claim is useful for satisfying the conditions of \Cref{lemma:ero:acquire_next_response_classification_weak} and our second assumption.
    
    \begin{claimcustom}{\ref{lemma:ero:under_certain_conditions_acquire_returns_found_weak}.1}\label{lemma:ero:under_certain_conditions_acquire_returns_found_weak:claim_zero}
        Consider any invocation $I^*$ of the AcquireNext procedure on \cref{line:ero:acquire_acquire_next} during $I$.
        Note that since $I$ exits $I^*$ exits.
        Let $T^{\ref{line:ero:acquire_next_read_curr_unique_pointer}}_{I^*}$ (resp. $T^{\ref{line:ero:acquire_next_linearization_changed_check}}_{I^*}$) be the last time $p$ executes \cref{line:ero:acquire_next_read_curr_unique_pointer} (resp. \cref{line:ero:acquire_next_linearization_changed_check}) during $I^*$ (these are well-defined since $I^*$ exits).
        Then, the following are true:
        \begin{compactitem}
            \item $\linearizationobject{}.\uniquerepositoryoperationlong{} = \uniquerepositoryoperationshort_\linearizationobject{}$ at $T^{\ref{line:ero:acquire_next_linearization_changed_check}}_{I^*}$; and
            \item $T^{\ref{line:ero:acquire_next_read_curr_unique_pointer}}_{I^*} \in [T_b, T^{\ref{line:ero:acquire_next_linearization_changed_check}}]$.
        \end{compactitem}
    \end{claimcustom}

    \begin{proof}
        First 1.
        Since by the first assumption of \Cref{lemma:ero:under_certain_conditions_acquire_returns_found_weak} $\linearizationobject{}.\uniquerepositoryoperationlong{} = \uniquerepositoryoperationshort_\linearizationobject{}$ at the end of $\mathcal{I}'$ for every prefix $\mathcal{I}'$ of $\mathcal{I}$ in $[T_b, T^{\ref{line:ero:acquire_next_linearization_changed_check}}]$, we have that $\linearizationobject{}.\uniquerepositoryoperationlong{} = \uniquerepositoryoperationshort_\linearizationobject{}$ throughout $[T_b, T^{\ref{line:ero:acquire_next_linearization_changed_check}}]$.
        Since $I$ was invoked at $T_b$, $I^*$ was invoked during $I$, $T^{\ref{line:ero:acquire_next_linearization_changed_check}}_{I^*}$ is the last time $p$ executes \cref{line:ero:acquire_next_linearization_changed_check} during $I^*$, and $T^{\ref{line:ero:acquire_next_linearization_changed_check}}$ is the last time $p$ executes \cref{line:ero:acquire_next_linearization_changed_check} during $I$, by transitivity, $T^{\ref{line:ero:acquire_next_linearization_changed_check}}_{I^*} \in [T_b, T^{\ref{line:ero:acquire_next_linearization_changed_check}}]$.
        Hence, since $\linearizationobject{}.\uniquerepositoryoperationlong{} = \uniquerepositoryoperationshort_\linearizationobject{}$ throughout $[T_b, T^{\ref{line:ero:acquire_next_linearization_changed_check}}]$, we have that $\linearizationobject{}.\uniquerepositoryoperationlong{} = \uniquerepositoryoperationshort_\linearizationobject{}$ at $T^{\ref{line:ero:acquire_next_linearization_changed_check}}_{I^*}$.
        Now 2.
        Since $I$ was invoked at $T_b$, $I^*$ was invoked during $I$, and $T^{\ref{line:ero:acquire_next_read_curr_unique_pointer}}_{I^*}$ is the time of a step during $I^*$, by transitivity, $T_b < T^{\ref{line:ero:acquire_next_read_curr_unique_pointer}}_{I^*}$.
        Hence, since $T^{\ref{line:ero:acquire_next_read_curr_unique_pointer}}_{I^*} < T^{\ref{line:ero:acquire_next_linearization_changed_check}}_{I^*}$ and $T^{\ref{line:ero:acquire_next_linearization_changed_check}}_{I^*} \in [T_b, T^{\ref{line:ero:acquire_next_linearization_changed_check}}]$, by transitivity, $T^{\ref{line:ero:acquire_next_read_curr_unique_pointer}}_{I^*} \in [T_b, T^{\ref{line:ero:acquire_next_linearization_changed_check}}]$.
        \qH{\Cref{lemma:ero:under_certain_conditions_acquire_returns_found_weak:claim_zero}}
    \end{proof}

    We now prove that $I$ ``traverses" $\List(\mathcal{I})$.
    By \Cref{def:ero:logical_list}, $\List(\mathcal{I}) = \cellpointershort_0, \ldots, \cellpointershort_{n + 1}$ for some $n$.
    Hence, by \Cref{lemma:ero:every_list_sequence_is_from_universe}, $\cellpointershort_0 = \&\headobject$, for every $i \in [1..n]$ $\cellpointershort_i \in \celluniverse{}$, and $\cellpointershort_{n + 1} = \nullconstant$.
    
    \begin{claimcustom}{\ref{lemma:ero:under_certain_conditions_acquire_returns_found_weak}.2}\label{lemma:ero:under_certain_conditions_acquire_returns_found_weak:claim_one}
        Consider any iteration of the loop on \cref{line:ero:acquire_loop_until} during $I$, denoted by $I'$, such that the local variable $\currentcellpointershort{} = \cellpointershort_i$ for some $i \in [0..n)$ at the start of $I'$.\footnote{The start of an iteration $I$ of a loop on line $X$ refers to the time line $X$ was executed during $I$.}
        If $\cellpointershort_i \neq \targetcellpointershort$, then $p$ executes \cref{line:ero:acquire_update_current_unique_pointer} at some time $T^{\ref{line:ero:acquire_update_current_unique_pointer}}$ during $I'$ and $\currentuniquecellpointershort{} = \cellpointershort_{i + 1}$ at $T^{\ref{line:ero:acquire_update_current_unique_pointer}}$.
    \end{claimcustom}

    \begin{proof}
        Since $\currentcellpointershort{} = \cellpointershort_i$ at the start of $I'$, and $\cellpointershort_i \neq \targetcellpointershort$, we have that $p$ finds the condition on \cref{line:ero:acquire_loop_until} to be true at the start of $I'$.
        Hence, since $p$ exits $I$, $p$ begins and exits the AcquireNext procedure on \cref{line:ero:acquire_acquire_next} during $I'$.
        Denote this invocation by $I^*$.
        
        We first prove that $I^*$'s response is $(\found, \cellpointershort_{i + 1})$ by satisfying the conditions of \Cref{lemma:ero:acquire_next_response_classification_weak}.
        Since the first parameter of $I$ is $\uniquerepositoryoperationshort_\linearizationobject{}$ and $\currentuniquecellpointershort{} = \cellpointershort_i$ at the start of $I'$, the parameters of $I^*$ are $(\uniquerepositoryoperationshort_\linearizationobject{}, \cellpointershort_i)$.
        Let $T^{\ref{line:ero:acquire_next_read_curr_unique_pointer}}_{I^*}$ and $T^{\ref{line:ero:acquire_next_linearization_changed_check}}_{I^*}$ by defined as in \Cref{lemma:ero:under_certain_conditions_acquire_returns_found_weak:claim_zero}, and so $\linearizationobject{}.\uniquerepositoryoperationlong{} = \uniquerepositoryoperationshort_\linearizationobject{}$ at $T^{\ref{line:ero:acquire_next_linearization_changed_check}}_{I^*}$, and $T^{\ref{line:ero:acquire_next_read_curr_unique_pointer}}_{I^*} \in [T_b, T^{\ref{line:ero:acquire_next_linearization_changed_check}}]$.
        Hence, there is a prefix of $\mathcal{I}$ during $[T_b, T^{\ref{line:ero:acquire_next_linearization_changed_check}}]$ up to and including $T^{\ref{line:ero:acquire_next_read_curr_unique_pointer}}_{I^*}$; say $\mathcal{I}'$.
        Thus, by the second assumption of \Cref{lemma:ero:under_certain_conditions_acquire_returns_found_weak}, the list of cells conforms to $\List(\mathcal{I})$ in $\mathcal{I}'$.
        So, since $\List(\mathcal{I}) = \cellpointershort_0, \ldots, \cellpointershort_{n + 1}$, and $i \in [0..n]$, by \Cref{def:ero:logical_list}, at the end of $\mathcal{I}'$ $(*\cellpointershort_i).\nextlong.\uniquecellpointercontentlong{} = \cellpointershort_{i + 1}$.
        Thus, since $\cellpointershort_{i + 1} \in \celluniverse{}$ (because $i + 1 \in [1..n]$), by \Cref{assumption:ero:head_and_null_not_in_cell_universe} $\cellpointershort_{i + 1} \neq \nullconstant$, and so $(*\cellpointershort_i).\nextlong.\uniquecellpointercontentlong{} = \cellpointershort_{i + 1} \neq \nullconstant$ at the end of $\mathcal{I}'$.
        Therefore, we have established the following: (1) $I^*$ has parameters $(\uniquerepositoryoperationshort_\linearizationobject{}, \cellpointershort_i)$; (2) $\linearizationobject{}.\uniquerepositoryoperationlong{} = \uniquerepositoryoperationshort_\linearizationobject{}$ at $T^{\ref{line:ero:acquire_next_linearization_changed_check}}_{I^*}$; and (3) $(*\cellpointershort_i).\nextlong.\uniquecellpointercontentlong{} = \cellpointershort_{i + 1} \neq \nullconstant$ at $T^{\ref{line:ero:acquire_next_read_curr_unique_pointer}}_{I^*}$ (this is equivalent to the end of $\mathcal{I}'$), and so by \Cref{lemma:ero:acquire_next_response_classification_weak}, $I^*$'s response is $(\found, \cellpointershort_{i + 1})$ as wanted.
        
        We now finish the proof of \Cref{lemma:ero:under_certain_conditions_acquire_returns_found_weak:claim_one}.
        Since $I^*$'s response is $(\found, \cellpointershort_{i + 1})$, and $p$ exits $I$, we have that $p$ finds the condition on \cref{line:ero:acquire_found_check} to be true during $I'$, and so $p$ executes \cref{line:ero:acquire_update_current_unique_pointer} during $I'$; say at time $T^{\ref{line:ero:acquire_update_current_unique_pointer}}$.
        Therefore, $\currentuniquecellpointershort{} = \cellpointershort_{i + 1}$ at $T^{\ref{line:ero:acquire_update_current_unique_pointer}}$ as wanted.
        \qH{\Cref{lemma:ero:under_certain_conditions_acquire_returns_found_weak:claim_one}}
    \end{proof}

    \begin{claimcustom}{\ref{lemma:ero:under_certain_conditions_acquire_returns_found_weak}.3}\label{lemma:ero:under_certain_conditions_acquire_returns_found_weak:claim_two}
        Suppose for some $i \in [1..n]$ for every $j \in [0..i - 1]$ $\cellpointershort_j \neq \targetcellpointershort$.
        Then, for every $j \in [1..i + 1]$, (1) $p$ executes \cref{line:ero:acquire_loop_until} $j$ times during $I$ and (2) at the time $p$ executes \cref{line:ero:acquire_loop_until} for the $j$th time during $I$ the local variable $\currentuniquecellpointershort{} = \cellpointershort_{j - 1}$.
    \end{claimcustom}

    \begin{proof}
        By induction on $j$.
        \begin{itemize}
            \item[] \hspace{0pt}\textbf{Base Case.} $j = 1$.

            In this case, (1) holds immediately since $p$ must execute \cref{line:ero:acquire_loop_until} at least once during $I$ as $p$ exits $I$.
            Let $T^{\ref{line:ero:acquire_loop_until}}_1$ be the time of $p$'s first execution of \cref{line:ero:acquire_loop_until} during $I$.
            For (2), since $\currentuniquecellpointershort{}$ at $T^{\ref{line:ero:acquire_loop_until}}_1$ is the value it was initialized to on \cref{line:ero:acquire_initial_current_pointer} during $I$, we have that $\currentuniquecellpointershort{} = \&\headobject$ at $T^{\ref{line:ero:acquire_loop_until}}_1$.
            Therefore, since $\cellpointershort_0 = \&\headobject$, we have that $\currentuniquecellpointershort{} = \cellpointershort_0$ at $T^{\ref{line:ero:acquire_loop_until}}_1$ as wanted.

            \item[] \hspace{0pt}\textbf{Inductive Case.} For every $j \in [1..i]$, if (1) and (2) hold for $j$, then (1) and (2) hold for $j + 1$.

            Suppose for any $j \in [1..i]$ (1) $p$ executes \cref{line:ero:acquire_loop_until} $j$ times during $I$ and (2) at the time $p$ executes \cref{line:ero:acquire_loop_until} for the $j$th time during $I$ $\currentuniquecellpointershort{} = \cellpointershort_{j - 1}$.
            This is the inductive hypothesis.
            Let $I_j$ be the $j$th iteration of the loop on \cref{line:ero:acquire_loop_until} during $I$, which is well-defined by (1) of the inductive hypothesis.
            Furthermore, let $T^{\ref{line:ero:acquire_loop_until}}_j$ be the time of $p$'s $j$th execution of \cref{line:ero:acquire_loop_until} during $I$ which is the start of $I_j$.
            Since by (2) of the inductive hypothesis $\currentuniquecellpointershort{} = \cellpointershort_{j - 1}$ at $T^{\ref{line:ero:acquire_loop_until}}_j$ where $j - 1 \in [0..n)$ (since $j \in [1..i]$ and $i \in [1..n]$), and by assumption $\cellpointershort_{j - 1} \neq \targetcellpointershort$ (since $j - 1 \in [0..i-1]$), by \Cref{lemma:ero:under_certain_conditions_acquire_returns_found_weak:claim_one}, $p$ executes \cref{line:ero:acquire_update_current_unique_pointer} at some time $T^{\ref{line:ero:acquire_update_current_unique_pointer}}_j$ during $I_j$ and $\currentuniquecellpointershort{} = \cellpointershort_j$ at $T^{\ref{line:ero:acquire_update_current_unique_pointer}}_j$.
            Hence, since $p$ exits $I$, it follows that $p$ executes \cref{line:ero:acquire_loop_until} one more time during $I$, so $p$ executes \cref{line:ero:acquire_loop_until} $j + 1$ times during $I$.
            Since $\currentuniquecellpointershort{} = \cellpointershort_{j}$ at $T^{\ref{line:ero:acquire_update_current_unique_pointer}}_j$, and the value of $\currentuniquecellpointershort{}$ does not change between $T^{\ref{line:ero:acquire_update_current_unique_pointer}}_j$ and the time of $p$'s $j + 1$th execution of \cref{line:ero:acquire_loop_until} during $I$, it follows that at the time $p$ executes \cref{line:ero:acquire_loop_until} for the $j + 1$th time during $I$ $\currentuniquecellpointershort{} = \cellpointershort_j$.
            Therefore, (1) and (2) hold for $j + 1$ as wanted.
            \qH{\Cref{lemma:ero:under_certain_conditions_acquire_returns_found_weak:claim_two}}
        \end{itemize}
    \end{proof}

    We now complete the proof of \Cref{lemma:ero:under_certain_conditions_acquire_returns_found_weak}.
    There are two cases.
    \begin{itemize}
        \item[] \hspace{0pt}\textbf{Case 1.} $\targetcellpointershort \in \List(\mathcal{I})$.

        Since $\targetcellpointershort \in \celluniverse{}$, by \Cref{assumption:ero:head_and_null_not_in_cell_universe}, $\targetcellpointershort \neq \&\headobject$ and $\targetcellpointershort \neq \nullconstant$.
        Hence, since  $\targetcellpointershort \in \List(\mathcal{I})$, $\List(\mathcal{I}) = \cellpointershort_0, \ldots, \cellpointershort_{n + 1}$, $\cellpointershort_0 = \&\headobject$, and $\cellpointershort_{n + 1} = \nullconstant$, we have that $\targetcellpointershort = \cellpointershort_i$ for some $i \in [1..n]$.
        Since $\mathcal{I}$ is a finite prefix of $\mathcal{I}^\mathcal{B}$, $\List(\mathcal{I}) = \cellpointershort_0, \ldots, \cellpointershort_{n + 1}$, and $P(\mathcal{I}^\mathcal{B})$ holds, by \Cref{lemma:ero:pointers_in_list_are_unique}, for every $j \in [0..n+1]$ if $j \neq i$, then $\cellpointershort_j \neq \cellpointershort_i$, so $\cellpointershort_j \neq \targetcellpointershort$.
        Hence, for every $j \in [0..i-1]$ $\cellpointershort_j \neq \targetcellpointershort$.
        Thus, by \Cref{lemma:ero:under_certain_conditions_acquire_returns_found_weak:claim_two}, $p$ executes \cref{line:ero:acquire_loop_until} $i + 1$ times during $I$ and at the time $p$ executes \cref{line:ero:acquire_loop_until} for the $i + 1$th time during $I$ $\currentcellpointershort{} = \cellpointershort_i$.
        Since $\targetcellpointershort = \cellpointershort_i$, this implies that $p$ finds the condition on \cref{line:ero:acquire_loop_until} to be false during $I$.
        Therefore, $p$ executes \cref{line:ero:acquire_return} during $I$, and so $\status = \found$ as wanted.
        
        \item[] \hspace{0pt}\textbf{Case 2.} $\targetcellpointershort \notin \List(\mathcal{I})$.

        Hence, for every $j \in [1..n]$ $\cellpointershort_j \neq \targetcellpointershort$.
        Thus, by \Cref{lemma:ero:under_certain_conditions_acquire_returns_found_weak:claim_two}, $p$ executes \cref{line:ero:acquire_loop_until} $n + 1$ times during $I$ and at the time $p$ executes \cref{line:ero:acquire_loop_until} for the $n + 1$th time during $I$ $\currentuniquecellpointershort{} = \cellpointershort_{n}$.
        Let $I_{n + 1}$ be the $n + 1$th iteration of the loop on \cref{line:ero:acquire_loop_until} during $I$.
        Since $p$ exits $I$, $p$ invokes and exits the AcquireNext procedure during $I_{n + 1}$.
        Denote this invocation by $I^*$.

        We first prove that $I^*$'s response is $(\notfound, \arbitraryvalue)$ by satisfying the conditions of \Cref{lemma:ero:acquire_next_response_classification_weak}.
        Since the first parameter of $I$ is $\uniquerepositoryoperationshort_\linearizationobject{}$ and $\currentuniquecellpointershort{} = \cellpointershort_n$ at the start of $I_{n + 1}$, the parameters of $I^*$ are $(\uniquerepositoryoperationshort_\linearizationobject{}, \cellpointershort_n)$.
        Let $T^{\ref{line:ero:acquire_next_read_curr_unique_pointer}}_{I^*}$ and $T^{\ref{line:ero:acquire_next_linearization_changed_check}}_{I^*}$ by defined as in \Cref{lemma:ero:under_certain_conditions_acquire_returns_found_weak:claim_zero}, and so $\linearizationobject{}.\uniquerepositoryoperationlong{} = \uniquerepositoryoperationshort_\linearizationobject{}$ at $T^{\ref{line:ero:acquire_next_linearization_changed_check}}_{I^*}$, and $T^{\ref{line:ero:acquire_next_read_curr_unique_pointer}}_{I^*} \in [T_b, T^{\ref{line:ero:acquire_next_linearization_changed_check}}]$.
        Hence, there is a prefix of $\mathcal{I}$ during $[T_b, T^{\ref{line:ero:acquire_next_linearization_changed_check}}]$ up to and including $T^{\ref{line:ero:acquire_next_read_curr_unique_pointer}}_{I^*}$; say $\mathcal{I}'$.
        Thus, by the second assumption of \Cref{lemma:ero:under_certain_conditions_acquire_returns_found_weak}, the list of cells conforms to $\List(\mathcal{I})$ in $\mathcal{I}'$.
        So, since $\List(\mathcal{I}) = \cellpointershort_0, \ldots, \cellpointershort_{n + 1}$, by \Cref{def:ero:logical_list}, at the end of $\mathcal{I}'$ $(*\cellpointershort_n).\nextlong.\uniquecellpointercontentlong{} = \cellpointershort_{n + 1}$.
        Hence, since $\cellpointershort_{n + 1} = \nullconstant$, we have that $(*\cellpointershort_n).\nextlong.\uniquecellpointercontentlong{} = \nullconstant$ at the end of $\mathcal{I}'$.
        Therefore, we have established the following: (1) $I^*$ has parameters $(\uniquerepositoryoperationshort_\linearizationobject{}, \cellpointershort_n)$; (2) $\linearizationobject{}.\uniquerepositoryoperationshort = \uniquerepositoryoperationshort_\linearizationobject{}$ at $T^{\ref{line:ero:acquire_next_linearization_changed_check}}_{I^*}$; and (3) $(*\cellpointershort_n).\nextlong.\uniquecellpointercontentlong{} = \nullconstant$ at $T^{\ref{line:ero:acquire_next_read_curr_unique_pointer}}_{I^*}$ (this is equivalent to the end of $\mathcal{I}'$), by \Cref{lemma:ero:acquire_next_response_classification_weak}, $I^*$'s response is $(\notfound, \arbitraryvalue)$ as wanted.
        
        We now finish the proof of Case 2.
        Since $I^*$'s response is $(\notfound, \arbitraryvalue)$, and $p$ exits $I$, we have that $p$ finds the condition on \cref{line:ero:acquire_not_found_check} to be true during $I_{n + 1}$, and so $p$ executes \cref{line:ero:acquire_early_exit} during $I_{n + 1}$.
        Therefore, given the response of $I^*$, it follows that $\status = \notfound$ as wanted.
        \qH{\Cref{lemma:ero:under_certain_conditions_acquire_returns_found_weak}}
    \end{itemize}
\end{proof}

\subsubsection{The \doaddcell{} procedure has the intended effect}

In this section, we prove that the \doaddcell{} procedure with parameters $(\arbitraryvalue, \cellpointershort_\linearizationobject{})$ has the intended effect: (1) once it exits there is a successful list-add attempt for $\cellpointershort_\linearizationobject{}$; and (2) once it exits there is a successful add-response-set attempt for $\cellpointershort_\linearizationobject{}$.

\begin{lemma}\label{lemma:ero:exit_do_add_implies_done_strong}
    Consider any invocation of the \doaddcell{} procedure with a second parameter of $\cellpointershort_\linearizationobject$ which exits the loop on \cref{line:ero:add_cell_while_loop} at some time $T^{exit}$ during $\mathcal{I}^\mathcal{B}$.
    If $P(\mathcal{I}^\mathcal{B})$, $Q(\mathcal{I}^\mathcal{B})$, and $R(\mathcal{I}^\mathcal{B})$ hold, then there is a successful list-add attempt for $\cellpointershort_\linearizationobject$ before $T^{exit}$.
\end{lemma}

\begin{proof}
    Suppose, for contradiction, there is an invocation $I$ of the \doaddcell{} procedure with parameters $(\uniquerepositoryoperationshort_\linearizationobject{}, \cellpointershort_\linearizationobject)$ which exits the loop on \cref{line:ero:add_cell_while_loop} at some time $T^{exit}$ during $\mathcal{I}^\mathcal{B}$ such that there is not a successful list-add attempt for $\cellpointershort_\linearizationobject$ before $T^{exit}$ in $\mathcal{I}^\mathcal{B}$.
    Let $p$ be the process that invoked $I$.
    Since $I$ has parameters $(\uniquerepositoryoperationshort_\linearizationobject{}, \cellpointershort_\linearizationobject)$, by \Cref{lemma:ero:l_event_corresponding_to_do_low_level_op}, there is an $L$-add event $e$ for $\cellpointershort_\linearizationobject$ before $I$ was invoked that set $\linearizationobject{}$ to $(\uniquerepositoryoperationshort_\linearizationobject{}, \cellpointershort_\linearizationobject)$.
    Hence, by \Cref{lemma:ero:every_l_event_is_for_pointer_from_universe} $\cellpointershort_\linearizationobject \in \celluniverse{}$.
    Furthermore, since $e$ is before $I$ was invoked and $T^{exit}$ is after $I$ was invoked, by transitivity, $e < T^{exit}$, thus all steps during the loop \cref{line:ero:add_cell_while_loop} during $I$ are during $(e, T^{exit}]$. 
    There are two cases.
    Suppose during $(e, T^{exit}]$ there is at least one $L$-event in $\mathcal{I}^\mathcal{B}$. 
    Let $e_a$ be the next $L$-event after $e$ in $\mathcal{I}^\mathcal{B}$.
    Hence, since $e$ and $e_a$ are successive $L$-events in $\mathcal{I}^\mathcal{B}$, and $e$ is an $L$-add event for $\cellpointershort_\linearizationobject$, by $R(\mathcal{I}^\mathcal{B})$, there is a successful list-add attempt for $\cellpointershort_\linearizationobject$ during $(e, e_a)$.
    Therefore, since $e_a < T^{exit}$, by transitivity, there is a successful list-add attempt for $\cellpointershort_\linearizationobject$ before $T^{exit}$.
    However, by our initial assumption of \Cref{lemma:ero:exit_do_add_implies_done_strong}, there are no successful list-add attempts for $\cellpointershort_\linearizationobject$ before $T^{exit}$, a contradiction.
    
    Now suppose during $(e, T^{exit}]$ there are no $L$-events.
    Hence, $e$ is the last $L$-event in $\mathcal{I}$ where $\mathcal{I}$ is any prefix of $\mathcal{I}^\mathcal{B}$ during $(e, T^{exit}]$.
    Thus, since $e$ set $\linearizationobject{}.\uniquerepositoryoperationlong{} = \uniquerepositoryoperationshort_\linearizationobject{}$, by \Cref{observation:ero:where_objects_change}, $\linearizationobject{}.\uniquerepositoryoperationlong{} = \uniquerepositoryoperationshort_\linearizationobject{}$ throughout $(e, T^{exit}]$ (*).
    We first show $(e, T^{exit}]$ is desolate in two other senses.

    \begin{claimcustom}{\ref{lemma:ero:exit_do_add_implies_done_strong}.1}\label{lemma:ero:exit_do_add_implies_done_strong_claim}
        There are no successful list-add or list-remove attempts during $(e, T^{exit}]$.
    \end{claimcustom}

    \begin{proof}
        Let $\mathcal{I}$ be the prefix of $\mathcal{I}^\mathcal{B}$ up to and including $T^{exit}$, so by (*) $e$ is the last $L$-event in $\mathcal{I}$.
        Hence, since $P(\mathcal{I}^\mathcal{B})$, $Q(\mathcal{I}^\mathcal{B})$, and $R(\mathcal{I}^\mathcal{B})$ hold, and the last $L$-event in $\mathcal{I}$, $e$, is an $L$-add event for $\cellpointershort_\linearizationobject$, by \Cref{lemma:ero:1_of_r_safety_holds}, from $e$ onwards in $\mathcal{I}$ there is at most one successful list-add attempt for $\cellpointershort_\linearizationobject$ and no other successful list-add or list-remove attempts for any other pointer.
        So, since $\mathcal{I}$ is the prefix of $\mathcal{I}^\mathcal{B}$ up to and including $T^{exit}$, during $(e, T^{exit}]$ there is at most one successful list-add attempt for $\cellpointershort_\linearizationobject$ and no other successful list-add or list-remove attempts for any other pointer.
        If during $(e, T^{exit}]$ there is a successful list-add attempt for $\cellpointershort_\linearizationobject$, there would be a successful list-add attempt for $\cellpointershort_\linearizationobject$ before $T^{exit}$, contradicting our initial assumption of \Cref{lemma:ero:exit_do_add_implies_done_strong}.
        Therefore, there are no successful list-add or list-remove attempts during $(e, T^{exit}]$ as wanted.
        \qH{\Cref{lemma:ero:exit_do_add_implies_done_strong_claim}}
    \end{proof}

    \begin{claimcustom}{\ref{lemma:ero:exit_do_add_implies_done_strong}.2}\label{lemma:ero:exit_do_add_implies_done_strong_zero_claim}
        For every prefix $\mathcal{I}$ of $\mathcal{I}^\mathcal{B}$ during $(e, T^{exit}]$, the list of cells conforms to $\List(\mathcal{I}^{exclude}_e)$ in $\mathcal{I}$ where $\mathcal{I}^{exclude}_e$ is the prefix of $\mathcal{I}^\mathcal{B}$ up to but excluding $e$.
    \end{claimcustom}

    \begin{proof}
        For the first part, consider any prefix $\mathcal{I}$ of $\mathcal{I}^\mathcal{B}$ during $(e, T^{exit}]$.
        Hence, by (*) $e$ is the last $L$-event in $\mathcal{I}$, and so the last $L$-event in $\mathcal{I}$ is an $L$-add event.
        Furthermore, since by \Cref{lemma:ero:exit_do_add_implies_done_strong_claim} there are no successful list-add or list-remove attempts during $(e, T^{exit}]$, we have that from $e$ onwards in $\mathcal{I}$ there are no successful list-add or list-remove attempts.
        Thus, since $\mathcal{I}$ is finite and by assumption $P(\mathcal{I}^\mathcal{B})$, $Q(\mathcal{I}^\mathcal{B})$, and $R(\mathcal{I}^\mathcal{B})$ hold, by \Cref{lemma:ero:conditional_classification_lemma}, the list of cells conforms to $\List(\mathcal{I}_e)$ in $\mathcal{I}$ where $\mathcal{I}_e$ is the prefix of $\mathcal{I}$ up to but excluding $e$.
        Therefore, since $\mathcal{I}$ is a prefix of $\mathcal{I}^\mathcal{B}$ after $e$, $\mathcal{I}_e = \mathcal{I}^{exclude}_e$, and so the list of cells conforms to $\List(\mathcal{I}^{exclude}_e)$ in $\mathcal{I}$ as wanted.
        \qH{\Cref{lemma:ero:exit_do_add_implies_done_strong_zero_claim}}
    \end{proof}

    We now prove that $I$ ``traverses" $\List(\mathcal{I}^{exclude}_e)$.
    The following three claims will be reminiscent of the proof of \Cref{lemma:ero:under_certain_conditions_acquire_returns_found_weak}.
    Let $\List(\mathcal{I}^{exclude}_e) = \cellpointershort_0, \ldots, \cellpointershort_{n + 1}$ for some $n \geq 0$.
    Hence, by \Cref{lemma:ero:every_list_sequence_is_from_universe}, $\cellpointershort_0 = \&\headobject$, for every $i \in [1..n]$ $\cellpointershort_i \in \celluniverse{}$, and $\cellpointershort_{n + 1} = \nullconstant$.
    Furthermore, since $P(\mathcal{I}^\mathcal{B})$ holds, by \Cref{lemma:ero:l_add_for_ptr_is_in_exclude_list}, $\cellpointershort_\linearizationobject \notin \List(\mathcal{I}^{exclude}_e)$, so for every $i \in [0..n+1]$ $\cellpointershort_i \neq \cellpointershort_\linearizationobject$ (**).

    \begin{claimcustom}{\ref{lemma:ero:exit_do_add_implies_done_strong}.3}\label{lemma:ero:exit_do_add_implies_done_strong:claim_acquire_next}
        Consider any invocation $I^*$ of the AcquireNext procedure on \cref{line:ero:add_cell_acquire_next} during $I$.
        Since $I$ exits the loop on \cref{line:ero:add_cell_while_loop} $I^*$ exits.
        Let $T^{\ref{line:ero:acquire_next_read_curr_unique_pointer}}_{I^*}$ (resp. $T^{\ref{line:ero:acquire_next_linearization_changed_check}}_{I^*}$) be the last time $p$ executes \cref{line:ero:acquire_next_read_curr_unique_pointer} (resp. \cref{line:ero:acquire_next_linearization_changed_check}) during $I^*$ (these are well-defined since $I^*$ exits).
        Then, the following are true:
        \begin{compactitem}
            \item $\linearizationobject{}.\uniquerepositoryoperationlong{} = \uniquerepositoryoperationshort_\linearizationobject{}$ at $T^{\ref{line:ero:acquire_next_linearization_changed_check}}_{I^*}$; and
            \item $T^{\ref{line:ero:acquire_next_read_curr_unique_pointer}}_{I^*} \in (e, T^{exit}]$.
        \end{compactitem}
    \end{claimcustom}

    \begin{proof}
        First 1.
        Since $I^*$ began and exited during the loop on \cref{line:ero:add_cell_while_loop} in $I$, all steps during the loop on \cref{line:ero:add_cell_while_loop} during $I$ are during $(e, T^{exit}]$, and by (*) $\linearizationobject{}.\uniquerepositoryoperationlong{} = \uniquerepositoryoperationshort_\linearizationobject{}$ throughout $(e, T^{exit}]$, we have that $\linearizationobject{}.\uniquerepositoryoperationlong{} = \uniquerepositoryoperationshort_\linearizationobject{}$ throughout $I^*$.
        Hence, since $T^{\ref{line:ero:acquire_next_linearization_changed_check}}_{I^*}$ is the time of a step during $I^*$, we have that $\linearizationobject{}.\uniquerepositoryoperationlong{} = \uniquerepositoryoperationshort_\linearizationobject{}$ at $T^{\ref{line:ero:acquire_next_linearization_changed_check}}_{I^*}$.
        Now 2.
        Since all steps during the loop on \cref{line:ero:add_cell_while_loop} during $I$ are during $(e, T^{exit}]$, and $T^{\ref{line:ero:acquire_next_read_curr_unique_pointer}}_{I^*}$ is the time of a step during $I^*$, we have that $T^{\ref{line:ero:acquire_next_read_curr_unique_pointer}}_{I^*} \in (e, T^{exit}]$.
        \qH{\Cref{lemma:ero:exit_do_add_implies_done_strong:claim_acquire_next}}
    \end{proof}

    \begin{claimcustom}{\ref{lemma:ero:exit_do_add_implies_done_strong}.4}\label{lemma:ero:exit_do_add_implies_done_strong_first_claim:claim_third}
        Consider any iteration of the loop on \cref{line:ero:add_cell_while_loop} during $I$, denoted by $I'$, such that the local variable $\currentuniquecellpointershort{} = \cellpointershort_i$ for some $i \in [0..n)$ at the start of $I'$.
        Then, $p$ executes \cref{line:ero:add_cell_update_current_pointer} at time $T^{\ref{line:ero:add_cell_update_current_pointer}}$ during $I'$ and the local variable $\currentuniquecellpointershort{} = \cellpointershort_{i + 1}$ at $T^{\ref{line:ero:add_cell_update_current_pointer}}$.
    \end{claimcustom}

    \begin{proof}
        Since by (**) for every $i \in [0..n+1]$ $\cellpointershort_i \neq \cellpointershort_\linearizationobject$, and by assumption $\currentuniquecellpointershort{} = \cellpointershort_i$ at the start of $I'$ for some $i \in [0..n)$, it follows that $p$ finds the condition on \cref{line:ero:add_cell_while_loop} to be true at the start of $I'$.
        Hence, since $p$ exits the loop on \cref{line:ero:add_cell_while_loop} during $I$, $p$ begins and exits the AcquireNext procedure on \cref{line:ero:add_cell_acquire_next} during $I'$.
        Denote this invocation by $I^*$.
        
        We first prove that $I^*$'s response is $(\found, \cellpointershort_{i + 1})$ by satisfying the conditions of \Cref{lemma:ero:acquire_next_response_classification_weak}.
        Since the first parameter of $I$ is $\uniquerepositoryoperationshort_\linearizationobject{}$ and $\currentuniquecellpointershort{} = \cellpointershort_i$ at the start of $I'$, the parameters of $I^*$ are $(\uniquerepositoryoperationshort_\linearizationobject{}, \cellpointershort_i)$.
        Let $T^{\ref{line:ero:acquire_next_read_curr_unique_pointer}}_{I^*}$ and $T^{\ref{line:ero:acquire_next_linearization_changed_check}}_{I^*}$ by defined as in \Cref{lemma:ero:exit_do_add_implies_done_strong:claim_acquire_next}, and so $\linearizationobject{}.\uniquerepositoryoperationlong{} = \uniquerepositoryoperationshort_\linearizationobject{}$ at $T^{\ref{line:ero:acquire_next_linearization_changed_check}}_{I^*}$, and $T^{\ref{line:ero:acquire_next_read_curr_unique_pointer}}_{I^*} \in (e, T^{exit}]$.       
        Hence, there is a prefix of $\mathcal{I}^\mathcal{B}$ during $(e, T^{exit}]$ up to and including $T^{\ref{line:ero:acquire_next_read_curr_unique_pointer}}_{I^*}$; say $\mathcal{I}$.
        Thus, by \Cref{lemma:ero:exit_do_add_implies_done_strong_zero_claim} the list of cells conforms to $\List(\mathcal{I}^{exclude}_e)$ in $\mathcal{I}$.
        So, since $\List(\mathcal{I}^{exclude}_e) = \cellpointershort_0, \ldots, \cellpointershort_{n + 1}$, and $i \in [0..n]$, by \Cref{def:ero:logical_list}, at the end of $\mathcal{I}$ $(*\cellpointershort_i).\nextlong.\uniquecellpointercontentlong{} = \cellpointershort_{i + 1}$.
        Thus, since $\cellpointershort_{i + 1} \in \celluniverse{}$ (because $i + 1 \in [1..n]$), by \Cref{assumption:ero:head_and_null_not_in_cell_universe} $\cellpointershort_{i + 1} \neq \nullconstant$, and so $(*\cellpointershort_i).\nextlong.\uniquecellpointercontentlong{} = \cellpointershort_{i + 1} \neq \nullconstant$ at the end of $\mathcal{I}$.
        Therefore, we have established the following: (1) $I^*$ has parameters $(\uniquerepositoryoperationshort_\linearizationobject{}, \cellpointershort_i)$; (2) $\linearizationobject{}.\uniquerepositoryoperationlong{} = \uniquerepositoryoperationshort_\linearizationobject{}$ at $T^{\ref{line:ero:acquire_next_linearization_changed_check}}_{I^*}$; and (3) $(*\cellpointershort_i).\nextlong.\uniquecellpointercontentlong{} = \cellpointershort_{i + 1} \neq \nullconstant$ at $T^{\ref{line:ero:acquire_next_read_curr_unique_pointer}}_{I^*}$ (equivalently, the end of $\mathcal{I}$), and so by \Cref{lemma:ero:acquire_next_response_classification_weak}, $I^*$'s response is $(\found, \cellpointershort_{i + 1})$.
        
        We now finish the proof of \Cref{lemma:ero:exit_do_add_implies_done_strong_first_claim:claim_third}.
        Since $p$ exits the loop on \cref{line:ero:add_cell_while_loop} during $I$ and $I^*$'s response is $(\found, \cellpointershort_{i + 1})$, we have that $p$ finds the condition on \cref{line:ero:add_cell_acquire_next_found} to be true, so $p$ executes \cref{line:ero:add_cell_update_current_pointer} during $I'$; say at time $T^{\ref{line:ero:add_cell_update_current_pointer}}$.
        Therefore, $\currentuniquecellpointershort{} = \cellpointershort_{i + 1}$ at $T^{\ref{line:ero:add_cell_update_current_pointer}}$ as wanted.
        \qH{\Cref{lemma:ero:exit_do_add_implies_done_strong_first_claim:claim_third}}
    \end{proof}

    \begin{claimcustom}{\ref{lemma:ero:exit_do_add_implies_done_strong}.5}\label{lemma:ero:exit_do_add_implies_done_strong_first_claim:claim_fourth}
        For every $i \in [1..n+1]$, (1) $p$ executes \cref{line:ero:add_cell_while_loop} $i$ times during $I$ and (2) at the time $p$ executes \cref{line:ero:add_cell_while_loop} for the $i$th time during $I$ the local variable $\currentuniquecellpointershort{} = \cellpointershort_{i - 1}$.
    \end{claimcustom}

    \begin{proof}
        By induction on $i$.
        \begin{itemize}
            \item[] \hspace{0pt}\textbf{Base Case.} $i = 1$.

            In this case, (1) holds immediately since $p$ must execute \cref{line:ero:add_cell_while_loop} at least once during $I$ as $p$ exits the loop on \cref{line:ero:add_cell_while_loop} during $I$.
            Let $T^{\ref{line:ero:add_cell_while_loop}}_1$ be the time of $p$'s first execution of \cref{line:ero:add_cell_while_loop} during $I$.
            For (2), since $\currentuniquecellpointershort{}$ at $T^{\ref{line:ero:add_cell_while_loop}}_1$ is the value it was initialized to on \cref{line:ero:add_cell_initial_current_pointer} during $I$, we have that $\currentuniquecellpointershort{} = \&\headobject$ at $T^{\ref{line:ero:add_cell_while_loop}}_1$.
            Therefore, since $\cellpointershort_0 = \&\headobject$, we have that $\currentuniquecellpointershort{} = \cellpointershort_0$ at $T^{\ref{line:ero:add_cell_while_loop}}_1$.

            \item[] \hspace{0pt}\textbf{Inductive Case.} For every $i \in [1..n]$, if (1) and (2) hold for $i$, then (1) and (2) hold for $i + 1$.

            Suppose for any $i \in [1..n]$ (1) $p$ executes \cref{line:ero:add_cell_while_loop} $i$ times during $I$ and (2) at the time $p$ executes \cref{line:ero:add_cell_while_loop} for the $i$th time during $I$ $\currentuniquecellpointershort{} = \cellpointershort_{i - 1}$.
            This is the inductive hypothesis.
            Let $I_i$ be the $i$th iteration of the loop on \cref{line:ero:add_cell_while_loop} during $I$, which is well-defined by (1) of the inductive hypothesis.
            Furthermore, let $T^{\ref{line:ero:add_cell_while_loop}}_i$ be the time of $p$'s $i$th execution of \cref{line:ero:add_cell_while_loop} during $I$ which is the start of $I_i$.
            Since by (2) of the inductive hypothesis $\currentuniquecellpointershort{} = \cellpointershort_{i - 1}$ at $T^{\ref{line:ero:add_cell_while_loop}}_i$ where $i - 1 \in [0..n)$, by \Cref{lemma:ero:exit_do_add_implies_done_strong_first_claim:claim_third}, $p$ executes \cref{line:ero:add_cell_update_current_pointer} at some time $T^{\ref{line:ero:add_cell_update_current_pointer}}_i$ during $I_i$ and $\currentuniquecellpointershort{} = \cellpointershort_i$ at $T^{\ref{line:ero:add_cell_update_current_pointer}}_i$.
            Hence, since $p$ exits the loop on \cref{line:ero:add_cell_while_loop} during $I$, it follows that $p$ executes \cref{line:ero:add_cell_while_loop} one more time during $I$, so $p$ executes \cref{line:ero:add_cell_while_loop} $i + 1$ times during $I$.
            Since $\currentuniquecellpointershort{} = \cellpointershort_{i}$ at $T^{\ref{line:ero:add_cell_update_current_pointer}}_i$, and the value of $\currentuniquecellpointershort{}$ does not change between $T^{\ref{line:ero:add_cell_update_current_pointer}}_i$ and the time of $p$'s $i + 1$th execution of \cref{line:ero:add_cell_while_loop} during $I$, it follows that at the time $p$ executes \cref{line:ero:add_cell_while_loop} for the $i + 1$th time during $I$ $\currentuniquecellpointershort{} = \cellpointershort_i$.
            Therefore, (1) and (2) hold for $i + 1$ as wanted.
            \qH{\Cref{lemma:ero:exit_do_add_implies_done_strong_first_claim:claim_fourth}}
        \end{itemize}
    \end{proof}

    Now that we have established $p$ ``traverses" to the end of $\List(\mathcal{I}^{exclude}_e)$ during $I$, we are ready to prove that $p$ performs a list-add attempt for $\cellpointershort_\linearizationobject{}$ during $I$.

    \begin{claimcustom}{\ref{lemma:ero:exit_do_add_implies_done_strong}.6}\label{lemma:ero:exit_do_add_implies_done_strong_first_claim}
        $p$ performs a list-add attempt for $\cellpointershort_\linearizationobject$ during $I$.
    \end{claimcustom}

    \begin{proof}
        By \Cref{lemma:ero:exit_do_add_implies_done_strong_first_claim:claim_fourth}, $p$ executes \cref{line:ero:add_cell_while_loop} $n + 1$ times during $I$ and at the time $p$ executes \cref{line:ero:add_cell_while_loop} for the $n + 1$th time during $I$ $\currentuniquecellpointershort{} = \cellpointershort_{n}$.
        Let $I_{n + 1}$ be the $n + 1$th iteration of the loop on \cref{line:ero:add_cell_while_loop} during $I$.
        Since by (**) for every $i \in [0..n+1]$ $\cellpointershort_i \neq \cellpointershort_\linearizationobject$, and $\currentuniquecellpointershort{} = \cellpointershort_n$ at the start of $I_{n + 1}$, it follows that $p$ finds the condition on \cref{line:ero:add_cell_while_loop} to be true at the start of $\mathcal{I}_{n + 1}$.
        Hence, since $p$ exits the loop on \cref{line:ero:add_cell_while_loop} during $I$, $p$ invokes and exits the AcquireNext procedure during $I_{n + 1}$.
        Denote this execution of the AcquireNext procedure by $I^*$.

        We prove that the response of $I^*$ is $(\notfound, \arbitraryvalue)$ by satisfying the conditions of \Cref{lemma:ero:acquire_next_response_classification_weak}.
        Since the first parameter of $I$ is $\uniquerepositoryoperationshort_\linearizationobject{}$ and $\currentuniquecellpointershort{} = \cellpointershort_n$ at the start of $I_{n + 1}$, the parameters of $I^*$ are $(\uniquerepositoryoperationshort_\linearizationobject{}, \cellpointershort_n)$.
        Let $T^{\ref{line:ero:acquire_next_read_curr_unique_pointer}}_{I^*}$ and $T^{\ref{line:ero:acquire_next_linearization_changed_check}}_{I^*}$ by defined as in \Cref{lemma:ero:exit_do_add_implies_done_strong:claim_acquire_next}, and so $\linearizationobject{}.\uniquerepositoryoperationlong{} = \uniquerepositoryoperationshort_\linearizationobject{}$ at $T^{\ref{line:ero:acquire_next_linearization_changed_check}}_{I^*}$, and $T^{\ref{line:ero:acquire_next_read_curr_unique_pointer}}_{I^*} \in (e, T^{exit}]$.       
        Hence, there is a prefix of $\mathcal{I}^\mathcal{B}$ during $(e, T^{exit}]$ up to and including $T^{\ref{line:ero:acquire_next_read_curr_unique_pointer}}_{I^*}$; say $\mathcal{I}$.
        Thus, by \Cref{lemma:ero:exit_do_add_implies_done_strong_zero_claim} the list of cells conforms to $\List(\mathcal{I}^{exclude}_e)$ in $\mathcal{I}$.
        So, since $\List(\mathcal{I}^{exclude}_e) = \cellpointershort_0, \ldots, \cellpointershort_{n + 1}$, by \Cref{def:ero:logical_list}, at the end of $\mathcal{I}$ $(*\cellpointershort_n).\nextlong.\uniquecellpointercontentlong{} = \cellpointershort_{n + 1}$.
        Thus, since $\cellpointershort_{n + 1} = \nullconstant$, we have that $(*\cellpointershort_n).\nextlong.\uniquecellpointercontentlong{} = \nullconstant$ at the end of $\mathcal{I}$.
        Therefore, we have established the following: (1) $I^*$ has parameters $(\uniquerepositoryoperationshort_\linearizationobject{},  \cellpointershort_n)$; (2) $\linearizationobject{}.\uniquerepositoryoperationlong{} = \uniquerepositoryoperationshort_\linearizationobject{}$ at $T^{\ref{line:ero:acquire_next_linearization_changed_check}}_{I^*}$; and (3) $(*\cellpointershort_n).\nextlong.\uniquecellpointercontentlong{} = \nullconstant$ at $T^{\ref{line:ero:acquire_next_read_curr_unique_pointer}}_{I^*}$ (equivalently, the end of $\mathcal{I}$), and so by \Cref{lemma:ero:acquire_next_response_classification_weak}, $I^*$'s response is $(\notfound, \arbitraryvalue)$ as wanted.

        We now finish the proof of \Cref{lemma:ero:exit_do_add_implies_done_strong_first_claim}.
        Since $p$ exits the loop on \cref{line:ero:add_cell_while_loop} during $I$ and $I^*$'s response is $(\notfound, \arbitraryvalue)$, we have that $p$ finds the condition on \cref{line:ero:add_cell_acquire_next_not_found} to be true during $I_{n + 1}$.
        Let $T^{\ref{line:ero:add_cell_before_updating_end_of_list_linearization_changed_check}}$ be the time $p$ executes \cref{line:ero:add_cell_before_updating_end_of_list_linearization_changed_check} during $I_{n + 1}$.
        Since all steps during the loop on \cref{line:ero:add_cell_while_loop} during $I$ are during $(e, T^{exit}]$, it follows that $T^{\ref{line:ero:add_cell_before_updating_end_of_list_linearization_changed_check}} \in (e, T^{exit}]$.
        Hence, since by (*) $\linearizationobject{}.\uniquerepositoryoperationlong{} = \uniquerepositoryoperationshort_\linearizationobject{}$ throughout $(e, T^{exit}]$, we have that $\linearizationobject{}.\uniquerepositoryoperationlong{} = \uniquerepositoryoperationshort_\linearizationobject{}$ at $T^{\ref{line:ero:add_cell_before_updating_end_of_list_linearization_changed_check}}$.
        Thus, $p$ finds the condition on \cref{line:ero:add_cell_before_updating_end_of_list_linearization_changed_check} to be false at $T^{\ref{line:ero:add_cell_before_updating_end_of_list_linearization_changed_check}}$, and so $p$ executes \cref{line:ero:add_cell_to_list} during $I_{n + 1}$.
        Therefore, since the second parameter of $I$ is $\cellpointershort_\linearizationobject$, by \Cref{def:ero:english}, $p$ performs a list-add attempt for $\cellpointershort_\linearizationobject$ during $I$ as wanted.
        \qH{\Cref{lemma:ero:exit_do_add_implies_done_strong_first_claim}}
    \end{proof}

    Suppose this list-add attempt is after some $\currentcellpointershort{}$, so by \Cref{lemma:ero:every_list_add_attempt_is_after_a_pointer_from_universe_or_head} $\currentcellpointershort{} \in \celluniverse{} \cup \{\&\headobject\}$.
    The remainder of the proof is dedicated to showing that this list-add attempt is successful.

    \begin{claimcustom}{\ref{lemma:ero:exit_do_add_implies_done_strong}.7}\label{lemma:ero:exit_do_add_implies_done_strong_third_claim}
        $(*\currentcellpointershort{}).\nextlong = (\arbitraryvalue, \false, 0, \nullconstant)$ throughout $(e, T^{exit}]$.
    \end{claimcustom}

    \begin{proof}
        Since $e$ is an $L$-add event for $\cellpointershort_\linearizationobject$ in $\mathcal{I}^\mathcal{B}$, by $P(\mathcal{I}^\mathcal{B})$, $e$ is the only $L$-add event for $\cellpointershort_\linearizationobject$ in $\mathcal{I}^\mathcal{B}$.
        Since $p$ performs a list-add attempt $a$ for $\cellpointershort_\linearizationobject$ after $\currentcellpointershort{}$ during $(e, T^{exit}]$, by $Q(\mathcal{I}^\mathcal{B})$, $e$ is the unique $L$-add event for $\cellpointershort_\linearizationobject$ preceding $a$ and $\currentcellpointershort{}$ is the second last pointer in $\List(\mathcal{I}^{exclude}_e)$, i.e., the one preceding $\nullconstant$ (so $\currentcellpointershort{} \in \List(\mathcal{I}^{exclude}_e)$).
        Thus, since by \Cref{lemma:ero:exit_do_add_implies_done_strong_zero_claim} at every prefix $\mathcal{I}$ of $\mathcal{I}^\mathcal{B}$ during $(e, T^{exit}]$, the list of cells conforms to $\List(\mathcal{I}^{exclude}_e)$ in $\mathcal{I}$, by \Cref{def:ero:logical_list}, at the end of $\mathcal{I}$ $(*\currentcellpointershort{}).\nextlong.\uniquecellpointercontentlong{} = \nullconstant$.
        Hence, since $\mathcal{I}$ is an arbitrary prefix of $\mathcal{I}^\mathcal{B}$ during $(e, T^{exit}]$ and both $e$ and $T^{exit}$ are in $\mathcal{I}^\mathcal{B}$, we have that $(*\currentcellpointershort{}).\nextlong = \arbitraryvalue, \arbitraryvalue, \arbitraryvalue, \nullconstant)$ throughout $(e, T^{exit}]$.
        Therefore, since $\currentcellpointershort{} \in \celluniverse{} \cup \{\&\headobject\}$, by \Cref{lemma:ero:if_next_upointer_is_initial_then_acquisitions_is_zero}, $(*\currentcellpointershort{}).\nextlong = (\arbitraryvalue, \arbitraryvalue, 0, \nullconstant)$ throughout $(e, T^{exit}]$.
        So, it suffices to prove $(*\currentcellpointershort{}).\nextlong.sealed = \false$ throughout $(e, T^{exit}]$.
        
        Suppose, for contradiction,$(*\currentcellpointershort{}).\nextlong.sealed \neq \false$ sometime during $(e, T^{exit}]$.
        Hence, since $\currentcellpointershort{} \in \celluniverse{} \cup \{\&\headobject\}$, $(*\currentcellpointershort{}).\nextlong.sealed$ is initialized to $\false$, and so the value of $(*\currentcellpointershort{}).\nextlong.sealed$ changed before $T^{exit}$.
        Thus, by \Cref{observation:ero:where_objects_change}, there is a successful list-sealed attempt for $\currentcellpointershort{}$ before $T^{exit}$ and so by \Cref{lemma:ero:every_list_add_seal_and_remove_attempt_is_for_ptr_from_universe} $\currentcellpointershort{} \in \celluniverse{}$.
        Let $s$ be this successful list-seal attempt, and let $q$ be the process that executed $s$.
        Since $s$ is for $\currentcellpointershort{}$, by \Cref{def:ero:english}, $q$ executed $s$ during an invocation of the \doremovecell{} procedure with a second parameter of $\currentcellpointershort{}$.
        Hence, by \Cref{lemma:ero:l_event_corresponding_to_do_low_level_op}, there is an $L$-remove event $e_1$ for $\currentcellpointershort{}$ before $q$ invoked this procedure, and so since $q$ executed $s$ during this procedure, we have that $e_1 < s$.
        Thus, since $s < T^{exit}$, by transitivity, $e_1 < T^{exit}$.
        Therefore, since $e_1$ is an $L$-remove event for $\currentcellpointershort{}$, by \Cref{lemma:ero:remove_events_are_preceeded_by_add_events}, there is an $L$-add event $e_2$ for $\currentcellpointershort{}$ before $e_1$ in $\mathcal{I}^\mathcal{B}$.
        
        We now prove that there is an $L$-add for $\currentcellpointershort{}$ other than $e_2$ in $\mathcal{I}^\mathcal{B}$.
        Since as we established above, $\currentcellpointershort{} \in \celluniverse{}$, by \Cref{assumption:ero:head_and_null_not_in_cell_universe}, $\currentcellpointershort{} \neq \&\headobject$ and $\currentcellpointershort{} \neq \nullconstant$.
        Thus, since $\currentcellpointershort{} \in \List(\mathcal{I}^{exclude}_e)$, by \Cref{def:ero:logical_list}, there is an $L$-add event $e_3$ for $\currentcellpointershort{}$ in $\mathcal{I}^{exclude}_e$ such that from $e_3$ onwards in $\mathcal{I}^{exclude}_e$ there are no $L$-remove events for $\currentcellpointershort{}$.
        % Since $\mathcal{I}^{exclude}_e$ is the prefix of $\mathcal{I}^\mathcal{B}$ and $e_3$ is in $\mathcal{I}^{exclude}_e$, $e_3$ is in $\mathcal{I}^\mathcal{B}$.
        Since $\mathcal{I}^{exclude}_e$ is the prefix of $\mathcal{I}^\mathcal{B}$ up to but excluding $e$, $e$ is an $L$-add event, and by (*) there are no $L$-events during $(e, T^{exit}]$, it follows that during $[e_3, T^{exit}]$ there are no $L$-remove events for $\currentcellpointershort{}$.
        Thus, since $e_1$ is a $L$-remove event for $\currentcellpointershort{}$ before $T^{exit}$, we have that $e_1 < e_3$.
        Hence, since $e_2 < e_1$, it follows that $e_2 < e_3$, and so $e_2 \neq e_3$.
        Therefore, since $e_2$ and $e_3$ are both $L$-add events for $\currentcellpointershort{}$ in $\mathcal{I}^\mathcal{B}$, there are two $L$-add events for $\currentcellpointershort{}$ in $\mathcal{I}^\mathcal{B}$.
        However, by $P(\mathcal{I}^\mathcal{B})$, there is at most one $L$-add event for $\currentcellpointershort{}$ in $\mathcal{I}^\mathcal{B}$, a contradiction.
        \qH{\Cref{lemma:ero:exit_do_add_implies_done_strong_third_claim}}
    \end{proof}

    \begin{claimcustom}{\ref{lemma:ero:exit_do_add_implies_done_strong}.8}\label{lemma:ero:exit_do_add_implies_done_strong_second_claim}
        $(*\currentcellpointershort{}).\nextlong$ is unchanged throughout $(e, T^{exit}]$.
    \end{claimcustom}

    \begin{proof}
        Suppose, for contradiction, $(*\currentcellpointershort{}).\nextlong$ changes during $(e, T^{exit}]$.
        Hence, by \Cref{observation:ero:where_objects_change}, during $(e, T^{exit}]$ there is either a successful list-add attempt after $\currentcellpointershort{}$, a successful list-seal attempt for $\currentcellpointershort{}$, a successful list-remove attempt between $\currentcellpointershort{}$ and some pointer, or a successful list-acquire-next attempt after $\currentcellpointershort{}$.
        Since by \Cref{lemma:ero:exit_do_add_implies_done_strong_claim} there are no successful list-add or list-remove attempts during $(e, T^{exit}]$, there is either a successful list-seal attempt for $\currentcellpointershort{}$ or a successful list-acquire-next attempt after $\currentcellpointershort{}$.

        \begin{itemize}
            \item[] \hspace{0pt}\textbf{Case 1.} There is a successful list-seal attempt for $\currentcellpointershort{}$ during $(e, T^{exit}]$.

            Hence, by \Cref{def:ero:english}, $(*\currentcellpointershort{}).\nextlong.sealed = \true$ sometime during $(e, T^{exit}]$.
            However, by \Cref{lemma:ero:exit_do_add_implies_done_strong_third_claim}, $(*\currentcellpointershort{}).\nextlong.sealed = \false$ at all times during $(e, T^{exit}]$, a contradiction.

            \item[] \hspace{0pt}\textbf{Case 2.} There is a successful list-acquire-next attempt after $\currentcellpointershort{}$ during $(e, T^{exit}]$.

            Let $a$ be this successful attempt and let $(*\currentcellpointershort{}).\nextlong.\acquisitions = acq$ at the step before $a$.
            Since $\currentcellpointershort{} \in \celluniverse{} \cup \{\&\headobject\}$, by \Cref{lemma:ero:acquisition_counter_always_non_negative}, $acq \geq 0$.
            Since $a$ is successful, by \Cref{def:ero:english}, $a$ sets $(*\currentcellpointershort{}).\nextlong.\acquisitions = acq + 1$.
            Hence, $(*\currentcellpointershort{}). \nextlong.\acquisitions > 0$ at $a$.
            Therefore, $(*\currentcellpointershort{}).\nextlong.\acquisitions > 0$ some time during $(e, T^{exit}]$.
            However, by \Cref{lemma:ero:exit_do_add_implies_done_strong_third_claim}, $(*\currentcellpointershort{}).\nextlong.\acquisitions = 0$ at all times during $(e, T^{exit}]$, a contradiction.
            \qH{\Cref{lemma:ero:exit_do_add_implies_done_strong_second_claim}}
        \end{itemize}
    \end{proof}

    We now return to the proof of \Cref{lemma:ero:exit_do_add_implies_done_strong}.
    By \Cref{lemma:ero:exit_do_add_implies_done_strong_first_claim} $p$ performs a list-add attempt $a$ for $\cellpointershort_\linearizationobject$.
    As defined afterwards, $a$ is after $\currentcellpointershort{}$.
    Since all steps during the loop on \cref{line:ero:add_cell_while_loop} during $I$ are during $(e, T^{exit}]$, we have that $a \in (e, T^{exit}]$, and since $a$ is neither $e$ or the last step in the loop, we have that the step before $a$ is in $(e, T^{exit}]$.
    Hence, by \Cref{lemma:ero:exit_do_add_implies_done_strong_third_claim} $(*\currentcellpointershort{}).\nextlong = (\arbitraryvalue, \false, 0, \nullconstant)$ at the step before $a$.
    Furthermore, since by \Cref{lemma:ero:exit_do_add_implies_done_strong_claim} there are no successful list-add or list-remove attempts during $(e, T^{exit}]$, we have that $a$ is unsuccessful.
    Hence, $(*\currentcellpointershort{}).\nextlong \neq (v, \false, 0, \nullconstant)$ at the step before $a$ where $(*\currentcellpointershort{}).\nextlong.\view = v$ at $p$'s last execution of \cref{line:ero:add_cell_read_end_of_list} before $a$; say at time $T^{\ref{line:ero:add_cell_read_end_of_list}}$.
    Since $T^{\ref{line:ero:add_cell_read_end_of_list}}$ is during the loop on \cref{line:ero:add_cell_while_loop} during $I$ and all steps during the loop on \cref{line:ero:add_cell_while_loop} during $I$ are during $(e, T^{exit}]$, we have that $T^{\ref{line:ero:add_cell_read_end_of_list}} \in (e, T^{exit}]$.
    Hence, since $a \in (e, T^{exit}]$ and $T^{\ref{line:ero:add_cell_read_end_of_list}} < a$, by transitivity, $[T^{\ref{line:ero:add_cell_read_end_of_list}}, a] \subseteq (e, T^{exit}]$.
    Thus, by \Cref{lemma:ero:exit_do_add_implies_done_strong_second_claim}, $(*\currentcellpointershort{}).\nextlong.\view$ is unchanged throughout $[T^{\ref{line:ero:add_cell_read_end_of_list}}, a]$.
    So, since $(*\currentcellpointershort{}).\nextlong.\view = v$ at $T^{\ref{line:ero:add_cell_read_end_of_list}}$, we have that $(*\currentcellpointershort{}).\nextlong.\view = v$ at the step before $a$.
    Therefore, since $(*\currentcellpointershort{}).\nextlong = (\arbitraryvalue, \false, 0, \nullconstant)$ at the step before $a$, we have that $(*\currentcellpointershort{}).\nextlong = (v, \false, 0, \nullconstant)$ at the step before $a$.
    However, as we established above, $(*\currentcellpointershort{}).\nextlong \neq (v, \false, 0, \nullconstant)$ at at the step before $a$, a contradiction.
    \qH{\Cref{lemma:ero:exit_do_add_implies_done_strong}}
\end{proof}

\begin{lemma}\label{lemma:ero:exit_add_implies_response_set}
    Consider any invocation of the \doaddcell{} procedure with a second parameter of $\cellpointershort_\linearizationobject$ which ends at some time $T^{exit}$ during $\mathcal{I}^\mathcal{B}$.
    If $P(\mathcal{I}^\mathcal{B})$, $Q(\mathcal{I}^\mathcal{B})$, and $R(\mathcal{I}^\mathcal{B})$ hold, then there is a successful add-response-set attempt for $\cellpointershort_\linearizationobject$ before $T^{exit}$.
\end{lemma}

\begin{proof}
    Consider any invocation $I$ of the \doaddcell{} procedure with parameters $(\uniquerepositoryoperationshort_\linearizationobject{}, \cellpointershort_\linearizationobject)$ which ends at some time $T^{exit}$ during $\mathcal{I}^\mathcal{B}$.
    Let $p$ be the process that invoked $I$.
    Since $I$ has parameters $(\uniquerepositoryoperationshort_\linearizationobject{}, \cellpointershort_\linearizationobject)$, by \Cref{lemma:ero:l_event_corresponding_to_do_low_level_op}, there is an $L$-add event $e$ for $\cellpointershort_\linearizationobject$ before $I$ was invoked that set $\linearizationobject{}$ to $(\uniquerepositoryoperationshort_\linearizationobject{}, \cellpointershort_\linearizationobject)$.
    Hence, by \Cref{lemma:ero:every_l_event_is_for_pointer_from_universe}, $\cellpointershort_\linearizationobject{} \in \celluniverse$.

    The proof strategy is to identify an add-response-set attempt for $\cellpointershort_\linearizationobject{}$ before $T^{exit}$.
    If this is successful, we are done, but if it is unsuccessful, then by \Cref{lemma:ero:unsuccessful_add_response_set_implies_successful_add_response_set}, there is a successful add-response-set attempt for $\cellpointershort_\linearizationobject{}$ beforehand, in which case we are also done.
   % We start by identifying the \setrepositoryoperationresponse{} procedure in which this attempt occurs.

    \begin{claimcustom}{\ref{lemma:ero:exit_add_implies_response_set}.1}\label{lemma:ero:exit_add_implies_response_set:claim_zero}
        Consider any invocation $I'$ of the \setrepositoryoperationresponse{} procedure invoked on \cref{line:ero:add_cell_set_response} with parameters $(\uniquerepositoryoperationshort_\linearizationobject{}, \cellpointershort_\linearizationobject, \done)$.
        Then, there is a successful list-add attempt $a$ for $\cellpointershort_\linearizationobject{}$ such that $a$ is before $I'$ was invoked and $a$ is before $e$.
    \end{claimcustom}

    \begin{proof}
        We first prove that there is a successful list-add attempt for $\cellpointershort_\linearizationobject{}$ before $I'$ was invoked.
        Let $q$ be the process that invoked $I'$.
        Since $q$ invoked $I'$ on \cref{line:ero:add_cell_set_response}, we have that $q$ invoked $I'$ during some invocation $I^*$ of the \doaddcell{} procedure, and since the parameter of $I'$ are $(\uniquerepositoryoperationshort_\linearizationobject{}, \cellpointershort_\linearizationobject, \done)$, it follows that the parameters of $I^*$ are $(\uniquerepositoryoperationshort_\linearizationobject{}, \cellpointershort_\linearizationobject)$.
        Furthermore, since $q$ invoked $I'$ on \cref{line:ero:add_cell_set_response} during $I^*$, we have that $q$ exited the loop on \cref{line:ero:add_cell_while_loop} at some time $T$ during $I^*$.
        Therefore, since $P(\mathcal{I}^\mathcal{B})$, $Q(\mathcal{I}^\mathcal{B})$, and $R(\mathcal{I}^\mathcal{B})$ hold, by \Cref{lemma:ero:exit_do_add_implies_done_strong}, there is a successful list-add attempt $a$ for $\cellpointershort_\linearizationobject{}$ before $T$ which is before $I'$ was invoked as wanted.
        We now prove that $e < a$.
        Suppose, for contradiction, $a \leq e$.
        Hence, since $a$ is a successful list-add attempt for $\cellpointershort_\linearizationobject{}$, by \Cref{lemma:ero:l_event_corresponding_to_do_low_level_op}, there is a $L$-add event $e'$ for $\cellpointershort_\linearizationobject{}$ before $a$.
        Thus, since $e' < a$ and $a \leq e$, by transitivity $e' < e$, so $e' \neq e$.
        Therefore, there are two $L$-add events for $\cellpointershort_\linearizationobject{}$ in $\mathcal{I}^\mathcal{B}$.
        However, by $P(\mathcal{I}^\mathcal{B})$, there is at most one $L$-add event for $\cellpointershort_\linearizationobject{}$ in $\mathcal{I}^\mathcal{B}$, a contradiction.
        \qH{\Cref{lemma:ero:exit_add_implies_response_set:claim_zero}}
    \end{proof}

    \begin{claimcustom}{\ref{lemma:ero:exit_add_implies_response_set}.2}\label{lemma:ero:exit_add_implies_response_set:claim_one}
        The \setrepositoryoperationresponse{} procedure was invoked with parameters $(\uniquerepositoryoperationshort_\linearizationobject{}, \cellpointershort_\linearizationobject, \done)$ at some time $T'_b$ and ends at some time $T'_e$ during $\mathcal{I}^\mathcal{B}$ such that: (1) $[T'_b, T'_e] \subseteq (e, T^{exit}]$; (2) $e$ is the last $L$-event in $\mathcal{I}$ where $\mathcal{I}$ is any prefix of $\mathcal{I}^\mathcal{B}$ during $[T'_b, T'_e]$; (3) $\linearizationobject.\uniquerepositoryoperationlong = \uniquerepositoryoperationshort_\linearizationobject{}$ at the end of $\mathcal{I}$ where $\mathcal{I}$ is as in (2); and (4) the list of cells conforms to $\List(\mathcal{I})$ in $\mathcal{I}$ where $\mathcal{I}$ is as in (2).
    \end{claimcustom}

    \begin{proof}
        There are two cases.
        \begin{itemize}
            \item[] \hspace{0pt}\textbf{Case 1.} During $(e, T^{exit}]$ there are no $L$-events.

            Since $p$ exited $I$, we have that $p$ invoked the \setrepositoryoperationresponse{} procedure on \cref{line:ero:add_cell_set_response} during $I$.
            Denote this invocation by $I'$.
            We prove that $I'$ is the desired invocation.
            Since $I$'s parameters are $(\uniquerepositoryoperationshort_\linearizationobject{}, \cellpointershort_\linearizationobject)$, and $I'$ is invoked during $I$, we have that $I'$'s parameters are $(\uniquerepositoryoperationshort_\linearizationobject{}, \cellpointershort_\linearizationobject, \done)$.
            Let $T'_b$ and $T'_e$ be the times that $p$ invokes and exits $I'$, respectively.
            We first prove (1).
            Since $e$ is before $I$ was invoked, and $I'$ was invoked during $I$, by transitivity, $e < T'_b$.
            Furthermore, since $I'$ exits before $I$ does, by transitivity, $T'_e < T^{exit}$.
            Together these imply $[T'_b, T'_e] \subseteq (e, T^{exit}]$.
            We now prove (2).
            Consider any prefix $\mathcal{I}$ of $\mathcal{I}^\mathcal{B}$ during $[T'_b, T'_e]$.
            Since $[T'_b, T'_e] \subseteq (e, T^{exit}]$, and by assumption there are no $L$-events during $(e, T^{exit}]$, we have that $e$ is the last $L$-event in $\mathcal{I}$.
            We now prove (3).
            Since by assumption of Case 1 there are no $L$-events during $(e, T^{exit}]$, and $e$ set $\linearizationobject{}.\uniquerepositoryoperationlong = \uniquerepositoryoperationshort_\linearizationobject{}$, by \Cref{observation:ero:where_objects_change}, $\linearizationobject.\uniquerepositoryoperationlong = \uniquerepositoryoperationshort_\linearizationobject{}$ throughout $(e, T^{exit}]$.
            Hence, since $[T'_b, T'_e] \subseteq (e, T^{exit}]$, we have that $\linearizationobject.\uniquerepositoryoperationlong = \uniquerepositoryoperationshort_\linearizationobject{}$ throughout $[T'_b, T'_e]$.
            This implies (3).
            We now prove (4).
            By \Cref{lemma:ero:exit_add_implies_response_set:claim_zero}, there is a successful list-add attempt $a$ for $\cellpointershort_\linearizationobject{}$ before $T'_b$ and $e < a$.
            Thus, $a$ is in $\mathcal{I}$.
            So, since $e$ is the last $L$-event in $\mathcal{I}$, and $e < a$, we have that from the last $L$-event in $\mathcal{I}$ onwards, there is a successful list-add attempt.
            Therefore, since $\mathcal{I}$ is finite, and $P(\mathcal{I}^\mathcal{B})$, $Q(\mathcal{I}^\mathcal{B})$, and $R(\mathcal{I}^\mathcal{B})$ hold, by \Cref{lemma:ero:conditional_classification_lemma}, the list of cells conforms to $\List(\mathcal{I})$ in $\mathcal{I}$.
    
            \item[] \hspace{0pt}\textbf{Case 2.} During $(e, T^{exit}]$ there is at least one $L$-event.
    
            % We identify an earlier invocation than the one during $I$ that satisfies the requirements of the claim.
            We identify an earlier invocation than the one during $I$.
            Let $e'$ be the next $L$-event after $e$ in $\mathcal{I}^\mathcal{B}$, so $e' \in (e, T^{exit}]$.
            Let $q$ be the process that executed $e'$.
            Hence, there are no $L$-events during $(e, e')$.
            Let $T^{\ref{line:ero:linearization_read}}_q$ be the time of $q$'s last execution of \cref{line:ero:linearization_read} before $e'$.
            Since $e$ and $e'$ are successive $L$-events and $T^{\ref{line:ero:linearization_read}}_q$ is the time of $q$'s last execution of \cref{line:ero:linearization_read} before $e'$, by \Cref{lemma:ero:successive_l_event_read_previous_l_event_value}, $q$ read the value that $e$ set $\linearizationobject{}$ to on \cref{line:ero:linearization_read} at $T^{\ref{line:ero:linearization_read}}_q$, and since $P(\mathcal{I}^\mathcal{B})$ holds, by \Cref{lemma:ero:successive_l_event_read_previous_l_event_value_after_it_happened}, $e < T^{\ref{line:ero:linearization_read}}_q$.
            Hence, since $e$ set $\linearizationobject{}$ to $(\uniquerepositoryoperationshort_\linearizationobject{}, \cellpointershort_\linearizationobject)$, we have that $q$ read $(\uniquerepositoryoperationshort_\linearizationobject{}, \cellpointershort_\linearizationobject)$ from $\linearizationobject$ on \cref{line:ero:linearization_read} at $T^{\ref{line:ero:linearization_read}}_q$.
            Thus, since $\uniquerepositoryoperationshort_\linearizationobject{} = (\arbitraryvalue, \addcell)$ (because $e$ set $\linearizationobject{}.\uniquerepositoryoperationlong{} = \uniquerepositoryoperationshort_\linearizationobject{}$ and $e$ is an $L$-add event), we have that between $T^{\ref{line:ero:linearization_read}}_q$ and $e'$, $q$ invoked and exited the \doaddcell{} procedure on \cref{line:ero:do_add_cell} with parameters $(\uniquerepositoryoperationshort_\linearizationobject{}, \cellpointershort_\linearizationobject)$.
            Denote this invocation by $I^*$ and the time $q$ exited it by $T^{exit}_*$, so $T^{exit}_* < e'$.
            Since $q$ exited $I^*$, we have that $q$ invoked the \setrepositoryoperationresponse{} procedure on \cref{line:ero:add_cell_set_response} during $I^*$.
            Denote this invocation by $I'$.
            We claim that $I'$ is the desired invocation.
            Since $I^*$'s parameters are $(\uniquerepositoryoperationshort_\linearizationobject{}, \cellpointershort_\linearizationobject)$, and $I'$ is invoked during $I^*$, we have that $I'$'s parameters are $(\uniquerepositoryoperationshort_\linearizationobject{}, \cellpointershort_\linearizationobject, \done)$.
            Let $T'_b$ and $T'_e$ be the times $q$ invokes and exits $I'$, respectively.
            We first prove (1)
            Since $e < T^{\ref{line:ero:linearization_read}}_q$, $T^{\ref{line:ero:linearization_read}}_q$ is before $I^*$ was invoked, and $I'$ was invoked during $I^*$, by transitivity, $e < T'_b$.
            Furthermore, since $T'_e < T^{exit}_*$ (because $I'$ exits during $I^*$), $T^{exit}_* < e'$, by transitivity, $T'_e < e'$.
            Together these imply $[T'_b, T'_e] \subseteq (e, e')$, and since $e' \leq T^{exit}$, by transitivity, $[T'_b, T'_e] \subseteq (e,  T^{exit}]$.
            We now prove (2).
            Consider any prefix $\mathcal{I}$ of $\mathcal{I}^\mathcal{B}$ during $[T'_b, T'_e]$.
            Since $[T'_b, T'_e] \subseteq (e, e')$, and there are no $L$-events during $(e, e')$, we have that $e$ is the last $L$-event in $\mathcal{I}$.
            We now prove (3).
            Since there are no $L$-events during $(e, e')$ and $e$ set $\linearizationobject.\uniquerepositoryoperationlong = \uniquerepositoryoperationshort_\linearizationobject{}$, by \Cref{observation:ero:where_objects_change}, $\linearizationobject.\uniquerepositoryoperationlong = \uniquerepositoryoperationshort_\linearizationobject{}$ throughout $(e, e')$.
            Hence, since $[T'_b, T'_e] \subseteq (e, e')$, we have that $\linearizationobject.\uniquerepositoryoperationlong = \uniquerepositoryoperationshort_\linearizationobject{}$ throughout $[T'_b, T'_e]$.
            This implies (3).
            We now prove (4).
            By \Cref{lemma:ero:exit_add_implies_response_set:claim_zero}, there is a successful list-add attempt $a$ for $\cellpointershort_\linearizationobject{}$ before $T'_b$ and $e < a$.
            Thus, $a$ is in $\mathcal{I}$.
            So, since $e$ is the last $L$-event in $\mathcal{I}$, and $e < a$, we have that from the last $L$-event in $\mathcal{I}$ onwards, there is a successful list-add attempt.
            Therefore, since $\mathcal{I}$ is finite, and $P(\mathcal{I}^\mathcal{B})$, $Q(\mathcal{I}^\mathcal{B})$, and $R(\mathcal{I}^\mathcal{B})$ hold, by \Cref{lemma:ero:conditional_classification_lemma}, the list of cells conforms to $\List(\mathcal{I})$ in $\mathcal{I}$.
            \qH{\Cref{lemma:ero:exit_add_implies_response_set:claim_one}}
        \end{itemize}
    \end{proof}

    \begin{claimcustom}{\ref{lemma:ero:exit_add_implies_response_set}.3}\label{lemma:ero:exit_add_implies_response_set:claim_two}
        There is an add-response-set attempt for $\cellpointershort_\linearizationobject{}$ before $T^{exit}$.
    \end{claimcustom}

    \begin{proof}
        Let $I'$ be the invocation of the \setrepositoryoperationresponse{} procedure identified in \Cref{lemma:ero:exit_add_implies_response_set:claim_one} and let $q$ be the process that executed $I'$.
        Furthermore, let $T'_b$ and $T'_e$ be the times during $\mathcal{I}^\mathcal{B}$ that $q$ begins and exits $I'$, respectively.
        Since $q$ exits $I'$, $q$ began and exited the Acquire procedure on \cref{line:ero:set_response_acquire} during $I'$.
        Denote this invocation of the Acquire procedure by $I^*$.
        Since the parameters of $I'$ are $(\uniquerepositoryoperationshort_\linearizationobject{}, \cellpointershort_\linearizationobject, \done)$, we have that the parameters of $I^*$ are $(\uniquerepositoryoperationshort_\linearizationobject{}, \cellpointershort_\linearizationobject)$.
        
        We now satisfy the conditions of \Cref{lemma:ero:under_certain_conditions_acquire_returns_found_weak}.
        As established above, $\cellpointershort_\linearizationobject{} \in \celluniverse{}$.
        Let $T_b$ be the time $q$ invoked $I^*$ and let $T^{\ref{line:ero:acquire_next_linearization_changed_check}}$ be the last time $q$ executes \cref{line:ero:acquire_next_linearization_changed_check} during $I^*$.
        Recall that $T^{\ref{line:ero:acquire_next_linearization_changed_check}}$ is well-defined by \Cref{lemma:ero:exit_acquire_implies_executing_105}.
        By assumption $P(\mathcal{I}^\mathcal{B})$ holds.
        Let $\mathcal{I}^{include}_e$ be the prefix of $\mathcal{I}^\mathcal{B}$ up to and including $e$.
        We plug in $\mathcal{I}^{include}_e$ for $\mathcal{I}$ in \Cref{lemma:ero:under_certain_conditions_acquire_returns_found_weak}.
        Consider any prefix $\mathcal{I}^*$ of $\mathcal{I}^\mathcal{B}$ during $[T_b, T^{\ref{line:ero:acquire_next_linearization_changed_check}}]$.
        We first satisfy condition 1.
        Since by definition all steps during $[T_b, T^{\ref{line:ero:acquire_next_linearization_changed_check}}]$ are during $I^*$, and all steps during $I^*$ are during $I'$, we have that $[T_b, T^{\ref{line:ero:acquire_next_linearization_changed_check}}] \subseteq [T'_b, T'_e]$.
        Hence, since by (3) of \Cref{lemma:ero:exit_add_implies_response_set:claim_one}, $\linearizationobject{}.\uniquerepositoryoperationlong{} = \uniquerepositoryoperationshort_\linearizationobject{}$ at the end of $\mathcal{I}'$ for any prefix $\mathcal{I}'$ of $\mathcal{I}^\mathcal{B}$ during $[T'_b, T'_e]$, we have that $\linearizationobject{}.\uniquerepositoryoperationlong{} = \uniquerepositoryoperationshort_\linearizationobject{}$ at the end of $\mathcal{I}^*$.
        We now satisfy condition 2.
        Since by (4) of \Cref{lemma:ero:exit_add_implies_response_set:claim_one} the list of cells conforms to $\List(\mathcal{I}')$ in $\mathcal{I}'$ for any prefix $\mathcal{I}'$ of $\mathcal{I}^\mathcal{B}$ during $[T'_b, T'_e]$, and $[T_b, T^{\ref{line:ero:acquire_next_linearization_changed_check}}] \subseteq [T'_b, T'_e]$, we have that that the list of cells conforms to $\List(\mathcal{I}^*)$ in $\mathcal{I}^*$.
        We now prove that $\List(\mathcal{I}^{include}_e) = \List(\mathcal{I}^*)$.
        Since $\mathcal{I}^{include}_e$ is the prefix of $\mathcal{I}^\mathcal{B}$ up to and including $e$, and $e$ is an $L$-event, it follows that $e$ is the last $L$-event in $\mathcal{I}^{include}_e$.
        Furthermore, since by (2) of \Cref{lemma:ero:exit_add_implies_response_set:claim_one} $e$ is the last $L$-event in $\mathcal{I}'$ for any prefix $\mathcal{I}'$ of $\mathcal{I}^\mathcal{B}$ during $[T'_b, T'_e]$, and $[T_b, T^{\ref{line:ero:acquire_next_linearization_changed_check}}] \subseteq [T'_b, T'_e]$, we have that $e$ is the last $L$-event in $\mathcal{I}^*$.
        Together, these imply that the sequence of $L$-events is the same in $\mathcal{I}^{include}_e$ and $\mathcal{I}^*$, and so by \Cref{def:ero:logical_list}, $\List(\mathcal{I}^{include}_e) = \List(\mathcal{I}^*)$.
        Therefore, since the list of cells conforms to $\List(\mathcal{I}^*)$ in $\mathcal{I}^*$, we have that the list of cells conforms to $\List(\mathcal{I}^{include}_e)$ in $\mathcal{I}^*$ as wanted.
        So, by \Cref{lemma:ero:under_certain_conditions_acquire_returns_found_weak} if $\cellpointershort_\linearizationobject \in \List(\mathcal{I}^{include}_e)$ then the response of $I^*$ is $\found$.

        We now finish the proof of \Cref{lemma:ero:exit_add_implies_response_set:claim_two}.
        Since $e$ is an $L$-add event for $\cellpointershort_\linearizationobject{}$, and $e$ is the last $L$-event in $\mathcal{I}^{include}_e$, we have that there are no $L$-remove events after $e$ for $\cellpointershort_\linearizationobject{}$ in $\mathcal{I}^{include}_e$.
        Hence, by \Cref{def:ero:logical_list}, $\cellpointershort_\linearizationobject \in \List(\mathcal{I}^{include}_e)$, so the response of $I^*$ is $\found$.
        Thus, since $q$ exits $I'$, we have that $q$ finds the condition on \cref{line:ero:set_response_found_check} to be true during $I'$.
        So, $q$ executes \cref{line:ero:responses_set_attempt} during $I'$.
        Since $I^*$'s parameters are $(\uniquerepositoryoperationshort_\linearizationobject{}, \cellpointershort_\linearizationobject, \done)$ and $\uniquerepositoryoperationshort_\linearizationobject{} = (\arbitraryvalue, \addcell)$ (because $e$ set $\linearizationobject{}.\uniquerepositoryoperationlong{} = \uniquerepositoryoperationshort_\linearizationobject{}$ and $e$ is an $L$-add event), by \Cref{def:ero:english}, this execution is an add-response-set attempt for $\cellpointershort_\linearizationobject$ during $I'$ (and thus $[T'_b, T'_e]$).
        Therefore, since by (1) of \Cref{lemma:ero:exit_add_implies_response_set:claim_one} $[T'_b, T'_e] \subseteq (e, T^{exit}]$, we have that there is an add-response-set attempt for $\cellpointershort_\linearizationobject{}$ before $T^{exit}$ as wanted.
        \qH{\Cref{lemma:ero:exit_add_implies_response_set:claim_two}}
    \end{proof}

    We now return to the proof of \Cref{lemma:ero:exit_add_implies_response_set}.
    Let $a$ be the add-response-set attempt for $\cellpointershort_\linearizationobject{}$ identified by \Cref{lemma:ero:exit_add_implies_response_set:claim_two}.
    Since $a$ is an add-response-set attempt for $\cellpointershort_\linearizationobject{}$ before $T^{exit}$, if $a$ is successful, we have satisfied the claim.
    If $a$ is unsuccessful, then by \Cref{lemma:ero:unsuccessful_add_response_set_implies_successful_add_response_set}, there is a successful add-response-set attempt for $\cellpointershort_\linearizationobject{}$ before $a$ (and thus before $T^{exit}$).
    Therefore, in either case, there is a successful add-response-set attempt for $\cellpointershort_\linearizationobject{}$ before $T^{exit}$ as wanted.
    \qH{\Cref{lemma:ero:exit_add_implies_response_set}}
\end{proof}

\subsubsection{The \doremovecell{} procedure has the intended effect}

In this section, we prove that the \doremovecell{} procedure with parameters $(\arbitraryvalue, \cellpointershort_\linearizationobject{})$ has the intended effect: (1) once it exits there is a successful list-remove attempt for $\cellpointershort_\linearizationobject{}$; and (2) once it exits there is a successful remove-response-set attempt for $\cellpointershort_\linearizationobject{}$.

\begin{lemma}\label{lemma:ero:exit_do_remove_implies_done}
    Consider any invocation $I$ of the \doremovecell{} procedure with a second parameter of $\cellpointershort_\linearizationobject$ which ends at some time $T^{exit}$ during $\mathcal{I}^\mathcal{B}$.
    If $P(\mathcal{I}^\mathcal{B})$, $Q(\mathcal{I}^\mathcal{B})$, and $R(\mathcal{I}^\mathcal{B})$ hold, then there is a successful list-remove attempt for $\cellpointershort_\linearizationobject$ before $T^{exit}$ in $\mathcal{I}^\mathcal{B}$.
\end{lemma}

\begin{proof}
    Suppose, for contradiction, there is an invocation $I$ of the \doremovecell{} procedure with parameters $(\uniquerepositoryoperationshort_\linearizationobject{}, \cellpointershort_\linearizationobject)$ which ends at some time $T^{exit}$ during $\mathcal{I}^\mathcal{B}$ such that there is not a successful list-remove attempt for $\cellpointershort_\linearizationobject$ before $T^{exit}$ in $\mathcal{I}^\mathcal{B}$.
    Let $p$ be the process that invoked $I$.
    Since $I$ has parameters $(\uniquerepositoryoperationshort_\linearizationobject{}, \cellpointershort_\linearizationobject)$, by \Cref{lemma:ero:l_event_corresponding_to_do_low_level_op}, there is an $L$-remove event $e$ for $\cellpointershort_\linearizationobject$ before $I$ was invoked that set $\linearizationobject{}$ to $(\uniquerepositoryoperationshort_\linearizationobject{}, \cellpointershort_\linearizationobject)$.
    Hence, by \Cref{lemma:ero:every_l_event_is_for_pointer_from_universe} $\cellpointershort_\linearizationobject \in \celluniverse{}$.
    Furthermore, since $e$ is before $I$ was invoked and $T^{exit}$ is after $I$ was invoked, by transitivity, $e < T^{exit}$, thus all steps during the loop \cref{line:ero:remove_cell_while_loop} during $I$ are during $(e, T^{exit}]$. 
    There are two cases.
    Suppose during $(e, T^{exit}]$ there is at least one $L$-event in $\mathcal{I}^\mathcal{B}$. 
    Let $e_a$ be the next $L$-event after $e$ in $\mathcal{I}^\mathcal{B}$.
    Hence, since $e$ and $e_a$ are successive $L$-events in $\mathcal{I}^\mathcal{B}$, and $e$ is an $L$-remove event for $\cellpointershort_\linearizationobject$, by $R(\mathcal{I}^\mathcal{B})$, there is a successful list-remove attempt for $\cellpointershort_\linearizationobject$ during $(e, e_a)$.
    Therefore, since $e_a < T^{exit}$, there is a successful list-remove attempt for $\cellpointershort_\linearizationobject$ before $T^{exit}$.
    However, by our initial assumption of \Cref{lemma:ero:exit_do_remove_implies_done}, there are no successful list-remove attempts for $\cellpointershort_\linearizationobject$ before $T^{exit}$, a contradiction.
    
    Now suppose during $(e, T^{exit}]$ there are no $L$-events in $\mathcal{I}$.
    Hence, $e$ is the last $L$-event in $\mathcal{I}$ where $\mathcal{I}$ is any prefix of $\mathcal{I}^\mathcal{B}$ during $(e, T^{exit}]$.
    Thus, since $e$ set $\linearizationobject{}.\uniquerepositoryoperationlong{} = \uniquerepositoryoperationshort_\linearizationobject{}$, by \Cref{observation:ero:where_objects_change}, $\linearizationobject{}.\uniquerepositoryoperationlong{} = \uniquerepositoryoperationshort_\linearizationobject{}$ throughout $(e, T^{exit}]$ (*).
    We first show $(e, T^{exit}]$ is desolate in two other senses.

    \begin{claimcustom}{\ref{lemma:ero:exit_do_remove_implies_done}.1}\label{lemma:ero:exit_do_remove_implies_done_claim}
        There are no successful list-add or list-remove attempts during $(e, T^{exit}]$.
    \end{claimcustom}

    \begin{proof}
        Let $\mathcal{I}$ be the prefix of $\mathcal{I}^\mathcal{B}$ up to and including $T^{exit}$, so by (*) $e$ is the last $L$-event in $\mathcal{I}$.
        Hence, since $P(\mathcal{I}^\mathcal{B})$, $Q(\mathcal{I}^\mathcal{B})$, and $R(\mathcal{I}^\mathcal{B})$ hold, and the last $L$-event in $\mathcal{I}$, $e$, is an $L$-remove event for $\cellpointershort_\linearizationobject$, by \Cref{lemma:ero:3_of_r_safety_holds}, from $e$ onwards in $\mathcal{I}$ there is at most one successful list-remove attempt for $\cellpointershort_\linearizationobject$ and no other successful list-add or list-remove attempts for any other pointer.
        Hence, since $\mathcal{I}$ is the prefix of $\mathcal{I}^\mathcal{B}$ up to and including $T^{exit}$, during $(e, T^{exit}]$ there is at most one successful list-remove attempt for $\cellpointershort_\linearizationobject$ and no other successful list-add or list-remove attempts for any other pointer.
        If during $(e, T^{exit}]$ there is a successful list-remove attempt for $\cellpointershort_\linearizationobject$, there would be a successful list-remove attempt for $\cellpointershort_\linearizationobject$ before $T^{exit}$, contradicting our initial assumption of \Cref{lemma:ero:exit_do_remove_implies_done}.
        Therefore, there are no successful list-add or list-remove attempts during $(e, T^{exit}]$.
        \qH{\Cref{lemma:ero:exit_do_remove_implies_done_claim}}
    \end{proof}

   \begin{claimcustom}{\ref{lemma:ero:exit_do_remove_implies_done}.2}\label{lemma:ero:exit_do_remove_implies_done_zero_claim}
        At every prefix $\mathcal{I}$ of $\mathcal{I}^\mathcal{B}$ during $(e, T^{exit}]$, the list of cells conforms to $\List(\mathcal{I}^{exclude}_e)$ in $\mathcal{I}$ where $\mathcal{I}^{exclude}_e$ is the prefix of $\mathcal{I}^\mathcal{B}$ up to but excluding $e$.
    \end{claimcustom}

    \begin{proof}
        For the first part, consider any prefix $\mathcal{I}$ of $\mathcal{I}^\mathcal{B}$ during $(e, T^{exit}]$.
        Hence, by (*) $e$ is the last $L$-event in $\mathcal{I}$, and so the last $L$-event in $\mathcal{I}$ is an $L$-remove event.
        Furthermore, since by \Cref{lemma:ero:exit_do_remove_implies_done_claim} there are no successful list-add or list-remove attempts during $(e, T^{exit}]$, we have that from $e$ onwards in $\mathcal{I}$ there are no successful list-add or list-remove attempts.
        Thus, since $\mathcal{I}$ is finite and by assumption $P(\mathcal{I}^\mathcal{B})$, $Q(\mathcal{I}^\mathcal{B})$, and $R(\mathcal{I}^\mathcal{B})$ hold, by \Cref{lemma:ero:conditional_classification_lemma}, the list of cells conforms to $\List(\mathcal{I}_e)$ in $\mathcal{I}$ where $\mathcal{I}_e$ is the prefix of $\mathcal{I}$ up to but excluding $e$.
        Therefore, since $\mathcal{I}$ is a prefix of $\mathcal{I}^\mathcal{B}$ after $e$, $\mathcal{I}_e = \mathcal{I}^{exclude}_e$, and so the list of cells conforms to $\List(\mathcal{I}^{exclude}_e)$ in $\mathcal{I}$.
        \qH{\Cref{lemma:ero:exit_do_remove_implies_done_zero_claim}}
    \end{proof}

    We now prove that $I$ ``traverses" $\List(\mathcal{I}^{exclude}_e)$.
    The following few claims will be reminiscent of the proof of \Cref{lemma:ero:under_certain_conditions_acquire_returns_found_weak}.
    Let $\List(\mathcal{I}^{exclude}_e) = \cellpointershort_0, \ldots, \cellpointershort_{n + 1}$ for some integer $n \geq 0$.
    Hence, by \Cref{lemma:ero:every_list_sequence_is_from_universe}, $\cellpointershort_0 = \&\headobject$, for every $i \in [1..n]$ $\cellpointershort_i \in \celluniverse{}$, and $\cellpointershort_{n + 1} = \nullconstant$.
    Furthermore, since $e$ is an $L$-remove event for $\cellpointershort_\linearizationobject{}$, and $P(\mathcal{I}^\mathcal{B})$ holds, by \Cref{lemma:ero:l_remove_for_ptr_is_uniquely_in_exclude_list}, there is exactly one $i \in [1..n]$ such that $\cellpointershort_i = \cellpointershort_\linearizationobject{}$.

    \begin{claimcustom}{\ref{lemma:ero:exit_do_remove_implies_done}.3}\label{lemma:ero:exit_do_remove_implies_done:claim_acquire_next}
        Consider any invocation $I^*$ of the AcquireNext procedure on \cref{line:ero:remove_cell_acquire_next} during $I$.
        Since $I$ exits $I^*$ exits.
        Let $T^{\ref{line:ero:acquire_next_read_curr_unique_pointer}}_{I^*}$ (resp. $T^{\ref{line:ero:acquire_next_linearization_changed_check}}_{I^*}$) be the last time $p$ executes \cref{line:ero:acquire_next_read_curr_unique_pointer} (resp. \cref{line:ero:acquire_next_linearization_changed_check}) during $I^*$ (these are well-defined since $I^*$ exits).
        Then, the following are true:
        \begin{compactenum}
            \item $\linearizationobject{}.\uniquerepositoryoperationlong{} = \uniquerepositoryoperationshort_\linearizationobject{}$ at $T^{\ref{line:ero:acquire_next_linearization_changed_check}}_{I^*}$; and
            \item $T^{\ref{line:ero:acquire_next_read_curr_unique_pointer}}_{I^*} \in (e, T^{exit}]$.
        \end{compactenum}
    \end{claimcustom}

    \begin{proof}
        First 1.
        Since $I^*$ began and exited during the loop on \cref{line:ero:remove_cell_while_loop} in $I$, all steps during the loop on \cref{line:ero:remove_cell_while_loop} during $I$ are during $(e, T^{exit}]$, and by (*) $\linearizationobject{}.\uniquerepositoryoperationlong{} = \uniquerepositoryoperationshort_\linearizationobject{}$ throughout $(e, T^{exit}]$, we have that $\linearizationobject{}.\uniquerepositoryoperationlong{} = \uniquerepositoryoperationshort_\linearizationobject{}$ throughout $I^*$.
        Hence, since $T^{\ref{line:ero:acquire_next_linearization_changed_check}}_{I^*}$ is the time of a step during $I^*$, we have that $\linearizationobject{}.\uniquerepositoryoperationlong{} = \uniquerepositoryoperationshort_\linearizationobject{}$ at $T^{\ref{line:ero:acquire_next_linearization_changed_check}}_{I^*}$.
        Now 2.
        Since all steps during the loop on \cref{line:ero:remove_cell_while_loop} during $I$ are during $(e, T^{exit}]$, and $T^{\ref{line:ero:acquire_next_read_curr_unique_pointer}}_{I^*}$ is the time of a step during $I^*$, we have that $T^{\ref{line:ero:acquire_next_read_curr_unique_pointer}}_{I^*} \in (e, T^{exit}]$.
        \qH{\Cref{lemma:ero:exit_do_remove_implies_done:claim_acquire_next}}
    \end{proof}

    \begin{claimcustom}{\ref{lemma:ero:exit_do_remove_implies_done}.4}\label{lemma:ero:exit_do_remove_implies_done_first_claim:claim_third}
        Consider any iteration of the loop on \cref{line:ero:remove_cell_while_loop} during $I$, denoted by $I'$, such that the local variable $\currentuniquecellpointershort{} = \cellpointershort_j$ for some $j \in [0..i)$ at the start of $I'$.
        Then, $p$ executes \cref{line:ero:remove_cell_update_pointers} at time $T^{\ref{line:ero:remove_cell_update_pointers}}$ during $I'$ and the local variable $\currentuniquecellpointershort{} = \cellpointershort_{j + 1}$ at $T^{\ref{line:ero:remove_cell_update_pointers}}$.
    \end{claimcustom}

    \begin{proof}
        Since $\cellpointershort_\linearizationobject = \cellpointershort_i$ for a unique $i \in [1..n]$, we have that every $j \in [0..i)$ $\cellpointershort_j \neq \cellpointershort_\linearizationobject$.
        Hence, since by assumption $\currentuniquecellpointershort{} = \cellpointershort_j$ at the start of $I'$ for some $j \in [0..i)$, it follows that $p$ finds the condition on \cref{line:ero:remove_cell_while_loop} to be true at the start of $I'$.
        Thus, since $p$ exits $I$, $p$ begins and exits the AcquireNext procedure on \cref{line:ero:remove_cell_acquire_next} during $I'$.
        Denote this invocation by $I^*$.
        
        We first prove that $I^*$'s response is $(\found, \cellpointershort_{j + 1})$ by satisfying the conditions of \Cref{lemma:ero:acquire_next_response_classification_weak}.
        Since the first parameter of $I$ is $\uniquerepositoryoperationshort_\linearizationobject{}$ and $\currentuniquecellpointershort{} = \cellpointershort_j$ at the start of $I'$, the parameters of $I^*$ are $(\uniquerepositoryoperationshort_\linearizationobject{}, \cellpointershort_j)$.
        Let $T^{\ref{line:ero:acquire_next_read_curr_unique_pointer}}_{I^*}$ and $T^{\ref{line:ero:acquire_next_linearization_changed_check}}_{I^*}$ by defined as in \Cref{lemma:ero:exit_do_remove_implies_done:claim_acquire_next}, and so $\linearizationobject{}.\uniquerepositoryoperationlong{} = \uniquerepositoryoperationshort_\linearizationobject{}$ at $T^{\ref{line:ero:acquire_next_linearization_changed_check}}_{I^*}$, and $T^{\ref{line:ero:acquire_next_read_curr_unique_pointer}}_{I^*} \in (e, T^{exit}]$.       
        Hence, there is a prefix of $\mathcal{I}^\mathcal{B}$ during $(e, T^{exit}]$ up to and including $T^{\ref{line:ero:acquire_next_read_curr_unique_pointer}}_{I^*}$; say $\mathcal{I}$.
        Thus, by \Cref{lemma:ero:exit_do_remove_implies_done_zero_claim} the list of cells conforms to $\List(\mathcal{I}^{exclude}_e)$ in $\mathcal{I}$.
        So, since $\List(\mathcal{I}^{exclude}_e) = \cellpointershort_0, \ldots, \cellpointershort_{n + 1}$, and $j \in [0..n]$, by \Cref{def:ero:logical_list}, at the end of $\mathcal{I}$ $(*\cellpointershort_j).\nextlong.\uniquecellpointercontentlong{} = \cellpointershort_{j + 1}$.
        Thus, since $\cellpointershort_{j + 1} \in \celluniverse{}$ (because $j + 1 \in [1..n]$), by \Cref{assumption:ero:head_and_null_not_in_cell_universe} $\cellpointershort_{j + 1} \neq \nullconstant$, and so $(*\cellpointershort_j).\nextlong.\uniquecellpointercontentlong{} = \cellpointershort_{j + 1} \neq \nullconstant$ at the end of $\mathcal{I}$.
        Therefore, we have established the following: (1) $I^*$ has parameters $(\uniquerepositoryoperationshort_\linearizationobject{}, \cellpointershort_j)$; (2) $\linearizationobject{}.\uniquerepositoryoperationlong{} = \uniquerepositoryoperationshort_\linearizationobject{}$ at $T^{\ref{line:ero:acquire_next_linearization_changed_check}}_{I^*}$; and (3) $(*\cellpointershort_j).\nextlong.\uniquecellpointercontentlong{} = \cellpointershort_{j + 1} \neq \nullconstant$ at $T^{\ref{line:ero:acquire_next_read_curr_unique_pointer}}_{I^*}$ (equivalently, the end of $\mathcal{I}$), and so by \Cref{lemma:ero:acquire_next_response_classification_weak}, $I^*$'s response is $(\found, \cellpointershort_{j + 1})$.
        
        We now finish the proof of \Cref{lemma:ero:exit_do_remove_implies_done_first_claim:claim_third}.
        Since $p$ exits $I$ and $I^*$'s response is $(\found, \cellpointershort_{j + 1})$, we have that $p$ finds the condition on \cref{line:ero:remove_cell_acquire_next_found} to be true and so $p$ executes \cref{line:ero:remove_cell_update_pointers} during $I'$; say at time $T^{\ref{line:ero:remove_cell_update_pointers}}$.
        Therefore, $\currentuniquecellpointershort{} = \cellpointershort_{j + 1}$ at $T^{\ref{line:ero:remove_cell_update_pointers}}$ as wanted.
        \qH{\Cref{lemma:ero:exit_do_remove_implies_done_first_claim:claim_third}}
    \end{proof}

    \begin{claimcustom}{\ref{lemma:ero:exit_do_remove_implies_done}.5}\label{lemma:ero:exit_do_remove_implies_done_first_claim:claim_fourth}
        For every $j \in [1..i+1]$, (1) $p$ executes \cref{line:ero:remove_cell_while_loop} $j$ times during $I$, and (2) at the time $p$ executes \cref{line:ero:remove_cell_while_loop} for the $j$th time during $I$ the local variable $\currentuniquecellpointershort{} = \cellpointershort_{j - 1}$.
    \end{claimcustom}

    \begin{proof}
        By induction on $j$.
        \begin{itemize}
            \item[] \hspace{0pt}\textbf{Base Case.} $j = 1$.

            In this case, (1) holds immediately since $p$ must execute \cref{line:ero:remove_cell_while_loop} at least once during $I$ as $p$ exits $I$.
            Let $T^{\ref{line:ero:remove_cell_while_loop}}_1$ be the time of $p$'s first execution of \cref{line:ero:remove_cell_while_loop} during $I$.
            For (2), since $\currentuniquecellpointershort{}$ at $T^{\ref{line:ero:remove_cell_while_loop}}_1$ is the value it was initialized to on \cref{line:ero:remove_cell_initialize_pointers} during $I$, we have that $\currentuniquecellpointershort{} = \&\headobject$ at $T^{\ref{line:ero:remove_cell_while_loop}}_1$.
            Therefore, since $\cellpointershort_0 = \&\headobject$, we have that $\currentuniquecellpointershort{} = \cellpointershort_0$ at $T^{\ref{line:ero:remove_cell_while_loop}}_1$.

            \item[] \hspace{0pt}\textbf{Inductive Case.} For every $j \in [1..i]$, if (1) and (2) hold for $j$, then (1) and (2) hold for $j + 1$.

            Suppose for any $j \in [1..i]$ (1) $p$ executes \cref{line:ero:remove_cell_while_loop} $j$ times during $I$ and (2) at the time $p$ executes \cref{line:ero:remove_cell_while_loop} for the $j$th time during $I$, $\currentuniquecellpointershort{} = \cellpointershort_{j - 1}$.
            This is the inductive hypothesis.
            Let $I_j$ be the $j$th iteration of the loop on \cref{line:ero:remove_cell_while_loop} during $I$, which is well-defined by (1) of the inductive hypothesis.
            Furthermore, let $T^{\ref{line:ero:remove_cell_while_loop}}_j$ be the time of $p$'s $j$th execution of \cref{line:ero:remove_cell_while_loop} during $I$ which is the start of $I_j$.
            Since by (2) of the inductive hypothesis $\currentuniquecellpointershort{} = \cellpointershort_{j - 1}$ at $T^{\ref{line:ero:remove_cell_while_loop}}_j$ where $j - 1 \in [0..i)$, by \Cref{lemma:ero:exit_do_remove_implies_done_first_claim:claim_third}, $p$ executes \cref{line:ero:remove_cell_update_pointers} at some time $T^{\ref{line:ero:remove_cell_update_pointers}}_j$ during $I_j$ and $\currentuniquecellpointershort{} = \cellpointershort_j$ at $T^{\ref{line:ero:remove_cell_update_pointers}}_j$.
            Hence, since $p$ exits $I$, it follows that $p$ executes \cref{line:ero:remove_cell_while_loop} one more time during $I$, so $p$ executes \cref{line:ero:remove_cell_while_loop} $j + 1$ times during $I$.
            % Therefore, $p$ executes \cref{line:ero:add_cell_while_loop} $i + 1$ times during $I$.
            % We now prove (2) holds for $i + 1$.
            Since $\currentuniquecellpointershort{} = \cellpointershort_{j}$ at $T^{\ref{line:ero:remove_cell_update_pointers}}_j$, and the value of $\currentuniquecellpointershort{}$ does not change between $T^{\ref{line:ero:remove_cell_update_pointers}}_j$ and the time of $p$'s $j + 1$th execution of \cref{line:ero:remove_cell_while_loop} during $I$, it follows that at the time $p$ executes \cref{line:ero:remove_cell_while_loop} for the $j + 1$th time during $I$ $\currentuniquecellpointershort{} = \cellpointershort_j$.
            Therefore, (1) and (2) hold for $j + 1$ as wanted.
            \qH{\Cref{lemma:ero:exit_do_remove_implies_done_first_claim:claim_fourth}}
        \end{itemize}
    \end{proof}

    Since $\cellpointershort_\linearizationobject{} = \cellpointershort_i$, this implies that $p$ will exit the first loop on \cref{line:ero:remove_cell_while_loop} during $I$.

    \begin{claimcustom}{\ref{lemma:ero:exit_do_remove_implies_done}.6}\label{lemma:ero:exit_do_remove_implies_done_first_claim:claim_seven}
        At the time $p$ executes \cref{line:ero:remove_cell_while_loop} for the $i +1$th time during $I$, which is well-defined by \Cref{lemma:ero:exit_do_remove_implies_done_first_claim:claim_fourth}, $p$ finds the condition on \cref{line:ero:remove_cell_while_loop} to be false.
    \end{claimcustom}

    \begin{proof}
        By \Cref{lemma:ero:exit_do_remove_implies_done_first_claim:claim_fourth} $\currentuniquecellpointershort{} = \cellpointershort_i$ at the time $p$ executes \cref{line:ero:remove_cell_while_loop} for the $i +1$th time, say $T$, and $\cellpointershort_i = \cellpointershort_\linearizationobject$, so $\currentuniquecellpointershort{} = \cellpointershort_\linearizationobject$ at $T$.
        Therefore, since the second parameter of $I$ is $\cellpointershort_\linearizationobject$, we have that $p$ finds the condition on  \cref{line:ero:remove_cell_while_loop} to be false at $T$.
        \qH{\Cref{lemma:ero:exit_do_remove_implies_done_first_claim:claim_seven}}
    \end{proof}

    The remainder of the proof will proceed as follows.
    Since $p$ exits the loop on \cref{line:ero:remove_cell_while_loop} during $I$, we know that $p$ will enter the loop on \cref{line:ero:remove_cell_remove_repeat_loop} during $I$.
    Hence, since $p$ exits $I$, it follows that $p$ will find the condition on either \cref{line:ero:remove_cell_before_removal_linearization_check} or \cref{line:ero:remove_cell_from_list} to be true.
    If the latter, then given $I$'s parameters, we are done, so the task is to prove that the former is impossible.
    We first prove that the value of the local variable $\previousuniquecellpointershort{} = \cellpointershort_{i - 1}$ during the loop on \cref{line:ero:remove_cell_remove_repeat_loop} in $I$, and then we prove this.

    \begin{claimcustom}{\ref{lemma:ero:exit_do_remove_implies_done}.7}\label{lemma:ero:exit_do_remove_implies_done_first_claim:claim_six}
        At the time $p$ executes \cref{line:ero:remove_cell_while_loop} for the $i +1$th time during $I$, which is well-defined by \Cref{lemma:ero:exit_do_remove_implies_done_first_claim:claim_fourth}, the local variable $\previousuniquecellpointershort{} = \cellpointershort_{i - 1}$.
    \end{claimcustom}

    \begin{proof}
        Since $i \in [1..n]$, by \Cref{lemma:ero:exit_do_remove_implies_done_first_claim:claim_fourth}, (1) $p$ executes \cref{line:ero:remove_cell_while_loop} $i$ \& $i + 1$ times during $I$ and (2) at the time $p$ executes \cref{line:ero:remove_cell_while_loop} for the $i$th time during $I$ $\currentuniquecellpointershort{} = \cellpointershort_{i - 1}$.
        Let $I_{i}$ be the $i$th iteration of the loop on \cref{line:ero:remove_cell_while_loop} during $I$.
        Since $p$ executes \cref{line:ero:remove_cell_while_loop} $i$ \& $i +1$ times during $I$ and the first parameter of the response from the AcquireNext procedure is either $\found$, $\notfound$, or $\timechange{}$, it follows that $\status = \found$ during $I_{i}$.
        Hence, $p$ found the condition on \cref{line:ero:remove_cell_acquire_next_found} to be true during $I_{i}$, and so $p$ executed \cref{line:ero:remove_cell_update_pointers} during $I_{i}$; say at time $T^{\ref{line:ero:remove_cell_update_pointers}}_{i}$.
        Thus, since at the time $p$ executes \cref{line:ero:remove_cell_while_loop} during $I_i$ $\currentuniquecellpointershort{} = \cellpointershort_{i - 1}$, we have that $p$ set $\previousuniquecellpointershort{} = \cellpointershort_{i - 1}$ at $T^{\ref{line:ero:remove_cell_update_pointers}}_{i}$.
        Therefore, since the value of $\previousuniquecellpointershort{}$ does not change between $T^{\ref{line:ero:remove_cell_update_pointers}}_{i}$ and the time of $p$'s $i +1$th execution of \cref{line:ero:remove_cell_while_loop} during $I$, at the time $p$ executes \cref{line:ero:remove_cell_while_loop} for the $i +1$th time during $I$ $\previousuniquecellpointershort{} = \cellpointershort_{i - 1}$.
        \qH{\Cref{lemma:ero:exit_do_remove_implies_done_first_claim:claim_six}}
    \end{proof}

    \begin{claimcustom}{\ref{lemma:ero:exit_do_remove_implies_done}.8}\label{lemma:ero:exit_do_remove_implies_done_zero_two_claim}
        $p$ never finds the left clause of \cref{line:ero:remove_cell_before_removal_linearization_check} to be true during $I$.
    \end{claimcustom}

    \begin{proof}
        Suppose, for contradiction, $p$ finds the left clause of \cref{line:ero:remove_cell_before_removal_linearization_check} to be true sometime during $I$.
        Hence, $\linearizationobject{}.\uniquerepositoryoperationlong{}$ doesn't equal the first parameter of $I$ sometime during $I$.
        Since the first parameter of $I$ is $\uniquerepositoryoperationshort_\linearizationobject{}$, we have that $\linearizationobject{}.\uniquerepositoryoperationlong{} \neq \uniquerepositoryoperationshort_\linearizationobject{}$ sometime during $I$.
        Therefore, since all steps during $I$ are during $(e, T^{exit}]$, we have that $\linearizationobject{}.\uniquerepositoryoperationlong{} \neq \uniquerepositoryoperationshort_\linearizationobject{}$  sometime during $(e, T^{exit}]$.
        However, by (*) $\linearizationobject{}.\uniquerepositoryoperationlong{} = \uniquerepositoryoperationshort_\linearizationobject{}$ throughout $(e, T^{exit}]$, a contradiction.
        \qH{\Cref{lemma:ero:exit_do_remove_implies_done_zero_two_claim}}
    \end{proof}

    \begin{claimcustom}{\ref{lemma:ero:exit_do_remove_implies_done}.9}\label{lemma:ero:exit_do_remove_implies_done_first_claim}
        $p$ never finds the condition on \cref{line:ero:remove_cell_before_removal_linearization_check} to be true during $I$.
    \end{claimcustom}

    \begin{proof}
        Let $T^{\ref{line:ero:remove_cell_while_loop}}_{i + 1}$ be the time of $p$'s $i + 1$th execution of \cref{line:ero:remove_cell_while_loop} during $I$.
        This time is well-defined by \Cref{lemma:ero:exit_do_remove_implies_done_first_claim:claim_fourth}.
        By \Cref{lemma:ero:exit_do_remove_implies_done_first_claim:claim_six}, at $T^{\ref{line:ero:remove_cell_while_loop}}_{i + 1}$ $\previousuniquecellpointershort{} = \cellpointershort_{i - 1}$.
        Hence, since by \Cref{lemma:ero:exit_do_remove_implies_done_first_claim:claim_seven} $p$ finds the condition on \cref{line:ero:remove_cell_while_loop} to be false at $T^{\ref{line:ero:remove_cell_while_loop}}_{i + 1}$ and $\previousuniquecellpointershort{}$ only changes on lines \ref{line:ero:remove_cell_initialize_pointers} and \ref{line:ero:remove_cell_update_pointers} during $I$, from $T^{\ref{line:ero:remove_cell_while_loop}}_{i + 1}$ onwards in $I$, $\previousuniquecellpointershort{} = \cellpointershort_{i - 1}$.
        Since $p$ exits the loop on \cref{line:ero:remove_cell_while_loop} via the condition on \cref{line:ero:remove_cell_while_loop} during $I$, and $p$ exits $I$, $p$ enters and exits the loop on \cref{line:ero:remove_cell_remove_seal_loop} during $I$, followed by entering the loop on \cref{line:ero:remove_cell_remove_repeat_loop} during $I$.
        Therefore, $p$ executes \cref{line:ero:remove_cell_before_removal_linearization_check} at least once during $I$.

        We now prove that $p$ finds the right clause to be false on its first execution of \cref{line:ero:remove_cell_before_removal_linearization_check} during $I$.
        Since $p$ executes \cref{line:ero:remove_cell_before_removal_linearization_check} at least once during $I$, we have that $p$ executes \cref{line:ero:remove_cell_read_pointer_to_remove} during $I$, and executes \cref{line:ero:remove_cell_read_previous_pointer} at least once during $I$.
        Let $T^{\ref{line:ero:remove_cell_read_pointer_to_remove}}$ be the time of $p$'s execution of \cref{line:ero:remove_cell_read_pointer_to_remove} during $I$ and let $T^{\ref{line:ero:remove_cell_read_previous_pointer}}$ be the times of $p$'s first executions of \cref{line:ero:remove_cell_read_previous_pointer} during $I$.
        Since the second parameter of $I$ is $\cellpointershort_\linearizationobject$, $p$ reads from $\cellpointershort_\linearizationobject$ at $T^{\ref{line:ero:remove_cell_read_pointer_to_remove}}$.
        Thus, since $\cellpointershort_i = \cellpointershort_\linearizationobject$, we have that $p$ reads from $\cellpointershort_i$ at $T^{\ref{line:ero:remove_cell_read_pointer_to_remove}}$.
        Furthermore, since $\previousuniquecellpointershort{} = \cellpointershort_{i - 1}$ from $T^{\ref{line:ero:remove_cell_while_loop}}_{i + 1}$ onwards in $I$, we have that $p$ reads from $\cellpointershort_{i - 1}$ at $T^{\ref{line:ero:remove_cell_read_previous_pointer}}$.
        Since $T^{\ref{line:ero:remove_cell_read_pointer_to_remove}}$ and $T^{\ref{line:ero:remove_cell_read_previous_pointer}}$ are during $I$ and all steps during $I$ are during $(e, T^{exit}]$, we have that $T^{\ref{line:ero:remove_cell_read_pointer_to_remove}}$ and 
        $T^{\ref{line:ero:remove_cell_read_previous_pointer}}$ are during $(e, T^{exit}]$.
        Hence, there is a prefix of $\mathcal{I}^\mathcal{B}$ during $(e, T^{exit}]$ up to and including $T^{\ref{line:ero:remove_cell_read_pointer_to_remove}}$ (resp. 
        $T^{\ref{line:ero:remove_cell_read_previous_pointer}}$); say $I^{\ref{line:ero:remove_cell_read_pointer_to_remove}}$ (resp. $I^{\ref{line:ero:remove_cell_read_previous_pointer}}$).
        Thus, by \Cref{lemma:ero:exit_do_remove_implies_done_zero_claim} the list of cells conforms to $\List(\mathcal{I}^{exclude}_e)$ in $I^{\ref{line:ero:remove_cell_read_pointer_to_remove}}$ and $I^{\ref{line:ero:remove_cell_read_previous_pointer}}$.
        So, since $\List(\mathcal{I}^{exclude}_e) = \cellpointershort_0, \ldots, \cellpointershort_{n + 1}$, and $i, i-1 \in [0..n]$, by \Cref{def:ero:logical_list}, at the end of $I^{\ref{line:ero:remove_cell_read_pointer_to_remove}}$ $(*\cellpointershort_i).\nextlong.\uniquecellpointercontentlong{} = \cellpointershort_{i + 1}$, and at the end of $I^{\ref{line:ero:remove_cell_read_previous_pointer}}$ $(*\cellpointershort_{i-1}).\nextlong.\uniquecellpointercontentlong{} = \cellpointershort_i$.
        Hence, by the definition of $I^{\ref{line:ero:remove_cell_read_pointer_to_remove}}$ and $I^{\ref{line:ero:remove_cell_read_previous_pointer}}$, we have that $p$ read $\cellpointershort_{i + 1}$ from $(*\cellpointershort_i).\nextlong.\uniquecellpointercontentlong{}$ at $T^{\ref{line:ero:remove_cell_read_pointer_to_remove}}$, and $p$ read $\cellpointershort_i$ from $(*\cellpointershort_{i - 1}).\nextlong.\uniquecellpointercontentlong{}$ at $T^{\ref{line:ero:remove_cell_read_previous_pointer}}$.
        Thus, since $\mathcal{I}^{exclude}_e$ is a finite prefix of $\mathcal{I}^\mathcal{B}$, $\List(\mathcal{I}^{exclude}_e) = \cellpointershort_0, \ldots, \cellpointershort_{n + 1}$, $P(\mathcal{I}^\mathcal{B})$ holds, $i,i+1 \in [0..n+1]$, and $i + 1 \neq i$, by \Cref{lemma:ero:pointers_in_list_are_unique}, $\cellpointershort_{i + 1} \neq \cellpointershort_{i}$.
        Therefore, $p$ finds the right clause to be false on its first execution of \cref{line:ero:remove_cell_before_removal_linearization_check} during $I$.

        We now finish the proof of \Cref{lemma:ero:exit_do_remove_implies_done_first_claim}.
        Suppose, for contradiction, that $p$ finds the condition on \cref{line:ero:remove_cell_before_removal_linearization_check} to be true sometime during $I$; say at time $T^{\ref{line:ero:remove_cell_before_removal_linearization_check}}$.
        Hence, since by \Cref{lemma:ero:exit_do_remove_implies_done_zero_two_claim} $p$ finds the left clause to be false at $T^{\ref{line:ero:remove_cell_before_removal_linearization_check}}$, we have that $p$ finds the right clause to be true at $T^{\ref{line:ero:remove_cell_before_removal_linearization_check}}$.
        Thus, since $p$ read $\cellpointershort_{i + 1}$ from $(*\cellpointershort_i).\nextlong.\uniquecellpointercontentlong{}$ at $T^{\ref{line:ero:remove_cell_read_pointer_to_remove}}$, and $\previouscellpointershort{} = \cellpointershort_{i - 1}$ from $T^{\ref{line:ero:remove_cell_while_loop}}_{i + 1}$ onwards in $I$, we have that $p$ read $\cellpointershort_{i + 1}$ from $(*\cellpointershort_{i - 1}).\nextlong.\uniquecellpointercontentlong{}$ on \cref{line:ero:remove_cell_read_previous_pointer} at some time $T$ during $I$.
        Since $p$ read $\cellpointershort_{i}$ from $(*\cellpointershort_{i - 1}).\nextlong.\uniquecellpointercontentlong{}$ on \cref{line:ero:remove_cell_read_previous_pointer} at time $T^{\ref{line:ero:remove_cell_read_previous_pointer}}$ and $\cellpointershort_{i + 1} \neq \cellpointershort_{i}$, we have that between $T^{\ref{line:ero:remove_cell_read_previous_pointer}}$ and $T$ (or vice versa), the value of $(*\cellpointershort_{i - 1}).\nextlong.\uniquecellpointercontentlong{}$ changed.
        Hence, since $T^{\ref{line:ero:remove_cell_read_previous_pointer}}$ and $T$ both occurred during $I$, and all steps during $I$ are during $(e, T^{exit}]$, we have that $(*\cellpointershort_{i - 1}).\nextlong.\uniquecellpointercontentlong{}$ changed during $(e, T^{exit}]$.
        Therefore, since $\cellpointershort_{i - 1} \in \celluniverse{}$ (because $i - 1 \in [0..n)$), by \Cref{observation:ero:where_objects_change}, there is a successful list-add or list-remove attempt during $(e, T^{exit}]$.
        However, by \Cref{lemma:ero:exit_do_remove_implies_done_claim}, there are no successful list-add or list-remove attempts during $(e, T^{exit}]$, a contradiction.
        \qH{\Cref{lemma:ero:exit_do_remove_implies_done_first_claim}}
    \end{proof}

    We now return to the proof of \Cref{lemma:ero:exit_do_remove_implies_done}.
    Since by \Cref{lemma:ero:exit_do_remove_implies_done_first_claim:claim_seven}, $p$ exits the loop on \cref{line:ero:remove_cell_while_loop} during $I$ by the finding the condition on \cref{line:ero:remove_cell_while_loop} to be false, and $p$ exits $I$, it follows that $p$ either finds the condition on  \cref{line:ero:remove_cell_before_removal_linearization_check} or \cref{line:ero:remove_cell_from_list} to be true during $I$.
    By \Cref{lemma:ero:exit_do_remove_implies_done_first_claim}, $p$ never finds the condition on \cref{line:ero:remove_cell_before_removal_linearization_check} to be true during $I$, so $p$ finds the condition on \cref{line:ero:remove_cell_from_list} to be true during $I$.
    Therefore, since all steps during $I$ are during $(e, T^{exit}]$, by \Cref{def:ero:english}, $p$ executes a successful list-remove attempt during $(e, T^{exit}]$.
    However, by \Cref{lemma:ero:exit_do_remove_implies_done_claim}, there are no successful list-remove attempts during $(e, T^{exit}]$, a contradiction.
    \qH{\Cref{lemma:ero:exit_do_remove_implies_done}}
\end{proof}

\begin{lemma}\label{lemma:ero:exit_remove_implies_response_set}
    Consider any invocation of the \doremovecell{} procedure with a second parameter of $\cellpointershort_\linearizationobject{}$ which ends at some time $T^{exit}$ during $\mathcal{I}^\mathcal{B}$.
    If $P(\mathcal{I}^\mathcal{B})$, $Q(\mathcal{I}^\mathcal{B})$, and $R(\mathcal{I}^\mathcal{B})$, hold then there is a successful remove-response-set attempt for $\cellpointershort_\linearizationobject$ before $T^{exit}$.
\end{lemma}

\begin{proof}
    Consider any invocation $I$ of the \doremovecell{} procedure with parameters $(\uniquerepositoryoperationshort_\linearizationobject{}, \cellpointershort_\linearizationobject)$ which ends at some time $T^{exit}$ during $\mathcal{I}^\mathcal{B}$.
    Let $p$ be the process that invoked $I$.
    Since $I$ has parameters $(\uniquerepositoryoperationshort_\linearizationobject{}, \cellpointershort_\linearizationobject)$, by \Cref{lemma:ero:l_event_corresponding_to_do_low_level_op}, there is an $L$-remove event $e$ for $\cellpointershort_\linearizationobject$ before $I$ was invoked that set $\linearizationobject{}$ to $(\uniquerepositoryoperationshort_\linearizationobject{}, \cellpointershort_\linearizationobject)$.
    Hence, by \Cref{lemma:ero:every_l_event_is_for_pointer_from_universe}, $\cellpointershort_\linearizationobject{} \in \celluniverse$.

    The proof strategy is to identify a remove-response-set attempt for $\cellpointershort_\linearizationobject{}$ before $T^{exit}$.
    If this is successful, we are done, but if it is unsuccessful, then by \Cref{lemma:ero:unsuccessful_remove_response_set_implies_successful_remove_response_set}, there is a successful remove-response-set attempt for $\cellpointershort_\linearizationobject{}$ beforehand, in which case we are also done.

    \begin{claimcustom}{\ref{lemma:ero:exit_remove_implies_response_set}.1}\label{lemma:ero:exit_remove_implies_response_set:claim_zero}
        There is a successful list-remove attempt $a$ for $\cellpointershort_\linearizationobject{}$ during $(e, T^{exit}]$ in $\mathcal{I}^\mathcal{B}$ such that during $(e, a)$ there are no $L$-events and successful list-add and list-remove attempts.
    \end{claimcustom}

    \begin{proof}
        Since $I$ is an invocation of the \doremovecell{} procedure with parameters $(\uniquerepositoryoperationshort_\linearizationobject{}, \cellpointershort_\linearizationobject)$ which ends at time $T^{exit}$ in $\mathcal{I}^\mathcal{B}$ and $P(\mathcal{I}^\mathcal{B})$, $Q(\mathcal{I}^\mathcal{B})$, and $R(\mathcal{I}^\mathcal{B})$ hold, by \Cref{lemma:ero:exit_do_remove_implies_done} there is a successful list-remove attempt $a$ for $\cellpointershort_\linearizationobject{}$ before $T^{exit}$ in $\mathcal{I}^\mathcal{B}$.
        Hence, by \Cref{lemma:ero:l_event_corresponding_to_do_low_level_op}, $a$'s corresponding $L$-event is an $L$-remove event $e'$ for $\cellpointershort_\linearizationobject{}$ before $a$ in $\mathcal{I}^\mathcal{B}$.
        Thus, since $e$ and $e'$ are both $L$-remove events for $\cellpointershort_\linearizationobject{}$, by $P(\mathcal{I}^\mathcal{B})$, $e = e'$.
        So, $e$ is $a$'s corresponding $L$-event and $e < a$.
        Since $a$ is a successful list-remove attempt in $\mathcal{I}^\mathcal{B}$, and $P(\mathcal{I}^\mathcal{B})$, $Q(\mathcal{I}^\mathcal{B})$, and $R(\mathcal{I}^\mathcal{B})$ hold, by \Cref{lemma:ero:any_list_attempt_outside_its_window_is_unsuccessful_alternate_statement}, there are no $L$-events during $(e, a)$.

        We now prove that there are no successful list-add and list-remove attempts during $(e, a)$.
        Let $\mathcal{I}^{include}_a$ be the prefix of $\mathcal{I}^\mathcal{B}$ up to and including $a$.
        Since $e < a$, we have that $e$ is in $\mathcal{I}^{include}_a$, and since there are no $L$-events during $(e, a)$, we have that $e$ is the last $L$-event in $\mathcal{I}^{include}_a$.
        Hence, since $P(\mathcal{I}^\mathcal{B})$, $Q(\mathcal{I}^\mathcal{B})$, and $R(\mathcal{I}^\mathcal{B})$ hold, and the last $L$-event in $\mathcal{I}^{include}_a$, $e$, is an $L$-remove event for $\cellpointershort_\linearizationobject{}$, by \Cref{lemma:ero:3_of_r_safety_holds}, from $e$ onwards in $\mathcal{I}^{include}_a$ there is at most one successful list-remove attempt for $\cellpointershort_\linearizationobject$ and no other successful list-remove or list-add attempt for any pointer.
        Therefore, there are no successful list-add and list-remove attempts during $(e, a)$.
        \qH{\Cref{lemma:ero:exit_remove_implies_response_set:claim_zero}}
    \end{proof}

    \begin{claimcustom}{\ref{lemma:ero:exit_remove_implies_response_set}.2}\label{lemma:ero:exit_remove_implies_response_set:claim_one}
        The \setrepositoryoperationresponse{} procedure was invoked with parameters $(\uniquerepositoryoperationshort_\linearizationobject{}, \cellpointershort_\linearizationobject, \done)$ at some time $T'_b$ and ends at some time $T'_e$ during $\mathcal{I}^\mathcal{B}$ such that: (1) $[T'_b, T'_e] \subseteq (e, T^{exit}]$; (2) $\linearizationobject.\uniquerepositoryoperationlong = \uniquerepositoryoperationshort_\linearizationobject{}$ at the end of $\mathcal{I}$ where $\mathcal{I}$ is any prefix of $\mathcal{I}^\mathcal{B}$ during $[T'_b, T'_e]$; and (3) the list of cells conforms to $\List(\mathcal{I}^{exclude}_e)$ in $\mathcal{I}$ where $\mathcal{I}^{exclude}_e$ is the prefix of $\mathcal{I}^\mathcal{B}$ up to but excluding $e$.
    \end{claimcustom}

    \begin{proof}
        By \Cref{lemma:ero:exit_remove_implies_response_set:claim_zero} there is a successful list-remove attempt $a$ for $\cellpointershort_\linearizationobject{}$ during $(e, T^{exit}]$ in $\mathcal{I}^\mathcal{B}$ such that during $(e, a)$ there are no $L$-events and successful list-add and list-remove attempts.
        Let $q$ be the process that executed $a$.
        Since $a$ is a list-remove attempt for $\cellpointershort_\linearizationobject{}$, we have that $q$ executed $a$ during some invocation $I^*$ of the \doremovecell{} procedure with parameters $(\uniquerepositoryoperationshort^*_\linearizationobject{}, \cellpointershort_\linearizationobject{})$.
        Hence, by \Cref{lemma:ero:l_event_corresponding_to_do_low_level_op}, there is an $L$-remove event $e^*$ for $\cellpointershort_\linearizationobject{}$ before $I^*$ was invoked that set $\linearizationobject{} = (\uniquerepositoryoperationshort^*_\linearizationobject{}, \cellpointershort_\linearizationobject{})$.
        Thus, since $e$ and $e^*$ are both $L$-remove events for $\cellpointershort_\linearizationobject{}$ in $\mathcal{I}^\mathcal{B}$, by $P(\mathcal{I}^\mathcal{B})$, $e = e^*$, and so $\uniquerepositoryoperationshort_\linearizationobject{} = \uniquerepositoryoperationshort^*_\linearizationobject{}$.
        So, the parameters of $I^*$ are $(\uniquerepositoryoperationshort_\linearizationobject{}, \cellpointershort_\linearizationobject{})$.
        Since $q$ executed $a$ during $I^*$, we have that $q$ invoked the \setrepositoryoperationresponse{} procedure on \cref{line:ero:remove_cell_set_response} during $I^*$.
        Denote this invocation by $I'$.
        We prove that $I'$ is the desired invocation.
        Since the parameters of $I^*$ are $(\uniquerepositoryoperationshort_\linearizationobject{}, \cellpointershort_\linearizationobject{})$, and $I'$ was invoked during $I^*$, we have that the parameters of $I'$ are $(\uniquerepositoryoperationshort_\linearizationobject{}, \cellpointershort_\linearizationobject, \done)$.
        Let $T'_b$ and $T'_e$ be the times that $q$ invokes and exits $I'$, respectively.
        We first prove (1).
        Since $e$ is before $I^*$ was invoked (because $e = e^*$), and $I'$ is invoked during $I^*$, by transitivity, $e < T'_b$.
        Hence, since $T'_e < a$ (because they are both executed during $I^*$), by transitivity, $[T'_b, T'_e] \subseteq (e, a)$, and since $a \leq T^{exit}$, by transitivity, $[T'_b, T'_e] \subseteq (e, T^{exit}]$.
        We now prove (2).
        Since there are no $L$-events during $(e, a)$, and $e$ set $\linearizationobject.\uniquerepositoryoperationlong = \uniquerepositoryoperationshort_\linearizationobject{}$, by \Cref{observation:ero:where_objects_change}, $\linearizationobject.\uniquerepositoryoperationlong = \uniquerepositoryoperationshort_\linearizationobject{}$ throughout $(e, a)$.
        Hence, since $[T'_b, T'_e] \subseteq (e, a)$, we have that $\linearizationobject.\uniquerepositoryoperationlong = \uniquerepositoryoperationshort_\linearizationobject{}$ throughout $[T'_b, T'_e]$.
        This implies (2).
        We now prove (3).
        Consider any prefix $\mathcal{I}$ of $\mathcal{I}^\mathcal{B}$ during $[T'_b, T'_e]$.
        Since there are no $L$-events during $(e, a)$, and $[T'_b, T'_e] \subseteq (e, a)$, it follows that $e$ is the last $L$-event in $\mathcal{I}$.
        Hence, since $e$ is an $L$-remove event, we have that the last $L$-event in $\mathcal{I}$ is an $L$-remove event.
        Since there are no successful list-add and list-remove attempts during $(e, a)$, and $[T'_b, T'_e] \subseteq (e, a)$, we have that there are no successful list-add or list-remove attempts during $[T'_b, T'_e]$.
        Thus, since $\mathcal{I}$ is a prefix of $\mathcal{I}^\mathcal{B}$ during $[T'_b, T'_e]$, we have that from $e$ onwards in $\mathcal{I}$ there are no successful list-add or list-remove attempts.
        So, since $\mathcal{I}$ is finite, and $P(\mathcal{I}^\mathcal{B})$, $Q(\mathcal{I}^\mathcal{B})$, and $R(\mathcal{I}^\mathcal{B})$ hold, by \Cref{lemma:ero:conditional_classification_lemma}, the list of cells conforms to $\List(\mathcal{I}_e)$ in $\mathcal{I}$ where $\mathcal{I}_e$ is the prefix of $\mathcal{I}$ up to but excluding $e$.
        Therefore, since $\mathcal{I}$ is a prefix of $\mathcal{I}^\mathcal{B}$ after $e$, $\mathcal{I}_e = \mathcal{I}^{exclude}_e$, and so the list of cells conforms to $\List(\mathcal{I}^{exclude}_e)$ in $\mathcal{I}$ as wanted.
        \qH{\Cref{lemma:ero:exit_remove_implies_response_set:claim_one}}
    \end{proof}

    \begin{claimcustom}{\ref{lemma:ero:exit_remove_implies_response_set}.3}\label{lemma:ero:exit_remove_implies_response_set:claim_two}
        There is a remove-response-set attempt for $\cellpointershort_\linearizationobject{}$ before $T^{exit}$.
    \end{claimcustom}

    \begin{proof}
        Let $I'$ be the invocation of the \setrepositoryoperationresponse{} procedure identified in \Cref{lemma:ero:exit_remove_implies_response_set:claim_one} and let $q$ be the process that executed $I'$.
        Furthermore, let $T'_b$ and $T'_e$ be the times during $\mathcal{I}^\mathcal{B}$ that $q$ begins and exits $I'$, respectively.
        Since $q$ exits $I'$, $q$ began and exited the Acquire procedure on \cref{line:ero:set_response_acquire} during $I'$.
        Denote this invocation of the Acquire procedure by $I^*$.
        Since the parameters of $I'$ are $(\uniquerepositoryoperationshort_\linearizationobject{}, \cellpointershort_\linearizationobject, \done)$, we have that the parameters of $I^*$ are $(\uniquerepositoryoperationshort_\linearizationobject{}, \cellpointershort_\linearizationobject)$.
        
        We now satisfy the conditions of \Cref{lemma:ero:under_certain_conditions_acquire_returns_found_weak}.
        By above, $\cellpointershort_\linearizationobject{} \in \celluniverse{}$.
        Let $T_b$ be the time $q$ invoked $I^*$ and let $T^{\ref{line:ero:acquire_next_linearization_changed_check}}$ be the last time $q$ executes \cref{line:ero:acquire_next_linearization_changed_check} during $I^*$.
        Recall that $T^{\ref{line:ero:acquire_next_linearization_changed_check}}$ is well-defined by \Cref{lemma:ero:exit_acquire_implies_executing_105}.
        By assumption $P(\mathcal{I}^\mathcal{B})$ holds.
        We plug in $\mathcal{I}^{exclude}_e$ for $\mathcal{I}$ in \Cref{lemma:ero:under_certain_conditions_acquire_returns_found_weak}.
        Consider any prefix $\mathcal{I}^*$ of $\mathcal{I}^\mathcal{B}$ during $[T_b, T^{\ref{line:ero:acquire_next_linearization_changed_check}}]$.
        We first satisfy condition 1.
        Since by definition all steps during $[T_b, T^{\ref{line:ero:acquire_next_linearization_changed_check}}]$ are during $I^*$, and all steps during $I^*$ are during $I'$, we have that $[T_b, T^{\ref{line:ero:acquire_next_linearization_changed_check}}] \subseteq [T'_b, T'_e]$.
        Hence, since by (2) of \Cref{lemma:ero:exit_remove_implies_response_set:claim_one}, $\linearizationobject{}.\uniquerepositoryoperationlong{} = \uniquerepositoryoperationshort_\linearizationobject{}$ at the end of $\mathcal{I}'$ for any prefix $\mathcal{I}'$ of $\mathcal{I}^\mathcal{B}$ during $[T'_b, T'_e]$, we have that $\linearizationobject{}.\uniquerepositoryoperationlong{} = \uniquerepositoryoperationshort_\linearizationobject{}$ at the end of $\mathcal{I}^*$.
        We now satisfy condition 2.
        Since by (3) of \Cref{lemma:ero:exit_remove_implies_response_set:claim_one} the list of cells conforms to $\List(\mathcal{I}')$ in $\mathcal{I}'$ for any prefix $\mathcal{I}'$ of $\mathcal{I}^\mathcal{B}$ during $[T'_b, T'_e]$, and $[T_b, T^{\ref{line:ero:acquire_next_linearization_changed_check}}] \subseteq [T'_b, T'_e]$, we have that that the list of cells conforms to $\List(\mathcal{I}^{exclude}_e)$ in $\mathcal{I}^*$.
        Therefore, by \Cref{lemma:ero:under_certain_conditions_acquire_returns_found_weak} if $\cellpointershort_\linearizationobject \in \List(\mathcal{I}^{exclude}_e)$ then the response of $I^*$ is $\found$ (*).

        We now prove that $\cellpointershort_\linearizationobject \in \List(\mathcal{I}^{exclude}_e)$.
        Suppose, for contradiction, $\cellpointershort_\linearizationobject \notin \List(\mathcal{I}^{exclude}_e)$.
        Since $e$ is an $L$-remove event for $\cellpointershort_\linearizationobject{}$ in $\mathcal{I}^\mathcal{B}$, by \Cref{lemma:ero:remove_events_are_preceeded_by_add_events}, there is an $L$-add event $e_{add}$ for $\cellpointershort_\linearizationobject{}$ before $e$ in $\mathcal{I}^\mathcal{B}$.
        Hence, since $\mathcal{I}^{exclude}_e$ is a prefix of $\mathcal{I}$ up to but excluding $e$, and $e_{add} < e$, we have that $e_{add}$ is in $\mathcal{I}^{exclude}_e$.
        Thus, since $\cellpointershort_\linearizationobject \notin \List(\mathcal{I}^{exclude}_e)$ and $e_{add}$ is an $L$-add event for $\cellpointershort_\linearizationobject$ in $\mathcal{I}^{exclude}_e$, by \Cref{def:ero:logical_list}, there is an $L$-remove event $e_{remove}$ for $\cellpointershort_\linearizationobject$ after $e_{add}$ in $\mathcal{I}^{exclude}_e$.
        Hence, since $\mathcal{I}^{exclude}_e$ is a prefix of $\mathcal{I}^\mathcal{B}$ up to but excluding $e$, we have that $e_{remove} < e$.
        Therefore, $e_{remove} \neq e$, and so there are two $L$-remove events for $\cellpointershort_\linearizationobject{}$ in $\mathcal{I}^\mathcal{B}$.
        However, by $P(\mathcal{I}^\mathcal{B})$, there is at most one $L$-remove event for $\cellpointershort_\linearizationobject{}$ in $\mathcal{I}^\mathcal{B}$, a contradiction.

        We now finish the proof of \Cref{lemma:ero:exit_remove_implies_response_set:claim_two}.
        Since $\cellpointershort_\linearizationobject \in \List(\mathcal{I}^{exclude}_e)$, by (*) the response of $I^*$ is $\found$.
        Thus, since $q$ exits $I'$, we have that $q$ finds the condition on \cref{line:ero:set_response_found_check} to be true during $I'$.
        So, $q$ executes \cref{line:ero:responses_set_attempt} during $I'$.
        Since $I^*$'s parameters are $(\uniquerepositoryoperationshort_\linearizationobject{}, \cellpointershort_\linearizationobject, \done)$ and $\uniquerepositoryoperationshort_\linearizationobject{} = (\arbitraryvalue, \removecell)$ (because $e$ set $\linearizationobject{}.\uniquerepositoryoperationlong{} = \uniquerepositoryoperationshort_\linearizationobject{}$ and $e$ is an $L$-remove event), by \Cref{def:ero:english}, this execution is a remove-response-set attempt for $\cellpointershort_\linearizationobject$ during $I'$ (and thus $[T'_b, T'_e]$).
        Therefore, since by (1) of \Cref{lemma:ero:exit_remove_implies_response_set:claim_one} $[T'_b, T'_e] \subseteq (e, T^{exit}]$, we have that there is an remove-response-set attempt for $\cellpointershort_\linearizationobject{}$ before $T^{exit}$ as wanted.
        \qH{\Cref{lemma:ero:exit_remove_implies_response_set:claim_two}}
    \end{proof}

    We now return to the proof of \Cref{lemma:ero:exit_remove_implies_response_set}.
    Let $a$ be the remove-response-set attempt for $\cellpointershort_\linearizationobject{}$ identified by \Cref{lemma:ero:exit_remove_implies_response_set:claim_two}.
    Since $a$ is a remove-response-set attempt for $\cellpointershort_\linearizationobject{}$ before $T^{exit}$, if $a$ is successful, we have satisfied the claim.
    If $a$ is unsuccessful, then by \Cref{lemma:ero:unsuccessful_remove_response_set_implies_successful_remove_response_set}, there is a successful remove-response-set attempt for $\cellpointershort_\linearizationobject{}$ before $a$ (and thus before $T^{exit}$).
    Therefore, in either case, there is a successful remove-response-set attempt for $\cellpointershort_\linearizationobject{}$ before $T^{exit}$.
    \qH{\Cref{lemma:ero:exit_remove_implies_response_set}}
\end{proof}

\subsubsection{The \doapplyandcopyresponse{} procedure has the intended effect}

In this section, we prove that the \doapplyandcopyresponse{} procedure with parameters $((\timeshort{}, \arbitraryvalue), \cellpointershort_\linearizationobject{})$ has the intended effect: (1) once it exits there is a successful $S$-attempt for timestamp $\timeshort{}$; and (2) once it exits there is a successful apply-response-set attempt for $\cellpointershort_\linearizationobject{}$.

\begin{lemma}\label{lemma:ero:conditional_exit_apply_implies_successful_s_attempt}
    Consider any invocation of the \doapplyandcopyresponse{} procedure with a first parameter of $\uniquerepositoryoperationshort_\linearizationobject = (\timeshort{}, \arbitraryvalue)$ which ends at some time $T^{exit}$ during $\mathcal{I}^\mathcal{B}$.
    If $P(\mathcal{I}^\mathcal{B})$ and $O(\mathcal{I}^\mathcal{B})$ hold, then there is a successful $S$-attempt for timestamp $\timeshort{}$ \mbox{before $T^{exit}$.}
\end{lemma}

\begin{proof}
    Suppose, for contradiction, there is an invocation $I$ of the \doapplyandcopyresponse{} procedure with a first parameter of $\uniquerepositoryoperationshort_\linearizationobject = (\timeshort{}, \arbitraryvalue)$ which ends at some time $T^{exit}$ during $\mathcal{I}^\mathcal{B}$ such that there is not a successful $S$-attempt for timestamp $\timeshort{}$ before $T^{exit}$ in $\mathcal{I}^\mathcal{B}$.
    Let $p$ be the process that invoked $I$.
    Since $I$ has a first parameter of $\uniquerepositoryoperationshort_\linearizationobject$, by \Cref{lemma:ero:l_event_corresponding_to_do_low_level_op}, there is an $L$-apply event $e$ before $I$ was invoked that set $\linearizationobject{}.\uniquerepositoryoperationlong{} = \uniquerepositoryoperationshort_\linearizationobject$.
    Hence, by \Cref{def:ero:english}, $e$ is for timestamp $\timeshort{}$.
    Furthermore, since $e$ is before $I$ was invoked and $T^{exit}$ is after $I$ was invoked, by transitivity, $e < T^{exit}$, thus all steps during $I$ are during $(e, T^{exit}]$. 
    There are two cases.
    Suppose during $(e, T^{exit}]$ there is at least one $L$-event in $\mathcal{I}^\mathcal{B}$. 
    Let $e_a$ be the next $L$-event after $e$ in $\mathcal{I}^\mathcal{B}$.
    Hence, since $e$ and $e_a$ are successive $L$-events in $\mathcal{I}^\mathcal{B}$, and $e$ is an $L$-apply event for timestamp $\timeshort{}$, by $O(\mathcal{I}^\mathcal{B})$, there is a successful $S$-attempt for timestamp $\timeshort{}$ during $(e, e_a)$.
    Therefore, since $e_a < T^{exit}$, there is a successful $S$-attempt for timestamp $\timeshort{}$ before $T^{exit}$.
    However, by our initial assumption of \Cref{lemma:ero:conditional_exit_apply_implies_successful_s_attempt}, there are no successful $S$-attempts for timestamp $\timeshort{}$ before $T^{exit}$ in $\mathcal{I}^\mathcal{B}$, a contradiction.

    Now suppose during $(e, T^{exit}]$ there are no $L$-events in $\mathcal{I}^\mathcal{B}$.
    Hence, $e$ is the last $L$-event in $\mathcal{I}$ where $\mathcal{I}$ is any prefix of $\mathcal{I}^\mathcal{B}$ during $(e, T^{exit}]$.
    Thus, since $e$ set $\linearizationobject{}.\uniquerepositoryoperationlong{} = \uniquerepositoryoperationshort_\linearizationobject{}$, by \Cref{observation:ero:where_objects_change}, $\linearizationobject{}.\uniquerepositoryoperationlong{} = \uniquerepositoryoperationshort_\linearizationobject{}$ throughout $(e, T^{exit}]$ (*).

    \begin{claimcustom}{\ref{lemma:ero:conditional_exit_apply_implies_successful_s_attempt}.1}\label{lemma:ero:conditional_exit_apply_implies_successful_s_attempt:claim_one}
        There are no successful $S$-attempts for timestamp $\timeshort{}$ during $(e, T^{exit}]$.
    \end{claimcustom}

    \begin{proof}
        Let $\mathcal{I}$ be the prefix of $\mathcal{I}^\mathcal{B}$ up to and including $T^{exit}$, so by (*) $e$ is the last $L$-event in $\mathcal{I}$.
        Hence, since $P(\mathcal{I}^\mathcal{B})$ and $O(\mathcal{I}^\mathcal{B})$ hold, and the last $L$-event in $\mathcal{I}$, $e$, is an $L$-apply event for timestamp $\timeshort{}$, by \Cref{lemma:ero:1_o_safety_holds}, from $e$ onwards in $\mathcal{I}$ there is at most one successful $S$-attempt for $\timeshort{}$ and no other successful $S$-attempts for any timestamp.
        So, since $\mathcal{I}$ is the prefix of $\mathcal{I}^\mathcal{B}$ up to and including $T^{exit}$, during $(e, T^{exit}]$ there is at most one successful $S$-attempt for $\timeshort{}$ and no other successful $S$-attempts for any timestamp.
        If during $(e, T^{exit}]$ there is a successful $S$-attempt for timestamp $\timeshort{}$, there would be a successful $S$-attempt for timestamp $\timeshort{}$ before $T^{exit}$, contradicting the initial assumption of \Cref{lemma:ero:conditional_exit_apply_implies_successful_s_attempt}.
        Therefore, there are no successful $S$-attempts for timestamp $\timeshort{}$ during $(e, T^{exit}$ as wanted.
        \qH{\Cref{lemma:ero:conditional_exit_apply_implies_successful_s_attempt:claim_one}}
    \end{proof}
    
    Let $(\uniquerepositoryoperationshort{}, \stateshort{}, \responseshort{})$ be the value $p$ read from $\stateobject$ on \cref{line:ero:state_read} during $I$; say at time $T^{\ref{line:ero:state_read}}$.

    \begin{claimcustom}{\ref{lemma:ero:conditional_exit_apply_implies_successful_s_attempt}.2}\label{lemma:ero:conditional_exit_apply_implies_successful_s_attempt:claim_two}
        $\uniquerepositoryoperationshort{} \neq \uniquerepositoryoperationshort{}_\linearizationobject{}$.
    \end{claimcustom}

    \begin{proof}
        Suppose, for contradiction, $\uniquerepositoryoperationshort{} = \uniquerepositoryoperationshort{}_\linearizationobject{}$.
        Since $e$ set $\linearizationobject{}.\uniquerepositoryoperationlong{} = \uniquerepositoryoperationshort{}_\linearizationobject{}$, by \Cref{lemma:ero:every_l_event_is_for_timestamp_other_than_zero}, $\uniquerepositoryoperationshort{}_\linearizationobject{} \neq (0, \arbitraryvalue)$.
        Furthermore, since $p$ read $\uniquerepositoryoperationshort{}$ from $\stateobject.\uniquerepositoryoperationlong{}$ at $T^{\ref{line:ero:state_read}}$, we have that $\stateobject.\uniquerepositoryoperationlong{} = \uniquerepositoryoperationshort{}_\linearizationobject{}$ at $T^{\ref{line:ero:state_read}}$.
        Hence, since $\uniquerepositoryoperationshort{}_\linearizationobject{} \neq (0, \arbitraryvalue)$ and the value of $\stateobject.\uniquerepositoryoperationlong{}$ is initially $(0, \noop)$, we have that some step set $\stateobject.\uniquerepositoryoperationlong{} = \uniquerepositoryoperationshort{}_\linearizationobject{}$ before $T^{\ref{line:ero:state_read}}$.
        Thus, by \Cref{observation:ero:where_objects_change}, there is a successful $S$-attempt $a$ that set $\stateobject.\uniquerepositoryoperationlong{} = \uniquerepositoryoperationshort{}_\linearizationobject{}$ before $T^{\ref{line:ero:state_read}}$.
        So, since $T^{\ref{line:ero:state_read}}$ is the time of a step during $I$, and $I$ exits at time $T^{exit}$, we have that $T^{\ref{line:ero:state_read}} \leq T^{exit}$, and thus $a < T^{exit}$.
        Let $e_a$ be $a$'s corresponding $L$-event, so $e_a < a$.
        Hence, since $a$ set $\stateobject.\uniquerepositoryoperationlong{} = \uniquerepositoryoperationshort{}_\linearizationobject{}$, by \Cref{lemma:ero:s_attempts_have_corresponding_l_events}, $e_a$ set $\linearizationobject{}.\uniquerepositoryoperationlong{} = \uniquerepositoryoperationshort{}_\linearizationobject{}$.
        Thus, since $e$ set $\linearizationobject{}.\uniquerepositoryoperationlong{} = \uniquerepositoryoperationshort{}_\linearizationobject{}$, $e$ and $e_a$ are both in $\mathcal{I}^\mathcal{B}$, and $P(\mathcal{I}^\mathcal{B})$ holds, by \Cref{lemma:ero:p_implies_unique_low_level_operations_in_linearization}, $e = e_a$.
        So, since $e_a < a$, we have that $e < a$, and since $a < T^{exit}$, we have that $a \in (e, T^{exit}]$.
        Therefore, there is a successful $S$-attempt during $(e, T^{exit}]$ (namely $a$).
        However, by \Cref{lemma:ero:conditional_exit_apply_implies_successful_s_attempt:claim_one}, there are no successful $S$-attempts during $(e, T^{exit}]$, a contradiction.
        \qH{\Cref{lemma:ero:conditional_exit_apply_implies_successful_s_attempt:claim_two}}
    \end{proof}

    We now finish the proof of \Cref{lemma:ero:conditional_exit_apply_implies_successful_s_attempt}.
    Since $\uniquerepositoryoperationshort{}_\linearizationobject{}$ is the first parameter of $I$, by (*) $\linearizationobject{}.\uniquerepositoryoperationlong{} = \uniquerepositoryoperationshort_\linearizationobject{}$ throughout $(e, T^{exit}]$, and all steps during $I$ are during $(e, T^{exit}]$, we have that $p$ finds the condition on \cref{line:ero:state_linearization_check} to be false during $I$.
    Hence, since $p$ exits $I$, $p$ executes \cref{line:ero:check_if_already_applied} during $I$.
    Since $p$ read $(\uniquerepositoryoperationshort{}, \stateshort{}, \responseshort{})$ from $\stateobject$ on \cref{line:ero:state_read} during $I$, $\uniquerepositoryoperationshort{}_\linearizationobject{}$ is the first parameter of $I$, and by \Cref{lemma:ero:conditional_exit_apply_implies_successful_s_attempt:claim_two} $\uniquerepositoryoperationshort{} \neq \uniquerepositoryoperationshort{}_\linearizationobject{}$, we have that $p$ finds the condition on \cref{line:ero:check_if_already_applied} to be true during $I$.
    Hence, since $p$ exits $I$, $p$ executes \cref{line:ero:state_cas} during $I$.
    Since by \Cref{lemma:ero:conditional_exit_apply_implies_successful_s_attempt:claim_one} there are no successful $S$-attempts during $(e, T^{exit}]$, and all steps during $I$ are during $(e, T^{exit}]$, we have that the value of $\stateobject$ is the same throughout $I$.
    Hence, since $p$ read $(\uniquerepositoryoperationshort{}, \stateshort{}, \responseshort{})$ from $\stateobject$ on \cref{line:ero:state_read} during $I$, we have that $p$'s \CASop{} operation on \cref{line:ero:state_cas} during $I$ is successful.
    Thus, by \Cref{def:ero:english}, $p$ performs a successful $S$-attempt during $I$.
    Therefore, since $I$ exits at $T^{exit}$, there is a successful $S$-attempt before $T^{exit}$ in $\mathcal{I}^\mathcal{B}$, contradicting the initial assumption of \Cref{lemma:ero:conditional_exit_apply_implies_successful_s_attempt}.
    \qH{\Cref{lemma:ero:conditional_exit_apply_implies_successful_s_attempt}}
\end{proof}

\begin{lemma}\label{lemma:ero:conditional_exit_apply_implies_response_set}
    Consider any invocation of the \doapplyandcopyresponse{} procedure with a second parameter of $\cellpointershort_\linearizationobject$ which ends at some time $T^{exit}$ during $\mathcal{I}^\mathcal{B}$.
    If $P(\mathcal{I}^\mathcal{B})$, $Q(\mathcal{I}^\mathcal{B})$, and $R(\mathcal{I}^\mathcal{B})$ hold, then there is a successful apply-response-set attempt for $\cellpointershort_\linearizationobject$ \mbox{before $T^{exit}$.}
\end{lemma}

\begin{proof}
    Consider any invocation $I$ of the \doapplyandcopyresponse{} procedure which ends at some time $T^{exit}$ during $\mathcal{I}^\mathcal{B}$.
    Let $(\uniquerepositoryoperationshort_\linearizationobject{}, \cellpointershort_\linearizationobject)$ be the parameters of $I$ and suppose $p$ is the process that invoked $I$.
    Hence, by \Cref{lemma:ero:l_event_corresponding_to_do_low_level_op}, there is an $L$-apply event $e$ for $\cellpointershort_\linearizationobject$ before $I$ was invoked that set $\linearizationobject{}$ to $(\uniquerepositoryoperationshort_\linearizationobject{}, \cellpointershort_\linearizationobject)$.
    Thus, by \Cref{lemma:ero:every_l_event_is_for_pointer_from_universe} $\cellpointershort_\linearizationobject \in \celluniverse{}$.

    The proof strategy is to identify an apply-response-set attempt for $\cellpointershort_\linearizationobject{}$ before $T^{exit}$.
    If this is successful, we are done, but if it is unsuccessful, then by \Cref{lemma:ero:unsuccessful_apply_response_set_implies_successful_apply_response_set}, there is a successful apply-response-set attempt for $\cellpointershort_\linearizationobject{}$ beforehand, in which case we are also done.

    \begin{claimcustom}{\ref{lemma:ero:conditional_exit_apply_implies_response_set}.1}\label{lemma:ero:conditional_exit_apply_implies_response_set:claim_one}
        The \doapplyandcopyresponse{} procedure was invoked with parameters $(\uniquerepositoryoperationshort_\linearizationobject{}, \cellpointershort_\linearizationobject)$ at some time $T'_b$ and ends at some time $T'_e$ during $\mathcal{I}^\mathcal{B}$ such that: (1) $[T'_b, T'_e] \subseteq (e, T^{exit}]$; (2) $e$ is the last $L$-event in $\mathcal{I}$ where $\mathcal{I}$ is any prefix of $\mathcal{I}^\mathcal{B}$ during $[T'_b, T'_e]$; (3) $\linearizationobject.\uniquerepositoryoperationlong = \uniquerepositoryoperationshort_\linearizationobject{}$ at the end of $\mathcal{I}$ where $\mathcal{I}$ is as in (2); and (4) the list of cells conforms to $\List(\mathcal{I})$ in $\mathcal{I}$ where $\mathcal{I}$ is as in (2).   
    \end{claimcustom}

    \begin{proof}
        There are two cases.
        \begin{itemize}
            \item[] \hspace{0pt}\textbf{Case 1.} During $(e, T^{exit}]$ there are no $L$-events.
    
            We prove $I$ is the desired invocation.
            By definition $I$'s parameters are $(\uniquerepositoryoperationshort_\linearizationobject{}, \cellpointershort_\linearizationobject)$.
            Let $T_b$ be the time $I$ was invoked.
            We first prove (1).
            Since $e < T_b$, we have that $[T_b, T^{exit}] \subseteq (e, T^{exit}]$.
            We now prove (2).
            Consider any prefix $\mathcal{I}$ of $\mathcal{I}^\mathcal{B}$ during $[T_b, T^{exit}]$.
            Since $[T_b, T^{exit}] \subseteq (e, T^{exit}]$, and by assumption there are no $L$-events during $(e, T^{exit}]$, we have that $e$ is the last $L$-event in $\mathcal{I}$.
            We now prove (3).
            Since by assumption of Case 1 there are no $L$-events during $(e, T^{exit}]$, and $e$ set $\linearizationobject{}.\uniquerepositoryoperationlong{} = \uniquerepositoryoperationshort_\linearizationobject{}$, by \Cref{observation:ero:where_objects_change}, $\linearizationobject.\uniquerepositoryoperationlong{} = \uniquerepositoryoperationshort_\linearizationobject{}$ throughout $(e, T^{exit}]$.
            Hence, since $[T_b, T^{exit}] \subseteq (e, T^{exit}]$, $\linearizationobject.\uniquerepositoryoperationlong{} = \uniquerepositoryoperationshort_\linearizationobject{}$ throughout $[T_b, T^{exit}]$.
            This implies (3).
            We now prove (4).
            Since $e$ is the last $L$-event in $\mathcal{I}$, and $e$ is an $L$-apply event, we have that the last $L$-event in $\mathcal{I}$ is not an $L$-add or $L$-remove event.
            Hence, since $\mathcal{I}$ is a finite and by assumption $P(\mathcal{I}^\mathcal{B})$, $Q(\mathcal{I}^\mathcal{B})$, and $R(\mathcal{I}^\mathcal{B})$ hold, by \Cref{lemma:ero:conditional_classification_lemma}, the list of cells conforms to $\List(\mathcal{I})$ in $\mathcal{I}$.
    
            \item[] \hspace{0pt}\textbf{Case 2.} During $(e, T^{exit}]$ there is at least one $L$-event.

            We identify an earlier invocation than $I$.
            Let $e'$ be the next $L$-event after $e$ in $\mathcal{I}^\mathcal{B}$, so $e' \in (e, T^{exit}]$.
            Let $q$ be the process that executed $e'$.
            Hence, there are no $L$-events during $(e, e')$.
            Let $T^{\ref{line:ero:linearization_read}}_q$ be the time of $q$'s last execution of \cref{line:ero:linearization_read} before $e'$.
            Since $e$ and $e'$ are successive $L$-events and $T^{\ref{line:ero:linearization_read}}_q$ is the time of $q$'s last execution of \cref{line:ero:linearization_read} before $e'$, by \Cref{lemma:ero:successive_l_event_read_previous_l_event_value}, $q$ read the value that $e$ set $\linearizationobject{}$ to on \cref{line:ero:linearization_read} at $T^{\ref{line:ero:linearization_read}}_q$, and since $P(\mathcal{I}^\mathcal{B})$ holds, by \Cref{lemma:ero:successive_l_event_read_previous_l_event_value_after_it_happened}, $e < T^{\ref{line:ero:linearization_read}}_q$.
            Hence, since $e$ set $\linearizationobject{}$ to $(\uniquerepositoryoperationshort_\linearizationobject{}, \cellpointershort_\linearizationobject)$, we have that $q$ read $(\uniquerepositoryoperationshort_\linearizationobject{}, \cellpointershort_\linearizationobject)$ from $\linearizationobject$ on \cref{line:ero:linearization_read} at $T^{\ref{line:ero:linearization_read}}_q$.
            Thus, since $\uniquerepositoryoperationshort_\linearizationobject{} = (\arbitraryvalue, \langle \doopandcopyresponse{}, \arbitraryvalue \rangle)$ (because $e$ set $\linearizationobject{}.\uniquerepositoryoperationlong{} = \uniquerepositoryoperationshort_\linearizationobject{}$ and $e$ is an $L$-apply event), we have that between $T^{\ref{line:ero:linearization_read}}_q$ and $e'$, $q$ invoked and exited the \doapplyandcopyresponse{} procedure on \cref{line:ero:do_apply_and_copy_response} with parameters $(\uniquerepositoryoperationshort_\linearizationobject{},  \cellpointershort_\linearizationobject)$.
            Denote this invocation by $I'$ and the time $q$ began and exited it by $T'_b$ and $T'_e$, respectively.
            We claim that $I'$ is the desired invocation.
            We already established $I'$ has the desired parameters, so first we prove (1).
            Since $e < T^{\ref{line:ero:linearization_read}}_q$, $T^{\ref{line:ero:linearization_read}}_q < T'_b$, $T'_b < T'_e$, and $T'_e < e'$, by transitivity, $[T'_b, T'_e] \subseteq (e, e')$, and since $e' \leq T^{exit}$, we have that $[T'_b, T'_e] \subseteq (e, T^{exit}]$.
            We now prove (2).
            Consider any prefix $\mathcal{I}$ of $\mathcal{I}^\mathcal{B}$ during $[T'_b, T'_e]$.
            Since $[T'_b, T'_e] \subseteq (e, e']$, and there are no $L$-events during $(e, e')$, we have that $e$ is the last $L$-event in $\mathcal{I}$.
            We now prove (3).
            Since there are no $L$-events during $(e, e')$, and $e$ set $\linearizationobject{}.\uniquerepositoryoperationlong{} = \uniquerepositoryoperationshort_\linearizationobject{}$, by \Cref{observation:ero:where_objects_change}, $\linearizationobject.\uniquerepositoryoperationlong{} = \uniquerepositoryoperationshort_\linearizationobject{}$ throughout $(e, e')$.
            Hence, since $[T'_b, T'_e] \subseteq (e, e')$, $\linearizationobject.\uniquerepositoryoperationlong{} = \uniquerepositoryoperationshort_\linearizationobject{}$ throughout $[T'_b, T'_e]$.
            This implies (3).
            We now prove (4).
            Since $e$ is the last $L$-event in $\mathcal{I}$, and $e$ is an $L$-apply event, we have that the last $L$-event in $\mathcal{I}$ is not an $L$-add or $L$-remove event.
            Hence, since $\mathcal{I}$ is a finite and by assumption $P(\mathcal{I}^\mathcal{B})$, $Q(\mathcal{I}^\mathcal{B})$, and $R(\mathcal{I}^\mathcal{B})$ hold, by \Cref{lemma:ero:conditional_classification_lemma}, the list of cells conforms to $\List(\mathcal{I})$ in $\mathcal{I}$.
            \qH{\Cref{lemma:ero:conditional_exit_apply_implies_response_set:claim_one}}
        \end{itemize}
    \end{proof}

    \begin{claimcustom}{\ref{lemma:ero:conditional_exit_apply_implies_response_set}.2}\label{lemma:ero:conditional_exit_apply_implies_response_set:claim_two}
        There is an apply-response-set attempt for $\cellpointershort_\linearizationobject{}$ before $T^{exit}$.
    \end{claimcustom}

    \begin{proof}
        Let $I'$ be the invocation of the \doapplyandcopyresponse{} procedure in \Cref{lemma:ero:conditional_exit_apply_implies_response_set:claim_one} and let $q$ be the process that executed $I'$.
        Furthermore, let $T'_b$ and $T'_e$ be the times during $\mathcal{I}^\mathcal{B}$ that $q$ begins and exits $I'$, respectively.
        Since $q$ exits $I'$, it follows that $q$ began and exited the \setrepositoryoperationresponse{} procedure invoked on \cref{line:ero:apply_set_response} during $I'$; denote this invocation by $I_r$.
        Hence, $q$ began and exited the Acquire procedure on \cref{line:ero:set_response_acquire} during $I_r$.
        Denote this invocation of the Acquire procedure by $I^*$.
        Since the parameters of $I'$ are $(\uniquerepositoryoperationshort_\linearizationobject{}, \cellpointershort_\linearizationobject)$, we have that the parameters of $I^*$ are $(\uniquerepositoryoperationshort_\linearizationobject{}, \cellpointershort_\linearizationobject)$.
        
        We now satisfy the conditions of \Cref{lemma:ero:under_certain_conditions_acquire_returns_found_weak}.
        As established above, $\cellpointershort_\linearizationobject{} \in \celluniverse{}$.
        Let $T_b$ be the time $q$ invoked $I^*$ and let $T^{\ref{line:ero:acquire_next_linearization_changed_check}}$ be the last time $q$ executes \cref{line:ero:acquire_next_linearization_changed_check} during $I^*$.
        Recall that $T^{\ref{line:ero:acquire_next_linearization_changed_check}}$ is well-defined by \Cref{lemma:ero:exit_acquire_implies_executing_105}.
        By assumption $P(\mathcal{I}^\mathcal{B})$ holds.
        Let $\mathcal{I}^{include}_e$ be the prefix of $\mathcal{I}^\mathcal{B}$ up to and including $e$.
        We plug in $\mathcal{I}^{include}_e$ for $\mathcal{I}$ in \Cref{lemma:ero:under_certain_conditions_acquire_returns_found_weak}.
        Consider any prefix $\mathcal{I}^*$ of $\mathcal{I}^\mathcal{B}$ during $[T_b, T^{\ref{line:ero:acquire_next_linearization_changed_check}}]$.
        We first satisfy condition 1.
        Since by definition all steps during $[T_b, T^{\ref{line:ero:acquire_next_linearization_changed_check}}]$ are during $I^*$, and all steps during $I^*$ are during $I'$, we have that $[T_b, T^{\ref{line:ero:acquire_next_linearization_changed_check}}] \subseteq [T'_b, T'_e]$.
        Hence, since by (3) of \Cref{lemma:ero:conditional_exit_apply_implies_response_set:claim_one}, $\linearizationobject{}.\uniquerepositoryoperationlong{} = \uniquerepositoryoperationshort_\linearizationobject{}$ at the end of $\mathcal{I}'$ for any prefix $\mathcal{I}'$ of $\mathcal{I}^\mathcal{B}$ during $[T'_b, T'_e]$, we have that $\linearizationobject{}.\uniquerepositoryoperationlong{} = \uniquerepositoryoperationshort_\linearizationobject{}$ at the end of $\mathcal{I}^*$.
        We now satisfy condition 2.
        Since by (4) of \Cref{lemma:ero:conditional_exit_apply_implies_response_set:claim_one} the list of cells conforms to $\List(\mathcal{I}')$ in $\mathcal{I}'$ for any prefix $\mathcal{I}'$ of $\mathcal{I}^\mathcal{B}$ during $[T'_b, T'_e]$, and $[T_b, T^{\ref{line:ero:acquire_next_linearization_changed_check}}] \subseteq [T'_b, T'_e]$, we have that that the list of cells conforms to $\List(\mathcal{I}^*)$ in $\mathcal{I}^*$.
        We now prove that $\List(\mathcal{I}^{include}_e) = \List(\mathcal{I}^*)$.
        Since $\mathcal{I}^{include}_e$ is the prefix of $\mathcal{I}^\mathcal{B}$ up to and including $e$, and $e$ is an $L$-event, it follows that $e$ is the last $L$-event in $\mathcal{I}^{include}_e$.
        Furthermore, since by (2) of \Cref{lemma:ero:conditional_exit_apply_implies_response_set:claim_one} $e$ is the last $L$-event in $\mathcal{I}'$ for any prefix $\mathcal{I}'$ of $\mathcal{I}^\mathcal{B}$ during $[T'_b, T'_e]$, and $[T_b, T^{\ref{line:ero:acquire_next_linearization_changed_check}}] \subseteq [T'_b, T'_e]$, we have that $e$ is the last $L$-event in $\mathcal{I}^*$.
        Together, these imply that the sequence of $L$-events is the same in $\mathcal{I}^{include}_e$ and $\mathcal{I}^*$, and so by \Cref{def:ero:logical_list}, $\List(\mathcal{I}^{include}_e) = \List(\mathcal{I}^*)$.
        Therefore, since the list of cells conforms to $\List(\mathcal{I}^*)$ in $\mathcal{I}^*$, we have that the list of cells conforms to $\List(\mathcal{I}^{include}_e)$ in $\mathcal{I}^*$ as wanted.
        So, by \Cref{lemma:ero:under_certain_conditions_acquire_returns_found_weak} if $\cellpointershort_\linearizationobject \in \List(\mathcal{I}^{include}_e)$ then the response of $I^*$ is $\found$ (*).

        We now prove that $\cellpointershort_\linearizationobject \in \List(\mathcal{I}^{include}_e)$.
        Suppose, for contradiction, $\cellpointershort_\linearizationobject \notin \List(\mathcal{I}^{include}_e)$.
        Since $e$ is an $L$-apply event for $\cellpointershort_\linearizationobject{}$ in $\mathcal{I}^\mathcal{B}$, by \Cref{lemma:ero:apply_events_are_preceeded_by_add_events}, there is an $L$-add event $e_{add}$ for $\cellpointershort_\linearizationobject$ before $e$ in $\mathcal{I}^\mathcal{B}$.
        Hence, since $\mathcal{I}^{include}_e$ is the prefix of $\mathcal{I}^\mathcal{B}$ up to and including $e$, and $e_{add} < e$, we have that $e_{add}$ is in $\mathcal{I}^{include}_e$.
        Thus, since $\cellpointershort_\linearizationobject \notin \List(\mathcal{I}^{include}_e)$ and $e_{add}$ is an $L$-add event for $\cellpointershort_\linearizationobject$ in $\mathcal{I}^{include}_e$, by \Cref{def:ero:logical_list}, there is an $L$-remove event $e_{remove}$ for $\cellpointershort_\linearizationobject$ after $e_{add}$ in $\mathcal{I}^{include}_e$.
        Hence, $e_{remove}$ is an $L$-remove event for $\cellpointershort_\linearizationobject$ in $\mathcal{I}^\mathcal{B}$, and so by \Cref{lemma:ero:remove_events_are_preceeded_by_apply_events}, there is an $L$-apply event $e'$ for $\cellpointershort_\linearizationobject$ before $e_{remove}$ in $\mathcal{I}^\mathcal{B}$.
        Since $e_{remove}$ is in $\mathcal{I}^{include}_e$, and $\mathcal{I}^{include}_e$ is the prefix of $\mathcal{I}^\mathcal{B}$ up to and including $e$, we have that $e_{remove} \leq e$.
        Hence, since $e' < e_{remove}$, by transitivity, $e' < e$, and so $e' \neq e$.
        Therefore, there are two $L$-apply events for $\cellpointershort_\linearizationobject$ in $\mathcal{I}^\mathcal{B}$.
        However, by $P(\mathcal{I}^\mathcal{B})$, there is at most one $L$-apply event for $\cellpointershort_\linearizationobject$ in $\mathcal{I}^\mathcal{B}$, a contradiction.
        
        We now finish the proof of \Cref{lemma:ero:conditional_exit_apply_implies_response_set:claim_two}.
        Since $\cellpointershort_\linearizationobject \in \List(\mathcal{I}^{include}_e)$, by (*), the response of $I^*$ is $\found$.
        Thus, since $q$ exits $I_r$, we have that $q$ finds the condition on \cref{line:ero:set_response_found_check} to be true during $I_r$.
        So, $q$ executes \cref{line:ero:responses_set_attempt} during $I_r$.
        Since $I^*$'s parameters are $(\uniquerepositoryoperationshort_\linearizationobject{}, \cellpointershort_\linearizationobject)$ and $\uniquerepositoryoperationshort_\linearizationobject{} = (\arbitraryvalue, \langle \doopandcopyresponse{}, \arbitraryvalue \rangle)$ (because $e$ set $\linearizationobject{}.\uniquerepositoryoperationlong{} = \uniquerepositoryoperationshort_\linearizationobject{}$ and $e$ is an $L$-apply event), by \Cref{def:ero:english}, this execution is an apply-response-set attempt for $\cellpointershort_\linearizationobject$ during $I_r$ (and thus $[T'_b, T'_e]$ because $I_r$ was invoked during $I'$).
        Therefore, since by (1) of \Cref{lemma:ero:conditional_exit_apply_implies_response_set:claim_one} $[T'_b, T'_e] \subseteq (e, T^{exit}]$, we have that there is an apply-response-set attempt for $\cellpointershort_\linearizationobject{}$ before $T^{exit}$ as wanted.
        \qH{\Cref{lemma:ero:conditional_exit_apply_implies_response_set:claim_two}}
    \end{proof}

    We now return to the proof of \Cref{lemma:ero:conditional_exit_apply_implies_response_set}.
    Let $a$ be the apply-response-set attempt for $\cellpointershort_\linearizationobject{}$ identified by \Cref{lemma:ero:conditional_exit_apply_implies_response_set:claim_two}.
    Since $a$ is an apply-response-set attempt for $\cellpointershort_\linearizationobject{}$ before $T^{exit}$, if $a$ is successful, we have satisfied the claim.
    If $a$ is unsuccessful, then by \Cref{lemma:ero:unsuccessful_apply_response_set_implies_successful_apply_response_set}, there is a successful apply-response-set attempt for $\cellpointershort_\linearizationobject{}$ before $a$ (and thus before $T^{exit}$).
    Therefore, in either case, there is a successful apply-response-set attempt for $\cellpointershort_\linearizationobject{}$ before $T^{exit}$.
    \qH{\Cref{lemma:ero:conditional_exit_apply_implies_response_set}}
\end{proof}

\subsubsection{The IsDone procedure has the intended effect}

Now that we have proven that the \doaddcell{}, \doremovecell{}, and \doapplyandcopyresponse{} procedures have the intended effect, we are ready to prove that the IsDone procedure has the intended effect: if a process $p$ invokes IsDone$(\arbitraryvalue, \uniquerepositoryoperationshort_\announceobject, \arbitraryvalue)$ on \cref{line:ero:check_if_announce_is_done}, where $\uniquerepositoryoperationshort_\announceobject$ is the unique low-level operation it read in $\announceobject$ on \cref{line:ero:announce_read}, and this invocation returns $\done$ (resp. $\notdone$) then $\uniquerepositoryoperationshort_\announceobject$ is written (resp. not written) in $\linearizationobject{}$.
The exact timing of this $L$-event (or its absence) is delicate, as we will see shortly.
We start by characterizing the behavior of the Acquire procedure during an invocation of the IsDone procedure.

\begin{lemma}\label{lemma:ero:after_helping_last_l_event_the_list_of_cells_is_stuck_weak}
    Consider any process $p$ and suppose during an iteration $I$ of the loop on \cref{line:ero:do_work_while_loop} $p$ does the following during $\mathcal{I}^\mathcal{B}$: (1) $p$ reads $(\arbitraryvalue, \cellpointershort)$ from $\announceobject$ on \cref{line:ero:announce_read} for some $\cellpointershort{} \in \celluniverse$; and (2) $p$ exits the Acquire procedure on \cref{line:ero:announce_acquire} with response $\status$.
    Let $\mathcal{I}$ be the prefix of $\mathcal{I}^\mathcal{B}$ up to and including the time $p$ executed \cref{line:ero:linearization_read} during $I$.
    If $P(\mathcal{I}^\mathcal{B})$, $Q(\mathcal{I}^\mathcal{B})$, and $R(\mathcal{I}^\mathcal{B})$ hold, and $\status \neq \timechange$, then the following are true:
    \begin{compactenum}
        \item if $\cellpointershort{} \in \List(\mathcal{I})$, then $\status = \found$; and
        \item if $\cellpointershort{} \notin \List(\mathcal{I})$, then $\status = \notfound$.
    \end{compactenum}
\end{lemma}

\begin{proof}
    Suppose $p$ read $\uniquerepositoryoperationshort_\linearizationobject{}$ from $\linearizationobject{}.\uniquerepositoryoperationlong$ on \cref{line:ero:linearization_read} during $I$; say at time $T^{\ref{line:ero:linearization_read}}$.
    Let $I'$ be the invocation of the Acquire procedure on \cref{line:ero:announce_acquire} during $I$.    
    We satisfy the conditions of \Cref{lemma:ero:under_certain_conditions_acquire_returns_found_weak} for $I'$.
    Since $p$ read $\uniquerepositoryoperationshort_\linearizationobject{}$ from $\linearizationobject{}.\uniquerepositoryoperationlong$ on \cref{line:ero:linearization_read} during $I$ and $p$ read $(\arbitraryvalue, \cellpointershort)$ from $\announceobject$ on \cref{line:ero:announce_read} during $I$, we have that the parameters of $I'$ are $(\uniquerepositoryoperationshort_\linearizationobject{}, \cellpointershort)$.
    Since $p$ exits $I'$, and by \Cref{lemma:ero:exit_acquire_implies_executing_105} $p$ executes \cref{line:ero:acquire_next_linearization_changed_check} at least once during $I'$, we have that $p$ executes \cref{line:ero:acquire_next_linearization_changed_check} for a final time during $I'$.
    Let $T_b$ be the time $p$ invoked $I'$ and let $T^{\ref{line:ero:acquire_next_linearization_changed_check}}$ be the last time $p$ executes \cref{line:ero:acquire_next_linearization_changed_check} during $I'$.
    Hence, since $T^{\ref{line:ero:linearization_read}}$ is the time $p$ executed \cref{line:ero:linearization_read} during $I$, by transitivity, $T^{\ref{line:ero:linearization_read}} < T_b$.
    Furthermore, since $T_b$ is the time $p$ invoked $I'$, and $T^{\ref{line:ero:acquire_next_linearization_changed_check}}$ is the last time $p$ executes \cref{line:ero:acquire_next_linearization_changed_check} during $I'$, we have that $T_b < T^{\ref{line:ero:acquire_next_linearization_changed_check}}$, and so $T^{\ref{line:ero:linearization_read}} < T_b < T^{\ref{line:ero:acquire_next_linearization_changed_check}}$.
    Now consider any prefix $\mathcal{I}^*$ of $\mathcal{I}^\mathcal{B}$ in $[T_b, T^{\ref{line:ero:acquire_next_linearization_changed_check}}]$.
    
    \begin{claimcustom}{\ref{lemma:ero:after_helping_last_l_event_the_list_of_cells_is_stuck_weak}.1}\label{lemma:ero:after_helping_last_l_event_the_list_of_cells_is_stuck_weak:claim_three}
        $\linearizationobject{}.\uniquerepositoryoperationlong = \uniquerepositoryoperationshort_\linearizationobject{}$ at the end of $\mathcal{I}^*$.
    \end{claimcustom}

    \begin{proof}
        Suppose, for contradiction, $\linearizationobject{}.\uniquerepositoryoperationlong \neq \uniquerepositoryoperationshort_\linearizationobject{}$ at the end of $\mathcal{I}^*$.
        Hence, since $\mathcal{I}^*$ is any prefix of $\mathcal{I}^\mathcal{B}$ in $[T_b, T^{\ref{line:ero:acquire_next_linearization_changed_check}}]$, we have that $\linearizationobject{}.\uniquerepositoryoperationlong \neq \uniquerepositoryoperationshort_\linearizationobject{}$ at some time $T \in [T_b, T^{\ref{line:ero:acquire_next_linearization_changed_check}}]$.
        Since $p$ read $\uniquerepositoryoperationshort_\linearizationobject{}$ from $\linearizationobject{}.\uniquerepositoryoperationlong$ at $T^{\ref{line:ero:linearization_read}}$, $\linearizationobject{}.\uniquerepositoryoperationlong \neq \uniquerepositoryoperationshort_\linearizationobject{}$ at $T$, and $T^{\ref{line:ero:linearization_read}} < T_b \leq T$, we have that between $T^{\ref{line:ero:linearization_read}}$ and $T$, the value of $\linearizationobject{}.\uniquerepositoryoperationlong$ changed.
        Thus, by \Cref{observation:ero:where_objects_change}, there is an $L$-event between $T^{\ref{line:ero:linearization_read}}$ and $T$.
        Therefore, since $T \leq T^{\ref{line:ero:acquire_next_linearization_changed_check}}$, we have that there is an $L$-event between $T^{\ref{line:ero:linearization_read}}$ and $T^{\ref{line:ero:acquire_next_linearization_changed_check}}$.
        However, by \Cref{lemma:ero:if_announce_acquire_does_not_return_time_change_then_no_l_events}, there are no $L$-events between $T^{\ref{line:ero:linearization_read}}$ and $T^{\ref{line:ero:acquire_next_linearization_changed_check}}$, a contradiction.
        \qH{\Cref{lemma:ero:after_helping_last_l_event_the_list_of_cells_is_stuck_weak:claim_three}}
    \end{proof}

    \begin{claimcustom}{\ref{lemma:ero:after_helping_last_l_event_the_list_of_cells_is_stuck_weak}.2}\label{lemma:ero:after_helping_last_l_event_the_list_of_cells_is_stuck_weak:claim_four}
        The list of cells conforms to $\List(\mathcal{I})$ in $\mathcal{I}^*$.
    \end{claimcustom}

    \begin{proof}
        There are two cases, the first of which is trivial.
        
        Suppose $\mathcal{I}^*$ has zero $L$-events.
        Then, since $\mathcal{I}^*$ is a finite prefix of $\mathcal{I}^\mathcal{B}$, and $P(\mathcal{I}^\mathcal{B})$, $Q(\mathcal{I}^\mathcal{B})$, and $R(\mathcal{I}^\mathcal{B})$ hold, by \Cref{lemma:ero:conditional_classification_lemma}, the list of cells conforms to $\List(\mathcal{I}^*)$ in $\mathcal{I}^*$ as wanted.
        
        Now suppose $\mathcal{I}^*$ has at least one $L$-event.
        Let $e_{last}$ be the last $L$-event in $\mathcal{I}^*$.
        Since $T^{\ref{line:ero:linearization_read}} < T_b < T^{\ref{line:ero:acquire_next_linearization_changed_check}}$, and by \Cref{lemma:ero:if_announce_acquire_does_not_return_time_change_then_no_l_events} there are no $L$-events throughout $[T^{\ref{line:ero:linearization_read}}, T^{\ref{line:ero:acquire_next_linearization_changed_check}}]$, it follows that $e_{last} < T^{\ref{line:ero:linearization_read}}$.
        Thus, since $\mathcal{I}$ is the prefix of $\mathcal{I}^\mathcal{B}$ up to and including $T^{\ref{line:ero:linearization_read}}$, we have that $e_{last}$ is the last $L$-event in $\mathcal{I}$.
        So, by \Cref{observation:ero:where_objects_change}, $\linearizationobject{}$ is unchanged from $e_{last}$ onwards in $\mathcal{I}$.
        Hence, since $e_{last} < T^{\ref{line:ero:linearization_read}}$ and $p$ read $\uniquerepositoryoperationshort_\linearizationobject{}$ from $\linearizationobject{}.\uniquerepositoryoperationlong$ on \cref{line:ero:linearization_read} at $T^{\ref{line:ero:linearization_read}}$, we have that $e_{last}$ set $\linearizationobject{}.\uniquerepositoryoperationlong = \uniquerepositoryoperationshort_\linearizationobject{}$.
        Suppose $e_{last}$ set $\linearizationobject{} = (\uniquerepositoryoperationshort_\linearizationobject{}, \uniquecellpointershort_\linearizationobject{})$.
        So, by \Cref{def:ero:english}, $e_{last}$ is an $L$-event for $\cellpointershort{}_\linearizationobject{}$, and so by \Cref{lemma:ero:every_l_event_is_for_pointer_from_universe} $\cellpointershort{}_\linearizationobject{} \in \celluniverse$.
        Hence, by \Cref{lemma:ero:every_l_event_is_add_apply_or_remove} $e_{last}$ is either an $L$-add event for $\cellpointershort{}_\linearizationobject{}$, an $L$-apply event for $\cellpointershort{}_\linearizationobject{}$, or an $L$-remove event for $\cellpointershort{}_\linearizationobject{}$.
        We consider each case separately and prove that the list of cells conforms to $\List(\mathcal{I}^*)$ in $\mathcal{I}^*$.
        We then prove that $\List(\mathcal{I}^*) = \List(\mathcal{I})$ afterwards.
        \begin{itemize}
            \item[] \hspace{0pt}\textbf{Case 1.} $e_{last}$ is an $L$-add event for $\cellpointershort{}_\linearizationobject{}$.
    
            Hence, since $e_{last}$ set $\linearizationobject{} = (\uniquerepositoryoperationshort_\linearizationobject{}, \uniquecellpointershort_\linearizationobject{})$, by \Cref{def:ero:english}, $\uniquerepositoryoperationshort_\linearizationobject{} = (\arbitraryvalue, \addcell)$.
            Thus, since $p$ read $(\uniquerepositoryoperationshort_\linearizationobject{}, \uniquecellpointershort_\linearizationobject{})$ from $\linearizationobject{}$ on \cref{line:ero:linearization_read} at $T^{\ref{line:ero:linearization_read}}$, and $p$ invokes $I'$ during $I$, we have that $p$ found the condition on \cref{line:ero:do_add_cell_condition} to be true during $I$, and so $p$ invoked and exited the \doaddcell{} procedure on \cref{line:ero:do_add_cell} with a second parameter of $\cellpointershort{}_\linearizationobject{}$ which ends at some time $T_e < T_b$ during $I$.
            Hence, since $P(\mathcal{I}^\mathcal{B})$, $Q(\mathcal{I}^\mathcal{B})$, and $R(\mathcal{I}^\mathcal{B})$ hold, by \Cref{lemma:ero:exit_do_add_implies_done_strong}, there is a successful list-add attempt $a$ for $\cellpointershort{}_\linearizationobject{}$ before $T_e$.
            Let $e_b$ be $a$'s corresponding $L$-event (see \Cref{lemma:ero:l_event_corresponding_to_do_low_level_op}), so  $e_b < a$ and $e_b$ is an $L$-add event for $\cellpointershort_\linearizationobject{}$.
            Hence, since $e_{last}$ is an $L$-add event for $\cellpointershort_\linearizationobject{}$, by $P(\mathcal{I}^\mathcal{B})$, $e_b = e_{last}$, and so $e_{last} < a$.
            Since $a < T_e$ and $T_e < T_b$, by transitivity, $a < T_b$, and so $a$ is in $\mathcal{I}^*$.
            Thus, since by definition $e_{last}$ is the last $L$-event in $\mathcal{I}^*$ and $e_{last} < a$, we have that from $e_{last}$ onwards in $\mathcal{I}^*$ there is a successful list-add attempt.
            Hence, since $\mathcal{I}^*$ is finite, and $P(\mathcal{I}^\mathcal{B})$, $Q(\mathcal{I}^\mathcal{B})$, and $R(\mathcal{I}^\mathcal{B})$ hold, by \Cref{lemma:ero:conditional_classification_lemma}, the list of cells conforms to $\List(\mathcal{I}^*)$ in $\mathcal{I}^*$ as wanted.
    
            \item[] \hspace{0pt}\textbf{Case 2.} $e_{last}$ is an $L$-apply event for $\cellpointershort{}_\linearizationobject{}$.
    
            Since by definition $e_{last}$ is the last $L$-event in $\mathcal{I}^*$, and $e_{last}$ is an $L$-apply event, we have that the last $L$-event in $\mathcal{I}^*$ is not an $L$-add or $L$-remove event.
            Hence, since $\mathcal{I}^*$ is finite, and $P(\mathcal{I}^\mathcal{B})$, $Q(\mathcal{I}^\mathcal{B})$, and $R(\mathcal{I}^\mathcal{B})$ hold, by \Cref{lemma:ero:conditional_classification_lemma}, the list of cells conforms to $\List(\mathcal{I}^*)$ in $\mathcal{I}^*$.

            \item[] \hspace{0pt}\textbf{Case 3.} $e_{last}$ is an $L$-remove event for $\cellpointershort{}_\linearizationobject{}$.

            The proof is the same as Case 1 by replacing the adds with the appropriate removes (i.e., list-add with list-remove and $\addcell$ with $\removecell$) and \Cref{lemma:ero:exit_do_add_implies_done_strong} with \Cref{lemma:ero:exit_do_remove_implies_done}.
        \end{itemize}

        We now prove that $\List(\mathcal{I}^*) = \List(\mathcal{I})$ which completes the proof.
        Since $e_{last}$ is the last $L$-event in $\mathcal{I}^*$ and $\mathcal{I}$, and $\mathcal{I}^*$ and $\mathcal{I}$ are both prefixes of $\mathcal{I}^\mathcal{B}$, we have that the sequence of $L$-events is the same in $\mathcal{I}^*$ and $\mathcal{I}$.
        Therefore, by \Cref{def:ero:logical_list}, $\List(\mathcal{I}) = \List(\mathcal{I}^*)$.
        \qH{\Cref{lemma:ero:after_helping_last_l_event_the_list_of_cells_is_stuck_weak:claim_four}}
    \end{proof}

    We now finish the proof of \Cref{lemma:ero:after_helping_last_l_event_the_list_of_cells_is_stuck_weak}.
    We have established: (1) $I'$ is an invocation of the Acquire procedure in $\mathcal{I}^\mathcal{B}$ with parameters $(\uniquerepositoryoperationshort_\linearizationobject{}, \cellpointershort)$ for some $\cellpointershort{} \in \celluniverse$ and exits with response $\status$; (2) $T_b$ is the time $p$ invoked $I'$ and $T^{\ref{line:ero:acquire_next_linearization_changed_check}}$ is the last time $p$ executes \cref{line:ero:acquire_next_linearization_changed_check} during $I'$; (3) $P(\mathcal{I}^\mathcal{B})$ holds; and (4) $\mathcal{I}$ is a finite prefix of $\mathcal{I}^\mathcal{B}$ such that for every prefix $\mathcal{I}^*$ of $\mathcal{I}^\mathcal{B}$ during $[T_b, T^{\ref{line:ero:acquire_next_linearization_changed_check}}]$ 1. by \Cref{lemma:ero:after_helping_last_l_event_the_list_of_cells_is_stuck_weak:claim_three} $\linearizationobject{}.\uniquerepositoryoperationlong = \uniquerepositoryoperationshort_\linearizationobject{}$ at the end of $\mathcal{I}^*$ and 2. by \Cref{lemma:ero:after_helping_last_l_event_the_list_of_cells_is_stuck_weak:claim_four} the list of cells conforms to $\List(\mathcal{I})$ in $\mathcal{I}^*$, and so by \Cref{lemma:ero:under_certain_conditions_acquire_returns_found_weak}, if $\cellpointershort{} \in \List(\mathcal{I})$, then $\status = \found$, and if $\cellpointershort{} \notin \List(\mathcal{I})$, then $\status = \notfound$ as wanted.
    \qH{\Cref{lemma:ero:after_helping_last_l_event_the_list_of_cells_is_stuck_weak}}
\end{proof}

We now prove that the IsDone procedure has the intended effect for each type of low-level operation that can be provided as input (excluding the initial value of $\announceobject$).

\begin{lemma}\label{lemma:ero:add_done_check_pass_implies_l_add_event}
    Consider any process $p$ and any iteration $I$ of the loop on \cref{line:ero:do_work_while_loop} by $p$ in $\mathcal{I}^\mathcal{B}$.
    Let $\mathcal{I}$ be the prefix of $\mathcal{I}^\mathcal{B}$ up to and including the time $p$ executed \cref{line:ero:linearization_read} during $I$.
    If $P(\mathcal{I}^\mathcal{B})$, $Q(\mathcal{I}^\mathcal{B})$, and $R(\mathcal{I}^\mathcal{B})$ hold, and $p$ read $((\arbitraryvalue, \addcell), \cellpointershort)$ from $\announceobject$ on \cref{line:ero:announce_read} during $I$, then:
    \begin{compactenum}
        \item If $p$ received $\notdone$ on \cref{line:ero:check_if_announce_is_done} during $I$, then there is no $L$-add event for $\cellpointershort{}$ in $\mathcal{I}$; and
        \item If $p$ received $\done$ on \cref{line:ero:check_if_announce_is_done} during $I$, then there is a $L$-add event for $\cellpointershort{}$ in $\mathcal{I}$.
    \end{compactenum}
\end{lemma}

\begin{proof}
    Since $p$ read $((\arbitraryvalue, \addcell), \cellpointershort)$ from $\announceobject$ during $I$, and $\announceobject$ is initially $((0, \noop), (0, \nullconstant))$, we have that $\announceobject$ was set to $((\arbitraryvalue, \addcell), \cellpointershort)$, and so by \Cref{observation:ero:where_objects_change} some $A$-event $a$ set $\announceobject = ((\arbitraryvalue, \addcell), \cellpointershort)$.
    Hence, by \Cref{def:ero:english}, this is an $A$-add event for $\cellpointershort$, so by \Cref{lemma:ero:every_a_event_is_for_pointer_from_universe}, $\cellpointershort \in \celluniverse{}$.

    We now prove 1.
    Suppose $p$ received $\notdone$ on \cref{line:ero:check_if_announce_is_done} during $I$.
    Since $p$ exited the IsDone procedure during $I$, we have that $p$ exited the Acquire procedure on \cref{line:ero:announce_acquire} during it.
    Denote this invocation by $I'$, and its response by $\status$.
    Since $p$ received $\notdone$ as a response on \cref{line:ero:check_if_announce_is_done} during $I$, it follows that $p$ found the condition on \cref{line:ero:announce_acquire_l_changed_check} to be false during $I$, and so $\status \neq \timechange$.
    Hence, since $P(\mathcal{I}^\mathcal{B})$, $Q(\mathcal{I}^\mathcal{B})$, and $R(\mathcal{I}^\mathcal{B})$ hold, by \Cref{lemma:ero:after_helping_last_l_event_the_list_of_cells_is_stuck_weak}, if $\status \neq \found$, then $\cellpointershort{} \notin \List(\mathcal{I})$.
    Therefore, since $\status \neq \timechange$ and $\status$ is either $\found$, $\notfound$, or $\timechange$, this is equivalent to: if $\status = \notfound$, then $\cellpointershort{} \notin \List(\mathcal{I})$ (*).
    We now prove that $\status = \notfound$.
    Since $p$ read $((\arbitraryvalue, \addcell), \cellpointershort)$ from $\announceobject$ on \cref{line:ero:announce_read} during $I$, we have that $p$ found the conditions on \cref{line:ero:remove_cell_done_check} and \cref{line:ero:write_and_read_done_check} to be false during $I$.
    Furthermore, since $p$ received $\notdone$ as a response on \cref{line:ero:check_if_announce_is_done} during $I$, we have that between the time $response$ was initialized to $\done$ on \cref{line:ero:done_initialization} during $I$ and the time $p$ exited the IsDone procedure during $I$, the value of $response$ changed.
    Together, these imply that $p$ found the condition on \cref{line:ero:add_cell_done_check} to be true during $I$.
    Therefore, $\status = \notfound$ as wanted.
    We now finish the proof of 1.
    Since $\status = \notfound$, by (*), $\cellpointershort{} \notin \List(\mathcal{I})$.
    Hence, since $\cellpointershort{} \in \celluniverse$, by \Cref{assumption:ero:head_and_null_not_in_cell_universe} $\cellpointershort{} \neq \&\headobject$ and $\cellpointershort{} \neq \nullconstant$, and so by \Cref{def:ero:logical_list}, either there is not a $L$-add event for $\cellpointershort$ in $\mathcal{I}$ or there is an $L$-add event for $\cellpointershort$ followed by an $L$-remove event for $\cellpointershort$ in $\mathcal{I}$.
    If the former, the claim is satisfied, so it suffices to consider the latter.
    We prove that the latter is impossible.
    Let $e$ be the $L$-remove event for $\cellpointershort$ in $\mathcal{I}$.
    Hence, by \Cref{lemma:ero:l_event_corresponding_a_event_type_and_pointer_relations}, there is an $A$-remove event $e'$ for $\cellpointershort$ before $e$ in $\mathcal{I}$, and so by \Cref{def:ero:english}, $\announceobject = ((\arbitraryvalue, \removecell), \cellpointershort)$ at $e'$.
    Thus, since $p$ read $((\arbitraryvalue, \addcell), \cellpointershort)$ from $\announceobject$ on \cref{line:ero:announce_read} during $I$, and this step is not in $\mathcal{I}$ by definition, we have that between $e'$ and this step, $\announceobject$ was set to $((\arbitraryvalue, \addcell), \cellpointershort)$.
    Therefore, by \Cref{observation:ero:where_objects_change} and \Cref{def:ero:english}, there is an $A$-add event for $\cellpointershort$ after $e'$ in $\mathcal{I}^\mathcal{B}$.
    However, since $e'$ is an $A$-remove event for $\cellpointershort$ in $\mathcal{I}^\mathcal{B}$, by \Cref{lemma:ero:no_add_announce_after_remove_announce}, there are no $A$-add events for $\cellpointershort$ from $e'$ onwards in $\mathcal{I}^\mathcal{B}$, a contradiction, and so the latter case is impossible.
    
    We now prove 2.
    Suppose $p$ received $\done$ on \cref{line:ero:check_if_announce_is_done} during $I$.
    Since $p$ exited the IsDone procedure during $I$, we have that $p$ exited the Acquire procedure on \cref{line:ero:announce_acquire} during it.
    Denote this invocation by $I'$, and its response by $\status$.
    Since $p$ received $\done$ as a response on \cref{line:ero:check_if_announce_is_done} during $I$, $p$ found the condition on \cref{line:ero:announce_acquire_l_changed_check} to be false during $I$, and so $\status \neq \timechange$.
    Therefore, since $P(\mathcal{I}^\mathcal{B})$, $Q(\mathcal{I}^\mathcal{B})$, and $R(\mathcal{I}^\mathcal{B})$ hold, by \Cref{lemma:ero:after_helping_last_l_event_the_list_of_cells_is_stuck_weak}, if $\status \neq \notfound$, then $\cellpointershort{} \in \List(\mathcal{I})$ (*).
    We now prove that $\status \neq \notfound$.
    Since $p$ received $\done$ as a response on \cref{line:ero:check_if_announce_is_done} during $I$, we have that between the time $response$ was initialized to $\done$ on \cref{line:ero:done_initialization} during $I$ and the time $p$ exited the IsDone procedure during $I$, the value of $response$ was not changed.
    Hence, $p$ found the condition on \cref{line:ero:add_cell_done_check} to be false during $I$.
    Thus, since $p$ read $((\arbitraryvalue, \addcell), \cellpointershort)$ from $\announceobject$ on \cref{line:ero:announce_read} during $I$, this implies that $p$ found the right condition on \cref{line:ero:add_cell_done_check} to be false during $I$.
    Therefore, $\status \neq \notfound$.
    We now finish the proof of 2.
    Since $\status \neq \notfound$, by (*), $\cellpointershort{} \in \List(\mathcal{I})$.
    Therefore, since $\cellpointershort{} \in \celluniverse$, by \Cref{assumption:ero:head_and_null_not_in_cell_universe} $\cellpointershort{} \neq \&\headobject$ and $\cellpointershort{} \neq \nullconstant$, and so by \Cref{def:ero:logical_list}, there is a $L$-add event for $\cellpointershort{}$ in $\mathcal{I}$ as wanted.
    \qH{\Cref{lemma:ero:add_done_check_pass_implies_l_add_event}}
\end{proof}

We now prove the same for apply low-level operations with one important difference: in the second case, we cannot conclude that there is an $L$-apply event for $\cellpointershort$ in $\mathcal{I}$, but only in $\mathcal{I}^\mathcal{B}$.
Roughly speaking, this is because there is some lag between when the Acquire completes on \cref{line:ero:announce_acquire} and when the response of $\cellpointershort$ is checked on \cref{line:ero:announce_op_response_check}.
As we will see, the second case is only needed during the proof of wait-freedom, and in the single location it is needed, identifying a single $L$-apply event (or $L$-add and $L$-remove event in the other two cases) for $\cellpointershort$ in $\mathcal{I}^\mathcal{B}$ is sufficient.

\begin{lemma}\label{lemma:ero:apply_done_check_pass_implies_l_apply_event_weak}
    Consider any process $p$ and any iteration $I$ of the loop on \cref{line:ero:do_work_while_loop} by $p$ in $\mathcal{I}^\mathcal{B}$.
    Let $\mathcal{I}$ be the prefix of $\mathcal{I}^\mathcal{B}$ up to and including the time $p$ executed \cref{line:ero:linearization_read} during $I$.
    If $P(\mathcal{I}^\mathcal{B})$, $Q(\mathcal{I}^\mathcal{B})$, and $R(\mathcal{I}^\mathcal{B})$ hold, and during $I$ $p$ read $((\arbitraryvalue, \langle \doopandcopyresponse, \arbitraryvalue \rangle), \cellpointershort)$ from $\announceobject$ on \cref{line:ero:announce_read}, then:
    \begin{compactenum}
        \item If $p$ received $\notdone$ on \cref{line:ero:check_if_announce_is_done} during $I$, then there is no $L$-apply event for $\cellpointershort{}$ in $\mathcal{I}$; and
        \item If $p$ received $\done$ on \cref{line:ero:check_if_announce_is_done} during $I$, then there is a $L$-apply event for $\cellpointershort{}$ in $\mathcal{I}^\mathcal{B}$.
    \end{compactenum}
\end{lemma}

\begin{proof}
    Since $p$ read $((\arbitraryvalue, \langle \doopandcopyresponse, \arbitraryvalue \rangle), \cellpointershort)$ from $\announceobject$ during $I$, and $\announceobject$ is initially $((0, \noop), (0, \nullconstant))$, we have that $\announceobject$ was set to $((\arbitraryvalue, \langle \doopandcopyresponse, \arbitraryvalue \rangle), \cellpointershort)$, and so by \Cref{observation:ero:where_objects_change} some $A$-event $a$ set $\announceobject = ((\arbitraryvalue, \langle \doopandcopyresponse, \arbitraryvalue \rangle), \cellpointershort)$.
    Hence, by \Cref{def:ero:english}, this is an $A$-apply event for $\cellpointershort$, so by \Cref{lemma:ero:every_a_event_is_for_pointer_from_universe}, $\cellpointershort \in \celluniverse{}$.

    We now prove 1.
    Suppose, for contradiction, $p$ received $\notdone$ on \cref{line:ero:check_if_announce_is_done} during $I$ and there is a $L$-apply event $e$ for $\cellpointershort{}$ in $\mathcal{I}$.

    \begin{claimcustom}{\ref{lemma:ero:apply_done_check_pass_implies_l_apply_event_weak}.1}\label{lemma:ero:apply_done_check_pass_implies_l_apply_event_weak:claim_one}
        Let $T^{\ref{line:ero:announce_read}}$ be the time $p$ executed \cref{line:ero:announce_read} during $I$.
        Then, from $T^{\ref{line:ero:announce_read}}$ onwards in $\mathcal{I}^\mathcal{B}$ $(*\cellpointershort).\lastrepositoryoperationresponse{} \neq ((\arbitraryvalue{}, \langle \doopandcopyresponse{}, \arbitraryvalue{} \rangle), \nullconstant)$.
    \end{claimcustom}

    \begin{proof}
        We first identify an invocation $I^*$ of the \doapplyandcopyresponse{} procedure with a second parameter of $\cellpointershort$ that exits at time $T^{exit}_* < T^{\ref{line:ero:announce_read}}$.
        There are two cases.
        \begin{itemize}
            \item[] \hspace{0pt}\textbf{Case 1.} $e$ is the last $L$-event in $\mathcal{I}$.

            Hence, since $\mathcal{I}$ is the prefix of $\mathcal{I}^\mathcal{B}$ up to and including the time $p$ executed \cref{line:ero:linearization_read} during $I$, say $T^{\ref{line:ero:linearization_read}}$, we have that $p$ read $((\arbitraryvalue, \langle \doopandcopyresponse, \arbitraryvalue \rangle), \cellpointershort)$ from $\linearizationobject{}$ at $T^{\ref{line:ero:linearization_read}}$.
            Thus, $p$ finds the condition on \cref{line:ero:do_apply_and_copy_response_condition} to be true during $I$, and so $p$ begins and exits the \doapplyandcopyresponse{} procedure on \cref{line:ero:do_apply_and_copy_response} with a second parameter of $\cellpointershort$ during $I$.
            Denote this invocation of the \doapplyandcopyresponse{} procedure by $I^*$ and let $T^{exit}_*$ be the time $I^*$ exits.
            Hence, $T^{exit}_* < T^{\ref{line:ero:announce_read}}$.

            \item[] \hspace{0pt}\textbf{Case 2.} $e$ is not the last $L$-event in $\mathcal{I}$.

            % Suppose, for contradiction, at some time at or after $T^{\ref{line:ero:announce_read}}_p$ this is not true.
            Let $e_a$ be the next $L$-event after $e$ in $\mathcal{I}$ and let $q$ be the process that executed $e_a$.
            Since $e_a$ is the next $L$-event after $e$, by \Cref{lemma:ero:successive_l_event_read_previous_l_event_value}, $q$ read the value that $e$ set $\linearizationobject{}$ to on its last execution of \cref{line:ero:linearization_read} before $e_a$; say at time $T^{\ref{line:ero:linearization_read}}_q$.
            Hence, since $e$ is an $L$-apply event for $\cellpointershort$, by \Cref{def:ero:english}, $e$ set $\linearizationobject{}$ to $((\arbitraryvalue, \langle \doopandcopyresponse{}, \arbitraryvalue\rangle), \cellpointershort)$.
            Thus, since $q$ read the value that $e$ set $\linearizationobject{}$ to on \cref{line:ero:linearization_read} at $T^{\ref{line:ero:linearization_read}}_q$, we have that $q$ read $((\arbitraryvalue, \langle \doopandcopyresponse{}, \arbitraryvalue\rangle), \cellpointershort)$ from $\linearizationobject{}$ at $T^{\ref{line:ero:linearization_read}}_q$.
            Hence, $q$ finds the condition on \cref{line:ero:do_apply_and_copy_response_condition} to be true between $T^{\ref{line:ero:linearization_read}}_q$ and $e_a$, and so $q$ begins and exits the \doapplyandcopyresponse{} procedure on \cref{line:ero:do_apply_and_copy_response} with a second parameter of $\cellpointershort$ between $T^{\ref{line:ero:linearization_read}}_q$ and $e_a$.
            Denote this invocation of the \doapplyandcopyresponse{} procedure by $I^*$ and let $T^{exit}_*$ be the time $I^*$ exits.
            Hence, $T^{exit}_* < e_a$, and since $e_a$ is in $\mathcal{I}$, we have that $T^{exit}_*$ is in $\mathcal{I}$, so $T^{exit}_* < T^{\ref{line:ero:announce_read}}$.
        \end{itemize}

        We now return to the proof of \Cref{lemma:ero:apply_done_check_pass_implies_l_apply_event_weak:claim_one}.
        As we just established, $I^*$ is an invocation of the \doapplyandcopyresponse{} procedure with a second parameter of $\cellpointershort$ that exits at time $T^{exit}_*$, and $P(\mathcal{I}^\mathcal{B})$, $Q(\mathcal{I}^\mathcal{B})$, and $R(\mathcal{I}^\mathcal{B})$ hold, by \Cref{lemma:ero:conditional_exit_apply_implies_response_set}, there is a successful apply-response-set attempt $a$ for $\cellpointershort$ before $T^{exit}_*$.
        Hence, by \Cref{lemma:ero:once_response_not_null_for_apply_never_null_for_apply_again}, from $a$ onwards in $\mathcal{I}^\mathcal{B}$ $(*\cellpointershort).\lastrepositoryoperationresponse{} \neq ((\arbitraryvalue{}, \langle \doopandcopyresponse{}, \arbitraryvalue{} \rangle), \nullconstant)$.
        Therefore, since $a < T^{exit}_*$ and $T^{exit}_* < T^{\ref{line:ero:announce_read}}$, from $T^{\ref{line:ero:announce_read}}$ onwards in $\mathcal{I}^\mathcal{B}$ $(*\cellpointershort).\lastrepositoryoperationresponse{} \neq ((\arbitraryvalue{}, \langle \doopandcopyresponse{}, \arbitraryvalue{} \rangle), \nullconstant)$.
        \qH{\Cref{lemma:ero:apply_done_check_pass_implies_l_apply_event_weak:claim_one}}
    \end{proof}

    We now finish the proof of 1.
    Since $p$ exited the IsDone procedure during $I$, we have that $p$ exited the Acquire procedure on \cref{line:ero:announce_acquire} during it.
    Denote this invocation by $I'$, and its response by $\status$.
    Since $p$ received $\done$ as a response on \cref{line:ero:check_if_announce_is_done} during $I$, $p$ found the condition on \cref{line:ero:announce_acquire_l_changed_check} to be false during $I$.
    Hence, since $p$ read $((\arbitraryvalue, \langle \doopandcopyresponse, \arbitraryvalue \rangle), \cellpointershort)$ from $\announceobject$ on \cref{line:ero:announce_read} during $I$, we have that $p$ found the conditions on \cref{line:ero:add_cell_done_check} and \cref{line:ero:remove_cell_done_check} to be false during $I$.
    Thus, since $p$ received $\notdone$ as a response on \cref{line:ero:check_if_announce_is_done} during $I$, we have that $p$ found the condition on \cref{line:ero:announce_op_response_check} to be true during $I$; say at time $T^{\ref{line:ero:announce_op_response_check}}$.
    So, $(*\cellpointershort).\lastrepositoryoperationresponse{} = ((\arbitraryvalue{}, \langle \doopandcopyresponse{}, \arbitraryvalue{} \rangle), \nullconstant)$ at $T^{\ref{line:ero:announce_op_response_check}}$.
    Furthermore, since $T^{\ref{line:ero:announce_read}}$ is the time $p$ executed \cref{line:ero:announce_read} during $I$, and $T^{\ref{line:ero:announce_op_response_check}}$ is the time $p$ executed \cref{line:ero:announce_op_response_check} during $I$, we have that $T^{\ref{line:ero:announce_read}} < T^{\ref{line:ero:announce_op_response_check}}$.
    Therefore, $(*\cellpointershort).\lastrepositoryoperationresponse{} = ((\arbitraryvalue{}, \langle \doopandcopyresponse{}, \arbitraryvalue{} \rangle), \nullconstant)$ after $T^{\ref{line:ero:announce_read}}$ in $\mathcal{I}^\mathcal{B}$.
    However, this contradicts \Cref{lemma:ero:apply_done_check_pass_implies_l_apply_event_weak:claim_one}.

    We now prove 2.
    Suppose, for contradiction, $p$ received $\done$ on \cref{line:ero:check_if_announce_is_done} during $I$ and there is not a $L$-apply event for $\cellpointershort{}$ in $\mathcal{I}^\mathcal{B}$.

    \begin{claimcustom}{\ref{lemma:ero:apply_done_check_pass_implies_l_apply_event_weak}.2}\label{lemma:ero:apply_done_check_pass_implies_l_apply_event_weak:claim_three}
        There is a successful apply-response-set attempt for $\cellpointershort{}$ in $\mathcal{I}^\mathcal{B}$.
    \end{claimcustom}

    \begin{proof}
        % Since, by \Cref{lemma:ero:apply_done_check_pass_implies_l_apply_event_weak:claim_two} $\cellpointershort \in \List(\mathcal{I})$, by (**), $\status = \found{}$.
        Since $p$ exited the IsDone procedure during $I$, we have that $p$ exited the Acquire procedure on \cref{line:ero:announce_acquire} during it.
        Denote this invocation by $I'$, and its response by $\status$.
        Since $p$ received $\done$ as a response on \cref{line:ero:check_if_announce_is_done} during $I$, $p$ found the condition on \cref{line:ero:announce_acquire_l_changed_check} to be false during $I$.
        Hence, since $p$ read $((\arbitraryvalue, \langle \doopandcopyresponse, \arbitraryvalue \rangle), \cellpointershort)$ from $\announceobject$ on \cref{line:ero:announce_read} during $I$, we have that $p$ found the conditions on \cref{line:ero:add_cell_done_check} and \cref{line:ero:remove_cell_done_check} to be false during $I$.
        Thus, since $p$ received $\done$ as a response on \cref{line:ero:check_if_announce_is_done} during $I$, we have that $p$ found the condition on \cref{line:ero:announce_op_response_check} to be false during $I$; say at time $T^{\ref{line:ero:announce_op_response_check}}$.
        Hence, $(*\cellpointershort{}).\lastrepositoryoperationresponse{} \neq ((\arbitraryvalue, \langle \doopandcopyresponse, \arbitraryvalue \rangle), \nullconstant)$ at $T^{\ref{line:ero:announce_op_response_check}}$.
        Since $p$ read $((\arbitraryvalue, \langle \doopandcopyresponse, \arbitraryvalue \rangle), \cellpointershort)$ from $\announceobject$ on \cref{line:ero:announce_read} during $I$, and $\announceobject$ is initially $((\arbitraryvalue, \noop), (\arbitraryvalue, \nullconstant))$, we have that $\announceobject$ was set to $((\arbitraryvalue, \langle \doopandcopyresponse, \arbitraryvalue \rangle), \cellpointershort)$ before $p$'s execution of \cref{line:ero:announce_read} during $I$.
        Hence, by \Cref{observation:ero:where_objects_change}, there is an $A$-event $e_{apply}$ that sets $\announceobject$ to this value before $p$'s execution of \cref{line:ero:announce_read} during $I$.
        Thus, it follows that the process $q$ that executed $e_{apply}$ did so during an invocation $I_q$ of the \doworkuntildone{} with parameters $(\langle \doopandcopyresponse, \arbitraryvalue \rangle, \cellpointershort)$.
        There are two cases.
    
        \begin{itemize}
            \item[] \hspace{0pt}\textbf{Case 1.} $q$ exits $I_q$.
    
            Hence, since $I_q$'s parameters are $(\langle \doopandcopyresponse, \arbitraryvalue \rangle, \cellpointershort)$, by \Cref{lemma:ero:successful_apply_response_set_before_apply_low_level_exits}, there is a successful apply-response-set attempt for $\cellpointershort{}$ in $\mathcal{I}^\mathcal{B}$.
    
            \item[] \hspace{0pt}\textbf{Case 2.} $q$ does not exit $I_q$.

            Since $q$ executed $e_{apply}$ during $I_q$ and by \Cref{def:ero:english} $e_{apply}$ is an execution of either \cref{line:ero:announce_gcas} or \cref{line:ero:announce_cas}, we have that $q$ executed \cref{line:ero:do_work_initialize_response} during $I_q$ before $e_{apply}$; say at time $T^{\ref{line:ero:do_work_initialize_response}}_q$, so $T^{\ref{line:ero:do_work_initialize_response}}_q < e_{apply}$.
            Hence, since $I_q$'s parameters are $(\langle \doopandcopyresponse, \arbitraryvalue \rangle, \cellpointershort)$, it follows that $q$ sets $(*\cellpointershort{}).\lastrepositoryoperationresponse{} = ((\arbitraryvalue, \langle \doopandcopyresponse, \arbitraryvalue \rangle), \nullconstant)$ at $T^{\ref{line:ero:do_work_initialize_response}}_q$.
            Since $e_{apply}$ is before $p$'s execution of \cref{line:ero:announce_read} during $I$, and $T^{\ref{line:ero:announce_op_response_check}}$ is the time of $p$'s execution of \cref{line:ero:announce_op_response_check} during $I$, we have that $e_{apply} < T^{\ref{line:ero:announce_op_response_check}}$.
            Hence, since $T^{\ref{line:ero:do_work_initialize_response}}_q < e_{apply}$, by transitivity, $T^{\ref{line:ero:do_work_initialize_response}}_q < T^{\ref{line:ero:announce_op_response_check}}$.
            Thus, since $(*\cellpointershort{}).\lastrepositoryoperationresponse{}$ equals $((\arbitraryvalue, \langle \doopandcopyresponse, \arbitraryvalue \rangle), \nullconstant)$ at $T^{\ref{line:ero:do_work_initialize_response}}_q$ and does not at $T^{\ref{line:ero:announce_op_response_check}}$, we have that between $T^{\ref{line:ero:do_work_initialize_response}}_q$ and $T^{\ref{line:ero:announce_op_response_check}}$ the value of $(*\cellpointershort{}).\lastrepositoryoperationresponse{}$ changed.
            Hence, since $\cellpointershort{} \in \celluniverse$, by \Cref{observation:ero:where_objects_change}, there is either a response-reset event for $\cellpointershort{}$ or a successful response-set attempt for $\cellpointershort{}$ between $T^{\ref{line:ero:do_work_initialize_response}}_q$ and $T^{\ref{line:ero:announce_op_response_check}}$.
            Let $e^*$ be the first such step between $T^{\ref{line:ero:do_work_initialize_response}}_q$ and $T^{\ref{line:ero:announce_op_response_check}}$.
            Hence, throughout $[T^{\ref{line:ero:do_work_initialize_response}}_q, e^*)$  $(*\cellpointershort{}).\lastrepositoryoperationresponse{} = ((\arbitraryvalue, \langle \doopandcopyresponse, \arbitraryvalue \rangle), \nullconstant)$.
            Since $I_q$'s parameters are $(\arbitraryvalue, \cellpointershort)$, by Case 2 $q$ does not exit $I_q$, and $T^{\ref{line:ero:do_work_initialize_response}}_q$ is the time $q$ executes \cref{line:ero:do_work_initialize_response} during $I_q$, by \Cref{lemma:ero:at_most_one_response_reset_for_ptr_during_low_level_op_never_exits}, there are no response-reset events for $\cellpointershort{}$ after $T^{\ref{line:ero:do_work_initialize_response}}_q$, and so $e^*$ is a successful response-set attempt for $\cellpointershort{}$.
            Therefore, since throughout $[T^{\ref{line:ero:do_work_initialize_response}}_q, e^*)$  $(*\cellpointershort{}).\lastrepositoryoperationresponse{} = ((\arbitraryvalue, \langle \doopandcopyresponse, \arbitraryvalue \rangle), \nullconstant)$ and by \Cref{def:ero:english} $e^*$ is a successful \CASop{} on \cref{line:ero:responses_set_attempt}, we have that the second parameter of $e^*$ is $((\arbitraryvalue, \langle \doopandcopyresponse, \arbitraryvalue \rangle), \nullconstant)$, and so by \Cref{def:ero:english} $e^*$ is a successful apply-response-set attempt for $\cellpointershort{}$ as wanted.
            \qH{\Cref{lemma:ero:apply_done_check_pass_implies_l_apply_event_weak:claim_three}}
        \end{itemize}
    \end{proof}

    We now finish the proof of 2.
    Since by \Cref{lemma:ero:apply_done_check_pass_implies_l_apply_event_weak:claim_three} there is a successful apply-response-set attempt $a$ for $\cellpointershort{}$, by \Cref{corollary:ero:response_set_attempts_and_corresponding_l_events_are_matching}, there is a $L$-apply event for $\cellpointershort$ in $\mathcal{I}^\mathcal{B}$.
    However, by our initial assumption, there are no $L$-apply events for $\cellpointershort{}$, a contradiction.
    \qH{\Cref{lemma:ero:apply_done_check_pass_implies_l_apply_event_weak}}
\end{proof}

We now prove the same for remove low-level operations

\begin{lemma}\label{lemma:ero:remove_done_check_pass_implies_l_remove_event_weak}
    Consider any process $p$ and any iteration $I$ of the loop on \cref{line:ero:do_work_while_loop} by $p$ in $\mathcal{I}^\mathcal{B}$.
    Let $\mathcal{I}$ be the prefix of $\mathcal{I}^\mathcal{B}$ up to and including the time $p$ executed \cref{line:ero:linearization_read} during $I$.
    If $P(\mathcal{I}^\mathcal{B})$, $Q(\mathcal{I}^\mathcal{B})$, and $R(\mathcal{I}^\mathcal{B})$ hold, and during $I$ $p$ read $((\arbitraryvalue, \removecell), \cellpointershort)$ from $\announceobject$ on \cref{line:ero:announce_read} then:
    \begin{compactenum}
        \item If $p$ received $\notdone$ on \cref{line:ero:check_if_announce_is_done} during $I$, then there is no $L$-remove event for $\cellpointershort{}$ in $\mathcal{I}$; and
        \item If $p$ received $\done$ on \cref{line:ero:check_if_announce_is_done} during $I$, then there is a $L$-remove event for $\cellpointershort{}$ in $\mathcal{I}$.
    \end{compactenum}
\end{lemma}

\begin{proof}
    Since $p$ read $((\arbitraryvalue, \removecell), \cellpointershort)$ from $\announceobject$ on \cref{line:ero:announce_read} during $I$, and the value of $\announceobject$ is initially $((\arbitraryvalue, \noop), (\arbitraryvalue, \nullconstant))$, we have that $\announceobject$ was set to $((\arbitraryvalue, \removecell), \cellpointershort)$.
    Hence, by \Cref{observation:ero:where_objects_change}, there is an $A$-event $e_{remove}$ that sets $\announceobject = ((\arbitraryvalue, \removecell), \cellpointershort)$ before $p$ read it on \cref{line:ero:announce_read} during $I$.
    Thus, by \Cref{def:ero:english}, $e_{remove}$ is an $A$-remove event for $\cellpointershort$, so by \Cref{lemma:ero:every_a_event_is_for_pointer_from_universe}, $\cellpointershort \in \celluniverse{}$.

    We now prove 1.
    Suppose, for contradiction, $p$ received $\notdone$ on \cref{line:ero:check_if_announce_is_done} during $I$, and there is an $L$-remove event $e$ for $\cellpointershort$ in $\mathcal{I}$.
    Since $p$ exited the IsDone procedure during $I$, we have that $p$ exited the Acquire procedure on \cref{line:ero:announce_acquire} during it.
    Denote this invocation by $I'$, and its response by $\status$.
    Since $p$ received $\notdone$ as a response on \cref{line:ero:check_if_announce_is_done} during $I$, $p$ found the condition on \cref{line:ero:announce_acquire_l_changed_check} to be false during $I$, and so $\status \neq \timechange$.
    Hence, since $P(\mathcal{I}^\mathcal{B})$, $Q(\mathcal{I}^\mathcal{B})$, and $R(\mathcal{I}^\mathcal{B})$ hold, by \Cref{lemma:ero:after_helping_last_l_event_the_list_of_cells_is_stuck_weak}, if $\status \neq \notfound$, then $\cellpointershort{} \in \List(\mathcal{I})$.
    Therefore, since $\status \neq \timechange$ and $\status$ is either $\found$, $\notfound$, or $\timechange$, this is equivalent to: if $\status = \found$, then $\cellpointershort{} \in \List(\mathcal{I})$ (*).
    We now prove that $\status = \found$.
    Since $p$ read $((\arbitraryvalue, \removecell), \cellpointershort)$ from $\announceobject$ on \cref{line:ero:announce_read} during $I$, we have that $p$ found the conditions on \cref{line:ero:add_cell_done_check} and \cref{line:ero:write_and_read_done_check} to be false during $I$.
    Furthermore, since $p$ received $\notdone$ as a response on \cref{line:ero:check_if_announce_is_done} during $I$, we have that between the time $response$ was initialized to $\done$ on \cref{line:ero:done_initialization} during $I$ and the time $p$ exited the IsDone procedure during $I$, the value of $response$ changed.
    Together, these imply that $p$ found the condition on \cref{line:ero:remove_cell_done_check} to be true during $I$.
    Therefore, $\status = \found$ as wanted.
    We now finish the proof of 1.
    Since $\status = \found$, by (*), $\cellpointershort{} \in \List(\mathcal{I})$.
    Hence, since $\cellpointershort \in \celluniverse{}$, by \Cref{assumption:ero:head_and_null_not_in_cell_universe}, $\cellpointershort \neq \&\headobject$ and $\cellpointershort \neq \nullconstant$, so by \Cref{def:ero:logical_list}, there is an $L$-add event $e'$ for $\cellpointershort$ in $\mathcal{I}$ without a subsequent $L$-remove event for $\cellpointershort$ in $\mathcal{I}$.
    Thus, $e < e'$.
    Therefore, since $e$ is an $L$-remove event for $\cellpointershort$, by \Cref{lemma:ero:remove_events_are_preceeded_by_add_events}, there is an $L$-add event for $\cellpointershort$ before $e$, and so there are two $L$-add events for $\cellpointershort$ in $\mathcal{I}^\mathcal{B}$.
    However, by $P(\mathcal{I}^\mathcal{B})$, there is at most one $L$-add event for $\cellpointershort$ in $\mathcal{I}^\mathcal{B}$, a contradiction.
    
    We now prove 2.
    Suppose, for contradiction, $p$ received $\done$ on \cref{line:ero:check_if_announce_is_done} during $I$ and there is not a $L$-remove event for $\cellpointershort{}$ in $\mathcal{I}$.
    Since $p$ exited the IsDone procedure during $I$, we have that $p$ exited the Acquire procedure on \cref{line:ero:announce_acquire} during $I$.
    Denote this invocation by $I'$, and its response by $\status$.
    Since $p$ received $\done$ as a response on \cref{line:ero:check_if_announce_is_done} during $I$, $p$ found the condition on \cref{line:ero:announce_acquire_l_changed_check} to be false during $I$, and so $\status \neq \timechange$.
    Let $T^{\ref{line:ero:linearization_read}}$ be the time $p$ executes \cref{line:ero:linearization_read} during $I$ and let $T^{\ref{line:ero:acquire_next_linearization_changed_check}}$ be the last time $p$ executes \cref{line:ero:acquire_next_linearization_changed_check} during $I'$ (this is well-defined by \Cref{lemma:ero:exit_acquire_implies_executing_105}).
    Hence, since $\status \neq \timechange$ and $P(\mathcal{I}^\mathcal{B})$ holds, by \Cref{lemma:ero:if_announce_acquire_does_not_return_time_change_then_no_l_events}, there are no $L$-events throughout $[T^{\ref{line:ero:linearization_read}}, T^{\ref{line:ero:acquire_next_linearization_changed_check}}]$.
    Furthermore, since $P(\mathcal{I}^\mathcal{B})$, $Q(\mathcal{I}^\mathcal{B})$, and $R(\mathcal{I}^\mathcal{B})$ hold, by \Cref{lemma:ero:after_helping_last_l_event_the_list_of_cells_is_stuck_weak}, if $\cellpointershort{} \in \List(\mathcal{I})$, then $\status = \found$ (*).

    \begin{claimcustom}{\ref{lemma:ero:remove_done_check_pass_implies_l_remove_event_weak}.1}\label{lemma:ero:remove_done_check_pass_implies_l_remove_event_weak:claim_one}
        $\status = \notfound$.
    \end{claimcustom}

    \begin{proof}
        Since $p$ received $\done$ as a response on \cref{line:ero:check_if_announce_is_done} during $I$, we have that between the time $response$ was initialized to $\done$ on \cref{line:ero:done_initialization} during $I$ and the time $p$ exited the IsDone procedure during $I$, the value of $response$ was not changed.
        Hence, $p$ found the condition on \cref{line:ero:remove_cell_done_check} to be false during $I$.
        Thus, since $p$ read $((\arbitraryvalue, \removecell), \cellpointershort)$ from $\announceobject$ on \cref{line:ero:announce_read} during $I$, this implies that $\status \neq \found$ on \cref{line:ero:remove_cell_done_check} during $I$.
        Hence, since $\status \neq \timechange$, the response of $I'$ is not $\timechange$ and not $\found$.
        Therefore, since the response of $I'$ is either $\found$, $\timechange$, or $\notfound$, we have that $\status = \notfound$ as wanted.
        \qH{\Cref{lemma:ero:remove_done_check_pass_implies_l_remove_event_weak:claim_one}}
    \end{proof}

    \begin{claimcustom}{\ref{lemma:ero:remove_done_check_pass_implies_l_remove_event_weak}.2}\label{lemma:ero:remove_done_check_pass_implies_l_remove_event_weak:claim_two}
        $\cellpointershort \in \List(\mathcal{I})$.
    \end{claimcustom}
    
    \begin{proof}
        Suppose, for contradiction, $\cellpointershort \notin \List(\mathcal{I})$.
        % Thus, by \Cref{def:ero:english}, $e_{remove}$ is an $A$-remove event for $\cellpointershort{}$, and so by \Cref{lemma:ero:every_a_event_is_for_pointer_from_universe} $\cellpointershort{} \in \celluniverse$.
        Since $e_{remove}$ is before $p$ executed \cref{line:ero:announce_read} during $I$, $I'$ is invoked on \cref{line:ero:announce_acquire} during $I$, and $T^{\ref{line:ero:acquire_next_linearization_changed_check}}$ is the last time $p$ executes \cref{line:ero:acquire_next_linearization_changed_check} during $I'$, by transitivity, $e_{remove} < T^{\ref{line:ero:acquire_next_linearization_changed_check}}$.
        Furthermore, the process $q$ that executed $e_{remove}$ did so during an invocation $I_1$ of the \doworkuntildone{} with parameters $(\removecell, \cellpointershort)$.
        Since the first parameter of $I_1$ is $\removecell$, $q$ invoked $I_1$ on \cref{line:ero:low_level_remove_cell} during some invocation $I^*$ of the \highleveloperation{} procedure.
        Hence, during $I^*$, $q$ began and exited an invocation $I_2$ the \doworkuntildone{} procedure on \cref{line:ero:low_level_add_cell}.
        Since $I_1$'s parameters are $(\removecell, \cellpointershort)$, and $I_1$ and $I_2$ are in during the same invocation $I^*$ of the \highleveloperation{} procedure, we have that $I_2$'s parameters are $(\addcell, \cellpointershort)$.
        Hence, since $P(\mathcal{I}^\mathcal{B})$ holds, by \Cref{lemma:ero:l_add_event_before_add_low_level_exits}, there is an $L$-add event $e_{add}$ for $\cellpointershort{}$ in $\mathcal{I}^\mathcal{B}$ before $q$ exited $I_2$.
        Since $e_{add}$ is before $q$ exited $I_2$, $q$ exited $I_2$ before $q$ invoked $I_1$, $q$ executed $e_{remove}$ during $I_1$, and $e_{remove} < T^{\ref{line:ero:acquire_next_linearization_changed_check}}$, by transitivity, we have that $e_{add} < T^{\ref{line:ero:acquire_next_linearization_changed_check}}$. 
        Thus, since $e_{add}$ is an $L$-event, and by (*) there are no $L$-events throughout $[T^{\ref{line:ero:linearization_read}}, T^{\ref{line:ero:acquire_next_linearization_changed_check}}]$, we have that $e_{add} < T^{\ref{line:ero:linearization_read}}$.
        Hence, since $\mathcal{I}$ is the prefix of $\mathcal{I}^\mathcal{B}$ up to and including $T^{\ref{line:ero:linearization_read}}$, we have that $e_{add}$ is in $\mathcal{I}$.
        Therefore, since by assumption $\cellpointershort \notin \List(\mathcal{I})$, by \Cref{def:ero:english}, there is an $L$-remove event for $\cellpointershort{}$ in $\mathcal{I}$.
        However, our initial assumption was that there is not an $L$-remove event for $\cellpointershort{}$ in $\mathcal{I}$, a contradiction.
        \qH{\Cref{lemma:ero:remove_done_check_pass_implies_l_remove_event_weak:claim_two}}
    \end{proof}

    We now finish the proof of 2.
    Since by \Cref{lemma:ero:remove_done_check_pass_implies_l_remove_event_weak:claim_two} $\cellpointershort \in \List(\mathcal{I})$, by (*), $\status = \found{}$.
    However, by \Cref{lemma:ero:remove_done_check_pass_implies_l_remove_event_weak:claim_one}, $\status = \notfound$, a contradiction.
    \qH{\Cref{lemma:ero:remove_done_check_pass_implies_l_remove_event_weak}}
\end{proof}

\subsubsection{The $L$-invariants hold}

The goal of this subsection is to prove the following lemma.

\begin{lemmablank}
    $P(\mathcal{I}^\mathcal{B})$, $Q(\mathcal{I}^\mathcal{B})$, $R(\mathcal{I}^\mathcal{B})$, and $O(\mathcal{I}^\mathcal{B})$ hold.
\end{lemmablank}

We first prove this claim for any finite implementation history of $\mathcal{B}$; the infinite case is then trivial, since if one of the invariants doesn't hold for an infinite $\mathcal{I}^\mathcal{B}$, we can identify a finite prefix of $\mathcal{I}^\mathcal{B}$ in which this invariant also doesn't hold.
For the entirety of this section, we fix a finite implementation history $\mathcal{I}$ of $\mathcal{B}$ and consider a one-step continuation $s$ of $\mathcal{I}$, denoted by $\mathcal{I} \circ s$.
% In other words, $s$ is the last step in $\mathcal{I} \circ s$ and $\mathcal{I}$ is the prefix of $\mathcal{I} \circ s$ up to but excluding $s$.

\subsubsection*{\Underline{Inductive Case for $P$}}

\begin{proposition}\label{lemma:ero:1_of_p_holds}
    If $P(\mathcal{I})$, $Q(\mathcal{I})$, and $R(\mathcal{I})$ hold, then $P(\mathcal{I} \circ s)$ holds.
\end{proposition}

\begin{proof}
    Suppose, for contradiction, $P(\mathcal{I} \circ s)$ does not hold.
    Since $P(\mathcal{I})$ holds, if $s$ is anything other than an $L$-event, $P(\mathcal{I} \circ s)$ holds by definition, so $s$ must be an $L$-event; say for some $\cellpointershort$ which by \Cref{lemma:ero:every_l_event_is_for_pointer_from_universe} is in $\celluniverse{}$.
    Since $P(\mathcal{I})$ holds and $P(\mathcal{I} \circ s)$ does not, there is some $L$-event in $\mathcal{I}$, denoted by $s'$, which is of the same form as $s$.
    Specifically, since by \Cref{lemma:ero:every_l_event_is_add_apply_or_remove} $s$ is either an $L$-add, $L$-apply, or $L$-remove event for $\cellpointershort$, then the following are true.
    If $s$ is an $L$-add event for $\cellpointershort$, then $s'$ is an $L$-add event for $\cellpointershort$, if $s$ is an $L$-apply event for $\cellpointershort$, then $s'$ is an $L$-apply event for $\cellpointershort$, and if $s$ is an $L$-remove event for $\cellpointershort$, then $s'$ is an $L$-remove event for $\cellpointershort$.
    Let $e$ be the last $L$-event before $s$ ($e$ exists because $s'$ does), so $e$ is in $\mathcal{I}$ and $s' \leq e$.
    Furthermore, let $T^{\ref{line:ero:linearization_read}}$ be the last time $p$ executed \cref{line:ero:linearization_read} before $s$, so $T^{\ref{line:ero:linearization_read}}$ is in $\mathcal{I}$.
    Since $e$ and $s$ are successive $L$-events in $\mathcal{I} \circ s$, $e$ is in $\mathcal{I}$, and $P(\mathcal{I})$ holds, by \Cref{lemma:ero:successive_l_event_read_previous_l_event_value_after_it_happened}, $e < T^{\ref{line:ero:linearization_read}}$.
    Let $I$ be the iteration of the loop on \cref{line:ero:do_work_while_loop} that $p$ executed \cref{line:ero:linearization_read} at $T^{\ref{line:ero:linearization_read}}$ during.
    Furthermore, let $\mathcal{I}^{\ref{line:ero:linearization_read}}$ be the prefix of $\mathcal{I}$ up to and including $T^{\ref{line:ero:linearization_read}}$ (this is well-defined because $T^{\ref{line:ero:linearization_read}}$ is in $\mathcal{I}$).
    Hence, since $s' \leq e$ and $e < T^{\ref{line:ero:linearization_read}}$, we have that $s'$ is in $\mathcal{I}^{\ref{line:ero:linearization_read}}$.
    Since $p$ executed \cref{line:ero:linearization_read} at $T^{\ref{line:ero:linearization_read}}$ during $I$, and $T^{\ref{line:ero:linearization_read}}$ is the last time $p$ executed \cref{line:ero:linearization_read} before $s$, we have that $p$ executed $s$ during $I$.
    Hence, $p$ received $\notdone$ on \cref{line:ero:check_if_announce_is_done} during $I$, and so since this is before $s$, we have that $p$ received $\notdone$ on \cref{line:ero:check_if_announce_is_done} during $I$ in $\mathcal{I}$.
    Since $s$ is either an $L$-add, $L$-apply, or $L$-remove event for $\cellpointershort$, we have that $p$ read either $((\arbitraryvalue, \addcell), \cellpointershort)$, $((\arbitraryvalue, \langle \doopandcopyresponse{}, \arbitraryvalue \rangle), \cellpointershort)$, or $((\arbitraryvalue, \removecell), \cellpointershort)$ from $\announceobject$ on \cref{line:ero:announce_read} during $I$, when $s$ is an $L$-add, $L$-apply, or $L$-remove event for $\cellpointershort$, respectively.
    Therefore, since $\mathcal{I}^{\ref{line:ero:linearization_read}}$ is the prefix of $\mathcal{I}$ up to and including $T^{\ref{line:ero:linearization_read}}$, and $P(\mathcal{I})$, $Q(\mathcal{I})$, and $R(\mathcal{I})$ hold, by 1. of \Cref{lemma:ero:add_done_check_pass_implies_l_add_event}, \Cref{lemma:ero:apply_done_check_pass_implies_l_apply_event_weak}, and \Cref{lemma:ero:remove_done_check_pass_implies_l_remove_event_weak}, we have that there is not an $L$-add, $L$-apply, or $L$-remove event for $\cellpointershort$ in $\mathcal{I}^{\ref{line:ero:linearization_read}}$, when $s$ is an $L$-add, $L$-apply, or $L$-remove event for $\cellpointershort$, respectively.\footnote{Observe that $\mathcal{I}^{\ref{line:ero:linearization_read}}$ is plugged in for ``$\mathcal{I}$", and $\mathcal{I}$ is plugged in for ``$\mathcal{I}^\mathcal{B}$" when applying these lemmas.}
    However, $s'$ is in $\mathcal{I}^{\ref{line:ero:linearization_read}}$ and is an $L$-add, $L$-apply, or $L$-remove event for $\cellpointershort$, when $s$ is an $L$-add, $L$-apply, or $L$-remove event for $\cellpointershort$, respectively, a contradiction.
    \qH{\Cref{lemma:ero:1_of_p_holds}}
\end{proof}

\subsubsection*{\Underline{Inductive Case for $Q$}}

\begin{proposition}\label{lemma:ero:1_of_q_holds}
    If $P(\mathcal{I})$, $Q(\mathcal{I})$, and $R(\mathcal{I})$ hold, then 1. of $Q(\mathcal{I} \circ s)$ holds.
\end{proposition}

\begin{proof}
    Since $Q(\mathcal{I})$ holds, if $s$ is anything other than a list-add attempt, 1. of $Q(\mathcal{I} \circ s)$ holds by definition, so it suffices to assume that $s$ is a list-add attempt for some $\cellpointershort$.
    Hence, by \Cref{lemma:ero:every_list_add_seal_and_remove_attempt_is_for_ptr_from_universe}, $\cellpointershort \in \celluniverse{}$.
    Furthermore, by \Cref{def:ero:english}, $p$ executed $s$ during some invocation $I$ of the \doaddcell{} procedure with parameters $(\uniquerepositoryoperationshort_\linearizationobject{}, \cellpointershort)$.
    Hence, by \Cref{lemma:ero:l_event_corresponding_to_do_low_level_op}, there is an $L$-add event $e$ for $\cellpointershort$ before $I$ was invoked that set $\linearizationobject{}$ to $(\uniquerepositoryoperationshort_\linearizationobject{}, \cellpointershort)$.
    Thus, $e < s$, so $e$ is in $\mathcal{I}$, and by $P(\mathcal{I})$, $e$ is the only $L$-add event for $\cellpointershort$ in $\mathcal{I}$.
    Therefore, $s$ is preceded by a unique $L$-add event for $\cellpointershort$ (namely $e$).
    Thus, what remains is to prove that if $\mathcal{I}^{exclude}_{e}$ is the prefix of $\mathcal{I} \circ s$ up to but excluding $e$, then $s$ is after the second last pointer in $\List(\mathcal{I}^{exclude}_{e})$.
    We start by establishing some basic facts for the proof.
    Since $e$ is before $p$ invoked $I$, we have that all steps during $I$ are after $e$. 
    Furthermore, since $p$ executed $s$ during $I$, and $s$ is a list-add attempt, $p$ executed \cref{line:ero:add_cell_before_updating_end_of_list_linearization_changed_check} before $s$ during $I$; say at time $T^{\ref{line:ero:add_cell_before_updating_end_of_list_linearization_changed_check}}$.
    Hence, since a process can only execute \cref{line:ero:add_cell_before_updating_end_of_list_linearization_changed_check} once during a single invocation of the \doaddcell{} procedure, $T^{\ref{line:ero:add_cell_before_updating_end_of_list_linearization_changed_check}}$ is the only time during $I$ that $p$ executes \cref{line:ero:add_cell_before_updating_end_of_list_linearization_changed_check}.
    Furthermore, since $T^{\ref{line:ero:add_cell_before_updating_end_of_list_linearization_changed_check}} < s$, it follows that the step at $T^{\ref{line:ero:add_cell_before_updating_end_of_list_linearization_changed_check}}$ is during $\mathcal{I}$.
    Lastly, since $e$ is before $p$ invoked $I$ and the step at $T^{\ref{line:ero:add_cell_before_updating_end_of_list_linearization_changed_check}}$ is executed during $I$, we have that $e < T^{\ref{line:ero:add_cell_before_updating_end_of_list_linearization_changed_check}}$, so $(e, T^{\ref{line:ero:add_cell_before_updating_end_of_list_linearization_changed_check}}]$ is during $\mathcal{I}$.
    We now prove that during $(e, T^{\ref{line:ero:add_cell_before_updating_end_of_list_linearization_changed_check}}]$ there are no $L$-events and the ``shape" of the list is in one of two states.

    \begin{claimcustom}{\ref{lemma:ero:1_of_q_holds}.1}\label{lemma:ero:1_of_q_holds_first_claim}
        There are no $L$-events during $(e, T^{\ref{line:ero:add_cell_before_updating_end_of_list_linearization_changed_check}}]$.
    \end{claimcustom}

    \begin{proof}
        Suppose, for contradiction, there is an $L$-event $e'$ during $(e, T^{\ref{line:ero:add_cell_before_updating_end_of_list_linearization_changed_check}}]$.
        Since $e < e'$, we have that $e \neq e'$.
        Furthermore, since $e' < T^{\ref{line:ero:add_cell_before_updating_end_of_list_linearization_changed_check}}$ and  $T^{\ref{line:ero:add_cell_before_updating_end_of_list_linearization_changed_check}} < s$, by transitivity, $e' < s$.
        Hence, since $s$ is the step after $\mathcal{I}$ in $\mathcal{I} \circ s$, we have that $e'$ is in $\mathcal{I}$.
        Therefore, since $\linearizationobject{}.\uniquerepositoryoperationlong{} = \uniquerepositoryoperationshort_\linearizationobject{}$ at $e$, $e'$ is an $L$-event other than $e$ in $\mathcal{I}$, and by assumption $P(\mathcal{I})$ holds, by \Cref{lemma:ero:p_implies_unique_low_level_operations_in_linearization}, $\linearizationobject{}.\uniquerepositoryoperationlong{} \neq \uniquerepositoryoperationshort_\linearizationobject{}$ at $e'$.
        
        We now prove that $\linearizationobject{}.\uniquerepositoryoperationlong \neq \uniquerepositoryoperationshort_\linearizationobject{}$ at $T^{\ref{line:ero:add_cell_before_updating_end_of_list_linearization_changed_check}}$ (*).
        Suppose, for contradiction, the value of $\linearizationobject{}.\uniquerepositoryoperationlong = \uniquerepositoryoperationshort_\linearizationobject{}$ at $T^{\ref{line:ero:add_cell_before_updating_end_of_list_linearization_changed_check}}$.
        Hence, since $\linearizationobject{}.\uniquerepositoryoperationlong{} \neq \uniquerepositoryoperationshort_\linearizationobject{}$ at $e'$, $\linearizationobject{}.\uniquerepositoryoperationlong = \uniquerepositoryoperationshort_\linearizationobject{}$ at $T^{\ref{line:ero:add_cell_before_updating_end_of_list_linearization_changed_check}}$, and $e' < T^{\ref{line:ero:add_cell_before_updating_end_of_list_linearization_changed_check}}$, we have that some step set $\linearizationobject{}.\uniquerepositoryoperationlong = \uniquerepositoryoperationshort_\linearizationobject{}$ during $(e', T^{\ref{line:ero:add_cell_before_updating_end_of_list_linearization_changed_check}})$.
        Thus, by \Cref{observation:ero:where_objects_change}, some $L$-event $e^*$ set $\linearizationobject{}.\uniquerepositoryoperationlong = \uniquerepositoryoperationshort_\linearizationobject{}$ during $(e', T^{\ref{line:ero:add_cell_before_updating_end_of_list_linearization_changed_check}})$.
        Hence, since $e^* < T^{\ref{line:ero:add_cell_before_updating_end_of_list_linearization_changed_check}}$ and $T^{\ref{line:ero:add_cell_before_updating_end_of_list_linearization_changed_check}} < s$, by transitivity, $e^* < s$, so $e^*$ is in $\mathcal{I}$.
        Furthermore, since $e < e'$ and $e' < e^*$, by transitivity, $e < e^*$, and so $e \neq e^*$.
        Therefore, since both $e$ and $e^*$ are in $\mathcal{I}$, and $\linearizationobject{}.\uniquerepositoryoperationlong = \uniquerepositoryoperationshort_\linearizationobject{}$ at both $e$ and $e^*$, there are two $L$-events in $\mathcal{I}$ (namely $e$ and $e^*$) which set $\linearizationobject{}.\uniquerepositoryoperationlong$ to the same value.
        However, since $P(\mathcal{I})$ holds, by \Cref{lemma:ero:p_implies_unique_low_level_operations_in_linearization} every $L$-event in $\mathcal{I}$ sets $\linearizationobject{}.\uniquerepositoryoperationlong$ to a different value, a contradiction.

        We now finish the proof of \Cref{lemma:ero:1_of_q_holds_first_claim}.
        Since $p$ executes $s$ during $I$, it follows that $p$ finds the condition on \cref{line:ero:add_cell_before_updating_end_of_list_linearization_changed_check} to be false at $T^{\ref{line:ero:add_cell_before_updating_end_of_list_linearization_changed_check}}$.
        Therefore, since the parameters of $I$ are $(\uniquerepositoryoperationshort_\linearizationobject{}, \uniquecellpointercontentshort)$, it follows that $\linearizationobject{}.\uniquerepositoryoperationlong{} = \uniquerepositoryoperationshort_\linearizationobject{}$ at $T^{\ref{line:ero:add_cell_before_updating_end_of_list_linearization_changed_check}}$.
        However, by (*), $\linearizationobject{}.\uniquerepositoryoperationlong \neq \uniquerepositoryoperationshort_\linearizationobject{}$ at $T^{\ref{line:ero:add_cell_before_updating_end_of_list_linearization_changed_check}}$, a contradiction.
        \qH{\Cref{lemma:ero:1_of_q_holds_first_claim}}
    \end{proof}

    \begin{claimcustom}{\ref{lemma:ero:1_of_q_holds}.2}\label{lemma:ero:1_of_q_holds_second_claim}
        Consider any prefix $\mathcal{I}'$ of $\mathcal{I}$ during $(e, T^{\ref{line:ero:add_cell_before_updating_end_of_list_linearization_changed_check}}]$.
        The list of cells conforms to either $\List(\mathcal{I}^{exclude}_{e})$ or $\List(\mathcal{I}^{include}_{e})$ in $\mathcal{I}'$ where $\mathcal{I}^{exclude}_{e}$ is the prefix of $\mathcal{I}$ up to but \Underline{excluding} $e$ and $\mathcal{I}^{include}_{e}$ is the prefix of $\mathcal{I}$ up to and \Underline{including} $e$.           
    \end{claimcustom}

    \begin{proof}
        Since $\mathcal{I}'$ is finite, and by assumption $P(\mathcal{I})$, $Q(\mathcal{I})$, and $R(\mathcal{I})$ hold, by \Cref{lemma:ero:conditional_classification_lemma}, the list of cells conforms to either $\List(\mathcal{I}'_{e})$ or $\List(\mathcal{I}')$ in $\mathcal{I}'$ where $\mathcal{I}'_{e}$ is the prefix of $\mathcal{I}'$ up to but excluding $e$.
        Since $\mathcal{I}'$ is a prefix of $\mathcal{I}$, $e$ is in $\mathcal{I}'$, and by definition $\mathcal{I}^{exclude}_{e}$ is the prefix of $\mathcal{I}$ up to be excluding $e$, it follows that $\mathcal{I}'_{e} = \mathcal{I}^{exclude}_{e}$.
        Furthermore, since $\mathcal{I}'$ is a prefix of $\mathcal{I}$ during $(e, T^{\ref{line:ero:add_cell_before_updating_end_of_list_linearization_changed_check}}]$ and by \Cref{lemma:ero:1_of_q_holds_first_claim} there are no $L$-events during $(e, T^{\ref{line:ero:add_cell_before_updating_end_of_list_linearization_changed_check}}]$, we have that the sequence of $L$-events in $\mathcal{I}'$ is exactly the sequence of $L$-events in $\mathcal{I}$ up to and including $e$, so by \Cref{def:ero:logical_list}, $\List(\mathcal{I}') = \List(\mathcal{I}^{include}_{e})$.
        Therefore, the list of cells conforms to either $\List(\mathcal{I}^{exclude}_{e})$ or $\List(\mathcal{I}^{include}_{e})$ in $\mathcal{I}'$ as wanted.
        \qH{\Cref{lemma:ero:1_of_q_holds_second_claim}}
    \end{proof}

    We now prove that $I$ ``traverses" the list.
    Let $\List(\mathcal{I}^{exclude}_{e}) = \cellpointershort_0, \ldots, \cellpointershort_{n + 1}$ for some $n \geq 0$.
    Hence, by \Cref{lemma:ero:every_list_sequence_is_from_universe}, $\cellpointershort_0 = \&\headobject$, for every $i \in [1..n]$ $\cellpointershort_i \in \celluniverse{}$, and $\cellpointershort_{n + 1} = \nullconstant$.
    Furthermore, since $\mathcal{I}^{include}_{e}$ is a single step more than $\mathcal{I}^{exclude}_{e}$ and this single step is $e$, which is an $L$-add event for $\cellpointershort$, by the expansion of $\List(\mathcal{I}^{exclude}_{e})$ along with \Cref{def:ero:logical_list}, it follows that $\List(\mathcal{I}^{include}_{e}) = \cellpointershort_0, \ldots, \cellpointershort_{n}, \cellpointershort, \cellpointershort_{n + 1}$.
    Since $P(\mathcal{I})$ holds, by \Cref{lemma:ero:l_add_for_ptr_is_in_exclude_list}, $\cellpointershort \notin \List(\mathcal{I}^{exclude}_{e})$, and so since $\List(\mathcal{I}^{exclude}_{e}) = \cellpointershort_0, \ldots, \cellpointershort_{n + 1}$, for every $i \in [0..n+1]$ $\cellpointershort \neq \cellpointershort_i$ (*).

    \begin{claimcustom}{\ref{lemma:ero:1_of_q_holds}.3}\label{lemma:ero:1_of_q_holds:claim_acquire_next}
        Consider any invocation $I^*$ of the AcquireNext procedure on \cref{line:ero:add_cell_acquire_next} during $I$.
        $I^*$ exits since $s$ is executed during $I$.
        Let $T^{\ref{line:ero:acquire_next_read_curr_unique_pointer}}_{I^*}$ (resp. $T^{\ref{line:ero:acquire_next_linearization_changed_check}}_{I^*}$) be the last time $p$ executes \cref{line:ero:acquire_next_read_curr_unique_pointer} (resp. \cref{line:ero:acquire_next_linearization_changed_check}) during $I^*$ (these are well-defined since $I^*$ exits).
        Then, the following are true:
        \begin{compactenum}
            \item $\linearizationobject{}.\uniquerepositoryoperationlong{} = \uniquerepositoryoperationshort_\linearizationobject{}$ at $T^{\ref{line:ero:acquire_next_linearization_changed_check}}_{I^*}$; and
            \item $T^{\ref{line:ero:acquire_next_read_curr_unique_pointer}}_{I^*} \in (e, T^{\ref{line:ero:add_cell_before_updating_end_of_list_linearization_changed_check}}]$.
        \end{compactenum}
    \end{claimcustom}

    \begin{proof}
        Since $I^*$ began and exited during $I$, and all steps during $I$ are after $e$, we have that all steps during $I^*$ are after $e$.
        Furthermore, since $T^{\ref{line:ero:add_cell_before_updating_end_of_list_linearization_changed_check}}$ is the only time $p$ executes \cref{line:ero:add_cell_before_updating_end_of_list_linearization_changed_check} during $I$, it follows that $I^*$ exited before $T^{\ref{line:ero:add_cell_before_updating_end_of_list_linearization_changed_check}}$, and so all steps during $I^*$ are during $(e, T^{\ref{line:ero:add_cell_before_updating_end_of_list_linearization_changed_check}}]$.
        Now 1.
        Since $e$ set $\linearizationobject{}.\uniquerepositoryoperationlong{} = \uniquerepositoryoperationshort_\linearizationobject{}$ and by \Cref{lemma:ero:1_of_q_holds_first_claim} there are no $L$-events during $(e, T^{\ref{line:ero:add_cell_before_updating_end_of_list_linearization_changed_check}}]$, we have that $\linearizationobject{}.\uniquerepositoryoperationlong{} = \uniquerepositoryoperationshort_\linearizationobject{}$ throughout $(e, T^{\ref{line:ero:add_cell_before_updating_end_of_list_linearization_changed_check}}]$.
        Hence, since all steps during $I^*$ are during $(e, T^{\ref{line:ero:add_cell_before_updating_end_of_list_linearization_changed_check}}]$, we have that $\linearizationobject{}.\uniquerepositoryoperationlong{} = \uniquerepositoryoperationshort_\linearizationobject{}$ throughout $I^*$.
        Therefore, since $T^{\ref{line:ero:acquire_next_linearization_changed_check}}_{I^*}$ is the time of a step during $I^*$, we have that $\linearizationobject{}.\uniquerepositoryoperationlong{} = \uniquerepositoryoperationshort_\linearizationobject{}$ at $T^{\ref{line:ero:acquire_next_linearization_changed_check}}_{I^*}$.
        Now 2.
        Since all steps during $I^*$ are during $(e, T^{\ref{line:ero:add_cell_before_updating_end_of_list_linearization_changed_check}}]$, and $T^{\ref{line:ero:acquire_next_read_curr_unique_pointer}}_{I^*}$ is the time of a step during $I^*$, we have that $T^{\ref{line:ero:acquire_next_read_curr_unique_pointer}}_{I^*} \in (e, T^{\ref{line:ero:add_cell_before_updating_end_of_list_linearization_changed_check}}]$.
        \qH{\Cref{lemma:ero:1_of_q_holds:claim_acquire_next}}
    \end{proof}

    \begin{claimcustom}{\ref{lemma:ero:1_of_q_holds}.4}\label{lemma:ero:1_of_q_holds_fifth_claim}
        Consider any iteration of the loop on \cref{line:ero:add_cell_while_loop} during $I$, denoted by $I'$, such that the local variable $\currentuniquecellpointershort{} = \cellpointershort_i$ for some $i \in [0..n)$ at the start of $I'$.
        Then, $p$ executes \cref{line:ero:add_cell_update_current_pointer} at time $T^{\ref{line:ero:add_cell_update_current_pointer}}$ during $I'$ and the local variable $\currentuniquecellpointershort{} = \cellpointershort_{i + 1}$ at $T^{\ref{line:ero:add_cell_update_current_pointer}}$.
    \end{claimcustom}

    \begin{proof}
        Since by (*) for every $i \in [0..n+1]$ $\cellpointershort \neq \cellpointershort_i$, and by assumption $\currentuniquecellpointershort{} =\cellpointershort_i$ at the start of $I'$ for some $i \in [0..n)$, it follows that $p$ finds the condition on \cref{line:ero:add_cell_while_loop} to be true at the start of $I'$.
        Hence, since $p$ executes $s$ during $I$, $p$ begins and exits the AcquireNext procedure on \cref{line:ero:add_cell_acquire_next} during $I'$.
        Denote this invocation by $I^*$.
        
        We first prove that $I^*$'s response is $(\found, \cellpointershort_{i + 1})$ by satisfying the conditions of \Cref{lemma:ero:acquire_next_response_classification_weak}.
        Since the first parameter of $I$ is $\uniquerepositoryoperationshort_\linearizationobject{}$ and $\currentuniquecellpointershort{} = \cellpointershort_i$ at the start of $I'$, the parameters of $I^*$ are $(\uniquerepositoryoperationshort_\linearizationobject{}, \cellpointershort_i)$.
        Let $T^{\ref{line:ero:acquire_next_read_curr_unique_pointer}}_{I^*}$ and $T^{\ref{line:ero:acquire_next_linearization_changed_check}}_{I^*}$ by defined as in \Cref{lemma:ero:1_of_q_holds:claim_acquire_next}, and so $\linearizationobject{}.\uniquerepositoryoperationlong{} = \uniquerepositoryoperationshort_\linearizationobject{}$ at $T^{\ref{line:ero:acquire_next_linearization_changed_check}}_{I^*}$, and $T^{\ref{line:ero:acquire_next_read_curr_unique_pointer}}_{I^*} \in (e, T^{\ref{line:ero:add_cell_before_updating_end_of_list_linearization_changed_check}}]$.
        Hence, there is a prefix of $\mathcal{I}$ during $(e, T^{\ref{line:ero:add_cell_before_updating_end_of_list_linearization_changed_check}}]$ up to and including $T^{\ref{line:ero:acquire_next_read_curr_unique_pointer}}_{I^*}$; say $\mathcal{I}'$.
        Thus, by \Cref{lemma:ero:1_of_q_holds_second_claim} the list of cells conforms to either $\List(\mathcal{I}^{exclude}_{e})$ or $\List(\mathcal{I}^{include}_{e})$ in $\mathcal{I}'$.
        So, since $\List(\mathcal{I}^{exclude}_{e}) = \cellpointershort_0, \ldots, \cellpointershort_{n + 1}$, $\List(\mathcal{I}^{include}_{e}) = \cellpointershort_0, \ldots, \cellpointershort_n, \cellpointershort, \cellpointershort_{n + 1}$ and $i \in [0..n)$, by \Cref{def:ero:logical_list}, at the end of $\mathcal{I}'$ $(*\cellpointershort_i).\nextlong.\uniquecellpointercontentlong{} = \cellpointershort_{i + 1}$.
        Hence, since $\cellpointershort_{i + 1} \in \celluniverse{}$ (because $i + 1 \in [1..n]$), by \Cref{assumption:ero:head_and_null_not_in_cell_universe} $\cellpointershort_{i + 1} \neq \nullconstant$, and so $(*\cellpointershort_i).\nextlong.\uniquecellpointercontentlong{} = \cellpointershort_{i + 1} \neq \nullconstant$ at $T^{\ref{line:ero:acquire_next_read_curr_unique_pointer}}$.
        Therefore, we have established the following: (1) $I^*$ has parameters $(\uniquerepositoryoperationshort_\linearizationobject{}, \cellpointershort_i)$; (2) $\linearizationobject{}.\uniquerepositoryoperationlong{} = \uniquerepositoryoperationshort_\linearizationobject{}$ at $T^{\ref{line:ero:acquire_next_linearization_changed_check}}_{I^*}$; and (3) $(*\cellpointershort_i).\nextlong.\uniquecellpointercontentlong{} = \cellpointershort_{i + 1} \neq \nullconstant$ at $T^{\ref{line:ero:acquire_next_read_curr_unique_pointer}}_{I^*}$ (equivalently, the end of $\mathcal{I}'$), and so by \Cref{lemma:ero:acquire_next_response_classification_weak}, $I^*$'s response is $(\found, \cellpointershort_{i + 1})$ as wanted.
        
        We now finish the proof of \Cref{lemma:ero:1_of_q_holds_fifth_claim}.
        Since $p$ executes $s$ during $I$ and $I^*$'s response is $(\found, \cellpointershort_{i + 1})$, we have that $p$ finds the condition on \cref{line:ero:add_cell_acquire_next_found} to be true and so $p$ executes \cref{line:ero:add_cell_update_current_pointer} during $I'$; say at time $T^{\ref{line:ero:add_cell_update_current_pointer}}$.
        Therefore, $\currentuniquecellpointershort{} = \cellpointershort_{i + 1}$ at $T^{\ref{line:ero:add_cell_update_current_pointer}}$ as wanted.
        \qH{\Cref{lemma:ero:1_of_q_holds_fifth_claim}}
    \end{proof}

    \begin{claimcustom}{\ref{lemma:ero:1_of_q_holds}.5}\label{lemma:ero:1_of_q_holds_sixth_claim}
        For every $i \in [1..n+1]$, (1) $p$ executes \cref{line:ero:add_cell_while_loop} $i$ times during $I$ and (2) at the time $p$ executes \cref{line:ero:add_cell_while_loop} for the $i$th time during $I$, the local variable $\currentuniquecellpointershort{} = \cellpointershort_{i - 1}$.
    \end{claimcustom}

    \begin{proof}
        By induction on $i$.
        \begin{itemize}
            \item[] \hspace{0pt}\textbf{Base Case.} $i = 1$.

            In this case, (1) holds immediately since $p$ must execute \cref{line:ero:add_cell_while_loop} at least once during $I$ as $p$ executes $s$ during $I$.
            Let $T^{\ref{line:ero:add_cell_while_loop}}_1$ be the time of $p$'s first execution of \cref{line:ero:add_cell_while_loop} during $I$.
            For (2), since $\currentuniquecellpointershort{}$ at $T^{\ref{line:ero:add_cell_while_loop}}_1$ is the value it was initialized to on \cref{line:ero:add_cell_initial_current_pointer} during $I$, we have that $\currentuniquecellpointershort{} = \&\headobject$ at $T^{\ref{line:ero:add_cell_while_loop}}_1$.
            Therefore, since $\cellpointershort_0 = \&\headobject$, we have that $\currentuniquecellpointershort{} = \cellpointershort_0$ at $T^{\ref{line:ero:add_cell_while_loop}}_1$.

            \item[] \hspace{0pt}\textbf{Inductive Case.} For every $i \in [1..n]$, if (1) and (2) hold for $i$, then (1) and (2) hold for $i + 1$.

            Suppose for any $i \in [1..n]$ (1) $p$ executes \cref{line:ero:add_cell_while_loop} $i$ times during $I$ and (2) at the time $p$ executes \cref{line:ero:add_cell_while_loop} for the $i$th time during $I$, $\currentuniquecellpointershort{} = \cellpointershort_{i - 1}$.
            This is the inductive hypothesis.
            Let $I_i$ be the $i$th iteration of the loop on \cref{line:ero:add_cell_while_loop} during $I$, which is well-defined by (1) of the inductive hypothesis.
            Furthermore, let $T^{\ref{line:ero:add_cell_while_loop}}_i$ be the time of $p$'s $i$th execution of \cref{line:ero:add_cell_while_loop} during $I$ which is the start of $I_i$.
            Since by (2) of the inductive hypothesis $\currentuniquecellpointershort{} = \cellpointershort_{i - 1}$ at $T^{\ref{line:ero:add_cell_while_loop}}_i$ where $i - 1 \in [0..n)$, by \Cref{lemma:ero:1_of_q_holds_fifth_claim}, $p$ executes \cref{line:ero:add_cell_update_current_pointer} at some time $T^{\ref{line:ero:add_cell_update_current_pointer}}_i$ during $I_i$ and $\currentuniquecellpointershort{} = \cellpointershort_i$ at $T^{\ref{line:ero:add_cell_update_current_pointer}}_i$.
            Hence, since $p$ executes $s$ during $I$, it follows that $p$ executes \cref{line:ero:add_cell_while_loop} one more time during $I$, so $p$ executes \cref{line:ero:add_cell_while_loop} $i + 1$ times during $I$.
            Since $\currentuniquecellpointershort{} = \cellpointershort_{i}$ at $T^{\ref{line:ero:add_cell_update_current_pointer}}_i$, and the value of $\currentuniquecellpointershort{}$ does not change between $T^{\ref{line:ero:add_cell_update_current_pointer}}_i$ and the time of $p$'s $i + 1$th execution of \cref{line:ero:add_cell_while_loop} during $I$, it follows that at the time $p$ executes \cref{line:ero:add_cell_while_loop} for the $i + 1$th time during $I$ $\currentuniquecellpointershort{} = \cellpointershort_i$.
            Therefore, (1) and (2) hold for $i + 1$ as wanted.
            \qH{\Cref{lemma:ero:1_of_q_holds_sixth_claim}}
        \end{itemize}
    \end{proof}

    What remains is to deal with the possibility that the ``shape" of the list changes during $I$.

    \begin{claimcustom}{\ref{lemma:ero:1_of_q_holds}.6}\label{lemma:ero:1_of_q_holds_seventh_claim}
        In the $n + 1$th iteration of the loop on \cref{line:ero:add_cell_while_loop} during $I$, which is well-defined by \Cref{lemma:ero:1_of_q_holds_sixth_claim}, either $p$ finds the condition on \cref{line:ero:add_cell_acquire_next_not_found} to be true, or $p$ executes \cref{line:ero:add_cell_while_loop} $n + 2$ times during $I$, and at the time $p$ executes \cref{line:ero:add_cell_while_loop} for the $n+2$th time during $I$, the local variable $\currentuniquecellpointershort{} = \uniquecellpointercontentshort$.
    \end{claimcustom}

    \begin{proof}
        By \Cref{lemma:ero:1_of_q_holds_sixth_claim}, at the time $p$ executes \cref{line:ero:add_cell_while_loop} for the $n + 1$th time during $I$, $\currentuniquecellpointershort{} = \cellpointershort_{n}$.
        Let $I_{n + 1}$ be the $n + 1$th iteration of the loop on \cref{line:ero:add_cell_while_loop} during $I$.
        Since by (*) for every $i \in [0..n+1]$ $\cellpointershort \neq \cellpointershort_i$, and $\currentuniquecellpointershort{} =\cellpointershort_n$ at the start of $I_{n + 1}$, it follows that $p$ finds the condition on \cref{line:ero:add_cell_while_loop} to be true at the start of $\mathcal{I}_{n + 1}$.
        Thus, since $p$ executes $s$ during $I$, $p$ invokes and exits the AcquireNext procedure during $I_{n + 1}$.
        Denote this execution by $I^*$.

        We first prove that $I^*$'s response is either $(\notfound, \arbitraryvalue)$ or $(\found, \uniquecellpointercontentshort)$ by satisfying the conditions of \Cref{lemma:ero:acquire_next_response_classification_weak}.
        Since the first parameter of $I$ is $\uniquerepositoryoperationshort_\linearizationobject{}$ and $\currentuniquecellpointershort{} = \cellpointershort_n$ at the start of $I_{n + 1}$, the parameters of $I^*$ are $(\uniquerepositoryoperationshort_\linearizationobject{}, \cellpointershort_n)$.
        Let $T^{\ref{line:ero:acquire_next_read_curr_unique_pointer}}_{I^*}$ and $T^{\ref{line:ero:acquire_next_linearization_changed_check}}_{I^*}$ by defined as in \Cref{lemma:ero:1_of_q_holds:claim_acquire_next}, and so $\linearizationobject{}.\uniquerepositoryoperationlong{} = \uniquerepositoryoperationshort_\linearizationobject{}$ at $T^{\ref{line:ero:acquire_next_linearization_changed_check}}_{I^*}$, and $T^{\ref{line:ero:acquire_next_read_curr_unique_pointer}}_{I^*} \in (e, T^{\ref{line:ero:add_cell_before_updating_end_of_list_linearization_changed_check}}]$.
        Hence, there is a prefix of $\mathcal{I}$ during $(e, T^{\ref{line:ero:add_cell_before_updating_end_of_list_linearization_changed_check}}]$ up to and including $T^{\ref{line:ero:acquire_next_read_curr_unique_pointer}}_{I^*}$; say $\mathcal{I}'$.
        Thus, by \Cref{lemma:ero:1_of_q_holds_second_claim}, the list of cells conforms to either $\List(\mathcal{I}^{exclude}_{e})$ or $\List(\mathcal{I}^{include}_{e})$ in $\mathcal{I}'$.
        So, since $\List(\mathcal{I}^{exclude}_{e}) = \cellpointershort_0, \ldots, \cellpointershort_{n + 1}$, and $\List(\mathcal{I}^{include}_{e}) = \cellpointershort_0, \ldots, \cellpointershort_n, \cellpointershort, \cellpointershort_{n + 1}$, by \Cref{def:ero:logical_list}, at the end of $\mathcal{I}'$ $(*\cellpointershort_n).\nextlong.\uniquecellpointercontentlong{}$ equals either $\cellpointershort_{n + 1}$ or $\cellpointershort$.
        Thus, since $\cellpointershort \in \celluniverse{}$, by \Cref{assumption:ero:head_and_null_not_in_cell_universe}, $\cellpointershort \neq \nullconstant$, and so since $\cellpointershort_{n + 1} = \nullconstant$, it follows that  $(*\cellpointershort_n).\nextlong.\uniquecellpointercontentlong{}$ equals either $\nullconstant$ or $\cellpointershort \neq \nullconstant$ at the end of $\mathcal{I}'$.
        Therefore, we have established the following: (1) $I^*$ has parameters $(\uniquerepositoryoperationshort_\linearizationobject{}, \cellpointershort_n)$; (2) $\linearizationobject{}.\uniquerepositoryoperationlong{} = \uniquerepositoryoperationshort_\linearizationobject{}$ at $T^{\ref{line:ero:acquire_next_linearization_changed_check}}_{I^*}$; and (3) $(*\cellpointershort_n).\nextlong.\uniquecellpointercontentlong{}$ is either $\nullconstant$ or $\cellpointershort \neq \nullconstant$ at $T^{\ref{line:ero:acquire_next_read_curr_unique_pointer}}_{I^*}$ (equivalently, the end of $\mathcal{I}'$), and so by \Cref{lemma:ero:acquire_next_response_classification_weak}, $I^*$'s response is either $(\notfound, \arbitraryvalue)$ or $(\found, \uniquecellpointercontentshort)$ as wanted.

        We now finish the proof of \Cref{lemma:ero:1_of_q_holds_seventh_claim}.
        Suppose $I^*$'s response is $(\notfound, \arbitraryvalue)$.
        Hence, since $p$ executes $s$ during $I$, it follows that $p$ finds the condition on \cref{line:ero:add_cell_acquire_next_not_found} to be true during $I_{n + 1}$.
        Now suppose $I^*$'s response is $(\found, \uniquecellpointercontentshort)$.
        Hence, since $p$ executes $s$ during $I$, it follows that $p$ finds the condition on \cref{line:ero:add_cell_acquire_next_found} to be true during $I_{n + 1}$, and so $p$ executes \cref{line:ero:add_cell_update_current_pointer} during $I_{n + 1}$; say at time $T^{\ref{line:ero:add_cell_update_current_pointer}}$.
        Thus, $\currentuniquecellpointershort{} = \uniquecellpointercontentshort$ at $T^{\ref{line:ero:add_cell_update_current_pointer}}$.
        Therefore, $p$ executes \cref{line:ero:add_cell_while_loop} $n + 2$ times during $I$, and at the time $p$ executes \cref{line:ero:add_cell_while_loop} for the $n + 2$th time during $I$, $\currentuniquecellpointershort{} = \uniquecellpointercontentshort$.
        \qH{\Cref{lemma:ero:1_of_q_holds_seventh_claim}}
    \end{proof}

    We now return to the proof of \Cref{lemma:ero:1_of_q_holds}.
    By \Cref{lemma:ero:1_of_q_holds_seventh_claim}, there are two cases.
    \begin{itemize}
        \item[] \hspace{0pt}\textbf{Case 1.} During the $n + 1$th iteration of the loop on \cref{line:ero:add_cell_while_loop} during $I$, $p$ finds the condition on \cref{line:ero:add_cell_acquire_next_not_found} to be true.

        Let $I_{n + 1}$ be the the $n + 1$th iteration of the loop on \cref{line:ero:add_cell_while_loop} during $I$.
        
        We first prove that $p$ executes $s$ during $I_{n + 1}$.
        Since $p$ executes $s$ during $I$, and $p$ finds the condition on \cref{line:ero:add_cell_acquire_next_not_found} to be true during $I_{n + 1}$, it follows that $p$ either executes the break on \cref{line:ero:add_cell_before_updating_end_of_list_linearization_changed_check} or \cref{line:ero:add_cell_to_list} during $I_{n + 1}$.
        If the former, $p$ would break out of the loop on \cref{line:ero:add_cell_while_loop} during $I$ before executing $s$, which is impossible (because $p$ executes $s$ during $I$), so $p$ executes \cref{line:ero:add_cell_to_list} during $I_{n + 1}$.
        Therefore, since $p$ executes \cref{line:ero:add_cell_to_list} at most once during $I$, $p$ executes $s$ during $I$, and $p$ executes \cref{line:ero:add_cell_to_list} during $I_{n + 1}$, we have that $p$ executes $s$ during $I_{n + 1}$ as wanted.

        We now finish the proof.
        Since by \Cref{lemma:ero:1_of_q_holds_sixth_claim} the local variable $\currentuniquecellpointershort{} = \cellpointershort_n$ at the start of $I_{n + 1}$, and $p$ executes $s$ during $I_{n + 1}$, we have that $s$ is a \CASop{} operation on $(*\cellpointershort_n).\nextlong$.
        Thus, since $s$ is a list-add attempt for $\cellpointershort$, by \Cref{def:ero:english}, $s$ is a list-add attempt for $\cellpointershort$ after $\cellpointershort_n$.
        Hence, since $\List(\mathcal{I}^{exclude}_{e}) = \cellpointershort_0, \ldots, \cellpointershort_{n + 1}$, we have that $\cellpointershort_n$ is the second last pointer in $\List(\mathcal{I}^{exclude}_{e})$.
        Therefore, $s$ is after the second last pointer in $\List(\mathcal{I}^{exclude}_{e})$ as wanted.

        \item[] \hspace{0pt}\textbf{Case 2.} $p$ executes \cref{line:ero:add_cell_while_loop} $n + 2$ times during $I$, and at the time $p$ executes \cref{line:ero:add_cell_while_loop} for the $n+2$th time during $I$, the local variable $\currentuniquecellpointershort{} = \uniquecellpointercontentshort$.

        Hence, since the second parameter of $I$ is $\uniquecellpointercontentshort$, it follows that $p$ finds the condition on \cref{line:ero:add_cell_while_loop} to be false on its $n+2$th execution of \cref{line:ero:add_cell_while_loop} during $I$.
        Therefore, $p$ does not execute \cref{line:ero:add_cell_to_list} during $I$ (otherwise, $p$ would break out of the loop on \cref{line:ero:add_cell_while_loop} during $I$ on \cref{line:ero:add_cell_not_found_break} and not \cref{line:ero:add_cell_while_loop}).
        However, $p$ executes $s$ during $I$, a contradiction, so this case is impossible.
        \qH{\Cref{lemma:ero:1_of_q_holds}}
    \end{itemize}
\end{proof}

\begin{proposition}\label{lemma:ero:2_of_q_holds}
    If $P(\mathcal{I})$, $Q(\mathcal{I})$, and $R(\mathcal{I})$ hold, then 2. of $Q(\mathcal{I} \circ s)$ holds.
\end{proposition}

\begin{proof}
    Since $Q(\mathcal{I})$ holds, if $s$ is anything other than a list-remove attempt, 2. of $Q(\mathcal{I} \circ s)$ holds by definition, so it suffices to assume that $s$ is a list-remove attempt for some $\cellpointershort$.
    Hence, by \Cref{lemma:ero:every_list_add_seal_and_remove_attempt_is_for_ptr_from_universe}, $\cellpointershort \in \celluniverse{}$.
    Furthermore, by \Cref{def:ero:english}, $p$ executed $s$ during some invocation $I$ of the \doremovecell{} procedure with parameters $(\uniquerepositoryoperationshort_\linearizationobject{}, \cellpointershort)$.
    Hence, by \Cref{lemma:ero:l_event_corresponding_to_do_low_level_op}, there is an $L$-remove event $e$ for $\cellpointershort$ before $I$ was invoked that set $\linearizationobject{}$ to $(\uniquerepositoryoperationshort_\linearizationobject{}, \cellpointershort)$.
    Thus, $e < s$, so $e$ is in $\mathcal{I}$, and by $P(\mathcal{I})$, $e$ is the only $L$-remove event for $\cellpointershort$ in $\mathcal{I}$.
    Therefore, $s$ is preceded by a unique $L$-remove event for $\cellpointershort$ (namely $e$).
    Thus, what remains is to prove that if $\mathcal{I}^{exclude}_{e}$ is the prefix of $\mathcal{I} \circ s$ up to but excluding $e$, then $\cellpointershort$ is in $\List(\mathcal{I}^{exclude}_{e})$ exactly once, and $s$ is between pointers preceding and succeeding $\cellpointershort$ in $\List(\mathcal{I}^{exclude}_{e})$.
    We start by establishing some basic facts for the proof.
    Since $e$ is before $p$ invoked $I$, we have that all steps during $I$ are after $e$. 
    Furthermore, since $p$ executed $s$ during $I$, and $s$ is a list-remove attempt, $p$ executed \cref{line:ero:remove_cell_before_removal_linearization_check} during $I$.
    Hence, it executed \cref{line:ero:remove_cell_before_removal_linearization_check} for a last time during $I$ before $s$; say at time $T^{\ref{line:ero:remove_cell_before_removal_linearization_check}}$.
    Since $T^{\ref{line:ero:remove_cell_before_removal_linearization_check}} < s$, it follows that the step at $T^{\ref{line:ero:remove_cell_before_removal_linearization_check}}$ is during $\mathcal{I}$.
    Lastly, since $e$ is before $p$ invoked $I$ and the step at $T^{\ref{line:ero:remove_cell_before_removal_linearization_check}}$ is executed during $I$, we have that $e < T^{\ref{line:ero:remove_cell_before_removal_linearization_check}}$, so $(e, T^{\ref{line:ero:remove_cell_before_removal_linearization_check}}]$ is during $\mathcal{I}$.
    We now prove that during $(e, T^{\ref{line:ero:remove_cell_before_removal_linearization_check}}]$ there are no $L$-events and the ``shape" of the list is in one of two states.

    \begin{claimcustom}{\ref{lemma:ero:2_of_q_holds}.1}\label{lemma:ero:2_of_q_holds_first_claim}
        There are no $L$-events during $(e, T^{\ref{line:ero:remove_cell_before_removal_linearization_check}}]$.
    \end{claimcustom}

    \begin{proof}
        Suppose, for contradiction, there is an $L$-event $e'$ during $(e, T^{\ref{line:ero:remove_cell_before_removal_linearization_check}}]$.
        Since $e < e'$, we have that $e \neq e'$.
        Furthermore, since $e' < T^{\ref{line:ero:remove_cell_before_removal_linearization_check}}$ and $T^{\ref{line:ero:remove_cell_before_removal_linearization_check}} < s$, by transitivity, $e' < s$.
        Hence, since $s$ is the step after $\mathcal{I}$ in $\mathcal{I} \circ s$, we have that $e'$ is in $\mathcal{I}$.
        Therefore, since $\linearizationobject{}.\uniquerepositoryoperationlong{} = \uniquerepositoryoperationshort_\linearizationobject{}$ at $e$, $e'$ is an $L$-event other than $e$ in $\mathcal{I}$, and by assumption $P(\mathcal{I})$ holds, by \Cref{lemma:ero:p_implies_unique_low_level_operations_in_linearization}, $\linearizationobject{}.\uniquerepositoryoperationlong{} \neq \uniquerepositoryoperationshort_\linearizationobject{}$ at $e'$.

        We now prove that $\linearizationobject{}.\uniquerepositoryoperationlong \neq \uniquerepositoryoperationshort_\linearizationobject{}$ at $T^{\ref{line:ero:remove_cell_before_removal_linearization_check}}$ (*).
        Suppose, for contradiction, the value of $\linearizationobject{}.\uniquerepositoryoperationlong = \uniquerepositoryoperationshort_\linearizationobject{}$ at $T^{\ref{line:ero:remove_cell_before_removal_linearization_check}}$.
        Hence, since $\linearizationobject{}.\uniquerepositoryoperationlong{} \neq \uniquerepositoryoperationshort_\linearizationobject{}$ at $e'$, $\linearizationobject{}.\uniquerepositoryoperationlong = \uniquerepositoryoperationshort_\linearizationobject{}$ at $T^{\ref{line:ero:remove_cell_before_removal_linearization_check}}$, and $e' < T^{\ref{line:ero:remove_cell_before_removal_linearization_check}}$, we have that some step set $\linearizationobject{}.\uniquerepositoryoperationlong = \uniquerepositoryoperationshort_\linearizationobject{}$ during $(e', T^{\ref{line:ero:remove_cell_before_removal_linearization_check}})$.
        Thus, by \Cref{observation:ero:where_objects_change}, some $L$-event $e^*$ set $\linearizationobject{}.\uniquerepositoryoperationlong = \uniquerepositoryoperationshort_\linearizationobject{}$ during $(e', T^{\ref{line:ero:remove_cell_before_removal_linearization_check}})$.
        Hence, since $e^* < T^{\ref{line:ero:remove_cell_before_removal_linearization_check}}$ and by definition $T^{\ref{line:ero:remove_cell_before_removal_linearization_check}} < s$, by transitivity, $e^* < s$, so $e^*$ is in $\mathcal{I}$.
        Furthermore, since $e < e'$ and $e' < e^*$, by transitivity, $e < e^*$, and so $e \neq e^*$.
        Therefore, since both $e$ and $e^*$ are in $\mathcal{I}$, and $\linearizationobject{}.\uniquerepositoryoperationlong = \uniquerepositoryoperationshort_\linearizationobject{}$ at both $e$ and $e^*$, there are two $L$-events in $\mathcal{I}$ (namely $e$ and $e^*$) which set $\linearizationobject{}.\uniquerepositoryoperationlong$ to the same value.
        However, since $P(\mathcal{I})$ holds, by \Cref{lemma:ero:p_implies_unique_low_level_operations_in_linearization} every $L$-event in $\mathcal{I}$ sets $\linearizationobject{}.\uniquerepositoryoperationlong$ to a different value, a contradiction.

        We now finish the proof of \Cref{lemma:ero:2_of_q_holds_first_claim}.
        Since $p$ executes $s$ during $I$, $p$ finds the condition on \cref{line:ero:remove_cell_before_removal_linearization_check} to be false at $T^{\ref{line:ero:remove_cell_before_removal_linearization_check}}$.
        Therefore, since the parameters of $I$ are $(\uniquerepositoryoperationshort_\linearizationobject{}, \uniquecellpointercontentshort)$, it follows that $\linearizationobject{}.\uniquerepositoryoperationlong{} = \uniquerepositoryoperationshort_\linearizationobject{}$ at $T^{\ref{line:ero:remove_cell_before_removal_linearization_check}}$.
        However, by (*), $\linearizationobject{}.\uniquerepositoryoperationlong \neq \uniquerepositoryoperationshort_\linearizationobject{}$ at $T^{\ref{line:ero:remove_cell_before_removal_linearization_check}}$, a contradiction.
        \qH{\Cref{lemma:ero:2_of_q_holds_first_claim}}
    \end{proof}
    
    \begin{claimcustom}{\ref{lemma:ero:2_of_q_holds}.2}\label{lemma:ero:2_of_q_holds_second_claim}
        Consider any prefix $\mathcal{I}'$ of $\mathcal{I}$ during $(e, T^{\ref{line:ero:remove_cell_before_removal_linearization_check}}]$.
        The list of cells conforms to either $\List(\mathcal{I}^{exclude}_{e})$ or $\List(\mathcal{I}^{include}_{e})$ in $\mathcal{I}'$ where $\mathcal{I}^{exclude}_{e}$ is the prefix of $\mathcal{I}$ up to but \Underline{excluding} $e$ and $\mathcal{I}^{include}_{e}$ is the prefix of $\mathcal{I}$ up to and \Underline{including} $e$.           
    \end{claimcustom}

    \begin{proof}
        Since $\mathcal{I}'$ is finite, and by assumption $P(\mathcal{I})$, $Q(\mathcal{I})$, and $R(\mathcal{I})$ hold, by \Cref{lemma:ero:conditional_classification_lemma}, the list of cells conforms to either $\List(\mathcal{I}'_{e})$ or $\List(\mathcal{I}')$ in $\mathcal{I}'$ where $\mathcal{I}'_{e}$ is the prefix of $\mathcal{I}'$ up to but excluding $e$.
        Since $\mathcal{I}'$ is a prefix of $\mathcal{I}$, $e$ is in $\mathcal{I}'$, and by definition $\mathcal{I}^{exclude}_{e}$ is the prefix of $\mathcal{I}$ up to be excluding $e$, it follows that $\mathcal{I}'_{e} = \mathcal{I}^{exclude}_{e}$.
        Furthermore, since $\mathcal{I}'$ is a prefix of $\mathcal{I}$ during $(e, T^{\ref{line:ero:remove_cell_before_removal_linearization_check}}]$ and by \Cref{lemma:ero:2_of_q_holds_first_claim} there are no $L$-events during $(e, T^{\ref{line:ero:remove_cell_before_removal_linearization_check}}]$, we have that the sequence of $L$-events in $\mathcal{I}'$ is exactly the sequence of $L$-events in $\mathcal{I}$ up to and including $e$, so by \Cref{def:ero:logical_list}, $\List(\mathcal{I}') = \List(\mathcal{I}^{include}_{e})$.
        Therefore, the list of cells conforms to either $\List(\mathcal{I}^{exclude}_{e})$ or $\List(\mathcal{I}^{include}_{e})$ in $\mathcal{I}'$ as wanted.
        \qH{\Cref{lemma:ero:2_of_q_holds_second_claim}}
    \end{proof}

    We now prove that $I$ ``traverses" the list.
    Let $\List(\mathcal{I}^{exclude}_{e}) = \cellpointershort_0, \ldots, \cellpointershort_{n + 1}$ for some integer $n \geq 0$.
    Hence, by \Cref{lemma:ero:every_list_sequence_is_from_universe}, $\cellpointershort_0 = \&\headobject$, for every $i \in [1..n]$ $\cellpointershort_i \in \celluniverse{}$, and $\cellpointershort_{n + 1} = \nullconstant$.
    Furthermore, since $e$ is an $L$-remove event for $\cellpointershort$, and $P(\mathcal{I})$ holds, by \Cref{lemma:ero:l_remove_for_ptr_is_uniquely_in_exclude_list}, there is exactly one $i \in [1..n]$ such that $\cellpointershort_i = \cellpointershort$.
    Hence, since $\mathcal{I}^{include}_{e}$ is a single step more than $\mathcal{I}^{exclude}_{e}$ and this single step is $e$, which is a $L$-remove event for $\cellpointershort$, by \Cref{def:ero:logical_list}, it follows that $\List(\mathcal{I}^{include}_{e}) = \cellpointershort_0, \ldots, \cellpointershort_{i-1}, \cellpointershort_{i+1}, \ldots, \cellpointershort_{n + 1}$.

    \begin{claimcustom}{\ref{lemma:ero:2_of_q_holds}.3}\label{lemma:ero:2_of_q_holds:claim_acquire_next}
        Consider any invocation $I^*$ of the AcquireNext procedure on \cref{line:ero:remove_cell_acquire_next} during $I$.
        $I^*$ exits since $s$ is executed during $I$.
        Let $T^{\ref{line:ero:acquire_next_read_curr_unique_pointer}}_{I^*}$ (resp. $T^{\ref{line:ero:acquire_next_linearization_changed_check}}_{I^*}$) be the last time $p$ executes \cref{line:ero:acquire_next_read_curr_unique_pointer} (resp. \cref{line:ero:acquire_next_linearization_changed_check}) during $I^*$ (these are well-defined since $I^*$ exits).
        Then, the following are true:
        \begin{compactenum}
            \item $\linearizationobject{}.\uniquerepositoryoperationlong{} = \uniquerepositoryoperationshort_\linearizationobject{}$ at $T^{\ref{line:ero:acquire_next_linearization_changed_check}}_{I^*}$; and
            \item $T^{\ref{line:ero:acquire_next_read_curr_unique_pointer}}_{I^*} \in (e, T^{\ref{line:ero:remove_cell_before_removal_linearization_check}}]$.
        \end{compactenum}
    \end{claimcustom}

    \begin{proof}
        Since $I^*$ began and exited during $I$, and all steps during $I$ are after $e$, we have that all steps during $I^*$ are after $e$.
        Furthermore, since $T^{\ref{line:ero:remove_cell_before_removal_linearization_check}}$ is the last time $p$ executes \cref{line:ero:remove_cell_before_removal_linearization_check} during $I$, it follows that $I^*$ exited before $T^{\ref{line:ero:remove_cell_before_removal_linearization_check}}$, and so all steps during $I^*$ are during $(e, T^{\ref{line:ero:remove_cell_before_removal_linearization_check}}]$.
        Now 1.
        Since $e$ set $\linearizationobject{}.\uniquerepositoryoperationlong{} = \uniquerepositoryoperationshort_\linearizationobject{}$ and by \Cref{lemma:ero:2_of_q_holds_first_claim} there are no $L$-events during $(e, T^{\ref{line:ero:remove_cell_before_removal_linearization_check}}]$, we have that $\linearizationobject{}.\uniquerepositoryoperationlong{} = \uniquerepositoryoperationshort_\linearizationobject{}$ throughout $(e, T^{\ref{line:ero:remove_cell_before_removal_linearization_check}}]$.
        Hence, since all steps during $I^*$ are during $(e, T^{\ref{line:ero:remove_cell_before_removal_linearization_check}}]$, we have that $\linearizationobject{}.\uniquerepositoryoperationlong{} = \uniquerepositoryoperationshort_\linearizationobject{}$ throughout $I^*$.
        Therefore, since $T^{\ref{line:ero:acquire_next_linearization_changed_check}}_{I^*}$ is the time of a step during $I^*$, we have that $\linearizationobject{}.\uniquerepositoryoperationlong{} = \uniquerepositoryoperationshort_\linearizationobject{}$ at $T^{\ref{line:ero:acquire_next_linearization_changed_check}}_{I^*}$.
        Now 2.
        Since all steps during $I^*$ are during $(e, T^{\ref{line:ero:remove_cell_before_removal_linearization_check}}]$, and $T^{\ref{line:ero:acquire_next_read_curr_unique_pointer}}_{I^*}$ is the time of a step during $I^*$, we have that $T^{\ref{line:ero:acquire_next_read_curr_unique_pointer}}_{I^*} \in (e, T^{\ref{line:ero:remove_cell_before_removal_linearization_check}}]$.
        \qH{\Cref{lemma:ero:2_of_q_holds:claim_acquire_next}}
    \end{proof}

    \begin{claimcustom}{\ref{lemma:ero:2_of_q_holds}.4}\label{lemma:ero:2_of_q_holds_fifth_claim}
        Consider any iteration of the loop on \cref{line:ero:remove_cell_while_loop} during $I$, denoted by $I'$, such that the local variable $\currentuniquecellpointershort{} = \cellpointershort_j$ for some $j \in [0..n) \setminus \{i - 1,i\}$ at the start of $I'$.
        Then, $p$ executes \cref{line:ero:remove_cell_update_pointers} at time $T^{\ref{line:ero:remove_cell_update_pointers}}$ during $I'$ and the local variable $\currentuniquecellpointershort{} = \cellpointershort_{j + 1}$ at $T^{\ref{line:ero:remove_cell_update_pointers}}$.\footnote{We exclude $i$ because if $\currentcellpointershort{} = \cellpointershort_i$ on \cref{line:ero:remove_cell_while_loop} then $p$ will find the condition to be false because the second parameter of $I$ is $\cellpointershort$ which is equal to $\cellpointershort_i$. We exclude $i - 1$ because $p$ will set $\currentuniquecellpointershort{}$ either to $\cellpointershort_{i}$ or $\cellpointershort_{i + 1}$ at $T^{\ref{line:ero:remove_cell_update_pointers}}$ depending on whether $\cellpointershort$ has been removed from the list or not (see \Cref{lemma:ero:2_of_q_holds_eigth_claim}).}
    \end{claimcustom}

    \begin{proof}
        Since $\cellpointershort = \cellpointershort_i$ for a unique $i \in [1..n]$, we have that every $j \in [0..n] \setminus \{i\}$ $\cellpointershort \neq \cellpointershort_j$.
        Hence, since by assumption $\currentuniquecellpointershort{} = \cellpointershort_j$ at the start of $I'$ for some $j \in [0..n) \setminus \{i-1,i\}$ and the second parameter of $I$ is $\cellpointershort$, it follows that $\cellpointershort \neq \cellpointershort_j$, so $p$ finds the condition on \cref{line:ero:remove_cell_while_loop} to be true at the start of $I'$.
        Thus, since $p$ executes $s$ during $I$, $p$ begins and exits the AcquireNext procedure on \cref{line:ero:remove_cell_acquire_next} during $I'$.
        Denote this invocation by $I^*$.
        
        We first prove that $I^*$'s response is $(\found, \cellpointershort_{j + 1})$ by satisfying the conditions of \Cref{lemma:ero:acquire_next_response_classification_weak}.
        Since the first parameter of $I$ is $\uniquerepositoryoperationshort_\linearizationobject{}$ and $\currentuniquecellpointershort{} = \cellpointershort_j$ at the start of $I'$, we have that the parameters of $I^*$ are $(\uniquerepositoryoperationshort_\linearizationobject{}, \cellpointershort_j)$.
        Let $T^{\ref{line:ero:acquire_next_read_curr_unique_pointer}}_{I^*}$ and $T^{\ref{line:ero:acquire_next_linearization_changed_check}}_{I^*}$ by defined as in \Cref{lemma:ero:2_of_q_holds:claim_acquire_next}, and so $\linearizationobject{}.\uniquerepositoryoperationlong{} = \uniquerepositoryoperationshort_\linearizationobject{}$ at $T^{\ref{line:ero:acquire_next_linearization_changed_check}}_{I^*}$, and $T^{\ref{line:ero:acquire_next_read_curr_unique_pointer}}_{I^*} \in (e, T^{\ref{line:ero:remove_cell_before_removal_linearization_check}}]$.
        Hence, there is a prefix of $\mathcal{I}$ during $(e, T^{\ref{line:ero:remove_cell_before_removal_linearization_check}}]$ up to and including $T^{\ref{line:ero:acquire_next_read_curr_unique_pointer}}_{I^*}$; say $\mathcal{I}'$.
        Thus, by \Cref{lemma:ero:2_of_q_holds_second_claim} the list of cells conforms to either $\List(\mathcal{I}^{exclude}_{e})$ or $\List(\mathcal{I}^{include}_{e})$ in $\mathcal{I}'$.
        So, since $\List(\mathcal{I}^{exclude}_{e}) = \cellpointershort_0, \ldots, \cellpointershort_{n + 1}$, $\List(\mathcal{I}^{include}_{e}) = \cellpointershort_0, \ldots, \cellpointershort_{i - 1}, \cellpointershort_{i + 1}, \ldots, \cellpointershort_{n + 1}$, and $j \in [0..n) \setminus \{i-1, i\}$, by \Cref{def:ero:logical_list}, at the end of $\mathcal{I}'$ $(*\cellpointershort_j).\nextlong.\uniquecellpointercontentlong{} = \cellpointershort_{j + 1}$.
        Hence, since $\cellpointershort_{j + 1} \in \celluniverse{}$ (because $j + 1 \in [1..n]$), by \Cref{assumption:ero:head_and_null_not_in_cell_universe} $\cellpointershort_{j + 1} \neq \nullconstant$, and so $(*\cellpointershort_j).\nextlong.\uniquecellpointercontentlong{} = \cellpointershort_{j + 1} \neq \nullconstant$ at $T^{\ref{line:ero:acquire_next_read_curr_unique_pointer}}$.
        Therefore, we have established the following: (1) $I^*$ has parameters $(\uniquerepositoryoperationshort_\linearizationobject{}, \cellpointershort_j)$; (2) $\linearizationobject{}.\uniquerepositoryoperationlong{} = \uniquerepositoryoperationshort_\linearizationobject{}$ at $T^{\ref{line:ero:acquire_next_linearization_changed_check}}_{I^*}$; and (3) $(*\cellpointershort_j).\nextlong.\uniquecellpointercontentlong{} = \cellpointershort_{j + 1} \neq \nullconstant$ at $T^{\ref{line:ero:acquire_next_read_curr_unique_pointer}}_{I^*}$ (equivalently, the end of $\mathcal{I}'$), and so by \Cref{lemma:ero:acquire_next_response_classification_weak}, $I^*$'s response is $(\found,\cellpointershort_{j + 1})$.
        
        We now finish the proof of \Cref{lemma:ero:2_of_q_holds_fifth_claim}.
        Since $p$ executes $s$ during $I$ and $I^*$'s response is $(\found, \cellpointershort_{j + 1})$, it follows that $p$ finds the condition on \cref{line:ero:remove_cell_acquire_next_found} to be true and so $p$ executes \cref{line:ero:remove_cell_update_pointers} during $I'$ say at time $T^{\ref{line:ero:remove_cell_update_pointers}}$.
        Therefore, $\currentuniquecellpointershort{} = \cellpointershort_{j + 1}$ at $T^{\ref{line:ero:remove_cell_update_pointers}}$ as wanted.
        \qH{\Cref{lemma:ero:2_of_q_holds_fifth_claim}}
    \end{proof}

    \begin{claimcustom}{\ref{lemma:ero:2_of_q_holds}.5}\label{lemma:ero:2_of_q_holds_sixth_claim}
    For every $j \in [1..i]$ (1) $p$ executes \cref{line:ero:remove_cell_while_loop} $j$ times during $I$, (2) at the time $p$ executes \cref{line:ero:remove_cell_while_loop} for the $j$th time during $I$, the local variable $\currentuniquecellpointershort{}$ is $\cellpointershort_{j - 1}$.
    \end{claimcustom}

    \begin{proof}
        By induction on $j$.
        \begin{itemize}
            \item[] \hspace{0pt}\textbf{Base Case.} $j = 1$.
            
            In this case, (1) holds immediately since $p$ must execute \cref{line:ero:remove_cell_while_loop} at least once during $I$ as $p$ executes $s$ during $I$.
            Let $T^{\ref{line:ero:remove_cell_while_loop}}_1$ be the time of $p$'s first execution of \cref{line:ero:remove_cell_while_loop} during $I$.
            For (2), since $\currentuniquecellpointershort{}$ at $T^{\ref{line:ero:remove_cell_while_loop}}_1$ is the value it was initialized to on \cref{line:ero:remove_cell_initialize_pointers} during $I$, we have that $\currentuniquecellpointershort{} = \&\headobject$ at $T^{\ref{line:ero:remove_cell_while_loop}}_1$.
            Therefore, since $\cellpointershort_0 = \&\headobject$, we have that $\currentuniquecellpointershort{} = \cellpointershort_0$ at $T^{\ref{line:ero:remove_cell_while_loop}}_1$.

            \item[] \hspace{0pt}\textbf{Inductive Case.} For every $j \in [1..i)$, if (1) and (2) hold for $j$, then (1) and (2) hold for $j + 1$.

            Suppose for any $j \in [1..i)$ (1) $p$ executes \cref{line:ero:remove_cell_while_loop} $j$ times during $I$ and (2) at the time $p$ executes \cref{line:ero:remove_cell_while_loop} for the $j$th time during $I$, $\currentuniquecellpointershort{} = \cellpointershort_{j - 1}$.
            This is the inductive hypothesis.
            Let $I_j$ be the $j$th iteration of the loop on \cref{line:ero:remove_cell_while_loop} during $I$, which is well-defined by (1) of the inductive hypothesis.
            Furthermore, let $T^{\ref{line:ero:remove_cell_while_loop}}_j$ be the time of $p$'s $j$th execution of \cref{line:ero:remove_cell_while_loop} during $I$ which is the start of $I_j$.
            Since by (2) of the inductive hypothesis $\currentuniquecellpointershort{} = \cellpointershort_{j - 1}$ at $T^{\ref{line:ero:remove_cell_while_loop}}_j$ where $j - 1 \in [0..i-1)$, by \Cref{lemma:ero:2_of_q_holds_fifth_claim}, $p$ executes \cref{line:ero:remove_cell_update_pointers} at some time $T^{\ref{line:ero:remove_cell_update_pointers}}_j$ during $I_j$ and $\currentuniquecellpointershort{} = \cellpointershort_j$ at $T^{\ref{line:ero:remove_cell_update_pointers}}_j$.
            Hence, since $p$ executes $s$ during $I$, it follows that $p$ executes \cref{line:ero:remove_cell_while_loop} one more time during $I$, and so $p$ executes \cref{line:ero:remove_cell_while_loop} $j + 1$ times during $I$.
            % Therefore, $p$ executes \cref{line:ero:add_cell_while_loop} $i + 1$ times during $I$.
            % We now prove (2) holds for $i + 1$.
            Since $\currentuniquecellpointershort{} = \cellpointershort_{j}$ at $T^{\ref{line:ero:remove_cell_update_pointers}}_j$, and the value of $\currentuniquecellpointershort{}$ does not change between $T^{\ref{line:ero:remove_cell_update_pointers}}_j$ and the time of $p$'s $j + 1$th execution of \cref{line:ero:remove_cell_while_loop} during $I$, it follows that at the time $p$ executes \cref{line:ero:remove_cell_while_loop} for the $j + 1$th time during $I$ $\currentuniquecellpointershort{} = \cellpointershort_j$.
            Therefore, (1) and (2) hold for $j + 1$ as wanted.
            \qH{\Cref{lemma:ero:2_of_q_holds_sixth_claim}}
        \end{itemize}
    \end{proof}

    What remains is to deal with the possibility that the ``shape" of the list changes during $I$.

    \begin{claimcustom}{\ref{lemma:ero:2_of_q_holds}.6}\label{lemma:ero:2_of_q_holds_eigth_claim}
        $p$ executes \cref{line:ero:remove_cell_while_loop} $i + 1$ times during $I$ and at the time $p$ executes \cref{line:ero:remove_cell_while_loop} for the $i + 1$th time during $I$, the local variable $\currentuniquecellpointershort{}$ is either $\uniquecellpointercontentshort$ or $\cellpointershort_{i + 1}$.
        Furthermore, if $\currentuniquecellpointershort{} = \cellpointershort_{i + 1}$, then $i < n$.
    \end{claimcustom}

    \begin{proof}
        By \Cref{lemma:ero:2_of_q_holds_sixth_claim}, $p$ executes \cref{line:ero:remove_cell_while_loop} $i$ times and at the time $p$ executes \cref{line:ero:remove_cell_while_loop} for the $i$th time during $I$, $\currentuniquecellpointershort{}=\cellpointershort_{i-1}$.
        Let $I_{i}$ be the $i$th iteration of the loop on \cref{line:ero:remove_cell_while_loop} during $I$.
        Since $\mathcal{I}^{exclude}_{e}$ is a prefix of $\mathcal{I}$, $\List(\mathcal{I}^{exclude}_{e}) = \cellpointershort_0, \ldots, \cellpointershort_{n + 1}$, by assumption $P(\mathcal{I})$ holds, and $i \in [1..n]$, by \Cref{lemma:ero:pointers_in_list_are_unique} for every $j \in [0..n+1]$, if $i \neq j$, then $\cellpointershort_i \neq \cellpointershort_j$.
        Hence, since $\cellpointershort_i = \cellpointershort$, for every $j \in [0..n+1]$, if $i \neq j$, then $\cellpointershort \neq \cellpointershort_j$.
        Thus, since $i - 1\neq i$, we have that $\cellpointershort \neq \cellpointershort_{i - 1}$.
        Hence, since $\currentuniquecellpointershort{} = \cellpointershort_{i - 1}$ at the start of $I_i$ and the second parameter of $I$ is $\cellpointershort$, we have that $\currentuniquecellpointershort{} \neq \uniquecellpointercontentshort$ at the start of $I'$.
        Hence, $p$ finds the condition on \cref{line:ero:remove_cell_while_loop} to be true at the start of $I_{i}$.
        Thus, since $p$ executes $s$ during $I$, $p$ invokes and exits the AcquireNext procedure on \cref{line:ero:remove_cell_acquire_next} during $I_{i}$.
        Denote this execution of the AcquireNext procedure by $I^*$.

        We first prove that $I^*$'s response is either $(\notfound, \arbitraryvalue)$, $(\found, \uniquecellpointercontentshort)$, or $(\found, \cellpointershort_{i + 1})$ by satisfying the conditions of \Cref{lemma:ero:acquire_next_response_classification_weak}.
        Moreover, if $I^*$'s response is $(\found, \cellpointershort_{i + 1})$, then $i < n$.
        Since the first parameter of $I$ is $\uniquerepositoryoperationshort_\linearizationobject{}$ and $\currentuniquecellpointershort{}$ is $\cellpointershort_{i - 1}$ at the start of $I_{i}$, the parameters of $I^*$ are $(\uniquerepositoryoperationshort_\linearizationobject{}, \cellpointershort_{i - 1})$.
        Let $T^{\ref{line:ero:acquire_next_read_curr_unique_pointer}}_{I^*}$ and $T^{\ref{line:ero:acquire_next_linearization_changed_check}}_{I^*}$ by defined as in \Cref{lemma:ero:2_of_q_holds:claim_acquire_next}, and so $\linearizationobject{}.\uniquerepositoryoperationlong{} = \uniquerepositoryoperationshort_\linearizationobject{}$ at $T^{\ref{line:ero:acquire_next_linearization_changed_check}}_{I^*}$, and $T^{\ref{line:ero:acquire_next_read_curr_unique_pointer}}_{I^*} \in (e, T^{\ref{line:ero:remove_cell_before_removal_linearization_check}}]$.
        Hence, there is a prefix of $\mathcal{I}$ during $(e, T^{\ref{line:ero:remove_cell_before_removal_linearization_check}}]$ up to and including $T^{\ref{line:ero:acquire_next_read_curr_unique_pointer}}_{I^*}$; say $\mathcal{I}'$.
        Thus, by \Cref{lemma:ero:2_of_q_holds_second_claim} the list of cells conforms to either $\List(\mathcal{I}^{exclude}_{e})$ or $\List(\mathcal{I}^{include}_{e})$ in $\mathcal{I}'$.
        So, since $\List(\mathcal{I}^{exclude}_{e}) = \cellpointershort_0, \ldots, \cellpointershort_{n + 1}$, and $\List(\mathcal{I}^{include}_{e}) = \cellpointershort_0, \ldots, \cellpointershort_{i - 1}, \cellpointershort_{i + 1}, \ldots, \cellpointershort_{n + 1}$, by \Cref{def:ero:logical_list}, at the end of $\mathcal{I}'$ $(*\cellpointershort_{i - 1}).\nextlong.\uniquecellpointercontentlong{}$ equals either $\cellpointershort_i$ or $\cellpointershort_{i + 1}$.
        Therefore, we have established the following: (1) $I^*$ has parameters $(\uniquerepositoryoperationshort_\linearizationobject{}, \cellpointershort_{i - 1})$; (2) $\linearizationobject{}.\uniquerepositoryoperationlong{} = \uniquerepositoryoperationshort_\linearizationobject{}$ at $T^{\ref{line:ero:acquire_next_linearization_changed_check}}_{I^*}$; and (3) $(*\cellpointershort_{i - 1}).\nextlong.\uniquecellpointercontentlong{}$ is either $\cellpointershort_i$ or $\cellpointershort_{i + 1}$ at $T^{\ref{line:ero:acquire_next_read_curr_unique_pointer}}_{I^*}$ (equivalently, the end of $\mathcal{I}'$), and so by \Cref{lemma:ero:acquire_next_response_classification_weak}, $I^*$'s response is either $(\notfound, \arbitraryvalue)$, $(\found, \cellpointershort_i)$ (equivalently $(\found, \cellpointershort)$ since $\cellpointershort = \cellpointershort_i$), or $(\found, \cellpointershort_{i + 1})$ as wanted.
        We now prove the ``moreover" part.
        Suppose, for contradiction, $I^*$'s response is $(\found, \cellpointershort_{i + 1})$ and $i \geq n$.
        Hence, by \Cref{lemma:ero:acquire_next_response_classification_weak}, $(*\cellpointershort_{i - 1}).\nextlong.\uniquecellpointercontentlong{} = \cellpointershort_{i + 1} \neq \nullconstant$ at the end of $\mathcal{I}'$.
        Furthermore, since $i \in [1..n]$, we have that $i + 1 = n + 1$.
        Therefore, $\cellpointershort_{n + 1} \neq \nullconstant$.
        However, $\cellpointershort_{n + 1} = \nullconstant$, a contradiction.

        We now finish the proof of \Cref{lemma:ero:2_of_q_holds_eigth_claim}.
        If $I^*$'s response is $(\notfound, \arbitraryvalue)$, then since $p$ executes $s$ during $I$, it follows that $p$ finds the condition on \cref{line:ero:remove_cell_check_acquire_status} to be true during $I_{i}$, so $p$ executes \cref{line:ero:exit_1} during $I_i$.
        Therefore, $p$ does not execute \cref{line:ero:remove_cell_from_list} during $I$.
        However, by assumption $p$ executes $s$ during $I$, so this case is impossible.
        Now suppose $I^*$'s response is either $(\found, \uniquecellpointercontentshort)$ or $(\found, \cellpointershort_{i + 1})$.
        Since $p$ executes $s$ during $I$, and $I^*$'s response is either $(\found, \uniquecellpointercontentshort)$ or $(\found, \cellpointershort_{i + 1})$, we have that $p$ finds the condition on \cref{line:ero:remove_cell_acquire_next_found} to be true and so $p$ executes \cref{line:ero:remove_cell_update_pointers} during $I_{i}$; say at time $T^{\ref{line:ero:remove_cell_update_pointers}}$.
        Hence, $\currentuniquecellpointershort{}$ is set to either $\uniquecellpointercontentshort$ or $\cellpointershort_{i + 1}$ at $T^{\ref{line:ero:remove_cell_update_pointers}}$.
        Therefore, since $p$ executes $s$ during $I$, it follows that $p$ executes \cref{line:ero:remove_cell_while_loop} $i + 1$ times during $I$, and at the time $p$ executes \cref{line:ero:remove_cell_while_loop} for the $i + 1$th time during $I$, the value of $\currentuniquecellpointershort{}$ is either $\uniquecellpointercontentshort$ or $\cellpointershort_{i + 1}$.

        For the furthermore part, if the value of  $\currentuniquecellpointershort{}$ is $\cellpointershort_{i + 1}$, then $I^*$'s response is $(\found, \cellpointershort_{i + 1})$, so by the ``moreover" part above, $i < n$ as required.
        \qH{\Cref{lemma:ero:2_of_q_holds_eigth_claim}}
    \end{proof}

    \begin{claimcustom}{\ref{lemma:ero:2_of_q_holds}.7}\label{lemma:ero:2_of_q_holds_ninth_claim}
        At the time $p$ executes \cref{line:ero:remove_cell_while_loop} for the $i + 1$th time during $I$, which is well-defined by \Cref{lemma:ero:2_of_q_holds_eigth_claim}, the local variable $\previousuniquecellpointershort{} = \cellpointershort_{i - 1}$.
    \end{claimcustom}

    \begin{proof}
        By Claims \ref{lemma:ero:2_of_q_holds_sixth_claim} and \ref{lemma:ero:2_of_q_holds_eigth_claim}, respectively, $p$ executes \cref{line:ero:remove_cell_while_loop} $i$ \& $i + 1$ times during $I$.
        Furthermore, by \Cref{lemma:ero:2_of_q_holds_sixth_claim}, at the time $p$ executes \cref{line:ero:remove_cell_while_loop} for the $i$th time during $I$, $\currentuniquecellpointershort{} = \cellpointershort_{i - 1}$.
        Let $I_{i}$ be the $i$th iteration of the loop on \cref{line:ero:remove_cell_while_loop} during $I$.
        Since $p$ executes \cref{line:ero:remove_cell_while_loop} $i$ \& $i + 1$ times during $I$ and the first parameter of the response of every invocation of the AcquireNext procedure is either $\found$, $\notfound$, or $\timechange{}$, the local variable $\status = \found$ during $I_{i}$.
        Hence, $p$ found the condition on \cref{line:ero:remove_cell_acquire_next_found} to be true during $I_{i}$, and so $p$ executed \cref{line:ero:remove_cell_update_pointers} during $I_{i}$; say at time $T^{\ref{line:ero:remove_cell_update_pointers}}_{i}$.
        Thus, since at the time $p$ executes \cref{line:ero:remove_cell_while_loop} for the $i$th time during $I$  $\currentuniquecellpointershort{}=\cellpointershort_{i - 1}$, we have that $p$ set $\previousuniquecellpointershort{}=\cellpointershort_{i - 1}$ at time $T^{\ref{line:ero:remove_cell_update_pointers}}_{i}$.
        Therefore, since the value of $\previousuniquecellpointershort{}$ does not change between $T^{\ref{line:ero:remove_cell_update_pointers}}_{i}$ and the time of $p$'s $i + 1$th execution of \cref{line:ero:remove_cell_while_loop} during $I$, at the time $p$ executes \cref{line:ero:remove_cell_while_loop} for the $i + 1$th time, $\previousuniquecellpointershort{} = \cellpointershort_{i - 1}$ as wanted.
        \qH{\Cref{lemma:ero:2_of_q_holds_ninth_claim}}
    \end{proof}

    \begin{claimcustom}{\ref{lemma:ero:2_of_q_holds}.8}\label{lemma:ero:2_of_q_holds_10_claim}
        $(*\cellpointershort_{i}).\nextlong.\uniquecellpointercontentlong{} = \cellpointershort_{i + 1}$ throughout $(e, T^{\ref{line:ero:remove_cell_before_removal_linearization_check}}]$.
    \end{claimcustom}

    \begin{proof}
        We first prove that $(*\cellpointershort_{i}).\nextlong.\uniquecellpointercontentlong{} = \cellpointershort_{i + 1}$ at $e$.
        Recall that $\mathcal{I}^{include}_{e}$ is the prefix of $\mathcal{I}$ up to and including $e$.
        Hence, $e$ is the last step in $\mathcal{I}^{include}_{e}$, and since $e$ is an $L$-event, we have $e$ is the last $L$-event in $\mathcal{I}^{include}_{e}$.
        Thus, from $e$ onwards in $\mathcal{I}^{include}_{e}$, there are no successful list-add or list-remove attempts.
        So, since $\mathcal{I}^{include}_{e}$ is finite, by assumption $P(\mathcal{I})$, $Q(\mathcal{I})$, and $R(\mathcal{I})$ hold, $\mathcal{I}^{include}_{e}$ has a last $L$-event (namely $e$), the last $L$-event in $\mathcal{I}^{include}_{e}$ is a $L$-remove event, and from $e$ onwards in $\mathcal{I}^{include}_{e}$, there are no successful list-add or list-remove attempts, by \Cref{lemma:ero:conditional_classification_lemma}, the list of cells conforms to $\List(\mathcal{I}')$ in $\mathcal{I}^{include}_{e}$ where $\mathcal{I}'$ is the prefix of $\mathcal{I}^{include}_{e}$ up to but excluding $e$.
        Since $\mathcal{I}^{include}_{e}$ is a prefix of $\mathcal{I}$, $e$ is in $\mathcal{I}^{include}_{e}$, and by definition $\mathcal{I}^{exclude}_{e}$ is the prefix of $\mathcal{I}$ up to be excluding $e$, it follows that $\mathcal{I}' = \mathcal{I}^{exclude}_{e}$.
        Hence, the list of cells conforms to $\List(\mathcal{I}^{exclude}_{e})$ in $\mathcal{I}^{include}_{e}$.
        Therefore, since $\List(\mathcal{I}^{exclude}_{e}) = \cellpointershort_0, \ldots, \cellpointershort_{n + 1}$, and $i \in [1..n]$, by \Cref{def:ero:logical_list}, at $e$ (equivalently, the end of $\mathcal{I}^{include}_{e}$) $(*\cellpointershort_{i}).\nextlong.\uniquecellpointercontentlong{} = \cellpointershort_{i + 1}$ as wanted.

        So, it suffices to prove that $(*\cellpointershort_{i}).\nextlong.\uniquecellpointercontentlong{}$ is unchanged throughout $(e, T^{\ref{line:ero:remove_cell_before_removal_linearization_check}}]$.
        Suppose, for contradiction, $(*\cellpointershort_{i}).\nextlong.\uniquecellpointercontentlong{}$ changes during $(e, T^{\ref{line:ero:remove_cell_before_removal_linearization_check}}]$.
        Hence, by \Cref{observation:ero:where_objects_change}, there is either a successful list-add attempt after $\cellpointershort_i$ or there is a successful list-remove attempt between $\cellpointershort_i$ and some pointer during $(e, T^{\ref{line:ero:remove_cell_before_removal_linearization_check}}]$.
        Since $e$ is an $L$-remove event, and by \Cref{lemma:ero:2_of_q_holds_first_claim} there are no $L$-events during $(e, T^{\ref{line:ero:remove_cell_before_removal_linearization_check}}]$, it follows that $e$ is the last $L$-event in the prefix $\mathcal{I}^{\ref{line:ero:remove_cell_before_removal_linearization_check}}$ of $\mathcal{I}$ up to and including $T^{\ref{line:ero:remove_cell_before_removal_linearization_check}}$.
        Hence, since $e$ is an $L$-remove event for $\cellpointershort$, and $P(\mathcal{I})$, $Q(\mathcal{I})$, and $R(\mathcal{I})$ hold, by \Cref{lemma:ero:3_of_r_safety_holds}, there is at most one successful list-remove attempt for $\cellpointershort$ and no other successful list-add or list-remove attempts for any pointer from $e$ onwards in $\mathcal{I}^{\ref{line:ero:remove_cell_before_removal_linearization_check}}$, or equivalently, during $(e, T^{\ref{line:ero:remove_cell_before_removal_linearization_check}}]$.
        Thus, since $a$ is during $(e, T^{\ref{line:ero:remove_cell_before_removal_linearization_check}}]$, we have that $a$ is a successful list-remove attempt for $\cellpointershort$ between $\cellpointershort_i$ and some pointer.
        So, since $\cellpointershort = \cellpointershort_i$, we have that $a$ is a list-remove attempt for $\cellpointershort_i$ between $\cellpointershort_i$ and some pointer.
        Since $a$ is before $T^{\ref{line:ero:remove_cell_before_removal_linearization_check}}$, and $T^{\ref{line:ero:remove_cell_before_removal_linearization_check}}$ is in $\mathcal{I}$, we have that $a$ is in $\mathcal{I}$.
        Hence, there is a list-remove attempt (namely $a$) for $\cellpointershort_i$ between $\cellpointershort_i$ and some pointer in $\mathcal{I}$.
        Therefore, since by assumption $P(\mathcal{I})$ and $Q(\mathcal{I})$ hold, by \Cref{lemma:ero:list_remove_attempt_has_different_next}, $\cellpointershort_i \neq \cellpointershort_i$.
        However, $\cellpointershort_i = \cellpointershort_i$, a contradiction.
        \qH{\Cref{lemma:ero:2_of_q_holds_10_claim}}
    \end{proof}

    \begin{claimcustom}{\ref{lemma:ero:2_of_q_holds}.9}\label{lemma:ero:2_of_q_holds_12_claim}
    If the local variable $\currentuniquecellpointershort{} = \cellpointershort_{i + 1}$ at the time $p$ executes \cref{line:ero:remove_cell_while_loop} for the $i + 1$th time during $I$, then for every $j \in [i + 1..n]$, (1) $p$ executes \cref{line:ero:remove_cell_while_loop} $j$ times during $I$, (2) at the time $p$ executes \cref{line:ero:remove_cell_while_loop} for the $j$th time during $I$, the local variable $\currentuniquecellpointershort{}=\cellpointershort_{j}$.
    \end{claimcustom}

    \begin{proof}
        By induction on $j$.
        \begin{itemize}
            \item[] \hspace{0pt}\textbf{Base Case.} $j = i + 1$.

            In this case, (1) holds by \Cref{lemma:ero:2_of_q_holds_eigth_claim} and (2) holds by assumption.

            \item[] \hspace{0pt}\textbf{Inductive Case.} For every $j \in [i + 1..n)$, if (1) and (2) hold for $j$, then (1) and (2) hold for $j + 1$.

            Suppose for any $j \in [i+1..n)$ (1) $p$ executes \cref{line:ero:remove_cell_while_loop} $j$ times during $I$ and (2) at the time $p$ executes \cref{line:ero:remove_cell_while_loop} for the $j$th time during $I$, $\currentuniquecellpointershort{} = \cellpointershort_{j}$.
            This is the inductive hypothesis.
            Let $I_j$ be the $j$th iteration of the loop on \cref{line:ero:remove_cell_while_loop} during $I$, which is well-defined by (1) of the inductive hypothesis.
            Furthermore, let $T^{\ref{line:ero:remove_cell_while_loop}}_j$ be the time of $p$'s $j$th execution of \cref{line:ero:remove_cell_while_loop} during $I$ which is the start of $I_j$.
            Since by (2) of the inductive hypothesis $\currentuniquecellpointershort{} = \cellpointershort_{j}$ at $T^{\ref{line:ero:remove_cell_while_loop}}_j$ where $j \in [i + 1..n)$, by \Cref{lemma:ero:2_of_q_holds_fifth_claim}, $p$ executes \cref{line:ero:remove_cell_update_pointers} at some time $T^{\ref{line:ero:remove_cell_update_pointers}}_j$ during $I_j$ and $\currentuniquecellpointershort{} = \cellpointershort_{j + 1}$ at $T^{\ref{line:ero:remove_cell_update_pointers}}_j$.
            Hence, since $p$ executes $s$ during $I$, it follows that $p$ executes \cref{line:ero:remove_cell_while_loop} one more time during $I$, and so $p$ executes \cref{line:ero:remove_cell_while_loop} $j + 1$ times during $I$.
            Since $\currentuniquecellpointershort{} = \cellpointershort_{j + 1}$ at $T^{\ref{line:ero:remove_cell_update_pointers}}_j$, and the value of $\currentuniquecellpointershort{}$ does not change between $T^{\ref{line:ero:remove_cell_update_pointers}}_j$ and the time of $p$'s $j + 1$th execution of \cref{line:ero:remove_cell_while_loop} during $I$, it follows that at the time $p$ executes \cref{line:ero:remove_cell_while_loop} for the $j + 1$th time during $I$ $\currentuniquecellpointershort{} = \cellpointershort_{j + 1}$.
            Therefore, (1) and (2) hold for $j + 1$ as wanted.
            \qH{\Cref{lemma:ero:2_of_q_holds_12_claim}}
        \end{itemize}
    \end{proof}

    We now return to the proof of \Cref{lemma:ero:2_of_q_holds}.
    Let $T^{\ref{line:ero:remove_cell_while_loop}}_{i + 1}$ be the time $p$ executes \cref{line:ero:remove_cell_while_loop} for the $i + 1$th time during $I$.
    By \Cref{lemma:ero:2_of_q_holds_eigth_claim}, there are two cases.

    \begin{itemize}
        \item[] \hspace{0pt}\textbf{Case 1.} The local variable $\currentuniquecellpointershort{} = \uniquecellpointercontentshort$ at $T^{\ref{line:ero:remove_cell_while_loop}}_{i + 1}$.

        Hence, since the second parameter of $I$ is $\uniquecellpointercontentshort$, $p$ finds the condition on \cref{line:ero:remove_cell_while_loop} to be false at $T^{\ref{line:ero:remove_cell_while_loop}}_{i + 1}$.
        Thus, since by \Cref{lemma:ero:2_of_q_holds_ninth_claim} $\previousuniquecellpointershort{} = \cellpointershort_{i - 1}$ at $T^{\ref{line:ero:remove_cell_while_loop}}_{i + 1}$, and $\previousuniquecellpointershort{}$ only changes on lines \ref{line:ero:remove_cell_initialize_pointers} and \ref{line:ero:remove_cell_update_pointers}, we have that $\previousuniquecellpointershort{} = \cellpointershort_{i - 1}$ from $T^{\ref{line:ero:remove_cell_while_loop}}_{i + 1}$ onwards in $I$.
        Hence, every \CASop{} operation on \cref{line:ero:remove_cell_from_list} during $I$ is on $(*\cellpointershort_{i - 1}).\nextlong$.
        Thus, since $s$ is an execution of \cref{line:ero:remove_cell_from_list} during $I$, we have that $s$ is a \CASop{} operation on $(*\cellpointershort_{i - 1}).\nextlong$.
        Let $T^{\ref{line:ero:remove_cell_read_pointer_to_remove}}$ be the time $p$ executed \cref{line:ero:remove_cell_read_pointer_to_remove} during $I$ (this is well-defined since $p$ executed $s$ during $I$).
        Hence, $T^{\ref{line:ero:remove_cell_read_pointer_to_remove}} < T^{\ref{line:ero:remove_cell_before_removal_linearization_check}}$.
        Furthermore, since all steps during $I$ are after $e$, we have that $e < T^{\ref{line:ero:remove_cell_read_pointer_to_remove}}$, and so $T^{\ref{line:ero:remove_cell_read_pointer_to_remove}} \in (e, T^{\ref{line:ero:remove_cell_before_removal_linearization_check}}]$.
        Hence, by \Cref{lemma:ero:2_of_q_holds_10_claim}, $(*\cellpointershort_{i}).\nextlong.\uniquecellpointercontentlong{} = \cellpointershort_{i + 1}$ at $T^{\ref{line:ero:remove_cell_read_pointer_to_remove}}$.
        Thus, since the second parameter of $I$ is $\cellpointershort$, and $\cellpointershort = \cellpointershort_i$, we have that $p$ read $\cellpointershort_{i + 1}$ from $(*\cellpointershort_i).\nextlong.\uniquecellpointercontentlong{}$ at $T^{\ref{line:ero:remove_cell_read_pointer_to_remove}}$.
        So, since $s$ is a \CASop{} operation on $(*\cellpointershort_{i - 1}).\nextlong$, we have that $s$ attempts to change the next field of $\cellpointershort_{i -1}$ to $(\arbitraryvalue, \arbitraryvalue, \arbitraryvalue, \cellpointershort_{i + 1})$.
        Hence, by \Cref{def:ero:english}, $s$ is a list-remove attempt between $\cellpointershort_{i - 1}$ and $\cellpointershort_{i + 1}$.
        Since $s$ is a list-remove attempt for $\cellpointershort$ and $\cellpointershort = \cellpointershort_i$, we have that $s$ is a list-remove attempt for $\cellpointershort_i$ between $\cellpointershort_{i - 1}$ and $\cellpointershort_{i + 1}$.
        Hence, since $i$ is unique, $i$ is in $[1..n]$, and $\List(\mathcal{I}^{exclude}_{e}) = \cellpointershort_0, \ldots, \cellpointershort_{n + 1}$, we have that $\cellpointershort_i$ appears in $\List(\mathcal{I}^{exclude}_{e})$ exactly once and $\cellpointershort_{i - 1}$ and $\cellpointershort_{i + 1}$ are the pointers preceding and succeeding $\cellpointershort_i$ in $\List(\mathcal{I}^{exclude}_{e})$.
        Therefore, we have established the following: $s$ is a list-remove attempt for $\cellpointershort_i$ between $\cellpointershort_{i - 1}$ and $\cellpointershort_{i + 1}$, $s$ is preceded by a unique $L$-remove event for $\cellpointershort$ (namely $e$), $\mathcal{I}^{exclude}_{e}$ is the prefix of $\mathcal{I}$ up to but excluding this $L$-remove event, $\cellpointershort_i$ appears in $\List(\mathcal{I}^{exclude}_{e})$ exactly once, and $\cellpointershort_{i - 1}$ and $\cellpointershort_{i + 1}$ are the pointers preceding and succeeding $\cellpointershort_i$ in $\List(\mathcal{I}^{exclude}_{e})$, and so 2. of $Q(\mathcal{I} \circ s)$ holds as wanted.
        
        \item[] \hspace{0pt}\textbf{Case 2.} The local variable $\currentuniquecellpointershort{} = \cellpointershort_{i + 1}$ at $T^{\ref{line:ero:remove_cell_while_loop}}_{i + 1}$.

        Hence, by \Cref{lemma:ero:2_of_q_holds_eigth_claim} $i < n$.
        Furthermore, by \Cref{lemma:ero:2_of_q_holds_12_claim} $p$ executes \cref{line:ero:remove_cell_while_loop} $n$ times during $I$, and at the time $p$ executes \cref{line:ero:remove_cell_while_loop} for the $n$th time during $I$ $\currentuniquecellpointershort{} = \cellpointershort_n$.
        Let $I_n$ be the $n$th iteration of the loop on \cref{line:ero:remove_cell_while_loop} during $I$.
        Since $\mathcal{I}^{exclude}_{e}$ is a prefix of $\mathcal{I}$, $\List(\mathcal{I}^{exclude}_{e}) = \cellpointershort_0, \ldots, \cellpointershort_{n + 1}$, by assumption $P(\mathcal{I})$ holds, and $i \in [1..n]$, by \Cref{lemma:ero:pointers_in_list_are_unique} for every $j \in [0..n+1]$, if $i \neq j$, then $\cellpointershort_i \neq \cellpointershort_j$.
        Hence, since $\cellpointershort_i = \cellpointershort$, for every $j \in [0..n+1]$, if $i \neq j$, then $\cellpointershort \neq \cellpointershort_j$.
        Thus, since $i < n$, we have that $i \neq n$, so $\cellpointershort \neq \cellpointershort_n$.
        Hence, since $\currentuniquecellpointershort{} = \cellpointershort_n$ at the start of $I_n$ and the second parameter of $I$ is $\cellpointershort$, we have that $p$ finds the condition on \cref{line:ero:remove_cell_while_loop} to be true at the start of $I_n$.
        Thus, since $p$ executes $s$ during $I$, $p$ invokes and exits the AcquireNext procedure during $I_n$.
        Denote this execution of the AcquireNext procedure by $I^*$.

        We first prove that the response of $I^*$ is $(\notfound, \arbitraryvalue)$ by satisfying the conditions of \Cref{lemma:ero:acquire_next_response_classification_weak}.
        Since the first parameter of $I$ is $\uniquerepositoryoperationshort_\linearizationobject{}$ and $\currentuniquecellpointershort{} = \cellpointershort_n$ at the start of $I_{n}$, the parameters of $I^*$ are $(\uniquerepositoryoperationshort_\linearizationobject{},\cellpointershort_n)$.
        Let $T^{\ref{line:ero:acquire_next_read_curr_unique_pointer}}_{I^*}$ and $T^{\ref{line:ero:acquire_next_linearization_changed_check}}_{I^*}$ by defined as in \Cref{lemma:ero:2_of_q_holds:claim_acquire_next}, and so $\linearizationobject{}.\uniquerepositoryoperationlong{} = \uniquerepositoryoperationshort_\linearizationobject{}$ at $T^{\ref{line:ero:acquire_next_linearization_changed_check}}_{I^*}$, and $T^{\ref{line:ero:acquire_next_read_curr_unique_pointer}}_{I^*} \in (e, T^{\ref{line:ero:remove_cell_before_removal_linearization_check}}]$.
        Hence, there is a prefix of $\mathcal{I}$ during $(e, T^{\ref{line:ero:remove_cell_before_removal_linearization_check}}]$ up to and including $T^{\ref{line:ero:acquire_next_read_curr_unique_pointer}}_{I^*}$; say $\mathcal{I}'$.
        Thus, by \Cref{lemma:ero:2_of_q_holds_second_claim} the list of cells conforms to either $\List(\mathcal{I}^{exclude}_{e})$ or $\List(\mathcal{I}^{include}_{e})$ in $\mathcal{I}'$.
        So, since $\List(\mathcal{I}^{exclude}_{e}) = \cellpointershort_0, \ldots, \cellpointershort_{n + 1}$, $\List(\mathcal{I}^{include}_{e}) = \cellpointershort_0, \ldots, \cellpointershort_{i - 1}, \cellpointershort_{i + 1}, \ldots, \cellpointershort_{n + 1}$, and $i < n$, by \Cref{def:ero:logical_list}, at the end of $\mathcal{I}'$ $(*\cellpointershort_{n}).\nextlong.\uniquecellpointercontentlong{} = \cellpointershort_{n + 1}$.
        Hence, since $\cellpointershort_{n + 1} = \nullconstant$, we have that $(*\cellpointershort_n).\nextlong.\uniquecellpointercontentlong{} = \nullconstant$ at the end of $\mathcal{I}'$.
        Therefore, we have established the following: since (1) $I^*$ has parameters $(\uniquerepositoryoperationshort_\linearizationobject{}, \cellpointershort_n)$; (2) $\linearizationobject{}.\uniquerepositoryoperationlong{} = \uniquerepositoryoperationshort_\linearizationobject{}$ at $T^{\ref{line:ero:acquire_next_linearization_changed_check}}_{I^*}$; and (3) $(*\cellpointershort_n).\nextlong.\uniquecellpointercontentlong{} = \nullconstant$ at $T^{\ref{line:ero:acquire_next_read_curr_unique_pointer}}_{I^*}$, and so by \Cref{lemma:ero:acquire_next_response_classification_weak}, $I^*$'s response is $(\notfound, \arbitraryvalue)$ as wanted.

        We now finish the proof of Case 2.
        Since $p$ executes $s$ during $I$ and $I^*$'s response is $(\notfound, \arbitraryvalue)$, $p$ finds the condition on \cref{line:ero:remove_cell_check_acquire_status} to be true during $I_{n}$ and so $p$ executes \cref{line:ero:exit_1} during $I_n$.
        Therefore, $p$ does not execute \cref{line:ero:remove_cell_from_list} during $I$.
        However, $p$ executes $s$ during $I$, which is an execution of \cref{line:ero:remove_cell_from_list}, a contradiction, so this case is impossible.
        \qH{\Cref{lemma:ero:2_of_q_holds}}
    \end{itemize}
\end{proof}

\subsubsection*{\Underline{Inductive Case for $R$}}

\begin{proposition}\label{lemma:ero:1_of_r_holds}
    If $P(\mathcal{I})$, $Q(\mathcal{I})$, and $R(\mathcal{I})$ hold, then 1. of $R(\mathcal{I} \circ s)$ holds.
\end{proposition}

\begin{proof}
    Since by assumption $R(\mathcal{I})$ holds, it suffices to consider the case where $\mathcal{I}$ has at least one $L$-event, the last $L$-event in $\mathcal{I}$ is an $L$-add event for $\cellpointershort$, and $s$ is an $L$-event, with the goal of proving that between $e$ and $s$ there is one successful list-add attempt for $\cellpointershort$ and no other successful list-add or list-remove attempts for any pointer.
    Let $e$ be the last $L$-event in $\mathcal{I}$, so $e$ is an $L$-add event for $\cellpointershort$.
    Hence, by $P(\mathcal{I})$, $e$ is the only $L$-add event for $\cellpointershort$ in $\mathcal{I}$.
    Furthermore, $e$ and $s$ are successive $L$-events in $\mathcal{I} \circ s$.
    Thus, by \Cref{lemma:ero:successive_l_event_read_previous_l_event_value} $p$ read the value $v$ that $e$ set $\linearizationobject{}$ to on its last execution of \cref{line:ero:linearization_read} before $s$; say at time $T^{\ref{line:ero:linearization_read}}$.
    Hence, since $e$ is an $L$-add event for $\cellpointershort$, by \Cref{def:ero:english}, $v = ((\arbitraryvalue, \addcell), \cellpointershort)$.
    Thus, between $T^{\ref{line:ero:linearization_read}}$ and $s$, $p$ finds the condition on \cref{line:ero:do_add_cell_condition} to be true, and so $p$ invokes and exits the \doaddcell{} procedure with a second parameter of $\cellpointershort$ between $T^{\ref{line:ero:linearization_read}}$ and $s$.
    Denote this invocation of the \doaddcell{} procedure by $I$.
    % Hence, by \Cref{lemma:ero:l_event_corresponding_to_do_low_level_op}, there is a $L$-add event for $\cellpointershort$ before $I$ was invoked.
    % Thus, since $e$ is the only $L$-add event for $\cellpointershort$ in $\mathcal{I}$, we have that $e$ occured before $I$ was invoked.
    Since $I$ exits before $s$, and $s$ is the step after $\mathcal{I}$ in $\mathcal{I} \circ s$, we have that $I$ exits at some time $T_e$ during $\mathcal{I}$.
    Hence, since $P(\mathcal{I})$, $Q(\mathcal{I})$, and $R(\mathcal{I})$ hold, by \Cref{lemma:ero:exit_do_add_implies_done_strong}, there is a successful list-add attempt $a$ for $\cellpointershort$ before $T_e$ in $\mathcal{I}$.
    Thus, since $T_e$ is in $\mathcal{I}$, it is before $s$, and so $a < s$.
    Furthermore, since $a$ is a successful list-add attempt for $\cellpointershort$, by \Cref{def:ero:english}, it was executed during a \doaddcell{} procedure with a second parameter of $\cellpointershort$, so by \Cref{lemma:ero:l_event_corresponding_to_do_low_level_op}, there is a $L$-add event for $\cellpointershort$ before $a$, which must be $e$ since it is the only $L$-add event for $\cellpointershort$ in $\mathcal{I}$.
    Therefore, since $e < a$ and $a < s$, we have that there is a successful list-add attempt for $\cellpointershort$ between $e$ and $s$.
    What remains is to show that there are no other successful list-add or list-remove attempts between $e$ and $s$.
    Since (1) $\mathcal{I}$ has at least one $L$-event, (2) $\mathcal{I}$ has a last $L$-event (namely $e$), (3) $P(\mathcal{I})$, $Q(\mathcal{I})$, and $R(\mathcal{I})$ hold and (4) $e$ is an $L$-add event for $\cellpointershort$, by \Cref{lemma:ero:1_of_r_safety_holds} from $e$ onwards in $\mathcal{I}$, there is at most one successful list-add attempt for $\cellpointershort$ and no other successful list-add or list-remove attempts for any pointer.
    \qH{\Cref{lemma:ero:1_of_r_holds}}
\end{proof}

\begin{proposition}\label{lemma:ero:2_of_r_holds}
    If $P(\mathcal{I})$, $Q(\mathcal{I})$, and $R(\mathcal{I})$ hold, then 2. of $R(\mathcal{I} \circ s)$ holds.
\end{proposition}

\begin{proof}
    Since $R(\mathcal{I})$ holds, it suffices to consider the case where $\mathcal{I}$ has at least one $L$-event, the last $L$-event in $\mathcal{I}$ is an $L$-apply event, and $s$ is an $L$-event.
    Let $e$ be the last $L$-event in $\mathcal{I}$ and consider any list-add or list-remove attempt $a$ during $(e, s)$ in $\mathcal{I}$.
    Since (1) $\mathcal{I}$ has at least one $L$-event, (2) $\mathcal{I}$ has a last $L$-event (namely $e$), (3) $P(\mathcal{I})$, $Q(\mathcal{I})$, and $R(\mathcal{I})$ hold, and (4) $e$ is an $L$-apply event, by \Cref{lemma:ero:2_of_r_safety_holds}, from $e$ onwards in $\mathcal{I}$ there are no successful list-add or list-remove attempts.
    Therefore, $a$ is unsuccessful as wanted.
    \qH{\Cref{lemma:ero:2_of_r_holds}}
\end{proof}

\begin{proposition}\label{lemma:ero:3_of_r_holds}
    If $P(\mathcal{I})$, $Q(\mathcal{I})$, and $R(\mathcal{I})$ hold, then 3. of $R(\mathcal{I} \circ s)$ holds.
\end{proposition}

\begin{proof}
    Since by assumption $R(\mathcal{I})$ holds, it suffices to consider the case where $\mathcal{I}$ has at least one $L$-event, the last $L$-event in $\mathcal{I}$ is an $L$-remove event for $\cellpointershort$, and $s$ is an $L$-event, with the goal of proving that between $e$ and $s$ there is one successful list-remove attempt for $\cellpointershort$ and no other successful list-add or list-remove attempts for any pointer.
    Let $e$ be the last $L$-event in $\mathcal{I}$, and so $e$ is an $L$-remove event for $\cellpointershort$.
    Hence, by $P(\mathcal{I})$, $e$ is the only $L$-remove event for $\cellpointershort$ in $\mathcal{I}$.
    Furthermore, $e$ and $s$ are successive $L$-events in $\mathcal{I} \circ s$.
    Thus, by \Cref{lemma:ero:successive_l_event_read_previous_l_event_value} $p$ read the value $v$ that $e$ set $\linearizationobject{}$ to on its last execution of \cref{line:ero:linearization_read} before $s$; say at time $T^{\ref{line:ero:linearization_read}}$.
    Hence, since $e$ is an $L$-remove event for $\cellpointershort$, by \Cref{def:ero:english}, $v$ is of the form $((\arbitraryvalue, \removecell), \cellpointershort)$.
    Thus, between $T^{\ref{line:ero:linearization_read}}$ and $s$, $p$ finds the condition on \cref{line:ero:do_remove_cell_condition} to be true, and so $p$ invokes and exits the \doremovecell{} procedure with a second parameter of $\cellpointershort$ between $T^{\ref{line:ero:linearization_read}}$ and $s$.
    Denote this invocation of the \doremovecell{} procedure by $I$.
    Since $I$ exits before $s$, and $s$ is the step after $\mathcal{I}$ in $\mathcal{I} \circ s$, we have that $I$ exits at some time $T_e$ during $\mathcal{I}$.
    Hence, since by assumption $P(\mathcal{I})$, $Q(\mathcal{I})$, and $R(\mathcal{I})$ hold, by \Cref{lemma:ero:exit_do_remove_implies_done}, there is a successful list-remove attempt $a$ for $\cellpointershort$ before $T_e$ in $\mathcal{I}$.
    Thus, since $T_e$ is in $\mathcal{I}$ it is before $s$, and so $a < s$.
    Furthermore, since $a$ is a successful list-remove attempt for $\cellpointershort$, by \Cref{def:ero:english}, it was executed during a \doremovecell{} procedure with a second parameter of $\cellpointershort$, so by \Cref{lemma:ero:l_event_corresponding_to_do_low_level_op}, there is a $L$-remove event for $\cellpointershort$ before $a$, which must be $e$ since it is the only $L$-remove event for $\cellpointershort$ in $\mathcal{I}$.
    Therefore, since $e < a$ and $a < s$, there is a successful list-remove attempt for $\cellpointershort$ between $e$ and $s$.
    What remains is to show that there are no other successful list-add or list-remove attempts between $e$ and $s$.
    Since (1) $\mathcal{I}$ has at least one $L$-event, (2) $\mathcal{I}$ has a last $L$-event (namely $e$), (3) by assumption $P(\mathcal{I})$, $Q(\mathcal{I})$, and $R(\mathcal{I})$ hold and (4) $e$ is an $L$-remove event for $\cellpointershort$, by \Cref{lemma:ero:3_of_r_safety_holds} from $e$ onwards in $\mathcal{I}$, there is at most one successful list-remove attempt for $\cellpointershort$ and no other successful list-add or list-remove attempts for any pointer.
    \qH{\Cref{lemma:ero:3_of_r_holds}}
\end{proof}

\subsubsection*{\Underline{Inductive Case for $O$}}

\begin{proposition}\label{lemma:ero:o_holds}
   If $P(\mathcal{I})$ and $O(\mathcal{I})$ hold, then $O(\mathcal{I} \circ s)$ holds.
\end{proposition}

\begin{proof}
    Since by assumption $O(\mathcal{I})$ holds, it suffices to consider the case where $\mathcal{I}$ has at least one $L$-event, and $s$ is an $L$-event.
    Let $e$ be the last $L$-event in $\mathcal{I}$.
    By \Cref{lemma:ero:every_l_event_is_add_apply_or_remove} $e$ is either an $L$-add, $L$-remove, or $L$-apply event.
    Suppose $e$ is an $L$-add or $L$-remove event.
    Hence, (1) $\mathcal{I}$ has a last $L$-event (namely $e$), (2) by assumption $P(\mathcal{I})$ and $O(\mathcal{I})$ hold, and (3) $e$ is an $L$-add or $L$-remove event, and so by \Cref{lemma:ero:2_o_safety_holds}, from $e$ onwards in $\mathcal{I}$ there are no successful $S$-attempts as wanted.
    Now suppose $e$ is an $L$-apply event for some timestamp $\timeshort{}$.
    Hence, since $P(\mathcal{I})$ holds, by \Cref{lemma:ero:every_l_event_has_a_unique_timestamp}, $e$ is the only $L$-apply event for timestamp $\timeshort{}$ in $\mathcal{I}$.
    Let $p$ be the process that executed $s$.
    Since $e$ and $s$ are successive $L$-events in $\mathcal{I} \circ s$, by \Cref{lemma:ero:successive_l_event_read_previous_l_event_value}, $p$ read the value that $e$ set $\linearizationobject{}$ to on its last execution of \cref{line:ero:linearization_read} before $s$; say at time $T^{\ref{line:ero:linearization_read}}$.
    Hence, since $e$ is an $L$-apply event for timestamp $t$, by \Cref{def:ero:english}, $p$ read a value of the form $(t, \langle \doopandcopyresponse{}, \arbitraryvalue \rangle)$ from $\linearizationobject{}.\uniquerepositoryoperationlong{}$ at $T^{\ref{line:ero:linearization_read}}$.
    Thus, $p$ finds the condition on \cref{line:ero:do_apply_and_copy_response_condition} to be true between $T^{\ref{line:ero:linearization_read}}$ and $s$, and therefore $p$ invokes and exits the \doapplyandcopyresponse{} procedure on \cref{line:ero:do_apply_and_copy_response} with a first parameter of $(t, \arbitraryvalue)$ between $T^{\ref{line:ero:linearization_read}}$ and $s$.
    Denote this invocation by $I$.
    Since $I$ exits before $s$, and $s$ is the step after $\mathcal{I}$ in $\mathcal{I} \circ s$, we have that $I$ exits at some time $T_e$ during $\mathcal{I}$.
    Hence, since by assumption $P(\mathcal{I})$ and $O(\mathcal{I})$ hold, by \Cref{lemma:ero:conditional_exit_apply_implies_successful_s_attempt}, there is a successful $S$-attempt $a$ for timestamp $t$ before $T_e$ in $\mathcal{I}$.
    Thus, since $T_e$ is in $\mathcal{I}$ it is before $s$, and so $a < s$.
    Furthermore, since $a$ is a successful $S$-attempt for timestamp $t$, by \Cref{corollary:ero:s_attempts_and_corresponding_l_events_are_for_matching_timestamps}, there is a $L$-apply event for timestamp $\timeshort{}$ before $a$, which must be $e$ since it is the only $L$-apply event for timestamp $t$ in $\mathcal{I}$.
    Therefore, since $e < a$ and $a < s$, there is a successful $S$-attempt for timestamp $\timeshort{}$ between $e$ and $s$.
    What remains is to show that there are no other successful $S$-attempts between $e$ and $s$.
    Since (1) $\mathcal{I}$ has a last $L$-event (namely $e$), (2) by assumption $P(\mathcal{I})$ and $O(\mathcal{I})$ hold, and (3) $e$ is an $L$-apply event for timestamp $t$, by \Cref{lemma:ero:1_o_safety_holds}, from $e$ onwards in $\mathcal{I}$, there is at most one successful $S$-attempt for $\timeshort{}$ and no other successful $S$-attempts for any timestamp.
    \qH{\Cref{lemma:ero:o_holds}}
\end{proof}

\subsubsection*{\Underline{The Finale}}

\begin{lemma}\label{lemma:ero:the_list_invariants_hold}
    $P(\mathcal{I}^\mathcal{B})$, $Q(\mathcal{I}^\mathcal{B})$, $R(\mathcal{I}^\mathcal{B})$, and $O(\mathcal{I}^\mathcal{B})$ hold.    
\end{lemma}

\begin{proof}
    % If $\mathcal{I}^\mathcal{B}$ is empty, $P(\mathcal{I}^\mathcal{B})$, $Q(\mathcal{I}^\mathcal{B})$, $R(\mathcal{I}^\mathcal{B})$, and $O(\mathcal{I}^\mathcal{B})$ hold trivially, so it suffices to consider the case where $\mathcal{I}^\mathcal{B}$ is non-empty.
    Let $\mathcal{P}(n)$ be the predicate: for every implementation history $\mathcal{I}_n$ of $\mathcal{B}$ comprised of $n$ steps, $P(\mathcal{I}_n)$, $Q(\mathcal{I}_n)$, $R(\mathcal{I}_n)$, and $O(\mathcal{I}_n)$ hold. 
    We prove $\mathcal{P}(n)$ by induction on $n$.

    \begin{itemize}
        \item[] \hspace{0pt}\textbf{Base Case.} $\mathcal{P}(0)$.

        Since $\mathcal{I}_0$ contains zero steps, and $P(\mathcal{I}_0)$, $Q(\mathcal{I}_0)$, $R(\mathcal{I}_0)$, and $O(\mathcal{I}_0)$, assert properties about certain steps in $\mathcal{I}_0$, they are vacuously true.

        \item[] \hspace{0pt}\textbf{Inductive Case.} $\forall n\ \mathcal{P}(n) \implies \mathcal{P}(n + 1)$.

        Suppose for some $n \geq 0$ $\mathcal{P}(n)$ holds and consider any implementation history $\mathcal{I}_{n + 1}$ of $\mathcal{B}$ comprised of $n + 1$ steps.
        Let $\mathcal{I}_n$ be the prefix of $\mathcal{I}_{n + 1}$ up to but excluding its last step, so $\mathcal{I}_n$ is an implementation history of $\mathcal{B}$ comprised of $n$ steps.
        Hence, since $\mathcal{P}(n)$ holds, we have that $P(\mathcal{I}_n)$, $Q(\mathcal{I}_n)$, $R(\mathcal{I}_n)$, and $O(\mathcal{I}_n)$ hold.
        Thus, by \Cref{lemma:ero:1_of_p_holds} $P(\mathcal{I}_{n + 1})$ holds.
        Furthermore, by Propositions \ref{lemma:ero:1_of_q_holds} and \ref{lemma:ero:2_of_q_holds} $Q(\mathcal{I}_{n + 1})$ holds.
        Moreover, by Propositions \ref{lemma:ero:1_of_r_holds}, \ref{lemma:ero:2_of_r_holds}, and \ref{lemma:ero:3_of_r_holds} $R(\mathcal{I}_{n + 1})$ holds.
        Finally, by \Cref{lemma:ero:o_holds}, $O(\mathcal{I}_{n + 1})$ holds.
    \end{itemize}
    Therefore, if $\mathcal{I}^\mathcal{B}$ is finite, then $P(\mathcal{I}^\mathcal{B})$, $Q(\mathcal{I}^\mathcal{B})$, $R(\mathcal{I}^\mathcal{B})$, and $O(\mathcal{I}^\mathcal{B})$ hold.

    What remains is the case where $\mathcal{I}^\mathcal{B}$ is infinite.
    Observe that, if $P(\mathcal{I}^\mathcal{B})$, $Q(\mathcal{I}^\mathcal{B})$, $R(\mathcal{I}^\mathcal{B})$, or $O(\mathcal{I}^\mathcal{B})$ did not hold, then there is a finite prefix $\mathcal{I}$ of $\mathcal{I}^\mathcal{B}$ where $P(\mathcal{I})$, $Q(\mathcal{I})$, $R(\mathcal{I})$ or $O(\mathcal{I})$ does not hold, a contradiction to what we just proved.
    Therefore, the lemma follows.
    \qH{\Cref{lemma:ero:the_list_invariants_hold}}
\end{proof}

\subsection{$\mathcal{B}$ is Linearizable}
\label{sec:b_is_linearizable}

In this section, we prove that $\mathcal{B}$ is linearizable.
Consider any implementation history $\mathcal{I}^\mathcal{B}$ of $\mathcal{B}$, let $\mathcal{H}$ be the object history obtained by removing all implementation steps from $\mathcal{I}^\mathcal{B}$, and let
\begin{align*}
    V = ((t_1, \langle \doopandcopyresponse{}, o_1 \rangle), s_1, r_1), ((t_2, \langle \doopandcopyresponse{}, o_2 \rangle), s_2, r_2), \ldots
\end{align*}
be the sequence of values written into the state object $\stateobject$ on \cref{line:ero:state_cas} during $\mathcal{I}^\mathcal{B}$.
To define our completion of $\mathcal{H}'$, we map entries of $V$ to operation executions in $\mathcal{I}^\mathcal{B}$ as follows.

\begin{lemma}\label{lemma:ero:mapping_between_v_and_opx_is_well_defined}
    For every index $i$ of $V$, there is a unique operation execution $opx_i$ in $\mathcal{I}^\mathcal{B}$ that received $t_i$ as a response on \cref{line:ero:operation_timestamp} during an invocation of the \doworkuntildone{} procedure invoked on \cref{line:ero:low_level_apply_and_copy_response}.
    Furthermore, $opx_i$ was invoked before $((t_i, \langle \doopandcopyresponse{}, o_i \rangle), s_i, r_i)$ was \Underline{first} written into $\stateobject$ on \cref{line:ero:state_cas} during $\mathcal{I}^\mathcal{B}$.
\end{lemma}

\begin{proof}
    Since $((t_i, \langle \doopandcopyresponse{}, o_i \rangle), s_i, r_i)$ is in $V$, some process $p$ set the value of $\stateobject$ to it on \cref{line:ero:state_cas} during $\mathcal{I}^\mathcal{B}$ for the first time; say $T^{\ref{line:ero:state_cas}}$.
    Hence, $p$ did so during an invocation of the \doapplyandcopyresponse{} procedure with a first parameter of $(t_i, \langle \doopandcopyresponse{}, o_i \rangle)$.
    Thus, by \Cref{lemma:ero:l_event_corresponding_to_do_low_level_op}, some $L$-event set $\linearizationobject{}.\uniquerepositoryoperationlong{} = (t_i, \langle \doopandcopyresponse{}, o_i \rangle)$ before $T^{\ref{line:ero:state_cas}}$.
    So, by \Cref{lemma:ero:l_events_have_corresponding_a_events}, some $A$-event $e'$ set $\announceobject{}.\uniquerepositoryoperationlong{} = (t_i, \langle \doopandcopyresponse{}, o_i \rangle)$ before $T^{\ref{line:ero:state_cas}}$.
    Let $q$ be the process that executed $e'$.
    Since $e'$ set $\announceobject{}.\uniquerepositoryoperationlong{} = (t_i, \langle \doopandcopyresponse{}, o_i \rangle)$, $q$ received $t_i$ as a response on \cref{line:ero:operation_timestamp}.
    Therefore, there is an operation execution that received $t_i$ as a response on \cref{line:ero:operation_timestamp} and so by definition, $opx_i$ exists in $\mathcal{I}^\mathcal{B}$, and this operation execution is unique since responses on \cref{line:ero:operation_timestamp} are unique (see \Cref{observation:ero:operation_timestamp_is_unique}).
    Furthermore, since $q$ received $t_i$ on \cref{line:ero:operation_timestamp} before $e'$, and $e' < T^{\ref{line:ero:state_cas}}$, by transitivity, $opx_i$ received $t_i$ as a response on \cref{line:ero:operation_timestamp} before $T^{\ref{line:ero:state_cas}}$.
    Therefore, since $q$ invoked $opx_i$ before it received $t_i$ as a response on \cref{line:ero:operation_timestamp}, and $p$ set $\stateobject$ to $((t_i, \langle \doopandcopyresponse{}, o_i \rangle), s_i, r_i)$ at $T^{\ref{line:ero:state_cas}}$, we have that $opx_i$ was invoked before $((t_i, \langle \doopandcopyresponse{}, o_i \rangle), s_i, r_i)$ was written into $\stateobject$ on \cref{line:ero:state_cas} as wanted.
    \qH{\Cref{lemma:ero:mapping_between_v_and_opx_is_well_defined}}
\end{proof}

Let
\begin{align*}
    Opx = opx_1, opx_2, \ldots
\end{align*}
be the corresponding sequence of operation executions to values in $V$.
Note that, as of now, there may be duplicate values in $V$ (and hence $Opx$).
Our first order of business in this section will be to prove that this is not the case.

We define the completion $\mathcal{H}'$ of $\mathcal{H}$ as follows.
Consider any incomplete operation execution $opx$ in $\mathcal{H}$.
If $opx$ appears in $Opx$ and the first index in which it appears is $i$, then the response step for $opx$ is appended at the end of $\mathcal{H}'$ with response $r_i$.
Otherwise, $opx$'s invocation step is removed from $\mathcal{H}'$.
We define a sequential object history $\mathcal{S}$ using $Opx$ as follows:
\begin{align*}
    invocation(opx_1, o_1), response(opx_1, r_1), invocation(opx_2, o_2), response(opx_2, r_2), \ldots
\end{align*}
where $invocation(opx_i, o_i)$ is the invocation step for $opx_i$ and $response(opx_i, r_i)$ is the response step for $opx_i$ which returned the response $r_i$.
The remainder of this section proves that $<_{\mathcal{H}'} \subseteq <_\mathcal{S}$, $\mathcal{S}$ is legal with respect to type $\mathcal{T}$, and $\mathcal{H}'$ is equivalent to $\mathcal{S}$.
The plan for doing so is as follows.
\begin{compactitem}
    \item First, we prove that the values in $V$ are pairwise distinct, implying that so are the operation executions in $Opx$.
    \item We then define the linearization point $\ell(opx)$ for $opx$ in $Opx$ to be the time of the $i$th successful \CASop{} operation on \cref{line:ero:state_cas} during $\mathcal{I}^\mathcal{B}$ where $i$ is the unique index $opx$ appears at in $Opx$.
    %Since $opx_i$ appears exactly once in $Opx$, $\ell(opx_i)$ is well-defined, and the operations in $Opx$ appear in increasing order of their linearization points.
    \item We then prove (a) every complete operation execution in $\mathcal{I}^\mathcal{B}$ is in $Opx$ and (b) the linearization point $\ell(opx)$ of every $opx$ in $Opx$ (whether complete in $\mathcal{I}^\mathcal{B}$ or not) is between $opx$'s invocation and response step in $\mathcal{I}^\mathcal{B}$ (if it exists).
    These two facts imply that $<_{\mathcal{H}'} \subseteq <_\mathcal{S}$.
    \item We then prove that for every index $i$ of $V$, $(s_i, r_i) = apply_\mathcal{T}(o_i, s_{i - 1})$.
    This implies that $\mathcal{S}$ is legal with respect to $\mathcal{T}$.
    \item Finally, we prove that if $opx$ is a complete operation execution in $\mathcal{I}^\mathcal{B}$ then its invocation step is for $o_i$ and its response is $r_i$ in $\mathcal{I}^\mathcal{B}$.
    This implies that $\mathcal{H}'$ is equivalent to $\mathcal{S}$.
\end{compactitem}

\subsubsection{Linearization points}

\begin{lemma}\label{lemma:ero:every_s_attempt_has_a_unique_timestamp}
    Every successful $S$-attempt in $\mathcal{I}^\mathcal{B}$ is for a unique timestamp.
\end{lemma}

\begin{proof}
    Suppose, for contradiction, there are two successful $S$-attempts in $\mathcal{I}^\mathcal{B}$ for the same timestamp $\timeshort{}$.
    Let $a_1$ and $a_2$ be these two attempts and let $e_1$ and $e_2$ be their corresponding $L$-events, respectively.
    Since $a_1$ (resp. $a_2$) is for timestamp $\timeshort{}$, by \Cref{corollary:ero:s_attempts_and_corresponding_l_events_are_for_matching_timestamps}, $e_1$ (resp. $e_2$) is for timestamp $\timeshort{}$.
    Hence, since by \Cref{lemma:ero:the_list_invariants_hold} $P(\mathcal{I}^\mathcal{B})$ holds, by \Cref{lemma:ero:every_l_event_has_a_unique_timestamp}, $e_1 = e_2 = e$.
    Thus, by \Cref{lemma:ero:l_event_corresponding_to_do_low_level_op}, $e$ is an $L$-apply event.
    There are two cases.

    \begin{itemize}
        \item[] \hspace{0pt}\textbf{Case 1.} $e$ is the last $L$-event in $\mathcal{I}^\mathcal{B}$.

        Hence, since $a_1$ and $a_2$ are after $e$, from $e$ onwards in $\mathcal{I}^\mathcal{B}$, there are two successful $S$-attempts for $\timeshort{}$.
        However, since $\mathcal{I}^\mathcal{B}$ has a last $L$-event (namely $e$), by \Cref{lemma:ero:the_list_invariants_hold} $P(\mathcal{I}^\mathcal{B})$ and $O(\mathcal{I}^\mathcal{B})$ holds, and $e$ is an $L$-apply event for timestamp $\timeshort{}$, by \Cref{lemma:ero:1_o_safety_holds}, from $e$ onwards in $\mathcal{I}^\mathcal{B}$ there is at most one successful $S$-attempt for $\timeshort{}$, a contradiction.

        \item[] \hspace{0pt}\textbf{Case 2.} $e$ is not the last $L$-event in $\mathcal{I}^\mathcal{B}$.

        Hence, there is next $L$-event after $e$ in $\mathcal{I}^\mathcal{B}$; say $e'$.
        Since $a_1$ and $a_2$ are successful $S$-attempts in $\mathcal{I}^\mathcal{B}$, $e$ is there corresponding $L$-event, and by \Cref{lemma:ero:the_list_invariants_hold} $P(\mathcal{I}^\mathcal{B})$ and $O(\mathcal{I}^\mathcal{B})$ hold, by \Cref{lemma:ero:any_s_attempt_outside_its_window_is_unsuccessful_alternate_statement}, there are no $L$-events during $(e, a_1)$ and $(e, a_2)$ in $\mathcal{I}^\mathcal{B}$.
        Hence, since $e'$ is the next $L$-event after $e$ in $\mathcal{I}^\mathcal{B}$, we have that $a_1$ and $a_2$ are before $e'$.
        Therefore, between $e$ and $e'$, there are two successful $S$-attempts for timestamp $\timeshort{}$.
        However, since $e$ and $e'$ are successive $L$-events in $\mathcal{I}^\mathcal{B}$, by $O(\mathcal{I}^\mathcal{B})$, there is at most one successful $S$-attempt between $e$ and $e'$, a contradiction.
        \qH{\Cref{lemma:ero:every_s_attempt_has_a_unique_timestamp}}
    \end{itemize}
\end{proof}

\begin{lemma}\label{lemma:ero:sequence_opx_is_pairwise_distinct}
    Every operation execution $opx$ in $\mathcal{I}^\mathcal{B}$ appears at most once in $Opx$.
\end{lemma}

\begin{proof}
    Suppose, for contradiction, there exists an operation execution $opx$ in $\mathcal{I}^\mathcal{B}$ that appears twice in $Opx$; say at indices $i$ and $j$, i.e., $opx_i = opx_j = opx$ for $i \neq j$.
    Hence, by \Cref{lemma:ero:mapping_between_v_and_opx_is_well_defined}, $opx$ received $t_i$ and $t_j$ as a response on \cref{line:ero:operation_timestamp} during an invocation of the \doworkuntildone{} procedure invoked on \cref{line:ero:low_level_apply_and_copy_response} during $\mathcal{I}^\mathcal{B}$.
    Thus, since \cref{line:ero:operation_timestamp} is executed at most once during an invocation of the \doworkuntildone{} procedure invoked on \cref{line:ero:low_level_apply_and_copy_response} by the process that executed $opx$ during $opx$, we have that $t_i = t_j$.
    Therefore, since $i \neq j$, and the $i$th (resp. $j$th) value written into $\stateobject$ on \cref{line:ero:state_cas} during $\mathcal{I}^\mathcal{B}$ is $((t_i, \arbitraryvalue), \arbitraryvalue, \arbitraryvalue)$ (resp. $((t_j, \arbitraryvalue), \arbitraryvalue, \arbitraryvalue)$), by \Cref{def:ero:english}, we have that there are two successful $S$-attempts for the same timestamp in $\mathcal{I}^\mathcal{B}$.
    However, this contradicts \Cref{lemma:ero:every_s_attempt_has_a_unique_timestamp}.
    \qH{\Cref{lemma:ero:sequence_opx_is_pairwise_distinct}}
\end{proof}

\begin{definition}\label{def:ero:linearization_point}
    Consider any $opx$ in $Opx$ in $\mathcal{I}^\mathcal{B}$.
    By \Cref{lemma:ero:sequence_opx_is_pairwise_distinct}, $opx$ appears exactly once in $Opx$, say at index $i$.
    We define the linearization point of $opx$, denoted by $\ell(opx)$, to be the time of the $i$th successful \CASop{} operation on \cref{line:ero:state_cas} during $\mathcal{I}^\mathcal{B}$.
\end{definition}

\subsubsection{The linearization respects the real-time order of operations}

\begin{lemma}\label{lemma:ero:every_complete_operation_with_timestamp_t_has_a_successful_s_attempt_for_t}
    (a) Every complete operation execution $opx$ in $\mathcal{I}^\mathcal{B}$ is in $Opx$.
    (b) The linearization point $\ell(opx)$ of every operation execution $opx$ in $Opx$ (whether complete in $\mathcal{I}^\mathcal{B}$ or not) is after $opx$'s invocation step in $\mathcal{I}^\mathcal{B}$ and before $opx$'s response step in $\mathcal{I}^\mathcal{B}$ if it exists.
\end{lemma}

\begin{proof}
    For part (a) let $\cellpointershort$ be the response on \cref{line:ero:allocate_cell} during $opx$.
    Let $I^{apply}$ and $I^{remove}$ be the invocations on \cref{line:ero:low_level_apply_and_copy_response} and \cref{line:ero:low_level_remove_cell}, respectively, during $opx$.
    Hence, $I^{apply}$ has parameters $(\langle \doopandcopyresponse{}, \arbitraryvalue\rangle, \cellpointershort)$ and $I^{remove}$ has parameters $(\removecell, \cellpointershort)$.
    Since $opx$ is complete, $I^{apply}$ begins and exits at times $T^{apply}_b$ and $T^{apply}_e$, respectively, and $I^{remove}$ begins and exits at times $T^{remove}_b$ and $T^{remove}_e$, respectively, such that $T^{apply}_b < T^{apply}_e < T^{remove}_b < T^{remove}_e$.
    Since by \Cref{lemma:ero:the_list_invariants_hold} $P(\mathcal{I}^\mathcal{B})$ holds, by \Cref{lemma:ero:l_apply_event_before_apply_low_level_exits}, there is an $L$-apply event $e^{apply}$ for $\cellpointershort$ between $T^{apply}_b$ and $T^{apply}_e$, and by \Cref{lemma:ero:l_remove_event_before_remove_low_level_exits}, there is an $L$-remove event $e^{remove}$ for $\cellpointershort$ between $T^{remove}_b$ and $T^{remove}_e$.
    Hence, since $T^{apply}_b < T^{apply}_e < T^{remove}_b < T^{remove}_e$, we have that $e^{apply} < e^{remove}$.
    Let $\timeshort{}$ be the response on \cref{line:ero:operation_timestamp} during $I^{apply}$.
    Hence, by \Cref{lemma:ero:l_apply_event_before_apply_low_level_exits_is_for_correct_timestamp}, $e^{apply}$ is for timestamp $t$.
    Since $e^{apply} < e^{remove}$, we have that there is a next $L$-event after $e^{apply}$ in $\mathcal{I}^\mathcal{B}$; say $e$.
    Hence, since $e^{apply}$ and $e$ are successive $L$-events, and $e^{apply}$ is an $L$-apply event for timestamp $\timeshort{}$, by $O(\mathcal{I}^\mathcal{B})$ (which holds by \Cref{lemma:ero:the_list_invariants_hold}), there is a successful $S$-attempt $a$ for timestamp $\timeshort{}$ during $(e^{apply}, e)$.
    Thus, by \Cref{def:ero:english}, $a$ wrote a value of the form $((\timeshort{}, \arbitraryvalue), \arbitraryvalue, \arbitraryvalue)$ into $\stateobject$ during $\mathcal{I}^\mathcal{B}$, and so $((\timeshort{}, \arbitraryvalue), \arbitraryvalue, \arbitraryvalue)$ appears in $V$, say at index $i$, so $\timeshort{} = t_i$.
    So, by \Cref{lemma:ero:mapping_between_v_and_opx_is_well_defined}, there is a unique operation execution $opx_i$ that received $t_i$ as a response on \cref{line:ero:operation_timestamp} during an invocation of the \doworkuntildone{} procedure invoked on \cref{line:ero:low_level_apply_and_copy_response} during $\mathcal{I}^\mathcal{B}$.
    Therefore, since $opx$ received $\timeshort{}$ as a response on \cref{line:ero:operation_timestamp} during $I^{apply}$ which is an invocation of the \doworkuntildone{} procedure invoked on \cref{line:ero:low_level_apply_and_copy_response}, we have that $opx = opx_i$, and so $opx$ is in $Opx$, which completes the proof of part (a).

    We now prove part (b) for $opx$.
    Since by \Cref{observation:ero:where_objects_change} only successful $S$-attempts change the value of $\stateobject$ in $\mathcal{I}^\mathcal{B}$, and by \Cref{lemma:ero:every_s_attempt_has_a_unique_timestamp} every successful $S$-attempt in $\mathcal{I}^\mathcal{B}$ is for a unique timestamp, it follows that for every index $i$ and $j$ of $V$ if $i \neq j$, then $t_i \neq t_j$.
    Hence, $((t_i, \arbitraryvalue), \arbitraryvalue, \arbitraryvalue)$ is written into $\stateobject$ in $\mathcal{I}^\mathcal{B}$ once, and by the $i$th successful \CASop{} operation on \cref{line:ero:state_cas} during $\mathcal{I}^\mathcal{B}$.
    Thus, since $a$ wrote a value of the form $((\timeshort{}, \arbitraryvalue), \arbitraryvalue, \arbitraryvalue)$ into $\stateobject$ during $\mathcal{I}^\mathcal{B}$, and $\timeshort{} = t_i$, we have that $a$ is the $i$th successful \CASop{} operation on \cref{line:ero:state_cas} during $\mathcal{I}^\mathcal{B}$.
    So, since $opx$ is in $Opx$ and appears at index $i$, by \Cref{def:ero:linearization_point}, $\ell(opx)$ is the time of $a$.
    We now position $a$ (and hence $\ell(opx)$) between $opx$'s invocation and response steps.
    Since $e$ is the next $L$-event after $e^{apply}$ in $\mathcal{I}^\mathcal{B}$, and $e^{apply} < e^{remove}$, it follows that $e \leq e^{remove}$.
    Hence, since the invocation step of $opx$ is before $T^{apply}_b$, $T^{apply}_b < e^{apply}$, $e^{apply} < a$, $a < e$, $e \leq e^{remove}$, $e^{remove} < T^{remove}_e$, and the response step of $opx$ is after $T^{remove}_e$, by transitivity, $a$ (and hence $\ell(opx)$) is between the invocation and response step of $opx$.
    Therefore, for every complete operation execution $opx$ in $\mathcal{I}^\mathcal{B}$, $\ell(opx)$ is is after $opx$'s invocation step in $\mathcal{I}^\mathcal{B}$ and before $opx$'s response step in $\mathcal{I}^\mathcal{B}$.

    To complete the proof of part (b), consider any operation execution $opx$ in $Opx$ that is incomplete in $\mathcal{I}^\mathcal{B}$.
    Hence, $opx$ does not have a response step in $\mathcal{I}^\mathcal{B}$, so it suffices to prove that $\ell(opx)$ is after $opx$'s invocation step in $\mathcal{I}^\mathcal{B}$.
    Suppose $opx = opx_i$.
    Hence, by \Cref{def:ero:linearization_point}, the step at time $\ell(opx)$ set $\stateobject$ to $((t_i, \langle \doopandcopyresponse{}, o_i \rangle), s_i, r_i)$.
    Therefore, by \Cref{lemma:ero:mapping_between_v_and_opx_is_well_defined}, $opx_i$ (and hence $opx$) was invoked before $\ell(opx)$, which completes the proof of part (b).
    \qH{\Cref{lemma:ero:every_complete_operation_with_timestamp_t_has_a_successful_s_attempt_for_t}}
\end{proof}

\begin{lemma}\label{lemma:ero:respects_real_time_order}
    $<_{\mathcal{H}'} \subseteq <_\mathcal{S}$
\end{lemma}

\begin{proof}
    Consider any two operation executions $opx$ and $opx'$ in $\mathcal{H}'$ such that $opx <_{\mathcal{H}'} opx'$.
    Thus $opx$'s response step in $\mathcal{H}'$ is before $opx'$'s invocation step in $\mathcal{H}'$.
    Hence, by the construction of $\mathcal{H}'$: $opx$ is complete in $\mathcal{H}$, $opx <_{\mathcal{H}} opx'$, and if $opx'$ is incomplete in $\mathcal{H}$, then $opx'$ is in $Opx$.
    Since $\mathcal{H}$ is the result of removing all implementation steps in $\mathcal{I}^\mathcal{B}$, this implies that $opx$ is complete in $\mathcal{I}^\mathcal{B}$ and that $opx$'s response step in $\mathcal{I}^\mathcal{B}$ is before $opx'$'s invocation step in $\mathcal{I}^\mathcal{B}$.
    Since $opx$ is complete in $\mathcal{I}^\mathcal{B}$, by \Cref{lemma:ero:every_complete_operation_with_timestamp_t_has_a_successful_s_attempt_for_t} (a), $opx$ is in $Opx$.
    Thus, by \Cref{lemma:ero:every_complete_operation_with_timestamp_t_has_a_successful_s_attempt_for_t} (b), $\ell(opx)$ is before $opx$'s response step in $\mathcal{I}^\mathcal{B}$ (which exists since $opx$ is complete in $\mathcal{I}^\mathcal{B}$).
    Likewise, if $opx'$ is complete in $\mathcal{H}$, then $opx'$ is complete in $\mathcal{I}^\mathcal{B}$, in which case by \Cref{lemma:ero:every_complete_operation_with_timestamp_t_has_a_successful_s_attempt_for_t} (a), $opx'$ is in $Opx$.
    Hence, since if $opx'$ is incomplete in $\mathcal{H}$, then $opx'$ is in $Opx$, we have that in any case, $opx'$ is in $Opx$.
    Thus, by \Cref{lemma:ero:every_complete_operation_with_timestamp_t_has_a_successful_s_attempt_for_t} (b), $\ell(opx')$ is after $opx'$'s invocation step in $\mathcal{I}^\mathcal{B}$.
    Therefore, we have established the following: (1) $\ell(opx)$ is before $opx$'s response step; (2) $opx$'s response step is before $opx'$'s invocation step; and (3) $opx'$'s invocation step is before $\ell(opx')$.
    So, by transitivity, $\ell(opx) < \ell(opx')$.
    Since $opx$ and $opx'$ are both in $Opx$, we have that $opx = opx_i$ and $opx' = opx_j$ for indices $i$ and $j$ in $Opx$.
    Hence, since $\ell(opx) < \ell(opx')$, by \Cref{def:ero:linearization_point}, $i < j$.
    Thus, $opx$ appears before $opx'$ in $Opx$.
    Therefore, by the construction of $\mathcal{S}$, $opx$'s response step is before $opx'$'s invocation step in $\mathcal{S}$, which implies $opx <_\mathcal{S} opx'$, completing the lemma.
    \qH{\Cref{lemma:ero:respects_real_time_order}}
\end{proof}

\subsubsection{The linearization respects the specification of the target object type}

We start by proving that $o_i$ comes from the proper domain.

\begin{lemma}\label{lemma:ero:o_i_in_op_set}
    Let $OP$ be the set of operations of type $\mathcal{T}$.
    Then, $o_i \in OP$.
\end{lemma}

\begin{proof}
    Consider the $i$th successful \CASop{} operation on $\stateobject{}$ on \cref{line:ero:state_cas} in $\mathcal{I}^\mathcal{B}$ and denote it by $a_i$.
    Hence, $a_i$ was executed during some invocation of the \doapplyandcopyresponse{} with a first parameter of $(t_i, \langle \doopandcopyresponse{}, o_i \rangle)$.
    Thus, by \Cref{lemma:ero:l_event_corresponding_to_do_low_level_op}, there is an $L$-event that set $\linearizationobject{}.\uniquerepositoryoperationlong{} = (t_i, \langle \doopandcopyresponse{}, o_i \rangle)$.
    So, by \Cref{lemma:ero:l_events_have_corresponding_a_events}, there is an $A$-event that set $\announceobject{}.\uniquerepositoryoperationlong{} = (t_i, \langle \doopandcopyresponse{}, o_i \rangle)$.
    Hence, it was executed during some invocation of the \doworkuntildone{} procedure with a first parameter of $\langle \doopandcopyresponse{}, o_i \rangle$.
    Thus, this \doworkuntildone{} procedure was invoked on \cref{line:ero:low_level_apply_and_copy_response}, and so $o_i$ was the first parameter of some invocation of the \highleveloperation{} procedure.
    Therefore, $o_i \in OP$ as wanted.
    \qH{\Cref{lemma:ero:o_i_in_op_set}}
\end{proof}

\begin{lemma}\label{lemma:ero:correct_execution}
    For every index $i$ of $Opx$, $(s_i, r_i) = apply_\mathcal{T}(o_i, s_{i - 1})$, where $s_0$ is the initial state of type $\mathcal{T}$.
\end{lemma}

\begin{proof}
    By definition, $s_i$ and $r_i$ are the values written in $\stateobject{}.\statelong{}$ and $\stateobject{}.\responselong{}$, respectively, by the $i$th successful \CASop{} on $\stateobject{}$ on \cref{line:ero:state_cas} in $\mathcal{I}^\mathcal{B}$.
    Let $a_i$ denote this \CASop{}, let $p$ be the process that performed $a_i$, and let $I$ be the invocation of the \doapplyandcopyresponse{} procedure that $q$ performed $a_i$ during.
    Since $a_i$ is the $i$th \CASop{} on $\stateobject{}$ on \cref{line:ero:state_cas} in $\mathcal{I}^\mathcal{B}$, we have that $a_i$ set $\stateobject$ to $((t_i, \langle \doopandcopyresponse{}, o_i \rangle), s_i, r_i)$.
    Hence, we have that the first parameter of $I$ is $(t_i, \langle \doopandcopyresponse{}, o_i \rangle)$.
    Thus, by \cref{line:ero:apply_op} and \ref{line:ero:unpack_operation_to_apply}, $(s_i, r_i) = apply_{\mathcal{T}}(o_i, s)$, where $s$ is the state in $\stateobject{}.\statelong{}$ that $p$ read on \cref{line:ero:state_read} during $I$.
    So, since $a_i$ is successful, it follows that $\stateobject{}.\statelong{} = s$ at the step before $a_i$ in $\mathcal{I}^\mathcal{B}$.
    Hence, by \Cref{observation:ero:where_objects_change}, $s$ is the value written in $\stateobject{}.\statelong{}$ by the $(i-1)$-th successful \CASop{} on $\stateobject{}$ on \cref{line:ero:state_cas} during $\mathcal{I}^\mathcal{B}$, or the initial value of $\stateobject$ if $i = 1$ (which is $s_0$), so $s = s_{i - 1}$.
    Therefore, $(s_i, r_i) = apply_{\mathcal{T}}(o_i, s_{i - 1})$, as wanted.
    \qH{\Cref{lemma:ero:correct_execution}}
\end{proof}

Since by \Cref{lemma:ero:o_i_in_op_set} $o_i$ is a valid operation of type $\mathcal{T}$, by a simple induction, \Cref{lemma:ero:correct_execution} implies that for every index $i$ of $Opx$ $s_i \in Q$ where $Q$ is the set of states of type $\mathcal{T}$ and $r_i \in RES$ where $RES$ is the set of responses of type $\mathcal{T}$.
Therefore, $o_i$, $s_i$, and $r_i$ are all valid operations, states, and responses of type $\mathcal{T}$, respectively, so by \Cref{lemma:ero:correct_execution} and the definition of $apply_\mathcal{T}$:

\begin{corollary}\label{corollary:ero:correct_execution}
    % For every index $i$ of $Opx$, $(s_{i - 1}, o_i, s_i, r_i) \in \delta$, where $s_0$ is the initial state of object $O$ and $\delta$ is the state transition relation of $\mathcal{T}$, so
    $\mathcal{S}$ is legal with respect to $\mathcal{T}$.
\end{corollary}

\subsubsection{The linearization is equivalent to the completed history}

\begin{proposition}\label{lemma:ero:every_operation_execution_in_h_is_in_opx}
    Every operation execution $opx$ in $\mathcal{H}'$ is in $Opx$.
\end{proposition}

\begin{proof}
    Since $\mathcal{H}'$ is derived from $\mathcal{H}$, and $\mathcal{H}$ is derived from $\mathcal{I}^\mathcal{B}$, we have that $opx$ is an operation execution in $\mathcal{I}^\mathcal{B}$.
    There are two cases.
    \begin{itemize}
        \item[] \hspace{0pt}\textbf{Case 1.} $opx$ is complete in $\mathcal{I}^\mathcal{B}$.

        Hence, by \Cref{lemma:ero:every_complete_operation_with_timestamp_t_has_a_successful_s_attempt_for_t} (a), $opx$ is in $Opx$.

        \item[] \hspace{0pt}\textbf{Case 2.} $opx$ is incomplete in $\mathcal{I}^\mathcal{B}$.

        Hence, $opx$ is incomplete in $\mathcal{H}$, and so by the definition of $\mathcal{H}'$, $opx$ is in $Opx$.
        \qH{\Cref{lemma:ero:every_operation_execution_in_h_is_in_opx}}
    \end{itemize}
\end{proof}

\begin{lemma}\label{lemma:ero:invocations_match_in_h_prime_and_s}
    Consider any operation execution $opx$ in $\mathcal{H}'$ whose invocation step is for operation $o$ and let $i$ be the unique index that $opx$ appears in $Opx$, i.e., $opx = opx_i$ ($i$ is well-defined by \Cref{lemma:ero:every_operation_execution_in_h_is_in_opx} and \Cref{lemma:ero:sequence_opx_is_pairwise_distinct}).
    Then, $o = o_i$.
\end{lemma}

\begin{proof}
    Since $opx$ is in $\mathcal{H}'$, by definition, it is in $\mathcal{I}^\mathcal{B}$.
    Furthermore, since $opx = opx_i$, by \Cref{lemma:ero:mapping_between_v_and_opx_is_well_defined}, $opx$ received $\timeshort{}_i$ as a response on \cref{line:ero:operation_timestamp} during an invocation $I$ of the \doworkuntildone{} procedure invoked on \cref{line:ero:low_level_apply_and_copy_response} during $\mathcal{I}^\mathcal{B}$.
    %and $((t_i, \langle \doopandcopyresponse{}, o_i \rangle), s_i, r_i)$ was written into $\stateobject$.
    Hence, since $I$ is an invocation of the \doworkuntildone{} procedure invoked on \cref{line:ero:low_level_apply_and_copy_response} during $opx$ and $opx$'s invocation step is for operation $o$, the first parameter of $I$ is $\langle \doopandcopyresponse{}, o \rangle$.
    Furthermore, $((t_i, \langle \doopandcopyresponse{}, o_i \rangle), s_i, r_i)$ was written into $\stateobject$ during $\mathcal{I}^\mathcal{B}$.
    Hence, by \Cref{observation:ero:where_objects_change}, a successful $S$-attempt set $\stateobject$ to \linebreak $((t_i, \langle \doopandcopyresponse{}, o_i \rangle), s_i, r_i)$ during $\mathcal{I}^\mathcal{B}$.
    Thus, by \Cref{lemma:ero:s_attempts_have_corresponding_l_events}, an $L$-event set $\linearizationobject{}.\uniquerepositoryoperationlong{} = (t_i, \langle \doopandcopyresponse{}, o_i \rangle)$.
    So, by \Cref{lemma:ero:l_events_have_corresponding_a_events}, an $A$-event $e$ set $\announceobject{}.\uniquerepositoryoperationlong{} = (t_i, \langle \doopandcopyresponse{}, o_i \rangle)$.
    Let $p$ be the process that executed $opx$, and let $q$ be the process that executed $e$.
    Since $q$ executed $e$ and $e$ set $\announceobject{}.\uniquerepositoryoperationlong{} = (t_i, \langle \doopandcopyresponse{}, o_i \rangle)$, we have that $q$ received $\timeshort{}_i$ as a response on \cref{line:ero:operation_timestamp}.
    Hence, since $p$ also received $\timeshort{}_i$ as a response on \cref{line:ero:operation_timestamp}, and by \Cref{observation:ero:operation_timestamp_is_unique} the responses on \cref{line:ero:operation_timestamp} are unique, it follows that $p = q$.
    Thus, $p$ performed $e$, and since $e$ set $\announceobject{}.\uniquerepositoryoperationlong{} = (t_i, \langle \doopandcopyresponse{}, o_i \rangle)$, we have that $p$ performed $e$ during an invocation $I'$ of the \doworkuntildone{} procedure in which $p$ received $\timeshort{}_i$ as a response on \cref{line:ero:operation_timestamp} and whose first parameter is $\langle \doopandcopyresponse{}, o_i \rangle$.
    So, since $p$ also received $\timeshort{}_i$ as a response on \cref{line:ero:operation_timestamp} during $I$, and the responses on \cref{line:ero:operation_timestamp} are unique, we have that $I = I'$.
    Therefore, since $I$'s first parameter is $\langle \doopandcopyresponse{}, o \rangle$, and $I'$'s first parameter is $\langle \doopandcopyresponse{}, o_i \rangle$, we have that $o = o_i$ as wanted.
    \qH{\Cref{lemma:ero:invocations_match_in_h_prime_and_s}}
\end{proof}

\begin{lemma}\label{lemma:ero:responses_match_in_h_prime_and_s}
    Consider any operation execution $opx$ in $\mathcal{H}'$ whose response step is for response $r$ in $\mathcal{H}'$ and let $i$ be the unique index that $opx$ appears in $Opx$, i.e., $opx = opx_i$ ($i$ is well-defined by \Cref{lemma:ero:every_operation_execution_in_h_is_in_opx} and \Cref{lemma:ero:sequence_opx_is_pairwise_distinct}).
    Then, $r = r_i$.
\end{lemma}

\begin{proof}
    By definition of $\mathcal{H}'$ and the fact that $i$ is unique, the lemma trivially holds when $opx$ is incomplete in $\mathcal{H}$.
    Hence, it suffices to consider the case where $opx$ is complete in $\mathcal{H}$.
    Since $\mathcal{H}$ is derived from $\mathcal{I}^\mathcal{B}$, $opx$ is complete in $\mathcal{I}^\mathcal{B}$.
    Hence, since $opx$ is in $Opx$, by \Cref{lemma:ero:every_complete_operation_with_timestamp_t_has_a_successful_s_attempt_for_t} (b), $\ell(opx)$ is after $opx$'s invocation step and before $opx$'s response step in $\mathcal{I}^\mathcal{B}$.
    Let $p$ be the process that executed $opx$.
    Since $opx$ is complete in $\mathcal{I}^\mathcal{B}$ $p$ did the following during $opx$: $p$ executed \cref{line:ero:allocate_cell} and got response $\cellpointershort \in \celluniverse$, $p$ invoked an invocation $I^{apply}$ of the \doworkuntildone{} procedure on \cref{line:ero:low_level_apply_and_copy_response} with parameters $(\langle \doopandcopyresponse{}, \arbitraryvalue \rangle, \cellpointershort)$ that began at time $T^{apply}_b$ and exited at some time $T^{apply}_e$ during $opx$, and $p$ invoked an invocation $I^{remove}$ of the \doworkuntildone{} procedure on \cref{line:ero:low_level_remove_cell} with parameters $(\removecell, \cellpointershort)$ that began at time $T^{remove}_b$ and exited at some time $T^{remove}_e$ during $opx$.    
    Hence, $T^{apply}_b < T^{apply}_e < T^{remove}_b < T^{remove}_e$.
    Let $T^{\ref{line:ero:do_work_initialize_response}}$ be the time $p$ executed \cref{line:ero:do_work_initialize_response} during $I^{apply}$.
    Then, by \Cref{lemma:ero:successful_apply_response_set_before_apply_low_level_exits}, between $T^{\ref{line:ero:do_work_initialize_response}}$ and $T^{apply}_e$, there is a successful apply-response-set attempt $a_R$ for $\cellpointershort$.
    Since by \Cref{lemma:ero:the_list_invariants_hold} $P(\mathcal{I}^\mathcal{B})$ holds, by \Cref{lemma:ero:l_apply_event_before_apply_low_level_exits}, there is an $L$-apply event $e$ for $\cellpointershort$ between $T^{apply}_b$ and $T^{apply}_e$.
    Likewise, by \Cref{lemma:ero:l_remove_event_before_remove_low_level_exits}, there is an $L$-remove $e'$ event for $\cellpointershort$ between $T^{remove}_b$ and $T^{remove}_e$.
    Hence, since $T^{apply}_b < T^{apply}_e < T^{remove}_b < T^{remove}_e$, we have that $e < e'$.
    Let $\timeshort{}$ be the response $p$ received on \cref{line:ero:operation_timestamp} during $I^{apply}$, so by \Cref{lemma:ero:l_apply_event_before_apply_low_level_exits_is_for_correct_timestamp} $e$ is for timestamp $\timeshort{}$.
    Furthermore, since $opx = opx_i$, by \Cref{lemma:ero:mapping_between_v_and_opx_is_well_defined}, $\timeshort{} = t_i$.
    Let $a_S$ be the successful $S$-attempt at $\ell(opx)$.
    Hence, since $opx = opx_i$, by \Cref{def:ero:linearization_point}, $a_S$ is a successful $S$-attempt for timestamp $t_i$, and since $\timeshort{} = t_i$, we have that $a_S$ is a successful $S$-attempt for timestamp $\timeshort{}$.

    \begin{claimcustom}{\ref{lemma:ero:responses_match_in_h_prime_and_s}.1}\label{lemma:ero:responses_match_in_h_prime_and_s:claim_one}
        $a_R$ is a successful apply-response-set attempt for $\cellpointershort$ to $r_i$.
    \end{claimcustom}

    \begin{proof}
        Let $q$ be the process that executed $a_R$ and let $I'$ be the invocation of the \setrepositoryoperationresponse{} procedure that $q$ executed $a_R$ during.
        Since $a_R$ is a successful apply-response-set attempt for $\cellpointershort$, by \Cref{observation:ero:response_set_attempt_invocation}, $q$ invoked $I'$ on \cref{line:ero:apply_set_response} during some invocation $I^*$ of the \doapplyandcopyresponse{} procedure.
        The remainder of the proof is split into two cases.
        
        \begin{itemize}
            \item[] \hspace{0pt}\textbf{Case 1.} $q$ found the condition on \cref{line:ero:check_if_already_applied} to be true during $I^*$.

            Hence, $q$ executed \cref{line:ero:apply_update_response} during $I^*$; say at time $T^{\ref{line:ero:apply_update_response}}$.
            Thus, since $q$ executed $a_R$ during the \setrepositoryoperationresponse{} procedure on \cref{line:ero:apply_set_response} during $I^*$, we have that $T^{\ref{line:ero:apply_update_response}} < a_R$.
            Since $e < e'$, it follows that there is a next $L$-event after $e$ in $\mathcal{I}^\mathcal{B}$; say $e_a$.
            Hence, since $e$ is an $L$-apply event for timestamp $\timeshort{}$, and by \Cref{lemma:ero:the_list_invariants_hold} $O(\mathcal{I}^\mathcal{B})$ holds, we have that there is exactly one successful $S$-attempt between $e$ and $e_a$ in $\mathcal{I}^\mathcal{B}$ and it is for timestamp $\timeshort{}$.
            Therefore, since $a_S$ is a successful $S$-attempt for timestamp $\timeshort{}$, by \Cref{lemma:ero:every_s_attempt_has_a_unique_timestamp}, $a_S$ is the single successful $S$-attempt between $e$ and $e_a$ in $\mathcal{I}^\mathcal{B}$.

            Let $e_S$ be $a_S$'s corresponding $L$-event (see \Cref{lemma:ero:s_attempts_have_corresponding_l_events}), so $e_S$ is before $I^*$ was invoked.
            We prove that $e = e_S$.
            Since $a_S$ is a successful $S$-attempt for timestamp $\timeshort{}$, by \Cref{corollary:ero:s_attempts_and_corresponding_l_events_are_for_matching_timestamps}, $e_S$ is an $L$-event for timestamp $\timeshort{}$.
            Hence, since $P(\mathcal{I}^\mathcal{B})$ holds, by \Cref{lemma:ero:every_l_event_has_a_unique_timestamp}, $e_S$ is the only $L$-event for timestamp $\timeshort{}$ in $\mathcal{I}^\mathcal{B}$.
            Therefore, since $e$ is an $L$-event for timestamp $\timeshort{}$, we have that $e = e_S$ as wanted.
            Hence, since $e_S$ is before $I^*$ was invoked, we have that $e$ is before $I^*$ was invoked.
            
            We now prove that $a_S < T^{\ref{line:ero:apply_update_response}}$.
            Suppose, for contradiction, that $T^{\ref{line:ero:apply_update_response}} < a_S$.
            Let $T^{\ref{line:ero:state_read}}$ and $T^{\ref{line:ero:state_cas}}$ be the times that $q$ executed lines \ref{line:ero:state_read} and \ref{line:ero:state_cas} during $I^*$, respectively.
            Since by definition $T^{\ref{line:ero:state_read}} < T^{\ref{line:ero:state_cas}} < T^{\ref{line:ero:apply_update_response}}$, and by assumption $T^{\ref{line:ero:apply_update_response}} < a_S$, by transitivity, $T^{\ref{line:ero:state_read}} < T^{\ref{line:ero:state_cas}} < T^{\ref{line:ero:apply_update_response}} < a_S$.
            Since $e$ is before $I^*$ is invoked, we have that $e < T^{\ref{line:ero:state_read}}$, and so $e < T^{\ref{line:ero:state_read}} < T^{\ref{line:ero:state_cas}} < T^{\ref{line:ero:apply_update_response}} < a_S$.
            Furthermore, since $a_S < e_a$, by transitivity, $e < T^{\ref{line:ero:state_read}} < T^{\ref{line:ero:state_cas}} < T^{\ref{line:ero:apply_update_response}} < a_S < e_a$.
            There are two cases.

            \begin{itemize}
                \item[] \hspace{0pt}\textbf{Case 1.1.} $q$'s \CASop{} operation at $T^{\ref{line:ero:state_cas}}$ is successful.

                Hence, by \Cref{observation:ero:where_objects_change}, $q$'s step at $T^{\ref{line:ero:state_cas}}$ is a successful $S$-attempt.
                Therefore, since $e < T^{\ref{line:ero:state_read}} < T^{\ref{line:ero:state_cas}} < T^{\ref{line:ero:apply_update_response}} < a_S < e_a$, we have that there are two successful $S$-attempts between $e$ and $e_a$.
                However, $a_S$ is the only successful $S$-attempt between $e$ and $e_a$, a contradiction.

                \item[] \hspace{0pt}\textbf{Case 1.2.} $q$'s \CASop{} operation at $T^{\ref{line:ero:state_cas}}$ is unsuccessful.

                Hence, the value of $\stateobject$ changed between $T^{\ref{line:ero:state_read}}$ and $T^{\ref{line:ero:state_cas}}$.
                Thus, by \Cref{observation:ero:where_objects_change}, there is a successful $S$-attempt between $T^{\ref{line:ero:state_read}}$ and $T^{\ref{line:ero:state_cas}}$.
                Therefore, since $e < T^{\ref{line:ero:state_read}} < T^{\ref{line:ero:state_cas}} < T^{\ref{line:ero:apply_update_response}} < a_S < e_a$, we have that there are two successful $S$-attempts between $e$ and $e_a$.
                However, $a_S$ is the only successful $S$-attempt between $e$ and $e_a$, a contradiction.
            \end{itemize}

            We now prove that $T^{\ref{line:ero:apply_update_response}} < e_a$.
            Let $r$ be the process that executed $e_a$.
            Since $e$ and $e_a$ are successive $L$-events, by \Cref{lemma:ero:successive_l_event_read_previous_l_event_value}, $r$ read the value that $e$ set $\linearizationobject{}$ to on its last execution of \cref{line:ero:linearization_read} before $e_a$; say time $T^{\ref{line:ero:linearization_read}}$.
            Since $e$ is an $L$-apply event for $\cellpointershort$, by \Cref{def:ero:english}, $r$ read a value of the form $((\arbitraryvalue, \langle \doopandcopyresponse{}, \arbitraryvalue \rangle), \cellpointershort)$ from $\linearizationobject{}$ at $T^{\ref{line:ero:linearization_read}}$.
            Hence, since $r$ executes $e_a$, $p$ finds the condition on \cref{line:ero:do_apply_and_copy_response} to be true after $T^{\ref{line:ero:linearization_read}}$, and so $p$ invokes the \doapplyandcopyresponse{} procedure with parameters $(\arbitraryvalue, \cellpointershort)$ after $T^{\ref{line:ero:linearization_read}}$ and exits it before $e_a$.
            Denote this invocation of \doapplyandcopyresponse{} procedure by $I$.
            Since $I$ has parameters $(\arbitraryvalue, \cellpointershort)$ and exits before $e_a$, and by \Cref{lemma:ero:the_list_invariants_hold} $P(\mathcal{I}^\mathcal{B})$, $Q(\mathcal{I}^\mathcal{B})$, and $R(\mathcal{I}^\mathcal{B})$ holds, by \Cref{lemma:ero:conditional_exit_apply_implies_response_set}, there is a successful apply-response-set attempt for $\cellpointershort$ before $e_a$.
            Hence, since $a_R$ is a successful apply-response-set attempt for $\cellpointershort$, and by \Cref{lemma:ero:at_most_one_response_set_per_type_and_pointer} there is at most one successful apply-response-set attempt for $\cellpointershort$ in $\mathcal{I}^\mathcal{B}$, it follows that $a_R < e_a$.
            Therefore, since $T^{\ref{line:ero:apply_update_response}} < a_R$, by transitivity, $T^{\ref{line:ero:apply_update_response}} < e_a$ as wanted.

            We now finish the proof of Case 1.
            So far we have established that $e < a_S < T^{\ref{line:ero:apply_update_response}} < e_a$ and $a_S$ is the only successful $S$-attempt between $e$ and $e_a$.
            Hence, since $a_S$ is at time $\ell(opx)$ and $opx = opx_i$, by \Cref{def:ero:linearization_point}, $a_S$ set $\stateobject$ to $((t_i, \langle \doopandcopyresponse{}, o_i \rangle), s_i, r_i)$, and so the value of $\stateobject$ is $((t_i, \langle \doopandcopyresponse{}, o_i \rangle), s_i, r_i)$ throughout $(a_S, e_a)$.
            Thus, since $T^{\ref{line:ero:apply_update_response}} \in (a_S, e_a)$, we have that $q$ read $r_i$ from $\stateobject.\responselong{}$ on \cref{line:ero:apply_update_response} at $T^{\ref{line:ero:apply_update_response}}$.
            So, since $q$ invoked $I'$ on \cref{line:ero:apply_set_response} during $I^*$, we have that the third parameter of $I'$ is $r_i$.
            Therefore, since $q$ executed $a_R$ during $I'$, by \Cref{def:ero:english}, $a_R$ is a successful apply-response-set attempt for $\cellpointershort$ to $r_i$ as wanted.

            \item[] \hspace{0pt}\textbf{Case 2.} $q$ found the condition on \cref{line:ero:check_if_already_applied} to be false during $I^*$.
            
            Let $(\uniquerepositoryoperationshort{}, \stateshort{}, \responseshort{})$ be the value $q$ read from $\stateobject{}$ on \cref{line:ero:state_read} during $I^*$, say at time $T^{\ref{line:ero:state_read}}$, and let $\uniquerepositoryoperationshort{}_\linearizationobject{}$ be the first parameter of $I^*$.
            Hence, since $q$ executed $I'$ during $I^*$, we have that the first parameter of $I'$ is $\uniquerepositoryoperationshort{}_\linearizationobject{}$.
            Thus, since $q$ executed $a_R$ during $I'$, and $a_R$ is an apply-response-set attempt for $\cellpointershort$, we have that $a_R$ tries to set $(*\cellpointershort).\lastrepositoryoperationresponse{}.\uniquerepositoryoperationlong{}$ to $\uniquerepositoryoperationshort{}_\linearizationobject{}$.
            So, by \Cref{lemma:ero:response_set_attempts_have_corresponding_l_events} and \Cref{corollary:ero:response_set_attempts_and_corresponding_l_events_are_matching}, an $L$-apply event $e_R$ set $\linearizationobject{} = (\uniquerepositoryoperationshort{}_\linearizationobject{}, \cellpointershort)$ before $a_R$.
            Since by \Cref{lemma:ero:the_list_invariants_hold} $P(\mathcal{I}^\mathcal{B})$ holds, we have that $e_R$ is the only $L$-apply event for $\cellpointershort$ in $\mathcal{I}^\mathcal{B}$, and so since $e$ is also an $L$-apply event for $\cellpointershort$ in $\mathcal{I}^\mathcal{B}$, it follows that $e = e_R$.
            Hence, $e$ set $\linearizationobject{} = (\uniquerepositoryoperationshort{}_\linearizationobject{}, \cellpointershort)$ and $e < a_R$.
            Thus, since $e$ is for timestamp $\timeshort{}$, by \Cref{def:ero:english}, $\uniquerepositoryoperationshort{}_\linearizationobject{} = (\timeshort{}, \arbitraryvalue)$.
            Since $q$ found the condition on \cref{line:ero:check_if_already_applied} to be false during $I^*$, it follows that $\uniquerepositoryoperationshort{} = \uniquerepositoryoperationshort{}_\linearizationobject{}$.
            Hence, since $\uniquerepositoryoperationshort{}_\linearizationobject{} = (\timeshort{}, \arbitraryvalue)$, we have that $\uniquerepositoryoperationshort{} = (\timeshort{}, \arbitraryvalue)$.
            Since $a_S$ is for timestamp $\timeshort{}$, by \Cref{lemma:ero:every_s_attempt_is_for_timestamp_other_than_zero}, $\timeshort{} > 0$.
            Hence, since $q$ read $(\uniquerepositoryoperationshort{}, \stateshort{}, \responseshort{})$ from $\stateobject{}$ on \cref{line:ero:state_read} at time $T^{\ref{line:ero:state_read}}$, and $\uniquerepositoryoperationshort{} = (\timeshort{}, \arbitraryvalue)$, we have that some step set $\stateobject = (\uniquerepositoryoperationshort{}, \stateshort{}, \responseshort{})$.
            Thus, by \Cref{observation:ero:where_objects_change}, some successful $S$-attempt set $\stateobject = (\uniquerepositoryoperationshort{}, \stateshort{}, \responseshort{})$, and since $\uniquerepositoryoperationshort{} = (\timeshort{}, \arbitraryvalue)$, by \Cref{def:ero:english}, this $S$-attempt is for timestamp $\timeshort{}$.
            So, since by \Cref{lemma:ero:every_s_attempt_has_a_unique_timestamp} every successful $S$-attempt has a unique timestamp, and $a_S$ is for timestamp $\timeshort{}$, we have that $a_S$ set $\stateobject = (\uniquerepositoryoperationshort{}, \stateshort{}, \responseshort{})$.
            Hence, since $a_S$ is the step at $\ell(opx)$, and $opx = opx_i$, by \Cref{def:ero:linearization_point}, $a_S$ set $\stateobject$ to $((t_i, \langle \doopandcopyresponse{}, o_i \rangle), s_i, r_i)$, and so $((t_i, \langle \doopandcopyresponse{}, o_i \rangle), s_i, r_i) = (\uniquerepositoryoperationshort{}, \stateshort{}, \responseshort{})$.
            Thus, $q$ read $((t_i, \langle \doopandcopyresponse{}, o_i \rangle), s_i, r_i)$ from $\stateobject{}$ on \cref{line:ero:state_read} at $T^{\ref{line:ero:state_read}}$, and so since $q$ found the condition on \cref{line:ero:check_if_already_applied} to be false during $I^*$, and $q$ invoked $I'$ during $I^*$, we have that the third parameter of $I'$ is $r_i$.
            Therefore, since $q$ executed $a_R$ during $I'$, by \Cref{def:ero:english}, $a_R$ is a successful apply-response-set attempt for $\cellpointershort$ to $r_i$ as wanted.
            \qH{\Cref{lemma:ero:responses_match_in_h_prime_and_s:claim_one}}
        \end{itemize}
    \end{proof}

     Let $T^{\ref{line:ero:copy_response_out_of_cell}}$ be the time $p$ executes \cref{line:ero:copy_response_out_of_cell} during $opx$ (this is well-defined since $opx$ is complete in $\mathcal{H}$).
     Since $a_R$ is before $T^{apply}_e$, $I^{apply}$ exits at $T^{apply}_e$, and $I^{apply}$ is the invocation of the \doworkuntildone{} procedure on \cref{line:ero:low_level_apply_and_copy_response} during $opx$, we have that $a_R < T^{\ref{line:ero:copy_response_out_of_cell}}$.

    \begin{claimcustom}{\ref{lemma:ero:responses_match_in_h_prime_and_s}.2}\label{lemma:ero:responses_match_in_h_prime_and_s:claim_two}
        The value of $(*\cellpointershort).\lastrepositoryoperationresponse{}$ is unchanged throughout $(a_R, T^{\ref{line:ero:copy_response_out_of_cell}}]$.
    \end{claimcustom}

    \begin{proof}
        Suppose, for contradiction, the value of $(*\cellpointershort).\lastrepositoryoperationresponse{}$ changes during $(a_R, T^{\ref{line:ero:copy_response_out_of_cell}}]$.
        Hence, by \Cref{observation:ero:where_objects_change}, there is a response-reset event for $\cellpointershort$ or a successful response-set attempt for $\cellpointershort$ during $(a_R, T^{\ref{line:ero:copy_response_out_of_cell}}]$.
        We consider each case separately.
        \begin{itemize}
            \item[] \hspace{0pt}\textbf{Case 1.} There is a successful response-set attempt $a$ for $\cellpointershort$ during $(a_R, T^{\ref{line:ero:copy_response_out_of_cell}}]$.

            Hence, by \Cref{lemma:ero:every_apply_response_set_attempt_is_to_a_not_null_response}, $(*\cellpointershort).\lastrepositoryoperationresponse{}.\responselong{} \neq \nullconstant{}$ at $a_R$.
            Hence, since by \Cref{def:ero:english}, $a$ is a \CASop{} operation on \cref{line:ero:responses_set_attempt} and $a$ is successful, we have that $(*\cellpointershort).\lastrepositoryoperationresponse{}.\responselong{} = \nullconstant{}$ at the step before $a$.
            Thus, $(*\cellpointershort).\lastrepositoryoperationresponse{}.\responselong{}$ was set to $\nullconstant{}$ between $a_R$ and $a$.
            Hence, by \Cref{observation:ero:where_objects_change}, either a response-reset event for $\cellpointershort$ or a successful response-set attempt for $\cellpointershort$ set $(*\cellpointershort).\lastrepositoryoperationresponse{}.\responselong{} = \nullconstant{}$ between $a_R$ and $a$.
            Therefore, since by \Cref{lemma:ero:every_apply_response_set_attempt_is_to_a_not_null_response}, every successful response-set attempt for $\cellpointershort$ sets $(*\cellpointershort).\lastrepositoryoperationresponse{}.\responselong{} \neq \nullconstant{}$, we have that there is a response-reset event for $\cellpointershort$ between $a_R$ and $a$.
            However, since $a \leq T^{\ref{line:ero:copy_response_out_of_cell}}$, there is a response-reset event for $\cellpointershort$ during $(a_R, T^{\ref{line:ero:copy_response_out_of_cell}}]$ and so this case reduces to the next one.

            \item[] \hspace{0pt}\textbf{Case 2.} There is a response-reset event $e$ for $\cellpointershort$ during $(a_R, T^{\ref{line:ero:copy_response_out_of_cell}}]$.

            Let $q$ be the process that executed $e$ and let $I'$ be the invocation of the \doworkuntildone{} procedure that $q$ executed $e$ during.
            Since $e$ is a response-reset event for $\cellpointershort$, by \Cref{def:ero:english}, the parameters of $I'$ are $(\arbitraryvalue, \cellpointershort)$.
            Hence, $q$ received $\cellpointershort$ as a response on \cref{line:ero:allocate_cell}.
            Thus, since by \Cref{alg:lazy_cell_manager_specification} every response on \cref{line:ero:allocate_cell} is unique and $p$ received $\cellpointershort$ as a response on \cref{line:ero:allocate_cell}, we have that $p = q$.
            Therefore, since $e$ is between $a_R$ and $T^{\ref{line:ero:copy_response_out_of_cell}}$, $a_R$ is between $T^{\ref{line:ero:do_work_initialize_response}}$ and $T^{apply}_e$, and $p$ executes \cref{line:ero:do_work_initialize_response} at $T^{\ref{line:ero:do_work_initialize_response}}$, we have that $p$ executes \cref{line:ero:do_work_initialize_response} twice during $[T^{\ref{line:ero:do_work_initialize_response}}, T^{\ref{line:ero:copy_response_out_of_cell}}]$.
            However, since $p$ is inside $I^{apply}$ throughout $[T^{\ref{line:ero:do_work_initialize_response}}, T^{apply}_e]$ and $I^{apply}$ was invoked on \cref{line:ero:low_level_apply_and_copy_response} during $opx$, there is at most one execution of \cref{line:ero:do_work_initialize_response} during $[T^{\ref{line:ero:do_work_initialize_response}}, T^{\ref{line:ero:copy_response_out_of_cell}}]$, a contradiction.
            \qH{\Cref{lemma:ero:responses_match_in_h_prime_and_s:claim_two}}
        \end{itemize}
    \end{proof}

    We now finish the proof of \Cref{lemma:ero:responses_match_in_h_prime_and_s}.
    Since by \Cref{lemma:ero:responses_match_in_h_prime_and_s:claim_one} $a_R$ is a successful apply-response-set attempt for $\cellpointershort$ to $\responseshort{}_i$, by \Cref{lemma:ero:responses_match_in_h_prime_and_s:claim_two} the value of $(*\cellpointershort).\lastrepositoryoperationresponse{}$ is unchanged during $(a_R, T^{\ref{line:ero:copy_response_out_of_cell}}]$, and $a_R < T^{\ref{line:ero:copy_response_out_of_cell}}$, we have that $(*\cellpointershort).\lastrepositoryoperationresponse{} = (\arbitraryvalue, \responseshort{}_i)$ at $T^{\ref{line:ero:copy_response_out_of_cell}}$.
    Hence, since $p$ received $\cellpointershort$ as a response on \cref{line:ero:allocate_cell} during $opx$, and $T^{\ref{line:ero:copy_response_out_of_cell}}$ is the time $p$ executes \cref{line:ero:copy_response_out_of_cell} during $opx$, the value of the local variable $\responselong{}$ on \cref{line:ero:copy_response_out_of_cell} during $opx$ is $\responseshort{}_i$.
    Therefore, since the local variable $\responselong{}$ on \cref{line:ero:copy_response_out_of_cell} is unchanged for the remainder of $opx$, the response of $opx$ on \cref{line:ero:response_step} is $\responseshort{}_i$ as wanted.
    \qH{\Cref{lemma:ero:responses_match_in_h_prime_and_s}}
\end{proof}

\begin{lemma}\label{lemma:ero:h_prime_and_s_are_equivalent}
    $\mathcal{H'}$ is equivalent to $\mathcal{S}$.
\end{lemma}

\begin{proof}
    We must prove that $\subhistory{\mathcal{H'}}{p} = \subhistory{\mathcal{S}}{p}$ for each process $p$.
    Since by \Cref{lemma:ero:every_operation_execution_in_h_is_in_opx} every operation execution in $\mathcal{H}'$ is in $Opx$, by \Cref{lemma:ero:sequence_opx_is_pairwise_distinct}, every operation execution in $\mathcal{H}'$ is in $Opx$ exactly once.
    Furthermore, since every operation execution in $Opx$ is an operation execution in $\mathcal{I}^\mathcal{B}$, and thus $\mathcal{H}'$, we have that every operation execution in $Opx$ is in $\mathcal{H}'$.
    Hence, since the sequence of operation executions in $\mathcal{S}$ is $Opx$, we have that every operation execution in $\mathcal{H}'$ is in $\mathcal{S}$ exactly once, and every operation execution in $\mathcal{S}$ is in $\mathcal{H}'$ exactly once.
    Now consider any operation execution $opx$ in $\mathcal{H}'$.
    Suppose $opx = opx_i$, and let $invocation(opx, o)$ and $response(opx, r)$ be its invocation and response steps in $\mathcal{H}'$, respectively.
    Hence, by \Cref{lemma:ero:invocations_match_in_h_prime_and_s} $o = o_i$ and by \Cref{lemma:ero:responses_match_in_h_prime_and_s} $r = r_i$, so $invocation(opx, o_i)$ and $response(opx, r_i)$ are the invocation and response steps of $opx$ in $\mathcal{H}'$.
    Likewise, since $opx = opx_i$, by the definition of $\mathcal{S}$, $invocation(opx, o_i)$ and $response(opx, r_i)$ are $opx$'s invocation and response steps in $\mathcal{S}$, respectively.
    Therefore, the invocation and response steps are the same for $opx$ in $\mathcal{H}'$ and $\mathcal{S}$.
    Since operation executions for each process $p$ appear sequentially in $\mathcal{I}^\mathcal{B}$ from which $\mathcal{H'}$ is derived, $\subhistory{\mathcal{H'}}{p}$ is a sequential object history.
    Thus $<_{\subhistory{\mathcal{H'}}{p}}$ is a total order over all operation executions in $\subhistory{\mathcal{H'}}{p}$.
    Likewise, since $\subhistory{\mathcal{S}}{p}$ is a sequential object object history, $<_{\subhistory{\mathcal{S}}{p}}$ is a total order over all operation executions in $\subhistory{\mathcal{S}}{p}$.
    Since (1) every operation execution in $\mathcal{H}'$ is in $\mathcal{S}$ exactly once, and every operation execution in $\mathcal{S}$ is in $\mathcal{H}'$ exactly once, (2) the invocation and response steps are the same for every operation execution $opx$ in $\mathcal{H}'$ and $\mathcal{S}$, (3) $<_{\subhistory{\mathcal{H'}}{p}}$ is a total order over all operation executions in $\subhistory{\mathcal{H'}}{p}$, (4) $<_{\subhistory{\mathcal{S}}{p}}$ is a total order over all operation executions in $\subhistory{\mathcal{S}}{p}$, and (5) by \Cref{lemma:ero:respects_real_time_order} $<_{\subhistory{\mathcal{H'}}{p}} \subseteq <_{\subhistory{\mathcal{S}}{p}}$, it follows that $\subhistory{\mathcal{H'}}{p} = \subhistory{\mathcal{S}}{p}$ as wanted.
    \qH{\Cref{lemma:ero:h_prime_and_s_are_equivalent}}
\end{proof}

\begin{theorem}\label{theorem:ero:b_is_linearizable}
    $\mathcal{B}$ is a linearizable with respect to $\mathcal{T}$.
\end{theorem}

\begin{proof}
    Since $\mathcal{H}'$ is a completion of $\mathcal{H}$, by \Cref{lemma:ero:h_prime_and_s_are_equivalent} $\mathcal{H'}$ is equivalent to $\mathcal{S}$, by \Cref{corollary:ero:correct_execution} $\mathcal{S}$ is legal with respect to $\mathcal{T}$, and by \Cref{lemma:ero:respects_real_time_order} $<_{\mathcal{H'}} \subseteq <_{\mathcal{S}}$, we have that $\mathcal{H}$ is linearizable with respect to $\mathcal{T}$.
    Hence, since $\mathcal{H}$ is the object history obtained by removing all implementation steps from $\mathcal{I}^\mathcal{B}$, we have that $\mathcal{I}^\mathcal{B}$ is linearizable with respect to $\mathcal{T}$.
    Therefore, since $\mathcal{I}^\mathcal{B}$ is any implementation history of $\mathcal{B}$, we have that $\mathcal{B}$ is a linearizable with respect to $\mathcal{T}$.
    \qH{\Cref{theorem:ero:b_is_linearizable}}
\end{proof}

\subsection{$\mathcal{B}$ is Wait-free}
\label{sec:b_is_wait_free}

In this section, we prove that $\mathcal{B}$ is wait-free.
The proof is by contradiction, so we start by assuming that there is an implementation history $\mathcal{I}^\mathcal{B}$ of $\mathcal{B}$ with an operation execution that is ``stuck":

\begin{definition}\label{def:ero:stuck}
    We call an operation execution $opx$ in $\mathcal{I}^\mathcal{B}$ \textit{stuck} when the process that executed $opx$ takes infinitely many steps during $opx$ without completing it.
    Let $\mathbf{S}$ be the set of operation executions in $\mathcal{I}^\mathcal{B}$ that are stuck.
\end{definition}

We note that this $\mathcal{I}^\mathcal{B}$ is \Underline{fixed} throughout the entire section, and it is assumed that $\mathbf{S} \neq \emptyset$.
We first note the following properties of stuck operation executions.

\begin{observation}\label{observation:ero:stuck_operation_basic_properties}
    For every operation execution $opx \in \mathbf{S}$, the process that executed $opx$ does the following during $opx$.
    \begin{compactenum}
        \item[(1)] Takes infinitely many steps inside exactly one invocation of the \doworkuntildone{} procedure.
        \item[(2)] Takes infinitely many steps inside exactly one instance of a loop.
    \end{compactenum}
\end{observation}

We now define an operation execution $opx_{\min} \in \mathbf{S}$ with the goal of showing that it is not stuck.
In short, we show that $opx_{\min}$ is not stuck by showing that it cannot get stuck in each loop.

\begin{definition}\label{def:ero:do_work_timestamp}
    Consider any invocation $I$ of the \doworkuntildone{} procedure in $\mathcal{I}^\mathcal{B}$.
    Let $t(I)$ denote the response of \cref{line:ero:operation_timestamp} during $I$ or $\infty$ if \cref{line:ero:operation_timestamp} was not executed during $I$.
\end{definition}

\begin{definition}\label{def:ero:opx_min}
    For every $opx \in \mathbf{S}$, let $I(opx)$ denote the invocation of the \doworkuntildone{} procedure identified by (1) of \Cref{observation:ero:stuck_operation_basic_properties}.
    % If $\mathbf{S} \neq \emptyset$, then let $opx_{\min}$ be the operation in $\mathbf{S}$ such that every $opx \in \mathbf{S}, t(I(opx_{\min})) \leq t(I(opx))$.
    Let $opx_{\min}$ be the operation in $\mathbf{S}$ such that every $opx \in \mathbf{S}, t(I(opx_{\min})) \leq t(I(opx))$.
\end{definition}

Throughout the remainder of the section, we define the following regarding $opx_{\min}$.
Let $p_{\min}$ be the process that executed $opx_{\min}$.
Since $opx_{\min} \in \mathbf{S}$, by \Cref{def:ero:stuck}, $p_{\min}$ takes infinitely many steps during $opx_{\min}$ in $\mathcal{I}^\mathcal{B}$ without completing it.
Furthermore, by (2) of \Cref{observation:ero:stuck_operation_basic_properties}, $p_{\min}$ takes infinitely many steps inside exactly one instance of a loop $\mathcal{L}_{\min}$ during $opx_{\min}$.

\subsubsection{Processes cannot get stuck in the loops on lines \ref{line:ero:add_cell_while_loop}, \ref{line:ero:remove_cell_while_loop}, and \ref{line:ero:acquire_loop_until}.}

This section shows that $p_{\min}$ cannot take infinitely many steps in the loops on lines \ref{line:ero:add_cell_while_loop}, \ref{line:ero:remove_cell_while_loop}, and \ref{line:ero:acquire_loop_until} during $opx_{\min}$.
If we suppose, for contradiction, that this is not the case, we have:

\begin{scenario}\label{scenario:ero:helper_loop_scenario}
    % Suppose $\mathbf{S} \neq \emptyset$, so by \Cref{def:ero:opx_min} $opx_{\min}$ is well-defined.
    Suppose $\mathcal{L}_{\min}$ is an instance of any loop except the loop on \cref{line:ero:do_work_while_loop}.
    Let $I_{\min}$ denote the invocation of the \doaddcell, \doremovecell{}, Acquire, or AcquireNext procedure that $\mathcal{L}_{\min}$ was executed during.
    Furthermore, let $\uniquerepositoryoperationshort_\linearizationobject{}$ be the first parameter of $I_{\min}$.
    By tracing backwards, we have that $p_{\min}$ read $(\uniquerepositoryoperationshort_\linearizationobject{}, \uniquecellpointershort_\linearizationobject{})$ from $\linearizationobject{}$ on its last execution of \cref{line:ero:linearization_read} before invoking $I_{\min}$ for some $\cellpointershort{}_\linearizationobject{}$; say at time $T^{\ref{line:ero:linearization_read}}_{\min}$.
\end{scenario}

The reason for this scenario being more general than stating that $\mathcal{L}_{\min}$ is an instance of a loop on lines \ref{line:ero:add_cell_while_loop}, \ref{line:ero:remove_cell_while_loop}, and \ref{line:ero:acquire_loop_until}, is that most of the facts we prove will be useful when showing that $p_{\min}$ cannot take infinitely many steps in the loops on lines \ref{line:ero:remove_cell_remove_seal_loop}, \ref{line:ero:remove_cell_remove_repeat_loop}, and \ref{line:ero:acquire_next_repeat_loop} during $opx_{\min}$.

The high-level argument for why $p_{\min}$ cannot take infinitely many steps in the loops on lines \ref{line:ero:add_cell_while_loop}, \ref{line:ero:remove_cell_while_loop}, and \ref{line:ero:acquire_loop_until} during $opx_{\min}$ is the following.
First, we prove that in \Cref{scenario:ero:helper_loop_scenario}, there is a last $L$-event $e_{\min}$ in $\mathcal{I}^\mathcal{B}$, that $e_{\min}$ set $\linearizationobject{} = (\uniquerepositoryoperationshort_\linearizationobject{}, \uniquecellpointershort_\linearizationobject{})$ (the value $p_{\min}$ read), and that $e_{\min} < T^{\ref{line:ero:linearization_read}}_{\min}$.
This implies that the ``shape" of the list is one of two finite lists from $e_{\min}$ onwards (in particular, it is either $\List(\mathcal{I}^{exclude}_{e_{\min}})$ or $\List(\mathcal{I}^{include}_{e_{\min}})$).
Second, we prove that in every iteration of $\mathcal{L}_{\min}$, $p_{\min}$ ``traverses" through a cell from one of these two finite lists (in particular, the response of every AcquireNext procedure is $(\found, \cellpointershort{})$ where $\cellpointershort{}$ is in $\List(\mathcal{I}^{exclude}_{e_{\min}})$ or $\List(\mathcal{I}^{include}_{e_{\min}})$).
Third, we prove that the pointers it traverses through are distinct (in particular, the response of every AcquireNext procedure is $(\found, \cellpointershort{})$ where $\cellpointershort{}$ is different than any pointer previously returned from the AcquireNext procedure during $\mathcal{L}_{\min}$).
The finale is then: since $p_{\min}$ takes infinitely many steps in $\mathcal{L}_{\min}$, it received infinitely many responses from the AcquireNext procedure, and since they are all for different pointers in one of these two lists, we have that there are infinitely many pointers between these two lists, contradicting the fact that they are both finite.

We start by proving that in \Cref{scenario:ero:helper_loop_scenario} $\linearizationobject{}$ is fixed from $T^{\ref{line:ero:linearization_read}}_{\min}$ onwards in $\mathcal{I}^\mathcal{B}$.

\begin{proposition}\label{lemma:ero:opx_is_stuck_implies_l_is_same_infinitely_often}
    In \Cref{scenario:ero:helper_loop_scenario}, $\linearizationobject{}.\uniquerepositoryoperationlong = \uniquerepositoryoperationshort_\linearizationobject{}$ infinitely often in $\mathcal{I}^\mathcal{B}$.
\end{proposition}

\begin{proof}
    Since $\mathcal{L}_{\min}$ is an instance of any loop except the loop on \cref{line:ero:do_work_while_loop}, there are six cases.

    \begin{itemize}
        \item[] \hspace{0pt}\textbf{Case 1.} $\mathcal{L}_{\min}$ is the loop on \cref{line:ero:add_cell_while_loop}.

        Since $p_{\min}$ takes infinitely many steps inside $\mathcal{L}_{\min}$, we have that $p_{\min}$ invokes and exits the AcquireNext procedure on \cref{line:ero:add_cell_acquire_next} infinitely often.
        Hence, by the condition on \cref{line:ero:add_cell_acquire_next_l_changed}, the response of every invocation of the AcquireNext procedure on \cref{line:ero:add_cell_acquire_next} during $\mathcal{L}_{\min}$ returns a value other than $\timechange$.
        Thus, $p_{\min}$ finds the condition on \cref{line:ero:acquire_next_linearization_changed_check} to be false infinitely often.
        Therefore, since $\mathcal{L}_{\min}$ was executed during $I_{\min}$, and the first parameter of $I_{\min}$ is $\uniquerepositoryoperationshort_\linearizationobject{}$, we have that the first parameter of every invocation of the AcquireNext procedure on \cref{line:ero:add_cell_acquire_next} during $\mathcal{L}_{\min}$ is also $\uniquerepositoryoperationshort_\linearizationobject{}$, and so $\linearizationobject{}.\uniquerepositoryoperationlong = \uniquerepositoryoperationshort_\linearizationobject{}$ infinitely often as wanted.

        \item[] \hspace{0pt}\textbf{Case 2.} $\mathcal{L}_{\min}$ is the loop on \cref{line:ero:remove_cell_while_loop}.

        Since $p_{\min}$ takes infinitely many steps inside $\mathcal{L}_{\min}$, we have that $p_{\min}$ invokes and exits the AcquireNext procedure on \cref{line:ero:remove_cell_acquire_next} infinitely often.
        Hence, by the condition on \cref{line:ero:remove_cell_check_acquire_status}, the response of every invocation of the AcquireNext procedure on \cref{line:ero:remove_cell_acquire_next} during $\mathcal{L}_{\min}$ returns a value other than $\timechange$.
        Thus, $p_{\min}$ finds the condition on \cref{line:ero:acquire_next_linearization_changed_check} to be false infinitely often.
        Therefore, since $\mathcal{L}_{\min}$ was executed during $I_{\min}$, and the first parameter of $I_{\min}$ is $\uniquerepositoryoperationshort_\linearizationobject{}$, we have that the first parameter of every invocation of the AcquireNext procedure on \cref{line:ero:remove_cell_acquire_next} during $\mathcal{L}_{\min}$ is also $\uniquerepositoryoperationshort_\linearizationobject{}$, and so $\linearizationobject{}.\uniquerepositoryoperationlong = \uniquerepositoryoperationshort_\linearizationobject{}$ infinitely often as wanted.

        \item[] \hspace{0pt}\textbf{Case 3.} $\mathcal{L}_{\min}$ is the loop on \cref{line:ero:remove_cell_remove_seal_loop}.

        Hence, $I_{\min}$ is an invocation of \doremovecell{} procedure with parameters $(\uniquerepositoryoperationshort_\linearizationobject{}, \uniquecellpointershort_\linearizationobject{})$.
        Thus, by \Cref{lemma:ero:l_event_corresponding_to_do_low_level_op} there is an $L$-remove $e$ for $\cellpointershort{}_\linearizationobject{}$ that set $\linearizationobject{} = (\uniquerepositoryoperationshort_\linearizationobject{}, \uniquecellpointershort_\linearizationobject{})$.
        So, by \Cref{lemma:ero:every_l_event_is_for_pointer_from_universe}, $\cellpointershort{}_\linearizationobject{} \in \celluniverse$.
        We claim that $e$ is the last $L$-event in $\mathcal{I}^\mathcal{B}$ which completes the proof for this case.
        Suppose, for contradiction, there is an $L$-event after $e$ in $\mathcal{I}^\mathcal{B}$.
        Let $e'$ be the next $L$-event after $e$ in $\mathcal{I}^\mathcal{B}$, so $e$ and $e'$ are successive $L$-events in $\mathcal{I}^\mathcal{B}$.
        Hence, since by \Cref{lemma:ero:the_list_invariants_hold} $R(\mathcal{I}^\mathcal{B})$ holds, and $e$ is an $L$-remove event for $\cellpointershort{}_\linearizationobject{}$, we have that there is a successful list-remove attempt for $\cellpointershort{}_\linearizationobject{}$ in $\mathcal{I}^\mathcal{B}$.
        Thus, by \Cref{lemma:ero:list_seal_before_list_remove}, there is a successful list-seal attempt for $\cellpointershort{}_\linearizationobject{}$ in $\mathcal{I}^\mathcal{B}$; say at time $T$.
        So, $(*\cellpointershort{}_\linearizationobject{}).\nextlong.sealed = \true$ at $T$.
        Therefore, since $\cellpointershort{}_\linearizationobject{} \in \celluniverse$, by \Cref{observation:ero:where_objects_change}, only successful list-seal attempts change the value of $(*\cellpointershort{}_\linearizationobject{}).\nextlong.sealed$, we have that from $T$ onwards  $(*\cellpointershort{}_\linearizationobject{}).\nextlong.sealed = \true$.
        However, since $p_{\min}$ takes infinitely many steps inside $\mathcal{L}_{\min}$, we have that $p_{\min}$ finds the condition on \cref{line:ero:remove_cell_remove_seal_loop} to be false infinitely often, and since $\cellpointershort{}_\linearizationobject{}$ is the second parameter of $I_{\min}$, it follows that $(*\cellpointershort{}_\linearizationobject{}).\nextlong.sealed = \false$ infinitely often, a contradiction.

        \item[] \hspace{0pt}\textbf{Case 4.} $\mathcal{L}_{\min}$ is the loop on \cref{line:ero:remove_cell_remove_repeat_loop}.

        Since $p_{\min}$ takes infinitely many steps inside $\mathcal{L}_{\min}$, we have that $p_{\min}$ finds the condition on \cref{line:ero:remove_cell_before_removal_linearization_check} to be false infinitely often.
        Thus, since $\mathcal{L}_{\min}$ was executed during $I_{\min}$, and the first parameter of $I_{\min}$ is $\uniquerepositoryoperationshort_\linearizationobject{}$, we have that $\linearizationobject{}.\uniquerepositoryoperationlong = \uniquerepositoryoperationshort_\linearizationobject{}$ infinitely often as wanted.

        \item[] \hspace{0pt}\textbf{Case 5.} $\mathcal{L}_{\min}$ is the loop on \cref{line:ero:acquire_loop_until}.

        Since $p_{\min}$ takes infinitely many steps inside $\mathcal{L}_{\min}$, we have that $p_{\min}$ invokes and exits the AcquireNext procedure on \cref{line:ero:acquire_acquire_next} infinitely often.
        Hence, by the condition on \cref{line:ero:acquire_loop_until}, the response of every invocation of the AcquireNext procedure on \cref{line:ero:acquire_acquire_next} during $\mathcal{L}_{\min}$ returns a value other than $\timechange$.
        Thus, $p_{\min}$ finds the condition on \cref{line:ero:acquire_next_linearization_changed_check} to be false infinitely often.
        Therefore, since $\mathcal{L}_{\min}$ was executed during $I_{\min}$, and the first parameter of $I_{\min}$ is $\uniquerepositoryoperationshort_\linearizationobject{}$, we have that the first parameter of every invocation of the AcquireNext procedure on \cref{line:ero:acquire_acquire_next} during $\mathcal{L}_{\min}$ is also $\uniquerepositoryoperationshort_\linearizationobject{}$, and so $\linearizationobject{}.\uniquerepositoryoperationlong = \uniquerepositoryoperationshort_\linearizationobject{}$ infinitely often as wanted.
    
        \item[] \hspace{0pt}\textbf{Case 6.} $\mathcal{L}_{\min}$ is the loop on \cref{line:ero:acquire_next_repeat_loop}.

        Since $p_{\min}$ takes infinitely many steps inside $\mathcal{L}_{\min}$, we have that $p_{\min}$ finds the condition on \cref{line:ero:acquire_next_linearization_changed_check} to be false infinitely often.
        Thus, since $\mathcal{L}_{\min}$ was executed during $I_{\min}$, and the first parameter of $I_{\min}$ is $\uniquerepositoryoperationshort_\linearizationobject{}$, we have that $\linearizationobject{}.\uniquerepositoryoperationlong = \uniquerepositoryoperationshort_\linearizationobject{}$ infinitely often as wanted.
        \qH{\Cref{lemma:ero:opx_is_stuck_implies_l_is_same_infinitely_often}}
    \end{itemize}
\end{proof}

\begin{proposition}\label{lemma:ero:l_is_fixed_if_opx_is_stuck}
    In \Cref{scenario:ero:helper_loop_scenario}, $\linearizationobject{}.\uniquerepositoryoperationlong = \uniquerepositoryoperationshort_\linearizationobject{}$ from $T^{\ref{line:ero:linearization_read}}_{\min}$ onwards in $\mathcal{I}^\mathcal{B}$.
\end{proposition}

\begin{proof}
    Suppose, for contradiction, at some time $T$ after $T^{\ref{line:ero:linearization_read}}_{\min}$ in $\mathcal{I}^\mathcal{B}$ that $\linearizationobject{}.\uniquerepositoryoperationlong \neq \uniquerepositoryoperationshort_\linearizationobject{}$.
    Hence, by \Cref{lemma:ero:opx_is_stuck_implies_l_is_same_infinitely_often}, $\linearizationobject{}.\uniquerepositoryoperationlong = \uniquerepositoryoperationshort_\linearizationobject{}$ some time after $T$; say $T'$.
    Thus, since $\linearizationobject{}.\uniquerepositoryoperationlong{} \neq \uniquerepositoryoperationshort_\linearizationobject{}$ at $T$ and $\linearizationobject{}.\uniquerepositoryoperationlong{} = \uniquerepositoryoperationshort_\linearizationobject{}$ at $T'$, we have that the value of $\linearizationobject{}.\uniquerepositoryoperationlong$ was set to $\uniquerepositoryoperationshort_\linearizationobject{}$ between $T$ and $T'$, and so by \Cref{observation:ero:where_objects_change}, some $L$-event $e'$ set $\linearizationobject{}.\uniquerepositoryoperationlong = \uniquerepositoryoperationshort_\linearizationobject{}$ between $T$ and $T'$.
    So, by \Cref{lemma:ero:l_events_set_llo_to_a_value_different_from_initial}, $\uniquerepositoryoperationshort_\linearizationobject{} \neq (0, \noop)$.
    Hence, since $\linearizationobject{}.\uniquerepositoryoperationlong = \uniquerepositoryoperationshort_\linearizationobject{}$ at $T^{\ref{line:ero:linearization_read}}$, it follows that $\linearizationobject{}.\uniquerepositoryoperationlong$ was set to $\uniquerepositoryoperationshort_\linearizationobject{}$ before $T^{\ref{line:ero:linearization_read}}$, and so by \Cref{observation:ero:where_objects_change}, some $L$-event $e$ set $\linearizationobject{}.\uniquerepositoryoperationlong = \uniquerepositoryoperationshort_\linearizationobject{}$ before $T^{\ref{line:ero:linearization_read}}$.
    Thus, since $e < T^{\ref{line:ero:linearization_read}}$, $T^{\ref{line:ero:linearization_read}} < T$, and $T < e'$, by transitivity, $e < e'$, so $e \neq e'$.
    Therefore, two different $L$-events in $\mathcal{I}^\mathcal{B}$ set $\linearizationobject{}.\uniquerepositoryoperationlong$ to the same value (namely $\uniquerepositoryoperationshort_\linearizationobject{}$).
    However, since by \Cref{lemma:ero:the_list_invariants_hold} $P(\mathcal{I}^\mathcal{B})$ holds, by \Cref{lemma:ero:p_implies_unique_low_level_operations_in_linearization}, every $L$-event in $\mathcal{I}^\mathcal{B}$ sets $\linearizationobject{}.\uniquerepositoryoperationlong$ to a unique value, a contradiction.
    \qH{\Cref{lemma:ero:l_is_fixed_if_opx_is_stuck}}
\end{proof}

\begin{proposition}\label{lemma:ero:if_read_initial_there_are_no_l_events}
    In \Cref{scenario:ero:helper_loop_scenario}, if $\uniquerepositoryoperationshort_\linearizationobject{} = (0, \noop)$, then there are no $L$-event in $\mathcal{I}^\mathcal{B}$.
\end{proposition}

\begin{proof}
    Suppose, for contradiction, $\uniquerepositoryoperationshort_\linearizationobject{} = (0, \noop)$ and there is an $L$-event $e$ in $\mathcal{I}^\mathcal{B}$.
    Hence, by \Cref{lemma:ero:l_events_set_llo_to_a_value_different_from_initial}, $e$ sets $\linearizationobject{}.\uniquerepositoryoperationlong \neq (0, \noop)$.
    Thus, since $\uniquerepositoryoperationshort_\linearizationobject{} = (0, \noop)$, by \Cref{lemma:ero:l_is_fixed_if_opx_is_stuck}, $\linearizationobject{}.\uniquerepositoryoperationlong = (0, \noop)$ from $T^{\ref{line:ero:linearization_read}}_{\min}$ onwards in $\mathcal{I}^\mathcal{B}$, and so $e < T^{\ref{line:ero:linearization_read}}_{\min}$.
    Since (1) $\linearizationobject{}.\uniquerepositoryoperationlong \neq (0, \noop)$ at $e$, (2) $\linearizationobject{}.\uniquerepositoryoperationlong = (0, \noop)$ at $T^{\ref{line:ero:linearization_read}}_{\min}$, and (3) $e < T^{\ref{line:ero:linearization_read}}_{\min}$, we have that $\linearizationobject{}.\uniquerepositoryoperationlong$ was set to $(0, \noop)$ between $e$ and $T^{\ref{line:ero:linearization_read}}_{\min}$.
    Therefore, by \Cref{observation:ero:where_objects_change}, an $L$-event set $\linearizationobject{}.\uniquerepositoryoperationlong = (0, \noop)$.
    However, by \Cref{lemma:ero:l_events_set_llo_to_a_value_different_from_initial}, every $L$-event sets $\linearizationobject{}.\uniquerepositoryoperationlong \neq (0, \noop)$, a contradiction.
    \qH{\Cref{lemma:ero:if_read_initial_there_are_no_l_events}}
\end{proof}

\begin{lemma}\label{lemma:ero:i_first_parameter_is_not_initial}
    In \Cref{scenario:ero:helper_loop_scenario}, $\uniquerepositoryoperationshort_\linearizationobject{} \neq (0, \noop)$.
\end{lemma}

\begin{proof}
    There are two cases.

    \begin{itemize}
        \item[] \hspace{0pt}\textbf{Case 1.} $\mathcal{L}_{\min}$ is the loop on line \ref{line:ero:add_cell_while_loop}, \ref{line:ero:remove_cell_while_loop}, \ref{line:ero:remove_cell_remove_seal_loop}, or \ref{line:ero:remove_cell_remove_repeat_loop}.

        Hence, $I_{\min}$ is an invocation of the \doaddcell{} or \doremovecell{} procedure.
        Thus, since $\uniquerepositoryoperationshort_\linearizationobject{}$ is the first parameter of $I_{\min}$, by the conditions on lines \ref{line:ero:do_add_cell_condition} and \ref{line:ero:do_remove_cell_condition}, we have that $\uniquerepositoryoperationshort_\linearizationobject{}$ equals $(\arbitraryvalue, \addcell)$ or $(\arbitraryvalue, \removecell)$.
        Therefore, the claim follows.

        \item[] \hspace{0pt}\textbf{Case 2.} $\mathcal{L}_{\min}$ is the loop on line  \ref{line:ero:acquire_loop_until} or \ref{line:ero:acquire_next_repeat_loop}.

        Suppose, for contradiction, $\uniquerepositoryoperationshort_\linearizationobject{} = (0, \noop)$.
        
        We first define an invocation $I$ of the AcquireNext procedure by $p_{\min}$ such that $p_{\min}$ finds the condition on \cref{line:ero:acquire_next_not_found_check} to be false some time during $I$. 
        If $\mathcal{L}_{\min}$ is the loop on \cref{line:ero:acquire_loop_until}, then since $p_{\min}$ takes infinitely many steps in $\mathcal{L}_{\min}$, we have that $p_{\min}$ invokes and exits the AcquireNext procedure on \cref{line:ero:acquire_acquire_next} during $\mathcal{L}_{\min}$ infinitely often.
        Let $I$ be any of these invocations.
        Since $p_{\min}$ invokes and exits the AcquireNext procedure on \cref{line:ero:acquire_acquire_next} during $\mathcal{L}_{\min}$ infinitely often, we have that $I$'s response is of the form $(\found, \arbitraryvalue)$ as otherwise, $p_{\min}$ would find the condition on \cref{line:ero:acquire_not_found_check} to be true during $\mathcal{L}_{\min}$ implying $p_{\min}$ would exit $\mathcal{L}_{\min}$.
        Hence, $p_{\min}$ finds the condition on \cref{line:ero:acquire_next_not_found_check} to be false some time during $I$ (otherwise it would return $(\notfound, \arbitraryvalue)$).
        If $\mathcal{L}_{\min}$ is the loop on \cref{line:ero:acquire_next_repeat_loop}, then $I_{\min}$ is an invocation of the AcquireNext procedure.
        We let $I = I_{\min}$.
        Since $p_{\min}$ takes infinitely many steps during $\mathcal{L}_{\min}$ in $I$, it immediately follows that $p_{\min}$ finds the condition on \cref{line:ero:acquire_next_not_found_check} to be false some time during $I$ (otherwise $p_{\min}$ would exit $\mathcal{L}_{\min}$).
        
        We now finish the proof for Case 2.
        Let $\currentcellpointershort{}$ be the second parameter of $I$.
        Hence, by \Cref{lemma:ero:acquire_next_is_for_pointer_from_universe_or_head}, $\currentcellpointershort{} \in \celluniverse \cup \{\&\headobject\}$.
        %Thus, the initial value of $(*\currentcellpointershort{}).\nextlong.\cellpointerlong{}$ is $\nullconstant$.
        Furthermore, since $p_{\min}$ finds the condition on \cref{line:ero:acquire_next_not_found_check} to be false sometime during $I$, we have that $(*\currentcellpointershort{}).\nextlong.\cellpointerlong{} \neq \nullconstant$ sometime during $\mathcal{I}^\mathcal{B}$.
        Hence, since $\currentcellpointershort{} \in \celluniverse \cup \{\&\headobject\}$, the value of $(*\currentcellpointershort{}).\nextlong.\cellpointerlong{}$ is initially $\nullconstant$, and so it changed during $\mathcal{I}^\mathcal{B}$.
        Thus, by \Cref{observation:ero:where_objects_change}, there is a successful list-add or list-remove attempt in $\mathcal{I}^\mathcal{B}$.
        Therefore, by \Cref{lemma:ero:l_event_corresponding_to_do_low_level_op}, there is an $L$-event in $\mathcal{I}^\mathcal{B}$.
        However, since $\uniquerepositoryoperationshort_\linearizationobject{} = (0, \noop)$, by \Cref{lemma:ero:if_read_initial_there_are_no_l_events}, there are no $L$-events in $\mathcal{I}^\mathcal{B}$, a contradiction.
        \qH{\Cref{lemma:ero:i_first_parameter_is_not_initial}}
    \end{itemize}
\end{proof}

\begin{lemma}\label{lemma:ero:i_is_proceeded_by_a_unique_l_event}
    In \Cref{scenario:ero:helper_loop_scenario}, there is an $L$-event $e_{\min}$ in $\mathcal{I}^\mathcal{B}$ which set $\linearizationobject{} = (\uniquerepositoryoperationshort_\linearizationobject{}, \uniquecellpointershort_\linearizationobject{})$ such that (1) $e_{\min}$ is the only $L$-event in $\mathcal{I}^\mathcal{B}$ which set $\linearizationobject{} = (\uniquerepositoryoperationshort_\linearizationobject{}, \uniquecellpointershort_\linearizationobject{})$ and (2) $e_{\min} < T^{\ref{line:ero:linearization_read}}_{\min}$.
\end{lemma}

\begin{proof}
    Since $p$ read $\linearizationobject{} = (\uniquerepositoryoperationshort_\linearizationobject{}, \uniquecellpointershort_\linearizationobject{})$ at $T^{\ref{line:ero:linearization_read}}_{\min}$, and by \Cref{lemma:ero:i_first_parameter_is_not_initial} $\uniquerepositoryoperationshort_\linearizationobject{} \neq (0, \noop)$, we have that $\linearizationobject{}$ was set to $(\uniquerepositoryoperationshort_\linearizationobject{}, \uniquecellpointershort_\linearizationobject{})$ before $T^{\ref{line:ero:linearization_read}}_{\min}$.
    Hence, by \Cref{observation:ero:where_objects_change}, some $L$-event $e_{\min}$ set $\linearizationobject{} = (\uniquerepositoryoperationshort_\linearizationobject{}, \uniquecellpointershort_\linearizationobject{})$ before $T^{\ref{line:ero:linearization_read}}_{\min}$.
    Thus, since by \Cref{lemma:ero:the_list_invariants_hold} $P(\mathcal{I}^\mathcal{B})$ holds, by \Cref{lemma:ero:p_implies_unique_values_in_linearization}, $e_{\min}$ is the only $L$-event in $\mathcal{I}^\mathcal{B}$ that sets  $\linearizationobject{} = (\uniquerepositoryoperationshort_\linearizationobject{}, \uniquecellpointershort_\linearizationobject{})$.
    \qH{\Cref{lemma:ero:i_is_proceeded_by_a_unique_l_event}}
\end{proof}

\begin{lemma}\label{lemma:ero:e_is_the_last_l_event}
    In \Cref{scenario:ero:helper_loop_scenario}, $e_{\min}$ is the last $L$-event in $\mathcal{I}^\mathcal{B}$.
\end{lemma}

\begin{proof}
    Suppose, for contradiction, there is an $L$-event $e$ after $e_{\min}$ in $\mathcal{I}^\mathcal{B}$.
    There are two cases.
    \begin{itemize}
        \item[] \hspace{0pt}\textbf{Case 1.} $e < T^{\ref{line:ero:linearization_read}}_{\min}$.

        Hence, $e_{\min} < e < T^{\ref{line:ero:linearization_read}}_{\min}$.
        Since by \Cref{lemma:ero:the_list_invariants_hold} $P(\mathcal{I}^\mathcal{B})$ holds, by \Cref{lemma:ero:p_implies_unique_low_level_operations_in_linearization}, $e$ sets $\linearizationobject{}.\uniquerepositoryoperationlong$ to a different value than $e_{\min}$.
        Hence, since by \Cref{lemma:ero:i_is_proceeded_by_a_unique_l_event} $e_{\min}$ sets $\linearizationobject{}.\uniquerepositoryoperationlong = \uniquerepositoryoperationshort_\linearizationobject{}$, we have that $\linearizationobject{}.\uniquerepositoryoperationlong \neq \uniquerepositoryoperationshort_\linearizationobject{}$ at $e$.
        Thus, since $\linearizationobject{}.\uniquerepositoryoperationlong = \uniquerepositoryoperationshort_\linearizationobject{}$ at $T^{\ref{line:ero:linearization_read}}_{\min}$, we have that $\linearizationobject{}.\uniquerepositoryoperationlong$ was set to $\uniquerepositoryoperationshort_\linearizationobject{}$ between $e$ and $T^{\ref{line:ero:linearization_read}}_{\min}$.
        So, by \Cref{observation:ero:where_objects_change}, some $L$-event $e'$ set $\linearizationobject{}.\uniquerepositoryoperationlong = \uniquerepositoryoperationshort_\linearizationobject{}$ between $e$ and $T^{\ref{line:ero:linearization_read}}_{\min}$.
        Therefore, since $e_{\min} < e$ and $e < e'$, we have that $e_{\min} \neq e'$, and so there are two $L$-events in $\mathcal{I}^\mathcal{B}$ which set $\linearizationobject{}.\uniquerepositoryoperationlong = \uniquerepositoryoperationshort_\linearizationobject{}$.
        However, since by \Cref{lemma:ero:the_list_invariants_hold} $P(\mathcal{I}^\mathcal{B})$ holds, by \Cref{lemma:ero:p_implies_unique_low_level_operations_in_linearization}, every $L$-event sets $\linearizationobject{}.\uniquerepositoryoperationlong$ to a unique value, a contradiction.

        \item[] \hspace{0pt}\textbf{Case 2.} $T^{\ref{line:ero:linearization_read}}_{\min} < e$.

        Hence, since by \Cref{lemma:ero:i_is_proceeded_by_a_unique_l_event} $e_{\min} < T^{\ref{line:ero:linearization_read}}_{\min}$, by transitivity, $e_{\min} < T^{\ref{line:ero:linearization_read}}_{\min} < e$.
        Since by \Cref{lemma:ero:the_list_invariants_hold} $P(\mathcal{I}^\mathcal{B})$ holds, by \Cref{lemma:ero:p_implies_unique_low_level_operations_in_linearization}, $e$ sets $\linearizationobject{}.\uniquerepositoryoperationlong$ to a different value than $e_{\min}$.
        Therefore, since by \Cref{lemma:ero:i_is_proceeded_by_a_unique_l_event} $e_{\min}$ sets $\linearizationobject{}.\uniquerepositoryoperationlong = \uniquerepositoryoperationshort_\linearizationobject{}$, we have that $\linearizationobject{}.\uniquerepositoryoperationlong \neq \uniquerepositoryoperationshort_\linearizationobject{}$ at $e$.
        However, by \Cref{lemma:ero:l_is_fixed_if_opx_is_stuck}, from $T^{\ref{line:ero:linearization_read}}_{\min}$ onwards $\linearizationobject{}.\uniquerepositoryoperationlong = \uniquerepositoryoperationshort_\linearizationobject{}$, a contradiction.
        \qH{\Cref{lemma:ero:e_is_the_last_l_event}}
    \end{itemize}
\end{proof}

\begin{proposition}\label{lemma:ero:from_e_onwards_the_list_is_one_of_two_structures}
    In \Cref{scenario:ero:helper_loop_scenario}, consider any finite prefix $\mathcal{I}$ of $\mathcal{I}^\mathcal{B}$ at or after $e_{\min}$.
    The list of cells conforms to either $\List(\mathcal{I}^{exclude}_{e_{\min}})$ or $\List(\mathcal{I}^{include}_{e_{\min}})$ in $\mathcal{I}$ where $\mathcal{I}^{exclude}_{e_{\min}}$ is the prefix of $\mathcal{I}^\mathcal{B}$ up to but excluding $e_{\min}$ and $\mathcal{I}^{include}_{e_{\min}}$ is the prefix of $\mathcal{I}^\mathcal{B}$ up to and including $e_{\min}$.
\end{proposition}

\begin{proof}
    By \Cref{lemma:ero:e_is_the_last_l_event} $e_{\min}$ is the last $L$-event in $\mathcal{I}^\mathcal{B}$ and since $\mathcal{I}$ is a prefix of $\mathcal{I}^\mathcal{B}$ at or after $e_{\min}$, we have that $e_{\min}$ is the last $L$-event in $\mathcal{I}$.
    Since $e_{\min}$ is the last $L$-event in $\mathcal{I}$, and by \Cref{lemma:ero:the_list_invariants_hold} $P(\mathcal{I}^\mathcal{B})$, $Q(\mathcal{I}^\mathcal{B})$, and $R(\mathcal{I}^\mathcal{B})$ hold, by \Cref{lemma:ero:conditional_classification_lemma}, the list of cells conforms to either $\List(\mathcal{I}^{exclude}_{e_{\min}})$ or $\List(\mathcal{I})$ in $\mathcal{I}$.
    Furthermore, since $\mathcal{I}$ is a prefix of $\mathcal{I}^\mathcal{B}$ at or after $e_{\min}$, $e_{\min}$ is the last $L$-event in $\mathcal{I}$, and $\mathcal{I}^{include}_{e_{\min}}$ is the prefix of $\mathcal{I}^\mathcal{B}$ up to and including $e_{\min}$, it follows that the sequence of $L$-events is identical in $\mathcal{I}$ and $\mathcal{I}^{include}_{e_{\min}}$, so by \Cref{def:ero:logical_list} $\List(\mathcal{I}) = \List(\mathcal{I}^{include}_{e_{\min}})$.
    Therefore, the list of cells conforms to either $\List(\mathcal{I}^{exclude}_{e_{\min}})$ or $\List(\mathcal{I}^{include}_{e_{\min}})$ in $\mathcal{I}$ as wanted.
    \qH{\Cref{lemma:ero:from_e_onwards_the_list_is_one_of_two_structures}}
\end{proof}

This completes the first part of the high-level argument for why $p_{\min}$ cannot take infinitely many steps in the loops on lines \ref{line:ero:add_cell_while_loop}, \ref{line:ero:remove_cell_while_loop}, and \ref{line:ero:acquire_loop_until} during $opx_{\min}$.
We now prove that in every iteration of $\mathcal{L}_{\min}$, $p_{\min}$ ``traverses" through a cell from one of these two finite lists (in particular, the response of every AcquireNext procedure is $(\found, \cellpointershort{})$ where $\cellpointershort{}$ is in $\List(\mathcal{I}^{exclude}_{e_{\min}})$ or $\List(\mathcal{I}^{include}_{e_{\min}})$).

\begin{lemma}\label{lemma:ero:if_q_read_e_s_value_then_its_after_e}
    In \Cref{scenario:ero:helper_loop_scenario}, if a process read $(\uniquerepositoryoperationshort_\linearizationobject{}, \uniquecellpointershort{}_\linearizationobject{})$ from $\linearizationobject{}$ at time $T$ in $\mathcal{I}^\mathcal{B}$, then \mbox{$e_{\min} < T$.}
\end{lemma}

\begin{proof}
    Since a process read $(\uniquerepositoryoperationshort_\linearizationobject{}, \uniquecellpointershort{}_\linearizationobject{})$ from $\linearizationobject{}$ at time $T$, $\linearizationobject{}.\uniquerepositoryoperationlong$ is initially $(0, \noop)$, and by \Cref{lemma:ero:i_first_parameter_is_not_initial} $\uniquerepositoryoperationshort_\linearizationobject{} \neq (0, \noop)$, we have that some step set $\linearizationobject{}$ to $(\uniquerepositoryoperationshort_\linearizationobject{}, \uniquecellpointershort{}_\linearizationobject{})$ before $T$.
    Hence, by \Cref{observation:ero:where_objects_change}, some $L$-event $e$ set $\linearizationobject{} = (\uniquerepositoryoperationshort_\linearizationobject{}, \uniquecellpointershort{}_\linearizationobject{})$ before $T$.
    Therefore, by \Cref{lemma:ero:i_is_proceeded_by_a_unique_l_event}, $e = e_{\min}$, and so $e_{\min} < T$ as wanted.
    \qH{\Cref{lemma:ero:if_q_read_e_s_value_then_its_after_e}}
\end{proof}

There are three claims that allow us to conclude that in every iteration of $\mathcal{L}_{\min}$, $p_{\min}$ ``traverses" through a cell from either $\List(\mathcal{I}^{exclude}_{e_{\min}})$ or $\List(\mathcal{I}^{include}_{e_{\min}})$.
% They are \Cref{lemma:ero:acquire_next_for_ptr_in_exclude_is_in_include}, \Cref{lemma:ero:acquire_next_for_ptr_not_in_exclude_implies_a_prefix_that_does_not_conform_to_exclude}, and \Cref{lemma:ero:if_the_list_is_only_include_after_e_it_is_always_include}.
The first of which is \Cref{lemma:ero:acquire_next_for_ptr_in_exclude_is_in_include}.
Roughly speaking, \Cref{lemma:ero:acquire_next_for_ptr_in_exclude_is_in_include} asserts that if $p_{\min}$ tries to acquire a pointer after some $\cellpointershort{}$ in $\List(\mathcal{I}^{exclude}_{e_{\min}})$, then $\cellpointershort{}$ is in $\List(\mathcal{I}^{include}_{e_{\min}})$.
This is useful for the following reason.
Suppose $\cellpointershort{}$ is in $\List(\mathcal{I}^{exclude}_{e_{\min}})$ but at the moment $p_{\min}$ reads $(*\cellpointershort{}).\nextlong.\cellpointerlong{}$ on \cref{line:ero:acquire_next_read_curr_unique_pointer}, the list of cells conforms to $\List(\mathcal{I}^{include}_{e_{\min}})$.
If $\cellpointershort{}$ is not in $\List(\mathcal{I}^{include}_{e_{\min}})$, then the fact that the list of cells conforms to $\List(\mathcal{I}^{include}_{e_{\min}})$ tells us nothing about the value that $p_{\min}$ read from $(*\cellpointershort{}).\nextlong.\cellpointerlong{}$ on \cref{line:ero:acquire_next_read_curr_unique_pointer}.
\Cref{lemma:ero:acquire_next_for_ptr_in_exclude_is_in_include} lets us avoid this problem entirely because the fact that $\cellpointershort{}$ is in $\List(\mathcal{I}^{exclude}_{e_{\min}})$ tells us that $\cellpointershort{}$ is in $\List(\mathcal{I}^{include}_{e_{\min}})$, and since by \Cref{lemma:ero:from_e_onwards_the_list_is_one_of_two_structures} the list of cells conforms to $\List(\mathcal{I}^{exclude}_{e_{\min}})$ or $\List(\mathcal{I}^{include}_{e_{\min}})$, we always know that the value that $p_{\min}$ read from $(*\cellpointershort{}).\nextlong.\cellpointerlong{}$ on \cref{line:ero:acquire_next_read_curr_unique_pointer} comes from $\List(\mathcal{I}^{exclude}_{e_{\min}})$ or $\List(\mathcal{I}^{include}_{e_{\min}})$.
We now prove \Cref{lemma:ero:acquire_next_for_ptr_in_exclude_is_in_include}.

\begin{proposition}\label{lemma:ero:acquire_next_pointer_is_from_include_list}
    In \Cref{scenario:ero:helper_loop_scenario}, consider any invocation $I$ of the \doaddcell, \doremovecell, or Acquire procedure and denote the time that $I$ was invoked in $\mathcal{I}^\mathcal{B}$ by $T_b$.
    Let $I_1, I_2, \ldots$ be the (possibly infinite) sequence of invocations of the AcquireNext procedure during $I$ in the order they are invoked.\footnote{More precisely, if $p$ is the process that invoked $I$, then $I_1, I_2, \ldots$ is the (possibly infinite) sequence of invocations of the AcquireNext procedure by $p$ during $I$ in the order they are invoked.}
    If for every finite prefix $\mathcal{I}$ of $\mathcal{I}^\mathcal{B}$ at or after $T_b$ the list of cells conforms to $\List(\mathcal{I}^{include}_{e_{\min}})$ in $\mathcal{I}$, the second parameter of $I_i$ is in $\List(\mathcal{I}^{include}_{e_{\min}})$ for every $i$ (assuming $I_i$ exists).
\end{proposition}

\begin{proof}
    By induction on $i$.

    \begin{itemize}
        \item[] \hspace{0pt}\textbf{Base Case.} $i = 1$.

        Hence, the second parameter of $I_1$ is $\&\headobject$
        By \Cref{def:ero:logical_list}, $\&\headobject$ is the first element of $\List(\mathcal{I})$ for every finite implementation history $\mathcal{I}$, so the first element of $\List(\mathcal{I}^{include}_{e_{\min}})$ is $\&\headobject$.
        Therefore, the second parameter of $I_i$ is in $\List(\mathcal{I}^{include}_{e_{\min}})$ as wanted.

        \item[] \hspace{0pt}\textbf{Inductive Case.} For every $i \geq 1$ if the second parameter of $I_i$ is in $\List(\mathcal{I}^{include}_{e_{\min}})$, then the second parameter of $I_{i + 1}$ is in $\List(\mathcal{I}^{include}_{e_{\min}})$ (assuming $I_{i + 1}$ exists).
        
        Suppose for some $i \geq 1$ that the second parameter of $I_i$ is in $\List(\mathcal{I}^{include}_{e_{\min}})$.
        This is the inductive hypothesis.
        Since $I_{i + 1}$ exists, it follows that the response of $I_i$ is $(\found, \cellpointershort{}_{i + 1})$, and the second parameter of $I_{i + 1}$ is $\cellpointershort{}_{i + 1}$.
        Let $\cellpointershort{}_i$ be the second parameter of $I_i$, so by the inductive hypothesis, $\cellpointershort{}_i$ is in $\List(\mathcal{I}^{include}_{e_{\min}})$.
        Furthermore, by \Cref{lemma:ero:acquire_next_is_for_pointer_from_universe_or_head}, $\cellpointershort{}_i \in \celluniverse \cup \{\&\headobject\}$.
        Hence, by \Cref{assumption:ero:head_and_null_not_in_cell_universe} $\cellpointershort{}_i \neq \nullconstant$, and so since by \Cref{def:ero:logical_list} $\nullconstant$ is the last element of  $\List(\mathcal{I}^{include}_{e_{\min}})$, we have that $\cellpointershort{}_i$ is not the last element of $\List(\mathcal{I}^{include}_{e_{\min}})$.
        Since $(\found, \cellpointershort{}_{i + 1})$ is the response of $I_{i}$, we have that $(*\currentcellpointershort{}_i).\nextlong.\cellpointerlong{} = \cellpointershort{}_{i + 1}$ at the time of the last execution of \cref{line:ero:acquire_next_read_curr_unique_pointer} during $I_i$; say at time $T^{\ref{line:ero:acquire_next_read_curr_unique_pointer}}$.
        Since $T^{\ref{line:ero:acquire_next_read_curr_unique_pointer}}$ is during $I_i$ and $I_i$ is during $I$, we have that $T^{\ref{line:ero:acquire_next_read_curr_unique_pointer}}$ is during $I$, and so $T^{\ref{line:ero:acquire_next_read_curr_unique_pointer}}$ is after $T_b$.
        Hence, there is a prefix of $\mathcal{I}^\mathcal{B}$ at or after $T_b$ and up to and including $T^{\ref{line:ero:acquire_next_read_curr_unique_pointer}}$; say $\mathcal{I}^{\ref{line:ero:acquire_next_read_curr_unique_pointer}}$.
        Thus, since $(*\cellpointershort{}_{i}).\nextlong.\cellpointerlong{} = \cellpointershort{}_{i + 1}$ at $T^{\ref{line:ero:acquire_next_read_curr_unique_pointer}}$, we have that $(*\cellpointershort{}_{i}).\nextlong.\cellpointerlong{} = \cellpointershort{}_{i + 1}$ at the end of $\mathcal{I}^{\ref{line:ero:acquire_next_read_curr_unique_pointer}}$.        
        Therefore, since by assumption the list of cells conforms to $\List(\mathcal{I}^{include}_{e_{\min}})$ in $\mathcal{I}^{\ref{line:ero:acquire_next_read_curr_unique_pointer}}$, $\cellpointershort{}_{i}$ is in $\List(\mathcal{I}^{include}_{e_{\min}})$, $\cellpointershort{}_{i}$ is not the last element of $\List(\mathcal{I}^{include}_{e_{\min}})$, and $(*\cellpointershort{}_{i}).\nextlong.\cellpointerlong{} = \cellpointershort{}_{i + 1}$ at the end of $\mathcal{I}^{\ref{line:ero:acquire_next_read_curr_unique_pointer}}$, by \Cref{def:ero:logical_list}, $\cellpointershort{}_{i + 1}$ is in $\List(\mathcal{I}^{include}_{e_{\min}})$ as wanted.
        \qH{\Cref{lemma:ero:acquire_next_pointer_is_from_include_list}}
    \end{itemize}
\end{proof}

\begin{lemma}\label{lemma:ero:if_e_is_a_remove_event_then_no_one_acquires_it}
    In \Cref{scenario:ero:helper_loop_scenario}, suppose $e_{\min}$ is an $L$-remove event for $\cellpointershort{}_\linearizationobject{}$.
    Consider any invocation $I$ of the AcquireNext procedure in $\mathcal{I}^\mathcal{B}$ such that the process which invoked $I$ read $(\uniquerepositoryoperationshort_\linearizationobject{}, \uniquecellpointershort_\linearizationobject{})$ from $\linearizationobject{}$ on its last execution of \cref{line:ero:linearization_read} before invoking $I$.
    Then, the second parameter of $I$ is not $\cellpointershort{}_\linearizationobject{}$.
\end{lemma}

\begin{proof}
    Suppose, for contradiction, that there is an invocation $I$ of the AcquireNext procedure such that the process $p$ which invoked $I$ read $(\uniquerepositoryoperationshort_\linearizationobject{}, \uniquecellpointershort_\linearizationobject{})$ from $\linearizationobject{}$ on its last execution of \cref{line:ero:linearization_read} before invoking $I$ and the second parameter of $I$ is $\cellpointershort{}_\linearizationobject{}$.
    Since $e_{\min}$ is an $L$-remove event for $\cellpointershort{}_\linearizationobject{}$, by \Cref{lemma:ero:every_l_event_is_for_pointer_from_universe}, $\cellpointershort{}_\linearizationobject{} \in \celluniverse$.
    Furthermore, since by \Cref{lemma:ero:i_is_proceeded_by_a_unique_l_event} $e_{\min}$ set $\linearizationobject{} = (\uniquerepositoryoperationshort_\linearizationobject{}, \cellpointershort{}_\linearizationobject{})$, by \Cref{def:ero:english}, $(\uniquerepositoryoperationshort_\linearizationobject{}, \uniquecellpointershort_\linearizationobject{})$ is of the form $((\arbitraryvalue, \removecell), \cellpointershort{}_\linearizationobject{})$.
    Hence, since $p$ read $(\uniquerepositoryoperationshort_\linearizationobject{}, \uniquecellpointershort_\linearizationobject{})$ from $\linearizationobject{}$ on its last execution of \cref{line:ero:linearization_read} before invoking $I$, we have that $p$ invoked $I$ either during an invocation of the \doremovecell{} procedure on \cref{line:ero:do_remove_cell} or during an invocation of the Acquire procedure on \cref{line:ero:announce_acquire}.
    In the first case, it follows that $p$ invoked $I$ either: (1) during an invocation of the Acquire procedure during an invocation of the \setrepositoryoperationresponse{} procedure on \cref{line:ero:remove_cell_set_response}; or (2) on \cref{line:ero:remove_cell_acquire_next}.
    Therefore, $I$ is invoked either: (1) during an invocation of the Acquire procedure during an invocation of the \setrepositoryoperationresponse{} procedure on \cref{line:ero:remove_cell_set_response}; (2) on \cref{line:ero:remove_cell_acquire_next}; or (3) during an invocation of the Acquire procedure on \cref{line:ero:announce_acquire}.
    We consider each separately.

    \begin{itemize}
        \item[] \hspace{0pt}\textbf{Case 1.} $I$ was invoked during an invocation $I_{parent}$ of the Acquire procedure during an invocation of the \setrepositoryoperationresponse{} procedure on \cref{line:ero:remove_cell_set_response}.

        Hence, since $p$ read $(\uniquerepositoryoperationshort_\linearizationobject{}, \uniquecellpointershort_\linearizationobject{})$ from $\linearizationobject{}$ on its last execution of \cref{line:ero:linearization_read} before invoking $I$, it follows that the parameters of $I_{parent}$ are $(\uniquerepositoryoperationshort_\linearizationobject{}, \uniquecellpointershort_\linearizationobject{})$.
        Furthermore, since $\uniquecellpointershort_\linearizationobject{}$ is the second parameter of $I$, it follows that  $\uniquecellpointershort_\linearizationobject{}$ is either $\&\headobject$ or $(\found, \uniquecellpointershort_\linearizationobject{})$ is the response of an invocation $I'$ of the AcquireNext procedure on \cref{line:ero:acquire_acquire_next} during $I_{parent}$ such that $I'$ exited before $I$ was invoked.
        However, since $\uniquecellpointershort_\linearizationobject{} \in \celluniverse$, by \Cref{assumption:ero:head_and_null_not_in_cell_universe}, $\uniquecellpointershort_\linearizationobject{} \neq \&\headobject$, so the latter is the only possibility.
        Hence, $p$ executed \cref{line:ero:acquire_loop_until} between when $I'$ exited and when $I$ was invoked.
        Let $T^{\ref{line:ero:acquire_found_check}}$ and $T^{\ref{line:ero:acquire_loop_until}}$ denote the next time $p$ execute lines \ref{line:ero:acquire_found_check} and \ref{line:ero:acquire_loop_until} after exiting $I'$ during $I_{parent}$.
        Since the response of $I'$ is $(\found, \uniquecellpointershort_\linearizationobject{})$, $p$ finds the condition on \cref{line:ero:acquire_found_check} to be true at $T^{\ref{line:ero:acquire_found_check}}$ and so $p$ sets its local variable $\currentcellpointershort{}$ to $\uniquecellpointershort_\linearizationobject{}$ on \cref{line:ero:acquire_update_current_unique_pointer}.
        Hence, since the parameters of $I_{parent}$ are $(\uniquerepositoryoperationshort_\linearizationobject{}, \uniquecellpointershort_\linearizationobject{})$, we have that $p$ finds the condition on \cref{line:ero:acquire_loop_until} to be false at $T^{\ref{line:ero:acquire_loop_until}}$.
        Therefore, $p$ exits the loop on \cref{line:ero:acquire_loop_until} during $I_{parent}$, and so $I'$ is the last invocation of the AcquireNext procedure during $I_{parent}$.
        However, since $I$ is an invocation of the AcquireNext procedure during $I_{parent}$ and $I'$ exited before $I$ was invoked, we have that there is an invocation of the AcquireNext procedure after $I'$ during $I_{parent}$, a contradiction.

        \item[] \hspace{0pt}\textbf{Case 2.} $I$ was invoked on \cref{line:ero:remove_cell_acquire_next}.

        Let $I_{parent}$ be the invocation of the \doremovecell{} procedure that $p$ invoked $I'$ during.
        Since $p$ read $(\uniquerepositoryoperationshort_\linearizationobject{}, \uniquecellpointershort_\linearizationobject{})$ from $\linearizationobject{}$ on its last execution of \cref{line:ero:linearization_read} before invoking $I$, we have that the parameters of $I_{parent}$ are $(\uniquerepositoryoperationshort_\linearizationobject{}, \uniquecellpointershort_\linearizationobject{})$.
        Furthermore, since $\uniquecellpointershort_\linearizationobject{}$ is the second parameter of $I$, it follows that $\uniquecellpointershort_\linearizationobject{}$ is either $\&\headobject$ or $(\found, \uniquecellpointershort_\linearizationobject{})$ is the response of an invocation $I'$ of the AcquireNext procedure on \cref{line:ero:remove_cell_acquire_next} during $I_{parent}$ such that $I'$ exited before $I$ was invoked.
        However, since $\uniquecellpointershort_\linearizationobject{} \in \celluniverse$, by \Cref{assumption:ero:head_and_null_not_in_cell_universe}, $\uniquecellpointershort_\linearizationobject{} \neq \&\headobject$, so the latter is the only possibility.
        Hence, $p$ executed \cref{line:ero:remove_cell_while_loop} between when $I'$ exited and when $I$ was invoked.
        Let $T^{\ref{line:ero:remove_cell_check_acquire_status}}$ and $T^{\ref{line:ero:remove_cell_while_loop}}$ denote the next time $p$ executes lines \ref{line:ero:remove_cell_check_acquire_status} and \ref{line:ero:remove_cell_while_loop} after exiting $I'$ during $I_{parent}$.
        Since the response of $I'$ is $(\found, \uniquecellpointershort_\linearizationobject{})$, $p$ finds the condition on \cref{line:ero:remove_cell_acquire_next_found} to be true at $T^{\ref{line:ero:remove_cell_acquire_next_found}}$ and so $p$ sets its local variable $\currentcellpointershort{}$ to $\uniquecellpointershort_\linearizationobject{}$ on \cref{line:ero:remove_cell_update_pointers}.
        Hence, since the parameters of $I_{parent}$ are $(\uniquerepositoryoperationshort_\linearizationobject{}, \uniquecellpointershort_\linearizationobject{})$, we have that $p$ finds the condition on \cref{line:ero:remove_cell_while_loop} to be false at $T^{\ref{line:ero:remove_cell_while_loop}}$.
        Therefore, $p$ exits the while loop on \cref{line:ero:remove_cell_while_loop} during $I_{parent}$, and so $I'$ is the last invocation of the AcquireNext procedure during $I_{parent}$.
        However, since $I$ is an invocation of the AcquireNext procedure during $I_{parent}$ and $I'$ exited before $I$ was invoked, we have that there is an invocation of the AcquireNext procedure after $I'$ during $I_{parent}$, a contradiction.

        \item[] \hspace{0pt}\textbf{Case 3.} $I$ was invoked during an invocation $I_{parent}$ of the Acquire procedure on \cref{line:ero:announce_acquire}.

        Let $T^{\ref{line:ero:linearization_read}}$ be the time of $p$'s last execution of \cref{line:ero:linearization_read} before invoking $I$.
        Since $p$ read $(\uniquerepositoryoperationshort_\linearizationobject{}, \uniquecellpointershort_\linearizationobject{})$ from $\linearizationobject{}$ at $T^{\ref{line:ero:linearization_read}}$, and $(\uniquerepositoryoperationshort_\linearizationobject{}, \uniquecellpointershort_\linearizationobject{})$ is of the form $((\arbitraryvalue, \removecell), \cellpointershort{}_\linearizationobject{})$, we have that between $T^{\ref{line:ero:linearization_read}}$ and the time $p$ invoked $I_{parent}$, $p$ invoked and exited the \doremovecell{} on \cref{line:ero:do_remove_cell} with parameters $(\uniquerepositoryoperationshort_\linearizationobject{}, \uniquecellpointershort_\linearizationobject{})$.
        Denote this invocation by $I'$.
        Since by \Cref{lemma:ero:the_list_invariants_hold} $P(\mathcal{I}^\mathcal{B})$, $Q(\mathcal{I}^\mathcal{B})$, and $R(\mathcal{I}^\mathcal{B})$ hold, by \Cref{lemma:ero:exit_do_remove_implies_done}, there is a successful list-remove attempt for $\uniquecellpointershort_\linearizationobject{}$ before $I'$ exits in $\mathcal{I}^\mathcal{B}$.
        Thus, since $I'$ exits before $I_{parent}$ is invoked, there is a successful list-remove attempt $a$ for $\uniquecellpointershort_\linearizationobject{}$ before $I_{parent}$ is invoked.

        We first prove that $e_{\min} < a$.
        Since $a$ is a successful list-remove for $\cellpointershort{}_\linearizationobject{}$, by \Cref{lemma:ero:l_event_corresponding_to_do_low_level_op}, there is an $L$-remove event $e$ for $\cellpointershort{}_\linearizationobject{}$ before $a$.
        Hence, since by \Cref{lemma:ero:the_list_invariants_hold} $P(\mathcal{I}^\mathcal{B})$ holds, we have that $e$ is the only $L$-remove event for $\cellpointershort{}_\linearizationobject{}$ in $\mathcal{I}^\mathcal{B}$.
        Therefore, since $e_{\min}$ is an $L$-remove event for $\cellpointershort{}_\linearizationobject{}$, we have that $e = e_{\min}$, and so $e_{\min} < a$ as wanted.

        We now prove that for every prefix $\mathcal{I}$ of $\mathcal{I}^\mathcal{B}$ at or after $a$, the list of cells conforms to $\List(\mathcal{I}^{include}_{e_{\min}})$ in $\mathcal{I}$.
        Since $e_{\min} < a$, we have that $\mathcal{I}$ is a prefix of $\mathcal{I}^\mathcal{B}$ at or after $e_{\min}$.
        Thus, since by \Cref{lemma:ero:e_is_the_last_l_event} $e_{\min}$ is the last $L$-event in $\mathcal{I}^\mathcal{B}$, it follows that $e_{\min}$ is also the last $L$-event in $\mathcal{I}$.
        Hence, since $e_{\min} < a$ and $a$ is a successful list-remove attempt, we have that there is a successful list-remove attempt after the last $L$-event in $\mathcal{I}$.
        Thus, since by \Cref{lemma:ero:the_list_invariants_hold} $P(\mathcal{I}^\mathcal{B})$, $Q(\mathcal{I}^\mathcal{B})$, and $R(\mathcal{I}^\mathcal{B})$ hold, by \Cref{lemma:ero:conditional_classification_lemma}, the list of cells conforms to $\List(\mathcal{I})$ in $\mathcal{I}$.
        Since $\mathcal{I}$ is a prefix of $\mathcal{I}^\mathcal{B}$ at or after $e_{\min}$, $e_{\min}$ is the last $L$-event in $\mathcal{I}$, and $\mathcal{I}^{include}_{e_{\min}}$ is the prefix of $\mathcal{I}^\mathcal{B}$ up to and including $e_{\min}$, it follows that the sequence of $L$-events is identical in $\mathcal{I}$ and $\mathcal{I}^{include}_{e_{\min}}$.
        Thus, by \Cref{def:ero:english} $\List(\mathcal{I}) = \List(\mathcal{I}^{include}_{e_{\min}})$.
        Therefore, the list of cells conforms to $\List(\mathcal{I}^{include}_{e_{\min}})$ in $\mathcal{I}$.

        We now finish the proof of Case 3.
        Since $a$ is before $I_{parent}$ is invoked and for every prefix $\mathcal{I}$ of $\mathcal{I}^\mathcal{B}$ at or after $a$, the list of cells conforms to $\List(\mathcal{I}^{include}_{e_{\min}})$ in $\mathcal{I}$, we have that for every prefix $\mathcal{I}$ of $\mathcal{I}^\mathcal{B}$ at or after the time $I_{parent}$ is invoked, the list of cells conforms to $\List(\mathcal{I}^{include}_{e_{\min}})$ in $\mathcal{I}$.
        Hence, by \Cref{lemma:ero:acquire_next_pointer_is_from_include_list}, the second parameter of $I$ is in $\List(\mathcal{I}^{include}_{e_{\min}})$.
        Therefore, since the second parameter of $I$ is $\uniquecellpointershort_\linearizationobject{}$, we have that $\uniquecellpointershort_\linearizationobject{}$ is in $\List(\mathcal{I}^{include}_{e_{\min}})$.
        However, since $e_{\min}$ is an $L$-remove event for $\uniquecellpointershort_\linearizationobject{}$ and $\mathcal{I}^{include}_{e_{\min}}$ is the prefix of $\mathcal{I}^\mathcal{B}$ up to and including $e_{\min}$, by \Cref{def:ero:logical_list}, we have that $\uniquecellpointershort_\linearizationobject{}$ is not in $\List(\mathcal{I}^{include}_{e_{\min}})$, a contradiction.
        \qH{\Cref{lemma:ero:if_e_is_a_remove_event_then_no_one_acquires_it}}
    \end{itemize}
\end{proof}

\begin{proposition}\label{lemma:ero:acquire_next_for_ptr_in_exclude_is_in_include}
    In \Cref{scenario:ero:helper_loop_scenario}, consider any invocation $I$ of the \doaddcell, \doremovecell, or Acquire procedure in $\mathcal{I}^\mathcal{B}$ such that the process which invoked $I$ read $(\uniquerepositoryoperationshort_\linearizationobject{}, \uniquecellpointershort_\linearizationobject{})$ from $\linearizationobject{}$ on its last execution of \cref{line:ero:linearization_read} before invoking $I$.
    Let $I_1, I_2, \ldots$ be the (possibly infinite) sequence of invocations of the AcquireNext procedure during $I$ in the order they are invoked, and let $\cellpointershort{}_i$ denote the second parameter of $I_i$.
    If $\cellpointershort{}_i$ is in $\List(\mathcal{I}^{exclude}_{e_{\min}})$, $\cellpointershort{}_i$ is in $\List(\mathcal{I}^{include}_{e_{\min}})$ (assuming $I_i$ exists).
\end{proposition}

\begin{proof}
    Suppose, for contradiction, that $\cellpointershort{}_{i}$ is in $\List(\mathcal{I}^{exclude}_{e_{\min}})$ and $\cellpointershort{}_{i}$ is not in $\List(\mathcal{I}^{include}_{e_{\min}})$.
    Hence, since by \Cref{lemma:ero:every_l_event_is_add_apply_or_remove} $e_{\min}$ is either an $L$-add, $L$-apply, or $L$-remove event, and the sequence of $L$-events in $\mathcal{I}^{exclude}_{e_{\min}}$ and $\mathcal{I}^{include}_{e_{\min}}$ are the same except the former excludes $e_{\min}$ and the latter includes it, by \Cref{def:ero:logical_list}, the only way that this is possible is if $e_{\min}$ is an $L$-remove event for $\cellpointershort{}_{i}$.
    Thus, since by \Cref{lemma:ero:i_is_proceeded_by_a_unique_l_event} $e_{\min}$ is an $L$-event for $\uniquecellpointershort_\linearizationobject{}$, we have that $\cellpointershort{}_{i} = \uniquecellpointershort_\linearizationobject{}$, so $e_{\min}$ is an $L$-remove event for $\cellpointershort{}_\linearizationobject{}$.
    Therefore, since $p$ read $(\uniquerepositoryoperationshort_\linearizationobject{}, \uniquecellpointershort_\linearizationobject{})$ from $\linearizationobject{}$ on its last execution of \cref{line:ero:linearization_read} before invoking $I_{i}$, by \Cref{lemma:ero:if_e_is_a_remove_event_then_no_one_acquires_it}, the second parameter of $I_{i}$ is not $\cellpointershort{}_\linearizationobject{}$.
    However, the second parameter of $I_{i}$ is $\cellpointershort{}_{i} = \uniquecellpointershort_\linearizationobject{}$, a contradiction.
    \qH{\Cref{lemma:ero:acquire_next_for_ptr_in_exclude_is_in_include}}
\end{proof}

We now prove the other two claims that allow us to conclude that in every iteration of $\mathcal{L}_{\min}$, $p_{\min}$ ``traverses" through a cell from either $\List(\mathcal{I}^{exclude}_{e_{\min}})$ or $\List(\mathcal{I}^{include}_{e_{\min}})$.
The first of which is \Cref{lemma:ero:acquire_next_for_ptr_not_in_exclude_implies_a_prefix_that_does_not_conform_to_exclude}, which, roughly speaking, asserts that if $p_{\min}$ tries to acquire a pointer after some $\cellpointershort{}$ not in $\List(\mathcal{I}^{exclude}_{e_{\min}})$, then the list of cells does not conform to $\List(\mathcal{I}^{exclude}_{e_{\min}})$ beforehand.
The second is \Cref{lemma:ero:if_the_list_is_only_include_after_e_it_is_always_include}, which, roughly speaking, asserts that if the list of cells does not conform to $\List(\mathcal{I}^{exclude}_{e_{\min}})$, then the list of cells conforms to $\List(\mathcal{I}^{include}_{e_{\min}})$ from then onwards.
These two facts together let us conclude that if $\cellpointershort{}$ is not in $\List(\mathcal{I}^{exclude}_{e_{\min}})$ then the list of cells conforms to $\List(\mathcal{I}^{include}_{e_{\min}})$ at the time $p_{\min}$ tries to acquire the pointer after $\cellpointershort{}$.

\begin{proposition}\label{lemma:ero:acquire_next_for_ptr_not_in_exclude_implies_a_prefix_that_does_not_conform_to_exclude}
    In \Cref{scenario:ero:helper_loop_scenario}, consider any invocation $I$ of the \doaddcell, \doremovecell, or Acquire procedure in $\mathcal{I}^\mathcal{B}$ such that the process which invoked $I$ read $(\uniquerepositoryoperationshort_\linearizationobject{}, \uniquecellpointershort_\linearizationobject{})$ from $\linearizationobject{}$ on its last execution of \cref{line:ero:linearization_read} before invoking $I$.
    Let $I_1, I_2, \ldots$ be the (possibly infinite) sequence of invocations of the AcquireNext procedure during $I$ in the order they are invoked, and let $\cellpointershort{}_i$ denote the second parameter of $I_i$.
    If $\cellpointershort{}_i$ is not in $\List(\mathcal{I}^{exclude}_{e_{\min}})$, then there is a finite prefix $\mathcal{I}$ of $\mathcal{I}^\mathcal{B}$ after $e_{\min}$ and before $I_i$ was invoked where the list of cells does not conform to $\List(\mathcal{I}^{exclude}_{e_{\min}})$ in $\mathcal{I}$ (assuming $I_i$ exists).
\end{proposition}

\begin{proof}
    By induction on $i$.
    \begin{itemize}
        \item[] \hspace{0pt}\textbf{Base Case.} $i = 1$.

        Suppose $I_1$ exists.
        By definition, there are no invocations of the AcquireNext procedure during $I$ before $I_1$.
        Hence, the second parameter of $I_1$ is $\&\headobject$ and so $\cellpointershort{}_1 = \&\headobject$.
        By \Cref{def:ero:logical_list}, $\&\headobject$ is the first element of $\List(\mathcal{I})$ for every finite implementation history $\mathcal{I}$.
        Hence, the first element of $\List(\mathcal{I}^{exclude}_{e_{\min}})$ is $\&\headobject$, and so $\cellpointershort{}_1$ is in $\List(\mathcal{I}^{exclude}_{e_{\min}})$.
        Therefore, the claim is vacuously true.

        \item[] \hspace{0pt}\textbf{Inductive Case.} For every $i \geq 1$ if the claim holds for $I_i$, then the claim holds for $I_{i + 1}$ (assuming $I_{i + 1}$ exists).

        Suppose for some $i \geq 1$ if $\cellpointershort{}_i$ is not in $\List(\mathcal{I}^{exclude}_{e_{\min}})$, then there is a prefix $\mathcal{I}$ of $\mathcal{I}^\mathcal{B}$ after $e_{\min}$ and before $I_i$ was invoked where the list of cells does not conform to $\List(\mathcal{I}^{exclude}_{e_{\min}})$ in $\mathcal{I}$.
        This is the inductive hypothesis.
        Suppose, for contradiction, $\cellpointershort{}_{i + 1}$ is not in $\List(\mathcal{I}^{exclude}_{e_{\min}})$ and for every prefix $\mathcal{I}$ of $\mathcal{I}^\mathcal{B}$ after $e_{\min}$ and before $I_{i + 1}$ was invoked the list of cells conforms to $\List(\mathcal{I}^{exclude}_{e_{\min}})$ in $\mathcal{I}$.
        Since $I_i$ is the invocation of the AcquireNext procedure which proceeded $I_{i + 1}$ during $I$, it follows that $(\found, \cellpointershort{}_{i + 1})$ is the response of $I_i$.
        Let $p$ be the process that invoked $I$ and let $T^{\ref{line:ero:linearization_read}}_p$ be the last time $p$ executed \cref{line:ero:linearization_read} before invoking $I$.
        Since $p$ read $(\uniquerepositoryoperationshort_\linearizationobject{}, \uniquecellpointershort_\linearizationobject{})$ from $\linearizationobject{}$ at $T^{\ref{line:ero:linearization_read}}_p$, by \Cref{lemma:ero:if_q_read_e_s_value_then_its_after_e}, $e_{\min} < T^{\ref{line:ero:linearization_read}}_p$.
        Since $\cellpointershort{}_i$ is the second parameter of $I_i$, by \Cref{lemma:ero:acquire_next_is_for_pointer_from_universe_or_head}, $\cellpointershort{}_i \in \celluniverse \cup \{\&\headobject\}$, and so by \Cref{assumption:ero:head_and_null_not_in_cell_universe} $\cellpointershort{}_i \neq \nullconstant$.
        Furthermore, since $(\found, \cellpointershort{}_{i + 1})$ is the response of $I_i$, it follows that $p$ read $\cellpointershort{}_{i + 1}$ from $(*\cellpointershort{}_i).\nextlong.\cellpointerlong{}$ on $p$'s final execution of \cref{line:ero:acquire_next_read_curr_unique_pointer} during $I_i$; say at time $T^{\ref{line:ero:acquire_next_read_curr_unique_pointer}}_i$.
        Since $e_{\min} < T^{\ref{line:ero:linearization_read}}_p$, $T^{\ref{line:ero:linearization_read}}_p$ is $p$'s last execution of \cref{line:ero:linearization_read} before invoking $I$, $I_i$ is invoked during $I$, and $T^{\ref{line:ero:acquire_next_read_curr_unique_pointer}}_i$ is during $I_i$, by transitivity, $e_{\min} < T^{\ref{line:ero:acquire_next_read_curr_unique_pointer}}_i$.
        Hence, there is a prefix $\mathcal{I}_i$ of $\mathcal{I}^\mathcal{B}$ after $e_{\min}$ that ends at time $T^{\ref{line:ero:acquire_next_read_curr_unique_pointer}}_i$.
        Thus, since $(*\cellpointershort{}_i).\nextlong.\cellpointerlong{} = \cellpointershort{}_{i + 1}$ at $T^{\ref{line:ero:acquire_next_read_curr_unique_pointer}}_i$, it follows that $(*\cellpointershort{}_i).\nextlong.\cellpointerlong{} = \cellpointershort{}_{i + 1}$ at the end of $\mathcal{I}_i$.
        There are two cases.
        \begin{itemize}
            \item[] \hspace{0pt}\textbf{Case 1.} $\cellpointershort{}_i$ is in $\List(\mathcal{I}^{exclude}_{e_{\min}})$.

            Since $T^{\ref{line:ero:acquire_next_read_curr_unique_pointer}}_i$ is during $I_i$, and $I_i$ exits before $I_{i + 1}$ was invoked, by transitivity, $T^{\ref{line:ero:acquire_next_read_curr_unique_pointer}}_i$ is before $I_{i + 1}$ was invoked.
            Hence, since $\mathcal{I}_i$ is a prefix of $\mathcal{I}^\mathcal{B}$ after $e_{\min}$ up to $T^{\ref{line:ero:acquire_next_read_curr_unique_pointer}}_i$, we have that $\mathcal{I}_i$ is a prefix of $\mathcal{I}^\mathcal{B}$  after $e_{\min}$ and before $I_{i + 1}$.
            Thus, the list of cells conforms to $\List(\mathcal{I}^{exclude}_{e_{\min}})$ in $\mathcal{I}_i$.
            Therefore, since $\cellpointershort{}_i \neq \nullconstant$, $\cellpointershort{}_i$ is in $\List(\mathcal{I}^{exclude}_{e_{\min}})$, and $(*\cellpointershort{}_i).\nextlong.\cellpointerlong{} = \cellpointershort{}_{i + 1}$ at the end of $\mathcal{I}_i$, by \Cref{def:ero:logical_list}, $\cellpointershort{}_{i + 1}$ is in $\List(\mathcal{I}^{exclude}_{e_{\min}})$.
            However, $\cellpointershort{}_{i + 1}$ is not in $\List(\mathcal{I}^{exclude}_{e_{\min}})$, a contradiction.

            \item[] \hspace{0pt}\textbf{Case 2.} $\cellpointershort{}_i$ is not in $\List(\mathcal{I}^{exclude}_{e_{\min}})$.

            Hence, by the inductive hypothesis, we have that there is a prefix $\mathcal{I}$ of $\mathcal{I}^\mathcal{B}$ after $e_{\min}$ and before $I_i$ was invoked where the list of cells does not conform to $\List(\mathcal{I}^{exclude}_{e_{\min}})$ in $\mathcal{I}$.
            Therefore, since $I_i$ is before $I_{i + 1}$, we have that there is a prefix $\mathcal{I}$ of $\mathcal{I}^\mathcal{B}$ after $e_{\min}$ and before $I_{i + 1}$ was invoked where the list of cells does not conform to $\List(\mathcal{I}^{exclude}_{e_{\min}})$ in $\mathcal{I}$.
            However, for every prefix $\mathcal{I}$ of $\mathcal{I}^\mathcal{B}$ after $e_{\min}$ and before $I_{i + 1}$ was invoked the list of cells conforms to $\List(\mathcal{I}^{exclude}_{e_{\min}})$ in $\mathcal{I}$, a contradiction.
            \qH{\Cref{lemma:ero:acquire_next_for_ptr_not_in_exclude_implies_a_prefix_that_does_not_conform_to_exclude}}
        \end{itemize}
    \end{itemize}
\end{proof}

\begin{lemma}\label{lemma:ero:if_the_list_is_only_include_after_e_it_is_always_include}
    In \Cref{scenario:ero:helper_loop_scenario}, if there is a finite prefix $\mathcal{I}$ of $\mathcal{I}^\mathcal{B}$ at or after $e_{\min}$ where the list of cells does not conform to $\List(\mathcal{I}^{exclude}_{e_{\min}})$ in $\mathcal{I}$, then for every finite prefix $\mathcal{I}'$ of $\mathcal{I}^\mathcal{B}$ such that $\mathcal{I}$ is a prefix of $\mathcal{I}'$ we have that the list of cells conforms to $\List(\mathcal{I}^{include}_{e_{\min}})$ in $\mathcal{I}'$.
\end{lemma}

\begin{proof}
    Suppose, that there is a prefix $\mathcal{I}$ of $\mathcal{I}^\mathcal{B}$ at or after $e_{\min}$ where the list of cells does not conform to $\List(\mathcal{I}^{exclude}_{e_{\min}})$ in $\mathcal{I}$.
    Since by \Cref{lemma:ero:e_is_the_last_l_event} $e_{\min}$ is the last $L$-event in $\mathcal{I}^\mathcal{B}$ and, $\mathcal{I}$ is a prefix of $\mathcal{I}^\mathcal{B}$ at or after $e_{\min}$, we have that $e_{\min}$ is the last $L$-event in $\mathcal{I}$.

    We first prove that there is a successful list-add or list-remove attempt after $e_{\min}$ in $\mathcal{I}$ (*).
    Suppose, for contradiction, there is not a successful list-add or list-remove attempt after $e_{\min}$ in $\mathcal{I}$.
    Hence, we have that (1) there is an $L$-event in $\mathcal{I}$; (2) $e_{\min}$ is the last $L$-event in $\mathcal{I}$; and (3) from $e_{\min}$ onwards in $\mathcal{I}$ there are no successful list-add or list-remove attempts.
    Therefore, since by \Cref{lemma:ero:the_list_invariants_hold} $P(\mathcal{I}^\mathcal{B})$, $Q(\mathcal{I}^\mathcal{B})$, and $R(\mathcal{I}^\mathcal{B})$ hold, by \Cref{lemma:ero:conditional_classification_lemma}, the list of cells conforms to $\List(\mathcal{I}^{exclude}_{e_{\min}})$ in $\mathcal{I}$.
    However, the list of cells does not conform to $\List(\mathcal{I}^{exclude}_{e_{\min}})$ in $\mathcal{I}$, a contradiction.

    Consider any prefix $\mathcal{I}'$ of $\mathcal{I}^\mathcal{B}$ such that $\mathcal{I}$ is a prefix of $\mathcal{I}'$.
    Hence, $\mathcal{I}'$ is a prefix of $\mathcal{I}^\mathcal{B}$ at or after $e_{\min}$.
    Thus, since $e_{\min}$ is the last $L$-event in $\mathcal{I}$, we have that $e_{\min}$ is the last $L$-event in $\mathcal{I}'$.
    Since by (*) there is a successful list-add or list-remove attempt after $e_{\min}$ in $\mathcal{I}$, and $\mathcal{I}$ is a prefix of $\mathcal{I}'$, there is a successful list-add or list-remove attempt after $e_{\min}$ in $\mathcal{I}'$.
    Hence, we have that (1) there is an $L$-event in $\mathcal{I}'$; (2) $e_{\min}$ is the last $L$-event in $\mathcal{I}'$; and (3) there is a successful list-add or list-remove attempt after $e_{\min}$ in $\mathcal{I}'$.
    Thus, since by \Cref{lemma:ero:the_list_invariants_hold} $P(\mathcal{I}^\mathcal{B})$, $Q(\mathcal{I}^\mathcal{B})$, and $R(\mathcal{I}^\mathcal{B})$ hold, by \Cref{lemma:ero:conditional_classification_lemma}, the list of cells conforms to $\List(\mathcal{I}')$ in $\mathcal{I}'$.
    So, since $\mathcal{I}^{include}_{e_{\min}}$ is the prefix of $\mathcal{I}^\mathcal{B}$ up to and including $e_{\min}$, and $e_{\min}$ is the last $L$-event in $\mathcal{I}'$, it follows that the sequence of $L$-events is identical in $\mathcal{I}'$ and $\mathcal{I}^{include}_{e_{\min}}$.
    Therefore, by \Cref{def:ero:logical_list}, $\List(\mathcal{I}') = \List(\mathcal{I}^{include}_{e_{\min}})$, and so the list of cells conforms to $\List(\mathcal{I}^{include}_{e_{\min}})$ in $\mathcal{I}'$ as wanted.
    \qH{\Cref{lemma:ero:if_the_list_is_only_include_after_e_it_is_always_include}}
\end{proof}

We now complete the proof of every iteration of $\mathcal{L}_{\min}$, $p_{\min}$ ``traverses" through a cell from either $\List(\mathcal{I}^{exclude}_{e_{\min}})$ or $\List(\mathcal{I}^{include}_{e_{\min}})$.

\begin{lemma}\label{lemma:ero:acquire_next_for_ptr_is_in_either_include_or_exclude}
    In \Cref{scenario:ero:helper_loop_scenario}, consider any invocation $I$ of the \doaddcell, \doremovecell, or Acquire procedure in $\mathcal{I}^\mathcal{B}$ such that the process which invoked $I$ read $(\uniquerepositoryoperationshort_\linearizationobject{}, \uniquecellpointershort_\linearizationobject{})$ from $\linearizationobject{}$ on its last execution of \cref{line:ero:linearization_read} before invoking $I$.
    Let $I_1, I_2, \ldots$ be the (possibly infinite) sequence of invocations of the AcquireNext procedure during $I$ in the order they are invoked, and let $\cellpointershort{}_i$ denote the second parameter of $I_i$.
    $\cellpointershort{}_i$ is in either $\List(\mathcal{I}^{exclude}_{e_{\min}})$ or $\List(\mathcal{I}^{include}_{e_{\min}})$ (assuming $I_i$ exists).
\end{lemma}

\begin{proof}
    By induction on $i$.
    \begin{itemize}
        \item[] \hspace{0pt}\textbf{Base Case.} $i = 1$.

        Suppose $I_1$ exists.
        By definition, there are no invocations of the AcquireNext procedure during $I$ before $I_1$.
        Hence, the second parameter of $I_1$ is $\&\headobject$ and so $\cellpointershort{}_1 = \&\headobject$.
        By \Cref{def:ero:logical_list}, $\&\headobject$ is the first element of $\List(\mathcal{I})$ for every finite implementation history $\mathcal{I}$.
        Hence, the first element of $\List(\mathcal{I}^{exclude}_{e_{\min}})$ is $\&\headobject$, and so $\cellpointershort{}_1$ is in $\List(\mathcal{I}^{exclude}_{e_{\min}})$.
        Therefore, the claim follows.

        \item[] \hspace{0pt}\textbf{Inductive Case.} For every $i \geq 1$ if $\cellpointershort{}_i$ is in either $\List(\mathcal{I}^{exclude}_{e_{\min}})$ or $\List(\mathcal{I}^{include}_{e_{\min}})$, then $\cellpointershort{}_{i + 1}$ is in either $\List(\mathcal{I}^{exclude}_{e_{\min}})$ or $\List(\mathcal{I}^{include}_{e_{\min}})$ (assuming $I_{i + 1}$ exists).

        Suppose for some $i \geq 1$ that $\cellpointershort{}_i$ is in either $\List(\mathcal{I}^{exclude}_{e_{\min}})$ or $\List(\mathcal{I}^{include}_{e_{\min}})$.
        This is the inductive hypothesis.
        Suppose $I_{i + 1}$ exists.
        The setup is identical to \Cref{lemma:ero:acquire_next_for_ptr_not_in_exclude_implies_a_prefix_that_does_not_conform_to_exclude}, which we repeat for completeness below.
        Since $I_i$ is the invocation of the AcquireNext procedure which proceeded $I_{i + 1}$ during $I$, it follows that $(\found, \cellpointershort{}_{i + 1})$ is the response of $I_i$.
        Let $p$ be the process that invoked $I$ and let $T^{\ref{line:ero:linearization_read}}_p$ be the last time $p$ executed \cref{line:ero:linearization_read} before invoking $I$.
        Since $p$ read $(\uniquerepositoryoperationshort_\linearizationobject{}, \uniquecellpointershort_\linearizationobject{})$ from $\linearizationobject{}$ at $T^{\ref{line:ero:linearization_read}}_p$, by \Cref{lemma:ero:if_q_read_e_s_value_then_its_after_e}, $e_{\min} < T^{\ref{line:ero:linearization_read}}_p$.
        Since $\cellpointershort{}_i$ is the second parameter of $I_i$, by \Cref{lemma:ero:acquire_next_is_for_pointer_from_universe_or_head}, $\cellpointershort{}_i \in \celluniverse \cup \{\&\headobject\}$, and so by \Cref{assumption:ero:head_and_null_not_in_cell_universe} $\cellpointershort{}_i \neq \nullconstant$.
        Furthermore, since $(\found, \cellpointershort{}_{i + 1})$ is the response of $I_i$, it follows that $p$ read $\cellpointershort{}_{i + 1}$ from $(*\cellpointershort{}_i).\nextlong.\cellpointerlong{}$ on $p$'s final execution of \cref{line:ero:acquire_next_read_curr_unique_pointer} during $I_i$; say at time $T^{\ref{line:ero:acquire_next_read_curr_unique_pointer}}_i$.
        Since $e_{\min} < T^{\ref{line:ero:linearization_read}}_p$, $T^{\ref{line:ero:linearization_read}}_p$ is $p$'s last execution of \cref{line:ero:linearization_read} before invoking $I$, $I_i$ is invoked during $I$, and $T^{\ref{line:ero:acquire_next_read_curr_unique_pointer}}_i$ is during $I_i$, by transitivity, $e_{\min} < T^{\ref{line:ero:acquire_next_read_curr_unique_pointer}}_i$.
        Hence, there is a prefix $\mathcal{I}_i$ of $\mathcal{I}^\mathcal{B}$ after $e_{\min}$ that ends at time $T^{\ref{line:ero:acquire_next_read_curr_unique_pointer}}_i$.
        Thus, since $(*\cellpointershort{}_i).\nextlong.\cellpointerlong{} = \cellpointershort{}_{i + 1}$ at $T^{\ref{line:ero:acquire_next_read_curr_unique_pointer}}_i$, it follows that $(*\cellpointershort{}_i).\nextlong.\cellpointerlong{} = \cellpointershort{}_{i + 1}$ at the end of $\mathcal{I}_i$.
        There are two cases.
        \begin{itemize}
            \item[] \hspace{0pt}\textbf{Case 1.} $\cellpointershort{}_i$ is in $\List(\mathcal{I}^{exclude}_{e_{\min}})$.
    
            Since $\mathcal{I}_i$ is a prefix of $\mathcal{I}^\mathcal{B}$ after $e_{\min}$, by \Cref{lemma:ero:from_e_onwards_the_list_is_one_of_two_structures}, the list of cells conforms to either $\List(\mathcal{I}^{exclude}_{e_{\min}})$ or $\List(\mathcal{I}^{include}_{e_{\min}})$ in $\mathcal{I}_i$.
            Furthermore, since $\cellpointershort{}_i$ is in $\List(\mathcal{I}^{exclude}_{e_{\min}})$, by \Cref{lemma:ero:acquire_next_for_ptr_in_exclude_is_in_include}, $\cellpointershort{}_i$ is in $\List(\mathcal{I}^{include}_{e_{\min}})$, and so $\cellpointershort{}_i$ is in both $\List(\mathcal{I}^{exclude}_{e_{\min}})$ and $\List(\mathcal{I}^{include}_{e_{\min}})$.
            Hence, since $\cellpointershort{}_i \neq \nullconstant$, and $(*\cellpointershort{}_i).\nextlong.\cellpointerlong{} = \cellpointershort{}_{i}$ at the end of $\mathcal{I}_i$, by \Cref{def:ero:logical_list}, $\cellpointershort{}_{i}$ is in either $\List(\mathcal{I}^{exclude}_{e_{\min}})$ or $\List(\mathcal{I}^{include}_{e_{\min}})$.
            Therefore, the claim follows.            
    
            \item[] \hspace{0pt}\textbf{Case 2.} $\cellpointershort{}_i$ is not in $\List(\mathcal{I}^{exclude}_{e_{\min}})$.

            Hence, by the inductive hypothesis, $\cellpointershort{}_i$ is in $\List(\mathcal{I}^{include}_{e_{\min}})$.
            Furthermore, by \Cref{lemma:ero:acquire_next_for_ptr_not_in_exclude_implies_a_prefix_that_does_not_conform_to_exclude}, there is a prefix $\mathcal{I}$ of $\mathcal{I}^\mathcal{B}$ after $e_{\min}$ and before $I_i$ was invoked where the list of cells does not conform to $\List(\mathcal{I}^{exclude}_{e_{\min}})$ in $\mathcal{I}$.
            Since the end of $\mathcal{I}$ is before $I_i$ is invoked and the end of $\mathcal{I}_i$ is after $I_i$ is invoked, we have that $\mathcal{I}$ is a prefix of $\mathcal{I}_i$.
            Hence, by \Cref{lemma:ero:if_the_list_is_only_include_after_e_it_is_always_include}, the list of cells conforms to  $\List(\mathcal{I}^{include}_{e_{\min}})$ in $\mathcal{I}_i$.
            Thus, since $\cellpointershort{}_i \neq \nullconstant$, $\cellpointershort{}_i$ is in $\List(\mathcal{I}^{include}_{e_{\min}})$, and $(*\cellpointershort{}_i).\nextlong.\cellpointerlong{} = \cellpointershort{}_{i}$ at the end of $\mathcal{I}_i$, by \Cref{def:ero:logical_list}, $\cellpointershort{}_{i}$ is in $\List(\mathcal{I}^{include}_{e_{\min}})$.
            Therefore, the claim follows.
            \qH{\Cref{lemma:ero:acquire_next_for_ptr_is_in_either_include_or_exclude}}
        \end{itemize}
    \end{itemize}
\end{proof}

This completes the second part of the high-level argument for why $p_{\min}$ cannot take infinitely many steps in the loops on lines \ref{line:ero:add_cell_while_loop}, \ref{line:ero:remove_cell_while_loop}, and \ref{line:ero:acquire_loop_until} during $opx_{\min}$.
We now prove the third and final part: that the pointers $p_{\min}$ traverses through are distinct.

\begin{lemma}\label{lemma:ero:all_acquire_next_during_traversal_are_for_different_pointers}
    In \Cref{scenario:ero:helper_loop_scenario}, consider any invocation $I$ of the \doaddcell, \doremovecell, or Acquire procedure in $\mathcal{I}^\mathcal{B}$ such that the process which invoked $I$ read $(\uniquerepositoryoperationshort_\linearizationobject{}, \uniquecellpointershort_\linearizationobject{})$ from $\linearizationobject{}$ on its last execution of \cref{line:ero:linearization_read} before invoking $I$.
    Let $I_1, I_2, \ldots$ be the (possibly infinite) sequence of invocations of the AcquireNext procedure during $I$ in the order they were invoked, and let $\cellpointershort{}_i$ denote the second parameter of $I_i$.
    Then, for every $i$ and $j$ such that $i \neq j$, $\cellpointershort{}_i \neq \cellpointershort{}_j$ (assuming $I_i$ and $I_j$ exist).
\end{lemma}

\begin{proof}
    Suppose, for contradiction, there is $I_i$ and $I_j$ such that $i \neq j$ and $\cellpointershort{}_i = \cellpointershort{}_j$.
    Without loss of generality suppose $i < j$ and $j$ is the first non-distinct pointer, i.e., for all $k, l \in [1..j)$ if $k \neq l$, then $\cellpointershort{}_k \neq \cellpointershort{}_l$.
    Let $p$ be the process that invoked $I$ and let $T^{\ref{line:ero:linearization_read}}_p$ be the last time $p$ executed \cref{line:ero:linearization_read} before invoking $I$.
    Since $p$ read $(\uniquerepositoryoperationshort_\linearizationobject{}, \uniquecellpointershort_\linearizationobject{})$ from $\linearizationobject{}$ at $T^{\ref{line:ero:linearization_read}}_p$, by \Cref{lemma:ero:if_q_read_e_s_value_then_its_after_e}, $e_{\min} < T^{\ref{line:ero:linearization_read}}_p$.
    % Consider any $i, j \in [1..n]$ such that $i \neq j$.
    % Without loss of generality, suppose $i < j$.
    Since $1 \leq i$ and $i < j$, by transitivity $1 < j$, and so $1 \leq j - 1$.
    Hence, $I_{j - 1}$ is well-defined.
    Since $\cellpointershort{}_{j - 1}$ is the second parameter of $I_{j - 1}$, by \Cref{lemma:ero:acquire_next_is_for_pointer_from_universe_or_head}, $\cellpointershort{}_{j - 1} \in \celluniverse \cup \{\&\headobject\}$, and so by \Cref{assumption:ero:head_and_null_not_in_cell_universe} $\cellpointershort{}_{j - 1} \neq \nullconstant$.
    Furthermore, since $I_{j - 1}$ is the invocation of the AcquireNext procedure preceding $I_j$ during $I$, we have that $(\found,  \cellpointershort{}_{j})$ is the response of $I_{j - 1}$.
    Hence, since $\cellpointershort{}_{j - 1}$ is the second parameter of $I_{j - 1}$, on $p$'s last execution of \cref{line:ero:acquire_next_read_curr_unique_pointer} during $I_{j - 1}$, say at time $T^{\ref{line:ero:acquire_next_read_curr_unique_pointer}}_{j - 1}$, $p$ read $\cellpointershort{}_{j}$ from $(*\cellpointershort{}_{j - 1}).\nextlong.\cellpointerlong{}$.

    \begin{claimcustom}{\ref{lemma:ero:all_acquire_next_during_traversal_are_for_different_pointers}.1}\label{lemma:ero:all_acquire_next_during_traversal_are_for_different_pointers:claim_one}
        $i > 1$ and so $I_{i - 1}$ is well-defined.
    \end{claimcustom}

    \begin{proof}
        Suppose, for contradiction, $i = 1$.
        Hence, $I_1$ is the first invocation of the AcquireNext procedure during $I$, and so there are no invocations of the AcquireNext procedure before $I_1$ during $I$.
        Thus, the second parameter of $I_1$ is $\&\headobject$ and so $\cellpointershort{}_1 = \&\headobject$.
        Hence, since $i = 1$ and $\cellpointershort{}_i = \cellpointershort{}_j$, we have that $\cellpointershort{}_j = \&\headobject$.
        Thus, since $(*\cellpointershort{}_{j - 1}).\nextlong.\cellpointerlong{} = \cellpointershort{}_{j}$ at $T^{\ref{line:ero:acquire_next_read_curr_unique_pointer}}_{j - 1}$, it follows that $(*\cellpointershort{}_{j - 1}).\nextlong.\cellpointerlong{} = \&\headobject$ at $T^{\ref{line:ero:acquire_next_read_curr_unique_pointer}}_{j - 1}$.
        Therefore, since by \Cref{lemma:ero:acquire_next_is_for_pointer_from_universe_or_head} $\cellpointershort{}_{j - 1} \in \celluniverse \cup \{\&\headobject\}$, by \Cref{lemma:ero:next_pointer_is_always_from_universe_or_null}, $\&\headobject \in \celluniverse \cup \{\nullconstant\}$.
        However, by \Cref{assumption:ero:head_and_null_not_in_cell_universe} $\&\headobject \notin \celluniverse$ and $\&\headobject \neq \nullconstant$, a contradiction.
        Therefore, $I_{i - 1}$ is well-defined as wanted.
        \qH{\Cref{lemma:ero:all_acquire_next_during_traversal_are_for_different_pointers:claim_one}}
    \end{proof}

    Since $\cellpointershort{}_{i - 1}$ is the second parameter of $I_{i - 1}$, by \Cref{lemma:ero:acquire_next_is_for_pointer_from_universe_or_head}, $\cellpointershort{}_{i - 1} \in \celluniverse \cup \{\&\headobject\}$, and so by \Cref{assumption:ero:head_and_null_not_in_cell_universe} $\cellpointershort{}_{i - 1} \neq \nullconstant$.
    Furthermore, since $I_{i - 1}$ is the invocation of the AcquireNext procedure preceding $I_i$ during $I$, we have that the response of $I_{i - 1}$ is $(\found,  \cellpointershort{}_{i})$.
    Hence, since $\cellpointershort{}_{i - 1}$ is the second parameter of $I_{i - 1}$, on $p$'s last execution of \cref{line:ero:acquire_next_read_curr_unique_pointer} during $I_{i - 1}$, say at time $T^{\ref{line:ero:acquire_next_read_curr_unique_pointer}}_{i - 1}$, $p$ read $\cellpointershort{}_{i}$ from $(*\cellpointershort{}_{i - 1}).\nextlong.\cellpointerlong{}$.
    Since $e_{\min} < T^{\ref{line:ero:linearization_read}}_p$, $T^{\ref{line:ero:linearization_read}}_p$ is before $p$ invoked $I$, $I_{i - 1}$ (resp. $I_{j - 1}$) is invoked during $I$, and $T^{\ref{line:ero:acquire_next_read_curr_unique_pointer}}_{i - 1}$ (resp. $T^{\ref{line:ero:acquire_next_read_curr_unique_pointer}}_{j - 1}$) is during $I_{j - 1}$ (resp. $I_{i - 1}$), by transitivity, we have that $e_{\min} < T^{\ref{line:ero:acquire_next_read_curr_unique_pointer}}_{i - 1}$ (resp. $e_{\min} < T^{\ref{line:ero:acquire_next_read_curr_unique_pointer}}_{j - 1}$).
    Hence, there is a prefix $\mathcal{I}_{i - 1}$ (resp. $\mathcal{I}_{j - 1}$) of $\mathcal{I}^\mathcal{B}$ at or after $e_{\min}$ and up to and including $T^{\ref{line:ero:acquire_next_read_curr_unique_pointer}}_{i - 1}$ (resp. $T^{\ref{line:ero:acquire_next_read_curr_unique_pointer}}_{j - 1}$).
    Therefore, since $p$ read $\cellpointershort{}_{i}$ (resp. $\cellpointershort{}_{j}$) from $(*\cellpointershort{}_{i - 1}).\nextlong.\cellpointerlong{}$ (resp. $(*\cellpointershort{}_{j - 1}).\nextlong.\cellpointerlong{}$) at $T^{\ref{line:ero:acquire_next_read_curr_unique_pointer}}_{i - 1}$ (resp. $T^{\ref{line:ero:acquire_next_read_curr_unique_pointer}}_{j - 1}$), we have that $(*\cellpointershort{}_{i - 1}).\nextlong.\cellpointerlong{} = \cellpointershort{}_{i}$ (resp. $(*\cellpointershort{}_{j - 1}).\nextlong.\cellpointerlong{} = \cellpointershort{}_{j}$) at the end of $\mathcal{I}_{i - 1}$ (resp. $\mathcal{I}_{j - 1}$) (*).
    We now perform a case reduction to simplify the finale.

    \begin{claimcustom}{\ref{lemma:ero:all_acquire_next_during_traversal_are_for_different_pointers}.2}\label{lemma:ero:all_acquire_next_during_traversal_are_for_different_pointers:claim_two}
        One of the following three scenarios must occur.
        \begin{compactenum}
            \item[(1)] $\cellpointershort{}_{i - 1}$ and $\cellpointershort{}_{j - 1}$ are both in $\List(\mathcal{I}^{exclude}_{e_{\min}})$ and the list of cells conforms to $\List(\mathcal{I}^{exclude}_{e_{\min}})$ in both $\mathcal{I}_{i - 1}$ and $\mathcal{I}_{j - 1}$,
            \item[(2)] $\cellpointershort{}_{i - 1}$ and $\cellpointershort{}_{j - 1}$ are both in $\List(\mathcal{I}^{include}_{e_{\min}})$ and the list of cells conforms to $\List(\mathcal{I}^{include}_{e_{\min}})$ in both $\mathcal{I}_{i - 1}$ and $\mathcal{I}_{j - 1}$, and
            \item[(3)] $\cellpointershort{}_{i - 1}$ is in $\List(\mathcal{I}^{exclude}_{e_{\min}})$, $\cellpointershort{}_{j - 1}$ is in $\List(\mathcal{I}^{include}_{e_{\min}})$, the list of cells conforms to $\List(\mathcal{I}^{exclude}_{e_{\min}})$ in $\mathcal{I}_{i - 1}$, and the list of cells conforms to $\List(\mathcal{I}^{include}_{e_{\min}})$ in $\mathcal{I}_{j - 1}$.
        \end{compactenum}
    \end{claimcustom}

    \begin{proof}
        There are four cases.
        \begin{itemize}
            \item[] \hspace{0pt}\textbf{Case 1.} $\cellpointershort{}_{i - 1}$ and $\cellpointershort{}_{j - 1}$ are both in $\List(\mathcal{I}^{exclude}_{e_{\min}})$.
    
            Since $\mathcal{I}_{i - 1}$ (resp. $\mathcal{I}_{j - 1}$) is a prefix of $\mathcal{I}^\mathcal{B}$ at or after $e_{\min}$, by \Cref{lemma:ero:from_e_onwards_the_list_is_one_of_two_structures}, the list of cells conforms to either $\List(\mathcal{I}^{exclude}_{e_{\min}})$ or $\List(\mathcal{I}^{include}_{e_{\min}})$ in $\mathcal{I}_{i - 1}$ (resp. $\mathcal{I}_{j - 1}$).
            We consider each combination. 
            \begin{itemize}
                \item[] \hspace{0pt}\textbf{Case 1.1.} the list of cells conforms to $\List(\mathcal{I}^{exclude}_{e_{\min}})$ in both $\mathcal{I}_{i - 1}$ and $\mathcal{I}_{j - 1}$.
    
                Hence, since $\cellpointershort{}_{i - 1}$ and $\cellpointershort{}_{j - 1}$ are both in $\List(\mathcal{I}^{exclude}_{e_{\min}})$ (1) is satisfied.
    
                \item[] \hspace{0pt}\textbf{Case 1.2.} the list of cells conforms to $\List(\mathcal{I}^{include}_{e_{\min}})$ in both $\mathcal{I}_{i - 1}$ and $\mathcal{I}_{j - 1}$.
    
                Since $\cellpointershort{}_{i - 1}$ and $\cellpointershort{}_{j - 1}$ are both in $\List(\mathcal{I}^{exclude}_{e_{\min}})$, by \Cref{lemma:ero:acquire_next_for_ptr_in_exclude_is_in_include}, $\cellpointershort{}_{i - 1}$ and $\cellpointershort{}_{j - 1}$ are both in $\List(\mathcal{I}^{include}_{e_{\min}})$.
                Therefore, (2) is satisfied.
    
                \item[] \hspace{0pt}\textbf{Case 1.3.} the list of cells conforms to $\List(\mathcal{I}^{exclude}_{e_{\min}})$ in $\mathcal{I}_{i - 1}$ and the list of cells conforms to $\List(\mathcal{I}^{include}_{e_{\min}})$ in $\mathcal{I}_{j - 1}$.
    
                Since $\cellpointershort{}_{j - 1}$ is in $\List(\mathcal{I}^{exclude}_{e_{\min}})$, by \Cref{lemma:ero:acquire_next_for_ptr_in_exclude_is_in_include}, $\cellpointershort{}_{j - 1}$ is in $\List(\mathcal{I}^{include}_{e_{\min}})$.
                Therefore, since $\cellpointershort{}_{i - 1}$ is in $\List(\mathcal{I}^{exclude}_{e_{\min}})$, $\cellpointershort{}_{j - 1}$ is in $\List(\mathcal{I}^{include}_{e_{\min}})$, the list of cells conforms to $\List(\mathcal{I}^{exclude}_{e_{\min}})$ in $\mathcal{I}_{i - 1}$, and the list of cells conforms to $\List(\mathcal{I}^{include}_{e_{\min}})$ in $\mathcal{I}_{j - 1}$, (3) is satisfied.
    
                \item[] \hspace{0pt}\textbf{Case 1.4.} the list of cells conforms to $\List(\mathcal{I}^{include}_{e_{\min}})$ in $\mathcal{I}_{i - 1}$ and the list of cells conforms to $\List(\mathcal{I}^{exclude}_{e_{\min}})$ in $\mathcal{I}_{j - 1}$.
    
                It suffices to assume that the list of cells does not conform to $\List(\mathcal{I}^{exclude}_{e_{\min}})$ in $\mathcal{I}_{i - 1}$, as otherwise, this case reduces to Case 1.1.
                Since $i - 1 < j - 1$, we have that $I_{i - 1}$ exits before $I_{j - 1}$ is invoked.
                Hence, since $T^{\ref{line:ero:acquire_next_read_curr_unique_pointer}}_{i - 1}$ is during $I_{i - 1}$ and $T^{\ref{line:ero:acquire_next_read_curr_unique_pointer}}_{j - 1}$ is during $I_{j - 1}$, we have that $T^{\ref{line:ero:acquire_next_read_curr_unique_pointer}}_{i - 1} < T^{\ref{line:ero:acquire_next_read_curr_unique_pointer}}_{j - 1}$.
                Thus, since $\mathcal{I}_{i - 1}$ is the prefix of $\mathcal{I}^\mathcal{B}$ up to and including $T^{\ref{line:ero:acquire_next_read_curr_unique_pointer}}_{i - 1}$, and $\mathcal{I}_{j - 1}$ is the prefix of $\mathcal{I}^\mathcal{B}$ up to and including $T^{\ref{line:ero:acquire_next_read_curr_unique_pointer}}_{j - 1}$, it follows that $\mathcal{I}_{i - 1}$ is a prefix of $\mathcal{I}_{j - 1}$.
                Thus, since the list of cells does not conform to $\List(\mathcal{I}^{exclude}_{e_{\min}})$ in $\mathcal{I}_{i - 1}$, and $\mathcal{I}_{j - 1}$ is the prefix of $\mathcal{I}^\mathcal{B}$ such that $\mathcal{I}_{i - 1}$ is a prefix of $\mathcal{I}_{j - 1}$, by \Cref{lemma:ero:if_the_list_is_only_include_after_e_it_is_always_include}, the list of cells conforms to $\List(\mathcal{I}^{include}_{e_{\min}})$ in $\mathcal{I}_{j - 1}$.
                So, the list of cells conforms to $\List(\mathcal{I}^{include}_{e_{\min}})$ in both  $\mathcal{I}_{i - 1}$ and $\mathcal{I}_{j - 1}$.
                Since $\cellpointershort{}_{i - 1}$ and $\cellpointershort{}_{j - 1}$ are both in $\List(\mathcal{I}^{exclude}_{e_{\min}})$, by \Cref{lemma:ero:acquire_next_for_ptr_in_exclude_is_in_include}, $\cellpointershort{}_{i - 1}$ and $\cellpointershort{}_{j - 1}$ are both in $\List(\mathcal{I}^{include}_{e_{\min}})$.
                Therefore, (2) is satisfied.
            \end{itemize}
    
            \item[] \hspace{0pt}\textbf{Case 2.} $\cellpointershort{}_{i - 1}$ is not in $\List(\mathcal{I}^{exclude}_{e_{\min}})$ and $\cellpointershort{}_{j - 1}$ is in $\List(\mathcal{I}^{exclude}_{e_{\min}})$.
    
            Hence, by \Cref{lemma:ero:acquire_next_for_ptr_is_in_either_include_or_exclude}, $\cellpointershort{}_{i - 1}$ is in $\List(\mathcal{I}^{include}_{e_{\min}})$, and by \Cref{lemma:ero:acquire_next_for_ptr_in_exclude_is_in_include}, $\cellpointershort{}_{j - 1}$ is in $\List(\mathcal{I}^{include}_{e_{\min}})$, so $\cellpointershort{}_{i - 1}$ and $\cellpointershort{}_{j - 1}$ are both in $\List(\mathcal{I}^{include}_{e_{\min}})$.
            Furthermore, by \Cref{lemma:ero:acquire_next_for_ptr_not_in_exclude_implies_a_prefix_that_does_not_conform_to_exclude} there is a prefix $\mathcal{I}$ of $\mathcal{I}^\mathcal{B}$ after $e_{\min}$ and before $I_{i - 1}$ was invoked such that the list of cells does not conform to $\List(\mathcal{I}^{exclude}_{e_{\min}})$ in $\mathcal{I}$.
            Since the end of $\mathcal{I}$ is before $I_{i - 1}$ is invoked and the end of $\mathcal{I}_{i - 1}$ (resp. $\mathcal{I}_{j - 1}$) is after $I_{i - 1}$ was invoked (the end of $\mathcal{I}_{i - 1}$ is during $I_{i - 1}$ and $i - 1 < j - 1$ implies $\mathcal{I}_{j - 1}$ is a prefix of $\mathcal{I}_{i - 1}$), we have that $\mathcal{I}$ is a prefix of $\mathcal{I}_{i - 1}$ (resp. $\mathcal{I}_{j - 1}$).
            Hence, by \Cref{lemma:ero:if_the_list_is_only_include_after_e_it_is_always_include}, the list of cells conforms to $\List(\mathcal{I}^{include}_{e_{\min}})$ in $\mathcal{I}_{i - 1}$ (resp. $\mathcal{I}_{j - 1}$).
            Therefore, (2) is satisfied.
    
            \item[] \hspace{0pt}\textbf{Case 3.} $\cellpointershort{}_{i - 1}$ is in $\List(\mathcal{I}^{exclude}_{e_{\min}})$ and $\cellpointershort{}_{j - 1}$ is not in $\List(\mathcal{I}^{exclude}_{e_{\min}})$.

            Hence, by \Cref{lemma:ero:acquire_next_for_ptr_is_in_either_include_or_exclude}, $\cellpointershort{}_{j - 1}$ is in $\List(\mathcal{I}^{include}_{e_{\min}})$, and by \Cref{lemma:ero:acquire_next_for_ptr_in_exclude_is_in_include}, $\cellpointershort{}_{i - 1}$ is in $\List(\mathcal{I}^{include}_{e_{\min}})$, so $\cellpointershort{}_{i - 1}$ and $\cellpointershort{}_{j - 1}$ are both in $\List(\mathcal{I}^{include}_{e_{\min}})$.
            Furthermore, by \Cref{lemma:ero:acquire_next_for_ptr_not_in_exclude_implies_a_prefix_that_does_not_conform_to_exclude} there is a prefix $\mathcal{I}$ of $\mathcal{I}^\mathcal{B}$ after $e_{\min}$ and before $I_{j - 1}$ was invoked such that the list of cells does not conform to $\List(\mathcal{I}^{exclude}_{e_{\min}})$ in $\mathcal{I}$.
            Since the end of $\mathcal{I}$ is before $I_{j - 1}$ is invoked and the end of $\mathcal{I}_{j - 1}$ is after $I_{j - 1}$ was invoked, we have that $\mathcal{I}$ is a prefix of $\mathcal{I}_{j - 1}$.
            Hence, by \Cref{lemma:ero:from_e_onwards_the_list_is_one_of_two_structures}, the list of cells conforms to $\List(\mathcal{I}^{include}_{e_{\min}})$ in $\mathcal{I}_{j - 1}$.
            Since $\mathcal{I}_{i - 1}$ is a prefix of $\mathcal{I}^\mathcal{B}$ at or after $e_{\min}$, by \Cref{lemma:ero:from_e_onwards_the_list_is_one_of_two_structures}, the list of cells conforms to either $\List(\mathcal{I}^{exclude}_{e_{\min}})$ or $\List(\mathcal{I}^{include}_{e_{\min}})$ in $\mathcal{I}_{i - 1}$.
            Suppose the list of cells conforms to $\List(\mathcal{I}^{exclude}_{e_{\min}})$ in $\mathcal{I}_{i - 1}$.
            Therefore, $\cellpointershort{}_{i - 1}$ is in $\List(\mathcal{I}^{exclude}_{e_{\min}})$, $\cellpointershort{}_{j - 1}$ is in $\List(\mathcal{I}^{include}_{e_{\min}})$, the list of cells conforms to $\List(\mathcal{I}^{exclude}_{e_{\min}})$ in $\mathcal{I}_{i - 1}$, and the list of cells conforms to $\List(\mathcal{I}^{include}_{e_{\min}})$ in $\mathcal{I}_{j - 1}$, and so (3) is satisfied.
            Now suppose the list of cells conforms to $\List(\mathcal{I}^{include}_{e_{\min}})$ in $\mathcal{I}_{i - 1}$.
            Hence, the list of cells conforms to $\List(\mathcal{I}^{include}_{e_{\min}})$ in $\mathcal{I}_{i - 1}$ and $\mathcal{I}_{j - 1}$.
            Therefore, since $\cellpointershort{}_{i - 1}$ and $\cellpointershort{}_{j - 1}$ are both in $\List(\mathcal{I}^{include}_{e_{\min}})$, (2) is satisfied.
    
            \item[] \hspace{0pt}\textbf{Case 4.} $\cellpointershort{}_{i - 1}$ and $\cellpointershort{}_{j - 1}$ are both not in $\List(\mathcal{I}^{exclude}_{e_{\min}})$.
    
            Hence, by \Cref{lemma:ero:acquire_next_for_ptr_is_in_either_include_or_exclude}, $\cellpointershort{}_{i - 1}$ and $\cellpointershort{}_{j - 1}$ are both in $\List(\mathcal{I}^{include}_{e_{\min}})$, and by \Cref{lemma:ero:acquire_next_for_ptr_not_in_exclude_implies_a_prefix_that_does_not_conform_to_exclude} there is a prefix $\mathcal{I}$ of $\mathcal{I}^\mathcal{B}$ at or after $e_{\min}$ and before $I_{i - 1}$ was invoked such that the list of cells does not conform to $\List(\mathcal{I}^{exclude}_{e_{\min}})$ in $\mathcal{I}$.
            Since the end of $\mathcal{I}$ is before $I_{i - 1}$ is invoked and the end of $\mathcal{I}_{i - 1}$ (resp. $\mathcal{I}_{j - 1}$) is after $I_{i - 1}$ was invoked, we have that $\mathcal{I}$ is a prefix of $\mathcal{I}_{i - 1}$ (resp. $\mathcal{I}_{j - 1}$).
            Hence, by \Cref{lemma:ero:from_e_onwards_the_list_is_one_of_two_structures}, the list of cells conforms to $\List(\mathcal{I}^{include}_{e_{\min}})$ in $\mathcal{I}_{i - 1}$ (resp. $\mathcal{I}_{j - 1}$).
            Therefore, (2) is satisfied.
            \qH{\Cref{lemma:ero:all_acquire_next_during_traversal_are_for_different_pointers:claim_two}}
        \end{itemize}
    \end{proof}

    We now finish the proof of \Cref{lemma:ero:all_acquire_next_during_traversal_are_for_different_pointers}.
    We first note that, since $\mathcal{I}^{exclude}_{e_{\min}}$ and $\mathcal{I}^{include}_{e_{\min}}$ are finite, and by \Cref{lemma:ero:the_list_invariants_hold} $P(\mathcal{I}^\mathcal{B})$ holds, by \Cref{lemma:ero:pointers_in_list_are_unique}, the values in $\List(\mathcal{I}^{exclude}_{e_{\min}})$ and $\List(\mathcal{I}^{include}_{e_{\min}})$ are unique.
    By \Cref{lemma:ero:all_acquire_next_during_traversal_are_for_different_pointers:claim_two}, there are three cases.
    \begin{itemize}
        \item[] \hspace{0pt}\textbf{Case 1.} $\cellpointershort{}_{i - 1}$ and $\cellpointershort{}_{j - 1}$ are both in $\List(\mathcal{I}^{exclude}_{e_{\min}})$ and the list of cells conforms to $\List(\mathcal{I}^{exclude}_{e_{\min}})$ in both $\mathcal{I}_{i - 1}$ and $\mathcal{I}_{j - 1}$.

        Hence, since $\cellpointershort{}_{i - 1} \neq \nullconstant$, $\cellpointershort{}_{j - 1} \neq \nullconstant$, $(*\cellpointershort{}_{i - 1}).\nextlong.\cellpointerlong{} = \cellpointershort{}_{i}$ at the end of $\mathcal{I}_{i - 1}$, and $(*\cellpointershort{}_{j - 1}).\nextlong.\cellpointerlong{} = \cellpointershort{}_{j}$ at the end of $\mathcal{I}_{j - 1}$, by \Cref{def:ero:logical_list}, $\cellpointershort{}_i$ and $\cellpointershort{}_j$ are the pointers after $\cellpointershort{}_{i - 1}$ and $\cellpointershort{}_{j - 1}$ in $\List(\mathcal{I}^{exclude}_{e_{\min}})$, respectively.
        Let $\cellpointershort{}_{i - 1}$ and $\cellpointershort{}_{j - 1}$ be the $k$th and $l$th pointers in $\List(\mathcal{I}^{exclude}_{e_{\min}})$, respectively.
        Hence, $\cellpointershort{}_{i}$ and $\cellpointershort{}_{j}$ are the $k + 1$th and $l + 1$th pointers in $\List(\mathcal{I}^{exclude}_{e_{\min}})$, respectively.
        
        We prove that $k = l$.
        Suppose, for contradiction, $k \neq l$.
        Hence, $k + 1 \neq l + 1$.
        Thus, since $\cellpointershort{}_{i}$ and $\cellpointershort{}_{j}$ are the $k + 1$th and $l + 1$th pointers in $\List(\mathcal{I}^{exclude}_{e_{\min}})$, we have that $\cellpointershort{}_i \neq \cellpointershort{}_j$.
        However, by our initial assumption $\cellpointershort{}_i = \cellpointershort{}_j$, a contradiction.

        Since $\cellpointershort{}_{i - 1}$ and $\cellpointershort{}_{j - 1}$ are the $k$th and $l$th pointers in $\List(\mathcal{I}^{exclude}_{e_{\min}})$, respectively, and $k = l$, we have that $\cellpointershort{}_{i - 1} = \cellpointershort{}_{j - 1}$.
        Therefore, since $1 < i < j$, we have that $i - 1$ and $j - 1$ are in $[1..j)$, $i - 1 \neq j - 1$, and $\cellpointershort{}_{i - 1} = \cellpointershort{}_{j - 1}$.
        However, this contradicts the minimality of $j$.

        \item[] \hspace{0pt}\textbf{Case 2.} $\cellpointershort{}_{i - 1}$ and $\cellpointershort{}_{j - 1}$ are both in $\List(\mathcal{I}^{include}_{e_{\min}})$ and the list of cells conforms to $\List(\mathcal{I}^{include}_{e_{\min}})$ in both $\mathcal{I}_{i - 1}$ and $\mathcal{I}_{j - 1}$.

        Hence, since $\cellpointershort{}_{i - 1} \neq \nullconstant$, $\cellpointershort{}_{j - 1} \neq \nullconstant$, $(*\cellpointershort{}_{i - 1}).\nextlong.\cellpointerlong{} = \cellpointershort{}_{i}$ at the end of $\mathcal{I}_{i - 1}$, and $(*\cellpointershort{}_{j - 1}).\nextlong.\cellpointerlong{} = \cellpointershort{}_{j}$ at the end of $\mathcal{I}_{j - 1}$, by \Cref{def:ero:logical_list}, $\cellpointershort{}_i$ and $\cellpointershort{}_j$ are the pointers after $\cellpointershort{}_{i - 1}$ and $\cellpointershort{}_{j - 1}$ in $\List(\mathcal{I}^{include}_{e_{\min}})$, respectively.
        Let $\cellpointershort{}_{i - 1}$ and $\cellpointershort{}_{j - 1}$ be the $k$th and $l$th pointers in $\List(\mathcal{I}^{include}_{e_{\min}})$, respectively.
        Hence, $\cellpointershort{}_{i}$ and $\cellpointershort{}_{j}$ are the $k + 1$th and $l + 1$th pointers in $\List(\mathcal{I}^{include}_{e_{\min}})$, respectively.
        
        We prove that $k = l$.
        Suppose, for contradiction, $k \neq l$.
        Hence, $k + 1 \neq l + 1$.
        Thus, since $\cellpointershort{}_{i}$ and $\cellpointershort{}_{j}$ are the $k + 1$th and $l + 1$th pointers in $\List(\mathcal{I}^{include}_{e_{\min}})$, we have that $\cellpointershort{}_i \neq \cellpointershort{}_j$.
        However, by our initial assumption $\cellpointershort{}_i = \cellpointershort{}_j$, a contradiction.

        Since $\cellpointershort{}_{i - 1}$ and $\cellpointershort{}_{j - 1}$ are the $k$th and $l$th pointers in $\List(\mathcal{I}^{include}_{e_{\min}})$, respectively, and $k = l$, we have that $\cellpointershort{}_{i - 1} = \cellpointershort{}_{j - 1}$.
        Therefore, since $1 < i < j$, we have that $i - 1$ and $j - 1$ are in $[1..j)$, $i - 1 \neq j - 1$, and $\cellpointershort{}_{i - 1} = \cellpointershort{}_{j - 1}$.
        However, this contradicts the minimality of $j$.
        
        \item[] \hspace{0pt}\textbf{Case 3.} $\cellpointershort{}_{i - 1}$ is in $\List(\mathcal{I}^{exclude}_{e_{\min}})$, $\cellpointershort{}_{j - 1}$ is in $\List(\mathcal{I}^{include}_{e_{\min}})$, the list of cells conforms to $\List(\mathcal{I}^{exclude}_{e_{\min}})$ in $\mathcal{I}_{i - 1}$, and the list of cells conforms to $\List(\mathcal{I}^{include}_{e_{\min}})$ in $\mathcal{I}_{j - 1}$.

        Hence, since $\cellpointershort{}_{i - 1} \neq \nullconstant$, $\cellpointershort{}_{j - 1} \neq \nullconstant$, $(*\cellpointershort{}_{i - 1}).\nextlong.\cellpointerlong{} = \cellpointershort{}_{i}$ at the end of $\mathcal{I}_{i - 1}$, and $(*\cellpointershort{}_{j - 1}).\nextlong.\cellpointerlong{} = \cellpointershort{}_{j}$ at the end of $\mathcal{I}_{j - 1}$, by \Cref{def:ero:logical_list}, $\cellpointershort{}_i$ is the pointer after $\cellpointershort{}_{i - 1}$ in $\List(\mathcal{I}^{exclude}_{e_{\min}})$, and $\cellpointershort{}_j$ is the pointer after $\cellpointershort{}_{j - 1}$ in $\List(\mathcal{I}^{include}_{e_{\min}})$.
        Since $\cellpointershort{}_{i - 1}$ and $\cellpointershort{}_{i}$ are both in $\List(\mathcal{I}^{exclude}_{e_{\min}})$, by \Cref{lemma:ero:acquire_next_for_ptr_in_exclude_is_in_include}, $\cellpointershort{}_{i - 1}$ and $\cellpointershort{}_{i}$ are both in $\List(\mathcal{I}^{include}_{e_{\min}})$.

        We first prove that $\cellpointershort{}_{i}$ is the pointer after $\cellpointershort{}_{i - 1}$ in $\List(\mathcal{I}^{include}_{e_{\min}})$.
        Suppose, for contradiction, $\cellpointershort{}_{i}$ is not the pointer after $\cellpointershort{}_{i - 1}$ in $\List(\mathcal{I}^{include}_{e_{\min}})$.
        Since $\cellpointershort{}_{i - 1} \neq \nullconstant$ and $\cellpointershort{}_{i - 1}$ is in $\List(\mathcal{I}^{include}_{e_{\min}})$, by \Cref{def:ero:logical_list}, there is a pointer after $\cellpointershort{}_{i - 1}$ in $\List(\mathcal{I}^{include}_{e_{\min}})$; say $\cellpointershort{}$.
        Since $\cellpointershort{}_{i}$ is the pointer after $\cellpointershort{}_{i - 1}$ in $\List(\mathcal{I}^{exclude}_{e_{\min}})$, $\cellpointershort{}_{i - 1}$ and $\cellpointershort{}_{i}$ are both in $\List(\mathcal{I}^{include}_{e_{\min}})$, and by definition $\mathcal{I}^{exclude}_{e_{\min}}$ contains every $L$-event in $\mathcal{I}^{include}_{e_{\min}}$ with the exception of $e_{\min}$, by \Cref{def:ero:logical_list}, $e_{\min}$ is an $L$-add event for $\cellpointershort{}$.
        Hence, since $\cellpointershort{}$ is after $\cellpointershort{}_{i - 1}$ in $\List(\mathcal{I}^{include}_{e_{\min}})$, by \Cref{def:ero:logical_list}, $\cellpointershort{}_{i - 1}$ is the third last element in $\List(\mathcal{I}^{include}_{e_{\min}})$.
        Thus, since $e_{\min}$ is an $L$-add event for $\cellpointershort{}$, it follows that $\cellpointershort{}_{i - 1}$ is the second last element in $\List(\mathcal{I}^{exclude}_{e_{\min}})$.
        Therefore, since $\cellpointershort{}_i$ is the pointer after $\cellpointershort{}_{i - 1}$ in $\List(\mathcal{I}^{exclude}_{e_{\min}})$, by \Cref{def:ero:logical_list}, $\cellpointershort{}_i = \nullconstant$.
        However, since by \Cref{lemma:ero:acquire_next_is_for_pointer_from_universe_or_head} $\cellpointershort{}_{i} \in \celluniverse \cup \{\&\headobject\}$, by \Cref{assumption:ero:head_and_null_not_in_cell_universe}, $\cellpointershort{}_{i} \neq \nullconstant$, a contradiction.

        Let $\cellpointershort{}_{i - 1}$ and $\cellpointershort{}_{j - 1}$ be the $k$th and $l$th pointers in $\List(\mathcal{I}^{include}_{e_{\min}})$, respectively.
        Hence, since $\cellpointershort{}_i$ and $\cellpointershort{}_j$ are the pointers after $\cellpointershort{}_{i - 1}$ and $\cellpointershort{}_{j - 1}$ in $\List(\mathcal{I}^{include}_{e_{\min}})$, respectively, we have that $\cellpointershort{}_{i}$ and $\cellpointershort{}_{j}$ are the $k + 1$th and $l + 1$th pointers in $\List(\mathcal{I}^{include}_{e_{\min}})$, respectively.
        
        We prove that $k = l$.
        Suppose, for contradiction, $k \neq l$.
        Hence, $k + 1 \neq l + 1$.
        Thus, since $\cellpointershort{}_{i}$ and $\cellpointershort{}_{j}$ are the $k + 1$th and $l + 1$th pointers in $\List(\mathcal{I}^{include}_{e_{\min}})$, we have that $\cellpointershort{}_i \neq \cellpointershort{}_j$.
        However, by our initial assumption $\cellpointershort{}_i = \cellpointershort{}_j$, a contradiction.

        Since $\cellpointershort{}_{i - 1}$ and $\cellpointershort{}_{j - 1}$ are the $k$th and $l$th pointers in $\List(\mathcal{I}^{include}_{e_{\min}})$, respectively, and $k = l$, we have that $\cellpointershort{}_{i - 1} = \cellpointershort{}_{j - 1}$.
        Therefore, since $1 < i < j$, we have that $i - 1$ and $j - 1$ are in $[1..j)$, $i - 1 \neq j - 1$, and $\cellpointershort{}_{i - 1} = \cellpointershort{}_{j - 1}$.
        However, this contradicts the minimality of $j$.
        \qH{\Cref{lemma:ero:all_acquire_next_during_traversal_are_for_different_pointers}}
    \end{itemize}
\end{proof}

This completes the third part of the high-level argument for why $p_{\min}$ cannot take infinitely many steps in the loops on lines \ref{line:ero:add_cell_while_loop}, \ref{line:ero:remove_cell_while_loop}, and \ref{line:ero:acquire_loop_until} during $opx_{\min}$.
We are now ready to prove the main claim of this section.

\begin{lemma}\label{lemma:ero:if_stuck_must_be_the_main_loop_or_cas_loops}
    $\mathcal{L}_{\min}$ is not a loop on line \ref{line:ero:add_cell_while_loop}, \ref{line:ero:remove_cell_while_loop}, or \ref{line:ero:acquire_loop_until}.
\end{lemma}

\begin{proof}
    Suppose, for contradiction, $\mathcal{L}_{\min}$ is a loop on either line \ref{line:ero:add_cell_while_loop}, \ref{line:ero:remove_cell_while_loop}, or \ref{line:ero:acquire_loop_until}.
    Hence, $\mathcal{L}_{\min}$ is not a loop on \cref{line:ero:do_work_while_loop}.
    Thus, this is \Cref{scenario:ero:helper_loop_scenario}.
    Since $\mathcal{L}_{\min}$ is a loop on either line \ref{line:ero:add_cell_while_loop}, \ref{line:ero:remove_cell_while_loop}, or \ref{line:ero:acquire_loop_until}, we have that $I_{\min}$ is an invocation of the \doaddcell{}, \doremovecell{}, or Acquire procedure.
    Since $p_{\min}$ takes infinitely many steps inside $\mathcal{L}_{\min}$, we have that $p$ invokes infinitely many invocations of the AcquireNext procedure during $I_{\min}$.
    Let $I_1, I_2, \ldots$ denote these invocations of the AcquireNext procedure during $I_{\min}$ in the order they were invoked and let $\cellpointershort{}_i$ denote the second parameter of $I_i$.
    Since $\mathcal{I}^{exclude}_{e_{\min}}$ and $\mathcal{I}^{include}_{e_{\min}}$ are finite, by \Cref{def:ero:logical_list}, $\List(\mathcal{I}^{exclude}_{e_{\min}})$ and $\List(\mathcal{I}^{include}_{e_{\min}})$ are finite, and so the union of $\List(\mathcal{I}^{exclude}_{e_{\min}})$ and $\List(\mathcal{I}^{include}_{e_{\min}})$ is finite.
    Therefore, since $p_{\min}$ read $(\uniquerepositoryoperationshort_\linearizationobject{}, \uniquecellpointershort_\linearizationobject{})$ from $\linearizationobject{}$ on its last execution of \cref{line:ero:linearization_read} before invoking $I_{\min}$, by \Cref{lemma:ero:acquire_next_for_ptr_is_in_either_include_or_exclude}, $\cellpointershort{}_i$ is in either $\List(\mathcal{I}^{exclude}_{e_{\min}})$ or $\List(\mathcal{I}^{include}_{e_{\min}})$, and so $\{\cellpointershort{}_i\ \vert\ \forall i\}$ is finite.
    However, by \Cref{lemma:ero:all_acquire_next_during_traversal_are_for_different_pointers}, $\cellpointershort{}_1, \cellpointershort{}_2,\ldots$ are distinct so $\{\cellpointershort{}_i\ \vert\ \forall i\}$ is infinite, a contradiction.
    \qH{\Cref{lemma:ero:if_stuck_must_be_the_main_loop_or_cas_loops}}
\end{proof}

\subsubsection{Processes cannot get stuck in the loops on lines \ref{line:ero:remove_cell_remove_seal_loop}, \ref{line:ero:remove_cell_remove_repeat_loop}, and \ref{line:ero:acquire_next_repeat_loop}.}

This section shows that $p_{\min}$ cannot take infinitely many steps in the loops on lines  \ref{line:ero:remove_cell_remove_seal_loop}, \ref{line:ero:remove_cell_remove_repeat_loop}, and \ref{line:ero:acquire_next_repeat_loop} during $opx_{\min}$.
The high-level argument for why is the following.
First, we prove that the value of the next object of some cell changes infinitely often in $\mathcal{I}^\mathcal{B}$.
Second, we prove that the next object of each cell changes finitely many times in $\mathcal{I}^\mathcal{B}$.
The first property follows from the fact that each of these loops repeatedly performs a \CASop{} operation on some pointer, and once a single one of these \CASop{} operations is successful, $p_{\min}$ exits the loop, as we will soon show.
The main technical difficulty is that some of these \CASop{} operations (in particular, those on \cref{line:ero:remove_cell_from_list} and \cref{line:ero:acquire_next_cell}) have fixed values in their first parameter.
So, to deduce that an unsuccessful \CASop{} operation implies the value changed, we have to show these fixed values are actually the value of the object at the time $p_{\min}$ when it was read.
We start with two basic facts and then prove this.

\begin{proposition}\label{lemma:ero:loop_pointer_is_never_sealed:claim_two}
    In \Cref{scenario:ero:helper_loop_scenario}, suppose $\mathcal{L}_{\min}$ is the loop on \cref{line:ero:remove_cell_remove_repeat_loop}.
    So, $I_{\min}$ is an invocation of the \doremovecell{} procedure.
    Let $\cellpointershort{}$ be the value of the $p_{\min}$'s local variable $\previouscellpointershort{}$ during $\mathcal{L}_{\min}$.
    Then, there is an invocation of the AcquireNext procedure on \cref{line:ero:remove_cell_acquire_next} during $I_{\min}$ whose second parameter is $\cellpointershort{}$.
\end{proposition}

\begin{proof}
    Since $(\uniquerepositoryoperationshort_\linearizationobject{}, \uniquecellpointershort_\linearizationobject{})$ was read from $\linearizationobject{}$ on $p_{\min}$'s last execution of \cref{line:ero:linearization_read} before invoking $I_{\min}$, we have that $(\uniquerepositoryoperationshort_\linearizationobject{}, \uniquecellpointershort_\linearizationobject{})$ are the parameters of $I_{\min}$.
    Hence, by \Cref{lemma:ero:l_event_corresponding_to_do_low_level_op}, there is an $L$-remove event for $\cellpointershort{}_\linearizationobject{}$, and so by \Cref{lemma:ero:every_l_event_is_for_pointer_from_universe} $\cellpointershort{}_\linearizationobject{} \in \celluniverse$.
    Thus, by \Cref{assumption:ero:head_and_null_not_in_cell_universe} $\cellpointershort{}_\linearizationobject{} \neq \&\headobject$, and so the first time $p_{\min}$ executes \cref{line:ero:remove_cell_while_loop} during $I_{\min}$ it finds the condition on \cref{line:ero:remove_cell_while_loop} to be true.
    Since $\mathcal{L}_{\min}$ is the loop on \cref{line:ero:remove_cell_remove_repeat_loop} during $I_{\min}$, we have that $p_{\min}$ exits the loop on \cref{line:ero:remove_cell_while_loop} during $I_{\min}$ by finding the condition on \cref{line:ero:remove_cell_while_loop} to be false.
    Thus, since the first time $p_{\min}$ executes \cref{line:ero:remove_cell_while_loop} during $I_{\min}$ it finds the condition on \cref{line:ero:remove_cell_while_loop} to be true, we have that $p_{\min}$ executes \cref{line:ero:remove_cell_while_loop} at least twice and at most finitely many times during $I_{\min}$.
    Suppose $p_{\min}$ executes \cref{line:ero:remove_cell_while_loop} exactly $n \geq 2$ times during $I_{\min}$.
    Since $p_{\min}$ exits the loop on \cref{line:ero:remove_cell_while_loop} during $I_{\min}$ by finding the condition on \cref{line:ero:remove_cell_while_loop} to be false, and $p_{\min}$ executes \cref{line:ero:remove_cell_while_loop} exactly $n$ times during $I_{\min}$, we have that $p_{\min}$ finds the condition on \cref{line:ero:remove_cell_while_loop} to be false on $p_{\min}$'s $n$th execution of \cref{line:ero:remove_cell_while_loop} during $I_{\min}$.
    Let $\previousuniquecellpointershort{}_i$ and $\currentuniquecellpointershort{}_i$ be the values of $p_{\min}$'s local variables $\previousuniquecellpointershort{}$ and $\currentuniquecellpointershort{}$, respectively, at the time of $p_{\min}$'s $i$th execution of \cref{line:ero:remove_cell_while_loop} during $I_{\min}$ where $i \in [1..n]$.
    Since $n \geq 2$, $p_{\min}$ executes \cref{line:ero:remove_cell_while_loop} $n - 1$ times during $I_{\min}$, and between $p_{\min}$'s $n - 1$th and $n$th execution of \cref{line:ero:remove_cell_while_loop} during $I_{\min}$, $p_{\min}$ invokes the AcquireNext procedure on \cref{line:ero:remove_cell_acquire_next}.
    Denote this invocation by $I_{n - 1}$.
    Since $I_{n - 1}$ is invoked just after $p_{\min}$'s $n - 1$th execution of \cref{line:ero:remove_cell_while_loop}, its second parameter is $\currentuniquecellpointershort{}_{n - 1}$.
    Let $\status$ be the left field of $I_{n - 1}$'s response.
    By the AcquireNext procedure $\status$ is either $\found$, $\timechange$, or $\notfound$.
    Suppose $\status$ is $\timechange$ or $\notfound$.
    Hence, $p_{\min}$ would find the condition on \cref{line:ero:remove_cell_check_acquire_status} to be true after exiting $I_{n - 1}$, and so $p_{\min}$ would execute the goto on \cref{line:ero:exit_1}.
    Thus, $p_{\min}$ would not execute \cref{line:ero:remove_cell_while_loop} for an $n$th time during $I_{\min}$, which is impossible, so $\status$ is $\found$.
    Hence, after exiting $I_{n - 1}$, $p_{\min}$ executes \cref{line:ero:remove_cell_update_pointers}.
    Since $\previousuniquecellpointershort{}$ and $\currentuniquecellpointershort{}$ are the same at this time as they were when $p_{\min}$ executes \cref{line:ero:remove_cell_while_loop} for the $n - 1$th time during $I_{\min}$, we have that $p_{\min}$ sets $\previousuniquecellpointershort{} = \currentuniquecellpointershort{}_{n - 1}$.
    Therefore, since $\previousuniquecellpointershort{}$ is unchanged from this time until $p_{\min}$ \cref{line:ero:remove_cell_while_loop} for the $n$th time during $I_{\min}$, we have that $\previouscellpointershort{}_n = \currentuniquecellpointershort{}_{n - 1}$.
    Since $p_{\min}$ finds the condition on \cref{line:ero:remove_cell_while_loop} to be false on $p_{\min}$'s $n$th execution of \cref{line:ero:remove_cell_while_loop} during $I_{\min}$, we have that $p_{\min}$'s local variable $\previouscellpointershort{} = \currentuniquecellpointershort{}_{n - 1}$ from the time $p_{\min}$ exits the loop on \cref{line:ero:remove_cell_while_loop} during $I_{\min}$ onwards in $I_{\min}$.
    Hence, since $\cellpointershort{}$ is the value of the local variable $\previouscellpointershort{}$ in $\mathcal{L}_{\min}$, we have that $\cellpointershort{} = \currentuniquecellpointershort{}_{n - 1}$.
    Therefore, since $I_{n - 1}$ is an invocation of the AcquireNext procedure during $I_{\min}$ whose second parameter is $\currentcellpointershort{}_{n - 1}$, the claim follows.
    \qH{\Cref{lemma:ero:loop_pointer_is_never_sealed:claim_two}}
\end{proof}

\begin{proposition}\label{lemma:ero:case_a_b_c_ptr_is_from_universe}
    In \Cref{scenario:ero:helper_loop_scenario}, suppose $\mathcal{L}_{\min}$ is either the loop on line \ref{line:ero:remove_cell_remove_seal_loop} (Case A), \ref{line:ero:remove_cell_remove_repeat_loop} (Case B), or \ref{line:ero:acquire_next_repeat_loop} (Case C).
    Let $\cellpointershort{}$ be the value of the local variable $\cellpointershort{}_\linearizationobject{}$ (Case A), $\previouscellpointershort{}$ (Case B), or $\currentcellpointershort{}$ (Case C) in $\mathcal{L}_{\min}$.
    Then, $\cellpointershort{} \in \celluniverse \cup \{\&\headobject\}$.
\end{proposition}

\begin{proof}
    We consider each case separately.

    \begin{itemize}
        \item[] \hspace{0pt}\textbf{Case A.}

        Hence, $I_{\min}$ is an invocation of the \doremovecell{} procedure, and the second parameter of $I_{\min}$ is $\cellpointershort{}$.
        Thus, by \Cref{lemma:ero:l_event_corresponding_to_do_low_level_op}, there is an $L$-event for $\cellpointershort{}$.
        Therefore, by \Cref{lemma:ero:every_l_event_is_for_pointer_from_universe} $\cellpointershort{} \in \celluniverse$.

        \item[] \hspace{0pt}\textbf{Case B.}

        By \Cref{lemma:ero:loop_pointer_is_never_sealed:claim_two}, the second parameter of an invocation of the AcquireNext procedure is $\cellpointershort{}$.
        Hence, by \Cref{lemma:ero:acquire_next_is_for_pointer_from_universe_or_head}, we have that $\cellpointershort{} \in \celluniverse \cup \{\&\headobject\}$.

        \item[] \hspace{0pt}\textbf{Case C.}

        Since $\cellpointershort{}$ is the the second parameter of an invocation of the AcquireNext procedure, by \Cref{lemma:ero:acquire_next_is_for_pointer_from_universe_or_head}, we have that $\cellpointershort{} \in \celluniverse \cup \{\&\headobject\}$.
        \qH{\Cref{lemma:ero:case_a_b_c_ptr_is_from_universe}}
    \end{itemize}
\end{proof}

We now prove that $p_{\min}$ correctly fixes a value of $\false$ for the sealed field on \cref{line:ero:remove_cell_from_list} and \ref{line:ero:acquire_next_cell}.

\begin{proposition}\label{lemma:ero:loop_pointer_is_never_sealed_case_b_and_c}
    In \Cref{scenario:ero:helper_loop_scenario}, suppose $\mathcal{L}_{\min}$ is the loop on line \ref{line:ero:remove_cell_remove_repeat_loop} (Case A), or \ref{line:ero:acquire_next_repeat_loop} (Case B).
    Let $\cellpointershort{}$ be the value of the local variable $\previouscellpointershort{}$ (Case A), or $\currentcellpointershort{}$ (Case B) in $\mathcal{L}_{\min}$.
    By \Cref{lemma:ero:case_a_b_c_ptr_is_from_universe} $\cellpointershort{} \in \celluniverse \cup \{\&\headobject\}$.
    Then, $(*\cellpointershort{}).\nextlong.sealed = \false$ throughout $\mathcal{I}^\mathcal{B}$.
\end{proposition}

\begin{proof}
    The proof is done in two cases.
    First suppose that $e_{\min}$ is an $L$-remove event for $\cellpointershort{}$.
    Hence, since by \Cref{lemma:ero:i_is_proceeded_by_a_unique_l_event} $e_{\min}$ is an $L$-event for $\cellpointershort{}_\linearizationobject{}$, we have that $\cellpointershort{} = \cellpointershort{}_\linearizationobject{}$ and $e_{\min}$ is an $L$-remove event for $\cellpointershort{}_{\linearizationobject{}}$.
    We define an invocation $I$ of the AcquireNext procedure whose second parameter is $\cellpointershort{}$.
    In Case A, let $I$ be the invocation of the AcquireNext procedure identified by \Cref{lemma:ero:loop_pointer_is_never_sealed:claim_two}.
    In Case B, let $I_{\min}$ be the invocation.
    Hence, since $\cellpointershort{} = \cellpointershort{}_\linearizationobject{}$, we have that the second parameter of $I$ is $\cellpointershort{}_\linearizationobject{}$.
    In Case A, $I$ is invoked during $I_{\min}$, and so since $p$ read $(\uniquerepositoryoperationshort_\linearizationobject{}, \uniquecellpointershort_\linearizationobject{})$ from $\linearizationobject{}$ on its last execution of \cref{line:ero:linearization_read} before invoking $I_{\min}$, we have that $p$ read $(\uniquerepositoryoperationshort_\linearizationobject{}, \uniquecellpointershort_\linearizationobject{})$ from $\linearizationobject{}$ on its last execution of \cref{line:ero:linearization_read} before invoking $I$.
    In Case B, this is immediate since $I = I_{\min}$.
    Since $e_{\min}$ is an $L$-remove event for $\cellpointershort{}_{\linearizationobject{}}$, and $I$ is an invocation of the AcquireNext procedure such that $p$ read $(\uniquerepositoryoperationshort_\linearizationobject{}, \uniquecellpointershort_\linearizationobject{})$ from $\linearizationobject{}$ on its last execution of \cref{line:ero:linearization_read} before invoking $I$, by \Cref{lemma:ero:if_e_is_a_remove_event_then_no_one_acquires_it}, the second parameter of $I$ is not $\cellpointershort{}_\linearizationobject{}$.
    However, the second parameter of $I$ is $\cellpointershort{}_\linearizationobject{}$, a contradiction.

    Now suppose that $e_{\min}$ is not an $L$-remove event for $\cellpointershort{}$ and suppose, for contradiction, that $(*\cellpointershort{}).\nextlong.sealed \neq \false$ at some time $T$ during $\mathcal{I}^\mathcal{B}$.
    This implies the following.

    \begin{claimcustom}{\ref{lemma:ero:loop_pointer_is_never_sealed_case_b_and_c}.1}\label{lemma:ero:loop_pointer_is_never_sealed:claim_six}
        There is an $L$-remove event for $\cellpointershort{}$ in $\mathcal{I}^\mathcal{B}$.
    \end{claimcustom}

    \begin{proof}
        Since $\cellpointershort{} \in \celluniverse \cup \{\&\headobject\}$, $(*\cellpointershort{}).\nextlong.sealed$ is initially $\false$.
        Thus, since by our initial assumption $(*\cellpointershort{}).\nextlong.sealed \neq \false$ at $T$, we have that $(*\cellpointershort{}).\nextlong.sealed$ changed in $\mathcal{I}^\mathcal{B}$.
        Hence, by \Cref{observation:ero:where_objects_change}, there is a successful list-sealed attempt $a$ for $\cellpointershort{}$ in $\mathcal{I}^\mathcal{B}$.
        Therefore, by \Cref{lemma:ero:l_event_corresponding_to_do_low_level_op}, there is an $L$-remove event for $\cellpointershort{}$ in $\mathcal{I}^\mathcal{B}$.
        \qH{\Cref{lemma:ero:loop_pointer_is_never_sealed:claim_six}}
    \end{proof}

    \begin{claimcustom}{\ref{lemma:ero:loop_pointer_is_never_sealed_case_b_and_c}.2}\label{lemma:ero:loop_pointer_is_never_sealed:claim_five}
        $\cellpointershort{}$ is in either $\List(\mathcal{I}^{exclude}_{e_{\min}})$ or $\List(\mathcal{I}^{include}_{e_{\min}})$.
    \end{claimcustom}

    \begin{proof}
        We first standardize the proof of both cases by defining an invocation $I_{parent}$ of either the \doaddcell{}, \doremovecell{}, or Acquire procedure, and an invocation $I$ of the AcquireNext procedure which occurs during $I_{parent}$ and whose second parameter is $\cellpointershort{}$.
        In Case A, $I_{\min}$ is an invocation of the \doremovecell{} procedure.
        Let $I_{parent} = I_{\min}$ and let $I$ be the invocation of the AcquireNext procedure during $I_{\min}$ identified by \Cref{lemma:ero:loop_pointer_is_never_sealed:claim_two}.
        In Case B, $I_{\min}$ is an invocation of the AcquireNext procedure.
        Let $I_{parent}$ be the invocation of the \doaddcell, \doremovecell, or Acquire procedure in which $p_{\min}$ invokes $I_{\min}$ during, and let $I = I_{\min}$.
    
        Since $T^{\ref{line:ero:linearization_read}}_{\min}$ is the time of $p_{\min}$'s last execution of \cref{line:ero:linearization_read} before invoking $I_{\min}$, it follows that $T^{\ref{line:ero:linearization_read}}_{\min}$ is also the time of $p_{\min}$'s last execution of \cref{line:ero:linearization_read} before invoking $I_{parent}$.
        Thus, since $p_{\min}$ read $(\uniquerepositoryoperationshort_\linearizationobject{}, \uniquecellpointershort_\linearizationobject{})$ from $\linearizationobject{}$ at $T^{\ref{line:ero:linearization_read}}_{\min}$, we have that $p_{\min}$ read $(\uniquerepositoryoperationshort_\linearizationobject{}, \uniquecellpointershort_\linearizationobject{})$ from $\linearizationobject{}$ on its last execution of \cref{line:ero:linearization_read} before invoking $I_{parent}$.
        Therefore, since $\cellpointershort{}$ is the second parameter of $I$, by \Cref{lemma:ero:acquire_next_for_ptr_is_in_either_include_or_exclude}, $\cellpointershort{}$ is in either $\List(\mathcal{I}^{exclude}_{e_{\min}})$ or $\List(\mathcal{I}^{include}_{e_{\min}})$ as wanted.
        \qH{\Cref{lemma:ero:loop_pointer_is_never_sealed:claim_five}}
    \end{proof}

    We now finish the proof of \Cref{lemma:ero:loop_pointer_is_never_sealed_case_b_and_c}.
    By \Cref{lemma:ero:loop_pointer_is_never_sealed:claim_five} $\cellpointershort{}$ is in either $\List(\mathcal{I}^{exclude}_{e_{\min}})$ or $\List(\mathcal{I}^{include}_{e_{\min}})$, so it suffices to prove that $\cellpointershort{} \notin \List(\mathcal{I}^{exclude}_{e_{\min}})$ and $\cellpointershort{} \notin \List(\mathcal{I}^{include}_{e_{\min}})$.
    Let $e_{remove}$ be the $L$-remove event for $\cellpointershort{}$ identified by \Cref{lemma:ero:loop_pointer_is_never_sealed:claim_six}.
    Since $e_{\min}$ is not an $L$-remove event for $\cellpointershort{}$ and $e_{remove}$ is an $L$-remove event for $\cellpointershort{}$, it follows that $e_{\min} \neq e_{remove}$.
    Hence, since by \Cref{lemma:ero:e_is_the_last_l_event} $e_{\min}$ is the last $L$-event in $\mathcal{I}^\mathcal{B}$, we have $e_{remove} \leq e_{\min}$, and since $e_{\min} \neq e_{remove}$, it follows that $e_{remove} < e_{\min}$.
    Since $e_{remove}$ is an $L$-remove event for $\cellpointershort{}$ in $\mathcal{I}^\mathcal{B}$, by \Cref{lemma:ero:remove_events_are_preceeded_by_add_events}, there is an $L$-add event $e_{add}$ for $\cellpointershort{}$ before $e_{remove}$ in $\mathcal{I}^\mathcal{B}$, and since by \Cref{lemma:ero:the_list_invariants_hold} $P(\mathcal{I}^\mathcal{B})$ holds, we have that $e_{add}$ is the only $L$-add event for $\cellpointershort{}$ in $\mathcal{I}^\mathcal{B}$.
    Hence, since $e_{add} < e_{remove}$ and $e_{remove} < e_{\min}$, by transitivity $e_{add} < e_{\min}$, and so $e_{add} \neq e_{\min}$.
    Thus, since $e_{add}$ is the only $L$-add event for $\cellpointershort{}$ in $\mathcal{I}^\mathcal{B}$, we have that $e_{\min}$ is not an $L$-add event for $\cellpointershort{}$.
    Since $\mathcal{I}^{exclude}_{e_{\min}}$ is the prefix of $\mathcal{I}^\mathcal{B}$ up to but excluding $e_{\min}$, and $\mathcal{I}^{include}_{e_{\min}}$ is the prefix of $\mathcal{I}^\mathcal{B}$ up to and including $e_{\min}$, it follows that the sequence of $L$-events in $\mathcal{I}^{exclude}_{e_{\min}}$ and $\mathcal{I}^{include}_{e_{\min}}$ are the same except the former excludes $e_{\min}$ and the latter includes $e_{\min}$.
    Hence, since $e_{\min}$ is not an $L$-add event for $\cellpointershort{}$, by \Cref{def:ero:logical_list}, if $\cellpointershort{} \notin \List(\mathcal{I}^{exclude}_{e_{\min}})$, then $\cellpointershort{} \notin \List(\mathcal{I}^{include}_{e_{\min}})$.
    Therefore, what remains is to prove that $\cellpointershort{} \notin \List(\mathcal{I}^{exclude}_{e_{\min}})$.
    Suppose, for contradiction, that $\cellpointershort{} \in \List(\mathcal{I}^{exclude}_{e_{\min}})$.
    Hence, by \Cref{def:ero:english}, we have that there is an $L$-add event $e$ for $\cellpointershort{}$ in $\mathcal{I}^{exclude}_{e_{\min}}$ such that there are no $L$-remove events for $\cellpointershort{}$ from $e$ onwards in $\mathcal{I}^{exclude}_{e_{\min}}$.
    Since $e_{add}$ is the only $L$-add event for $\cellpointershort{}$ in $\mathcal{I}^\mathcal{B}$, and $e$ is an $L$-add event for $\cellpointershort{}$ in $\mathcal{I}^\mathcal{B}$, we have that $e_{add} = e$.
    Hence, there are no $L$-remove events for $\cellpointershort{}$ from $e_{add}$ onwards in $\mathcal{I}^{exclude}_{e_{\min}}$.
    Therefore, since $\mathcal{I}^{exclude}_{e_{\min}}$ is the prefix of $\mathcal{I}^\mathcal{B}$ up to but excluding $e_{\min}$, there are no $L$-remove events for $\cellpointershort{}$ between $e_{add}$ and $e_{\min}$.
    However, since $e_{remove}$ is an $L$-remove event for $\cellpointershort{}$, $e_{add} < e_{remove}$, and $e_{remove} < e_{\min}$, we have that there is an $L$-remove event for $\cellpointershort{}$ between $e_{add}$ and $e_{\min}$, a contradiction.
    \qH{\Cref{lemma:ero:loop_pointer_is_never_sealed_case_b_and_c}}
\end{proof}

We now prove that $p_{\min}$ correctly fixes a value of $\cellpointershort{}_\linearizationobject{}$ for the $\cellpointerlong{}$ field on \cref{line:ero:remove_cell_from_list}.

\begin{proposition}\label{lemma:ero:remove_loop_next_is_always_same}
    In \Cref{scenario:ero:helper_loop_scenario}, suppose $\mathcal{L}_{\min}$ is the loop on \cref{line:ero:remove_cell_remove_repeat_loop} so $I_{\min}$ is an invocation of the \doremovecell{} procedure.
    Let $\cellpointershort{}$ be the value of the local variable of $\previouscellpointershort{}$ in $\mathcal{L}_{\min}$.
    Then, $\cellpointershort{} \in \celluniverse \cup \{\&\headobject\}$ and from $e_{\min}$ onwards in $\mathcal{I}^\mathcal{B}$ $(*\cellpointershort{}).\nextlong.\cellpointerlong{} = \uniquecellpointershort_\linearizationobject{}$.
\end{proposition}

\begin{proof}
    Since $\mathcal{L}_{\min}$ is the loop on \cref{line:ero:remove_cell_remove_repeat_loop}, we have that $I_{\min}$ is an invocation of the \doremovecell{} procedure.
    Hence, since $p_{\min}$ read $(\uniquerepositoryoperationshort_\linearizationobject{}, \uniquecellpointershort_\linearizationobject{})$ from $\linearizationobject{}$ on its last execution of \cref{line:ero:linearization_read} before invoking $I_{\min}$, we have that the parameters of $I_{\min}$ are $(\uniquerepositoryoperationshort_\linearizationobject{}, \uniquecellpointershort_\linearizationobject{})$.
    Furthermore, by \cref{line:ero:do_remove_cell_condition}, $\uniquerepositoryoperationshort_\linearizationobject{} = (\arbitraryvalue, \removecell)$.
    Therefore, since by \Cref{lemma:ero:i_is_proceeded_by_a_unique_l_event} $e_{\min}$ set $\linearizationobject = (\uniquerepositoryoperationshort_\linearizationobject{}, \uniquecellpointershort_\linearizationobject{})$, by \Cref{def:ero:english}, $e_{\min}$ is an $L$-remove event for $\cellpointershort{}_\linearizationobject{}$, so by \Cref{lemma:ero:every_l_event_is_for_pointer_from_universe} $\cellpointershort{}_\linearizationobject{} \in \celluniverse$.

    \begin{claimcustom}{\ref{lemma:ero:remove_loop_next_is_always_same}.1}\label{lemma:ero:remove_loop_next_is_always_same:claim_one}
        $\cellpointershort{}_\linearizationobject{}$ is in $\List(\mathcal{I}^{exclude}_{e_{\min}})$ exactly once, and $\cellpointershort{}$ is the pointer preceding $\cellpointershort{}_\linearizationobject{}$ in $\List(\mathcal{I}^{exclude}_{e_{\min}})$.
        Furthermore, $\cellpointershort{} \in \celluniverse \cup \{\&\headobject\}$.
    \end{claimcustom}

    \begin{proof}
        Since $p_{\min}$ takes infinitely many steps in $\mathcal{L}_{\min}$, we have that $p_{\min}$ executes \cref{line:ero:remove_cell_from_list} infinitely often.
        Hence, since $\cellpointershort{}_\linearizationobject{}$ is the second parameter of $I_{\min}$, and $\cellpointershort{}$ is the value of the local variable of $\previouscellpointershort{}$ in $\mathcal{L}_{\min}$, by \Cref{def:ero:english}, $p_{\min}$ performs infinitely many list-remove attempts for $\cellpointershort{}_\linearizationobject{}$ between $\cellpointershort{}$ and some pointer.
        Let $a$ be the first list-remove attempt in $\mathcal{L}_{\min}$.
        Since by \Cref{lemma:ero:the_list_invariants_hold} $Q(\mathcal{I}^\mathcal{B})$ holds, we have that before $a$ there is a unique $L$-remove event $e$ for $\cellpointershort{}_\linearizationobject{}$ and if $\mathcal{I}$ is the prefix of $\mathcal{I}^\mathcal{B}$ up to but excluding $e$, $\cellpointershort{}_\linearizationobject{}$ is in $\List(\mathcal{I})$ exactly once, and $\cellpointershort{}$ is the pointer preceding $\cellpointershort{}_\linearizationobject{}$ in $\List(\mathcal{I})$.
        Since by \Cref{lemma:ero:the_list_invariants_hold} $P(\mathcal{I}^\mathcal{B})$ holds, we have that there is at most one $L$-remove event for $\cellpointershort{}_\linearizationobject{}$ in $\mathcal{I}^\mathcal{B}$.
        Hence, since $e_{\min}$ and $e$ are both $L$-remove events for $\cellpointershort{}_\linearizationobject{}$, we have that $e = e_{\min}$.
        Thus, since $\mathcal{I}$ is the prefix of $\mathcal{I}^\mathcal{B}$ up to but excluding $e$, we have that $\mathcal{I}$ is the prefix of $\mathcal{I}^\mathcal{B}$ up to but excluding $e_{\min}$, and so $\mathcal{I} = \mathcal{I}^{exclude}_{e_{\min}}$.
        Therefore, since $\cellpointershort{}_\linearizationobject{}$ is in $\List(\mathcal{I})$ exactly once, and $\cellpointershort{}$ is the pointer preceding $\cellpointershort{}_\linearizationobject{}$ in $\List(\mathcal{I})$, we have that $\cellpointershort{}_\linearizationobject{}$ is in $\List(\mathcal{I}^{exclude}_{e_{\min}})$ exactly once, and $\cellpointershort{}$ is the pointer preceding $\cellpointershort{}_\linearizationobject{}$ in $\List(\mathcal{I}^{exclude}_{e_{\min}})$.
        Furthermore, since $\cellpointershort{}_\linearizationobject{}$ is in $\List(\mathcal{I}^{exclude}_{e_{\min}})$ exactly once, and $\cellpointershort{}$ is the pointer preceding $\cellpointershort{}_\linearizationobject{}$ in $\List(\mathcal{I}^{exclude}_{e_{\min}})$, it follows that $\cellpointershort{}$ is not the last pointer in $\List(\mathcal{I}^{exclude}_{e_{\min}})$, and so by \Cref{def:ero:logical_list}, $\cellpointershort{} \in \celluniverse \cup \{\&\headobject\}$.
        \qH{\Cref{lemma:ero:remove_loop_next_is_always_same:claim_one}}
    \end{proof}

    \begin{claimcustom}{\ref{lemma:ero:remove_loop_next_is_always_same}.2}\label{lemma:ero:remove_loop_next_is_always_same:claim_two}
        There is a pointer $\nextcellpointershort{}$ succeeding $\cellpointershort{}_\linearizationobject{}$ in $\List(\mathcal{I}^{exclude}_{e_{\min}})$.
        Furthermore, from $e_{\min}$ onwards in $\mathcal{I}^\mathcal{B}$ $(*\cellpointershort{}_\linearizationobject{}).\nextlong.\cellpointerlong{} = \nextcellpointershort{}$.
    \end{claimcustom}

    \begin{proof}
        Since $\cellpointershort{}_\linearizationobject{} \in \celluniverse$, by \Cref{assumption:ero:head_and_null_not_in_cell_universe}, $\cellpointershort{}_\linearizationobject{} \neq \nullconstant$.
        Hence, since by \Cref{lemma:ero:remove_loop_next_is_always_same:claim_one} $\cellpointershort{}_\linearizationobject{}$ is in $\List(\mathcal{I}^{exclude}_{e_{\min}})$ exactly once, by \Cref{def:ero:logical_list}, there is a pointer succeeding $\cellpointershort{}_\linearizationobject{}$ in $\List(\mathcal{I}^{exclude}_{e_{\min}})$; say $\nextcellpointershort{}$.
        We now prove the furthermore part.
        Suppose, for contradiction, $(*\cellpointershort{}_\linearizationobject{}).\nextlong.\cellpointerlong{} \neq \nextcellpointershort{}$ at some time $T \geq e_{\min}$ in $\mathcal{I}^\mathcal{B}$.
        We first prove that $(*\cellpointershort{}_\linearizationobject{}).\nextlong.\cellpointerlong{} = \nextcellpointershort{}$ at $e_{\min}$.
        Since by \Cref{lemma:ero:e_is_the_last_l_event} $e_{\min}$ is the last $L$-event in $\mathcal{I}^\mathcal{B}$, and by definition $\mathcal{I}^{include}_{e_{\min}}$ is the prefix of $\mathcal{I}^\mathcal{B}$ up to and including $e_{\min}$, we have that $e_{\min}$ is the last $L$-event in $\mathcal{I}^{include}_{e_{\min}}$.
        Hence, since $e_{\min}$ is the last step in $\mathcal{I}^{include}_{e_{\min}}$, we have that from $e_{\min}$ onwards in $\mathcal{I}^{include}_{e_{\min}}$ there are no successful list-add or list-remove attempts.
        Thus, since by \Cref{lemma:ero:the_list_invariants_hold}, $P(\mathcal{I}^\mathcal{B})$, $Q(\mathcal{I}^\mathcal{B})$, and $R(\mathcal{I}^\mathcal{B})$ hold, by \Cref{lemma:ero:conditional_classification_lemma}, the list of cells conforms to $\List(\mathcal{I}^{exclude}_{e_{\min}})$ in $\mathcal{I}^{include}_{e_{\min}}$.
        Hence, since $\nextcellpointershort{}$ is the pointer after $\cellpointershort{}_\linearizationobject{}$ in $\List(\mathcal{I}^{exclude}_{e_{\min}})$, and the the list of cells conforms to $\List(\mathcal{I}^{exclude}_{e_{\min}})$ in $\mathcal{I}^{include}_{e_{\min}}$, by \Cref{def:ero:logical_list}, at the end of $\mathcal{I}^{include}_{e_{\min}}$, $(*\cellpointershort{}_\linearizationobject{}).\nextlong.\cellpointerlong{} = \nextcellpointershort{}$.
        Therefore, since $e_{\min}$ is the last step in $\mathcal{I}^{include}_{e_{\min}}$, we have that $(*\cellpointershort{}_\linearizationobject{}).\nextlong.\cellpointerlong{} = \nextcellpointershort{}$ at $e_{\min}$ as wanted.
        Since $(*\cellpointershort{}_\linearizationobject{}).\nextlong.\cellpointerlong{} \neq \nextcellpointershort{}$ at some time $T \geq e_{\min}$, this implies that $T > e_{\min}$.
        Hence, since $(*\cellpointershort{}_\linearizationobject{}).\nextlong.\cellpointerlong{} = \nextcellpointershort{}$ at $e_{\min}$, and $(*\cellpointershort{}_\linearizationobject{}).\nextlong.\cellpointerlong{} \neq \nextcellpointershort{}$ at $T > e_{\min}$, we have that $(*\cellpointershort{}_\linearizationobject{}).\nextlong.\cellpointerlong{}$ changed between $e_{\min}$ and $T$.
        Thus, since $\cellpointershort{}_\linearizationobject{} \in \celluniverse$, by \Cref{observation:ero:where_objects_change}, there is a successful list-add attempt after $\cellpointershort{}_\linearizationobject{}$ or there is a successful list-remove attempt between $\cellpointershort{}_\linearizationobject{}$ and some pointer after $e_{\min}$ in $\mathcal{I}^\mathcal{B}$; say $a$.
        Since by \Cref{lemma:ero:e_is_the_last_l_event} $e_{\min}$ is the last $L$-event in $\mathcal{I}^\mathcal{B}$, $e_{\min}$ is an $L$-remove event for $\cellpointershort{}_\linearizationobject{}$, and by \Cref{lemma:ero:the_list_invariants_hold}, $P(\mathcal{I}^\mathcal{B})$, $Q(\mathcal{I}^\mathcal{B})$, and $R(\mathcal{I}^\mathcal{B})$ hold, by \Cref{lemma:ero:3_of_r_safety_holds}, from $e_{\min}$ onwards in $\mathcal{I}^\mathcal{B}$ there is at most one successful list-remove attempt for $\cellpointershort{}_\linearizationobject{}$ and no other successful list-remove or list-add attempts for any pointer.
        Hence, since $a$ is a successful list-attempt after $e_{\min}$, we have that $a$ is a successful list-remove attempt for $\cellpointershort{}_\linearizationobject{}$.
        Thus, since $a$ is either a successful list-add attempt after $\cellpointershort{}_\linearizationobject{}$ or a successful list-remove attempt between $\cellpointershort{}_\linearizationobject{}$ and some pointer, we have that $a$ is a successful list-remove attempt for $\cellpointershort{}_\linearizationobject{}$ between $\cellpointershort{}_\linearizationobject{}$ and some pointer.
        Therefore, since by \Cref{lemma:ero:the_list_invariants_hold} $P(\mathcal{I}^\mathcal{B})$ and $Q(\mathcal{I}^\mathcal{B})$ hold, by \Cref{lemma:ero:list_remove_attempt_has_different_next}, $\cellpointershort{}_\linearizationobject{} \neq \cellpointershort{}_\linearizationobject{}$.
        However, $\cellpointershort{}_\linearizationobject{} = \cellpointershort{}_\linearizationobject{}$, a contradiction.
        \qH{\Cref{lemma:ero:remove_loop_next_is_always_same:claim_two}}
    \end{proof}

    \begin{claimcustom}{\ref{lemma:ero:remove_loop_next_is_always_same}.3}\label{lemma:ero:remove_loop_next_is_always_same:claim_three}
        $\cellpointershort{} \neq \cellpointershort{}_\linearizationobject{} \neq \nextcellpointershort{}$, $\cellpointershort{}$ is in $\List(\mathcal{I}^{include}_{e_{\min}})$ exactly once, and $\nextcellpointershort{}$ is the pointer succeeding $\cellpointershort{}$ in $\List(\mathcal{I}^{include}_{e_{\min}})$.
    \end{claimcustom}

    \begin{proof}
        Since $\mathcal{I}^{exclude}_{e_{\min}}$ is finite, and by \Cref{lemma:ero:the_list_invariants_hold} $P(\mathcal{I}^\mathcal{B})$ holds, by \Cref{lemma:ero:pointers_in_list_are_unique}, the pointers in $\List(\mathcal{I}^{exclude}_{e_{\min}})$ are pairwise distinct.
        Hence, since by \Cref{lemma:ero:remove_loop_next_is_always_same:claim_one} $\cellpointershort{}_\linearizationobject{}$ is in $\List(\mathcal{I}^{exclude}_{e_{\min}})$ exactly once and $\cellpointershort{}$ precedes $\cellpointershort{}_\linearizationobject{}$ in $\List(\mathcal{I}^{exclude}_{e_{\min}})$, and by \Cref{lemma:ero:remove_loop_next_is_always_same:claim_two} $\nextcellpointershort{}$ succeeds $\cellpointershort{}_\linearizationobject{}$ in $\List(\mathcal{I}^{exclude}_{e_{\min}})$, we have that $\cellpointershort{} \neq \cellpointershort{}_\linearizationobject{} \neq \nextcellpointershort{}$.
        Since $e_{\min}$ is an $L$-remove event for $\cellpointershort{}_\linearizationobject{}$, and by definition $\List(\mathcal{I}^{include}_{e_{\min}})$ is a one step extension of $\List(\mathcal{I}^{exclude}_{e_{\min}})$ that includes $e_{\min}$, by \Cref{def:ero:logical_list}, we have that $\List(\mathcal{I}^{include}_{e_{\min}})$ and $\List(\mathcal{I}^{exclude}_{e_{\min}})$ are identical with the exception that $\cellpointershort{}_\linearizationobject{}$ is in $\List(\mathcal{I}^{exclude}_{e_{\min}})$ but not in $\List(\mathcal{I}^{include}_{e_{\min}})$.
        Therefore, since $\cellpointershort{} \neq \cellpointershort{}_\linearizationobject{}$ and $\cellpointershort{}$ is in $\List(\mathcal{I}^{exclude}_{e_{\min}})$ exactly once, we have that $\cellpointershort{}$ is in $\List(\mathcal{I}^{include}_{e_{\min}})$ exactly once.
        Likewise, since $\nextcellpointershort{} \neq \cellpointershort{}_\linearizationobject{}$, we have that $\nextcellpointershort{}$ is in $\List(\mathcal{I}^{include}_{e_{\min}})$.
        Therefore, since $\List(\mathcal{I}^{include}_{e_{\min}})$ and $\List(\mathcal{I}^{exclude}_{e_{\min}})$ are identical with the exception that $\cellpointershort{}_\linearizationobject{}$ is in $\List(\mathcal{I}^{exclude}_{e_{\min}})$ but not in $\List(\mathcal{I}^{include}_{e_{\min}})$, $\cellpointershort{}$ is the pointer preceding $\cellpointershort{}_\linearizationobject{}$ in $\List(\mathcal{I}^{exclude}_{e_{\min}})$, $\nextcellpointershort{}$ is the pointer succeeding $\cellpointershort{}_\linearizationobject{}$ in $\List(\mathcal{I}^{exclude}_{e_{\min}})$, $\cellpointershort{}$ is in $\List(\mathcal{I}^{include}_{e_{\min}})$ exactly once, and  $\nextcellpointershort{}$ is in $\List(\mathcal{I}^{include}_{e_{\min}})$, we have that $\nextcellpointershort{}$ is the pointer succeeding $\cellpointershort{}$ in $\List(\mathcal{I}^{include}_{e_{\min}})$ as wanted.
        \qH{\Cref{lemma:ero:remove_loop_next_is_always_same:claim_three}}
    \end{proof}

    \begin{claimcustom}{\ref{lemma:ero:remove_loop_next_is_always_same}.4}\label{lemma:ero:remove_loop_next_is_always_same:claim_four}
        For every finite prefix $\mathcal{I}$ of $\mathcal{I}^\mathcal{B}$ at or after $e_{\min}$ the list of cells conforms to $\List(\mathcal{I}^{exclude}_{e_{\min}})$ in $\mathcal{I}$.
    \end{claimcustom}

    \begin{proof}
        Suppose, for contradiction, there is a prefix $\mathcal{I}$ of $\mathcal{I}^\mathcal{B}$ at or after $e_{\min}$ such that the list of cells does not conform to $\List(\mathcal{I}^{exclude}_{e_{\min}})$ in $\mathcal{I}$.
        % Since $\mathcal{I}$ is a prefix of $\mathcal{I}^\mathcal{B}$ at or after $e_{\min}$, by \Cref{lemma:ero:from_e_onwards_the_list_is_one_of_two_structures}, the list of cells conforms to either $\List(\mathcal{I}^{exclude}_{e_{\min}})$ or $\List(\mathcal{I}^{include}_{e_{\min}})$ in $\mathcal{I}$.
        % Hence, the list of cells conforms to $\List(\mathcal{I}^{include}_{e_{\min}})$ in $\mathcal{I}$.
        Since $p_{\min}$ takes infinitely many steps in $\mathcal{L}_{\min}$, we have that $p_{\min}$ executes \cref{line:ero:remove_cell_read_previous_pointer} infinitely often in $\mathcal{L}_{\min}$.
        Hence, $p_{\min}$ executes \cref{line:ero:remove_cell_read_previous_pointer} in $\mathcal{L}_{\min}$ after the end of $\mathcal{I}$.
        Let $T^{\ref{line:ero:remove_cell_read_previous_pointer}}$ be the first time $p_{\min}$ executes \cref{line:ero:remove_cell_read_previous_pointer} in $\mathcal{L}_{\min}$ after the end of $\mathcal{I}$, and let $T^{\ref{line:ero:remove_cell_read_pointer_to_remove}}$ be the time of $p_{\min}$'s execution of \cref{line:ero:remove_cell_read_pointer_to_remove} immediately preceding $T^{\ref{line:ero:remove_cell_read_previous_pointer}}$.
        Since by \Cref{lemma:ero:i_is_proceeded_by_a_unique_l_event}, $e_{\min} < T^{\ref{line:ero:linearization_read}}_{\min}$, $T^{\ref{line:ero:linearization_read}}_{\min}$ is the time of $p_{\min}$ last execution of \cref{line:ero:linearization_read} before $I_{\min}$, and $T^{\ref{line:ero:remove_cell_read_pointer_to_remove}}$ and $T^{\ref{line:ero:remove_cell_read_previous_pointer}}$ are times of steps executing during $I_{\min}$, we have that $e_{\min} < T^{\ref{line:ero:remove_cell_read_pointer_to_remove}} < T^{\ref{line:ero:remove_cell_read_previous_pointer}}$.
        We first show what value $p_{\min}$ read at $T^{\ref{line:ero:remove_cell_read_pointer_to_remove}}$.
        Since the parameters of $I_{\min}$ are $(\uniquerepositoryoperationshort_\linearizationobject{}, \uniquecellpointershort_\linearizationobject{})$, and $e_{\min} < T^{\ref{line:ero:remove_cell_read_pointer_to_remove}}$, by \Cref{lemma:ero:remove_loop_next_is_always_same:claim_two},  $p_{\min}$ read $\nextuniquecellpointershort$ from $(*\cellpointershort{}_\linearizationobject{}).\nextlong.\cellpointerlong{}$ on \cref{line:ero:remove_cell_read_pointer_to_remove} at $T^{\ref{line:ero:remove_cell_read_pointer_to_remove}}$.
        We now show what value $p_{\min}$ read at $T^{\ref{line:ero:remove_cell_read_previous_pointer}}$.
        Let $\mathcal{I}^{\ref{line:ero:remove_cell_read_previous_pointer}}$ be the prefix of $\mathcal{I}^\mathcal{B}$ up to and including $T^{\ref{line:ero:remove_cell_read_previous_pointer}}$.
        Since by definition $T^{\ref{line:ero:remove_cell_read_previous_pointer}}$ is the first time $p_{\min}$ executes \cref{line:ero:remove_cell_read_previous_pointer} in $\mathcal{L}_{\min}$ after the end of $\mathcal{I}$, we have that $\mathcal{I}$ is a prefix of $\mathcal{I}^{\ref{line:ero:remove_cell_read_previous_pointer}}$.
        Hence, since the list of cells does not conform to $\List(\mathcal{I}^{exclude}_{e_{\min}})$ in $\mathcal{I}$, by \Cref{lemma:ero:if_the_list_is_only_include_after_e_it_is_always_include}, the list of cells conforms to $\List(\mathcal{I}^{include}_{e_{\min}})$ in $\mathcal{I}^{\ref{line:ero:remove_cell_read_previous_pointer}}$.
        Thus, since by \Cref{lemma:ero:remove_loop_next_is_always_same:claim_three} $\cellpointershort{}$ is in $\List(\mathcal{I}^{include}_{e_{\min}})$ and $\nextcellpointershort{}$ is the pointer succeeding $\cellpointershort{}$ in $\List(\mathcal{I}^{include}_{e_{\min}})$, by \Cref{def:ero:logical_list}, $(*\cellpointershort{}).\nextlong.\cellpointerlong{} = \nextcellpointershort{}$ at the end of $\mathcal{I}^{\ref{line:ero:remove_cell_read_previous_pointer}}$.
        Therefore, since $\cellpointershort{}$ is the value of the local variable of $\previouscellpointershort{}$ in $\mathcal{L}_{\min}$, we have that $p_{\min}$ read $\nextuniquecellpointershort$ from $(*\cellpointershort{}).\nextlong.\cellpointerlong{}$ on \cref{line:ero:remove_cell_read_previous_pointer} at $T^{\ref{line:ero:remove_cell_read_previous_pointer}}$.
        We now finish the proof.
        Since $(*\cellpointershort{}_\linearizationobject{}).\nextlong.\cellpointerlong{} = \nextuniquecellpointershort$ at $T^{\ref{line:ero:remove_cell_read_pointer_to_remove}}$, and $(*\cellpointershort{}).\nextlong.\cellpointerlong{} = \nextuniquecellpointershort$ at $T^{\ref{line:ero:remove_cell_read_previous_pointer}}$, we have that $p_{\min}$ finds the condition on \cref{line:ero:remove_cell_before_removal_linearization_check} to be true after $T^{\ref{line:ero:remove_cell_read_previous_pointer}}$.
        Therefore, $p_{\min}$ exits $\mathcal{L}_{\min}$, and so $p_{\min}$ takes finitely many steps in $\mathcal{L}_{\min}$.
        However, $p_{\min}$ takes infinitely many steps in $\mathcal{L}_{\min}$, a contradiction.
        \qH{\Cref{lemma:ero:remove_loop_next_is_always_same:claim_four}}
    \end{proof}

    We now finish the proof of \Cref{lemma:ero:remove_loop_next_is_always_same}.
    Consider any prefix $\mathcal{I}$ of $\mathcal{I}^\mathcal{B}$ at or after $e_{\min}$.
    By \Cref{lemma:ero:remove_loop_next_is_always_same:claim_four}, the list of cells conforms to $\List(\mathcal{I}^{exclude}_{e_{\min}})$ in $\mathcal{I}$.
    Since by \Cref{lemma:ero:remove_loop_next_is_always_same:claim_one} $\cellpointershort{}$ precedes $\cellpointershort{}_\linearizationobject{}$ in $\List(\mathcal{I}^{exclude}_{e_{\min}})$ in $\mathcal{I}$, by \Cref{def:ero:logical_list}, $(*\cellpointershort{}).\nextlong.\cellpointerlong{} =  \cellpointershort{}_\linearizationobject{}$ at the end of $\mathcal{I}$. 
    Therefore, since $\mathcal{I}$ is any prefix of $\mathcal{I}^\mathcal{B}$ at or after $e_{\min}$, the claim follows.
    \qH{\Cref{lemma:ero:remove_loop_next_is_always_same}}
\end{proof}

We are now ready to finish the first part of this subsection: that the value of the next object of some cell changes infinitely in $\mathcal{I}^\mathcal{B}$.

\begin{proposition}\label{lemma:ero:infinite_cas_loops_imply_a_pointer_changes_infinitely_often}
    In \Cref{scenario:ero:helper_loop_scenario}, suppose $\mathcal{L}_{\min}$ is either the loop on line \ref{line:ero:remove_cell_remove_seal_loop} (Case A), \ref{line:ero:remove_cell_remove_repeat_loop} (Case B), or \ref{line:ero:acquire_next_repeat_loop} (Case C).
    Let $\cellpointershort{}$ be the value of the local variable $\cellpointershort{}_\linearizationobject{}$ (Case A), $\previouscellpointershort{}$ (Case B), or $\currentcellpointershort{}$ (Case C) in $\mathcal{L}_{\min}$.
    Then, $(*\cellpointershort{}).\nextlong$ changes infinitely often during $\mathcal{I}^\mathcal{B}$.
\end{proposition}

\begin{proof}
    We consider each case separately.

    \begin{itemize}
        \item[] \hspace{0pt}\textbf{Case A.}

        Hence, $\mathcal{L}_{\min}$ is the loop on \cref{line:ero:remove_cell_remove_seal_loop}.
        Since $p_{\min}$ takes infinitely many steps inside $\mathcal{L}_{\min}$, we have that $p_{\min}$ executes infinitely many unsuccessful \CASop{} operations on \cref{line:ero:seal_cell} during $\mathcal{L}_{\min}$.
        Hence, since the first parameter of each of these \CASop{} operations is the value read from $(*\cellpointershort{}).\nextlong$ on the line before, we have that $(*\cellpointershort{}).\nextlong$ changes infinitely often as wanted.

        \item[] \hspace{0pt}\textbf{Case B.}

        Hence, $\mathcal{L}_{\min}$ is the loop on \cref{line:ero:remove_cell_remove_repeat_loop}.
        Thus, $I_{\min}$ is an invocation of the \doremovecell{} procedure.
        Since $p_{\min}$ read $(\uniquerepositoryoperationshort_\linearizationobject{}, \uniquecellpointershort{}_\linearizationobject{})$ from $\linearizationobject{}$ on its last execution of \cref{line:ero:linearization_read} before invoking $I_{\min}$, we have that $(\uniquerepositoryoperationshort_\linearizationobject{}, \uniquecellpointershort{}_\linearizationobject{})$ are the parameters of $I_{\min}$.
        Furthermore, by \Cref{lemma:ero:if_q_read_e_s_value_then_its_after_e}, $p_{\min}$ invoked $I_{\min}$ after $e_{\min}$.
        Since $p_{\min}$ takes infinitely many steps inside $\mathcal{L}_{\min}$, we have that $p_{\min}$ executes infinitely many unsuccessful \CASop{} operations on \cref{line:ero:remove_cell_from_list} during $\mathcal{L}_{\min}$.
        Consider any of these unsuccessful \CASop{} operations and denote it by $o$.
        It suffices to prove that between $p_{\min}$'s last execution of  \cref{line:ero:remove_cell_read_previous_pointer} before $o$ and $o$, the value of $(*\cellpointershort{}).\nextlong$ changed.
        Let $T^{\ref{line:ero:remove_cell_read_previous_pointer}}$ be time of $p_{\min}$'s last execution of  \cref{line:ero:remove_cell_read_previous_pointer} before $o$.
        Since $o$ is executed inside $\mathcal{L}_{\min}$ which is executed inside $I_{\min}$, and the second parameter of $I_{\min}$ is $\uniquecellpointershort{}_\linearizationobject{}$, we have that the first parameter of $o$ is of the form $(\view', \false, a', \uniquecellpointershort_\linearizationobject{})$.
        Hence, by \cref{line:ero:remove_cell_read_previous_pointer}, $(*\cellpointershort{}).\nextlong = (\view', \arbitraryvalue, a', \arbitraryvalue)$ at $T^{\ref{line:ero:remove_cell_read_previous_pointer}}$.
        Furthermore, by \Cref{lemma:ero:loop_pointer_is_never_sealed_case_b_and_c}, $(*\cellpointershort{}).\nextlong.sealed = \false$ at $T^{\ref{line:ero:remove_cell_read_previous_pointer}}$.
        Lastly, since $e_{\min}$ is before $p_{\min}$ invoked $I_{\min}$ and $T^{\ref{line:ero:remove_cell_read_previous_pointer}}$ is in $I_{\min}$, by transitivity, $e_{\min} < T^{\ref{line:ero:remove_cell_read_previous_pointer}}$, and so by \Cref{lemma:ero:remove_loop_next_is_always_same}, $(*\cellpointershort{}).\nextlong.\cellpointerlong{} = \uniquecellpointershort_\linearizationobject{}$ at $T^{\ref{line:ero:remove_cell_read_previous_pointer}}$.
        Hence, $(*\cellpointershort{}).\nextlong = (\view', \false, a', \uniquecellpointershort_\linearizationobject{})$ at $T^{\ref{line:ero:remove_cell_read_previous_pointer}}$.
        Therefore, since the first parameter of $o$ equals the value of $(*\cellpointershort{}).\nextlong$ at the time of $p_{\min}$'s last execution of \cref{line:ero:remove_cell_read_previous_pointer} before $o$, and $o$ was unsuccessful, we have that the value of $(*\cellpointershort{}).\nextlong$ changed, as wanted.

        \item[] \hspace{0pt}\textbf{Case C.}

        Hence, $\mathcal{L}_{\min}$ is the loop on \cref{line:ero:acquire_next_repeat_loop}.
        Since $p_{\min}$ takes infinitely many steps inside $\mathcal{L}_{\min}$, we have that $p_{\min}$ executes infinitely many unsuccessful \CASop{} operations on \cref{line:ero:acquire_next_cell} during $\mathcal{L}_{\min}$.
        Consider any of these unsuccessful \CASop{} operations and denote it by $o$.
        It suffices to prove that between $p_{\min}$'s last execution of  \cref{line:ero:acquire_next_read_curr_unique_pointer} before $o$ and $o$, the value of $(*\cellpointershort{}).\nextlong$ changed.
        Let $(\view, \false, a, \nextuniquecellpointershort)$ be the first parameter of $o$.
        Hence, by \cref{line:ero:acquire_next_read_curr_unique_pointer}, $(*\cellpointershort{}).\nextlong = (\view, \arbitraryvalue, a, \nextuniquecellpointershort)$ on $p_{\min}$'s last execution of  \cref{line:ero:acquire_next_read_curr_unique_pointer} before $o$.
        Furthermore, by \Cref{lemma:ero:loop_pointer_is_never_sealed_case_b_and_c}, $(*\cellpointershort{}).\nextlong.sealed = \false$ throughout $\mathcal{I}^\mathcal{B}$.
        Hence, $(*\cellpointershort{}).\nextlong = (\view, \false, a, \nextuniquecellpointershort)$ on $p_{\min}$'s last execution of  \cref{line:ero:acquire_next_read_curr_unique_pointer} before $o$.
        Therefore, since the first parameter of $o$ equals the value of $(*\cellpointershort{}).\nextlong$ at the time of $p_{\min}$'s last execution of \cref{line:ero:acquire_next_read_curr_unique_pointer} before $o$, and $o$ was unsuccessful, we have that the value of $(*\cellpointershort{}).\nextlong$ changed, as wanted.
        \qH{\Cref{lemma:ero:infinite_cas_loops_imply_a_pointer_changes_infinitely_often}}
    \end{itemize}
\end{proof}

This completes the first part of the high-level argument for why $p_{\min}$ cannot take infinitely many steps in the loops on lines  \ref{line:ero:remove_cell_remove_seal_loop}, \ref{line:ero:remove_cell_remove_repeat_loop}, and \ref{line:ero:acquire_next_repeat_loop} during $opx_{\min}$.
We now prove the second part of this subsection: the next object of each cell changes finitely many times in $\mathcal{I}^\mathcal{B}$.
The high-level argument for why this is true is that: (1) each process performs a finite number of successful \CASop{} operations on the next object of each cell; and (2) finitely many processes take steps in $\mathcal{I}^\mathcal{B}$.
The idea for proving (1) is that if a process performing infinitely many successful \CASop{} operations on the next object of some cell, then it must perform an $L$-event after $e_{\min}$ in $\mathcal{I}^\mathcal{B}$, contradicting \Cref{lemma:ero:e_is_the_last_l_event}.
We start by proving two facts that relate $\announceobject$ and $\linearizationobject{}$.

\begin{lemma}\label{lemma:ero:a_events_which_get_into_l_eventually_get_out_of_a}
    Suppose there is an $L$-event $e$ in $\mathcal{I}^\mathcal{B}$ that sets $\linearizationobject{} = v$.
    Then, there are finitely many executions of \cref{line:ero:announce_gcas} or \cref{line:ero:announce_cas} in $\mathcal{I}^\mathcal{B}$ that try to set $\announceobject = v$.
\end{lemma}

\begin{proof}
    Suppose, for contradiction, there are infinitely many executions of \cref{line:ero:announce_gcas} or \cref{line:ero:announce_cas} in $\mathcal{I}^\mathcal{B}$ that try to set $\announceobject = v$.
    Hence, by \Cref{lemma:ero:try_to_set_a_to_same_value_by_same_process}, some process $p$ performs infinitely many executions of \cref{line:ero:announce_gcas} or \cref{line:ero:announce_cas} that try to set $\announceobject = v$ during some invocation $I$ of the \doworkuntildone{} procedure.
    Since $e$ is an $L$-event, by \Cref{def:ero:english}, $e$ sets $\linearizationobject{} = v = ((\arbitraryvalue, \repositoryoperationshort), \cellpointershort{})$.
    Hence, by \Cref{def:ero:english}, $e$ is an $L$-event for $\cellpointershort{}$ so by 
    \Cref{lemma:ero:every_l_event_is_add_apply_or_remove} $\repositoryoperationshort$ is either $\addcell$, $\langle \doopandcopyresponse, \arbitraryvalue \rangle$, or $\removecell$, and by \Cref{lemma:ero:every_l_event_is_for_pointer_from_universe} $\cellpointershort{} \in \celluniverse$.
    Since $p$ performs infinitely many executions of \cref{line:ero:announce_gcas} or \cref{line:ero:announce_cas} that try to set $\announceobject = v$ during $I$ and $v = ((\arbitraryvalue, \repositoryoperationshort), \cellpointershort{})$, we have that $(\repositoryoperationshort, \cellpointershort{})$ are of the parameters of $I$.
    Furthermore, $p$ executes \cref{line:ero:do_work_while_loop} and \cref{line:ero:linearization_read} infinitely often during $I$, and so $(*\cellpointershort{}).\lastrepositoryoperationresponse{} = ((\arbitraryvalue, \repositoryoperationshort), \nullconstant{})$ infinitely often during $\mathcal{I}^\mathcal{B}$ (*).
    There are three cases.
    We note that the proofs of each case are essentially the same, but they rely on different lemmas.
    
    \begin{itemize}
        \item[] \hspace{0pt}\textbf{Case 1.} $\repositoryoperationshort = \addcell$.

        There are two cases.

        \begin{itemize}
            \item[] \hspace{0pt}\textbf{Case 1.1.} $\linearizationobject{} = ((\arbitraryvalue, \addcell), \cellpointershort{})$ from $e$ onwards.

            Since $p$ executes \cref{line:ero:linearization_read} infinitely often during $I$, it follows that $p$ executes \cref{line:ero:linearization_read} at some time $T^{\ref{line:ero:linearization_read}}_p$ after $e$, and so $p$ reads $((\arbitraryvalue, \addcell), \cellpointershort{})$ from $\linearizationobject{}$ at $T^{\ref{line:ero:linearization_read}}_p$.
            Hence, since $p$ executes \cref{line:ero:linearization_read} again after $T^{\ref{line:ero:linearization_read}}_p$, it follows that $p$ invokes the \doaddcell{} procedure with a second parameter of $\cellpointershort{}$ on \cref{line:ero:do_add_cell} immediately after $T^{\ref{line:ero:linearization_read}}_p$ and $p$ exits this invocation of the \doaddcell{} procedure; say at time $T^{\ref{line:ero:do_add_cell}}_p$.
            Hence, since by \Cref{lemma:ero:the_list_invariants_hold} $P(\mathcal{I}^\mathcal{B})$, $Q(\mathcal{I}^\mathcal{B})$, and $R(\mathcal{I}^\mathcal{B})$ hold, by \Cref{lemma:ero:exit_add_implies_response_set}, there is a successful add-response-set attempt $a$ for $\cellpointershort{}$ before $T^{\ref{line:ero:do_add_cell}}_p$.
            Therefore, since $\cellpointershort{} \in \celluniverse$, by \Cref{lemma:ero:once_response_not_null_for_add_never_null_for_add_again}, from $a$ onwards $(*\cellpointershort{}).\lastrepositoryoperationresponse{} \neq ((\arbitraryvalue, \addcell), \nullconstant)$.
            However, since by (*) $(*\cellpointershort{}).\lastrepositoryoperationresponse{} = ((\arbitraryvalue, \repositoryoperationshort), \nullconstant{})$ infinitely often during $\mathcal{I}^\mathcal{B}$ and $\repositoryoperationshort = \addcell$, we have that $(*\cellpointershort{}).\lastrepositoryoperationresponse{} = ((\arbitraryvalue, \addcell), \nullconstant)$ some time after $a$, a contradiction.

            \item[] \hspace{0pt}\textbf{Case 1.2.} $\linearizationobject{} \neq ((\arbitraryvalue, \addcell), \cellpointershort{})$ some time after $e$.

            Hence, since $e$ sets $\linearizationobject{} = ((\arbitraryvalue, \repositoryoperationshort), \cellpointershort{})$ and $\repositoryoperationshort = \addcell$, the value of $\linearizationobject{}$ changed after $e$, and so by \Cref{observation:ero:where_objects_change} there is an $L$-event after $e$.
            Let $e'$ be the next $L$-event after $e$ and let $q$ be the process that executed $e'$.
            Hence, by \Cref{lemma:ero:successive_l_event_read_previous_l_event_value} $q$ read the value that $e$ set $\linearizationobject{}$ to on its last execution of \cref{line:ero:linearization_read} before $e'$; say at time $T^{\ref{line:ero:linearization_read}}_{q}$.
            Since $e$ sets $\linearizationobject{} = ((\arbitraryvalue, \addcell), \cellpointershort{})$, we have that $q$ read $((\arbitraryvalue, \addcell), \cellpointershort{})$ from $\linearizationobject{}$ at $T^{\ref{line:ero:linearization_read}}_q$.
            Hence, since $q$ executes $e'$ after $T^{\ref{line:ero:linearization_read}}_q$, it follows that $q$ invokes the \doaddcell{} procedure with parameters $\cellpointershort$ on \cref{line:ero:do_add_cell} immediately after $T^{\ref{line:ero:linearization_read}}_q$ and $q$ exits this invocation of the \doaddcell{} procedure; say at time $T^{\ref{line:ero:do_add_cell}}_q$.
            Thus, since by \Cref{lemma:ero:the_list_invariants_hold} $P(\mathcal{I}^\mathcal{B})$, $Q(\mathcal{I}^\mathcal{B})$, and $R(\mathcal{I}^\mathcal{B})$ hold, by \Cref{lemma:ero:exit_add_implies_response_set}, there is a successful add-response-set attempt $a$ for $\cellpointershort{}$ before $T^{\ref{line:ero:do_add_cell}}_q$.
            Therefore, since $\cellpointershort{} \in \celluniverse$, by \Cref{lemma:ero:once_response_not_null_for_add_never_null_for_add_again}, from $a$ onwards $(*\cellpointershort{}).\lastrepositoryoperationresponse{} \neq ((\arbitraryvalue, \addcell), \nullconstant)$.
            However, since by (*) $(*\cellpointershort{}).\lastrepositoryoperationresponse{} = ((\arbitraryvalue, \repositoryoperationshort), \nullconstant{})$ infinitely often during $\mathcal{I}^\mathcal{B}$ and $\repositoryoperationshort = \addcell$, we have that $(*\cellpointershort{}).\lastrepositoryoperationresponse{} = ((\arbitraryvalue, \addcell), \nullconstant)$ some time after $a$, a contradiction.
        \end{itemize}

        \item[] \hspace{0pt}\textbf{Case 2.} $\repositoryoperationshort = \langle \doopandcopyresponse, \arbitraryvalue \rangle$.
        
        There are two cases.

        \begin{itemize}
            \item[] \hspace{0pt}\textbf{Case 2.1.} $\linearizationobject{} = ((\arbitraryvalue, \langle \doopandcopyresponse, \arbitraryvalue \rangle), \cellpointershort{})$ from $e$ onwards.

            Since $p$ executes \cref{line:ero:linearization_read} infinitely often during $I$, it follows that $p$ executes \cref{line:ero:linearization_read} at some time $T^{\ref{line:ero:linearization_read}}_p$ after $e$, and so $p$ reads $((\arbitraryvalue, \langle \doopandcopyresponse, \arbitraryvalue \rangle), \cellpointershort{})$ from $\linearizationobject{}$ at $T^{\ref{line:ero:linearization_read}}_p$.
            Hence, since $p$ executes \cref{line:ero:linearization_read} again after $T^{\ref{line:ero:linearization_read}}_p$, it follows that $q$ invokes the \doapplyandcopyresponse{} procedure with a second parameter of $\cellpointershort$ on \cref{line:ero:do_apply_and_copy_response} immediately after $T^{\ref{line:ero:linearization_read}}_p$ and $p$ exits this invocation of the \doapplyandcopyresponse{} procedure; say at time $T^{\ref{line:ero:do_apply_and_copy_response}}_p$.
            Hence, since by \Cref{lemma:ero:the_list_invariants_hold} $P(\mathcal{I}^\mathcal{B})$, $Q(\mathcal{I}^\mathcal{B})$, and $R(\mathcal{I}^\mathcal{B})$ hold, by \Cref{lemma:ero:conditional_exit_apply_implies_response_set}, there is a successful apply-response-set attempt $a$ for $\cellpointershort{}$ before $T^{\ref{line:ero:do_apply_and_copy_response}}_p$.
            Therefore, since $\cellpointershort{} \in \celluniverse$, by \Cref{lemma:ero:once_response_not_null_for_apply_never_null_for_apply_again}, from $a$ onwards $(*\cellpointershort{}).\lastrepositoryoperationresponse{} \neq ((\arbitraryvalue, \langle \doopandcopyresponse, \arbitraryvalue \rangle), \nullconstant)$.
            However, since by (*) $(*\cellpointershort{}).\lastrepositoryoperationresponse{} = ((\arbitraryvalue, \repositoryoperationshort), \nullconstant{})$ infinitely often during $\mathcal{I}^\mathcal{B}$ and $\repositoryoperationshort = \langle \doopandcopyresponse, \arbitraryvalue \rangle$, we have that $(*\cellpointershort{}).\lastrepositoryoperationresponse{} = ((\arbitraryvalue, \langle \doopandcopyresponse, \arbitraryvalue \rangle), \nullconstant)$ some time after $a$, a contradiction.

            \item[] \hspace{0pt}\textbf{Case 2.2.} $\linearizationobject{} \neq ((\arbitraryvalue, \langle \doopandcopyresponse, \arbitraryvalue \rangle), \cellpointershort{})$ some time after $e$.

            Hence, since $e$ sets $\linearizationobject{} = ((\arbitraryvalue, \repositoryoperationshort), \cellpointershort{})$ and $\repositoryoperationshort = \langle \doopandcopyresponse, \arbitraryvalue \rangle$, the value of $\linearizationobject{}$ changed after $e$, and so by \Cref{observation:ero:where_objects_change} there is an $L$-event after $e$.
            Let $e'$ be the next $L$-event after $e$ and let $q$ be the process that executed $e'$.
            Hence, by \Cref{lemma:ero:successive_l_event_read_previous_l_event_value} $q$ read the value that $e$ set $\linearizationobject{}$ to on its last execution of \cref{line:ero:linearization_read} before $e'$; say at time $T^{\ref{line:ero:linearization_read}}_q$.
            Since $e$ sets $\linearizationobject{} = ((\arbitraryvalue, \langle \doopandcopyresponse, \arbitraryvalue \rangle), \cellpointershort{})$, we have that $q$ read $((\arbitraryvalue, \langle \doopandcopyresponse, \arbitraryvalue \rangle), \cellpointershort{})$ from $\linearizationobject{}$ at $T^{\ref{line:ero:linearization_read}}_q$.
            Hence, since $q$ executes $e'$ after $T^{\ref{line:ero:linearization_read}}_q$, $q$ invokes the \doapplyandcopyresponse{} procedure with a second parameter of $\cellpointershort$ on \cref{line:ero:do_apply_and_copy_response} immediately after $T^{\ref{line:ero:linearization_read}}_q$ and $q$ exits this invocation of the \doapplyandcopyresponse{} procedure; say at time $T^{\ref{line:ero:do_apply_and_copy_response}}_q$.
            Thus, since by \Cref{lemma:ero:the_list_invariants_hold} $P(\mathcal{I}^\mathcal{B})$, $Q(\mathcal{I}^\mathcal{B})$, and $R(\mathcal{I}^\mathcal{B})$ hold, by \Cref{lemma:ero:conditional_exit_apply_implies_response_set}, there is a successful apply-response-set attempt $a$ for $\cellpointershort{}$ before $T^{\ref{line:ero:do_apply_and_copy_response}}_q$.
            Therefore, since $\cellpointershort{} \in \celluniverse$, by \Cref{lemma:ero:once_response_not_null_for_apply_never_null_for_apply_again}, from $a$ onwards $(*\cellpointershort{}).\lastrepositoryoperationresponse{} \neq ((\arbitraryvalue, \langle \doopandcopyresponse, \arbitraryvalue \rangle), \nullconstant)$.
            However, since by (*) $(*\cellpointershort{}).\lastrepositoryoperationresponse{} = ((\arbitraryvalue, \repositoryoperationshort), \nullconstant{})$ infinitely often during $\mathcal{I}^\mathcal{B}$ and $\repositoryoperationshort = \langle \doopandcopyresponse, \arbitraryvalue \rangle$, we have that $(*\cellpointershort{}).\lastrepositoryoperationresponse{} = ((\arbitraryvalue, \langle \doopandcopyresponse, \arbitraryvalue \rangle), \nullconstant)$ some time after $a$, a contradiction.
        \end{itemize}

        \item[] \hspace{0pt}\textbf{Case 3.} $\repositoryoperationshort = \removecell$.

        There are two cases.

        \begin{itemize}
            \item[] \hspace{0pt}\textbf{Case 3.1.} $\linearizationobject{} = ((\arbitraryvalue, \removecell), \cellpointershort{})$ from $e$ onwards.

            Since $p$ executes \cref{line:ero:linearization_read} infinitely often during $I$, it follows that $p$ executes \cref{line:ero:linearization_read} at some time $T^{\ref{line:ero:linearization_read}}_p$ after $e$, and so $p$ reads $((\arbitraryvalue, \removecell), \cellpointershort{})$ from $\linearizationobject{}$ at $T^{\ref{line:ero:linearization_read}}_p$.
            Hence, since $p$ executes \cref{line:ero:linearization_read} again after $T^{\ref{line:ero:linearization_read}}_p$, it follows that $p$ invokes the \doremovecell{} procedure with a second parameter of $\cellpointershort$ on \cref{line:ero:do_remove_cell} immediately after $T^{\ref{line:ero:linearization_read}}_p$ and $p$ exits this invocation of the \doremovecell{} procedure; say at time $T^{\ref{line:ero:do_remove_cell}}_p$.
            Hence, since by \Cref{lemma:ero:the_list_invariants_hold} $P(\mathcal{I}^\mathcal{B})$, $Q(\mathcal{I}^\mathcal{B})$, and $R(\mathcal{I}^\mathcal{B})$ hold, by \Cref{lemma:ero:exit_remove_implies_response_set}, there is a successful remove-response-set attempt $a$ for $\cellpointershort{}$ before $T^{\ref{line:ero:do_remove_cell}}_p$.
            Therefore, since $\cellpointershort{} \in \celluniverse$, by \Cref{lemma:ero:once_response_not_null_for_remove_never_null_for_remove_again}, from $a$ onwards $(*\cellpointershort{}).\lastrepositoryoperationresponse{} \neq ((\arbitraryvalue, \removecell), \nullconstant)$.
            However, since by (*) $(*\cellpointershort{}).\lastrepositoryoperationresponse{} = ((\arbitraryvalue, \repositoryoperationshort), \nullconstant{})$ infinitely often during $\mathcal{I}^\mathcal{B}$ and $\repositoryoperationshort = \removecell$, we have that $(*\cellpointershort{}).\lastrepositoryoperationresponse{} = ((\arbitraryvalue, \removecell), \nullconstant)$ some time after $a$, a contradiction.

            \item[] \hspace{0pt}\textbf{Case 3.2.} $\linearizationobject{} \neq ((\arbitraryvalue, \removecell), \cellpointershort{})$ some time after $e$.

            Hence, since $e$ sets $\linearizationobject{} = ((\arbitraryvalue, \repositoryoperationshort), \cellpointershort{})$ and $\repositoryoperationshort = \removecell$, the value of $\linearizationobject{}$ changed after $e$, and so by \Cref{observation:ero:where_objects_change} there is an $L$-event after $e$.
            Let $e'$ be the next $L$-event after $e$ and let $q$ be the process that executed $e'$.
            Hence, by \Cref{lemma:ero:successive_l_event_read_previous_l_event_value} $q$ read the value that $e$ set $\linearizationobject{}$ to on its last execution of \cref{line:ero:linearization_read} before $e'$; say at time $T^{\ref{line:ero:linearization_read}}_q$.
            Since $e$ sets $\linearizationobject{} = ((\arbitraryvalue, \removecell), \cellpointershort{})$, we have that $q$ read $((\arbitraryvalue, \removecell), \cellpointershort{})$ from $\linearizationobject{}$ at $T^{\ref{line:ero:linearization_read}}_q$.
            Hence, since $q$ executes $e'$ after $T^{\ref{line:ero:linearization_read}}_q$, it follows that $q$ invokes the \doremovecell{} procedure with a second parameter of $\cellpointershort$ on \cref{line:ero:do_remove_cell} immediately after $T^{\ref{line:ero:linearization_read}}_q$ and $q$ exits this invocation of the \doremovecell{} procedure; say at time $T^{\ref{line:ero:do_remove_cell}}_q$.
            Thus, since by \Cref{lemma:ero:the_list_invariants_hold} $P(\mathcal{I}^\mathcal{B})$, $Q(\mathcal{I}^\mathcal{B})$, and $R(\mathcal{I}^\mathcal{B})$ hold, by \Cref{lemma:ero:exit_remove_implies_response_set}, there is a successful remove-response-set attempt $a$ for $\cellpointershort{}$ before $T^{\ref{line:ero:do_remove_cell}}_q$.
            Therefore, since $\cellpointershort{} \in \celluniverse$, by \Cref{lemma:ero:once_response_not_null_for_remove_never_null_for_remove_again}, from $a$ onwards $(*\cellpointershort{}).\lastrepositoryoperationresponse{} \neq ((\arbitraryvalue, \removecell), \nullconstant)$.
            However, since by (*) $(*\cellpointershort{}).\lastrepositoryoperationresponse{} = ((\arbitraryvalue, \repositoryoperationshort), \nullconstant{})$ infinitely often during $\mathcal{I}^\mathcal{B}$ and $\repositoryoperationshort = \removecell$, we have that $(*\cellpointershort{}).\lastrepositoryoperationresponse{} = ((\arbitraryvalue, \removecell), \nullconstant)$ some time after $a$, a contradiction.
            \qH{\Cref{lemma:ero:a_events_which_get_into_l_eventually_get_out_of_a}}
        \end{itemize}
    \end{itemize}
\end{proof}

\begin{lemma}\label{lemma:ero:if_p_reads_unfinished_value_from_a_it_cannot_fail_the_done_check}
    Consider any process $p$ and iteration $I$ of the loop on \cref{line:ero:do_work_while_loop} by $p$ in $\mathcal{I}^\mathcal{B}$.
    If $p$ reads a value $v$ from $\announceobject$ on \cref{line:ero:announce_read} during $I$ such that $\linearizationobject{} \neq v$ throughout $\mathcal{I}^\mathcal{B}$, then $p$ does not receive $\done$ on \cref{line:ero:check_if_announce_is_done} during $I$.
\end{lemma}

\begin{proof}
    Suppose, for contradiction, $p$ received $\done$ on \cref{line:ero:check_if_announce_is_done} during $I$.    
    Since $\announceobject$ and $\linearizationobject{}$ are both initially $((0, \noop), \nullconstant)$ and $\linearizationobject{} \neq v$ throughout $\mathcal{I}^\mathcal{B}$, it follows that $v \neq ((0, \noop), \nullconstant)$.
    Hence, $\announceobject$ was set to $v$ at some time, so by \Cref{observation:ero:where_objects_change}, there is an $A$-event $e_A$ that set $\announceobject = v$.
    Suppose $e_A$ is an $A$-event for $\cellpointershort{}$.
    Hence, by \Cref{lemma:ero:every_a_event_is_add_apply_or_remove} $e_A$ is an $A$-add, $A$-apply, or $A$-remove event for $\cellpointershort{}$.
    We consider each case separately.
    We note that the proofs of each case are essentially the same, but they rely on different lemmas.

    \begin{itemize}
        \item[] \hspace{0pt}\textbf{Case 1.} $e_A$ is an $A$-add event for $\cellpointershort{}$.

        Hence, by \Cref{def:ero:english}, $v = ((\arbitraryvalue, \addcell), \cellpointershort)$, and so $p$ read $((\arbitraryvalue, \addcell), \cellpointershort)$ from $\announceobject$ on \cref{line:ero:announce_read} during $I$.
        Thus, since by \Cref{lemma:ero:the_list_invariants_hold} $P(\mathcal{I}^\mathcal{B})$, $Q(\mathcal{I}^\mathcal{B})$, and $R(\mathcal{I}^\mathcal{B})$ hold, and $p$ received $\done$ on \cref{line:ero:check_if_announce_is_done} during $I$, by \Cref{lemma:ero:add_done_check_pass_implies_l_add_event}, there is an $L$-add event $e_L$ for $\cellpointershort{}$ in $\mathcal{I}^\mathcal{B}$.
        Therefore, since $e_A$ is an $A$-add event for $\cellpointershort{}$, $e_L$ is an an $L$-add event for $\cellpointershort{}$, and $e_A$ sets $\announceobject = v$, by \Cref{lemma:ero:a_and_l_events_of_same_type_for_same_pointer_set_same_value}, $e_L$ sets $\linearizationobject{} = v$.
        However, by assumption $\linearizationobject{} \neq v$ throughout $\mathcal{I}^\mathcal{B}$, a contradiction.

        \item[] \hspace{0pt}\textbf{Case 2.} $e_A$ is an $A$-apply event for $\cellpointershort{}$.

        Hence, by \Cref{def:ero:english}, $v = ((\arbitraryvalue, \langle \doapplyandcopyresponse{}, \arbitraryvalue \rangle), \cellpointershort)$, and so we have that $p$ read $((\arbitraryvalue, \langle \doapplyandcopyresponse{}, \arbitraryvalue \rangle), \cellpointershort)$ from $\announceobject$ on \cref{line:ero:announce_read} during $I$.
        Thus, since by \Cref{lemma:ero:the_list_invariants_hold} $P(\mathcal{I}^\mathcal{B})$, $Q(\mathcal{I}^\mathcal{B})$, and $R(\mathcal{I}^\mathcal{B})$ hold, and $p$ received $\done$ on \cref{line:ero:check_if_announce_is_done} during $I$, by \Cref{lemma:ero:apply_done_check_pass_implies_l_apply_event_weak}, there is an $L$-apply event $e_L$ for $\cellpointershort{}$ in $\mathcal{I}^\mathcal{B}$.
        Therefore, since $e_A$ is an $A$-apply event for $\cellpointershort{}$, $e_L$ is an an $L$-apply event for $\cellpointershort{}$, and $e_A$ sets $\announceobject = v$, by \Cref{lemma:ero:a_and_l_events_of_same_type_for_same_pointer_set_same_value}, $e_L$ sets $\linearizationobject{} = v$.
        However, by assumption $\linearizationobject{} \neq v$ throughout $\mathcal{I}^\mathcal{B}$, a contradiction.

        \item[] \hspace{0pt}\textbf{Case 3.} $e_A$ is an $A$-remove event for $\cellpointershort{}$.

        Hence, by \Cref{def:ero:english}, $v = ((\arbitraryvalue, \removecell), \cellpointershort)$, and so $p$ read $((\arbitraryvalue, \removecell), \cellpointershort)$ from $\announceobject$ on \cref{line:ero:announce_read} during $I$.
        Thus, since by \Cref{lemma:ero:the_list_invariants_hold} $P(\mathcal{I}^\mathcal{B})$, $Q(\mathcal{I}^\mathcal{B})$, and $R(\mathcal{I}^\mathcal{B})$ hold, and $p$ received $\done$ on \cref{line:ero:check_if_announce_is_done} during $I$, by \Cref{lemma:ero:remove_done_check_pass_implies_l_remove_event_weak}, there is an $L$-remove event $e_L$ for $\cellpointershort{}$ in $\mathcal{I}^\mathcal{B}$.
        Therefore, since $e_A$ is an $A$-remove event for $\cellpointershort{}$, $e_L$ is an an $L$-remove event for $\cellpointershort{}$, and $e_A$ sets $\announceobject = v$, by \Cref{lemma:ero:a_and_l_events_of_same_type_for_same_pointer_set_same_value}, $e_L$ sets $\linearizationobject{} = v$.
        However, by assumption $\linearizationobject{} \neq v$ throughout $\mathcal{I}^\mathcal{B}$, a contradiction.
        \qH{\Cref{lemma:ero:if_p_reads_unfinished_value_from_a_it_cannot_fail_the_done_check}}
    \end{itemize}
\end{proof}

We now prove (1).

\begin{proposition}\label{lemma:ero:each_process_does_finitely_many_successful_cas_on_ptr_next}
    In \Cref{scenario:ero:helper_loop_scenario}, consider any process $p$ and any pointer $\cellpointershort{} \in \celluniverse \cup \{\&\headobject\}$.
    Then, $p$ performs a finite number of successful \CASop{} operations on $(*\cellpointershort{}).\nextlong$ in $\mathcal{I}^\mathcal{B}$.
\end{proposition}

\begin{proof}
    Suppose, for contradiction, $p$ performs an infinite number of successful \CASop{} operations on $(*\cellpointershort{}).\nextlong$ in $\mathcal{I}^\mathcal{B}$.
    Hence, $p$ takes infinitely many steps in $\mathcal{I}^\mathcal{B}$.

    \begin{claimcustom}{\ref{lemma:ero:each_process_does_finitely_many_successful_cas_on_ptr_next}.1}\label{lemma:ero:each_process_does_finitely_many_successful_cas_on_ptr_next:claim_one}
        Consider any incarnation $\mathcal{L}$ of the loops on lines \ref{line:ero:add_cell_while_loop}, \ref{line:ero:remove_cell_while_loop}, \ref{line:ero:remove_cell_remove_seal_loop}, \ref{line:ero:remove_cell_remove_repeat_loop}, \ref{line:ero:acquire_loop_until}, and \ref{line:ero:acquire_next_repeat_loop} by $p$ during $\mathcal{I}^\mathcal{B}$.
        Then, $p$ takes finitely many steps during $\mathcal{L}$.
    \end{claimcustom}

    \begin{proof}
        Suppose, for contradiction, $p$ takes infinitely many steps during $\mathcal{L}$.
        There are two cases.
        \begin{itemize}
            \item[] \hspace{0pt}\textbf{Case 1.}  $\mathcal{L}$ is the loop on line \ref{line:ero:remove_cell_remove_seal_loop} (Case A), \ref{line:ero:remove_cell_remove_repeat_loop} (Case B), or \ref{line:ero:acquire_next_repeat_loop} (Case C).
      
            There are two cases.
        
            \begin{itemize}
                \item[] \hspace{0pt}\textbf{Case 1.1.} The value of $\cellpointercontentshort{}_\linearizationobject{}$ (Case A), $\previouscellpointershort{}$ (Case B), and $\currentcellpointershort{}$ (Case C) in $\mathcal{L}$ is $\cellpointershort{}$.
        
                Hence, since $p$ takes infinitely many steps during $\mathcal{L}$, we have that $p$ never performs a successful \CASop{} operation on $(*\cellpointershort{}).\nextlong$ on \cref{line:ero:seal_cell} (Case A), \cref{line:ero:remove_cell_from_list} (Case B), and \cref{line:ero:acquire_next_cell} (Case C) during $\mathcal{L}$.
                Thus, there is a time after which $p$ never performs another another successful \CASop{} operation on $(*\cellpointershort{}).\nextlong$ during $\mathcal{I}^\mathcal{B}$.
                Therefore, $p$ performs a finite number of successful \CASop{} operations on $(*\cellpointershort{}).\nextlong$ in $\mathcal{I}^\mathcal{B}$.
                However, by our initial assumption of \Cref{lemma:ero:each_process_does_finitely_many_successful_cas_on_ptr_next}, $p$ performs an infinite number of successful \CASop{} operations on $(*\cellpointershort{}).\nextlong$ in $\mathcal{I}^\mathcal{B}$, a contradiction.
        
                \item[] \hspace{0pt}\textbf{Case 1.2.} The value of $\cellpointercontentshort{}_\linearizationobject{}$ (Case A), $\previouscellpointershort{}$ (Case B), and $\currentcellpointershort{}$ (Case C) in $\mathcal{L}$ is not $\cellpointershort{}$.
        
                Hence, since these are the only values that $p$ performs \CASop{} operations on in $\mathcal{L}$, and $p$ takes infinitely many steps during $\mathcal{L}$, there is a time after which $p$ never performs another \CASop{} operation on $(*\cellpointershort{}).\nextlong$ during $\mathcal{I}^\mathcal{B}$.
                Therefore, $p$ performs a finite number of successful \CASop{} operations on $(*\cellpointershort{}).\nextlong$ in $\mathcal{I}^\mathcal{B}$.
                However, by our initial assumption of \Cref{lemma:ero:each_process_does_finitely_many_successful_cas_on_ptr_next}, $p$ performs an infinite number of successful \CASop{} operations on $(*\cellpointershort{}).\nextlong$ in $\mathcal{I}^\mathcal{B}$, a contradiction.
            \end{itemize}
    
            \item[] \hspace{0pt}\textbf{Case 2.} $\mathcal{L}$ is the loop on \cref{line:ero:add_cell_while_loop}, \ref{line:ero:remove_cell_while_loop}, or \ref{line:ero:acquire_loop_until}.
    
            Let $I$ be the invocation of the \doaddcell{}, \doremovecell{}, or Acquire procedure in which $\mathcal{L}$ is in.
            Since, as proven in Case 1, $p$ exits every invocation of the AcquireNext procedure in $\mathcal{I}^\mathcal{B}$, we have that $p$ invokes infinitely many invocations of the AcquireNext procedure during $I$.
            Let $I_1, I_2, \ldots$ denote these invocations of the AcquireNext procedure during $I$ in the order they were invoked.

            \begin{claimcustom}{\ref{lemma:ero:each_process_does_finitely_many_successful_cas_on_ptr_next}.1.1}\label{lemma:ero:each_process_does_finitely_many_successful_cas_on_ptr_next:claim_one:claim_one}
                $p$ read $(\uniquerepositoryoperationshort_\linearizationobject{}, \uniquecellpointershort_\linearizationobject{})$ from $\linearizationobject{}$ on its last execution of \cref{line:ero:linearization_read} before invoking $I$.
            \end{claimcustom}

            \begin{proof}
                Suppose, for contradiction, $p$ read $(\uniquerepositoryoperationshort_\linearizationobject{}', \uniquecellpointershort_\linearizationobject{}') \neq (\uniquerepositoryoperationshort_\linearizationobject{}, \uniquecellpointershort_\linearizationobject{})$ from $\linearizationobject{}$ on its last execution of \cref{line:ero:linearization_read} before invoking $I$; say at time $T^{\ref{line:ero:linearization_read}}_p$.
                Since by \Cref{lemma:ero:i_is_proceeded_by_a_unique_l_event} $e_{\min}$ set $\linearizationobject{} = (\uniquerepositoryoperationshort_\linearizationobject{}, \uniquecellpointershort_\linearizationobject{})$ and by \Cref{lemma:ero:e_is_the_last_l_event} $e_{\min}$ is the last $L$-event in $\mathcal{I}^\mathcal{B}$, we have that from $e_{\min}$ onwards in $\mathcal{I}^\mathcal{B}$ $\linearizationobject{} = (\uniquerepositoryoperationshort_\linearizationobject{}, \uniquecellpointershort_\linearizationobject{})$, and so given the value $p$ read at $T^{\ref{line:ero:linearization_read}}_p$ we have that $T^{\ref{line:ero:linearization_read}}_p < e_{\min}$.

                We claim that $\uniquerepositoryoperationshort_\linearizationobject{}' \neq \uniquerepositoryoperationshort_\linearizationobject{}$ (*).
                Suppose, for contradiction, that $\uniquerepositoryoperationshort_\linearizationobject{}' = \uniquerepositoryoperationshort_\linearizationobject{}$.
                Since by \Cref{lemma:ero:i_first_parameter_is_not_initial} $\uniquerepositoryoperationshort_\linearizationobject{} \neq (0, \noop)$, we have that $\uniquerepositoryoperationshort_\linearizationobject{}' \neq (0, \noop)$.
                Hence, since $p$ read $(\uniquerepositoryoperationshort_\linearizationobject{}', \uniquecellpointershort_\linearizationobject{}')$ from $\linearizationobject{}$ at $T^{\ref{line:ero:linearization_read}}_p$, we have that  $\linearizationobject{}.\uniquerepositoryoperationlong$ was set to $\uniquerepositoryoperationshort_\linearizationobject{}'$ before $T^{\ref{line:ero:linearization_read}}_p$.
                Thus, by \Cref{observation:ero:where_objects_change}, some $L$-event $e$ set $\linearizationobject{}.\uniquerepositoryoperationlong = \uniquerepositoryoperationshort_\linearizationobject{}'$ before $T^{\ref{line:ero:linearization_read}}_p$.
                So, since $e < T^{\ref{line:ero:linearization_read}}_p$ and $T^{\ref{line:ero:linearization_read}}_p < e_{\min}$, by transitivity, $e < e_{\min}$, and so $e \neq e_{\min}$.
                Therefore, since $\uniquerepositoryoperationshort_\linearizationobject{}' = \uniquerepositoryoperationshort_\linearizationobject{}$, we have that there are two $L$-events in $\mathcal{I}^\mathcal{B}$ which set $\linearizationobject{}.\uniquerepositoryoperationlong = \uniquerepositoryoperationshort_\linearizationobject{}$ (namely $e_{\min}$ and $e$).
                However, since by \Cref{lemma:ero:the_list_invariants_hold} $P(\mathcal{I}^\mathcal{B})$ holds, by \Cref{lemma:ero:p_implies_unique_low_level_operations_in_linearization} every $L$-event in $\mathcal{I}^\mathcal{B}$ sets $\linearizationobject{}.\uniquerepositoryoperationlong{}$ to a unique value, a contradiction.
                
                We now return to the proof of \Cref{lemma:ero:each_process_does_finitely_many_successful_cas_on_ptr_next:claim_one:claim_one}.
                Since $p$ read $(\uniquerepositoryoperationshort_\linearizationobject{}', \uniquecellpointershort_\linearizationobject{}')$ from $\linearizationobject{}$ on its last execution of \cref{line:ero:linearization_read} before invoking $I$, the first parameter of $I$ is $\uniquerepositoryoperationshort_\linearizationobject{}'$.
                Hence, since $p$ invokes $I_1, I_2, \ldots$ during $I$, it follows that the first parameter of $I_1, I_2, \ldots$ is also $\uniquerepositoryoperationshort_\linearizationobject{}'$.
                Since $p$ invokes infinitely many invocations of the AcquireNext procedure during $I$, there exists some invocation $I_i$ of the AcquireNext procedure during $I$ which is invoked after $e_{\min}$.
                Hence, since from $e_{\min}$ onwards in $\mathcal{I}^\mathcal{B}$ $\linearizationobject{} = (\uniquerepositoryoperationshort_\linearizationobject{}, \uniquecellpointershort_\linearizationobject{})$, we have that $\linearizationobject{} = (\uniquerepositoryoperationshort_\linearizationobject{}, \uniquecellpointershort_\linearizationobject{})$ throughout $I_i$.
                Since $p$ eventually exits $I_i$, it executes \cref{line:ero:acquire_next_linearization_changed_check} during $I_i$ at least once.
                Hence, since the first parameter of $I_i$ is $\uniquerepositoryoperationshort_\linearizationobject{}'$, by (*) $\uniquerepositoryoperationshort_\linearizationobject{}' \neq \uniquerepositoryoperationshort_\linearizationobject{}$, and $\linearizationobject{} = (\uniquerepositoryoperationshort_\linearizationobject{}, \uniquecellpointershort_\linearizationobject{})$ throughout $I_i$, we have that $p$ finds the condition on \cref{line:ero:acquire_next_linearization_changed_check} to be true during $I_i$.
                Thus, since $p$ exits $I_i$, it exits on \cref{line:ero:acquire_next_linearization_changed_return}, and returns $(\timechange{}, \arbitraryvalue{})$.
                Therefore, the value of $\status$ during $I$ is $\timechange$, and so $p$ only takes finitely many steps in $\mathcal{L}$.
                However, by assumption, $p$ takes infinitely many steps in $\mathcal{L}$, a contradiction.
                \qH{\Cref{lemma:ero:each_process_does_finitely_many_successful_cas_on_ptr_next:claim_one:claim_one}}
            \end{proof}
        
            We now finish the proof of Case 2.
            Let $\cellpointershort{}_i$ denote the second parameter of $I_i$.
            Since $\mathcal{I}^{exclude}_{e_{\min}}$ and $\mathcal{I}^{include}_{e_{\min}}$ are finite, by \Cref{def:ero:logical_list}, $\List(\mathcal{I}^{exclude}_{e_{\min}})$ and $\List(\mathcal{I}^{include}_{e_{\min}})$ are finite, and so the union of $\List(\mathcal{I}^{exclude}_{e_{\min}})$ and $\List(\mathcal{I}^{include}_{e_{\min}})$ is finite.
            Therefore, since by \Cref{lemma:ero:each_process_does_finitely_many_successful_cas_on_ptr_next:claim_one:claim_one} $p$ read $(\uniquerepositoryoperationshort_\linearizationobject{}, \uniquecellpointershort_\linearizationobject{})$ from $\linearizationobject{}$ on its last execution of \cref{line:ero:linearization_read} before invoking $I$, by \Cref{lemma:ero:acquire_next_for_ptr_is_in_either_include_or_exclude}, $\cellpointershort{}_i$ is in either $\List(\mathcal{I}^{exclude}_{e_{\min}})$ or $\List(\mathcal{I}^{include}_{e_{\min}})$, and so $\{\cellpointershort{}_i\ \vert\ \forall i\}$ is finite.
            However, by \Cref{lemma:ero:all_acquire_next_during_traversal_are_for_different_pointers}, $\cellpointershort{}_1, \cellpointershort{}_2,\ldots$ are distinct so $\{\cellpointershort{}_i\ \vert\ \forall i\}$ is infinite, a contradiction.
            \qH{\Cref{lemma:ero:each_process_does_finitely_many_successful_cas_on_ptr_next:claim_one}}
        \end{itemize}
    \end{proof}

    \begin{claimcustom}{\ref{lemma:ero:each_process_does_finitely_many_successful_cas_on_ptr_next}.2}\label{lemma:ero:each_process_does_finitely_many_successful_cas_on_ptr_next:claim_two}
        $p$ executes \cref{line:ero:linearization_read} infinitely often in $\mathcal{I}^\mathcal{B}$.
        Furthermore, $p$ reads $(\uniquerepositoryoperationshort_\linearizationobject{}, \uniquecellpointershort_\linearizationobject{})$ from $\linearizationobject{}$ on every execution of \cref{line:ero:linearization_read} at or after $e_{\min}$.
    \end{claimcustom}

    \begin{proof}
        Since $p$ takes infinitely many steps in $\mathcal{I}^\mathcal{B}$, and by \Cref{lemma:ero:each_process_does_finitely_many_successful_cas_on_ptr_next:claim_one} $p$ exits every incarnation of the loops on lines \ref{line:ero:add_cell_while_loop}, \ref{line:ero:remove_cell_while_loop}, \ref{line:ero:remove_cell_remove_seal_loop}, \ref{line:ero:remove_cell_remove_repeat_loop}, \ref{line:ero:acquire_loop_until}, and \ref{line:ero:acquire_next_repeat_loop} during $\mathcal{I}^\mathcal{B}$, we have that $p$ exits every invocation of every procedure other than \highleveloperation{} and \doworkuntildone{} in $\mathcal{I}^\mathcal{B}$.
        Hence, since $p$ takes infinitely many steps in $\mathcal{I}^\mathcal{B}$ and there are no loops in the \highleveloperation{} procedure, we have that $p$ takes infinitely many steps inside the \doworkuntildone{} procedure in $\mathcal{I}^\mathcal{B}$.
        Thus, since there is only a single loop inside the \doworkuntildone{} procedure, and $p$ exits every invocation of every procedure invoked inside the \doworkuntildone{} procedure, we have that $p$ executes \cref{line:ero:linearization_read} infinitely often in $\mathcal{I}^\mathcal{B}$.
        Since by \Cref{lemma:ero:i_is_proceeded_by_a_unique_l_event} $e_{\min}$ set $\linearizationobject{} = (\uniquerepositoryoperationshort_\linearizationobject{}, \uniquecellpointershort_\linearizationobject{})$ and by \Cref{lemma:ero:e_is_the_last_l_event} $e_{\min}$ is the last $L$-event in $\mathcal{I}^\mathcal{B}$, we have that from $e_{\min}$ onwards in $\mathcal{I}^\mathcal{B}$ $\linearizationobject{} = (\uniquerepositoryoperationshort_\linearizationobject{}, \uniquecellpointershort_\linearizationobject{})$.
        Hence, $p$ reads $(\uniquerepositoryoperationshort_\linearizationobject{}, \uniquecellpointershort_\linearizationobject{})$ from $\linearizationobject{}$ on every execution of \cref{line:ero:linearization_read} at or after $e_{\min}$.
        \qH{\Cref{lemma:ero:each_process_does_finitely_many_successful_cas_on_ptr_next:claim_two}}
    \end{proof}

    \begin{claimcustom}{\ref{lemma:ero:each_process_does_finitely_many_successful_cas_on_ptr_next}.3}\label{lemma:ero:each_process_does_finitely_many_successful_cas_on_ptr_next:claim_four}
        There exists an iteration $I$ of the loop on \cref{line:ero:do_work_while_loop} by $p$ during $\mathcal{I}^\mathcal{B}$ such that $p$'s execution of \cref{line:ero:linearization_read} during $I$ is at or after $e_{\min}$ and $p$ executes \cref{line:ero:linearization_cas} during $I$.
    \end{claimcustom}

    \begin{proof}
        Suppose, for contradiction, every iteration $I$ of the loop on \cref{line:ero:do_work_while_loop} by $p$ during $\mathcal{I}^\mathcal{B}$ either (1) $p$'s execution of \cref{line:ero:linearization_read} during $I$ is before $e_{\min}$ or (2) $p$ does not execute \cref{line:ero:linearization_cas} during $I$.

        \begin{claimcustom}{\ref{lemma:ero:each_process_does_finitely_many_successful_cas_on_ptr_next}.3.1}\label{lemma:ero:each_process_does_finitely_many_successful_cas_on_ptr_next:claim_four:claim_one}
            There are infinitely many $A$-events in $\mathcal{I}^\mathcal{B}$.
        \end{claimcustom}
        
        \begin{proof}
            Suppose, for contradiction, there are finitely many $A$-events in $\mathcal{I}^\mathcal{B}$.
            Since by \Cref{lemma:ero:e_is_the_last_l_event} $e_{\min}$ is in $\mathcal{I}^\mathcal{B}$, by \Cref{lemma:ero:l_events_have_corresponding_a_events}, there is an $A$-event in $\mathcal{I}^\mathcal{B}$, and so there is a last $A$-event in $\mathcal{I}^\mathcal{B}$; say $e_{last}$.
            Hence, since by \Cref{observation:ero:where_objects_change} only $A$-events change the value of $\announceobject$, from $e_{last}$ onwards the value of $\announceobject$ does not change.
            Since by \Cref{lemma:ero:each_process_does_finitely_many_successful_cas_on_ptr_next:claim_two} $p$ executes \cref{line:ero:linearization_read} infinitely often in $\mathcal{I}^\mathcal{B}$, $p$ executes \cref{line:ero:linearization_read} after $e_{last}$ and $e_{\min}$.
            Let $I$ be any iteration of the loop on \cref{line:ero:do_work_while_loop} in which $p$ executes \cref{line:ero:linearization_read} after $e_{last}$ and $e_{\min}$.
            Hence, since by \Cref{lemma:ero:e_is_the_last_l_event} $e_{\min}$ is the last $L$-event in $\mathcal{I}^\mathcal{B}$, by \Cref{lemma:ero:if_p_reads_last_l_event_then_announce_acquire_never_fails}, $p$ does not find the condition on \cref{line:ero:announce_acquire_l_changed_check} to be true during $I$.
            Thus, since $p$ takes infinitely many steps during $\mathcal{I}^\mathcal{B}$, we have that $p$'s response on \cref{line:ero:check_if_announce_is_done} is either $\notdone$ or $\done$, and so $p$ either executes \cref{line:ero:linearization_cas} or \cref{line:ero:announce_cas} during $I$.
            So, since $I$ is an iteration of the loop on \cref{line:ero:do_work_while_loop} by $p$ during $\mathcal{I}^\mathcal{B}$ and $p$'s execution of \cref{line:ero:linearization_read} during $I$ is after $e_{\min}$, by our initial assumption of \Cref{lemma:ero:each_process_does_finitely_many_successful_cas_on_ptr_next:claim_four}, we have that $p$ does not execute \cref{line:ero:linearization_cas} during $I$, and so $p$ executes \cref{line:ero:announce_cas} during $I$.
            Denote this execution of \cref{line:ero:announce_cas} by $e_A$.
            If $e_A$ is successful, then by \Cref{def:ero:english}, $e_A$ is an $A$-event.
            Since $I$ is chosen such that $p$'s execution of \cref{line:ero:linearization_read} during $I$ is after $e_{last}$, we have that $e_{last} < e_A$, and so there is an $A$-event after $e_{last}$.
            However, this is impossible since $e_{last}$ is by definition the last $A$-event in $\mathcal{I}^\mathcal{B}$.
            Hence, $e_A$ is unsuccessful.
            Let $T^{\ref{line:ero:announce_read}}_p$ be the time of $p$'s last execution of \cref{line:ero:announce_read} before $e_A$.
            Since the first parameter of $e_A$ is the value $p$ read from $\announceobject$ at $T^{\ref{line:ero:announce_read}}_p$ and $e_A$ is unsuccessful, we have that the value of $\announceobject$ changed between $T^{\ref{line:ero:announce_read}}_p$ and $e_A$.
            Thus, by \Cref{observation:ero:where_objects_change}, there is an $A$-event between $T^{\ref{line:ero:announce_read}}_p$ and $e_A$.
            Since $I$ is chosen such that $p$'s execution of \cref{line:ero:linearization_read} during $I$ is after $e_{last}$, we have that $e_{last} < T^{\ref{line:ero:announce_read}}_p$, and so there is an $A$-event after $e_{last}$.
            However, this is impossible since $e_{last}$ is by definition the last $A$-event in $\mathcal{I}^\mathcal{B}$.
            Therefore, all cases are impossible, so there are infinitely many $A$-events in $\mathcal{I}^\mathcal{B}$.
            \qH{\Cref{lemma:ero:each_process_does_finitely_many_successful_cas_on_ptr_next:claim_four:claim_one}}
        \end{proof}

        \begin{claimcustom}{\ref{lemma:ero:each_process_does_finitely_many_successful_cas_on_ptr_next}.3.2}\label{lemma:ero:each_process_does_finitely_many_successful_cas_on_ptr_next:claim_four:claim_three}
            There is a time $T_1$ where for all times $T \geq T_1$ $\announceobject = v$ at $T$ for some value $v$ such that $\linearizationobject{} \neq v$ throughout $\mathcal{I}^\mathcal{B}$.
        \end{claimcustom}

        \begin{proof}
            By \Cref{lemma:ero:e_is_the_last_l_event} there are finitely many $L$-events in $\mathcal{I}^\mathcal{B}$.
            Let $l_1, l_2, \ldots, l_n$ denote these $L$-events and suppose $l_i$ sets $\linearizationobject{} = v_i$.
            Hence, by \Cref{lemma:ero:a_events_which_get_into_l_eventually_get_out_of_a}, there are finitely many executions of \cref{line:ero:announce_gcas} or \cref{line:ero:announce_cas} in $\mathcal{I}^\mathcal{B}$ that try to set $\announceobject = v_i$, and so by \Cref{observation:ero:where_objects_change}, there are finitely many $A$-events in $\mathcal{I}^\mathcal{B}$ that set $\announceobject = v_i$.        
            Thus, since by \Cref{lemma:ero:e_is_the_last_l_event} $e_{\min}$ is in $\mathcal{I}^\mathcal{B}$, by \Cref{lemma:ero:l_events_have_corresponding_a_events} there is an $A$-event in $\mathcal{I}^\mathcal{B}$, and so there is a last $A$-event in $\mathcal{I}^\mathcal{B}$ that set $\announceobject = v_i$ for any $i \in [1..n]$; say $e_{last}$.
            Since by \Cref{lemma:ero:each_process_does_finitely_many_successful_cas_on_ptr_next:claim_four:claim_one} there are infinitely many $A$-events in $\mathcal{I}^\mathcal{B}$, there is an $A$-event after $e_{last}$; say $e_{next}$.
            Since $e_{next}$ is an $A$-event after $e_{last}$, we have that $e_{next}$ sets $\announceobject = v$ such that $v \neq v_i$ for all $i \in [1..n]$. 
            We claim that $e_{next}$ is the time $T_1$ listed in the statement of \Cref{lemma:ero:each_process_does_finitely_many_successful_cas_on_ptr_next:claim_four:claim_three}.

            Suppose, for contradiction, at some time $T \geq e_{next}$ $\announceobject = v'$ at $T$ for some value $v'$ such that $\linearizationobject{} = v'$ at sometime during $\mathcal{I}^\mathcal{B}$.
            We first prove that $v' \neq ((0, \noop), \nullconstant)$.
            Suppose, for contradiction, $v' = ((0, \noop), \nullconstant)$.
            Since $e_{next}$ is an $A$-event that set $\announceobject = v$, by \Cref{lemma:ero:every_a_event_is_add_apply_or_remove}, $v \neq ((0, \noop), \nullconstant)$.
            Hence, $v \neq v'$, and so since $\announceobject = v$ at $e_{next}$ and $\announceobject = v'$ at $T \geq e_{next}$, we have that $\announceobject$ was set to $v'$.
            Therefore, by \Cref{observation:ero:where_objects_change}, an $A$-event set $\announceobject = v'$, and so by \Cref{lemma:ero:every_a_event_is_add_apply_or_remove}, $v' \neq ((0, \noop), \nullconstant)$.
            However, $v' = ((0, \noop), \nullconstant)$, a contradiction.
            Since $\linearizationobject{}$ is initially $((0, \noop), \nullconstant)$, $v' \neq ((0, \noop), \nullconstant)$, and $\linearizationobject{} = v'$ at sometime during $\mathcal{I}^\mathcal{B}$, we have that $\linearizationobject{}$ was set to $v'$ during $\mathcal{I}^\mathcal{B}$, and so by \Cref{observation:ero:where_objects_change}, an $L$-event $e$ set $\linearizationobject{} = v'$ during $\mathcal{I}^\mathcal{B}$.
            Hence, since $l_1, l_2, \ldots, l_n$ are the only $L$-events in $\mathcal{I}^\mathcal{B}$, we have that $e = l_i$ and $v' = v_i$ for some $i \in [1..n]$.
            Thus, since $v \neq v_i$ for all $i \in [1..n]$, we have that $v \neq v'$.
            So, since $\announceobject = v$ at $e_{next}$ and $\announceobject = v'$ at $T \geq e_{next}$, we have that $\announceobject$ was set to $v'$ after $e_{next}$.
            Hence, by \Cref{observation:ero:where_objects_change}, an $A$-event $e'$ after $e_{next}$ set $\announceobject = v'$.
            Therefore, since $e_{last} < e_{next}$ and $e_{next} < e'$, by transitivity, $e_{last} < e'$, and so there is an $A$-event after $e_{last}$ that set $\announceobject = v' = v_i$.
            However, $e_{last}$ is the last $A$-event in $\mathcal{I}^\mathcal{B}$ that set $\announceobject = v_i$ for any $i \in [1..n]$, a contradiction.
            \qH{\Cref{lemma:ero:each_process_does_finitely_many_successful_cas_on_ptr_next:claim_four:claim_three}}
        \end{proof}

        We now finish the proof of \Cref{lemma:ero:each_process_does_finitely_many_successful_cas_on_ptr_next:claim_four}.
        By \Cref{lemma:ero:each_process_does_finitely_many_successful_cas_on_ptr_next:claim_two}, there is an iteration $I$ of the loop on \cref{line:ero:do_work_while_loop} by $p$ during $\mathcal{I}^\mathcal{B}$ such that $p$'s execution of \cref{line:ero:linearization_read} during $I$ is after $\max(e_{\min}, T_1)$ and $p$ reads $(\uniquerepositoryoperationshort_\linearizationobject{}, \uniquecellpointershort_\linearizationobject{})$ from $\linearizationobject{}$ on \cref{line:ero:linearization_read} during $I$.
        Hence, since by \Cref{lemma:ero:e_is_the_last_l_event} $e_{\min}$ is the last $L$-event in $\mathcal{I}^\mathcal{B}$, by \Cref{lemma:ero:if_p_reads_last_l_event_then_announce_acquire_never_fails}, $p$ does not find the condition on \cref{line:ero:announce_acquire_l_changed_check} to be true during $I$.
        Thus, since $p$ takes infinitely many steps during $\mathcal{I}^\mathcal{B}$, we have that $p$ receives $\notdone$ or $\done$ on \cref{line:ero:check_if_announce_is_done} during $I$.
        So, since $I$ is an iteration of the loop on \cref{line:ero:do_work_while_loop} by $p$ during $\mathcal{I}^\mathcal{B}$ and $p$'s execution of \cref{line:ero:linearization_read} during $I$ is after $e_{\min}$, by our initial assumption of \Cref{lemma:ero:each_process_does_finitely_many_successful_cas_on_ptr_next:claim_four}, we have that $p$ receives $\done$ on \cref{line:ero:check_if_announce_is_done} during $I$ (*).
        We now satisfy the condition of \Cref{lemma:ero:if_p_reads_unfinished_value_from_a_it_cannot_fail_the_done_check}.
        Suppose $p$ read $v$ from $\announceobject$ on \cref{line:ero:announce_read} during $I$; say at time $T^{\ref{line:ero:announce_read}}$.
        Hence, since $p$'s execution of \cref{line:ero:linearization_read} during $I$ is after $\max(e_{\min}, T_1)$, we have that $T^{\ref{line:ero:announce_read}} \geq T_1$, and so by \Cref{lemma:ero:each_process_does_finitely_many_successful_cas_on_ptr_next:claim_four:claim_three}, $\linearizationobject{} \neq v$ throughout $\mathcal{I}^\mathcal{B}$.
        Therefore, by \Cref{lemma:ero:if_p_reads_unfinished_value_from_a_it_cannot_fail_the_done_check}, $p$ does does not receive $\done$ on \cref{line:ero:check_if_announce_is_done} during $I$.
        However, by (*), $p$ receives $\done$ on \cref{line:ero:check_if_announce_is_done} during $I$, a contradiction.
        \qH{\Cref{lemma:ero:each_process_does_finitely_many_successful_cas_on_ptr_next:claim_four}}
    \end{proof}

    We now finish the proof of \Cref{lemma:ero:each_process_does_finitely_many_successful_cas_on_ptr_next}.
    Since by \Cref{lemma:ero:each_process_does_finitely_many_successful_cas_on_ptr_next:claim_four} there is an iteration $I$ of the loop on \cref{line:ero:do_work_while_loop} by $p$ during $\mathcal{I}^\mathcal{B}$ such that $p$'s execution of \cref{line:ero:linearization_read} during $I$ is at or after $e_{\min}$ and $p$ executes \cref{line:ero:linearization_cas} during $I$.
    There are two cases.

    \begin{itemize}
        \item[] \hspace{0pt}\textbf{Case 1.} $p$'s execution of \cref{line:ero:linearization_cas} during $I$ is successful.

        Hence, since $p$'s execution of \cref{line:ero:linearization_read} during $I$ is at or after $e_{\min}$, we have that $p$'s execution of \cref{line:ero:linearization_cas} during $I$ is after $e_{\min}$.
        Therefore, by \Cref{def:ero:english}, there is an $L$-event after $e_{\min}$ in $\mathcal{I}^\mathcal{B}$.
        However, by \Cref{lemma:ero:e_is_the_last_l_event}, $e_{\min}$ is the last $L$-event in $\mathcal{I}^\mathcal{B}$, a contradiction.

        \item[] \hspace{0pt}\textbf{Case 2.} $p$'s execution of \cref{line:ero:linearization_cas} during $I$ is unsuccessful.

        Hence, between $p$'s execution of \cref{line:ero:linearization_read} and \cref{line:ero:linearization_cas} during $I$, the value of $\linearizationobject{}$ changed.
        Thus, by \Cref{observation:ero:where_objects_change}, there is an $L$-event after $p$'s execution of \cref{line:ero:linearization_read} during $I$.
        Therefore, since $p$'s execution of \cref{line:ero:linearization_read} during $I$ is at or after $e_{\min}$, we have that there is an $L$-event after $e_{\min}$ in $\mathcal{I}^\mathcal{B}$.
        However, by \Cref{lemma:ero:e_is_the_last_l_event}, $e_{\min}$ is the last $L$-event in $\mathcal{I}^\mathcal{B}$, a contradiction.
        \qH{\Cref{lemma:ero:each_process_does_finitely_many_successful_cas_on_ptr_next}}
    \end{itemize}
\end{proof}

We now prove (2).

\begin{proposition}\label{lemma:ero:only_finitely_many_processes_in_stuck_run}
    In \Cref{scenario:ero:helper_loop_scenario}, finitely many processes take steps in $\mathcal{I}^\mathcal{B}$.
\end{proposition}

\begin{proof}
    Suppose, for contradiction, infinitely many processes take steps in $\mathcal{I}^\mathcal{B}$.
    Since by \Cref{assumption:ero:bounded_concurrency} $\mathcal{I}^\mathcal{B}$ has bounded concurrency\footnote{This is the only place in the entire proof where we rely on this assumption.}, this implies infinitely many operations complete in $\mathcal{I}^\mathcal{B}$.
    Let $opx_1, opx_2, \ldots$ denote this infinite sequence of complete operation executions and let $\cellpointershort{}_i$ denote the response the process that executed $opx_i$ received on \cref{line:ero:allocate_cell} during $opx_i$.
    We prove that there is an $L$-add event for $\cellpointershort{}_i$ in $\mathcal{I}^\mathcal{B}$.
    Since $opx_i$ completes and received $\cellpointershort{}_i$ on \cref{line:ero:allocate_cell}, we have that the process that executed $opx_i$ invokes and exits the \doworkuntildone{} procedure on \cref{line:ero:low_level_add_cell} with parameters $(\addcell, \cellpointershort{}_i)$.
    Therefore, since by \Cref{lemma:ero:the_list_invariants_hold} $P(\mathcal{I}^\mathcal{B})$ holds, by \Cref{lemma:ero:l_add_event_before_add_low_level_exits}, there is an $L$-add event for $\cellpointershort{}_i$ in $\mathcal{I}^\mathcal{B}$.
    We now finish the proof.
    Since by \Cref{alg:lazy_cell_manager_specification} the responses on \cref{line:ero:allocate_cell} are unique, we have that $\cellpointershort{}_i \neq \cellpointershort{}_j$ for every $i \neq j$.
    Hence, since there is an $L$-add event for $\cellpointershort{}_i$ for every $i$ in $\mathcal{I}^\mathcal{B}$, we have that there are infinitely many $L$-events in $\mathcal{I}^\mathcal{B}$.
    However, by \Cref{lemma:ero:e_is_the_last_l_event}, there are finitely many $L$-events in $\mathcal{I}^\mathcal{B}$, a contradiction.
    \qH{\Cref{lemma:ero:only_finitely_many_processes_in_stuck_run}}
\end{proof}

This completes the second part of the high-level argument for why $p_{\min}$ cannot take infinitely many steps in the loops on lines  \ref{line:ero:remove_cell_remove_seal_loop}, \ref{line:ero:remove_cell_remove_repeat_loop}, and \ref{line:ero:acquire_next_repeat_loop} during $opx_{\min}$.
We are now ready to prove the main claim of this section.

\begin{lemma}\label{lemma:ero:if_stuck_must_be_the_main_loop}
    $\mathcal{L}_{\min}$ is not a loop on line \ref{line:ero:remove_cell_remove_seal_loop}, \ref{line:ero:remove_cell_remove_repeat_loop}, or \ref{line:ero:acquire_next_repeat_loop}.
\end{lemma}

\begin{proof}
    Suppose, for contradiction, $\mathcal{L}_{\min}$ is a loop on line \ref{line:ero:remove_cell_remove_seal_loop} (Case A), \ref{line:ero:remove_cell_remove_repeat_loop} (Case B), or \ref{line:ero:acquire_next_repeat_loop} (Case C).
    Hence, $\mathcal{L}_{\min}$ is not a loop on \cref{line:ero:do_work_while_loop}.
    Thus, this is \Cref{scenario:ero:helper_loop_scenario}.
    Let $\cellpointershort{}$ be the value of the local variable $\cellpointershort{}_\linearizationobject{}$ (Case A), $\previouscellpointershort{}$ (Case B), or $\currentcellpointershort{}$ (Case C) in $\mathcal{L}_{\min}$.
    Since by \Cref{lemma:ero:case_a_b_c_ptr_is_from_universe} $\cellpointershort{} \in \celluniverse \cup \{\&\headobject\}$, and by \Cref{lemma:ero:infinite_cas_loops_imply_a_pointer_changes_infinitely_often}, $(*\cellpointershort{}).\nextlong$ changes infinitely often during $\mathcal{I}^\mathcal{B}$, by \Cref{observation:ero:where_objects_change} and \Cref{def:ero:english}, we have that there are infinitely many successful \CASop{} operations on $(*\cellpointershort{}).\nextlong$ in $\mathcal{I}^\mathcal{B}$.
    However, since $\cellpointershort{} \in \celluniverse \cup \{\&\headobject\}$, by \Cref{lemma:ero:each_process_does_finitely_many_successful_cas_on_ptr_next} each process performs a finite number of successful \CASop{} operations on $(*\cellpointershort{}).\nextlong$ in $\mathcal{I}^\mathcal{B}$, and by \Cref{lemma:ero:only_finitely_many_processes_in_stuck_run} only finitely many processes take steps in $\mathcal{I}^\mathcal{B}$, we have that there are finitely many successful \CASop{} operations on $(*\cellpointershort{}).\nextlong$ in $\mathcal{I}^\mathcal{B}$, a contradiction.
    \qH{\Cref{lemma:ero:if_stuck_must_be_the_main_loop}}
\end{proof}

By \Cref{lemma:ero:if_stuck_must_be_the_main_loop_or_cas_loops} and \Cref{lemma:ero:if_stuck_must_be_the_main_loop} we have the following.

\begin{corollary}\label{lemma:ero:stuck_in_main_loop}
    $\mathcal{L}_{\min}$ is the loop on \cref{line:ero:do_work_while_loop}.
\end{corollary}

\subsubsection{Processes cannot get stuck in the loop on \cref{line:ero:do_work_while_loop}}

In this section, we show that $p_{\min}$ does not take infinitely many steps in $\mathcal{L}_{\min}$.
Let $I_{\min}$ denote the invocation of the \doworkuntildone{} procedure that $\mathcal{L}_{\min}$ was executed during.
Furthermore, let $v_{\min} = ((t(I(opx_{\min})), \repositoryoperationshort_{\min}), \uniquecellpointershort_{\min})$ where $(\repositoryoperationshort_{\min}, \uniquecellpointershort_{\min})$ are the parameters of $I_{\min}$.

\begin{observation}\label{observation:ero:every_a_event_in_i_min_tries_to_set_a_to_v_min}
    Every execution of \cref{line:ero:announce_gcas} or \cref{line:ero:announce_cas} during $I_{\min}$ tries to set $\announceobject = v_{\min}$.
\end{observation}

The high-level argument for why $p_{\min}$ does not take infinitely many steps in $\mathcal{L}_{\min}$ follows closely from the wait-freedom proof of \Cref{alg:wait-free-simple}.

\begin{proposition}\label{lemma:ero:v_min_is_never_in_l}
    $\linearizationobject{} \neq v_{\min}$ throughout $\mathcal{I}^\mathcal{B}$.    
\end{proposition}

\begin{proof}
    Suppose, for contradiction, $\linearizationobject{} = v_{\min}$ at some time during $\mathcal{I}^\mathcal{B}$.
    Hence, since $\linearizationobject{}$ is initially $((0, \noop), \nullconstant)$, $v_{\min} = ((t(I(opx_{\min})), \repositoryoperationshort_{\min}), \uniquecellpointershort_{\min})$, and $\repositoryoperationshort_{\min} \neq \noop$, we have that $\linearizationobject{}$ was set to $v_{\min}$.
    Thus, by \Cref{observation:ero:where_objects_change}, there is an $L$-event $e$ that set $\linearizationobject{} = v_{\min}$.
    Since $p_{\min}$ takes infinitely many steps in $\mathcal{L}_{\min}$, and by \Cref{lemma:ero:stuck_in_main_loop} $\mathcal{L}_{\min}$ is the loop on \cref{line:ero:do_work_while_loop} during $I_{\min}$, we have that $p_{\min}$ executes \cref{line:ero:announce_gcas} infinitely many times in $I_{\min}$.
    Therefore, by \Cref{observation:ero:every_a_event_in_i_min_tries_to_set_a_to_v_min}, $p_{\min}$ performs infinitely many executions of \cref{line:ero:announce_gcas} that try to set $\announceobject = v_{\min}$.
    However, since $e$ is an $L$-event that set $\linearizationobject{} = v_{\min}$ in $\mathcal{I}^\mathcal{B}$, by \Cref{lemma:ero:a_events_which_get_into_l_eventually_get_out_of_a}, there are only finitely many executions of \cref{line:ero:announce_gcas} that try to set $\announceobject = v_{\min}$, a contradiction.
    \qH{\Cref{lemma:ero:v_min_is_never_in_l}}
\end{proof}

\begin{proposition}\label{lemma:ero:stuck_in_a_implies_finitely_many_l_events}
    Suppose $\announceobject = v$ from some time $T$ onwards in $\mathcal{I}^\mathcal{B}$.
    Then, there are finitely many $L$-events in $\mathcal{I}^\mathcal{B}$.
\end{proposition}

\begin{proof}
    Suppose, for contradiction, there are infinitely many $L$-events in $\mathcal{I}^\mathcal{B}$.
    Let $l_1, l_2, \ldots$ denote the infinitely many $L$-events in $\mathcal{I}^\mathcal{B}$ in the order they occur, and suppose $l_i$ sets $\linearizationobject{} = v_i$.
    Hence, by \Cref{lemma:ero:l_events_have_corresponding_a_events}, there is an $A$-event that set $\announceobject = v_i$.
    Furthermore, since by \Cref{lemma:ero:the_list_invariants_hold} $P(\mathcal{I}^\mathcal{B})$ holds, by \Cref{lemma:ero:p_implies_unique_values_in_linearization}, we have that $v_i \neq v_j$ for every $i \neq j$.
    Hence, since there is an $A$-event that set $\announceobject = v_i$, we have that there are infinitely many $A$-events that set $\announceobject$ to different values in $\mathcal{I}^\mathcal{B}$.
    Thus, for every time, there is a later $A$-event that sets $\announceobject$ to a value that $\announceobject$ has never been set to before.
    Therefore, since $\announceobject = v$ at $T$, there is an $A$-event after $T$ that sets $\announceobject$ to a value other than $v$.
    However, by assumption, $\announceobject = v$ from $T$ onwards, a contradiction.
    \qH{\Cref{lemma:ero:stuck_in_a_implies_finitely_many_l_events}}
\end{proof}

\begin{proposition}\label{lemma:ero:stuck_in_a_implies_written_into_l}
    Suppose $\announceobject = v$ from some time $T$ onwards in $\mathcal{I}^\mathcal{B}$.
    Then, $\linearizationobject{} = v$ at some time in $\mathcal{I}^\mathcal{B}$.
\end{proposition}

\begin{proof}
    Suppose, for contradiction, from some time $T$ onwards $\announceobject = v$ and $\linearizationobject{} \neq v$ throughout $\mathcal{I}^\mathcal{B}$.
    Hence, since $\linearizationobject{}$ is initially $((0, \noop), \nullconstant)$, it follows that $v \neq ((0, \noop), \nullconstant)$.
    Thus, since $\announceobject = v$ at $T$, and $\announceobject{}$ is initially $((0, \noop), \nullconstant)$, it follows that $\announceobject$ was set to $v$, and so by \Cref{observation:ero:where_objects_change}, there is an $A$-event $e$ that set $\announceobject = v$.
    Suppose $e$ is an $A$-event for $\cellpointershort{}$.
    Hence, by \Cref{def:ero:english}, $v = (\arbitraryvalue, \cellpointershort)$, and by \Cref{lemma:ero:every_a_event_is_for_pointer_from_universe} $\cellpointershort{} \in \celluniverse$.
    Since by assumption $\announceobject = v$ from $T$ onwards in $\mathcal{I}^\mathcal{B}$, by \Cref{lemma:ero:stuck_in_a_implies_finitely_many_l_events}, there are finitely many $L$-events in $\mathcal{I}^\mathcal{B}$.
    Hence, there is a last $L$-event in $\mathcal{I}^\mathcal{B}$; say $e_{last}$.

    \begin{claimcustom}{\ref{lemma:ero:stuck_in_a_implies_written_into_l}.1}\label{lemma:ero:stuck_in_a_implies_written_into_l:claim_one}
        Suppose $p_{\min}$ executes \cref{line:ero:linearization_read} at some time after $\max(T, e_{last})$ in $\mathcal{I}^\mathcal{B}$.
        Then, in the same iteration of the loop on \cref{line:ero:do_work_while_loop}, $p_{\min}$ executes \cref{line:ero:linearization_cas} and tries to set $\linearizationobject{} = v$.
    \end{claimcustom}

    \begin{proof}
        Suppose $p_{\min}$ executes \cref{line:ero:linearization_read} at some time $T^{\ref{line:ero:linearization_read}}$ after $\max(T, e_{last})$.
        Let $I$ be the iteration of the loop on \cref{line:ero:do_work_while_loop} that $p_{\min}$ executes \cref{line:ero:linearization_read} at $T^{\ref{line:ero:linearization_read}}$.
        Since $p_{\min}$ takes infinitely many steps in $\mathcal{L}_{\min}$, we have that $p_{\min}$ executes \cref{line:ero:announce_read} during $I$; say at time  $T^{\ref{line:ero:announce_read}}$.
        Hence, $\max(T, e_{last}) < T^{\ref{line:ero:linearization_read}} < T^{\ref{line:ero:announce_read}}$.
        Since by assumption $\announceobject = v$ from $T$ onwards, and $T < T^{\ref{line:ero:announce_read}}$, we have that $p_{\min}$ read $v$ from $\announceobject$ at $T^{\ref{line:ero:announce_read}}$.
        Hence, if $p_{\min}$ executes \cref{line:ero:linearization_cas} during $I$, we have that $p_{\min}$ tries to set $\linearizationobject{} = v$.
        Thus, it suffices to prove that $p_{\min}$ executes \cref{line:ero:linearization_cas} during $I$.

        Suppose, for contradiction, $p_{\min}$ does not execute \cref{line:ero:linearization_cas} during $I$.
        Since $e_{last}$ is the last $L$-event in $\mathcal{I}^\mathcal{B}$, and $p_{\min}$'s execution of \cref{line:ero:linearization_read} during $I$ is after $e_{last}$, by \Cref{lemma:ero:if_p_reads_last_l_event_then_announce_acquire_never_fails}, $p_{\min}$ does not find the condition on \cref{line:ero:announce_acquire_l_changed_check} to be true during $I$.
        Thus, since $p_{\min}$ takes infinitely many steps in  $\mathcal{L}_{\min}$, we have that $p_{\min}$ receives $\notdone$ or $\done$ on \cref{line:ero:check_if_announce_is_done} during $I$.
        Therefore, since by assumption $p_{\min}$ does not execute \cref{line:ero:linearization_cas} during $I$, we have that $p_{\min}$ receives $\done$ on \cref{line:ero:check_if_announce_is_done} during $I$.
        However, since $p_{\min}$ reads $v$ from $\announceobject$ on \cref{line:ero:announce_read} during $I$ and $\linearizationobject{} \neq v$ throughout $\mathcal{I}^\mathcal{B}$, by \Cref{lemma:ero:if_p_reads_unfinished_value_from_a_it_cannot_fail_the_done_check}, $p_{\min}$ does not receive $\done$ on \cref{line:ero:check_if_announce_is_done} during $I$, a contradiction.
        \qH{\Cref{lemma:ero:stuck_in_a_implies_written_into_l:claim_one}}
    \end{proof}

    \begin{claimcustom}{\ref{lemma:ero:stuck_in_a_implies_written_into_l}.2}\label{lemma:ero:stuck_in_a_implies_written_into_l:claim_two}
        There is a time $T'$ after which all executions of \cref{line:ero:linearization_cas} try to set $\linearizationobject{} = v$.
    \end{claimcustom}

    \begin{proof}
        Suppose, for contradiction, that for all times there exists a later time where an execution of \cref{line:ero:linearization_cas} tries to set $\linearizationobject{}$ to a value other than $v$.
        Thus, since every execution of \cref{line:ero:linearization_cas} tries to set $\linearizationobject{}$ to a value read from $\announceobject$, we have that there are infinitely many executions of \cref{line:ero:announce_read} that read a value other than $v$ from $\announceobject$.
        However, from $T$ onwards $\announceobject = v$, a contradiction.
        \qH{\Cref{lemma:ero:stuck_in_a_implies_written_into_l:claim_two}}
    \end{proof}

    We now complete the proof of \Cref{lemma:ero:stuck_in_a_implies_written_into_l}.
    Since $p_{\min}$ takes infinitely many steps in $\mathcal{L}_{\min}$ and by \Cref{lemma:ero:stuck_in_main_loop} $\mathcal{L}_{\min}$ is the loop on \cref{line:ero:do_work_while_loop}, we have that $p_{\min}$ executes \cref{line:ero:linearization_read} at some time after $\max(T, e_{last}, T')$.
    Consider the first iteration of the loop on \cref{line:ero:do_work_while_loop} during $\mathcal{L}_{\min}$ that $p_{\min}$ executes \cref{line:ero:linearization_read} after $\max(T, e_{last}, T')$.
    Denote this loop iteration by $I$ and let $T^{\ref{line:ero:linearization_read}}$ be the time $p_{\min}$ executed \cref{line:ero:linearization_read} during $I$.
    By \Cref{lemma:ero:stuck_in_a_implies_written_into_l:claim_one}, $p_{\min}$ executes \cref{line:ero:linearization_cas} during $I$ and tries to set $\linearizationobject{} = v$; say at time $T^{\ref{line:ero:linearization_cas}}$.
    If $p_{\min}$'s execution of \cref{line:ero:linearization_cas} at $T^{\ref{line:ero:linearization_cas}}$ is successful, then $\linearizationobject{} = v$ at $T^{\ref{line:ero:linearization_cas}}$.
    However, by assumption $\linearizationobject{} \neq v$ at all times in $\mathcal{I}^\mathcal{B}$, so $p_{\min}$'s execution of \cref{line:ero:linearization_cas} at $T^{\ref{line:ero:linearization_cas}}$ is unsuccessful.
    Thus, since the first parameter of $p_{\min}$'s execution of \cref{line:ero:linearization_cas} at $T^{\ref{line:ero:linearization_cas}}$ is the value that $p_{\min}$ read from $\linearizationobject{}$ on \cref{line:ero:linearization_read} at $T^{\ref{line:ero:linearization_read}}$, we have that between $T^{\ref{line:ero:linearization_read}}$ and $T^{\ref{line:ero:linearization_cas}}$ the value of $\linearizationobject{}$ changed, and so by \Cref{observation:ero:where_objects_change}, there is an $L$-event $e'$ after $T^{\ref{line:ero:linearization_read}}$.
    Hence, since $T^{\ref{line:ero:linearization_read}}$ is after $T'$, by transitivity, $T' < e$.
    Therefore, since by \Cref{def:ero:english} $e'$ is an execution of \cref{line:ero:linearization_cas}, by \Cref{lemma:ero:stuck_in_a_implies_written_into_l:claim_two}, $e'$ set $\linearizationobject{} = v$.
    However, by assumption $\linearizationobject{} \neq v$ at all times in $\mathcal{I}^\mathcal{B}$, a contradiction.
    \qH{\Cref{lemma:ero:stuck_in_a_implies_written_into_l}}
\end{proof}

\begin{proposition}\label{lemma:ero:timestamps_smaller_than_t_min_take_finitely_many_steps}
    Suppose some process $p$ during some invocation $I$ of the \doworkuntildone{} procedure in $\mathcal{I}^\mathcal{B}$ received $t < t(I(opx_{\min}))$ as a response on \cref{line:ero:operation_timestamp} during $I$.
    Then, $p$ takes finitely many steps in $I$.
\end{proposition}

\begin{proof}
    Suppose, for contradiction, $p$ takes infinitely many steps in $I$.
    Let $opx$ be the operation execution $p$ is executing $I$ during.
    Hence, since $p$ takes infinitely many steps in $I$ during $opx$, we have that $p$ never executes \cref{line:ero:response_step} during $opx$, and so $p$ takes infinitely many steps during $opx$ without completing it.
    Thus, by \Cref{def:ero:stuck}, $opx$ is stuck, and so $opx \in \mathbf{S}$.
    Since $p$ takes infinitely many steps inside $I$ during $opx$ and $I$ is an invocation of the \doworkuntildone{} procedure, it follows that $I$ is the invocation of the \doworkuntildone{} procedure identified in (1) of \Cref{observation:ero:stuck_operation_basic_properties}.
    Hence, by \Cref{def:ero:opx_min}, $I(opx) = I$.
    Thus, since $p$ received $t$ as a response on \cref{line:ero:operation_timestamp} during $I$, by \Cref{def:ero:do_work_timestamp}, $t(I) = t$, and since $I(opx) = I$, we have that $t(I(opx)) = t$.
    Therefore, since $opx \in \mathbf{S}$, by \Cref{def:ero:opx_min}, $t(I(opx_{\min})) \leq t(I(opx))$, and so $t(I(opx_{\min})) \leq t$.
    However, by definition $t < t(I(opx_{\min}))$, a contradiction.
    \qH{\Cref{lemma:ero:timestamps_smaller_than_t_min_take_finitely_many_steps}}
\end{proof}

\begin{proposition}\label{lemma:ero:v_min_is_stuck_in_a}
    $A = v_{\min}$ from some time $T$ onwards in $\mathcal{I}^\mathcal{B}$.
\end{proposition}

\begin{proof}
    Suppose, for contradiction, that for all times there exists a later time when $\announceobject \neq v_{\min}$.

    \begin{claimcustom}{\ref{lemma:ero:v_min_is_stuck_in_a}.1}\label{lemma:ero:v_min_is_stuck_in_a:claim_one}
        There is a time $T_1$ after which $\announceobject \neq ((t, \arbitraryvalue), \arbitraryvalue)$ for every $t \in [0..t(I(opx_{\min})))$.
    \end{claimcustom}

    \begin{proof}
        Suppose, for contradiction, that for all times there is a later time when $\announceobject = ((t, \arbitraryvalue), \arbitraryvalue)$ for some $t \in [0..t(I(opx_{\min})))$.
        There are two cases.
        \begin{itemize}
            \item[] \hspace{0pt}\textbf{Case 1.} The value of $\announceobject$ changes infinitely often.

            Since $[0..t(I(opx_{\min})))$ is finite, we have that $\announceobject$ is set to $((t, \arbitraryvalue), \arbitraryvalue)$ for some $t \in [0..t(I(opx_{\min})))$ infinitely often.
            Hence, by \Cref{observation:ero:where_objects_change}, there are infinitely many $A$-events that set $\announceobject$ to $((t, \arbitraryvalue), \arbitraryvalue)$.
            Thus, by \Cref{def:ero:english}, there are infinitely many executions of \cref{line:ero:announce_gcas} or \cref{line:ero:announce_cas} that try to set $\announceobject = ((t, \arbitraryvalue), \arbitraryvalue)$.
            Therefore, by \Cref{lemma:ero:try_to_set_a_to_same_timestamp_by_same_process}, all of these executions are by the same process $p$ and inside the same invocation $I$ of the \doworkuntildone{} procedure where $p$ received $t$ as a response on \cref{line:ero:operation_timestamp} during $I$. 
            However, since $t < t(I(opx_{\min}))$, by \Cref{lemma:ero:timestamps_smaller_than_t_min_take_finitely_many_steps}, $p$ takes finitely many steps in $I$, a contradiction.
            
            \item[] \hspace{0pt}\textbf{Case 2.} The value in $\announceobject$ changes finitely often.

            Since infinitely often $A = ((t, \arbitraryvalue), \arbitraryvalue)$ for some $t \in [0..t(I(opx_{\min})))$, it follows that there exists a $t \in [0..t(I(opx_{\min})))$ such that $A = ((t, \arbitraryvalue), \arbitraryvalue)$ from some time $T$ onwards in $\mathcal{I}^\mathcal{B}$.
            Hence, by \Cref{lemma:ero:stuck_in_a_implies_finitely_many_l_events}, there are finitely many $L$-events in $\mathcal{I}^\mathcal{B}$; say $e_{last}$ is the last $L$-event in $\mathcal{I}^\mathcal{B}$.
            Since $p_{\min}$ takes infinitely many steps in $\mathcal{L}_{\min}$, we have that $p_{\min}$ executes \cref{line:ero:linearization_read} after $\max(T, e_{last})$.
            Consider the first time $p_{\min}$ does so and let $I$ be the iteration of the loop on \cref{line:ero:do_work_while_loop} that $p_{\min}$ does so in.
            Hence, since $e_{last}$ is the last $L$-event in $\mathcal{I}^\mathcal{B}$, by \Cref{lemma:ero:if_p_reads_last_l_event_then_announce_acquire_never_fails}, $p_{\min}$ does not find the condition on \cref{line:ero:announce_acquire_l_changed_check} to be true during $I$.
            Thus, since $p_{\min}$ takes infinitely many steps in $\mathcal{L}_{\min}$, we have that $p_{\min}$ receives $\notdone$ or $\done$ on \cref{line:ero:check_if_announce_is_done} during $I$, and so $p_{\min}$ executes either \cref{line:ero:linearization_cas} or \cref{line:ero:announce_cas} during $I$.
            We consider each case separately.
            
            \begin{itemize}
                \item[] \hspace{0pt}\textbf{Case 2.1.} $p_{\min}$ executes \cref{line:ero:linearization_cas} during $I$.

                Let $T^{\ref{line:ero:linearization_cas}}$ be the time of  $p_{\min}$'s execution of \cref{line:ero:linearization_cas} during $I$.
                Since $p_{\min}$ executes \cref{line:ero:linearization_read} after $e_{last}$ during $I$, we have that $T^{\ref{line:ero:linearization_cas}} > e_{last}$.
                Thus, if $p_{\min}$'s execution of \cref{line:ero:linearization_cas} at $T^{\ref{line:ero:linearization_cas}}$ is successful, by \Cref{def:ero:english}, there is an $L$-event after $e_{last}$ in $\mathcal{I}^\mathcal{B}$.
                However, $e_{last}$ is the last $L$-event in $\mathcal{I}^\mathcal{B}$ so this is impossible.
                Hence, $p_{\min}$'s execution of \cref{line:ero:linearization_cas} at $T^{\ref{line:ero:linearization_cas}}$ is unsuccessful.
                Suppose $e_{last}$ set $\linearizationobject{} = (\uniquerepositoryoperationshort, \uniquecellpointershort)$.
                Hence, since $e_{last}$ is the last $L$-event in $\mathcal{I}^\mathcal{B}$, by \Cref{observation:ero:where_objects_change}, $\linearizationobject{} = (\uniquerepositoryoperationshort, \uniquecellpointershort)$ from $e_{last}$ onwards in $\mathcal{I}^\mathcal{B}$.
                Thus, since $p_{\min}$ executes \cref{line:ero:linearization_read} after $e_{last}$ during $I$, we have that $p_{\min}$ read $(\uniquerepositoryoperationshort, \uniquecellpointershort)$ from $\linearizationobject{}$ on \cref{line:ero:linearization_read} during $I$.
                So, since $p_{\min}$ executes \cref{line:ero:linearization_cas} during $I$ at $T^{\ref{line:ero:linearization_cas}}$, we have that its first parameter is $(\uniquerepositoryoperationshort, \uniquecellpointershort)$.
                Therefore, since $p_{\min}$'s execution of \cref{line:ero:linearization_cas} at $T^{\ref{line:ero:linearization_cas}}$ is unsuccessful, we have that $\linearizationobject{} \neq (\uniquerepositoryoperationshort, \uniquecellpointershort)$ at $T^{\ref{line:ero:linearization_cas}}$.
                However, since $T^{\ref{line:ero:linearization_cas}} > e_{last}$ and $\linearizationobject{} = (\uniquerepositoryoperationshort, \uniquecellpointershort)$ from $e_{last}$ onwards in $\mathcal{I}^\mathcal{B}$, we have that $\linearizationobject{} = (\uniquerepositoryoperationshort, \uniquecellpointershort)$ at $T^{\ref{line:ero:linearization_cas}}$, a contradiction.

                \item[] \hspace{0pt}\textbf{Case 2.2.} $p_{\min}$ executes \cref{line:ero:announce_cas} during $I$.

                Let $T^{\ref{line:ero:announce_cas}}$ be the time of $p_{\min}$'s execution of \cref{line:ero:announce_cas} during $I$.                
                Since $p_{\min}$ executes \cref{line:ero:linearization_read} during $I$ after $T$, we have that $p_{\min}$ executes \cref{line:ero:announce_read} during $I$ after $T$ and $T < T^{\ref{line:ero:announce_cas}}$.
                Hence, since $\announceobject = v$ from $T$ onwards in $\mathcal{I}^\mathcal{B}$, we have that $p_{\min}$ read $v$ from $\announceobject$ on \cref{line:ero:announce_read} during $I$.
                Thus, the first parameter of $p_{\min}$'s execution of \cref{line:ero:announce_cas} during $I$ at $T^{\ref{line:ero:announce_cas}}$ is $v$.
                If this execution is unsuccessful, it follows that $\announceobject \neq v$ at $T^{\ref{line:ero:announce_cas}}$, and since $T < T^{\ref{line:ero:announce_cas}}$, we have that $\announceobject \neq v$ some time after $T$.
                However, $\announceobject = v$ from $T$ onwards in $\mathcal{I}^\mathcal{B}$, and so this is impossible.
                Hence, $p_{\min}$'s execution of \cref{line:ero:announce_cas} during $I$ at $T^{\ref{line:ero:announce_cas}}$ is successful.
                Since this execution is during $I_{\min}$, by \Cref{observation:ero:every_a_event_in_i_min_tries_to_set_a_to_v_min}, it sets $\announceobject = v_{\min}$.
                Hence, $\announceobject = v_{\min}$ at $T^{\ref{line:ero:announce_cas}}$, and since $T < T^{\ref{line:ero:announce_cas}}$, we have that $\announceobject = v_{\min}$ some time after $T$.
                Therefore, since $v = ((t, \arbitraryvalue), \arbitraryvalue)$ for some $t < t(I(opx_{\min}))$, and $v_{\min} = ((t(I(opx_{\min})), \arbitraryvalue), \arbitraryvalue)$, we have that $v \neq v_{\min}$, and so $\announceobject \neq v$ some time after $T$.
                However, $\announceobject = v$ from $T$ onwards in $\mathcal{I}^\mathcal{B}$, a contradiction.
                \qH{\Cref{lemma:ero:v_min_is_stuck_in_a:claim_one}}
            \end{itemize}
        \end{itemize}
    \end{proof}

    \begin{claimcustom}{\ref{lemma:ero:v_min_is_stuck_in_a}.2}\label{lemma:ero:v_min_is_stuck_in_a:claim_two}
        $\announceobject = v_{\min}$ at some time $T_2 \geq T_1$.
    \end{claimcustom}

    \begin{proof}
        Since $p_{\min}$ takes infinitely many steps in $\mathcal{L}_{\min}$, we have that $p_{\min}$ executes \cref{line:ero:announce_gcas} in $\mathcal{L}_{\min}$ at some time $T_2 \geq T_1$.
        Hence, since $p_{\min}$'s execution of \cref{line:ero:announce_gcas} at $T_2$ is in $\mathcal{L}_{\min}$, and $\mathcal{L}_{\min}$ is in $I_{\min}$, by \Cref{observation:ero:every_a_event_in_i_min_tries_to_set_a_to_v_min}, $p_{\min}$'s execution of \cref{line:ero:announce_gcas} at $T_2$ tries to set $\announceobject = v_{\min}$.
        If this \GCASop{} returns true, then $\announceobject = v_{\min}$ at $T_2$ as wanted.
        Otherwise, this \GCASop{} returns false, so $\announceobject = v$ at $T_2$ such that $v \leq v_{\min}$.
        However, since this \GCASop{} is at $T_2 \geq T_1$, $v_{\min} = ((t(I(opx_{\min})), \arbitraryvalue), \arbitraryvalue)$, and the left component of $\announceobject.\uniquerepositoryoperationlong$ is always an integer greater than $0$ (because it is initially zero, and is only set to the response of \cref{line:ero:operation_timestamp}), by \Cref{lemma:ero:v_min_is_stuck_in_a:claim_one}, $v \geq v_{\min}$.
        Hence, $v = ((t, \arbitraryvalue), \arbitraryvalue)$ for some $t$ such that $t \leq t(I(opx_{\min}))$ and $t \geq t(I(opx_{\min}))$, and so $t = t(I(opx_{\min}))$.
        Since $\announceobject = v$ at $T_2$, it suffices to prove that $v = v_{\min}$.
        Since $t(I(opx_{\min}))$ is the response $p_{\min}$ received on \cref{line:ero:operation_timestamp} during $I_{\min}$, by the initialization of $\clockobject{}$, we have that $t(I(opx_{\min})) > 0$, and so $t > 0$.
        Thus, since $\announceobject$ is initially $((0, \noop), \nullconstant)$ and $\announceobject = v = ((t, \arbitraryvalue), \arbitraryvalue)$ at $T_2$, we have that $\announceobject$ was set to $v$ before $T_2$.
        Hence, by \Cref{observation:ero:where_objects_change}, some $A$-event set $\announceobject = v$ before $T_2$, and so by \Cref{def:ero:english} some execution $e_1$ of \cref{line:ero:announce_gcas} or \cref{line:ero:announce_cas} set $\announceobject = v$.
        Let $e_2$ be $p_{\min}$'s execution of \cref{line:ero:announce_gcas} at $T_2$ which tries to set $\announceobject = v_{\min}$.
        Since $v = ((t, \arbitraryvalue), \arbitraryvalue)$, $v_{\min} = ((t(I(opx_{\min})), \arbitraryvalue), \arbitraryvalue)$, and $t = t(I(opx_{\min}))$, by \Cref{lemma:ero:try_to_set_a_to_same_timestamp_by_same_process}, we have that $e_1$ and $e_2$ are executed by the same process during the same invocation of the \doworkuntildone{} procedure.
        Hence, since $e_2$ is executed by $p_{\min}$ during $I_{\min}$, we have that $e_1$ is also executed by $p_{\min}$ during $I_{\min}$.
        Thus, by \Cref{observation:ero:every_a_event_in_i_min_tries_to_set_a_to_v_min}, $e_1$ sets $\announceobject = v_{\min}$.
        Therefore, since by definition $e_1$ sets $\announceobject = v$, we have that $v = v_{\min}$ as required.
        \qH{\Cref{lemma:ero:v_min_is_stuck_in_a:claim_two}}
    \end{proof}

    We now complete the proof of \Cref{lemma:ero:v_min_is_stuck_in_a}.
    Since by assumption for all times there exists a later time when $\announceobject \neq v_{\min}$, we have that there is a time $T_3 > T_2$ where $\announceobject \neq v_{\min}$ at $T_3$.
    Without loss of generality, suppose this is the first time after $T_2$ where $\announceobject \neq v_{\min}$.
    For $\announceobject \neq v_{\min}$ at $T_3$, a process $p$ performed a successful execution $e$ of \cref{line:ero:announce_gcas} or \cref{line:ero:announce_cas} during some invocation $I$ of the \doworkuntildone{} which set $\announceobject = v$ for some $v \neq v_{\min}$ at $T_3$.
    Since $p$ executed $e$ during $I$, we have that $v = ((t, \arbitraryvalue), \arbitraryvalue)$ where $t$ was the response $p$ received on \cref{line:ero:operation_timestamp} during $I$.
    Hence, $t > 0$.
    Thus, since $T_3 > T_2 \geq T_1$, and $\announceobject = ((t, \arbitraryvalue), \arbitraryvalue)$ at $T_3$, by \Cref{lemma:ero:v_min_is_stuck_in_a:claim_one}, $t \geq t(I(opx_{\min}))$, and so $v \geq v_{\min}$.
    So, since $T_3$ is the first time after $T_2$ where $\announceobject \neq v_{\min}$, it follows that $e$ could not have been on \cref{line:ero:announce_gcas}.
    Hence, $e$ was on \cref{line:ero:announce_cas}.
    Let $I'$ be the iteration of the loop on \cref{line:ero:do_work_while_loop} that $p$ executed $e$ during.
    Since $T_3$ is the first time after $T_2$ where $\announceobject \neq v_{\min}$ and $e$ is a successful execution of \cref{line:ero:announce_cas}, we have that $p$ read $v_{\min}$ from $\announceobject$ on \cref{line:ero:announce_read} during $I'$; say at time $T^{\ref{line:ero:announce_read}}$.
    Therefore, since $e$ is an execution of \cref{line:ero:announce_cas} during $I'$, we have that $p$ received $\done$ on \cref{line:ero:check_if_announce_is_done} during $I'$.
    However, since $p$ read $v_{\min}$ from $\announceobject$ on \cref{line:ero:announce_read} during $I'$ and by \Cref{lemma:ero:v_min_is_never_in_l} $\linearizationobject{} \neq v_{\min}$ throughout $\mathcal{I}^\mathcal{B}$, by \Cref{lemma:ero:if_p_reads_unfinished_value_from_a_it_cannot_fail_the_done_check}, $p$ does not received $\done$ on \cref{line:ero:check_if_announce_is_done} during $I'$, a contradiction.
    \qH{\Cref{lemma:ero:v_min_is_stuck_in_a}}
\end{proof}

\begin{theorem}\label{theorem:ero:b_is_wait_free}
    $\mathcal{B}$ is wait-free.
\end{theorem}

\begin{proof}
    Suppose, for contradiction, $\mathcal{B}$ is not wait-free.
    Hence, there is an implementation history of $\mathcal{B}$ with an operation execution that is stuck.
    Let $\mathcal{I}^\mathcal{B}$, the history defined at the beginning of \Cref{sec:b_is_wait_free}, be this history.
    Consider the value $v_{\min}$.
    By \Cref{lemma:ero:v_min_is_stuck_in_a} $\announceobject = v_{\min}$ from some time $T$ onwards in $\mathcal{I}^\mathcal{B}$, and so by \Cref{lemma:ero:stuck_in_a_implies_written_into_l}, $\linearizationobject{} = v_{\min}$ at some time in $\mathcal{I}^\mathcal{B}$.
    However, by \Cref{lemma:ero:v_min_is_never_in_l}, $\linearizationobject{} \neq v_{\min}$ throughout $\mathcal{I}^\mathcal{B}$, a contradiction.
    \qH{\Cref{theorem:ero:b_is_wait_free}}
\end{proof}

\subsection{$\mathcal{B}$ Correctly Manages Cells and is Space-Efficient}
\label{sec:b_memory_management_and_space_efficiency}

Throughout this section, $\mathcal{I}^\mathcal{B}$ refers to an arbitrary implementation history, i.e., all statements that refer to $\mathcal{I}^\mathcal{B}$ begin with ``for every implementation history $\mathcal{I}^\mathcal{B}$ of $\mathcal{B}$" which is omitted for brevity.
The goal of this section is to prove the following two theorems.

\begin{theoremblank}[$\mathcal{B}$ Correctly Manages Cells]
    For every $\cellpointershort \in \celluniverse$ the following are true.
    \begin{compactenum}
        \item There is at most one $\allocatecelloperation{}$ operation whose response is $\cellpointershort$, and at most one $\freecelloperation{}(\cellpointershort)$ operation in $\mathcal{I}^\mathcal{B}$.
        \item If there is a $\freecelloperation{}(\cellpointershort)$ operation in $\mathcal{I}^\mathcal{B}$, then it is after an $\allocatecelloperation{}$ operation whose response is $\cellpointershort$.
        \item Every operation on an object of the cell pointed to by $\cellpointershort$ in $\mathcal{I}^\mathcal{B}$ is after an $\allocatecelloperation{}$ operation whose response is $\cellpointershort$, and is before any $\freecelloperation{}(\cellpointershort)$ operation.
    \end{compactenum}
\end{theoremblank}

\begin{theoremblank}[$\mathcal{B}$ is Space-Efficient]
    Suppose $\mathcal{I}^\mathcal{B}$ is finite.
    Let $Allocate(\mathcal{I}^\mathcal{B})$ be the set of pointers which have been allocated in $\mathcal{I}^\mathcal{B}$, i.e., $\cellpointershort \in Allocate(\mathcal{I}^\mathcal{B})$ if and only if there is an $\allocatecelloperation{}$ operation in $\mathcal{I}^\mathcal{B}$ with response $\cellpointershort$.
    Likewise, let $Free(\mathcal{I}^\mathcal{B})$ be the set of pointers which have been freed in $\mathcal{I}^\mathcal{B}$, i.e., $\cellpointershort \in Free(\mathcal{I}^\mathcal{B})$ if and only if there is a $\freecelloperation{}(\cellpointershort)$ operation in $\mathcal{I}^\mathcal{B}$.
    Then, $|Allocate(\mathcal{I}^\mathcal{B}) \setminus Free(\mathcal{I}^\mathcal{B})| \leq 6c + 1$ where $c$ is the point contention in $\mathcal{I}^\mathcal{B}$.
\end{theoremblank}

\subsubsection{At most one \freecelloperation{} operation per pointer}

The main goal of this section is to prove the first two bullets of the $\mathcal{B}$ correctly manages cells theorem.
We begin by proving some basic properties of successful list-add and list-remove attempts for a given pointer, which are used extensively throughout this section.

\begin{lemma}\label{lemma:ero:at_most_one_successful_list_add_attempt}
    There is at most one successful list-add attempt for $\cellpointershort$ in $\mathcal{I}^\mathcal{B}$.
\end{lemma}

\begin{proof}
    Suppose, for contradiction, there are at least two successful list-add attempts for $\cellpointershort$ in $\mathcal{I}^\mathcal{B}$; say $a_1$ and $a_2$ such that $a_1 < a_2$.
    Hence, there are two successful list-add attempts for $\cellpointershort$ in $\mathcal{I}^\mathcal{B}$.
    Let $e_1$ and $e_2$ be there corresponding $L$-events, so $e_1 < a_1$.
    Hence, by \Cref{lemma:ero:l_event_corresponding_to_do_low_level_op}, $e_1$ and $e_2$ are both $L$-add events for $\cellpointershort$.
    Thus, since by \Cref{lemma:ero:the_list_invariants_hold} $P(\mathcal{I}^\mathcal{B})$ holds, we have that $e_1 = e_2 = e$.
    Hence, since $e_1 < a_1$ and $a_1 < a_2$, by transitivity, $e < a_1 < a_2$.
    Furthermore, $e$ is the corresponding $L$-event for $a_2$.
    Hence, since by \Cref{lemma:ero:the_list_invariants_hold} $P(\mathcal{I}^\mathcal{B})$, $Q(\mathcal{I}^\mathcal{B})$, and $R(\mathcal{I}^\mathcal{B})$ hold, by \Cref{lemma:ero:any_list_attempt_outside_its_window_is_unsuccessful_alternate_statement}, $e$ is the last $L$-event before $a_2$ in $\mathcal{I}^\mathcal{B}$, and so $e$ is the last $L$-event in $\mathcal{I}^{include}_{a_2}$: the prefix of $\mathcal{I}^\mathcal{B}$ up to and including $a_2$.
    Therefore, since $e$ is an $L$-add event for $\cellpointershort$ (because $e_1 = e_2 = e$), and $P(\mathcal{I}^\mathcal{B})$, $Q(\mathcal{I}^\mathcal{B})$, and $R(\mathcal{I}^\mathcal{B})$ hold, by \Cref{lemma:ero:1_of_r_safety_holds}, there is at most one successful list-add attempt for $\cellpointershort$ from $e$ onwards in $\mathcal{I}^{include}_{a_2}$.
    However, since $e < a_1 < a_2$, there are two successful list-add attempts for $\cellpointershort$ from $e$ onwards in $\mathcal{I}^{include}_{a_2}$, a contradiction.
    \qH{\Cref{lemma:ero:at_most_one_successful_list_add_attempt}}
\end{proof}

The next lemma and its proof are the same as this one, except it's for list-remove attempts.

\begin{lemma}\label{lemma:ero:at_most_one_successful_list_remove_attempt}
    There is at most one successful list-remove attempt for $\cellpointershort$ in $\mathcal{I}^\mathcal{B}$.
\end{lemma}

\begin{proof}
    Suppose, for contradiction, there are at least two successful list-remove attempts for $\cellpointershort$ in $\mathcal{I}^\mathcal{B}$; say $a_1$ and $a_2$ such that $a_1 < a_2$.
    Hence, there are two successful list-remove attempts for $\cellpointershort$ in $\mathcal{I}^\mathcal{B}$.
    Let $e_1$ and $e_2$ be there corresponding $L$-events, so $e_1 < a_1$.
    Hence, by \Cref{lemma:ero:l_event_corresponding_to_do_low_level_op}, $e_1$ and $e_2$ are both $L$-remove events for $\cellpointershort$.
    Thus, since by \Cref{lemma:ero:the_list_invariants_hold} $P(\mathcal{I}^\mathcal{B})$ holds, we have that $e_1 = e_2 = e$.
    Hence, since $e_1 < a_1$ and $a_1 < a_2$, by transitivity, $e < a_1 < a_2$.
    Furthermore, $e$ is the corresponding $L$-event for $a_2$.
    Hence, since by \Cref{lemma:ero:the_list_invariants_hold} $P(\mathcal{I}^\mathcal{B})$, $Q(\mathcal{I}^\mathcal{B})$, and $R(\mathcal{I}^\mathcal{B})$ hold, by \Cref{lemma:ero:any_list_attempt_outside_its_window_is_unsuccessful_alternate_statement}, $e$ is the last $L$-event before $a_2$ in $\mathcal{I}^\mathcal{B}$, and so $e$ is the last $L$-event in $\mathcal{I}^{include}_{a_2}$: the prefix of $\mathcal{I}^\mathcal{B}$ up to and including $a_2$.
    Therefore, since $e$ is an $L$-remove event for $\cellpointershort$ (because $e_1 = e_2 = e$), and $P(\mathcal{I}^\mathcal{B})$, $Q(\mathcal{I}^\mathcal{B})$, and $R(\mathcal{I}^\mathcal{B})$ hold, by \Cref{lemma:ero:3_of_r_safety_holds}, there is at most one successful list-remove attempt for $\cellpointershort$ from $e$ onwards in $\mathcal{I}^{include}_{a_2}$.
    However, since $e < a_1 < a_2$, there are two successful list-remove attempts for $\cellpointershort$ from $e$ onwards in $\mathcal{I}^{include}_{a_2}$, a contradiction.
    \qH{\Cref{lemma:ero:at_most_one_successful_list_remove_attempt}}
\end{proof}

\begin{lemma}\label{lemma:ero:successful_list_add_before_successful_list_remove}
    If there is a successful list-add attempt $a_{add}$ for $\cellpointershort$ in $\mathcal{I}^\mathcal{B}$ and a successful list-remove attempt $a_{remove}$ for $\cellpointershort$ in $\mathcal{I}^\mathcal{B}$, then $a_{add}$ is before $a_{remove}$.
\end{lemma}

\begin{proof}
    Since by \Cref{lemma:ero:the_list_invariants_hold} $R(\mathcal{I}^\mathcal{B})$ holds, by \Cref{lemma:ero:list_add_before_list_remove}, there is a successful list-add attempt for $\cellpointershort$ before $a_{remove}$, which must be $a_{add}$ by \Cref{lemma:ero:at_most_one_successful_list_add_attempt}.
    \qH{\Cref{lemma:ero:successful_list_add_before_successful_list_remove}}
\end{proof}

We now prove the first bullet of the $\mathcal{B}$ correctly manages cells theorem, i.e., for every $\cellpointershort \in \celluniverse$, there is at most one $\freecelloperation{}(\cellpointershort)$ operation in $\mathcal{I}^\mathcal{B}$.
The strategy for doing so is to prove that there is at most one revocation event for $\cellpointershort$ whose response is $-1$ in $\mathcal{I}^\mathcal{B}$.
As we will see, this is a consequence of the following lemma.

\begin{lemma}\label{lemma:ero:at_most_one_acquisition_copy}
    There is at most one acquire-copy event for $\cellpointershort$ in $\mathcal{I}^\mathcal{B}$. 
\end{lemma}

\begin{proof}
    Suppose, for contradiction, there are two acquire-copy events for $\cellpointershort$ in $\mathcal{I}^\mathcal{B}$; say $e_1$ and $e_2$.
    Let $p_1$ (resp. $p_2$) be the process that executed $e_1$ (resp. $e_2$) and let $I_1$ (resp. $I_2$) be the invocation of the \doremovecell{} procedure that $e_1$ (resp. $e_2$) was executed during.
    Hence, since by \Cref{def:ero:english} $e_1$ and $e_2$ are distinct executions of \cref{line:ero:copy_acquisitions_to_revocations} and \cref{line:ero:copy_acquisitions_to_revocations} is executed at most one per invocation of the \doremovecell{} procedure, we have that $I_1 \neq I_2$.
    By \Cref{lemma:ero:before_acquire_copy_is_successful}, $p_1$ (resp. $p_2$) performed a successful list-remove attempt $a_1$ (resp. $a_2$) for $\cellpointershort$ during $I_1$ (resp. $I_2$).
    Since $a_1$ was executed during $I_1$, $a_2$ was executed during $I_2$, and $I_1 \neq I_2$, we have that $a_1 \neq a_2$.
    Therefore, there are two successful list-remove attempts for $\cellpointershort$ in $\mathcal{I}^\mathcal{B}$.
    However, by \Cref{lemma:ero:at_most_one_successful_list_remove_attempt}, there is at most one successful list-remove attempt for $\cellpointershort$ in $\mathcal{I}^\mathcal{B}$, a contradiction.
    \qH{\Cref{lemma:ero:at_most_one_acquisition_copy}}
\end{proof}

\begin{proposition}\label{lemma:ero:at_most_one_revocation_event_response_is_negative_one}
    There is at most one revocation event for $\cellpointershort$ in $\mathcal{I}^\mathcal{B}$ whose \mbox{response is $-1$.}
\end{proposition}

\begin{proof}
    Suppose, for contradiction, there are at least two revocation events for $\cellpointershort$ in $\mathcal{I}^\mathcal{B}$ whose responses are $-1$; say $e_1$ and $e_2$ such that $e_1 < e_2$.
    Hence, by \Cref{lemma:ero:revocation_event_is_for_pointer_from_universe} $\cellpointershort \in \celluniverse$, and so  $(*\cellpointershort).\revocations$ is initially 0.
    Furthermore, by \Cref{observation:ero:where_objects_change}, the only steps that change the value of $(*\cellpointershort).\revocations$ are acquire-copy events for $\cellpointershort$ and revocation events for $\cellpointershort$.
    Hence, since $(*\cellpointershort).\revocations$ is initially 0, each revocation event for $\cellpointershort$ increases the value of $(*\cellpointershort).\revocations$ by 1, and the response of $e_1$ is $-1$, we have that there is an acquire-copy event $e'_1$ for $\cellpointershort$ before $e_1$.
    Since the response of $e_1$ is $-1$ and $e_1$ is a revocation event for $\cellpointershort$, we have that $(*\cellpointershort).\revocations = 0$ at $e_1$.
    Thus, since $e_1 < e_2$, each revocation event for $\cellpointershort$ increases the value of $(*\cellpointershort).\revocations$ by 1, and the response of $e_2$ is $-1$, we have that there is an acquire-copy event $e'_2$ for $\cellpointershort$ between $e_1$ and $e_2$.
    Therefore, since $e'_1 < e_1$ and $e_1 < e'_2$, we have that $e'_1 \neq e'_2$, and so there are two acquire-copy events for $\cellpointershort$ in $\mathcal{I}^\mathcal{B}$.
    However, by \Cref{lemma:ero:at_most_one_acquisition_copy}, there is at most one acquire-copy event for $\cellpointershort$ in $\mathcal{I}^\mathcal{B}$, a contradiction.
    \qH{\Cref{lemma:ero:at_most_one_revocation_event_response_is_negative_one}}
\end{proof}

\begin{lemma}\label{lemma:ero:at_most_one_free_per_pointer}
    There is at most one $\freecelloperation{}(\cellpointershort)$ operation in $\mathcal{I}^\mathcal{B}$.
\end{lemma}

\begin{proof}
    Suppose, for contradiction, there are at least two $\freecelloperation{}(\cellpointershort)$ operations in $\mathcal{I}^\mathcal{B}$; say at time $T_1$ and $T_2$.
    Let $p_1$ (resp. $p_2$) be the process that executed the $\freecelloperation{}$ operation at $T_1$ (resp. $T_2$) and let $I_1$ (resp. $I_2$) be the invocation of the Relinquish procedure that $p_1$ (resp. $p_2$) executed the step at $T_1$ (resp. $T_2$) during.
    Hence, since $T_1 \neq T_2$, and there is at most one execution of the \freecelloperation{} operation during an invocation of the Relinquish procedure, we have that $I_1 \neq I_2$.
    Furthermore, $p_1$ (resp. $p_2$) found the condition on \cref{line:ero:free_cell} during $I_1$ (resp. $I_2$) to be true.
    Let $e_1$ (resp. $e_2$) be the execution of \cref{line:ero:relinquish_revocations} during $I_1$ (resp. $I_2$).
    Since the $\freecelloperation{}$ operation at $T_1$ (resp. $T_2$) has parameter $\cellpointershort$, it follows that $e_1$ (resp. $e_2$) is of the form $\FAop{}((*\cellpointershort).\revocations, 1)$, so by \Cref{def:ero:english}, $e_1$ (resp. $e_2$) is a revocation event for $\cellpointershort$.
    Hence, since $p_1$ (resp. $p_2$) found the condition on \cref{line:ero:free_cell} to be true at $e_1$ (resp. $e_2$), we have that the response of $e_1$ (resp. $e_2$) is $-1$.
    Since $e_1$ is an execution of \cref{line:ero:relinquish_revocations} during $I_1$, $e_2$ is an execution of \cref{line:ero:relinquish_revocations} during $I_2$, and $I_1 \neq I_2$, we have that $e_1 \neq e_2$.
    Therefore, there are two revocation events for $\cellpointershort$ in $\mathcal{I}^\mathcal{B}$ whose response is $-1$.
    However, by \Cref{lemma:ero:at_most_one_revocation_event_response_is_negative_one}, there is at most one revocation event for $\cellpointershort$ in $\mathcal{I}^\mathcal{B}$ whose response is $-1$, a contradiction.
    \qH{\Cref{lemma:ero:at_most_one_free_per_pointer}}
\end{proof}

%\subsubsection{Free After Allocate}

We now prove the second bullet of the $\mathcal{B}$ correctly manages cells theorem.

\begin{lemma}\label{lemma:ero:every_free_ptr_is_after_an_allocate_ptr}
    If there is a $\freecelloperation{}(\cellpointershort)$ operation in $\mathcal{I}^\mathcal{B}$, then it is after an $\allocatecelloperation{}$ operation whose response is $\cellpointershort$.
\end{lemma}

\begin{proof}
    Consider a $\freecelloperation{}(\cellpointershort)$ operation in $\mathcal{I}^\mathcal{B}$ at time $T$.
    Hence, by \Cref{lemma:ero:free_cell_operations_are_preceeded_by_remove_event}, there is an $L$-remove event $e$ for $\cellpointershort$ before $T$.
    Thus, by \Cref{lemma:ero:l_event_corresponding_a_event_type_and_pointer_relations}, there is an $A$-remove event $e'$ for $\cellpointershort$ before $e$.
    So, by \Cref{def:ero:english}, this $A$-remove event was executed during some invocation of the \doworkuntildone{} procedure with a second parameter of $\cellpointershort$.
    Hence, there is an $\allocatecelloperation{}$ operation whose response is $\cellpointershort$ before $e'$.
    Therefore, since $e' < e$, and $e < T$, by transitivity, there is an $\allocatecelloperation{}$ operation whose response is $\cellpointershort$ before $T$ as wanted.
    \qH{\Cref{lemma:ero:every_free_ptr_is_after_an_allocate_ptr}}
\end{proof}

\subsubsection{Tracking acquisitions and revocations per operation execution}

Over the next few sections, we prove the third bullet of the $\mathcal{B}$ correctly manages cells theorem and then prove that $\mathcal{B}$ is space-efficient.
Both of these theorems require proving some properties about the number of acquisitions and revocations performed by the process that executed $opx$ during some operation execution $opx$.
These properties are informally stated below.
\begin{compactitem}
    \item The number of successful list-acquire-next attempts for $\cellpointershort$ is larger than the number of revocation events for $\cellpointershort$ (\Cref{lemma:ero:reference_count_is_non_negative}).
    \item A process only performs an operation on an object of a cell when it has the right to use it (\Cref{lemma:ero:operation_acquisition_invariant}).
    \item The number of successful list-acquire-next attempts for $\cellpointershort$ is equal to the number of revocation events for $\cellpointershort$ at the time an operation execution completes (\Cref{lemma:ero:reference_count_of_complete_operation_execution}).
    \item Each process has the right to use at most three cells at all times (\Cref{lemma:ero:total_reference_count_is_bounded}).
\end{compactitem}

We note that the first two properties are used in the proof of the third bullet of the $\mathcal{B}$ correctly manages cells theorem, and all four properties are used in the proof of the $\mathcal{B}$ is space-efficient theorem.
Formally, all four of these properties are stated with respect to $R$ defined below.

\begin{definition}\label{def:ero:reference_count}
    For every operation execution $opx$ in $\mathcal{I}^\mathcal{B}$, we define $R(\mathcal{I}^\mathcal{B}, opx, \cellpointershort)$ as the number of successful list-acquire-next attempts for $\cellpointershort$ minus the number of revocation events for $\cellpointershort$ performed by the process that executed $opx$ during $opx$ in $\mathcal{I}^\mathcal{B}$.
\end{definition}

Note that, since $\mathcal{B}$ is wait-free by \Cref{theorem:ero:b_is_wait_free}, we have that the process that executed $opx$ performed a finite number of successful list-acquire-next attempts for $\cellpointershort$ and revocation for $\cellpointershort$ during $opx$ (otherwise the process that executed $opx$ would perform infinitely many steps during $opx$ without completing it, implying $\mathcal{B}$ is not wait-free), so $R(\mathcal{I}^\mathcal{B}, opx, \cellpointershort)$ is always an integer.

The main utility we get by proving bounds on $R$ is that it implies bounds on the total difference between successful list-acquire-next attempts and revocation events in $\mathcal{I}^\mathcal{B}$.
We formalize this below.

\begin{definition}\label{def:ero:number_of_successful_acquires}
    Let $A(\mathcal{I}^\mathcal{B}, \cellpointershort)$ (resp. $X(\mathcal{I}^\mathcal{B}, \cellpointershort)$) denote the number of successful list-acquire-next attempts (resp. revocation events) for $\cellpointershort$ in $\mathcal{I}^\mathcal{B}$.
\end{definition}

Note that unlike $R$, $A$ and $X$ may be infinite when $\mathcal{I}^\mathcal{B}$ is infinite, in which case they equal $\infty$. By \Cref{def:ero:reference_count} and \Cref{def:ero:number_of_successful_acquires}, we have the following.

\begin{observation}\label{lemma:ero:relate_a_x_and_r}
    If $\mathcal{I}^\mathcal{B}$ is finite, then
    \begin{align*}
        A(\mathcal{I}^\mathcal{B}, \cellpointershort) - X(\mathcal{I}^\mathcal{B}, \cellpointershort) &= \sum_{\text{$opx$ is an operation execution in $\mathcal{I}^\mathcal{B}$}} R(\mathcal{I}^\mathcal{B}, opx, \cellpointershort).
    \end{align*}
\end{observation}

We now prove the four properties mentioned at the start of the section.
We start with a few observations regarding the sequence of successful list-acquire-next attempts and revocation events performed during an invocation $I$ of the Acquire, \doremovecell{}, and \doaddcell{} procedures.
These observations are a consequence of the order in which the AcquireNext procedure and the Relinquish procedure are invoked during $I$.
Furthermore, these sequences are necessarily finite by the fact that $\mathcal{B}$ is wait-free by \Cref{theorem:ero:b_is_wait_free}. 

\begin{observation}\label{lemma:ero:acquire_acquire_revocation_sequence}
    Consider any invocation $I$ of the Acquire procedure by process $p$ in $\mathcal{I}^\mathcal{B}$.
    At any time $T$ in $\mathcal{I}^\mathcal{B}$, the sequence of successful list-acquire-next attempts and revocation events performed by $p$ during $I$ is some prefix of the following sequence, and is the entire sequence if $p$ exited $I$ by time $T$.
    First is a successful list-acquire-next attempt for $\cellpointershort_1$.
    Then, the following pattern occurs for $i = 1, \ldots, n$ where $n \geq 0$: a successful list-acquire-next attempt for $\cellpointershort_{i + 1}$ followed by a revocation event for $\cellpointershort_{i}$.
    Finally, if $p$ exits $I$ with response $\found$, then $\cellpointershort_{n + 1}$ is the second parameter of $I$, and otherwise, the last element of this sequence is a revocation event for $\cellpointershort_{n + 1}$.
\end{observation}

\begin{observation}\label{lemma:ero:do_remove_cell_acquire_revocation_sequence}
    Consider any invocation $I$ of the \doremovecell{} procedure by process $p$.
    At any time $T$ in $\mathcal{I}^\mathcal{B}$, the sequence of successful list-acquire-next attempts and revocation events performed by $p$ during $I$, other than those performed during the \setrepositoryoperationresponse{} procedure on \cref{line:ero:remove_cell_set_response}, is some prefix of the following sequence, and is the entire sequence if $p$ exited $I$ by time $T$.
    First is a successful list-acquire-next attempt for $\cellpointershort_1$.
    Then, either (1) the next is a revocation event for $\cellpointershort_1$, and there are no other successful list-acquire-next attempts and revocation events performed by $p$ during $I$, or (2) the next is a successful list-acquire-next attempt for $\cellpointershort_2$ and the following pattern occurs for $i = 1, \ldots, n$ where $n \geq 0$: a successful list-acquire-next attempt for $\cellpointershort_{i + 2}$ followed by a revocation event for $\cellpointershort_{i}$; finally, the last two elements of this sequence are a revocation event for $\cellpointershort_{n + 1}$ and a revocation event for $\cellpointershort_{n + 2}$.
\end{observation}

\begin{observation}\label{lemma:ero:do_add_cell_acquire_revocation_sequence}
    Consider any invocation $I$ of the \doaddcell{} procedure by process $p$.
    At any time $T$ in $\mathcal{I}^\mathcal{B}$, the sequence of successful list-acquire-next attempts and revocation events performed by $p$ during $I$, other than those performed during the \setrepositoryoperationresponse{} procedure on \cref{line:ero:add_cell_set_response}, is some prefix of the following sequence, and is the entire sequence if $p$ exited $I$ by time $T$.
    First is a successful list-acquire-next attempt for $\cellpointershort_1$.
    Then, the following pattern occurs for $i = 1, \ldots, n$ where $n \geq 0$: a successful list-acquire-next attempt for $\cellpointershort_{i + 1}$ followed by a revocation event for $\cellpointershort_{i}$.
    Finally, the last element of this sequence is a revocation event for $\cellpointershort_{n + 1}$.
\end{observation}

We now prove the first property of this section, which is, with the exception of the revocation event performed during the Relinquish procedure invoked on \cref{line:ero:owner_relinquish}, $R$ is always non-negative.
We first note an immediate consequence of \Cref{def:ero:reference_count}.

\begin{observation}\label{observation:ero:initially_no_acquisitions}
    For every operation execution $opx$ in $\mathcal{I}^\mathcal{B}$ if the only step by the process that executed $opx$ during $opx$ in $\mathcal{I}^\mathcal{B}$ is the invocation step of $opx$, then $R(\mathcal{I}^\mathcal{B}, opx, \cellpointershort) = 0$ for every $\cellpointershort$.
\end{observation}

\begin{lemma}\label{lemma:ero:reference_count_is_non_negative_before_owner_relinquish}
    Consider any operation execution $opx$ in $\mathcal{I}^\mathcal{B}$ such that the process that executed $opx$ has not executed \cref{line:ero:owner_relinquish} during $opx$ in $\mathcal{I}^\mathcal{B}$.
    Then, $R(\mathcal{I}^\mathcal{B}, opx, \cellpointershort) \geq 0$ for every $\cellpointershort$.
\end{lemma}

\begin{proof}
    Let $p$ be the process that executed $opx$.
    Suppose, for contradiction, $R(\mathcal{I}^\mathcal{B}, opx, \cellpointershort) < 0$ for some $\cellpointershort$.    
    Hence, by \Cref{def:ero:reference_count}, $p$ performed a revocation event for $\cellpointershort$ during $opx$ in $\mathcal{I}^\mathcal{B}$, so there is an invocation step for $opx$ in $\mathcal{I}^\mathcal{B}$.
    Thus, by \Cref{observation:ero:initially_no_acquisitions} $R(\mathcal{I}^{invoke}, opx, \cellpointershort) = 0$ where $\mathcal{I}^{invoke}$ is the prefix of $\mathcal{I}^\mathcal{B}$ up to and including the invocation step of $opx$.
    So, since $R(\mathcal{I}^\mathcal{B}, opx, \cellpointershort) < 0$, it follows that there is a finite prefix $\mathcal{I}$ of $\mathcal{I}^\mathcal{B}$ where $R(\mathcal{I}, opx, \cellpointershort) < 0$ and for every proper prefix $\mathcal{I}'$ of $\mathcal{I}$ $R(\mathcal{I}', opx, \cellpointershort) \geq 0$.
    Let $\mathcal{I}^-$ be the prefix of $\mathcal{I}$ up to but excluding the last step of $\mathcal{I}$.
    Hence, $R(\mathcal{I}^-, opx, \cellpointershort) \geq 0$, and since $R(\mathcal{I}, opx, \cellpointershort) < 0$, we have that $R(\mathcal{I}^-, opx, \cellpointershort) = 0$.
    So, the last step of $\mathcal{I}$ is a revocation event for $\cellpointershort$ by $p$ during $opx$.
    Hence, by \Cref{def:ero:english}, the last step of $\mathcal{I}$ is an execution of \cref{line:ero:relinquish_revocations} during an invocation $I$ of the Relinquish procedure by $p$.
    Thus, since $\mathcal{I}^-$ is a prefix of $\mathcal{I}$ excluding the last step, we have that $p$ invoked $I$ during $\mathcal{I}^-$.
    Since $p$ has not executed \cref{line:ero:owner_relinquish} during $opx$ in $\mathcal{I}^\mathcal{B}$, and $\mathcal{I}$ is a prefix of $\mathcal{I}^\mathcal{B}$, we have that $p$ invoked $I$ during an invocation $I^+$ of one of the following procedures: \doaddcell{}, \doremovecell{}, \setrepositoryoperationresponse{}, IsDone, or Acquire.

    \begin{itemize}
        \item[] \hspace{0pt}\textbf{Case 1.} $I^+$ is an invocation of the \doaddcell{}, \doremovecell{}, or Acquire procedure.

        Hence, since the last step of $\mathcal{I}$ is a revocation event for $\cellpointershort$ by $p$ during $I$ (and thus $I^+$), by Observations \ref{lemma:ero:do_add_cell_acquire_revocation_sequence}, \ref{lemma:ero:do_remove_cell_acquire_revocation_sequence}, and \ref{lemma:ero:acquire_acquire_revocation_sequence}, 
        the number of successful list-acquire-next attempts for $\cellpointershort$ minus the number of revocation events for $\cellpointershort$ by $p$ during $I^+$ in $\mathcal{I}$ is non-negative. 
        Therefore, since $R(\mathcal{I}, opx, \cellpointershort) < 0$, it follows that $R(\mathcal{I}', opx, \cellpointershort) < 0$ for some proper prefix $\mathcal{I}'$ of $\mathcal{I}$.
        However, for every proper prefix $\mathcal{I}'$ of $\mathcal{I}$ $R(\mathcal{I}', opx, \cellpointershort) \geq 0$, a contradiction.

        \item[] \hspace{0pt}\textbf{Case 2.} $I^+$ is an invocation of the \setrepositoryoperationresponse{} or IsDone procedure.

        Hence, since the last step of $\mathcal{I}$ is a revocation event for $\cellpointershort$ by $p$ during $I$, we have that $p$ invoked $I$ on either \cref{line:ero:set_response_relinquish} or \cref{line:ero:announce_acquire_relinquish} depending on which procedure $I^+$ is.
        Thus, $p$ invoked the Acquire procedure with a second parameter of $\cellpointershort$ during $I^+$ before the end of $\mathcal{I}$; denote this invocation by $I^{acq}$.
        Furthermore, by the conditions on lines \ref{line:ero:set_response_found_check} and \ref{line:ero:announce_acquire_relinquish}, we have that the response of $I^{acq}$ is $\found$.
        So, by \Cref{lemma:ero:acquire_acquire_revocation_sequence}, there is a successful list-acquire-next attempt $a$ for $\cellpointershort$ by $p$ during $I^{acq}$ such that there is no revocation event for $\cellpointershort$ after $a$ by $p$ during $I^{acq}$.
        Therefore, since there are no revocation events by $p$ during $I^+$ between the end of $I^{acq}$ and the last step of $\mathcal{I}$, we have that from $a$ onwards in $\mathcal{I}$ there are no revocation events for $\cellpointershort$ by $p$.

        We now claim that $R(\mathcal{I}^{include}_a, opx, \cellpointershort) \geq 1$ where $\mathcal{I}^{include}_a$ is the prefix of $\mathcal{I}$ up to and including $a$.
        Suppose, for contradiction, $R(\mathcal{I}^{include}_a, opx, \cellpointershort) < 1$.
        Hence, since $\mathcal{I}^{include}_a$ is a proper prefix of $\mathcal{I}$, by the minimality of $\mathcal{I}$, $R(\mathcal{I}^{include}_a, opx, \cellpointershort) \geq 0$, and so $R(\mathcal{I}^{include}_a, opx, \cellpointershort) = 0$.
        Thus, since $a$ is a successful list-acquire-next attempt for $\cellpointershort$ by $p$ during $opx$, we have that $R( \mathcal{I}^{exclude}_a, opx, \cellpointershort) = -1$ where $\mathcal{I}^{exclude}_a$ is the prefix of $\mathcal{I}$ up to but excluding $a$.
        However, since $\mathcal{I}^{exclude}_a$ is a proper prefix of $\mathcal{I}$, by the minimality of $\mathcal{I}$, $R(\mathcal{I}^{exclude}_a, opx, \cellpointershort) \geq 0$, a contradiction.
        
        We now finish the proof of Case 2.
        Since from $a$ onwards in $\mathcal{I}$ there are no revocation events for $\cellpointershort$ by $p$ and $R(\mathcal{I}^{include}_a, opx, \cellpointershort) \geq 1$, we have that $R(\mathcal{I}^-, opx, \cellpointershort) \geq 1$.
        However, $R(\mathcal{I}^-, opx, \cellpointershort) = 0$, a contradiction.
        \qH{\Cref{lemma:ero:reference_count_is_non_negative_before_owner_relinquish}}
    \end{itemize}
\end{proof}

\Cref{lemma:ero:reference_count_is_non_negative_before_owner_relinquish} implies the first property, which is stated formally below.

\begin{corollary}\label{lemma:ero:reference_count_is_non_negative}
    Consider any operation execution $opx$ in $\mathcal{I}^\mathcal{B}$.
    The following are true.
    \begin{compactenum}
        \item If the process that executed $opx$ executed \cref{line:ero:allocate_cell} during $opx$ with response $\cellpointershort_{opx}$, then for every $\cellpointershort \neq \cellpointershort_{opx}$ $R(\mathcal{I}^\mathcal{B}, opx, \cellpointershort) \geq 0$, and $R(\mathcal{I}^\mathcal{B}, opx, \cellpointershort_{opx}) \geq -1$.
        \item Otherwise, $R(\mathcal{I}^\mathcal{B}, opx, \cellpointershort) \geq 0$ for every $\cellpointershort$.
    \end{compactenum}
\end{corollary}

We now prove the second property of this section, which is, roughly speaking, that the process that executed $opx$ only performs an operation on an object of the cell when it has the right to use it.
We start with a few observations regarding $R$ at the beginning of every iteration of the loops on lines \ref{line:ero:add_cell_while_loop}, \ref{line:ero:remove_cell_while_loop}, and \ref{line:ero:acquire_loop_until}.
By a straightforward induction, each of these observations follows.

\begin{observation}\label{observation:ero:add_cell_acquisition_loop_invariant}
    If the last step of $\mathcal{I}^\mathcal{B}$ is an execution of \cref{line:ero:add_cell_while_loop} during some invocation $I$ of the \doaddcell{} procedure during some operation execution $opx$, and the value $\cellpointershort$ of the local variable $\currentcellpointershort$ in $I$ at the end of $\mathcal{I}^\mathcal{B}$ is in $\celluniverse$, then $R(\mathcal{I}^\mathcal{B}, opx, \cellpointershort) \geq 1$.
\end{observation}

\begin{observation}\label{observation:ero:remove_cell_acquisition_loop_invariant}
    If the last step of $\mathcal{I}^\mathcal{B}$ is an execution of \cref{line:ero:remove_cell_while_loop} during some invocation $I$ of the \doremovecell{} procedure during some operation execution $opx$, and the value $\cellpointershort$ of the local variable $\currentcellpointershort$ (resp. $\previouscellpointershort$) in $I$ at the end of $\mathcal{I}^\mathcal{B}$ is in $\celluniverse$, then $R(\mathcal{I}^\mathcal{B}, opx, \cellpointershort) \geq 1$.
\end{observation}

\begin{observation}\label{observation:ero:acquire_acquisition_loop_invariant}
    If the last step of $\mathcal{I}^\mathcal{B}$ is an execution of \cref{line:ero:acquire_loop_until} during some invocation $I$ of the Acquire procedure during some operation execution $opx$, and the value $\cellpointershort$ of the local variable $\currentcellpointershort$ in $I$ at the end of $\mathcal{I}^\mathcal{B}$ is in $\celluniverse$, then $R(\mathcal{I}^\mathcal{B}, opx, \cellpointershort) \geq 1$.
\end{observation}

Before proving the second property, we need one more fact about the special case of \cref{line:ero:announce_op_response_check}.

\begin{proposition}\label{lemma:ero:announce_op_check_means_the_acquire_was_successful}
    If a process $p$ executes \cref{line:ero:announce_op_response_check} during $\mathcal{I}^\mathcal{B}$, then $p$ received response $\found$ on \cref{line:ero:announce_acquire} during the same invocation of the IsDone procedure.
\end{proposition}

\begin{proof}
    Suppose, for contradiction, some process $p$ executes \cref{line:ero:announce_op_response_check} during some invocation $I$ of the IsDone procedure, and received a response other than $\found$ on \cref{line:ero:announce_acquire} during $I$.
    Since $p$ executes \cref{line:ero:announce_op_response_check} during $I$, by the condition on \cref{line:ero:announce_acquire_l_changed_check}, we have that $p$ did not receive $\timechange$ on \cref{line:ero:announce_acquire} during $I$.
    Hence, since the Acquire procedure returns either $\found$, $\timechange$, or $\notfound$, we have that $p$ received $\notfound$ on \cref{line:ero:announce_acquire} during $I$.
    Let $\uniquerepositoryoperationshort_\announceobject$ (resp. $\cellpointershort_\announceobject$) be the second (resp. third) parameter of $I$.
    Hence, $p$ read $(\uniquerepositoryoperationshort_\announceobject, \cellpointershort_\announceobject)$ from $\announceobject$ on \cref{line:ero:announce_read} during some iteration $I^{\ref{line:ero:do_work_while_loop}}$ of the loop on \cref{line:ero:do_work_while_loop}; say at time $T^{\ref{line:ero:announce_read}}$.
    Furthermore, since $p$ executes \cref{line:ero:announce_op_response_check} during $I$, by the condition on \cref{line:ero:write_and_read_done_check}, $\uniquerepositoryoperationshort_\announceobject = (\arbitraryvalue{}, \langle \doopandcopyresponse{}, \arbitraryvalue{}\rangle)$.
    Thus, since $\announceobject$ is initially $((0, \noop), \nullconstant)$, we have that $\announceobject$ was set to $(\uniquerepositoryoperationshort_\announceobject, \cellpointershort_\announceobject)$ before $T^{\ref{line:ero:announce_read}}$, and so by \Cref{observation:ero:where_objects_change}, some $A$-event set $\announceobject = (\uniquerepositoryoperationshort_\announceobject, \cellpointershort_\announceobject)$ before $T^{\ref{line:ero:announce_read}}$.
    So, by \Cref{lemma:ero:every_a_event_is_for_pointer_from_universe}, $\cellpointershort_\announceobject \in \celluniverse$.
    Therefore, since by \Cref{lemma:ero:the_list_invariants_hold} $P(\mathcal{I}^\mathcal{B})$, $Q(\mathcal{I}^\mathcal{B})$, and $R(\mathcal{I}^\mathcal{B})$ hold and $p$ received $\notfound$ on \cref{line:ero:announce_acquire} during $I$ (or equivalently on \cref{line:ero:announce_acquire} during $I^{\ref{line:ero:do_work_while_loop}}$), by \Cref{lemma:ero:after_helping_last_l_event_the_list_of_cells_is_stuck_weak}, $\cellpointershort_\announceobject{} \notin \List(\mathcal{I})$ where $\mathcal{I}$ is the prefix of $\mathcal{I}^\mathcal{B}$ up to and including the $p$'s execution of \cref{line:ero:linearization_read} during $I^{\ref{line:ero:do_work_while_loop}}$.
    Let $e_A$ be the last $A$-event that set $\announceobject = (\uniquerepositoryoperationshort_\announceobject, \cellpointershort_\announceobject)$ before $T^{\ref{line:ero:announce_read}}$.
    Since $\uniquerepositoryoperationshort_\announceobject = (\arbitraryvalue{}, \langle \doopandcopyresponse{}, \arbitraryvalue{}\rangle)$, by \Cref{def:ero:english}, $e_A$ is an $A$-apply event.
    Furthermore, since $p$ read $(\uniquerepositoryoperationshort_\announceobject, \cellpointershort_\announceobject)$ from $\announceobject$ at $T^{\ref{line:ero:announce_read}}$, we have that $e_A$ is the last $A$-event before $T^{\ref{line:ero:announce_read}}$.
    %Since $e < T^{\ref{line:ero:announce_read}}$, and 
    By \Cref{def:ero:english}, $e_A$ was executed by some process $q$ during an invocation of the \doworkuntildone{} procedure with parameters $(\langle \doopandcopyresponse{}, \arbitraryvalue{}\rangle, \cellpointershort_\announceobject)$.
    Hence, before $e_A$, $q$ exited the \doworkuntildone{} procedure with parameters $(\addcell, \cellpointershort_\announceobject)$.
    Thus, since by \Cref{lemma:ero:the_list_invariants_hold} $P(\mathcal{I}^\mathcal{B})$ holds, by \Cref{lemma:ero:l_add_event_before_add_low_level_exits}, there is an $L$-add event $e_L$ for $\cellpointershort_\announceobject$ before $q$ exited this invocation of the \doworkuntildone{} procedure (and thus $e_A$).
    Therefore, since $e_A < T^{\ref{line:ero:announce_read}}$, by transitivity, $e_L < T^{\ref{line:ero:announce_read}}$.
    Let $T^{\ref{line:ero:linearization_read}}$ be the time $p$ executes \cref{line:ero:linearization_read} during $I^{\ref{line:ero:do_work_while_loop}}$ and let $T^{\ref{line:ero:acquire_next_linearization_changed_check}}$ be the last time $p$ executes \cref{line:ero:acquire_next_linearization_changed_check} during the Acquire procedure on \cref{line:ero:announce_acquire} during $I$ (since $p$ exits the Acquire procedure on \cref{line:ero:announce_acquire} during $I$, by \Cref{lemma:ero:exit_acquire_implies_executing_105}, $T^{\ref{line:ero:acquire_next_linearization_changed_check}}$ is well-defined), so $T^{\ref{line:ero:announce_read}} \in [T^{\ref{line:ero:linearization_read}}, T^{\ref{line:ero:acquire_next_linearization_changed_check}}]$.
    Hence, since $P(\mathcal{I}^\mathcal{B})$ holds and the Acquire procedure on \cref{line:ero:announce_acquire} during $I$ (equivalently $I^{\ref{line:ero:do_work_while_loop}}$) exits with response $\notfound$, by \Cref{lemma:ero:if_announce_acquire_does_not_return_time_change_then_no_l_events}, there are no $L$-events throughout $[T^{\ref{line:ero:linearization_read}}, T^{\ref{line:ero:acquire_next_linearization_changed_check}}]$ during $\mathcal{I}^\mathcal{B}$.
    Thus, since $e_L < T^{\ref{line:ero:announce_read}}$ and $T^{\ref{line:ero:announce_read}} \in [T^{\ref{line:ero:linearization_read}}, T^{\ref{line:ero:acquire_next_linearization_changed_check}}]$, we have that $e_L < T^{\ref{line:ero:linearization_read}}$.
    Therefore, since $\mathcal{I}$ is the prefix of $\mathcal{I}^\mathcal{B}$ up to and including $p$'s execution of \cref{line:ero:linearization_read} during $I^{\ref{line:ero:do_work_while_loop}}$, we have that $e_L$ is in $\mathcal{I}$.
    Since $e_L$ is an $L$-add event for $\cellpointershort_\announceobject$, $e_L$ is in $\mathcal{I}$, and $\cellpointershort_\announceobject \notin \List(\mathcal{I})$, by \Cref{def:ero:logical_list}, there is an $L$-remove event $e$ for $\cellpointershort_\announceobject$ in $\mathcal{I}$.
    Hence, by \Cref{lemma:ero:l_event_corresponding_a_event_type_and_pointer_relations}, there is an $A$-remove event $e'$ for $\cellpointershort_\announceobject$ before $e$.
    Since $e$ is in $\mathcal{I}$, and the time of the last step of $\mathcal{I}$ is $T^{\ref{line:ero:linearization_read}}$, we have that $e \leq T^{\ref{line:ero:linearization_read}}$, and so since $e' < e$, by transitivity, $e' < T^{\ref{line:ero:linearization_read}}$.
    Hence, since $T^{\ref{line:ero:linearization_read}} < T^{\ref{line:ero:announce_read}}$, by transitivity, $e' < T^{\ref{line:ero:announce_read}}$.
    Thus, since $e_A$ is the last $A$-event before $T^{\ref{line:ero:announce_read}}$, we have that $e' \leq e_A$, and since $e_A$ is an $A$-apply event and $e'$ is an $A$-remove event, we have that $e' \neq e_A$, and so $e' < e_A$.
    Therefore, there is an $A$-remove event for $\cellpointershort_\announceobject$ in $\mathcal{I}^\mathcal{B}$ such that after $e'$ there is an $A$-apply event for $\cellpointershort_\announceobject$.
    However, by \Cref{lemma:ero:no_apply_announce_after_remove_announce}, there are no $A$-apply events for $\cellpointershort_\announceobject$ from $e'$ onwards in $\mathcal{I}^\mathcal{B}$, a contradiction.
    \qH{\Cref{lemma:ero:announce_op_check_means_the_acquire_was_successful}}
\end{proof}

The second property is formally stated below.

\begin{lemma}\label{lemma:ero:operation_acquisition_invariant}
    If the last step of $\mathcal{I}^\mathcal{B}$ is an operation on an object of the cell pointed to by $\cellpointershort \in \celluniverse$ during some operation execution $opx$, then the following are true:
    \begin{compactenum}
        \item if the last step of $\mathcal{I}^\mathcal{B}$ is on line \ref{line:ero:copy_response_out_of_cell}, \ref{line:ero:do_work_initialize_response}, \ref{line:ero:do_work_while_loop}, or \ref{line:ero:relinquish_revocations} during an invocation of the Relinquish procedure invoked on \cref{line:ero:owner_relinquish}, then $R(\mathcal{I}^-, opx, \cellpointershort) \geq 0$;
        \item otherwise, $R(\mathcal{I}^-, opx, \cellpointershort) \geq 1$ 
    \end{compactenum}
    % \begin{compactenum}
    %     \item if the owner of $opx$ executed \cref{line:ero:allocate_cell} during $opx$ with response $\cellpointershort$, then $R(opx, \cellpointershort, \mathcal{I}^-) \geq 0$;
    %     \item otherwise, $R(opx, \cellpointershort, \mathcal{I}^-) \geq 1$ 
    % \end{compactenum}
    where $\mathcal{I}^-$ is the prefix of $\mathcal{I}^\mathcal{B}$ excluding the last step.
\end{lemma}

\begin{proof}
    Let $p$ be the process that executed $opx$.
    Observe that the last step of $\mathcal{I}^\mathcal{B}$ is either on line \ref{line:ero:copy_response_out_of_cell}, \ref{line:ero:do_work_initialize_response}, \ref{line:ero:do_work_while_loop}, \ref{line:ero:add_cell_read_end_of_list}, \ref{line:ero:add_cell_to_list}, \ref{line:ero:remove_cell_remove_seal_loop}, \ref{line:ero:remove_cell_read_pointer_to_remove_before_seal}, \ref{line:ero:seal_cell}, \ref{line:ero:remove_cell_read_pointer_to_remove}, \ref{line:ero:remove_cell_read_previous_pointer}, \ref{line:ero:remove_cell_from_list}, \ref{line:ero:copy_acquisitions_to_revocations}, \ref{line:ero:responses_set_attempt}, \ref{line:ero:announce_op_response_check}, \ref{line:ero:acquire_next_read_curr_unique_pointer}, \ref{line:ero:acquire_next_cell}, and \ref{line:ero:relinquish_revocations}.
    The proof is by cases.

    \begin{itemize}
        \item[] \hspace{0pt}\textbf{Case 1.} The last step of $\mathcal{I}^\mathcal{B}$ is either on line \ref{line:ero:copy_response_out_of_cell}, \ref{line:ero:do_work_initialize_response}, or \ref{line:ero:do_work_while_loop}.

        Hence, $p$ executed \cref{line:ero:allocate_cell} during $opx$ with response $\cellpointershort$, so we must show that $R(\mathcal{I}^-, opx, \cellpointershort) \geq 0$.
        This follows from \Cref{lemma:ero:reference_count_is_non_negative_before_owner_relinquish}.

        \item[] \hspace{0pt}\textbf{Case 2.} The last step of $\mathcal{I}^\mathcal{B}$ is either on line \ref{line:ero:add_cell_read_end_of_list} or \ref{line:ero:add_cell_to_list}.

        Hence, $p$ performed the last step of $\mathcal{I}^\mathcal{B}$ during some invocation $I$ of the \doaddcell{} procedure, and $\cellpointershort$ is the value of the local variable $\currentcellpointershort$ in $I$ at the end of $\mathcal{I}^\mathcal{B}$.
        Let $\mathcal{I}'$ be the prefix of $\mathcal{I}^\mathcal{B}$ up to and including $p$'s last execution of \cref{line:ero:add_cell_while_loop}.
        Hence, this execution is during $I$, and the value of the local variable $\currentcellpointershort$ at the end of $\mathcal{I}'$ is $\cellpointershort$.
        Thus, since $\cellpointershort \in \celluniverse$, by \Cref{observation:ero:add_cell_acquisition_loop_invariant}, $R( \mathcal{I}', opx, \currentcellpointershort) \geq 1$.
        So, since there are no successful list-acquire-next attempts by $p$ between the end of $\mathcal{I}'$ and $\mathcal{I}^\mathcal{B}$, we have that $R(\mathcal{I}^-, opx, \cellpointershort) \geq 1$ as wanted.

        \item[] \hspace{0pt}\textbf{Case 3.} The last step of $\mathcal{I}^\mathcal{B}$ is either on line \ref{line:ero:remove_cell_remove_seal_loop}, \ref{line:ero:remove_cell_read_pointer_to_remove_before_seal}, \ref{line:ero:seal_cell}, \ref{line:ero:remove_cell_read_pointer_to_remove}, or \ref{line:ero:copy_acquisitions_to_revocations}.

        Let $\mathcal{I}'$ be the prefix of $\mathcal{I}^\mathcal{B}$ up to and including $p$'s last execution of \cref{line:ero:remove_cell_while_loop} before the last step of $\mathcal{I}$.
        Since $\cellpointershort \in \celluniverse$, by \Cref{observation:ero:remove_cell_acquisition_loop_invariant}, if $\cellpointershort$ is the value of $\currentcellpointershort$ at the end of $\mathcal{I}'$, then $R(\mathcal{I}', opx, \cellpointershort) \geq 1$.
        Since this is the last execution of \cref{line:ero:remove_cell_while_loop} before the last step of $\mathcal{I}^\mathcal{B}$, and the last step of $\mathcal{I}^\mathcal{B}$ is either on line \ref{line:ero:remove_cell_remove_seal_loop}, \ref{line:ero:remove_cell_read_pointer_to_remove_before_seal}, \ref{line:ero:seal_cell}, \ref{line:ero:remove_cell_read_pointer_to_remove}, or \ref{line:ero:copy_acquisitions_to_revocations}, we have that $p$ found the condition on \cref{line:ero:remove_cell_while_loop} at the end of $\mathcal{I}'$ to be false and $\currentcellpointershort = \cellpointershort$ at the end of $\mathcal{I}'$.
        Hence, $R(\mathcal{I}', opx, \cellpointershort) \geq 1$.
        So, since there are no successful list-acquire-next attempts by $p$ between the end of $\mathcal{I}'$ and $\mathcal{I}^\mathcal{B}$, we have that $R(\mathcal{I}^-, opx, \cellpointershort) \geq 1$ as wanted.

        \item[] \hspace{0pt}\textbf{Case 4.} The last step of $\mathcal{I}^\mathcal{B}$ is either on line \ref{line:ero:remove_cell_read_previous_pointer} or \ref{line:ero:remove_cell_from_list}.

        Let $\mathcal{I}'$ be the prefix of $\mathcal{I}^\mathcal{B}$ up to and including $p$'s last execution of \cref{line:ero:remove_cell_while_loop} before the last step of $\mathcal{I}$.
        Since $\cellpointershort \in \celluniverse$, by \Cref{observation:ero:remove_cell_acquisition_loop_invariant}, if $\cellpointershort$ is the value of $\previouscellpointershort$ at the end of $\mathcal{I}'$, then $R(\mathcal{I}', opx, \cellpointershort) \geq 1$.
        Since this is the last execution of \cref{line:ero:remove_cell_while_loop} before the last step of $\mathcal{I}^\mathcal{B}$, and the last step of $\mathcal{I}^\mathcal{B}$ is either on line \ref{line:ero:remove_cell_read_previous_pointer} or \ref{line:ero:remove_cell_from_list}, we have that $p$ found the condition on \cref{line:ero:remove_cell_while_loop} at the end of $\mathcal{I}'$ to be false and $\previouscellpointershort = \cellpointershort$ at the end of $\mathcal{I}'$.
        Hence, $R(\mathcal{I}', opx, \cellpointershort) \geq 1$.
        So, since there are no successful list-acquire-next attempts by $p$ between the end of $\mathcal{I}'$ and $\mathcal{I}^\mathcal{B}$, we have that $R(\mathcal{I}^-, opx, \cellpointershort) \geq 1$ as wanted.

        \item[] \hspace{0pt}\textbf{Case 5.} The last step of $\mathcal{I}^\mathcal{B}$ is on \cref{line:ero:responses_set_attempt}.

        Let $I$ be the invocation of the \setrepositoryoperationresponse{} procedure that $p$ performed the last step of $\mathcal{I}^\mathcal{B}$ during.
        Since the last step of $\mathcal{I}$ is on \cref{line:ero:responses_set_attempt} and is an operation on an object of the cell pointed to by $\cellpointershort$, we have that the response of the Acquire procedure during $I$ is $\found$ and the second parameter of $I$ is $\cellpointershort$.
        Hence, by \Cref{lemma:ero:acquire_acquire_revocation_sequence}, $p$ performed a successful list-acquire-next attempt $a$ for $\cellpointershort$ such that $p$ did not perform a revocation event for $\cellpointershort$ after $a$ during the Acquire procedure during $I$.
        Thus, by \Cref{lemma:ero:reference_count_is_non_negative_before_owner_relinquish}, $R(\mathcal{I}_a, opx, \cellpointershort) \geq 0$ where $\mathcal{I}_a$ is the prefix of $\mathcal{I}^\mathcal{B}$ up to and including $a$, and since $a$ is a successful list-acquire-next attempt for $\cellpointershort$, it follows that $R(\mathcal{I}_a, opx, \cellpointershort) \geq 1$ (otherwise $R$ would be less than $0$ at the step before $a$).
        Therefore, since $p$ did not perform a revocation event for $\cellpointershort$ after $a$ during the Acquire procedure during $I$, and $p$ does not perform any successful list-acquire-next attempts or revocation events during $I$, other than those performed during the Acquire procedure during $I$, we have that $R(\mathcal{I}^-, opx, \cellpointershort) \geq 1$ as wanted.

        \item[] \hspace{0pt}\textbf{Case 6.} The last step of $\mathcal{I}^\mathcal{B}$ is on \cref{line:ero:announce_op_response_check}.

        Let $I$ be the invocation of the IsDone procedure that $p$ performed the last step of $\mathcal{I}^\mathcal{B}$ during.
        Since the last step of $\mathcal{I}^\mathcal{B}$ is on \cref{line:ero:announce_op_response_check} and is an operation on an object of the cell pointed to by $\cellpointershort$, by \Cref{lemma:ero:announce_op_check_means_the_acquire_was_successful}, the response of the Acquire procedure during $I$ is $\found$.
        Furthermore, the third parameter of $I$ is $\cellpointershort$.
        Hence, by \Cref{lemma:ero:acquire_acquire_revocation_sequence}, $p$ performed a successful list-acquire-next attempt $a$ for $\cellpointershort$ such that $p$ did not perform a revocation event for $\cellpointershort$ after $a$ during the Acquire procedure during $I$.
        Since $a$ is a successful list-acquire-next attempt for $\cellpointershort$, by \Cref{lemma:ero:reference_count_is_non_negative_before_owner_relinquish}, $R(\mathcal{I}_a, opx, \cellpointershort) \geq 1$ where $\mathcal{I}_a$ is the prefix of $\mathcal{I}^\mathcal{B}$ up to and including $a$.
        Therefore, since $p$ did not perform a revocation event for $\cellpointershort$ after $a$ during the Acquire procedure during $I$, and $p$ does not perform any successful list-acquire-next attempts or revocation events during $I$, other than those performed during the Acquire procedure during $I$, we have that $R(\mathcal{I}^-, opx, \cellpointershort) \geq 1$.
        
        \item[] \hspace{0pt}\textbf{Case 7.} The last step of $\mathcal{I}^\mathcal{B}$ is either on line \ref{line:ero:acquire_next_read_curr_unique_pointer} or \ref{line:ero:acquire_next_cell}.

        Let $I$ be the invocation of the AcquireNext procedure that $p$ performed the last step of $\mathcal{I}$ during.
        Since the last step of $\mathcal{I}^\mathcal{B}$ is an operation on an object of the cell pointed to by $\cellpointershort$, we have that the second parameter of $I$ is $\cellpointershort$.
        Hence, since $I$ is invoked on either line \ref{line:ero:add_cell_acquire_next}, \ref{line:ero:remove_cell_acquire_next}, or \ref{line:ero:acquire_acquire_next}, we have that $p$ executed line \ref{line:ero:add_cell_while_loop}, \ref{line:ero:remove_cell_while_loop}, or \ref{line:ero:acquire_loop_until}, respectively, immediately before invoking $I$.
        Let $\mathcal{I}'$ be the prefix of $\mathcal{I}^\mathcal{B}$ up to and including $p$'s execution of this line.
        Hence, since the second parameter of $I$ is $\cellpointershort$, by Observation \ref{observation:ero:add_cell_acquisition_loop_invariant}, \ref{observation:ero:remove_cell_acquisition_loop_invariant}, and \ref{observation:ero:acquire_acquisition_loop_invariant}, we have that $R(\mathcal{I}', opx, \cellpointershort) \geq 1$.
        Therefore, since $p$ does not perform any revocation events during $I$, we have that $R(\mathcal{I}^-, opx, \cellpointershort) \geq 1$ as wanted.
        
        \item[] \hspace{0pt}\textbf{Case 8.} The last step of $\mathcal{I}^\mathcal{B}$ is on \cref{line:ero:relinquish_revocations}.

        Hence, since the last step of $\mathcal{I}^\mathcal{B}$ is an operation on an object of the cell pointed to by $\cellpointershort$, we have that the last step of $\mathcal{I}^\mathcal{B}$ is a revocation event for $\cellpointershort$ by $p$ during $opx$.
        Let $I$ be the invocation of the Relinquish procedure that $p$ performed the last step of $\mathcal{I}^\mathcal{B}$ during.
        First suppose that $p$ did not invoke $I$ on \cref{line:ero:owner_relinquish}, so we must prove that $R(\mathcal{I}^-, opx, \cellpointershort) \geq 1$.
        Suppose, for contradiction, $R(\mathcal{I}^-, opx, \cellpointershort) < 1$.
        Since $p$ did not invoke $I$ on \cref{line:ero:owner_relinquish}, by \Cref{lemma:ero:reference_count_is_non_negative_before_owner_relinquish}, $R(\mathcal{I}^-, opx, \cellpointershort) \geq 0$.
        Hence, since $R(\mathcal{I}^-, opx, \cellpointershort) < 1$, we have that $R(\mathcal{I}^-, opx, \cellpointershort) = 0$.
        Therefore, since $\mathcal{I}^-$ is the prefix of $\mathcal{I}^\mathcal{B}$ excluding the last step, and the last step of $\mathcal{I}^\mathcal{B}$ is a revocation event for $\cellpointershort$ by $p$ during $opx$, we have that $R(\mathcal{I}^\mathcal{B}, opx, \cellpointershort) = -1$.
        However, since $p$ has not invoked the Relinquish procedure on \cref{line:ero:owner_relinquish} during $opx$ in $\mathcal{I}^\mathcal{B}$, by \Cref{lemma:ero:reference_count_is_non_negative_before_owner_relinquish}, $R(\mathcal{I}^\mathcal{B}, opx, \cellpointershort) \geq 0$, a contradiction.
        Now suppose that $p$ invoked $I$ on \cref{line:ero:owner_relinquish}, so we must prove that $R(\mathcal{I}^-, opx, \cellpointershort) \geq 0$.
        Suppose, for contradiction, $R(\mathcal{I}^-, opx, \cellpointershort) < 0$.
        By \Cref{lemma:ero:reference_count_is_non_negative} $R(\mathcal{I}^-, opx, \cellpointershort) \geq -1$, so $R(\mathcal{I}^-, opx, \cellpointershort) = -1$.
        Therefore, since $\mathcal{I}^-$ is the prefix of $\mathcal{I}^\mathcal{B}$ excluding the last step, and the last step of $\mathcal{I}^\mathcal{B}$ is a revocation event for $\cellpointershort$ by $p$ during $opx$, we have that $R(\mathcal{I}^\mathcal{B}, opx, \cellpointershort) = -2$.
        However, by \Cref{lemma:ero:reference_count_is_non_negative} $R(\mathcal{I}^\mathcal{B}, opx, \cellpointershort) \geq -1$, a contradiction.
        \qH{\Cref{lemma:ero:operation_acquisition_invariant}}
    \end{itemize}
\end{proof}

We now prove the third property of this section, which is, roughly speaking, with the exception of the pointer allocated on \cref{line:ero:allocate_cell} during $opx$, $R$ is $0$ for every pointer once $opx$ is complete.
We start by proving that individual procedures ``clean up" all the pointers that they allocated. 

\begin{proposition}\label{lemma:ero:if_no_acquisitons_before_then_no_acquisitions_after}
    Consider any operation execution $opx$ in $\mathcal{I}^\mathcal{B}$ and any invocation $I$ of the \doaddcell{}, \doremovecell{}, \doapplyandcopyresponse{}, \setrepositoryoperationresponse{}, or IsDone procedure that the process that executed $opx$ exited during $opx$.
    Let $\mathcal{I}^{invoke}$ (resp. $\mathcal{I}^{exit}$) be the prefix of $\mathcal{I}^\mathcal{B}$ up to and including the first (resp. last) step of $I$.
    If $R(\mathcal{I}^{invoke}, opx, \cellpointershort) = 0$ for every $\cellpointershort$, then $R(\mathcal{I}^{exit}, opx, \cellpointershort) = 0$ for every $\cellpointershort$.
\end{proposition}

\begin{proof}
    Let $p$ be the process that executed $opx$.
    We consider each case of $I$ separately.
    \begin{itemize}
        \item[] \hspace{0pt}\textbf{Case 1.} $I$ is an invocation of the \setrepositoryoperationresponse{} or IsDone procedure.

        Since $p$ exits $I$, we have that $p$ exited the Acquire procedure on \cref{line:ero:set_response_acquire} or \cref{line:ero:announce_acquire} during $I$ with response $\status$ (depending on $I$).
        Denote this invocation of the Acquire procedure by $I^{acq}$.
        If $\status \neq \found$, then by \Cref{lemma:ero:acquire_acquire_revocation_sequence}, the number of successful list-acquire-next attempts for $\cellpointershort$ by $p$ during $I^{acq}$ is equal to the number of revocation events for $\cellpointershort$ by $p$ for every $\cellpointershort$ during $I^{acq}$.
        Furthermore, since $\status \neq \found$, observe that $p$ does not perform any successful list-acquire-next attempts or revocation events for any pointer during $I$, other than those performed during $I^{acq}$.
        Therefore, since $R(\mathcal{I}^{invoke}, opx, \cellpointershort) = 0$ for every $\cellpointershort$, we have that for $R(\mathcal{I}^{exit}, opx, \cellpointershort) = 0$ for every $\cellpointershort$ as wanted.
        Now suppose $\status = \found$ and let $\targetcellpointershort$ be the second parameter of $I^{acq}$.        
        By \Cref{lemma:ero:acquire_acquire_revocation_sequence}, for every $\cellpointershort \neq \targetcellpointershort$ the number of successful list-acquire-next attempts for $\cellpointershort$ by $p$ during $I^{acq}$ is equal to the number of revocation events for $\cellpointershort$ by $p$ during $I^{acq}$, and the number of successful list-acquire-next attempts for $\targetcellpointershort$ is one greater than the number of revocation events for $\targetcellpointershort$ by $p$ during $I^{acq}$.
        Since there is a successful list-acquire-next attempt for $\targetcellpointershort$, by \Cref{lemma:ero:list_acquire_next_attempt_for_pointer_from_universe_and_after_pointer_from_universe_or_head}, $\targetcellpointershort \in \celluniverse$.
        Hence, since $\status = \found$, $p$ does not perform any successful list-acquire-next attempts for any pointer during $I$, other than those performed during $I^{acq}$, and $p$ only performs a single revocation event for $\targetcellpointershort$ during $I$, other than those performed during $I^{acq}$.
        Therefore, since $R(\mathcal{I}^{invoke}, opx, \cellpointershort) = 0$ for every $\cellpointershort$, we have that $R(\mathcal{I}^{exit}, opx, \cellpointershort) = 0$ for every $\cellpointershort$ as wanted.

        \item[] \hspace{0pt}\textbf{Case 2.} $I$ is an invocation of the \doapplyandcopyresponse{} procedure.

        Observe that $p$ does not perform any successful list-acquire-next attempts or revocation events other than those performed during the \setrepositoryoperationresponse{} procedure on \cref{line:ero:apply_set_response} during $I$.
        Therefore, this case reduces to Case 1.

        \item[] \hspace{0pt}\textbf{Case 3.} $I$ is an invocation of the \doremovecell{} procedure.

        Since $R(\mathcal{I}^{invoke}, opx, \cellpointershort) = 0$ for every $\cellpointershort$, by Case 1, $R(\mathcal{I}, opx, \cellpointershort) = 0$ for every $\cellpointershort$ where $\mathcal{I}$ is the prefix of $\mathcal{I}^\mathcal{B}$ up to and including the last step by $p$ of the \setrepositoryoperationresponse{} procedure on \cref{line:ero:remove_cell_set_response} during $I$.
        Then, by \Cref{lemma:ero:do_remove_cell_acquire_revocation_sequence}, the number of successful list-acquire-next attempts for $\cellpointershort$ by $p$ during $I$ is equal to the number of revocation events for $\cellpointershort$ by $p$ during $I$ for every $\cellpointershort$.
        Therefore, $R(\mathcal{I}^{exit}, opx, \cellpointershort) = 0$ for every $\cellpointershort$ as wanted.

        \item[] \hspace{0pt}\textbf{Case 4.} $I$ is an invocation of the \doaddcell{} procedure.

        % Let $I'$ be the invocation of the \setrepositoryoperationresponse{} procedure on \cref{line:ero:add_cell_set_response} by $p$ during $I$.
        Since $R(\mathcal{I}^{invoke}, opx, \cellpointershort) = 0$ for every $\cellpointershort$, and by \Cref{lemma:ero:do_add_cell_acquire_revocation_sequence} the number of successful list-acquire-next attempts for $\cellpointershort$ by $p$ during $I$ is equal to the number of revocation events for $\cellpointershort$ by $p$ during $I$ for every $\cellpointershort$, other than those performed during the \setrepositoryoperationresponse{} procedure on \cref{line:ero:add_cell_set_response} during $I$, we have that $R(\mathcal{I}, opx, \cellpointershort) = 0$ for every $\cellpointershort$ where $\mathcal{I}$ is the prefix of $\mathcal{I}^\mathcal{B}$ up to and including the first step by $p$ of the \setrepositoryoperationresponse{} procedure on \cref{line:ero:add_cell_set_response} during $I$.
        Therefore, by Case 1, we have that $R(\mathcal{I}^{exit}, opx, \cellpointershort) = 0$ for every $\cellpointershort$ as wanted.
        \qH{\Cref{lemma:ero:if_no_acquisitons_before_then_no_acquisitions_after}}
    \end{itemize}
\end{proof}

Since the \doworkuntildone{} procedure only invokes the \doaddcell{}, \doapplyandcopyresponse{}, \doremovecell{}, and IsDone procedures, \Cref{lemma:ero:if_no_acquisitons_before_then_no_acquisitions_after} implies the following corollary.

\begin{corollary}\label{lemma:ero:if_no_acquisitions_before_then_no_acquisitions_after_do_low_level_op}
    Consider any operation execution $opx$ in $\mathcal{I}^\mathcal{B}$ and any invocation $I$ of the \doworkuntildone{} procedure that the process that executed $opx$ exited during $opx$.
    Let $\mathcal{I}^{invoke}$ (resp. $\mathcal{I}^{exit}$) be the prefix of $\mathcal{I}^\mathcal{B}$ up to and including the first (resp. last) step of $I$.
    If $R(\mathcal{I}^{invoke}, opx, \cellpointershort) = 0$ for every $\cellpointershort$, then $R(\mathcal{I}^{exit}, opx, \cellpointershort) = 0$ for every $\cellpointershort$.
\end{corollary}

Hence, since by \Cref{observation:ero:initially_no_acquisitions} every operation execution $opx$ starts with no successful list-acquire-next attempts and no revocations events, after the third \doworkuntildone{} procedure exits on \cref{line:ero:low_level_remove_cell}, there is a single revocation event for the pointer returned on \cref{line:ero:allocate_cell} during $opx$, and after $opx$ completes there are no successful list-acquire-next attempts or revocation events during $opx$, it follows that \Cref{lemma:ero:if_no_acquisitions_before_then_no_acquisitions_after_do_low_level_op} implies the third property which is formally stated below.

\begin{corollary}\label{lemma:ero:reference_count_of_complete_operation_execution}
    Consider any complete operation execution $opx$ in $\mathcal{I}^\mathcal{B}$.
    Hence, the process that executed $opx$ executed \cref{line:ero:allocate_cell} during $opx$; say $\cellpointershort_{opx}$ is its response.
    Then, $R(\mathcal{I}^\mathcal{B}, opx, \cellpointershort) = 0$ for every $\cellpointershort \neq \cellpointershort_{opx}$ and $R(\mathcal{I}^\mathcal{B}, opx, \cellpointershort_{opx}) = -1$.
\end{corollary}

We now prove the fourth and final property of this section, which is, roughly speaking, that the process that executed $opx$ has ``acquired" the right to use at most three cells at all times.

\begin{proposition}\label{lemma:ero:add_remove_acquire_no_acqusitions_at_invocation}
    Consider any operation execution $opx$ in $\mathcal{I}^\mathcal{B}$ and any invocation $I$ of the \doaddcell{}, \doremovecell{}, Acquire, and \setrepositoryoperationresponse{} procedure that the process that executed $opx$ invoked during $opx$.
    Let $\mathcal{I}^{invoke}$ be the prefix of $\mathcal{I}^\mathcal{B}$ up to and including the first step of $I$.
    Then, $R(\mathcal{I}^{invoke}, opx, \cellpointershort) = 0$ for every $\cellpointershort$.
\end{proposition}

\begin{proof}
    Let $p$ be the process that executed $opx$.
    By \Cref{observation:ero:initially_no_acquisitions} and \Cref{lemma:ero:if_no_acquisitions_before_then_no_acquisitions_after_do_low_level_op} we have the following.
    Consider any invocation $I$ of the \doworkuntildone{} procedure that $p$ invoked during $opx$.
    Let $\mathcal{I}^{invoke}$ be the prefix of $\mathcal{I}^\mathcal{B}$ up to and including the first step of $I$.
    Then, $R(\mathcal{I}^{invoke}, opx, \cellpointershort) = 0$ for every $\cellpointershort$.
    Hence, since the \doworkuntildone{} procedure only invokes the \doaddcell{}, \doapplyandcopyresponse{}, \doremovecell{}, and IsDone procedures, and the \linebreak \doworkuntildone{} does not perform any successful list-acquire-next attempts or revocation events except for those performed within these procedures, \Cref{lemma:ero:if_no_acquisitons_before_then_no_acquisitions_after} implies the following.
    Consider any invocation $I$ of the \doaddcell{}, \doapplyandcopyresponse{}, \doremovecell{}, and IsDone procedure that $p$ invoked during $opx$.
    Let $\mathcal{I}^{invoke}$ be the prefix of $\mathcal{I}^\mathcal{B}$ up to and including the first step of $I$.
    Then, $R(\mathcal{I}^{invoke}, opx, \cellpointershort) = 0$ for every $\cellpointershort$ (*).
    So, what remains is to justify the claim for the Acquire and \setrepositoryoperationresponse{} procedures.
    Since the Acquire procedure is only invoked on lines \ref{line:ero:set_response_acquire} and \ref{line:ero:announce_acquire}, and we know that the claim holds at the start of every invocation of the IsDone procedure, it suffices to prove that the claim for the \setrepositoryoperationresponse{} procedure.
    Consider any invocation $I$ of the \setrepositoryoperationresponse{} procedure by $p$ during $opx$ and let $\mathcal{I}^{invoke}$ be the prefix of $\mathcal{I}^\mathcal{B}$ up to and including the first step of $I$.
    Observe that $I$ was invoked on either line \ref{line:ero:add_cell_set_response}, \ref{line:ero:remove_cell_set_response}, or \ref{line:ero:apply_set_response}.
    In the first case, let $I'$ be the invocation of the \doaddcell{} procedure that $I$ was invoked during.
    By (*) $R(\mathcal{I}', opx, \cellpointershort) = 0$ for every $\cellpointershort$ where $\mathcal{I}'$ is the prefix of $\mathcal{I}^\mathcal{B}$ up to and including the first step of $I'$.
    Furthermore, since by \Cref{lemma:ero:do_add_cell_acquire_revocation_sequence} the number of successful list-acquire-next attempts and revocation events for every pointer is equal before invoking \cref{line:ero:add_cell_set_response} during $I'$, we have that $R(\mathcal{I}^{invoke}, opx, \cellpointershort) = 0$ for every $\cellpointershort$ as wanted.
    In the second case, let $I'$ be the invocation of the \doremovecell{} procedure that $I$ was invoked during.
    By (*) $R(\mathcal{I}', opx, \cellpointershort) = 0$ for every $\cellpointershort$ where $\mathcal{I}'$ is the prefix of $\mathcal{I}^\mathcal{B}$ up to and including the first step of $I'$.
    This immediately implies $R(\mathcal{I}^{invoke}, opx, \cellpointershort) = 0$ for every $\cellpointershort$ because $\mathcal{I}^{invoke}$ is a single step after $\mathcal{I}'$.
    In the third and final case, let $I'$ be the invocation of the \doapplyandcopyresponse{} procedure that $I$ was invoked during.
    By (*) $R(\mathcal{I}', opx, \cellpointershort) = 0$ for every $\cellpointershort$ where $\mathcal{I}'$ is the prefix of $\mathcal{I}^\mathcal{B}$ up to and including the first step of $I'$.
    This immediately implies $R(\mathcal{I}^{invoke}, opx, \cellpointershort) = 0$ for every $\cellpointershort$ because $p$ does not perform any successful list-acquire-next attempts or revocation events during $I'$ other than those during $I$.
    \qH{\Cref{lemma:ero:add_remove_acquire_no_acqusitions_at_invocation}}
\end{proof}

The fourth property is formalized below.

\begin{lemma}\label{lemma:ero:total_reference_count_is_bounded}
    For every operation execution $opx$ in $\mathcal{I}^\mathcal{B}$
    \begin{align*}
        \sum_{\cellpointershort} R(\mathcal{I}^\mathcal{B}, opx, \cellpointershort) \leq 3.
    \end{align*}
\end{lemma}

\begin{proof}
    Let $p$ be the process that executed $opx$.
    Suppose, for contradiction, \mbox{$\sum_{\cellpointershort} R(\mathcal{I}^\mathcal{B}, opx, \cellpointershort) > 3$.}
    Hence, by \Cref{def:ero:reference_count}, $p$ performed a successful list-acquire-next attempt for $\cellpointershort$ during $opx$ in $\mathcal{I}^\mathcal{B}$, so there is an invocation step for $opx$ in $\mathcal{I}^\mathcal{B}$.
    Thus, by \Cref{observation:ero:initially_no_acquisitions} $R(\mathcal{I}^{invoke}, opx, \cellpointershort) = 0$ where $\mathcal{I}^{invoke}$ is the prefix of $\mathcal{I}^\mathcal{B}$ up to and including the invocation step of $opx$.
    So, since $R(\mathcal{I}^\mathcal{B}, opx, \cellpointershort) > 3$, it follows that there is a finite prefix $\mathcal{I}$ of $\mathcal{I}^\mathcal{B}$ where $R(\mathcal{I}, opx, \cellpointershort) > 3$ and for every proper prefix $\mathcal{I}'$ of $\mathcal{I}$ $R(\mathcal{I}', opx, \cellpointershort) \leq 3$.
    Hence, the last step of $\mathcal{I}$ is a successful list-acquire-next attempt by $p$.
    Observe that $p$ performed this step during an invocation of the AcquireNext procedure, which was performed during an invocation $I$ of either the Acquire, \doremovecell{}, or \doaddcell{} procedure.
    By \Cref{lemma:ero:add_remove_acquire_no_acqusitions_at_invocation}, $R(\mathcal{I}^{invoke}, opx, \cellpointershort) = 0$ for every $\cellpointershort$ where $\mathcal{I}^{invoke}$ is the prefix of $\mathcal{I}^\mathcal{B}$ up to and including the first step of $I$.
    Therefore, since the last step during $\mathcal{I}$ is during an invocation of the AcquireNext procedure invoked during $I$ and $\sum_{\cellpointershort} R(\mathcal{I}, opx, \cellpointershort) > 3$, we have that $p$ performed more than three successful list-acquire-next attempts during $I$ such that between then and the end of $\mathcal{I}$, $p$ did not perform revocation events for any of these pointers.
    However, since $I$ is an invocation of either the Acquire, \doremovecell{}, or \doaddcell{} procedure, this is impossible by \Cref{lemma:ero:acquire_acquire_revocation_sequence}, \ref{lemma:ero:do_remove_cell_acquire_revocation_sequence}, and \ref{lemma:ero:do_add_cell_acquire_revocation_sequence}, respectively, a contradiction.
    \qH{\Cref{lemma:ero:total_reference_count_is_bounded}}
\end{proof}

\subsubsection{Acquire-copy events copy the total number of acquisitions}

The main ingredient for proving the third bullet of the $\mathcal{B}$ correctly manages cells theorem, and the $\mathcal{B}$ is space-efficient theorem, is the following lemma.

\begin{lemmablank}
    For any acquire-copy event $e$ for $\cellpointershort$ in $\mathcal{I}^\mathcal{B}$ the following are true:
    \begin{compactenum}
        \item $e = \FAop{}((*\cellpointershort).\revocations, -(A(\mathcal{I}^\mathcal{B}, \cellpointershort) + 1))$; and
        \item if $\mathcal{I}$ is the prefix of $\mathcal{I}^\mathcal{B}$ up to and including $e$ then $A(\mathcal{I}, \cellpointershort) = A(\mathcal{I}^\mathcal{B}, \cellpointershort)$.
    \end{compactenum}
\end{lemmablank}

Since by \Cref{lemma:ero:at_most_one_acquisition_copy} there is at most one acquire-copy event for $\cellpointershort$, (1) of this lemma states that the process that performs an acquire-copy event $e$ for $\cellpointershort$ ``knows" the total number of successful list-acquire-next attempts that will ever happen for $\cellpointershort$, and (2) states that all of these successful list-acquire-next attempts have already happened by the time of $e$.
To get a sense of why this lemma is useful, we note that, roughly speaking, in conjunction with the facts proved in the last section, this lemma allows us to conclude: (1) for all times at and after a $\freecelloperation{}(\cellpointershort)$ operation, no process has the right to access $\cellpointershort$; and (2) every pointer with an acquire-copy event gets freed by some time except at most some number linear in the point contention at that time.
Since by \Cref{lemma:ero:operation_acquisition_invariant} processes only perform operations on an object of the cell pointed to by $\cellpointershort$ if it has the right to access it, (1) implies that all operations on an object of the cell pointed to $\cellpointershort$ are before any $\freecelloperation{}(\cellpointershort)$ operation.
This implies the third bullet of the $\mathcal{B}$ correctly manages cells theorem because the other requirement is proved by a simple tracing argument.
Furthermore, (2) is used to complete the $\mathcal{B}$ is space-efficient theorem.

As we will see, proving this lemma is almost entirely accomplished by proving that the acquisition counter of the cell pointed to by $\cellpointershort$ is semantically correct, i.e., it is equal to the number of successful-list-acquire-next attempts for $\cellpointershort$ plus one (\Cref{lemma:ero:acquisition_count_is_semantically_correct}).
Since the acquisition counter of the cell pointed to by $\cellpointershort$ is stored in the cell preceding it in the list, we have to deal with the fact that the acquisition counter for $\cellpointershort$ is only defined at some times (in particular, times when $\cellpointershort$ is in the list).
We deal with this issue by defining the notion of an \emph{active} pointer at a particular time, and prove that if $\cellpointershort$ is active at some time, then $\cellpointershort$ is in the list at that time.

\begin{definition}\label{def:ero:active}
    We say $\cellpointershort$ is active in $\mathcal{I}^\mathcal{B}$ if and only if there is a single successful list-add attempt for $\cellpointershort$ in $\mathcal{I}^\mathcal{B}$ and no successful list-remove attempt for $\cellpointershort$ in $\mathcal{I}^\mathcal{B}$.
\end{definition}

Hence, by \Cref{lemma:ero:every_list_add_seal_and_remove_attempt_is_for_ptr_from_universe}, we have the following.

\begin{corollary}\label{lemma:ero:active_implies_from_universe}
    If $\cellpointershort$ is active in $\mathcal{I}^\mathcal{B}$, then $\cellpointershort \in \celluniverse$.
\end{corollary}

We now prove that if $\cellpointershort$ is active at some time, then $\cellpointershort$ is in the list at that time.

\begin{lemma}\label{lemma:ero:l_add_event_in_active_prefix}
    Suppose $\cellpointershort$ is active in $\mathcal{I}^\mathcal{B}$.
    Let $a$ be the successful list-add attempt for $\cellpointershort$ in $\mathcal{I}^\mathcal{B}$ which exists by \Cref{def:ero:active}.
    Then, there is an $L$-add event for $\cellpointershort$ before $a$ in $\mathcal{I}^\mathcal{B}$.
\end{lemma}

\begin{proof}
    Let $e$ be the corresponding $L$-event of $a$, so $e < a$.
    Hence, $e$ is in $\mathcal{I}^\mathcal{B}$.
    Furthermore, since $a$ is a list-add attempt for $\cellpointershort$, by \Cref{lemma:ero:l_event_corresponding_to_do_low_level_op}, $e$ is an $L$-add event for $\cellpointershort$.
    %Therefore, there is an $L$-add event for $\cellpointershort$ before $a$ in $\mathcal{I}$ as required.
    \qH{\Cref{lemma:ero:l_add_event_in_active_prefix}}
\end{proof}

This implies the following.

\begin{corollary}\label{corollary:ero:active_prefix_has_last_l_event}
    If $\mathcal{I}^\mathcal{B}$ is finite and $\cellpointershort$ is active in $\mathcal{I}^\mathcal{B}$, then there is a last $L$-event in $\mathcal{I}^\mathcal{B}$.
    % Consider any $\cellpointershort \in \celluniverse$ and any finite prefix $\mathcal{I}$ of $\mathcal{I}^\mathcal{B}$ such that $\cellpointershort$ is active in $\mathcal{I}$.
    % There is a last $L$-event in $\mathcal{I}$.
\end{corollary}

\begin{proposition}\label{lemma:ero:if_from_last_l_event_onwards_no_list_change_then_l_add_event_in_i}
    Suppose $\mathcal{I}^\mathcal{B}$ is finite and $\cellpointershort$ is active in $\mathcal{I}^\mathcal{B}$.
    Let $e_{last}$ be the last $L$-event in $\mathcal{I}^\mathcal{B}$ which is well-defined by \Cref{corollary:ero:active_prefix_has_last_l_event}.
    If from $e_{last}$ onwards in $\mathcal{I}^\mathcal{B}$ there are no successful list-add or list-remove attempts, then there is an $L$-add event for $\cellpointershort$ in $\mathcal{I}^{exclude}_{e_{last}}$ where $\mathcal{I}^{exclude}_{e_{last}}$ is the prefix of $\mathcal{I}^\mathcal{B}$ up to but excluding $e_{last}$.
\end{proposition}

\begin{proof}
    Suppose, for contradiction, there are no $L$-add events for $\cellpointershort$ in $\mathcal{I}^{exclude}_{e_{last}}$.
    Since $\cellpointershort$ is active in $\mathcal{I}^\mathcal{B}$, by \Cref{def:ero:active}, there is a successful list-add attempt $a$ for $\cellpointershort$ in $\mathcal{I}^\mathcal{B}$.
    By \Cref{lemma:ero:l_add_event_in_active_prefix}, there is an $L$-add event $e_{add}$ for $\cellpointershort$ before $a$ in $\mathcal{I}^\mathcal{B}$.
    There are two cases.
    \begin{itemize}
        \item[] \hspace{0pt}\textbf{Case 1.} $e_{add} \neq e_{last}$.
    
        Hence, since $e_{add}$ is an $L$-event in $\mathcal{I}^\mathcal{B}$ and $e_{last}$ is the last $L$-event in $\mathcal{I}^\mathcal{B}$, we have that $e_{add} \leq e_{last}$, and so $e_{add} < e_{last}$.
        Thus, since $\mathcal{I}^{exclude}_{e_{last}}$ is the the prefix of $\mathcal{I}^\mathcal{B}$ up to but excluding $e_{last}$, we have that $e_{add}$ is in $\mathcal{I}^{exclude}_{e_{last}}$.
        Therefore, there is an $L$-add event for $\cellpointershort$ in $\mathcal{I}^{exclude}_{e_{last}}$.
        However, by assumption there are no $L$-add events for $\cellpointershort$ in $\mathcal{I}^{exclude}_{e_{last}}$, a contradiction.
    
        \item[] \hspace{0pt}\textbf{Case 2.} $e_{last} = e_{add}$.
    
        Hence, since by assumption from $e_{last}$ onwards in $\mathcal{I}^\mathcal{B}$ there are no successful list-add or list-remove attempts, we have that from $e_{add}$ onwards in $\mathcal{I}^\mathcal{B}$ there are no successful list-add or list-remove attempts.
        However, since $e_{add} < a$, $a$ is in $\mathcal{I}^\mathcal{B}$, and $a$ is a successful list-add attempt, we have that from $e_{add}$ onwards in $\mathcal{I}^\mathcal{B}$ there is a successful list-add attempt, a contradiction.
        \qH{\Cref{lemma:ero:if_from_last_l_event_onwards_no_list_change_then_l_add_event_in_i}}
    \end{itemize}
\end{proof}

\begin{proposition}\label{lemma:ero:if_from_last_l_event_onwards_no_list_change_then_ptr_in_list}
    Suppose $\mathcal{I}^\mathcal{B}$ is finite and $\cellpointershort$ is active in $\mathcal{I}^\mathcal{B}$.
    Let $e_{last}$ be the last $L$-event in $\mathcal{I}^\mathcal{B}$ which is well-defined by \Cref{corollary:ero:active_prefix_has_last_l_event}.
    If from $e_{last}$ onwards in $\mathcal{I}^\mathcal{B}$ there are no successful list-add or list-remove attempts, then $\cellpointershort$ is in $\List(\mathcal{I}^{exclude}_{e_{last}})$ exactly once where $\mathcal{I}^{exclude}_{e_{last}}$ is the prefix of $\mathcal{I}^\mathcal{B}$ up to but excluding $e_{last}$.
\end{proposition}

\begin{proof}
    It suffices to prove that $\cellpointershort$ is in $\List(\mathcal{I}^{exclude}_{e_{last}})$ at least once because by \Cref{lemma:ero:the_list_invariants_hold} $P(\mathcal{I}^{exclude}_{e_{last}})$ holds, and so by \Cref{lemma:ero:pointers_in_list_are_unique} the elements of $\List(\mathcal{I}^{exclude}_{e_{last}})$ are pairwise distinct.
    Suppose, for contradiction, $\cellpointershort \notin \List(\mathcal{I}^{exclude}_{e_{last}})$.
    Hence, since by \Cref{lemma:ero:if_from_last_l_event_onwards_no_list_change_then_l_add_event_in_i} there is an $L$-add event $e_{add}$ for $\cellpointershort$ in $\mathcal{I}^{exclude}_{e_{last}}$, by \Cref{def:ero:logical_list}, there is a subsequent $L$-remove event $e_{remove}$ for $\cellpointershort$ after $e_{add}$ in $\mathcal{I}^{exclude}_{e_{last}}$.
    Since $e_{remove}$ is in $\mathcal{I}^{exclude}_{e_{last}}$ and $\mathcal{I}^{exclude}_{e_{last}}$ is a prefix of $\mathcal{I}^\mathcal{B}$, we have that $e_{remove}$ is in $\mathcal{I}^\mathcal{B}$.
    Furthermore, since $\mathcal{I}^{exclude}_{e_{last}}$ is the prefix of $\mathcal{I}^\mathcal{B}$ up to but excluding $e_{last}$, we have that there is a next $L$-event after $e_{remove}$ in $\mathcal{I}^\mathcal{B}$; say $e$.
    Therefore, since $e_{remove}$ is an $L$-remove event for $\cellpointershort$ in $\mathcal{I}^\mathcal{B}$ and by \Cref{lemma:ero:the_list_invariants_hold} $R(\mathcal{I}^\mathcal{B})$ holds, we have that there is a successful list-remove attempt for $\cellpointershort$ in $\mathcal{I}^\mathcal{B}$.
    However, since $\cellpointershort$ is active in $\mathcal{I}^\mathcal{B}$, by \Cref{def:ero:active}, there are no successful list-remove attempts for $\cellpointershort$ in $\mathcal{I}^\mathcal{B}$, a contradiction.
    \qH{\Cref{lemma:ero:if_from_last_l_event_onwards_no_list_change_then_ptr_in_list}}
\end{proof}

\begin{proposition}\label{lemma:ero:if_from_last_l_event_onwards_list_change_then_ptr_in_list}
    Suppose $\mathcal{I}^\mathcal{B}$ is finite and $\cellpointershort$ is active in $\mathcal{I}^\mathcal{B}$.
    Let $e_{last}$ be the last $L$-event in $\mathcal{I}^\mathcal{B}$ which is well-defined by \Cref{corollary:ero:active_prefix_has_last_l_event}.
    If $e_{last}$ is not an $L$-add nor an $L$-remove event or from $e_{last}$ onwards in $\mathcal{I}^\mathcal{B}$ there is a successful list-add or list-remove attempt, then $\cellpointershort$ is in $\List(\mathcal{I}^\mathcal{B})$ exactly once.
\end{proposition}

\begin{proof}
    It suffices to prove that $\cellpointershort$ is in $\List(\mathcal{I}^\mathcal{B})$ at least once because by \Cref{lemma:ero:the_list_invariants_hold} $P(\mathcal{I}^\mathcal{B})$ holds, and so by \Cref{lemma:ero:pointers_in_list_are_unique} the elements of $\List(\mathcal{I}^\mathcal{B})$ are pairwise distinct.
    Suppose, for contradiction, $\cellpointershort \notin \List(\mathcal{I}^\mathcal{B})$.
    Hence, since by \Cref{lemma:ero:l_add_event_in_active_prefix} there is an $L$-add event $e_{add}$ for $\cellpointershort$ in $\mathcal{I}^\mathcal{B}$, by \Cref{def:ero:logical_list}, there is a subsequent $L$-remove event $e_{remove}$ for $\cellpointershort$ after $e_{add}$ in $\mathcal{I}^\mathcal{B}$.
    There are two cases.
    \begin{itemize}
        \item[] \hspace{0pt}\textbf{Case 1.} $e_{remove} \neq e_{last}$.

        Hence, since $e_{remove}$ is an $L$-event in $\mathcal{I}^\mathcal{B}$ and $e_{last}$ is the last $L$-event in $\mathcal{I}^\mathcal{B}$, we have that $e_{remove} \leq e_{last}$, and so $e_{remove} < e_{last}$.
        Thus, there is a next $L$-event after $e_{remove}$ in $e_{last}$; say $e$.
        Therefore, since $e_{remove}$ is an $L$-remove event for $\cellpointershort$ in $\mathcal{I}^\mathcal{B}$ and by \Cref{lemma:ero:the_list_invariants_hold} $R(\mathcal{I}^\mathcal{B})$ holds, we have that there is a successful list-remove attempt for $\cellpointershort$ in $\mathcal{I}^\mathcal{B}$.
        However, since $\cellpointershort$ is active in $\mathcal{I}^\mathcal{B}$, by \Cref{def:ero:active}, there are no successful list-remove attempts for $\cellpointershort$ in $\mathcal{I}^\mathcal{B}$, a contradiction.

        \item[] \hspace{0pt}\textbf{Case 2.} $e_{remove} = e_{last}$.

        Hence, $e_{last}$ is an $L$-remove event, and so from $e_{last}$ onwards in $\mathcal{I}^\mathcal{B}$ there is a successful list-add or list-remove attempt.
        Thus, since $e_{remove}$ is the last $L$-event in $\mathcal{I}^\mathcal{B}$, $e_{remove}$ is an $L$-remove event for $\cellpointershort$, and by \Cref{lemma:ero:the_list_invariants_hold} $P(\mathcal{I}^\mathcal{B})$, $Q(\mathcal{I}^\mathcal{B})$, and $R(\mathcal{I}^\mathcal{B})$ hold, by \Cref{lemma:ero:3_of_r_safety_holds}, from $e_{remove}$ onwards in $\mathcal{I}^\mathcal{B}$ there is at most one successful list-remove attempt for $\cellpointershort$ and no other successful list-remove or list-add attempt for any pointer.
        Therefore, since from $e_{remove}$ onwards in $\mathcal{I}^\mathcal{B}$ there is a successful list-add or list-remove attempt, we have that from $e_{remove}$ onwards in $\mathcal{I}^\mathcal{B}$ there is a successful list-remove attempt for $\cellpointershort$.
        However, since $\cellpointershort$ is active in $\mathcal{I}^\mathcal{B}$, by \Cref{def:ero:active}, there are no successful list-remove attempts for $\cellpointershort$ in $\mathcal{I}^\mathcal{B}$, a contradiction.
        \qH{\Cref{lemma:ero:if_from_last_l_event_onwards_list_change_then_ptr_in_list}}
    \end{itemize}
\end{proof}

\Cref{lemma:ero:if_from_last_l_event_onwards_no_list_change_then_ptr_in_list} and \Cref{lemma:ero:if_from_last_l_event_onwards_list_change_then_ptr_in_list} cover the two possible cases of what the list may conform to (see \Cref{lemma:ero:conditional_classification_lemma}), so we have our desired conclusion: if $\cellpointershort$ is active at some time, then $\cellpointershort$ is in the list at that time.
This allows us to identify the cell that stores the acquisition counter for the cell pointed to by $\cellpointershort$: the cell before the cell pointed to by $\cellpointershort$ in the list. 

\begin{definition}\label{def:ero:acquisition_pointer}
    Suppose $\mathcal{I}^\mathcal{B}$ is finite and $\cellpointershort$ is active in $\mathcal{I}^\mathcal{B}$.
    We define $prev(\mathcal{I}^\mathcal{B}, \cellpointershort)$ as follows.
    Let $e_{last}$ be the last $L$-event in $\mathcal{I}^\mathcal{B}$ which is well-defined by \Cref{corollary:ero:active_prefix_has_last_l_event} and let $\mathcal{I}^{exclude}_{e_{last}}$ be the prefix of $\mathcal{I}^\mathcal{B}$ up to but excluding $e_{last}$.
    Then,
    \begin{compactitem}
        \item if $e_{last}$ is an $L$-add or $L$-remove event and from $e_{last}$ onwards in $\mathcal{I}^\mathcal{B}$ there are no successful list-add or list-remove attempts, then by \Cref{lemma:ero:if_from_last_l_event_onwards_no_list_change_then_ptr_in_list}, $\cellpointershort$ is in $\List(\mathcal{I}^{exclude}_{e_{last}})$ exactly once, and we define $prev(\mathcal{I}^\mathcal{B}, \cellpointershort)$ as the pointer immediately before $\cellpointershort$ in $\List(\mathcal{I}^{exclude}_{e_{last}})$; and
        \item otherwise, by \Cref{lemma:ero:if_from_last_l_event_onwards_list_change_then_ptr_in_list}, $\cellpointershort$ is in $\List(\mathcal{I}^\mathcal{B})$ exactly once, and we define $prev(\mathcal{I}^\mathcal{B}, \cellpointershort)$ as the pointer immediately before $\cellpointershort$ in $\List(\mathcal{I}^\mathcal{B})$.
    \end{compactitem}
    $prev(\mathcal{I}^\mathcal{B}, \cellpointershort)$ is well-defined in both cases since $\cellpointershort \neq \&\headobject{}$ (because by \Cref{lemma:ero:active_implies_from_universe} $\cellpointershort \in \celluniverse$ and \Cref{assumption:ero:head_and_null_not_in_cell_universe}) and by \Cref{def:ero:logical_list} $\&\headobject{}$ is the first element of $\List(\mathcal{I}^{exclude}_{e_{last}})$ and $\List(\mathcal{I}^\mathcal{B})$.
\end{definition}

Before continuing, we record a simple fact about $prev(\mathcal{I}^\mathcal{B}, \cellpointershort)$.

\begin{proposition}\label{lemma:ero:acquisition_pointer_in_universe_or_head}
    If $\mathcal{I}^\mathcal{B}$ is finite and $\cellpointershort$ is active in $\mathcal{I}^\mathcal{B}$, then $prev(\mathcal{I}^\mathcal{B}, \cellpointershort) \in \celluniverse \cup \{\&\headobject{}\}$.
\end{proposition}

\begin{proof}
    By \Cref{def:ero:acquisition_pointer} there is a prefix $\mathcal{I}$ of $\mathcal{I}^\mathcal{B}$ where $\cellpointershort$ is in $\List(\mathcal{I})$ exactly once, and $prev(\mathcal{I}^\mathcal{B}, \cellpointershort)$ is immediately before $\cellpointershort$ in $\List(\mathcal{I})$.
    Hence, $prev(\mathcal{I}^\mathcal{B}, \cellpointershort)$ is not the last element of $\List(\mathcal{I})$.
    Thus, by \Cref{def:ero:logical_list}, $prev(\mathcal{I}^\mathcal{B}, \cellpointershort)$ is either $\&\headobject{}$ or there is an $L$-add event for $prev(\mathcal{I}^\mathcal{B}, \cellpointershort)$ in $\mathcal{I}$.
    If the latter, then by \Cref{lemma:ero:every_l_event_is_for_pointer_from_universe}, $prev(\mathcal{I}^\mathcal{B}, \cellpointershort) \in \celluniverse$.
    Therefore, $prev(\mathcal{I}^\mathcal{B}, \cellpointershort) \in \celluniverse \cup \{\&\headobject{}\}$ as wanted.
    \qH{\Cref{lemma:ero:acquisition_pointer_in_universe_or_head}}
\end{proof}

Now that we have defined the cell that contains the acquisition counter for the cell pointed to by $\cellpointershort$, we need to prove that the acquisition counter of this cell is semantically correct, i.e., it is equal to the number of successful-list-acquire-next attempts for $\cellpointershort$ plus one (\Cref{lemma:ero:acquisition_count_is_semantically_correct}).
To do so, we prove two facts: (1) every successful list-acquire-next attempt for $\cellpointershort$ is after its previous cell (\Cref{lemma:ero:successful_acquire_attempt_for_ptr_is_after_prev_ptr}); and (2) every successful list-acquire-next attempt after its previous cell is for $\cellpointershort$ (\Cref{lemma:ero:successful_acquire_attempt_after_prev_ptr_is_for_ptr}).
These two facts together imply that during periods when the cell that contains the acquisition counter for the cell pointed to by $\cellpointershort$ remains the same, the acquisition counter for $\cellpointershort$ changes proportionally to the number of successful list-acquire-next attempts for $\cellpointershort$.
In other words, in the special case where the cell that contains the acquisition counter for the cell pointed to by $\cellpointershort$ is always the same, (1) and (2) imply that the acquisition counter for $\cellpointershort$ is semantically correct.
We start the proof of (1) and (2) by proving that $\cellpointershort$ is active at the time of any successful list-acquire-next attempt for $\cellpointershort$, implying that our definition of the acquisition counter for the cell pointed to by $\cellpointershort$ (\Cref{def:ero:acquisition_pointer}) is well-defined.

\begin{proposition}\label{lemma:ero:acquisition_after_implies_active_after}
    If the last step in $\mathcal{I}^\mathcal{B}$ is a successful list-acquire-next attempt after $\previouscellpointershort$ such that $\previouscellpointershort \neq \&\headobject{}$, then $\previouscellpointershort$ is active in $\mathcal{I}^\mathcal{B}$.
\end{proposition}

\begin{proof}
    Let $a_{acquire}$ be the last step in $\mathcal{I}^\mathcal{B}$.
    By \Cref{def:ero:active}, we must prove that there is a single successful list-add attempt for $\cellpointershort$ in $\mathcal{I}^\mathcal{B}$ and no successful list-remove attempts for $\cellpointershort$ in $\mathcal{I}^\mathcal{B}$.
    Since $a_{acquire}$ is a list-acquire-next attempt after $\previouscellpointershort$ in $\mathcal{I}^\mathcal{B}$, by \Cref{lemma:ero:list_acquire_next_attempt_for_pointer_from_universe_and_after_pointer_from_universe_or_head}, $\previouscellpointershort \in \celluniverse \cup \{\&\headobject{}\}$, and since $\previouscellpointershort \neq \&\headobject{}$ we have that $\previouscellpointershort \in \celluniverse$ and so by \Cref{assumption:ero:head_and_null_not_in_cell_universe} $\previouscellpointershort \neq \nullconstant$.
    Furthermore, by \Cref{lemma:ero:before_any_list_acquire_is_a_successful_list_add_or_list_remove_attempt}, there is either a successful list-add attempt after $\previouscellpointershort$ or a successful list-remove attempt between $\previouscellpointershort$ and some pointer in $\mathcal{I}^\mathcal{B}$.
    Let $a$ be this successful list-add or list-remove attempt.
    Since by \Cref{lemma:ero:the_list_invariants_hold} $Q(\mathcal{I}^\mathcal{B})$ holds, we have that there is an $L$-event $e$ before $a$ such that if $\mathcal{I}^{exclude}_e$ is the prefix of $\mathcal{I}^\mathcal{B}$ up to but excluding $e$ then $\previouscellpointershort \in \List(\mathcal{I}^{exclude}_e)$.
    Hence, since $\previouscellpointershort \neq \&\headobject{}$ and $\previouscellpointershort \neq \nullconstant$, by \Cref{def:ero:logical_list}, there is an $L$-add event $e_{add}$ for $\previouscellpointershort$ in $\mathcal{I}^{exclude}_e$.
    Since $e_{add}$ is in $\mathcal{I}^{exclude}_e$, we have that $e_{add} < e$, and so there is a next $L$-event after $e_{add}$ in $\mathcal{I}^\mathcal{B}$; say $e'$.
    Hence, since $e_{add}$ is an $L$-add event for $\previouscellpointershort$ in $\mathcal{I}^\mathcal{B}$ and by \Cref{lemma:ero:the_list_invariants_hold} $R(\mathcal{I}^\mathcal{B})$ holds, we have that there is a successful list-add attempt $a_{add}$ for $\previouscellpointershort$ in $\mathcal{I}^\mathcal{B}$.
    Therefore, by \Cref{lemma:ero:at_most_one_successful_list_add_attempt}, $a_{add}$ is the only successful list-add attempt for $\previouscellpointershort$ in $\mathcal{I}^\mathcal{B}$ as wanted.
    What remains is to prove that there are no successful list-remove attempts for $\cellpointershort$ in $\mathcal{I}^\mathcal{B}$.
    This follows from \Cref{lemma:ero:no_list_remove_before_acquire} because $a_{acquire}$ is a successful list-acquire-next attempt after $\previouscellpointershort$ and is the last step of $\mathcal{I}^\mathcal{B}$.
    \qH{\Cref{lemma:ero:acquisition_after_implies_active_after}}
\end{proof}

\begin{lemma}\label{lemma:ero:acquisition_for_implies_active_for}
    If the last step in $\mathcal{I}^\mathcal{B}$ is a successful list-acquire-next attempt for $\cellpointershort$, then $\cellpointershort$ is active in $\mathcal{I}^\mathcal{B}$.
\end{lemma}

\begin{proof}
    Let $a_{acquire}$ be the last step in $\mathcal{I}^\mathcal{B}$ and suppose $a_{acquire}$ is a successful list-acquire-next attempt for $\cellpointershort$ after $\previouscellpointershort$.
    This setup yields the following three facts.
    First, by \Cref{lemma:ero:list_acquire_next_attempt_for_pointer_from_universe_and_after_pointer_from_universe_or_head}, $\cellpointershort \in \celluniverse$ and $\previouscellpointershort \in \celluniverse \cup \{\&\headobject{}\}$.
    Second, $(*\previouscellpointershort).\nextlong{}.\cellpointerlong = \cellpointershort$ at $a_{acquire}$.
    Third, by \Cref{lemma:ero:before_any_list_acquire_is_a_successful_list_add_or_list_remove_attempt}, there is a successful list-add or list-remove attempt before $a_{acquire}$, and so by \Cref{lemma:ero:l_event_corresponding_to_do_low_level_op}, there is an $L$-event before $a_{acquire}$.
    Hence, there is a last $L$-event before $a_{acquire}$; say $e_{last}$.
    Thus, since $a_{acquire}$ is the last step in $\mathcal{I}^\mathcal{B}$, we have that $e_{last}$ is the last $L$-event in $\mathcal{I}^\mathcal{B}$.
    Let $\mathcal{I}^{exclude}_{e_{last}}$ be the prefix of $\mathcal{I}^\mathcal{B}$ up to but excluding $e_{last}$.
    There are two cases.

    \begin{itemize}
        \item[] \hspace{0pt}\textbf{Case 1.} $e_{last}$ is an $L$-add or $L$-remove event and from $e_{last}$ onwards in $\mathcal{I}^\mathcal{B}$ there are no successful list-add or list-remove attempts.

        Hence, since $\mathcal{I}^\mathcal{B}$ is finite, by \Cref{lemma:ero:the_list_invariants_hold} $P(\mathcal{I}^\mathcal{B})$, $Q(\mathcal{I}^\mathcal{B})$, and $R(\mathcal{I}^\mathcal{B})$ hold, $e_{last}$ is the last $L$-event in $\mathcal{I}^\mathcal{B}$, $e_{last}$ is an $L$-add or $L$-remove event, and from $e_{last}$ onwards in $\mathcal{I}^\mathcal{B}$ there are no successful list-add or list-remove attempts, by \Cref{lemma:ero:conditional_classification_lemma}, the list of cells conforms to $\List(\mathcal{I}^{exclude}_{e_{last}})$ in $\mathcal{I}^\mathcal{B}$.
        
        We now prove that $\previouscellpointershort \in \List(\mathcal{I}^{exclude}_{e_{last}})$.
        Since $a_{acquire}$ is a successful list-acquire-next attempt after $\previouscellpointershort$, by \Cref{lemma:ero:acquisition_after_implies_active_after}, $\previouscellpointershort$ is active in $\mathcal{I}^\mathcal{B}$.
        Therefore, since $\mathcal{I}^\mathcal{B}$ is finite, $e_{last}$ is the last $L$-event in $\mathcal{I}^\mathcal{B}$, and from $e_{last}$ onwards in $\mathcal{I}^\mathcal{B}$ there are no successful list-add or list-remove attempts, by \Cref{lemma:ero:if_from_last_l_event_onwards_no_list_change_then_ptr_in_list}, $\previouscellpointershort \in \List(\mathcal{I}^{exclude}_{e_{last}})$.

        We now prove that $\cellpointershort \in \List(\mathcal{I}^{exclude}_{e_{last}})$.
        Since $\previouscellpointershort \in \celluniverse \cup \{\&\headobject{}\}$, by \Cref{assumption:ero:head_and_null_not_in_cell_universe}, $\previouscellpointershort \neq \nullconstant$.
        Furthermore, since as established above $(*\previouscellpointershort).\nextlong{}.\cellpointerlong = \cellpointershort$ at $a_{acquire}$, and $a_{acquire}$ is the last step in $\mathcal{I}^\mathcal{B}$, we have that $(*\previouscellpointershort).\nextlong{}.\cellpointerlong = \cellpointershort$ at the end of $\mathcal{I}^\mathcal{B}$.
        Since the list of cells conforms to $\List(\mathcal{I}^{exclude}_{e_{last}})$ in $\mathcal{I}^\mathcal{B}$, $\previouscellpointershort \in \List(\mathcal{I}^{exclude}_{e_{last}})$, and $\previouscellpointershort \neq \nullconstant$, by \Cref{def:ero:logical_list}, $\cellpointershort \in \List(\mathcal{I}^{exclude}_{e_{last}})$ as wanted.

        We now finish the proof of Case 1.
        We must prove that there is exactly one successful list-add attempt for $\cellpointershort$ in $\mathcal{I}^\mathcal{B}$ and no successful list-remove attempts for $\cellpointershort$ in $\mathcal{I}^\mathcal{B}$.
        Since $\cellpointershort \in \celluniverse$, by \Cref{assumption:ero:head_and_null_not_in_cell_universe}, $\cellpointershort \neq \&\headobject{}$ and $\cellpointershort \neq \nullconstant$.
        Hence, since $\cellpointershort \in \List(\mathcal{I}^{exclude}_{e_{last}})$, by \Cref{def:ero:logical_list}, there is an $L$-add event $e_{add}$ for $\cellpointershort$ in $\mathcal{I}^{exclude}_{e_{last}}$.
        Hence, $e_{add} < e_{last}$, and so there is a next $L$-event after $e_{add}$ in $\mathcal{I}^\mathcal{B}$; say $e$.
        Thus, since $e_{add}$ is an $L$-add event for $\cellpointershort$ in $\mathcal{I}^\mathcal{B}$ and by \Cref{lemma:ero:the_list_invariants_hold} $R(\mathcal{I}^\mathcal{B})$ holds, we have that there is a successful list-add attempt $a_{add}$ for $\cellpointershort$ in $\mathcal{I}^\mathcal{B}$.
        Therefore, by \Cref{lemma:ero:at_most_one_successful_list_add_attempt}, $a_{add}$ is the only successful list-add attempt for $\cellpointershort$ in $\mathcal{I}^\mathcal{B}$ as wanted.
        What remains is to prove that there are no successful list-remove attempts for $\cellpointershort$ in $\mathcal{I}^\mathcal{B}$.
        Since by \Cref{lemma:ero:the_list_invariants_hold} $P(\mathcal{I}^\mathcal{B})$ holds, and $\cellpointershort \in \List(\mathcal{I}^{exclude}_{e_{last}})$, by \Cref{lemma:ero:if_in_list_then_no_l_remove_events_for_in_history}, there are no list-remove attempts for $\cellpointershort$ in $\mathcal{I}^{exclude}_{e_{last}}$.
        Therefore, since $e_{last}$ is not a list-remove attempt, and from $e_{last}$ onwards in $\mathcal{I}^\mathcal{B}$ there are no successful list-remove attempts, we have that there are no successful list-remove attempts for $\cellpointershort$ in $\mathcal{I}^\mathcal{B}$ as wanted.

        \item[] \hspace{0pt}\textbf{Case 2.} $e_{last}$ is not an $L$-add nor an $L$-remove event or from $e_{last}$ onwards in $\mathcal{I}^\mathcal{B}$ there is a successful list-add or list-remove attempt.

        Hence, since $\mathcal{I}^\mathcal{B}$ is finite, by \Cref{lemma:ero:the_list_invariants_hold} $P(\mathcal{I}^\mathcal{B})$, $Q(\mathcal{I}^\mathcal{B})$, and $R(\mathcal{I}^\mathcal{B})$ hold, by \Cref{lemma:ero:conditional_classification_lemma}, the list of cells conforms to $\List(\mathcal{I}^\mathcal{B})$ in $\mathcal{I}^\mathcal{B}$.

        We now prove that $\previouscellpointershort \in \List(\mathcal{I}^\mathcal{B})$.
        Since $a_{acquire}$ is a successful list-acquire-next attempt after $\previouscellpointershort$, by \Cref{lemma:ero:acquisition_after_implies_active_after}, $\previouscellpointershort$ is active in $\mathcal{I}^\mathcal{B}$.
        Therefore, since $\mathcal{I}^\mathcal{B}$ is finite, $e_{last}$ is the last $L$-event in $\mathcal{I}^\mathcal{B}$, $e_{last}$ is not an $L$-add nor an $L$-remove event or from $e_{last}$ onwards in $\mathcal{I}^\mathcal{B}$ there is a successful list-add or list-remove attempt, by \Cref{lemma:ero:if_from_last_l_event_onwards_list_change_then_ptr_in_list}, $\previouscellpointershort \in \List(\mathcal{I}^\mathcal{B})$.

        We now prove that $\cellpointershort \in \List(\mathcal{I}^\mathcal{B})$.
        Since $\previouscellpointershort \in \celluniverse \cup \{\&\headobject{}\}$, by \Cref{assumption:ero:head_and_null_not_in_cell_universe}, $\previouscellpointershort \neq \nullconstant$.
        Furthermore, since as established above $(*\previouscellpointershort).\nextlong{}.\cellpointerlong = \cellpointershort$ at $a_{acquire}$, and $a_{acquire}$ is the last step in $\mathcal{I}^\mathcal{B}$, we have that $(*\previouscellpointershort).\nextlong{}.\cellpointerlong = \cellpointershort$ at the end of $\mathcal{I}^\mathcal{B}$.
        Since the list of cells conforms to $\List(\mathcal{I}^\mathcal{B})$ in $\mathcal{I}^\mathcal{B}$, $\previouscellpointershort \in \List(\mathcal{I}^\mathcal{B})$, and $\previouscellpointershort \neq \nullconstant$, by \Cref{def:ero:logical_list}, $\cellpointershort \in \List(\mathcal{I}^\mathcal{B})$ as wanted.

        We now prove that there is a successful list-add attempt for $\cellpointershort$ in $\mathcal{I}^\mathcal{B}$.
        Since $a_{acquire}$ is a list-acquire-next attempt for $\cellpointershort$, by \Cref{lemma:ero:before_any_list_acquire_is_a_successful_list_add_or_list_remove_attempt}, there is either a successful list-add attempt for $\cellpointershort$ or there is a successful list-remove attempt between some pointer and $\cellpointershort$ before $a_{acquire}$.
        Let $a$ be this successful list-add or list-remove attempt.
        Hence, if $a$ is a list-add attempt for $\cellpointershort$, we are done, so suppose $a$ is a successful list-remove attempt between some pointer and $\cellpointershort$.
        Thus, since by \Cref{lemma:ero:the_list_invariants_hold} $Q(\mathcal{I}^\mathcal{B})$ holds, we have that there is an $L$-event $e$ before $a$ such that if $\mathcal{I}^{exclude}_e$ is the prefix of $\mathcal{I}^\mathcal{B}$ up to but excluding $e$, then $\cellpointershort \in \List(\mathcal{I}^{exclude}_e)$.
        Since $\cellpointershort \in \celluniverse$, by \Cref{assumption:ero:head_and_null_not_in_cell_universe}, $\cellpointershort \neq \&\headobject{}$ and $\cellpointershort \neq \nullconstant$.
        Hence, since $\cellpointershort \in \List(\mathcal{I}^{exclude}_e)$, by \Cref{def:ero:logical_list}, there is an $L$-add event $e_{add}$ for $\cellpointershort$ in $\mathcal{I}^{exclude}_e$, so $e_{add} < e$.
        Thus, there is a next $L$-event after $e_{add}$ in $\mathcal{I}^\mathcal{B}$; say $e'$.
        Therefore, since $e_{add}$ is an $L$-add event for $\cellpointershort$ in $\mathcal{I}^\mathcal{B}$ and by \Cref{lemma:ero:the_list_invariants_hold} $R(\mathcal{I}^\mathcal{B})$ holds, we have that there is a successful list-add attempt $a_{add}$ for $\cellpointershort$ in $\mathcal{I}^\mathcal{B}$ as wanted.
        
        We now finish the proof of Case 2.
        We must prove that there is exactly one successful list-add attempt for $\cellpointershort$ in $\mathcal{I}^\mathcal{B}$ and no successful list-remove attempts for $\cellpointershort$ in $\mathcal{I}^\mathcal{B}$.
        Since there is a successful list-add attempt for $\cellpointershort$ in $\mathcal{I}^\mathcal{B}$, by \Cref{lemma:ero:at_most_one_successful_list_add_attempt}, there is exactly one successful list-add attempt for $\cellpointershort$ in $\mathcal{I}^\mathcal{B}$.
        What remains is to prove that there are no successful list-remove attempts for $\cellpointershort$ in $\mathcal{I}^\mathcal{B}$.
        This follows from \Cref{lemma:ero:if_in_list_then_no_l_remove_events_for_in_history} since $P(\mathcal{I}^\mathcal{B})$ holds and $\cellpointershort \in \List(\mathcal{I}^\mathcal{B})$.
        \qH{\Cref{lemma:ero:acquisition_for_implies_active_for}}
    \end{itemize}
\end{proof}

We are now ready to prove (1) and (2), i.e., (1) every successful list-acquire-next attempt for $\cellpointershort$ is after its previous cell (\Cref{lemma:ero:successful_acquire_attempt_for_ptr_is_after_prev_ptr}), and (2) every successful list-acquire-next attempt after its previous cell is for $\cellpointershort$ (\Cref{lemma:ero:successful_acquire_attempt_after_prev_ptr_is_for_ptr}).
We start by recording a few useful facts.

\begin{proposition}\label{lemma:ero:list_add_format_ordering_in_list}
    If the last step of $\mathcal{I}^\mathcal{B}$ is a successful list-add attempt for $\cellpointershort$ after $\previouscellpointershort$, then $\previouscellpointershort$ is immediately before $\cellpointershort$ in $\List(\mathcal{I}^\mathcal{B})$.
\end{proposition}

\begin{proof}
    Let $a$ be the last step of $\mathcal{I}^\mathcal{B}$.
    Hence, since by \Cref{lemma:ero:the_list_invariants_hold} $Q(\mathcal{I}^\mathcal{B})$ holds, we have that $a$ is preceded by a unique $L$-add event $e$ for $\cellpointershort$ and if $\mathcal{I}^{exclude}_{e}$ is the prefix of $\mathcal{I}^\mathcal{B}$ up to but excluding $e$ then $\previouscellpointershort$ is the second last pointer in $\List(\mathcal{I}^{exclude}_{e})$.
    Since $a$ is a list-add attempt for $\cellpointershort$, by \Cref{lemma:ero:l_event_corresponding_to_do_low_level_op}, there is an $L$-add event $e'$ for $\cellpointershort$ before $a$.
    Hence, since $e$ and $e'$ are both $L$-add events for $\cellpointershort$ in $\mathcal{I}^\mathcal{B}$ and by \Cref{lemma:ero:the_list_invariants_hold} $P(\mathcal{I}^\mathcal{B})$ holds, we have that $e' = e$.
    Thus, $e$ is $a$'s corresponding $L$-event, and so since by \Cref{lemma:ero:the_list_invariants_hold} $P(\mathcal{I}^\mathcal{B})$, $Q(\mathcal{I}^\mathcal{B})$, and $R(\mathcal{I}^\mathcal{B})$ hold, by \Cref{lemma:ero:any_list_attempt_outside_its_window_is_unsuccessful_alternate_statement}, $e$ is the last $L$-event in $\mathcal{I}^\mathcal{B}$. 
    So, since $\mathcal{I}^{exclude}_{e}$ is the prefix of $\mathcal{I}^\mathcal{B}$ up to but excluding $e$, we have that the sequences of $L$-events in $\mathcal{I}^{exclude}_{e}$ and $\mathcal{I}^\mathcal{B}$ are the same except $\mathcal{I}^{exclude}_{e}$ excludes $e$ and $\mathcal{I}^\mathcal{B}$ includes $e$.
    Therefore, since $\previouscellpointershort$ is the second last pointer in $\List(\mathcal{I}^{exclude}_{e})$, and $e$ is an $L$-add event for $\cellpointershort$, by \Cref{def:ero:logical_list}, $\previouscellpointershort$ is the third last pointer in $\List(\mathcal{I}^\mathcal{B})$ and $\cellpointershort$ is the second last pointer in $\List(\mathcal{I}^\mathcal{B})$ as wanted.
    \qH{\Cref{lemma:ero:list_add_format_ordering_in_list}}
\end{proof}

\begin{proposition}\label{lemma:ero:list_remove_format_ordering_in_list}
    If the last step of $\mathcal{I}^\mathcal{B}$ is a successful list-remove attempt between $\previouscellpointershort$ and $\nextcellpointershort$, then $\previouscellpointershort$ is immediately before $\nextcellpointershort$ in $\List(\mathcal{I}^\mathcal{B})$.
\end{proposition}

\begin{proof}
    Let $a$ be the last step of $\mathcal{I}^\mathcal{B}$ and suppose $a$ is for $\cellpointershort$.
    Hence, since by \Cref{lemma:ero:the_list_invariants_hold} $Q(\mathcal{I}^\mathcal{B})$ holds, we have that $a$ is preceded by a unique $L$-remove event $e$ for $\cellpointershort$ in $\mathcal{I}^\mathcal{B}$ and if $\mathcal{I}^{exclude}_{e}$ is the prefix of $\mathcal{I}^\mathcal{B}$ up to but excluding $e$ then $\cellpointershort$ is in $\List(\mathcal{I}^{exclude}_{e})$ exactly once and $\previouscellpointershort{}$ and $\nextuniquecellpointershort{}$ are the pointers preceding and succeeding $\cellpointershort$ in $\List(\mathcal{I}^{exclude}_{e})$.
    Since $a$ is a list-remove attempt for $\cellpointershort$, by, \Cref{lemma:ero:l_event_corresponding_to_do_low_level_op}, there is an $L$-remove event $e'$ for $\cellpointershort$ before $a$.
    Hence, since $e$ and $e'$ are both $L$-remove events for $\cellpointershort$ in $\mathcal{I}^\mathcal{B}$ and by \Cref{lemma:ero:the_list_invariants_hold} $P(\mathcal{I}^\mathcal{B})$ holds, we have that $e' = e$.
    Thus, $e$ is $a$'s corresponding $L$-event, and so since by \Cref{lemma:ero:the_list_invariants_hold} $P(\mathcal{I}^\mathcal{B})$, $Q(\mathcal{I}^\mathcal{B})$, and $R(\mathcal{I}^\mathcal{B})$ hold, by \Cref{lemma:ero:any_list_attempt_outside_its_window_is_unsuccessful_alternate_statement}, $e$ is the last $L$-event in $\mathcal{I}^\mathcal{B}$. 
    So, since $\mathcal{I}^{exclude}_{e}$ is the prefix of $\mathcal{I}^\mathcal{B}$ up to but excluding $e$, we have that the sequences of $L$-events in $\mathcal{I}^{exclude}_{e}$ and $\mathcal{I}^\mathcal{B}$ are the same except $\mathcal{I}^{exclude}_{e}$ excludes $e$ and $\mathcal{I}^\mathcal{B}$ includes $e$.
    Therefore, since $\previouscellpointershort{}$ and $\nextuniquecellpointershort{}$ are the pointers preceding and succeeding $\cellpointershort$ in $\List(\mathcal{I}^{exclude}_{e})$ and $e$ is an $L$-remove event for $\cellpointershort$, by \Cref{def:ero:logical_list}, $\previouscellpointershort$ is the pointer preceding $\nextcellpointershort$ in $\List(\mathcal{I}^\mathcal{B})$ as wanted.
    \qH{\Cref{lemma:ero:list_remove_format_ordering_in_list}}
\end{proof}

\begin{proposition}\label{lemma:ero:nice_list_property_of_successful_list_add_and_remove_attempt}
    If the last step of $\mathcal{I}^\mathcal{B}$ is a successful list-add attempt for $\cellpointershort$ after $\previouscellpointershort$ or a successful list-remove attempt between $\previouscellpointershort$ and $\cellpointershort$, then (1) the list of cells conforms to $\List(\mathcal{I}^\mathcal{B})$ in $\mathcal{I}^\mathcal{B}$ and (2) $\previouscellpointershort$ is immediately before $\cellpointershort$ in $\List(\mathcal{I}^\mathcal{B})$.
\end{proposition}

\begin{proof}
    Since the last step of $\mathcal{I}^\mathcal{B}$ is a successful list-add or list-remove attempt, by \Cref{lemma:ero:l_event_corresponding_to_do_low_level_op}, there is an $L$-event in $\mathcal{I}^\mathcal{B}$, so there is a last $L$-event in $\mathcal{I}^\mathcal{B}$; say $e_{last}$.
    Hence, since the last step of $\mathcal{I}^\mathcal{B}$ is a successful list-add or list-remove attempt, we have that from $e_{last}$ onwards in $\mathcal{I}^\mathcal{B}$ there is a successful list-add or list-remove attempt.
    Therefore, since $\mathcal{I}^\mathcal{B}$ is finite, and by \Cref{lemma:ero:the_list_invariants_hold} $P(\mathcal{I}^\mathcal{B})$, $Q(\mathcal{I}^\mathcal{B})$, and $R(\mathcal{I}^\mathcal{B})$ hold, by \Cref{lemma:ero:conditional_classification_lemma}, the list of cells conforms to $\List(\mathcal{I}^\mathcal{B})$ in $\mathcal{I}^\mathcal{B}$ satisfying (1).
    Furthermore, (2) follows from \Cref{lemma:ero:list_add_format_ordering_in_list} and \Cref{lemma:ero:list_remove_format_ordering_in_list}.
    \qH{\Cref{lemma:ero:nice_list_property_of_successful_list_add_and_remove_attempt}}
\end{proof}

We now prove (1).

\begin{proposition}\label{lemma:ero:successful_acquire_attempt_for_ptr_is_after_prev_ptr}
    If the last step in $\mathcal{I}^\mathcal{B}$ is a successful list-acquire-next attempt for $\cellpointershort$, then it is after $prev(\mathcal{I}^\mathcal{B}, \cellpointershort)$.
\end{proposition}

\begin{proof}
    %By \Cref{lemma:ero:acquisition_for_implies_active_for} $\cellpointershort$ is active in $\mathcal{I}^\mathcal{B}$.
    Let $a_{acquire}$ be the last step in $\mathcal{I}^\mathcal{B}$, $p$ be the process that executed $a_{acquire}$, and suppose $a_{acquire}$ is for $\cellpointershort$ after $\previouscellpointershort$.
    Hence, by \Cref{lemma:ero:list_acquire_next_attempt_for_pointer_from_universe_and_after_pointer_from_universe_or_head}, $\cellpointershort \in \celluniverse$ and $\previouscellpointershort \in \celluniverse \cup \{\&\headobject{}\}$.
    Furthermore, by \Cref{lemma:ero:before_any_list_acquire_is_a_successful_list_add_or_list_remove_attempt}, there is a successful list-add attempt for $\cellpointershort$ after $\previouscellpointershort$ or a successful list-remove attempt between $\previouscellpointershort$ and $\cellpointershort$ before $a_{acquire}$ in $\mathcal{I}^\mathcal{B}$; say $a$.
    Hence, by \Cref{lemma:ero:nice_list_property_of_successful_list_add_and_remove_attempt}, the list of cells conforms to $\List(\mathcal{I}^{include}_a)$ in $\mathcal{I}^{include}_a$ and $\previouscellpointershort$ is immediately before $\cellpointershort$ in $\List(\mathcal{I}^{include}_a)$ where $\mathcal{I}^{include}_a$ is the prefix of $\mathcal{I}^\mathcal{B}$ up to and including $a$ (*).

    Since the last step in $\mathcal{I}^\mathcal{B}$ is a successful list-acquire-next attempt for $\cellpointershort$, by \Cref{lemma:ero:acquisition_for_implies_active_for}, $\cellpointershort$ is active in $\mathcal{I}^\mathcal{B}$, and so by \Cref{corollary:ero:active_prefix_has_last_l_event}, there is a last $L$-event in $\mathcal{I}^\mathcal{B}$; say $e_{last}$. 

    \begin{itemize}
        \item[] \hspace{0pt}\textbf{Case 1.} $e_{last} < a$.
        
        Hence, since $a$ is a successful list-add or list-remove attempt, from $e_{last}$ onwards in $\mathcal{I}^\mathcal{B}$ there is a successful list-add or list-remove attempt.
        Thus, by \Cref{def:ero:acquisition_pointer}, $prev(\mathcal{I}^\mathcal{B}, \cellpointershort)$ is immediately before $\cellpointershort$ in $\List(\mathcal{I}^\mathcal{B})$.
        % We now prove that the sequences of $L$-events are the same in $\mathcal{I}^{include}_a$ and $\mathcal{I}^\mathcal{B}$ which implies $\List(\mathcal{I}^{include}_a) = \List(\mathcal{I}^\mathcal{B})$ by \Cref{def:ero:logical_list}.
        Since $e_{last} < a$, we have that $e_{last}$ is in $\mathcal{I}^{include}_a$.
        Hence, since $e_{last}$ is the last $L$-event in $\mathcal{I}^\mathcal{B}$ and $\mathcal{I}^{include}_a$ is a prefix of $\mathcal{I}^\mathcal{B}$, we have that the sequences of $L$-events are the same in $\mathcal{I}^{include}_a$ and $\mathcal{I}^\mathcal{B}$.
        Thus, by \Cref{def:ero:logical_list}, $\List(\mathcal{I}^{include}_a) = \List(\mathcal{I}^\mathcal{B})$.
        So, since by (*) $\previouscellpointershort$ is immediately before $\cellpointershort$ in $\List(\mathcal{I}^{include}_a)$, we have that $\previouscellpointershort$ is immediately before $\cellpointershort$ in $\List(\mathcal{I}^\mathcal{B})$.
        Therefore, since $prev(\mathcal{I}^\mathcal{B}, \cellpointershort)$ is immediately before $\cellpointershort$ in $\List(\mathcal{I}^\mathcal{B})$, we have that $prev(\mathcal{I}^\mathcal{B}, \cellpointershort) = \previouscellpointershort$ as wanted.

        \item[] \hspace{0pt}\textbf{Case 2.} $a < e_{last}$.

        There are two cases.

        \begin{itemize}
            \item[] \hspace{0pt}\textbf{Case 2.1.} $e_{last}$ is a $L$-remove event for $\previouscellpointershort$.

            Hence, by \Cref{lemma:ero:every_l_event_is_for_pointer_from_universe}, $\previouscellpointershort \in \celluniverse$, so by \Cref{assumption:ero:head_and_null_not_in_cell_universe} $\previouscellpointershort \neq \&\headobject{}$ and $\nullconstant$.

            We first prove that from $e_{last}$ onwards in $\mathcal{I}^\mathcal{B}$ there are no successful list-add or list-remove attempts.
            Since $e_{last}$ is the last $L$-event in $\mathcal{I}^\mathcal{B}$, $e_{last}$ is an $L$-remove event for $\previouscellpointershort$, and by \Cref{lemma:ero:the_list_invariants_hold}, $P(\mathcal{I}^\mathcal{B})$, $Q(\mathcal{I}^\mathcal{B})$, and $R(\mathcal{I}^\mathcal{B})$ hold, by \Cref{lemma:ero:3_of_r_safety_holds}, from $e_{last}$ onwards in $\mathcal{I}^\mathcal{B}$ there is at most one successful list-remove attempt for $\previouscellpointershort$ and no other successful list-remove or list-add attempt for any pointer.
            Furthermore, since $a_{acquire}$ is a successful list-acquire-next attempt after $\previouscellpointershort$ in $\mathcal{I}^\mathcal{B}$, by \Cref{lemma:ero:no_list_remove_before_acquire}, there are no successful list-remove attempts for $\previouscellpointershort$ before $a_{acquire}$ in $\mathcal{I}^\mathcal{B}$, and so since $a_{acquire}$ is the last step in $\mathcal{I}^\mathcal{B}$, we have that there are no successful list-remove attempts for $\previouscellpointershort$ in $\mathcal{I}^\mathcal{B}$.
            These two facts together imply that from $e_{last}$ onwards in $\mathcal{I}^\mathcal{B}$ there are no successful list-add or list-remove attempts as wanted.

            Since $e_{last}$ is an $L$-remove event and from $e_{last}$ onwards in $\mathcal{I}^\mathcal{B}$ there are no successful list-add or list-remove attempts, by \Cref{def:ero:acquisition_pointer}, 
            $\cellpointershort$ is in $\List(\mathcal{I}^{exclude}_{e_{last}})$ exactly once and
            $prev(\mathcal{I}^\mathcal{B}, \cellpointershort)$ is immediately before $\cellpointershort$ in $\List(\mathcal{I}^{exclude}_{e_{last}})$ where $\mathcal{I}^{exclude}_{e_{last}}$ is the prefix of $\mathcal{I}^\mathcal{B}$ up to but excluding $e_{last}$.
            Since $a < e_{last}$, we have that $\mathcal{I}^{include}_a$ is a prefix of $\mathcal{I}^{exclude}_{e_{last}}$.

            We now prove that $\previouscellpointershort \in \List(\mathcal{I}^{exclude}_{e_{last}})$.
            Suppose, for contradiction, $\previouscellpointershort \notin \List(\mathcal{I}^{exclude}_{e_{last}})$.
            Since by (*) $\previouscellpointershort$ is in $\List(\mathcal{I}^{include}_a)$, and $\previouscellpointershort \neq \&\headobject{}$ and $\previouscellpointershort \neq \nullconstant$, by \Cref{def:ero:logical_list}, there is an $L$-add event $e_{add}$ for $\previouscellpointershort$ in $\mathcal{I}^{include}_a$.
            Hence, since $\mathcal{I}^{include}_a$ is a prefix of $\mathcal{I}^{exclude}_{e_{last}}$, we have that $e_{add}$ is in $\mathcal{I}^{exclude}_{e_{last}}$.
            Thus, since $\previouscellpointershort \notin \List(\mathcal{I}^{exclude}_{e_{last}})$, by \Cref{def:ero:logical_list}, there is an $L$-remove event $e_{remove}$ for $\previouscellpointershort$ in $\mathcal{I}^{exclude}_{e_{last}}$.
            So, since $\mathcal{I}^{exclude}_{e_{last}}$ is the prefix of $\mathcal{I}^\mathcal{B}$ up to but excluding $e_{last}$, we have that $e_{remove} < e_{last}$, and thus $e_{remove} \neq e_{last}$.
            Therefore, there are two $L$-remove events for $\previouscellpointershort$ in $\mathcal{I}^\mathcal{B}$ (namely $e_{remove}$ and $e_{last}$).
            However, since by \Cref{lemma:ero:the_list_invariants_hold} $P(\mathcal{I}^\mathcal{B})$ holds, there is at most one $L$-remove event for $\previouscellpointershort$ in $\mathcal{I}^\mathcal{B}$, a contradiction.

            We now finish the proof of Case 2.1.
            Since $\mathcal{I}^{include}_a$ is a prefix of $\mathcal{I}^{exclude}_{e_{last}}$, 
            by (*) $\previouscellpointershort \in \celluniverse$ is immediately before $\cellpointershort \in \celluniverse$ in $\List(\mathcal{I}^{include}_a)$, $\previouscellpointershort$ is in $\List(\mathcal{I}^{exclude}_{e_{last}})$, $\cellpointershort$ is in $\List(\mathcal{I}^{exclude}_{e_{last}})$, and by \Cref{lemma:ero:the_list_invariants_hold} $P(\mathcal{I}^{exclude}_{e_{last}})$ holds, by \Cref{lemma:ero:list_neighbours_are_the_same_if_they_contained_in_an_extended_list}, $\previouscellpointershort$ is immediately before $\cellpointershort$ in $\List(\mathcal{I}^{exclude}_{e_{last}})$.
            Therefore, since $\cellpointershort$ is in $\List(\mathcal{I}^{exclude}_{e_{last}})$ exactly once and $prev(\mathcal{I}^\mathcal{B}, \cellpointershort)$ is immediately before $\cellpointershort$ in $\List(\mathcal{I}^{exclude}_{e_{last}})$, we have that $prev(\mathcal{I}^\mathcal{B}, \cellpointershort) = \previouscellpointershort$ as wanted.

            \item[] \hspace{0pt}\textbf{Case 2.2.} $e_{last}$ is not a $L$-remove event for $\previouscellpointershort$.

            By \Cref{def:ero:acquisition_pointer}, either: (A) $\cellpointershort$ is in $\List(\mathcal{I}^{exclude}_{e_{last}})$ exactly once and $prev(\mathcal{I}^\mathcal{B}, \cellpointershort)$ is immediately before $\cellpointershort$ in $\List(\mathcal{I}^{exclude}_{e_{last}})$ where $\mathcal{I}^{exclude}_{e_{last}}$ is the prefix of $\mathcal{I}^\mathcal{B}$ up to but excluding $e_{last}$; or (B) $\cellpointershort$ is in $\List(\mathcal{I}^\mathcal{B})$ exactly once and $prev(\mathcal{I}^\mathcal{B}, \cellpointershort)$ is immediately before $\cellpointershort$ in $\List(\mathcal{I}^\mathcal{B})$.
            Since $a < e_{last}$, we have that $\mathcal{I}^{include}_a$ is a prefix of $\mathcal{I}^{exclude}_{e_{last}}$.

            We first prove that there are no $L$-remove events for $\previouscellpointershort$ in $\mathcal{I}^{exclude}_{e_{last}}$ and $\mathcal{I}^\mathcal{B}$.
            Since $a_{acquire}$ is the last step in $\mathcal{I}^\mathcal{B}$, $a_{acquire}$ is a successful list-acquire-next attempt after $\previouscellpointershort$, $e_{last}$ is the last $L$-event in $\mathcal{I}^\mathcal{B}$, $e_{last}$ is not a $L$-remove event for $\previouscellpointershort$, and by \Cref{lemma:ero:the_list_invariants_hold} $R(\mathcal{I}^\mathcal{B})$ holds, by \Cref{lemma:ero:l_remove_event_is_last_before_acquire}, there are no $L$-remove events for $\previouscellpointershort$ in $\mathcal{I}^\mathcal{B}$.
            Therefore, since $\mathcal{I}^{exclude}_{e_{last}}$ is a prefix of $\mathcal{I}^\mathcal{B}$, we have that there are no $L$-remove events for $\previouscellpointershort$ in $\mathcal{I}^{exclude}_{e_{last}}$ and $\mathcal{I}^\mathcal{B}$ as wanted.

            We now prove that $\previouscellpointershort$ is in $\List(\mathcal{I}^{exclude}_{e_{last}})$ and $\List(\mathcal{I}^\mathcal{B})$.
            If $\previouscellpointershort = \&\headobject{}$, this immediately follows by \Cref{def:ero:logical_list}, so since $\previouscellpointershort \in \celluniverse \cup \{\&\headobject{}\}$, it remains to consider the case where $\previouscellpointershort \in \celluniverse$.
            Since $\previouscellpointershort \in \celluniverse$, by \Cref{assumption:ero:head_and_null_not_in_cell_universe}, $\previouscellpointershort \neq \&\headobject{}$ and $\previouscellpointershort \neq \nullconstant$.
            Hence, since by (*) $\previouscellpointershort$ is in $\List(\mathcal{I}^{include}_a)$, by \Cref{def:ero:logical_list}, there is an $L$-add event $e_{add}$ for $\previouscellpointershort$ in $\mathcal{I}^{include}_a$.
            Thus, since $\mathcal{I}^{include}_a$ is a prefix of $\mathcal{I}^{exclude}_{e_{last}}$ and $\mathcal{I}^\mathcal{B}$, we have that $e_{add}$ is in $\mathcal{I}^{exclude}_{e_{last}}$ and $\mathcal{I}^\mathcal{B}$.
            Therefore, since as we just proved there are no $L$-remove events for $\previouscellpointershort$ in $\mathcal{I}^{exclude}_{e_{last}}$ and $\mathcal{I}^\mathcal{B}$, by \Cref{def:ero:logical_list}, $\previouscellpointershort$ is in $\List(\mathcal{I}^{exclude}_{e_{last}})$ and $\List(\mathcal{I}^\mathcal{B})$ as wanted.

            We now finish the proof of Case 2.2.
            First, consider (A).
            Hence, $\cellpointershort$ is in $\List(\mathcal{I}^{exclude}_{e_{last}})$ exactly once and $prev(\mathcal{I}^\mathcal{B}, \cellpointershort)$ is immediately before $\cellpointershort$ in $\List(\mathcal{I}^{exclude}_{e_{last}})$.
            Since $\mathcal{I}^{include}_a$ is a prefix of $\mathcal{I}^{exclude}_{e_{last}}$, 
            by (*) $\previouscellpointershort \in \celluniverse \cup \{\&\headobject{}\}$ is immediately before $\cellpointershort \in \celluniverse$ in $\List(\mathcal{I}^{include}_a)$, $\previouscellpointershort$ is in $\List(\mathcal{I}^{exclude}_{e_{last}})$, $\cellpointershort$ is in $\List(\mathcal{I}^{exclude}_{e_{last}})$, and by \Cref{lemma:ero:the_list_invariants_hold} $P(\mathcal{I}^{exclude}_{e_{last}})$ holds, by \Cref{lemma:ero:list_neighbours_are_the_same_if_they_contained_in_an_extended_list}, $\previouscellpointershort$ is immediately before $\cellpointershort$ in $\List(\mathcal{I}^{exclude}_{e_{last}})$.
            Therefore, $prev(\mathcal{I}^\mathcal{B}, \cellpointershort) = \previouscellpointershort$ as wanted.
            Now consider (B).
            Hence, $\cellpointershort$ is in $\List(\mathcal{I}^\mathcal{B})$ exactly once, and $prev(\mathcal{I}^\mathcal{B}, \cellpointershort)$ is immediately before $\cellpointershort$ in $\List(\mathcal{I}^\mathcal{B})$.
            Since $\mathcal{I}^{include}_a$ is a prefix of $\mathcal{I}^\mathcal{B}$, 
            by (*) $\previouscellpointershort \in \celluniverse \cup \{\&\headobject{}\}$ is immediately before $\cellpointershort \in \celluniverse$ in $\List(\mathcal{I}^{include}_a)$, $\previouscellpointershort$ is in $\mathcal{I}^\mathcal{B}$, $\cellpointershort$ is in $\mathcal{I}^\mathcal{B}$, and by \Cref{lemma:ero:the_list_invariants_hold} $P(\mathcal{I}^\mathcal{B})$ holds, by \Cref{lemma:ero:list_neighbours_are_the_same_if_they_contained_in_an_extended_list}, $\previouscellpointershort$ is immediately before $\cellpointershort$ in $\List(\mathcal{I}^\mathcal{B})$.
            Therefore, $prev(\mathcal{I}^\mathcal{B}, \cellpointershort) = \previouscellpointershort$ as wanted.
            \qH{\Cref{lemma:ero:successful_acquire_attempt_for_ptr_is_after_prev_ptr}}
        \end{itemize}
    \end{itemize}
\end{proof}

We now prove (2).

\begin{proposition}\label{lemma:ero:successful_acquire_attempt_after_prev_ptr_is_for_ptr}
    Suppose $\mathcal{I}^\mathcal{B}$ is finite and $\cellpointershort$ is active in $\mathcal{I}^\mathcal{B}$.
    If the last step in $\mathcal{I}^\mathcal{B}$ is a successful list-acquire-next attempt after $prev(\mathcal{I}^\mathcal{B}, \cellpointershort)$, then it is for $\cellpointershort$.
\end{proposition}

\begin{proof}
    Let $a_{acquire}$ be the last step of $\mathcal{I}^\mathcal{B}$ and suppose it is for $\cellpointershort'$ after $prev(\mathcal{I}^\mathcal{B}, \cellpointershort)$.
    Hence, by \Cref{lemma:ero:acquisition_for_implies_active_for}, $\cellpointershort'$ is active in $\mathcal{I}^\mathcal{B}$, and so by \Cref{lemma:ero:successful_acquire_attempt_for_ptr_is_after_prev_ptr}, $a_{acquire}$ is a successful list-acquire-next attempt for $\cellpointershort'$ after $prev(\mathcal{I}^\mathcal{B}, \cellpointershort')$, implying $prev(\mathcal{I}^\mathcal{B}, \cellpointershort) = prev(\mathcal{I}^\mathcal{B}, \cellpointershort')$.
    Let $e_{last}$ be the last $L$-event in $\mathcal{I}^\mathcal{B}$ which is well-defined by \Cref{corollary:ero:active_prefix_has_last_l_event} and let $\mathcal{I}^{exclude}_{e_{last}}$ be the prefix of $\mathcal{I}^\mathcal{B}$ up to but excluding $e_{last}$.
    We must prove that $\cellpointershort = \cellpointershort'$.
    There are two cases.

    \begin{itemize}
        \item[] \hspace{0pt}\textbf{Case 1.} $e_{last}$ is an $L$-add or $L$-remove event and from $e_{last}$ onwards in $\mathcal{I}^\mathcal{B}$ there are no successful list-add or list-remove attempts.

        Hence, since $\cellpointershort$ (resp. $\cellpointershort'$) is active in $\mathcal{I}^\mathcal{B}$, by \Cref{def:ero:acquisition_pointer}, $\cellpointershort$ (resp. $\cellpointershort'$) is in $\List(\mathcal{I}^{exclude}_{e_{last}})$ exactly once, and $prev(\mathcal{I}^\mathcal{B}, \cellpointershort)$ (resp. $prev(\mathcal{I}^\mathcal{B}, \cellpointershort')$) is immediately before $\cellpointershort$ (resp. $\cellpointershort'$) in $\List(\mathcal{I}^{exclude}_{e_{last}})$.
        Therefore, since $prev(\mathcal{I}^\mathcal{B}, \cellpointershort) = prev(\mathcal{I}^\mathcal{B}, \cellpointershort')$, we have that $\cellpointershort = \cellpointershort'$.

        \item[] \hspace{0pt}\textbf{Case 2.} $e_{last}$ is not an $L$-add nor an $L$-remove event or from $e_{last}$ onwards in $\mathcal{I}^\mathcal{B}$ there is a successful list-add or list-remove attempt.

        Hence, since $\cellpointershort$ (resp. $\cellpointershort'$) is active in $\mathcal{I}^\mathcal{B}$, by \Cref{def:ero:acquisition_pointer}, $\cellpointershort$ (resp. $\cellpointershort'$) is in $\List(\mathcal{I}^\mathcal{B})$ exactly once, and $prev(\mathcal{I}^\mathcal{B}, \cellpointershort)$ (resp. $prev(\mathcal{I}^\mathcal{B}, \cellpointershort')$) is immediately before $\cellpointershort$ (resp. $\cellpointershort'$) in $\List(\mathcal{I}^\mathcal{B})$.
        Therefore, since $prev(\mathcal{I}^\mathcal{B}, \cellpointershort) = prev(\mathcal{I}^\mathcal{B}, \cellpointershort')$, we have that $\cellpointershort = \cellpointershort'$.
        \qH{\Cref{lemma:ero:successful_acquire_attempt_after_prev_ptr_is_for_ptr}}
    \end{itemize}
\end{proof}

As we mentioned before, in the special case where the cell that contains the acquisition counter for the cell pointed to by $\cellpointershort$ is always the same, \Cref{lemma:ero:successful_acquire_attempt_for_ptr_is_after_prev_ptr} and \Cref{lemma:ero:successful_acquire_attempt_after_prev_ptr_is_for_ptr} imply that the acquisition counter for $\cellpointershort$ is semantically correct, i.e., it is equal to the number of successful list-acquire-next attempts for $\cellpointershort$ plus one (\Cref{lemma:ero:acquisition_count_is_semantically_correct}).
So, what remains is to deal with the case where the cell that contains the acquisition counter for the cell pointed to by $\cellpointershort$ changes.
The challenge is ensuring that the acquisition counter for $\cellpointershort$ is correctly copied when the cell it is stored in changes.
Ultimately, this concern is addressed by the sealing mechanism, but to leverage this, we must first show that when the cell that stores the acquisition counter for $\cellpointershort$ changes, it is a particular list-remove attempt that does so (\Cref{lemma:ero:if_acquisition_pointer_changes_then_its_because_of_a_specific_list_remove_attempt}).

\begin{proposition}\label{lemma:ero:every_prefix_between_i_first_and_i_is_active}
    Suppose $\cellpointershort$ is active in $\mathcal{I}^\mathcal{B}$.
    By \Cref{def:ero:active}, there is a single successful list-add attempt $a_{add}$ for $\cellpointershort$ in $\mathcal{I}^\mathcal{B}$.
    For every prefix $\mathcal{I}$ of $\mathcal{I}^\mathcal{B}$ if $a_{add}$ is in $\mathcal{I}$, then $\cellpointershort$ is active in $\mathcal{I}$.
\end{proposition}

\begin{proof}
    Suppose, for contradiction, there is a prefix $\mathcal{I}$ of $\mathcal{I}^\mathcal{B}$ where $a_{add}$ is in $\mathcal{I}$ and $\cellpointershort$ is not active in $\mathcal{I}$.
    Hence, since $a_{add}$ is a successful list-add attempt for $\cellpointershort$, by \Cref{lemma:ero:at_most_one_successful_list_add_attempt}, $a_{add}$ is the only successful list-add attempt for $\cellpointershort$ in $\mathcal{I}$.
    Thus, since $\cellpointershort$ is not active in $\mathcal{I}$, by \Cref{def:ero:active}, there is a successful list-remove attempt $a_{remove}$ for $\cellpointershort$ in $\mathcal{I}$.
    Therefore, since $\mathcal{I}$ is a prefix of $\mathcal{I}^\mathcal{B}$, we have that $a_{remove}$ is in $\mathcal{I}^\mathcal{B}$, and so there is a successful list-remove attempt for $\cellpointershort$ in $\mathcal{I}^\mathcal{B}$.
    However, since $\cellpointershort$ is active in $\mathcal{I}^\mathcal{B}$, by \Cref{def:ero:active}, there are no successful list-remove attempts for $\cellpointershort$ in $\mathcal{I}^\mathcal{B}$, a contradiction.
    \qH{\Cref{lemma:ero:every_prefix_between_i_first_and_i_is_active}}
\end{proof}

\begin{proposition}\label{lemma:ero:if_acquisition_pointer_changes_then_its_because_of_a_specific_list_remove_attempt}
    Suppose $\cellpointershort$ is active in $\mathcal{I}^\mathcal{B}$.
    By \Cref{def:ero:active}, there is a single successful list-add attempt $a_{add}$ for $\cellpointershort$ in $\mathcal{I}^\mathcal{B}$.
    Consider any proper prefix $\mathcal{I}$ of $\mathcal{I}^\mathcal{B}$ such that $a_{add}$ is in $\mathcal{I}$ and let $s$ be the step after $\mathcal{I}$ in $\mathcal{I}^\mathcal{B}$.
    By \Cref{lemma:ero:every_prefix_between_i_first_and_i_is_active}, $\cellpointershort$ is active in both $\mathcal{I}$ and $\mathcal{I} \circ s$, and so $prev(\mathcal{I}, \cellpointershort)$ and $prev(\mathcal{I} \circ s, \cellpointershort)$ are well-defined.
    If $prev(\mathcal{I}, \cellpointershort) \neq prev(\mathcal{I} \circ s, \cellpointershort)$, then $s$ is a successful list-remove attempt for $prev(\mathcal{I}, \cellpointershort)$ between $prev(\mathcal{I} \circ s, \cellpointershort)$ and $\cellpointershort$.
\end{proposition}

\begin{proof}
    Suppose $prev(\mathcal{I}, \cellpointershort) \neq prev(\mathcal{I} \circ s, \cellpointershort)$.
    Since $\cellpointershort$ is active in $\mathcal{I}$ (resp. $\mathcal{I} \circ s$), by \Cref{corollary:ero:active_prefix_has_last_l_event}, there is a last $L$-event in $\mathcal{I}$ (resp. $\mathcal{I} \circ s$); say $e^{\mathcal{I}}_{last}$ (resp. $e^{\mathcal{I} \circ s}_{last}$).
    There are two cases.
    \begin{itemize}
        \item[] \hspace{0pt}\textbf{Case 1.} $e^{\mathcal{I}}_{last} \neq e^{\mathcal{I} \circ s}_{last}$.

        Hence, since $\mathcal{I} \circ s$ is a one step continuation of $\mathcal{I}$, we have that $s = e^{\mathcal{I} \circ s}_{last}$.

        We first prove that $\cellpointershort$ is in $\List(\mathcal{I})$ exactly once, and $prev(\mathcal{I} \circ s, \cellpointershort)$ is immediately before $\cellpointershort$ in $\List(\mathcal{I})$.
        Since $e^{\mathcal{I} \circ s}_{last}$ is an $L$-event, by \Cref{lemma:ero:every_l_event_is_add_apply_or_remove}, $e^{\mathcal{I} \circ s}_{last}$ is either an $L$-add, $L$-apply, or $L$-remove event.
        There are two cases.
        \begin{itemize}
            \item[] \hspace{0pt}\textbf{Case 1.1.} $e^{\mathcal{I} \circ s}_{last}$ is an $L$-add or $L$-remove event.

            Hence, since $s = e^{\mathcal{I} \circ s}_{last}$, we have that from $e^{\mathcal{I} \circ s}_{last}$ onwards in $\mathcal{I} \circ s$ there are no successful list-add or list-remove attempts, and so by \Cref{def:ero:acquisition_pointer}, $\cellpointershort$ is in $\List(\mathcal{I}^*)$ exactly once, and $prev(\mathcal{I} \circ s, \cellpointershort)$ is immediately before $\cellpointershort$ in $\List(\mathcal{I}^*)$ where $\mathcal{I}^*$ is the prefix of $\mathcal{I} \circ s$ up to but excluding $e^{\mathcal{I} \circ s}_{last}$.
            Thus, since $s = e^{\mathcal{I} \circ s}_{last}$, we have that $\mathcal{I}^* = \mathcal{I}$.
            Therefore, $\cellpointershort$ is in $\List(\mathcal{I})$ exactly once, $prev(\mathcal{I} \circ s, \cellpointershort)$ is immediately before $\cellpointershort$ in $\List(\mathcal{I})$ as wanted.

            \item[] \hspace{0pt}\textbf{Case 1.2.} $e^{\mathcal{I} \circ s}_{last}$ is an $L$-apply event.

            Hence, by \Cref{def:ero:acquisition_pointer}, $\cellpointershort$ is in $\List(\mathcal{I} \circ s)$ exactly once, and $prev(\mathcal{I} \circ s, \cellpointershort)$ is immediately before $\cellpointershort$ in $\List(\mathcal{I} \circ s)$.
            Since $s = e^{\mathcal{I} \circ s}_{last}$ and $e^{\mathcal{I} \circ s}_{last}$ is an $L$-apply event, by \Cref{def:ero:logical_list}, $\List(\mathcal{I}) = \List(\mathcal{I} \circ s)$.
            Therefore, $\cellpointershort$ is in $\List(\mathcal{I})$ exactly once, $prev(\mathcal{I} \circ s, \cellpointershort)$ is immediately before $\cellpointershort$ in $\List(\mathcal{I})$ as wanted.
        \end{itemize}

        We now prove that $prev(\mathcal{I}, \cellpointershort)$ is immediately before $\cellpointershort$ in $\List(\mathcal{I})$.
        Since $e^{\mathcal{I}}_{last}$ is an $L$-event, by \Cref{lemma:ero:every_l_event_is_add_apply_or_remove}, $e^{\mathcal{I}}_{last}$ is either an $L$-add, $L$-apply, or $L$-remove event.
        If $e^{\mathcal{I}}_{last}$ is an $L$-apply event, by \Cref{def:ero:acquisition_pointer}, the claim follows.
        Otherwise, $e^{\mathcal{I}}_{last}$ is an $L$-add or $L$-remove event.
        We prove that from $e^{\mathcal{I}}_{last}$ onwards in $\mathcal{I}$ there is a successful list-add or list-remove attempt.
        Since $e^{\mathcal{I}}_{last}$ is in $\mathcal{I}$, we have that $e^{\mathcal{I}}_{last}$ is in $\mathcal{I} \circ s$.
        Hence, since $s = e^{\mathcal{I}' \circ s}_{last}$, we have that $e^{\mathcal{I}}_{last} \leq e^{\mathcal{I} \circ s}_{last}$, and since $e^{\mathcal{I}}_{last}$ is in $\mathcal{I}$, it follows that $e^{\mathcal{I}}_{last} < e^{\mathcal{I}' \circ s}_{last}$.
        Thus, there is a next $L$-event after $e^{\mathcal{I}}_{last}$ in $\mathcal{I} \circ s$; say $e$.
        So, since $e^{\mathcal{I}}_{last}$ is an $L$-add or $L$-remove event in $\mathcal{I} \circ s$, and by \Cref{lemma:ero:the_list_invariants_hold} $R(\mathcal{I} \circ s)$ holds, we have that between 
        $e^{\mathcal{I}}_{last}$ and $e$ in $\mathcal{I} \circ s$, there is either a successful list-add or list-remove attempt, say $a$, so $e^{\mathcal{I}}_{last} < a < e$.
        Since $e$ is in $\mathcal{I} \circ s$, we have that $e \leq s$, so by transitivity, $a < s$.
        Hence, since $a$ is in $\mathcal{I} \circ s$, we have that $a$ is in $\mathcal{I}$.
        Therefore, since $e^{\mathcal{I}}_{last} < a$ and $a$ is either a successful list-add or list-remove attempt, we have that from $e^{\mathcal{I}}_{last}$ onwards in $\mathcal{I}$ there is a successful list-add or list-remove attempt.
        Since $e^{\mathcal{I}}_{last}$ is an $L$-add or $L$-remove event, and from $e^{\mathcal{I}}_{last}$ onwards in $\mathcal{I}$ there is a successful list-add or list-remove attempt, by \Cref{def:ero:acquisition_pointer}, the claim follows.

        We now finish the proof of Case 1.
        So far we have proved that $\cellpointershort$ is in $\List(\mathcal{I})$ exactly once, $prev(\mathcal{I} \circ s, \cellpointershort)$ is immediately before $\cellpointershort$ in $\List(\mathcal{I})$, and $prev(\mathcal{I}, \cellpointershort)$ is immediately before $\cellpointershort$ in $\List(\mathcal{I})$.
        Together, these facts imply $prev(\mathcal{I}, \cellpointershort) = prev(\mathcal{I} \circ s, \cellpointershort)$.
        However, our initial assumption is that $prev(\mathcal{I}, \cellpointershort) \neq prev(\mathcal{I} \circ s, \cellpointershort)$, so this case is impossible.

        \item[] \hspace{0pt}\textbf{Case 2.} $e^{\mathcal{I}}_{last} = e^{\mathcal{I} \circ s}_{last}$.

        Hence, the last $L$-event in $\mathcal{I}$ and $\mathcal{I} \circ s$ is the same, and so for brevity we drop the superscript and denote it as $e_{last}$.
        Furthermore, $s$ is not an $L$-event.
        Hence, the sequence of $L$-events in $\mathcal{I}$ and $\mathcal{I} \circ s$ are the same, so by \Cref{def:ero:logical_list} $\List(\mathcal{I}) = \List(\mathcal{I} \circ s)$.
        % Since $e_{last}$ is an $L$-event, by \Cref{lemma:ero:every_l_event_is_add_apply_or_remove}, $e_{last}$ is either an $L$-add, $L$-apply, or $L$-remove event.
        There are two cases.

        \begin{itemize}
            \item[] \hspace{0pt}\textbf{Case 2.1.} $e_{last}$ is not an $L$-add nor an $L$-remove event, or from $e_{last}$ onwards in $\mathcal{I}$ there is a successful list-add or list-remove attempt.

            Hence, $e_{last}$ is not an $L$-add nor an $L$-remove event, or from $e_{last}$ onwards in $\mathcal{I} \circ s$ there is a successful list-add or list-remove attempt, so by \Cref{def:ero:acquisition_pointer}, $\cellpointershort$ is in $\List(\mathcal{I} \circ s)$ exactly once, $prev(\mathcal{I} \circ s, \cellpointershort)$ is immediately before $\cellpointershort$ in $\List(\mathcal{I} \circ s)$, and $prev(\mathcal{I}, \cellpointershort)$ is immediately before $\cellpointershort$ in $\List(\mathcal{I})$.
            Thus, since $\List(\mathcal{I}) = \List(\mathcal{I} \circ s)$, we have that $\cellpointershort$ is in $\List(\mathcal{I})$ exactly once, $prev(\mathcal{I} \circ s, \cellpointershort)$ is immediately before $\cellpointershort$ in $\List(\mathcal{I})$, and $prev(\mathcal{I}, \cellpointershort)$ is immediately before $\cellpointershort$ in $\List(\mathcal{I})$.
            Together, these facts imply $prev(\mathcal{I}, \cellpointershort) = prev(\mathcal{I} \circ s, \cellpointershort)$.
            However, our initial assumption is that $prev(\mathcal{I}, \cellpointershort) \neq prev(\mathcal{I} \circ s, \cellpointershort)$, so this case is impossible.
            
            \item[] \hspace{0pt}\textbf{Case 2.2.} $e_{last}$ is an $L$-add or $L$-remove event, and from $e_{last}$ onwards in $\mathcal{I}$ there are no successful list-add or list-remove attempts.

            Hence, by \Cref{def:ero:acquisition_pointer}, $\cellpointershort$ is in $\List(\mathcal{I}^{exclude}_{e_{last}})$ exactly once and $prev(\mathcal{I}, \cellpointershort)$ is immediately before $\cellpointershort$ in $\List(\mathcal{I}^{exclude}_{e_{last}})$ where $\mathcal{I}^{exclude}_{e_{last}}$ is the prefix of $\mathcal{I}$ up to but excluding $e_{last}$.
            Since $e_{last}$ is in $\mathcal{I} \circ s$, we have that $\mathcal{I}^{exclude}_{e_{last}}$ is the prefix of $\mathcal{I} \circ s$ up to but excluding $e_{last}$.
            There are two cases.

            \begin{itemize}
                \item[] \hspace{0pt}\textbf{Case 2.2.1.} $s$ is not a successful list-add nor list-remove attempt.

                Hence, since from $e_{last}$ onwards in $\mathcal{I}$ there are no successful list-add or list-remove attempts, we have that from $e_{last}$ onwards in $\mathcal{I} \circ s$ there are no successful list-add or list-remove attempts.
                Thus, by \Cref{def:ero:acquisition_pointer}, $prev(\mathcal{I} \circ s, \cellpointershort)$ is immediately before $\cellpointershort$ in $\List(\mathcal{I}^{exclude}_{e_{last}})$.
                Therefore, since $\cellpointershort$ is in $\List(\mathcal{I}^{exclude}_{e_{last}})$ exactly once and $prev(\mathcal{I}, \cellpointershort)$ is immediately before $\cellpointershort$ in $\List(\mathcal{I}^{exclude}_{e_{last}})$, we have that $prev(\mathcal{I}, \cellpointershort) = prev(\mathcal{I} \circ s, \cellpointershort)$.
                However, our initial assumption is that $prev(\mathcal{I}, \cellpointershort) \neq prev(\mathcal{I} \circ s, \cellpointershort)$, so this case is impossible.

                \item[] \hspace{0pt}\textbf{Case 2.2.2.} $s$ is a successful list-add or list-remove attempt.

                Hence, by \Cref{def:ero:acquisition_pointer}, $\cellpointershort$ is in $\List(\mathcal{I} \circ s)$ exactly once and $prev(\mathcal{I} \circ s, \cellpointershort)$ is immediately before $\cellpointershort$ in $\List(\mathcal{I} \circ s)$.
                Since $s$ is a successful list-add or list-remove attempt, and by \Cref{lemma:ero:the_list_invariants_hold} $P(\mathcal{I}^\mathcal{B})$, $Q(\mathcal{I}^\mathcal{B})$, and $R(\mathcal{I}^\mathcal{B})$ hold, by \Cref{lemma:ero:any_list_attempt_outside_its_window_is_unsuccessful_alternate_statement}, $s$' corresponding $L$-event $e$, is the last $L$-event before $s$ in $\mathcal{I}^\mathcal{B}$.
                Hence, since $\mathcal{I} \circ s$ is the prefix of $\mathcal{I}^\mathcal{B}$ up to and including $s$, we have that $e$ is the last $L$-event in $\mathcal{I} \circ s$.
                Therefore, since $e_{last}$ is the last $L$-event in $\mathcal{I} \circ s$, we have that $e = e_{last}$, and so $e_{last}$ is $s$' corresponding $L$-event.
                There are three cases.

                \begin{itemize}
                    \item[] \hspace{0pt}\textbf{Case 2.2.2.1.} $s$ is a successful list-add attempt for any pointer.

                    Hence, since $e_{last}$ is $s$' corresponding $L$-event, by \Cref{lemma:ero:l_event_corresponding_to_do_low_level_op}, $e_{last}$ is an $L$-add event.
                    Since $e_{last}$ is the last $L$-event in $\mathcal{I} \circ s$ and $\mathcal{I}^{exclude}_{e_{last}}$ is the prefix of $\mathcal{I} \circ s$ up to but excluding $e_{last}$, we have that the sequences of $L$-events is the same in $\mathcal{I}^{exclude}_{e_{last}}$ and $\mathcal{I} \circ s$ except $e_{last}$ is not in $\mathcal{I}^{exclude}_{e_{last}}$ and $e_{last}$ is in $\mathcal{I} \circ s$.
                    Thus, since $e_{last}$ is an $L$-add event, by \Cref{def:ero:logical_list}, every element in $\List(\mathcal{I}^{exclude}_{e_{last}})$ is in $\List(\mathcal{I} \circ s)$.
                    So, since by the beginning of Case 2.2 $prev(\mathcal{I}, \cellpointershort) \in \List(\mathcal{I}^{exclude}_{e_{last}})$, we have that $prev(\mathcal{I}, \cellpointershort) \in \List(\mathcal{I} \circ s)$.
                    Since $\mathcal{I}^{exclude}_{e_{last}}$ is a prefix of $\mathcal{I} \circ s$, by \Cref{lemma:ero:acquisition_pointer_in_universe_or_head} $prev(\mathcal{I}, \cellpointershort) \in \celluniverse \cup \{\&\headobject{}\}$, by \Cref{lemma:ero:active_implies_from_universe} $\cellpointershort \in \celluniverse$, by the beginning of Case 2.2 $prev(\mathcal{I}, \cellpointershort)$ is immediately before $\cellpointershort$ in $\List(\mathcal{I}^{exclude}_{e_{last}})$, $prev(\mathcal{I}, \cellpointershort) \in \List(\mathcal{I} \circ s)$, by the beginning of Case 2.2.2 $\cellpointershort \in \List(\mathcal{I} \circ s)$, and by \Cref{lemma:ero:the_list_invariants_hold} $P(\mathcal{I} \circ s)$ holds, by \Cref{lemma:ero:list_neighbours_are_the_same_if_they_contained_in_an_extended_list}, $prev(\mathcal{I}, \cellpointershort)$ is immediately before $\cellpointershort$ in $\List(\mathcal{I} \circ s)$.
                    Therefore, since by the beginning of Case 2.2.2 $\cellpointershort$ is in $\List(\mathcal{I} \circ s)$ exactly once and $prev(\mathcal{I} \circ s, \cellpointershort)$ is immediately before $\cellpointershort$ in $\List(\mathcal{I} \circ s)$, we have that $prev(\mathcal{I}, \cellpointershort) = prev(\mathcal{I} \circ s, \cellpointershort)$.
                    However, our initial assumption is that $prev(\mathcal{I}, \cellpointershort) \neq prev(\mathcal{I} \circ s, \cellpointershort)$, so this case is impossible.

                    \item[] \hspace{0pt}\textbf{Case 2.2.2.2.} $s$ is a successful list-remove attempt for a pointer other than $prev(\mathcal{I}, \cellpointershort)$.

                    Hence, since $e_{last}$ is $s$' corresponding $L$-event, by \Cref{lemma:ero:l_event_corresponding_to_do_low_level_op}, $e_{last}$ is an $L$-remove event for $v \neq prev(\mathcal{I}, \cellpointershort)$.
                    Since $e_{last}$ is the last $L$-event in $\mathcal{I} \circ s$ and $\mathcal{I}^{exclude}_{e_{last}}$ is the prefix of $\mathcal{I} \circ s$ up to but excluding $e_{last}$, we have that the sequences of $L$-events is the same in $\mathcal{I}^{exclude}_{e_{last}}$ and $\mathcal{I} \circ s$ except $e_{last}$ is not in $\mathcal{I}^{exclude}_{e_{last}}$ and $e_{last}$ is in $\mathcal{I} \circ s$.
                    Thus, since $e_{last}$ is an $L$-remove event for $v \neq prev(\mathcal{I}, \cellpointershort)$, and by the beginning of Case 2.2 $prev(\mathcal{I}, \cellpointershort) \in \List(\mathcal{I}^{exclude}_{e_{last}})$, by \Cref{def:ero:logical_list}, $prev(\mathcal{I}, \cellpointershort) \in \List(\mathcal{I} \circ s)$.
                    Since $\mathcal{I}^{exclude}_{e_{last}}$ is a prefix of $\mathcal{I} \circ s$, by \Cref{lemma:ero:acquisition_pointer_in_universe_or_head} $prev(\mathcal{I}, \cellpointershort) \in \celluniverse \cup \{\&\headobject{}\}$, by \Cref{lemma:ero:active_implies_from_universe} $\cellpointershort \in \celluniverse$, by the beginning of Case 2.2 $prev(\mathcal{I}, \cellpointershort)$ is immediately before $\cellpointershort$ in $\List(\mathcal{I}^{exclude}_{e_{last}})$, $prev(\mathcal{I}, \cellpointershort) \in \List(\mathcal{I} \circ s)$, by the beginning of Case 2.2.2 $\cellpointershort \in \List(\mathcal{I} \circ s)$, and by \Cref{lemma:ero:the_list_invariants_hold} $P(\mathcal{I} \circ s)$ holds, by \Cref{lemma:ero:list_neighbours_are_the_same_if_they_contained_in_an_extended_list}, $prev(\mathcal{I}, \cellpointershort)$ is immediately before $\cellpointershort$ in $\List(\mathcal{I} \circ s)$.
                    Therefore, since by the beginning of Case 2.2.2 $\cellpointershort$ is in $\List(\mathcal{I} \circ s)$ exactly once and $prev(\mathcal{I} \circ s, \cellpointershort)$ is immediately before $\cellpointershort$ in $\List(\mathcal{I} \circ s)$, we have that $prev(\mathcal{I}, \cellpointershort) = prev(\mathcal{I} \circ s, \cellpointershort)$.
                    However, our initial assumption is that $prev(\mathcal{I}, \cellpointershort) \neq prev(\mathcal{I} \circ s, \cellpointershort)$, so this case is impossible.

                    \item[] \hspace{0pt}\textbf{Case 2.2.2.3.} $s$ is a successful list-remove attempt for $prev(\mathcal{I}, \cellpointershort)$.

                    Let $s$ be a successful list-remove attempt for $prev(\mathcal{I}, \cellpointershort)$ between $\previouscellpointershort$ and $\nextcellpointershort$.
                    We will prove that $\nextcellpointershort = \cellpointershort$ and $\previouscellpointershort = prev(\mathcal{I} \circ s, \cellpointershort)$.
                    Since $s$ is a successful list-remove attempt for $prev(\mathcal{I}, \cellpointershort)$ between $\previouscellpointershort$ and $\nextcellpointershort$ in $\mathcal{I} \circ s$ and by \Cref{lemma:ero:the_list_invariants_hold} $Q(\mathcal{I} \circ s)$ holds, there is an $L$-remove event $e$ for $prev(\mathcal{I}, \cellpointershort)$ before $s$ in $\mathcal{I} \circ s$ such that if $\mathcal{I}^{exclude}_e$ is the prefix of $\mathcal{I} \circ s$ up to but excluding $e$, then $prev(\mathcal{I}, \cellpointershort)$ is in $\List(\mathcal{I}^{exclude}_e)$ exactly once and $\previouscellpointershort$ and $\nextcellpointershort$ are the pointers preceding and succeeding $prev(\mathcal{I}, \cellpointershort)$ in $\List(\mathcal{I}^{exclude}_e)$, respectively.
                    Since $s$ is a successful list-remove attempt for $prev(\mathcal{I}, \cellpointershort)$ and $e_{last}$ is $s$' corresponding $L$-event, by \Cref{lemma:ero:l_event_corresponding_to_do_low_level_op}, $e_{last}$ is an $L$-remove event for $prev(\mathcal{I}, \cellpointershort)$.
                    Hence, since by \Cref{lemma:ero:the_list_invariants_hold} $P(\mathcal{I} \circ s)$ holds, and $e$ and $e_{last}$ are both $L$-remove events for $prev(\mathcal{I}, \cellpointershort)$ in $\mathcal{I} \circ s$, we have that $e = e_{last}$.
                    Thus, $\mathcal{I}^{exclude}_e = \mathcal{I}^{exclude}_{e_{last}}$, and so $prev(\mathcal{I}, \cellpointershort)$ is in $\List(\mathcal{I}^{exclude}_{e_{last}})$ exactly once and $\previouscellpointershort$ and $\nextcellpointershort$ are the pointers preceding and succeeding $prev(\mathcal{I}, \cellpointershort)$ in $\List(\mathcal{I}^{exclude}_{e_{last}})$, respectively.
                    So, since by the beginning of Case 2.2 $prev(\mathcal{I}, \cellpointershort)$ is immediately before $\cellpointershort$ in $\List(\mathcal{I}^{exclude}_{e_{last}})$, we have that $\nextcellpointershort = \cellpointershort$.
                    What remains is to prove that $\previouscellpointershort = prev(\mathcal{I} \circ s, \cellpointershort)$.
                    Since $prev(\mathcal{I}, \cellpointershort)$ is in $\List(\mathcal{I}^{exclude}_{e_{last}})$ exactly once, $\previouscellpointershort$ and $\nextcellpointershort$ are the pointers preceding and succeeding $prev(\mathcal{I}, \cellpointershort)$ in $\List(\mathcal{I}^{exclude}_{e_{last}})$, respectively, and $\nextcellpointershort = \cellpointershort$, by \Cref{def:ero:logical_list}, $\List(\mathcal{I}^{exclude}_{e_{last}}) = \&\headobject{}, \ldots, \previouscellpointershort, prev(\mathcal{I}, \cellpointershort), \cellpointershort, \ldots, \nullconstant$ where the dots represent zero or more pointers.
                    Hence, since $e_{last}$ is the last $L$-event in $\mathcal{I} \circ s$ and $\mathcal{I}^{exclude}_{e_{last}}$ is the prefix of $\mathcal{I} \circ s$ up to but excluding $e_{last}$, we have that the sequences of $L$-events is the same in $\mathcal{I}^{exclude}_{e_{last}}$ and $\mathcal{I} \circ s$ except $e_{last}$ is not in $\mathcal{I}^{exclude}_{e_{last}}$ and $e_{last}$ is in $\mathcal{I} \circ s$, and so since $e_{last}$ is an $L$-remove event for $prev(\mathcal{I}, \cellpointershort)$, by \Cref{def:ero:logical_list}, $\List(\mathcal{I} \circ s) = \&\headobject{}, \ldots, \previouscellpointershort, \cellpointershort, \ldots, \nullconstant$.
                    Thus, $\previouscellpointershort$ is immediately before $\cellpointershort$ in $\List(\mathcal{I} \circ s)$.
                    Therefore, since by the beginning of Case 2.2.2 $\cellpointershort$ is in $\List(\mathcal{I} \circ s)$ exactly once and $prev(\mathcal{I} \circ s, \cellpointershort)$ is immediately before $\cellpointershort$ in $\List(\mathcal{I} \circ s)$, we have that $\previouscellpointershort = prev(\mathcal{I} \circ s, \cellpointershort)$ as wanted.
                    \qH{\Cref{lemma:ero:if_acquisition_pointer_changes_then_its_because_of_a_specific_list_remove_attempt}}
                \end{itemize}                
            \end{itemize}
        \end{itemize}
    \end{itemize}
\end{proof}

We are now ready to prove that the acquisition counter for $\cellpointershort$ is semantically correct.

\begin{proposition}\label{lemma:ero:acquisition_count_is_semantically_correct}
    Suppose $\mathcal{I}^\mathcal{B}$ is finite and $\cellpointershort$ is active in $\mathcal{I}^\mathcal{B}$.
    Then, at the end of $\mathcal{I}^\mathcal{B}$
    \begin{align*}
        (*prev(\mathcal{I}^\mathcal{B}, \cellpointershort)).\nextlong{}.\acquisitions{} = A(\mathcal{I}^\mathcal{B}, \cellpointershort) + 1.
    \end{align*}
\end{proposition}

\begin{proof}
    Since $\cellpointershort$ is active in $\mathcal{I}^\mathcal{B}$, by \Cref{def:ero:active}, there is a single successful list-add attempt $a_{add}$ for $\cellpointershort$ in $\mathcal{I}^\mathcal{B}$.
    To prove this claim, it suffices to prove the following statement.
    Consider any prefix $\mathcal{I}$ of $\mathcal{I}^\mathcal{B}$ where $a_{add}$ is in $\mathcal{I}$.
    By \Cref{lemma:ero:every_prefix_between_i_first_and_i_is_active}, $\cellpointershort$ is active in $\mathcal{I}$ so $prev(\mathcal{I}, \cellpointershort)$ is well-defined.
    Then, at the end of $\mathcal{I}$, $(*prev(\mathcal{I}, \cellpointershort)).\nextlong{}.\acquisitions{} = A(\mathcal{I}, \cellpointershort) + 1$.
    We prove this statement by induction on the step number in $\mathcal{I}^\mathcal{B}$ starting from $a_{add}$ to the end of $\mathcal{I}^\mathcal{B}$.

    \begin{itemize}
        \item[] \hspace{0pt}\textbf{Base Case.} $\mathcal{I}$ is the prefix of $\mathcal{I}^\mathcal{B}$ up to and including $a_{add}$.

        We claim that there are no successful list-acquire-next attempts for $\cellpointershort$ before $a_{add}$ in $\mathcal{I}^\mathcal{B}$.
        Suppose, for contradiction, there is a list-acquire-next attempt $a_{acquire}$ for $\cellpointershort$ before $a_{add}$.
        Let $\mathcal{I}^*$ be the prefix of $\mathcal{I}^\mathcal{B}$ up to and including $a_{acquire}$.
        Hence, by \Cref{lemma:ero:acquisition_for_implies_active_for}, $\cellpointershort$ is active in $\mathcal{I}^*$.
        Thus, by \Cref{def:ero:active}, there is a successful list-add attempt $a$ for $\cellpointershort$ in $\mathcal{I}^*$, so $a < a_{acquire}$.
        Therefore, since $a_{acquire} < a_{add}$, by transitivity $a < a_{add}$, and so there are two successful list-add attempts for $\cellpointershort$ in $\mathcal{I}^\mathcal{B}$.
        However, by \Cref{lemma:ero:at_most_one_successful_list_add_attempt}, there is at most one successful list-add attempt for $\cellpointershort$ in $\mathcal{I}^\mathcal{B}$, a contradiction.
        Since there are no successful list-acquire-next attempts for $\cellpointershort$ before $a_{add}$ in $\mathcal{I}^\mathcal{B}$, and $a_{add}$ is the last step in $\mathcal{I}$, we have that there are no successful list-acquire-next attempts for $\cellpointershort$ in $\mathcal{I}$.
        Therefore, by \Cref{def:ero:number_of_successful_acquires}, $A(\mathcal{I}, \cellpointershort) = 0$, and so we must prove that at the end of $\mathcal{I}$, $(*prev(\mathcal{I}, \cellpointershort)).\nextlong{}.\acquisitions{} = 1$.

        Since $\cellpointershort$ is active in $\mathcal{I}$, by \Cref{corollary:ero:active_prefix_has_last_l_event}, there is a last $L$-event in $\mathcal{I}$; say $e_{last}$.
        Hence, since $\mathcal{I}$ is the prefix of $\mathcal{I}^\mathcal{B}$ up to and including $a_{add}$, we have that from $e_{last}$ onwards in $\mathcal{I}$ there is a successful list-add attempt (namely $a_{add}$).
        Thus, since $\cellpointershort$ is active in $\mathcal{I}$, by \Cref{def:ero:acquisition_pointer}, $\cellpointershort$ is in $\List(\mathcal{I})$ exactly once and $prev(\mathcal{I}, \cellpointershort)$ is the pointer preceding $\cellpointershort$ in $\List(\mathcal{I})$.
        Therefore, since $a_{add}$ is the last step of $\mathcal{I}$, $a_{add}$ is a successful list-add attempt for $\cellpointershort$ after some $\previouscellpointershort$, by \Cref{lemma:ero:nice_list_property_of_successful_list_add_and_remove_attempt}, $\previouscellpointershort$ is immediately before $\cellpointershort$ in $\List(\mathcal{I})$, and so $prev(\mathcal{I}, \cellpointershort) = \previouscellpointershort$.
        Hence, $a_{add}$ is a successful list-add attempt for $\cellpointershort$ after $prev(\mathcal{I}, \cellpointershort)$.
        Thus, by \cref{line:ero:add_cell_to_list}, $(*prev(\mathcal{I}, \cellpointershort)).\nextlong{}.\acquisitions{} = 1$ at $a_{add}$.
        Therefore, since $a_{add}$ is the last step in $\mathcal{I}$, we have that $(*prev(\mathcal{I}, \cellpointershort)).\nextlong{}.\acquisitions{} = 1$ at the end of $\mathcal{I}$ as required.
        
        \item[] \hspace{0pt}\textbf{Inductive Case.} If the claim holds for some proper prefix $\mathcal{I}$ of $\mathcal{I}^\mathcal{B}$ then it holds for $\mathcal{I} \circ s$ where $s$ is the step after $\mathcal{I}$ in $\mathcal{I}^\mathcal{B}$ ($s$ is well-defined since $\mathcal{I}$ is a proper prefix of $\mathcal{I}^\mathcal{B}$).

        Suppose for some proper prefix $\mathcal{I}$ of $\mathcal{I}^\mathcal{B}$ that $(*prev(\mathcal{I}, \cellpointershort)).\nextlong{}.\acquisitions{} = A(\mathcal{I}, \cellpointershort) + 1$ at the end of $\mathcal{I}$.
        This is the inductive hypothesis.
        We will prove that $(*prev(\mathcal{I} \circ s, \cellpointershort)).\nextlong{}.\acquisitions{} = A(\mathcal{I} \circ s, \cellpointershort) + 1$ at the end of $\mathcal{I} \circ s$ where $s$ is the step after $\mathcal{I}$ in $\mathcal{I}^\mathcal{B}$.    
        Since $\cellpointershort$ is active in $\mathcal{I}$ and $\mathcal{I} \circ s$, by \Cref{lemma:ero:acquisition_pointer_in_universe_or_head}, $prev(\mathcal{I}, \cellpointershort)$ and $prev(\mathcal{I} \circ s, \cellpointershort)$ are in $\celluniverse \cup \{\&\headobject{}\}$.  
        There are two cases.

        \begin{itemize}
            \item[] \hspace{0pt}\textbf{Case 1.} $prev(\mathcal{I}, \cellpointershort) = prev(\mathcal{I} \circ s, \cellpointershort)$.

            There are two more cases.

            \begin{itemize}
                \item[] \hspace{0pt}\textbf{Case 1.1.} $s$ is a successful list-acquire-next attempt for $\cellpointershort$.

                Hence, by \Cref{def:ero:number_of_successful_acquires}, $A(\mathcal{I} \circ s, \cellpointershort) = A(\mathcal{I}, \cellpointershort) + 1$.
                Furthermore, by \Cref{lemma:ero:successful_acquire_attempt_for_ptr_is_after_prev_ptr}, $s$ is a successful list-acquire-next attempt for $\cellpointershort$ after $prev(\mathcal{I} \circ s, \cellpointershort)$.
                Hence, by \Cref{def:ero:english}, $s$ is an execution of the form $\CASop{}((*prev(\mathcal{I} \circ s, \cellpointershort)).\nextlong{}, (\arbitraryvalue, \arbitraryvalue, a, \cellpointershort), (\arbitraryvalue, \arbitraryvalue, a + 1, \cellpointershort))$.
                Thus, since $s$ is successful we have that (1) $(*prev(\mathcal{I} \circ s, \cellpointershort)).\nextlong{}.\acquisitions = a$ at the step before $s$ (the end of $\mathcal{I}$) and (2) $(*prev(\mathcal{I} \circ s, \cellpointershort)).\nextlong{}.\acquisitions = a + 1$ at $s$ (the end of $\mathcal{I} \circ s$).
                Since $prev(\mathcal{I}, \cellpointershort) = prev(\mathcal{I} \circ s, \cellpointershort)$, by (1), we have that $(*prev(\mathcal{I}, \cellpointershort)).\nextlong{}.\acquisitions = a$ at the end of $\mathcal{I}$.
                Hence, since by the inductive hypothesis $(*prev(\mathcal{I}, \cellpointershort)).\nextlong{}.\acquisitions{} = A(\mathcal{I}, \cellpointershort) + 1$ at the end of $\mathcal{I}$, we have that $a = A(\mathcal{I}, \cellpointershort) + 1$.
                Thus, by (2), we have that $(*prev(\mathcal{I} \circ s, \cellpointershort)).\nextlong{}.\acquisitions = A(\mathcal{I}, \cellpointershort) + 2$ at the end of $\mathcal{I} \circ s$.
                Therefore, since $A(\mathcal{I} \circ s, \cellpointershort) = A(\mathcal{I}, \cellpointershort) + 1$, we have that $(*prev(\mathcal{I} \circ s, \cellpointershort)).\nextlong{}.\acquisitions{} = A(\mathcal{I} \circ s, \cellpointershort) + 1$ at the end of $\mathcal{I} \circ s$ as required.
            
                \item[] \hspace{0pt}\textbf{Case 1.2.} $s$ is not a successful list-acquire-next attempt for $\cellpointershort$.

                Hence, by \Cref{def:ero:number_of_successful_acquires}, $A(\mathcal{I}, \cellpointershort) = A(\mathcal{I} \circ s, \cellpointershort)$.
                Furthermore, since $\cellpointershort$ is active in $\mathcal{I} \circ s$, by \Cref{lemma:ero:successful_acquire_attempt_after_prev_ptr_is_for_ptr}, $s$ is not a successful list-acquire-next attempt after $prev(\mathcal{I} \circ s, \cellpointershort)$.
                
                We now prove that $s$ is not a successful list-add attempt after $prev(\mathcal{I} \circ s, \cellpointershort)$ and $s$ is not a successful list-remove attempt between $prev(\mathcal{I} \circ s, \cellpointershort)$ and some pointer.
                Suppose, for contradiction, that $s$ is a successful list-add attempt for $\nextcellpointershort$ after $prev(\mathcal{I} \circ s, \cellpointershort)$ or a successful list-remove attempt between $prev(\mathcal{I} \circ s, \cellpointershort)$ and $\nextcellpointershort$.
                Hence, by \Cref{lemma:ero:nice_list_property_of_successful_list_add_and_remove_attempt}, $prev(\mathcal{I} \circ s, \cellpointershort)$ is immediately before $\nextcellpointershort$ in $\List(\mathcal{I} \circ s)$.
                Furthermore, the last step in $\mathcal{I} \circ s$ is a successful list-add or list-remove attempt, and so since $\cellpointershort$ is active in $\mathcal{I} \circ s$, by \Cref{def:ero:acquisition_pointer}, $prev(\mathcal{I} \circ s, \cellpointershort)$ is immediately before $\cellpointershort$ in $\List(\mathcal{I} \circ s)$.
                Hence, $\cellpointershort = \nextcellpointershort$.
                Therefore, $s$ is either a successful list-add attempt for $\cellpointershort$ after $prev(\mathcal{I} \circ s, \cellpointershort)$ or a successful list-remove attempt between $prev(\mathcal{I} \circ s, \cellpointershort)$ and $\cellpointershort$.
                In the former case, since $a_{add}$ is in $\mathcal{I}$, and $a_{add}$ is a successful list-add attempt for $\cellpointershort$, this implies there are two successful list-add attempts for $\cellpointershort$, a contradiction to \Cref{lemma:ero:at_most_one_successful_list_add_attempt}.
                In the latter case, since $s$ is a successful list-remove attempt for $\cellpointershort$, by \Cref{def:ero:active}, $\cellpointershort$ is not active in $\mathcal{I} \circ s$, a contradiction to the fact that $\cellpointershort$ is active in $\mathcal{I} \circ s$.

                We now finish the proof of Case 1.2.
                Since $prev(\mathcal{I} \circ s, \cellpointershort) \in \celluniverse \cup \{\&\headobject{}\}$, $s$ is not a successful list-add attempt after $prev(\mathcal{I} \circ s, \cellpointershort)$, $s$ is not a successful list-remove attempt between $prev(\mathcal{I} \circ s, \cellpointershort)$ and some pointer, and $s$ is not a successful list-acquire-next attempt after $prev(\mathcal{I} \circ s, \cellpointershort)$, by \Cref{observation:ero:where_objects_change}, $(*prev(\mathcal{I} \circ s, \cellpointershort)).\nextlong.\acquisitions$ is the same at the end of $\mathcal{I}$ and at the end of $\mathcal{I} \circ s$.
                Thus, since $prev(\mathcal{I}, \cellpointershort) = prev(\mathcal{I} \circ s, \cellpointershort)$, $(*prev(\mathcal{I}, \cellpointershort)).\nextlong.\acquisitions$ at the end of $\mathcal{I}$ is equal to $(*prev(\mathcal{I} \circ s, \cellpointershort)).\nextlong.\acquisitions$ at the end of $\mathcal{I} \circ s$.
                Hence, since by the inductive hypothesis, $(*prev(\mathcal{I}, \cellpointershort)).\nextlong{}.\acquisitions{} = A(\mathcal{I}, \cellpointershort) + 1$ at the end of $\mathcal{I}$, we have that $(*prev(\mathcal{I} \circ s, \cellpointershort)).\nextlong{}.\acquisitions{} = A(\mathcal{I}, \cellpointershort) + 1$ at the end of $\mathcal{I} \circ s$.
                Therefore, since $A(\mathcal{I}, \cellpointershort) = A(\mathcal{I} \circ s, \cellpointershort)$, we have that $(*prev(\mathcal{I} \circ s, \cellpointershort)).\nextlong{}.\acquisitions{} = A(\mathcal{I} \circ s, \cellpointershort) + 1$ at the end of $\mathcal{I} \circ s$ as required.
            \end{itemize}

            \item[] \hspace{0pt}\textbf{Case 2.} $prev(\mathcal{I}, \cellpointershort) \neq prev(\mathcal{I} \circ s, \cellpointershort)$.

            Hence, by \Cref{lemma:ero:if_acquisition_pointer_changes_then_its_because_of_a_specific_list_remove_attempt}, $s$ is a successful list-remove attempt for $prev(\mathcal{I}, \cellpointershort)$ between $prev(\mathcal{I} \circ s, \cellpointershort)$ and $\cellpointershort$.
            Thus, by \Cref{def:ero:number_of_successful_acquires}, $A(\mathcal{I}, \cellpointershort) = A(\mathcal{I} \circ s, \cellpointershort)$.
            Let $p$ be the process that executed $s$ and let $T^{\ref{line:ero:remove_cell_read_pointer_to_remove}}$ be the time of $p$'s last execution of \cref{line:ero:remove_cell_read_pointer_to_remove} before $s$.
            Hence, by \Cref{lemma:ero:list_seal_before_list_remove}, there is a successful list-seal attempt $a_{seal}$ for $prev(\mathcal{I}, \cellpointershort)$ before $T^{\ref{line:ero:remove_cell_read_pointer_to_remove}}$ in $\mathcal{I}^\mathcal{B}$.
            We now prove that there are no successful list-add attempts after $prev(\mathcal{I}, \cellpointershort)$, successful list-remove attempts between $prev(\mathcal{I}, \cellpointershort)$ and some pointer, and successful list-acquire-next attempts after $prev(\mathcal{I}, \cellpointershort)$ from $T^{\ref{line:ero:remove_cell_read_pointer_to_remove}}$ onwards in $\mathcal{I}^\mathcal{B}$, implying from $T^{\ref{line:ero:remove_cell_read_pointer_to_remove}}$ onwards in $\mathcal{I}^\mathcal{B}$ $(*prev(\mathcal{I}, \cellpointershort)).\nextlong{}.\acquisitions$ is unchanged.

            We first prove that there are no successful list-add attempts after $prev(\mathcal{I}, \cellpointershort)$ from $T^{\ref{line:ero:remove_cell_read_pointer_to_remove}}$ onwards in $\mathcal{I}^\mathcal{B}$.
            Suppose, for contradiction, there is a successful list-add attempt $a_{add}$ after $prev(\mathcal{I}, \cellpointershort)$ at or after $T^{\ref{line:ero:remove_cell_read_pointer_to_remove}}$ in $\mathcal{I}^\mathcal{B}$.
            Therefore, by \Cref{lemma:ero:no_list_seal_before_add}, there are no successful list-seal attempts for $prev(\mathcal{I}, \cellpointershort)$ before $a_{add}$ in $\mathcal{I}^\mathcal{B}$.
            However, since $a_{seal} <  T^{\ref{line:ero:remove_cell_read_pointer_to_remove}}$, and $T^{\ref{line:ero:remove_cell_read_pointer_to_remove}} \leq a_{add}$, by transitivity, $a_{seal} < a_{add}$, so there is a successful list-seal attempt for $prev(\mathcal{I}, \cellpointershort)$ before $a_{add}$ in $\mathcal{I}^\mathcal{B}$, a contradiction.

            We now prove that there are no successful list-remove attempts between $prev(\mathcal{I}, \cellpointershort)$ and some pointer from $T^{\ref{line:ero:remove_cell_read_pointer_to_remove}}$ onwards in $\mathcal{I}^\mathcal{B}$.
            Suppose, for contradiction, there is a successful list-remove attempt $a_{remove}$ between $prev(\mathcal{I}, \cellpointershort)$ and some pointer at or after $T^{\ref{line:ero:remove_cell_read_pointer_to_remove}}$ in $\mathcal{I}^\mathcal{B}$.
            Therefore, by \Cref{lemma:ero:no_list_seal_before_remove}, there are no successful list-seal attempts for $prev(\mathcal{I}, \cellpointershort)$ before $a_{add}$ in $\mathcal{I}^\mathcal{B}$.
            However, since $a_{seal} <  T^{\ref{line:ero:remove_cell_read_pointer_to_remove}}$, and $T^{\ref{line:ero:remove_cell_read_pointer_to_remove}} \leq a_{remove}$, by transitivity, $a_{seal} < a_{remove}$, so there is a successful list-seal attempt for $prev(\mathcal{I}, \cellpointershort)$ before $a_{remove}$ in $\mathcal{I}^\mathcal{B}$, a contradiction.

            We now prove that there are no successful list-acquire-next attempts after $prev(\mathcal{I}, \cellpointershort)$ from $T^{\ref{line:ero:remove_cell_read_pointer_to_remove}}$ onwards in $\mathcal{I}^\mathcal{B}$.
            Suppose, for contradiction, there is a successful list-acquire-next attempt $a_{acquire}$ after $prev(\mathcal{I}, \cellpointershort)$ at or after $T^{\ref{line:ero:remove_cell_read_pointer_to_remove}}$ in $\mathcal{I}^\mathcal{B}$.
            Therefore, by \Cref{lemma:ero:no_list_seal_before_acquire}, there are no successful list-seal attempts for $prev(\mathcal{I}, \cellpointershort)$ before $a_{acquire}$ in $\mathcal{I}^\mathcal{B}$.
            However, since $a_{seal} <  T^{\ref{line:ero:remove_cell_read_pointer_to_remove}}$, and $T^{\ref{line:ero:remove_cell_read_pointer_to_remove}} \leq a_{acquire}$, by transitivity, $a_{seal} < a_{acquire}$, so there is a successful list-seal attempt for $prev(\mathcal{I}, \cellpointershort)$ before $a_{acquire}$ in $\mathcal{I}^\mathcal{B}$, a contradiction.

            We now finish the proof of Case 2.
            Since $prev(\mathcal{I}, \cellpointershort) \in \celluniverse \cup \{\&\headobject{}\}$, and there are no successful list-add attempts after $prev(\mathcal{I}, \cellpointershort)$, successful list-remove attempts between $prev(\mathcal{I}, \cellpointershort)$ and some pointer, and successful list-acquire-next attempts after $prev(\mathcal{I}, \cellpointershort)$ from $T^{\ref{line:ero:remove_cell_read_pointer_to_remove}}$ onwards in $\mathcal{I}^\mathcal{B}$, by \Cref{observation:ero:where_objects_change}, from $T^{\ref{line:ero:remove_cell_read_pointer_to_remove}}$ onwards in $\mathcal{I}^\mathcal{B}$ $(*prev(\mathcal{I}, \cellpointershort)).\nextlong{}.\acquisitions$ is unchanged.
            Let $(*prev(\mathcal{I}, \cellpointershort)).\nextlong{}.\acquisitions = a$ at $T^{\ref{line:ero:remove_cell_read_pointer_to_remove}}$, so $(*prev(\mathcal{I}, \cellpointershort)).\nextlong{}.\acquisitions = a$ from $T^{\ref{line:ero:remove_cell_read_pointer_to_remove}}$ onwards in $\mathcal{I}^\mathcal{B}$.
            Hence, since by the inductive hypothesis $(*prev(\mathcal{I}, \cellpointershort)).\nextlong{}.\acquisitions{} = A(\mathcal{I}, \cellpointershort) + 1$ at the end of $\mathcal{I}$, and $T^{\ref{line:ero:remove_cell_read_pointer_to_remove}}$ is in $\mathcal{I}$ (because $T^{\ref{line:ero:remove_cell_read_pointer_to_remove}} < s$), we have that $a = A(\mathcal{I}, \cellpointershort) + 1$.
            Since $T^{\ref{line:ero:remove_cell_read_pointer_to_remove}}$ is the time of $p$'s last execution of \cref{line:ero:remove_cell_read_pointer_to_remove} before $s$, $(*prev(\mathcal{I}, \cellpointershort)).\nextlong{}.\acquisitions = a$ at $T^{\ref{line:ero:remove_cell_read_pointer_to_remove}}$, and $s$ is a successful list-remove attempt for $prev(\mathcal{I}, \cellpointershort)$ between $prev(\mathcal{I} \circ s, \cellpointershort)$ and $\cellpointershort$, we have that $p$ read $a$ from $(*prev(\mathcal{I}, \cellpointershort)).\nextlong{}.\acquisitions$ at $T^{\ref{line:ero:remove_cell_read_pointer_to_remove}}$ and set $(*prev(\mathcal{I} \circ s, \cellpointershort)).\nextlong{}.\acquisitions = a$ at $s$ (equivalently, the end of $\mathcal{I} \circ s$).
            Hence, since $a = A(\mathcal{I}, \cellpointershort) + 1$, we have that $(*prev(\mathcal{I} \circ s, \cellpointershort)).\nextlong{}.\acquisitions = A(\mathcal{I}, \cellpointershort) + 1$ at the end of $\mathcal{I} \circ s$.
            Therefore, since $A(\mathcal{I}, \cellpointershort) = A(\mathcal{I} \circ s, \cellpointershort)$, we have that $(*prev(\mathcal{I} \circ s, \cellpointershort)).\nextlong{}.\acquisitions = A(\mathcal{I} \circ s, \cellpointershort) + 1$ at the end of $\mathcal{I} \circ s$ as required.
            \qH{\Cref{lemma:ero:acquisition_count_is_semantically_correct}}
        \end{itemize}
    \end{itemize}
\end{proof}

We are now ready to prove the main claim of this section.

\begin{lemma}\label{lemma:ero:acquire_copy_copies_the_final_number_of_acquisitions}
    For any acquire-copy event $e$ for $\cellpointershort$ in $\mathcal{I}^\mathcal{B}$ the following are true:
    \begin{compactenum}
        \item $e = \FAop{}((*\cellpointershort).\revocations, -(A(\mathcal{I}^\mathcal{B}, \cellpointershort) + 1))$; and
        \item if $\mathcal{I}$ is the prefix of $\mathcal{I}^\mathcal{B}$ up to and including $e$ then $A(\mathcal{I}, \cellpointershort) = A(\mathcal{I}^\mathcal{B}, \cellpointershort)$.
    \end{compactenum}
\end{lemma}

\begin{proof}
    Let $p$ be the process that executed $e$ and let $I$ be the invocation of the \doremovecell{} procedure that $e$ was executed during.
    Hence, by \Cref{lemma:ero:every_acquire_copy_event_is_for_pointer_from_universe} $\cellpointershort \in \celluniverse$ and by \Cref{def:ero:english} the second parameter of $I$ is $\cellpointershort$.
    Furthermore, by \Cref{lemma:ero:before_acquire_copy_is_successful}, $p$ performed a successful list-remove attempt $a_{remove}$ for $\cellpointershort$ before $e$ during $I$. 
    Since $e$ is for $\cellpointershort$, by \Cref{def:ero:english}, $e$ is of the form $\FAop{}((*\cellpointershort).\revocations, \arbitraryvalue)$, and so the remainder of the proof is dedicated to showing that the second parameter of $e$ is $-(A(\mathcal{I}^\mathcal{B}, \cellpointershort) + 1)$ (2 is proved along the way).

    We start with some definitions and basic facts.
    Let $T^{\ref{line:ero:remove_cell_read_previous_pointer}}$ be the time of $p$'s last execution of \cref{line:ero:remove_cell_read_previous_pointer} before $a_{remove}$, so the step at time $T^{\ref{line:ero:remove_cell_read_previous_pointer}}$ is during $I$.
    Hence, since $a_{remove}$ is successful, we have that $T^{\ref{line:ero:remove_cell_read_previous_pointer}}$ is the last time $p$ executes \cref{line:ero:remove_cell_read_previous_pointer} during $I$.
    Since $a_{remove}$ is a successful list-remove attempt for $\cellpointershort$ in $\mathcal{I}^\mathcal{B}$, by \Cref{lemma:ero:at_most_one_successful_list_remove_attempt}, $a_{remove}$ is the only successful list-remove attempt for $\cellpointershort$ in $\mathcal{I}^\mathcal{B}$.
    Furthermore, by \Cref{lemma:ero:l_event_corresponding_to_do_low_level_op}, $a_{remove}$'s corresponding $L$-event $e_{remove}$, is an $L$-remove event for $\cellpointershort$ before $p$ invoked $I$, and so $e_{remove} < T^{\ref{line:ero:remove_cell_read_previous_pointer}}$.
    Hence, since by \Cref{lemma:ero:the_list_invariants_hold} $R(\mathcal{I}^\mathcal{B})$ holds, by \Cref{lemma:ero:list_add_before_list_remove}, there is a successful list-add attempt $a_{add}$ for $\cellpointershort$ before $e_{remove}$ in $\mathcal{I}^\mathcal{B}$.
    Thus, by \Cref{lemma:ero:at_most_one_successful_list_add_attempt}, $a_{add}$ is the only successful list-add attempt for $\cellpointershort$ in $\mathcal{I}^\mathcal{B}$.
    Finally, since $e_{remove}$ is $a_{remove}$'s corresponding $L$-event, and by \Cref{lemma:ero:the_list_invariants_hold} $P(\mathcal{I}^\mathcal{B})$, $Q(\mathcal{I}^\mathcal{B})$, and $R(\mathcal{I}^\mathcal{B})$ hold, by \Cref{lemma:ero:any_list_attempt_outside_its_window_is_unsuccessful_alternate_statement}, $e_{remove}$ is the last $L$-event before $a_{remove}$ in $\mathcal{I}^\mathcal{B}$.
    Hence, $e_{remove}$ is the last $L$-event in $\mathcal{I}^{exclude}_{a_{remove}}$ where $\mathcal{I}^{exclude}_{a_{remove}}$ is the prefix of $\mathcal{I}^\mathcal{B}$ up to but excluding $a_{remove}$.

    \begin{claimcustom}{\ref{lemma:ero:acquire_copy_copies_the_final_number_of_acquisitions}.1}\label{lemma:ero:acquire_copy_copies_the_final_number_of_acquisitions:claim_one}
        $\cellpointershort$ is active in $\mathcal{I}^{exclude}_{a_{remove}}$.
    \end{claimcustom}

    \begin{proof}
        Since $a_{add} < e_{remove}$ and $e_{remove} < a_{remove}$, by transitivity, $a_{add} < a_{remove}$, and so $a_{add}$ is in $\mathcal{I}^{exclude}_{a_{remove}}$.
        Hence, since $a_{add}$ is the only successful list-add attempt for $\cellpointershort$ in $\mathcal{I}^\mathcal{B}$ and $\mathcal{I}^{exclude}_{a_{remove}}$ is the prefix of $\mathcal{I}^\mathcal{B}$, it follows there is a single successful list-add attempt for $\cellpointershort$ in $\mathcal{I}^{exclude}_{a_{remove}}$.
        Furthermore, since $a_{remove}$ is not in $\mathcal{I}^{exclude}_{a_{remove}}$, $a_{remove}$ is the only successful list-remove attempt for $\cellpointershort$ in $\mathcal{I}^\mathcal{B}$, and $\mathcal{I}^{exclude}_{a_{remove}}$ is the prefix of $\mathcal{I}^\mathcal{B}$, we have that there are no successful list-remove attempts for $\cellpointershort$ in $\mathcal{I}^{exclude}_{a_{remove}}$.
        Therefore, by \Cref{def:ero:active}, $\cellpointershort$ is active in $\mathcal{I}^{exclude}_{a_{remove}}$ as wanted.
        \qH{\Cref{lemma:ero:acquire_copy_copies_the_final_number_of_acquisitions:claim_one}}
    \end{proof}

    \begin{claimcustom}{\ref{lemma:ero:acquire_copy_copies_the_final_number_of_acquisitions}.2}\label{lemma:ero:acquire_copy_copies_the_final_number_of_acquisitions:claim_two}
        From $e_{remove}$ onwards in $\mathcal{I}^{exclude}_{a_{remove}}$ there are no successful list-add or list-remove attempts.
    \end{claimcustom}

    \begin{proof}
        Since $e_{remove}$ is the last $L$-event in $\mathcal{I}^{exclude}_{a_{remove}}$, by \Cref{lemma:ero:the_list_invariants_hold} $P(\mathcal{I}^{exclude}_{a_{remove}})$, $Q(\mathcal{I}^{exclude}_{a_{remove}})$, and $R(\mathcal{I}^{exclude}_{a_{remove}})$ hold, and $e_{remove}$ is an $L$-remove event for $\cellpointershort$, by \Cref{lemma:ero:3_of_r_safety_holds}, from $e_{remove}$ onwards in $\mathcal{I}^{exclude}_{a_{remove}}$ there is at most one successful list-remove attempt for $\cellpointershort$ and no other successful list-remove or list-add attempt for any pointer.
        Therefore, since $a_{remove}$ is the only successful list-remove attempt for $\cellpointershort$ in $\mathcal{I}^\mathcal{B}$ and $a_{remove}$ is not in $\mathcal{I}^{exclude}_{a_{remove}}$, we have that from $e_{remove}$ onwards in $\mathcal{I}^{exclude}_{a_{remove}}$ there are no successful list-add or list-remove attempt as wanted.
        \qH{\Cref{lemma:ero:acquire_copy_copies_the_final_number_of_acquisitions:claim_two}}
    \end{proof}

    \begin{claimcustom}{\ref{lemma:ero:acquire_copy_copies_the_final_number_of_acquisitions}.3}\label{lemma:ero:acquire_copy_copies_the_final_number_of_acquisitions:claim_three}
        $a_{remove}$ is a successful list-remove attempt for $\cellpointershort$ between $prev(\mathcal{I}^{exclude}_{a_{remove}}, \cellpointershort)$ and some pointer.
    \end{claimcustom}

    \begin{proof}
        Recall that $a_{remove}$ is a successful list-remove attempt for $\cellpointershort$.
        Let $a_{remove}$ be a successful list-remove attempt for $\cellpointershort$ between $\previouscellpointershort$ and some pointer.
        We must prove that $\previouscellpointershort = prev(\cellpointershort, \mathcal{I}^{exclude}_{a_{remove}})$.
        Since $\mathcal{I}^{exclude}_{a_{remove}}$ is finite, by \Cref{lemma:ero:acquire_copy_copies_the_final_number_of_acquisitions:claim_one} $\cellpointershort$ is active in $\mathcal{I}^{exclude}_{a_{remove}}$, $e_{remove}$ is the last $L$-event in $\mathcal{I}^{exclude}_{a_{remove}}$, $e_{remove}$ is an $L$-remove event, and by \Cref{lemma:ero:acquire_copy_copies_the_final_number_of_acquisitions:claim_two} from $e_{remove}$ onwards in $\mathcal{I}^{exclude}_{a_{remove}}$ there are no successful list-add or list-remove events, by \Cref{def:ero:acquisition_pointer}, $\cellpointershort$ is in $\List(\mathcal{I}^{exclude}_{e_{remove}})$ exactly once and $prev(\mathcal{I}^{exclude}_{a_{remove}}, \cellpointershort)$ is immediately before $\cellpointershort$ in $\List(\mathcal{I}^{exclude}_{e_{remove}})$ where $\mathcal{I}^{exclude}_{e_{remove}}$ is the prefix of $\mathcal{I}^{exclude}_{a_{remove}}$ up to but excluding $e_{remove}$.
        Note that since $\mathcal{I}^{exclude}_{a_{remove}}$ is a prefix of $\mathcal{I}^\mathcal{B}$, $\mathcal{I}^{exclude}_{e_{remove}}$ is also the prefix of $\mathcal{I}^\mathcal{B}$ up to but excluding $e_{remove}$.
        Since $a_{remove}$ is a list-remove attempt for $\cellpointershort$ between $\previouscellpointershort$ and some pointer in $\mathcal{I}^\mathcal{B}$ and by \Cref{lemma:ero:the_list_invariants_hold} $Q(\mathcal{I}^\mathcal{B})$ holds, we have that before $a_{remove}$ in $\mathcal{I}^\mathcal{B}$ there is a $L$-remove event $e$ for $\cellpointershort$ such that if $\mathcal{I}^{exclude}_e$ is the prefix of $\mathcal{I}^\mathcal{B}$ up to but excluding $e$, then $\cellpointershort$ is in $\List(\mathcal{I}^{exclude}_e)$ exactly once and $\previouscellpointershort$ is the pointer preceding $\cellpointershort$ in $\List(\mathcal{I}^{exclude}_e)$.
        Since $e$ and $e_{remove}$ are both $L$-remove events for $\cellpointershort$ in $\mathcal{I}^\mathcal{B}$, and by \Cref{lemma:ero:the_list_invariants_hold} $P(\mathcal{I}^\mathcal{B})$ holds, we have that $e = e_{remove}$.
        Hence, $\mathcal{I}^{exclude}_e$ is the prefix of $\mathcal{I}^\mathcal{B}$ up to but excluding $e_{remove}$, and so $\mathcal{I}^{exclude}_e = \mathcal{I}^{exclude}_{e_{remove}}$.
        Thus, since $\previouscellpointershort$ is the pointer preceding $\cellpointershort$ in $\List(\mathcal{I}^{exclude}_e)$, we have that $\previouscellpointershort$ is the pointer preceding $\cellpointershort$ in $\List(\mathcal{I}^{exclude}_{e_{remove}})$.
        Therefore, since $\cellpointershort$ is in $\List(\mathcal{I}^{exclude}_{e_{remove}})$ exactly once and $prev(\mathcal{I}^{exclude}_{a_{remove}}, \cellpointershort)$ is immediately before $\cellpointershort$ in $\List(\mathcal{I}^{exclude}_{e_{remove}})$, we have that $\previouscellpointershort = prev(\mathcal{I}^{exclude}_{a_{remove}}, \cellpointershort)$ as wanted.
        \qH{\Cref{lemma:ero:acquire_copy_copies_the_final_number_of_acquisitions:claim_three}}
    \end{proof}

    \begin{claimcustom}{\ref{lemma:ero:acquire_copy_copies_the_final_number_of_acquisitions}.4}\label{lemma:ero:acquire_copy_copies_the_final_number_of_acquisitions:claim_four}
        For every prefix $\mathcal{I}$ of $\mathcal{I}^{exclude}_{a_{remove}}$ if $T^{\ref{line:ero:remove_cell_read_previous_pointer}}$ is in $\mathcal{I}$, then $\cellpointershort$ is active in $\mathcal{I}$ and $prev(\mathcal{I}, \cellpointershort) = prev(\mathcal{I}^{exclude}_{a_{remove}}, \cellpointershort)$.
    \end{claimcustom}

    \begin{proof}
        We first prove that $\cellpointershort$ is active in $\mathcal{I}$.
        Since by \Cref{lemma:ero:acquire_copy_copies_the_final_number_of_acquisitions:claim_one} $\cellpointershort$ is active in $\mathcal{I}^{exclude}_{a_{remove}}$, by \Cref{def:ero:active}, there is a single successful list-add attempt $a$ for $\cellpointershort$ in $\mathcal{I}^{exclude}_{a_{remove}}$.
        Since $a_{add}$ is the only successful list-add attempt for $\cellpointershort$ in $\mathcal{I}^\mathcal{B}$ and $\mathcal{I}^{exclude}_{a_{remove}}$ is a prefix of $\mathcal{I}^\mathcal{B}$, we have that $a = a_{add}$.
        Hence, since $a_{add} < e_{remove}$ and $e_{remove} < T^{\ref{line:ero:remove_cell_read_previous_pointer}}$, by transitivity, $a_{add} < T^{\ref{line:ero:remove_cell_read_previous_pointer}}$, and so $a_{add}$ is in $\mathcal{I}$.
        Therefore, by \Cref{lemma:ero:every_prefix_between_i_first_and_i_is_active}, $\cellpointershort$ is active in $\mathcal{I}$ as wanted.
        So, $prev(\mathcal{I}, \cellpointershort)$ is well-defined.

        We now prove that $prev(\mathcal{I}, \cellpointershort) = prev(\mathcal{I}^{exclude}_{a_{remove}}, \cellpointershort)$.
        Since $e_{remove} < T^{\ref{line:ero:remove_cell_read_previous_pointer}}$, we have that $e_{remove}$ is in $\mathcal{I}$.
        Hence, since $\mathcal{I}$ is a prefix of $\mathcal{I}^{exclude}_{a_{remove}}$ and $e_{remove}$ is the last $L$-event in $\mathcal{I}^{exclude}_{a_{remove}}$, we have that $e_{remove}$ is the last $L$-event in $\mathcal{I}$.
        Furthermore, since by \Cref{lemma:ero:acquire_copy_copies_the_final_number_of_acquisitions:claim_two} from $e_{remove}$ onwards in $\mathcal{I}^{exclude}_{a_{remove}}$ there are no successful list-add or list-remove events, we have that from $e_{remove}$ onwards in $\mathcal{I}$ there are no successful list-add or list-remove events.
        Since $\mathcal{I}$ is finite, $\cellpointershort$ is active in $\mathcal{I}$, $e_{remove}$ is the last $L$-event in $\mathcal{I}$, $e_{remove}$ is an $L$-remove event, and from $e_{remove}$ onwards in $\mathcal{I}$ there are no successful list-add or list-remove events, by \Cref{def:ero:acquisition_pointer}, $\cellpointershort$ is in $\List(\mathcal{I}')$ exactly once and $prev(\mathcal{I}, \cellpointershort)$ is immediately before $\cellpointershort$ in $\List(\mathcal{I}')$ where $\mathcal{I}'$ is the prefix of $\mathcal{I}$ up to but excluding $e_{remove}$.
        Note that since $\mathcal{I}$ is a prefix of $\mathcal{I}^{exclude}_{a_{remove}}$, $\mathcal{I}'$ is also the prefix of $\mathcal{I}^{exclude}_{a_{remove}}$ up to but excluding $e_{remove}$.
        Since $\mathcal{I}^{exclude}_{a_{remove}}$ is finite, $\cellpointershort$ is active in $\mathcal{I}^{exclude}_{a_{remove}}$, $e_{remove}$ is the last $L$-event in $\mathcal{I}^{exclude}_{a_{remove}}$, $e_{remove}$ is an $L$-remove event, and from $e_{remove}$ onwards in $\mathcal{I}^{exclude}_{a_{remove}}$ there are no successful list-add or list-remove attempts, by \Cref{def:ero:acquisition_pointer}, $\cellpointershort$ is in $\List(\mathcal{I}^*)$ exactly once and $prev(\mathcal{I}^{exclude}_{a_{remove}}, \cellpointershort)$ is immediately before $\cellpointershort$ in $\List(\mathcal{I}^*)$ where $\mathcal{I}^*$ is the prefix of $\mathcal{I}^{exclude}_{a_{remove}}$ up to but excluding $e_{remove}$.
        Hence, since $\mathcal{I}'$ is the prefix of $\mathcal{I}^{exclude}_{a_{remove}}$ up to but excluding $e_{remove}$, we have that $\mathcal{I}' = \mathcal{I}^*$.
        Thus, since $prev(\mathcal{I}^{exclude}_{a_{remove}}, \cellpointershort)$ is immediately before $\cellpointershort$ in $\List(\mathcal{I}^*)$, we have that $prev(\mathcal{I}^{exclude}_{a_{remove}}, \cellpointershort)$ is immediately before $\cellpointershort$ in $\List(\mathcal{I}')$.
        Therefore, since $\cellpointershort$ is in $\List(\mathcal{I}')$ exactly once and $prev(\mathcal{I}, \cellpointershort)$ is immediately before $\cellpointershort$ in $\List(\mathcal{I}')$, we have that $prev(\mathcal{I}, \cellpointershort) = prev(\mathcal{I}^{exclude}_{a_{remove}}, \cellpointershort)$ as wanted.
        \qH{\Cref{lemma:ero:acquire_copy_copies_the_final_number_of_acquisitions:claim_four}}
    \end{proof}

    \begin{claimcustom}{\ref{lemma:ero:acquire_copy_copies_the_final_number_of_acquisitions}.5}\label{lemma:ero:acquire_copy_copies_the_final_number_of_acquisitions:claim_five}
        There are no successful list-acquire-next attempts for $\cellpointershort$ after $T^{\ref{line:ero:remove_cell_read_previous_pointer}}$ in $\mathcal{I}^\mathcal{B}$.
    \end{claimcustom}

    \begin{proof}
        Suppose, for contradiction, there is a successful list-acquire-next attempt $a_{acquire}$ for $\cellpointershort$ after $T^{\ref{line:ero:remove_cell_read_previous_pointer}}$ in $\mathcal{I}^\mathcal{B}$.
        There are two cases.
        
        \begin{itemize}
            \item[] \hspace{0pt}\textbf{Case 1.} $a_{remove} < a_{acquire}$.

            Let $\mathcal{I}$ be the prefix of $\mathcal{I}^\mathcal{B}$ up to and including $a_{acquire}$.
            Hence, since $a_{acquire}$ is a successful list-acquire-next attempt for $\cellpointershort$, by \Cref{lemma:ero:acquisition_for_implies_active_for}, $\cellpointershort$ is active in $\mathcal{I}$.
            Thus, by \Cref{def:ero:active}, there are no successful list-remove attempts for $\cellpointershort$ in $\mathcal{I}$.
            Therefore, since $\mathcal{I}$ is the prefix of $\mathcal{I}^\mathcal{B}$ up to and including $a_{acquire}$, we have that there are no successful list-remove attempts for $\cellpointershort$ before $a_{acquire}$ in $\mathcal{I}^\mathcal{B}$.
            However, since $a_{remove} < a_{acquire}$, we have that there is a successful list-remove attempt for $\cellpointershort$ before $a_{acquire}$ in $\mathcal{I}^\mathcal{B}$, a contradiction.

            \item[] \hspace{0pt}\textbf{Case 2.} $a_{acquire} < a_{remove}$.

            Hence, since $T^{\ref{line:ero:remove_cell_read_previous_pointer}} < a_{acquire}$, we have that $T^{\ref{line:ero:remove_cell_read_previous_pointer}} < a_{acquire} < a_{remove}$.
            Thus, there is a prefix $\mathcal{I}$ of $\mathcal{I}^{exclude}_{a_{remove}}$ up to and including $a_{acquire}$ such that $T^{\ref{line:ero:remove_cell_read_previous_pointer}}$ is in $\mathcal{I}$.
            So, by \Cref{lemma:ero:acquire_copy_copies_the_final_number_of_acquisitions:claim_four}, $prev(\mathcal{I}, \cellpointershort) = prev(\mathcal{I}^{exclude}_{a_{remove}}, \cellpointershort)$.
            Hence, since the last step of $\mathcal{I}$, $a_{acquire}$, is a successful list-acquire-next attempt for $\cellpointershort$, by \Cref{lemma:ero:successful_acquire_attempt_for_ptr_is_after_prev_ptr}, $a_{acquire}$ is a successful list-acquire-next attempt for $\cellpointershort$ after $prev(\mathcal{I}, \cellpointershort)$.
            Thus, since $prev(\mathcal{I}, \cellpointershort) = prev(\mathcal{I}^{exclude}_{a_{remove}}, \cellpointershort)$, we have that $a_{acquire}$ is a successful list-acquire-next attempt for $\cellpointershort$ after $prev(\mathcal{I}^{exclude}_{a_{remove}}, \cellpointershort)$.
            Since $\mathcal{I}^{exclude}_{a_{remove}}$ is finite, and by \Cref{lemma:ero:acquire_copy_copies_the_final_number_of_acquisitions:claim_one} $\cellpointershort$ is active in $\mathcal{I}^{exclude}_{a_{remove}}$, by \Cref{lemma:ero:acquisition_pointer_in_universe_or_head}, $prev(\mathcal{I}^{exclude}_{a_{remove}}, \cellpointershort) \in \celluniverse \cup \{\&\headobject{}\}$.
            Hence, by \Cref{def:ero:english}, $a_{acquire}$ is of the form \CASop{}$((*prev(\mathcal{I}^{exclude}_{a_{remove}}, \cellpointershort)).\nextlong, (v, \arbitraryvalue, \arbitraryvalue, \arbitraryvalue), (v + 1, \arbitraryvalue, \arbitraryvalue, \arbitraryvalue))$ for some view $v$.
            Thus, since $a_{acquire}$ is successful, $(*prev(\mathcal{I}^{exclude}_{a_{remove}}, \cellpointershort)).\nextlong.\view = v$ at the step before $a_{acquire}$, and $(*prev(\mathcal{I}^{exclude}_{a_{remove}}, \cellpointershort)).\nextlong.\view = v + 1$ at $a_{acquire}$.
            So, since $prev(\mathcal{I}^{exclude}_{a_{remove}}, \cellpointershort) \in \celluniverse \cup \{\&\headobject{}\}$, by \Cref{observation:ero:views_are_monotonic}, $(*prev(\mathcal{I}^{exclude}_{a_{remove}}, \cellpointershort)).\nextlong.\view$ is monotonically increasing, and $T^{\ref{line:ero:remove_cell_read_previous_pointer}} < a_{acquire} < a_{remove}$, it follows that (1) $(*prev(\mathcal{I}^{exclude}_{a_{remove}}, \cellpointershort)).\nextlong.\view \leq v$ at $T^{\ref{line:ero:remove_cell_read_previous_pointer}}$, and (2) $(*prev(\mathcal{I}^{exclude}_{a_{remove}}, \cellpointershort)).\nextlong.\view > v$ at the step before $a_{remove}$.
            Since by \Cref{lemma:ero:acquire_copy_copies_the_final_number_of_acquisitions:claim_three} $a_{remove}$ is a list-remove attempt for $\cellpointershort$ between $prev(\mathcal{I}^{exclude}_{a_{remove}}, \cellpointershort)$ and some pointer, by \Cref{def:ero:english} $a_{remove}$ is of the form \CASop{}$((*prev(\mathcal{I}^{exclude}_{a_{remove}}, \cellpointershort)).\nextlong, (v', \arbitraryvalue, \arbitraryvalue, \arbitraryvalue), (v' + 1, \arbitraryvalue, \arbitraryvalue, \arbitraryvalue))$ for some view $v'$.
            Hence, since $T^{\ref{line:ero:remove_cell_read_previous_pointer}}$ is the time of $p$'s last execution of \cref{line:ero:remove_cell_read_previous_pointer} before $a_{remove}$, we have that $p$ saw that $(*prev(\mathcal{I}^{exclude}_{a_{remove}}, \cellpointershort)).\nextlong.\view = v'$ at $T^{\ref{line:ero:remove_cell_read_previous_pointer}}$. 
            Thus, since by (1) $(*prev(\mathcal{I}^{exclude}_{a_{remove}}, \cellpointershort)).\nextlong.\view \leq v$ at $T^{\ref{line:ero:remove_cell_read_previous_pointer}}$, we have that $v' \leq v$.
            Therefore, since $a_{remove}$ is successful, we have that $(*prev(\mathcal{I}^{exclude}_{a_{remove}}, \cellpointershort)).\nextlong.\view = v'$ at the step before $a_{remove}$, and since $v' \leq v$, we have that $(*prev(\mathcal{I}^{exclude}_{a_{remove}}, \cellpointershort)).\nextlong.\view \leq v$ at the step before $a_{remove}$.
            However, by (2) $(*prev(\mathcal{I}^{exclude}_{a_{remove}}, \cellpointershort)).\nextlong.\view > v$ at the step before $a_{remove}$, a contradiction.
            \qH{\Cref{lemma:ero:acquire_copy_copies_the_final_number_of_acquisitions:claim_five}}
        \end{itemize}
    \end{proof}

    We now finish the proof of \Cref{lemma:ero:acquire_copy_copies_the_final_number_of_acquisitions}.
    Let $\mathcal{I}^{\ref{line:ero:remove_cell_read_previous_pointer}}$ be the prefix of $\mathcal{I}^\mathcal{B}$ up to and including $T^{\ref{line:ero:remove_cell_read_previous_pointer}}$.
    Hence, since by \Cref{lemma:ero:acquire_copy_copies_the_final_number_of_acquisitions:claim_five} there are no successful list-acquire-next attempts for $\cellpointershort$ after $T^{\ref{line:ero:remove_cell_read_previous_pointer}}$ in $\mathcal{I}^\mathcal{B}$, by \Cref{def:ero:number_of_successful_acquires}, $A(\mathcal{I}^{\ref{line:ero:remove_cell_read_previous_pointer}}, \cellpointershort) = A(\mathcal{I}^\mathcal{B}, \cellpointershort)$.
    Furthermore, since $T^{\ref{line:ero:remove_cell_read_previous_pointer}}$ is in $\mathcal{I}^{\ref{line:ero:remove_cell_read_previous_pointer}}$ and $T^{\ref{line:ero:remove_cell_read_previous_pointer}} < a_{remove}$, we have that $\mathcal{I}^{\ref{line:ero:remove_cell_read_previous_pointer}}$ is also the prefix of $\mathcal{I}^{exclude}_{a_{remove}}$ up to and including $T^{\ref{line:ero:remove_cell_read_previous_pointer}}$.
    Thus, by \Cref{lemma:ero:acquire_copy_copies_the_final_number_of_acquisitions:claim_four}, $\cellpointershort$ is active in $\mathcal{I}^{\ref{line:ero:remove_cell_read_previous_pointer}}$ and $prev(\mathcal{I}^{\ref{line:ero:remove_cell_read_previous_pointer}}, \cellpointershort) = prev(\mathcal{I}^{exclude}_{a_{remove}}, \cellpointershort)$.
    Since $\mathcal{I}^{\ref{line:ero:remove_cell_read_previous_pointer}}$ is finite, and $\cellpointershort$ is active in $\mathcal{I}^{\ref{line:ero:remove_cell_read_previous_pointer}}$, by \Cref{lemma:ero:acquisition_count_is_semantically_correct}, $(*prev(\mathcal{I}^{\ref{line:ero:remove_cell_read_previous_pointer}}, \cellpointershort)).\nextlong{}.\acquisitions{} = A(\mathcal{I}^{\ref{line:ero:remove_cell_read_previous_pointer}}, \cellpointershort) + 1$ at the end of $\mathcal{I}^{\ref{line:ero:remove_cell_read_previous_pointer}}$.
    Hence, since $prev(\mathcal{I}^{\ref{line:ero:remove_cell_read_previous_pointer}}, \cellpointershort) = prev(\mathcal{I}^{exclude}_{a_{remove}}, \cellpointershort)$, and $A(\mathcal{I}^{\ref{line:ero:remove_cell_read_previous_pointer}}, \cellpointershort) = A(\mathcal{I}^\mathcal{B}, \cellpointershort)$, we have that $(*prev(\mathcal{I}^{exclude}_{a_{remove}}), \cellpointershort).\nextlong{}.\acquisitions{} = A(\mathcal{I}^\mathcal{B}, \cellpointershort) + 1$ at the end of $\mathcal{I}^{\ref{line:ero:remove_cell_read_previous_pointer}}$ (equivalently, $T^{\ref{line:ero:remove_cell_read_previous_pointer}}$).
    Thus, since $T^{\ref{line:ero:remove_cell_read_previous_pointer}}$ is the time of $p$'s last execution of \cref{line:ero:remove_cell_read_previous_pointer} before $a_{remove}$, and by \Cref{lemma:ero:acquire_copy_copies_the_final_number_of_acquisitions:claim_three} $a_{remove}$ is a successful list-remove attempt for $\cellpointershort$ between $prev(\mathcal{I}^{exclude}_{a_{remove}}, \cellpointershort)$ and some pointer, we have that $p$ saw that $(*prev(\mathcal{I}^{exclude}_{a_{remove}}, \cellpointershort)).\nextlong.\acquisitions = A(\mathcal{I}^\mathcal{B}, \cellpointershort) + 1$ on \cref{line:ero:remove_cell_read_previous_pointer} at $T^{\ref{line:ero:remove_cell_read_previous_pointer}}$.
    Hence, since $T^{\ref{line:ero:remove_cell_read_previous_pointer}}$ is the last time $p$ executes \cref{line:ero:remove_cell_read_previous_pointer} before $a_{remove}$ during $I$, and $p$ executes $e$ on \cref{line:ero:copy_acquisitions_to_revocations} immediately after $a_{remove}$ on \cref{line:ero:remove_cell_from_list} during $I$, it follows that the second parameter of $e$ is $-(A(\mathcal{I}^\mathcal{B}, \cellpointershort) + 1)$.
    Therefore, $e = \FAop{}((*\cellpointershort).\revocations, -(A(\mathcal{I}^\mathcal{B}, \cellpointershort) + 1))$ as wanted.
    We now prove 2.
    Let $\mathcal{I}$ be the prefix of $\mathcal{I}^\mathcal{B}$ up to and including $e$.
    Since $T^{\ref{line:ero:remove_cell_read_previous_pointer}} < a_{remove}$ and $a_{remove} < e$, by transitivity, $T^{\ref{line:ero:remove_cell_read_previous_pointer}} < e$, and so $\mathcal{I}^{\ref{line:ero:remove_cell_read_previous_pointer}}$ is a prefix of $\mathcal{I}$.
    Therefore, since $A(\mathcal{I}^{\ref{line:ero:remove_cell_read_previous_pointer}}, \cellpointershort) = A(\mathcal{I}^\mathcal{B}, \cellpointershort)$, by \Cref{def:ero:number_of_successful_acquires}, $A(\mathcal{I}, \cellpointershort) = A(\mathcal{I}^\mathcal{B}, \cellpointershort)$ as wanted.
    \qH{\Cref{lemma:ero:acquire_copy_copies_the_final_number_of_acquisitions}}
\end{proof}

\subsubsection{$\mathcal{B}$ correctly manages cells}

In this section, we finish the proof of the $\mathcal{B}$ correctly manages cells theorem.
Recall that all that remains is to prove the third bullet, i.e., every operation on an object of the cell pointed to by $\cellpointershort$ in $\mathcal{I}^\mathcal{B}$ is after an $\allocatecelloperation{}$ operation whose response is $\cellpointershort$, and is before any $\freecelloperation{}(\cellpointershort)$ operation.
We start by proving the first half.

\begin{proposition}\label{lemma:ero:allocate_for_ptr_before_operation_on_ptr}
    For every $\cellpointershort \in \celluniverse$ and operation $o$ on and object of the cell pointed to by $\cellpointershort$ in $\mathcal{I}^\mathcal{B}$, there is an $\allocatecelloperation{}$ operation whose response is $\cellpointershort$ before $o$ in $\mathcal{I}^\mathcal{B}$.
\end{proposition}

\begin{proof}
    Consider any operation $o$ on $\cellpointershort$ in $\mathcal{I}^\mathcal{B}$.
    Let $opx$ be the operation execution that $o$ was performed during.
    Since there are no operations on any pointer before \cref{line:ero:allocate_cell} in $opx$, we have that the process that executed $opx$ performed an $\allocatecelloperation{}$ operation $o_A$ on \cref{line:ero:allocate_cell} before $o$ during $opx$ in $\mathcal{I}^\mathcal{B}$.
    Let $p$ be the process that executed $opx$.
    There are two cases.
    
    \begin{itemize}
        \item[] \hspace{0pt}\textbf{Case 1.} $o_A$'s response is $\cellpointershort$.

        Hence, since $o_A$ is before $o$ in $\mathcal{I}^\mathcal{B}$, there is an $\allocatecelloperation{}$ operation whose response is $\cellpointershort$ before $o$ in $\mathcal{I}^\mathcal{B}$ as wanted.

        \item[] \hspace{0pt}\textbf{Case 2.} $o_A$'s response is not $\cellpointershort$.

        Let $\mathcal{I}$ be the prefix of $\mathcal{I}^\mathcal{B}$ up to and including $o$.
        Since $o_A$'s response is not $\cellpointershort$, we have that $o$ is not on line \ref{line:ero:copy_response_out_of_cell}, \ref{line:ero:do_work_initialize_response}, \ref{line:ero:do_work_while_loop}, or \ref{line:ero:relinquish_revocations} during an invocation of the Relinquish procedure invoked on \cref{line:ero:owner_relinquish}.
        Hence, since the last step of $\mathcal{I}$ is $p$ performing an operation on an object of the cell pointed to by $\cellpointershort \in \celluniverse$, by \Cref{lemma:ero:operation_acquisition_invariant}, $R(\mathcal{I}^-, opx, \cellpointershort) \geq 1$ where $\mathcal{I}^-$ is the prefix of $\mathcal{I}$ excluding the last step.
        Thus, by \Cref{def:ero:reference_count}, $p$ performed a successful list-acquire-next attempt $a_{acquire}$ for $\cellpointershort$ before $o$ in $\mathcal{I}^\mathcal{B}$.
        Let $\mathcal{I}'$ be the prefix of $\mathcal{I}^\mathcal{B}$ up to and including $a_{acquire}$.
        Since the last step of $\mathcal{I}'$, $a_{acquire}$, is a successful list-acquire-next attempt for $\cellpointershort$, by \Cref{lemma:ero:acquisition_for_implies_active_for}, $\cellpointershort$ is active in $\mathcal{I}'$, and so by \Cref{def:ero:active}, there is a successful list-add attempt for $\cellpointershort$ in $\mathcal{I}'$.
        Hence, by \Cref{lemma:ero:l_event_corresponding_to_do_low_level_op}, there is an $L$-add event for $\cellpointershort$ in $\mathcal{I}'$.
        Thus, since $\mathcal{I}'$ is the prefix of $\mathcal{I}^\mathcal{B}$ up to and including $a_{acquire}$, and $a_{acquire}$ is before $o$, we have that there is an $L$-add event for $\cellpointershort$ before $o$ in $\mathcal{I}^\mathcal{B}$.
        So, by \Cref{lemma:ero:l_event_corresponding_a_event_type_and_pointer_relations}, there is an $A$-add event $e_{add}$ for $\cellpointershort$ before $o$.
        Let $q$ be the process that executed $e_{add}$.
        Hence, by \Cref{def:ero:english}, $q$ executed $e_{add}$ during an invocation $I$ of the \doworkuntildone{} with parameters $(\addcell, \cellpointershort)$.
        Thus, $q$ invoked $I$ on \cref{line:ero:low_level_add_cell} during an invocation $I'$ of the \highleveloperation{} procedure, and so $p$ executed \cref{line:ero:allocate_cell} with response $\cellpointershort$ during $I'$.
        Therefore, since $p$'s execution of \cref{line:ero:allocate_cell} during $I'$ is before $p$ invoked $I$, $p$ executed $e_{add}$ during $I$, and $e_{add}$ is before $o$, by transitivity, there is an $\allocatecelloperation{}$ operation whose response is $\cellpointershort$ before $o$ in $\mathcal{I}^\mathcal{B}$ as wanted.
        \qH{\Cref{lemma:ero:allocate_for_ptr_before_operation_on_ptr}}
    \end{itemize}    
\end{proof}

Now all that remains is to prove the second half of the third bullet of the $\mathcal{B}$ correctly manages cells theorem.
Recall from the last section that our strategy for doing so is to prove that for all times at and after a $\freecelloperation{}(\cellpointershort)$ operation, no process has the right to access $\cellpointershort$ (\Cref{lemma:ero:fixed_references_at_free_operation}).

\begin{proposition}\label{lemma:ero:there_is_at_most_one_more_revocation_than_acquisition}
    If $\mathcal{I}^\mathcal{B}$ is finite, then $A(\mathcal{I}^\mathcal{B}, \cellpointershort) - X(\mathcal{I}^\mathcal{B}, \cellpointershort) \geq -1$.
\end{proposition}

\begin{proof}
    Consider any $\cellpointershort$.
    Since by \Cref{alg:lazy_cell_manager_specification} the responses on \cref{line:ero:allocate_cell} are unique, by \Cref{lemma:ero:reference_count_is_non_negative}, there is at most one operation execution in $\mathcal{I}^\mathcal{B}$, say $opx_{\cellpointershort}$, such that $R(\mathcal{I}^\mathcal{B}, opx_{\cellpointershort}, \cellpointershort) \geq -1$, and every other operation execution $opx$ in $\mathcal{I}^\mathcal{B}$ has $R(\mathcal{I}^\mathcal{B}, opx, \cellpointershort) \geq 0$.
    Therefore, 
    \begin{align*}
        \sum_{\text{$opx$ is an operation execution in $\mathcal{I}^\mathcal{B}$}} R(\mathcal{I}^\mathcal{B}, opx, \cellpointershort) \geq -1
    \end{align*}
    and so by \Cref{lemma:ero:relate_a_x_and_r} $A(\mathcal{I}^\mathcal{B}, \cellpointershort) - X(\mathcal{I}^\mathcal{B}, \cellpointershort) \geq -1$ as wanted.
    \qH{\Cref{lemma:ero:there_is_at_most_one_more_revocation_than_acquisition}}
\end{proof}

\begin{lemma}\label{lemma:ero:fixed_references_at_free_operation}
    Consider any $\cellpointershort \in \celluniverse$, any $\freecelloperation{}(\cellpointershort)$ operation $o_F$ in $\mathcal{I}^\mathcal{B}$, and finite prefix $\mathcal{I}$ of $\mathcal{I}^\mathcal{B}$ such that $o_F$ is in $\mathcal{I}$.
    By \Cref{lemma:ero:every_free_ptr_is_after_an_allocate_ptr}, there is an $\allocatecelloperation{}$ operation $o_{A}$ whose response is $\cellpointershort$ before $o_F$ in $\mathcal{I}^\mathcal{B}$ which is unique by \Cref{alg:lazy_cell_manager_specification}, so $o_A$ is in $\mathcal{I}$.
    Let $opx_{\cellpointershort}$ be the operation execution that $o_A$ was performed during, so $opx_{\cellpointershort}$ was invoked in $\mathcal{I}$.
    Then, $R(\mathcal{I}, opx_{\cellpointershort}, \cellpointershort) = -1$ and $R(\mathcal{I}, opx, \cellpointershort) = 0$ for every operation execution $opx \neq opx_{\cellpointershort}$ in $\mathcal{I}$.
\end{lemma}

\begin{proof}
    Let $p$ be the process that executed $o_F$ and let $I$ be the invocation of the Relinquish procedure that $p$ executed $o_F$ during.
    Hence, $p$ found the condition on \cref{line:ero:free_cell} to be true during $I$.
    Let $e_{revocation}$ be the execution of \cref{line:ero:relinquish_revocations} during $I$.
    Hence, since $o_F$ has parameters $\cellpointershort$, $e_{revocation}$ is of the form $\FAop{}((*\cellpointershort).\revocations, 1)$, so by \Cref{def:ero:english}, $e_{revocation}$ is a revocation event for $\cellpointershort$.
    Furthermore, since $p$ found the condition on \cref{line:ero:free_cell} to be true during $I$, we have that the response of $e_{revocation}$ is $-1$.
    Since $\cellpointershort \in \celluniverse$, by definition, $(*\cellpointershort).\revocations$ is initially 0.
    Furthermore, by \Cref{observation:ero:where_objects_change}, the only steps that change the value of $(*\cellpointershort).\revocations$ are acquire-copy events for $\cellpointershort$ and revocation events for $\cellpointershort$.
    Hence, since $(*\cellpointershort).\revocations$ is initially 0, each revocation event for $\cellpointershort$ increases the value of $(*\cellpointershort).\revocations$ by 1, and the response of $e_{revocation}$ is $-1$, we have that there is an acquire-copy event $e_{acquire-copy}$ for $\cellpointershort$ before $e_{revocation}$ in $\mathcal{I}^\mathcal{B}$.
    Thus, by \Cref{lemma:ero:at_most_one_acquisition_copy}, $e_{acquire-copy}$ is the only acquire-copy event for $\cellpointershort$ in $\mathcal{I}^\mathcal{B}$.
    Furthermore, by \Cref{lemma:ero:acquire_copy_copies_the_final_number_of_acquisitions}, $e_{acquire-copy}$ is $\FAop{}((*\cellpointershort).\revocations, -(A(\mathcal{I}^\mathcal{B}, \cellpointershort) + 1))$.
    Since the response of $e_{revocation}$ is $-1$, we have that the $(*\cellpointershort).\revocations = 0$ at $e_{revocation}$.
    Hence, since $(*\cellpointershort).\revocations$ is initially 0, by \Cref{observation:ero:where_objects_change} the only steps that change the value of $(*\cellpointershort).\revocations$ are acquire-copy events for $\cellpointershort$ and revocation events for $\cellpointershort$, $e_{acquire-copy}$ is the only acquire-copy event for $\cellpointershort$ in $\mathcal{I}^\mathcal{B}$, $e_{acquire-copy}$ is $\FAop{}((*\cellpointershort).\revocations, -(A(\mathcal{I}^\mathcal{B}, \cellpointershort) + 1))$, each revocation event for $\cellpointershort$ increases the value of $(*\cellpointershort).\revocations$ by 1, and $e_{acquire-copy} < e_{revocation}$, we have that there are exactly $A(\mathcal{I}^\mathcal{B}, \cellpointershort) + 1$ revocation events for $\cellpointershort$ before or at $e_{revocation}$ in $\mathcal{I}^\mathcal{B}$.
    Let $\mathcal{I}^{include}_{e_{revocation}}$ be the prefix of $\mathcal{I}^\mathcal{B}$ up to and including $e_{revocation}$.
    Hence, by \Cref{def:ero:number_of_successful_acquires}, $X(\mathcal{I}^{include}_{e_{revocation}}, \cellpointershort) = A(\mathcal{I}^\mathcal{B}, \cellpointershort) + 1$.
    Thus, since $\mathcal{I}^{include}_{e_{revocation}}$ is finite, by \Cref{lemma:ero:there_is_at_most_one_more_revocation_than_acquisition}, $A(\mathcal{I}^{include}_{e_{revocation}}, \cellpointershort) - X(\mathcal{I}^{include}_{e_{revocation}}, \cellpointershort) \geq -1$, and so $A(\mathcal{I}^{include}_{e_{revocation}}, \cellpointershort) \geq A(\mathcal{I}^\mathcal{B}, \cellpointershort)$.
    Hence, since $\mathcal{I}^{include}_{e_{revocation}}$ is a prefix of $\mathcal{I}^\mathcal{B}$, by \Cref{def:ero:number_of_successful_acquires}, $A(\mathcal{I}^{include}_{e_{revocation}}, \cellpointershort) \leq A(\mathcal{I}^\mathcal{B}, \cellpointershort)$.
    Therefore, $A(\mathcal{I}^{include}_{e_{revocation}}, \cellpointershort) = A(\mathcal{I}^\mathcal{B}, \cellpointershort)$.

    We now prove that there are no successful list-acquire-next attempts or revocation events for $\cellpointershort$ after $e_{revocation}$ in $\mathcal{I}^\mathcal{B}$.
    Since $A(\mathcal{I}^{include}_{e_{revocation}}, \cellpointershort) = A(\mathcal{I}^\mathcal{B}, \cellpointershort)$ and $\mathcal{I}^{include}_{e_{revocation}}$ is the prefix of $\mathcal{I}^\mathcal{B}$ up to and including $e_{revocation}$, by \Cref{def:ero:number_of_successful_acquires}, there are no more successful list-acquire-next attempts for $\cellpointershort$ after $e_{revocation}$ in $\mathcal{I}^\mathcal{B}$.
    Now suppose, for contradiction, there is a revocation event for $\cellpointershort$ after $e_{revocation}$ in $\mathcal{I}^\mathcal{B}$.
    Let $e$ be the first revocation event for $\cellpointershort$ after $e_{revocation}$ in $\mathcal{I}^\mathcal{B}$ and let $\mathcal{I}^{include}_{e}$ be the prefix of $\mathcal{I}^\mathcal{B}$ up to and including $e$.
    Since $e$ is the first revocation event for $\cellpointershort$ after $e_{revocation}$ and $\mathcal{I}^{include}_{e_{revocation}}$ is the prefix of $\mathcal{I}^\mathcal{B}$ up to and including $e_{revocation}$, by \Cref{def:ero:number_of_successful_acquires}, $X(\mathcal{I}^{include}_{e}, \cellpointershort) = X(\mathcal{I}^{include}_{e_{revocation}}, \cellpointershort) + 1$.
    Hence, since $X(\mathcal{I}^{include}_{e_{revocation}}, \cellpointershort) = A(\mathcal{I}^\mathcal{B}, \cellpointershort) + 1$, we have that $X(\mathcal{I}^{include}_{e}, \cellpointershort) = A(\mathcal{I}^\mathcal{B}, \cellpointershort) + 2$.
    Since there are no successful list-acquire-next attempts for $\cellpointershort$ after $e_{revocation}$ in $\mathcal{I}^\mathcal{B}$, by \Cref{def:ero:number_of_successful_acquires}, $A(\mathcal{I}^{include}_{e}, \cellpointershort) = A(\mathcal{I}^{include}_{e_{revocation}}, \cellpointershort)$, and so since $A(\mathcal{I}^{include}_{e_{revocation}}, \cellpointershort) = A(\mathcal{I}^\mathcal{B}, \cellpointershort)$, we have that $A(\mathcal{I}^{include}_{e}, \cellpointershort) = A(\mathcal{I}^\mathcal{B}, \cellpointershort)$.
    Therefore, $A(\mathcal{I}^{include}_{e}, \cellpointershort) - X(\mathcal{I}^{include}_{e}, \cellpointershort) = -2$.
    However, since $\mathcal{I}^{include}_{e}$ is finite, by \Cref{lemma:ero:there_is_at_most_one_more_revocation_than_acquisition}, $A(\mathcal{I}^{include}_{e}, \cellpointershort) - X(\mathcal{I}^{include}_{e}, \cellpointershort) \geq -1$, a contradiction.
    
    Since $\mathcal{I}^{include}_{e_{revocation}}$ is the prefix of $\mathcal{I}^\mathcal{B}$ up to and including $e_{revocation}$, $e_{revocation} < o_F$, and $\mathcal{I}$ is a finite prefix of $\mathcal{I}^\mathcal{B}$ such that $o_F$ is in $\mathcal{I}$, we have that $\mathcal{I}^{include}_{e_{revocation}}$ is a prefix of $\mathcal{I}$.
    Hence, since there are no successful list-acquire-next attempts or revocation events for $\cellpointershort$ after $e_{revocation}$ in $\mathcal{I}^\mathcal{B}$, by \Cref{def:ero:number_of_successful_acquires},  $A(\mathcal{I}^{include}_{e_{revocation}}, \cellpointershort) = A(\mathcal{I}, \cellpointershort)$ and $X(\mathcal{I}^{include}_{e_{revocation}}, \cellpointershort) = X(\mathcal{I}, \cellpointershort)$.
    Thus, since $A(\mathcal{I}^{include}_{e_{revocation}}, \cellpointershort) = A(\mathcal{I}^\mathcal{B}, \cellpointershort)$ and $X(\mathcal{I}^{include}_{e_{revocation}}, \cellpointershort) = A(\mathcal{I}^\mathcal{B}, \cellpointershort) + 1$, we have that $A(\mathcal{I}, \cellpointershort) = A(\mathcal{I}^\mathcal{B}, \cellpointershort)$ and $X(\mathcal{I}, \cellpointershort) = A(\mathcal{I}^\mathcal{B}, \cellpointershort) + 1$, and so $A(\mathcal{I}, \cellpointershort) - X(\mathcal{I}, \cellpointershort) = -1$.
    Thus, since $\mathcal{I}$ is finite, by \Cref{lemma:ero:relate_a_x_and_r}
    \begin{align*}
        \sum_{\text{$opx$ is an operation execution in $\mathcal{I}$}} R(\mathcal{I}, opx, \cellpointershort) = -1.
    \end{align*}
    Since $opx_{\cellpointershort}$ executed $o_A$ during $opx_{\cellpointershort}$ in $\mathcal{I}$ and $o_A$'s response is $\cellpointershort$, by \Cref{lemma:ero:reference_count_is_non_negative}, we have that $R(opx_{\cellpointershort}, \cellpointershort, \mathcal{I}) \geq -1$.
    Furthermore, since by \Cref{alg:lazy_cell_manager_specification} there is at most one $\allocatecelloperation{}$ operation with response $\cellpointershort$ in $\mathcal{I}^\mathcal{B}$, we have that $opx_{\cellpointershort}$ is the only operation execution in $\mathcal{I}$ that receive $\cellpointershort$ as a response on \cref{line:ero:allocate_cell}.
    Hence, for every operation execution $opx \neq opx_{\cellpointershort}$ in $\mathcal{I}$ if the process that executed $opx$ executed \cref{line:ero:allocate_cell} during $opx$ in $\mathcal{I}$, then its response is not $\cellpointershort$.
    Thus, by \Cref{lemma:ero:reference_count_is_non_negative}, $R(\mathcal{I}, opx, \cellpointershort) \geq 0$.
    Therefore, since (1) $\sum R(\mathcal{I}, opx, \cellpointershort) = -1$, (2) $opx_{\cellpointershort}$ is an operation execution in $\mathcal{I}$ (3) $R(\mathcal{I}, opx_{\cellpointershort}, \cellpointershort) \geq -1$, and (4) for every operation execution $opx \neq opx_{\cellpointershort}$ in $\mathcal{I}$ $R(\mathcal{I}, opx, \cellpointershort) \geq 0$, we have that $R(\mathcal{I}, opx_{\cellpointershort}, \cellpointershort) = -1$ and $R(\mathcal{I}, opx, \cellpointershort) = 0$ for every operation execution $opx \neq opx_{\cellpointershort}$ in $\mathcal{I}$ as wanted.
    \qH{\Cref{lemma:ero:fixed_references_at_free_operation}}
\end{proof}

\begin{proposition}\label{lemma:ero:no_free_for_ptr_before_operation_on_ptr}
    For every $\cellpointershort \in \celluniverse$ and operation $o$ on an object of the cell pointed to by $\cellpointershort$ in $\mathcal{I}^\mathcal{B}$, there are no $\freecelloperation{}(\cellpointershort)$ operations before $o$ in $\mathcal{I}^\mathcal{B}$.
\end{proposition}

\begin{proof}
    Suppose, for contradiction, there is an operation $o$ on an object of the cell pointed to by $\cellpointershort$ in $\mathcal{I}^\mathcal{B}$ and there is a $\freecelloperation{}(\cellpointershort)$ operation $o_F$ before $o$ in $\mathcal{I}^\mathcal{B}$.
    Let $opx$ be the operation execution that $o$ was performed during, and let $\mathcal{I}^{exclude}_{o}$ be the prefix of $\mathcal{I}^\mathcal{B}$ up to but excluding $o$.
    Hence, since $o_F$ is before $o$, we have that $o_F$ is in $\mathcal{I}^{exclude}_{o}$.
    % Hence, since the last step of $opx$ is not an operation on a pointer in $\celluniverse$, $opx$ is pending at the end of $\mathcal{I}$.
    There are two cases.
    \begin{itemize}
        \item[] \hspace{0pt}\textbf{Case 1.} The process that executed $opx$ executed \cref{line:ero:allocate_cell} during $opx$ in $\mathcal{I}^{exclude}_{o}$ with response $\cellpointershort$.

        Let $o_A$ be this $\allocatecelloperation{}$ operation.
        Since $o_A$'s response is $\cellpointershort$, by \Cref{alg:lazy_cell_manager_specification}, $o_A$ is the only $\allocatecelloperation{}$ operation in $\mathcal{I}^\mathcal{B}$ whose response is $\cellpointershort$.
        Therefore, since $\mathcal{I}^{exclude}_{o}$ is a finite prefix of $\mathcal{I}^\mathcal{B}$ such that $o_F$ is in $\mathcal{I}^{exclude}_{o}$, by \Cref{lemma:ero:fixed_references_at_free_operation}, $R(\mathcal{I}^{exclude}_{o}, opx, \cellpointershort) = -1$.
        However, since $\mathcal{I}^{exclude}_{o}$ is a finite prefix of $\mathcal{I}^\mathcal{B}$ such that the step after $\mathcal{I}^{exclude}_{o}$ in $\mathcal{I}^\mathcal{B}$ is the process that executed $opx$ performing an operation on an object of the cell pointed to by $\cellpointershort \in \celluniverse$ during $opx$, by \Cref{lemma:ero:operation_acquisition_invariant}, $R(\mathcal{I}^{exclude}_{o}, opx, \cellpointershort) \geq 0$, a contradiction.

        \item[] \hspace{0pt}\textbf{Case 2.} Otherwise.

        Hence, $o$ was not executed on line \ref{line:ero:copy_response_out_of_cell}, \ref{line:ero:do_work_initialize_response}, \ref{line:ero:do_work_while_loop}, or \ref{line:ero:relinquish_revocations} during an invocation of the Relinquish procedure invoked on \cref{line:ero:owner_relinquish}.
        Therefore, since $\mathcal{I}^{exclude}_{o}$ is a finite prefix of $\mathcal{I}^\mathcal{B}$ such that the step after $\mathcal{I}^{exclude}_{o}$ in $\mathcal{I}^\mathcal{B}$ is the process that executed $opx$ performing an operation on an object of the cell pointed to by $\cellpointershort \in \celluniverse$ during $opx$, by \Cref{lemma:ero:operation_acquisition_invariant}, $R(\mathcal{I}^{exclude}_{o}, opx, \cellpointershort) \geq 1$.
        However, since $\mathcal{I}^{exclude}_{o}$ is a finite prefix of $\mathcal{I}^\mathcal{B}$ such that $o_F$ is in $\mathcal{I}^{exclude}_{o}$, by \Cref{lemma:ero:fixed_references_at_free_operation}, $R(\mathcal{I}^{exclude}_{o}, opx, \cellpointershort) \leq 0$, a contradiction.
        \qH{\Cref{lemma:ero:no_free_for_ptr_before_operation_on_ptr}}
    \end{itemize}
\end{proof}

\begin{theorem}\label{thm:reduction:algorithm_b_is_well_behaved}
    For every $\cellpointershort \in \celluniverse$ the following are true.
    \begin{compactenum}
        \item There is at most one $\allocatecelloperation{}$ operation whose response is $\cellpointershort$, and at most one $\freecelloperation{}(\cellpointershort)$ operation in $\mathcal{I}^\mathcal{B}$.
        \item If there is a $\freecelloperation{}(\cellpointershort)$ operation in $\mathcal{I}^\mathcal{B}$, then it is after an $\allocatecelloperation{}$ operation whose response is $\cellpointershort$.
        \item Every operation on an object of the cell pointed to by $\cellpointershort$ in $\mathcal{I}^\mathcal{B}$ is after an $\allocatecelloperation{}$ operation whose response is $\cellpointershort$, and is before any $\freecelloperation{}(\cellpointershort)$ operation.
    \end{compactenum}
\end{theorem}

\begin{proof}
    \Cref{alg:lazy_cell_manager_specification} and \Cref{lemma:ero:at_most_one_free_per_pointer} imply 1, \Cref{lemma:ero:every_free_ptr_is_after_an_allocate_ptr} implies 2, and Propositions \ref{lemma:ero:allocate_for_ptr_before_operation_on_ptr} and \ref{lemma:ero:no_free_for_ptr_before_operation_on_ptr} imply 3.
    \qH{\Cref{thm:reduction:algorithm_b_is_well_behaved}}
\end{proof}

\subsubsection{$\mathcal{B}$ is space-efficient}

This section proves the $\mathcal{B}$ is space-efficient theorem (\Cref{thm:ero:b_is_space_efficient}).

\begin{proposition}\label{lemma:ero:number_of_acquire_pointers_per_operation_is_at_most_four}
    For every operation execution $opx$ in $\mathcal{I}^\mathcal{B}$ define the set $Acquired(\mathcal{I}^\mathcal{B}, opx)$ as $\cellpointershort \in Acquired(\mathcal{I}^\mathcal{B}, opx)$ if and only if $R(\mathcal{I}^\mathcal{B}, opx, \cellpointershort) > 0$.
    Then, $|Acquire(\mathcal{I}^\mathcal{B}, opx)| \leq 3$.
\end{proposition}

\begin{proof}
    % By \Cref{lemma:ero:total_reference_count_is_bounded} $\sum_{\cellpointershort} R(\mathcal{I}^\mathcal{B}, opx, \cellpointershort) \leq 3$.
    There are two cases.
    \begin{itemize}
        \item[] \hspace{0pt}\textbf{Case 1.} The process that executed $opx$ has not executed \cref{line:ero:owner_relinquish} during $opx$ in $\mathcal{I}^\mathcal{B}$.

        Hence, by \Cref{lemma:ero:reference_count_is_non_negative_before_owner_relinquish}, $R(\mathcal{I}^\mathcal{B}, opx, \cellpointershort) \geq 0$ for every $\cellpointershort$.
        Therefore, since by \Cref{lemma:ero:total_reference_count_is_bounded} $\sum_{\cellpointershort} R(\mathcal{I}^\mathcal{B}, opx, \cellpointershort) \leq 3$, we have that $|Acquire(\mathcal{I}^\mathcal{B}, opx)| \leq 3$ as wanted.
    
        \item[] \hspace{0pt}\textbf{Case 2.} The process that executed $opx$ has executed \cref{line:ero:owner_relinquish} during $opx$ in $\mathcal{I}^\mathcal{B}$.

        Let $\mathcal{I}$ be the prefix of $\mathcal{I}^\mathcal{B}$ up to but excluding the time that the process that executed $opx$ executed \cref{line:ero:owner_relinquish} during $opx$ in $\mathcal{I}^\mathcal{B}$.
        Hence, the process that executed $opx$ has not executed \cref{line:ero:owner_relinquish} during $opx$ in $\mathcal{I}$.
        Thus, by \Cref{lemma:ero:reference_count_is_non_negative_before_owner_relinquish}, $R(\mathcal{I}, opx, \cellpointershort) \geq 0$ for every $\cellpointershort$.
        So, since by \Cref{lemma:ero:total_reference_count_is_bounded} $\sum_{\cellpointershort} R(\mathcal{I}, opx, \cellpointershort) \leq 3$, we have that $|Acquire(\mathcal{I}, opx)| \leq 3$.
        Observe that, after $\mathcal{I}$ in $\mathcal{I}^\mathcal{B}$, the process that executed $opx$ performs zero successful list-acquire-next attempts and at most one revocation event during $opx$.
        Hence, $Acquire(\mathcal{I}^\mathcal{B}, opx) \subseteq Acquire(\mathcal{I}, opx)$.
        Therefore, since $|Acquire(\mathcal{I}, opx)| \leq 3$, we have that $|Acquire(\mathcal{I}^\mathcal{B}, opx)| \leq 3$ as wanted.
        \qH{\Cref{lemma:ero:number_of_acquire_pointers_per_operation_is_at_most_four}}
    \end{itemize}
\end{proof}

\begin{theorem}\label{thm:ero:b_is_space_efficient}
    Suppose $\mathcal{I}^\mathcal{B}$ is finite.
    Let $Allocate(\mathcal{I}^\mathcal{B})$ be the set of pointers which have been allocated in $\mathcal{I}^\mathcal{B}$, i.e., $\cellpointershort \in Allocate(\mathcal{I}^\mathcal{B})$ if and only if there is an $\allocatecelloperation{}$ operation in $\mathcal{I}^\mathcal{B}$ with response $\cellpointershort$.
    Likewise, let $Free(\mathcal{I}^\mathcal{B})$ be the set of pointers which have been freed in $\mathcal{I}^\mathcal{B}$, i.e., $\cellpointershort \in Free(\mathcal{I}^\mathcal{B})$ if and only if there is a $\freecelloperation{}(\cellpointershort)$ operation in $\mathcal{I}^\mathcal{B}$.
    Then, $|Allocate(\mathcal{I}^\mathcal{B}) \setminus Free(\mathcal{I}^\mathcal{B})| \leq 6c + 1$ where $c$ is the point contention in $\mathcal{I}^\mathcal{B}$.
\end{theorem}

\begin{proof}
    We note that since by \Cref{alg:lazy_cell_manager_specification} the response of every $\allocatecelloperation{}$ operation is in $\celluniverse$, $Allocate(\mathcal{I}^\mathcal{B}) \subseteq \celluniverse$.
    Furthermore, since $\mathcal{I}^\mathcal{B}$ is finite, we have that $Allocate(\mathcal{I}^\mathcal{B})$ is finite.
    Let $Pending(\mathcal{I}^\mathcal{B})$ be defined as follows.
    $\cellpointershort \in Pending(\mathcal{I}^\mathcal{B})$ if and only if there is a pending operation execution $opx$ in $\mathcal{I}^\mathcal{B}$ such that the process that executed $opx$ executed \cref{line:ero:allocate_cell} during $opx$ in $\mathcal{I}^\mathcal{B}$ and received response $\cellpointershort$.
    Since by definition there are $c$ pending operation executions in $\mathcal{I}^\mathcal{B}$, we have that $|Pending(\mathcal{I}^\mathcal{B})| \leq c$.
    Furthermore, since by \Cref{alg:lazy_cell_manager_specification} every response to an $\allocatecelloperation{}$ operation is in $\celluniverse$, we have that $Pending(\mathcal{I}^\mathcal{B}) \subseteq \celluniverse$.
    $Pending(\mathcal{I}^\mathcal{B})$ is useful for the following reason.

    \begin{claimcustom}{\ref{thm:ero:b_is_space_efficient}.1}\label{thm:ero:b_is_space_efficient:claim_one}
        For every $\cellpointershort \in Allocate(\mathcal{I}^\mathcal{B}) \setminus Pending(\mathcal{I}^\mathcal{B})$ there is a unique operation execution $opx_{\cellpointershort}$ in $\mathcal{I}^\mathcal{B}$ such that the process that executed $opx_{\cellpointershort}$ executed \cref{line:ero:allocate_cell} during $opx_{\cellpointershort}$ in $\mathcal{I}^\mathcal{B}$ and received response $\cellpointershort$.
        Furthermore, $opx_{\cellpointershort}$ is complete in $\mathcal{I}^\mathcal{B}$.
    \end{claimcustom}

    \begin{proof}
        Consider any $\cellpointershort \in Allocate(\mathcal{I}^\mathcal{B}) \setminus Pending(\mathcal{I}^\mathcal{B})$.
        Since $\cellpointershort \in Allocate(\mathcal{I}^\mathcal{B})$, we have that there is an operation execution $opx_{\cellpointershort}$ in $\mathcal{I}^\mathcal{B}$ such that the process that executed $opx_{\cellpointershort}$ executed \cref{line:ero:allocate_cell} during $opx_{\cellpointershort}$ in $\mathcal{I}^\mathcal{B}$ and received response $\cellpointershort$.
        Hence, by \Cref{alg:lazy_cell_manager_specification}, this execution of \cref{line:ero:allocate_cell} is the only $\allocatecelloperation{}$ operation whose response is $\cellpointershort$ in $\mathcal{I}^\mathcal{B}$, and so $opx_{\cellpointershort}$ is unique.
        Since $\cellpointershort \notin Pending(\mathcal{I}^\mathcal{B})$, by the definition of $Pending(\mathcal{I}^\mathcal{B})$, for every pending operation execution $opx$ in $\mathcal{I}^\mathcal{B}$, the process that executed $opx$ received a response other than $\cellpointershort$ on \cref{line:ero:allocate_cell} during $opx$ in $\mathcal{I}^\mathcal{B}$.
        Hence, $opx_{\cellpointershort}$ is not a pending operation execution in $\mathcal{I}^\mathcal{B}$.
        Therefore, since $opx_{\cellpointershort}$ is an operation execution in $\mathcal{I}^\mathcal{B}$, we have that $opx_{\cellpointershort}$ is complete in $\mathcal{I}^\mathcal{B}$ as wanted.
        \qH{\Cref{thm:ero:b_is_space_efficient:claim_one}}
    \end{proof}

    We now define another useful set of pointers, $Acquired(\mathcal{I}^\mathcal{B})$, as $\cellpointershort \in Acquired(\mathcal{I}^\mathcal{B})$ if and only if there is a pending operation execution $opx$ in $\mathcal{I}^\mathcal{B}$ with $R(\mathcal{I}^\mathcal{B}, opx, \cellpointershort) > 0$.
    Hence,
    \begin{align*}
        Acquired(\mathcal{I}^\mathcal{B}) = \bigcup_{\text{$opx$ is a pending operation execution in $\mathcal{I}^\mathcal{B}$}} Acquire(\mathcal{I}^\mathcal{B}, opx)
    \end{align*}
    where $Acquire(\mathcal{I}^\mathcal{B}, opx)$ is defined in \Cref{lemma:ero:number_of_acquire_pointers_per_operation_is_at_most_four}.
    Thus, since there are $c$ pending operation executions in $\mathcal{I}^\mathcal{B}$ and by \Cref{lemma:ero:number_of_acquire_pointers_per_operation_is_at_most_four} $|Acquire(\mathcal{I}^\mathcal{B}, opx)| \leq 3$, we have that $|Acquired(\mathcal{I}^\mathcal{B})| \leq 3c$.
    Since $|Pending(\mathcal{I})| \leq c$ and $|Acquired(\mathcal{I})| \leq 3c$, we have that $|Pending(\mathcal{I}) \cup Acquired(\mathcal{I})| \leq 4c$.
    For brevity, let $E_1 = Pending(\mathcal{I}) \cup Acquired(\mathcal{I})$ where ``$E$" stands for exceptions, so $|E_1| \leq 4c$.

    \begin{claimcustom}{\ref{thm:ero:b_is_space_efficient}.2}\label{thm:ero:b_is_space_efficient:claim_two}
        For every $\cellpointershort \in Allocate(\mathcal{I}^\mathcal{B}) \setminus E_1$ there is one $L$-remove event $\cellpointershort$ in $\mathcal{I}^\mathcal{B}$.
    \end{claimcustom}

    \begin{proof}
        Consider any $\cellpointershort \in Allocate(\mathcal{I}^\mathcal{B}) \setminus E_1$.
        Hence, $\cellpointershort \in Allocate(\mathcal{I}^\mathcal{B}) \setminus Pending(\mathcal{I}^\mathcal{B})$, so by \Cref{thm:ero:b_is_space_efficient:claim_one}, there is a unique operation execution $opx_{\cellpointershort}$ in $\mathcal{I}^\mathcal{B}$ such that the process that executed $opx_{\cellpointershort}$ executed \cref{line:ero:allocate_cell} during $opx_{\cellpointershort}$ in $\mathcal{I}^\mathcal{B}$ and received response $\cellpointershort$, and $opx_{\cellpointershort}$ is complete in $\mathcal{I}^\mathcal{B}$.
        Thus, the process that executed $opx_{\cellpointershort}$, say $p$, invoked and exited the \doworkuntildone{} procedure on \cref{line:ero:low_level_remove_cell} during $opx_{\cellpointershort}$ in $\mathcal{I}^\mathcal{B}$.
        Let $I$ denote the invocation of this procedure.
        Since $p$ received $\cellpointershort$ as a response on \cref{line:ero:allocate_cell} during $opx_{\cellpointershort}$ we have that the parameters of $I$ are $(\removecell, \cellpointershort)$.
        Hence, since by \Cref{lemma:ero:the_list_invariants_hold} $P(\mathcal{I}^\mathcal{B})$ holds, by \Cref{lemma:ero:l_remove_event_before_remove_low_level_exits}, there is a $L$-remove event for $\cellpointershort$ in $\mathcal{I}^\mathcal{B}$.
        Therefore, by $P(\mathcal{I}^\mathcal{B})$, this is the only $L$-remove event for $\cellpointershort$ in $\mathcal{I}^\mathcal{B}$ as wanted.
        \qH{\Cref{thm:ero:b_is_space_efficient:claim_two}}
    \end{proof}

    \begin{claimcustom}{\ref{thm:ero:b_is_space_efficient}.3}\label{thm:ero:b_is_space_efficient:claim_three}
        For every set $\mathcal{S}$ with $|\mathcal{S}| \leq 4c + 1$ if $|Allocate(\mathcal{I}^\mathcal{B}) \setminus \mathcal{S}| \leq 2c$, then $|Allocate(\mathcal{I}^\mathcal{B}) \setminus Free(\mathcal{I}^\mathcal{B})| \leq 6c + 1$.
    \end{claimcustom}

    \begin{proof}
        If $|Allocate(\mathcal{I}^\mathcal{B}) \setminus \mathcal{S}| \leq 2c$, then since $|\mathcal{S}| \leq 4c + 1$, we have that $|Allocate(\mathcal{I}^\mathcal{B})| \leq 6c + 1$, and so $|Allocate(\mathcal{I}^\mathcal{B}) \setminus Free(\mathcal{I}^\mathcal{B})| \leq 6c + 1$ as required.
        \qH{\Cref{thm:ero:b_is_space_efficient:claim_three}}
    \end{proof}

    For every $\cellpointershort \in Allocate(\mathcal{I}^\mathcal{B}) \setminus E_1$ let $L(\cellpointershort)$ be the unique $L$-remove event for $\cellpointershort$ in $\mathcal{I}^\mathcal{B}$ identified by \Cref{thm:ero:b_is_space_efficient:claim_two}.
    Since $|Pending(\mathcal{I}^\mathcal{B}) \cup Acquired(\mathcal{I}^\mathcal{B})| \leq 4c$, by \Cref{thm:ero:b_is_space_efficient:claim_three}, the theorem holds if $|Allocate(\mathcal{I}^\mathcal{B}) \setminus E_1| = 0$, so it suffices to assume that $Allocate(\mathcal{I}^\mathcal{B}) \setminus E_1 \neq \emptyset$.
    Hence, there is at least one element in $Allocate(\mathcal{I}^\mathcal{B}) \setminus E_1$.
    Thus, since $Allocate(\mathcal{I}^\mathcal{B})$ is finite, we have that there is a last $L$-remove event for some pointer in $Allocate(\mathcal{I}^\mathcal{B}) \setminus E_1$ in $\mathcal{I}^\mathcal{B}$.
    Let $\cellpointershort_{last} \in Allocate(\mathcal{I}^\mathcal{B}) \setminus E_1$ be this pointer, i.e., and for every $\cellpointershort \in Allocate(\mathcal{I}^\mathcal{B}) \setminus E_1$ $L(\cellpointershort) \leq L(\cellpointershort_{last})$.
    Note that since $|E_1| \leq 4c$, we have that $|E_1 \cup \{\cellpointershort_{last}\}| \leq 4c + 1$.
    As with $E_1$, we define $E_2 = E_1 \cup \{\cellpointershort_{last}\}$ or equivalently $E_2 = Pending(\mathcal{I}^\mathcal{B}) \cup Acquired(\mathcal{I}^\mathcal{B}) \cup \{\cellpointershort_{last}\}$, so $|E_2| \leq 4c + 1$.

    \begin{claimcustom}{\ref{thm:ero:b_is_space_efficient}.4}\label{thm:ero:b_is_space_efficient:claim_four}
        For every $\cellpointershort \in Allocate(\mathcal{I}^\mathcal{B}) \setminus E_2$ there is a successful list-remove attempt for $\cellpointershort$ in $\mathcal{I}^\mathcal{B}$.
    \end{claimcustom}

    \begin{proof}
        Consider any $\cellpointershort \in Allocate(\mathcal{I}^\mathcal{B}) \setminus E_2$.
        Since $\cellpointershort \neq \cellpointershort_{last}$, we have that $L(\cellpointershort) \neq L(\cellpointershort_{last})$.
        Hence, since for every $\cellpointershort \in Allocate(\mathcal{I}^\mathcal{B}) \setminus E_1$ $L(\cellpointershort) \leq L(\cellpointershort_{last})$, we have that $L(\cellpointershort) < L(\cellpointershort_{last})$.
        Thus, there is a next $L$-event after $L(\cellpointershort)$ in $\mathcal{I}^\mathcal{B}$; say $e$.
        Therefore, since by \Cref{lemma:ero:the_list_invariants_hold} $R(\mathcal{I}^\mathcal{B})$ holds, we have that there is a successful list-remove attempt for $\cellpointershort$ in $\mathcal{I}^\mathcal{B}$ as wanted.
        \qH{\Cref{thm:ero:b_is_space_efficient:claim_four}}
    \end{proof}

    Since $Allocate(\mathcal{I}^\mathcal{B})$ is finite, we have that $|Allocate(\mathcal{I}^\mathcal{B}) \setminus E_2| = n$ for some non-negative integer $n$.
    Since $|E_2| \leq 4c + 1$, by \Cref{thm:ero:b_is_space_efficient:claim_three}, the theorem holds if $|Allocate(\mathcal{I}^\mathcal{B}) \setminus E_2| \leq 2c$ and so we way assume that $n > 2c$.
    Hence, since $c \geq 0$, we have that $n$ is a positive integer.

    \begin{claimcustom}{\ref{thm:ero:b_is_space_efficient}.5}\label{thm:ero:b_is_space_efficient:claim_five}
        There is an acquire-copy event for every $\cellpointershort \in Allocate(\mathcal{I}^\mathcal{B}) \setminus E_2$ except at most $c$ in $\mathcal{I}^\mathcal{B}$.
        Let $Coalesced(\mathcal{I}^\mathcal{B})$ be the maximal subset of $Allocate(\mathcal{I}^\mathcal{B}) \setminus E_2$ such that every pointer in $Coalesced(\mathcal{I}^\mathcal{B})$ has an acquire-copy event for it in $\mathcal{I}^\mathcal{B}$, so $|Coalesced(\mathcal{I}^\mathcal{B})| \geq n - c$.
    \end{claimcustom}

    \begin{proof}
        Suppose, for contradiction, for some set $\{\cellpointershort_1, \ldots, \cellpointershort_{c + 1}\} \subseteq Allocate(\mathcal{I}^\mathcal{B}) \setminus E_2$ of size $c + 1$ there is not an acquire-copy event for every $\cellpointershort_i$ in $\mathcal{I}^\mathcal{B}$.
        Consider an $1 \leq i \leq c + 1$.
        By \Cref{thm:ero:b_is_space_efficient:claim_four}, there is a successful list-remove attempt $a_i$ for $\cellpointershort_i$ in $\mathcal{I}^\mathcal{B}$.
        Let $p_i$ be the process that executed this successful list-remove attempt.
        Hence, since $a_i$ is a list-remove attempt for $\cellpointershort_i$, by \Cref{def:ero:english}, $p_i$ executed $a_i$ during an invocation $I_i$ of the \doremovecell{} procedure with a second parameter of $\cellpointershort_i$.
        Thus, since $a_i$ is a successful list-remove attempt for $\cellpointershort_i$, it follows that $p_i$ will execute \cref{line:ero:copy_acquisitions_to_revocations} during $I_i$ on its next step.
        So, if $p_i$ executes this step, then since $I_i$'s second parameter is $\cellpointershort_i$, by \Cref{def:ero:english}, it would be an acquire-copy event for $\cellpointershort_i$.
        Hence, since by assumption there is not an acquire-copy event for $\cellpointershort_i$ in $\mathcal{I}^\mathcal{B}$, we have that $p_i$ does not execute this step during $\mathcal{I}^\mathcal{B}$.
        Thus, $p_i$ does not exit $I_i$ in $\mathcal{I}^\mathcal{B}$, and so if $opx_i$ is the operation execution $p_i$ executed $a_i$ during, then $opx_i$ is pending in $\mathcal{I}^\mathcal{B}$.
        Since there is at most one successful list-remove attempt per invocation of the \doremovecell{} procedure, we have that for every $1 \leq i, j \leq c + 1$ if $i \neq j$, then $a_i$ and $a_j$ are executed during different invocations of the \doremovecell{} procedure.
        Hence, since $a_i$ is executed during $I_i$ and $a_j$ is executed during $I_j$, we have that $I_i \neq I_j$.
        Thus, since each process is executing at most one invocation of the \doremovecell{} procedure at a time and for every $1 \leq i \leq c + 1$ $p_i$ does not exit $I_i$ in $\mathcal{I}^\mathcal{B}$, it follows that for every $1 \leq i, j \leq c + 1$ if $i \neq j$, then $p_i \neq p_j$.
        So, since $opx_i$ is the operation execution $p_i$ executed $a_i$ during, we have that for every $1 \leq i, j \leq c + 1$ if $i \neq j$, then $opx_i \neq opx_j$.
        Therefore, since for every $1 \leq i \leq c + 1$ $opx_i$ is pending in $\mathcal{I}^\mathcal{B}$, we have that there are $c + 1$ pending operation executions in $\mathcal{I}^\mathcal{B}$.
        However, by the definition of $c$, there are $c$ pending operation executions in $\mathcal{I}^\mathcal{B}$, a contradiction.
        \qH{\Cref{thm:ero:b_is_space_efficient:claim_five}}
    \end{proof}

    \begin{claimcustom}{\ref{thm:ero:b_is_space_efficient}.6}\label{thm:ero:b_is_space_efficient:claim_six}
        For every $\cellpointershort \in Coalesced(\mathcal{I}^\mathcal{B})$ there is a revocation event for $\cellpointershort$ with response $-1$ in $\mathcal{I}^\mathcal{B}$.
    \end{claimcustom}

    \begin{proof}
        Since $\cellpointershort \in Coalesced(\mathcal{I}^\mathcal{B})$, by \Cref{thm:ero:b_is_space_efficient:claim_five}, $\cellpointershort \in Allocate(\mathcal{I}^\mathcal{B}) \setminus E_2$.
        Hence, since $E_2 = Pending(\mathcal{I}^\mathcal{B}) \cup Acquired(\mathcal{I}^\mathcal{B}) \cup \{\cellpointershort_{last}\}$, we have that $\cellpointershort \in Allocate(\mathcal{I}^\mathcal{B}) \setminus Pending(\mathcal{I}^\mathcal{B})$.
        Thus, by \Cref{thm:ero:b_is_space_efficient:claim_one}, there is a unique operation execution $opx_{\cellpointershort}$ in $\mathcal{I}^\mathcal{B}$ such that the process that executed $opx_{\cellpointershort}$ executed \cref{line:ero:allocate_cell} during $opx_{\cellpointershort}$ in $\mathcal{I}^\mathcal{B}$ and received response $\cellpointershort$, and $opx_{\cellpointershort}$ is complete in $\mathcal{I}^\mathcal{B}$.
        Furthermore, since $\cellpointershort \in Coalesced(\mathcal{I}^\mathcal{B})$, by \Cref{thm:ero:b_is_space_efficient:claim_five}, there is an acquire-copy event $e$ for $\cellpointershort$ in $\mathcal{I}^\mathcal{B}$.
        Hence, by \Cref{lemma:ero:acquire_copy_copies_the_final_number_of_acquisitions}, $e$ is $\FAop{}((*\cellpointershort).\revocations, -(A(\mathcal{I}^\mathcal{B}, \cellpointershort) + 1))$ and if $\mathcal{I}^{include}_{e}$ is the prefix of $\mathcal{I}^\mathcal{B}$ up to and including $e$, then $A(\mathcal{I}^{include}_{e}, \cellpointershort) = A(\mathcal{I}^\mathcal{B}, \cellpointershort)$.
        Let $\mathcal{I}^{exclude}_{e}$ be the prefix of $\mathcal{I}^\mathcal{B}$ up to but excluding $e$.
        Hence, since the last step of $\mathcal{I}^{include}_{e}$ is not a successful list-acquire-next attempt, by \Cref{def:ero:number_of_successful_acquires}, $A(\mathcal{I}^{exclude}_{e}, \cellpointershort) = A(\mathcal{I}^{include}_{e}, \cellpointershort)$ and so $A(\mathcal{I}^{exclude}_{e}, \cellpointershort) = A(\mathcal{I}^\mathcal{B}, \cellpointershort)$.
        
        We claim that $A(\mathcal{I}^\mathcal{B}, \cellpointershort) - X(\mathcal{I}^\mathcal{B}, \cellpointershort) = -1$. 
        For every operation execution $opx$ in $\mathcal{I}^\mathcal{B}$ $opx$ is either complete or pending in $\mathcal{I}^\mathcal{B}$.
        We first consider complete operation executions in $\mathcal{I}^\mathcal{B}$.
        Since $opx_{\cellpointershort}$ is complete in $\mathcal{I}^\mathcal{B}$ and the process that executed $opx_{\cellpointershort}$ executed \cref{line:ero:allocate_cell} during $opx_{\cellpointershort}$ and received response $\cellpointershort$, by \Cref{lemma:ero:reference_count_of_complete_operation_execution}, $R(\mathcal{I}^\mathcal{B}, opx_{\cellpointershort}, \cellpointershort) = -1$.
        Now consider any complete operation execution $opx$ in $\mathcal{I}^\mathcal{B}$ other than $opx_{\cellpointershort}$.
        Since $opx_{\cellpointershort}$ is the only operation execution in $\mathcal{I}^\mathcal{B}$ such that the process that executed $opx_{\cellpointershort}$ received $\cellpointershort$ on \cref{line:ero:allocate_cell} during $opx_{\cellpointershort}$ in $\mathcal{I}^\mathcal{B}$, we have that the process that executed $opx$ received a different response on \cref{line:ero:allocate_cell} during $opx$, and so by \Cref{lemma:ero:reference_count_of_complete_operation_execution}, $R(\mathcal{I}^\mathcal{B}, opx, \cellpointershort) = 0$.
        Now consider any pending operation execution $opx$ in $\mathcal{I}^\mathcal{B}$.
        Hence, since $opx_{\cellpointershort}$ is complete in $\mathcal{I}^\mathcal{B}$, we have that $opx \neq opx_{\cellpointershort}$.
        Since $\cellpointershort \notin Acquired(\mathcal{I}^\mathcal{B})$, by the definition of $Acquire(\mathcal{I}^\mathcal{B})$, we have that $R(\mathcal{I}^\mathcal{B}, opx, \cellpointershort) \leq 0$.
        Furthermore, since $opx_{\cellpointershort}$ is the only operation execution in $\mathcal{I}^\mathcal{B}$ such that the process that executed $opx_{\cellpointershort}$ received $\cellpointershort$ on \cref{line:ero:allocate_cell} during $opx_{\cellpointershort}$ in $\mathcal{I}^\mathcal{B}$, we have that if the process that executed $opx$ executed \cref{line:ero:allocate_cell} during $opx$, then it received a response other than $\cellpointershort$, and so by \Cref{lemma:ero:reference_count_is_non_negative}, $R(\mathcal{I}^\mathcal{B}, opx, \cellpointershort) \geq 0$.
        Together, these imply that $R(\mathcal{I}^\mathcal{B}, opx, \cellpointershort) = 0$.
        Hence, since (1) $R(\mathcal{I}^\mathcal{B}, opx_{\cellpointershort}, \cellpointershort) = -1$, (2) for every complete operation execution $opx$ in $\mathcal{I}^\mathcal{B}$ other than $opx_{\cellpointershort}$ $R(\mathcal{I}^\mathcal{B}, opx, \cellpointershort) = 0$, and (3) for every pending operation execution $opx$ in $\mathcal{I}^\mathcal{B}$ $R(\mathcal{I}^\mathcal{B}, opx, \cellpointershort) = 0$, we have that
        \begin{align*}
            \sum_{\text{$opx$ is an operation execution in $\mathcal{I}^\mathcal{B}$}} R(\mathcal{I}^\mathcal{B}, opx, \cellpointershort) = -1.
        \end{align*}
        Therefore, since $\mathcal{I}^\mathcal{B}$ is finite, by \Cref{lemma:ero:relate_a_x_and_r}, $A(\mathcal{I}^\mathcal{B}, \cellpointershort) - X(\mathcal{I}^\mathcal{B}, \cellpointershort) = -1$.

        Since $A(\mathcal{I}^\mathcal{B}, \cellpointershort) - X(\mathcal{I}^\mathcal{B}, \cellpointershort) = -1$, we have that $X(\mathcal{I}^\mathcal{B}, \cellpointershort) = A(\mathcal{I}^\mathcal{B}, \cellpointershort) + 1$.
        Let $e_{revocation}$ be the $A(\cellpointershort, \mathcal{I}^\mathcal{B}) + 1$th revocation event for $\cellpointershort$ in $\mathcal{I}^\mathcal{B}$, so by \Cref{def:ero:number_of_successful_acquires} $e_{revocation}$ is the last revocation event for $\cellpointershort$ in $\mathcal{I}^\mathcal{B}$.
        We claim that $e < e_{revocation}$.
        Suppose, for contradiction, $e_{revocation} \leq e$.
        Hence, since $e_{revocation}$ is the last revocation event for $\cellpointershort$ in $\mathcal{I}^\mathcal{B}$, we have that there are no revocation events for $\cellpointershort$ after $e$ in $\mathcal{I}^\mathcal{B}$.
        Let $q$ be the process that performed $e$ and suppose $q$ did so during an operation execution $opx$.
        % Since $\mathcal{I}^{exclude}_{e}$ is the prefix of $\mathcal{I}^\mathcal{B}$, up to but excluding $e$, we have that $opx$ is pending in $\mathcal{I}^{exclude}_{e}$.
        Hence, since the step after $\mathcal{I}^{exclude}_{e}$ in $\mathcal{I}^\mathcal{B}$ is $q$ performing an acquire-copy event for $\cellpointershort \in \celluniverse$ (because $\cellpointershort \in Allocate(\mathcal{I}^\mathcal{B})$), by \Cref{lemma:ero:operation_acquisition_invariant}, $R(\mathcal{I}^{exclude}_{e}, opx, \cellpointershort) \geq 1$.
        Since $opx$ is an operation execution in $\mathcal{I}^\mathcal{B}$, as proved above, $R(\mathcal{I}^\mathcal{B}, opx, \cellpointershort)$ is $0$ or $-1$, and so $R(\mathcal{I}^\mathcal{B}, opx, \cellpointershort) < 1$.
        Thus, since $R(\mathcal{I}^{exclude}_{e}, opx, \cellpointershort) \geq 1$, and $R(\mathcal{I}^\mathcal{B}, opx, \cellpointershort) < 1$, by \Cref{def:ero:number_of_successful_acquires}, there is revocation event for $\cellpointershort$ in $\mathcal{I}^\mathcal{B}$ that is not in $\mathcal{I}^{exclude}_{e}$.
        Therefore, since $\mathcal{I}^{exclude}_{e}$ is the prefix of $\mathcal{I}^\mathcal{B}$ up to but excluding $e$, we have that there is a revocation event for $\cellpointershort$ after $e$ in $\mathcal{I}^\mathcal{B}$.
        However, there are no revocation events for $\cellpointershort$ after $e$ in $\mathcal{I}^\mathcal{B}$, a contradiction.

        We now finish the proof of \Cref{thm:ero:b_is_space_efficient:claim_six}.
        Since $\cellpointershort \in \celluniverse$, by \Cref{observation:ero:where_objects_change}, only acquire-copy events for $\cellpointershort$ and revocation events for $\cellpointershort$ change $(*\cellpointershort).\revocations$.
        Furthermore, since $e$ is an acquire-copy event for $\cellpointershort$ in $\mathcal{I}^\mathcal{B}$, by \Cref{lemma:ero:at_most_one_acquisition_copy}, $e$ is the only acquire-copy event for $\cellpointershort$ in $\mathcal{I}^\mathcal{B}$.
        Hence, since $e_{revocation}$ is the last revocation event for $\cellpointershort$ in $\mathcal{I}^\mathcal{B}$, and $e < e_{revocation}$, we have that $e_{revocation}$ is the last operation on $(*\cellpointershort).\revocations$ in $\mathcal{I}^\mathcal{B}$.
        Since $\cellpointershort \in \celluniverse$, we have that $(*\cellpointershort).\revocations$ is initially 0.
        Hence, since there are exactly $A(\mathcal{I}^\mathcal{B}, \cellpointershort) + 1$ revocation events for $\cellpointershort$ in $\mathcal{I}^\mathcal{B}$ (because $X(\mathcal{I}^\mathcal{B}, \cellpointershort) = A(\mathcal{I}^\mathcal{B}, \cellpointershort) + 1$), each revocation event for $\cellpointershort$ increases $(*\cellpointershort).\revocations$ by 1, $e$ decreases $(*\cellpointershort).\revocations$ by $A(\mathcal{I}^\mathcal{B}, \cellpointershort) + 1$, and $e$ is the only acquire-copy event for $\cellpointershort$ in $\mathcal{I}^\mathcal{B}$, it follows that $(*\cellpointershort).\revocations = 0$ at the end of $\mathcal{I}^\mathcal{B}$.
        Thus, since $e_{revocation}$ is the last operation on $(*\cellpointershort).\revocations$ in $\mathcal{I}^\mathcal{B}$, we have that $(*\cellpointershort).\revocations = 0$ at $e_{revocation}$.
        Therefore, since $e_{revocation}$ is a revocation event for $\cellpointershort$, by \Cref{def:ero:english}, $e_{revocation}$ is $\FAop{}((*\cellpointershort).\revocations, 1)$, and so $e_{revocation}$'s response is $-1$ as wanted.
        \qH{\Cref{thm:ero:b_is_space_efficient:claim_six}}
    \end{proof}

    Recall that by \Cref{thm:ero:b_is_space_efficient:claim_five}, $Coalesced(\mathcal{I}^\mathcal{B}) \subseteq Allocate(\mathcal{I}^\mathcal{B}) \setminus E_2$ and $|Coalesced(\mathcal{I}^\mathcal{B})| \geq n - c$.
    Since $Allocate(\mathcal{I}^\mathcal{B})$ is finite, we have that $|Coalesced(\mathcal{I}^\mathcal{B})| = m$ for some non-negative integer, so $m \geq n - c$.
    Since $n > 2c$, we have that $m > c$, and since $c$ is a non-negative integer, we have that $m > 0$.
    Hence, since $m$ is a non-negative integer, $m$ is a positive integer.

    \begin{claimcustom}{\ref{thm:ero:b_is_space_efficient}.7}\label{thm:ero:b_is_space_efficient:claim_seven}
        There is a $\freecelloperation{}(\cellpointershort)$ operation for every $\cellpointershort \in Coalesced(\mathcal{I}^\mathcal{B})$ except at most $c$ in $\mathcal{I}^\mathcal{B}$.
        Let $Freed(\mathcal{I}^\mathcal{B})$ be the maximal subset of $Coalesced(\mathcal{I}^\mathcal{B})$ such that every pointer $\cellpointershort \in Freed(\mathcal{I}^\mathcal{B})$ has a $\freecelloperation{}(\cellpointershort)$ operation in $\mathcal{I}^\mathcal{B}$, so $|Freed(\mathcal{I}^\mathcal{B})| \geq m - c$.
    \end{claimcustom}

    \begin{proof}
        Suppose, for contradiction, for some set $\{\cellpointershort_1, \ldots, \cellpointershort_{c + 1}\}\subseteq Coalesced(\mathcal{I}^\mathcal{B})$ of size $c + 1$ there is no $\freecelloperation{}(\cellpointershort_i)$ operation in $\mathcal{I}^\mathcal{B}$.
        Consider an integer $1 \leq i \leq c + 1$.
        By \Cref{thm:ero:b_is_space_efficient:claim_six}, 
        there is a revocation event $e_i$ for $\cellpointershort_i$ with response $-1$ in $\mathcal{I}^\mathcal{B}$ by process $p_i$.
        Hence, since $e_i$ is a revocation event for $\cellpointershort_i$, by \Cref{def:ero:english}, $p_i$ executed $e_i$ during an invocation $I_i$ of the Relinquish procedure with parameter $\cellpointershort_i$.
        Thus, since $e_i$ is a revocation event for $\cellpointershort_i$ whose response is $-1$, $p_i$ will execute \cref{line:ero:free_cell} during $I_i$ on its next step.
        So, if $p_i$ executes this step, then since $I_i$'s second parameter is $\cellpointershort_i$, it would be a $\freecelloperation{}(\cellpointershort_i)$ operation.
        Hence, since by assumption there is not a $\freecelloperation{}(\cellpointershort_i)$ operation in $\mathcal{I}^\mathcal{B}$, we have that $p_i$ does not execute this step during $\mathcal{I}^\mathcal{B}$.
        Thus, $p_i$ does not exit $I_i$ in $\mathcal{I}^\mathcal{B}$, and so if $opx_i$ is the operation execution $p_i$ executed $e_i$ during, then $opx_i$ is pending in $\mathcal{I}^\mathcal{B}$.
        Since there is at most one revocation event per invocation of the Relinquish procedure, we have that for every $1 \leq i, j \leq c + 1$ if $i \neq j$, then $e_i$ and $e_j$ are executed during different invocations of the Relinquish procedure.
        Hence, since $e_i$ is executed during $I_i$ and $e_j$ is executed during $I_j$, we have that $I_i \neq I_j$.
        Thus, since each process is executing at most one invocation of the Relinquish procedure at a time and for every $1 \leq i \leq c + 1$ $p_i$ does not exit $I_i$ in $\mathcal{I}^\mathcal{B}$, it follows that for every $1 \leq i, j \leq c + 1$ if $i \neq j$, then $p_i \neq p_j$.
        So, since $opx_i$ is the operation execution $p_i$ executed $e_i$ during, we have that for every $1 \leq i, j \leq c + 1$ if $i \neq j$, then $opx_i \neq opx_j$.
        Therefore, since for every $1 \leq i \leq c + 1$ $opx_i$ is pending in $\mathcal{I}^\mathcal{B}$, we have that there are $c + 1$ pending operation executions in $\mathcal{I}^\mathcal{B}$.
        However, by the definition of $c$, there are $c$ pending operation executions in $\mathcal{I}^\mathcal{B}$, a contradiction.
        \qH{\Cref{thm:ero:b_is_space_efficient:claim_seven}}
    \end{proof}

    We now finish the proof of \Cref{thm:ero:b_is_space_efficient}.        
    By \Cref{thm:ero:b_is_space_efficient:claim_seven} $Freed(\mathcal{I}^\mathcal{B}) \subseteq Coalesced(\mathcal{I}^\mathcal{B})$ with $|Freed(\mathcal{I}^\mathcal{B})| \geq m - c$ such that for every $\cellpointershort \in Freed(\mathcal{I}^\mathcal{B})$ there is a $\freecelloperation{}(\cellpointershort)$ operation in $\mathcal{I}^\mathcal{B}$.
    Hence, by the definition of $Free(\mathcal{I}^\mathcal{B})$, we have that $Freed(\mathcal{I}^\mathcal{B}) \subseteq Free(\mathcal{I}^\mathcal{B})$.
    Furthermore, since $m \geq n - c$, we have that $|Freed(\mathcal{I}^\mathcal{B})| \geq n - 2c$.
    Hence, since $|Allocate(\mathcal{I}^\mathcal{B}) \setminus E_2| = n$ and $|E_2| \leq 4c + 1$, we have that $|Allocate(\mathcal{I}^\mathcal{B})| \leq n + 4c + 1$, and so $|Allocate(\mathcal{I}^\mathcal{B})| - |Freed(\mathcal{I}^\mathcal{B})| \leq 6c + 1$.
    Since $Freed(\mathcal{I}^\mathcal{B}) \subseteq Coalesced(\mathcal{I}^\mathcal{B})$, and by \Cref{thm:ero:b_is_space_efficient:claim_five} $Coalesced(\mathcal{I}^\mathcal{B}) \subseteq Allocate(\mathcal{I}^\mathcal{B})$, by transitivity, $Freed(\mathcal{I}^\mathcal{B}) \subseteq Allocate(\mathcal{I}^\mathcal{B})$.
    Hence, since $Allocate(\mathcal{I}^\mathcal{B})$ is finite, we have that $|Allocate(\mathcal{I}^\mathcal{B}) \setminus Freed(\mathcal{I}^\mathcal{B})| = |Allocate(\mathcal{I}^\mathcal{B})| - |Freed(\mathcal{I}^\mathcal{B})|$.
    Thus, since $|Allocate(\mathcal{I}^\mathcal{B})| - |Freed(\mathcal{I}^\mathcal{B})| \leq 6c + 1$ we have that $|Allocate(\mathcal{I}^\mathcal{B}) \setminus Freed(\mathcal{I}^\mathcal{B})| \leq 6c + 1$.
    Therefore, since $Freed(\mathcal{I}^\mathcal{B}) \subseteq Free(\mathcal{I}^\mathcal{B})$, we have that $|Allocate(\mathcal{I}^\mathcal{B}) \setminus Free(\mathcal{I}^\mathcal{B})| \leq 6c + 1$ as wanted.
    \qH{\Cref{thm:ero:b_is_space_efficient}}
\end{proof}

\subsection{$\mathcal{A}$ is Linearizable, Wait-free, and Space-Efficient}
\label{sec:mapping_a_to_b}

In this section, we show that $\mathcal{B}$ being linearizable, wait-free, and having space complexity linear in the point contention implies $\mathcal{A}$ has these properties as well.
Recall from \Cref{def:reduction:algorithm_a} and \Cref{def:reduction:algorithm_b} that the difference between $\mathcal{B}$ and $\mathcal{A}$ is the following: (1) $\mathcal{B}$ allocates a pointer at most once (in contrast to $\mathcal{A}$ which can reallocate a pointer arbitrarily many times); (2) all operations on any object of any cell in $\celluniverse$ respects the semantics of its type (in contrast to $\mathcal{A}$ where the response of an operation on an object of a cell which is not allocated is arbitrary); and (3) an $\allocatecelloperation{}$ operation whose response is $\cellpointershort$ does not change the state assigned to the objects of the cell pointed to by $\cellpointershort$ (in contrast to $\mathcal{A}$ where an $\allocatecelloperation{}$ operation whose response is $\cellpointershort$ sets the state of each object of the cell pointed to by $\cellpointershort$ to its initial state).
The strategy for resolving these differences is by mapping each implementation history $\mathcal{I}^\mathcal{A}$ of $\mathcal{A}$ to an implementation history $\mathcal{I}^\mathcal{B}$ of $\mathcal{B}$ such that: (A) the object histories obtained by removing all implementation steps from of $\mathcal{I}^\mathcal{A}$ and $\mathcal{I}^\mathcal{B}$, respectively, are the same; (B) the program counter of each process is the same in the $i$th configuration of $\mathcal{I}^\mathcal{A}$ and $\mathcal{I}^\mathcal{B}$; (C) the number of allocated cells is the same in the $i$th configuration of $\mathcal{I}^\mathcal{A}$ and $\mathcal{I}^\mathcal{B}$.
(A) is the property that lets us prove that $\mathcal{A}$ is linearizable because it allows us to reuse the linearization function of $\mathcal{B}$.
(B) is the property that lets us prove that $\mathcal{A}$ is wait-free because any supposed operation execution in $\mathcal{A}$ that takes in infinitely many steps without completing would be an operation execution in $\mathcal{B}$ that takes infinitely many steps without completing, contradicting the fact that $\mathcal{B}$ is wait-free.
(C) is the property that lets us prove that the space complexity of $\mathcal{A}$ is linear in the point contention, because at any supposed time $t$ where the number of allocated cells is larger than $6c + 1$, where $c$ is the point contention at $t$, is a time in $\mathcal{B}$ where the number of allocated cells is larger than $6c + 1$, contradicting the space bound of $\mathcal{B}$.

We now sketch how we will build the implementation history $\mathcal{I}^\mathcal{B}$ from $\mathcal{I}^\mathcal{A}$, and why it resolves differences (1)-(3).
The main difficulty in building $\mathcal{I}^\mathcal{B}$ is dealing with (1).
To see this, consider algorithm $\mathcal{A}'$ which is the same as $\mathcal{A}$ except it uses the memory manager given in \Cref{alg:lazy_cell_manager_specification} instead of \Cref{alg:cell_manager_specification}, so $\mathcal{A}'$ and $\mathcal{B}$ are the same except for differences (2) and (3).
Mapping implementation histories of $\mathcal{A}'$ to $\mathcal{B}$ is trivial: \emph{every implementation history of $\mathcal{A}'$ is an implementation history of $\mathcal{B}$.}
To see why, we provide a proof sketch for resolving (2) and (3) between $\mathcal{A'}$ and $\mathcal{B}$.
Consider any implementation history $\mathcal{I}$ of $\mathcal{A}$, and suppose that the prefix of $\mathcal{I}$ up to and including the $n$th step, denoted by $\mathcal{I}_n$, is an implementation history of $\mathcal{B}$.
We sketch why the prefix of $\mathcal{I}$ up to and including the $n + 1$th step is an implementation history of $\mathcal{B}$ by resolving (2) and (3).
For (2), it suffices to suppose that the $n + 1$th step in $\mathcal{I}$ executes an operation on an object of a cell in $\celluniverse$ which is not allocated; let $p_{n + 1}$ be the process that executed this step.
% In $\mathcal{B}$, the response of this operation respects the semantics of the type of the object, but in $\mathcal{A}'$ the response could be arbitrary.
Since $\mathcal{I}_n$ is an implementation history of $\mathcal{B}$, we have that a one step continuation of $\mathcal{I}_n$ by $p_{n + 1}$ would yield an implementation history of $\mathcal{B}$ where a process executes an operation on an object of a cell in $\celluniverse$ which is not allocated, contradicting 3 of \Cref{thm:reduction:algorithm_b_is_well_behaved}.
For (3), it suffices to suppose that the $n + 1$th step in $\mathcal{I}$ executes an $\allocatecelloperation{}$ operation with response $\cellpointershort$, and the state of one of the objects of the cell pointed to by $\cellpointershort$ in the $n + 1$th configuration in $\mathcal{I}$ differs from a one-step continuation by the same process in $\mathcal{B}$ from $\mathcal{I}_n$.
Since $\mathcal{A}'$ sets the state of each object of the cell pointed to by $\cellpointershort$ to its initial state in the $n + 1$th configuration in $\mathcal{I}$, this implies that some step in $\mathcal{I}_n$ executed an operation on this object.
Hence, since $\mathcal{I}_n$ is an implementation history of $\mathcal{B}$, by 3 of \Cref{thm:reduction:algorithm_b_is_well_behaved}, there is an $\allocatecelloperation{}$ operation whose response is $\cellpointershort$ in $\mathcal{I}_n$.
Therefore, there are two $\allocatecelloperation{}$ operations in $\mathcal{I}$ with the same response, which is impossible by \Cref{alg:lazy_cell_manager_specification}.

To deal with difference (1) between $\mathcal{A}$ and $\mathcal{B}$, we need to consistently ``rename'' the response of each $\allocatecelloperation{}$ operation in $\mathcal{I}^\mathcal{A}$ when building $\mathcal{I}^\mathcal{B}$ so that the response of each $\allocatecelloperation{}$ operation in $\mathcal{I}^\mathcal{B}$ is unique.
Our approach for doing so is simple: \emph{use the step number as a source of uniqueness to pick a pointer from $\celluniverse$.}
More precisely, we define an injective function $\mathcal{M}$ from $\mathbb{N}$ to $\celluniverse$ (such a function exists because $\celluniverse$ is infinite), and define the response of an $\allocatecelloperation{}$ operation during the $n$th step of $\mathcal{I}^\mathcal{B}$ as $\mathcal{M}(n)$.
The injectivity of $\mathcal{M}$ yields the desired uniqueness of responses to $\allocatecelloperation{}$ operations in $\mathcal{I}^\mathcal{B}$.
Our task now is two-fold: (I) how do we make sure these changes in the responses to $\allocatecelloperation{}$ operations reflect in the subsequent configurations in $\mathcal{I}^\mathcal{B}$; and (II) how do we assign states to objects of cells.
To see why (I) and (II) are delicate, we give some examples.
For (I), if a process $p$ receives $\mathcal{M}(n)$ as a response to an $\allocatecelloperation{}$ operation during the $n$th step of $\mathcal{I}^\mathcal{B}$ it must be that the local variable $\cellpointershort$ at each step inside the same invocation of the \highleveloperation{} procedure is also $\mathcal{M}(n)$.
For (II), in $\mathcal{I}^\mathcal{A}$ it could be the case that every $\allocatecelloperation{}$ operation returns the same response (this could happen when only a single process takes steps in $\mathcal{I}^\mathcal{A}$), but in $\mathcal{I}^\mathcal{B}$ we use infinitely many cells, so after a given step in $\mathcal{I}^\mathcal{A}$, how do we decide what state to assign each of these cells in $\mathcal{I}^\mathcal{B}$?

We solve (I) by ``tracking'' the dissemination of a response from an $\allocatecelloperation{}$ operation.
More precisely, we \emph{watermark} the $j$th value of each object or local variable $O$ in the $i$th configuration of $\mathcal{I}^\mathcal{A}$ (for our purposes, a value is the smallest unit in the state of an object or a local variable), if it is in $\celluniverse$ as follows.
If the $i$th step of $\mathcal{I}^\mathcal{A}$ does not change the $j$th value of $O$, then the $i$th watermark of the $j$th value of $O$ is the same as the $i - 1$th watermark of the $j$th value of $O$.
If the $i$th step of $\mathcal{I}^\mathcal{A}$ changes the $j$th value of $O$ to a value originating from the $k$th value of some object or local variable $O'$, then the $i$th watermark of the $j$th value of $O$ is the watermark of the $i - 1$th watermark of the $k$th value of $O'$.
Lastly, if the $i$th step of $\mathcal{I}^\mathcal{A}$ is an $\allocatecelloperation{}$ operation, then the $i$th watermark of the $j$th value of $O$ is $i$.
This watermarking strategy lets us map states of objects and local variables in the $i$th configuration of $\mathcal{I}^\mathcal{A}$ to the $i$th configuration of $\mathcal{I}^\mathcal{B}$ by swapping every value with its watermarked counterpart.
More precisely, the $j$th value of any object or local variable $O$ in the $i$th configuration of $\mathcal{I}^\mathcal{B}$ is swapped to the output of $\mathcal{M}$ on the $i$th watermark of the $j$th value of $O$ if it is well-defined, and is the same as the value in $\mathcal{I}^\mathcal{A}$ otherwise.
% This swapping strategy takes care of the problem of determining what state a local variable should be in a configuration of $\mathcal{I}^\mathcal{B}$, but what remains is to determine what state to assign each object to.

To solve (II), each statically allocated base object or local variable is assigned to its mapped version of the state as described above.
This suffices because the objects and local variables are not ``renamed''.
The case of objects of cells is more delicate because an object $O$ of a cell pointed to $\cellpointershort$ may map to many different objects in $\mathcal{I}^\mathcal{B}$.
For example, the response of multiple $\allocatecelloperation{}$ operations may be $\cellpointershort$ in $\mathcal{I}^\mathcal{A}$, and since each $\allocatecelloperation{}$ operation is unique in $\mathcal{I}^\mathcal{B}$, the object $O$ corresponds to multiple different objects in $\mathcal{I}^\mathcal{B}$.
To deal with this ambiguity, we map the current state of $O$, using the mapping above, to the ``latest'' version of $O$ in $\mathcal{I}^\mathcal{B}$, and all other versions of $O$ use the mapping above on the configuration when they were the latest version.
More precisely, let $W$ be the set of step numbers up to and including the $i$th step of $\mathcal{I}^\mathcal{A}$ which perform $\allocatecelloperation{}$ operations.
For every $\cellpointershort \in \celluniverse \setminus \mathcal{M}[W]$\footnote{This is the function image of a subset, i.e., for a function $f\colon X \to Y$ and $S \subseteq X$ $f[S] = \{f(s)\ \vert\ s \in S\}$.}, the $i$th configuration of $\mathcal{I}^\mathcal{A}$ assigns every object of the cell pointed to by $\cellpointershort$ to its initial state.
Now consider any $w \in W$.
Let $\cellpointershort$ be the response of the $\allocatecelloperation{}$ operation performed during the $w$th step of $\mathcal{I}^\mathcal{A}$.
If for all $w < j \leq i$ the $j$th step of $\mathcal{I}^\mathcal{A}$ does not perform an $\allocatecelloperation{}$ operation whose response is $\cellpointershort$, then the $i$th configuration of $\mathcal{I}^\mathcal{B}$ assigns state $\mathcal{S}_i((*\cellpointershort).f)$ (the mapped version of $(*\cellpointershort).f$) to $(*\mathcal{M}(w)).f$ for every $f$ equal to $\lastrepositoryoperationresponse{}$, $\revocations$, or $\nextlong$.
Otherwise, let $j$ be the minimum $w < j \leq i$ such that the $j$th step of $\mathcal{I}^\mathcal{A}$ performs an $\allocatecelloperation{}$ cell operation whose response is $\cellpointershort$.
Then, the $i$th configuration of $\mathcal{I}^\mathcal{B}$ assigns state $\mathcal{S}_{j - 1}((*\cellpointershort).f)$ to $(*\mathcal{M}(w)).f$.

\textbf{Roadmap.} We start by introducing some notation and a basic fact about $\mathcal{A}$.
We then define the mapping sketched above from the implementation histories of $\mathcal{A}$ to those of $\mathcal{B}$ and prove some basic facts about it.
We then prove that this mapping actually yields implementation histories of $\mathcal{B}$; this is the majority of the work in this section.
Finally, we prove that $\mathcal{A}$ is linearizable, wait-free, and has space complexity linear in the point contention using this mapping.

\begin{definition}\label{def:reduction:shared_objects}
    We define the set $\celluniverse_O$ of objects of cells as $O \in \celluniverse_O$ if and only if for some $\cellpointershort \in \celluniverse \cup \{\&\headobject\}$ $O$ equals either $(*\cellpointershort).\lastrepositoryoperationresponse{}$, $(*\cellpointershort).\revocations$, or $(*\cellpointershort).\nextlong$.
    The set of base objects of $\mathcal{A}$ are $\{\clockobject{}, \announceobject, \linearizationobject, \stateobject\} \cup \celluniverse_O$ and the memory manager given in \Cref{alg:cell_manager_specification}.
    The set of base objects of $\mathcal{B}$ are $\{\clockobject{}, \announceobject, \linearizationobject, \stateobject\} \cup \celluniverse_O$ and the memory manager given in \Cref{alg:lazy_cell_manager_specification}.
\end{definition}

\begin{lemma}\label{lemma:reduction:timestamp_of_announce_object_in_a_determine_state_of_announce_object}
    Consider any implementation history $\mathcal{I}^\mathcal{A}$ of $\mathcal{A}$.
    If two configurations of $\mathcal{I}^\mathcal{A}$ assign states of the form $((t, \arbitraryvalue), \arbitraryvalue)$ to $\announceobject$, then they assign the same state to $\announceobject$.
\end{lemma}

\begin{proof}
    Suppose, for contradiction, the $i$th and $j$th configurations of $\mathcal{I}^\mathcal{A}$ assign states of the form $((t, \arbitraryvalue), \arbitraryvalue)$ to $\announceobject$, but they assign different states to $\announceobject$.
    Let $C^\mathcal{A}_i$ (resp. $C^\mathcal{A}_j$) be the $i$th (resp. $j$th) configuration of $\mathcal{I}^\mathcal{A}$.
    Furthermore, let $s_i$ (resp. $s_j$) be the states they assign to $\announceobject$.
    By assumption, $s_i$ and $s_j$ are of the form $((t, \arbitraryvalue), \arbitraryvalue)$ but $s_i \neq s_j$.
    Hence, $i \neq j$.
    Without loss of generality, assume $i < j$.
    Hence, since $s_i \neq s_j$, we have that some process $p_j$ set the state of $\announceobject$ to $s_j$ on the $k_j$th step of $\mathcal{I}^\mathcal{A}$ where $k_j \in (i, j]$.
    Thus, since the state of $\announceobject$ is only changed on \cref{line:ero:announce_gcas} or \ref{line:ero:announce_cas}, $p_j$ executed \cref{line:ero:announce_gcas} or \ref{line:ero:announce_cas} on the $k_j$th step of $\mathcal{I}^\mathcal{A}$ with a third parameter of $s_j$; say during some invocation $I_j$ of the \doworkuntildone{} procedure.
    Since $s_j$ is of the form $((t, \arbitraryvalue), \arbitraryvalue)$, we have that $p_j$ set the state of $\announceobject = ((t, \arbitraryvalue), \arbitraryvalue)$ on the $k_j$th step of $\mathcal{I}^\mathcal{A}$.
    Therefore, $p_j$ received $t$ as a response to its execution of \cref{line:ero:operation_timestamp} during $I_j$.
    Since the state of $\clockobject{}$ is initially 1, by the definition of \FIop{}, all responses on \cref{line:ero:operation_timestamp} are bigger than 0, and so $t \neq 0$.
    Hence, since the state of $\announceobject$ is initially $((0, \noop), \nullconstant)$, and $s_i$ is of the form $((t, \arbitraryvalue), \arbitraryvalue)$, we have that some process $p_i$ set the state of $\announceobject$ to $s_i$ on the $k_i$th step of $\mathcal{I}^\mathcal{A}$ where $k_i \in [1.. i]$.
    Thus, since the state of $\announceobject$ is only changed on \cref{line:ero:announce_gcas} or \ref{line:ero:announce_cas}, $p_i$ executed \cref{line:ero:announce_gcas} or \ref{line:ero:announce_cas} on the $k_i$th step of $\mathcal{I}^\mathcal{A}$ with a third parameter of $s_i$; say during some invocation $I_i$ of the \doworkuntildone{} procedure.
    Since $s_i$ is of the form $((t, \arbitraryvalue), \arbitraryvalue)$, we have that $p_i$ set the state of $\announceobject = ((t, \arbitraryvalue), \arbitraryvalue)$ on the $k_i$th step of $\mathcal{I}^\mathcal{A}$.
    Therefore, $p_i$ received $t$ as a response to its execution of \cref{line:ero:operation_timestamp} during $I_i$.
    Since $p_i$ and $p_j$ both received $t$ as a response to an execution of \cref{line:ero:operation_timestamp}, by the definition of \FIop{}, we have that $p_i = p_j$.
    Furthermore, since $p_i$ (resp. $p_j$) performed this executions of \cref{line:ero:operation_timestamp} during $I_i$ (resp. $I_j$), we have that $I_i = I_j$.
    Let $p_i = p_j = p$, and let $I_i = I_j = I$.
    Hence, $p$ performed the $k_i$th and $k_j$th step during $I$.
    Let $(\myrepositoryoperationshort, \cellpointershort)$ be the parameters of $I$.
    Since $p$ received $t$ as a response on \cref{line:ero:operation_timestamp} during $I$, it follows that every execution of \cref{line:ero:announce_gcas} and \ref{line:ero:announce_cas} during $I$ has a third parameter of $((t, \myrepositoryoperationshort), \cellpointershort)$.
    Therefore, since $p$ executed the $k_i$th and $k_j$th step during $I$, and the third parameter of the $k_i$th (resp. $k_j$th) step is $s_i$ (resp. $s_j$), we have that $s_i = s_j$.
    However, $s_i \neq s_j$, a contradiction.
    \qH{\Cref{lemma:reduction:timestamp_of_announce_object_in_a_determine_state_of_announce_object}}
\end{proof}

\subsubsection{A correctness-preserving mapping from implementation histories of $\mathcal{A}$ to $\mathcal{B}$}

In this section, we define our mapping of implementation histories of $\mathcal{A}$ to $\mathcal{B}$ and prove some basic facts about it.
We start by defining what a value is.
For our purposes, a value is the smallest unit in the state of an object or a local variable, as defined below.

\begin{observation}\label{observation:reduction:finite_sequence_of_values}
    In both algorithms $\mathcal{A}$ and $\mathcal{B}$, the state of every local variable or base object other than the memory manager is a finite sequence of values.
    For example,
    \begin{compactitem}
        \item The state of $\clockobject{}$ is a single value $n$, so the sequence is $n$.
        \item The state of $\announceobject$ and $\linearizationobject$ is of the form $((t, llo), ptr)$ so the sequences is $t, llo, ptr$. 
        %When $llo = \langle \doopandcopyresponse{}, hlo \rangle$, the sequence is $t, \doopandcopyresponse{}, hlo, ptr$.
        \item The state of $\stateobject$ is of the form $((t, llo), s, r)$ so the sequence is $t, llo, s, r$.
        %In the special case when $llo = \langle \doopandcopyresponse{}, hlo \rangle$, the sequence is $t, \doopandcopyresponse{}, hlo, s, r$.
        \item The state of a cell's $\lastrepositoryoperationresponse{}$ object is of the form $((t, llo), r)$ so the sequence is $t, llo, r$.
        \item The state of a cell's $\revocations$ object is a single value $n$, so the sequence is $n$.
        \item The state of a cell's $\nextlong$ object is of the form $(v, s, a, ptr)$ so the sequence is $v, s, a, ptr$.
    \end{compactitem}
    In all the cases above, when $llo = \langle \doopandcopyresponse{}, hlo \rangle$, $llo$ in the above sequence is replaced with $\doopandcopyresponse{}, hlo$.
    Furthermore, in the same fashion as the cases above, the input and output to each operation performed on a local variable or base object is a finite sequence of values.
\end{observation}

We now define the watermarking scheme we described at the beginning of the section.
We note that the phrase ``process $p$ set the value (or some index) of some local variable or base object $O$ during some step $s$'' means that $p$ performs a write or \CASop{} operation on $O$ during $s$.

\begin{definition}[Watermarks]\label{def:reduction:watermarks}
    Consider any implementation history $\mathcal{I}^\mathcal{A} = C^\mathcal{A}_0, p_1, C^\mathcal{A}_1, \ldots$ of $\mathcal{A}$.
    For every $C^\mathcal{A}_i$, local variable or base object $O$ other than the memory manager of $\mathcal{A}$, and $j$th index of the state assigned to $O$ in $C^\mathcal{A}_i$, we define a watermarking function $\mathcal{W}_i(O, j)$ as follows.
    Let $v_1, v_2, \ldots$ be the state assigned to $O$ in $C^\mathcal{A}_i$ (see \Cref{observation:reduction:finite_sequence_of_values}). Suppose $i > 0$, if $O$ is a base object (not a local variable), then $O \in \{\announceobject, \linearizationobject\} \cup \{(*\cellpointershort).\nextlong\ \vert\ \cellpointershort \in \celluniverse \cup \{\&\headobject\}\}$, and $v_j \in \celluniverse$.
    \begin{compactenum}
        % \item If the $j$th index of the state assigned to $O$ in $C^\mathcal{A}_{i - 1}$ is $v_j$, then $\mathcal{W}_i(O, j) = \mathcal{W}_{i - 1}(O, j)$.
        \item If $p_i$ does not set the $j$th index of $O$ during the $i$th step of $\mathcal{I}^\mathcal{A}$, then $\mathcal{W}_i(O, j) = \mathcal{W}_{i - 1}(O, j)$.\footnote{Note that this is not equivalent to saying that the $j$th index of $O$ is the same in $C^\mathcal{A}_{i - 1}$ and $C^\mathcal{A}_{i}$.}
        \item If $O$ is a local variable of $p_i$ other than its program counter, then:
        \begin{compactenum}
            \item If $O$ is the local variable $\cellpointershort$ of $p_i$ on \cref{line:ero:allocate_cell}, and $p_i$ 
            sets the $j$th index of $O$ to $v_j$ during the $i$th step of $\mathcal{I}^\mathcal{A}$ because $p_i$ performs an $\allocatecelloperation{}$ operation whose response is $v_j$, then $\mathcal{W}_i(O, j) = i$.
            \item If $p_i$ sets the $j$th index of $O$ to $v_j$ during the $i$th step of $\mathcal{I}^\mathcal{A}$ because $p_i$ performs a read operation during the $i$th step of $\mathcal{I}^\mathcal{A}$ on a base object $O'$ whose $k$th index of its response is $v_j$ which is also the $k$th index of its state in $C^\mathcal{A}_{i - 1}$, then $\mathcal{W}_i(O, j) = \mathcal{W}_{i - 1}(O', k)$.
            % $O' \in \{\announceobject, \linearizationobject \} \cup \{(*\cellpointershort).\nextlong\ \vert\ \cellpointershort \in \celluniverse \cup \{\&\headobject\}\}$ 
            \item If $p_i$ sets the $j$th index of $O$ to $v_j$ during the $i$th step of $\mathcal{I}^\mathcal{A}$ because the $k$th index of one of $p_i$'s local variables $O'$ in $C^\mathcal{A}_{i - 1}$ is $v_j$, then $\mathcal{W}_i(O, j) = \mathcal{W}_{i - 1}(O', k)$.
        \end{compactenum}
        \item If $O$ is a base object, and $p_i$ sets the $j$th index of $O$ to $v_j$ during the $i$th step of $\mathcal{I}^\mathcal{A}$ because the $k$th index of one of $p_i$'s local variables $O'$ in $C^\mathcal{A}_{i - 1}$ is $v_j$, then $\mathcal{W}_i(O, j) = \mathcal{W}_{i - 1}(O', k)$.
    \end{compactenum}
    In all other cases $\mathcal{W}_i(O, j) = \bot$.
    So, by definition, $\mathcal{W}_i(O, j) \in \mathbb{N} \cup \{\bot\}$.
\end{definition}

For convenience, it is useful to ``rename'' pointers in $\mathcal{A}$ to ``fresh'' pointers in $\mathcal{B}$, i.e., pointers that were not used in the implementation history of $\mathcal{A}$ that we are mapping to $\mathcal{B}$.
To define these fresh pointers, we define the set of pointers used in an implementation $\mathcal{I}^\mathcal{A}$ of $\mathcal{A}$ below.

\begin{definition}\label{def:reduction:used_pointers_in_run_of_a}
    Consider any implementation history $\mathcal{I}^\mathcal{A}$ of $\mathcal{A}$.
    Let $\celluniverse(\mathcal{I}^\mathcal{A})$ be defined as $\cellpointershort \in \celluniverse(\mathcal{I}^\mathcal{A})$ if and only if $\cellpointershort \in \celluniverse$ and there is a configuration $C^\mathcal{A}_i$ of $\mathcal{I}^\mathcal{A}$ where $C^\mathcal{A}_i$ assigns state $s$ to a local variable or base object and $\cellpointershort$ is an element of $s$ (because $s$ is a sequence \Cref{observation:reduction:finite_sequence_of_values}).
\end{definition}

We define a set of fresh pointers that is large enough.
Note that this set trivially exists when $\celluniverse$ is uncountable, and it can be shown that it exists even when $\celluniverse$ is countable by reasoning about the gaps between $\allocatecelloperation{}$ operations in any implementation history of $\mathcal{A}$.

\begin{observation}\label{observation:reduction:unused_pointers_in_run_of_a}
    For every implementation history $\mathcal{I}^\mathcal{A}$ of $\mathcal{A}$ there exists a subset $\celluniverse^0(\mathcal{I}^\mathcal{A}) \subseteq \celluniverse$ such that (a) $\celluniverse^0(\mathcal{I}^\mathcal{A})$ and $\celluniverse(\mathcal{I}^\mathcal{A})$ are disjoint, and (b) $\celluniverse^0(\mathcal{I}^\mathcal{A})$ is countably infinite.
\end{observation}

We are now ready to define how we map states of objects and local variables in $\mathcal{A}$ to $\mathcal{B}$.

\begin{definition}[Swapping Function]\label{def:reduction:swapper}
    Consider any implementation history $\mathcal{I}^\mathcal{A} = C^\mathcal{A}_0, p_1, C^\mathcal{A}_1, \ldots$ of $\mathcal{A}$.
    For every $C^\mathcal{A}_i$, local variable or base object $O$ other than the memory manager of $\mathcal{A}$, and $j$th index of the state assigned to $O$ in $C^\mathcal{A}_i$, we define a swapping function $\mathcal{S}_i(O, j)$ as follows.
    Let $s = v_1, v_2, \ldots$ be the state assigned to $O$ in $C^\mathcal{A}_i$.
    \begin{align*}
        \mathcal{S}_i(O, j) &=
        \begin{cases} 
          \mathcal{M}(\mathcal{W}_i(O, j)) & \text{if}\ \mathcal{W}_i(O, j) \neq \bot \\
          v_j & \text{otherwise} 
       \end{cases}
    \end{align*}
    where $\mathcal{M}$ is an injective function from $\mathbb{N}$ to $\celluniverse^0(\mathcal{I}^\mathcal{A})$.
    This function exists because $\celluniverse^0(\mathcal{I}^\mathcal{A})$ is countably infinite (see \Cref{observation:reduction:unused_pointers_in_run_of_a}).
    For convenience, the notation $\mathcal{S}_i(O)$ means the sequence $\mathcal{S}_i(O, 1), \mathcal{S}_i(O, 2), \ldots$ for each index of $s$ (see \Cref{observation:reduction:finite_sequence_of_values}).
\end{definition}

Observe that $\mathcal{M}$ is defined only after we fix an implementation history $\mathcal{I}^\mathcal{A}$ of $\mathcal{A}$ (because its co-domain is $\celluniverse^0(\mathcal{I}^\mathcal{A})$), so $\mathcal{M}$ is dependent on $\mathcal{I}^\mathcal{A}$.
Throughout the proof, it will always be clear from context which $\mathcal{I}^\mathcal{A}$ we are referring to when using $\mathcal{M}$, so we drop any reference to it.

We now define the mapping from $\mathcal{A}$ to $\mathcal{B}$.

\begin{definition}[$\mathcal{A}$ to $\mathcal{B}$ Mapping]\label{def:reduction:run_mapping}
    Let $\mathcal{I}^\mathcal{A} = C^\mathcal{A}_0, p_1, C^\mathcal{A}_1, \ldots$ be any implementation history of $\mathcal{A}$.
    We define $\mathcal{I} = C_0, p_1, C_1, \ldots$ as follows.
    Consider any configuration $C^\mathcal{A}_i$ in $\mathcal{I}^\mathcal{A}$.
    For every local variable or base object $O$ other than the memory manager in $\mathcal{A}$ such that $O \notin \celluniverse_O$ or $O$ is in $\headobject$, $C_i$ assigns state $\mathcal{S}_i(O)$ to $O$.
    $C_i$ assigns states to objects of cells as follows.
    Let $W$ be the set of step numbers up to and including the $i$th step of $\mathcal{I}^\mathcal{A}$ which perform $\allocatecelloperation{}$ operations.
    For every $\cellpointershort \in \celluniverse \setminus \mathcal{M}[W]$, $C_i$ assigns every object of the cell pointed to by $\cellpointershort$ to its initial state (as defined in \Cref{alg:efficient_algo}).
    Consider any $w \in W$.
    Let $\cellpointershort$ be the response of the $\allocatecelloperation{}$ operation performed during the $w$th step of $\mathcal{I}^\mathcal{A}$.
    If for all $w < j \leq i$ the $j$th step of $\mathcal{I}^\mathcal{A}$ does not perform an $\allocatecelloperation{}$ operation whose response is $\cellpointershort$, then $C_i$ assigns state $\mathcal{S}_i((*\cellpointershort).f)$ to $(*\mathcal{M}(w)).f$ for every $f$ equal to $\lastrepositoryoperationresponse{}$, $\revocations$, or $\nextlong$.
    Otherwise, let $j$ be the minimum $w < j \leq i$ such that the $j$th step of $\mathcal{I}^\mathcal{A}$ performs an $\allocatecelloperation{}$ cell operation whose response is $\cellpointershort$.
    Then, $C_i$ assigns state $\mathcal{S}_{j - 1}((*\cellpointershort).f)$ to $(*\mathcal{M}(w)).f$.
    Finally, $C_i$ assigns state $\mathcal{M}[W]$ to the memory manager.
\end{definition}

We now prove some basic facts about this mapping.

\begin{lemma}\label{lemma:reduction:same_program_counters_in_mapped_run}
    Let $\mathcal{I}^\mathcal{A} = C^\mathcal{A}_0, p_1, C^\mathcal{A}_1, \ldots$ be any implementation history of $\mathcal{A}$ and let $\mathcal{I}$ be the sequence $C_0, p_1, C_1, \ldots$ defined in \Cref{def:reduction:run_mapping}.
    For every $i$, the program counter of each process is the same in $C^\mathcal{A}_{i}$ and $C_{i}$.
\end{lemma}

\begin{proof}
    Suppose, for contradiction, the program counter $pc$ of some process $p$ is different in $C^\mathcal{A}_{i}$ and $C_{i}$.
    Hence, since $pc$ stores a single value, by \Cref{def:reduction:run_mapping}, $\mathcal{S}_i(pc, 1) \neq pc$, and so by \Cref{def:reduction:swapper}, $\mathcal{W}_i(pc, 1) \neq \bot$.
    However, since $pc$ is a program counter, by \Cref{def:reduction:watermarks}, $\mathcal{W}_i(pc, 1) = \bot$, a contradiction.
    \qH{\Cref{lemma:reduction:same_program_counters_in_mapped_run}}
\end{proof}

\begin{lemma}\label{lemma:reduction:allocate_response_in_b_is_swapped}
    Let $\mathcal{I}^\mathcal{A} = C^\mathcal{A}_0, p_1, C^\mathcal{A}_1, \ldots$ be any implementation history of $\mathcal{A}$ and let $\mathcal{I} = C_0, p_1, C_1, \ldots$ be the sequence defined in \Cref{def:reduction:run_mapping}.
    Suppose the $i$th step of $\mathcal{I}^\mathcal{A}$ performs an $\allocatecelloperation{}$ operation and $\mathcal{I}_{i - 1} = C_0, p_1, C_1, \ldots C_{i - 1}$ is an implementation history of $\mathcal{B}$.
    Let $\mathcal{I}^\mathcal{B}_i = C_0, p_1, C_1, \ldots C_{i - 1}, p_i, C^\mathcal{B}_i$ be a one step continuation of $\mathcal{I}_{i - 1}$ by $p_i$.
    If the state of the memory manager is the same in $C_i$ and $C^\mathcal{B}_i$, then the $i$th step of $\mathcal{I}^\mathcal{B}_i$ performs an $\allocatecelloperation{}$ operation whose response is $\mathcal{M}(i)$.
\end{lemma}

\begin{proof}
    Suppose the $i$th step of $\mathcal{I}^\mathcal{A}$ performs an $\allocatecelloperation{}$ operation.
    Let $A_{i - 1}$ (resp. $A_i$) be the state of the memory manager in $C_{i - 1}$ (resp. $C_i$) and let $W$ be the set of step numbers up to and including the $i$th step of $\mathcal{I}^\mathcal{A}$ which perform $\allocatecelloperation{}$ operations.
    Since the $i$th step of $\mathcal{I}^\mathcal{A}$ is an $\allocatecelloperation{}$ operation, we have that $i \in W$.
    Hence, by \Cref{def:reduction:run_mapping}, $A_{i - 1} = \mathcal{M}[W\setminus \{i\}]$ and $A_{i} = \mathcal{M}[W]$.
    Thus, either $A_i \setminus A_{i - 1} = \emptyset$ or $A_i \setminus A_{i - 1} = \{\mathcal{M}(i)\}$.
    
    We now prove that $A_i \setminus A_{i - 1} \neq \emptyset$, which implies that $A_i \setminus A_{i - 1} = \{\mathcal{M}(i)\}$.
    Suppose, for contradiction, $A_i \setminus A_{i - 1} = \emptyset$.
    Since $A_{i} = \mathcal{M}[W]$ and $A_{i - 1} = \mathcal{M}[W\setminus \{i\}]$, we have that $A_i \setminus A_{i - 1} = \mathcal{M}[W] \setminus \mathcal{M}[W\setminus \{i\}]$, and so $\mathcal{M}[W] \setminus \mathcal{M}[W\setminus \{i\}] = \emptyset$.
    Furthermore, since $i \in W$, we have that $\mathcal{M}(i) \in \mathcal{M}[W]$.
    Hence, since $\mathcal{M}[W] \setminus \mathcal{M}[W\setminus \{i\}] = \emptyset$, we have that $\mathcal{M}(i) \in \mathcal{M}[W\setminus \{i\}]$.
    Thus, for some $j \in W\setminus \{i\}$, we have that $\mathcal{M}(i) = \mathcal{M}(j)$.
    Therefore, since $j \in W\setminus \{i\}$, we have that $i \neq j$ and $\mathcal{M}(i) = \mathcal{M}(j)$.
    However, $\mathcal{M}$ is injective, so $i \neq j$ implies $\mathcal{M}(i) \neq \mathcal{M}(j)$, a contradiction.
    
    We now finish the proof of \Cref{lemma:reduction:allocate_response_in_b_is_swapped}.
    Since $p_i$ takes the $i$th step of $\mathcal{I}^\mathcal{A}$ and $\mathcal{I}^\mathcal{B}_i$, and by \Cref{lemma:reduction:same_program_counters_in_mapped_run} the program counter of $p_{i}$ is the same in $C^\mathcal{A}_{i - 1}$ and $C_{i - 1}$, we have that the $i$th step of $\mathcal{I}^\mathcal{B}_i$ performs an $\allocatecelloperation{}$ operation.
    Hence, since $\mathcal{I}^\mathcal{B}_i$ is an implementation history of $\mathcal{B}$, by \Cref{alg:lazy_cell_manager_specification}, the response of the $\allocatecelloperation{}$ operation performed during the $i$th step of $\mathcal{I}^\mathcal{B}_i$ is the pointer in the state of the memory manager in $C^\mathcal{B}_i$ which is not in the state of the memory manager in $C_{i - 1}$.
    Therefore, since by assumption the state of the memory manager is the same in $C_i$ and $C^\mathcal{B}_i$, and $A_i$ is the state of the memory manager in $C_i$, we have that $A_i$ is the state of the memory manager in $C^\mathcal{B}_i$, and since $A_{i - 1}$ is the state of the memory manager in $C_{i - 1}$, and $A_i \setminus A_{i - 1} = \{\mathcal{M}(i)\}$, we have that the $i$th step of $\mathcal{I}^\mathcal{B}_i$ performs an $\allocatecelloperation{}$ operation whose response is $\mathcal{M}(i)$ as wanted.
    \qH{\Cref{lemma:reduction:allocate_response_in_b_is_swapped}}
\end{proof}

\begin{lemma}\label{lemma:reduction:free_input_in_b_is_swapped}
    Let $\mathcal{I}^\mathcal{A} = C^\mathcal{A}_0, p_1, C^\mathcal{A}_1, \ldots$ be any implementation history of $\mathcal{A}$ and let $\mathcal{I} = C_0, p_1, C_1, \ldots$ be the sequence defined in \Cref{def:reduction:run_mapping}.
    Suppose the $i$th step of $\mathcal{I}^\mathcal{A}$ performs a $\freecelloperation{}(\cellpointershort)$ operation and suppose $\mathcal{I}_{i - 1} = C_0, p_1, C_1, \ldots C_{i - 1}$ is an implementation history of $\mathcal{B}$.
    Let $\mathcal{I}^\mathcal{B}_i = C_0, p_1, C_1, \ldots C_{i - 1}, p_i, C^\mathcal{B}_i$ be a one step continuation of $\mathcal{I}_{i - 1}$ by $p_i$.
    Furthermore, let $O$ be the local variable $\currentcellpointershort$ of $p_i$ in the Relinquish procedure.
    Then, $C^\mathcal{A}_{i - 1}$ assigns state $\cellpointershort$ to $O$ and the $i$th step of $\mathcal{I}^\mathcal{B}_{i}$ performs a $\freecelloperation{}(\mathcal{S}_{i - 1}(O))$ operation.
\end{lemma}

\begin{proof}
    Since the $i$th step of $\mathcal{I}^\mathcal{A}$ performs a $\freecelloperation{}(\cellpointershort)$ operation, we know that $p_{i}$ performed this operation because it saw $O$ to be assigned $\cellpointershort$ in $C^\mathcal{A}_{i - 1}$.
    Hence, the value of $O$ determines the input to $p_{i}$'s $\freecelloperation{}$ operation in the $i$th step of $\mathcal{I}^\mathcal{A}$.
    Thus, since by \Cref{lemma:reduction:same_program_counters_in_mapped_run} the program counter of $p_{i}$ is the same in $C^\mathcal{A}_{i-1}$ and $C_{i-1}$, we have that $i$th step of $\mathcal{I}^\mathcal{B}_{i}$ performs a $\freecelloperation{}(v)$ operation where $v$ is the value of $O$ in $C_{i - 1}$.
    Since $O$ is assigned to $\cellpointershort$ in $C^\mathcal{A}_{i - 1}$, by \Cref{def:reduction:run_mapping}, $O$ is assigned to $ \mathcal{S}_{i - 1}(O)$ in $C_{i - 1}$.
    Therefore, since $i$th step of $\mathcal{I}^\mathcal{B}_{i}$ performs a $\freecelloperation{}(v)$ operation where $v$ is the value of $O$ in $C_{i - 1}$, we have that the $i$th step of $\mathcal{I}^\mathcal{B}_{i}$ performs a $\freecelloperation{}(\mathcal{S}_{i - 1}(O))$ operation as wanted.
    \qH{\Cref{lemma:reduction:free_input_in_b_is_swapped}}
\end{proof}

\begin{lemma}\label{lemma:reduction:associated_pointer}
    Consider any implementation history $\mathcal{I}^\mathcal{A} = C^\mathcal{A}_0, p_1, C^\mathcal{A}_1, \ldots$ of $\mathcal{A}$.
    For every configuration $C^\mathcal{A}_i$ of $\mathcal{I}^\mathcal{A}$, local variable or base object $O$ other than the memory manager of $\mathcal{A}$, and $j$th index of the state assigned to $O$ in $C^\mathcal{A}_i$, if $\mathcal{W}_i(O, j) = k \neq \bot$, then $k \leq i$ and $p_k$ performed an $\allocatecelloperation{}$ operation during the $k$th step of $\mathcal{I}^\mathcal{A}$ with response $v_j$ where $v_j$ is the value of the $j$th index of the state assigned to $O$ in $C^\mathcal{A}_i$.
\end{lemma}

\begin{proof}
    By induction on $i$.
    
    \begin{itemize}
        \item[] \hspace{0pt}\textbf{Base Case.} $i = 0$.

        Hence, by \Cref{def:reduction:watermarks}, $\mathcal{W}_0(O, j) = \bot$ for every local variable or base object $O$ other than the memory manager of $\mathcal{A}$ and $j$th index of the state of $O$, so the claim vacuously holds.
        
        \item[] \hspace{0pt}\textbf{Inductive Case.} for every $i$, if the claim holds for $C^\mathcal{A}_i$, then the claim holds for $C^\mathcal{A}_{i + 1}$.

        Suppose for some $i$, the claim holds for $C^\mathcal{A}_{i}$.
        This is the inductive hypothesis.
        Consider $C^\mathcal{A}_{i + 1}$,  local variable or base object $O$ other than the memory manager of $\mathcal{A}$, and $j$th index of the state assigned to $O$ in $C^\mathcal{A}_{i + 1}$.
        Suppose $\mathcal{W}_{i + 1}(O, j) = k \neq \bot$, we will prove that $k \leq i + 1$ and $p_k$ performed an $\allocatecelloperation{}$ operation during the $k$th step of $\mathcal{I}^\mathcal{A}$ with response $v_j$ where $v_j$ is the value of the $j$th index of the state assigned to $O$ in $C^\mathcal{A}_{i + 1}$.
        Hence, by \Cref{def:reduction:watermarks}, $k = i + 1$ (2.a) or $k = \mathcal{W}_i(O', l)$ for some local variable or base object $O'$ other than the memory manager of $\mathcal{A}$ and the $l$th index of the state assigned to $O'$ in $C^\mathcal{A}_i$ (1, 2.b, 2.c, and 3).
        We consider each case separately.
        \begin{itemize}
            \item[] \hspace{0pt}\textbf{Case 1.} $k = i + 1$.

            Hence, $k \leq i + 1$ and by \Cref{def:reduction:watermarks}, during the $i + 1$th step of $\mathcal{I}^\mathcal{A}$ $p_{i + 1}$ performs an $\allocatecelloperation{}$ operation whose response is $v_j$.
            Therefore, since $k = i + 1$, we have that $p_k$ performed an $\allocatecelloperation{}$ operation during the $k$th step of $\mathcal{I}^\mathcal{A}$ with response $v_j$ as wanted.

            \item[] \hspace{0pt}\textbf{Case 2.} $k = \mathcal{W}_i(O', l)$ for some local variable or base object $O'$ other than the memory manager of $\mathcal{A}$ and the $l$th index of the state assigned to $O'$ in $C^\mathcal{A}_i$.

            Hence, by \Cref{def:reduction:watermarks}, the $j$th index of $O$ is $v_j$ in $C^\mathcal{A}_{i + 1}$ because the $l$th index of $O'$ in $C^\mathcal{A}_i$ is $v_j$.
            Since $\mathcal{W}_i(O', l) = k \neq \bot$, by the inductive hypothesis, $k \leq i$ and $p_k$ performed an $\allocatecelloperation{}$ operation during the $k$th step of $\mathcal{I}^\mathcal{A}$ with response $v_l$ where $v_l$ is the value of the $l$th index of the state assigned to $O'$ in $C^\mathcal{A}_i$.
            Hence, since the $l$th index of $O'$ in $C^\mathcal{A}_i$ is $v_j$, we have that $p_k$ performed an $\allocatecelloperation{}$ operation during the $k$th step of $\mathcal{I}^\mathcal{A}$ with response $v_j$ as wanted.
            \qH{\Cref{lemma:reduction:associated_pointer}}
        \end{itemize}
    \end{itemize}
\end{proof}

\begin{lemma}\label{lemma:ero:psi_is_basically_injective}
    Consider any implementation history $\mathcal{I}^\mathcal{A}$ of $\mathcal{A}$, configuration $C^\mathcal{A}_i$ of $\mathcal{I}^\mathcal{A}$, local variable or base object $O$ (resp. $O'$) other than the memory manager of $\mathcal{A}$, and value $v_j$ (resp. $v'_k$) in the $j$th (resp. $k$th) index of the state assigned to $O$ (resp. $O'$) in $C^\mathcal{A}_i$.
    Suppose if $v_j \in \celluniverse$ (resp. $v'_k \in \celluniverse$), then $v_j \in \celluniverse(\mathcal{I}^\mathcal{A})$ (resp. $v'_k \in \celluniverse(\mathcal{I}^\mathcal{A})$).
    If $v_j \neq v'_k$, then $\mathcal{S}_i(O, j) \neq \mathcal{S}_i(O', k)$.
\end{lemma}

\begin{proof}
    There are four cases.

    \begin{itemize}
        \item[] \hspace{0pt}\textbf{Case 1.} $\mathcal{S}_i(O, j) = v_j$ and $\mathcal{S}_i(O', k) = v'_k$.

        Hence, since $v_j \neq v'_k$, we have that $\mathcal{S}_i(O, j) \neq \mathcal{S}_i(O', k)$.

        \item[] \hspace{0pt}\textbf{Case 2.} $\mathcal{S}_i(O, j) = v_j$ and $\mathcal{S}_i(O', k) \neq v'_k$.

        Hence, by \Cref{def:reduction:swapper} $\mathcal{S}_i(O', k) \in C^0(\mathcal{I}^\mathcal{A})$, and so by \Cref{observation:reduction:unused_pointers_in_run_of_a} $\mathcal{S}_i(O', k) \in \celluniverse$. If $v_j \notin \celluniverse$, then $v_j \neq \mathcal{S}_i(O', k)$, and so $\mathcal{S}_i(O, j) \neq \mathcal{S}_i(O', k)$.
        Otherwise, if $v_j \in \celluniverse$, then by assumption $v_j \in \celluniverse(\mathcal{I}^\mathcal{A})$.
        Hence, since by \Cref{observation:reduction:unused_pointers_in_run_of_a} $\celluniverse^0(\mathcal{I}^\mathcal{A})$ and $\celluniverse(\mathcal{I}^\mathcal{A})$ are disjoint, $\mathcal{S}_i(O', k) \in C^0(\mathcal{I}^\mathcal{A})$, and $v_j \in \celluniverse(\mathcal{I}^\mathcal{A})$, we have that $v_j \neq \mathcal{S}_i(O', k)$, and so $\mathcal{S}_i(O, j) \neq \mathcal{S}_i(O', k)$.

        \item[] \hspace{0pt}\textbf{Case 3.} $\mathcal{S}_i(O, j) \neq v_j$ and $\mathcal{S}_i(O', k) = v'_k$.

        The proof is symmetrical to Case 2 and is included below for completeness.
        Since $\mathcal{S}_i(O, j) \neq v_j$ and $\mathcal{S}_i(O', k) = v'_k$, by \Cref{def:reduction:swapper}, $\mathcal{S}_i(O, j) \in C^0(\mathcal{I}^\mathcal{A})$, and so by \Cref{observation:reduction:unused_pointers_in_run_of_a} $\mathcal{S}_i(O, j) \in \celluniverse$. If $v'_k \notin \celluniverse$, then $v'_k \neq \mathcal{S}_i(O, j)$, and so $\mathcal{S}_i(O, j) \neq \mathcal{S}_i(O', k)$.
        Otherwise, if $v'_k \in \celluniverse$, then by assumption $v'_k \in \celluniverse(\mathcal{I}^\mathcal{A})$.
        Hence, since by \Cref{observation:reduction:unused_pointers_in_run_of_a} $\celluniverse^0(\mathcal{I}^\mathcal{A})$ and $\celluniverse(\mathcal{I}^\mathcal{A})$ are disjoint, $\mathcal{S}_i(O, j) \in C^0(\mathcal{I}^\mathcal{A})$, and $v'_k \in \celluniverse(\mathcal{I}^\mathcal{A})$, we have that $v'_k \neq \mathcal{S}_i(O, j)$, and so $\mathcal{S}_i(O, j) \neq \mathcal{S}_i(O', k)$ as wanted.

        \item[] \hspace{0pt}\textbf{Case 4.} $\mathcal{S}_i(O, j) \neq v_j$ and $\mathcal{S}_i(O', k) \neq v'_k$.

        Hence, by \Cref{def:reduction:swapper}, $\mathcal{S}_i(O, j) = \mathcal{M}(\mathcal{W}_i(O, j))$ and $\mathcal{S}_i(O', k) = \mathcal{M}(\mathcal{W}_i(O', k))$, and so by \Cref{observation:reduction:unused_pointers_in_run_of_a} $\mathcal{S}_i(O, j) \in \celluniverse$ and $\mathcal{S}_i(O', k) \in \celluniverse$.
        Furthermore, $\mathcal{W}_i(O, j) = w \neq \bot$ and $\mathcal{W}_i(O', k) = w' \neq \bot$.
        Thus, by \Cref{lemma:reduction:associated_pointer}, the $w$th (resp. $w'$th) step of $\mathcal{I}^\mathcal{A}$ is an $\allocatecelloperation{}$ operation whose response is $v_j$ (resp. $v'_k$).
        Since $v_j \neq v'_k$, this implies that $w \neq w'$.
        Hence, $\mathcal{W}_i(O, j) \neq \mathcal{W}_i(O', k)$, and so since $\mathcal{M}$ is injective, we have that $\mathcal{M}(\mathcal{W}_i(O, j)) \neq \mathcal{M}(\mathcal{W}_i(O', k))$.
        Therefore, $\mathcal{S}_i(O, j) \neq \mathcal{S}_i(O', k)$ as wanted.
        \qH{\Cref{lemma:ero:psi_is_basically_injective}}
    \end{itemize}
\end{proof}

\subsubsection{The mapping produces implementation histories of $\mathcal{B}$}

In this section, we prove that the mapping is an implementation of $\mathcal{B}$.
We start with an observation about how a step decides what object to perform a step on, and then prove this claim. 

\begin{observation}\label{observation:reduction:source_of_pointer}
    In any implementation history $\mathcal{I}^\mathcal{A} = C^\mathcal{A}_0, p_1, C^\mathcal{A}_1, \ldots$ of $\mathcal{A}$ if $p_{i}$ performs an operation on a base object $O \in \celluniverse_O$ during the $i$th step of $\mathcal{I}^\mathcal{A}$, then by \Cref{def:reduction:shared_objects} $O = (*\cellpointershort).f$ for some $\cellpointershort \in \celluniverse \cup \{\&\headobject\}$ where $f$ is either $\lastrepositoryoperationresponse{}$, $\revocations$, or $\nextlong$, because one of $p_i$'s local variables $O_s$ was assigned state $\cellpointershort$ in $C^\mathcal{A}_{i - 1}$.
    We call $O_s$ the source of $O$.
\end{observation}

\begin{lemma}\label{lemma:reduction:mapped_history_is_for_b}
    For every implementation history $\mathcal{I}^\mathcal{A} = C^\mathcal{A}_0, p_1, C^\mathcal{A}_1, \ldots$ of algorithm $\mathcal{A}$, $\mathcal{I} = C_0, p_1, C_1, \ldots$ as defined in \Cref{def:reduction:run_mapping} is an implementation history of $\mathcal{B}$.
\end{lemma}

\begin{proof}
    The claim follows by proving the following predicate.
    Let $\mathcal{P}(n)$ be the predicate: for every implementation history $\mathcal{I}^\mathcal{A}_n = C^\mathcal{A}_0, p_1, C^\mathcal{A}_1, \ldots, C^\mathcal{A}_n$ of $\mathcal{A}$, $\mathcal{I}_n = C_0, p_1, C_1, \ldots, C_n$ as defined in \Cref{def:reduction:run_mapping} is an implementation history of $\mathcal{B}$.
    We prove $\mathcal{P}(n)$ by induction on $n$.

    \begin{itemize}
        \item[] \textbf{Base Case.} $\mathcal{P}(0)$.

        Consider any local variable or base object $O$ other than the memory manager such that $O \notin \celluniverse_O$.
        Hence, by \Cref{def:reduction:run_mapping}, $C_0$ assigns state $\mathcal{S}_0(O)$ to $O$.
        Since by \Cref{def:reduction:watermarks} the output of $\mathcal{W}_0$ for every input is $\bot$, by \Cref{def:reduction:swapper}, $\mathcal{S}_0(O)$ is the state assigned to $O$ in $\mathcal{C}^\mathcal{A}_0$.
        We now consider objects of cells and the memory manager.
        Since no steps have been performed in $C^\mathcal{A}_0$, we have that no $\allocatecelloperation{}$ operations have been performed in $C^\mathcal{A}_0$, and so the $W$ defined in \Cref{def:reduction:run_mapping} is empty.
        Hence, $\mathcal{M}[W]$ is empty, implying (1) for every $\cellpointershort \in \celluniverse$ $C_0$ assigns every object of the cell pointed to by $\cellpointershort$ to its initial state; and (2) $C_0$ assigns the empty state to the memory manager.
        Thus, $C_0$ assigns the same state to (1) every object of every cell; and (2) the memory manager as $C^\mathcal{A}_0$.
        Therefore, since $C_0$ assigns the same state to every local variable and every statically allocated object, we have that $C^\mathcal{A}_0 = C_0$, and so since the initial configurations of $\mathcal{A}$ and $\mathcal{B}$ are the same, we have that $C_0$ is an implementation history of $\mathcal{B}$ as wanted.

        \item[] \textbf{Inductive Case.} $\forall n\ \mathcal{P}(n) \implies \mathcal{P}(n + 1)$.

        Suppose for some $n$ $\mathcal{P}(n)$ holds.
        This is the inductive hypothesis.
        Consider any implementation history $\mathcal{I}^\mathcal{A}_{n + 1} = C^\mathcal{A}_0, p_1, C^\mathcal{A}_1, \ldots, C^\mathcal{A}_{n + 1}$ of $\mathcal{A}$ and let $\mathcal{I}_{n + 1} = C_0, p_1, C_1, \ldots, C_{n + 1}$ be the sequence defined in \Cref{def:reduction:run_mapping}.
        Let $\mathcal{I}_n$ be the prefix of $\mathcal{I}_{n + 1}$ up to and including the $n$th step, i.e., $\mathcal{I}_n = C_0, p_1, C_1, \ldots, C_{n}$.
        By the inductive hypothesis, we have that $\mathcal{I}_n$ is an implementation history of $\mathcal{B}$.
        We must prove that $C_{n + 1}$ is reachable after a single step of $p_{n + 1}$ after $\mathcal{I}_n$.
        We start by identifying a reachable configuration $C^\mathcal{B}_{n + 1}$ after a single step of $p_{n + 1}$ after $\mathcal{I}_n$ which we will prove is equal to $C_{n + 1}$.
    
        \begin{claimcustom}{\ref{lemma:reduction:mapped_history_is_for_b}.1}\label{claim:reduction:specificy_b_i+1}
            Since $\mathcal{I}_n$ is an implementation history of $\mathcal{B}$, a single step of $p_{n + 1}$ after $\mathcal{I}_n$ leads to a set of possible configurations denoted by $\mathbf{C} = \{\mathbf{c}_1, \mathbf{c}_2, \ldots\}$.
            If $p_{n + 1}$ performs an $\allocatecelloperation{}$ operation after $\mathcal{I}_n$, then some $\mathbf{c}_j \in \mathbf{C}$ assigns $A_n \cup \{\mathcal{M}(n + 1)\}$ to the memory manager where $A_n$ is the state of the memory manager in $C_n$.
        \end{claimcustom}
    
        \begin{proof}
            Suppose, for contradiction, $p_{n + 1}$ performs an $\allocatecelloperation{}$ operation after $\mathcal{I}_n$ and every $\mathbf{c} \in \mathbf{C}$ does not assign $A_n \cup \{\mathcal{M}(n + 1)\}$ to the memory manager.
            By \Cref{alg:lazy_cell_manager_specification}, the state of the memory manager after an $\allocatecelloperation{}$ operation starting from $C_n$ can be any set $A_{n + 1}$ so long as $A_{n + 1} \setminus A_n = \{x\}$ for $x \in \celluniverse$.
            Hence, since by assumption every $\mathbf{c} \in \mathbf{C}$ does not assign $A_n \cup \{\mathcal{M}(n + 1)\}$ to the memory manager, we have either $(A_n \cup \{\mathcal{M}(n + 1)\}) \setminus A_n = \emptyset$ or $\mathcal{M}(n + 1) \notin \celluniverse$.
            Since the co-domain of $\mathcal{M}$ is $\celluniverse^0(\mathcal{I}^\mathcal{A})$, which by \Cref{observation:reduction:unused_pointers_in_run_of_a} is a subset of $\celluniverse$, the latter is impossible, so $(A_n \cup \{\mathcal{M}(n + 1)\}) \setminus A_n = \emptyset$.
            Hence, $\mathcal{M}(n + 1) \in A_n$, so by \Cref{def:reduction:run_mapping}, for some $i \leq n$ $\mathcal{M}(i) = \mathcal{M}(n + 1)$.
            Therefore, $i \neq n + 1$ and $\mathcal{M}(i) = \mathcal{M}(n + 1)$.
            However, since $\mathcal{M}$ is injective, $i \neq n+1$ implies $\mathcal{M}(i) \neq \mathcal{M}(n + 1)$, a contradiction.
            \qH{\Cref{claim:reduction:specificy_b_i+1}}
        \end{proof}
    
        We now define a configuration $C^\mathcal{B}_{n + 1}$ reachable by a single step of $p_{n + 1}$ after $\mathcal{I}_n$.
        Since each line of code except \cref{line:ero:invocation_step} and \ref{line:ero:apply_op} are deterministic\footnote{\Cref{line:ero:invocation_step} is non-deterministic because, given a configuration $C$ of $\mathcal{B}$ where a process $p$'s program counter is one in $C$, a one step continuation from $C$ by $p$ leads to a different configuration depending on the operation $p$ invokes (these configurations are different because $p$ assigns its local variable $\operationshort{}$ in the \highleveloperation{} procedure to the operation it invokes). Conversely, \cref{line:ero:apply_op} is non-deterministic when the implemented object type $\mathcal{T}$ is non-deterministic.}, every base object other than the memory manager is deterministic, and for every $v$ the $\freecelloperation{}(v)$ operation on the lazy memory manager is deterministic, we have that if $p_{n + 1}$ does not execute \cref{line:ero:invocation_step} nor \ref{line:ero:apply_op} nor perform an $\allocatecelloperation{}$ operation after $\mathcal{I}_n$, then there is only a single reachable configuration $C$ by a single step of $p_{n + 1}$ after $\mathcal{I}_n$.
        We now consider these three exceptions.
        Since by \Cref{lemma:reduction:same_program_counters_in_mapped_run} the program counter of $p_{n + 1}$ is the same in $C^\mathcal{A}_n$ and $C_n$, if $p_{n + 1}$ executes line $\ell$ during the $n + 1$th step of $\mathcal{I}^\mathcal{A}_{n + 1}$, then $p_{n + 1}$ executes line $\ell$ after $\mathcal{I}_n$.
        If $p_{n + 1}$ executes \cref{line:ero:invocation_step} during the $n + 1$th step of $\mathcal{I}^\mathcal{A}_{n + 1}$, then $C$ is the configuration where $p_{n + 1}$ performs an invocation step for the same operation after $\mathcal{I}_n$.
        Otherwise, if $p_{n + 1}$ executes \cref{line:ero:apply_op} during the $n + 1$th step of $\mathcal{I}^\mathcal{A}_{n + 1}$, then $C$ is the configuration where $p_{n + 1}$ sets its local variables $s'$ and $r'$ to the same values as in $C^\mathcal{A}_{n + 1}$.
        Note that this is a possible configuration because: (1) by tracing backward $\operationshort{}$ is an operation of type $\mathcal{T}$; and (2) by a simple induction $s$ is a state of type $\mathcal{T}$.
        Lastly, if $p_{n + 1}$ performs an $\allocatecelloperation{}$ operation during the $n + 1$th step of $\mathcal{I}^\mathcal{A}_{n + 1}$, then $p_{n + 1}$ performs an $\allocatecelloperation{}$ operation after $\mathcal{I}_n$, so by \Cref{claim:reduction:specificy_b_i+1}, there is a reachable configuration $C$ by a single step of $p_{n + 1}$ after $\mathcal{I}_n$ that assigns $A_n \cup \{\mathcal{M}(n + 1)\}$ to the memory manager.
        In all cases, we let $C^\mathcal{B}_{n + 1} = C$ (where the $C$ chosen is dependent on the case).
        
        We must prove that $C^\mathcal{B}_{n + 1} = C_{n + 1}$.
        Let $\mathcal{I}^\mathcal{B}_{n + 1} = C_0, p_1, C_1, \ldots, C_{n}, p_{n + 1}, C^\mathcal{B}_{n + 1}$.
        Since $\mathcal{I}_n$ is an implementation history of $\mathcal{B}$ and $C^\mathcal{B}_{n + 1}$ is reachable by a single step of $p_{n + 1}$ after $\mathcal{I}_n$, we have that $\mathcal{I}^\mathcal{B}_{n + 1}$ is an implementation history of $\mathcal{B}$.
        We start by proving that every line of code executed during $\mathcal{I}^\mathcal{A}_{n + 1}$ that intends to perform an operation on an object of a cell actually does.

        \begin{claimcustom}{\ref{lemma:reduction:mapped_history_is_for_b}.2}\label{claim:reduction:steps_in_a_are_not_on_bad_pointers}
            Every execution of line \ref{line:ero:copy_response_out_of_cell}, \ref{line:ero:do_work_initialize_response}, \ref{line:ero:do_work_while_loop}, \ref{line:ero:add_cell_read_end_of_list}, \ref{line:ero:add_cell_to_list}, \ref{line:ero:remove_cell_remove_seal_loop}, \ref{line:ero:remove_cell_read_pointer_to_remove_before_seal}, \ref{line:ero:seal_cell}, \ref{line:ero:remove_cell_read_pointer_to_remove}, \ref{line:ero:remove_cell_read_previous_pointer}, \ref{line:ero:remove_cell_from_list}, \ref{line:ero:copy_acquisitions_to_revocations}, \ref{line:ero:responses_set_attempt}, \ref{line:ero:announce_op_response_check}, \ref{line:ero:acquire_next_read_curr_unique_pointer}, \ref{line:ero:acquire_next_cell}, and \ref{line:ero:relinquish_revocations} in $\mathcal{I}^\mathcal{A}_{n + 1}$ performs an operation on an object of a cell in $\celluniverse{} \cup \{\&\headobject\}$.
        \end{claimcustom}

        \begin{proof}
            Suppose, for contradiction, that an execution of one of these lines in $\mathcal{I}^\mathcal{A}_{n + 1}$ does not perform an operation on an object of a cell in $\celluniverse{} \cup \{\&\headobject\}$; say the $i$th step.
            Hence, since $p_i$ takes the $i$th step in $\mathcal{I}^\mathcal{A}_{n + 1}$ and $\mathcal{I}^\mathcal{B}_{n + 1}$, and by \Cref{lemma:reduction:same_program_counters_in_mapped_run} the program counter of $p_i$ is the same in $C^\mathcal{A}_{i - 1}$ and $C_{i - 1}$, it follows that there is an execution of one of the lines listed in the claim in $\mathcal{I}^\mathcal{B}_{n + 1}$ that does not perform an operation on an object of a cell in $\celluniverse{} \cup \{\&\headobject\}$.
            However, since $\mathcal{I}^\mathcal{B}_{n + 1}$ is an implementation history of $\mathcal{B}$, this contradicts \Cref{lemma:ero:b_never_performs_an_op_on_a_bad_pointer}.
            \qH{\Cref{claim:reduction:steps_in_a_are_not_on_bad_pointers}}
        \end{proof}

        We now prove that the state of the memory manager is the same in $C^\mathcal{B}_{n + 1}$ and $C_{n + 1}$.
        This is useful because it allows us to satisfy the conditions of \Cref{lemma:reduction:allocate_response_in_b_is_swapped}, enabling us to prove that if the $i$th step of $\mathcal{I}^\mathcal{A}_{n + 1}$ performs an $\allocatecelloperation{}$ operation, then the $i$th step of $\mathcal{I}^\mathcal{B}_{n + 1}$ performs an $\allocatecelloperation{}$ operation whose response is $\mathcal{M}(i)$ (see \Cref{claim:reduction:allocate_response_in_b_is_swapped}).
    
        \begin{claimcustom}{\ref{lemma:reduction:mapped_history_is_for_b}.3}\label{claim:reduction:c_b_i+1_and_c_i+1_same_cell_manager}
            The state of the memory manager is the same in $C^\mathcal{B}_{n + 1}$ and $C_{n + 1}$.
        \end{claimcustom}
    
        \begin{proof}
            There are two cases.
    
            \begin{itemize}
                \item[] \hspace{0pt}\textbf{Case 1.} $p_{n + 1}$ does not perform an $\allocatecelloperation{}$ operation during the $n + 1$th step of $\mathcal{I}^\mathcal{A}_{n + 1}$.
    
                Hence, by \Cref{def:reduction:run_mapping}, the state of the memory manager is the same in $C_{n}$ and $C_{n + 1}$.
                Furthermore, since by \Cref{lemma:reduction:same_program_counters_in_mapped_run} the program counter of $p_{n + 1}$ is the same in $C^\mathcal{A}_n$ and $C_n$, and $p_{n + 1}$ takes the $n + 1$th step of $\mathcal{I}^\mathcal{B}_{n + 1}$, we have that $p_{n + 1}$ does not perform an $\allocatecelloperation{}$ operation during the $n + 1$th step of $\mathcal{I}^\mathcal{B}_{n + 1}$.
                Hence, the state of the memory manager is the same in $C_n$ and $C^\mathcal{B}_{n + 1}$.
                Therefore, since the state of the memory manager is the same in $C_{n}$ and $C_{n + 1}$, we have that the state of the memory manager is the same in $C^\mathcal{B}_{n + 1}$ and $C_{n + 1}$ as wanted.
    
                \item[] \hspace{0pt}\textbf{Case 2.} $p_{n + 1}$ performs an $\allocatecelloperation{}$ operation during the $n + 1$th step of $\mathcal{I}^\mathcal{A}_{n + 1}$.
    
                Hence, since by \Cref{lemma:reduction:same_program_counters_in_mapped_run} the program counter of $p_{n + 1}$ is the same in $C^\mathcal{A}_n$ and $C_n$, and $p_{n + 1}$ takes the $n + 1$th step of $\mathcal{I}^\mathcal{B}_{n + 1}$, we have that $p_{n + 1}$ perform an $\allocatecelloperation{}$ operation during the $n + 1$th step of $\mathcal{I}^\mathcal{B}_{n + 1}$.
                Thus, by the definition of $C^\mathcal{B}_{n + 1}$, we have that $C^\mathcal{B}_{n + 1}$ assigns state $A_n \cup \{\mathcal{M}(n + 1)\}$ to the memory manager where $C_n$ assigns state $A_n$ to the memory manager.
                By \Cref{def:reduction:run_mapping}, $C_n$ (resp. $C_{n + 1}$) assigns state $\mathcal{M}[W_{n}]$ (resp. $\mathcal{M}[W_{n + 1}]$) where $W_n$ (resp. $W_{n + 1}$) is the set of step numbers up to and including the $n$th (resp. $n + 1$th) step of $\mathcal{I}^\mathcal{A}_{n + 1}$ which perform $\allocatecelloperation{}$ operations, so $A_n = \mathcal{M}[W_n]$.
                Hence, since $p_{n + 1}$ performs an $\allocatecelloperation{}$ operation during the $n + 1$th step of $\mathcal{I}^\mathcal{A}_{n + 1}$, we have that $n + 1 \notin W_n$ and $n + 1 \in W_{n + 1}$, and so it follows that $W_{n + 1} = W_n \cup \{n + 1\}$.
                Thus, $\mathcal{M}[W_{n + 1}] = \mathcal{M}[W_n \cup \{n + 1\}]$ which simplifies to $\mathcal{M}[W_n] \cup \{\mathcal{M}(n + 1)\}$.
                So, since $A_n = \mathcal{M}[W_n]$, we have that $\mathcal{M}[W_{n + 1}] = A_n \cup \{\mathcal{M}(n + 1)\}$.
                Therefore, $C_{n + 1}$ assigns state $A_n \cup \{\mathcal{M}(n + 1)\}$ to the memory manager, and so the state of the memory manager is the same in $C^\mathcal{B}_{n + 1}$ and $C_{n + 1}$.
                \qH{\Cref{claim:reduction:c_b_i+1_and_c_i+1_same_cell_manager}}
            \end{itemize}
        \end{proof}
    
        \begin{claimcustom}{\ref{lemma:reduction:mapped_history_is_for_b}.4}\label{claim:reduction:allocate_response_in_b_is_swapped}
            If the $i$th step of $\mathcal{I}^\mathcal{A}_{n + 1}$ performs an $\allocatecelloperation{}$ operation, then the $i$th step of $\mathcal{I}^\mathcal{B}_{n + 1}$ performs an $\allocatecelloperation{}$ operation whose response is $\mathcal{M}(i)$.
        \end{claimcustom}
    
        \begin{proof}
            %Suppose the $i$th step of $\mathcal{I}^\mathcal{A}_{n + 1}$ is an $\allocatecelloperation{}$ operation.
            First, suppose that $i \leq n$.
            Hence, by the inductive hypothesis, $\mathcal{I}_{i  - 1} = C_0, p_1, C_1, \ldots, C_{i - 1}$ is an implementation history of $\mathcal{B}$, $\mathcal{I}_i = C_0, p_1, C_1, \ldots, C_{i - 1}, p_i, C_i$ is a one step continuation of $\mathcal{I}_{i - 1}$ by $p_i$, and the state of the memory manager is by definition the same in $C_i$ and $C_i$, and so by \Cref{lemma:reduction:allocate_response_in_b_is_swapped}, the $i$th step of $\mathcal{I}_{i}$ (and thus $\mathcal{I}^\mathcal{B}_{n + 1}$) performs an $\allocatecelloperation{}$ operation whose response is $\mathcal{M}(i)$.
            Now suppose $i = n + 1$.
            By the inductive hypothesis $\mathcal{I}_n = C_0, p_1, C_1, \ldots, C_n$ is an implementation history of $\mathcal{B}$, and by definition $\mathcal{I}^\mathcal{B}_{n + 1} = C_0, p_1, C_1, \ldots, C_i, p_{n + 1}, C^\mathcal{B}_{n + 1}$ is a one step continuation of $\mathcal{I}_n$ by $p_{n + 1}$.
            Therefore, since by \Cref{claim:reduction:c_b_i+1_and_c_i+1_same_cell_manager} the state of the memory manager is the same in $C_{n + 1}$ and $C^\mathcal{B}_{n + 1}$, the claim follows by \Cref{lemma:reduction:allocate_response_in_b_is_swapped}.
            \qH{\Cref{claim:reduction:allocate_response_in_b_is_swapped}}
        \end{proof}

        We now prove a similar claim for $\freecelloperation{}$ operations in $\mathcal{I}^\mathcal{B}_{n + 1}$.
        We first prove a simple claim asserting that the response of every $\allocatecelloperation{}$ operation in $\mathcal{I}^\mathcal{B}_{n + 1}$ is not in $\celluniverse(\mathcal{I}^\mathcal{A}_{n + 1})$.
        This is one place where we make use of the fact that we ``rename'' pointers in $\mathcal{A}$ to ``fresh'' pointers in $\mathcal{B}$.
    
        \begin{claimcustom}{\ref{lemma:reduction:mapped_history_is_for_b}.5}\label{claim:reduction:no_allocate_for_used_pointer_in_b}
            The response of every $\allocatecelloperation{}$ operation in $\mathcal{I}^\mathcal{B}_{n + 1}$ is not in $\celluniverse(\mathcal{I}^\mathcal{A}_{n + 1})$.
        \end{claimcustom}
    
        \begin{proof}
            Consider any $\allocatecelloperation{}$ operation during $\mathcal{I}^\mathcal{B}_{n + 1}$; say it is performed during the $i$th step.
            Hence, since by \Cref{lemma:reduction:same_program_counters_in_mapped_run} the program counter of $p_{i}$ is the same in $C^\mathcal{A}_{i - 1}$ and $C_{i - 1}$ and $p_i$ performs the $i$th step of $\mathcal{I}^\mathcal{A}_{n + 1}$ and $\mathcal{I}^\mathcal{B}_{n + 1}$, we have that $p_i$ performed an $\allocatecelloperation{}$ operation during the $i$th step of $\mathcal{I}^\mathcal{A}_{n + 1}$.
            Thus, by \Cref{claim:reduction:allocate_response_in_b_is_swapped}, the $i$th step of $\mathcal{I}^\mathcal{B}_{n + 1}$ performs an $\allocatecelloperation{}$ operation whose response is $\mathcal{M}(i)$, and so by \Cref{def:reduction:swapper} it is in $\celluniverse^0(\mathcal{I}^\mathcal{A}_{n + 1})$.
            Therefore, since by \Cref{observation:reduction:unused_pointers_in_run_of_a} $\celluniverse(\mathcal{I}^\mathcal{A})$ and $\celluniverse^0(\mathcal{I}^\mathcal{A}_{n + 1})$ are disjoint, we have that the response of every $\allocatecelloperation{}$ operation in $\mathcal{I}^\mathcal{B}_{n + 1}$ is not in $\celluniverse(\mathcal{I}^\mathcal{A}_{n + 1})$ as wanted.
            \qH{\Cref{claim:reduction:no_allocate_for_used_pointer_in_b}}
        \end{proof}
    
        \begin{claimcustom}{\ref{lemma:reduction:mapped_history_is_for_b}.6}\label{claim:reduction:frees_in_b_are_swapped}
             Consider any $\cellpointershort \in \celluniverse$.
             If the $i$th step of $\mathcal{I}^\mathcal{A}_{n + 1}$ performs a $\freecelloperation{}(\cellpointershort)$ operation and $\mathcal{W}_{i - 1}(O, 1) = w$ where $O$ is the local variable $\currentcellpointershort$ of $p_i$ in the Relinquish procedure, then $w \neq \bot$ and the $i$th step of $\mathcal{I}^\mathcal{B}_{n + 1}$ performs a $\freecelloperation{}(\mathcal{M}(w))$ operation.
        \end{claimcustom}
    
        \begin{proof}
            Consider any $\cellpointershort \in \celluniverse$ and suppose the $i$th step of $\mathcal{I}^\mathcal{A}_{n + 1}$ performs a $\freecelloperation{}(\cellpointershort)$ operation and $\mathcal{W}_{i - 1}(O, 1) = w$ where $O$ is the local variable $\currentcellpointershort$ of $p_i$ in the Relinquish procedure.
            Hence, by \Cref{lemma:reduction:free_input_in_b_is_swapped}, $C^\mathcal{A}_{i - 1}$ assigns state $\cellpointershort$ to $O$ and the $i$th step of $\mathcal{I}^\mathcal{B}_{n + 1}$ performs a $\freecelloperation{}(\mathcal{S}_{i - 1}(O))$ operation.
        
            We first prove that $\mathcal{W}_{i - 1}(O, 1) \neq \bot$.
            Suppose, for contradiction, $\mathcal{W}_{i - 1}(O, 1) = \bot$.
            Hence, since $C^\mathcal{A}_{i - 1}$ assigns $\cellpointershort$ to $O$, by \Cref{def:reduction:swapper}, $\mathcal{S}_{i - 1}(O) = \cellpointershort$.
            Thus, since the $i$th step of $\mathcal{I}^\mathcal{B}_{n + 1}$ performs a $\freecelloperation{}(\mathcal{S}_{i - 1}(O))$ operation, the $i$th step of $\mathcal{I}^\mathcal{B}_{n + 1}$ is a $\freecelloperation{}(\cellpointershort)$ operation.
            Since $C^\mathcal{A}_{i - 1}$ assigns $\cellpointershort$ to $O$, by \Cref{def:reduction:used_pointers_in_run_of_a}, $\cellpointershort \in \celluniverse(\mathcal{I}^\mathcal{A}_{n + 1})$.
            Hence, by \Cref{claim:reduction:no_allocate_for_used_pointer_in_b} the response of every $\allocatecelloperation{}$ operation in $\mathcal{I}^\mathcal{B}_{n + 1}$ is not $\cellpointershort$.
            Therefore, in $\mathcal{I}^\mathcal{B}_{n + 1}$ there is a $\freecelloperation{}(\cellpointershort)$ such that there is no $\allocatecelloperation{}$ operation whose response is $\cellpointershort$ before it.
            However, since $\mathcal{I}^\mathcal{B}_{n + 1}$ is an implementation history of $\mathcal{B}$ and $\cellpointershort \in \celluniverse$, by 2 of \Cref{thm:reduction:algorithm_b_is_well_behaved}, every $\freecelloperation{}(\cellpointershort)$ operation in $\mathcal{I}^\mathcal{B}_{n + 1}$ is after an $\allocatecelloperation{}$ operation whose response is $\cellpointershort$, a contradiction.
    
            We now prove that the $i$th step of $\mathcal{I}^\mathcal{B}_{n + 1}$ performs a $\freecelloperation{}(\mathcal{M}(w))$ operation.
            Since $\mathcal{W}_{i - 1}(O, 1) = w \neq \bot$, by \Cref{def:reduction:swapper}, $\mathcal{S}_{i - 1}(O, 1) = \mathcal{M}(w)$.
            Hence, since the state of $O$ is a single value, $\mathcal{S}_{i - 1}(O) = \mathcal{M}(w)$.
            Therefore, since the $i$th step of $\mathcal{I}^\mathcal{B}_{n + 1}$ performs a $\freecelloperation{}(\mathcal{S}_{i - 1}(O))$ operation, we have that the $i$th step of $\mathcal{I}^\mathcal{B}_{n + 1}$ performs a $\freecelloperation{}(\mathcal{M}(w))$ operation as wanted.
            \qH{\Cref{claim:reduction:frees_in_b_are_swapped}}
        \end{proof}

        % \textcolor{red}{The following claim should be thought of as, if a pointer is allocated and then freed, there is a matching free operation for the allocate in $\mathcal{B}$.}

        We now show that between any two $\allocatecelloperation{}$ and $\freecelloperation{}$ operations for $\cellpointershort$ in $\mathcal{I}^\mathcal{A}_{n + 1}$ there is a $\freecelloperation{}$ operation for a ``matching'' pointer in $\mathcal{I}^\mathcal{B}_{n + 1}$.
    
        \begin{claimcustom}{\ref{lemma:reduction:mapped_history_is_for_b}.7}\label{claim:reduction:allocate_followed_by_free_in_a_implies_free_for_same_watermark}
            Consider any $\cellpointershort \in \celluniverse$.
            If the $i$th step of $\mathcal{I}^\mathcal{A}_{n + 1}$ performs an $\allocatecelloperation{}$ operation whose response is $\cellpointershort$ and the $j$th step of $\mathcal{I}^\mathcal{A}_{n + 1}$ performs a $\freecelloperation{}(\cellpointershort)$ operation where $j \in (i..n + 1]$, then the $k$th step of $\mathcal{I}^\mathcal{A}_{n + 1}$ performs a $\freecelloperation{}(\cellpointershort)$ operation where $k \in (i..j]$, and $\mathcal{W}_{k - 1}(O_k, 1) = i$ where $O_k$ is the local variable $\currentcellpointershort$ of $p_k$ in the Relinquish procedure.
        \end{claimcustom}
    
        \begin{proof}
            By induction on $i$.
    
            \begin{itemize}
                \item[] \hspace{0pt}\textbf{Base Case.} $i = 1$.
    
                Since every $\allocatecelloperation{}$ operation is performed on \cref{line:ero:allocate_cell}, we have that there is at least one step before any $\allocatecelloperation{}$ operation, and so the first step of $\mathcal{I}^\mathcal{A}_{n + 1}$ cannot perform an $\allocatecelloperation{}$ operation.
                Therefore, the claim for $i = 1$ is vacuously true.
    
                \item[] \hspace{0pt}\textbf{Inductive Case.} $\forall i \in [1.. n]$ if the claim holds for all $j \in [1..i]$, then the claim holds for $i + 1$.
    
                Suppose for some $i \in [1..n]$ and every $j \in [1..i]$ that the claim holds for $j$.
                This is the inductive hypothesis.
                We must prove that the claim holds for $i + 1$.
                Suppose, for contradiction, that the $i + 1$th step of $\mathcal{I}^\mathcal{A}_{n + 1}$ performs an $\allocatecelloperation{}$ operation whose response is $\cellpointershort$, for some $i + 1 < j \leq n + 1$ the $j$th step of $\mathcal{I}^\mathcal{A}_{n + 1}$ performs a $\freecelloperation{}(\cellpointershort)$ operation, and for all $i + 1 < k \leq j$ the $k$th step of $\mathcal{I}^\mathcal{A}_{n + 1}$ does not perform a $\freecelloperation{}(\cellpointershort)$ operation or $\mathcal{W}_{k - 1}(O_k, 1) \neq i + 1$.
                Without loss of generality, suppose $j$ is the smallest such step number, i.e., for every $i + 1 < j' < j$ if the $j'$th step of $\mathcal{I}^\mathcal{A}_{n + 1}$ performs a $\freecelloperation{}(\cellpointershort)$ operation, then for some $i + 1 < k' \leq j'$ the $k'$th step of $\mathcal{I}^\mathcal{A}_{n + 1}$ performs a $\freecelloperation{}(\cellpointershort)$ operation, and $\mathcal{W}_{k' - 1}(O_{k'}, 1) = i + 1$ (*).
                
                Since $p_j$ in the $j$th step of $\mathcal{I}^\mathcal{A}_{n + 1}$ performs a $\freecelloperation{}(\cellpointershort)$ operation, by \Cref{claim:reduction:frees_in_b_are_swapped}, $\mathcal{W}_{j - 1}(O_j, 1) = w \neq \bot$ and the $j$th step of $\mathcal{I}^\mathcal{B}_{n + 1}$ performs a $\freecelloperation{}(\mathcal{M}(w))$ operation.
                Hence, by \Cref{lemma:reduction:associated_pointer}, $w < j$ and the $w$th step of $\mathcal{I}^\mathcal{A}_{n + 1}$ performed an $\allocatecelloperation{}$ operation with response $v_1$ where $v_1$ is the value of the $1$st index of the state assigned to $O_j$ in $C^\mathcal{A}_{j - 1}$.
                Thus, since $O_j$ is the local variable $\currentcellpointershort$ of $p_j$ in the Relinquish procedure, and the $j$th step of $\mathcal{I}^\mathcal{A}_{n + 1}$ performs a $\freecelloperation{}(\cellpointershort)$ operation, we have that $C^\mathcal{A}_{j - 1}$ assigns state $\cellpointershort$ to $O_j$, and so the $w$th step of $\mathcal{I}^\mathcal{A}_{n + 1}$ performed an $\allocatecelloperation{}$ operation with response $\cellpointershort$.
                There are two cases.
    
                \begin{itemize}
                    \item[] \hspace{0pt}\textbf{Case 1.} $i + 1 < w$.
    
                    Hence, since both the $i + 1$th and $w$th step of $\mathcal{I}^\mathcal{A}_{n + 1}$ perform an $\allocatecelloperation{}$ operation whose response is $\cellpointershort$, by \Cref{alg:cell_manager_specification}, for some $i + 1 < j' < w$ the $j'$th step of $\mathcal{I}^\mathcal{A}_{n + 1}$ performs a $\freecelloperation{}(\cellpointershort)$ operation.
                    Since $w < j$, this implies that for some $i + 1 < j' < j$ the $j'$th step of $\mathcal{I}^\mathcal{A}_{n + 1}$ performs a $\freecelloperation{}(\cellpointershort)$ operation.
                    Therefore, by (*), for some $i + 1 < k' \leq j'$ (or equivalently $i + 1 < k' \leq j$ since $j' < j$)  the $k'$th step of $\mathcal{I}^\mathcal{A}_{n + 1}$ performs a $\freecelloperation{}(\cellpointershort)$ operation, and $\mathcal{W}_{k' - 1}(O_{k'}, 1) = i + 1$.
                    However, by our initial assumption, for all $i + 1 < k \leq j$ the $k$th step of $\mathcal{I}^\mathcal{A}_{n + 1}$ does not perform a $\freecelloperation{}(\cellpointershort)$ operation or $\mathcal{W}_{k - 1}(O_k, 1) \neq i + 1$, a contradiction.
    
                    \item[] \hspace{0pt}\textbf{Case 2.} $w \leq i + 1$.
    
                    We first prove that the claim holds for $w$.
                    Since, by assumption, the $j$th step of $\mathcal{I}^\mathcal{A}_{n + 1}$ performs a $\freecelloperation{}(\cellpointershort)$ operation, and for all $i + 1 < k \leq j$ the $k$th step of $\mathcal{I}^\mathcal{A}_{n + 1}$ does not perform a $\freecelloperation{}(\cellpointershort)$ operation or $\mathcal{W}_{k - 1}(O_k, 1) \neq i + 1$, we have that $\mathcal{W}_{j - 1}(O_j, 1) \neq i + 1$.
                    Hence, since $\mathcal{W}_{j - 1}(O_j, 1) = w$, we have that $w \neq i + 1$.
                    Thus, since $w \leq i + 1$, we have that $w < i + 1$, and so $w \leq i$.
                    Therefore, since $1 \leq w$, by the inductive hypothesis, the claim holds for $w$.
    
                    We now prove that, roughly speaking, there is another $\freecelloperation{}$ operation whose watermark is the same as $j$'s.
                    Since both the $w$th and $i + 1$th step of $\mathcal{I}^\mathcal{A}_{n + 1}$ perform an $\allocatecelloperation{}$ operation whose response is $\cellpointershort$, by \Cref{alg:cell_manager_specification}, for some $w < j' < i + 1$ the $j'$th step of $\mathcal{I}^\mathcal{A}_{n + 1}$ performs a $\freecelloperation{}(\cellpointershort)$ operation.
                    Hence, since $i + 1 \leq n + 1$, by transitivity, we have that $w < j' < n + 1$.
                    Thus, since the $w$th step of $\mathcal{I}^\mathcal{A}_{n + 1}$ perform an $\allocatecelloperation{}$ operation whose response is $\cellpointershort$, the $j'$th step of $\mathcal{I}^\mathcal{A}_{n + 1}$ performs a $\freecelloperation{}(\cellpointershort)$ operation, and the claim holds for $w$, we have that for some $w < k \leq j'$ the $k$th step of $\mathcal{I}^\mathcal{A}_{n + 1}$ performs a $\freecelloperation{}(\cellpointershort)$ operation, and $\mathcal{W}_{k - 1}(O_k, 1) = w$.
    
                    We now finish the proof of Case 2.
                    Since the $k$th step of $\mathcal{I}^\mathcal{A}_{n + 1}$ performs a $\freecelloperation{}(\cellpointershort)$ operation and $\mathcal{W}_{k - 1}(O_k, 1) = w$, by \Cref{claim:reduction:frees_in_b_are_swapped}, the $k$th step of $\mathcal{I}^\mathcal{B}_{n + 1}$ performs a $\freecelloperation{}(\mathcal{M}(w))$ operation.
                    Furthermore, since $k \leq j'$, $j' < i + 1$, and $i + 1 < j$, by transitivity, $k < j$, and so $k \neq j$.
                    Therefore, since both the $k$th and $j$th step of $\mathcal{I}^\mathcal{B}_{n + 1}$ perform a $\freecelloperation{}(\mathcal{M}(w))$ operation, we have that there are two $\freecelloperation{}(\mathcal{M}(w))$ operations during $\mathcal{I}^\mathcal{B}_{n + 1}$.
                    However, since $\mathcal{I}^\mathcal{B}_{n + 1}$ is an implementation history of $\mathcal{B}$, and by \Cref{def:reduction:swapper} $\mathcal{M}(w) \in \celluniverse$, by 1 of \Cref{thm:reduction:algorithm_b_is_well_behaved}, there is at most one $\freecelloperation{}(\mathcal{M}(w))$ operation in $\mathcal{I}^\mathcal{B}_{n + 1}$, a contradiction.
                    \qH{\Cref{claim:reduction:allocate_followed_by_free_in_a_implies_free_for_same_watermark}}
                \end{itemize}
            \end{itemize}
        \end{proof}

        We now prove that the $n + 1$th step of $\mathcal{I}^\mathcal{A}_{n + 1}$ cannot perform an operation on an object of a cell which is unallocated.
        We will do this over the next few claims.
    
        \begin{claimcustom}{\ref{lemma:reduction:mapped_history_is_for_b}.8}\label{claim:reduction:if_i+1_performs_op_on_o_in_a_then_i+1_performs_op_on_swapped_o_in_b}
            If the $n + 1$th step of $\mathcal{I}^\mathcal{A}_{n + 1}$ performs an operation on a base object $O$ other than the memory manager, then the $n + 1$th step of $\mathcal{I}^\mathcal{B}_{n + 1}$ performs an operation on $O^*$ defined as follows.
            Let $O^*$ be $O$ if $O \notin \celluniverse_O$ and otherwise $O$ is the object $f$ of some cell where $f$ is either $\lastrepositoryoperationresponse{}$, $\revocations$, or $\nextlong$, and $O^* = (*\mathcal{S}_{n}(O_s)).f$ where $O_s$ is the source of $O$ (see \Cref{observation:reduction:source_of_pointer}).
        \end{claimcustom}
    
        \begin{proof}
            %Suppose the $i + 1$th step of $\mathcal{I}^\mathcal{A}$ performs an operation on a base object $O$.
            There are two cases.
            \begin{itemize}
                \item[] \hspace{0pt}\textbf{Case 1.} $O \notin \celluniverse_O$.
    
                Hence, since $O$ is not the memory manager, by \Cref{def:reduction:shared_objects}, $O \in \{\clockobject{}, \announceobject, \linearizationobject, \stateobject\}$.
                Thus, $p_{n + 1}$ performed an operation $O$ during the $n + 1$th step of $\mathcal{I}^\mathcal{A}_{n + 1}$ because of the line of code it executed (as opposed to because of the state of its local variables).
                Therefore, since by \Cref{lemma:reduction:same_program_counters_in_mapped_run} the program counter of $p_{n + 1}$ is the same in $C^\mathcal{A}_{n}$ and $C_{n}$, and $p_{n + 1}$ performs the $n + 1$th step of $\mathcal{I}^\mathcal{B}_{n + 1}$, we have that the $n + 1$th step of $\mathcal{I}^\mathcal{B}_{n + 1}$ performs an operation on $O$ as wanted.
    
                \item[] \hspace{0pt}\textbf{Case 2.} $O \in \celluniverse_O$.
    
                Hence, by \Cref{def:reduction:shared_objects}, for some $\cellpointershort \in \celluniverse \cup \{\&\headobject\}$ $O$ equals $(*\cellpointershort).f$ where $f$ is defined in the claim.
                Observe that the line of code $p_{n + 1}$ executed determines $f$, but a local variable determines $\cellpointershort$.
                Since $O_s$ is the source of $O$, by \Cref{observation:reduction:source_of_pointer}, $p_{n + 1}$ performed an operation on $O = (*\cellpointershort).f$ during the $n + 1$th step of $\mathcal{I}^\mathcal{A}$ because $O_s$ was assigned to state $\cellpointershort$ in $C^\mathcal{A}_{n}$.
                Thus, since by \Cref{lemma:reduction:same_program_counters_in_mapped_run} the program counter of $p_{n + 1}$ is the same in $C^\mathcal{A}_{n}$ and $C_{n}$ and $p_{n + 1}$ performs the $n + 1$th step of $\mathcal{I}^\mathcal{B}_{n + 1}$, $p_{n + 1}$ performs an operation on $(*\cellpointershort').f$ where  $\cellpointershort'$ is the state assigned to $O_s$ in $C_n$.
                Since $O_s = \cellpointershort$ in $C^\mathcal{A}_{n}$, by \Cref{def:reduction:run_mapping}, $O_s = \mathcal{S}_n(O_s)$ in $C_{n}$.
                Therefore, $p_{n + 1}$ performs an operation on $(*\mathcal{S}_n(O_s)).f$ during the $n + 1$th step of $\mathcal{I}^\mathcal{B}_{n + 1}$ as wanted.
                \qH{\Cref{claim:reduction:if_i+1_performs_op_on_o_in_a_then_i+1_performs_op_on_swapped_o_in_b}}
            \end{itemize}
        \end{proof}
    
        \begin{claimcustom}{\ref{lemma:reduction:mapped_history_is_for_b}.9}\label{claim:reduction:op_on_ptr_after_allocate_in_a}
            If the $n + 1$th step of $\mathcal{I}^\mathcal{A}_{n + 1}$ performs an operation on a base object $O \in \celluniverse_O$, $O$ is not in $\headobject$, and $\mathcal{W}_n(O_s, 1) = w$ where $O_s$ is the source of $O$ (see \Cref{observation:reduction:source_of_pointer}), then (a) $w \neq \bot$ (b) the $w$th step of $\mathcal{I}^\mathcal{A}_{n + 1}$ performs an $\allocatecelloperation{}$ operation whose response is $\cellpointershort$ where $O$ is an object of the cell pointed to by $\cellpointershort$ and (c) $\mathcal{S}_n(O_s) = \mathcal{M}(w)$.
        \end{claimcustom}
    
        \begin{proof}
            We first prove (a).
            Suppose, for contradiction, the $n + 1$th step of $\mathcal{I}^\mathcal{A}_{n + 1}$ performs an operation on a base object $O\in \celluniverse_O$, $O$ is not in $\headobject$, and $\mathcal{W}_n(O_s, 1) = \bot$.
            Hence, by \Cref{def:reduction:shared_objects}, $O = (*\cellpointershort).f$ where $\cellpointershort \in \celluniverse$ and $f$ is either $\lastrepositoryoperationresponse{}$, $\revocations$, or $\nextlong$.
            Since $O_s$ is the source of $O$, by \Cref{observation:reduction:source_of_pointer}, $C^\mathcal{A}_n$ assigned state $\cellpointershort$ to $O_s$, and so by \Cref{def:reduction:used_pointers_in_run_of_a}, $\cellpointershort \in \celluniverse(\mathcal{I}^\mathcal{A})$.
            Furthermore, by \Cref{claim:reduction:if_i+1_performs_op_on_o_in_a_then_i+1_performs_op_on_swapped_o_in_b}, the $n + 1$th step of $\mathcal{I}^\mathcal{B}_{n + 1}$ performs an operation on $(*\mathcal{S}_n(O_s)).f$.
            Since $C^\mathcal{A}_n$ assigned state $\cellpointershort$ to $O_s$ and $\mathcal{W}_n(O_s, 1) = \bot$, by $\mathcal{S}_n(O_s) = \cellpointershort$.
            Hence, since the $n + 1$th step of $\mathcal{I}^\mathcal{B}_{n + 1}$ performs an operation on $(*\mathcal{S}_n(O_s)).f$, we have that the $n + 1$th step of $\mathcal{I}^\mathcal{B}_{n + 1}$ performs an operation on $(*\cellpointershort).f$.
            Therefore, since $\cellpointershort \in \celluniverse(\mathcal{I}^\mathcal{A})$, by \Cref{claim:reduction:no_allocate_for_used_pointer_in_b}, the response of every $\allocatecelloperation{}$ operation in $\mathcal{I}^\mathcal{B}_{n + 1}$ is not $\cellpointershort$, and so in $\mathcal{I}^\mathcal{B}_{n + 1}$ the $n + 1$th step performs an operation on an object of the cell pointed to by $\cellpointershort$ and there are no $\allocatecelloperation{}$ operations whose response is $\cellpointershort$ in $\mathcal{I}^\mathcal{B}_{n + 1}$.
            However, since $\mathcal{I}^\mathcal{B}_{n + 1}$ is an implementation history of $\mathcal{B}$ and $\cellpointershort \in \celluniverse$, by 3. of \Cref{thm:reduction:algorithm_b_is_well_behaved}, every operation on an object of the cell pointed to by $\cellpointershort$ is after an $\allocatecelloperation{}$ operation whose response is $\cellpointershort$, a contradiction.
    
            We now prove (b) and (c).
            Since $\mathcal{W}_n(O_s, 1) = w \neq \bot$, by \Cref{lemma:reduction:associated_pointer}, the $w$th step of $\mathcal{I}^\mathcal{A}_{n + 1}$ performs an $\allocatecelloperation{}$ operation during the $w$th step of $\mathcal{I}^\mathcal{A}_{n + 1}$ with response $v_1$ where $v_1$ is the value of the $1$st index of the state assigned to $O_s$ in $C^\mathcal{A}_n$.
            Thus, since the $n + 1$th step of $\mathcal{I}^\mathcal{A}_{n + 1}$ performs an operation on $O$, $O$ is an object of the cell pointed to by $\cellpointershort$, and $O_s$ is the source of $O$, by \Cref{observation:reduction:source_of_pointer}, $C^\mathcal{A}_n$ assigns state $\cellpointershort$ to $O_s$, and so the $w$th step of $\mathcal{I}^\mathcal{A}_{n + 1}$ performs an $\allocatecelloperation{}$ operation with response $\cellpointershort$.
            Furthermore, since $C^\mathcal{A}_n$ assigns a single value (namely $\cellpointershort$) to $O_s$, and $\mathcal{W}_n(O_s, 1) = w \neq \bot$, by \Cref{def:reduction:swapper}, $\mathcal{S}_n(O_s) = \mathcal{M}(w)$.
            \qH{\Cref{claim:reduction:op_on_ptr_after_allocate_in_a}}
        \end{proof}
        
        \begin{claimcustom}{\ref{lemma:reduction:mapped_history_is_for_b}.10}\label{claim:reduction:never_undefined}
            If the $n + 1$th step of $\mathcal{I}^\mathcal{A}_{n + 1}$ performs an operation on an object of the cell pointed to by $\cellpointershort \in \celluniverse$, then $\cellpointershort$ is in the state of the memory manager in $C^\mathcal{A}_n$.
        \end{claimcustom}
        \begin{proof}
            Suppose, for contradiction, the $n + 1$th step of $\mathcal{I}^\mathcal{A}_{n + 1}$ performs an operation on an object $O$ of the cell pointed to by $\cellpointershort \in \celluniverse$, and $\cellpointershort$ is not in the state of the memory manager in $C^\mathcal{A}_n$.
            Hence, $O = (*\cellpointershort).f$ where $f$ is either $\lastrepositoryoperationresponse{}$, $\revocations$, or $\nextlong$, and so, by \Cref{def:reduction:shared_objects}, $O \in \celluniverse_O$.
            Let $O_s$ be the source of $O$.
            Hence, since the $n + 1$th step of $\mathcal{I}^\mathcal{A}_{n + 1}$ performs an operation on the cell pointed to by $\cellpointershort \in \celluniverse$, by \Cref{claim:reduction:if_i+1_performs_op_on_o_in_a_then_i+1_performs_op_on_swapped_o_in_b}, the $n + 1$th step of $\mathcal{I}^\mathcal{B}_{n + 1}$ performs an operation on $(*\mathcal{S}_n(O_s)).f$.
            Furthermore, by \Cref{claim:reduction:op_on_ptr_after_allocate_in_a}, $\mathcal{W}_n(O_s, 1) = w \neq \bot$, the $w$th step of $\mathcal{I}^\mathcal{A}_{n + 1}$ performs an $\allocatecelloperation{}$ operation whose response is $\cellpointershort$, and $\mathcal{S}_n(O_s) = \mathcal{M}(w)$.
            Therefore, the $n + 1$th step of $\mathcal{I}^\mathcal{B}_{n + 1}$ performs an operation on $(*\mathcal{M}(w)).f$.
    
            We now identify a $\freecelloperation{}(\mathcal{M}(w))$ operation in $\mathcal{I}^\mathcal{B}_{n + 1}$.
            Since the $w$th step of $\mathcal{I}^\mathcal{A}_{n + 1}$ performs an $\allocatecelloperation{}$ operation whose response is $\cellpointershort$, by \Cref{def:reduction:used_pointers_in_run_of_a} $\cellpointershort \in \celluniverse(\mathcal{I}^\mathcal{A}_{n + 1})$.
            Hence, since by assumption $\cellpointershort$ is not in the state of the memory manager in $C^\mathcal{A}_n$, by \Cref{alg:cell_manager_specification}, for some $w < j \leq n$ the $j$th step of $\mathcal{I}^\mathcal{A}_{n + 1}$ performs a $\freecelloperation{}(\cellpointershort)$ operation.
            Thus, by \Cref{claim:reduction:allocate_followed_by_free_in_a_implies_free_for_same_watermark}, for some $w < k \leq j$ the $k$th step of $\mathcal{I}^\mathcal{A}_{n + 1}$ performs a $\freecelloperation{}(\cellpointershort)$ operation and $\mathcal{W}_{k - 1}(O_k, 1) = w$ where $O_j$ is the local variable $\currentcellpointershort$ of $p_k$ in the Relinquish procedure.
            Therefore, by \Cref{claim:reduction:frees_in_b_are_swapped}, the $k$th step of $\mathcal{I}^\mathcal{B}_{n + 1}$ performs a $\freecelloperation{}(\mathcal{M}(w))$ operation.
    
            We now finish the proof \Cref{claim:reduction:never_undefined}.
            Therefore, since $k < n + 1$, there is an operation on an object of the cell pointed to by $\mathcal{M}(w)$ during the $n + 1$th step of $\mathcal{I}^\mathcal{B}_{n + 1}$ which is after a $\freecelloperation{}(\mathcal{M}(w))$ operation during the $k$th step of $\mathcal{I}^\mathcal{B}_{n + 1}$.
            However, since $\mathcal{I}^\mathcal{B}_{n + 1}$ is an implementation history of $\mathcal{B}$ and $\mathcal{M}(w) \in \celluniverse$, by 3. of \Cref{thm:reduction:algorithm_b_is_well_behaved}, every operation on an object the cell pointed to by $\mathcal{M}(w)$ is before any $\freecelloperation{}(\mathcal{M}(w))$ operation, a contradiction.
            \qH{\Cref{claim:reduction:never_undefined}}
        \end{proof}

        %\textcolor{red}{stopped here.}

        The next claim should be thought of as: if the $n + 1$th step of $\mathcal{I}^\mathcal{A}_{n + 1}$ performs an operation on an object of the cell, then when mapped to $\mathcal{B}$, it is the latest version of that cell.
    
        \begin{claimcustom}{\ref{lemma:reduction:mapped_history_is_for_b}.11}\label{claim:reduction:if_performed_op_on_cell_it_is_the_latest_version}
            Suppose the $n + 1$th step of $\mathcal{I}^\mathcal{A}_{n + 1}$ performs an operation on a base object $O \in \celluniverse_O$ and $O$ is not in $\headobject$.
            By \Cref{claim:reduction:op_on_ptr_after_allocate_in_a}, $\mathcal{W}_n(O_s, 1) = w \neq \bot$ where $O_s$ is the source of $O$ (see \Cref{observation:reduction:source_of_pointer}).
            Then, the $i$th step of $\mathcal{I}^\mathcal{A}_{n + 1}$ does not perform an $\allocatecelloperation{}$ operation whose response is $\cellpointershort$, where $i \in (w..n+1]$ and $\cellpointershort$ is the pointer to the cell that $O$ is an object of.
        \end{claimcustom}
    
        \begin{proof}
            Suppose, for contradiction, $O \in \celluniverse_O$, $O$ is not in $\headobject$, $\mathcal{W}_n(O_s, 1) = w$, and for some $i \in (w..n+1]$ the $i$th step of $\mathcal{I}^\mathcal{A}_{n + 1}$ performs an $\allocatecelloperation{}$ operation whose response is $\cellpointershort$.
            Hence, by \Cref{claim:reduction:op_on_ptr_after_allocate_in_a}, the $w$th step of $\mathcal{I}^\mathcal{A}_{n + 1}$ performs an $\allocatecelloperation{}$ operation whose response is $\cellpointershort$.
            Since the $w$th and $i$th step of $\mathcal{I}^\mathcal{A}_{n + 1}$ perform an $\allocatecelloperation{}$ operation with response $\cellpointershort$, and $w < i$, by \Cref{alg:cell_manager_specification}, for some $w < j < i$ the $j$th step of $\mathcal{I}^\mathcal{A}_{n + 1}$ performs a $\freecelloperation{}(\cellpointershort)$ operation.
            Hence, by \Cref{claim:reduction:allocate_followed_by_free_in_a_implies_free_for_same_watermark}, for some $w < k \leq j$ the $k$th step of $\mathcal{I}^\mathcal{A}_{n + 1}$ performs a $\freecelloperation{}(\cellpointershort)$ operation, and $\mathcal{W}_{k - 1}(O_k, 1) = w$ where $O_k$ is the local variable $\currentcellpointershort$ of $p_k$ in the Relinquish procedure.
            Since $k \leq j$, $j < i$, and $i \leq n + 1$, by transitivity, $k < n + 1$.
            Furthermore, since the $k$th step of $\mathcal{I}^\mathcal{A}_{n + 1}$ performs a $\freecelloperation{}(\cellpointershort)$ operation and $\mathcal{W}_{k - 1}(O_k, 1) = w$, by \Cref{claim:reduction:frees_in_b_are_swapped}, the $k$th step of $\mathcal{I}^\mathcal{B}_{n + 1}$ performs a $\freecelloperation{}(\mathcal{M}(w))$ operation.
            Since the $n + 1$th step of $\mathcal{I}^\mathcal{A}_{n + 1}$ performs an operation on $O \in \celluniverse_O$ and $O$ is an object of the cell pointed to by $\cellpointershort$, by \Cref{claim:reduction:if_i+1_performs_op_on_o_in_a_then_i+1_performs_op_on_swapped_o_in_b}, the $n + 1$th step of $\mathcal{I}^\mathcal{B}_{n + 1}$ performs an operation on an object of the cell pointed to by $\mathcal{S}_n(O_s)$.
            Hence, since $\mathcal{W}_n(O_s, 1) = w$, by \Cref{claim:reduction:op_on_ptr_after_allocate_in_a}, $\mathcal{S}_n(O_s) = \mathcal{M}(w)$, and so the $n + 1$th step of $\mathcal{I}^\mathcal{B}_{n + 1}$ performs an operation on the cell pointed to by $\mathcal{M}(w)$.
            Therefore, since $k < n + 1$, in $\mathcal{I}^\mathcal{B}_{n + 1}$, there is an operation on an object of the cell pointed to by $\mathcal{M}(w)$ after a $\freecelloperation{}(\mathcal{M}(w))$ operation.
            However, since $\mathcal{I}^\mathcal{B}_{n + 1}$ is an implementation history of $\mathcal{B}$ and $\mathcal{M}(w) \in \celluniverse$, by 3 of \Cref{thm:reduction:algorithm_b_is_well_behaved}, every operation on an object of the cell pointed to by $\mathcal{M}(w)$ is before any $\freecelloperation{}(\mathcal{M}(w))$ operation, a contradiction.
            \qH{\Cref{claim:reduction:if_performed_op_on_cell_it_is_the_latest_version}}
        \end{proof}

        % \textcolor{red}{remark about the task now.}

        We now have all the facts we need to do the majority of the work to prove \Cref{lemma:reduction:mapped_history_is_for_b}.
        The rest of the proof will go as follows.
        First, we will prove that if the $n + 1$th step of $\mathcal{I}^\mathcal{A}_{n + 1}$ performs an operation on a base object $O$ other than the memory manager, then the state of the ``corresponding'' object in $\mathcal{C}^\mathcal{B}_{n + 1}$ is the same as $C_{n + 1}$.
        We will then prove that the state of every local variable other than the program counters is the same in $C^\mathcal{B}_{n + 1}$ and $C_{n + 1}$.
        Then, we will prove that the state of every program counter is the same in $C^\mathcal{B}_{n + 1}$ and $C_{n + 1}$.
    
        \begin{claimcustom}{\ref{lemma:reduction:mapped_history_is_for_b}.12}\label{claim:reduction:b_gets_the_right_response}
            Suppose the $n + 1$th step of $\mathcal{I}^\mathcal{A}_{n + 1}$ performs an operation $o^\mathcal{A}_{n + 1}$ on some base object $O$ other than the memory manager with response $r^\mathcal{A}_{n + 1}$.
            Let $O^*$ be $O$ if $O \notin \celluniverse_O$ and otherwise $O$ is the object $f$ of some cell where $f$ is either $\lastrepositoryoperationresponse{}$, $\revocations$, or $\nextlong$, and $O^* = (*\mathcal{S}_{n}(O_s)).f$ where $O_s$ is the source of $O$.
            Then, the $n + 1$th step of $\mathcal{I}^\mathcal{B}_{n + 1}$ performs an operation on $O^*$ with response $r_{n + 1}$ and $C^\mathcal{B}_{n + 1}$ assigns state $\mathcal{S}_{n + 1}(O)$ to $O^*$ where $r_{n + 1}$ is defined as follows.
            If $o^\mathcal{A}_{n + 1}$ is a read operation, then $r_{n + 1} = \mathcal{S}_n(O)$; Otherwise, $r_{n + 1} = r^\mathcal{A}_{n + 1}$.
        \end{claimcustom}
        
        \begin{proof}
            Suppose the $n + 1$th step of $\mathcal{I}^\mathcal{A}_{n + 1}$ performs an operation $o^\mathcal{A}_{n + 1}$ on some base object $O$ other than the memory manager with response $r^\mathcal{A}_{n + 1}$.
            Hence, by \Cref{claim:reduction:if_i+1_performs_op_on_o_in_a_then_i+1_performs_op_on_swapped_o_in_b}, the $n + 1$th step of $\mathcal{I}^\mathcal{B}_{n + 1}$ performs an operation $o_{n + 1}$ on $O^*$; say with response $r_{n + 1}$.
            Since $O$ is a base object other than the memory manager, $O$ is either a $\FAop$, $\CASop$, or $\GCASop$ object, so $O$ is deterministic.
            Let $\delta$ be the state transition function of $O$.
            Furthermore, let $s^\mathcal{A}_{n}$ (resp. $s^\mathcal{A}_{n + 1}$) be the state that $C^\mathcal{A}_{n}$ (resp. $C^\mathcal{A}_{n + 1}$) assigns to $O$.
            Hence, since the $n + 1$th step of $\mathcal{I}^\mathcal{A}_{n + 1}$ performs an operation $o^\mathcal{A}_{n + 1}$ on $O$ with response $r^\mathcal{A}_{n + 1}$, if $O \notin \celluniverse_O$, then we have that $\delta(s^\mathcal{A}_n, o^\mathcal{A}_{n + 1}) = (s^\mathcal{A}_{n + 1}, r^\mathcal{A}_{n + 1})$.
            Otherwise, if $O \in \celluniverse_O$, then by \Cref{def:reduction:shared_objects}, $O$ is an object of some cell, say the one pointed to by $\cellpointershort \in \celluniverse$, and so by \Cref{claim:reduction:never_undefined}, $\cellpointershort$ is in the memory manager in $C^\mathcal{A}_n$.
            Thus, by \Cref{def:reduction:algorithm_a}, we have that $\delta(s^\mathcal{A}_n, o^\mathcal{A}_{n + 1}) = (s^\mathcal{A}_{n + 1}, r^\mathcal{A}_{n + 1})$.
            Therefore, in all cases, $\delta(s^\mathcal{A}_n, o^\mathcal{A}_{n + 1}) = (s^\mathcal{A}_{n + 1}, r^\mathcal{A}_{n + 1})$.
    
            We first prove that $C_n$ assigns $\mathcal{S}_n(O)$ to $O^*$ (*).
            If $O \notin \celluniverse_O$ or $O$ is in $\headobject$, then $O^* = O$, and so by \Cref{def:reduction:run_mapping} $C_n$ assigns $\mathcal{S}_n(O)$ to $O$.
            If $O \in \celluniverse_O$ and $O$ is not in $\headobject$, then by \Cref{def:reduction:shared_objects}, $O = (*\cellpointershort).f$ for some $\cellpointershort \in \celluniverse$ and $f$ which is either $\lastrepositoryoperationresponse{}$, $\revocations$, or $\nextlong$.
            Let $\mathcal{W}_{n}(O_s, 1) = w$.
            Hence, by \Cref{claim:reduction:op_on_ptr_after_allocate_in_a}, the $w$th step of $\mathcal{I}^\mathcal{A}_{n + 1}$ performs an $\allocatecelloperation{}$ operation whose response is $\cellpointershort$, and $\mathcal{S}_n(O_s) = \mathcal{M}(w)$.
            Furthermore, by \Cref{claim:reduction:if_performed_op_on_cell_it_is_the_latest_version} for all $w < i \leq n + 1$ the $i$th step of $\mathcal{I}^\mathcal{A}_{n + 1}$ does not perform an $\allocatecelloperation{}$ operation whose response is $\cellpointershort$.
            Hence, by \Cref{def:reduction:run_mapping}, $C_n$ assigns state $\mathcal{S}_n((*\cellpointershort).f)$ to $(*\mathcal{M}(w)).f$.
            Thus, since $\mathcal{S}_n(O_s) = \mathcal{M}(w)$, we have that $C_n$ assigns state $\mathcal{S}_n((*\cellpointershort).f)$ to $(*\mathcal{S}_n(O_s)).f$.
            Therefore, since $O = (*\cellpointershort).f$, and $O^* = (*\mathcal{S}_{n}(O_s)).f$, we have that $C_n$ assigns state $\mathcal{S}_n(O)$ to $O^*$ as wanted.

            The remainder of the proof is by cases depending on the type of $o^\mathcal{A}_{n + 1}$.
            
            \begin{itemize}
                \item[] \hspace{0pt}\textbf{Case 1.} $o^\mathcal{A}_{n + 1}$ is a read operation.
    
                % Hence, since $\delta(o^\mathcal{A}_{n + 1}, s^\mathcal{A}_n) = (s^\mathcal{A}_{n + 1}, r^\mathcal{A}_{n + 1})$, we have that $s^\mathcal{A}_{n + 1} = s^\mathcal{A}_n$ and $r^\mathcal{A}_{n + 1} = s^\mathcal{A}_n$.
                Hence, the state of $O$ is the same in $C^\mathcal{A}_n$ and $C^\mathcal{A}_{n + 1}$.
                Furthermore, since by \Cref{lemma:reduction:same_program_counters_in_mapped_run} the program counter of $p_{n + 1}$ is the same in $C^\mathcal{A}_{n}$ and $C_{n}$, and $p_{n + 1}$ takes the $n + 1$th step of $\mathcal{I}^\mathcal{B}_{n + 1}$, we have that $o_{n + 1}$ is a read operation.
                Thus, since $o_{n + 1}$ is an operation on $O^*$, and by (*) $C_n$ assigns $\mathcal{S}_n(O)$ to $O^*$, we have that $r_{n + 1} = \mathcal{S}_n(O)$ and $C^\mathcal{B}_{n + 1}$ assigns state $\mathcal{S}_n(O)$ to $O^*$.
                What remains is to show that $\mathcal{S}_{n + 1}(O) = \mathcal{S}_n(O)$.
                Let $v_j$ be the value of the $j$th index of the state of $O$ in $C^\mathcal{A}_{n + 1}$.
                Hence, since the state of $O$ is the same in $C^\mathcal{A}_n$ and $C^\mathcal{A}_{n + 1}$, it follows that $v_j$ is the value of the $j$th index of the state of $O$ in $C^\mathcal{A}_n$.
                Furthermore, since $n$ is a non-negative integer, $n + 1 > 0$, so $n \geq 0$.
                There are two cases.

                \begin{itemize}
                    \item[] \hspace{0pt}\textbf{Case 1.1.} if $O$ is a base object, then $O \in \{\announceobject, \linearizationobject\} \cup \{(*\cellpointershort).\nextlong\ \vert\ \cellpointershort \in \celluniverse \cup \{\&\headobject\}\}$, and $v_j \in \celluniverse$.
    
                    Since $o^\mathcal{A}_{n + 1}$ is a read operation, we have that $p_{n + 1}$ does not set any index of $O$ during the $n + 1$th step of $\mathcal{I}^\mathcal{A}_{n + 1}$, and so by 1 of \Cref{def:reduction:watermarks}, $\mathcal{W}_{n + 1}(O, j) = \mathcal{W}_n(O, j)$.
                    Hence, since the state of $O$ is the same in $C^\mathcal{A}_n$ and $C^\mathcal{A}_{n + 1}$, by \Cref{def:reduction:swapper}, $\mathcal{S}_{n + 1}(O, j) = \mathcal{S}_n(O, j)$.

                    \item[] \hspace{0pt}\textbf{Case 1.2.} otherwise.

                    Hence, by \Cref{def:reduction:watermarks}, $\mathcal{W}_{n + 1}(O, j) = \bot$.
                    We now show that $\mathcal{W}_{n}(O, j) = \bot$.
                    Recall that $n \geq 0$.
                    If $n = 0$, then by \Cref{def:reduction:watermarks}, $\mathcal{W}_{n}(O, j) = \bot$, as wanted, so suppose $n > 0$.
                    Hence, by assumption if $O$ is a base object, then $O \notin \{\announceobject, \linearizationobject\} \cup \{(*\cellpointershort).\nextlong\ \vert\ \cellpointershort \in \celluniverse \cup \{\&\headobject\}\}$, or $v_j \notin \celluniverse$.
                    Thus, since $v_j$ is the value of the $j$th index of the state of $O$ in $C^\mathcal{A}_n$, in either case, by \Cref{def:reduction:watermarks}, $\mathcal{W}_{n}(O, j) = \bot$.
                    Therefore, since in all cases $\mathcal{W}_{n + 1}(O, j) = \mathcal{W}_{n}(O, j)$, and the state of $O$ is the same in $C^\mathcal{A}_n$ and $C^\mathcal{A}_{n + 1}$, by \Cref{def:reduction:swapper}, $\mathcal{S}_{n + 1}(O, j) = \mathcal{S}_n(O, j)$.
                \end{itemize}

                We now finish the proof of Case 1.
                Since $\mathcal{S}_{n + 1}(O, j) = \mathcal{S}_n(O, j)$ for every index $j$ of $O$, by \Cref{def:reduction:swapper}, $\mathcal{S}_{n + 1}(O) = \mathcal{S}_n(O)$.
                Therefore, since $C^\mathcal{B}_{n + 1}$ assigns state $\mathcal{S}_n(O)$ to $O^*$, we have that $C^\mathcal{B}_{n + 1}$ assigns state $\mathcal{S}_{n + 1}(O)$ to $O^*$ as wanted.
    
                \item[] \hspace{0pt}\textbf{Case 2.} $o^\mathcal{A}_{n + 1}$ is a write operation.
    
                Observe that only \cref{line:ero:do_work_initialize_response} executes a write operation.
                Hence, $p_{n + 1}$ executes \cref{line:ero:do_work_initialize_response} during the $n + 1$th step of $\mathcal{I}^\mathcal{A}_{n + 1}$.
                Thus, $o^\mathcal{A}_{n + 1}$ is a write operation for a value $v_1, v_2, \nullconstant$ where $v_1, v_2$ is the state of $p_{n + 1}$'s local variable $\myuniquerepositoryoperationshort{}$ in $C^\mathcal{A}_n$.
                So, $s^\mathcal{A}_{n + 1} = v_1, v_2, \nullconstant$.
                Since $p_{n + 1}$ performs a write operation on $O$ during the $n + 1$th step of $\mathcal{I}^\mathcal{A}_{n + 1}$, and $p_{n + 1}$ executes \cref{line:ero:do_work_initialize_response} during the $n + 1$th step of $\mathcal{I}^\mathcal{A}_{n + 1}$, it follows that $O = (*\cellpointershort).\lastrepositoryoperationresponse{}$, and so $O \notin \{\announceobject, \linearizationobject\} \cup \{(*\cellpointershort).\nextlong\ \vert\ \cellpointershort \in \celluniverse \cup \{\&\headobject\}\}$.
                Hence, by \Cref{def:reduction:watermarks}, $\mathcal{W}_{n + 1}(O, 1) = \bot$ and $\mathcal{W}_{n}(O, 2) = \bot$.
                Furthermore, since $p_{n + 1}$ sets the third index of $O$ to $\nullconstant$ and by \Cref{assumption:ero:head_and_null_not_in_cell_universe} $\nullconstant \notin \celluniverse$, by \Cref{def:reduction:watermarks}, $\mathcal{W}_{n + 1}(O, 3) = \bot$.
                Thus, by \Cref{def:reduction:swapper}, $\mathcal{S}_{n + 1}(O) = v_1, v_2, \nullconstant$.
                Now observe that the values of $p_{n + 1}$'s local variable $\myuniquerepositoryoperationshort{}$ on \cref{line:ero:do_work_initialize_response} does not originate from an $\allocatecelloperation{}$ operation on \cref{line:ero:allocate_cell} (because $\timeshort{}$ originates from \cref{line:ero:operation_timestamp} and $\myrepositoryoperationshort$ is a fixed input on either line \ref{line:ero:low_level_add_cell}, \ref{line:ero:low_level_apply_and_copy_response}, or \ref{line:ero:low_level_remove_cell}).
                Hence, by \Cref{def:reduction:watermarks}, $\mathcal{W}_{n}(\myuniquerepositoryoperationshort{}, 1) = \mathcal{W}_{n}(\myuniquerepositoryoperationshort{}, 2) = \bot$.
                Thus, by \Cref{def:reduction:swapper}, $\mathcal{S}_{n}(\myuniquerepositoryoperationshort{}, 1) = v_1$, and $\mathcal{S}_{n}(\myuniquerepositoryoperationshort{}, 1) = v_2$.
                Therefore, since $\mathcal{S}_{n + 1}(O) = v_1, v_2, \nullconstant$, we have that $\mathcal{S}_{n + 1}(O) = \mathcal{S}_{n}(\myuniquerepositoryoperationshort{}, 1), \mathcal{S}_{n}(\myuniquerepositoryoperationshort{}, 2), \nullconstant$.
    
                Since $o^\mathcal{A}_{n + 1}$ is a write operation for a value $v_1, v_2, \nullconstant$ where $v_1, v_2$ is the state of the local variable $\myuniquerepositoryoperationshort{}$ in $C^\mathcal{A}_n$, by \Cref{lemma:reduction:same_program_counters_in_mapped_run} the program counter of $p_{n + 1}$ is the same in $C^\mathcal{A}_{n}$ and $C_{n}$, and $p_{n + 1}$ takes the $n + 1$th step of $\mathcal{I}^\mathcal{B}_{n + 1}$, we have that $o_{n + 1}$ is a write operation for a value $v'_1, v'_2, \nullconstant$ where $v'_1, v'_2$ is the state of the local variable $\myuniquerepositoryoperationshort{}$ in $C_n$.
                Hence, since by \Cref{def:reduction:run_mapping}, $C_n$ assigns state $\mathcal{S}_n(\myuniquerepositoryoperationshort{})$ to $\myuniquerepositoryoperationshort{}$, we have that $\mathcal{S}_n(\myuniquerepositoryoperationshort{}) = v'_1, v'_2$.
                Thus, $o_{n + 1}$ is a write operation for a value $\mathcal{S}_{n}(\myuniquerepositoryoperationshort{}, 1), \mathcal{S}_{n}(\myuniquerepositoryoperationshort{}, 2), \nullconstant$.
                Therefore, since $\mathcal{S}_{n + 1}(O) = \mathcal{S}_{n}(\myuniquerepositoryoperationshort{}, 1), \mathcal{S}_{n}(\myuniquerepositoryoperationshort{}, 2), \nullconstant$ and $o_{n + 1}$ is an operation on $O^*$, we have that $C^\mathcal{B}_{n + 1}$ assigns state $\mathcal{S}_{n + 1}(O)$ to $O^*$.
                Furthermore, since $o^\mathcal{A}_{n + 1}$ and $o_{n + 1}$ are both write operations, their responses are both $\done$.
                
                \item[] \hspace{0pt}\textbf{Case 3.} $o^\mathcal{A}_{n + 1}$ is a \FAop{} operation.
    
                Observe that only \cref{line:ero:operation_timestamp}, \ref{line:ero:copy_acquisitions_to_revocations}, and \ref{line:ero:relinquish_revocations} execute \FAop{} operations.
                Hence, $O$ is $\clockobject{}$ or $(*\cellpointershort).\revocations$ for some $\cellpointershort \in \celluniverse \cup \{\&\headobject\}$.
                Thus, by \Cref{def:reduction:watermarks}, $\mathcal{W}_{n + 1}(O, 1) = \mathcal{W}_{n}(O, 1) = \bot$.
                So, by \Cref{def:reduction:swapper}, $\mathcal{S}_n(O) = s^\mathcal{A}_n$ and $\mathcal{S}_{n + 1}(O) = s^\mathcal{A}_n + a$ where $a$ is the input of $o^\mathcal{A}_{n + 1}$ (in the case where $o^\mathcal{A}_{n + 1}$ is an \FIop{} operation, $a = 1$).
                Furthermore, the response of $o^\mathcal{A}_{n + 1}$, i.e., $r^\mathcal{A}_{n + 1}$, is $s^\mathcal{A}_n$.
                In the single case where $p_{n + 1}$ read $a$ from one of its local variables, say $pr$, on \cref{line:ero:copy_acquisitions_to_revocations} it can be seen that $a$ did not originate from the response of an $\allocatecelloperation{}$ operation, and so by \Cref{def:reduction:watermarks} $\mathcal{W}_n(pr, 1) = \bot$, and so by \Cref{def:reduction:swapper} $\mathcal{S}_n(pr) = a$.
                Since $o^\mathcal{A}_{n + 1}$ is a \FAop{} operation, by \Cref{lemma:reduction:same_program_counters_in_mapped_run} the program counter of $p_{n + 1}$ is the same in $C^\mathcal{A}_{n}$ and $C_{n}$, and $p_{n + 1}$ takes the $n + 1$th step of $\mathcal{I}^\mathcal{B}_{n + 1}$, we have that $o_{n + 1}$ is a \FAop{} operation.
                Let $a'$ be the input to $o_{n + 1}$.
                Hence, $a'$ is either one, or a value $p_{n + 1}$ read from $pr$.
                Since by \Cref{def:reduction:run_mapping} $C_n$ assigns state $\mathcal{S}_n(pr)$ to $pr$, and $\mathcal{S}_n(pr) = a$, we have that $a' = a$.
                Hence, the input to $o_{n + 1}$ is $a$.
                Thus, since by (*) $C_n$ assigns $\mathcal{S}_n(O)$ to $O^*$, we have that the response of $o_{n + 1}$ is $\mathcal{S}_n(O)$, and $C^\mathcal{B}_{n + 1}$ assigns state $\mathcal{S}_n(O) + a$ to $O^*$.
                Therefore, since $\mathcal{S}_n(O) = s^\mathcal{A}_n$ and $\mathcal{S}_{n + 1}(O) = s^\mathcal{A}_n + a$, we have that $C^\mathcal{B}_{n + 1}$ assigns state $\mathcal{S}_{n + 1}(O)$ to $O^*$ and $r_{n + 1} = s^\mathcal{A}_n$.
    
                \item[] \hspace{0pt}\textbf{Case 4.} $o^\mathcal{A}_{n + 1}$ is a \CASop{} operation.
    
                Hence, since by \Cref{lemma:reduction:same_program_counters_in_mapped_run} the program counter of $p_{n + 1}$ is the same in $C^\mathcal{A}_{n}$ and $C_{n}$, and $p_{n + 1}$ takes the $n + 1$th step of $\mathcal{I}^\mathcal{B}_{n + 1}$, we have that $o_{n + 1}$ is a \CASop{} operation.
                Observe that only \cref{line:ero:linearization_cas}, \ref{line:ero:announce_cas}, \ref{line:ero:add_cell_to_list}, \ref{line:ero:seal_cell}, \ref{line:ero:remove_cell_from_list}, \ref{line:ero:state_cas}, \ref{line:ero:responses_set_attempt}, and \ref{line:ero:acquire_next_cell} execute \CASop{} operations.
                Let $old^\mathcal{A}_{n + 1}$ (resp. $old_{n + 1}$) be the first parameter of $o^\mathcal{A}_{n + 1}$ (resp. $o_{n + 1}$).
                Furthermore, let $new^\mathcal{A}_{n + 1}$ (resp. $new_{n + 1}$) be the second parameter of $o^\mathcal{A}_{n + 1}$ (resp. $o_{n + 1}$).
                Let $\ell$ be the line of code $p_{n + 1}$ executed during the $n + 1$th step of $\mathcal{I}^\mathcal{A}_{n + 1}$.
                There are two cases.
                \begin{itemize}
                    \item[] \hspace{0pt}\textbf{Case 4.1.} $r^\mathcal{A}_{i + 1} = \false$.

                    Hence, since $\delta(s^\mathcal{A}_n, o^\mathcal{A}_{n + 1}) = (s^\mathcal{A}_{n + 1}, r^\mathcal{A}_{n + 1})$, we have that $old^\mathcal{A}_{n + 1} \neq s^\mathcal{A}_n$ and $s^\mathcal{A}_{n + 1} = s^\mathcal{A}_n$ (i.e., the state of $O$ is the same in $C^\mathcal{A}_n$ and $C^\mathcal{A}_{n + 1}$).
                    Thus, $p_{n + 1}$ does not set any index of $O$ during the $n + 1$th step of $\mathcal{I}^\mathcal{A}_{n + 1}$.
                    
                    We first prove that $\mathcal{S}_{n + 1}(O) = \mathcal{S}_n(O)$.
                    Let $v_j$ be the value of the $j$th index of the state of $O$ in $C^\mathcal{A}_{n + 1}$.
                    Hence, since the state of $O$ is the same in $C^\mathcal{A}_n$ and $C^\mathcal{A}_{n + 1}$, it follows that $v_j$ is the value of the $j$th index of the state of $O$ in $C^\mathcal{A}_n$.
                    Furthermore, since $n$ is a non-negative integer, $n + 1 > 0$, so $n \geq 0$.
                    There are two cases.
    
                    \begin{itemize}
                        \item[] \hspace{0pt}\textbf{Case 4.1.1.} if $O$ is a base object, then $O \in \{\announceobject, \linearizationobject\} \cup \{(*\cellpointershort).\nextlong\ \vert\ \cellpointershort \in \celluniverse \cup \{\&\headobject\}\}$, and $v_j \in \celluniverse$.
        
                        Hence, since $p_{n + 1}$ does not set any index of $O$ during the $n + 1$th step of $\mathcal{I}^\mathcal{A}_{n + 1}$, by 1 of \Cref{def:reduction:watermarks}, $\mathcal{W}_{n + 1}(O, j) = \mathcal{W}_n(O, j)$.
                        Therefore, since the state of $O$ is the same in $C^\mathcal{A}_n$ and $C^\mathcal{A}_{n + 1}$, by \Cref{def:reduction:swapper}, $\mathcal{S}_{n + 1}(O, j) = \mathcal{S}_n(O, j)$.
    
                        \item[] \hspace{0pt}\textbf{Case 4.1.2.} otherwise.
    
                        Hence, by \Cref{def:reduction:watermarks}, $\mathcal{W}_{n + 1}(O, j) = \bot$.
                        We now show that $\mathcal{W}_{n}(O, j) = \bot$.
                        Recall that $n \geq 0$.
                        If $n = 0$, then by \Cref{def:reduction:watermarks}, $\mathcal{W}_{n}(O, j) = \bot$, as wanted, so suppose $n > 0$.
                        Hence, by assumption if $O$ is a base object, then $O \notin \{\announceobject, \linearizationobject\} \cup \{(*\cellpointershort).\nextlong\ \vert\ \cellpointershort \in \celluniverse \cup \{\&\headobject\}\}$, or $v_j \notin \celluniverse$.
                        Thus, since $v_j$ is the value of the $j$th index of the state of $O$ in $C^\mathcal{A}_n$, in either case, by \Cref{def:reduction:watermarks}, $\mathcal{W}_{n}(O, j) = \bot$.
                        Therefore, since in all cases $\mathcal{W}_{n + 1}(O, j) = \mathcal{W}_{n}(O, j)$, and the state of $O$ is the same in $C^\mathcal{A}_n$ and $C^\mathcal{A}_{n + 1}$, by \Cref{def:reduction:swapper}, $\mathcal{S}_{n + 1}(O, j) = \mathcal{S}_n(O, j)$.
                    \end{itemize}

                    Since $\mathcal{S}_{n + 1}(O, j) = \mathcal{S}_n(O, j)$ for every index $j$ of $O$, by \Cref{def:reduction:swapper}, $\mathcal{S}_{n + 1}(O) = \mathcal{S}_n(O)$.
                    This completes the proof that $\mathcal{S}_{n + 1}(O) = \mathcal{S}_n(O)$.

                    We now prove that $old_{n + 1} \neq \mathcal{S}_n(O)$.
                    Since $old^\mathcal{A}_{n + 1} \neq s^\mathcal{A}_n$, we have that some index of $old^\mathcal{A}_{n + 1}$ and $s^\mathcal{A}_n$ differ; let this be the $i$th index.
                    Let $o^\mathcal{A}_i$ (resp. $o_i$) be the value of the $i$th index of $old^\mathcal{A}_{n + 1}$ (resp. $old_{n + 1}$).
                    Observe that $o^\mathcal{A}_i$ is either (a) a value dictated by $\ell$ or (b) $p_{n + 1}$ read $o^\mathcal{A}_i$ from the $j$th index of one of its local variable, say $pr$, in $C^\mathcal{A}_n$.
                    Hence, if (a), then $o_i = o^\mathcal{A}_i$ and, if (b), then by \Cref{def:reduction:run_mapping} $o_i = \mathcal{S}_n(pr, j)$.
                    Let $s^\mathcal{A}_i$ be the value of the $i$th index of $s^\mathcal{A}_{n}$, so $o^\mathcal{A}_i \neq s^\mathcal{A}_i$.
                    Hence, by \Cref{def:reduction:used_pointers_in_run_of_a}, if $s^\mathcal{A}_i \in \celluniverse$, then $s^\mathcal{A}_i \in \celluniverse(\mathcal{I}^\mathcal{A}_{n + 1})$.
                    We consider cases (a) and (b) separately.

                    \begin{itemize}
                        \item[] \textbf{Case (a).}

                        Hence, since $o^\mathcal{A}_i$ is a static value determined by $\ell$, it follows that $o^\mathcal{A}_i \notin \celluniverse$.
                        Furthermore, $o_i = o^\mathcal{A}_i$.
                        Thus, since $o^\mathcal{A}_i \neq s^\mathcal{A}_i$, we have that $o_i \neq s^\mathcal{A}_i$, and since $o^\mathcal{A}_i \notin\celluniverse$, we have that $o_i \notin \celluniverse$.
                        Let $\mathcal{W}_n(O, i) = w$.
                        If $w = \bot$, then by \Cref{def:reduction:swapper} $\mathcal{S}_n(O, i) = s^\mathcal{A}_i$, and so since $o_i \neq s^\mathcal{A}_i$, we have that $o_i \neq \mathcal{S}_n(O, i)$.
                        If $w \neq \bot$, then \Cref{def:reduction:swapper} $\mathcal{S}_n(O, i) = \mathcal{M}(w) \in \celluniverse$, and so since $o_i \notin \celluniverse$, we have that $o_i \neq \mathcal{S}_n(O, i)$.
                        Therefore, $old_{n + 1} \neq \mathcal{S}_n(O)$ as wanted.

                        \item[] \textbf{Case (b).}

                        Hence, $o_i = \mathcal{S}_n(pr, j)$ and $o^\mathcal{A}_i$ is the value of the $j$th index of $pr$ in $C^\mathcal{A}_n$.
                        Thus, by \Cref{def:reduction:used_pointers_in_run_of_a}, if $o^\mathcal{A}_i \in \celluniverse$, then $o^\mathcal{A}_i \in \celluniverse(\mathcal{I}^\mathcal{A}_{n + 1})$.
                        So, since $o^\mathcal{A}_i \neq s^\mathcal{A}_i$, by \Cref{lemma:ero:psi_is_basically_injective}, $\mathcal{S}_n(pr, j) \neq \mathcal{S}_n(O, i)$.
                        Therefore, since $o_i = \mathcal{S}_n(pr, j)$, we have that $o_i \neq \mathcal{S}_n(O, i)$, and so $old_{n + 1} \neq \mathcal{S}_n(O)$ as wanted.
                    \end{itemize}
                    
                    This completes the proof that $old_{n + 1} \neq \mathcal{S}_n(O)$.
                    
                    We now finish the proof of Case 4.1.
                    Since by (*) $C_n$ assigns $\mathcal{S}_n(O)$ to $O^*$, $o_{n + 1}$ is an operation on $O^*$, and $old_{n + 1} \neq \mathcal{S}_n(O)$, we have that $o_{n + 1}$ is unsuccessful, so $r_{n + 1} = \false$ and $C^\mathcal{B}_{n + 1}$ assigns state $\mathcal{S}_n(O)$ to $O^*$.
                    Therefore, since $\mathcal{S}_{n + 1}(O) = \mathcal{S}_n(O)$, we have that $C^\mathcal{B}_{n + 1}$ assigns state $\mathcal{S}_{n + 1}(O)$ to $O^*$, and $r_{n + 1} = r^\mathcal{A}_{n + 1}$ as wanted.
            
                    \item[] \hspace{0pt}\textbf{Case 4.2.} $r^\mathcal{A}_{n + 1} = \true$.
    
                    Hence, since $\delta(s^\mathcal{A}_n, o^\mathcal{A}_{n + 1}) = (s^\mathcal{A}_{n + 1}, r^\mathcal{A}_{n + 1})$, we have that $old^\mathcal{A}_{n + 1} = s^\mathcal{A}_n$ and $s^\mathcal{A}_{n + 1} = new^\mathcal{A}_{n + 1}$.
                    
                    We first prove that $old_{n + 1} = \mathcal{S}_n(O)$.
                    There are seven cases.

                    \begin{itemize}
                        \item[] \hspace{0pt}\textbf{Case 4.2.1} $\ell$ is \ref{line:ero:state_cas}.

                        Hence, $O = \stateobject$.
                        Thus, $old^\mathcal{A}_{n + 1} = s^\mathcal{A}_n$ is $\timeshort{}, \myrepositoryoperationshort, \stateshort{}, \responseshort{}$ where one of $p_{n + 1}$'s local variables, say $\myuniquerepositoryoperationshort{}$, is assigned to $\timeshort{}, \myrepositoryoperationshort$ in $C^\mathcal{A}_n$ and two other local variables, say $pr_\stateshort{}$ and $pr_\responseshort{}$, are assigned to $\stateshort{}$ and $\responseshort{}$ in $C^\mathcal{A}_n$, respectively.
                        Since $\stateobject \notin \{\announceobject, \linearizationobject\} \cup \{(*\cellpointershort).\nextlong\ \vert\ \cellpointershort \in \celluniverse \cup \{\&\headobject\}\}$, by \Cref{def:reduction:watermarks}, every index of its state is not watermarked.
                        Hence, since $s^\mathcal{A}_n = \timeshort{}, \myrepositoryoperationshort, \stateshort{}, \responseshort{}$, and $O = \stateobject$, by \Cref{def:reduction:swapper}, $\mathcal{S}_n(O) = \timeshort{}, \myrepositoryoperationshort, \stateshort{}, \responseshort{}$.
                        Furthermore, since the contents of $\myuniquerepositoryoperationshort{}$, $pr_\stateshort{}$, and $pr_\responseshort{}$, originated from $\stateobject$, by \Cref{def:reduction:watermarks}, every index of their state is not watermarked.
                        Hence, by \Cref{def:reduction:swapper}, $\mathcal{S}_n(\myuniquerepositoryoperationshort{}) = \timeshort{}, \myrepositoryoperationshort$, $\mathcal{S}_n(pr_\stateshort{}) = \stateshort{}$, and $\mathcal{S}_n(pr_\responseshort{}) = \responseshort{}$.
                        Thus, by \Cref{def:reduction:run_mapping} $C_n$ assigns state $\timeshort{}, \myrepositoryoperationshort$ to $\myuniquerepositoryoperationshort{}$, $\stateshort{}$ to $pr_\stateshort$, and $\responseshort{}$ to $pr_\responseshort{}$.
                        Therefore, $old_{n + 1} = \timeshort{}, \myrepositoryoperationshort, \stateshort{}, \responseshort{}$ which is equal to $\mathcal{S}_n(O)$ as wanted.
        
                        \item[] \hspace{0pt}\textbf{Case 4.2.2} $\ell$ is \ref{line:ero:responses_set_attempt}.

                        Hence, $O = (*\cellpointershort).\lastrepositoryoperationresponse{}$ for some $\cellpointershort \in \celluniverse \cup \{\&\headobject\}$.
                        Thus, $old^\mathcal{A}_{n + 1} = s^\mathcal{A}_n$ is $\timeshort{}, \myrepositoryoperationshort, \nullconstant$, where one of $p_{n + 1}$'s local variables, say $\myuniquerepositoryoperationshort{}$, is assigned to $\timeshort{}, \myrepositoryoperationshort$ in $C^\mathcal{A}_n$.
                        Observe that the content of $\linearizationobject.\uniquerepositoryoperationlong$ never originates from the response of an $\allocatecelloperation{}$ operation, so by \Cref{def:reduction:watermarks}, every index of its state is not watermarked.
                        Hence, since the contents of $\myuniquerepositoryoperationshort{}$ originated from $\linearizationobject.\uniquerepositoryoperationlong$, by \Cref{def:reduction:watermarks}, every index of their state is not watermarked.
                        Thus, by \Cref{def:reduction:swapper}, $\mathcal{S}_n(\myuniquerepositoryoperationshort{}) = \timeshort{}, \myrepositoryoperationshort$.
                        Therefore, $old_{n + 1} = \timeshort{}, \myrepositoryoperationshort, \nullconstant$.
                        Since $O = (*\cellpointershort).\lastrepositoryoperationresponse{}$ for some $\cellpointershort \in \celluniverse \cup \{\&\headobject\}$, we have that $O \notin \{\announceobject, \linearizationobject\} \cup \{(*\cellpointershort).\nextlong\ \vert\ \cellpointershort \in \celluniverse \cup \{\&\headobject\}\}$, and so by \Cref{def:reduction:watermarks}, every index of its state is not watermarked.
                        Thus, since $s^\mathcal{A}_n = \timeshort{}, \myrepositoryoperationshort, \nullconstant$, by \Cref{def:reduction:swapper}, $\mathcal{S}_n(O) = \timeshort{}, \myrepositoryoperationshort, \nullconstant$.
                        Therefore, since $old_{n + 1} = \timeshort{}, \myrepositoryoperationshort, \nullconstant$, we have that $old_{n + 1} = \mathcal{S}_n(O)$.

                        \item[] \hspace{0pt}\textbf{Case 4.2.3} $\ell$ is \ref{line:ero:add_cell_to_list}.

                        Hence, $O = (*\currentcellpointershort).\nextlong$ for some $\currentcellpointershort \in \celluniverse \cup \{\&\headobject\}$.
                        Thus, it follows that $old^\mathcal{A}_{n + 1} = s^\mathcal{A}_n$ is $\view, \false, 0, \nullconstant$, where one of $p_{n + 1}$'s local variables, say $pr_{\view}$, is assigned to $\view$ in $C^\mathcal{A}_n$.
                        Observe that the contents of the first three indices of the state of $(*\currentcellpointershort).\nextlong$ do not originate from the response of an $\allocatecelloperation{}$ operation, so by \Cref{def:reduction:watermarks}, the first three indices of $(*\currentcellpointershort).\nextlong$ are not watermarked.
                        Hence, since $s^\mathcal{A}_n = \view, \false, 0, \nullconstant$, by \Cref{def:reduction:swapper}, $\mathcal{S}_n(O) = \view, \false, 0, \mathcal{S}_n(O, 4)$.
                        Furthermore, since the contents of $pr_{\view}$ originated from $(*\currentcellpointershort).\nextlong.\view$, by \Cref{def:reduction:watermarks}, every index of their state is not watermarked.
                        Thus, by \Cref{def:reduction:swapper}, $\mathcal{S}_n(pr_{\view}) = \view$.
                        Therefore, $old_{n + 1} = \view, \false, 0, \nullconstant$.
                        Since the fourth index of $s^\mathcal{A}_n$ is $\nullconstant$, and by \Cref{assumption:ero:head_and_null_not_in_cell_universe} $\nullconstant \notin \celluniverse$, by \Cref{def:reduction:watermarks}, $\mathcal{W}_n(O, 4) = \bot$.
                        Hence, by \Cref{def:reduction:swapper}, $\mathcal{S}_n(O, 4) = \nullconstant$.
                        Therefore, since $\mathcal{S}_n(O) = \view, \false, 0, \mathcal{S}_n(O, 4)$, we have that $\mathcal{S}_n(O) = \view, \false, 0, \nullconstant$, and so $old_{n + 1} = \mathcal{S}_n(O)$ as wanted.

                        \item[] \hspace{0pt}\textbf{Case 4.2.4} $\ell$ is \ref{line:ero:seal_cell}.

                        Hence, $O = (*\cellpointershort).\nextlong$ for some $\cellpointershort \in \celluniverse \cup \{\&\headobject\}$, so by \Cref{def:reduction:shared_objects}, $O \in \celluniverse_O$.
                        Thus, it follows that $old^\mathcal{A}_{n + 1} = s^\mathcal{A}_n$ is $\view, sealed, \acquisitions, \nextcellpointershort$, where one of $p_{n + 1}$'s local variables, say $pr_{\view}$, is assigned to $\view$ in $C^\mathcal{A}_n$, another, say $pr_{sealed}$, is assigned $sealed$ in $C^\mathcal{A}_n$, another, say $pr_{\acquisitions}$, is assigned $\acquisitions$ in $C^\mathcal{A}_n$, and another, say $pr_{\nextcellpointershort}$, is assigned $\nextcellpointershort$ in $C^\mathcal{A}_n$.
                        Observe that the contents of the first three indices of the state of $(*\cellpointershort).\nextlong$ do not originate from the response of an $\allocatecelloperation{}$ operation, so by \Cref{def:reduction:watermarks}, the first three indices of $(*\cellpointershort).\nextlong$ are not watermarked.
                        Hence, since $s^\mathcal{A}_n = \view, sealed, \acquisitions, \nextcellpointershort$, by \Cref{def:reduction:swapper}, $\mathcal{S}_n(O) = \view, sealed, \acquisitions, \mathcal{S}_n(O, 4)$.
                        Furthermore, since the contents of $pr_{\view}$, $pr_{sealed}$, and $pr_{\acquisitions}$ originated from the first three indices of $(*\cellpointershort).\nextlong$, respectively, by \Cref{def:reduction:watermarks}, every index of their state is not watermarked.
                        Thus, by \Cref{def:reduction:swapper}, $\mathcal{S}_n(pr_{\view}) = \view$, $\mathcal{S}_n(pr_{sealed}) = sealed$, and $\mathcal{S}_n(pr_{\acquisitions}) = \acquisitions$.
                        Therefore, $old_{n + 1} = \view, sealed, \acquisitions, \mathcal{S}_n(pr_{\nextcellpointershort})$.
                        
                        Since by (*) $C_n$ assigns $\mathcal{S}_n(O)$ to $O^*$, what remains is to prove that $\mathcal{S}_n(pr_{\nextcellpointershort}) = \mathcal{S}_n(O, 4)$.
                        Since $O \in \celluniverse_O$, it follows that $O^* = (*\mathcal{S}_n(O_s)).\nextlong$.
                        Hence, if $O$ is in $\headobject$, then $O^* = \headobject.\nextlong$, and otherwise, by \Cref{claim:reduction:op_on_ptr_after_allocate_in_a}, $\mathcal{W}_n(O_s, 1) = w \neq \bot$ and $\mathcal{S}_n(O_s) = \mathcal{M}(w)$, so $O^* = (*\mathcal{M}(w)).\nextlong$.
                        Thus, since $O^* = (*\mathcal{S}_n(O_s)).\nextlong$, and $O^*$ is either $\headobject.\nextlong$ or $(*\mathcal{M}(w)).\nextlong$, we have that $\mathcal{S}_n(O_s) \in \celluniverse \cup \{\&\headobject\}$.
                        Since $p_{n + 1}$ performs a \CASop{} operation on $O^*$ during the $n + 1$th step of $\mathcal{I}^\mathcal{B}_{n + 1}$, and $old_{n + 1} = \view, sealed, \acquisitions, \mathcal{S}_n(pr_{\nextcellpointershort})$, we have that $p_{n + 1}$ read the value $\view, sealed, \acquisitions, \mathcal{S}_n(pr_{\nextcellpointershort})$ from $O^*$ on its last execution of \cref{line:ero:remove_cell_read_pointer_to_remove_before_seal} before the $n + 1$th step of $\mathcal{I}^\mathcal{B}_{n + 1}$; say during the $i$th step of $\mathcal{I}^\mathcal{B}_{n + 1}$.
                        Hence, $(*\mathcal{S}_n(O_s)).\nextlong.\view = \view$ in $C_{i - 1}$.
                        Furthermore, since by (*) $C_n$ assigns $\mathcal{S}_n(O)$ to $O^*$, and $\mathcal{S}_n(O) = \view, sealed, \acquisitions, \mathcal{S}_n(O, 4)$, we have that $(*\mathcal{S}_n(O_s)).\nextlong.\view = \view$ in $C_{n}$.
                        Therefore, $(*\mathcal{S}_n(O_s)).\nextlong.\view$ is the same in $C_{i - 1}$ and $C_n$. 

                        We prove that for every $i < j < n + 1$ the $j$th step of $\mathcal{I}^\mathcal{B}_{n + 1}$ does not set the value of $O^*$.
                        Suppose, for contradiction, for some $i < j < n + 1$ the $j$th step of $\mathcal{I}^\mathcal{B}_{n + 1}$ sets the value of $O^*$.
                        Since the $j$th step of $\mathcal{I}^\mathcal{B}_{n + 1}$ sets the value of $O^*$, and $O^* = (*\mathcal{S}_n(O_s)).\nextlong$, by \Cref{observation:ero:where_objects_change}, the $j$th step either performs a successful list-add attempt after $\mathcal{S}_n(O_s)$, a successful list-seal attempt for $\mathcal{S}_n(O_s)$, a successful list-remove attempt between $\mathcal{S}_n(O_s)$ and some pointer, or a successful list-acquire-next attempt after $\mathcal{S}_n(O_s)$.
                        Hence, by \Cref{def:ero:english}, the $j$th step performs a successful $\CASop$ operation of the form $\CASop((*\mathcal{S}_n(O_s)).\nextlong, (v, \arbitraryvalue, \arbitraryvalue, \arbitraryvalue), (v + 1, \arbitraryvalue, \arbitraryvalue, \arbitraryvalue))$ for some view $v$.
                        Thus, $(*\mathcal{S}_n(O_s)).\nextlong.\view = v$ in $C_{j - 1}$ and $(*\mathcal{S}_n(O_s)).\nextlong.\view = v + 1$ in $C_{j}$.
                        Therefore, since $\mathcal{S}_n(O_s) \in \celluniverse{} \cup \{\&\headobject\}$ by \Cref{observation:ero:views_are_monotonic} $(*\mathcal{S}_n(O_s)).\nextlong.\view$ is monotonically increasing, and $i < j < n + 1$, it follows that (1) $(*\mathcal{S}_n(O_s)).\nextlong.\view \leq v$ in $C_{i - 1}$, and (2) $(*\mathcal{S}_n(O_s)).\nextlong.\view > v$ in $C_n$, and so $(*\mathcal{S}_n(O_s)).\nextlong.\view$ is different in $C_{i - 1}$ and $C_n$.
                        However, as we established above, $(*\mathcal{S}_n(O_s)).\nextlong.\view$ is the same in $C_{i - 1}$ and $C_n$, a contradiction.

                        We now finish the proof of Case 4.2.4.            
                        Since $p_{n + 1}$ reads $O^*$ during the $i$th step of $\mathcal{I}^\mathcal{B}_{n + 1}$, and for every $i < j < n + 1$ the $j$th step of $\mathcal{I}^\mathcal{B}_{n + 1}$ does not set the value of $O^*$, we have that the state of $O^*$ is the same in $C_{i - 1}$ and $C_n$.
                        Therefore, since $p_{n + 1}$ read $\view, sealed, \acquisitions, \mathcal{S}_n(pr_{\nextcellpointershort})$ during the $i$th step of $\mathcal{I}^\mathcal{B}_{n + 1}$, and $C_n$ assigns state $\mathcal{S}_n(O) = \view, sealed, \acquisitions, \mathcal{S}_n(O, 4)$ to $O^*$, we have that $\mathcal{S}_n(pr_{\nextcellpointershort}) = \mathcal{S}_n(O, 4)$ as wanted.
                        
                        \item[] \hspace{0pt}\textbf{Case 4.2.5} $\ell$ is \ref{line:ero:acquire_next_cell}.

                        The proof is essentially the same as Case 4.2.4.
                        Since $\ell$ is \ref{line:ero:acquire_next_cell}, we have that $O = (*\cellpointershort).\nextlong$ for some $\cellpointershort \in \celluniverse \cup \{\&\headobject\}$, so by \Cref{def:reduction:shared_objects}, $O \in \celluniverse_O$.
                        Thus, it follows that $old^\mathcal{A}_{n + 1} = s^\mathcal{A}_n$ is the sequence $\view, \false, \acquisitions, \nextcellpointershort$, where one of $p_{n + 1}$'s local variables, say $pr_{\view}$, is assigned to $\view$ in $C^\mathcal{A}_n$, another, say $pr_{\acquisitions}$, is assigned to $\acquisitions$ in $C^\mathcal{A}_n$, and another, say $pr_{\nextcellpointershort}$, is assigned to $\nextcellpointershort$ in $C^\mathcal{A}_n$.
                        Observe that the first three indices of the state of $(*\cellpointershort).\nextlong$ do not originate from the response of an $\allocatecelloperation{}$ operation, so by \Cref{def:reduction:watermarks}, the first three indices of $(*\cellpointershort).\nextlong$ are not watermarked.
                        Hence, since $s^\mathcal{A}_n = \view, \false, \acquisitions, \nextcellpointershort$, by \Cref{def:reduction:swapper}, $\mathcal{S}_n(O) = \view, \false, \acquisitions, \mathcal{S}_n(O, 4)$.
                        Furthermore, since the contents of $pr_{\view}$ and $pr_{\acquisitions}$ originated from the first and third index of $(*\cellpointershort).\nextlong$, respectively, by \Cref{def:reduction:watermarks}, their state is not watermarked.
                        Thus, by \Cref{def:reduction:swapper}, $\mathcal{S}_n(pr_{\view}) = \view$ and $\mathcal{S}_n(pr_{\acquisitions}) = \acquisitions$.
                        Therefore, $old_{n + 1} = \view, \false, \acquisitions, \mathcal{S}_n(pr_{\nextcellpointershort})$.
            
                        Since by (*) $C_n$ assigns $\mathcal{S}_n(O)$ to $O^*$, what remains is to prove that $\mathcal{S}_n(pr_{\nextcellpointershort}) = \mathcal{S}_n(O, 4)$.
                        Since $O \in \celluniverse_O$, it follows that $O^* = (*\mathcal{S}_n(O_s)).\nextlong$.
                        Hence, if $O$ is in $\headobject$, then $O^* = \headobject.\nextlong$, and otherwise, by \Cref{claim:reduction:op_on_ptr_after_allocate_in_a}, $\mathcal{W}_n(O_s, 1) = w \neq \bot$ and $\mathcal{S}_n(O_s) = \mathcal{M}(w)$, so $O^* = (*\mathcal{M}(w)).\nextlong$.
                        Thus, since $O^* = (*\mathcal{S}_n(O_s)).\nextlong$, and $O^*$ is either $\headobject.\nextlong$ or $(*\mathcal{M}(w)).\nextlong$, we have that $\mathcal{S}_n(O_s) \in \celluniverse \cup \{\&\headobject\}$.
                        Since $p_{n + 1}$ performs a \CASop{} operation on $O^*$ during the $n + 1$th step of $\mathcal{I}^\mathcal{B}_{n + 1}$, and $old_{n + 1} = \view, \false, \acquisitions, \mathcal{S}_n(pr_{\nextcellpointershort})$, we have that $p_{n + 1}$ read the value $\view, \unusedvalue, \acquisitions, \mathcal{S}_n(pr_{\nextcellpointershort})$ from $O^*$ on its last execution of \cref{line:ero:acquire_next_read_curr_unique_pointer} before the $n + 1$th step of $\mathcal{I}^\mathcal{B}_{n + 1}$; say during the $i$th step of $\mathcal{I}^\mathcal{B}_{n + 1}$.
                        Hence, $(*\mathcal{S}_n(O_s)).\nextlong.\view = \view$ in $C_{i - 1}$.
                        Furthermore, since by (*) $C_n$ assigns $\mathcal{S}_n(O)$ to $O^*$, and $\mathcal{S}_n(O) = \view, \false, \acquisitions, \mathcal{S}_n(O, 4)$, we have that $(*\mathcal{S}_n(O_s)).\nextlong.\view = \view$ in $C_{n}$.
                        Therefore, $(*\mathcal{S}_n(O_s)).\nextlong.\view$ is the same in $C_{i - 1}$ and $C_n$. 
                        
                        We prove that for every $i < j < n + 1$ the $j$th step of $\mathcal{I}^\mathcal{B}_{n + 1}$ does not set the value of $O^*$.
                        Suppose, for contradiction, for some $i < j < n + 1$ the $j$th step of $\mathcal{I}^\mathcal{B}_{n + 1}$ sets the value of $O^*$.
                        Since the $j$th step of $\mathcal{I}^\mathcal{B}_{n + 1}$ sets the value of $O^*$, and $O^* = (*\mathcal{S}_n(O_s)).\nextlong$, by \Cref{observation:ero:where_objects_change}, the $j$th step either performs a successful list-add attempt after $\mathcal{S}_n(O_s)$, a successful list-seal attempt for $\mathcal{S}_n(O_s)$, a successful list-remove attempt between $\mathcal{S}_n(O_s)$ and some pointer, or a successful list-acquire-next attempt after $\mathcal{S}_n(O_s)$.
                        Hence, by \Cref{def:ero:english}, the $j$th step performs a successful $\CASop$ operation of the form $\CASop((*\mathcal{S}_n(O_s)).\nextlong, (v, \arbitraryvalue, \arbitraryvalue, \arbitraryvalue), (v + 1, \arbitraryvalue, \arbitraryvalue, \arbitraryvalue))$ for some view $v$.
                        Thus, $(*\mathcal{S}_n(O_s)).\nextlong.\view = v$ in $C_{j - 1}$ and $(*\mathcal{S}_n(O_s)).\nextlong.\view = v + 1$ in $C_{j}$.
                        Therefore, since $\mathcal{S}_n(O_s) \in \celluniverse{} \cup \{\&\headobject\}$ by \Cref{observation:ero:views_are_monotonic} $(*\mathcal{S}_n(O_s)).\nextlong.\view$ is monotonically increasing, and $i < j < n + 1$, it follows that (1) $(*\mathcal{S}_n(O_s)).\nextlong.\view \leq v$ in $C_{i - 1}$, and (2) $(*\mathcal{S}_n(O_s)).\nextlong.\view > v$ in $C_n$, and so $(*\mathcal{S}_n(O_s)).\nextlong.\view$ is different in $C_{i - 1}$ and $C_n$.
                        However, $(*\mathcal{S}_n(O_s)).\nextlong.\view$ is the same in $C_{i - 1}$ and $C_n$, a contradiction.

                        We now finish the proof of Case 4.2.5.            
                        Since $p_{n + 1}$ reads $O^*$ during the $i$th step of $\mathcal{I}^\mathcal{B}_{n + 1}$, and for every $i < j < n + 1$ the $j$th step of $\mathcal{I}^\mathcal{B}_{n + 1}$ does not set the value of $O^*$, we have that the state of $O^*$ is the same in $C_{i - 1}$ and $C_n$.
                        Hence, since $p_{n + 1}$ read $\view, \unusedvalue, \acquisitions, \mathcal{S}_n(pr_{\nextcellpointershort})$ during the $i$th step of $\mathcal{I}^\mathcal{B}_{n + 1}$, and $C_n$ assigns state $\mathcal{S}_n(O) = \view, \false, \acquisitions, \mathcal{S}_n(O, 4)$ to $O^*$, we have that $\mathcal{S}_n(pr_{\nextcellpointershort}) = \mathcal{S}_n(O, 4)$ as wanted.

                        \item[] \hspace{0pt}\textbf{Case 4.2.6} $\ell$ is \ref{line:ero:remove_cell_from_list}.

                        Hence, $O = (*\cellpointershort).\nextlong$ for some $\cellpointershort \in \celluniverse \cup \{\&\headobject\}$, so by \Cref{def:reduction:shared_objects}, $O \in \celluniverse_O$.
                        Thus, it follows that $old^\mathcal{A}_{n + 1} = s^\mathcal{A}_n$ is $\view, \false, \acquisitions, \cellpointershort'$, where one of $p_{n + 1}$'s local variables, say $pr_{\view}$, is assigned to $\view$ in $C^\mathcal{A}_n$, another, say $pr_{\acquisitions}$, is assigned $\acquisitions$ in $C^\mathcal{A}_n$, and another, say $pr_{\cellpointershort'}$, is assigned $\cellpointershort'$ in $C^\mathcal{A}_n$.
                        Observe that the contents of the first three indices of the state of $(*\cellpointershort).\nextlong$ do not originate from the response of an $\allocatecelloperation{}$ operation, so by \Cref{def:reduction:watermarks}, the first three indices of $(*\cellpointershort).\nextlong$ are not watermarked.
                        Hence, since $s^\mathcal{A}_n = \view, \false, \acquisitions, \cellpointershort'$, by \Cref{def:reduction:swapper}, $\mathcal{S}_n(O) = \view, \false, \acquisitions, \mathcal{S}_n(O, 4)$.
                        Furthermore, since the contents of $pr_{\view}$ and $pr_{\acquisitions}$ originated from the first and third index of $(*\cellpointershort).\nextlong$, respectively, by \Cref{def:reduction:watermarks}, every index of their state is not watermarked.
                        Thus, by \Cref{def:reduction:swapper}, $\mathcal{S}_n(pr_{\view}) = \view$ and $\mathcal{S}_n(pr_{\acquisitions}) = \acquisitions$.
                        Therefore, $old_{n + 1} = \view, \false, \acquisitions, \mathcal{S}_n(pr_{\cellpointershort'})$.
                        
                        Since by (*) $C_n$ assigns $\mathcal{S}_n(O)$ to $O^*$, what remains is to prove that $\mathcal{S}_n(pr_{\cellpointershort'}) = \mathcal{S}_n(O, 4)$.
                        We start by proving that the state of $O^*$ is the same at the last time $p_{n + 1}$ read it and $C_n$.
                        The argument is essentially the same as Case 4.2.5.
                        Since $O \in \celluniverse_O$, it follows that $O^* = (*\mathcal{S}_n(O_s)).\nextlong$.
                        Hence, if $O$ is in $\headobject$, then $O^* = \headobject.\nextlong$, and otherwise, by \Cref{claim:reduction:op_on_ptr_after_allocate_in_a}, $\mathcal{W}_n(O_s, 1) = w \neq \bot$ and $\mathcal{S}_n(O_s) = \mathcal{M}(w)$, so $O^* = (*\mathcal{M}(w)).\nextlong$.
                        Thus, since $O^* = (*\mathcal{S}_n(O_s)).\nextlong$, and $O^*$ is either $\headobject.\nextlong$ or $(*\mathcal{M}(w)).\nextlong$, we have that $\mathcal{S}_n(O_s) \in \celluniverse \cup \{\&\headobject\}$.
                        Since $p_{n + 1}$ performs a \CASop{} operation on $O^*$ during the $n + 1$th step of $\mathcal{I}^\mathcal{B}_{n + 1}$, and $old_{n + 1} = \view, \false, \acquisitions, \mathcal{S}_n(pr_{\cellpointershort'})$, we have that $p_{n + 1}$ read the value $\view, \unusedvalue, \acquisitions, \unusedvalue$ from $O^*$ on its last execution of \cref{line:ero:remove_cell_read_previous_pointer} in $\mathcal{I}^\mathcal{B}_{n + 1}$; say during the $i$th step of $\mathcal{I}^\mathcal{B}_{n + 1}$.
                        Hence, $(*\mathcal{S}_n(O_s)).\nextlong.\view = \view$ in $C_{i - 1}$.
                        Furthermore, since by (*) $C_n$ assigns $\mathcal{S}_n(O)$ to $O^*$, and $\mathcal{S}_n(O) = \view, \false, \acquisitions, \mathcal{S}_n(O, 4)$, we have that $(*\mathcal{S}_n(O_s)).\nextlong.\view = \view$ in $C_{n}$.
                        Therefore, $(*\mathcal{S}_n(O_s)).\nextlong.\view$ is the same in $C_{i - 1}$ and $C_n$. 
                        
                        We prove that for every $i < j < n + 1$ the $j$th step of $\mathcal{I}^\mathcal{B}_{n + 1}$ does not set the value of $O^*$.
                        Suppose, for contradiction, for some $i < j < n + 1$ the $j$th step of $\mathcal{I}^\mathcal{B}_{n + 1}$ sets the value of $O^*$.
                        Since the $j$th step of $\mathcal{I}^\mathcal{B}_{n + 1}$ sets the value of $O^*$, and $O^* = (*\mathcal{S}_n(O_s)).\nextlong$, by \Cref{observation:ero:where_objects_change}, the $j$th step either performs a successful list-add attempt after $\mathcal{S}_n(O_s)$, a successful list-seal attempt for $\mathcal{S}_n(O_s)$, a successful list-remove attempt between $\mathcal{S}_n(O_s)$ and some pointer, or a successful list-acquire-next attempt after $\mathcal{S}_n(O_s)$.
                        Hence, by \Cref{def:ero:english}, the $j$th step performs a successful $\CASop$ operation of the form $\CASop((*\mathcal{S}_n(O_s)).\nextlong, (v, \arbitraryvalue, \arbitraryvalue, \arbitraryvalue), (v + 1, \arbitraryvalue, \arbitraryvalue, \arbitraryvalue))$ for some view $v$.
                        Thus, $(*\mathcal{S}_n(O_s)).\nextlong.\view = v$ in $C_{j - 1}$ and $(*\mathcal{S}_n(O_s)).\nextlong.\view = v + 1$ in $C_{j}$.
                        Therefore, since $\mathcal{S}_n(O_s) \in \celluniverse{} \cup \{\&\headobject\}$ by \Cref{observation:ero:views_are_monotonic} $(*\mathcal{S}_n(O_s)).\nextlong.\view$ is monotonically increasing, and $i < j < n + 1$, it follows that (1) $(*\mathcal{S}_n(O_s)).\nextlong.\view \leq v$ in $C_{i - 1}$, and (2) $(*\mathcal{S}_n(O_s)).\nextlong.\view > v$ in $C_n$, and so $(*\mathcal{S}_n(O_s)).\nextlong.\view$ is different in $C_{i - 1}$ and $C_n$.
                        However, as established above $(*\mathcal{S}_n(O_s)).\nextlong.\view$ is the same in $C_{i - 1}$ and $C_n$, a contradiction.
         
                        This implies that $O^*$ is in the same state in $C_{i - 1}$ and $C_n$ as wanted.
                        The difference between this case and the last two is that $p_{n + 1}$'s fourth value in $old_{n + 1}$ was not read from $O^*$ during the $i$th step of $\mathcal{I}^\mathcal{B}_{n + 1}$ but was read much earlier from $\linearizationobject$.
                        We now prove that the value $p_{n + 1}$ read from $(*\mathcal{S}_n(O_s)).\nextlong.\cellpointerlong$ during the $i$th step is the fourth value in $old_{n + 1}$, i.e., $\mathcal{S}_n(pr_{\cellpointershort'})$.

                        Since $p_{n + 1}$ performs a \CASop{} operation on $O^*$ during the $n + 1$th step of $\mathcal{I}^\mathcal{B}_{n + 1}$, we have that $p_{n + 1}$ found the condition on \cref{line:ero:remove_cell_before_removal_linearization_check} to be false between the $i$th and $n + 1$th step; say the $j$th step.
                        Hence, $\linearizationobject.\uniquerepositoryoperationlong = \uniquerepositoryoperationshort_\linearizationobject$ in $C_j$ where $(\uniquerepositoryoperationshort_\linearizationobject, \cellpointershort_\linearizationobject)$ are the parameters of the invocation $I$ of the \doremovecell{} procedure that $p_{n + 1}$ executed the $n + 1$th step during.
                        Thus, by \Cref{lemma:ero:l_event_corresponding_to_do_low_level_op}, there is an $L$-remove $e$ that set $\linearizationobject = (\uniquerepositoryoperationshort_\linearizationobject, \cellpointershort_\linearizationobject)$ before $I$ was invoked in $\mathcal{I}^\mathcal{B}_{n + 1}$.
                        So, by \Cref{lemma:ero:every_l_event_is_for_pointer_from_universe}, $\cellpointershort_\linearizationobject \in \celluniverse$.
                        Since the $n + 1$th step of $\mathcal{I}^\mathcal{B}_{n + 1}$ performs a \CASop{} operation on $O^* = (*\mathcal{S}_n(O_s)).\nextlong$ on \cref{line:ero:remove_cell_from_list}, and this is during $I$ whose second parameter is $\cellpointershort_\linearizationobject$, by \Cref{def:ero:english}, the $n + 1$th step of $\mathcal{I}^\mathcal{B}_{n + 1}$ is a list-remove attempt for $\cellpointershort_\linearizationobject$ between $\mathcal{S}_n(O_s)$ and some pointer $\nextcellpointershort$.
                        Hence, since $old_{n + 1}$ is the first parameter of the \CASop{} operation performed during the $n + 1$th step of $\mathcal{I}^\mathcal{B}_{n + 1}$, and $old_{n + 1} = \view, \false, \acquisitions, \mathcal{S}_n(pr_{\cellpointershort'})$, it follows that $\mathcal{S}_n(pr_{\cellpointershort'}) = \cellpointershort_\linearizationobject$.
                        
                        We now prove that $e$ is the last $L$-event before the $j$th step in $\mathcal{I}^\mathcal{B}_{n + 1}$.
                        Suppose, for contradiction, there is an $L$-event after $e$ and before the $j$th step in $\mathcal{I}^\mathcal{B}_{n + 1}$.
                        Let $e_{last}$ be the last $L$-event before the $j$th step in $\mathcal{I}^\mathcal{B}_{n + 1}$, so $e < e_{last}$.
                        Hence, since $\linearizationobject.\uniquerepositoryoperationlong = \uniquerepositoryoperationshort_\linearizationobject$ in $C_j$, we have that $e_{last}$ set $\linearizationobject.\uniquerepositoryoperationlong = \uniquerepositoryoperationshort_\linearizationobject$.
                        Therefore, since $e < e_{last}$, and $e$ set $\linearizationobject.\uniquerepositoryoperationlong = \uniquerepositoryoperationshort_\linearizationobject$, we have that two $L$-events in $\mathcal{I}^\mathcal{B}_{n + 1}$ set $\linearizationobject.\uniquerepositoryoperationlong$ to the same value.
                        However, since $\mathcal{I}^\mathcal{B}_{n + 1}$ is an implementation history of $\mathcal{B}$, by \Cref{lemma:ero:the_list_invariants_hold}, $P(\mathcal{I}^\mathcal{B}_{n + 1})$ holds, and so by \Cref{lemma:ero:p_implies_unique_low_level_operations_in_linearization}, every $L$-event in $\mathcal{I}^\mathcal{B}_{n + 1}$ sets $\linearizationobject.\uniquerepositoryoperationlong$ to a unique value, a contradiction.

                        We now prove that there are no successful list-add or list-remove attempts between $e$ and the $i$th step of $\mathcal{I}^\mathcal{B}_{n + 1}$.
                        Suppose, for contradiction, there is a successful list-add or list-remove attempt between $e$ and the $i$th step of $\mathcal{I}^\mathcal{B}_{n + 1}$.
                        The plan is to show that this implies $p_{n + 1}$ must have found the condition on \cref{line:ero:remove_cell_before_removal_linearization_check} to be true at step $j$, contradicting the fact that it found the condition to be false.
                        Let $k$ be the step of $p_{n + 1}$'s last execution of \cref{line:ero:remove_cell_read_pointer_to_remove} in $\mathcal{I}^\mathcal{B}_{n + 1}$, so $k < i$.
                        Hence, $p_{n + 1}$ executed the $k$th step during $I$, and so since $e$ occurred before $I$ was invoked, it follows that $e < k$.
                        Thus, since $k < i$, and $i < j$, by transitivity, $k < j$, and so since $e$ is the last $L$-event before the $j$th step in $\mathcal{I}^\mathcal{B}_{n + 1}$, we have that $e$ is the last $L$-event before the $k$th step in $\mathcal{I}^\mathcal{B}_{n + 1}$.
                        Let $\mathcal{I}_k$ be the prefix of $\mathcal{I}^\mathcal{B}_{n + 1}$ up to and including the $k$th step, so $e$ is the last $L$-event in $\mathcal{I}_k$.
                        Hence, since $\mathcal{I}_k$ is an implementation history of $\mathcal{B}$, by \Cref{lemma:ero:the_list_invariants_hold}, $P(\mathcal{I}_k)$, $Q(\mathcal{I}_k)$, and $R(\mathcal{I}_k)$ hold, and so since $e$ is an $L$-remove event for $\cellpointershort_\linearizationobject$, by \Cref{lemma:ero:3_of_r_safety_holds}, from $e$ onwards in $\mathcal{I}_k$ there is at most one successful list-remove attempt for $\cellpointershort_\linearizationobject$ and no other successful list-add or list-remove attempt for any pointer.
                        First, suppose from $e$ onwards in $\mathcal{I}_k$ there are no successful list-remove attempts for $\cellpointershort_\linearizationobject$.
                        Hence, from $e$ onwards in $\mathcal{I}_k$ there are no successful list-add or list-remove attempts, so since $\cellpointershort_\linearizationobject \in \celluniverse$, by \Cref{observation:ero:where_objects_change}, $(*\cellpointershort_\linearizationobject).\nextlong.\cellpointerlong$ is unchanged from $e$ onwards in $\mathcal{I}_k$.
                        Now suppose from $e$ onwards in $\mathcal{I}_k$ there is a successful list-remove attempt for $\cellpointershort_\linearizationobject$.
                        Hence, from $e$ onwards in $\mathcal{I}_k$ there is exactly one list-remove attempt for $\cellpointershort_\linearizationobject$, say $a$, and no other successful list-add or list-remove attempt for any pointer.
                        Let $a$ be between some pointer $\previouscellpointershort$ and some pointer.
                        Hence, since $a$ is for $\cellpointershort_\linearizationobject$, and $P(\mathcal{I}_k)$ and $Q(\mathcal{I}_k)$ hold, by \Cref{lemma:ero:list_remove_attempt_has_different_next}, $\previouscellpointershort \neq \cellpointershort_\linearizationobject$.
                        Thus, since $\cellpointershort_\linearizationobject \in \celluniverse$, by \Cref{observation:ero:where_objects_change}, $(*\cellpointershort_\linearizationobject).\nextlong.\cellpointerlong$ is unchanged from $e$ onwards in $\mathcal{I}_k$.
                        Therefore, in all cases, $(*\cellpointershort_\linearizationobject).\nextlong.\cellpointerlong$ is unchanged from $e$ onwards in $\mathcal{I}_k$ (A).
                        
                        Let $\mathcal{I}^{include}_e$ be the prefix of $\mathcal{I}^\mathcal{B}_{n + 1}$ up to and including $e$.
                        Hence, $e$ is the last step (and thus $L$-event) in $\mathcal{I}^{include}_e$, so there is no successful list-add or list-remove attempts from $e$ onwards $\mathcal{I}^{include}_e$.
                        Thus, since $e$ is an $L$-remove event and by \Cref{lemma:ero:the_list_invariants_hold}, $P(\mathcal{I}^\mathcal{B}_{n + 1})$, $Q(\mathcal{I}^\mathcal{B}_{n + 1})$, and $R(\mathcal{I}^\mathcal{B}_{n + 1})$ hold, by \Cref{lemma:ero:conditional_classification_lemma}, the list of cells conforms to $\List(\mathcal{I}^{exclude}_e)$ in $\mathcal{I}^{include}_e$ where $\mathcal{I}^{include}_e$ be the prefix of $\mathcal{I}^\mathcal{B}_{n + 1}$ up to but excluding $e$ (B).
                        
                        Let $\mathcal{I}_i$ be the prefix of $\mathcal{I}^\mathcal{B}_{n + 1}$ up to and including the $i$th step.
                        Since $i < j$, and $e$ is the last $L$-event before the $j$th step in $\mathcal{I}^\mathcal{B}_{n + 1}$, we have that $e$ is the last $L$-event in $\mathcal{I}_i$.
                        Hence, since by assumption there is a successful list-add or list-remove attempt between $e$ and the $i$th step of $\mathcal{I}^\mathcal{B}_{n + 1}$, we have that from $e$ onwards in $\mathcal{I}_i$ there is a successful list-add or list-remove attempt.
                        Thus, since $\mathcal{I}_i$ is a finite implementation history of $\mathcal{B}$, by \Cref{lemma:ero:the_list_invariants_hold}, $P(\mathcal{I}_i)$, $Q(\mathcal{I}_i)$, and $R(\mathcal{I}_i)$ hold, and so by \Cref{lemma:ero:conditional_classification_lemma}, the list of cells conforms to $\List(\mathcal{I}_i)$ in $\mathcal{I}_i$ (C).
                        
                        Since $\mathcal{I}^\mathcal{B}_{n + 1}$ is an implementation history of $\mathcal{B}$, by \Cref{lemma:ero:the_list_invariants_hold}, $Q(\mathcal{I}^\mathcal{B}_{n + 1})$ holds, and so since the $n + 1$th step of $\mathcal{I}^\mathcal{B}_{n + 1}$ performs a list-remove attempt for $\cellpointershort_\linearizationobject$ between $\mathcal{S}_n(O_s)$ and $\nextcellpointershort$, it is preceded by a unique $L$-remove event for $\cellpointershort_\linearizationobject$ such that if $\mathcal{I}$ is the prefix of $\mathcal{I}^\mathcal{B}_{n + 1}$ up to but excluding that $L$-event, then $\cellpointershort_\linearizationobject$ appears in $\List(\mathcal{I})$ exactly once and $\mathcal{S}_n(O_s)$ and $\nextcellpointershort$ are the pointers preceding and succeeding $\cellpointershort_\linearizationobject$ in $\List(\mathcal{I})$.
                        Thus, since $e$ is an $L$-remove event for $\cellpointershort_\linearizationobject$ in $\mathcal{I}^\mathcal{B}_{n + 1}$, it follows that $\mathcal{I} = \mathcal{I}^{exclude}_e$, so $\cellpointershort_\linearizationobject$ appears in $\List(\mathcal{I}^{exclude}_e)$ exactly once, and $\mathcal{S}_n(O_s)$ and $\nextcellpointershort$ are the pointers preceding and succeeding $\cellpointershort_\linearizationobject$ in $\List(\mathcal{I}^{exclude}_e)$ (D).
                        
                        Since $e$ is before the $i$th step of $\mathcal{I}^\mathcal{B}_{n + 1}$, it follows that $\mathcal{I}^{exclude}_e$ is the prefix of $\mathcal{I}_i$ up to but excluding $e$.
                        So, since $e$ is the last $L$-event in $\mathcal{I}_i$, we have that the sequence of $L$-events in $\mathcal{I}^{exclude}_e$ and $\mathcal{I}_i$ are the same except the former excludes $e$ and the latter includes $e$.
                        Hence, since $e$ is an $L$-remove event for $\cellpointershort_\linearizationobject$, $\cellpointershort_\linearizationobject$ appears in $\List(\mathcal{I}^{exclude}_e)$ exactly once, and $\mathcal{S}_n(O_s)$ and $\nextcellpointershort$ are the pointers preceding and succeeding $\cellpointershort_\linearizationobject$ in $\List(\mathcal{I}^{exclude}_e)$, by \Cref{def:ero:logical_list}, $\mathcal{S}_n(O_s)$ appears immediately before $\nextcellpointershort$ in $\List(\mathcal{I}_i)$ (E).
                        
                        We now put everything together.
                        Since by (B) the the list of cells conforms to $\List(\mathcal{I}^{exclude}_e)$ in $\mathcal{I}^{include}_e$, and by (D) $\cellpointershort_\linearizationobject$ appears in $\List(\mathcal{I}^{exclude}_e)$ exactly once and $\nextcellpointershort$ is the pointer succeeding $\cellpointershort_\linearizationobject$ in $\List(\mathcal{I}^{exclude}_e)$, by \Cref{def:ero:logical_list}, $(*\cellpointershort_\linearizationobject).\nextlong.\cellpointerlong = \nextcellpointershort$ at $e$.
                        Hence, since by (A) $(*\cellpointershort_\linearizationobject).\nextlong.\cellpointerlong$ is unchanged from $e$ onwards in $\mathcal{I}_k$, we have that $(*\cellpointershort_\linearizationobject).\nextlong.\cellpointerlong = \nextcellpointershort$ in $C_k$.
                        Thus, since $p_{n + 1}$ performs a list-remove attempt for $\cellpointershort_\linearizationobject$ during the $n + 1$th step of $\mathcal{I}^\mathcal{B}_{n + 1}$, and $k$ is $p_{n + 1}$'s last execution of \cref{line:ero:remove_cell_read_pointer_to_remove} in $\mathcal{I}^\mathcal{B}_{n + 1}$, we have that $p_{n + 1}$ read $\nextcellpointershort$ from $(*\cellpointershort_\linearizationobject).\nextlong.\cellpointerlong$ during the $k$th step of $\mathcal{I}^\mathcal{B}_{n + 1}$.
                        Since by (C) the list of cells conforms to $\List(\mathcal{I}_i)$ in $\mathcal{I}_i$, and by (E) $\mathcal{S}_n(O_s)$ appears immediately before $\nextcellpointershort$ in $\List(\mathcal{I}_i)$, by \Cref{def:ero:logical_list}, $(*\mathcal{S}_n(O_s)).\nextlong.\cellpointerlong = \nextcellpointershort$ in $C_i$.
                        Hence, since $p_{n + 1}$ performs a list-remove attempt between $\mathcal{S}_n(O_s)$ and $\nextcellpointershort$ during the $n + 1$th step of $\mathcal{I}^\mathcal{B}_{n + 1}$, and $i$ is $p_{n + 1}$'s last execution of \cref{line:ero:remove_cell_read_previous_pointer} in $\mathcal{I}^\mathcal{B}_{n + 1}$, we have that $p_{n + 1}$ read $\nextcellpointershort$ from $(*\mathcal{S}_n(O_s)).\nextlong.\cellpointerlong$ during the $i$th step of $\mathcal{I}^\mathcal{B}_{n + 1}$.
                        Therefore, since $k$ (resp. $i$) is $p_{n + 1}$'s last execution of \cref{line:ero:remove_cell_read_pointer_to_remove} (resp. \cref{line:ero:remove_cell_read_previous_pointer}) in $\mathcal{I}^\mathcal{B}_{n + 1}$, and they both read $\nextcellpointershort$ from the $\cellpointerlong$ field of a $\nextlong$ object of a cell during these steps, we have that $p_{n + 1}$ finds the condition on \cref{line:ero:remove_cell_before_removal_linearization_check} to be true on its last execution of it in $\mathcal{I}^\mathcal{B}_{n + 1}$, so $p_{n + 1}$ finds the condition on \cref{line:ero:remove_cell_before_removal_linearization_check} to be true during the $j$th step of $\mathcal{I}^\mathcal{B}_{n + 1}$.
                        However, $p_{n + 1}$ finds the condition on \cref{line:ero:remove_cell_before_removal_linearization_check} to be false during the $j$th step of $\mathcal{I}^\mathcal{B}_{n + 1}$, a contradiction.

                        We now finish the proof of Case 4.2.6.
                        Let $\mathcal{I}_i$ be the prefix of $\mathcal{I}^\mathcal{B}_{n + 1}$ up to and including the $i$th step.
                        Since $i < j$, and $e$ is the last $L$-event before the $j$th step in $\mathcal{I}^\mathcal{B}_{n + 1}$, we have that $e$ is the last $L$-event in $\mathcal{I}_i$.
                        Furthermore, since there are no successful list-add or list-remove attempts between $e$ and the $i$th step of $\mathcal{I}^\mathcal{B}_{n + 1}$, we have that from $e$ onwards in $\mathcal{I}_i$ there are no successful list-add or list-remove attempts.
                        Hence, since $P(\mathcal{I}_i)$, $Q(\mathcal{I}_i)$, and $R(\mathcal{I}_i)$ hold, by \Cref{lemma:ero:conditional_classification_lemma}, the list of cells conforms to $\List(\mathcal{I}^{exclude}_e)$ in $\mathcal{I}_i$ where $\mathcal{I}^{exclude}_e$ is the prefix of $\mathcal{I}^\mathcal{B}_{n + 1}$ up to but excluding $e$.
                        Since $Q(\mathcal{I}^\mathcal{B}_{n + 1})$ holds, and the $n + 1$th step of $\mathcal{I}^\mathcal{B}_{n + 1}$ performs a list-remove attempt for $\cellpointershort_\linearizationobject$ between $\mathcal{S}_n(O_s)$ and $\nextcellpointershort$, it is preceded by a unique $L$-remove event for $\cellpointershort_\linearizationobject$ such that if $\mathcal{I}$ is the prefix of $\mathcal{I}^\mathcal{B}_{n + 1}$ up to but excluding that $L$-event, then $\cellpointershort_\linearizationobject$ appears in $\List(\mathcal{I})$ exactly once and $\mathcal{S}_n(O_s)$ and $\nextcellpointershort$ are the pointers preceding and succeeding $\cellpointershort_\linearizationobject$ in $\List(\mathcal{I})$.
                        Thus, since $e$ is an $L$-remove event for $\cellpointershort_\linearizationobject$, it follows that $\mathcal{I} = \mathcal{I}^{exclude}_e$, so $\cellpointershort_\linearizationobject$ appears in $\List(\mathcal{I}^{exclude}_e)$ exactly once, and $\mathcal{S}_n(O_s)$ and $\nextcellpointershort$ are the pointers preceding and succeeding $\cellpointershort_\linearizationobject$ in $\List(\mathcal{I}^{exclude}_e)$.
                        Since the list of cells conforms to $\List(\mathcal{I}^{exclude}_e)$ in $\mathcal{I}_i$, and $\mathcal{S}_n(O_s)$ appears immediately before $\cellpointershort_\linearizationobject$ in $\List(\mathcal{I}^{exclude}_e)$, by \Cref{def:ero:logical_list}, $(*\mathcal{S}_n(O_s)).\nextlong.\cellpointerlong = \cellpointershort_\linearizationobject$ in $C_i$.
                        Hence, since $p_{n + 1}$ performs a list-remove attempt between $\mathcal{S}_n(O_s)$ and $\nextcellpointershort$ during the $n + 1$th step of $\mathcal{I}^\mathcal{B}_{n + 1}$, and $i$ is $p_{n + 1}$'s last execution of \cref{line:ero:remove_cell_read_previous_pointer} in $\mathcal{I}^\mathcal{B}_{n + 1}$, we have that $p_{n + 1}$ read $\cellpointershort_\linearizationobject$ from $(*\mathcal{S}_n(O_s)).\nextlong.\cellpointerlong$ during the $i$th step of $\mathcal{I}^\mathcal{B}_{n + 1}$.
                        So, since $O^* = (*\mathcal{S}_n(O_s)).\nextlong$, the fourth value of $O^*$ in $C_{i - 1}$ is $\cellpointershort_\linearizationobject$.
                        Hence, since as proved above $O^*$ is in the same state in $C_{i - 1}$ and $C_n$, we have that the fourth value of $O^*$ in $C_{n}$ is $\cellpointershort_\linearizationobject$.
                        Thus, since by (*) $C_n$ assigns $\mathcal{S}_n(O)$ to $O^*$, we have that $\mathcal{S}_n(O, 4) = \cellpointershort_\linearizationobject$.
                        Therefore, since $\mathcal{S}_n(pr_{\cellpointershort'}) = \cellpointershort_\linearizationobject$, we have that $\mathcal{S}_n(pr_{\cellpointershort'}) = \mathcal{S}_n(O, 4)$ as wanted.

                        \item[] \hspace{0pt}\textbf{Case 4.2.7} $\ell$ is either \ref{line:ero:linearization_cas} or \ref{line:ero:announce_cas}.

                        Hence, $O$ is either $\linearizationobject$ or $\announceobject$.
                        Thus, $old^\mathcal{A}_{n + 1} = s^\mathcal{A}_n = \timeshort{}, \myrepositoryoperationshort, \cellpointershort$, where one of $p_{n + 1}$'s local variables, say $\myuniquerepositoryoperationshort{}$, is assigned to $\timeshort{}, \myrepositoryoperationshort$ in $C^\mathcal{A}_n$, and another, say $pr$, is assigned to $\cellpointershort$ in $C^\mathcal{A}_n$.
                        Observe that, the contents of $\myuniquerepositoryoperationshort{}$ did not originate from the response of an $\allocatecelloperation{}$ operation, and so by \Cref{def:reduction:watermarks}, $\mathcal{W}_n(\myuniquerepositoryoperationshort{}, 1) = \mathcal{W}_n(\myuniquerepositoryoperationshort{}, 2) = \bot$.
                        Hence, by \Cref{def:reduction:swapper}, $\mathcal{S}_n(\myuniquerepositoryoperationshort{}) = \timeshort{}, \myrepositoryoperationshort$.
                        Thus, by \Cref{def:reduction:run_mapping} $C_n$ assigns state $\timeshort{}, \myrepositoryoperationshort$ to $\myuniquerepositoryoperationshort{}$, so $old_{n + 1} = \timeshort{}, \myrepositoryoperationshort, \mathcal{S}_n(pr)$.
                        Likewise, observe that the contents of $\announceobject.\uniquerepositoryoperationlong{}$ (resp. $\linearizationobject.\uniquerepositoryoperationlong$) do not originate from the response of an $\allocatecelloperation{}$ operation, and so by \Cref{def:reduction:watermarks}, $\mathcal{W}_n(\announceobject, 1) = \mathcal{W}_n(\announceobject, 2) = \bot$ (resp. $\mathcal{W}_n(\linearizationobject, 1) = \mathcal{W}_n(\linearizationobject, 2) = \bot$).
                        Hence, by \Cref{def:reduction:swapper}, $\mathcal{S}_n(O) = \timeshort{}, \myrepositoryoperationshort, \mathcal{S}_n(O, 3)$.
                        There are two cases.

                        \textbf{Case A.} $\mathcal{W}_n(pr, 1) = \bot$ and $\mathcal{W}_n(O, 3) = \bot$.
                        
                        Hence, since $pr$ is assigned to $\cellpointershort$ in $C^\mathcal{A}_n$ and the third index of $O$ is $\cellpointershort$, by \Cref{def:reduction:swapper}, $\mathcal{S}_n(pr) = \cellpointershort$ and $\mathcal{S}_n(O, 3) = \cellpointershort$, and so $\mathcal{S}_n(pr) = \mathcal{S}_n(O, 3)$.
                        Therefore, since $old_{n + 1} = \timeshort{}, \myrepositoryoperationshort, \mathcal{S}_n(pr)$ and $\mathcal{S}_n(O) = \timeshort{}, \myrepositoryoperationshort, \mathcal{S}_n(O, 3)$, we have that $old_{n + 1} = \mathcal{S}_n(O)$, as wanted.
                        
                        \textbf{Case B.} $\mathcal{W}_n(pr, 1)$ or $\mathcal{S}_n(O, 3)$ is not $\bot$.
                        
                        Hence, since $pr$ and the third index of $O$ is assigned to $\cellpointershort$ in $C^\mathcal{A}_n$, by \Cref{def:reduction:watermarks}, we have that $\cellpointershort \in \celluniverse$, so by \Cref{assumption:ero:head_and_null_not_in_cell_universe}, $\cellpointershort \neq \nullconstant$.
                        Thus, since $pr$ is a value read from the third index of $O$, and the third index of $O$ is initially $\nullconstant$, we have that some process set the third index of $O$ to $\cellpointershort$ in $\mathcal{I}^\mathcal{A}_{n + 1}$.
                        Since, as we can see, all values written into the third index of $O$ originate from the response of an $\allocatecelloperation{}$ operation, both $\mathcal{W}_n(pr, 1)$ and $\mathcal{W}_n(O, 3)$ are not $\bot$.
                        Let $\mathcal{W}_n(pr, 1) = w \neq \bot$ and $\mathcal{W}_n(O, 3) = w' \neq \bot$.
                        Hence, by \Cref{def:reduction:swapper}, $\mathcal{S}_n(pr) = \mathcal{M}(w)$ and $\mathcal{S}_n(O, 3) = \mathcal{M}(w')$.
                        Thus, $old_{n + 1} = \timeshort{}, \myrepositoryoperationshort, \mathcal{M}(w)$ and $\mathcal{S}_n(O) = \timeshort{}, \myrepositoryoperationshort, \mathcal{M}(w')$.
                        Therefore, if $\mathcal{M}(w) = \mathcal{M}(w')$, $old_{n + 1} = \mathcal{S}_n(O)$, so it suffices to consider the case where $\mathcal{M}(w) \neq \mathcal{M}(w')$.

                        We show that $\mathcal{M}(w) \neq \mathcal{M}(w')$ leads to a contradiction, so this case is impossible.
                        Since  $old_{n + 1} = \timeshort{}, \myrepositoryoperationshort, \mathcal{M}(w)$, we have that $p_{n + 1}$ read $\timeshort{}, \myrepositoryoperationshort, \mathcal{M}(w)$ from $O$ before the $n + 1$th step of $\mathcal{I}^\mathcal{B}_{n + 1}$; say during the $i$th step of $\mathcal{I}^\mathcal{B}_{n + 1}$.
                        Hence, since the third index of $O$ is initially $\nullconstant$ and by \Cref{assumption:ero:head_and_null_not_in_cell_universe} $\mathcal{M}(w) \neq \nullconstant$, we have that some process set $O$ to $\timeshort{}, \myrepositoryoperationshort, \mathcal{M}(w)$ before the $i$th step; say during the $j$th step of $\mathcal{I}^\mathcal{B}_{n + 1}$.
                        Since $\mathcal{S}_n(O) = \timeshort{}, \myrepositoryoperationshort, \mathcal{M}(w')$, by \Cref{def:reduction:run_mapping}, $C_n$ assigns $\timeshort{}, \myrepositoryoperationshort, \mathcal{M}(w')$ to $O$.
                        Hence, since $p_{n + 1}$ read $\mathcal{M}(w)$ during the $i$th step of $\mathcal{I}^\mathcal{B}_{n + 1}$, and $\mathcal{M}(w) \neq \mathcal{M}(w')$, we have that $O$ was set to $\timeshort{}, \myrepositoryoperationshort, \mathcal{M}(w')$ during the $k$th step of $\mathcal{I}^\mathcal{B}_{n + 1}$ where $i < k < n + 1$.
                        Thus, since $j < i$, by transitivity, $j < k$, so $j \neq k$.
                        There are two cases.

                        \textbf{Case B.1.} $O = \linearizationobject$.
                        
                        Therefore, the $j$th (resp. $k$th) step of $\mathcal{I}^\mathcal{B}_{n + 1}$ perform distinct $L$-events that set $\linearizationobject.\uniquerepositoryoperationlong$ to the same value (namely, $\timeshort{}, \myrepositoryoperationshort$).
                        However, since $\mathcal{I}^\mathcal{B}_{n + 1}$ is an implementation history of $\mathcal{B}$, by \Cref{lemma:ero:the_list_invariants_hold}, $P(\mathcal{I}^\mathcal{B}_{n + 1})$ holds, so by \Cref{lemma:ero:p_implies_unique_low_level_operations_in_linearization}, every $L$-event in $\mathcal{I}^\mathcal{B}_{n + 1}$ sets $\linearizationobject.\uniquerepositoryoperationlong$ to a distinct value, a contradiction, so Case B.1. is impossible.

                        \textbf{Case B.2.} $O = \announceobject$.

                        Since the $j$th (resp. $k$th) step of $\mathcal{I}^\mathcal{B}_{n + 1}$ set $\announceobject$ to $\timeshort{}, \myrepositoryoperationshort, \mathcal{M}(w)$ (resp. $\timeshort{}, \myrepositoryoperationshort, \mathcal{M}(w')$), and $\mathcal{I}^\mathcal{B}_{n + 1}$ is an implementation history of $\mathcal{B}$, by \Cref{lemma:ero:matching_timestamp_in_a_implies_matching_value_in_a}, $\mathcal{M}(w) = \mathcal{M}(w')$.
                        However, $\mathcal{M}(w) \neq \mathcal{M}(w')$, a contradiction, so Case B.2 is impossible.

                        This completes the proof that $old_{n + 1} = \mathcal{S}_n(O)$.
                    \end{itemize}

                    We now return to the proof of Case 4.2.
                    Since by (*) $C_n$ assigns $\mathcal{S}_n(O)$ to $O^*$, $o_{n + 1}$ is an operation on $O^*$, and $old_{n + 1} = \mathcal{S}_n(O)$, we have that $o_{n + 1}$ is successful, so $r_{n + 1} = \true$ and $C^\mathcal{B}_{n + 1}$ assigns state $new_{n + 1}$ to $O^*$.
                    Recall that $s^\mathcal{A}_{n + 1} = new^\mathcal{A}_{n + 1}$, so $C^\mathcal{A}_{n + 1}$ assigns state $new^\mathcal{A}_{n + 1}$ to $O$.
                    Let $new^\mathcal{A}_{n + 1} = v_1, v_2, \ldots$ and $new_{n + 1} = v'_1, v'_2, \ldots$.
                    Observe that $v_i$ is either (a) a value dictated by $\ell$ or (b) $p_{n + 1}$ read $v_i$ from the $j_i$th index of its local variable $pr_i$ in $C^\mathcal{A}_n$.
                    We consider each case separately.
                    \begin{itemize}
                        \item[] \hspace{0pt}\textbf{Case (a).}

                        Hence, 3 of \Cref{def:reduction:watermarks} is not satisfied so $\mathcal{W}_{n + 1}(O, i) = \bot$, and so by \Cref{def:reduction:swapper} $\mathcal{S}_{n + 1}(O, i) = v_i$.
                        Furthermore, since $p_{n + 1}$ executes line $\ell$ during the $n + 1$th step of $\mathcal{I}^\mathcal{B}_{n + 1}$, it follows that $v'_i = v_i$.
                        Therefore $\mathcal{S}_{n + 1}(O, i) = v'_i$.

                        \item[] \hspace{0pt}\textbf{Case (b).}

                        Hence, since $p_{n + 1}$ executes line $\ell$ during the $n + 1$th step of $\mathcal{I}^\mathcal{B}_{n + 1}$, by \Cref{def:reduction:run_mapping}, $v'_i = \mathcal{S}_n(pr_i, j_i)$.
                        Furthermore, 3 of \Cref{def:reduction:watermarks} is satisfied so $\mathcal{W}_{n + 1}(O, i) = \mathcal{W}_n(pr_i, j_i)$ or $\mathcal{W}_{n + 1}(O, i) = \bot$.
                        First suppose that $\mathcal{W}_{n + 1}(O, i) = \mathcal{W}_n(pr_i, j_i)$.
                        Hence, since the $i$th index of $O$ is $v_i$ in $C^\mathcal{A}_{n + 1}$ and the $j_i$th index of $pr_i$ is $v_i$ in $C^\mathcal{A}_n$, by \Cref{def:reduction:swapper} $\mathcal{S}_{n + 1}(O, i) = \mathcal{S}_{n}(pr, j_i)$.
                        Therefore, since $v'_i = \mathcal{S}_n(pr_i, j_i)$, we have that $\mathcal{S}_{n + 1}(O, i) = v'_i$.
                        Now suppose that $\mathcal{W}_{n + 1}(O, i) = \bot$.
                        Hence, since 3 of \Cref{def:reduction:watermarks} is satisfied, it follows that $v_i \notin \celluniverse$.
                        Thus, since the $j_i$th index of $pr_i$ is $v_i$ in $C^\mathcal{A}_n$, by \Cref{def:reduction:watermarks}, $\mathcal{W}_{n}(pr, j_i) = \bot$.
                        So, since $\mathcal{W}_{n + 1}(O, i) = \bot$ (resp. $\mathcal{W}_{n}(pr, j_i) = \bot$) and the value of the $i$th (resp. $j_i$th) index of $O$ (resp. $pr$) in $C^\mathcal{A}_{n + 1}$ (resp. $C^\mathcal{A}_{n}$) is $v_i$, by \Cref{def:reduction:swapper}, we have that $\mathcal{S}_{n + 1}(O, i) = \mathcal{S}_{n}(pr, j_i) = v_i$.
                        Therefore, since $v'_i = \mathcal{S}_n(pr_i, j_i)$, we have that $\mathcal{S}_{n + 1}(O, i) = v'_i$.
                    \end{itemize}

                    We now finish the proof of Case 4.2.
                    Since $\mathcal{S}_{n + 1}(O, i) = v'_i$ for each index $i$ of the state of $O$ in $C^\mathcal{A}_{n + 1}$, we have that $v'_1, v'_2, \ldots = \mathcal{S}_{n + 1}(O, 1), \mathcal{S}_{n + 1}(O, 2), \ldots$, so $new_{n + 1} = \mathcal{S}_{n + 1}(O)$.
                    Therefore, $C^\mathcal{B}_{n + 1}$ assigns $\mathcal{S}_{n + 1}(O)$ to $O^*$ and $r_{n + 1} = r^\mathcal{A}_{n + 1}$ as wanted.
                \end{itemize}
    
                \item[] \hspace{0pt}\textbf{Case 5.} $o^\mathcal{A}_{n + 1}$ is \GCASop{} operation.
    
                Hence, since by \Cref{lemma:reduction:same_program_counters_in_mapped_run} the program counter of $p_{n + 1}$ is the same in $C^\mathcal{A}_{n}$ and $C_{n}$, and $p_{n + 1}$ takes the $n + 1$th step of $\mathcal{I}^\mathcal{B}_{n + 1}$, we have that $o_{n + 1}$ is a \GCASop{} operation.
                If the comparator of $o^\mathcal{A}_{n + 1}$ is $=$, $o^\mathcal{A}_{n + 1}$ is simply a \CASop{} operation, which we covered in Case 4.
                Hence, since $\mathcal{A}$ only performs $\GCASop(=)$ and $\GCASop(>)$ operations, it suffices to consider the case where the comparator of $o^\mathcal{A}_{n + 1}$ is $>$.
                Observe that the only $\GCASop{}(>)$ operation is on \cref{line:ero:announce_gcas}.
                Hence, $O = \announceobject$.
                Furthermore, the second and third parameter of $o^\mathcal{A}_{n + 1}$ (resp. $o_{n + 1}$) are the same.
                Let $val^\mathcal{A}_{n + 1}$ (resp. $val_{n + 1}$) be the second and third parameter of $o^\mathcal{A}_{n + 1}$ (resp. $o_{n + 1}$).
                Hence, $val^\mathcal{A}_{n + 1} = \timeshort{}^\mathcal{A}_{n + 1}, \myrepositoryoperationshort^\mathcal{A}_{n + 1}, \cellpointershort^\mathcal{A}_{n + 1}$ where one of $p_{n + 1}$'s local variables, say $\myuniquerepositoryoperationshort{}$, is assigned to $\timeshort{}^\mathcal{A}_{n + 1}, \myrepositoryoperationshort^\mathcal{A}_{n + 1}$ in $C^\mathcal{A}_n$, and another, say $pr$, is assigned to $\cellpointershort^\mathcal{A}_{n + 1}$ in $C^\mathcal{A}_n$.
                Observe that the contents of $\myuniquerepositoryoperationshort{}$ do not originate from the response of an \allocatecelloperation{} operation, so by \Cref{def:reduction:watermarks}, $\mathcal{W}_n(\myuniquerepositoryoperationshort{}, 1) = \mathcal{W}_n(\myuniquerepositoryoperationshort{}, 2) = \bot$.
                Hence, by \Cref{def:reduction:swapper} $\mathcal{S}_n(\myuniquerepositoryoperationshort{}) = \timeshort{}^\mathcal{A}_{n + 1}, \myrepositoryoperationshort^\mathcal{A}_{n + 1}$.
                Furthermore, observe that the contents of $pr$ originated from the response of an $\allocatecelloperation{}$ operation, so by \Cref{alg:cell_manager_specification} $\cellpointershort^\mathcal{A}_{n + 1} \in \celluniverse$, and by \Cref{def:reduction:watermarks}, $\mathcal{W}_n(pr, 1) = w \neq \bot$.
                Hence, by \Cref{def:reduction:swapper}, $\mathcal{S}_n(pr, 1) = \mathcal{M}(w)$.
                Therefore, by \Cref{def:reduction:run_mapping}, $val_{n + 1} = \timeshort{}^\mathcal{A}_{n + 1}, \myrepositoryoperationshort^\mathcal{A}_{n + 1}, \mathcal{M}(w)$.

                Since $C^\mathcal{A}_n$ assigns state $s^\mathcal{A}_n$ to $O$, and $O = \announceobject$, we have that $s^\mathcal{A}_n = \timeshort^\mathcal{A}_n, \myrepositoryoperationshort^\mathcal{A}_n, \cellpointershort^\mathcal{A}_n$.
                Observe that the contents of $\announceobject.\uniquerepositoryoperationlong{}$ do not originate from the response of an $\allocatecelloperation{}$ operation, and so by \Cref{def:reduction:watermarks}, $\mathcal{W}_n(\announceobject, 1) = \mathcal{W}_n(\announceobject, 2) = \bot$.
                Hence, since $O = \announceobject$, by \Cref{def:reduction:swapper}, $\mathcal{S}_n(O) = \timeshort^\mathcal{A}_n, \myrepositoryoperationshort^\mathcal{A}_n, \mathcal{S}_n(O, 3)$.
                There are two cases.
                \begin{itemize}
                    \item[] \hspace{0pt}\textbf{Case 5.1.} $r^\mathcal{A}_{n + 1} = \false$.
    
                    Hence, since $\delta(s^\mathcal{A}_n, o^\mathcal{A}_{n + 1}) = (s^\mathcal{A}_{n + 1}, r^\mathcal{A}_{n + 1})$, and $o^\mathcal{A}_{n + 1}$ is a \GCASop{}($>$) operation, we have that $s^\mathcal{A}_n \leq val^\mathcal{A}_{n + 1}$ and $s^\mathcal{A}_{n + 1} = s^\mathcal{A}_n$ (i.e., the state of $O$ is the same in $C^\mathcal{A}_n$ and $C^\mathcal{A}_{n + 1}$).
                    Thus, $p_{n + 1}$ did not set any index of $O$ during the $n + 1$th step of $\mathcal{I}^\mathcal{A}_{n + 1}$.

                    We first prove that $\mathcal{S}_{n + 1}(O) = \mathcal{S}_n(O)$.
                    Let $v_j$ be the value of the $j$th index of the state of $O$ in $C^\mathcal{A}_{n + 1}$.
                    Hence, since the state of $O$ is the same in $C^\mathcal{A}_n$ and $C^\mathcal{A}_{n + 1}$, it follows that $v_j$ is the value of the $j$th index of the state of $O$ in $C^\mathcal{A}_n$.
                    Furthermore, since $n$ is a non-negative integer, $n + 1 > 0$, so $n \geq 0$.
                    There are two cases.
    
                    \begin{itemize}
                        \item[] \hspace{0pt}\textbf{Case 5.1.1.} $v_j \in \celluniverse$.
        
                        Hence, since $O = \announceobject$, and $p_{n + 1}$ does not set any index of $O$ during the $n + 1$th step of $\mathcal{I}^\mathcal{A}_{n + 1}$, by 1 of \Cref{def:reduction:watermarks}, $\mathcal{W}_{n + 1}(O, j) = \mathcal{W}_n(O, j)$.
                        Therefore, since the state of $O$ is the same in $C^\mathcal{A}_n$ and $C^\mathcal{A}_{n + 1}$, by \Cref{def:reduction:swapper}, $\mathcal{S}_{n + 1}(O, j) = \mathcal{S}_n(O, j)$.
    
                        \item[] \hspace{0pt}\textbf{Case 5.1.2.} $v_j \notin \celluniverse$.
    
                        Hence, by \Cref{def:reduction:watermarks}, $\mathcal{W}_{n + 1}(O, j) = \bot$.
                        We now show that $\mathcal{W}_{n}(O, j) = \bot$.
                        Recall that $n \geq 0$.
                        If $n = 0$, then by \Cref{def:reduction:watermarks}, $\mathcal{W}_{n}(O, j) = \bot$, as wanted, so suppose $n > 0$.
                        Hence, since by assumption $v_j \notin \celluniverse$, and $v_j$ is the value of the $j$th index of the state of $O$ in $C^\mathcal{A}_n$, by \Cref{def:reduction:watermarks}, $\mathcal{W}_{n}(O, j) = \bot$.
                        Therefore, since in all cases $\mathcal{W}_{n + 1}(O, j) = \mathcal{W}_{n}(O, j)$, and the state of $O$ is the same in $C^\mathcal{A}_n$ and $C^\mathcal{A}_{n + 1}$, by \Cref{def:reduction:swapper}, $\mathcal{S}_{n + 1}(O, j) = \mathcal{S}_n(O, j)$.
                    \end{itemize}
                    
                    Since $\mathcal{S}_{n + 1}(O, j) = \mathcal{S}_n(O, j)$ for every index $j$ of $O$, by \Cref{def:reduction:swapper}, $\mathcal{S}_{n + 1}(O) = \mathcal{S}_n(O)$.
                    This completes the proof that $\mathcal{S}_{n + 1}(O) = \mathcal{S}_n(O)$.

                    We now finish the proof of Case 5.1.
                    Since $s^\mathcal{A}_n = \timeshort^\mathcal{A}_n, \myrepositoryoperationshort^\mathcal{A}_n, \cellpointershort^\mathcal{A}_n$, $val^\mathcal{A}_{n + 1} = \timeshort^\mathcal{A}_{n + 1}, \myrepositoryoperationshort^\mathcal{A}_{n + 1}, \cellpointershort^\mathcal{A}_{n + 1}$, and $s^\mathcal{A}_n \leq val^\mathcal{A}_{n + 1}$, we have that $\timeshort^\mathcal{A}_n \leq \timeshort^\mathcal{A}_{n + 1}$.
                    Hence, since $\mathcal{S}_n(O) = \timeshort^\mathcal{A}_n, \myrepositoryoperationshort^\mathcal{A}_n, \mathcal{S}_n(O, 3)$, and $val_{n + 1} = \timeshort^\mathcal{A}_{n + 1}, \myrepositoryoperationshort^\mathcal{A}_{n + 1}, \mathcal{M}(w)$, we have that $\mathcal{S}_n(O) \leq val_{n + 1}$.
                    Thus, $o_{n + 1}$ is unsuccessful.
                    So, $r_{n + 1} = \false$ and $C^\mathcal{B}_{n + 1}$ assigns the same state to $O^*$ as $C_n$.
                    Therefore, since by (*) $C_n$ assigns state $\mathcal{S}_n(O)$ to $O^*$, and $\mathcal{S}_{n + 1}(O) = \mathcal{S}_n(O)$, we have that $C^\mathcal{B}_{n + 1}$ assigns state $\mathcal{S}_{n + 1}(O)$ to $O^*$ and $r_{n + 1} = r^\mathcal{A}_{n + 1}$.
                
                    \item[] \hspace{0pt}\textbf{Case 5.2.} $r^\mathcal{A}_{n + 1} = \true$.
    
                    Hence, since $\delta(s^\mathcal{A}_n, o^\mathcal{A}_{n + 1}) = (s^\mathcal{A}_{n + 1}, r^\mathcal{A}_{n + 1})$, and $o^\mathcal{A}_{n + 1}$ is a \GCASop{}($>$) operation, we have that $s^\mathcal{A}_n > val^\mathcal{A}_{n + 1}$ and $s^\mathcal{A}_{n + 1} = val^\mathcal{A}_{n + 1}$.
                    Hence, since $val^\mathcal{A}_{n + 1} = \timeshort^\mathcal{A}_{n + 1}, \myrepositoryoperationshort^\mathcal{A}_{n + 1}, \cellpointershort^\mathcal{A}_{n + 1}$, we have that $s^\mathcal{A}_{n + 1} = \timeshort^\mathcal{A}_{n + 1}, \myrepositoryoperationshort^\mathcal{A}_{n + 1}, \cellpointershort^\mathcal{A}_{n + 1}$.
                    Thus, since $p_{n + 1}$ set the first and second index of $O$ to $\timeshort^\mathcal{A}_{n + 1}$ and $\myrepositoryoperationshort^\mathcal{A}_{n + 1}$ during the $n + 1$th step of $\mathcal{I}^\mathcal{A}_{n + 1}$ because the first and second index of $\myuniquerepositoryoperationshort{}$ are $\timeshort^\mathcal{A}_{n + 1}$ and $\myrepositoryoperationshort^\mathcal{A}_{n + 1}$ in $C^\mathcal{A}_n$, respectively, by 3 of \Cref{def:reduction:watermarks}, $\mathcal{W}_{n + 1}(O, 1) = \mathcal{W}_n(\myuniquerepositoryoperationshort{}, 1)$ and $\mathcal{W}_{n + 1}(O, 2) =  \mathcal{W}_n(\myuniquerepositoryoperationshort{}, 2)$, or $\mathcal{W}_{n + 1}(O, 1) = \mathcal{W}_{n + 1}(O, 2) = \bot$.
                    So, since $\mathcal{W}_n(\myuniquerepositoryoperationshort{}, 1) = \mathcal{W}_n(\myuniquerepositoryoperationshort{}, 2) = \bot$, in either case, by \Cref{def:reduction:swapper}, $\mathcal{S}_{n + 1}(O, 1) = \timeshort^\mathcal{A}_{n + 1}$ and $\mathcal{S}_{n + 1}(O, 2) = \myrepositoryoperationshort^\mathcal{A}_{n + 1}$.
                    Similarly, since $\cellpointershort^\mathcal{A}_{n + 1}\in \celluniverse$ and $p_{n + 1}$ set the third index of $O$ to $\cellpointershort^\mathcal{A}_{n + 1}$ during the $n + 1$th step of $\mathcal{I}^\mathcal{A}_{n + 1}$ because the first index of $pr$ is assigned to $\cellpointershort^\mathcal{A}_{n + 1}$ in $C^\mathcal{A}_n$, by 3 of \Cref{def:reduction:watermarks}, $\mathcal{W}_{n + 1}(O, 3) = \mathcal{W}_n(pr, 1)$.
                    Hence, since $\mathcal{W}_n(pr, 1) = w \neq \bot$, by \Cref{def:reduction:swapper}, $\mathcal{S}_{n + 1}(O, 3) = \mathcal{M}(w)$.
                    Therefore, $\mathcal{S}_{n + 1}(O) = \timeshort^\mathcal{A}_{n + 1}, \myrepositoryoperationshort^\mathcal{A}_{n + 1}, \mathcal{M}(w)$, and so $val_{n + 1} = \mathcal{S}_{n + 1}(O)$.
                    We now prove that $o_{n + 1}$ sets $O^*$ to $val_{n + 1}$.
                    
                    We first prove that $\timeshort^\mathcal{A}_n > \timeshort^\mathcal{A}_{n + 1}$.
                    Suppose, for contradiction, that $\timeshort^\mathcal{A}_n \leq \timeshort^\mathcal{A}_{n + 1}$.
                    If $\timeshort^\mathcal{A}_n < \timeshort^\mathcal{A}_{n + 1}$, then since $s^\mathcal{A}_n = \timeshort^\mathcal{A}_n, \myrepositoryoperationshort^\mathcal{A}_n, \cellpointershort^\mathcal{A}_n$ and $val_{n + 1} = \timeshort{}^\mathcal{A}_{n + 1}, \myrepositoryoperationshort^\mathcal{A}_{n + 1}, \mathcal{M}(w)$, we have that $s^\mathcal{A}_n < val^\mathcal{A}_{n + 1}$.
                    However, as established above, $s^\mathcal{A}_n > val^\mathcal{A}_{n + 1}$, a contradiction, so $\timeshort^\mathcal{A}_n \geq \timeshort^\mathcal{A}_{n + 1}$.
                    Thus, since by assumption $\timeshort^\mathcal{A}_n \leq \timeshort^\mathcal{A}_{n + 1}$, we have that $\timeshort^\mathcal{A}_n = \timeshort^\mathcal{A}_{n + 1}$.
                    Hence, since $s^\mathcal{A}_n = \timeshort^\mathcal{A}_n, \myrepositoryoperationshort^\mathcal{A}_n, \cellpointershort^\mathcal{A}_n$ and $s^\mathcal{A}_{n + 1} = \timeshort^\mathcal{A}_{n + 1}, \myrepositoryoperationshort^\mathcal{A}_{n + 1}, \cellpointershort^\mathcal{A}_{n + 1}$, by \Cref{lemma:reduction:timestamp_of_announce_object_in_a_determine_state_of_announce_object}, $s^\mathcal{A}_{n + 1} = s^\mathcal{A}_n$.
                    Therefore, since $s^\mathcal{A}_{n + 1} = val^\mathcal{A}_{n + 1}$, we have that $s^\mathcal{A}_{n} = val^\mathcal{A}_{n + 1}$.
                    However, $s^\mathcal{A}_n > val^\mathcal{A}_{n + 1}$, a contradiction.

                    We now finish the proof of Case 5.2.
                    Since $\timeshort^\mathcal{A}_n > \timeshort^\mathcal{A}_{n + 1}$, $\mathcal{S}_n(O) = \timeshort^\mathcal{A}_n, \myrepositoryoperationshort^\mathcal{A}_n, \mathcal{S}_n(O, 3)$, and $val_{n + 1} = \timeshort^\mathcal{A}_{n + 1}, \myrepositoryoperationshort^\mathcal{A}_{n + 1}, \mathcal{M}(w)$, we have that $\mathcal{S}_n(O) > val_{n + 1}$.
                    Hence, $o_{n + 1}$ is successful.
                    Thus, $r_{n + 1} = \true$ and $C^\mathcal{B}_{n + 1}$ assigns state $val_{n + 1}$ to $O^*$.
                    Therefore, since $val_{n + 1} = \mathcal{S}_{n + 1}(O)$, we have that $C^\mathcal{B}_{n + 1}$ assigns state $\mathcal{S}_{n + 1}(O)$ to $O^*$ and $r_{n + 1} = r^\mathcal{A}_{n + 1}$ as wanted.
                    \qH{\Cref{claim:reduction:b_gets_the_right_response}}
                \end{itemize}
            \end{itemize}
        \end{proof}
    
        \begin{claimcustom}{\ref{lemma:reduction:mapped_history_is_for_b}.13}\label{claim:reduction:b_updates_private_registers_correctly}
            Consider any local variable $pr$ other than the program counters in $C^\mathcal{A}_{n + 1}$.
            Then, $C^\mathcal{B}_{n + 1}$ assigns state $\mathcal{S}_{n + 1}(pr)$ to $pr$.
        \end{claimcustom}
    
        \begin{proof}        
            Let $s^\mathcal{A}_n$ (resp. $s^\mathcal{A}_{n + 1}$) be the state assigned to $pr$ in $C^\mathcal{A}_n$ (resp. $C^\mathcal{A}_{n + 1}$) and let $s_n$ (resp. $s_{n + 1}$) be the state assigned to $pr$ in $C_n$ (resp. $C^\mathcal{B}_{n + 1}$).
            Consider any index $i$ of $s^\mathcal{A}_{n + 1}$.
            We will prove that the $i$th index of $s_{n + 1}$ is $\mathcal{S}_{n + 1}(pr, i)$.
            There are two cases.
            
            \begin{itemize}
                \item[] \hspace{0pt}\textbf{Case 1.} the line of code $\ell$ executed by $p_{n + 1}$ during the $n + 1$th step of $\mathcal{I}^\mathcal{A}_{n + 1}$ does not set the $i$th index of $pr$.

                Hence, the $i$th index of $s^\mathcal{A}_n$ and $s^\mathcal{A}_{n + 1}$ are the same.
                Furthermore, since by \Cref{lemma:reduction:same_program_counters_in_mapped_run}, the program counter of $p_{n + 1}$ is the same in $C^\mathcal{A}_n$ and $C_n$, and $p_{n + 1}$ performs the $n + 1$th step of $\mathcal{I}^\mathcal{B}_{n + 1}$, we have that $p_{n + 1}$ executes $\ell$ during the $n + 1$th step of $\mathcal{I}^\mathcal{B}_{n + 1}$, and so the $i$th index of $pr$ is the same in $C_n$ and $C^\mathcal{B}_{n + 1}$.
                So, the $i$th index of $s_n$ and $s_{n + 1}$ are the same.
                Hence, since by \Cref{def:reduction:run_mapping} the $i$th index of $s_n$ is $\mathcal{S}_n(pr, i)$, we have that the $i$th index of $s_{n + 1}$ is $\mathcal{S}_n(pr, i)$.
                Since $p_{n + 1}$ does not set the $i$th index of $pr$ during the $n + 1$th step of $\mathcal{I}^\mathcal{A}_{n + 1}$, we have that 1 of \Cref{def:reduction:watermarks} is satisfied, so  $\mathcal{W}_n(pr, i) = \mathcal{W}_{n + 1}(pr, i)$ if the $i$th index of $pr$ in $C^\mathcal{A}_n$ is in $\celluniverse$ and $\mathcal{W}_{n + 1} = \bot$ if the $i$th index of $pr$ in $C^\mathcal{A}_n$ is not in $\celluniverse$.
                Hence, in the latter case, by \Cref{def:reduction:watermarks}, $\mathcal{W}_{n} = \bot$, and so in all cases $\mathcal{W}_n(pr, i) = \mathcal{W}_{n + 1}(pr, i)$.
                Thus, since the $i$th index of $pr$ is the same in $C^\mathcal{A}_n$ and $C^\mathcal{A}_{n + 1}$, by \Cref{def:reduction:swapper}, $\mathcal{S}_n(pr, i) = \mathcal{S}_{n + 1}(pr, i)$.
                Therefore, since the $i$th index of $s_{n + 1}$ is $\mathcal{S}_n(pr, i)$, we have that the $i$th index of $s_{n + 1}$ is $\mathcal{S}_{n + 1}(pr, i)$ as wanted.
    
                \item[] \hspace{0pt}\textbf{Case 2.} the line of code $\ell$ executed by $p_{n + 1}$ during the $n + 1$th step of $\mathcal{I}^\mathcal{A}_{n + 1}$ sets the $i$th index of $pr$.

                We first deal with the special case of $\ell = \ref{line:ero:invocation_step}$.
                In this case, $pr$ is the local variable $hlo$ of $p_{n + 1}$ and $i = 1$.
                Notice that $pr$ is set ``externally" because $p_{n + 1}$ performs an invocation step during the $n + 1$th step of $\mathcal{I}^\mathcal{A}_{n + 1}$.
                Hence, the contents of $pr$ did not originate from the response of $\allocatecelloperation{}$ operation, and so by \Cref{def:reduction:watermarks}, $\mathcal{W}_{n + 1}(pr, 1) = \bot$.
                Thus, by \Cref{def:reduction:swapper}, $\mathcal{S}_{n + 1}(pr)$ is the value of $pr$ in $C^\mathcal{A}_{n + 1}$.
                Recall from the definition of $C^\mathcal{B}_{n + 1}$ that if $p_{n + 1}$ executes \cref{line:ero:invocation_step} during the $n + 1$th step of $\mathcal{I}^\mathcal{A}_{n + 1}$, then $C^\mathcal{B}_{n + 1}$ is choosen such that $p_{n + 1}$ performs the same invocation step during the $n + 1$th step of $\mathcal{I}^\mathcal{B}_{n + 1}$.
                Hence, the value of $pr$ is the same in $C^\mathcal{A}_{n + 1}$ and $C^\mathcal{B}_{n + 1}$.
                Therefore, since $i = 1$, we have that the $i$th index of $s_{n + 1}$ is $\mathcal{S}_{n + 1}(pr, i)$ as wanted.

                Now suppose that $\ell = \ref{line:ero:apply_op}$.
                Hence, $pr$ is either $\stateshort'$ or $\responseshort{}'$, so $s^\mathcal{A}_{n + 1}$ is a single value and $i = 1$.
                Thus, based on how we defined $C^\mathcal{B}_{n + 1}$, we have that $C^\mathcal{B}_{n + 1}$ assigns state $s^\mathcal{A}_{n + 1}$ to $pr$.
                Since in this case $pr$ does not satisfy the conditions of 2 of \Cref{def:reduction:watermarks}, we have that $\mathcal{W}_{n + 1}(pr, 1) = \bot$.
                Hence, since $C^\mathcal{A}_{n + 1}$ assigns $s^\mathcal{A}_{n + 1}$ to $pr$, by \Cref{def:reduction:swapper} $\mathcal{S}_{n + 1}(pr, 1) = s^\mathcal{A}_{n + 1}$.
                Therefore, since $C^\mathcal{B}_{n + 1}$ assigns state $s^\mathcal{A}_{n + 1}$ to $pr$, we have that $C^\mathcal{B}_{n + 1}$ assigns state $\mathcal{S}_{n + 1}(pr, 1)$ to $pr$ as wanted.
                
                Now suppose $\ell$ is neither \ref{line:ero:invocation_step} nor \ref{line:ero:apply_op}.
                Observe that the $i$th index of $s^\mathcal{A}_{n + 1}$ is either (a) the $j$th index of one of $p_{n + 1}$'s local variables $pr'$ in $C^\mathcal{A}_{n}$ or (b) the $j$th index of the response $p_{n + 1}$ received from an operation $o$ it performed during the $n + 1$th step of $\mathcal{I}^\mathcal{A}_{n + 1}$.
                In Case (b), by \Cref{claim:reduction:steps_in_a_are_not_on_bad_pointers}, it follows that $o$ is an operation on a base object.
                We consider each case separately.

                \begin{itemize}
                    \item[] \hspace{0pt}\textbf{Case (a).}

                    Hence, since by \Cref{lemma:reduction:same_program_counters_in_mapped_run}, the program counter of $p_{n + 1}$ is the same in $C^\mathcal{A}_n$ and $C_n$, and $p_{n + 1}$ performs the $n + 1$th step of $\mathcal{I}^\mathcal{B}_{n + 1}$, the $i$th index of $pr$ in $C^\mathcal{B}_{n + 1}$ is the $j$th index of $pr'$ in $C_{n}$.
                    Thus, the $i$th index of $s_{n + 1}$ is the $j$th index of $pr'$ in $C_{n}$.
                    Let $s$ be the state assigned to $pr'$ in $C^\mathcal{A}_n$.
    
                    We first prove that $\mathcal{S}_{n + 1}(pr, i) = \mathcal{S}_{n}(pr', j)$.
                    Let $v_j$ be the value of the $j$th index of $s$.
                    Hence, since the $i$th index of $s^\mathcal{A}_{n + 1}$ is the $j$th index of $pr'$ in $C^\mathcal{A}_{n}$, we have that the $i$th index of $s^\mathcal{A}_{n + 1}$ is $v_j$.
                    Since $s^\mathcal{A}_{n + 1}$ (resp. $s$) is the state assigned to $pr$ (resp. $pr'$) in $C^\mathcal{A}_{n + 1}$ (resp. $C^\mathcal{A}_n$), and the $i$th (resp. $j$th) index of $s^\mathcal{A}_{n + 1}$ (resp. $s$) is $v_j$, we have that if $v_j \notin \celluniverse$, then by \Cref{def:reduction:watermarks} $\mathcal{W}_{n + 1}(pr, i) = \bot$ (resp. $\mathcal{W}_{n}(pr', j) = \bot$).
                    Otherwise, since $pr$ is a local variable of $p_{n + 1}$ other than its program counter, and $p_{n + 1}$ sets the $i$th index of $pr$ to $v_j$ during the $n + 1$th step of $\mathcal{I}^\mathcal{A}_{n + 1}$ because the $j$th index of $pr'$ is $v_j$ in $C^\mathcal{A}_n$, by 2.3 of \Cref{def:reduction:watermarks}, $\mathcal{W}_{n + 1}(pr, i) = \mathcal{W}_{n}(pr', j)$.
                    So, in all cases, $\mathcal{W}_{n + 1}(pr, i) = \mathcal{W}_{n}(pr', j)$.
                    Therefore, since the $i$th (resp. $j$th) index of $s^\mathcal{A}_{n + 1}$ (resp. $s$) is $v_j$, by \Cref{def:reduction:swapper}, $\mathcal{S}_{n + 1}(pr, i) = \mathcal{S}_n(pr', j)$ as wanted.
    
                    We now finish the proof of Case (a).
                    Since by \Cref{def:reduction:run_mapping} the $j$th index of $pr'$ is $\mathcal{S}_n(pr', j)$ in $C_n$, and the $i$th index of $s_{n + 1}$ is the $j$th index of $pr'$ in $C_n$, we have that the $i$th index of $s_{n + 1}$ is $\mathcal{S}_{n}(pr', j)$.
                    Therefore, since $\mathcal{S}_{n + 1}(pr, i) = \mathcal{S}_{n}(pr', j)$, the $i$th index of $s_{n + 1}$ is $\mathcal{S}_{n + 1}(pr, i)$ as wanted.
                
                    \item[] \hspace{0pt}\textbf{Case (b).}

                    Hence, $p_{n + 1}$ performed an operation $o^\mathcal{A}_{n + 1}$ on a base object $O$ with response $r^\mathcal{A}_{n + 1}$ during the $n + 1$th step of $\mathcal{I}^\mathcal{A}_{n + 1}$ and the $i$th index of $s^\mathcal{A}_{n + 1}$ is the $j$th index of $r^\mathcal{A}_{n + 1}$.
                    Hence, since by \Cref{lemma:reduction:same_program_counters_in_mapped_run}, the program counter of $p_{n + 1}$ is the same in $C^\mathcal{A}_n$ and $C_n$, and $p_{n + 1}$ performs the $n + 1$th step of $\mathcal{I}^\mathcal{B}_{n + 1}$, we have that the $i$th index of $pr$ in $C^\mathcal{B}_{n + 1}$ is the $j$th index of the response $r_{n + 1}$ $p_{n + 1}$ received during the $n + 1$th step of $\mathcal{I}^\mathcal{B}_{n + 1}$.
                    There are three cases.

                    \begin{itemize}
                        \item[] \hspace{0pt}\textbf{Case (b).1.} $O$ is the memory manager.

                        Hence, $o^\mathcal{A}_{n + 1}$ is either an $\allocatecelloperation{}$ or $\freecelloperation{}$ operation.
                        However, since the only line of code that performs a $\freecelloperation{}$ operation is \cref{line:ero:free_cell}, and its response is not stored in any local variable, we have that $o^\mathcal{A}_{n + 1}$ is an $\allocatecelloperation{}$ operation whose response is $v_1$.    
                        Hence, since the only line of code that performs an $\allocatecelloperation{}$ operation is \cref{line:ero:allocate_cell}, we have that $p_{n + 1}$ executed \cref{line:ero:allocate_cell} during the $n + 1$th step of $\mathcal{I}^\mathcal{A}_{n + 1}$.
                        Thus, $pr$ is the local variable $\cellpointershort$ of $p_{n + 1}$ on \cref{line:ero:allocate_cell} and $i = 1$.
                        Since $o^\mathcal{A}_{n + 1}$ is an $\allocatecelloperation{}$ operation whose response is $v_1$, we have that $p_{n + 1}$ sets the $1$st index of $pr$ to $v_1$ during the $n + 1$th step of $\mathcal{I}^\mathcal{A}_{n + 1}$.
                        Hence, $s^\mathcal{A}_{n + 1} = v_1$.
                        Furthermore, by 2.1 of \Cref{def:reduction:watermarks}, $\mathcal{W}_{n + 1}(pr, 1) = n + 1$, and so by \Cref{def:reduction:swapper}, $\mathcal{S}_{n + 1}(pr, 1) = \mathcal{M}(n + 1)$.
                        Since the $n + 1$th step of $\mathcal{I}^\mathcal{A}_{n + 1}$ performs an $\allocatecelloperation{}$ operation, by \Cref{claim:reduction:allocate_response_in_b_is_swapped}, the $n + 1$th step of $\mathcal{I}^\mathcal{B}_{n + 1}$ performs an $\allocatecelloperation{}$ operation whose response is $\mathcal{M}(n + 1)$.
                        Hence, since $pr$ is the local variable $\cellpointershort$ of $p_{n + 1}$ on \cref{line:ero:allocate_cell}, we have that $s_{n + 1} = \mathcal{M}(n + 1)$.
                        Thus, since $\mathcal{S}_{n + 1}(pr, 1) = \mathcal{M}(n + 1)$, we have that $s_{n + 1} = \mathcal{S}_{n + 1}(pr, 1)$.
                        Therefore, since $i = 1$, the $i$th index of $s_{n + 1}$ is $\mathcal{S}_{n + 1}(pr, i)$ as wanted.

                        \item[] \hspace{0pt}\textbf{Case (b).2.} $O$ is not the memory manager and $o^\mathcal{A}_{n + 1}$ is a read operation.

                        We first prove that the $i$th index of $s^\mathcal{A}_{n + 1}$ is the $j$th index of the state $s$ assigned to $O$ in $C^\mathcal{A}_n$.
                        %We first deal with the problem of misbehaving cells.
                        If $O \notin \celluniverse_O$, then $r^\mathcal{A}_{n + 1} = s$.
                        Otherwise, if $O \in \celluniverse_O$, then by \Cref{claim:reduction:never_undefined}, $r^\mathcal{A}_{n + 1} = s$.
                        So, in all cases, $r^\mathcal{A}_{n + 1} = s$.
                        Therefore, since the $i$th index of $s^\mathcal{A}_{n + 1}$ is the $j$th index of $r^\mathcal{A}_{n + 1}$, we have that the $i$th index of $s^\mathcal{A}_{n + 1}$ is the $j$th index of $s$ as wanted.
                        
                        We now prove that $\mathcal{S}_{n + 1}(pr, i) = \mathcal{S}_n(O, j)$.
                        Let $v_j$ be the $j$th index of $s$, so the $i$th index of $s^\mathcal{A}_{n + 1}$ is $v_j$.
                        Since $s^\mathcal{A}_{n + 1}$ (resp. $s$) is the state assigned to $pr$ (resp. $O$) in $C^\mathcal{A}_{n + 1}$ (resp. $C^\mathcal{A}_n$), and the $i$th (resp. $j$th) index of $s^\mathcal{A}_{n + 1}$ (resp. $s$) is $v_j$, we have that if $v_j \notin \celluniverse$, then by \Cref{def:reduction:watermarks} $\mathcal{W}_{n + 1}(pr, i) = \bot$ (resp. $\mathcal{W}_{n}(O, j) = \bot$).
                        Otherwise, since $pr$ is a local variable of $p_{n + 1}$ other than its program counter, and $p_{n + 1}$ sets the $i$th index of $pr$ to $v_j$ during the $n + 1$th step of $\mathcal{I}^\mathcal{A}_{n + 1}$ because $p_{n + 1}$ performs a read operation on $O$ whose $j$th index of its response is $v_j$ which, as we just proved, is also the $j$th index of its state in $C^\mathcal{A}_{n}$, by 2.2 of \Cref{def:reduction:watermarks}, $\mathcal{W}_{n + 1}(pr, i) = \mathcal{W}_{n}(O, j)$.
                        So, in all cases, $\mathcal{W}_{n + 1}(pr, i) = \mathcal{W}_{n}(O, j)$.
                        Therefore, since the $i$th (resp. $j$th) index of $s^\mathcal{A}_{n + 1}$ (resp. $s$) is $v_j$, by \Cref{def:reduction:swapper}, $\mathcal{S}_{n + 1}(pr, i) = \mathcal{S}_n(O, j)$ as wanted.
        
                        We now finish the proof of Case (b).2.
                        Since $o^\mathcal{A}_{n + 1}$ is a read operation on $O$, by \Cref{claim:reduction:b_gets_the_right_response}, $r_{n + 1} = \mathcal{S}_n(O)$.
                        Hence, since the $i$th index of $s_{n + 1}$ is the $j$th index of $r_{n + 1}$, we have that the $i$th index of $s_{n + 1}$ is the $j$th index of $\mathcal{S}_n(O)$, or equivalently $\mathcal{S}_n(O, j)$.
                        Therefore, since $\mathcal{S}_{n + 1}(pr, i) = \mathcal{S}_n(O, j)$, we have the $i$th index of $s_{n + 1}$ is $\mathcal{S}_{n + 1}(pr, i)$ as wanted.

                        \item[] \hspace{0pt}\textbf{Case (b).3.} $O$ is not the memory manager and $o^\mathcal{A}_{n + 1}$ is not a read operation.

                        We first prove that $\mathcal{S}_{n + 1}(pr, i) = v_j$ where $v_j$ is the $j$th index of $r^\mathcal{A}_{n + 1}$.
                        Since the $i$th index of $s^\mathcal{A}_{n + 1}$ is the $j$th index of $r^\mathcal{A}_{n + 1}$, we have that the $i$th index of $s^\mathcal{A}_{n + 1}$ is $v_j$.
                        Furthermore, since $pr$ is a local variable of $p_{n + 1}$ other than its program counter, and $p_{i + 1}$ performs a non-read operation $o^\mathcal{A}_{n + 1}$ on $O$ during the $n + 1$th step of $\mathcal{I}^\mathcal{A}_{n + 1}$, none of the conditions of \Cref{def:reduction:watermarks} are satisfied, so $\mathcal{W}_{n + 1}(pr, i) = \bot$.
                        Therefore, since the $i$th index of $s^\mathcal{A}_{n + 1}$ is $v_j$, by \Cref{def:reduction:swapper}, $\mathcal{S}_{n + 1}(pr, i) = v_j$.
        
                        We now finish the proof of Case (b).3.
                        Since $o^\mathcal{A}_{n + 1}$ is not a read operation, by \Cref{claim:reduction:b_gets_the_right_response}, $r_{n + 1} = r^\mathcal{A}_{n + 1}$.
                        Hence, since the $i$th index of $s_{n + 1}$ is the $j$th index of $r_{n + 1}$, we have that the the $i$th index of $s_{n + 1}$ is the $j$th index of $r^\mathcal{A}_{n + 1}$, or equivalently $v_j$.
                        Therefore, since $\mathcal{S}_{n + 1}(pr, i) = v_j$, we have the $i$th index of $s_{n + 1}$ is $\mathcal{S}_{n + 1}(pr, i)$ as wanted.
                    \end{itemize}
                \end{itemize}
            \end{itemize}

            We now finish the proof of \Cref{claim:reduction:b_updates_private_registers_correctly}.
            Since the $i$th index of $s_{n + 1}$ is $\mathcal{S}_{n + 1}(pr, i)$, by \Cref{def:reduction:swapper}, $s_{n + 1} = \mathcal{S}_{n + 1}(pr)$.
            Therefore, $C^\mathcal{B}_{n + 1}$ assigns $\mathcal{S}_{n + 1}(pr)$ to $pr$ as wanted.
            \qH{\Cref{claim:reduction:b_updates_private_registers_correctly}}
        \end{proof}

        We now prove that the program counters are the same in $C^\mathcal{B}_{n + 1}$ and $C_{n + 1}$.
        We start by dealing with the difficult cases in which a process compares two pointers and updates its program counter based on the comparison's outcome.
    
        \begin{claimcustom}{\ref{lemma:reduction:mapped_history_is_for_b}.14}\label{claim:reduction:b_does_traversal_checks_correctly}
            If $p_{n + 1}$ executes line $\ell$ where $\ell$ is either \ref{line:ero:add_cell_while_loop}, \ref{line:ero:remove_cell_while_loop}, or \ref{line:ero:acquire_loop_until} during the $n + 1$th step of $\mathcal{I}^\mathcal{A}_{n + 1}$ and finds the condition on line $\ell$ to be false, then $p_{n + 1}$ executes line $\ell$ during the $n + 1$th step of $\mathcal{I}^\mathcal{B}_{n + 1}$ and finds the condition on line $\ell$ to be false.
        \end{claimcustom}
    
        \begin{proof}
            Since by \Cref{lemma:reduction:same_program_counters_in_mapped_run} the program counter of $p_{n + 1}$ is the same in $C^\mathcal{A}_n$ and $C_n$, and $p_{n + 1}$ takes the $n + 1$th step of $\mathcal{I}^\mathcal{B}_{n + 1}$, we have that $p_{n + 1}$ executes line $\ell$ during the $n + 1$th step of $\mathcal{I}^\mathcal{B}_{n + 1}$.
            Furthermore, since $p_{n + 1}$ finds the condition on line $\ell$ to be false during the $n + 1$th step of $\mathcal{I}^\mathcal{A}_{n + 1}$, it follows that $p_{n + 1}$ finds $\currentcellpointershort = \cellpointershort_\linearizationobject$ in $C^\mathcal{A}_n$ on \cref{line:ero:add_cell_while_loop} or \cref{line:ero:remove_cell_while_loop}, and $\currentcellpointershort = \targetcellpointershort$ in $C^\mathcal{A}_n$ on \cref{line:ero:acquire_loop_until}.
            For uniformity, we let $\cellpointershort$ be $\cellpointershort_\linearizationobject$ or $\targetcellpointershort$ depending on $\ell$.
            Hence, $\currentcellpointershort = \cellpointershort$ in $C^\mathcal{A}_n$.
            Observe that the state of $\cellpointershort$ and $\currentcellpointershort$ is a single value.
            For $\cellpointershort$, this is because it is the last value of $\linearizationobject$ or $\announceobject$, and for $\currentcellpointershort$, this is because it is either $\&\headobject$, or it is the last value of the $\nextlong$ object of some cell.
            Hence, by \Cref{def:reduction:swapper}, $\mathcal{S}_n(\cellpointershort) = \mathcal{S}_n(\cellpointershort, 1)$ and $\mathcal{S}_n(\currentcellpointershort) = \mathcal{S}_n(\currentcellpointershort, 1)$.
            There are four cases.
            \begin{itemize}
                \item[] \hspace{0pt}\textbf{Case 1.} $\mathcal{W}_n(\cellpointershort, 1) = \bot$ and $\mathcal{W}_n(\currentcellpointershort, 1) = \bot$.
    
                Hence, by \Cref{def:reduction:swapper}, $\mathcal{S}_n(\cellpointershort, 1)$ (resp. $\mathcal{S}_n(\currentcellpointershort, 1)$) is the same as the first value of $\cellpointershort$ (resp. $\currentcellpointershort$) in $C^\mathcal{A}_n$.
                Thus, since $\mathcal{S}_n(\cellpointershort) = \mathcal{S}_n(\cellpointershort, 1)$ (resp. $\mathcal{S}_n(\currentcellpointershort) = \mathcal{S}_n(\currentcellpointershort, 1)$), we have that $\mathcal{S}_n(\cellpointershort)$ (resp. $\mathcal{S}_n(\currentcellpointershort)$) is the same as the value of $\cellpointershort$ (resp. $\currentcellpointershort$) in $C^\mathcal{A}_n$.
                So, by \Cref{def:reduction:run_mapping}, the value of $\cellpointershort$ (resp. $\currentcellpointershort$) is the same in $C^\mathcal{A}_n$ and $C_n$.
                Therefore, since $\currentcellpointershort = \cellpointershort$ in $C^\mathcal{A}_n$, we have that $\currentcellpointershort = \cellpointershort$ in $C_n$, and so $p_{n + 1}$ finds the condition on line $\ell$ to be false during the $n + 1$th step of $\mathcal{I}^\mathcal{B}_{n + 1}$ as wanted.
    
                \item[] \hspace{0pt}\textbf{Case 2.} $\mathcal{W}_n(\cellpointershort, 1) \neq \bot$ and $\mathcal{W}_n(\currentcellpointershort, 1) = \bot$.
    
                Hence, since $\mathcal{S}_n(\cellpointershort) = \mathcal{S}_n(\cellpointershort, 1)$, by \Cref{def:reduction:watermarks}, the value of $\cellpointershort$ in $C^\mathcal{A}_n$ is in $\celluniverse$.
                Thus, since $\currentcellpointershort = \cellpointershort$ in $C^\mathcal{A}_n$, we have that the value of $\currentcellpointershort$ in $C^\mathcal{A}_n$ is in $\celluniverse$; say $v$.
                So, by \Cref{def:reduction:used_pointers_in_run_of_a}, $v \in \celluniverse(\mathcal{I}^\mathcal{A}_{n + 1})$.
                Furthermore, since $\mathcal{S}_n(\currentcellpointershort) = \mathcal{S}_n(\currentcellpointershort, 1)$ and $\mathcal{W}_n(\currentcellpointershort, 1) = \bot$, by \Cref{def:reduction:run_mapping}, the value of $\currentcellpointershort$ in $C_n$ is $v$.
                Since $v \in \celluniverse$, by \Cref{assumption:ero:head_and_null_not_in_cell_universe}, $v \neq \&\headobject$, and so since $\currentcellpointershort$ is initialized to $\&\headobject$ on either \cref{line:ero:add_cell_initial_current_pointer}, \cref{line:ero:remove_cell_initialize_pointers}, or \cref{line:ero:acquire_initial_current_pointer} depending on $\ell$ in $\mathcal{I}^\mathcal{B}_{n + 1}$, we have that $p_{n + 1}$ set $\currentcellpointershort$ to $v$ on either \cref{line:ero:add_cell_update_current_pointer}, \cref{line:ero:remove_cell_update_pointers}, or \cref{line:ero:acquire_update_current_unique_pointer} depending on $\ell$ in $\mathcal{I}^\mathcal{B}_{n + 1}$.
                Thus, $p_{n + 1}$ performed a list-acquire-next attempt $a$ for $v$ in $\mathcal{I}^\mathcal{B}_{n + 1}$.
                Let $\mathcal{I}^\mathcal{B}_a$ be the prefix of $\mathcal{I}^\mathcal{B}_{n + 1}$ up to and including $a$.
                Hence, by \Cref{lemma:ero:acquisition_for_implies_active_for}, $v$ is active in $\mathcal{I}^\mathcal{B}_a$, and so by \Cref{lemma:ero:l_add_event_in_active_prefix}, there is an $L$-event $e$ for $v$ in $\mathcal{I}^\mathcal{B}_a$ (and thus $\mathcal{I}^\mathcal{B}_{n + 1}$).
                Thus, by \Cref{lemma:ero:l_event_corresponding_a_event_type_and_pointer_relations}, there is an $A$-event $e'$ for $v$ before $e$ in $\mathcal{I}^\mathcal{B}_{n + 1}$.
                Hence, the process $p$ that executed $e'$ did so during an invocation of the \doworkuntildone{} procedure with a second parameter of $v$.
                Thus, $p$ performed an $\allocatecelloperation{}$ operation on \cref{line:ero:allocate_cell} whose response is $v$ in $\mathcal{I}^\mathcal{B}_{n + 1}$.
                Therefore, since $v \in \celluniverse(\mathcal{I}^\mathcal{A}_{n + 1})$, there is an $\allocatecelloperation{}$ operation in $\mathcal{I}^\mathcal{B}_{n + 1}$ whose response is in $\celluniverse(\mathcal{I}^\mathcal{A}_{n + 1})$.
                However, by \Cref{claim:reduction:no_allocate_for_used_pointer_in_b}, the response of every $\allocatecelloperation{}$ operation in $\mathcal{I}^\mathcal{B}_{n + 1}$ is not in $\celluniverse(\mathcal{I}^\mathcal{A}_{n + 1})$, so Case 2 is impossible.

                \item[] \hspace{0pt}\textbf{Case 3.} $\mathcal{W}_n(\cellpointershort, 1) = \bot$ and $\mathcal{W}_n(\currentcellpointershort, 1) \neq \bot$.
    
                Hence, since $\mathcal{S}_n(\currentcellpointershort) = \mathcal{S}_n(\currentcellpointershort, 1)$, the value of $\currentcellpointershort$ in $C^\mathcal{A}_n$ is in $\celluniverse$.
                Thus, since $\currentcellpointershort = \cellpointershort$ in $C^\mathcal{A}_n$, we have that the value of $\cellpointershort$ in $C^\mathcal{A}_n$ is in $\celluniverse$; say $v$.
                So, by \Cref{def:reduction:used_pointers_in_run_of_a}, $v \in \celluniverse(\mathcal{I}^\mathcal{A}_{n + 1})$.
                Furthermore, since $\mathcal{S}_n(\cellpointershort) = \mathcal{S}_n(\cellpointershort, 1)$ and $\mathcal{W}_n(\cellpointershort, 1) = \bot$, by \Cref{def:reduction:run_mapping}, the value of $\cellpointershort$ is $v$ in $C_n$.
                First consider the case where $\ell$ is either \ref{line:ero:add_cell_while_loop}, \ref{line:ero:remove_cell_while_loop}, or \ref{line:ero:acquire_loop_until} during an invocation of the Acquire procedure invoked on \cref{line:ero:set_response_acquire}.
                In these cases, $p_{n + 1}$ read $v$ from $\linearizationobject.\cellpointerlong$.
                Since $v \in \celluniverse$, by \Cref{assumption:ero:head_and_null_not_in_cell_universe}, $v \neq \nullconstant$.
                Hence, since $\linearizationobject.\cellpointerlong$ is initially $\nullconstant$, we have that $\linearizationobject.\cellpointerlong$ was set to $v$.
                Thus, by \Cref{def:ero:english}, there is an $L$-event for $v$ in $\mathcal{I}^\mathcal{B}_{n + 1}$.
                Therefore, by \Cref{lemma:ero:l_event_corresponding_a_event_type_and_pointer_relations}, there is an $A$-event for $v$ in $\mathcal{I}^\mathcal{B}_{n + 1}$.
                Now consider the case where $\ell$ is \ref{line:ero:acquire_loop_until} during an invocation of the Acquire procedure invoked on \cref{line:ero:announce_acquire}.
                In this case, $p_{n + 1}$ read $v$ from $\announceobject.\cellpointerlong$.
                Since $v \neq \nullconstant$, and $\announceobject.\cellpointerlong$ is initially $\nullconstant$, we have that $\announceobject.\cellpointerlong$ was set to $v$.
                Thus, by \Cref{def:ero:english}, there is an $A$-event for $v$ in $\mathcal{I}^\mathcal{B}_{n + 1}$.
                Therefore, in all cases, there is an $A$-event for $v$ in $\mathcal{I}^\mathcal{B}_{n + 1}$.
                Let $p$ be the process that executed this $A$-event for $v$.
                Hence, $p$ did so during an invocation of the \doworkuntildone{} procedure with a second parameter of $v$.
                Thus, $p$ performed an $\allocatecelloperation{}$ operation on \cref{line:ero:allocate_cell} whose response is $v$.
                Therefore, since $v \in \celluniverse(\mathcal{I}^\mathcal{A}_{n + 1})$, we have that there is an $\allocatecelloperation{}$ operation in $\mathcal{I}^\mathcal{B}_{n + 1}$ whose response is in $\celluniverse(\mathcal{I}^\mathcal{A}_{n + 1})$.
                However, by \Cref{claim:reduction:no_allocate_for_used_pointer_in_b}, the response of every $\allocatecelloperation{}$ operation in $\mathcal{I}^\mathcal{B}_{n + 1}$ is not in $\celluniverse(\mathcal{I}^\mathcal{A}_{n + 1})$, so Case 3 is impossible.
                
                \item[] \hspace{0pt}\textbf{Case 4.} $\mathcal{W}_n(\cellpointershort, 1) \neq \bot$ and $\mathcal{W}_n(\currentcellpointershort, 1) \neq \bot$.
    
                Let $\mathcal{W}_n(\cellpointershort, 1) = w \neq \bot$ and $\mathcal{W}_n(\currentcellpointershort, 1) = w' \neq \bot$.
                Hence, by \Cref{def:reduction:swapper}, $\mathcal{S}_n(\cellpointershort, 1) = \mathcal{M}(w)$ and $\mathcal{S}_n(\currentcellpointershort, 1) = \mathcal{M}(w')$.
                Therefore, since $\mathcal{S}_n(\cellpointershort) = \mathcal{S}_n(\cellpointershort, 1)$ and $\mathcal{S}_n(\currentcellpointershort) = \mathcal{S}_n(\currentcellpointershort, 1)$, we have that $\mathcal{S}_n(\cellpointershort) = \mathcal{M}(w)$ and $\mathcal{S}_n(\currentcellpointershort) = \mathcal{M}(w')$.
                So, by \Cref{def:reduction:run_mapping}, $\cellpointershort$ (resp. $\currentcellpointershort$) is assigned to $\mathcal{M}(w)$ (resp. $\mathcal{M}(w')$) in $C_n$.
                Hence, since $p_{n + 1}$ checks whether $\currentcellpointershort \neq \cellpointershort$ during the $n + 1$th step of $\mathcal{I}^\mathcal{A}_{n + 1}$ and $p_{n + 1}$ executes line $\ell$ during the $n + 1$th step of $\mathcal{I}^\mathcal{B}_{n + 1}$, we have that $p_{n + 1}$ checks whether $\currentcellpointershort \neq \cellpointershort$ (and thus whether $\mathcal{M}(w) \neq \mathcal{M}(w')$) during the $n + 1$th step of $\mathcal{I}^\mathcal{B}_{n + 1}$.
                Thus, if $w = w'$, then $\mathcal{M}(w) = \mathcal{M}(w')$, so it follows that $p_{n + 1}$ finds the condition on line $\ell$ to be false during the $n + 1$th step of $\mathcal{I}^\mathcal{B}_{n + 1}$.
                Therefore, it suffices to consider the case where $w \neq w'$.
                
                Since $\mathcal{W}_n(\cellpointershort, 1) = w \neq \bot$ (resp. $\mathcal{W}_n(\currentcellpointershort, 1) = w' \neq \bot$), by \Cref{lemma:reduction:associated_pointer}, the $w$th (resp. $w'$th) step of $\mathcal{I}^\mathcal{A}_{n + 1}$ performs an $\allocatecelloperation{}$ operation whose response is the value of $\cellpointershort$ (resp. $\currentcellpointershort$) in $C^\mathcal{A}_n$.
                Hence, since $\currentcellpointershort = \cellpointershort$ in $C^\mathcal{A}_n$, we have that this is the same value, say $v$, and so the $w$th and $w'$th step of $\mathcal{I}^\mathcal{A}_{n + 1}$ perform an $\allocatecelloperation{}$ operation whose response is $v$.
                Furthermore, by \Cref{claim:reduction:allocate_response_in_b_is_swapped}, the $w$th (resp. $w'$th) step of $\mathcal{I}^\mathcal{B}_{n + 1}$ performs an $\allocatecelloperation{}$ operation whose response is $\mathcal{M}(w)$ (resp. $\mathcal{M}(w')$). 
                Since $w \neq w'$, there are two cases.
    
                \begin{itemize}
                    \item[] \hspace{0pt}\textbf{Case 4.1.} $w' < w$.
    
                    Hence, since the $w'$th and $w$th step of $\mathcal{I}^\mathcal{A}_{n + 1}$ perform an $\allocatecelloperation{}$ operation whose response is $v$, by \Cref{alg:cell_manager_specification}, $v \in \celluniverse$ and for some $w' < j < w$ the $j$th step of $\mathcal{I}^\mathcal{A}_{n + 1}$ performs a $\freecelloperation{}(v)$ operation.
                    Thus, by \Cref{claim:reduction:allocate_followed_by_free_in_a_implies_free_for_same_watermark}, for some $w' < k \leq j$ the $k$th step of $\mathcal{I}^\mathcal{A}_{n + 1}$ performs a $\freecelloperation{}(v)$, and $\mathcal{W}_{k - 1}(O_k, 1) = w'$ where $O_k$ is the local variable $\currentcellpointershort$ of $p_k$ in the Relinquish procedure.
                    So, by \Cref{claim:reduction:frees_in_b_are_swapped}, the $k$th step of $\mathcal{I}^\mathcal{B}_{n + 1}$ performs a $\freecelloperation{}(\mathcal{M}(w'))$ operation.
                    Hence, by \Cref{lemma:ero:free_cell_operations_are_preceeded_by_successful_list_remove_attempt}, there is a successful list-remove attempt $a_{rmv}$ for $\mathcal{M}(w')$ before the $k$th step of $\mathcal{I}^\mathcal{B}_{n + 1}$.
                    Therefore, since $k \leq j$ and $j < w$, by transitivity, $a_{rmv} < w$ (*).

                    Since $\ell$ is either \ref{line:ero:add_cell_while_loop}, \ref{line:ero:remove_cell_while_loop}, or \ref{line:ero:acquire_loop_until}, $p_{n + 1}$ performs the $n + 1$th step of $\mathcal{I}^\mathcal{B}_{n + 1}$ in the context of an invocation $I$ of either a \doaddcell{}, \doremovecell{}, or Acquire procedure.
                    Hence, since $p_{n + 1}$ checks whether $\currentcellpointershort \neq \cellpointershort$ during the $n + 1$th step of $\mathcal{I}^\mathcal{B}_{n + 1}$, and $\cellpointershort$ is assigned to $\mathcal{M}(w)$ in $C_n$, we have that $p_{n + 1}$ read $\mathcal{M}(w)$ from either $\linearizationobject.\cellpointerlong$ or $\announceobject.\cellpointerlong$ before invoking $I$.
                    We now prove that $I$ was invoked after the $w$th step of $\mathcal{I}^\mathcal{B}_{n + 1}$ (**).
                    We start by proving that there is an $A$-event for $\mathcal{M}(w)$ before $p_{n + 1}$ invoked $I$.
                    Since $\mathcal{M}(w) \in \celluniverse$, by \Cref{assumption:ero:head_and_null_not_in_cell_universe}, $\mathcal{M}(w) \neq \nullconstant$.
                    First suppose $p_{n + 1}$ read $\mathcal{M}(w)$ from $\linearizationobject.\cellpointerlong$ before invoking $I$.
                    Hence, since $\linearizationobject.\cellpointerlong$ is initially $\nullconstant$, and $\mathcal{M}(w) \neq \nullconstant$, we have that $\linearizationobject.\cellpointerlong$ was set to $\mathcal{M}(w)$ before $I$ was invoked.
                    Thus, by \Cref{def:ero:english}, there is an $L$-event for $\mathcal{M}(w)$ before $T^{\ref{line:ero:linearization_read}}$.
                    Therefore, by \Cref{lemma:ero:l_event_corresponding_a_event_type_and_pointer_relations}, there is an $A$-event for $\mathcal{M}(w)$ before $I$ was invoked as wanted.
                    Now suppose that $p_{n + 1}$ read $\mathcal{M}(w)$ from $\announceobject.\cellpointerlong$ before invoking $I$.
                    Hence, since $\announceobject.\cellpointerlong$ is initially $\nullconstant$, and $\mathcal{M}(w) \neq \nullconstant$, we have that $\announceobject.\cellpointerlong$ was set to $\mathcal{M}(w)$ before $I$ was invoked.
                    Therefore, by \Cref{def:ero:english}, there is an $A$-event for $\mathcal{M}(w)$ before $I$ was invoked as wanted.
                    Let $e$ be this $A$-event for $\mathcal{M}(w)$ before $p_{n + 1}$ invoked $I$.
                    By \Cref{def:ero:english}, the process $q$ that executed $e$ did so during an invocation of the \doworkuntildone{} procedure whose second parameter is $\mathcal{M}(w)$.
                    Hence, $q$ performed an $\allocatecelloperation{}$ operation whose response is $\mathcal{M}(w)$ before executing $e$.
                    Since by \Cref{alg:lazy_cell_manager_specification} only a single $\allocatecelloperation{}$ operation has response $\mathcal{M}(w)$, and the $w$th step of $\mathcal{I}^\mathcal{B}_{n + 1}$ performs an $\allocatecelloperation{}$ operation has response $\mathcal{M}(w)$, we have that $w < e$.
                    Hence, since $p_{n + 1}$ invoked $I$ after $e$, we have that $p_{n + 1}$ invoked $I$ after the $w$th step of $\mathcal{I}^\mathcal{B}_{n + 1}$.

                    We now finish the proof of Case 4.1.
                    Since $\mathcal{M}(w') \in \celluniverse$, by \Cref{assumption:ero:head_and_null_not_in_cell_universe}, $\mathcal{M}(w') \neq \&\headobject$.
                    Hence, since $p_{n + 1}$ initializes $\currentcellpointershort$ to $\&\headobject$ on \cref{line:ero:add_cell_initial_current_pointer}, \ref{line:ero:remove_cell_initialize_pointers}, or \ref{line:ero:acquire_initial_current_pointer} depending on $\ell$ during $I$, and finds that $\currentcellpointershort = \mathcal{M}(w')$ on \cref{line:ero:add_cell_while_loop}, \cref{line:ero:remove_cell_while_loop}, or \cref{line:ero:acquire_loop_until} depending on $\ell$ during $I$, we have that $p_{n + 1}$ set $\currentcellpointershort = \mathcal{M}(w')$ on either \cref{line:ero:add_cell_update_current_pointer}, \cref{line:ero:remove_cell_update_pointers}, or \cref{line:ero:acquire_update_current_unique_pointer} depending on $\ell$ during $I$.
                    Hence, $p_{n + 1}$ received response $(\found, \mathcal{M}(w'))$ from an invocation of the AcquireNext procedure during $I$.
                    Thus, $p_{n + 1}$ performed a successful list-acquire-next attempt $a_{acq}$ for $\mathcal{M}(w')$ during $I$.
                    Let $\mathcal{I}^\mathcal{B}_{acq}$ be the prefix of $\mathcal{I}^\mathcal{B}_{n + 1}$ up to and including $a_{acq}$.
                    Since the last step of $\mathcal{I}^\mathcal{B}_{acq}$ is a successful list-acquire-next attempt for $\mathcal{M}(w')$, by \Cref{lemma:ero:acquisition_for_implies_active_for}, $\mathcal{M}(w')$ is active in $\mathcal{I}^\mathcal{B}_{acq}$.
                    Hence, by \Cref{def:ero:active}, there are no successful list-remove attempts for $\mathcal{M}(w')$ in $\mathcal{I}^\mathcal{B}_{acq}$, or equivalently, before $a_{acq}$ in $\mathcal{I}^\mathcal{B}_{n + 1}$.
                    Therefore, since $p_{n + 1}$ executed $a_{acq}$ during $I$, and by (**) $p_{n + 1}$ invoked $I$ after the $w$th step of $\mathcal{I}^\mathcal{B}_{n + 1}$, we have that $w < a_{acq}$, so there are no successful list-remove attempts for $\mathcal{M}(w')$ before the $w$th step of $\mathcal{I}^\mathcal{B}_{n + 1}$.
                    However, by (*) $a_{rmv} < w$, so there is a successful list-remove attempt for $\mathcal{M}(w')$ before the $w$th step of $\mathcal{I}^\mathcal{B}_{n + 1}$, a contradiction, so Case 4.1 is impossible.
                
                    \item[] \hspace{0pt}\textbf{Case 4.2.} $w < w'$.
    
                    Hence, since the $w$th and $w'$th step of $\mathcal{I}^\mathcal{A}_{n + 1}$ perform an $\allocatecelloperation{}$ operation whose response is $v$, by \Cref{alg:cell_manager_specification}, $v \in \celluniverse$ and for some $w < j < w'$ the $j$th step of $\mathcal{I}^\mathcal{A}_{n + 1}$ performs a $\freecelloperation{}(v)$ operation.
                    Thus, by \Cref{claim:reduction:allocate_followed_by_free_in_a_implies_free_for_same_watermark}, for some $w < k \leq j$ the $k$th step of $\mathcal{I}^\mathcal{A}_{n + 1}$ performs a $\freecelloperation{}(v)$, and $\mathcal{W}_{k - 1}(O_k, 1) = w$ where $O_k$ is the local variable $\currentcellpointershort$ of $p_k$ in the Relinquish procedure.
                    Therefore, by \Cref{claim:reduction:frees_in_b_are_swapped}, the $k$th step of $\mathcal{I}^\mathcal{B}_{n + 1}$ performs a $\freecelloperation{}(\mathcal{M}(w))$ operation.
                    Since $\ell$ is either \ref{line:ero:add_cell_while_loop}, \ref{line:ero:remove_cell_while_loop}, or \ref{line:ero:acquire_loop_until}, $p_{n + 1}$ performs the $n + 1$ step of $\mathcal{I}^\mathcal{B}_{n + 1}$ in the context of an invocation $I$ of either a \doaddcell{}, \doremovecell{}, or Acquire procedure.
                    Let $v'$ be the value $p_{n + 1}$ read from $\linearizationobject$ on its last execution of \cref{line:ero:linearization_read} before invoking $I$; say at time $T^{\ref{line:ero:linearization_read}}$.
                    There are two cases.
    
                    \begin{itemize}
                        \item[] \hspace{0pt}\textbf{Case 4.2.1.} $v'$ is the initial value of $\linearizationobject$, i.e., $v' = ((0, \noop), \nullconstant)$.
    
                        Hence, $p_{n + 1}$ does not find the condition on \cref{line:ero:do_add_cell_condition}, \cref{line:ero:do_remove_cell_condition}, or \cref{line:ero:do_apply_and_copy_response_condition} to be true on its next execution of these lines after $T^{\ref{line:ero:linearization_read}}$, and so $I$ is not an invocation of either \doaddcell{} or \doremovecell{}, or the Acquire procedure on \cref{line:ero:set_response_acquire}.
                        Thus, $I$ must be an invocation of the Acquire procedure on \cref{line:ero:announce_acquire}, and so $\ell = \ref{line:ero:acquire_loop_until}$.
                        Since $p_{n + 1}$ read $v'$ from $\linearizationobject$ on its last execution of \cref{line:ero:linearization_read} before invoking $I$, we have that the first parameter of $I$ is $(0, \noop)$.
                        Furthermore, by \Cref{def:reduction:swapper} $\mathcal{M}(w') \in \celluniverse$, so by \Cref{assumption:ero:head_and_null_not_in_cell_universe}, $\mathcal{M}(w') \neq \&\headobject$ and $\mathcal{M}(w') \neq \nullconstant$.
                        Hence, since $p_{n + 1}$ initializes $\currentcellpointershort$ to $\&\headobject$ on \cref{line:ero:acquire_initial_current_pointer} during $I$ and finds that $\currentcellpointershort = \mathcal{M}(w')$ on \cref{line:ero:acquire_loop_until} during $I$, we have that $p_{n + 1}$ set $\currentcellpointershort = \mathcal{M}(w')$ on \cref{line:ero:acquire_update_current_unique_pointer} during $I$.
                        Thus, $p_{n + 1}$ received response $(\found, \mathcal{M}(w'))$ from an invocation $I'$ of the AcquireNext procedure during $I$.
                        Let $\previouscellpointershort$ be the second parameter of $I'$, so by \Cref{lemma:ero:acquire_next_is_for_pointer_from_universe_or_head}, $\previouscellpointershort \in \celluniverse \cup \{\&\headobject\}$.
                        Since the first parameter of $I$ is $(0, \noop)$ and $p_{n + 1}$ invoked $I'$ during $I$, we have that the first parameter of $I'$ is $(0, \noop)$.
                        Hence, the parameters of $I'$ are exactly $((0, \noop), \previouscellpointershort)$.
                        Since the response of $I'$ is $(\found, \mathcal{M}(w'))$, we have that (a) $p_{n + 1}$ read $(*\previouscellpointershort).\nextlong = (\arbitraryvalue, \arbitraryvalue, \arbitraryvalue, \mathcal{M}(w'))$ on \cref{line:ero:acquire_next_read_curr_unique_pointer}; say at time $T^{\ref{line:ero:acquire_next_read_curr_unique_pointer}}$ and (b) $p_{n + 1}$ found that $\linearizationobject{}.\uniquerepositoryoperationlong{} = (0, \noop)$ on its next execution of \cref{line:ero:acquire_next_linearization_changed_check} during $I'$; say at time $T^{\ref{line:ero:acquire_next_linearization_changed_check}}$.
                        Hence, since $\previouscellpointershort \in \celluniverse \cup \{\&\headobject\}$, $(*\previouscellpointershort).\nextlong$ is initially $(\arbitraryvalue, \arbitraryvalue, \arbitraryvalue, \nullconstant)$, and so since $\mathcal{M}(w') \neq \nullconstant$ and $(*\previouscellpointershort).\nextlong = (\arbitraryvalue, \arbitraryvalue, \arbitraryvalue, \mathcal{M}(w'))$ at $T^{\ref{line:ero:acquire_next_read_curr_unique_pointer}}$, we have that $(*\previouscellpointershort).\nextlong$ was set to $(\arbitraryvalue, \arbitraryvalue, \arbitraryvalue, \mathcal{M}(w'))$ before $T^{\ref{line:ero:acquire_next_read_curr_unique_pointer}}$.
                        Thus, by \Cref{observation:ero:where_objects_change}, there is either a successful list-add or list-remove attempt before $T^{\ref{line:ero:acquire_next_read_curr_unique_pointer}}$, and so by \Cref{lemma:ero:l_event_corresponding_to_do_low_level_op}, there is a successful $L$-event before $T^{\ref{line:ero:acquire_next_read_curr_unique_pointer}}$.
                        Therefore, since $T^{\ref{line:ero:acquire_next_read_curr_unique_pointer}} < T^{\ref{line:ero:acquire_next_linearization_changed_check}}$, we have that is an $L$-event before $T^{\ref{line:ero:acquire_next_linearization_changed_check}}$.
                        However, since $\linearizationobject{}.\uniquerepositoryoperationlong{} = (0, \noop)$ at $T^{\ref{line:ero:acquire_next_linearization_changed_check}}$ and by \Cref{lemma:ero:l_events_set_llo_to_a_value_different_from_initial} every $L$-event sets $\linearizationobject{}.\uniquerepositoryoperationlong \neq (0, \noop)$, we have that there are no $L$-events before $T^{\ref{line:ero:acquire_next_linearization_changed_check}}$, a contradiction, so Case 4.2.1 is impossible.
    
                        \item[] \hspace{0pt}\textbf{Case 4.2.2.} $v'$ is not the initial value of $\linearizationobject$, i.e., $v' \neq ((0, \noop), \nullconstant)$.
    
                        Hence, $\linearizationobject$ was set to $v'$ before $T^{\ref{line:ero:linearization_read}}$, and so by \Cref{def:ero:english}, some $L$-event $e < T^{\ref{line:ero:linearization_read}}$ set $\linearizationobject = v'$.
    
                        \textbf{Claim:} There are no $A$-events for $\mathcal{M}(w')$ before $e$ in $\mathcal{I}^\mathcal{B}_{n + 1}$.
                        
                        We first prove that every $A$-event for $\mathcal{M}(w')$ is after the $w'$th step of $\mathcal{I}^\mathcal{B}_{n + 1}$ (*) .
                        Suppose, for contradiction, there is an $A$-event $e'$ for $\mathcal{M}(w')$ before or at the $w'$th step of $\mathcal{I}^\mathcal{B}_{n + 1}$.
                        Let $p$ be the process that executed $e'$.
                        Hence, by \Cref{def:ero:english}, $p$ executed $e'$ during an invocation of the \doworkuntildone{} procedure with a second parameter of $\mathcal{M}(w')$.
                        Thus, before invoking this procedure, and thus before executing $e'$, $p$ performed an $\allocatecelloperation{}$ operation whose response is $\mathcal{M}(w')$.
                        Therefore, since the $w'$th step of $\mathcal{I}^\mathcal{B}_{n + 1}$ performs an $\allocatecelloperation{}$ operation whose response is $\mathcal{M}(w')$, we have that there are two \allocatecelloperation{} operations in $\mathcal{I}^\mathcal{B}_{n + 1}$ whose response is $\mathcal{M}(w')$.
                        However, by \Cref{alg:lazy_cell_manager_specification}, the response of every $\allocatecelloperation{}$ is unique, a contradiction.
    
                        The rest of the proof of \textbf{Claim} is done in two cases.
    
                        \textbf{Case A.} $\ell$ is either \ref{line:ero:add_cell_while_loop}, \ref{line:ero:remove_cell_while_loop}, or \ref{line:ero:acquire_loop_until} during an invocation of the Acquire procedure invoked on \cref{line:ero:set_response_acquire}.
                        
                        Hence, since $v'$ is the value $p_{n + 1}$ read from $\linearizationobject$ on \cref{line:ero:linearization_read} at $T^{\ref{line:ero:linearization_read}}$ and $\cellpointershort$ is assigned to $\mathcal{M}(w)$ in $C_n$, it follows that $v' = (\arbitraryvalue, \mathcal{M}(w))$, and so by \Cref{def:ero:english}, $e$ is an $L$-event for $\mathcal{M}(w)$.
                        Since $\mathcal{I}^\mathcal{B}_{n + 1}$ is an implementation history of $\mathcal{B}$, by \Cref{lemma:ero:the_list_invariants_hold}, $P(\mathcal{I}^\mathcal{B}_{n + 1})$ holds.
                        Hence, since $\mathcal{M}(w) \in \celluniverse$, by $P(\mathcal{I}^\mathcal{B}_{n + 1})$, there are at-most three $L$-events for $\mathcal{M}(w)$ in $\mathcal{I}^\mathcal{B}_{n + 1}$.
                        
                        We now prove that there are three $L$-events for $\mathcal{M}(w)$ before the $k$th step of $\mathcal{I}^\mathcal{B}_{n + 1}$.
                        Recall that the $k$th step of $\mathcal{I}^\mathcal{B}_{n + 1}$ performs a $\freecelloperation{}(\mathcal{M}(w))$ operation.
                        Hence, by \Cref{lemma:ero:free_cell_operations_are_preceeded_by_remove_event}, there is an $L$-remove event for $\mathcal{M}(w)$ before the $k$th step of $\mathcal{I}^\mathcal{B}_{n + 1}$.
                        Thus, by \Cref{lemma:ero:remove_events_are_preceeded_by_apply_events} and \Cref{lemma:ero:remove_events_are_preceeded_by_add_events}, there is an $L$-apply event for $\mathcal{M}(w)$ and an $L$-add event for $\mathcal{M}(w)$ before the $k$th step of $\mathcal{I}^\mathcal{B}_{n + 1}$.
    
                        We now return to the proof of Case A.
                        Since $e$ is an $L$-event for $\mathcal{M}(w)$ in $\mathcal{I}^\mathcal{B}_{n + 1}$, there are three $L$-events for $\mathcal{M}(w)$ before the $k$th step of $\mathcal{I}^\mathcal{B}_{n + 1}$, and there are at-most three $L$-events for $\mathcal{M}(w)$ in $\mathcal{I}^\mathcal{B}_{n + 1}$, we have that $e$ is before the $k$th step of $\mathcal{I}^\mathcal{B}_{n + 1}$.
                        Hence, since $k < w'$, we have that $e$ is before the $w'$th step of $\mathcal{I}^\mathcal{B}_{n + 1}$.
                        Therefore, since by (*) every $A$-event for $\mathcal{M}(w')$ is after the $w'$th step of $\mathcal{I}^\mathcal{B}_{n + 1}$, we have that there are no $A$-events for $\mathcal{M}(w')$ before $e$ in $\mathcal{I}^\mathcal{B}_{n + 1}$ as wanted
    
                        \textbf{Case B.} $\ell$ is \ref{line:ero:acquire_loop_until} during an invocation of the Acquire procedure invoked on \cref{line:ero:announce_acquire}.
    
                        We first prove that every $A$-event for $\mathcal{M}(w)$ is before the $k$th step of $\mathcal{I}^\mathcal{B}_{n + 1}$ (**).
                        Suppose, for contradiction, there is an $A$-event for $\mathcal{M}(w)$ after the $k$th step of $\mathcal{I}^\mathcal{B}_{n + 1}$; say during the $i$th step of $\mathcal{I}^\mathcal{B}_{n + 1}$.
                        Let $p$ be the process that performed this $A$-event, suppose $p$ did so during an operation execution $opx$, and let $\mathcal{I}^\mathcal{B}_i$ be the prefix of $\mathcal{I}^\mathcal{B}_{n + 1}$ up to and including the $i$th step, so $opx$ was invoked in $\mathcal{I}^\mathcal{B}_i$.
                        % Since $k < i$, we have that the $k$th step is included in $\mathcal{I}^\mathcal{B}_i$.
                        Since $p$ performed an $A$-event for $\mathcal{M}(w)$ during the $i$th step of $\mathcal{I}^\mathcal{B}_{n + 1}$, by \Cref{def:ero:english}, it did so during some invocation $I'$ of the \doworkuntildone{} procedure whose second parameter is $\mathcal{M}(w)$.
                        Hence, $p$ performed an $\allocatecelloperation{}$ operation whose response is $\mathcal{M}(w)$ during $opx$.
                        Furthermore, $p$ did not execute \cref{line:ero:owner_relinquish} during $opx$ in $\mathcal{I}^\mathcal{B}_i$.
                        Thus, by \Cref{lemma:ero:reference_count_is_non_negative_before_owner_relinquish}, $R(\mathcal{I}^\mathcal{B}_i, opx, \mathcal{M}(w)) \geq 0$.
                        To reach the contradiction, recall that the $k$th step of $\mathcal{I}^\mathcal{B}_{n + 1}$ performs a $\freecelloperation{}(\mathcal{M}(w))$ operation.
                        Hence, since $k < i$, we have that the $k$th step is included in $\mathcal{I}^\mathcal{B}_i$, and so there is a $\freecelloperation{}(\mathcal{M}(w))$ operation during $\mathcal{I}^\mathcal{B}_{i}$.
                        Thus, since $p$ performed an $\allocatecelloperation{}$ operation whose response is $\mathcal{M}(w)$ during $opx$, by \Cref{lemma:ero:fixed_references_at_free_operation}, $R(\mathcal{I}^\mathcal{B}_i, opx, \mathcal{M}(w)) = -1$.
                        However, $R(\mathcal{I}^\mathcal{B}_i, opx, \mathcal{M}(w)) \geq 0$, a contradiction.
    
                        We now finish the proof of Case B.
                        Since $\ell$ is \ref{line:ero:acquire_loop_until} during an invocation of the Acquire procedure invoked on \cref{line:ero:announce_acquire}, and $\cellpointershort$ is assigned to $\mathcal{M}(w)$ in $C_n$, we have that $p_{n + 1}$ read $\mathcal{M}(w)$ from $\announceobject.\cellpointerlong$ on its last execution of \cref{line:ero:announce_read} before the $n + 1$th step of $\mathcal{I}^\mathcal{B}_{n + 1}$; say at time $T^{\ref{line:ero:announce_read}}$.
                        Thus, $T^{\ref{line:ero:linearization_read}} < T^{\ref{line:ero:announce_read}}$, and since $e < T^{\ref{line:ero:linearization_read}}$, by transitivity, $e < T^{\ref{line:ero:announce_read}}$.
                        Suppose, for contradiction, that there is an $A$-event $e'$ for $\mathcal{M}(w')$ before $e$ in $\mathcal{I}^\mathcal{B}_{n + 1}$.
                        Hence, since $e' < e$ and $e < T^{\ref{line:ero:announce_read}}$, by transitivity, $e' < T^{\ref{line:ero:announce_read}}$.
                        Since $w \neq w'$, and by \Cref{def:reduction:swapper} $\mathcal{M}$ is injective, we have that $\mathcal{M}(w) \neq \mathcal{M}(w')$.
                        Furthermore, since $e'$ is an $A$-event for $\mathcal{M}(w')$, by \Cref{def:ero:english}, $\announceobject.\cellpointerlong = \mathcal{M}(w')$ at $e'$.
                        Hence, since $p_{n + 1}$ read $\mathcal{M}(w)$ from $\announceobject.\cellpointerlong$ at $T^{\ref{line:ero:announce_read}}$, $\mathcal{M}(w) \neq \mathcal{M}(w')$, and $e' < T^{\ref{line:ero:announce_read}}$, we have that $\announceobject.\cellpointerlong$ was set to $\mathcal{M}(w)$ after $e'$.
                        Thus, by \Cref{observation:ero:where_objects_change}, there is an $A$-event $e^*$ for $\mathcal{M}(w)$ after $e'$.
                        Therefore, since by (*) every $A$-event for $\mathcal{M}(w')$ is after the $w'$th step of $\mathcal{I}^\mathcal{B}_{n + 1}$, and $e'$ is an $A$-event for $\mathcal{M}(w')$, by transitivity, $w' < e'$, and so since $e' < e^*$, by transitivity, $w' < e^*$.
                        However, since by (**) every $A$-event for $\mathcal{M}(w)$ is before the $k$th step of $\mathcal{I}^\mathcal{B}_{n + 1}$, we have that $e^* < k$, and so since $k \leq j$ and $j < w'$, by transitivity, we have that $e^* < w'$, a contradiction.
                        This completes the proof of \textbf{Claim}.
    
                        We now complete the proof of Case 4.2.2.
                        The plan is to show that there is an $A$-event for $\mathcal{M}(w')$ before $e$ in $\mathcal{I}^\mathcal{B}_{n + 1}$, contradicting \textbf{Claim}.
                        Since $p_{n + 1}$ read $v'$ from $\linearizationobject$ on its last execution of \cref{line:ero:linearization_read} before invoking $I$, we have that the first parameter of $I$ is $\uniquerepositoryoperationshort_\linearizationobject$ where $v' = (\uniquerepositoryoperationshort_\linearizationobject, \arbitraryvalue)$.
                        Furthermore, by \Cref{def:reduction:swapper} $\mathcal{M}(w') \in \celluniverse$, so by \Cref{assumption:ero:head_and_null_not_in_cell_universe}, $\mathcal{M}(w') \neq \&\headobject$ and $\mathcal{M}(w') \neq \nullconstant$.
                        Hence, since $p_{n + 1}$ initializes $\currentcellpointershort$ to $\&\headobject$ on \cref{line:ero:add_cell_initial_current_pointer}, \ref{line:ero:remove_cell_initialize_pointers}, or \ref{line:ero:acquire_initial_current_pointer} depending on $\ell$ during $I$ and finds that $\currentcellpointershort = \mathcal{M}(w')$ on \cref{line:ero:add_cell_while_loop}, \cref{line:ero:remove_cell_while_loop}, or \cref{line:ero:acquire_loop_until} depending on $\ell$ during $I$, we have that $p_{n + 1}$ set $\currentcellpointershort = \mathcal{M}(w')$ on either \cref{line:ero:add_cell_update_current_pointer}, \cref{line:ero:remove_cell_update_pointers}, or \cref{line:ero:acquire_update_current_unique_pointer} depending on $\ell$ during $I$.
                        Thus, $p_{n + 1}$ received response $(\found, \mathcal{M}(w'))$ from an invocation $I'$ of the AcquireNext procedure during $I$.
                        Let $\previouscellpointershort$ be the second parameter of $I'$, so by \Cref{lemma:ero:acquire_next_is_for_pointer_from_universe_or_head}, $\previouscellpointershort \in \celluniverse \cup \{\&\headobject\}$.
                        Since the first parameter of $I$ is $\uniquerepositoryoperationshort_\linearizationobject$ and $p_{n + 1}$ invoked $I'$ during $I$, we have that the first parameter of $I'$ is $\uniquerepositoryoperationshort_\linearizationobject$.
                        Hence, the parameters of $I'$ are exactly $(\uniquerepositoryoperationshort_\linearizationobject, \previouscellpointershort)$.
                        Since the response of $I'$ is $(\found, \mathcal{M}(w'))$, we have that (a) $p_{n + 1}$ read $(*\previouscellpointershort).\nextlong = (\arbitraryvalue, \arbitraryvalue, \arbitraryvalue, \mathcal{M}(w'))$ on \cref{line:ero:acquire_next_read_curr_unique_pointer} during $I'$; say at time $T^{\ref{line:ero:acquire_next_read_curr_unique_pointer}}$ and (b) $p_{n + 1}$ found that $\linearizationobject{}.\uniquerepositoryoperationlong{} = \uniquerepositoryoperationshort_\linearizationobject$ on its next execution of \cref{line:ero:acquire_next_linearization_changed_check} during $I'$; say at time $T^{\ref{line:ero:acquire_next_linearization_changed_check}}$.
                        Hence, since $\previouscellpointershort \in \celluniverse \cup \{\&\headobject\}$, $(*\previouscellpointershort).\nextlong$ is initially $(\arbitraryvalue, \arbitraryvalue, \arbitraryvalue, \nullconstant)$, and so since $\mathcal{M}(w') \neq \nullconstant$ and $(*\previouscellpointershort).\nextlong = (\arbitraryvalue, \arbitraryvalue, \arbitraryvalue, \mathcal{M}(w'))$ at $T^{\ref{line:ero:acquire_next_read_curr_unique_pointer}}$, we have that $(*\previouscellpointershort).\nextlong$ was set to $(\arbitraryvalue, \arbitraryvalue, \arbitraryvalue, \mathcal{M}(w'))$ before $T^{\ref{line:ero:acquire_next_read_curr_unique_pointer}}$.
                        Thus, by \Cref{observation:ero:where_objects_change}, there is either a successful list-add attempt for $\mathcal{M}(w')$ or a successful list-remove attempt between  $\previouscellpointershort$ and $\mathcal{M}(w')$ before $T^{\ref{line:ero:acquire_next_read_curr_unique_pointer}}$.
                        Let $a$ denote this successful list attempt, so $a < T^{\ref{line:ero:acquire_next_read_curr_unique_pointer}}$.
                        If $a$ is a successful list-add attempt for $\mathcal{M}(w')$, by \Cref{lemma:ero:l_event_corresponding_to_do_low_level_op}, there is an $L$-event for $\mathcal{M}(w')$ before $a$.
                        Furthermore, if $a$ is a successful list-remove attempt between $\previouscellpointershort$ and $\mathcal{M}(w')$, then since $\mathcal{I}^\mathcal{B}_{n + 1}$ is an implementation history of $\mathcal{B}$, by \Cref{lemma:ero:the_list_invariants_hold}, $Q(\mathcal{I}^\mathcal{B}_{n + 1})$ holds, and so $v \in \List(\mathcal{I}')$ where $\mathcal{I}'$ is a prefix of $\mathcal{I}^\mathcal{B}_{n + 1}$ before $a$.
                        Hence, since $\mathcal{M}(w') \neq \nullconstant$ and $\mathcal{M}(w') \neq \&\headobject$, by \Cref{def:ero:logical_list}, there is an $L$-event for $\mathcal{M}(w')$ in $\mathcal{I}'$.
                        Therefore, in all cases, there is an $L$-event for $\mathcal{M}(w')$ before $a$ in $\mathcal{I}^\mathcal{B}_{n + 1}$.
                        Since $v' = (\uniquerepositoryoperationshort_\linearizationobject, \arbitraryvalue)$ and $e$ set $\linearizationobject = v'$, we have that $e$ set $\linearizationobject.\uniquerepositoryoperationlong = \uniquerepositoryoperationshort_\linearizationobject$.
                        Furthermore, since $\mathcal{I}^\mathcal{B}_{n + 1}$ is an implementation history of $\mathcal{B}$, by \Cref{lemma:ero:the_list_invariants_hold}, $P(\mathcal{I}^\mathcal{B}_{n + 1})$ holds.
                        Hence, since $e < T^{\ref{line:ero:acquire_next_linearization_changed_check}}$ and $\linearizationobject{}.\uniquerepositoryoperationlong{} = \uniquerepositoryoperationshort_\linearizationobject$ at $T^{\ref{line:ero:acquire_next_linearization_changed_check}}$, by \Cref{lemma:ero:p_implies_unique_low_level_operations_in_linearization}, we have that $e$ is the last $L$-event before $T^{\ref{line:ero:acquire_next_linearization_changed_check}}$ in $\mathcal{I}^\mathcal{B}_{n + 1}$.
                        Thus, since there is an $L$-event for $\mathcal{M}(w')$ before $a$ in $\mathcal{I}^\mathcal{B}_{n + 1}$, $a < T^{\ref{line:ero:acquire_next_read_curr_unique_pointer}}$, and $T^{\ref{line:ero:acquire_next_read_curr_unique_pointer}} < T^{\ref{line:ero:acquire_next_linearization_changed_check}}$, we have that there is an $L$-event for $\mathcal{M}(w')$ before or at $e$ in $\mathcal{I}^\mathcal{B}_{n + 1}$.
                        Therefore, by \Cref{lemma:ero:l_event_corresponding_a_event_type_and_pointer_relations}, there is an $A$-event for $\mathcal{M}(w')$ before $e$ in $\mathcal{I}^\mathcal{B}_{n + 1}$.
                        However, by \textbf{Claim}, there are no $A$-events for $\mathcal{M}(w')$ before $e$ in $\mathcal{I}^\mathcal{B}_{n + 1}$, a contradiction, so Case 4.2.2 is impossible.
                        \qH{\Cref{claim:reduction:b_does_traversal_checks_correctly}}
                    \end{itemize}
                \end{itemize}
            \end{itemize}
        \end{proof}
    
        \begin{claimcustom}{\ref{lemma:reduction:mapped_history_is_for_b}.15}\label{claim:reduction:b_does_already_remove_check_correctly}
            If $p_{n + 1}$ executes \cref{line:ero:remove_cell_before_removal_linearization_check} during the $n + 1$th step of $\mathcal{I}^\mathcal{A}_{n + 1}$ and finds the condition on \cref{line:ero:remove_cell_before_removal_linearization_check} to be true, then $p_{n + 1}$ executes \cref{line:ero:remove_cell_before_removal_linearization_check} during the $n + 1$th step of $\mathcal{I}^\mathcal{B}_{n + 1}$ and finds the condition on \cref{line:ero:remove_cell_before_removal_linearization_check} to be true.
        \end{claimcustom}
    
        \begin{proof}
            Since by \Cref{lemma:reduction:same_program_counters_in_mapped_run} the program counter of $p_{n + 1}$ is the same in $C^\mathcal{A}_n$ and $C_n$, and $p_{n + 1}$ takes the $n + 1$th step of $\mathcal{I}^\mathcal{B}_{n + 1}$, we have that $p_{n + 1}$ executes \cref{line:ero:remove_cell_before_removal_linearization_check} during the $n + 1$th \mbox{step of $\mathcal{I}^\mathcal{B}_{n + 1}$.}
            
            First observe that, if $p_{n + 1}$ finds the first clause of \cref{line:ero:remove_cell_before_removal_linearization_check} to be true in $\mathcal{I}^\mathcal{A}_{n + 1}$, then so does $p_{n + 1}$ in $\mathcal{I}^\mathcal{B}_{n + 1}$.
            This is because $\linearizationobject{}.\uniquerepositoryoperationlong{}$ is not watermarked, and so by \Cref{def:reduction:run_mapping}, the value of $\linearizationobject{}.\uniquerepositoryoperationlong{}$ is the same in $C^\mathcal{A}_n$ and $C_n$.
            Hence, it suffices to consider the case where $p_{n + 1}$ finds the first clause to be false and the second clause to be true in $\mathcal{I}^\mathcal{A}_{n + 1}$, i.e., $p_{n + 1}$ finds $\nextcellpointershort' = \nextcellpointershort$ in $C^\mathcal{A}_n$.
            Thus, since the value of $\linearizationobject{}.\uniquerepositoryoperationlong{}$ is the same in $C^\mathcal{A}_n$ and $C_n$, we have that $p_{n + 1}$ finds the first clause to be false in $\mathcal{I}^\mathcal{B}_{n + 1}$, and we must prove that the second clause is true.
    
            We start with a few basic facts.
            Since $\nextcellpointershort'$ and $\nextcellpointershort$ are a single value, by \Cref{def:reduction:swapper}, $\mathcal{S}_n(\nextcellpointershort') = \mathcal{S}_n(\nextcellpointershort', 1)$ and $\mathcal{S}_n(\nextcellpointershort) = \mathcal{S}_n(\nextcellpointershort, 1)$.
            Furthermore, since $p_{n + 1}$ execute \cref{line:ero:remove_cell_before_removal_linearization_check} during the $n + 1$th step of $\mathcal{I}^\mathcal{B}_{n + 1}$, it did so during an invocation $I$ of the \doremovecell{} procedure with parameters $(\uniquerepositoryoperationshort_\linearizationobject, \cellpointershort_\linearizationobject)$.
            Hence, by \Cref{lemma:ero:l_event_corresponding_to_do_low_level_op}, there is an $L$-remove event $e$ for $\cellpointershort_\linearizationobject$ before $p_{n + 1}$ invoked $I$ that set $\linearizationobject = (\uniquerepositoryoperationshort_\linearizationobject, \cellpointershort_\linearizationobject)$.
            Thus, by \Cref{lemma:ero:every_l_event_is_for_pointer_from_universe} $\cellpointershort_\linearizationobject \in \celluniverse$, and so by \Cref{assumption:ero:head_and_null_not_in_cell_universe}, $\cellpointershort_\linearizationobject \neq \&\headobject$.
            So, since $\cellpointershort_\linearizationobject$ is the second parameter of $I$, and the local variable $\currentcellpointershort$ is initially $\&\headobject$ (see \cref{line:ero:remove_cell_initialize_pointers}), we have that $p_{n + 1}$ finds the condition on \cref{line:ero:remove_cell_while_loop} to true on its first execution of \cref{line:ero:remove_cell_while_loop} during $I$.
            Furthermore, since $p_{n + 1}$ executes \cref{line:ero:remove_cell_before_removal_linearization_check} during $I$, we have that $p_{n + 1}$ finds the condition on \cref{line:ero:remove_cell_while_loop} to be false during $I$.
            These facts together imply $p_{n + 1}$ executes \cref{line:ero:remove_cell_update_pointers} at least once during $I$; let $T$ be the first time $p_{n + 1}$ does so.
            Hence, by Lemmas \ref{lemma:ero:remove_cell_current_pointer_always_in_universe_or_head} and \ref{lemma:ero:remove_cell_previous_pointer_always_in_universe_or_head}, from $T$ onwards in $I$  $\currentcellpointershort \in \celluniverse \cup \{\&\headobject\}$ and $\previouscellpointershort \in \celluniverse \cup \{\&\headobject\}$.
            
            There are four cases.
            \begin{itemize}
                \item[] \hspace{0pt}\textbf{Case 1.} $\mathcal{W}_n(\nextcellpointershort, 1) = \bot$ and $\mathcal{W}_n(\nextcellpointershort', 1) = \bot$.
    
                Hence, by \Cref{def:reduction:swapper}, $\mathcal{S}_n(\nextcellpointershort, 1)$ (resp. $\mathcal{S}_n(\nextcellpointershort', 1)$) is the same as the first value of $\nextcellpointershort$ (resp. $\nextcellpointershort'$) in $C^\mathcal{A}_n$.
                Thus, since $\mathcal{S}_n(\nextcellpointershort) = \mathcal{S}_n(\nextcellpointershort, 1)$ (resp. $\mathcal{S}_n(\nextcellpointershort') = \mathcal{S}_n(\nextcellpointershort', 1)$), we have that $\mathcal{S}_n(\nextcellpointershort)$ (resp. $\mathcal{S}_n(\nextcellpointershort')$) is the same as the value of $\nextcellpointershort$ (resp. $\nextcellpointershort'$) in $C^\mathcal{A}_n$.
                So, by \Cref{def:reduction:run_mapping}, the value of $\nextcellpointershort$ (resp. $\nextcellpointershort'$) is the same in $C^\mathcal{A}_n$ and $C_n$.
                Therefore, since $\nextcellpointershort = \nextcellpointershort'$ in $C^\mathcal{A}_n$, we have that $\nextcellpointershort = \nextcellpointershort'$ in $C_n$, and so $p_{n + 1}$ finds the condition on \cref{line:ero:remove_cell_before_removal_linearization_check} to be true \mbox{during the $n + 1$th step of $\mathcal{I}^\mathcal{B}_{n + 1}$.}
    
                \item[] \hspace{0pt}\textbf{Case 2.} $\mathcal{W}_n(\nextcellpointershort, 1) \neq \bot$ and $\mathcal{W}_n(\nextcellpointershort', 1) = \bot$.
    
                Hence, since $\mathcal{S}_n(\nextcellpointershort) = \mathcal{S}_n(\nextcellpointershort, 1)$, by \Cref{def:reduction:watermarks}, the value of $\nextcellpointershort$ in $C^\mathcal{A}_n$ is in $\celluniverse$.
                Thus, since $\nextcellpointershort = \nextcellpointershort'$ in $C^\mathcal{A}_n$, we have that the value of $\nextcellpointershort'$ in $C^\mathcal{A}_n$ is in $\celluniverse$; say $v$.
                So, by \Cref{def:reduction:used_pointers_in_run_of_a}, $v \in \celluniverse(\mathcal{I}^\mathcal{A}_{n + 1})$.
                Furthermore, since $\mathcal{S}_n(\nextcellpointershort') = \mathcal{S}_n(\nextcellpointershort', 1)$ and $\mathcal{W}_n(\nextcellpointershort', 1) = \bot$, by \Cref{def:reduction:run_mapping}, the value of $\nextcellpointershort'$ is $v$ in $C_n$.
                We now show that there is an $\allocatecelloperation{}$ operation in $\mathcal{I}^\mathcal{B}_{n + 1}$ whose response is $v$.
                Since $C_n$ assigns state $v$ to $\nextcellpointershort'$, we have that $p_{n + 1}$ saw $(*\previouscellpointershort).\nextlong = (\arbitraryvalue, \arbitraryvalue, \arbitraryvalue, v)$ on its last execution of \cref{line:ero:remove_cell_read_previous_pointer} during $I$.
                Hence, since $\mathcal{I}^\mathcal{B}_{n + 1}$ is an implementation history of $\mathcal{B}$, by \Cref{lemma:ero:the_list_invariants_hold} $Q(\mathcal{I}^\mathcal{B}_{n + 1})$ holds, and so since $\previouscellpointershort \in \celluniverse \cup \{\&\headobject\}$ and $v \in \celluniverse$, by \Cref{lemma:ero:if_next_pointer_not_null_there_is_a_l_event_for_it}, there is an $L$-event $e$ for $v$ in $\mathcal{I}^\mathcal{B}_{n + 1}$.
                Thus, by \Cref{lemma:ero:l_event_corresponding_a_event_type_and_pointer_relations}, there is an $A$-event $e'$ for $v$ before $e$.
                So, by \Cref{def:ero:english}, the process $p$ that executed $e'$ did so during an invocation of the \doworkuntildone{} procedure with a second parameter of $v$.
                Thus, $p$ performed an $\allocatecelloperation{}$ operation on \cref{line:ero:allocate_cell} whose response is $v$.
                Therefore, since $v \in \celluniverse(\mathcal{I}^\mathcal{A}_{n + 1})$, there is an $\allocatecelloperation{}$ operation in $\mathcal{I}^\mathcal{B}_{n + 1}$ whose response is in $\celluniverse(\mathcal{I}^\mathcal{A}_{n + 1})$.
                However, by \Cref{claim:reduction:no_allocate_for_used_pointer_in_b}, the response of every $\allocatecelloperation{}$ operation in $\mathcal{I}^\mathcal{B}_{n + 1}$ is not in $\celluniverse(\mathcal{I}^\mathcal{A}_{n + 1})$, a contradiction, so Case 2 is impossible.

                \item[] \hspace{0pt}\textbf{Case 3.} $\mathcal{W}_n(\nextcellpointershort, 1) = \bot$ and $\mathcal{W}_n(\nextcellpointershort', 1) \neq \bot$.

                The proof is essentially the same as Case 2.
                Since $\mathcal{W}_n(\nextcellpointershort', 1) \neq \bot$, and $\mathcal{S}_n(\nextcellpointershort') = \mathcal{S}_n(\nextcellpointershort', 1)$, the value of $\nextcellpointershort'$ in $C^\mathcal{A}_n$ is in $\celluniverse$.
                Thus, since $\nextcellpointershort = \nextcellpointershort'$ in $C^\mathcal{A}_n$, we have that the value of $\nextcellpointershort$ in $C^\mathcal{A}_n$ is in $\celluniverse$; say $v$.
                So, by \Cref{def:reduction:used_pointers_in_run_of_a}, $v \in \celluniverse(\mathcal{I}^\mathcal{A}_{n + 1})$.
                Furthermore, since $\mathcal{S}_n(\nextcellpointershort) = \mathcal{S}_n(\nextcellpointershort, 1)$ and $\mathcal{W}_n(\nextcellpointershort, 1) = \bot$, by \Cref{def:reduction:run_mapping}, the value of $\nextcellpointershort$ is $v$ in $C_n$.
                We now show that there is an $\allocatecelloperation{}$ operation in $\mathcal{I}^\mathcal{B}_{n + 1}$ whose response is $v$.
                Since $C_n$ assigns state $v$ to $\nextcellpointershort$, we have that $p_{n + 1}$ saw $(*\cellpointershort_\linearizationobject).\nextlong = (\arbitraryvalue, \arbitraryvalue, \arbitraryvalue, v)$ on its last execution of \cref{line:ero:remove_cell_read_pointer_to_remove} during $I$.
                Hence, since $\mathcal{I}^\mathcal{B}_{n + 1}$ is an implementation history of $\mathcal{B}$, by \Cref{lemma:ero:the_list_invariants_hold} $Q(\mathcal{I}^\mathcal{B}_{n + 1})$ holds, and so since $\cellpointershort_\linearizationobject \in \celluniverse$ and $v \in \celluniverse$, by \Cref{lemma:ero:if_next_pointer_not_null_there_is_a_l_event_for_it}, there is an $L$-event $e$ for $v$ in $\mathcal{I}^\mathcal{B}_{n + 1}$.
                Thus, by \Cref{lemma:ero:l_event_corresponding_a_event_type_and_pointer_relations}, there is an $A$-event $e'$ for $v$ before $e$.
                So, by \Cref{def:ero:english}, the process $p$ that executed $e'$ did so during an invocation of the \doworkuntildone{} procedure with a second parameter of $v$.
                Thus, $p$ performed an $\allocatecelloperation{}$ operation on \cref{line:ero:allocate_cell} whose response is $v$.
                Therefore, since $v \in \celluniverse(\mathcal{I}^\mathcal{A}_{n + 1})$, there is an $\allocatecelloperation{}$ operation in $\mathcal{I}^\mathcal{B}_{n + 1}$ whose response is in $\celluniverse(\mathcal{I}^\mathcal{A}_{n + 1})$.
                However, by \Cref{claim:reduction:no_allocate_for_used_pointer_in_b}, the response of every $\allocatecelloperation{}$ operation in $\mathcal{I}^\mathcal{B}_{n + 1}$ is not in $\celluniverse(\mathcal{I}^\mathcal{A}_{n + 1})$, a contradiction, so Case 3 is impossible.
                
                \item[] \hspace{0pt}\textbf{Case 4.} $\mathcal{W}_n(\nextcellpointershort, 1) \neq \bot$ and $\mathcal{W}_n(\nextcellpointershort', 1) \neq \bot$.
    
                Let $\mathcal{W}_n(\nextcellpointershort, 1) = w$ and $\mathcal{W}_n(\nextcellpointershort', 1) = w'$.
                Hence, by \Cref{def:reduction:swapper}, $\mathcal{S}_n(\nextcellpointershort, 1) = \mathcal{M}(w)$ and $\mathcal{S}_n(\nextcellpointershort', 1) = \mathcal{M}(w')$.
                Therefore, since by above $\mathcal{S}_n(\nextcellpointershort) = \mathcal{S}_n(\nextcellpointershort, 1)$ and $\mathcal{S}_n(\nextcellpointershort') = \mathcal{S}_n(\nextcellpointershort', 1)$, we have that $\mathcal{S}_n(\nextcellpointershort) = \mathcal{M}(w)$ and $\mathcal{S}_n(\nextcellpointershort') = \mathcal{M}(w')$.
                So, by \Cref{def:reduction:run_mapping}, $\nextcellpointershort$ (resp. $\nextcellpointershort'$) is assigned to $\mathcal{M}(w)$ (resp. $\mathcal{M}(w')$) in $C_n$.
                Hence, since $p_{n + 1}$ checks whether $\nextcellpointershort = \nextcellpointershort'$ during the $n + 1$th step of $\mathcal{I}^\mathcal{B}_{n + 1}$, we have that $p_{n + 1}$ checks whether $\mathcal{M}(w) = \mathcal{M}
                (w')$ during the $n + 1$th step of $\mathcal{I}^\mathcal{B}_{n + 1}$.
    
                \begin{itemize}
                    \item[] \textbf{Case 4.1.} $w = w'$.
    
                    Hence, $\mathcal{M}(w) = \mathcal{M}(w')$.
                    Therefore, since $p_{n + 1}$ checks whether $\mathcal{M}(w) = \mathcal{M}(w')$ during the $n + 1$th step of $\mathcal{I}^\mathcal{B}_{n + 1}$, $p_{n + 1}$ finds the condition on \cref{line:ero:remove_cell_before_removal_linearization_check} to be true \mbox{during the $n + 1$th step of $\mathcal{I}^\mathcal{B}_{n + 1}$.}
    
                    \item[] \textbf{Case 4.2.} $w \neq w'$.
    
                    Hence, since $\mathcal{M}$ is injective, we have that $\mathcal{M}(w) \neq \mathcal{M}(w')$.
                    Thus, $p_{n + 1}$ finds the second clause of \cref{line:ero:remove_cell_before_removal_linearization_check} to be false during the $n + 1$th step of $\mathcal{I}^\mathcal{B}_{n + 1}$.
                    Therefore, since $p_{n + 1}$ finds the first clause of \cref{line:ero:remove_cell_before_removal_linearization_check} to be false during the $n + 1$th step of $\mathcal{I}^\mathcal{B}_{n + 1}$, we have that $p_{n + 1}$ finds the condition on \cref{line:ero:remove_cell_before_removal_linearization_check} to be false during the $n + 1$th step of $\mathcal{I}^\mathcal{B}_{n + 1}$, and so $p_{n + 1}$ is poised to execute \cref{line:ero:remove_cell_from_list} during its next step.
                    Let $\mathcal{I}^\mathcal{B}_{n + 2}$ be the one step of continuation of $\mathcal{I}^\mathcal{B}_{n +1}$ by $p_{n + 1}$, so $p_{n + 1}$ executes \cref{line:ero:remove_cell_from_list} during the last step of $\mathcal{I}^\mathcal{B}_{n + 2}$.
                    Since $\mathcal{I}^\mathcal{B}_{n + 1}$ is an implementation history of $\mathcal{B}$, we have that $\mathcal{I}^\mathcal{B}_{n + 2}$ is an implementation history of $\mathcal{B}$.
                    Furthermore, since $\mathcal{M}(w)$ is the value of $\nextcellpointershort$ in $C_n$, it follows that $p_{n + 1}$ executes \cref{line:ero:remove_cell_from_list} during the last step of $\mathcal{I}^\mathcal{B}_{n + 2}$ of the form $\CASop((*\previouscellpointershort).\nextlong, (\arbitraryvalue, \arbitraryvalue, \arbitraryvalue, \cellpointershort_\linearizationobject), (\arbitraryvalue, \arbitraryvalue, \arbitraryvalue, \mathcal{M}(w)))$.
                    Thus, by \Cref{def:ero:english}, the last step of $\mathcal{I}^\mathcal{B}_{n + 2}$ is a list-remove attempt for $\cellpointershort_\linearizationobject$ between $\previouscellpointershort$ and $\mathcal{M}(w)$.
                    Let $a$ denote this list-remove attempt.
                    Since $\mathcal{I}^\mathcal{B}_{n + 2}$ is an implementation history of $\mathcal{B}$, by \Cref{lemma:ero:the_list_invariants_hold}, $P(\mathcal{I}^\mathcal{B}_{n + 2})$, $Q(\mathcal{I}^\mathcal{B}_{n + 2})$, and $R(\mathcal{I}^\mathcal{B}_{n + 2})$ hold.
                    Hence, by $Q(\mathcal{I}^\mathcal{B}_{n + 2})$, $a$ is preceded by a unique $L$-remove event for $\cellpointershort_\linearizationobject$, and if $\mathcal{I}'$ is the prefix of $\mathcal{I}^\mathcal{B}_{n + 2}$ up to but excluding that $L$-remove event, $\cellpointershort_\linearizationobject$ is in $\List(\mathcal{I}')$ exactly once and $\previouscellpointershort$ and $\mathcal{M}(w)$ are the pointers preceding and succeeding $\cellpointershort_\linearizationobject$ in $\List(\mathcal{I}')$.
                    Therefore, since $e$ is an $L$-remove event for $\cellpointershort_\linearizationobject$ in $\mathcal{I}^\mathcal{B}_{n + 1}$ (as defined at the start of the proof), it follows that $\mathcal{I}'$ is the prefix of $\mathcal{I}^\mathcal{B}_{n + 2}$ up to but excluding $e$.           
        
                    We now prove that there are no $L$-events after $e$ in $\mathcal{I}^\mathcal{B}_{n + 2}$ (*).
                    Since $p_{n + 1}$ finds the first clause of \cref{line:ero:remove_cell_before_removal_linearization_check} to be false during the $n + 1$th step of $\mathcal{I}^\mathcal{B}_{n + 2}$, and the first parameter of $I$ is $\uniquerepositoryoperationshort_\linearizationobject$, we have that $\linearizationobject.\uniquerepositoryoperationlong = \uniquerepositoryoperationshort_\linearizationobject$ in $C^\mathcal{B}_{n + 1}$.
                    Hence, since $e$ is an $L$-event that set $\linearizationobject.\uniquerepositoryoperationlong = \uniquerepositoryoperationshort_\linearizationobject$ during $\mathcal{I}^\mathcal{B}_{n + 2}$ and $P(\mathcal{I}^\mathcal{B}_{n + 2})$ holds, by \Cref{lemma:ero:p_implies_unique_low_level_operations_in_linearization}, there are no $L$-events after $e$ in $\mathcal{I}^\mathcal{B}_{n + 2}$.
        
                    We now prove that $\cellpointershort_\linearizationobject = \mathcal{M}(w')$ (**).
                    Let $\mathcal{I}^{\ref{line:ero:remove_cell_read_previous_pointer}}$ be the prefix of $\mathcal{I}^\mathcal{B}_{n + 2}$ up to and including $p_{n + 1}$'s last execution of \cref{line:ero:remove_cell_read_previous_pointer} in $\mathcal{I}^\mathcal{B}_{n + 2}$.
                    Since $\nextcellpointershort'$ is assigned to $\mathcal{M}(w')$ in $C_n$, and $p_{n + 1}$ executes \cref{line:ero:remove_cell_before_removal_linearization_check} and \cref{line:ero:remove_cell_from_list} during the $n + 1$th and $n + 2$th step of $\mathcal{I}^\mathcal{B}_{n + 2}$, we have that $(*\previouscellpointershort).\nextlong = (\arbitraryvalue, \arbitraryvalue, \arbitraryvalue, \mathcal{M}(w'))$ at the end of $\mathcal{I}^{\ref{line:ero:remove_cell_read_previous_pointer}}$.
                    Furthermore, notice that $e$ is in $\mathcal{I}^{\ref{line:ero:remove_cell_read_previous_pointer}}$.
                    Hence, since by (*) there are no $L$-events after $e$ in $\mathcal{I}^\mathcal{B}_{n + 2}$, we have that $e$ is the last $L$-event in $\mathcal{I}^{\ref{line:ero:remove_cell_read_previous_pointer}}$.
                    Thus, since $P(\mathcal{I}^\mathcal{B}_{n + 2})$, $Q(\mathcal{I}^\mathcal{B}_{n + 2})$, and $R(\mathcal{I}^\mathcal{B}_{n + 2})$ hold, by \Cref{lemma:ero:conditional_classification_lemma}, the list of cells conforms to either $\List(\mathcal{I}')$ or $\List(\mathcal{I}^{\ref{line:ero:remove_cell_read_previous_pointer}})$ in $\mathcal{I}^{\ref{line:ero:remove_cell_read_previous_pointer}}$.
                    In the first case, since $\cellpointershort_\linearizationobject$ is in $\List(\mathcal{I}')$ exactly once and $\previouscellpointershort$ is the pointer preceding $\cellpointershort_\linearizationobject$ in $\List(\mathcal{I}')$, by \Cref{def:ero:logical_list}, $(*\previouscellpointershort).\nextlong = (\arbitraryvalue, \arbitraryvalue, \arbitraryvalue, \cellpointershort_\linearizationobject)$ at the end of $\mathcal{I}^{\ref{line:ero:remove_cell_read_previous_pointer}}$.
                    Hence, since $(*\previouscellpointershort).\nextlong = (\arbitraryvalue, \arbitraryvalue, \arbitraryvalue, \mathcal{M}(w'))$ at the end of $\mathcal{I}^{\ref{line:ero:remove_cell_read_previous_pointer}}$, we have that $\cellpointershort_\linearizationobject = \mathcal{M}(w')$ as wanted.
                    In the second case, since (a) the sequence of $L$-events in $\List(\mathcal{I}')$ and $\List(\mathcal{I}^{\ref{line:ero:remove_cell_read_previous_pointer}})$ are the same with the exception that $e$ is not in $\List(\mathcal{I}')$ and $e$ is in $\List(\mathcal{I}^{\ref{line:ero:remove_cell_read_previous_pointer}})$, (b) $e$ is an $L$-remove event for $\cellpointershort_\linearizationobject$, and (c) $\cellpointershort_\linearizationobject$ is in $\List(\mathcal{I}')$ exactly once and $\previouscellpointershort$ and $\mathcal{M}(w)$ are the pointers preceding and following $\cellpointershort_\linearizationobject$ in $\List(\mathcal{I}')$, by \Cref{def:ero:logical_list}, we have that $\previouscellpointershort$ is preceding $\mathcal{M}(w)$ in $\List(\mathcal{I}^{\ref{line:ero:remove_cell_read_previous_pointer}})$.
                    Hence, since the list of cells conforms to $\List(\mathcal{I}^{\ref{line:ero:remove_cell_read_previous_pointer}})$ in $\mathcal{I}^{\ref{line:ero:remove_cell_read_previous_pointer}}$, by \Cref{def:ero:logical_list}, $(*\previouscellpointershort).\nextlong = (\arbitraryvalue, \arbitraryvalue, \arbitraryvalue, \mathcal{M}(w))$ at the end of $\mathcal{I}^{\ref{line:ero:remove_cell_read_previous_pointer}}$.
                    Therefore, since $(*\previouscellpointershort).\nextlong = (\arbitraryvalue, \arbitraryvalue, \arbitraryvalue, \mathcal{M}(w'))$ at the end of $\mathcal{I}^{\ref{line:ero:remove_cell_read_previous_pointer}}$, we have that $\mathcal{M}(w) = \mathcal{M}(w')$.
                    However, $\mathcal{M}(w) \neq \mathcal{M}(w')$, a contradiction, so the second case is impossible.
        
                    We now return the proof of Case 4.2.
                    Since $e$ is an $L$-remove event for $\cellpointershort_\linearizationobject$, and by (**) $\cellpointershort_\linearizationobject = \mathcal{M}(w')$, we have that $e$ is an $L$-remove event for $\mathcal{M}(w')$.
                    Furthermore, since $\mathcal{W}_n(\nextcellpointershort, 1) = w \neq \bot$ (resp. $\mathcal{W}_n(\nextcellpointershort', 1) = w' \neq \bot$), by \Cref{lemma:reduction:associated_pointer}, the $w$th (resp. $w'$th) step of $\mathcal{I}^\mathcal{A}_{n + 1}$ performs an $\allocatecelloperation{}$ operation whose response is the value of $\nextcellpointershort$ (resp. $\nextcellpointershort'$) in $C^\mathcal{A}_n$.
                    Hence, since $\nextcellpointershort = \nextcellpointershort'$ in $C^\mathcal{A}_n$, we have that this is the same value, say $v$, and so the $w$th and $w'$th step of $\mathcal{I}^\mathcal{A}_{n + 1}$ perform an $\allocatecelloperation{}$ operation whose response is $v$.
                    Furthermore, by \Cref{claim:reduction:allocate_response_in_b_is_swapped}, the $w$th (resp. $w'$th) step of $\mathcal{I}^\mathcal{B}_{n + 1}$ performs an $\allocatecelloperation{}$ operation whose response is $\mathcal{M}(w)$ (resp. $\mathcal{M}(w')$).
                    There are two cases.
        
                    \begin{itemize}
                        \item[] \hspace{0pt}\textbf{Case 4.2.1.} $w < w'$.
    
                        Hence, since the $w$th and $w'$th step of $\mathcal{I}^\mathcal{A}_{n + 1}$ perform an $\allocatecelloperation{}$ operation whose response is $v$, by \Cref{alg:cell_manager_specification}, $v \in \celluniverse$ and for some $w < j < w'$ the $j$th step of $\mathcal{I}^\mathcal{A}_{n + 1}$ performs a $\freecelloperation{}(v)$ operation.
                        Thus, by \Cref{claim:reduction:allocate_followed_by_free_in_a_implies_free_for_same_watermark}, for some $w < k \leq j$ the $k$th step of $\mathcal{I}^\mathcal{A}_{n + 1}$ performs a $\freecelloperation{}(v)$, and $\mathcal{W}_{k - 1}(O_k, 1) = w$ where $O_k$ is the local variable $\currentcellpointershort$ of $p_k$ in the Relinquish procedure.
                        Therefore, by \Cref{claim:reduction:frees_in_b_are_swapped}, the $k$th step of $\mathcal{I}^\mathcal{B}_{n + 1}$ performs a $\freecelloperation{}(\mathcal{M}(w))$ operation, so by \Cref{lemma:ero:free_cell_operations_are_preceeded_by_remove_event}, there is an $L$-remove event for $\mathcal{M}(w)$ before the $k$th step of $\mathcal{I}^\mathcal{B}_{n + 1}$.
    
                        We now prove that $w' < e$.
                        Since $e$ is an $L$-event for $\mathcal{M}(w')$, by \Cref{lemma:ero:l_event_corresponding_a_event_type_and_pointer_relations}, there is an $A$-event for $\mathcal{M}(w')$ before $e$ in $\mathcal{I}^\mathcal{B}_{n + 2}$.
                        Hence, by \Cref{def:ero:english}, some process executed this $A$-event during some invocation of the \doworkuntildone{} with a second parameter $\mathcal{M}(w')$, and so it performed an $\allocatecelloperation{}$ operation with response $\mathcal{M}(w')$ before $e$.
                        Therefore, since the $w'$th step of $\mathcal{I}^\mathcal{B}_{n + 2}$ performs an $\allocatecelloperation{}$ operation with response $\mathcal{M}(w')$, and by \Cref{alg:lazy_cell_manager_specification} the response of each $\allocatecelloperation{}$ operation is unique, we have that $w' < e$.
    
                        We now finish the proof of Case 4.2.1.
                        Since $\mathcal{M}(w) \in \celluniverse$ and $\mathcal{M}(w) \in \List(\mathcal{I}')$, by \Cref{def:ero:logical_list}, there are no $L$-remove events for $\mathcal{M}(w)$ in $\mathcal{I}'$.
                        Therefore, since $\mathcal{I}'$ is the prefix of $\mathcal{I}^\mathcal{B}_{n + 2}$ up to but excluding $e$, we have that there are no $L$-remove events for $\mathcal{M}(w)$ before $e$ in $\mathcal{I}^\mathcal{B}_{n + 2}$.
                        However, as proved above, there is an $L$-remove event for $\mathcal{M}(w)$ before the $k$th step of $\mathcal{I}^\mathcal{B}_{n + 1}$, and so since $k \leq j$, $j < w'$, and as proved above $w' < e$, by transitivity, we have that there is an $L$-remove event for $\mathcal{M}(w)$ before $e$ in $\mathcal{I}^\mathcal{B}_{n + 2}$, a contradiction, so Case 4.2.1 is impossible.
        
                        \item[] \hspace{0pt}\textbf{Case 4.2.2.} $w' < w$.
    
                        Hence, since the $w'$th and $w$th step of $\mathcal{I}^\mathcal{A}_{n + 1}$ perform an $\allocatecelloperation{}$ operation whose response is $v$, by \Cref{alg:cell_manager_specification}, $v \in \celluniverse$ and for some $w' < j < w$ the $j$th step of $\mathcal{I}^\mathcal{A}_{n + 1}$ performs a $\freecelloperation{}(v)$ operation.
                        Thus, by \Cref{claim:reduction:allocate_followed_by_free_in_a_implies_free_for_same_watermark}, for some $w' < k \leq j$ the $k$th step of $\mathcal{I}^\mathcal{A}_{n + 1}$ performs a $\freecelloperation{}(v)$, and $\mathcal{W}_{k - 1}(O_k, 1) = w'$ where $O_k$ is the local variable $\currentcellpointershort$ of $p_k$ in the Relinquish procedure.
                        So, by \Cref{claim:reduction:frees_in_b_are_swapped}, the $k$th step of $\mathcal{I}^\mathcal{B}_{n + 1}$ performs a $\freecelloperation{}(\mathcal{M}(w'))$ operation, and thus by \Cref{lemma:ero:free_cell_operations_are_preceeded_by_remove_event}, there is an $L$-remove event for $\mathcal{M}(w')$ before the $k$th step of $\mathcal{I}^\mathcal{B}_{n + 1}$.
                        Therefore, since $e$ is an $L$-remove event for $\mathcal{M}(w')$, by $P(\mathcal{I}^\mathcal{B}_{n + 2})$, $e$ is the only $L$-remove event for $\mathcal{M}(w')$ in $\mathcal{I}^\mathcal{B}_{n + 2}$, and so $e < k$.
    
                        We now prove that there is an $L$-event for $\mathcal{M}(w)$ in $\mathcal{I}^\mathcal{B}_{n + 2}$.
                        Since $\nextcellpointershort$ is assigned to $\mathcal{M}(w)$ in $C_n$, and $p_{n + 1}$ executes \cref{line:ero:remove_cell_before_removal_linearization_check} and \cref{line:ero:remove_cell_from_list} during the $n + 1$th and $n + 2$th step of $\mathcal{I}^\mathcal{B}_{n + 2}$, we have that $p_{n + 1}$ read $(\arbitraryvalue, \arbitraryvalue, \arbitraryvalue, \mathcal{M}(w))$ from $\cellpointershort_\linearizationobject$ at the time of $p_{n + 1}$'s last execution of \cref{line:ero:remove_cell_read_pointer_to_remove} during $\mathcal{I}^\mathcal{B}_{n + 2}$.
                        Hence, since $\mathcal{I}^\mathcal{B}_{n + 2}$ is an implementation history of $\mathcal{B}$, by \Cref{lemma:ero:the_list_invariants_hold} $Q(\mathcal{I}^\mathcal{B}_{n + 2})$ holds, and so since $\cellpointershort_\linearizationobject \in \celluniverse$ and $\mathcal{M}(w) \in \celluniverse$, by \Cref{lemma:ero:if_next_pointer_not_null_there_is_a_l_event_for_it}, there is an $L$-event for $\mathcal{M}(w)$ in $\mathcal{I}^\mathcal{B}_{n + 2}$.
                        
                        We now finish the proof of Case 4.2.2.
                        Let $e'$ be the $L$-event for $\mathcal{M}(w)$ we just identified.
                        Since $e'$ is an $L$-event for $\mathcal{M}(w)$, by \Cref{lemma:ero:l_event_corresponding_a_event_type_and_pointer_relations}, there is an $A$-event for $\mathcal{M}(w)$ before $e'$ in $\mathcal{I}^\mathcal{B}_{n + 2}$.
                        Hence, by \Cref{def:ero:english}, some process executed this $A$-event during some invocation of the \doworkuntildone{} with a second parameter $\mathcal{M}(w)$, and so it performed an $\allocatecelloperation{}$ operation with response $\mathcal{M}(w)$ before $e'$.
                        Thus, since the $w$th step of $\mathcal{I}^\mathcal{B}_{n + 2}$ performs an $\allocatecelloperation{}$ operation with response $\mathcal{M}(w)$, and by \Cref{alg:lazy_cell_manager_specification} the response of each $\allocatecelloperation{}$ operation is unique, we have that $w < e'$.
                        Since by (*) $e$ is the last $L$-event in $\mathcal{I}^\mathcal{B}_{n + 2}$ and $e'$ is an $L$-event in $\mathcal{I}^\mathcal{B}_{n + 2}$, we have that $e' \leq e$.
                        Therefore, since $w < e'$ and as proved above $e < k$, by transitivity, $w < k$.
                        However, since $k \leq j$ and $j < w$, by transitivity, $k < w$, a contradiction, so Case 4.2.2 is impossible.
                        \qH{\Cref{claim:reduction:b_does_already_remove_check_correctly}}
                    \end{itemize}
                \end{itemize}
            \end{itemize}
        \end{proof}

        %\textcolor{red}{stopped here.}
    
        \begin{claimcustom}{\ref{lemma:reduction:mapped_history_is_for_b}.16}\label{claim:reduction:b_updates_program_counter_correctly}
            Let $pr$ be a program counter in $C^\mathcal{A}_{n + 1}$.
            Then, $C^\mathcal{B}_{n + 1}$ assigns $\mathcal{S}_{n + 1}(pr)$ to $pr$.
        \end{claimcustom}
    
        \begin{proof}
            Suppose $pr$ is the program counter for process $p$.
            Let $s^\mathcal{A}_n$ (resp. $s^\mathcal{A}_{n + 1}$) be the state assigned to $pr$ in $C^\mathcal{A}_n$ (resp. $C^\mathcal{A}_{n + 1}$) and let $s_n$ (resp. $s_{n + 1}$) be the state assigned to $pr$ in $C_n$ (resp. $C^\mathcal{B}_{n + 1}$).
            We prove that $s^\mathcal{A}_{n + 1} = s_{n + 1}$.
            There are two cases.
            \begin{itemize}
                \item[] \hspace{0pt}\textbf{Case 1.} $s^\mathcal{A}_n = s^\mathcal{A}_{n + 1}$.
    
                Hence, $p$ did not take the $n + 1$th step of $\mathcal{I}^\mathcal{A}_{n + 1}$, and so $p \neq p_{n + 1}$.
                Thus, since $p_{n + 1}$ takes the $n + 1$th step of $\mathcal{I}^\mathcal{B}_{n + 1}$, we have that $p$ also does not take the $n + 1$th step of $\mathcal{I}^\mathcal{B}_{n + 1}$, and so $s_n = s_{n + 1}$.
                Since by \Cref{lemma:reduction:same_program_counters_in_mapped_run}, the program counter of $p$ is the same in $C^\mathcal{A}_n$ and $C_n$, we have that $s^\mathcal{A}_n = s_n$.
                Therefore, since $s^\mathcal{A}_n = s^\mathcal{A}_{n + 1}$ and $s_n = s_{n + 1}$, we have that $s^\mathcal{A}_{n + 1} = s_{n + 1}$.
    
                \item[] \hspace{0pt}\textbf{Case 2.} $s^\mathcal{A}_n \neq s^\mathcal{A}_{n + 1}$.
    
                Hence, $p$ takes the $n + 1$th step of $\mathcal{I}^\mathcal{A}_{n + 1}$, and so $p = p_{n + 1}$.
                Thus, since $p_{n + 1}$ takes the $n + 1$th step of $\mathcal{I}^\mathcal{B}_{n + 1}$, we have that $p$ also takes the $n + 1$th step of $\mathcal{I}^\mathcal{B}_{n + 1}$, and so $s_n \neq s_{n + 1}$.
                Suppose $p$ executes the line of code $\ell$ during the $n + 1$th step of $\mathcal{I}^\mathcal{A}_{n + 1}$.
                Since by \Cref{lemma:reduction:same_program_counters_in_mapped_run}, the program counter of $p$ is the same in $C^\mathcal{A}_n$ and $C_n$, we have that $p$ executes the line of code $\ell$ during the $n + 1$th step of  $\mathcal{I}^\mathcal{B}_{n + 1}$, so $s^\mathcal{A}_n = s_n$.
                Observe that $s^\mathcal{A}_{n + 1}$ is either (a) solely determined based on $\ell$ (i.e., it increases by one, or deterministically jumps to a new line number because $\ell$ invokes a procedure, is a goto statement, or is a response step) or (b) depends on the the response $p$ received from an operation on a base object it performed during the $i + 1$th step of $\mathcal{I}^\mathcal{A}_{n + 1}$ and/or the state of $p$'s local variables in $C^\mathcal{A}_n$ (i.e., $\ell$ is an if, while, or until statement).
                We consider each case separately.
                
                \begin{itemize}
                    \item[] \hspace{0pt}\textbf{Case (a).}
    
                    Hence, since $p$ executes the line of code $\ell$ during the $n + 1$th step of $\mathcal{I}^\mathcal{B}_{n + 1}$, we have that $s_{n + 1} = s^\mathcal{A}_{n + 1}$.
    
                    \item[] \hspace{0pt}\textbf{Case (b).}
    
                    Hence, $\ell$ is an execution of either an if, while, or until statement.
                    We consider each line.
    
                    \begin{itemize}
                        % A unique low-level operations & status
                        \item[] \textbf{Case (b).1.} $\ell$ is either \ref{line:ero:add_cell_done_check}, \ref{line:ero:remove_cell_done_check}, or \ref{line:ero:write_and_read_done_check}.
    
                        Hence, $p_{n + 1}$ is comparing a unique low-level operation read from $\announceobject$ with a static value and possibly comparing the response of an invocation of the Acquire procedure with a static value.
                        By tracing backwards, we can see that the values of these local variables do not originate from the response of an $\allocatecelloperation{}$ operation, and so by \Cref{def:reduction:watermarks}, they are not watermarked.
                        Thus, by \Cref{def:reduction:run_mapping}, their state is the same in $C^\mathcal{A}_n$ and $C_n$.
                        Therefore, $p_{n + 1}$ finds the condition on line $\ell$ to be true during the $n + 1$th step of $\mathcal{I}^\mathcal{A}_{n + 1}$ if and only if $p_{n + 1}$ finds the condition on line $\ell$ to be true during the $n + 1$th step of $\mathcal{I}^\mathcal{B}_{n + 1}$, so $s_{n + 1} = s^\mathcal{A}_{n + 1}$ as wanted.
    
                        % L unique low-level operations & status
                        \item[] \textbf{Case (b).2.} $\ell$ is either \ref{line:ero:do_add_cell_condition}, \ref{line:ero:do_remove_cell_condition},  \ref{line:ero:do_apply_and_copy_response_condition}, \ref{line:ero:add_cell_before_updating_end_of_list_linearization_changed_check}, \ref{line:ero:state_linearization_check}, \ref{line:ero:check_if_already_applied}, or \ref{line:ero:acquire_next_linearization_changed_check}.
    
                        Hence, $p_{n + 1}$ compares a unique low-level operation read from $\linearizationobject$ with either a static value, the current unique low-level operation in $\linearizationobject$, or a unique low-level operation read from $\stateobject$.
                        Since any unique low-level operation written into $\stateobject$ was read from $\linearizationobject$, and any unique low-level operation written into $\linearizationobject$ was read from $\announceobject$, we have that any possible values $p_{n + 1}$ compares do not originate from the response of an $\allocatecelloperation{}$ operation, and so by \Cref{def:reduction:watermarks}, they are not watermarked.
                        Thus, by \Cref{def:reduction:run_mapping}, their state is the same in $C^\mathcal{A}_n$ and $C_n$.
                        Therefore, $p_{n + 1}$ finds the condition on line $\ell$ to be true during the $n + 1$th step of $\mathcal{I}^\mathcal{A}_{n + 1}$ if and only if $p_{n + 1}$ finds the condition on line $\ell$ to be true during the $n + 1$th step of $\mathcal{I}^\mathcal{B}_{n + 1}$, so $s_{n + 1} = s^\mathcal{A}_{n + 1}$ as wanted.
    
                        % Responses
                        \item[] \textbf{Case (b).3.} $\ell$ is either \ref{line:ero:do_work_while_loop} or \ref{line:ero:announce_op_response_check}.
    
                        Hence, by \Cref{claim:reduction:steps_in_a_are_not_on_bad_pointers}, $p_{n + 1}$ reads the current value of the response object $O$ of some cell and compares it to a unique low-level operation and $\nullconstant$.
                        Thus, by \Cref{claim:reduction:b_gets_the_right_response}, $p_{n + 1}$ reads $\mathcal{S}_n(O)$ in $\mathcal{I}^\mathcal{B}_{n + 1}$.
                        Since $O \neq (*\cellpointershort).\nextlong$ for any $\cellpointershort \in \celluniverse \cup \{\&\headobject\}$, and by tracing backward, we can see that any possible value $p_{n + 1}$ compares the state of $O$ to do not originate from the response of an $\allocatecelloperation{}$ operation, by \Cref{def:reduction:watermarks}, they are not watermarked.
                        Hence, by \Cref{def:reduction:swapper}, $\mathcal{S}_n(O)$ is the state assigned to $O$ in $C^\mathcal{A}_n$, and by \Cref{def:reduction:run_mapping}, the relevant local variables of $p_{n + 1}$ are the same in $C^\mathcal{A}_n$ and $C_n$.
                        Therefore, $p_{n + 1}$ finds the condition on line $\ell$ to be true during the $n + 1$th step of $\mathcal{I}^\mathcal{A}_{n + 1}$ if and only if $p_{n + 1}$ finds the condition on line $\ell$ to be true during the $n + 1$th step of $\mathcal{I}^\mathcal{B}_{n + 1}$, so $s_{n + 1} = s^\mathcal{A}_{n + 1}$ as wanted.
    
                        % comparisions to status (done, notdone, found, notfound, lchanged).
                        \item[] \textbf{Case (b).4.} $\ell$ is either \ref{line:ero:done_check}, \ref{line:ero:not_done_check}, \ref{line:ero:add_cell_acquire_next_l_changed}, \ref{line:ero:add_cell_acquire_next_not_found}, \ref{line:ero:add_cell_acquire_next_found}, \ref{line:ero:remove_cell_check_acquire_status}, \ref{line:ero:remove_cell_acquire_next_found}, \ref{line:ero:set_response_found_check}, \ref{line:ero:announce_acquire_l_changed_check}, \ref{line:ero:announce_acquire_relinquish}, \ref{line:ero:acquire_not_found_check}, or \ref{line:ero:acquire_found_check}.
    
                        Hence, $p_{n + 1}$ compares the response of an invocation of the Acquire procedure with a static value.
                        Since these values do not originate from the response of an $\allocatecelloperation{}$ operation, by \Cref{def:reduction:watermarks}, they are not watermarked.
                        Thus, by \Cref{def:reduction:run_mapping}, their state is the same in $C^\mathcal{A}_n$ and $C_n$.
                        Therefore, $p_{n + 1}$ finds the condition on line $\ell$ to be true during the $n + 1$th step of $\mathcal{I}^\mathcal{A}_{n + 1}$ if and only if $p_{n + 1}$ finds the condition on line $\ell$ to be true during the $n + 1$th step of $\mathcal{I}^\mathcal{B}_{n + 1}$, so $s_{n + 1} = s^\mathcal{A}_{n + 1}$ as wanted.

                        \item[] \textbf{Case (b).5.} $\ell$ is \ref{line:ero:remove_cell_remove_seal_loop}.

                        Hence, by \Cref{claim:reduction:steps_in_a_are_not_on_bad_pointers}, $p_{n + 1}$ compares the response of $O = (*\cellpointershort).\nextlong.sealed$ to $\false$ for some $\cellpointershort \in \celluniverse \cup \{\&\headobject\}$.
                        Thus, by \Cref{claim:reduction:b_gets_the_right_response}, $p_{n + 1}$ receives response $\mathcal{S}_n(O, 3)$ in $\mathcal{I}^\mathcal{B}_{n + 1}$.
                        Since the $sealed$ field of $(*\cellpointershort).\nextlong$ is initially $\false$ and is only set to $\true$, by \Cref{def:reduction:watermarks}, it is not watermarked, so $\mathcal{S}_n(O, 3)$ is the same as the value of $(*\cellpointershort).\nextlong.sealed$ in $C^\mathcal{A}_n$.
                        Therefore, $p_{n + 1}$ finds the condition on line $\ell$ to be true during the $n + 1$th step of $\mathcal{I}^\mathcal{A}_{n + 1}$ if and only if $p_{n + 1}$ finds the condition on line $\ell$ to be true during the $n + 1$th step of $\mathcal{I}^\mathcal{B}_{n + 1}$, so $s_{n + 1} = s^\mathcal{A}_{n + 1}$ as wanted.

                        % true or false
                        \item[] \textbf{Case (b).6.} $\ell$ is either \ref{line:ero:remove_cell_from_list} or \ref{line:ero:acquire_next_cell}.
    
                        Hence, by \Cref{claim:reduction:steps_in_a_are_not_on_bad_pointers}, $p_{n + 1}$ compares the response of a \CASop{} operation to $\true$.
                        Since by \Cref{claim:reduction:b_gets_the_right_response}, $p_{n + 1}$ receives the same response during the $n + 1$th step of $\mathcal{I}^\mathcal{B}_{n + 1}$, we have that $p_{n + 1}$ finds the condition on line $\ell$ to be true during the $n + 1$th step of $\mathcal{I}^\mathcal{A}_{n + 1}$ if and only if $p_{n + 1}$ finds the condition on line $\ell$ to be true during the $n + 1$th step of $\mathcal{I}^\mathcal{B}_{n + 1}$, so $s_{n + 1} = s^\mathcal{A}_{n + 1}$ as wanted.
    
                        % pointer to static comparison.
                        \item[] \textbf{Case (b).7.} $\ell$ is \ref{line:ero:free_cell}.
    
                        Hence, by \Cref{claim:reduction:steps_in_a_are_not_on_bad_pointers}, $p_{n + 1}$ compares the response of a \FAop{} operation to -1.
                        Since by \Cref{claim:reduction:b_gets_the_right_response}, $p_{n + 1}$ receives the same response during the $n + 1$th step of $\mathcal{I}^\mathcal{B}_{n + 1}$, we have that $p_{n + 1}$ finds the condition on line $\ell$ to be true during the $n + 1$th step of $\mathcal{I}^\mathcal{A}_{n + 1}$ if and only if $p_{n + 1}$ finds the condition on line $\ell$ to be true during the $n + 1$th step of $\mathcal{I}^\mathcal{B}_{n + 1}$, so $s_{n + 1} = s^\mathcal{A}_{n + 1}$ as wanted.
    
                        % pointer to static comparison.
                        \item[] \textbf{Case (b).8.} $\ell$ is either \ref{line:ero:acquire_next_not_found_check} or \ref{line:ero:early_exit}.
    
                        Hence, $p_{n + 1}$ compares a pointer with $\nullconstant$ or $\&\headobject$.
                        Let $pr$ be this local variable and suppose $C^\mathcal{A}_n$ assigns state $v$ to $pr$.
                        If $p_{n + 1}$ finds this comparison to be true, then $v$ is either $\nullconstant$ or $\&\headobject$.
                        Hence, by \Cref{def:reduction:watermarks}, $\mathcal{W}_n(pr, 1) = \bot$, and so by \Cref{def:reduction:swapper}, $\mathcal{S}_n(pr) = v$.
                        Thus, by \Cref{def:reduction:run_mapping}, $C_n$ assigns state $v$ to $pr$, and so the state of $pr$ is the same in $C^\mathcal{A}_{n}$ and $C_n$.
                        Therefore, $p_{n + 1}$ finds the condition on line $\ell$ to be true during the $n + 1$th step of $\mathcal{I}^\mathcal{B}_{n + 1}$.
                        On the other hand, if $p_{n + 1}$ finds the comparison on line $\ell$ to be false, then $v$ is not $\nullconstant$ nor $\&\headobject$.
                        Note that by \Cref{def:reduction:swapper} $\mathcal{S}_n(pr)$ is either $v$ or $\mathcal{M}(w)$ for some $w$.
                        If $\mathcal{S}_n(pr) = v$, then by \Cref{def:reduction:run_mapping}, $C_n$ assigns state $v$ to $pr$, and so the state of $pr$ is the same in $C^\mathcal{A}_{n}$ and $C_n$.
                        Hence, since $v$ is not $\nullconstant$ nor $\&\headobject$, it follows that $p_{n + 1}$ finds the condition on line $\ell$ to be false during the $n + 1$th step of $\mathcal{I}^\mathcal{B}_{n + 1}$.
                        If $\mathcal{S}_n(pr) = \mathcal{M}(w)$, then by \Cref{def:reduction:run_mapping}, $C_n$ assigns state $\mathcal{M}(w)$ to $pr$.
                        Hence, since $\mathcal{M}(w) \in \celluniverse$, by \Cref{assumption:ero:head_and_null_not_in_cell_universe}, $\mathcal{M}(w) \neq \&\headobject$ and $\mathcal{M}(w) \neq \nullconstant$.
                        Thus, $p_{n + 1}$ finds the condition on line $\ell$ to be false during the $n + 1$th step of $\mathcal{I}^\mathcal{B}_{n + 1}$.
                        Therefore $p_{n + 1}$ finds the condition on line $\ell$ to be true during the $n + 1$th step of $\mathcal{I}^\mathcal{A}_{n + 1}$ if and only if $p_{n + 1}$ finds the condition on line $\ell$ to be true during the $n + 1$th step of $\mathcal{I}^\mathcal{B}_{n + 1}$, so $s_{n + 1} = s^\mathcal{A}_{n + 1}$ as wanted.
    
                        % pointer to pointer comparison.
                        \item[] \textbf{Case (b).9.} $\ell$ is either \ref{line:ero:add_cell_while_loop}, \ref{line:ero:remove_cell_while_loop}, or \ref{line:ero:acquire_loop_until}.
    
                        \Cref{claim:reduction:b_does_traversal_checks_correctly} covers the case where $p_{n + 1}$ finds the condition on $\ell$ to be false during the $n + 1$th step of $\mathcal{I}^\mathcal{A}_{n + 1}$.
                        So, suppose $p_{n + 1}$ finds the condition on $\ell$ to be true during the $n + 1$th step of $\mathcal{I}^\mathcal{A}_{n + 1}$.
                        In this case, $\ell$ compares the value of two local variables, say $pr_1$ and $pr_2$.
                        Observe that, in all cases, the state of $pr_1$ and $pr_2$ is a single value, so by \Cref{def:reduction:swapper}, $\mathcal{S}_n(pr_1) = \mathcal{S}_n(pr_1, 1)$ and $\mathcal{S}_n(pr_2) = \mathcal{S}_n(pr_2, 1)$.
                        Let $v_1$ (resp. $v_2$) be the state of $pr_1$ (resp. $pr_2$) in $C^\mathcal{A}_n$.
                        Hence, since $p_{n + 1}$ finds the condition on $\ell$ to be true during the $n + 1$th step of $\mathcal{I}^\mathcal{A}_{n + 1}$, we have that $v_1 \neq v_2$.
                        Furthermore, since $v_1$ (resp. $v_2$) is the state of a local variable in $C^\mathcal{A}_n$, by \Cref{def:reduction:used_pointers_in_run_of_a}, if $v_1 \in \celluniverse$ (resp. $v_2 \in \celluniverse$), then $v_1 \in \celluniverse(\mathcal{I}^\mathcal{A}_{n + 1})$ (resp. $v_2 \in \celluniverse(\mathcal{I}^\mathcal{A}_{n + 1})$).
                        Hence, by \Cref{lemma:ero:psi_is_basically_injective}, $\mathcal{S}_n(pr_1, 1) \neq \mathcal{S}_n(pr_2, 1)$.
                        Thus, since $\mathcal{S}_n(pr_1) = \mathcal{S}_n(pr_1, 1)$ and $\mathcal{S}_n(pr_2) = \mathcal{S}_n(pr_2, 1)$, we have that $\mathcal{S}_n(pr_1) \neq \mathcal{S}_n(pr_2)$.
                        So, by \Cref{def:reduction:run_mapping}, the state of $pr_1$ and $pr_2$ differ in $C_n$, implying $p_{n + 1}$ finds the condition on $\ell$ to be true during the $n + 1$th step of $\mathcal{I}^\mathcal{B}_{n + 1}$.
                        Therefore $p_{n + 1}$ finds the condition on line $\ell$ to be true during the $n + 1$th step of $\mathcal{I}^\mathcal{A}_{n + 1}$ if and only if $p_{n + 1}$ finds the condition on line $\ell$ to be true during the $n + 1$th step of $\mathcal{I}^\mathcal{B}_{n + 1}$, so $s_{n + 1} = s^\mathcal{A}_{n + 1}$ as wanted.

                        % pointer to pointer comparison.
                        \item[] \textbf{Case (b).10.} $\ell$ is \ref{line:ero:remove_cell_before_removal_linearization_check}.

                        \Cref{claim:reduction:b_does_already_remove_check_correctly} covers the case where $p_{n + 1}$ finds the condition on $\ell$ to be true during the $n + 1$th step of $\mathcal{I}^\mathcal{A}_{n + 1}$.
                        So, suppose $p_{n + 1}$ finds the condition on $\ell$ to be false during the $n + 1$th step of $\mathcal{I}^\mathcal{A}_{n + 1}$.
                        In this case, $\ell$ compares a unique low-level operation read from $\linearizationobject$ with the unique low-level operation currently in $\linearizationobject$ and the value of two local variables, say $pr_1$ and $pr_2$.
                        The former is covered by Case (b).3, so it suffices to consider the case where $p_{n + 1}$ finds the condition on $\ell$ to be false during the $n + 1$th step of $\mathcal{I}^\mathcal{A}_{n + 1}$, because the values of $pr_1$ and $pr_2$ differ in $C^\mathcal{A}_n$.
                        The argument is the same as the previous case, which is provided below for completeness.
                        Observe that, in all cases, the state of $pr_1$ and $pr_2$ is a single value, so by \Cref{def:reduction:swapper}, $\mathcal{S}_n(pr_1) = \mathcal{S}_n(pr_1, 1)$ and $\mathcal{S}_n(pr_2) = \mathcal{S}_n(pr_2, 1)$.
                        Let $v_1$ (resp. $v_2$) be the state of $pr_1$ (resp. $pr_2$) in $C^\mathcal{A}_n$.
                        Hence, $v_1 \neq v_2$.
                        Furthermore, since $v_1$ (resp. $v_2$) is the state of a local variable in $C^\mathcal{A}_n$, by \Cref{def:reduction:used_pointers_in_run_of_a}, if $v_1 \in \celluniverse$ (resp. $v_2 \in \celluniverse$), then $v_1 \in \celluniverse(\mathcal{I}^\mathcal{A}_{n + 1})$ (resp. $v_2 \in \celluniverse(\mathcal{I}^\mathcal{A}_{n + 1})$).
                        Hence, by \Cref{lemma:ero:psi_is_basically_injective}, $\mathcal{S}_n(pr_1, 1) \neq \mathcal{S}_n(pr_2, 1)$.
                        Thus, since $\mathcal{S}_n(pr_1) = \mathcal{S}_n(pr_1, 1)$ and $\mathcal{S}_n(pr_2) = \mathcal{S}_n(pr_2, 1)$, we have that $\mathcal{S}_n(pr_1) \neq \mathcal{S}_n(pr_2)$.
                        So, by \Cref{def:reduction:run_mapping}, the state of $pr_1$ and $pr_2$ differ in $C_n$, implying $p_{n + 1}$ finds the condition on $\ell$ to be false during the $n + 1$th step of $\mathcal{I}^\mathcal{B}_{n + 1}$.
                        Therefore $p_{n + 1}$ finds the condition on line $\ell$ to be true during the $n + 1$th step of $\mathcal{I}^\mathcal{A}_{n + 1}$ if and only if $p_{n + 1}$ finds the condition on line $\ell$ to be true during the $n + 1$th step of $\mathcal{I}^\mathcal{B}_{n + 1}$, so $s_{n + 1} = s^\mathcal{A}_{n + 1}$ as wanted.
                    \end{itemize}
                \end{itemize}
            \end{itemize}
            We now finish the proof of \Cref{claim:reduction:b_updates_program_counter_correctly}.
            Since $pr$ is a program counter, by \Cref{def:reduction:watermarks}, $\mathcal{W}_{n + 1}(pr, 1) = \bot$.
            Hence, since $s^\mathcal{A}_{n + 1}$ is the state assigned to $pr$ in $C^\mathcal{A}_{n + 1}$, by \Cref{def:reduction:swapper}, $\mathcal{S}_{n + 1}(pr) = \mathcal{S}_{n + 1}(pr, 1) = s^\mathcal{A}_{n + 1}$.
            Thus, since $s_{n + 1} = s^\mathcal{A}_{n + 1}$, we have that $s_{n + 1} = \mathcal{S}_{n + 1}(pr)$.
            Therefore, since $C^\mathcal{B}_{n + 1}$ assigns state $s_{n + 1}$ to $pr$, we have that $C^\mathcal{B}_{n + 1}$ assigns state $\mathcal{S}_{n + 1}(pr)$ to $pr$.
            \qH{\Cref{claim:reduction:b_updates_program_counter_correctly}}
        \end{proof}
    
        We now finish the inductive case of \Cref{lemma:reduction:mapped_history_is_for_b}.
        Recall we must prove that $C^\mathcal{B}_{n + 1} = C_{n + 1}$.
        Since by \Cref{claim:reduction:b_updates_private_registers_correctly} and \Cref{claim:reduction:b_updates_program_counter_correctly}, $C^\mathcal{B}_{n + 1}$ assigns $\mathcal{S}_{n + 1}(pr)$ to $pr$ where $pr$ is any local variable.
        Therefore, by \Cref{def:reduction:run_mapping}, $C^\mathcal{B}_{n + 1}$ and $C_{n + 1}$ assign every local variable to the same state.
        What remains is to prove that $C^\mathcal{B}_{n + 1}$ and $C_{n + 1}$ assign every base object to the same state.
        There are three cases.
        \begin{itemize}
            \item[] \hspace{0pt}\textbf{Case 1.} The $n + 1$th step of $\mathcal{I}^\mathcal{A}_{n + 1}$ does not perform an operation on a base object.
    
            Hence, the state of every base object is the same in $C^\mathcal{A}_n$ and $C^\mathcal{A}_{n + 1}$, and $p_{n + 1}$ does not set any index of any base object $O$ during the $n + 1$th step of $\mathcal{I}^\mathcal{A}_{n + 1}$, so by \Cref{def:reduction:watermarks}, $\mathcal{W}_{n + 1}(O, i) = \mathcal{W}_n(O, i)$.
            Therefore, since the state of $O$ is the same in $C^\mathcal{A}_n$ and $C^\mathcal{A}_{n + 1}$, by \Cref{def:reduction:swapper}, $\mathcal{S}_{n + 1}(O, i) = \mathcal{S}_n(O, i)$, and so $\mathcal{S}_{n + 1}(O) = \mathcal{S}_n(O)$ for any base object $O$ other than the memory manager (*).
    
            \textbf{Claim:} The state of every base object is the same in $C_{n}$ and $C_{n + 1}$.
            
            Consider any base object $O$ other than the memory manager.
    
            \textbf{Case A.} $O \notin \celluniverse_O$ or $O$ is in $\headobject$.
            
            Hence, by \Cref{def:reduction:run_mapping}, $C_n$ (resp. $C_{n + 1}$) assigns state $\mathcal{S}_n(O)$ (resp. $\mathcal{S}_{n + 1}(O)$) to $O$.
            Therefore, since by (*) $\mathcal{S}_{n + 1}(O) = \mathcal{S}_n(O)$, we have that $C_n$ and $C_{n + 1}$ assigns the same state to $O$.
    
            \textbf{Case B.} $O \in \celluniverse_O$ and $O$ is not in $\headobject$.
            
            Hence, by \Cref{def:reduction:shared_objects}, $O = (*\cellpointershort).f$ for some $\cellpointershort \in \celluniverse$ and $f$ which is either $\lastrepositoryoperationresponse{}$, $\revocations$, or $\nextlong$.
            Let $W_n$ (resp. $W_{n + 1}$) be the set of step numbers up to and including the $n$th (resp. $n + 1$th) step of $\mathcal{I}^\mathcal{A}_{n + 1}$ which perform $\allocatecelloperation{}$ operations.
            Since the $n + 1$th step of $\mathcal{I}^\mathcal{A}_{n + 1}$ does not perform an operation on a base object, we have that the $n + 1$th step of $\mathcal{I}^\mathcal{A}_{n + 1}$ does not perform an $\allocatecelloperation{}$ operation, and so $W_n = W_{n + 1} = W$.
    
            \textbf{Case B.1.} $\cellpointershort \notin \mathcal{M}[W]$.
            
            Hence, by \Cref{def:reduction:run_mapping}, $C_n$ and $C_{n + 1}$ assign $O$ to its initial state.
    
            \textbf{Case B.2.} $\cellpointershort \in \mathcal{M}[W]$.
            
            Hence, $\mathcal{M}(w) = \cellpointershort$ for some $w \in W$.
            Thus, since $O = (*\cellpointershort).f$, we have that $O = (*\mathcal{M}(w)).f$.
            Let $\cellpointershort'$ be the response of the $\allocatecelloperation{}$ operation performed during the $w$th step of $\mathcal{I}^\mathcal{A}_{n + 1}$.
    
            \textbf{Case B.2.1.} For all $w < j \leq n$ the $j$th step of $\mathcal{I}^\mathcal{A}_{n + 1}$ does not perform an $\allocatecelloperation{}$ operation whose response is $\cellpointershort'$.
            
            Since the $n + 1$th step of $\mathcal{I}^\mathcal{A}_{n + 1}$ does not perform an $\allocatecelloperation{}$ operation, we have that for all $w < j \leq n + 1$ the $j$th step of $\mathcal{I}^\mathcal{A}_{n + 1}$ does not perform an $\allocatecelloperation{}$ operation whose response is $\cellpointershort'$.
            Hence, by \Cref{def:reduction:run_mapping}, $C_n$ (resp. $C_{n + 1}$) assigns state $\mathcal{S}_n((*\cellpointershort').f)$ (resp. $\mathcal{S}_{n + 1}((*\cellpointershort').f)$) to $(*\mathcal{M}(w)).f$.
            Thus, since by \Cref{alg:cell_manager_specification} $\cellpointershort' \in \celluniverse$, $(*\cellpointershort').f$ is a base object other than the memory manager, and so by (*) $\mathcal{S}_{n + 1}((*\cellpointershort').f) = \mathcal{S}_n((*\cellpointershort').f)$.
            So, $C_n$ and $C_{n + 1}$ assign the same state to $(*\mathcal{M}(w)).f$.
            Therefore, since $O = (*\mathcal{M}(w)).f$, $C_n$ and $C_{n + 1}$ assign the same state to $O$.

            \textbf{Case B.2.2.} For some $w < j \leq n$ the $j$th step of $\mathcal{I}^\mathcal{A}_{n + 1}$ performs an $\allocatecelloperation{}$ operation whose response is $\cellpointershort'$.
    
            Let $j$ be the minimum $w < j \leq n$ such that the $j$th step of $\mathcal{I}^\mathcal{A}_{n + 1}$ performs an $\allocatecelloperation{}$ cell operation whose response is $\cellpointershort'$.
            Hence, $j$ is also the minimum $w < j \leq n + 1$ such that the $j$th step of $\mathcal{I}^\mathcal{A}_{n + 1}$ performs an $\allocatecelloperation{}$ cell operation whose response is $\cellpointershort'$.
            Thus, by \Cref{def:reduction:run_mapping}, $C_n$ and $C_{n + 1}$ assigns state $\mathcal{S}_{j - 1}((*\cellpointershort').f)$ to $(*\mathcal{M}(w)).f$.
            Therefore, since $O = (*\mathcal{M}(w)).f$, we have that $C_n$ and $C_{n + 1}$ assign the same state to $O$.

            This completes the proof of Case B.
    
            Finally, consider the memory manager.
            By \Cref{def:reduction:run_mapping}, $C_n$ (resp. $C_{n + 1}$) assigns state $\mathcal{M}[W_n]$ (resp. $\mathcal{M}[W_{n + 1}]$) to the memory manager, and since $W_n = W_{n + 1}$, we have that $\mathcal{M}[W_n] = \mathcal{M}[W_{n + 1}]$, and so $C_n$ and $C_{n + 1}$ assign the same state to the memory manager.
    
            This completes the proof of \textbf{Claim}.
    
            We now finish the proof of Case 1.
            Since the $n + 1$th step of $\mathcal{I}^\mathcal{A}_{n + 1}$ does not perform an operation on a base object, and by \Cref{lemma:reduction:same_program_counters_in_mapped_run} the program counter of every process is the same in $C^\mathcal{A}_n$ and $C_n$, we have that the $n + 1$th step of $\mathcal{I}^\mathcal{B}_{n + 1}$ does not perform an operation on a base object.
            Hence, the state of every base object is the same in $C_n$ and $C^\mathcal{B}_{n + 1}$.
            Therefore, since by \textbf{Claim} the state of every base object is the same in $C_{n}$ and $C_{n + 1}$, we have that the state of every base object is the same in $C^\mathcal{B}_{n + 1}$ and $C_{n + 1}$ as wanted.
    
            \item[] \hspace{0pt}\textbf{Case 2.} The $n + 1$th step of $\mathcal{I}^\mathcal{A}_{n + 1}$ performs an operation on the memory manager.

            The main challenge to be dealt with in this case compared to the last case is that in $\mathcal{A}$, $\allocatecelloperation{}$ operations set the state of each object of the cell they return to their initial state while in $\mathcal{B}$ $\allocatecelloperation{}$ operations do not change the state of the objects of the cell they return.
    
            \textbf{Claim 1:} For every base object $O$ in $\mathcal{A}$ such that $O$ is not the memory manager, and $O \notin \celluniverse_O$ or $O$ is in $\headobject$, $\mathcal{S}_{n + 1}(O) = \mathcal{S}_n(O)$.
    
            Since the $n + 1$th step of $\mathcal{I}^\mathcal{A}_{n + 1}$ performs an operation on the memory manager, the state of $O$ is the same in $C^\mathcal{A}_n$ and $C^\mathcal{A}_{n + 1}$, and $p_{n + 1}$ does not set any index of $O$ during the $n + 1$th step of $\mathcal{I}^\mathcal{A}_{n + 1}$.
            Thus, for any index $i$ of the state of $O$, by \Cref{def:reduction:watermarks}, $\mathcal{W}_{n + 1}(O, i) = \mathcal{W}_n(O, i)$.
            Therefore, since the state of $O$ is the same in $C^\mathcal{A}_n$ and $C^\mathcal{A}_{n + 1}$, by \Cref{def:reduction:swapper}, $\mathcal{S}_{n + 1}(O, i) = \mathcal{S}_n(O, i)$, and so $\mathcal{S}_{n + 1}(O) = \mathcal{S}_n(O)$ as wanted.
    
            \textbf{Claim 2:} If the $n + 1$th step of $\mathcal{I}^\mathcal{A}_{n + 1}$ does not perform an $\allocatecelloperation{}$ operation whose response is $\cellpointershort$, then for every base object $O$ of the cell pointed to by $\cellpointershort$ $\mathcal{S}_{n + 1}(O) = \mathcal{S}_n(O)$.
    
            Consider any base object $O$ of the cell pointed to by $\cellpointershort$.
            Since the $n + 1$th step of $\mathcal{I}^\mathcal{A}_{n + 1}$ performs an operation on the memory manager and the $n + 1$th step of $\mathcal{I}^\mathcal{A}_{n + 1}$ does not perform an $\allocatecelloperation{}$ operation whose response is $\cellpointershort$, the state of $O$ is the same in $C^\mathcal{A}_n$ and $C^\mathcal{A}_{n + 1}$, and $p_{n + 1}$ does not set any index of $O$ during the $n + 1$th step of $\mathcal{I}^\mathcal{A}_{n + 1}$.
            Thus, for any index $i$ of the state of $O$, by \Cref{def:reduction:watermarks}, $\mathcal{W}_{n + 1}(O, i) = \mathcal{W}_n(O, i)$.
            Therefore, since the state of $O$ is the same in $C^\mathcal{A}_n$ and $C^\mathcal{A}_{n + 1}$, by \Cref{def:reduction:swapper}, $\mathcal{S}_{n + 1}(O, i) = \mathcal{S}_n(O, i)$, and so $\mathcal{S}_{n + 1}(O) = \mathcal{S}_n(O)$ as wanted.
            
            \textbf{Claim 3:} The state of every base object other than the memory manager is the same in $C_{n}$ and $C_{n + 1}$.
            
            Consider any base object $O$ other than the memory manager.
    
            \textbf{Case A.} $O \notin \celluniverse_O$ or $O$ is in $\headobject$.
            
            Hence, by \Cref{def:reduction:run_mapping}, $C_n$ (resp. $C_{n + 1}$) assigns state $\mathcal{S}_n(O)$ (resp. $\mathcal{S}_{n + 1}(O)$) to $O$.
            Therefore, since by \textbf{Claim 1} $\mathcal{S}_{n + 1}(O) = \mathcal{S}_n(O)$, $C_n$ and $C_{n + 1}$ assigns the same state to $O$.
    
            \textbf{Case B.} $O \in \celluniverse_O$ and $O$ is not in $\headobject$.
            
            Hence, by \Cref{def:reduction:shared_objects}, $O = (*\cellpointershort).f$ for some $\cellpointershort \in \celluniverse$ and $f$ which is either $\lastrepositoryoperationresponse{}$, $\revocations$, or $\nextlong$.
            Let $W_n$ (resp. $W_{n + 1}$) be the set of step numbers up to and including the $n$th (resp. $n + 1$th) step of $\mathcal{I}^\mathcal{A}_{n + 1}$ which perform $\allocatecelloperation{}$ operations.
            Since the $n + 1$th step of $\mathcal{I}^\mathcal{A}_{n + 1}$ performs an operation on the memory manager, we have that $W_{n + 1} = W_n \cup \{n + 1\}$.
            There are two cases.
    
            \textbf{Case B.1.} $\cellpointershort \notin \mathcal{M}[W_n]$.
            
            Hence, by \Cref{def:reduction:run_mapping}, $C_n$ assign $O$ to its initial state.
            There are two more cases.
    
            \textbf{Case B.1.1.} $\cellpointershort \notin \mathcal{M}[W_{n + 1}]$.
            
            Hence, by \Cref{def:reduction:run_mapping}, $C_{n + 1}$ assign $O$ to its initial state, which is the same as $C_n$.
    
            \textbf{Case B.1.2.} $\cellpointershort \in \mathcal{M}[W_{n + 1}]$.
            
            Hence, since $\cellpointershort \notin \mathcal{M}[W_n]$, we have that $\cellpointershort = \mathcal{M}(w)$ for some $w \in W_{n + 1}\setminus W_n$.
            Thus, since $W_{n + 1} = W_n \cup \{n + 1\}$, we have that $w = n + 1$, so $\cellpointershort = \mathcal{M}(n + 1)$.
            Therefore, since $O = (*\cellpointershort).f$, we have that $O = (*\mathcal{M}(n + 1)).f$.
            Let $\cellpointershort'$ be the response of the $\allocatecelloperation{}$ performed during the $n + 1$th step of $\mathcal{I}^\mathcal{A}_{n + 1}$.
            Hence, the set of steps satisfying $n + 1 < j \leq n + 1$ is empty, so by \Cref{def:reduction:run_mapping}, $C_{n + 1}$ assigns state $\mathcal{S}_{n + 1}((*\cellpointershort').f)$ to $((*\mathcal{M}(n + 1)).f)$.
            Since the $n + 1$th step of $\mathcal{I}^\mathcal{A}_{n + 1}$ performs an $\allocatecelloperation{}$ operation with response $\cellpointershort'$, by \Cref{def:reduction:algorithm_a}, $C^\mathcal{A}_{n + 1}$ assigns $(*\cellpointershort').f$ to its initial state.
            Hence, since every value of every index of the initial state of $(*\cellpointershort').f$ is not in $\celluniverse$, by \Cref{def:reduction:watermarks}, $\mathcal{W}_{n + 1}((*\cellpointershort').f, i) = \bot$ for any index $i$ of the state of $(*\cellpointershort').f$.
            Thus, by \Cref{def:reduction:swapper}, $\mathcal{S}_{n + 1}((*\cellpointershort').f)$ is the initial state of $(*\cellpointershort').f$.
            So, since $C_{n + 1}$ assigns state $\mathcal{S}_{n + 1}((*\cellpointershort').f)$ to $((*\mathcal{M}(n + 1)).f)$, we have that $C_{n + 1}$ assigns $(*\mathcal{M}(n + 1)).f$ to its initial state.
            Therefore, since $O = (*\mathcal{M}(n + 1)).f$, we have that $C_{n + 1}$ assign $O$ to its initial state, which is the same as $C_n$ as wanted.
    
            \textbf{Case B.2.} $\cellpointershort \in \mathcal{M}[W_n]$.
            
            Hence, $\mathcal{M}(w) = \cellpointershort$ for some $w \in W_n$, so since $O = (*\cellpointershort).f$, we have that $O = (*\mathcal{M}(w)).f$.
            Thus, since $W_{n + 1} = W_n \cup \{n + 1\}$, we have that $w \in W_{n + 1}$, and so $\cellpointershort \in \mathcal{M}[W_{n + 1}]$.
            Let $\cellpointershort'$ be the response of the $\allocatecelloperation{}$ operation performed during the $w$th step of $\mathcal{I}^\mathcal{A}_{n + 1}$.
            There are two cases.
    
            \textbf{Case B.2.1.} For all $w < j \leq n$ the $j$th step of $\mathcal{I}^\mathcal{A}_{n + 1}$ does not perform an $\allocatecelloperation{}$ operation whose response is $\cellpointershort'$.
            
            Hence, by \Cref{def:reduction:run_mapping}, $C_n$ assigns state $\mathcal{S}_n((*\cellpointershort').f)$ to $(*\mathcal{M}(w)).f$.
            Thus, since $O = (*\mathcal{M}(w)).f$, we have that $C_n$ assigns state $\mathcal{S}_n((*\cellpointershort').f)$ to $O$.
            There are two more cases.
    
            \textbf{Case B.2.1.1.} For all $w < j \leq n + 1$ the $j$th step of $\mathcal{I}^\mathcal{A}_{n + 1}$ does not perform an $\allocatecelloperation{}$ operation whose response is $\cellpointershort'$.

            Hence, by \textbf{Claim 2}, $\mathcal{S}_{n + 1}((*\cellpointershort').f) = \mathcal{S}_n((*\cellpointershort').f)$.
            Furthermore, by \Cref{def:reduction:run_mapping}, $C_{n + 1}$ assigns state $\mathcal{S}_{n + 1}((*\cellpointershort').f)$ to $(*\mathcal{M}(w)).f$.
            Thus, since $\mathcal{S}_{n + 1}((*\cellpointershort').f) = \mathcal{S}_n((*\cellpointershort').f)$, we have that $C_{n + 1}$ assigns state $\mathcal{S}_{n}((*\cellpointershort').f)$ to $(*\mathcal{M}(w)).f$.
            Therefore, since $O = (*\mathcal{M}(w)).f$, we have that $C_{n + 1}$ assigns state $\mathcal{S}_{n}((*\cellpointershort').f)$ to $O$ which is the same as $C_n$ as wanted.
    
            \textbf{Case B.2.1.2.} For some $w < j \leq n + 1$ the $j$th step of $\mathcal{I}^\mathcal{A}_{n + 1}$ performs an $\allocatecelloperation{}$ operation whose response is $\cellpointershort'$.
            
            Hence, since for all $w < j \leq n$ the $j$th step of $\mathcal{I}^\mathcal{A}_{n + 1}$ does not perform an $\allocatecelloperation{}$ operation whose response is $\cellpointershort'$, we have that the $n + 1$th step of $\mathcal{I}^\mathcal{A}_{n + 1}$ performs an $\allocatecelloperation{}$ operation whose response is $\cellpointershort'$.
            Thus, $j = n + 1$ is the minimum $w < j \leq n + 1$ such that the $j$th step of $\mathcal{I}^\mathcal{A}_{n + 1}$ performs an $\allocatecelloperation{}$ operation whose response is $\cellpointershort'$.
            Therefore, by \Cref{def:reduction:run_mapping}, $C_{n + 1}$ assigns state $\mathcal{S}_{n}((*\cellpointershort').f)$ to $(*\mathcal{M}(w)).f = O$, which is the same as $C_n$.
    
            \textbf{Case B.2.2.} For some $w < j \leq n$ the $j$th step of $\mathcal{I}^\mathcal{A}_{n + 1}$ performs an $\allocatecelloperation{}$ operation whose response is $\cellpointershort'$.
    
            Let $j$ be the minimum $w < j \leq n$ such that the $j$th step of $\mathcal{I}^\mathcal{A}_{n + 1}$ performs an $\allocatecelloperation{}$ cell operation whose response is $\cellpointershort'$.
            Hence, $j$ is also the minimum $w < j \leq n + 1$ such that the $j$th step of $\mathcal{I}^\mathcal{A}_{n + 1}$ performs an $\allocatecelloperation{}$ cell operation whose response is $\cellpointershort'$.
            Thus, by \Cref{def:reduction:run_mapping}, $C_n$ and $C_{n + 1}$ assigns state $\mathcal{S}_{j - 1}((*\cellpointershort').f)$ to $(*\mathcal{M}(w)).f$.
            Therefore, since $O = (*\mathcal{M}(w)).f$, we have that $C_n$ and $C_{n + 1}$ assign the same state to $O$ as wanted.
            
            This completes the proof of \textbf{Claim 3.}
    
            We now finish the proof of Case 2.
            Since the $n + 1$th step of $\mathcal{I}^\mathcal{A}_{n + 1}$ performs an operation on the memory manager, and by \Cref{lemma:reduction:same_program_counters_in_mapped_run} the program counter of every process is the same in $C^\mathcal{A}_n$ and $C_n$, we have that the $n + 1$th step of $\mathcal{I}^\mathcal{B}_{n + 1}$ performs an operation on the memory manager.
            Hence, the state of every base object other than the memory manager is the same in $C_n$ and $C^\mathcal{B}_{n + 1}$.
            Thus, since by \textbf{Claim 3} the state of every base object other than the memory manager is the same in $C_{n}$ and $C_{n + 1}$, we have that the state of every base object other than the memory manager is the same in $C^\mathcal{B}_{n + 1}$ and $C_{n + 1}$.
            Therefore, since by \Cref{claim:reduction:c_b_i+1_and_c_i+1_same_cell_manager} the state of the memory manager is the same in $C^\mathcal{B}_{n + 1}$ and $C_{n + 1}$, we have that the state of every base object is the same in $C^\mathcal{B}_{n + 1}$ and $C_{n + 1}$ as wanted.
    
            \item[] \hspace{0pt}\textbf{Case 3.} The $n + 1$th step of $\mathcal{I}^\mathcal{A}_{n + 1}$ performs an operation on a base object $O$ other than the memory manager.
    
            Hence, the state of every base object except $O$ is the same in $C^\mathcal{A}_n$ and $C^\mathcal{A}_{n + 1}$, and $p_{n + 1}$ does not set any index of any base object except $O$ during the $n + 1$th step of $\mathcal{I}^\mathcal{A}_{n + 1}$.
            Thus, for any base object $O'$ other than $O$ and the memory manager and index $i$ of the state of $O'$, by \Cref{def:reduction:watermarks}, $\mathcal{W}_{n + 1}(O', i) = \mathcal{W}_n(O', i)$.
            Therefore, since the state of every base object except $O$ is the same in $C^\mathcal{A}_n$ and $C^\mathcal{A}_{n + 1}$, by \Cref{def:reduction:swapper}, $\mathcal{S}_{n + 1}(O', i) = \mathcal{S}_n(O', i)$, and so $\mathcal{S}_{n + 1}(O') = \mathcal{S}_n(O')$ (*).
    
            \textbf{Claim:} The state of every base object except $O^*$ is the same in $C_{n}$ and $C_{n + 1}$ where $O^*$ is defined as: $O$ if $O \notin \celluniverse_O$ and otherwise $O$ is the object $f$ of some cell where $f$ is either $\lastrepositoryoperationresponse{}$, $\revocations$, or $\nextlong$, and $O^* = (*\mathcal{S}_{n}(O_s)).f$ where $O_s$ is the source of $O$ (see \Cref{observation:reduction:source_of_pointer}).
            
            Consider any base object $O'$ other than $O^*$ and the memory manager.
            
            \textbf{Case A.} $O' \notin \celluniverse_O$ or $O'$ is in $\headobject$.
            
            We first prove that $O' \neq O$.
            If $O \notin \celluniverse_O$ or $O$ is in $\headobject$, then $O^* = O$, and so since $O' \neq O^*$, we have that $O' \neq O$.
            Otherwise, if $O \in \celluniverse_O$ and $O$ is not in $\headobject$, then since $O' \notin \celluniverse_O$ or $O'$ is in $\headobject$, we have that $O' \neq O$.
            Therefore, in all cases, $O' \neq O$ as wanted.
            
            We now finish the proof of Case A. 
            Since $O' \notin \celluniverse_O$ or $O'$ is in $\headobject$, by \Cref{def:reduction:run_mapping}, $C_n$ (resp. $C_{n + 1}$) assigns state $\mathcal{S}_n(O')$ (resp. $\mathcal{S}_{n + 1}(O')$) to $O'$.
            Hence, since $O'$ is a base object other than $O$ and the memory manager, by (*), $\mathcal{S}_{n + 1}(O') = \mathcal{S}_n(O')$.
            Therefore, $C_n$ and $C_{n + 1}$ assigns the same state to $O'$ as wanted.
            
            \textbf{Case B.} $O' \in \celluniverse_O$ and $O'$ is not in $\headobject$.
            
            Hence, by \Cref{def:reduction:shared_objects}, $O' = (*\cellpointershort').f'$ for some $\cellpointershort' \in \celluniverse$ and $f'$ which is either $\lastrepositoryoperationresponse{}$, $\revocations$, or $\nextlong$.        
            Let $W_n$ (resp. $W_{n + 1}$) be the set of step numbers up to and including the $n$th (resp. $n + 1$th) step of $\mathcal{I}^\mathcal{A}_{n + 1}$ which perform $\allocatecelloperation{}$ operations.
            Since the $n + 1$th step of $\mathcal{I}^\mathcal{A}_{n + 1}$ performs an operation on a base object other than the memory manager, the $n + 1$th step of $\mathcal{I}^\mathcal{A}_{n + 1}$ does not perform an $\allocatecelloperation{}$ operation.
            Hence, $W_n = W_{n + 1} = W$.
            There are two cases.
    
            \textbf{Case B.1.} $\cellpointershort' \notin \mathcal{M}[W]$.
            
            Hence, by \Cref{def:reduction:run_mapping}, $C_n$ and $C_{n + 1}$ assign $O'$ to its initial state.
            
            \textbf{Case B.2.} $\cellpointershort' \in \mathcal{M}[W]$.
            
            Hence, for some $w' \in W$ $\mathcal{M}(w') = \cellpointershort'$.
            Thus, since $O' = (*\cellpointershort').f'$, we have that $O' = (*\mathcal{M}(w')).f'$.
            Let $\cellpointershort$ be the response of the $\allocatecelloperation{}$ operation performed during the $w'$th step of $\mathcal{I}^\mathcal{A}_{n + 1}$, so by \Cref{alg:cell_manager_specification} $\cellpointershort \in \celluniverse$.
            There are two more cases.
            
            \textbf{Case B.2.1.} For all $w' < j \leq n$ the $j$th step of $\mathcal{I}^\mathcal{A}_{n + 1}$ does not perform an $\allocatecelloperation{}$ operation whose response is $\cellpointershort$.
            
            Since the $n + 1$th step of $\mathcal{I}^\mathcal{A}_{n + 1}$ does not perform an $\allocatecelloperation{}$ operation, we have that for all $w' < j \leq n + 1$ the $j$th step of $\mathcal{I}^\mathcal{A}_{n + 1}$ does not perform an $\allocatecelloperation{}$ operation whose response is $\cellpointershort$.
            Hence, by \Cref{def:reduction:run_mapping}, $C_n$ (resp. $C_{n + 1}$) assigns state $\mathcal{S}_n((*\cellpointershort).f')$ (resp. $\mathcal{S}_{n + 1}((*\cellpointershort).f')$) to $(*\mathcal{M}(w')).f'$.
    
            We now prove that $O \neq (*\cellpointershort).f'$.
            Suppose, for contradiction, $O = (*\cellpointershort).f'$.
            Hence, since $\cellpointershort \in \celluniverse$, by \Cref{def:reduction:shared_objects}, $O \in \celluniverse_O$.
            Thus, by the definition of $O^*$, we have that $O^* = (*\mathcal{S}_{n}(O_s)).f'$.
            Let $\mathcal{W}_n(O_s, 1) = w$.
            Since $O = (*\cellpointershort).f'$ where $\cellpointershort \in \celluniverse$, by \Cref{claim:reduction:op_on_ptr_after_allocate_in_a}, $w$th step of $\mathcal{I}^\mathcal{A}_{n + 1}$ performs an $\allocatecelloperation{}$ operation whose response is $\cellpointershort$ and $\mathcal{S}_{n}(O_s) = \mathcal{M}(w)$.
            Hence, since $O^* = (*\mathcal{S}_{n}(O_s)).f'$, we have that $O^* = (*\mathcal{M}(w)).f'$.
            Furthermore, by \Cref{claim:reduction:if_performed_op_on_cell_it_is_the_latest_version}, for all $w < i \leq n + 1$ the $i$th step of $\mathcal{I}^\mathcal{A}_{n + 1}$ does not perform an $\allocatecelloperation{}$ operation whose response is $\cellpointershort$ where $O$ is an object of the cell pointed to by $\cellpointershort$.
            Since both the $w'$th and $w$th step of $\mathcal{I}^\mathcal{A}_{n + 1}$ perform an $\allocatecelloperation{}$ operation whose response is $\cellpointershort$, for all $w' < j \leq n + 1$ the $j$th step of $\mathcal{I}^\mathcal{A}_{n + 1}$ does not perform an $\allocatecelloperation{}$ operation whose response is $\cellpointershort$, and for all $w < i \leq n + 1$ the $i$th step of $\mathcal{I}^\mathcal{A}_{n + 1}$ does not perform an $\allocatecelloperation{}$ operation whose response is $\cellpointershort$, it follows that $w' = w$.
            % In other words, since both the $w'$ and $w$th step are the last step in $\mathcal{I}^\mathcal{A}_{n + 1}$ to perform $\allocatecelloperation{}$ operations whose response is $\cellpointershort$, $w' = w$.
            Therefore, since $O' = (*\mathcal{M}(w')).f'$, we have that $O' = (*\mathcal{M}(w)).f'$, and so $O' = O^*$.
            However, $O'$ was chosen to be different from $O^*$, a contradiction.
    
            We now finish the proof of Case B.2.1.
            Recall $C_n$ (resp. $C_{n + 1}$) assigns state $\mathcal{S}_n((*\cellpointershort).f')$ (resp. $\mathcal{S}_{n + 1}((*\cellpointershort).f')$) to $(*\mathcal{M}(w')).f'$.
            Hence, since $\cellpointershort \in \celluniverse$ and $O \neq (*\cellpointershort).f'$, we have that $(*\cellpointershort).f'$ is a base object other than $O$ and the memory manager, and so by (*) $\mathcal{S}_{n + 1}((*\cellpointershort).f') = \mathcal{S}_n((*\cellpointershort).f')$.
            Thus, $C_n$ and $C_{n + 1}$ assign the same state to $(*\mathcal{M}(w')).f'$.
            Therefore, since $O' = (*\mathcal{M}(w')).f'$, we have that $C_n$ and $C_{n + 1}$ assign the same state to $O'$ as wanted.
            
            \textbf{Case B.2.2.} For some $w' < j \leq n$ the $j$th step of $\mathcal{I}^\mathcal{A}_{n + 1}$ performs an $\allocatecelloperation{}$ operation whose response is $\cellpointershort'$.
    
            Let $j$ be the minimum $w' < j \leq n$ such that the $j$th step of $\mathcal{I}^\mathcal{A}_{n + 1}$ performs an $\allocatecelloperation{}$ cell operation whose response is $\cellpointershort'$.
            Hence, $j$ is also the minimum $w' < j \leq n + 1$ such that the $j$th step of $\mathcal{I}^\mathcal{A}_{n + 1}$ performs an $\allocatecelloperation{}$ cell operation whose response is $\cellpointershort$.
            Thus, by \Cref{def:reduction:run_mapping}, $C_n$ and $C_{n + 1}$ assigns state $\mathcal{S}_{j - 1}((*\cellpointershort).f)$ to $(*\mathcal{M}(w')).f'$.
            Therefore, since $O' = (*\mathcal{M}(w')).f'$, we have that $C_n$ and $C_{n + 1}$ assign the same state to $O'$.

            This completes the proof of Case B.
    
            Finally, consider the memory manager.
            By \Cref{def:reduction:run_mapping}, $C_n$ (resp. $C_{n + 1}$) assigns state $\mathcal{M}[W_n]$ (resp. $\mathcal{M}[W_{n + 1}]$) to the memory manager, and since $W_n = W_{n + 1}$, we have that $\mathcal{M}[W_n] = \mathcal{M}[W_{n + 1}]$, and so $C_n$ and $C_{n + 1}$ assign the same state to the memory manager.
            Therefore, the state of every base object except $O^*$ is the same in $C_{n}$ and $C_{n + 1}$.
            
            This completes the proof of \textbf{Claim}.
    
            We now finish the proof of Case 3.
            Since the $n + 1$th step of $\mathcal{I}^\mathcal{A}_{n + 1}$ performs an operation on a base object $O$ other than the memory manager, by \Cref{claim:reduction:if_i+1_performs_op_on_o_in_a_then_i+1_performs_op_on_swapped_o_in_b}, the $n + 1$th step of $\mathcal{I}^\mathcal{B}_{n + 1}$ performs an operation on $O^*$.
            Hence, the state of every base object except $O^*$ is the same in $C_n$ and $C^\mathcal{B}_{n + 1}$.
            Thus, since by \textbf{Claim} the state of every base object except $O^*$ is the same in $C_{n}$ and $C_{n + 1}$, we have that the state of every base object except $O^*$ is the same in $C^\mathcal{B}_{n + 1}$ and $C_{n + 1}$.
            What remains is to show that $C^\mathcal{B}_{n + 1}$ and $C_{n + 1}$ assign the same state to $O^*$.
            By \Cref{claim:reduction:b_gets_the_right_response} $C^{\mathcal{B}}_{n + 1}$ assigns state $\mathcal{S}_{n + 1}(O)$ to $O^*$.
            First, suppose $O \notin \celluniverse_O$ or $O$ is in $\headobject$.
            Hence, by \Cref{def:reduction:run_mapping}, $C_{n + 1}$ assigns state $\mathcal{S}_{n + 1}(O)$ to $O$.
            Thus, since $O \notin \celluniverse_O$ or $O$ is in $\headobject$, we have that $O^* = O$, so $C_{n + 1}$ assigns state $\mathcal{S}_{n + 1}(O)$ to $O^*$.
            Therefore, $C^\mathcal{B}_{n + 1}$ and $C_{n + 1}$ assign the same state to $O^*$ as wanted.
            Now suppose $O \in \celluniverse_O$ and $O$ is not in $\headobject$.
            Hence, by \Cref{def:reduction:shared_objects}, $O = (*\cellpointershort).f$ where $f$ is either $\lastrepositoryoperationresponse{}$, $\revocations$, or $\nextlong$.
            Let $\mathcal{W}_n(O_s, 1) = w$.
            By \Cref{claim:reduction:op_on_ptr_after_allocate_in_a} the $w$th step of $\mathcal{I}^\mathcal{A}_{n + 1}$ performs an $\allocatecelloperation{}$ operation whose response is $\cellpointershort$ and $\mathcal{S}_n(O_s) = \mathcal{M}(w)$.
            Hence, since $O = (*\cellpointershort).f$, by the definition of $O^*$, $O^* = (*\mathcal{S}_n(O_s)).f$, and so $O^* = (*\mathcal{M}(w)).f$.
            Furthermore, by \Cref{claim:reduction:if_performed_op_on_cell_it_is_the_latest_version} for all $w < i \leq n + 1$ the $i$th step of $\mathcal{I}^\mathcal{A}_{n + 1}$ does not perform an $\allocatecelloperation{}$ operation whose response is $\cellpointershort$.
            Hence, since the $w$th step of $\mathcal{I}^\mathcal{A}_{n + 1}$ performs an $\allocatecelloperation{}$ operation whose response is $\cellpointershort$, by \Cref{def:reduction:run_mapping}, $C_{n + 1}$ assigns state $\mathcal{S}_{n + 1}((*\cellpointershort).f)$ to $(*\mathcal{M}(w)).f$.
            Thus, since $O = (*\cellpointershort).f$ and $O^* = (*\mathcal{M}(w)).f$, we have that $C_{n + 1}$ assigns state $\mathcal{S}_{n + 1}(O)$ to $O^*$.
            Therefore, $C^\mathcal{B}_{n + 1}$ and $C_{n + 1}$ assign the same state to $O^*$ as wanted.
        \end{itemize}
        This completes the proof of the inductive case, and therefore the lemma.
        \qH{\Cref{lemma:reduction:mapped_history_is_for_b}}
    \end{itemize}
\end{proof}

\subsubsection{The mapping is correctness-preserving}

In this section, we show that $\mathcal{A}$ is linearizable, wait-free, and has space complexity linear in the point contention by using the mapping from $\mathcal{A}$ to $\mathcal{B}$ and the fact that $\mathcal{B}$ has these properties.

\begin{proposition}\label{lemma:reduction:a_is_linearizable}
    $\mathcal{A}$ is linearizable with respect to $\mathcal{T}$.
\end{proposition}

\begin{proof}
    Consider any implementation history $\mathcal{I}^\mathcal{A} = C^\mathcal{A}_0, p_1, C^\mathcal{A}_1, \ldots$ of $\mathcal{A}$.
    By \Cref{lemma:reduction:mapped_history_is_for_b}, $\mathcal{I} = C_0, p_1, C_1, \ldots$ as defined in \Cref{def:reduction:run_mapping} is an implementation history of $\mathcal{B}$.
    Let $\mathcal{H}^\mathcal{A}$ (resp. $\mathcal{H}$) be the object history of $\mathcal{I}^\mathcal{A}$ (resp. $\mathcal{I}$).
    We claim that $\mathcal{H}^\mathcal{A} = \mathcal{H}$.
    Since by \Cref{lemma:reduction:same_program_counters_in_mapped_run}, the program counter of each process is the same in the $i$th configuration of $\mathcal{I}^\mathcal{A}$ and $\mathcal{I}$, we have that the subsequence of invocation and response steps (ignoring the operation or response they are for) in $\mathcal{I}^\mathcal{A}$ and $\mathcal{I}$ are the same, so all that remains is to show that each invocation (resp. response) step is for the same operation (resp. response) in $\mathcal{I}^\mathcal{A}$ and $\mathcal{I}$.
    Consider any invocation step of $\mathcal{I}^\mathcal{A}$; say the $i$th step, and for some operation $o$.
    Let $p_i$ be the process that performed the $i$th step of $\mathcal{I}^\mathcal{A}$.
    Hence, some local variable $pr$ of $p_{i}$ is assigned the state of $o$ in $C^\mathcal{A}_{i}$.
    Thus, since the value of $pr$ in $C^\mathcal{A}_i$ did not originate from the response of an $\allocatecelloperation{}$ response operation, by \Cref{def:reduction:watermarks}, $\mathcal{W}_i(pr, 1) = \bot$, and so by \Cref{def:reduction:swapper} $\mathcal{S}_i(pr) = o$.
    So, by \Cref{def:reduction:run_mapping}, $C_{i}$ assigns state $o$ to $pr$.
    Therefore, since by definition $p_i$ performs the $i$th step of $\mathcal{I}$ and by \Cref{lemma:reduction:same_program_counters_in_mapped_run} the program counter of $p_i$ is the same in $C^\mathcal{A}_{i - 1}$ and $C_{i - 1}$, we have that the $i$th step of $\mathcal{I}$ is also an invocation step for $o$.
    Now consider any response step of $\mathcal{I}^\mathcal{A}$; say the $i$th step and for some response $r$.
    Let $p_i$ be the process that performed the $i$th step of $\mathcal{I}^\mathcal{A}$.
    Hence, some local variable $pr$ of $p_{i}$ is assigned the state of $r$ in $C^\mathcal{A}_{i - 1}$.
    Thus, since the value of $pr$ in $C^\mathcal{A}_i$ did not originate from the response of an $\allocatecelloperation{}$ response operation, by \Cref{def:reduction:watermarks}, $\mathcal{W}_i(pr, 1) = \bot$, and so by \Cref{def:reduction:swapper} $\mathcal{S}_i(pr) = r$.
    So, by \Cref{def:reduction:run_mapping}, $C_{i}$ assigns state $r$ to $pr$.
    Therefore, since by definition $p_i$ performs the $i$th step of $\mathcal{I}$ and by \Cref{lemma:reduction:same_program_counters_in_mapped_run} the program counter of $p_i$ is the same in $C^\mathcal{A}_{i - 1}$ and $C_{i - 1}$, we have that the $i$th step of $\mathcal{I}$ is also a response step for $r$.
    This completes the proof of that $\mathcal{H}^\mathcal{A} = \mathcal{H}$.
    Since $\mathcal{H}$ is the object history of $\mathcal{I}$, and $\mathcal{I}$ is an implementation history of $\mathcal{B}$, by \Cref{theorem:ero:b_is_linearizable}, $\mathcal{H}$ is linearizable with respect to $\mathcal{T}$.
    Hence, since $\mathcal{H}^\mathcal{A} = \mathcal{H}$, we have that $\mathcal{H}^\mathcal{A}$ is linearizable with respect to $\mathcal{T}$.
    Thus, since $\mathcal{H}^\mathcal{A}$ is the object history of $\mathcal{I}^\mathcal{A}$, we have that $\mathcal{I}^\mathcal{A}$ is linearizable with respect to $\mathcal{T}$.
    Therefore, every implementation history of $\mathcal{A}$ is linearizable with respect to $\mathcal{T}$, and so $\mathcal{A}$ is linearizable with respect to $\mathcal{T}$ as wanted.
    \qH{\Cref{lemma:reduction:a_is_linearizable}}
\end{proof}

\begin{proposition}\label{lemma:reduction:a_is_wait_free}
    $\mathcal{A}$ is wait-free.
\end{proposition}

\begin{proof}
    Suppose, for contradiction, $\mathcal{A}$ is not wait-free.
    Hence, there is an implementation history $\mathcal{I}^\mathcal{A} =C^\mathcal{A}_0, p_1, C^\mathcal{A}_1, \ldots$ of $\mathcal{A}$ such that there is an operation execution $opx$ in $\mathcal{I}^\mathcal{A}$ of $\mathcal{A}$ where the process that executed $opx$ takes infinitely many steps during $opx$ without completing it.
    Let $p$ be the process that executed $opx$ and let $s_1, s_2, \ldots$ be the infinite sequence of step numbers where $s_i$ denotes the step number of the $i$th step $p$ performed during $opx$ in $\mathcal{I}^\mathcal{A}$.
    Let $\mathcal{I} = C_0, p_1, C_1, \ldots$ be the implementation history defined in \Cref{def:reduction:run_mapping} from $\mathcal{I}^\mathcal{A}$, and so by \Cref{lemma:reduction:mapped_history_is_for_b}, $\mathcal{I}$ is an implementation history of $\mathcal{B}$.
    Since by definition the process that performs each step of $\mathcal{I}^\mathcal{A}$ and $\mathcal{I}$ is the same, and by \Cref{lemma:reduction:same_program_counters_in_mapped_run} the program counter of $p$ is the same in $C^\mathcal{A}_{i - 1}$ and $C_{i - 1}$ for the $i$th step of $\mathcal{I}^\mathcal{A}$, we have that if $p$ executes the line of code $\ell_i$ during the $s_i$th step of $\mathcal{I}^\mathcal{A}$, then $p$ executes the line of code $\ell_i$ during the $s_i$th step of $\mathcal{I}$.
    Hence, since $s_1$ is the first step $p$ performs during $opx$ in $\mathcal{I}^\mathcal{A}$, we have that $\ell_1 = \ref{line:ero:invocation_step}$, and so $p$ performs an invocation step during the $s_1$th step of $\mathcal{I}$.
    Furthermore, since by assumption $p$ never completes $opx$ in $\mathcal{I}^\mathcal{A}$, we have that $\ell_i \neq \ref{line:ero:response_step}$ for every $i$, and so $p$ never performs a revocation step from the $s_1$th step onwards in $\mathcal{I}$.
    Hence, since $p$ performs a step of $\mathcal{I}$ at step numbers $s_1, s_2, \ldots$, $p$ performs an invocation step at $s_1$, and $p$ never performs a revocation step from the $s_1$th step onwards in $\mathcal{I}$, we have that $p$ takes infinitely many steps during a single operation execution in $\mathcal{I}$ without completing it.
    Therefore, since $\mathcal{I}$ is an implementation history of $\mathcal{B}$, by the definition of wait-freedom, $\mathcal{B}$ is not wait-free.
    However, by \Cref{theorem:ero:b_is_wait_free}, $\mathcal{B}$ is wait-free, a contradiction.
    \qH{\Cref{lemma:reduction:a_is_wait_free}}
\end{proof}

\begin{proposition}\label{lemma:reduction:a_is_space_efficient}
    For every implementation history $\mathcal{I}^\mathcal{A}$ of $\mathcal{A}$, the space complexity at any time $t$ in $\mathcal{I}^\mathcal{A}$ is linear in the point contention at $t$.
\end{proposition}

\begin{proof}
    Since by \Cref{def:space_complexity} the space complexity at any time $t$ in $\mathcal{I}^\mathcal{A}$ is the number of statically allocated base objects plus the number of dynamically allocated base objects at time $t$ in $\mathcal{I}^\mathcal{A}$, and $\mathcal{A}$ uses 7 statically allocated base objects (i.e., $\clockobject{}$, $\announceobject$, $\linearizationobject$, $\stateobject$, and the three objects in $\headobject$), it suffices to prove that the number of dynamically allocated base objects at time $t$ in $\mathcal{I}^\mathcal{A}$ is linear in the point contention at $t$.
    Furthermore, since each cell in $\mathcal{A}$ is comprised of three base objects, it suffices to prove that the number of allocated cells at time $t$ in $\mathcal{I}^\mathcal{A}$ is linear in the point contention at $t$.
    We prove the following stronger claim.
    For every implementation history $\mathcal{I}^\mathcal{A}$ of $\mathcal{A}$, the number of allocated cells at any time $t$ in $\mathcal{I}^\mathcal{A}$ is at most $6c + 1$ where $c$ is the point contention at $t$ in $\mathcal{I}^\mathcal{A}$.
    In the same terminology used in \Cref{thm:ero:b_is_space_efficient}, this is equivalent to the following statement.
    For every finite implementation history $\mathcal{I}^\mathcal{A}$ of $\mathcal{A}$, the number of allocated cells at the end of $\mathcal{I}^\mathcal{A}$ is at most $6c + 1$ where $c$ is the point contention in $\mathcal{I}^\mathcal{A}$.
    
    Suppose, for contradiction, there is a finite implementation history $\mathcal{I}^\mathcal{A} = C^\mathcal{A}_0, p_1, C^\mathcal{A}_1, \ldots, C^\mathcal{A}_n$ of $\mathcal{A}$ such that the number of allocated cells at the end of $\mathcal{I}^\mathcal{A}$ (i.e., in $\mathcal{C}^\mathcal{A}_n$) is larger than $6c + 1$.
    Let $M$ be the state of the memory manager in $C^\mathcal{A}_n$, so $|M| > 6c + 1$.
    Furthermore, let $\mathcal{I} = C_0, p_1, C_1, \ldots, C_n$ be the implementation history defined in \Cref{def:reduction:run_mapping} from $\mathcal{I}^\mathcal{A}$, and so by \Cref{lemma:reduction:mapped_history_is_for_b}, $\mathcal{I}$ is an implementation history of $\mathcal{B}$.
    Consider the $i$th step of $\mathcal{I}^\mathcal{A}$.
    Since the same process $p_i$ takes the $i$th step of $\mathcal{I}^\mathcal{A}$ and $\mathcal{I}$, and by \Cref{lemma:reduction:same_program_counters_in_mapped_run} the program counter of $p_i$ is the same in $C^\mathcal{A}_{i - 1}$ and $C_{i - 1}$, we have that if the $i$th step of $\mathcal{I}^\mathcal{A}$ is an invocation step, then the $i$th step of $\mathcal{I}$ is an invocation step, and if the $i$th step of $\mathcal{I}^\mathcal{A}$ is an revocation step, then the $i$th step of $\mathcal{I}$ is an revocation step.
    Therefore, the subsequence of invocation and response steps (ignoring the operation or response they are for) in $\mathcal{I}^\mathcal{A}$ and $\mathcal{I}$ are the same, so the point contention in $\mathcal{I}$ is also $c$ (A).
    
    Since $|M| > 6c + 1$, we have that there is a set $\{\cellpointershort_1, \cellpointershort_2, \ldots, \cellpointershort_{6c+2}\}$ of $6c + 2$ pointers in $M$.
    Hence, for every $i \in [1.. 6c+2]$, by \Cref{alg:cell_manager_specification}, for some $s_i \leq n$ the $s_i$th step of $\mathcal{I}^\mathcal{A}$ performs an $\allocatecelloperation{}$ operation whose response is $\cellpointershort_i$ such that for all $s_i \leq j \leq n$ the $j$th step of $\mathcal{I}^\mathcal{A}$ does not perform a $\freecelloperation{}(\cellpointershort_i)$ operation (B).
    Hence, by \Cref{alg:cell_manager_specification}, $\cellpointershort_i \in \celluniverse$.
    Furthermore, by \Cref{lemma:reduction:allocate_response_in_b_is_swapped}, the $s_i$th step of $\mathcal{I}$ performs an $\allocatecelloperation{}$ operation whose response is $\mathcal{M}(s_i) \in \celluniverse^0(\mathcal{I}^\mathcal{A})$ (C).
    Lastly, since $|\{\cellpointershort_1, \cellpointershort_2, \ldots, \cellpointershort_{6c+2}\}| = 6c + 2$ and the $s_i$th step of $\mathcal{I}^\mathcal{A}$ performs an $\allocatecelloperation{}$ operation whose response is $\cellpointershort_i$, we have that $|\{s_1, s_2, \ldots, s_{6c+2}\}| = 6c + 2$.
    Therefore, since $\mathcal{M}$ is injective, we have that $|\{\mathcal{M}(s_1), \mathcal{M}(s_2), \ldots, \mathcal{M}(s_{6c+2})\}| = 6c + 2$ (D).

    We claim that every step of $\mathcal{I}$ does not perform a $\freecelloperation{}(\mathcal{M}(s_i))$ operation (E).
    Suppose, for contradiction, the $j$th step of $\mathcal{I}$ where $j \leq n$ does.
    \begin{itemize}
        \item[] \hspace{0pt}\textbf{Case 1.} $j \leq s_i$.

        Since $\mathcal{I}$ is an implementation history of $\mathcal{B}$, $\mathcal{M}(s_i) \in \celluniverse$, and by assumption the $j$th step of step of $\mathcal{I}$ performs a $\freecelloperation{}(\mathcal{M}(s_i))$ operation, by 2 of \Cref{thm:reduction:algorithm_b_is_well_behaved}, there is an $\allocatecelloperation{}$ operation whose response is $\mathcal{M}(s_i)$ before the $j$th step of $\mathcal{I}$.
        Therefore, since by (C) the $s_i$th step of $\mathcal{I}$ is an \allocatecelloperation{} operation whose response is  $\mathcal{M}(s_i)$, and $j \leq s_i$, we have that there are two $\allocatecelloperation{}$ operations whose response is $\mathcal{M}(s_i)$ during $\mathcal{I}$.
        However, by \Cref{alg:lazy_cell_manager_specification}, there is at most one $\allocatecelloperation{}$ operation whose response is $\mathcal{M}(s_i)$ in $\mathcal{I}$, a contradiction.

        \item[] \hspace{0pt}\textbf{Case 2.} $s_i < j$.

        Hence, since $j \leq n$, we have that $s_i < j \leq n$.
        Since the $j$th step of $\mathcal{I}$ performs a $\freecelloperation{}(\mathcal{M}(s_i))$ operation, we have that $C_{j - 1}$ assigns state $\mathcal{M}(s_i)$ to $p_j$'s local variable $\currentcellpointershort$ in the Relinquish procedure; denote this local variable by $pr$.
        Hence, by \Cref{def:reduction:run_mapping}, $\mathcal{S}_{j - 1}(pr) = \mathcal{M}(s_i)$.
        Thus, since the state assigned to $\currentcellpointershort$ is a single value, $\mathcal{S}_{j - 1}(pr) = \mathcal{S}_{j - 1}(pr, 1)$, and so by \Cref{def:reduction:swapper}, $C^\mathcal{A}_{j - 1}$ either assigns state $\mathcal{M}(s_i)$ to $pr$ or $\mathcal{W}_{j - 1}(pr, 1) = s_i$.
        If the former, then by \Cref{def:reduction:used_pointers_in_run_of_a}, $\mathcal{M}(s_i) \in \celluniverse(\mathcal{I}^\mathcal{A})$.
        However, $\mathcal{M}(s_i) \in \celluniverse^0(\mathcal{I}^\mathcal{A})$, and by \Cref{observation:reduction:unused_pointers_in_run_of_a} $\celluniverse(\mathcal{I}^\mathcal{A})$ and $\celluniverse^0(\mathcal{I}^\mathcal{A})$ are disjoint, so this case is impossible.
        Hence, $\mathcal{W}_{j - 1}(pr, 1) = s_i$.
        Thus, by \Cref{lemma:reduction:associated_pointer}, the $s_i$th step of $\mathcal{I}^\mathcal{A}$ performs an $\allocatecelloperation{}$ operation with response $v_1$ where $v_1$ is the value of the $1$st index of the state assigned to $pr$ in $C^\mathcal{A}_{j - 1}$.
        So, since by (B) the $s_i$th step of $\mathcal{I}^\mathcal{A}$ performs an $\allocatecelloperation{}$ operation whose response is $\cellpointershort_i$, we have that $v_1 = \cellpointershort_i$, so $C^\mathcal{A}_{j - 1}$ assigns state $\cellpointershort_i$ to $pr$.
        Since the $j$th step of $\mathcal{I}$ performs a \freecelloperation{} operation, the same process $p_j$ takes the $j$th step of $\mathcal{I}^\mathcal{A}$ and $\mathcal{I}$, and by \Cref{lemma:reduction:same_program_counters_in_mapped_run} the program counter of $p_j$ is the same in $C^\mathcal{A}_{j - 1}$ and $C_{j - 1}$, we have that the $j$th step of $\mathcal{I}^\mathcal{A}$ performs a $\freecelloperation{}$ operation.
        Therefore, since $C^\mathcal{A}_{j - 1}$ assigns state $\cellpointershort_i$ to $pr$ and $pr$ is $p_j$'s local variable $\currentcellpointershort$ in the Relinquish procedure, we have that the $j$th step of $\mathcal{I}^\mathcal{A}$ performs a $\freecelloperation{}(\cellpointershort_i)$ operation.
        However, by (B), for all $s_i \leq j \leq n$ the $j$th step of $\mathcal{I}^\mathcal{A}$ does not perform a $\freecelloperation{}(\cellpointershort_i)$, a contradiction.
    \end{itemize}

    We now finish the proof of \Cref{lemma:reduction:a_is_space_efficient}.
    Let $Allocate(\mathcal{I})$ be the set of pointers which have been allocated in $\mathcal{I}$, i.e., $\cellpointershort \in Allocate(\mathcal{I})$ if and only if there is an $\allocatecelloperation{}$ operation in $\mathcal{I}$ whose response is $\cellpointershort$.
    Likewise, let $Free(\mathcal{I})$ be the set of pointers which have been freed in $\mathcal{I}$, i.e., $\cellpointershort \in Free(\mathcal{I})$ if and only if there is an $\freecelloperation{}(\cellpointershort)$ operation in $\mathcal{I}$.
    Consider any $i \in [1.. 6c+2]$.
    Since by (C) the $s_i$th step of $\mathcal{I}$ is an $\allocatecelloperation{}$ operation whose response is $\mathcal{M}(s_i)$, we have that $\mathcal{M}(s_i) \in Allocate(\mathcal{I})$.
    Hence, since by (E) every step of $\mathcal{I}$ does not perform a $\freecelloperation{}(\mathcal{M}(s_i))$ operation, we have that $\mathcal{M}(s_i) \in Allocate(\mathcal{I}) \setminus Free(\mathcal{I})$.
    Thus, for every $i \in [1..6c+2]$ $\mathcal{M}(s_i) \in Allocate(\mathcal{I}) \setminus Free(\mathcal{I})$.
    Therefore, by (D) $|Allocate(\mathcal{I}) \setminus Free(\mathcal{I})| \geq 6c + 2 > 6c + 1$.
    However, since $\mathcal{I}$ is a finite implementation history of $\mathcal{B}$, and by (A) the point contention in $\mathcal{I}$ is $c$, by \Cref{thm:ero:b_is_space_efficient}, $|Allocate(\mathcal{I}) \setminus Free(\mathcal{I})| \leq 6c + 1$, a contradiction.
    \qH{\Cref{lemma:reduction:a_is_space_efficient}}
\end{proof}

This completes the main result of \Cref{sec:algo_2_proof}.

\begin{theorem}[\Cref{thm:algo_2} restated]\label{thm:real_algo_2}
    \Cref{alg:efficient_algo} is a wait-free universal construction for the infinite-arrival model with bounded concurrency.
    Its space complexity at time $t$ is linear in the point contention at $t$.
\end{theorem}

\begin{proof}
    By Propositions \ref{lemma:reduction:a_is_linearizable}, \ref{lemma:reduction:a_is_wait_free}, and \ref{lemma:reduction:a_is_space_efficient}.
    \qH{\Cref{thm:real_algo_2}}
\end{proof}

\end{document}